\documentclass[twoside,headsepline]{scrbook}
\pdfoutput=1
\KOMAoptions{twocolumn=false}
\KOMAoptions{paper=letter}
\KOMAoptions{titlepage=true}
\KOMAoptions{toc=graduated}
\KOMAoptions{listof=flat}
\usepackage{scrlayer-scrpage}
\usepackage{import}    
\usepackage{array}     
\clearplainofpairofpagestyles
\ihead{LZ Technical Design Report}
\ifoot[]{}
\ofoot[]{}

\automark[section]{chapter}
\usepackage[T1]{fontenc} 
\usepackage[utf8]{inputenc}
\usepackage{chapterbib}
\usepackage{calc}
\usepackage{typearea}
\usepackage{graphicx}
\usepackage{subfig}
\usepackage{wrapfig}
\usepackage{booktabs}
\usepackage{longtable}
\usepackage{multirow}
\usepackage{amsmath}
\usepackage{amssymb}
\usepackage{appendix}
\usepackage{eurosym}
\usepackage{afterpage} 
\usepackage{placeins}
\usepackage{flafter}
\usepackage{float}
\usepackage{multicol}
\usepackage{caption}
\usepackage[figuresright]{rotating}

\usepackage{balance}

\usepackage[dvipsnames]{xcolor}
\definecolor{referenceblue}{rgb}{0.0, 0.25, 0.44}
\definecolor{referencered}{rgb}{0.6, 0.0, 0.0}
\usepackage{colortbl}

\usepackage{mathptmx}  
\usepackage{wallpaper} 
\usepackage[numbers,square,comma,sort&compress,merge]{natbib}  
\usepackage{siunitx}
\usepackage{./LZtdrSupport/hepparticles} 
\usepackage[italic]{heppennames}

\usepackage{hyperref}         
\usepackage[all]{hypcap}

\usepackage{tablefootnote}

\hypersetup{
  pdftitle = {LZ Technical Design Report},
  colorlinks=true,
  breaklinks=true,
  linkcolor=referencered,
  anchorcolor=referenceblue,
  citecolor=referencered,
  filecolor=referencered,
  menucolor=referenceblue,
  urlcolor=referenceblue,
  linktocpage=true,
}

\input{LZsym}

\newcommand{\ChapLabel}{chap:empty}
\newcommand{\BibLabel}{empty:bib}
\newcommand{\SecPre}{empty}
\newcommand{\FloatPre}{empty}

\newcommand{\tdrchap}[2][\ChapLabel]
{
   \ifx#1\ChapLabel
     \chapter{#2\ifKL\space\texttt{#1}\fi}
     \label{#1}
   \else
    \renewcommand{\ChapLabel}{chap:#1} 
    \renewcommand{\BibLabel}{#1S:bib} 
    \renewcommand{\SecPre}{#1}
    \renewcommand{\FloatPre}{#1}
    \chapter{#2\ifKL\space\texttt{\ChapLabel}\fi}
    \label{\ChapLabel}
   \fi
}

\newcommand{\tdrsec}[2][\Alph{section}]
{
   \section{#2\ifKL\space\texttt{\SecPre S:#1}\fi}
   \label{\SecPre S:#1}
}

\newcommand{\tdrsubsec}[2][\Alph{section}\alph{subsection}]
{
   \subsection{#2\ifKL\space\texttt{\SecPre Ss:#1}\fi}
   \label{\SecPre Ss:#1}
}

\newcommand{\tdrsubsubsec}[2][\Alph{section}\alph{subsection}\Roman{subsubsection}]
{
   \subsubsection{#2\ifKL\space\texttt{\SecPre Sss:#1}\fi}
   \label{\SecPre Sss:#1}   
}

\newcommand{\tdrtcaption}[3][\arabic{section}.\arabic{subsection}.\arabic{table}]
{
   \caption[#2\ifKL\space\texttt{\FloatPre t:#1}\fi]{#3\ifKL\space\texttt{\FloatPre t:#1}\fi}
   \label{\FloatPre t:#1}
}

\newcommand{\tdrfcaption}[3][\arabic{section}.\arabic{subsection}.\arabic{figure}]
{
   \caption[#2\ifKL\space\texttt{\FloatPre f:#1}\fi]{#3\ifKL\space\texttt{\FloatPre f:#1}\fi}
   \label{\FloatPre f:#1}
}

\long\def\ignore#1{\relax}

\makeatletter
\renewcommand*\bib@heading{%
  \section{\bibname}
  \@mkboth{\bibname}{\bibname}
}
\makeatother

\makeatletter
\renewcommand*\l@figure{\@dottedtocline{1}{1.5em}{3.5em}} 
\let\l@table\l@figure
\makeatother

\makeatletter
\DeclareRobustCommand\bfseries{%
  \not@math@alphabet\bfseries\mathbf
  \fontseries\bfdefault\selectfont\boldmath}

\makeatother

\setcapindent{0pt} 

\setkomafont{captionlabel}{\normalfont\sffamily\bfseries}
\setkomafont{caption}{\normalfont\sffamily}
\setkomafont{pagefoot}{\normalfont\sffamily}
\setkomafont{pagenumber}{\normalfont\sffamily}
\setkomafont{pagehead}{\normalfont\sffamily}

\newcommand{\ltbcp}{lngtab}

\areaset[12mm]{165mm}{240mm}
\usepackage[pages=some,scale=1,angle=0,opacity=0.45]{background}
\backgroundsetup{firstpage=true,contents={\includegraphics[width=0.85\textwidth]{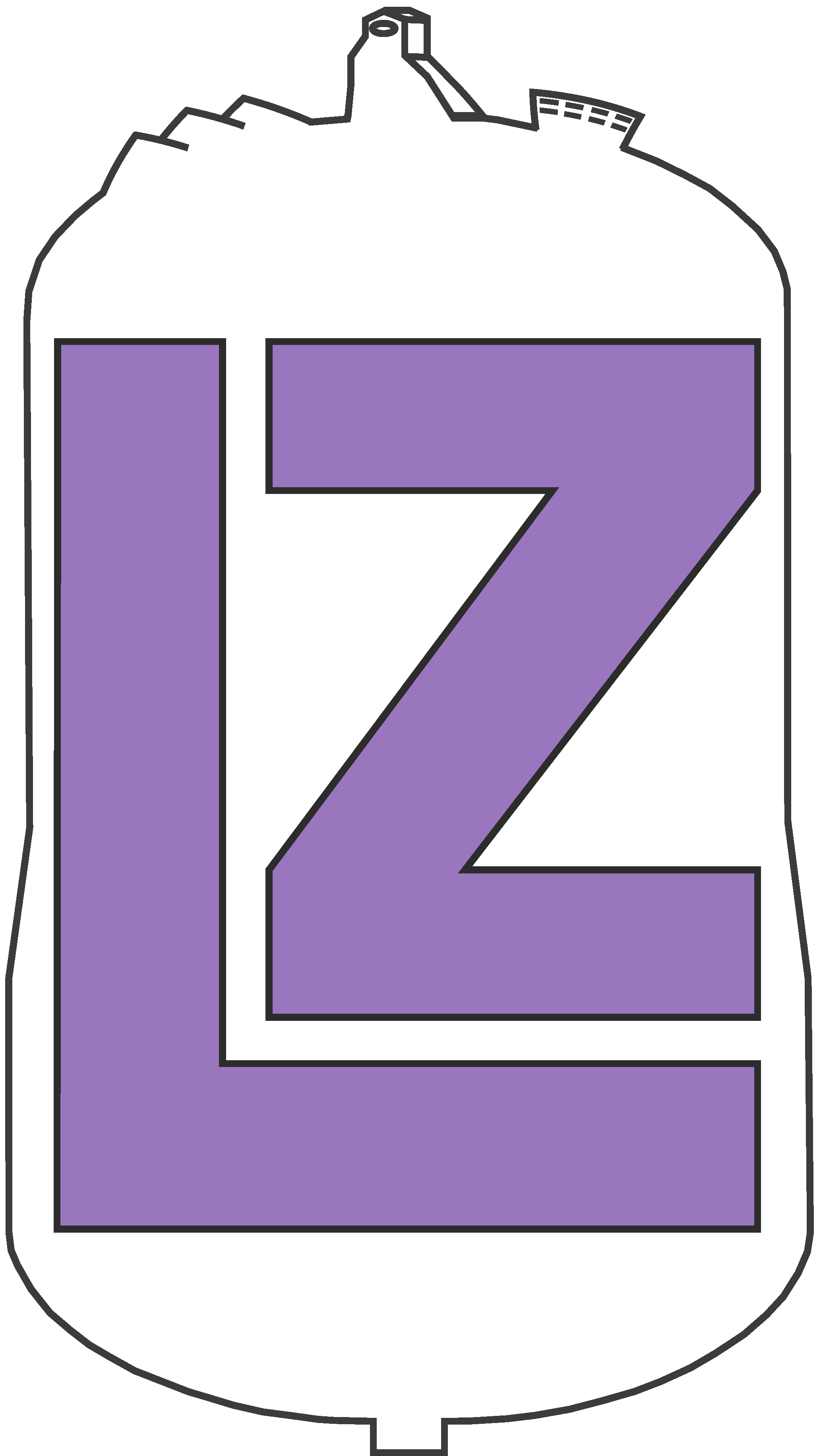}}}
\numberwithin{figure}{section}
\numberwithin{table}{section}
\numberwithin{equation}{section}

\newif\ifKL
\KLfalse

\begin{document}
\setcounter{chapter}{0}

\frontmatter
\pagestyle{scrplain}
%
%

\begin{flushright}
{\Large LBNL-1007256}
\end{flushright}
\vspace*{0.003\textheight}

\begin{center}
  {\href{http://lz.lbl.gov}{\normalfont\sffamily\LARGE\bfseries\fontsize{25}{30.0}\selectfont The LUX-ZEPLIN (LZ)\\[0.2cm] Technical Design Report}} \\*
\end{center}
~\\[11.5cm]



\begin{center}
\begin{minipage}{0.67\textwidth}
In this Technical Design Report (TDR) we describe the LZ detector
to be built at the Sanford Underground Research Facility (SURF).
The LZ dark matter experiment is designed to achieve sensitivity 
to a WIMP-nucleon spin-independent cross section of 
\SI{3e-48}{\cm\squared}.
\end{minipage}
\end{center}

\cleardoublepage

\pagenumbering{roman}
\begin{center}
  {\href{http://lz.lbl.gov}{\normalfont\sffamily\LARGE\bfseries\fontsize{16}{19.2}\selectfont The LUX-ZEPLIN (LZ) Collaboration}} \\*
\end{center}
%
%
\begin{center}
 B.J.~Mount \\* 
 {\href{http://www.bhsu.edu/Academics/ProgramsMajors/NaturalSciences/Physics.aspx}{\em \small Black Hills State University, School of Natural Sciences, 1200 University Street, Spearfish, SD 57799-0002, USA}}
\end{center}
%
%
\begin{center}
 S.~Hans, 
 R.~Rosero, 
 M.~Yeh \\*  
 {\href{https://www.bnl.gov/chemistry/}{\em \small Brookhaven National Laboratory (BNL) P.O. Box 5000, Upton, NY 11973-5000, USA}}
\end{center}
%
%
\begin{center}
 C.~Chan, 
 R.J.~Gaitskell, 
 D.Q.~Huang, 
 J.~Makkinje, 
 D.C.~Malling,\footnote{Now at: \href{https://www.ll.mit.edu}{Lincoln Laboratory, 244 Wood Street, Massachusetts Institute of Technology, Lexington, MA 02421-6426, USA}} 
 M.~Pangilinan,\footnote{Now at: \href{https://www.samba.tv}{Samba TV, 301 Brannan Street, Floor 6, San Francisco, CA 94107-3816, USA}} 
 C.A.~Rhyne, 
 W.C.~Taylor, 
 J.R.~Verbus \footnote{Now at: \href{https://www.linkedin.com/company/linkedin}{LinkedIn Corporation, 1000 West Maude Avenue, Sunnyvale, CA 94085, USA\label{linkedin}}} \\* 
 {\href{http://pa.brown.edu}{\em \small Brown University, Department of Physics, 182 Hope Street, Providence, RI 02912-9037, USA}}
 \end{center}
%
%
\begin{center}
 Y.D.~Kim, 
 H.S.~Lee, 
 J.~Lee, 
 D.S.~Leonard, 
 J.~Li \\* 
 {\href{http://cupweb.ibs.re.kr/html/cup_en/}{\em \small IBS Center for Underground Physics (CUP), 70, Yuseong-daero 1689-gil, Yuseong-gu, Daejeon, Korea}}
 \end{center}
%
%
\begin{center}
  J.~Belle,\footnote{Also at: Belle Aerospace Corporation, 2237 Warrenville Ave., Wheaton, IL 60189, USA} 
  A.~Cottle, 
 W.H.~Lippincott, 
 D.J.~Markley, 
 T.J.~Martin, 
 M.~Sarychev, 
 T.E.~Tope, 
 M.~Utes, 
 R.~Wang, 
 I.~Young \\* 
 {\href{http://astro.fnal.gov}{\em \small Fermi National Accelerator Laboratory (FNAL), Batavia, IL 60510-0500, USA}}
 \end{center}
%
%
\begin{center}
 H.M.~Ara\'ujo, 
 A.J.~Bailey,\footnote{Now at: \href{http://webific.ific.uv.es/web/en}{University of Valencia, Instituto de Física Corpuscular, Parc Científic de la Universitat de Val\`{e}ncia, C/ Catedrático José Beltrán, 2, E-46980 Paterna, Spain}} 
 D.~Bauer, 
 D.~Colling, 
 A.~Currie,\footnote{Now at: \href{https://www.gov.uk/government/organisations/hm-revenue-customs}{HM Revenue and Customs,  100 Parliament Street, London, SW1A 2BQ, UK}} 
 S.~Fayer, 
 F.~Froborg, 
 S.~Greenwood, 
 W.G.~Jones, 
 V.~Kasey, 
 M.~Khaleeq, 
 I.~Olcina, 
 B.~L\'opez Paredes, 
 A.~Richards, 
 T.J.~Sumner, 
 A.~Tom\'as, 
 A.~Vacheret\\* 
 {\href{http://www3.imperial.ac.uk/highenergyphysics/research/experiments/zeplin}{\em \small Imperial College London, Physics Department, Blackett Laboratory, Prince Consort Road, London, SW7 2BW, UK}}
 \end{center}
%
%
\begin{center}
 P.~Br\'{a}s, 
 A.~Lindote, 
 M.I.~Lopes, 
 F.~Neves, 
 J.P.~Rodrigues, 
 C.~Silva, 
 V.N.~Solovov \\* 
 {\href{http://coimbra.lip.pt/index.php?id=3&prid=36&lg=en}{\em \small Laborat\'orio de Instrumenta\c{c}\~ao e F\'isica Experimental de Part\'iculas (LIP), Department of Physics, University of Coimbra, Rua Larga, 3004-516, Coimbra, Portugal}}
 \end{center}
%
%
\begin{center}
 M.J.~Barry, 
 A.~Cole,  
 A.~Dobi,\footnote{Now at: \href{http://www.zenysis.com}{Zenysis Technology, 535 Mission Street Floor 14, San Franciso, CA 94105-3253, USA}} 
 W.R.~Edwards, 
 C.H.~Faham,\footref{linkedin} 
 S.~Fiorucci, 
 N.J.~Gantos, 
 V.M.~Gehman,\footnote{Now at: \href{http://www.cainthus.com}{Cainthus, Otherlab, 701 Alabama Street, San Francisco, CA 94110-2022, USA}} 
 M.G.D.~Gilchriese, 
 K.~Hanzel, 
 M.D.~Hoff, 
K.~Kamdin,\footnote{\label{UCBerk}Also at: \href{http://physics.berkeley.edu/research/particle-physics}{University of California (UC), Berkeley, Department of Physics, 366 LeConte Hall MC 7300, Berkeley, CA 94720-7300, USA}} 
 K.T.~Lesko, 
 C.T.~McConnell, 
 K.~O'Sullivan,\footnote{Now at: \href{http://insightdatascience.com}{Insight Data Science, 260 Sheridan Avenue Suite 310, Palo Alto, CA 94306-2010, USA}} 
 K.C.~Oliver-Mallory,\footref{UCBerk} 
 S.J.~Patton, 
 J.S.~Saba, 
 P.~Sorensen, 
 K.J.~Thomas,\footnote{Also at: \href{http://www.nuc.berkeley.edu}{University of California (UC), Berkeley, Department of Nuclear Engineering, 4155 Etcheverry Hall MC 1730,  CA 94720-1730, USA}}$^{,}$\footnote{Now at: \href{http://www.llnl.gov/}{Lawrence Livermore National Laboratory (LLNL), 7000 East Avenue, Livermore, CA 94550-9698, USA}} 
 C.E.~Tull, 
 W.L.~Waldron, 
 M.S.~Witherell\footref{UCBerk} \\* 
 {\href{https://commons.lbl.gov/display/physics/LUX}{\em \small Lawrence Berkeley National Laboratory (LBNL), 1 Cyclotron Road, Berkeley, CA 94720-8156, USA}}
 \end{center}
%
%
\begin{center}
 A.~Bernstein, 
 K.~Kazkaz, 
 J.~Xu \\* 
 {\href{http://www.llnl.gov/}{\em \small Lawrence Livermore National Laboratory (LLNL), 7000 East Avenue, Livermore, CA 94550-9698, USA}}
 \end{center}
%
%
\begin{center}
 D.~Yu.~Akimov, 
 A.I.~Bolozdynya, 
 A.V.~Khromov, 
 A.M.~Konovalov, 
 A.V.~Kumpan, 
 V.V.~Sosnovtsev \\* 
 {\href{http://enpl.mephi.ru/index.php?is_eng=1}{\em \small National Research Nuclear University MEPhI (NRNU MEPhI), 31 Kashirskoe shosse, Moscow, 115409, RUS}}
 \end{center}
%
%
\begin{center}  
 C.E.~Dahl,\footnote{Also at: \href{http://astro.fnal.gov}{Fermi National Accelerator Laboratory (FNAL), Batavia, IL 60510-0500, USA}} 
 D.~Temples\\* 
  {\href{http://www.physics.northwestern.edu}{\em \small Northwestern University, Department of Physics \& Astronomy, 2145 Sheridan Road, Evanston, IL 60208-3112, USA}}
\end{center}
%
%
\begin{center}
 M.C.~Carmona-Benitez, 
 L.~de~Viveiros \\* 
{\href{http://www.phys.psu.edu}{\em \small Pennsylvania State University, Department of Physics, 104 Davey Lab, University Park, PA 16802-6300, USA}} \\[0.2cm]
\end{center}
%
%
\begin{center}
 D.S.~Akerib,$^{ba}$ 
 H.~Auyeung,$^{b}$ 
 T.P.~Biesiadzinski,$^{ba}$ 
 M.~Breidenbach,$^{b}$ 
 R.~Bramante,$^{ba}$ 
 R.~Conley,$^{b}$ 
 W.W.~Craddock,$^{b}$ 
 A.~Fan,$^{ba}$ 
 A.~Hau,$^{b}$ 
 C.M.~Ignarra,$^{ba}$ 
 W.~Ji,$^{ba}$ 
 H.J.~Krebs,$^{b}$ 
 R.~Linehan,$^{ba}$ 
 C.~Lee,$^{ba,}$\footnote{Now at: \href{http://cupweb.ibs.re.kr/html/cup_en/}{IBS Center for Underground Physics (CUP), 70, Yuseong-daero 1689-gil, Yuseong-gu, Daejeon, Korea}} 
 S.~Luitz,$^{b}$ 
 E.~Mizrachi,$^{b}$ 
 M.E.~Monzani,$^{ba}$ 
 F.G.~O'Neill,$^{b}$ 
 S.~Pierson,$^{b}$ 
 M.~Racine,$^{b}$ 
 B.N.~Ratcliff,$^{b}$ 
 G.W.~Shutt,$^{b}$ 
 T.A.~Shutt,$^{ba}$ 
 K.~Skarpaas,$^{b}$ 
 K.~Stifter,$^{ba}$ 
 W.H.~To,$^{ba,}$\footnote{Now at: \href{http://physics.csustan.edu}{California State University, Stanislaus, Department of Physics, 1 University Circle, Turlock, CA 95382-3200, USA}} 
 J.~Va'vra,$^{b}$ 
 T.J.~Whitis,$^{ba}$ 
 W.J.~Wisniewski$^{b}$ \\* 
{\em \small {}\href{http://kipac.stanford.edu/kipac/}{$^a$Kavli Institute for Particle Astrophysics and Cosmology (KIPAC), 452 Lomita Mall, Stanford University, Stanford, CA 94305, USA}; {}$^b$\href{https://ppa.slac.stanford.edu/node/100}{SLAC National Accelerator Laboratory, 2575 Sand Hill Road, Menlo Park, CA 94205-7015, USA}}
\end{center}
%
%
\begin{center}
 X.~Bai, 
 R.~Bunker,\footnote{Now at: \href{https://www.pnl.gov}{Pacific Northwest National Laboratory (PNNL), 902 Battelle Blvd., Richland WA 99354-1793, USA}} 
 R.~Coughlen, 
 C.~Hjemfelt, 
 R.~Leonard, 
 E.H.~Miller, 
 E.~Morrison, 
 J.~Reichenbacher, 
 R.W.~Schnee, 
 M.R.~Stark,\footnote{Now at: \href{https://www.northeastern.edu/}{Northeastern University, 360 Huntington Ave. 111 Dana Research Center, Boston, MA 02115, USA}} 
 K.~Sundarnath, 
 D.R.~Tiedt, 
 M.~Timalsina\\* 
 {\href{http://www.sdsmt.edu/PHYS/}{\em \small South Dakota School of Mines and Technology, 501 East Saint Joseph Street, Rapid City, SD 57701-3901, USA}}
 \end{center}
%
%
\begin{center}
 P.~Bauer, 
 B.~Carlson, 
 M.~Horn, 
 M.~Johnson, 
 J.~Keefner, 
 C.~Maupin, 
 D.J.~Taylor \\* 
{\href{http://sanfordlab.org/sdsta}{\em \small South Dakota Science and Technology Authority (SDSTA), Sanford Underground Research Facility, 630 East Summit Street, Lead, SD 57754-1700, USA}}
\end{center}
%
%
\begin{center}
 S.~Balashov, 
 P.~Ford, 
 V.~Francis, 
 E.~Holtom, 
 A.~Khazov, 
 A.~Kaboth,\footnote{Also at: \href{https://pure.royalholloway.ac.uk/portal/en/organisations/department-of-physics(54da7e90-0544-4dbe-bd6d-4e85fa8f7465).html}{Royal Holloway, University of London, Department of Physics, Egham Hill, Egham, Surry, TW20 0EX, UK}} 
 P.~Majewski, 
 J.A.~Nikkel, 
 J.~O'Dell, 
 R.M.~Preece, 
 M.G.D.~van der Grinten, 
 S.D.~Worm\footnote{Now at: \href{http://www.ep.ph.bham.ac.uk/index.php?page=welcome}{University of Birmingham, Particle Physics Group, School of Physics and Astronomy, Edgbaston, Birmingham, B15 2TT, UK}}\\* 
 {\href{http://www.ppd.stfc.ac.uk/ppd/dark-matter-people.html}{\em \small STFC Rutherford Appleton Laboratory (RAL), Didcot, OX11 0QX, UK}}
 \end{center}
%
%
\begin{center}
 R.L.~Mannino, 
 T.M.~Stiegler, 
 P.A.~Terman, 
 R.C.~Webb \\* 
 {\href{http://physics.tamu.edu}{\em \small Texas A\&M University, Department of Physics and Astronomy, 4242 TAMU, College Station, TX 77843-4242, USA}}
 \end{center}
%
%
\begin{center}
C.~Levy, 
 J.~Mock,\footnote{\label{LBNLfoot}Also at: \href{https://commons.lbl.gov/display/physics/LUX}{Lawrence Berkeley National Laboratory (LBNL), 1 Cyclotron Road, Berkeley, CA 94720-8156 USA}}$^{,}$\footnote{Now at: \href{http://www.cosylab.com}{Cosylab USA, Inc., 1530 Page Mill Road, Suite 200, Palo Alto, CA 94304-1140, USA}} 
 M.~Szydagis \\* 
 {\href{http://www.albany.edu/physics/index.shtml}{\em \small University at Albany (SUNY), Department of Physics, 1400 Washington Avenue, Albany, NY 12222-1000, USA}}
 \end{center}
%
%
\begin{center}
 J.K.~Busenitz, 
 M.~Elnimr,\footnote{Now at: \href{https://www.physics.uci.edu}{University of California (UC), Irvine, Department of Physics \& Astronomy, 4129 Frederick Reines Hall, Irvine, CA 92697-4575, USA}} 
 J.Y-K.~Hor, 
 Y.~Meng, 
 A.~Piepke, 
 I.~Stancu \\* 
{\href{http://physics.ua.edu}{\em \small University of Alabama, Department of Physics \& Astronomy, 206 Gallalee Hall, 514 University Boulevard, Tuscaloosa, AL 34587-0324, USA}} \\[0.2cm]
\end{center}
%
%
\begin{center}
L.~Kreczko, 
B.~Krikler, 
B.~Penning \\* 
{\href{http://www.bristol.ac.uk/physics/}{\em \small University of Bristol, School of Physics, Bristol, BS8 1TL, UK}} \\[0.2cm]
\end{center}
%
%
\begin{center}
  E.P.~Bernard, \footref{LBNLfoot} 
 R.G.~Jacobsen, 
 D.N.~McKinsey,\footref{LBNLfoot} 
 R.~Watson\footref{LBNLfoot} \\* 
 {\href{http://physics.berkeley.edu/research/particle-physics}{\em \small University of California (UC), Berkeley, Department of Physics, 366 LeConte Hall MC 7300, Berkeley, CA 94720-7300, USA}}
 \end{center}
%
%
\begin{center}
  J.E.~Cutter, 
  S.~El-Jurf, 
  R.M.~Gerhard, 
  D.~Hemer, 
  S.~Hillbrand, 
 B.~Holbrook, 
 B.G.~Lenardo,\footnote{Also at: \href{http://www.llnl.gov/}{Lawrence Livermore National Laboratory (LLNL), 7000 East Avenue, Livermore, CA 94550-9698, USA}} 
 A.G.~Manalaysay, 
 J.A.~Morad, 
 S.~Stephenson,\footnote{Now at: \href{https://www.deepgram.com}{Deepgram, 148 Townsend Street, San Francisco, CA, 94107-1919, USA}} 
 J.A.~Thomson, 
 M.~Tripathi, 
 S.~Uvarov \\* 
{\href{http://www.physics.ucdavis.edu}{\em \small University of California (UC), Davis, Department of Physics, One Shields Avenue, Davis, CA 95616-5270, USA}} \\*
\end{center}
%
%
\begin{center}
 S.J.~Haselschwardt, 
 S.~Kyre, 
 C.~Nehrkorn, 
 H.N.~Nelson, 
 M.~Solmaz, 
 D.T.~White \\* 
 {\href{http://hep.ucsb.edu/}{\em \small University of California (UC), Santa Barbara, Department of Physics, Broida Hall, Santa Barbara, CA 93106-9530, USA}}
 \end{center}
%
%
\begin{center}
 M.~Cascella, 
 J.E.Y.~Dobson, 
 C.~Ghag, 
 X.~Liu, 
 L.~Manenti, 
 L.~Reichhart,\footnote{Now at: \href{http://www.ims.co.at}{IMS Nanofabrication AG, Wolfholzgasse 20-22, 2345 Brunn am Gebirge, AUT}} 
 S.~Shaw,\footnote{Now at: \href{http://hep.ucsb.edu/}{University of California (UC), Santa Barbara, Department of Physics, Broida Hall, Santa Barbara, CA 93106-9530, USA}} 
 U.~Utku \\* 
 {\href{http://www.hep.ucl.ac.uk}{\em \small University College London (UCL), Department of Physics and Astronomy, Gower Street, London, WC1E 6BT, UK}}
 \end{center}
%
%
\begin{center}
 P.~Beltrame, 
 T.J.R.~Davison, 
 M.F.~Marzioni, 
 A.St.J.~Murphy, 
 A.~Nilima \\* 
 {\href{http://www2.ph.ed.ac.uk/particle-physics-experiment/}{\em \small University of Edinburgh, SUPA, School of Physics and Astronomy, Edinburgh, EH9 3FD, UK}}
 \end{center}
%
%
\begin{center}
 B.~Boxer,\footnote{Also at: \href{http://www.ppd.stfc.ac.uk/ppd/dark-matter-people.html}{STFC Rutherford Appleton Laboratory (RAL), Didcot, OX11 0QX, UK}} 
 S.~Burdin, 
 A.~Greenall, 
 S.~Powell, 
 H.J.~Rose, 
 P.~Sutcliffe \\* 
 {\href{https://www.liv.ac.uk/physics/research/particle-physics/}{\em \small University of Liverpool, Department of Physics, Liverpool, L69 7ZE, UK}}
 \end{center}
%
%
\begin{center}
 J.~Balajthy, 
 T.K.~Edberg, 
 C.R.~Hall, 
 J.S.~Silk \\* 
 {\href{http://umdphysics.umd.edu}{\em \small University of Maryland, Department of Physics, College Park, MD 20742-4111, USA}}
 \end{center}
%
%
\begin{center}
S.~Hertel \\* 
 {\href{https://www.physics.umass.edu}{\em \small University of Massachusetts, Department of Physics, 1126 Lederle Graduate Research Tower (LGRT), Amherst, MA 01003-9337, USA}}
\end{center}
%
%
\begin{center}
  C.W.~Akerlof, 
  M.~Arthurs, 
 W.~Lorenzon, 
 K.~Pushkin, 
 M.~Schubnell \\* 
 {\href{https://www.lsa.umich.edu/physics/research/researchareas/elementaryparticlephysics}{\em \small University of Michigan, Randall Laboratory of Physics, 450 Church Street, Ann Arbor, MI 48109-1040, USA}}
 \end{center}
%
%
\begin{center}
 K.E.~Boast, 
 C.~Carels, 
 T.~Fruth, 
 H.~Kraus, 
 F.-T.~Liao, 
 J.~Lin,\footnote{Now  at: \href{http://physics.berkeley.edu/research/particle-physics}{University of California (UC), Berkeley, Department of Physics, 366 LeConte Hall MC 7300, Berkeley, CA 94720-7300, USA}} 
 P.R.~Scovell \\* 
 {\href{https://users.physics.ox.ac.uk/~kraus/research/DarkMatter.htm}{\em \small University of Oxford, Department of Physics, Oxford, OX1 3RH, UK}}
 \end{center}
%
%
\begin{center}
 E.~Druszkiewicz, 
 D.~Khaitan, 
 M.~Koyuncu, 
 W.~Skulski, 
 F.L.H.~Wolfs, 
 J.~Yin \\* 
 {\href{http://www.pas.rochester.edu/index.html}{\em \small University of Rochester, Department of Physics and Astronomy, Rochester, NY 14627-0171, USA}}
 \end{center}
\begin{center}
 E.V.~Korolkova, 
 V.A.~Kudryavtsev, 
 P.~Rossiter, 
 D.~Woodward \\* 
 {\href{https://www.hep.shef.ac.uk/research/dm/lz.php}{\em \small University of Sheffield, Department of Physics and Astronomy, Sheffield, S3 7RH, UK}}
 \end{center}
%
%
\begin{center}
 A.A.~Chiller, 
 C.~Chiller, 
 D.-M.~Mei, 
 L.~Wang, 
 W.-Z.~Wei, 
 M.~While, 
 C.~Zhang \\* 
 {\href{http://www.usd.edu/arts-and-sciences/physics}{\em \small University of South Dakota, Department of Physics, 414 East Clark Street, Vermillion, SD 57069-2307, USA}}
 \end{center}
%
%
\begin{center}
 S.K.~Alsum, 
 T.~Benson, 
 D.L.~Carlsmith, 
 J.J.~Cherwinka, 
 S.~Dasu, 
 G.~Gregerson, 
 B.~Gomber, 
 A.~Pagac, 
 K.J.~Palladino, 
 C.O.~Vuosalo, 
 Q.~Xiao \\* 
 {\href{https://www.physics.wisc.edu}{\em \small University of Wisconsin-Madison, Department of Physics, 1150 University Avenue Room 2320, Chamberlin Hall, Madison, WI 53706-1390, USA}}
 \end{center}
%
%
\begin{center}
 J.H.~Buckley, 
 V.V.~Bugaev, 
 M.A.~Olevitch \\* 
 {\href{https://physics.wustl.edu}{\em \small Washington University in St. Louis, Department of Physics, One Brookings Drive, St. Louis, MO 63130-4862, USA}}
 \end{center}
%
%
\begin{center}
 E.M.~Boulton,\footref{UCBerk}$^{,}$\footref{LBNLfoot} 
 W.T.~Emmet, 
 T.W.~Hurteau, 
 N.A.~Larsen,\footnote{Now at: \href{https://kicp.uchicago.edu/index.php}{The University of Chicago, The Kavli Institute for Cosmological Physics, William Eckhardt Research Center (ERC) - Suite 499, 5640 South Ellis Avenue, Chicago, IL 60637-1433, USA}} 
 E.K.~Pease,\footref{UCBerk}$^{,}$\footref{LBNLfoot} 
 B.P.~Tennyson, 
 L.~Tvrznikova\footref{UCBerk}$^{,}$\footref{LBNLfoot} \\* 
 {\href{http://physics.yale.edu}{\em \small Yale University, Department of Physics, 217 Prospect Street, New Haven, CT 06511-8499, USA}}
\end{center}

\setcounter{footnote}{0}
\cleardoublepage

\begin{center}
{\large\bfseries Acknowledgements}
\end{center}
\setlength{\emergencystretch}{3em}

This work was partially supported by the U.S. Department of Energy (DOE) under award numbers 
DESC0012704, 
DE-SC0010010, 
DE-AC02-05CH11231, 
DE-SC0012161, 
DE-SC0014223, 
DE-FG02-13ER42020, 
DE-FG02-91ER40674, DE-NA0000979, 
DE-SC0011702, 
DESC0012034, DESC0010072, DESC0006572, 
DE-SC0006605, 
DE-FG02-10ER46709, 
DE-AC02-07CH11359, 
DE-AC02-76SF00515, 
DE-AC52-07NA27344, 
DE-FG02-01ER41166, 
DE-FG02-91ER40628, 
DE-SC0015535, 
DE-SC0015708, and DE-SC0220187; 
by the U.S. National Science Foundation (NSF) under award numbers 
NSF PHY-110447, 
NSF PHY-1506068, NSF PHY-1312561, 
NSF PHY-1406943, 
NSF PHY-1642619, 
and NSF PHY-1429544, 
by the U.K. Science \& Technology Facilities Council under award numbers 
ST/K006428/1, ST/M003655/1, 
ST/M003981/1, 
ST/M003744/1, 
ST/M003639/1, 
ST/M003604/1, 
ST/M003469/1, 
ST/M003779/1, 
and ST/N000250/1; 
by the Portuguese Foundation for Science and Technology (FCT) under award numbers 
CERN/FP/123610/2011 and PTDC/FIS-NUC/1525/2014; 
by the National Research Nuclear University MEPhI 
(Moscow Engineering Physics Institute) in the 
framework of the MEPhI Academic Excellence Project (contract 
02.a03.21.0005, 27.08.2013); 
and by the Institute for Basic Science, Korea (budget numbers
IBS-R016-D1, and IBS-R016-S1).

University College London and Lawrence 
Berkeley National Laboratory thank the U.K. Royal Society for travel 
funds under the International Exchange Scheme (IE141517). We acknowledge 
additional support from the Boulby Underground Laboratory in the U.K., 
the University of Wisconsin for grant UW PRJ82AJ, the GridPP Collaboration, 
in particular at Imperial College London, 
and from the South Dakota Science and Technology Authority, 
and the State of South Dakota. The University of 
Edinburgh is a charitable body, registered in Scotland, with the 
registration number SC005336.

\begin{center}
{\large\bfseries Disclaimers}
\end{center}

This document was prepared as an account of work sponsored by the 
United States Government. While this document is believed to contain 
correct information, neither the United States Government nor any agency 
thereof, nor the Regents of the University of California, nor any of 
their employees, makes any warranty, express or implied, or assumes any 
legal responsibility for the accuracy, completeness, or usefulness of 
any information, apparatus, product, or process disclosed, or represents 
that its use would not infringe privately owned rights. Reference herein 
to any specific commercial product, process, or service by its trade name, 
trademark, manufacturer, or otherwise, does not necessarily constitute 
or imply its endorsement, recommendation, or favoring by the United 
States Government or any agency thereof, or the Regents of the University 
of California. The views and opinions of authors expressed herein do not 
necessarily state or reflect those of the United States Government or any 
agency thereof or the Regents of the University of California.

\begin{center}
{\large\bfseries Copyright Notice}
\end{center}

This manuscript has been authored by an author at Lawrence Berkeley 
National Laboratory under Contract No. DE-AC02-05CH11231 with the U.S. 
Department of Energy. The U.S. Government retains, and the publisher, by 
accepting the article for publication, acknowledges, that the U.S. 
Government retains a non-exclusive, paid-up, irrevocable, world-wide 
license to publish or reproduce the published form of this manuscript, 
or allow others to do so, for U.S. Government purposes.

\tableofcontents
\listoffigures
\listoftables
\cleardoublepage


\mainmatter
\pagestyle{scrheadings}

\renewcommand{\ChapLabel}{chap:IDO} 
\renewcommand{\BibLabel}{IDOS:bib} 
\renewcommand{\SecPre}{IDO}
\renewcommand{\FloatPre}{IDO}
\tdrchap{Overview}

\tdrsec[DirectDDM]{Direct Detection of Dark Matter}

During the past two decades, a standard cosmological picture of the
universe (the Lambda Cold Dark Matter or LCDM model) has emerged,
which includes a detailed breakdown of the main constituents of
the energy density of the universe. This theoretical framework is
now on a firm empirical footing, given the remarkable agreement
of a diverse set of astrophysical data~\cite{Spergel:2006hy,Percival:2006gt}.
Recent results by Planck largely confirm the earlier Wilkinson Microwave
Anisotropy Probe (WMAP) conclusions and confirm that the universe is spatially
flat, with an acceleration in the rate of expansion and an energy budget
comprising approximately \SI{5}{\percent} baryonic matter,
\SI{26}{\percent} cold dark matter (CDM), and roughly \SI{69}{\percent}
dark energy~\cite{Komatsu:2010fb,Ade:2013zuv,Ade:2015xua}.
With the generation-2 (G2) dark matter experiments, we are now
in a position to identify this dark matter through sensitive terrestrial
\emph{direct detection} experiments. Failing to detect a signal in G2, 
or in subsequent generation (G3), experiments would rule out most of 
the natural parameter space that describes weakly interacting massive
particles (WIMPs), forcing us to reassess the WIMP paradigm and look
for new detection techniques. In the following sections, we introduce
the cosmological and particle physics evidence pointing to the
hypothesis that the dark matter is composed of WIMPs, detectable 
through nuclear recoil (NR) interactions in low-background 
experiments. We then give the motivation for a massive liquid xenon
(LXe) detector as the logical next step in the direct detection of 
dark matter.

%
%

\tdrsubsec[Cosmo]{Cosmology and Complementarity}

While the Large Hadron Collider (LHC) experiments continue to verify
the Standard Model of particle physics to ever-greater precision,
the nature of the particles and fields that constitute dark energy
and dark matter remain elusive.
The gravitational effects of dark matter are evident throughout 
the cosmos, dominating gravitational interactions of objects as 
small as dwarf satellites of the Milky Way, up to galaxy clusters 
and superclusters.
Application of Kepler's laws leads to the inescapable conclusion 
that our own galaxy, and all others, are held together by the 
gravitational pull of a dark halo that outweighs the combined mass 
of stars and gas by an order of magnitude, and appears to form 
an extended halo beyond the distribution of luminous matter.

At the same time, very weakly interacting CDM, particles or compact
objects that were moving non-relativistically at the time of decoupling,
appear to be an essential ingredient in the evolution of structure in
the universe. N-body simulations of CDM can explain much of the
structure, ranging from objects made of tens of thousands of stars to
galaxy clusters. In the past few years, more realistic simulations,
including both baryonic matter (gas and stars) and dark matter, are
beginning to reveal how galaxy-like objects can arise from the primordial
perturbations in the early universe~\cite{Frenk:2012ph,Governato:2012fa}.

While we know much about the impact of dark matter on a variety of 
astrophysical phenomena, we know very little about its nature. 
An attractive conjecture is that dark matter particles were in equilibrium
with ordinary matter in the hot early universe. We note, however, that
there are viable dark matter candidates, including axions, where the
conjecture of thermal equilibrium is not made~\cite{Kusenko:2013saa}.
Thermal equilibrium describes the balance between annihilation of
dark matter into ordinary particle-antiparticle pairs, and vice versa.
As the universe expanded and cooled, the reaction rates (the product of
number density, cross section, and relative velocity) eventually fell
below the level required for thermal equilibrium, leaving behind a
relic abundance of dark matter. The lower the annihilation cross section
of dark matter into ordinary matter, the higher the relic abundance of
dark matter. An annihilation cross-section characteristic of the weak
interaction results in a dark matter energy consistent with that observed
by cosmological measurements~\cite{Lee:1977ua}.

\begin{wrapfigure}{o}[0pt]{0.60\textwidth}
\includegraphics[width=0.55\textwidth]{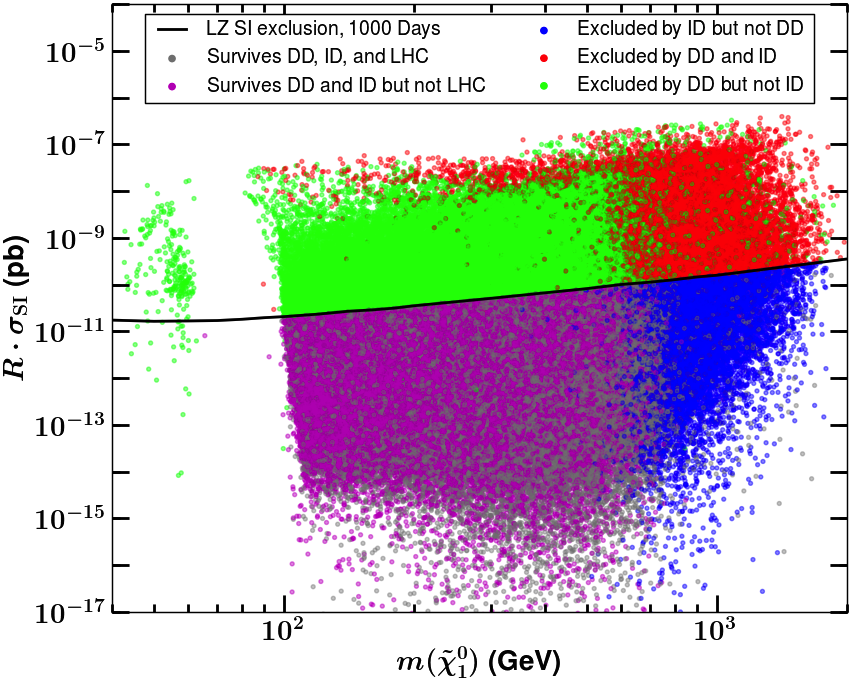}
\tdrfcaption[Complementarity]{
Scan of pMSSM parameter space and complementarity}
{Scan of pMSSM parameter space and complementarity.
Each point represents a SUSY model and is plotted as WIMP-nucleon
scattering cross section scaled by abundance versus mass of the WIMP.
The colors show models that can be tested by the three techniques of
detection: direct detection (DD), LHC, and indirect detection (ID),
and their combinations. The black curve shows the expected LZ
sensitivity based on an early (2013) LZ estimate, weakened by a
factor of four~\cite{CahillRowley:2014boa}.}
\end{wrapfigure} 

Models of supersymmetry (SUSY) predict the neutralino, a new 
particle that has properties appropriate to be a WIMP. SUSY
posits a fermion-like partner for every Standard Model boson, and a
boson-like partner for every Standard Model fermion. A principal
feature of SUSY is its natural means for achieving cancellations in
quantum field theory amplitudes that could cause the Higgs mass to
be much higher than the \SI{125}{\GeVpct} recently
observed~\cite{Aad:2012tfa,Chatrchyan:2012xdj}.
The neutralino is a coherent quantum state formed from the SUSY
partners of the photon, the \PZz, and Higgs boson \PHz, and is 
a ``Majorana'' particle, meaning it is its own antiparticle.

Astrophysical measurements show that dark matter behaves like a
particle and not like a modification of gravity. Gravitational
lensing of distant galaxies by foreground galactic clusters can
provide a map of the total gravitational mass, showing that this
mass far exceeds that of ordinary baryonic matter. By combining the
distribution of the total gravitational mass (from lensing) with the
distribution of the dominant component of baryonic matter (evident
in the X-ray-emitting cluster gas) one can see whether the dark mass
follows the distribution of baryonic matter. For a number of galaxy 
clusters, in particular the Bullet cluster~\cite{Clowe:2006eq}, 
the total gravitational mass (dominated by dark matter) follows 
the distribution of other non-interacting test particles (stars) 
rather than the dominant component of baryonic matter in the 
cluster gas. Combining this evidence with other observations 
of stellar distributions and velocity measurements for galaxies 
with a wide range of mass-to-light ratios, it appears that the 
total gravitational mass does not follow the distribution of 
baryonic matter as one would expect for modified gravity, but 
behaves like a second, dark component of relatively weakly 
interacting particles.

There are three complementary signals of WIMP dark matter. The dark matter
of the Milky Way can interact with atomic nuclei, resulting in NRs that
are the basis of direct detection (DD) experiments like LZ. At the LHC,
the dark matter should be produced as a stable, non-interacting particle 
that appears as missing energy and momentum. Out in the cosmos, dark matter
collects at the centers of galaxies and in the sun, where pairs of
dark matter particles will annihilate with one another, if the dark matter
is a Majorana particle, as expected in SUSY theories. The annihilations will
produce secondary particles, including positrons, antiprotons, neutrinos,
and gamma rays, providing the basis for ``indirect'' detection (ID) by gamma
ray, cosmic ray, and neutrino telescopes.

In Figure~\ref{IDOf:Complementarity}, we show the results of a 
recent analysis of the complementarity of the three signals from 
WIMP dark matter. The LHC has already provided constraints on the 
simplest SUSY parameter space, and the Higgs mass is in some 
tension with the most constrained versions of SUSY, requiring 
theorists to relax simplifying assumptions. One slightly less
restrictive choice of parameters is the so-called phenomenological 
minimal supersymmetric standard model (pMSSM) 
model~\cite{Djouadi:1998di,Berger:2008cq}.
In Figure~\ref{IDOf:Complementarity}, each point represents a 
choice of pMSSM parameters that satisfies all known physics and
astrophysics constraints~\cite{CahillRowley:2014boa}.
The color of the points show which experiments have adequate 
sensitivity to test whether that point is valid. 
The black curve shows the expected LZ sensitivity based on an 
early (2013) LZ estimate, weakened by a factor of four.
The three 
experimental tests are: DD with the proposed LZ experiment, 
current LHC data, and ID from a proposed ID experiment; an 
enhanced version of the Cherenkov Telescope Array\cite{Acharya:2013sxa}
with twice the number of telescopes compared with the current baseline.
Each of the three experimental techniques tends to be most sensitive to
one region in Figure~\ref{IDOf:Complementarity}, although
there are regions of overlapping sensitivity.

The most important goal for the G2 program is to produce the best
constraints over the natural mass range for dark matter. The LZ experiment
accomplishes this goal. A broader goal of dark matter research is to use the
three complementary approaches to establish the validity of any signal, and
to identify the properties of the dark matter particle. 

%
%

\tdrsubsec[DDExperiments]{Direct Detection Experiments}
\begin{wrapfigure}{o}{0.55\textwidth}
\begin{center}
\includegraphics[width=0.55\textwidth]{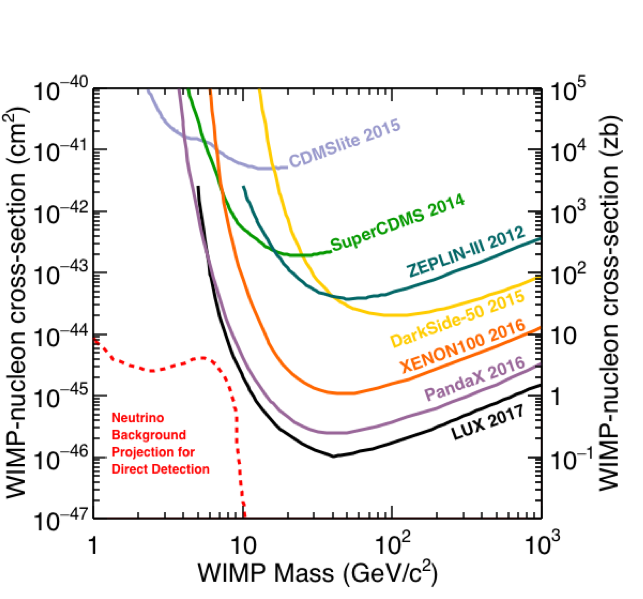}
\end{center}
\tdrfcaption[Spin-indep-2016]{Compilation of current 
WIMP-nucleon SI cross-section upper limits (\SI{90}{\percent} CL)}{A 
compilation of current WIMP-nucleon SI cross-section upper
limits at \SI{90}{\percent CL} confidence level (solid colored curves, 
labeled by experiment names).  The dotted region at the lower left 
indicates the region where solar neutrinos contribute background.}
\end{wrapfigure}

The direct detection of dark matter in earthbound experiments depends on
the local properties of the Milky Way's dark matter and on the properties
of the dark matter particles themselves. The local properties of the Milky
Way's dark halo are determined by astrophysical studies, and include the
local dark matter mass density as well as the distribution of the velocities
of dark matter particles. The conjecture that the dark matter particles
are WIMPs implies that their scattering with nuclei is non-relativistic
two-body scattering; in LZ, we seek to observe the xenon nuclei that recoil
after having been struck by an incoming WIMP. The mass assumed for the WIMP
determines the kinematics of the scattering, and the rate of WIMP-nucleus
scatters seen in a WIMP detector further depends on the exposure
defined as the product of the target mass and the live time, the 
WIMP-nucleus cross section, and the energy threshold for detection of 
the NR.

The dark matter halo of the Milky Way is harder to quantify than that of
other galaxies. The density of all matter, including dark matter, in
galaxies is quantified with rotation curves, which describe the average
circular velocity of matter in orbit about the galactic center as a function
of radius from the galactic center. The rotation curve of the Milky Way
for radii larger than \SI{8.0}{kpc}, that of our sun, is a 
challenge to quantify.
Estimates of the local dark matter density range from
\SI{0.235+-0.030}{\GeV\per\cm\cubed} to
\SI{0.389+-0.025}{\GeV\per\cm\cubed}~\cite{Sofue:2011kw,Bovy:2012tw,Catena:2009mf}.
Most DD experiments adopt, for ease of
inter-comparison, a standard value for the local dark matter mass density
of $\rho_0$ = \SI{0.3}{\GeV\per\cm\cubed}.
The experiments also adopt a standard distribution
function for the velocity of dark matter particles, characterized by a
Maxwell-Boltzmann distribution with solar circular velocity
$v_0$ = \SI{220}{\km\per\second}, which is cut off at the galactic escape velocity of
$\vesc$ = \SI{544}{\km\per\second}, and with proper accounting for the sun's peculiar
velocity and the periodic annual motion of the
Earth~\cite{Freese:1987wu,Savage:2006qr}. The minimum WIMP mass detectable with a particular DD experiment depends
on the maximum velocity in the galactic WIMP spectrum, the atomic mass $A$
of the target nucleus, and the energy threshold \Emin\ for NR detection in
that experiment. Straightforward kinematics gives the minimum detectable
WIMP mass as
$\MWIMP(\si{\GeV})\approx(1/4)\sqrt{\Emin(\si{\keV})A}$
for a maximum velocity of $\vesc$ = \SI{544}{\km\per\second}.
This minimum \MWIMP\ is \SI{1.6}{\GeV} for the recent CDMSlite
Ge-target result~\cite{Agnese:2015nto},
and \SI{4}{\GeV} for the recent LUX result~\cite{Akerib:2015rjg} using LXe.

These results are shown in Figure~\ref{IDOf:Spin-indep-2016},
along with the current experimental situation for the spin-\-independ\-ent
(SI) WIMP-nucleon cross section. We discuss the details of spin 
independence and other types of WIMP-nucleon interactions in
Chapter~\ref{chap:SIP}. The SI cross section is the standard 
benchmark, and would result from a WIMP coupling to the Standard
Model Higgs.
As the presumed \MWIMP\ rises above this value, the portion of 
the WIMP velocity distribution function that permits NR above 
the detectable threshold rises rapidly. This rapid rise drives 
the improvement in sensitivities as \MWIMP\ rises above \SI{5}{\GeV}, 
as shown in Figure~\ref{IDOf:Spin-indep-2016}.
As \MWIMP\ approaches the mass of the atom used in the target, the
kinematics of energy transfer to the target nuclei becomes most 
efficient, and experiments reach their maximum sensitivity. 

\begin{wrapfigure}{o}{0.6\textwidth}
\begin{center}
\includegraphics[width=0.59\textwidth]{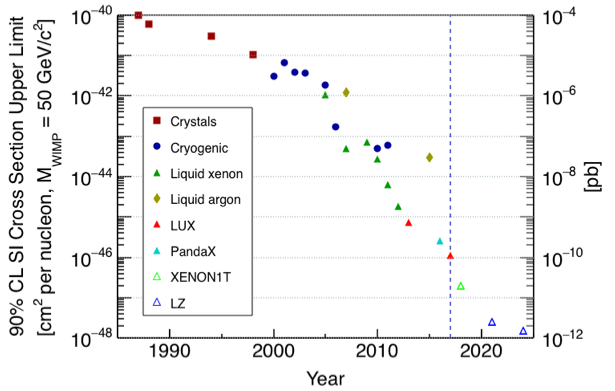}
\end{center}
\tdrfcaption[history1]{Evolution of cross-section limits 
for 50-GeV WIMPs as function of time}
{The evolution of cross-section limit for 50-GeV WIMPs as 
a function of time. Past points are published results. 
Future points are from the Snowmass 
meetings~\protect\cite{Cushman:2013zza}.}
\end{wrapfigure} 

This region, with $\MWIMP$ > \SI{10}{\GeV} roughly, is the region 
most likely for a WIMP, as other weak-interaction particles, 
including the \PZz, the \PW, and the Higgs particles, have 
masses in that region. As \Emin\ grows yet larger, the number
density of WIMPs implied by the astrophysical mass density,
$\rho_0$~=~(\SI{0.3}{\GeV}/\MWIMP)\,\si{\per\cm\cubed}
falls, reducing the sensitivity of direct-detection
experiments, as shown in Figure~\ref{IDOf:Spin-indep-2016}.
The maximum WIMP mass consistent with thermal equilibrium 
in the early universe is \SI{340}{\TeVpct}, above which
unitarity can no longer be satisfied~\cite{Griest:1989wd}.

The history of and future projections for WIMP sensitivity 
are shown in Figure~\ref{IDOf:history1}.
There are three distinct eras: (1) 1986-1996; (2) 2000-2010;
and (3) post-2010. In the first era, low-background Ge and 
NaI crystals dominated. In these experiments, discrimination 
between the dominant background of electron recoils (ERs) 
from gamma rays and NRs was not available. The cross-section 
sensitivity per nucleon achieved,
\SI{e-41}{\cm\squared}, 
was sufficient to rule out the most straightforward WIMP implementation: 
that the WIMP is a heavy Dirac neutrino, which was the original 
suggestion of Ref.~\cite{Lee:1977ua}.
In the second era, cryogenic Ge detectors with the ability 
to distinguish NR from the ER background dominated, and
achieved a cross-section per nucleon
sensitivity of \SI{5e-44}{\cm\squared}.
This sensitivity began to probe dark matter that is a 
Majorana fermion, which couples to nucleons via the Higgs 
particle. In the third era, still under way, LXe time 
projection chambers (TPCs) have been most sensitive, and
with the LUX experiment have achieved an upper limit for
the WIMP-nucleon spin-independent cross section of
\SI{1.1e-46}{\cm\squared}(=\SI{1.1e-10}{\pico\barn})
~\cite{Akerib:2016vxi}.
The LXe TPC has the ability to discriminate
ER and NR, and can be expanded to large, homogeneous volumes. 
It is possible to make accurate predictions for the 
backgrounds in the central LXe TPC region, particularly 
with the added power of outer detectors that can
characterize the radiation field in the TPC vicinity.

A more detailed portrayal of the variety of experiments, both from the past
and projected into the future, is given in Figure~\ref{IDOf:Snowmass}
~\cite{Cushman:2013zza}. In the region \MWIMP \SI{>10}{\GeVpct}, most 
likely for a WIMP due to proximity to the masses of the weak bosons, 
LZ will be the most sensitive experiment. Figure~\ref{IDOf:Snowmass}
also shows the ``neutrino floor'', where NRs from coherent neutrino 
scattering, a process that has not yet been observed, will greatly 
influence progress in sensitivity to WIMP interactions. A \num{1000}-day 
run of the LZ experiment will just begin to touch this background. 
Searches for direct interactions of dark matter with exposure 
substantially greater than LZ will see many candidate events from NRs 
in response to, primarily, atmospheric neutrinos.

\begin{figure}[tbp]
\begin{center}
\includegraphics[width=0.9\textwidth]{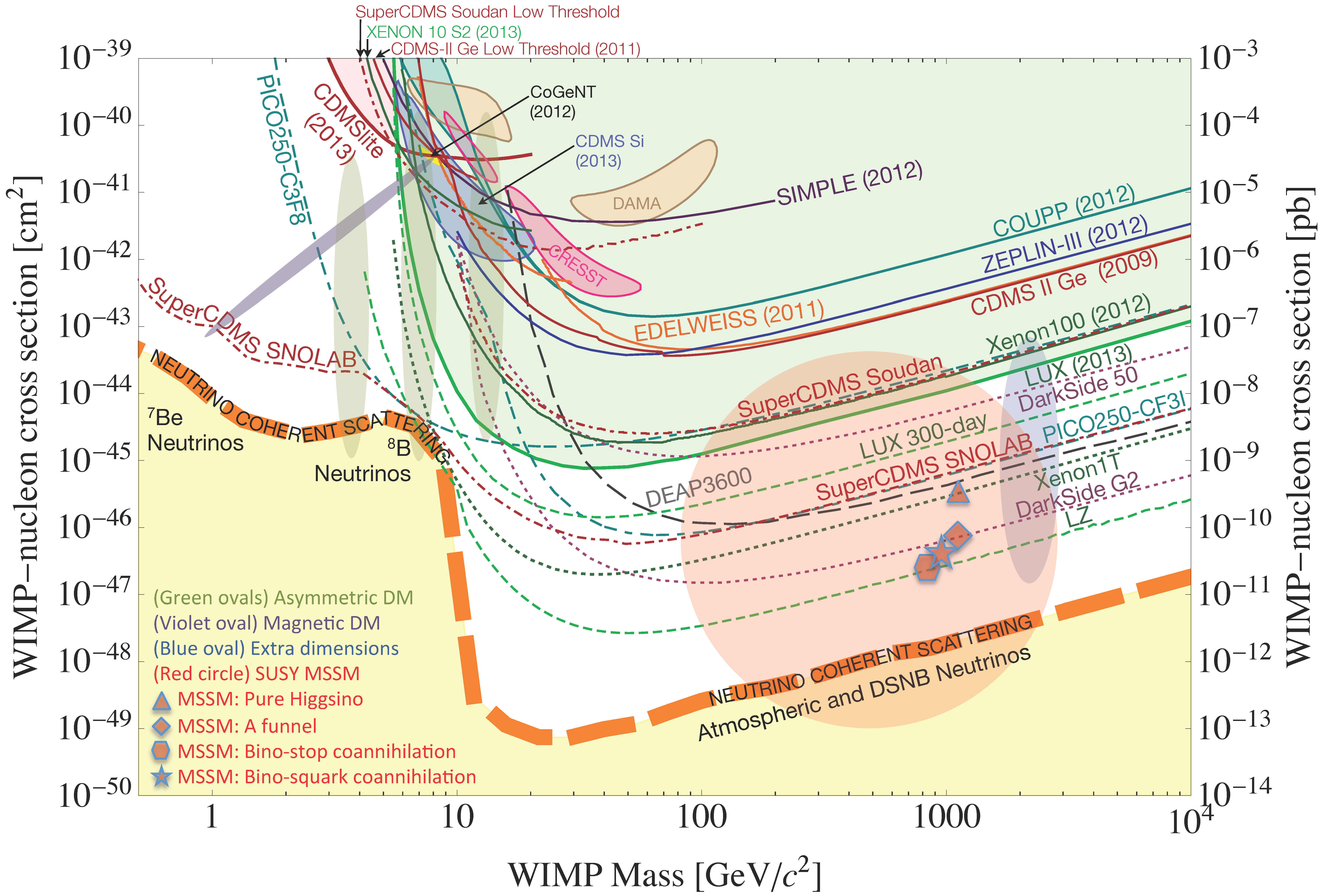}
\end{center}
\tdrfcaption[Snowmass]{A compilation of WIMP-nucleon SI 
cross-section sensitivity}
{A compilation of WIMP-nucleon SI cross-section sensitivity (solid
curves), hints for WIMP signals (shaded closed contours), and
projections (dot and dot-dash curves) for DD experiments of the past
and projected into the future. This figure captures the experimental
situation as of the end of 2013 and was reported in the 2013 Snowmass
CF1 summary. Also shown is an approximate band where
the coherent nuclear scattering of \IBe solar neutrinos, atmospheric
neutrinos, and diffuse supernova neutrinos will limit the sensitivity
of DD experiments to WIMPs. Finally, a suite of theoretical model
predictions is indicated by the shaded regions, with model references
included~\protect\cite{Cushman:2013zza}.}
\end{figure}

\clearpage
\tdrsec[InstOverview]{Instrument Overview}
The core of the LZ experiment is a two-phase xenon (Xe) time projection 
chamber (TPC) containing \num{7} fully active tonnes of LXe. 
Scattering events in LXe create both a prompt scintillation signal (S1) and 
free electrons. Electric fields are employed to drift the electrons 
to the liquid surface, extract them into the gas phase above, and accelerate 
them to create a proportional scintillation signal (S2). Both signals are 
detected by arrays of photomultiplier tubes (PMTs) above and below the central 
region. The difference in time of arrival between the signals measures the 
position of the event in $z$, while the $x$, $y$ position is determined from the 
pattern of S2 light in the top PMT array. Events with an S2 signal but no S1 
are also recorded. A 3-D model of the LZ detector located in a large water 
tank is shown in Figure~\ref{IDOf:LZSolid}. The water tank is located at the 
\num{4850}-foot level (4850L) of the Sanford Underground Research Facility (SURF). 
The heart of the LZ detector (including the inner titanium [Ti] cryostat) 
will be assembled on the surface at SURF, lowered in the Yates shaft to 
the 4850L of SURF, and deployed in the existing water tank in the Davis 
Cavern (where LUX is currently located). The LZ experiment’s principal 
parameters are given in Table~\ref{IDOt:WBS}, along with the Work 
Breakdown Structure (WBS) for the LZ Project.

\begin{figure}[htbp]
\begin{center}
\includegraphics[width=0.9\textwidth]{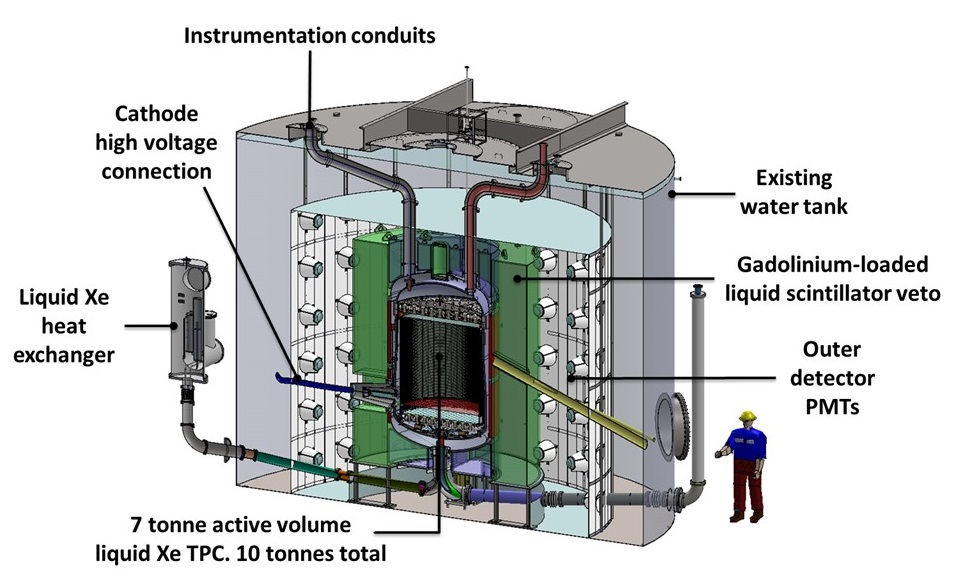}
\end{center}
\tdrfcaption[LZSolid]{The LZ detector concept}{The LZ detector concept.}
\end{figure} 

The LZ detector includes several added capabilities beyond the successfully 
demonstrated LUX and ZEPLIN designs. The most important addition is a nearly 
hermetic liquid organic scintillator (gadolinium-loaded linear alkyl benzene 
[LAB]) outer detector, which surrounds the central cryostat vessels and the TPC. 
The outer detector and the active Xe ``skin'' layer, the Xe between the inner 
cryostat wall and the outer wall of the TPC, operate as an integrated veto 
system, which has several benefits. The first is rejecting gammas and neutrons 
generated internally (e.g., in the PMTs) that scatter a single time in the 
fully active region of the TPC and would otherwise escape without detection; 
this could mimic a weakly interacting massive particle (WIMP) signal. As 
these internally generated backgrounds interact primarily at the outer regions 
of the detector, the veto thus allows an increase in the fiducial volume.

\begin{center}
\sffamily
\renewcommand{\ltbcp}{WBS}
\renewcommand*{\arraystretch}{1.2}
\setlength{\extrarowheight}{2pt}
\begin{longtable}{ | m{0.9cm} | m{8.4cm} | m{5.6cm} | }
\caption[Principal parameters and WBS \ifKL\space\texttt{\FloatPre t:\ltbcp}\fi]{Principal parameters of the LZ detector organized to reflect the responsible work package under the Work Breakdown Structure (WBS). \ifKL\space\texttt{\FloatPre t:\ltbcp}\fi\label{\FloatPre t:\ltbcp}}\\
\hline
\rowcolor{mrocol}
{\bfseries WBS} & {\bfseries Description} & {\bfseries Quantity} \\ \hline
\endfirsthead
\caption[]{\sfs (continued)}\\
\hline
\rowcolor{mrocol}
{\bfseries WBS} & {\bfseries Description} & {\bfseries Quantity} \\ \hline
\endhead
\rowcolor{lrocol}
\multicolumn{3}{|c|}{\sfs\Tstrut (continued on next page)}\\
\hline
\endfoot
\endlastfoot
\hline 
{\bfseries 1.1} & {\bfseries Xenon Procurement} & \\ \hline
    & Approximate total mass  & \SI{10}{\tonnesl} \\ \hline
    & Mass inside TPC active region & \SI{7}{\tonnesl} \\ \hline
    & Approximate fiducial mass & \SI{5.6}{\tonnesl} \\ \hline
{\bfseries 1.2} & {\bfseries Xe Vessel(Cryostat) }  & \\ \hline
    & Inner cryostat---inside diameter (tapered), height, wall thickness & \SI{1.58}-\SI{1.66}{\m}, \SI{2.59}{\m}, \SI{8}-\SI{11}{\mm} \\ \hline
    & Outer cryostat---inside diameter, height, wall thickness & \SI{1.83}{\m}, \SI{3.04}{\m}, \SI{7}-\SI{14}{\mm} \\ \hline
    & Approximate cryostat weights---inner, outer & \SI{0.95}{\tonnel}, \SI{1.12}{\tonnel} \\ \hline
{\bfseries 1.3} & {\bfseries Cryogenic System} & \\ \hline
          & Cooling power  & \SI{1}{\kW} @ \SI{80}{\kelvin} \\ \hline
          & Input electrical power & \SI{11}{\kW} \\ \hline
{\bfseries 1.4} & {\bfseries Xenon Purification} & \\ \hline
          & Krypton content & \SI{<0.015}{\ppt} (\si{g/g}) \\ \hline
          & Allowed air content (Kr equivalent) & \SI{<40} cc \\ \hline
          & \IRnttt content in active Xe & \SI{<2}{\micro\becquerel/kg} \\ \hline
          & Recirculation rate & \SI{500}{\slpm} \\ \hline
          & Electron lifetime & \SI{>0.8}{\ms} \\ \hline
          & Charge attenuation length & \SI{>1.5}{\m} \\ \hline
{\bfseries 1.5} & {\bfseries Xenon Detector} & \\ \hline
          & Top (Bottom) TPC 3-inch PMT array & 253 (241), 494 total tubes \\ \hline
          & ``Side skin'' PMT and ``dome'' PMT & 113 and 18, 131 (93 1-inch,38 2-inch) total \\ \hline
          & Nominal (Design) cathode operating voltage & \num{50} (\num{100}) \si{\kV} \\ \hline
          & Reverse field region (cathode to bottom tube shield) & \SI[group-separator={}]{0.1375}{\m} \\ \hline
         & TPC height (cathode to gate grid) & \SI{1.456}{\m} \\ \hline
         & TPC effective diameter & \SI{1.456}{\m} \\ \hline
{\bfseries 1.6} & {\bfseries Outer Detector System} & \\ \hline
          & Weight of Gd-loaded LAB scintillator & \SI{17.5}{\tonnesl} \\ \hline
          & Number of acrylic vessels, total acrylic mass & 9 vessels, \SI{3.1}{\tonnesl} \\ \hline
          & Number of 8-inch PMTs & 120 \\ \hline
          & Minimum thickness of scintillator & \SI{0.61}{\m} \\ \hline
          & Diameter of water tank & \SI{7.62}{\m} \\ \hline
          & Height of water tank & \SI{5.92}{\m} \\ \hline
          & Approximate weight of water & \SI{228}{\tonnesl} \\ \hline
{\bfseries 1.7} & {\bfseries Calibration System} & \\ \hline
          & Number of source deployment tubes & 3 \\ \hline
          & Number of neutron tubes & 3 \\ \hline
          & Other calibration tools & \IKreTm, \Pn-generator, tritiated \CCHf, \IArTS, \IXemes \\ \hline
{\bfseries 1.8} & {\bfseries Electronics, DAQ, Controls, and Computing} & \\ \hline
          & Trigger rate (all energies, \SIrange{0}{40}{\keV}) & \SI{40}{\Hz}, \SI{0.4}{\Hz} \\ \hline
          & Average event size (noncalibration, uncompressed) & \SIrange{0.2}{1.0}{\mega\byte} \\ \hline
          & Data volume per year & \SIrange{350}{850}{\tera\byte} \\ \hline
{\bfseries 1.9} & {\bfseries Integration and Installation} & \\ \hline
{\bfseries 1.10} & {\bfseries Cleanliness and Screening} & \\ \hline
{\bfseries 1.11} & {\bfseries Offline Computing} & \\ \hline
{\bfseries 1.12} & {\bfseries Project Management} & \\ \hline
\end{longtable}
\end{center}
Additionally, this direct vetoing is an important means of risk mitigation 
against one of the detector components (e.g., the cryostat materials or the 
PMTs) having a higher-than-expected background. A second benefit of the 
outer detector is that the combination of outer detector and segmented Xe 
detector will form a nearly hermetic detection system for all internal 
radioactivity. This will not only directly measure the internal backgrounds 
with a very high level of detail and completeness, but also help provide an 
understanding of the detector’s response to those backgrounds that can be 
included in our analysis for dark matter signals.

As shown in Figure~\ref{IDOf:LZSolid}, the liquid scintillator volume is
confined in segmented, clear acrylic vessels encapsulating the Ti cryostat 
of the central LZ detector. PMTs mounted on ladders in the outer water 
shield simultaneously view the light from both the scintillator and inner 
water volumes. The design of the acrylic vessels is advanced and industrial 
fabrication of the vessels has started. The production of 
the low-radioactivity liquid scintillator is based on the successful 
production of a similar scintillator for the Daya Bay experiment, albeit 
with lower backgrounds. The layer of LXe located between the PTFE 
(polytetrafluoroethylene, or Teflon\textregistered) structure that surrounds 
the fully 
active region and the cryostat wall, as well as the LXe region beneath 
the bottom PMT array forms the ``skin'' element of the veto system. 
A skin of some ($\sim$few \si{\cm}, more in the bottom region of the detector) 
thickness is difficult to avoid given the TPC geometry, the need for 
high-voltage (HV) standoff, and the strong mismatch in thermal expansion 
between the PTFE panels and Ti vessels. The skin readout alone has a limited 
veto efficiency, but has sufficient gamma-stopping power to augment the 
scintillator veto. The combination of skin readout and outer detector 
creates a highly efficient integrated veto system. Scintillation light 
from the skin region is observed by a sparse PMT array dedicated to this 
region. Thin PTFE panels are attached to the inner cryostat wall and 
bottom (dome) region to enhance light collection.

For detector calibration, neutron and gamma-ray sources will be brought 
next to the wall of the inner cryostat via an array of three source tubes 
that penetrate the water and organic scintillator. The principal calibrations, 
metastable krypton and xenon and tritiated methane, will be introduced directly into 
the LXe via the Xe gas-handling system to allow in situ calibration. An external 
neutron generator that produces neutrons through a deuterium-deuterium fusion 
reaction will also be employed, and the neutrons penetrate through the water 
and scintillator veto through dedicated tubes. Additional neutron calibration 
sources will also be employed.

Another key determinant of the sensitivity of the experiment is the level of 
discrimination of electron recoil (ER) backgrounds from nuclear recoils (NRs). 
This depends on the electric field established in the TPC. The nominal operating 
voltage (cathode-to-anode) is \SI{50}{\kV} but all components will be designed to a 
voltage of \SI{100}{\kV} (and in general tested to higher voltages) to have sufficient 
operating margin.

The LZ TPC detector will employ Hamamatsu R11410-20 3-inch-diameter PMTs 
with a demonstrated low level of radioactive contamination and high quantum 
efficiency. Materials for these PMTs have been procured and screened to 
demonstrate low radioactivity. Tube production and testing began in 2016. Care will 
be taken in the design of the highly reflective PTFE TPC structure to isolate 
light produced in the central volume from the skin region, and vice-versa. 
Additional 1-inch PMTs will be utilized in the cylindrical skin region as well as 
a few 2-inch tubes (from LUX)  and 2-inch tubes used in the dome region.

The design of TPC mechanical and HV systems is advanced. Small prototypes 
of the high-electric field region of the TPC have been fabricated and tested 
in liquid argon (LAr) and successfully in a LXe test facility at SLAC National 
Accelerator Laboratory (SLAC). A scaled-prototype of the TPC and all grids 
that shape the electric fields is being tested at SLAC. HV systems, including the 
cathode HV delivery system, will be tested in LAr at LBNL in a dedicated 
facility. Assembly of the TPC structures and HV elements will occur primarily 
at LBNL-SLAC. The low-radioactivity Ti mechanical supports for the 3-inch 
PMTs will be industrially manufactured and the PMTs loaded and tested at 
Brown University.

LZ will employ an array of liquid nitrogen (LN)-cooled thermosyphons to 
control the detector temperature and minimize thermal gradients. The 
purification system is a scaled-up version of the LUX room-temperature 
gas-phase purification system, and will exploit the liquid/gas heat-exchanger 
technology developed for LUX to minimize LN consumption. A cryocooler will 
be used underground to manufacture LN, and LN storage vessels used by LUX 
will be retained as backup. The Xe circulation and purification system is 
similar conceptually to that used in LUX but with a much higher capacity. 
The 10 tonnes of Xe will be circulated and purified in \numrange{2}{3} days. The Xe 
purity will be monitored carefully by a dedicated system, expanding and 
improving on a similar system used in LUX. Radon will be removed from 
selected regions of the detector that contain Xe gas using a dedicated 
system underground, to lower the radon content.

The acquisition of the \SI{10}{\tonnesl} of Xe has started. The Xe used in LUX 
and in other devices will be reused. Procurement began 
in 2015 and is planned to end by mid-2018. All 
gaseous Xe is to be delivered to SLAC, where krypton will be removed through 
a chromatographic process using scaled-up techniques already demonstrated 
successfully for LUX. 

The electronics front-end, trigger and data acquisition, and slow controls 
are based on the LUX experience but will be significantly expanded and 
improved. A test of the PMT analog-amplifier-digitizer system, with full 
cable lengths, has been successfully completed. In addition, a few PMTs in
the LUX experiment were operated successfully with prototype electronics systems after 
the completion of LUX data taking in mid-2016. 
Software and computing systems for LZ are based on the successful operation 
of LUX, analysis of LUX data, and the experience of other experiments at LBNL 
and collaborating institutions. LZ data will be buffered locally at SURF and 
then transmitted to primary data storage at LBNL that will be mirrored in the 
U.K. The analysis framework to be used by the LZ collaboration has been defined 
and the simulation of the LZ detector and its response is already well advanced. 
The results are given in other chapters of this Technical Design Report. 

Components of LZ will be manufactured at a number of locations and brought 
to SURF. Assembly of the Xe detector and the inner cryostat will occur in 
a dedicated and upgraded cleanroom at SURF that includes an augmented air 
handling system to reduce radon during the assembly process. The construction 
of the facilities to house the upgraded cleanroom began in September 2016. 
The reduced-radon air handling system is under construction as is a 
reduced-radon cleanroom. Work on improved surface facilities will be 
completed by mid-2017. A comprehensive system for tracking 
components, ensuring cleanliness and a very low dust environment, will be 
employed to minimize radioactive backgrounds. The inner cryostat with the 
Xe detector inside will be lowered as a unit down the Yates shaft and transported 
to the Davis Cavern water tank. Another system will be built underground to 
provide low-radon air to the water tank during the connection of the 
inner cryostat to other systems and for other uses. All other components 
will be staged at SURF, lowered via the Yates shaft, and similarly transported. 
This includes the segmented acrylic vessels for the outer detector system 
and other large components. The gaseous Xe will be stored in specially 
designed cylinders in a dedicated room underground close to the Davis 
Cavern. 

\tdrsec[DesignDrivers]{Design Drivers for WIMP Identification}
Having established the motivation to perform direct searches for WIMP dark
matter, we introduced in the previous section the configuration of LZ.
Searching for events that are rare (\SI{\lesssim0.1}{\pertd})
and that involve very small energy transfers (\SI{\lesssim100}{\keV}) is
extremely challenging. This section focuses on the more salient features 
of the experiment and the detection medium, and how these will contribute 
to the identification of a galactic WIMP signal with low systematic 
uncertainty. The detailed design and its technical implementation are 
described in later sections; here, we address the key requirements that 
drive the technical design.

%
%

\tdrsubsec[ExperimentStrategy]{Overview of the Experimental Strategy}
Xenon has long been recognized as a very attractive WIMP target 
material \cite{Belli:1990bx,Davies:1994bz,Benetti:1993203}.
Its high atomic mass provides a good kinematic match to intermediate 
WIMP masses of O(\SI{100}{\GeVpct}) and the largest spin-independent 
scattering cross section among the available detector technologies, 
as illustrated in Figure~\ref{IDOf:Xe-WIMP}. Sensitivity to lighter 
WIMPs, with masses of O(\SI{10}{\GeVpct}), can be also be achieved, 
given the excellent low-energy scintillation and ionization yields 
in the liquid phase \cite{Akerib:2013tjd}. Xenon contains neither 
long-lived radioactive isotopes with troublesome decays nor 
activation products that remain significant after the first few months
of underground deployment. It is also sensitive to spin-dependent 
interactions via the odd-neutron isotopes \IXeotn and \IXeoTo, 
which account for approximately half of the natural isotopic abundance. 
If a WIMP discovery were made, the properties of the new particle 
could be studied by altering the isotopic composition of the target. 
This broad WIMP sensitivity confers maximum discovery potential to LZ.

\begin{figure}[htbp]
\centering
\begin{minipage}{.5\textwidth}
  \centering
\includegraphics[width=7cm]{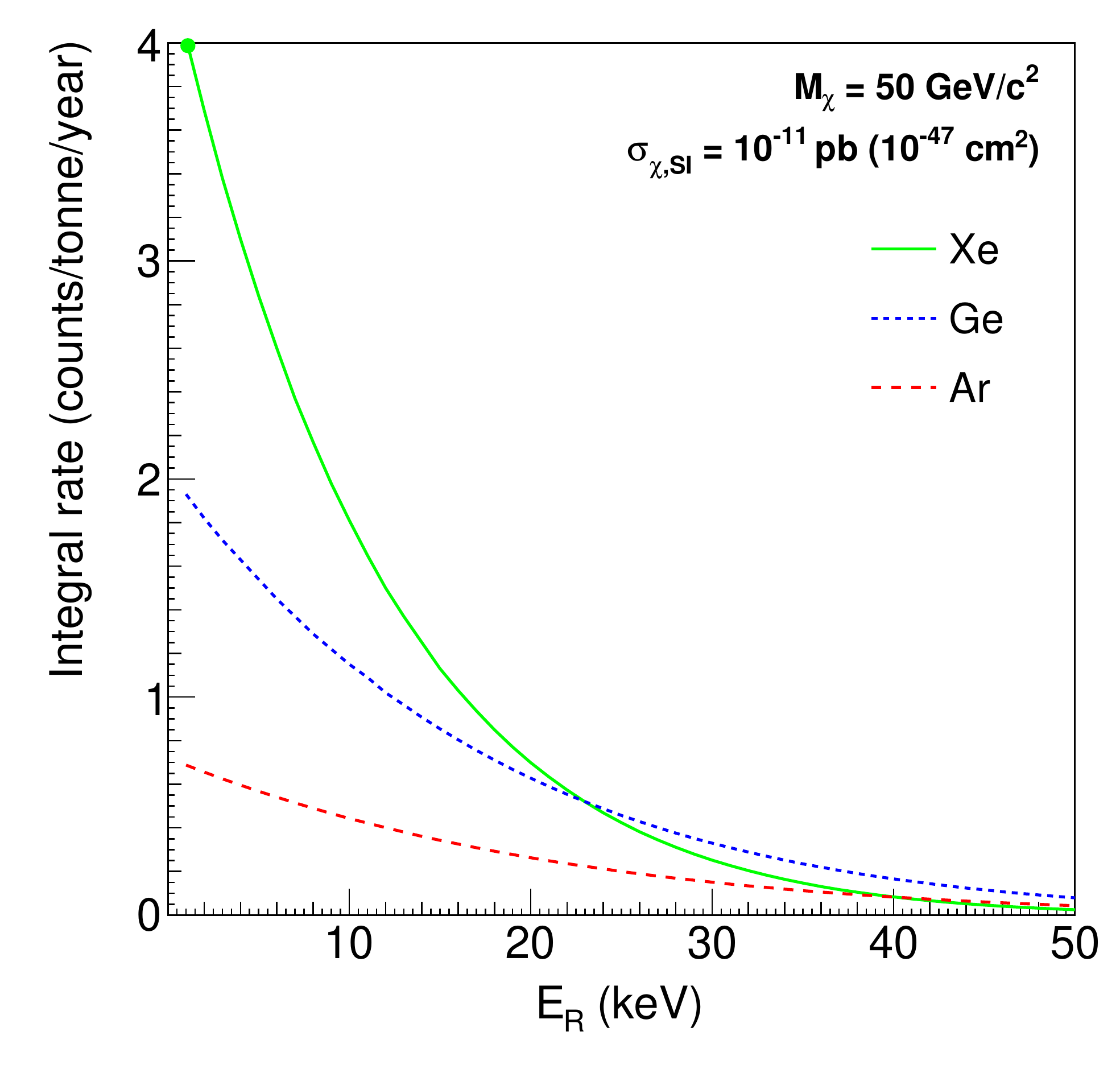}
\label{IDO:Xe-WIMPa}
\end{minipage}%
\begin{minipage}{.5\textwidth}
  \centering
\includegraphics[width=7cm]{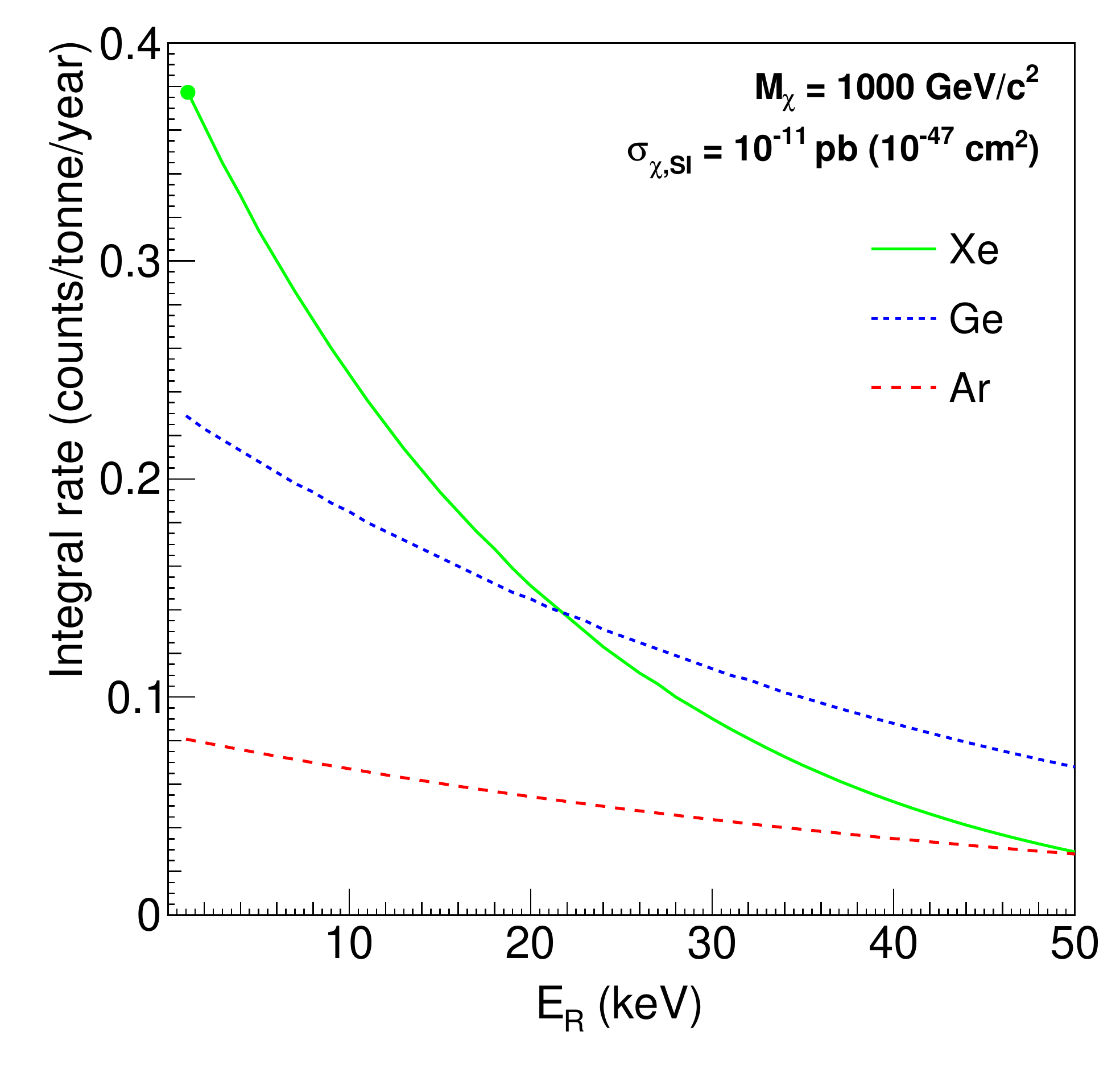}
\label{IDO:Xe-WIMPb}
\end{minipage}
\tdrfcaption[Xe-WIMP]{Integrated rates above threshold for various targets}
{Integrated rate above threshold per tonne-year of exposure for 
WIMP elastic scattering on Xe, Ge, and Ar targets for 
\SI{50}{\GeVpct} and \SI{1}{\TeVpct} WIMP masses and 
\SI{e-47}{\cm\squared} interaction cross section 
per nucleon. The green marker indicates the \SI{1.1}{\keVnr} 
WIMP-search threshold in LUX with nominal ER/NR discrimination
\protect\cite{Akerib:2013tjd}.
CDMS II searched above \SI{\sim5}{\keV} in their Ge target 
\protect\cite{Agnese:2015ywx}; 
selected SuperCDMS detectors allowed a \SI{1.6}{\keVnr} 
threshold with lower discrimination 
\protect\cite{Agnese:2014aze}; CDMSlite search 
for low-mass WIMPs with an electron recoil threshold down to 
\SI{56}{\eV} \protect\cite{Agnese:2015nto}.
In LAr, the DarkSide-50 experiment has recently conducted a 
WIMP search above \SI{13}{\keVnr} \protect\cite{Agnes:2015ftt}.}
\end{figure}

\begin{figure}[ht]
\begin{center}
\includegraphics[width=0.48\textwidth]{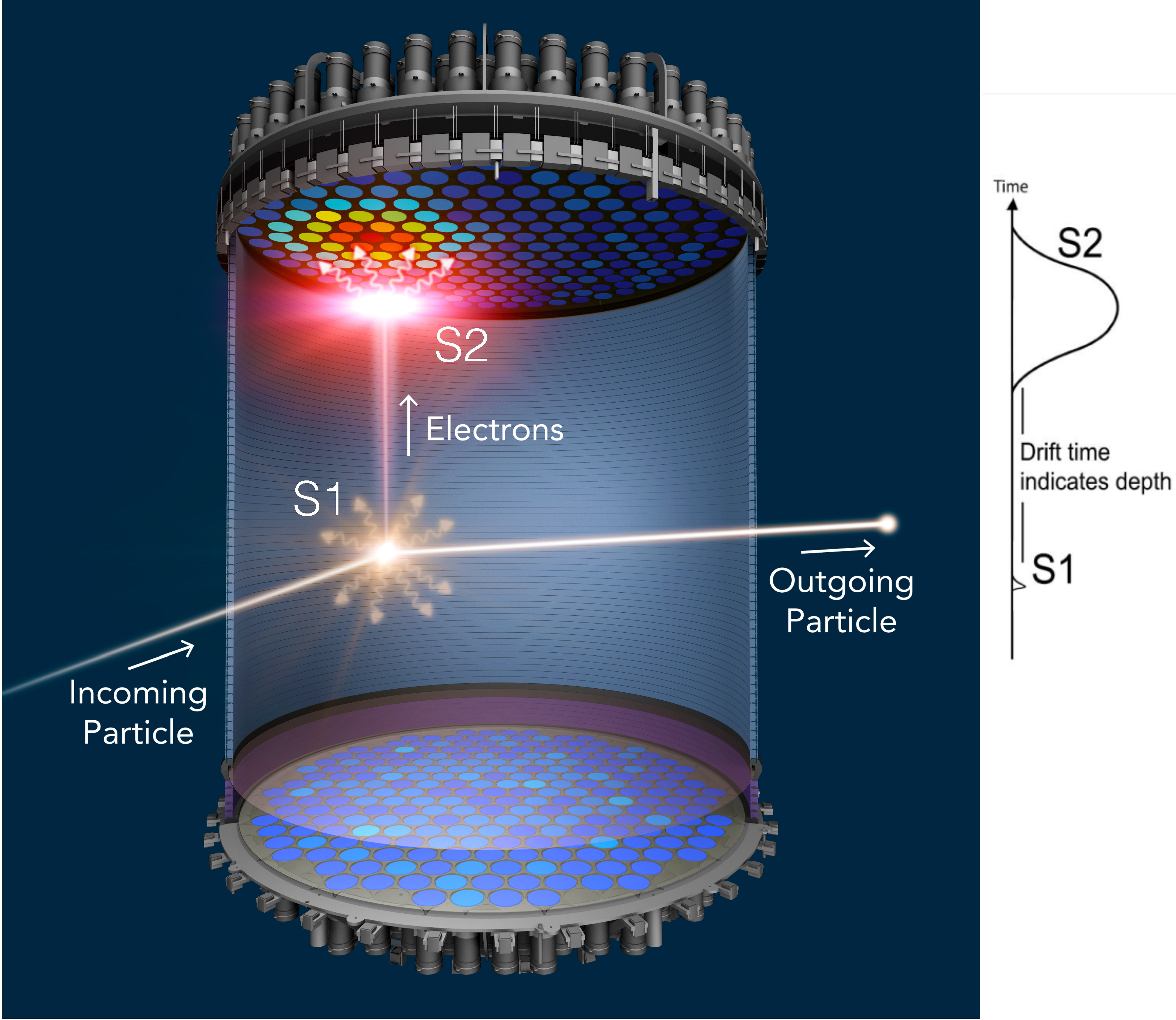}
\end{center}
\tdrfcaption[TPC-config-dpd]{Operating principle of the double-phase Xe TPC}
{Operating principle of the double-phase Xe TPC. Each particle 
interaction in the LXe (the WIMP target) produces two signatures: 
one from prompt scintillation (S1) and a second, delayed one from 
ionization, via electroluminescence in the vapor phase (S2). This 
allows precise vertex location in three dimensions and discrimination 
between nuclear and electron recoils.}
\end{figure}
\clearpage

The liquid phase is preferred over the gas phase due to its high 
density (\SI{3}{g\per\cm\cubed}) and high scintillation yield, and because 
its charge quenching of NRs provides a powerful particle ID mechanism. 
Early experiments such as ZEPLIN-I~\cite{Alner:2005pa} exploited 
simple pulse shape discrimination (PSD) of the scintillation
signal to reject electronic backgrounds; however, this achieved 
modest rejection efficiencies and only at relatively high recoil 
energies. When the first double-phase Xe detectors 
were deployed  for dark matter searches, in the 
ZEPLIN-II/III~\cite{Alner:2007ja,Akimov:2006qw} 
and XENON10~\cite{Aprile:2010bt} experiments, the 
increase in engineering complexity soon paid off 
in sensitivity, and this technique has been at the 
forefront of the field ever since. Comprehensive reviews 
on the application of the noble liquids to rare-event 
searches can be found in the literature~
\cite{Chepel:2012sj,Aprile:2009dv}.

The TPC configuration at the core of double-phase detectors, 
illustrated in Figure~\ref{IDOf:TPC-config-dpd}, has several
notable advantages for WIMP searches, in that two signatures 
are detected for every interaction: a prompt scintillation
signal (S1) and the delayed ionization response, detected 
via electroluminescence in a thin gaseous phase above the 
liquid (S2). These permit precise event localization in 
three dimensions (to within a few \si{\mm} 
\cite{Solovov:2011aa}) and discrimination between 
electron and nuclear recoil events (potentially 
reaching \SI{99.99}{\percent} 
rejection \cite{Lebedenko:2008gb}).

\begin{wrapfigure}{o}[0pt]{0.60\textwidth}
\begin{center}
\includegraphics[width=0.52\textwidth]{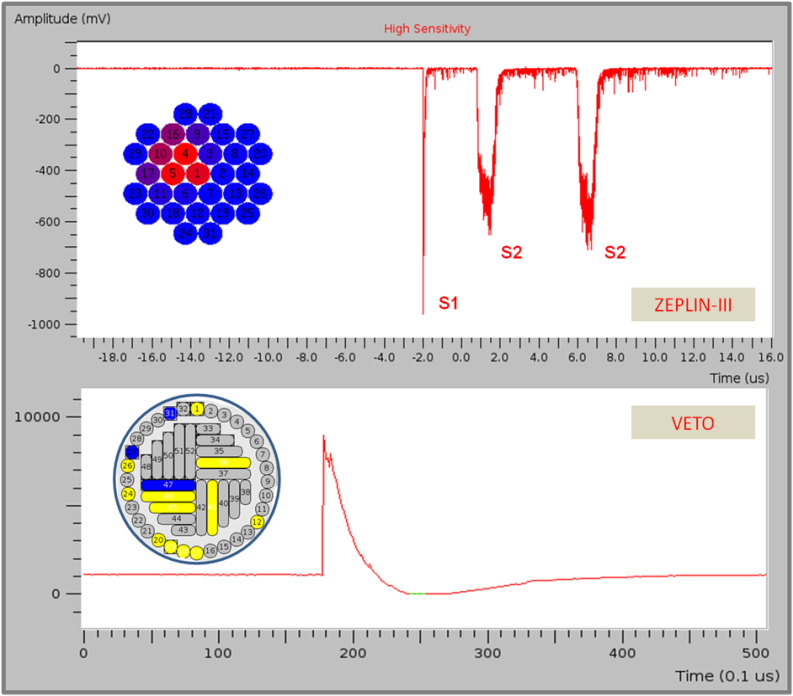}
\end{center}
\tdrfcaption[ZEPIII-nu]{A double-scatter neutron event recorded in ZEPLIN-III}
{A double-scatter neutron event recorded in ZEPLIN-III. The 
upper panel shows two elastic vertices clearly 
resolved in drift time (two S2 pulses, representing different 
vertical coordinates), although both have similar horizontal 
positions. The lower panel shows the summed waveform from the 
52-module veto detector which surrounded the main instrument,
indicating radiative capture of this neutron some \SI{17}{\us} 
after interacting in the LXe target. Recording additional 
particle scatters (either in the WIMP target or in an ancillary 
veto detector) provides a powerful rejection of backgrounds.}
\end{wrapfigure}

Both channels are sensitive to very low NR energies. The 
S2 response enables detection of single ionization 
electrons extracted from the liquid surface due to the high
photon yield that can be achieved with proportional
scintillation in the 
gas~\cite{Edwards:2007nj,Sorensen:2008a,Santos:2011ju}. 
In LUX we have demonstrated sufficient S1 light collection 
to achieve a NR energy threshold below 
\SI{1.1}{\keVnr}~\cite{Akerib:2016mzi}.
The combination of accurate 3-D imaging capability within 
a monolithic volume of a readily purifiable, highly 
self-shielding liquid is nearly an ideal architecture for
minimizing backgrounds. It allows optimal exploitation of 
the powerful attenuation of external gamma rays and 
neutrons into LXe, distinguishes multiply-scattered 
backgrounds from single-site signals, and precisely tags 
events on the surrounding surfaces. This latter feature is
important, given the difficulty of achieving 
contamination-free surfaces. The low surface-to-volume
ratio of the large, homogeneous TPC lowers surface 
backgrounds in comparison to signal, and stands in stark
contrast to the high surface-to-volume ratio of segmented
detectors.

These concepts are illustrated in Figure~\ref{IDOf:ZEPIII-nu}, 
which shows neutron interactions occurring just a few
millimeters apart in the ZEPLIN-III detector. The S1 
signals are essentially time-coincident, but the S2 pulses 
have different time delays corresponding to different 
vertical coordinates, making the rejection of such multiple
scatters extremely efficient. The figure shows also a pulse
observed in delayed coincidence in the surrounding veto
detector, indicating radiative capture of this neutron on 
the gadolinium-loaded plastic installed around the WIMP 
target. LZ will utilize a similar anticoincidence 
detection technique to characterize the radiation 
environment around the Xe detector and to further 
reduce backgrounds.

Nevertheless, when the first tonne-scale Xe experiments were 
proposed just over a decade ago, it was unclear whether LXe 
technology could be monolithically scaled as now proposed for 
LZ, or if it would be necessary to replicate smaller devices 
with target masses of a few hundred kilograms each. The latter 
option, while conceptually simple, fails to fully exploit the 
power of self-shielding. Since then, several aspects of the 
double-phase TPC technique have been further developed to 
make LZ technologically feasible. First, the ionization and
scintillation yields of LXe and their dependence on energy, 
electric field, and particle type have now been established 
down to a few keV by a comprehensive development program 
carried out around the globe, including substantial work by
members of the LZ Collaboration. Second, good acceptance for 
the primary scintillation light must be maintained as the 
detector becomes larger, and the remarkably high reflectance 
\SI{>95}{\percent} of PTFE at the \SI{178}{\nm} LXe scintillation
wavelength has made this practical. Third, considerable
control over electronegative impurities is required to 
drift charge over a distance of a meter or more, and 
commercial purification technology and new screening and 
detection methods developed by LZ scientists have made this
routinely achievable.
Fourth, the extraordinary self-shielding of an LZ-class 
instrument requires the use of internal calibration sources, 
and these have now been developed and deployed within LUX by 
LZ groups. Fifth, radioactive impurities such as Kr must
be reliably removed from the Xe, and other sources of 
internal radioactivity such as radon must be tightly 
controlled. Finally, a large detector requires a
substantial cathode voltage, or else fluctuations in the 
charge recombination near the interaction site will degrade 
the recoil discrimination. Very recently, LUX has demonstrated 
a rejection efficiency of \SI{99.6}{\percent} 
(for \SI{50}{\percent} NR acceptance)
even at a modest field of 
\SI{180}{\V\per\cm}~\cite{Akerib:2013tjd},
matching already the baseline assumption for LZ.

On the whole, the progressive nature of our program has 
contributed to an increase in the readiness level of this 
technology: 
LZ entails a twentyfold scale-up from LUX, the latter 
being also an order of magnitude or so larger than the 
ZEPLIN and XENON10 targets. Besides having the favorable 
properties of the WIMP target material and the proven
sensitivity of the technology to small energy deposits, 
a successful experiment must achieve very low background 
rates over a significant fraction of its active medium. 
Indeed, it is worth noting that LZ will be a factor of 
\num{e4} times more sensitive than current limits from 
the EDELWEISS and Cryogenic Dark Matter Search (CDMS)
experiments, which led in sensitivity only one decade 
ago \cite{Benoit:2001zu,Abrams:2002nb}. 
This implies a corresponding reduction in the background 
rate. This is achieved to first order by the power of 
self-shielding of local radioactivity, in combination 
with an outer layer of instrumented LXe and a hermetic
gadolinium-loaded scintillator ``veto'' shield capable of 
tagging neutrons and gamma rays with high efficiency.
The construction of a veto instrument at the required scale 
builds on two decades of development work in the field of 
reactor neutrino physics, and its development within LZ is 
led by scientists with considerable expertise in this area. 
Three other important developments, again pioneered 
by LZ groups, have also made this possible: the development, 
in collaboration with Hamamatsu, of very-low-background PMTs 
compatible with LXe \cite{Akerib:2012da}; the identification 
via the LUX program and LZ R\&D program of radio-clean titanium for cryostat 
fabrication \cite{Akerib:2011rr,Akerib:2017iwt}; and the development of 
krypton-removal and -screening technology capable of delivering 
sub-ppt concentrations \cite{Bolozdynya:2007zz,Leonard:2010zt}.

This strategy leads to a WIMP-search background of order \num{6} events 
in \num{1000} days of live exposure for a \num{5.6}-tonne fiducial mass. 
The remaining component is dominated by events from radon decay in the 
fiducial volume, a small fraction of which mimic NRs due to the finite 
S2/S1 discrimination power. Coherent scattering of atmospheric neutrinos 
from Xe nuclei (CNS) will constitute an even smaller, but irreducible,
background. These rates are well understood and background expectations 
are calculable with small systematic uncertainty (e.g., these events are 
spatially uniform and their energy spectra are well known). 
With its pioneering capability, LZ will be sensitive to 
these ultra-rare processes.

%
%

\tdrsubsec[Self-shielding]{Self-shielding in Liquid Xenon}
\begin{figure}[htbp]
\centering
\begin{minipage}{.5\textwidth}
  \centering
\includegraphics[width=7cm]{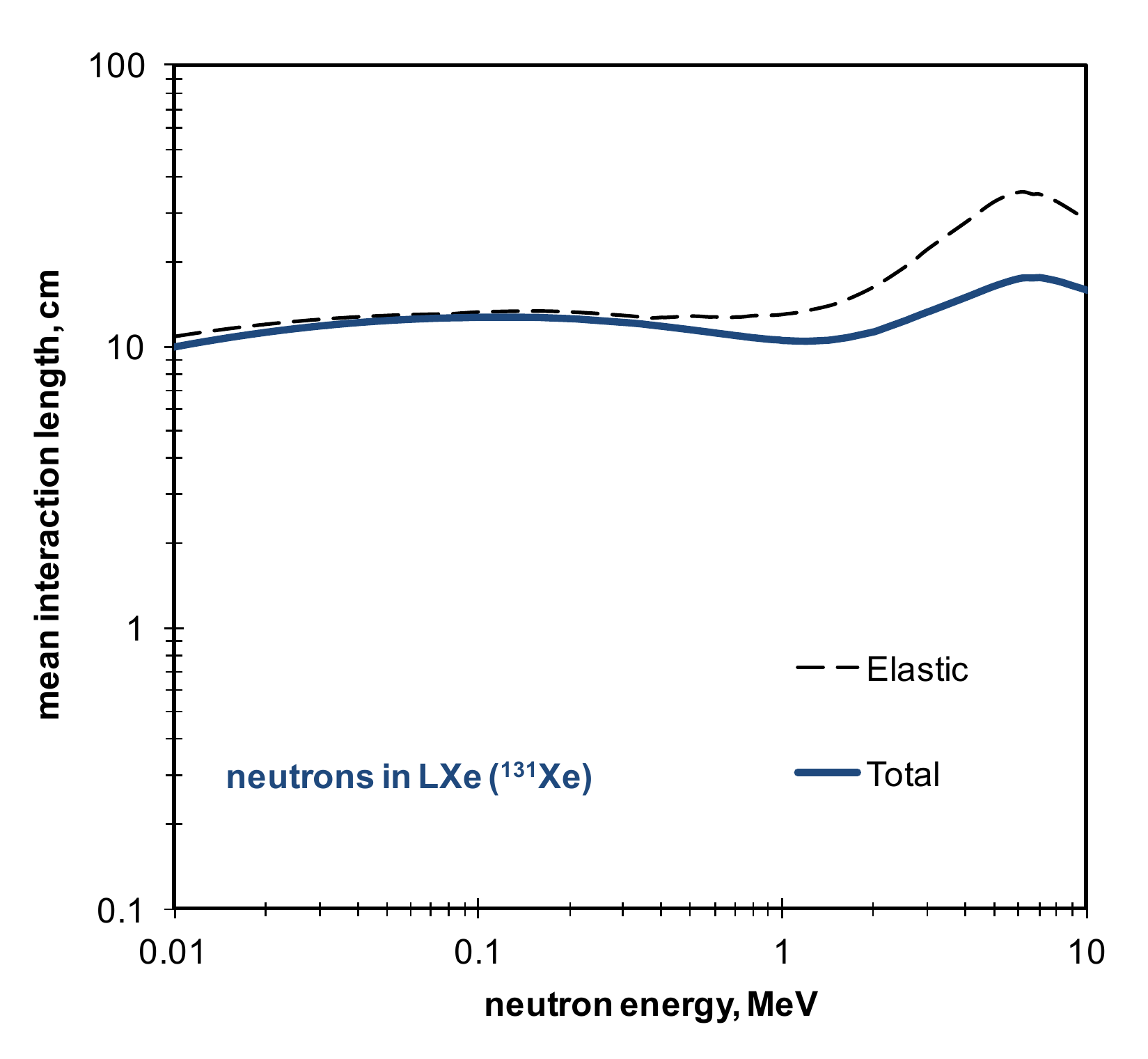}
\end{minipage}%
\begin{minipage}{.5\textwidth}
  \centering
\includegraphics[width=7cm]{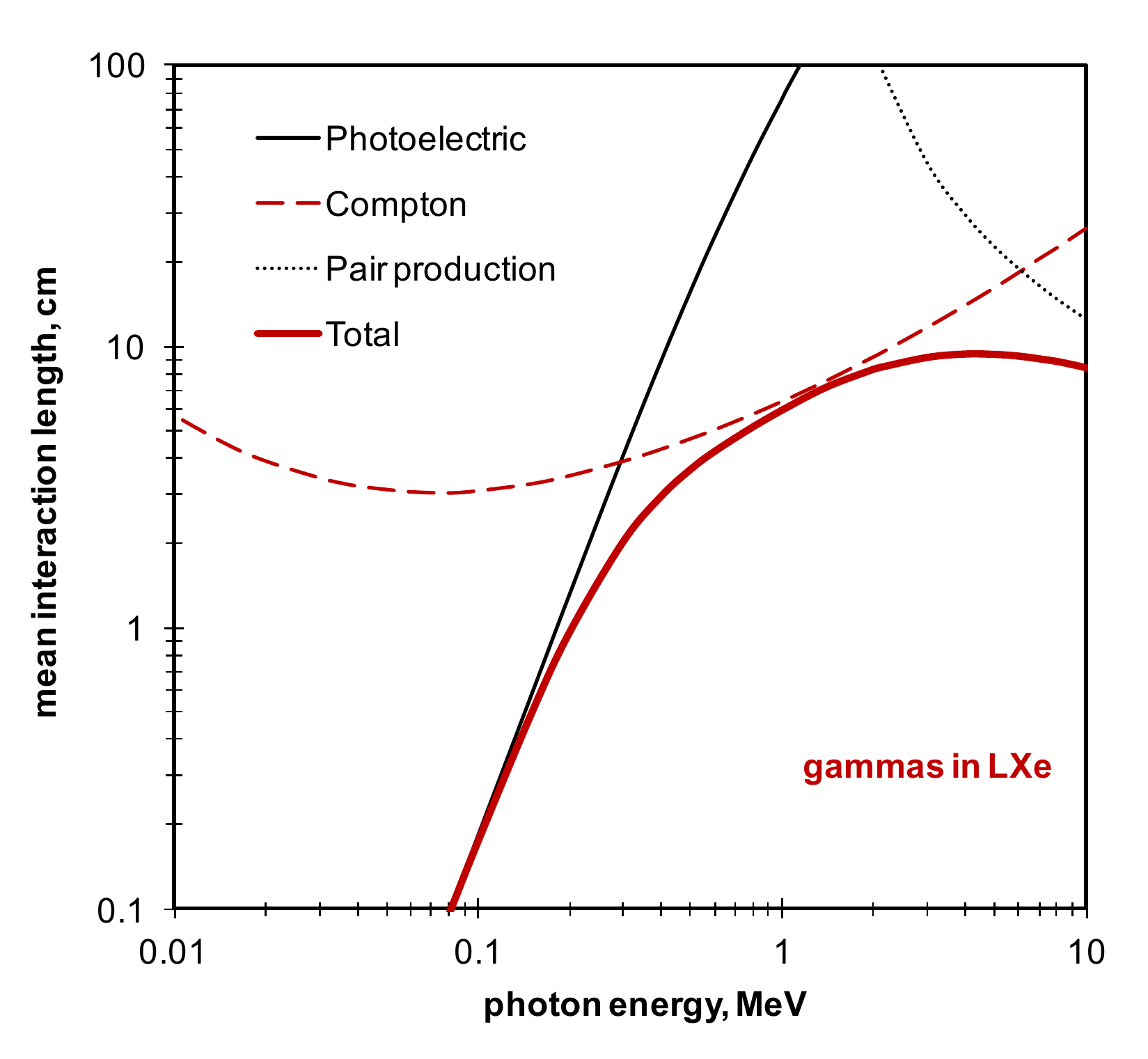}
\end{minipage}
\tdrfcaption[LXe-mean]{Mean interaction lengths for neutrons and gamma rays}
{Mean interaction lengths for neutrons
\protect\cite{Chadwick20112887} and gamma rays 
\protect\cite{xcom_2010} in LXe.}
\end{figure}

At the core of any WIMP search experiment is a substantial 
screening and materials-selection program (Chapter~\ref{chap:MAS})
that controls the trace radioactivity of the detector
components. In the case of LZ, however, backgrounds from
detector radioactivity will also be rejected to 
unprecedented levels by the combination of self-shielding 
of external particles and operation in anticoincidence 
with outer veto detectors (Chapter~\ref{chap:OD}). This 
will render external gamma rays and neutrons less 
problematic than in other experiments.

\begin{figure}[htbp]
\begin{center}
\includegraphics[width=0.7\textwidth]{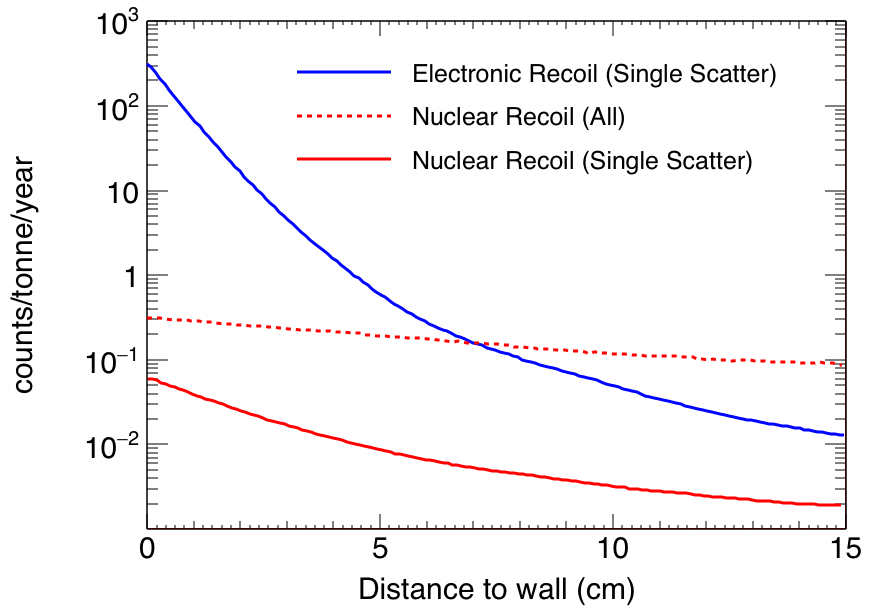}
\end{center}
\tdrfcaption[self-shield-LXe]{Self-shielding of external 
neutrons and gamma rays in LXe}{Self-shielding of external 
neutrons and gamma rays in LXe. The red lines indicate the 
number of elastic neutron scatters creating 
\SIrange{6}{30}{\keVnr}
NRs as a function of distance to the lateral TPC wall; the 
continuous line shows single scatters only, while the dashed 
line includes all multiplicities adding up to the same total 
energy; the input spectrum is that from (\Pga,\Pn) neutron 
production in PTFE, an important background near the TPC walls. 
The blue line represents single-site ER interactions from U/Th
gamma rays from PTFE with energy \SIrange{1.5}{6.5}{\keVee}. 
A tenfold decrease is achieved at \SI{\approx2}{\cm} and 
\SI{\approx6}{\cm} from the wall for gamma rays and neutrons,
respectively.}
\end{figure}

The self-shielding strategy, in particular, relies on the 
combination of a large, dense, high-Z and continuous detection 
medium with the ability to resolve interaction sites in three
dimensions with high precision. An outer layer of the target 
can therefore be defined (in data analysis) that shields 
a fiducial region with extremely low background at the center 
of the active medium.
The nonfiducial layer will be only a few centimeters thick. 
Because the size of the LZ detector is much larger than the 
interaction lengths for MeV gamma rays and neutrons, as shown 
in Figure~\ref{IDOf:LXe-mean}, when these particles penetrate 
more than a few cm they will scatter multiple times and be 
rejected (Figure~\ref{IDOf:self-shield-LXe}). X-rays, with 
energies similar to WIMP events, penetrate only a few \si{\mm} 
into the LXe.

Double-phase Xe detectors implement this strategy very 
successfully, and this is reflected in their present dominance 
in WIMP sensitivity---with LUX being a prime example of 
this concept. In LZ, a fiducial mass of nearly \SI{6}{\tonnesl} 
will be practically free of external gamma-ray or neutron 
backgrounds, which represents \SI{85}{\percent} of active 
mass within the TPC, compared to about \SI{50}{\percent} 
for LUX \cite{Akerib:2013tjd} and XENON100~\cite{Aprile:2012nq}. 
In addition to the required high density and high-Z of the 
detection medium, we highlight the importance of the precise 
spatial resolution that can be achieved in these detectors, 
which is of the order of \SI{1}{\cm} or better at threshold.

%
%

\tdrsubsec[ParticleDetection]{Low-energy Particle Detection in 
Liquid Xenon}

\begin{figure}[H]
\begin{minipage}[T]{0.46\textwidth}
\includegraphics[width=0.99\textwidth]{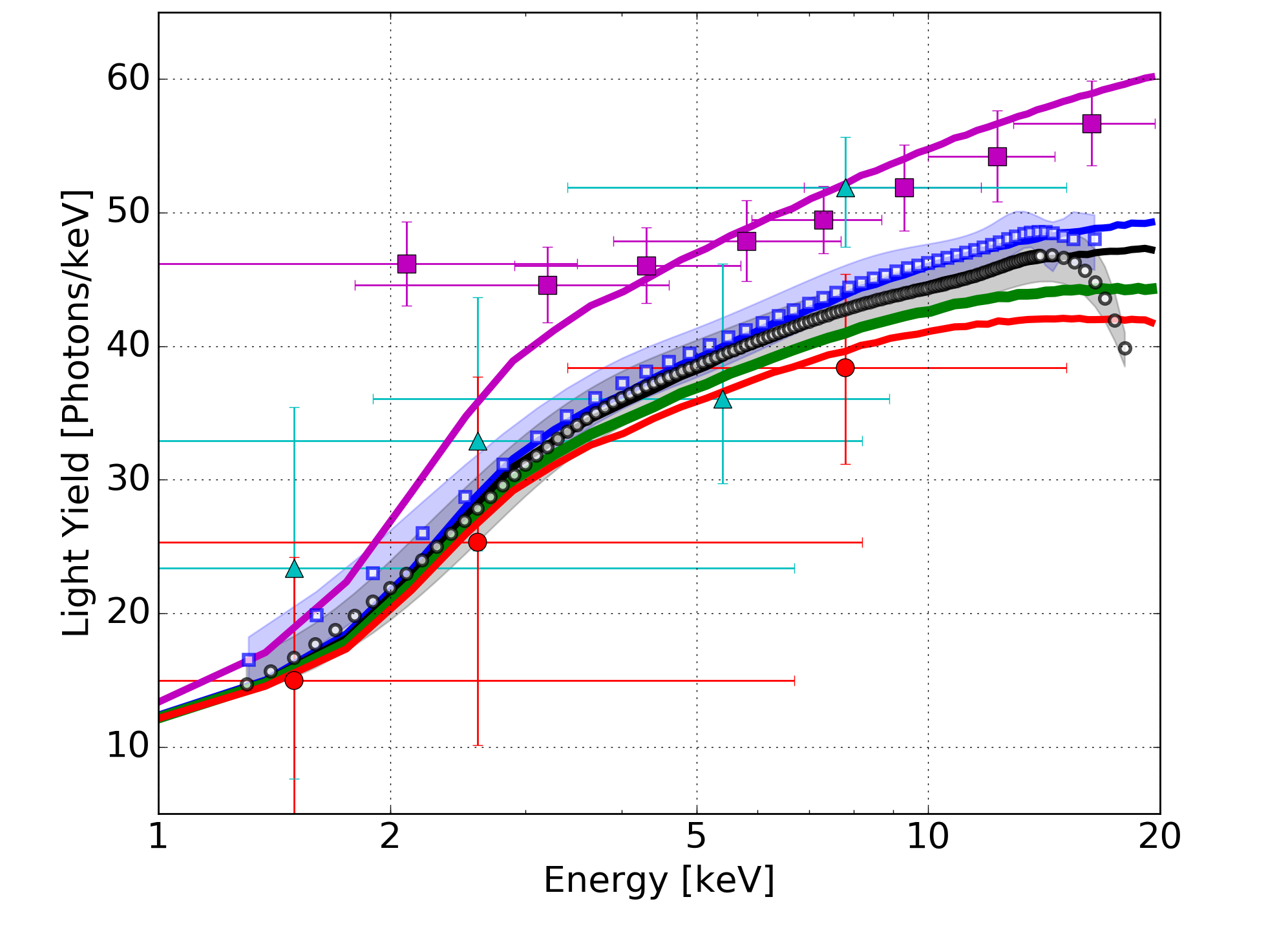}
\tdrfcaption[LY]{Absolute ER scintillation yield in LXe}
{Absolute ER scintillation yield in LXe. Purple squares are 
from Compton-scattering measurements at Columbia~\cite{Aprile:2012an}
and cyan triangles from~\cite{Baudis:2013cca} -- both at zero field.
Blue/black squares/circles are from LUX tritium beta 
emission measurements in situ at \num{105} and 
\SI{180}{\V\per\cm} respectively~\cite{Akerib:2015wdi}. 
Red circles are from Compton-scattering measurements 
performed in Z\"urich at \SI{450}{\V\per\cm}~\cite{Baudis:2013cca}. 
The NEST model (updated from~\cite{Szydagis:2011tk,Szydagis:2013sih} but 
using the same framework and formulae still) is shown in 
purple, blue, black, green, and red for \num{0}, \num{105}, 
\num{180}, \num{310} (LZ baseline), and \SI{450}{\V\per\cm},
respectively. }
\end{minipage}
\hfill
\begin{minipage}[T]{0.49\textwidth}
\includegraphics[width=0.99\textwidth]{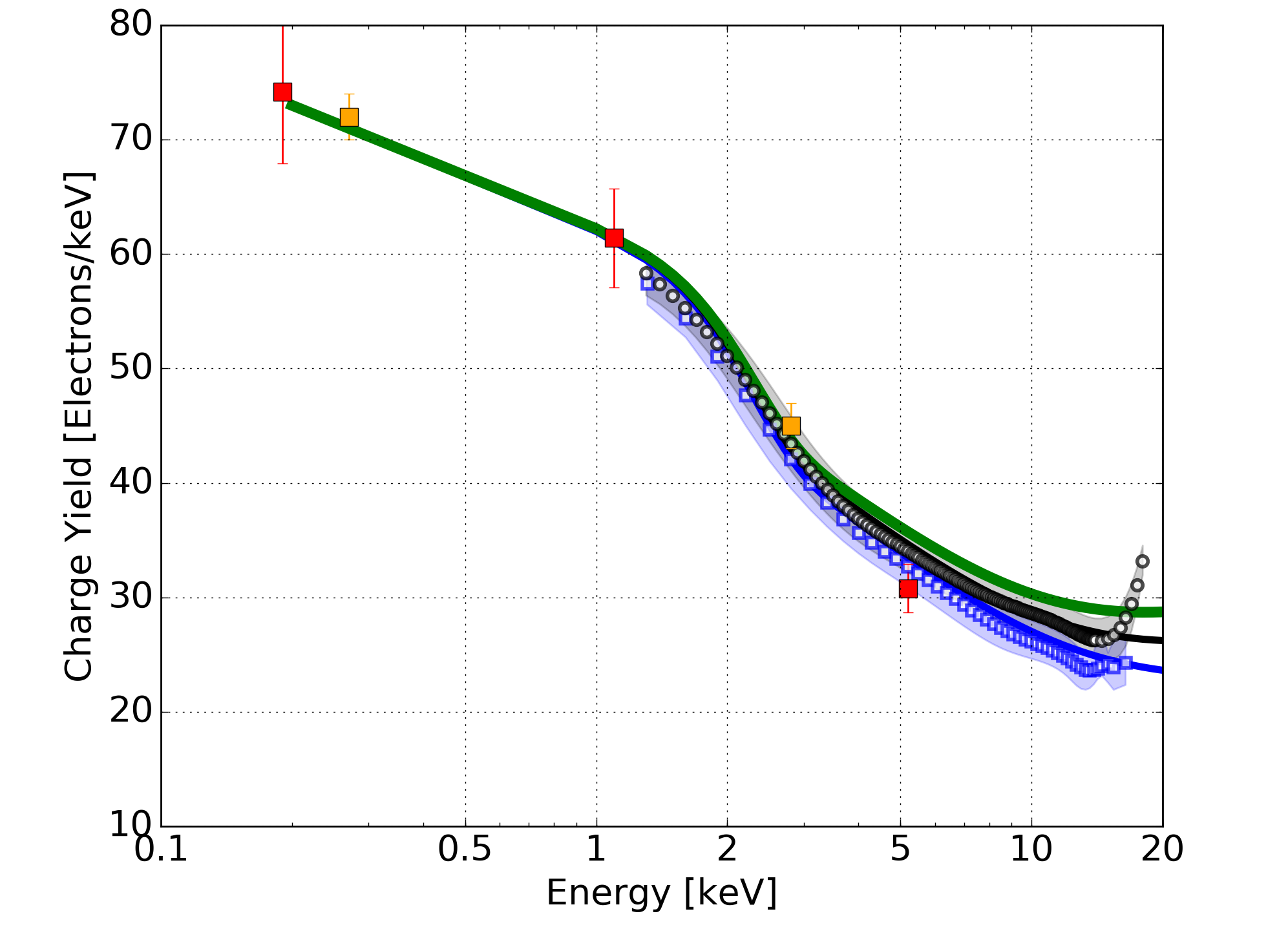}
\tdrfcaption[QY]{ER ionization yield in LXe}
{ER ionization yield in LXe. Data are as follows: blue 
squares from LUX tritium data beta spectrum matching at 
\SI{105}{\V\per\cm}~\cite{Akerib:2015wdi}; red and orange squares 
from \IXeotS activation lines in LUX and \IArTS in PIXeY, respectively, 
all at \SI{180}{\V\per\cm}~[F,G]; 
black circles from LUX at \SI{180}{\V\per\cm} once again~\cite{Akerib:2015wdi}. 
The NEST model is shown in blue, black, and green for 
\num{105}, \num{180}, and \SI{310}{\V\per\cm} (LZ baseline),
respectively. The green curve in each plot is what is used 
for all of the ER background modeling in LZ. Low-energy 
beta particles and gamma-rays are treated equivalently 
as both generating ER in the context of the simulation. 
Increasing the magnitude of the drift electric field 
reduces the recombination probability, thus raising the 
charge yield at the expense of light, as with NR.}
\end{minipage}
\end{figure}

The potential of this medium for particle detection was 
recognized in the mid-20th century, when the combination of 
good scintillation and ionization properties was first noted 
(see \cite{Chepel:2012sj} and references therein). In the 1970s,
the first double-phase detectors were demonstrated, originally 
using argon \cite{Dolgoshien:1970a}. Initially, our 
understanding of the mechanisms involved in generating the
scintillation and ionization responses in the noble liquids
progressed slowly, especially regarding the response to 
low-energy nuclear and electron recoils. However, great steps 
have been taken in the past decade, with LZ collaborators 
taking a central role. This effort continues around the world.

In this section, we summarize those LXe properties that 
affect the detection of low-energy nuclear and electronic 
recoils via scintillation and ionization; the next section 
discusses how to discriminate between them. The response of 
LXe to electron and nuclear recoils is now well understood 
over the energy range of interest for ``standard WIMP'' 
searches \SI{\gtrsim1.1}{\keVnr}~\cite{Akerib:2016mzi}. Significant progress
has equally been made in modeling its behavior as a function 
of incident particle species, energy, and electric field, 
in order to optimize detector design and the physics analyses.
Naturally, the increasing WIMP scattering rates with decreasing 
recoil energy and the need to probe lighter dark matter 
candidates mean that pushing further down in threshold is 
a perennial concern for any detection technology.

\begin{wrapfigure}{o}[0pt]{0.60\textwidth}
\begin{center}
\includegraphics[width=0.55\textwidth]{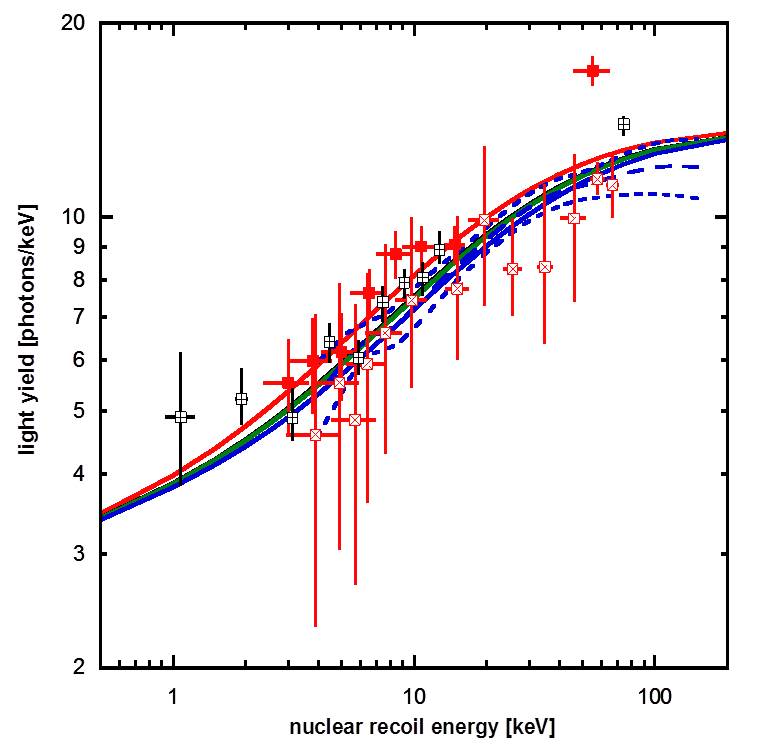}
\end{center}
\tdrfcaption[absNR-scintyield]{Absolute NR scintillation yield in LXe}
{Absolute NR scintillation yield in LXe. 
Hollow red markers are from neutron-beam measurements
at Yale \protect\cite{Manzur:2009hp} and filled markers 
from \protect\cite{Plante:2011hw}---both at zero field.
Black squares are from LUX D-D neutron gun measurements 
in situ at \SI{180}{\V\per\cm} ~\cite{Akerib:2015rjg}. 
Blue dashed lines are the combined mean and 1-$\sigma$
curves from two in situ measurements with Am-Be neutron
sources via fitting to MC simulation from ZEPLIN-III
\protect\cite{Horn:2011wz} (\SI{3650}{\V\per\cm}). 
The NEST model (updated from \cite{Lenardo:2014cva} 
but using the same framework and formulae) is shown 
in red, black, green, and blue for 0, 180, 310 
(LZ baseline), and \SI{3650}{\V\per\cm}, respectively. 
The green curve is used for LZ sensitivity calculations.
}
\end{wrapfigure}

Scintillation and ionization yields for 
ERs in LXe are 
shown in Figures~\ref{IDOf:LY}~and~\ref{IDOf:QY}, with predictions by the 
Noble Element Simulation Technique (NEST) model (see 
\cite{Szydagis:2013sih,Aprile:2012an,Baudis:2013cca} 
and the brief description in Chapter \ref{chap:SRD}); 
data for Compton electrons now reach down to 
\SI{1.5}{\keVee} \cite{Aprile:2012an,Lenardo:2014cva}, 
and NEST shows good agreement with these results 
\cite{Szydagis:2011tk}. LXe compares favorably to the 
best scintillators and is also a good ionization medium. 
For example, the maximum photon yield at a few tens of 
keV is some \SI{40}{\percent} higher than that of liquid 
argon. This is important for a number of reasons: It 
reduces the variance of the ER response, which is 
important for particle discrimination; it permits 
effective detector calibration; and it is directly relevant 
to some leptophilic dark matter searches.

As Figure~\ref{IDOf:LY} suggests, the scintillation 
yield is suppressed with increasing electric-field strength, 
while the ionization yield improves by the same amount. 
This behavior is also observed for individual events: A
fraction of the photon yield comes from recombination 
luminescence, whereby VUV photons are generated from 
electron-ion recombination occurring near the interaction 
site, and therefore some electrons contribute either to 
scintillation (S1) or to ionization (S2), but not to both. 
This event-by-event anticorrelation of the two signatures 
can be exploited very effectively at higher energies in
double-phase detectors to obtain good energy resolution 
for the spectroscopy of ERs, with application to gamma-ray 
background studies and searches for \Dznbb decay.

For NRs, both data and modeling have progressed markedly 
in recent years. The picture here is more complex than for 
electron interactions. Most of the energy deposited by 
electrons is shared between ionization and excitation of 
the medium, making much of it observable. In NR interactions, 
a larger fraction is spent in atomic collisions, which is 
dissipated as heat and not detected. However, data
obtained by scattering experiments at neutron beams ex situ, 
e.g. \cite{Manzur:2009hp,Plante:2011hw}, are now in 
good agreement with those from indirect in situ techniques 
\cite{Horn:2011wz}, down to \SI{\sim3}{\keV}. 

\begin{figure}[ht]
\begin{center}
\includegraphics[width=0.55\textwidth]{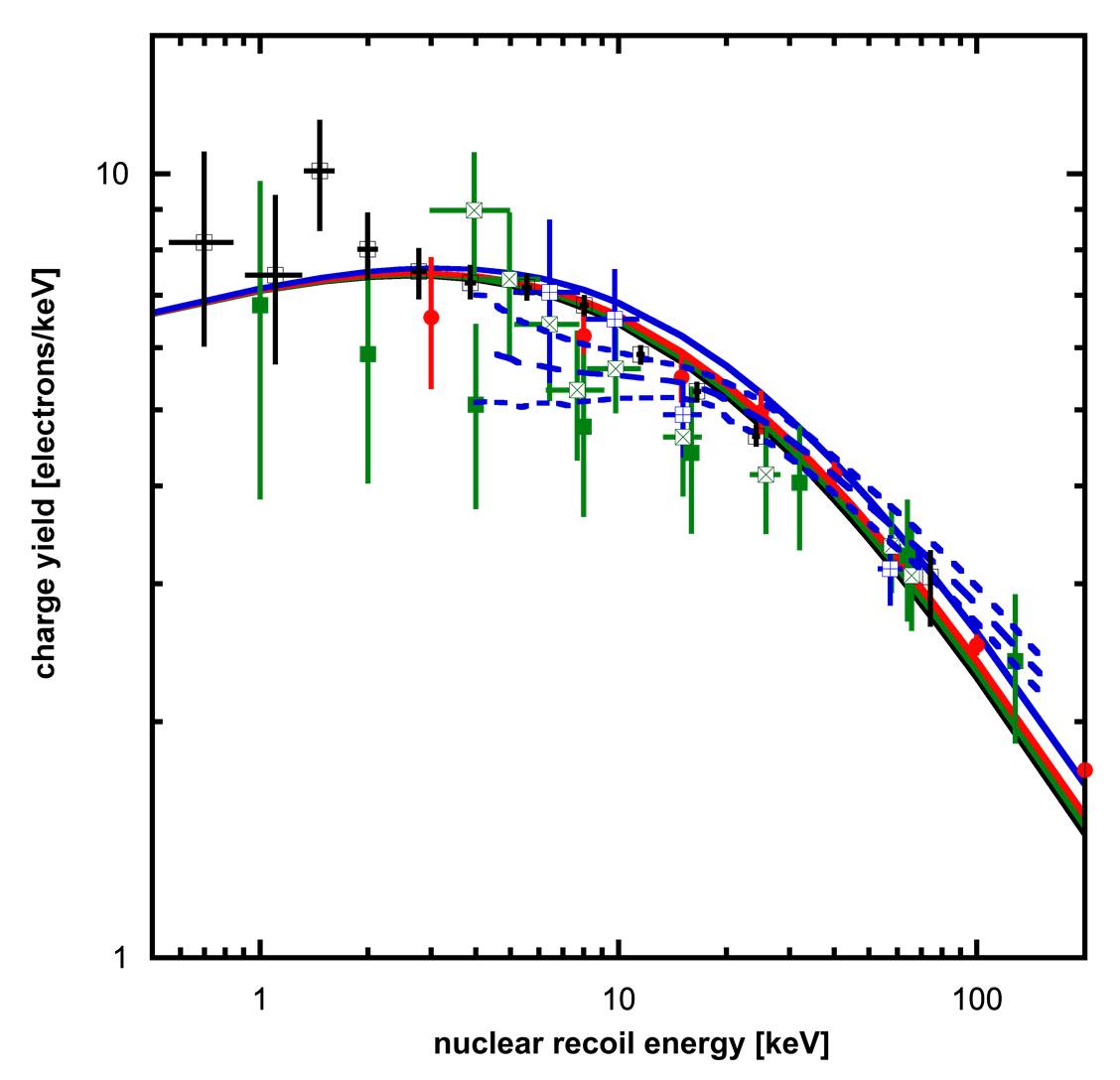}
\end{center}
\tdrfcaption[absNR-ionyield]{Absolute NR ionization yield in LXe}
{Absolute NR ionization yield in LXe. Data are as 
follows: blue and green hollow squares from neutron 
beam data from Yale at \SI{4}{\kilo\V\per\cm} and 
\SI{1}{\kilo\V\per\cm}, respectively 
\protect\cite{Manzur:2009hp}; dashed blue curves 
from MC matching from ZEPLIN-III \protect\cite{Horn:2011wz} 
at \SI{3650}{\V\per\cm}; solid green squares from XENON10 
at \SI{730}{V/cm}\protect\cite{Sorensen:2010hv}; red 
markers from XENON100 at \SI{530 }{\V\per\cm} 
\protect\cite{Aprile:2013teh}; black squares from LUX 
D-D neutron gun measurements in situ at 
\SI{180}{\V\per\cm}~\cite{Akerib:2015rjg}. The NEST model 
is shown in black, green, red, and blue for 180, 310 
(LZ baseline), 530, and \SI{3650}{\V\per\cm}, respectively. 
The green curve is used for LZ sensitivity calculations.
}
\end{figure}

As the NR scintillation yield declines gently at least down 
to \SI{3}{\keV}, the ionization yield increases accordingly, 
as shown in Figure~\ref{IDOf:absNR-ionyield}, with the current
state of the art for in-situ measurement coming from 
LUX. This measurement, for the first time, constrains the 
ionization yield below \SI{1}{\keV} (\SI{0.7}{\keV})~\cite{Akerib:2015rjg}.
We note that the experimental data suggest a remarkably 
high yield even for \num{1}-keV recoils, with several electrons 
being released per interaction. As with the scintillation 
yield, the electric field dependence is modest.

\tdrsubsubsec[LowEnergy-LowMass]{Low Energy and Low Mass Sensitivity}
The high ionization yield allied to the ability to detect 
single electrons with high efficiency is a very attractive 
feature of this technology: Not only does it provide a 
low-threshold channel for light WIMP searches but, from a 
practical standpoint, it allows very high triggering efficiency 
(on S2) for the lowest-energy events, which are associated 
with very small S1 pulses.

Double-phase Xe detectors achieve the best NR energy 
threshold among the leading WIMP-search technologies, while
maintaining discrimination and good vertex location. 
Of all such detectors operated so far, LUX can claim the 
lowest 50\% NR threshold of approximately \SI{4}{\keV}. 
WIMP masses down 
to \SI{\sim10}{\GeVpct} are directly accessible to an 
instrument such as LZ operating in the ``normal'' TPC mode,
requiring one S1 pulse and one S2 pulse.
A low-energy ER interaction (\SI{\approx1.5}{\keVee}) is 
shown in Figure~\ref{IDOf:double-phaseX3} -- 3-D position 
resolution and discrimination are fully effective even at 
these energies.

At the smallest NR energies (\SI{\lesssim4}{\keVnr}), it is 
clear that the S1 signal is often absent but S2 is still 
easily detectable, so that LZ can recover sensitivity in
this regime by performing an ``S2-only'' analysis 
\cite{Angle:2011th}. Discrimination based on S2/S1 ratio 
is not possible in this instance, but the detector retains
the ability to reject edge events in $(x,y)$. A more 
limited but still useful degree of $z$ position reconstruction 
is possible based on the broadening of the S2 pulse due to 
longitudinal diffusion of electrons as they drift in the
liquid. As a result, an S2-only search can still exploit 
the extremely radio-quiet inner region of the WIMP target, 
and place upper limits on the dark matter scattering cross 
section. Naturally, a thorough understanding of backgrounds is
required for this type of analysis; several background 
mechanisms create single S2 electrons, while the two-electron 
random coincidence rate might still be significant. This 
technique is particularly applicable to particle masses lower
than about \SI{10}{\GeVpct}.

One class of NR event that inevitably will be visible below 
the (3-phe) S1 threshold is due to coherent elastic scattering 
of \IBe solar neutrinos off Xe nuclei. The electron counting 
technique (S2-only) was in fact suggested a decade ago to 
allow a first observation of this process \cite{Hagmann:2004uv}. 
Due to energy resolution broadening of the scintillation signal, 
some events will register both S1 and S2---from this and other 
neutrino fluxes. Despite significant interest in this signal 
per se, coherent neutrino-nucleus scattering is also a
fundamental background for dark matter searches, which is 
quantified in Chapter~\ref{SIPSss:CENNS}.

\begin{figure}
\begin{center}
\includegraphics[width=0.59\textwidth]{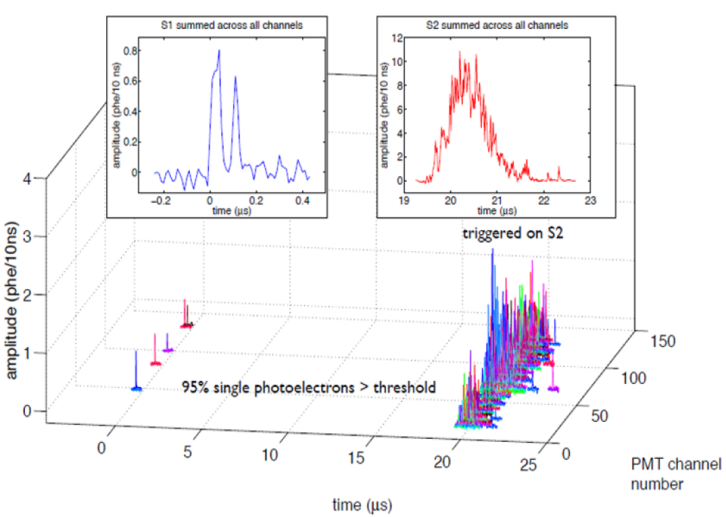}
\includegraphics[width=0.59\textwidth]{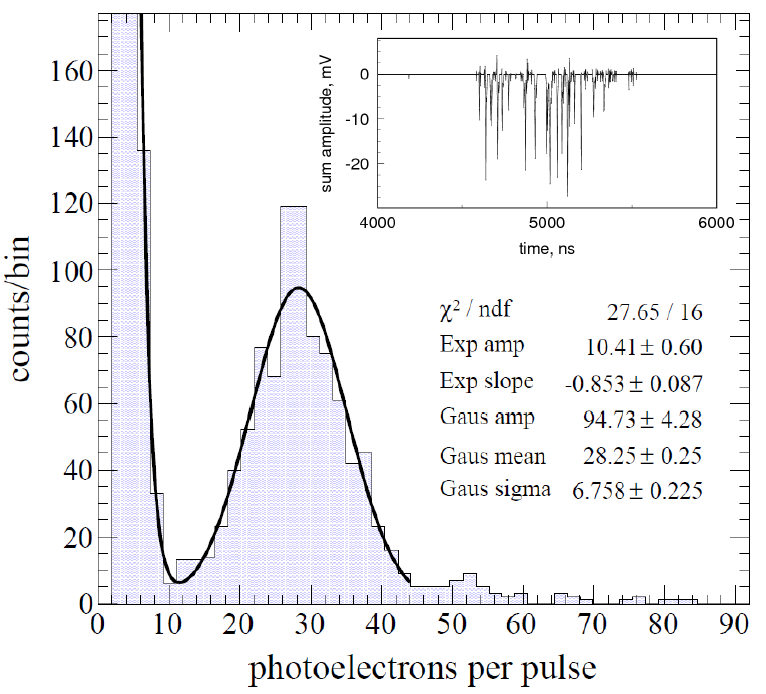}
\end{center}
\tdrfcaption[double-phaseX3]{Low-energy performance of 
double-phase Xe detectors}
{Low-energy performance of double-phase Xe detectors.
Top: A \SI{1.5}{\keVee} electron interaction in LUX 
\protect\cite{Akerib:2012ak}, showing a fivefold 
coincidence for S1 and the corresponding (much larger) S2
delayed by \SI{20}{\us}. Bottom: Pulse size distribution 
of single electrons measured by electroluminescence in 
ZEPLIN-III, showing a mean of 28 photoelectrons per 
emitted electron (one such waveform is shown in inset)
\protect\cite{Santos:2011ju}.}
\end{figure}

%
%

\tdrsubsec[Discrimination]{Electron/Nuclear Recoil Discrimination}
Discrimination of ER/NR is key to the positive identification 
of a WIMP signal, both by directly reducing the effect of the 
dominant electronic backgrounds in the detector, and by 
confirming a NR origin. The physical basis for discrimination
is the difference in the ratio of ionization electrons to 
scintillation photons that emerge from the interaction site 
and subsequently create the measured S2 and S1 signals, 
respectively. In a plot of the logarithm of S2/S1 as a 
function of S1, as in Figure~\ref{IDOf:discrimination}, electron 
and nuclear recoils each form a distinct band, with NRs having 
a lower average charge/light ratio.

A figure of merit for discrimination is commonly the ER leakage past 
the median of the NR population (i.e., retaining a flat 
\SI{50}{\percent} NR acceptance). Previous values are
between \SI{99.5}{\percent} in XENON10 \cite{Aprile:2010bt} 
and \SI{99.99}{\percent} in ZEPLIN-III
\cite{Lebedenko:2008gb}. 
We assume a baseline discrimination value of 
\SI{99.5}{\percent}, a conservative assumption given the
performance already obtained in LUX, as discussed below.

Electron/nuclear recoil discrimination is determined by the 
separation of the bands as well as their widths, and in 
particular the ``low tail'' in $\log_{10}$(S2/S1) of the ER 
band. Remarkably, the bands are mostly Gaussian when binned 
in slices of S1. 
Some skewness in the shape of the ER band has been observed 
in both the ZEPLIN-III and LUX experiments. In ZEPLIN-III, 
this skewness was observed using an external gamma-ray source 
whilst running at a high field~\cite{Lebedenko:2008gb}. 
Conversely, in LUX, the effect was seen when running an 
internal tritium calibration at a much lower electric field
\cite{Akerib:2015wdi}.

The physics determining both the position of the bands 
and their widths has been studied and we are increasingly 
able to model it successfully \cite{Dahl:2009nta}.
The overall separation of the bands is mostly due to NRs 
producing less initial ionization and more direct excitation 
(leading to scintillation) than do ERs. In turn, the 
band widths depend strongly on the physics of electron-ion
recombination at the interaction site. A recombination 
episode generates an excited Xe atom that de-excites through
scintillation (via the Xe$_2^*$ state).
Therefore, initial ionization is either measured as charge 
(via S2) or light (via S1), and event-by-event fluctuations 
in the amount of recombination are one of the primary sources 
of band broadening. These fluctuations increase with recoil 
energy over the range of interest.

At the lowest energies, however, the distributions 
``flare up'' due to statistical fluctuations in the S1 signal. 
This broadening is therefore reduced in chambers with higher 
light yield, improving discrimination. In fact, ER rejection 
in this technology is better just above threshold 
(\SIrange{\sim5}{15}{\keV}), where WIMP-induced recoil 
rates are highest, than at intermediate energies 
(\SIrange{\sim15}{40}{\keV}), and then improves
dramatically beyond \SI{\sim40}{\keV}. The excellent ER 
discrimination at low energies is due in part to a decrease 
of recombination fluctuations, but is also caused by the 
curvature of both bands at low energy, as shown in 
Figure~\ref{IDOf:discrimination}: The bands are largely
parallel to the direction of S1 fluctuations, sharply 
reducing their impact on discrimination.

In addition to light collection, the drift field is also 
expected to affect discrimination. Although largely determined 
by the amount of initial ionization, the positions of the band 
medians have a residual dependence on the different amounts 
of field-dependent recombination for the two recoil species, 
and their separation increases at higher fields 
\cite{Dahl:2009nta}. 
This may explain the world-best discrimination observed in 
ZEPLIN-III, which operated with close to \SI{4}{\kilo\V\per\cm}
drift field, and is an important driver of the LZ design. 
The band width should, in principle, also have some field 
dependence, though this has not been well measured. 
The model developed in \cite{Dahl:2009nta} has been 
incorporated in the NEST Monte Carlo package 
\cite{Lenardo:2014cva,Szydagis:2013sih},
which has informed the sensitivity projections for this report. 
However, caution must be exercised when comparing values from 
different experiments, since instrumental effects other than 
electric field and light yield can impact discrimination very 
severely. A case in point is the degraded discrimination measured 
in the second science run (SSR) of ZEPLIN-III \cite{Akimov:2011tj} 
relative to the first science run (FSR) \cite{Lebedenko:2008gb}, 
benchmarked essentially in the same detector and with the same 
software, but following upgrade of the TPC with underperforming
photomultipliers.

\begin{figure}[htbp]
\begin{center}
\includegraphics[width=0.8\textwidth]{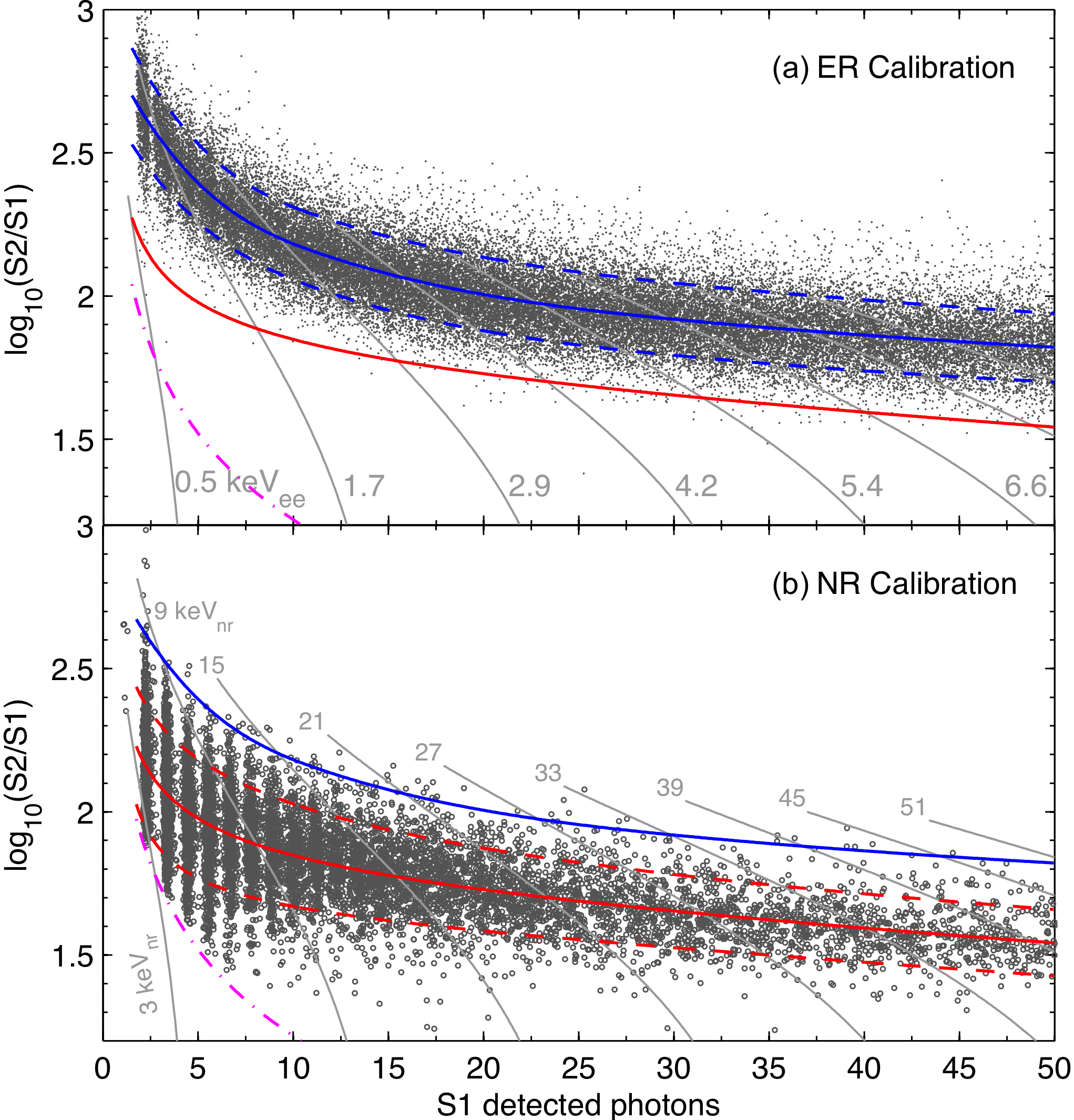}
\end{center}
\tdrfcaption[discrimination]{Discrimination in LUX with calibrations}{Discrimination parameter $\log_{10}$(S2/S1) as a function of S1 signal
obtained with LUX calibration \protect\cite{Akerib:2015wdi}. 
(a) ER band calibrated with beta decays from a dispersed 
\IHT source; the median is shown in blue, with \SI{80}{\percent}
population contours indicated by the dashed blue lines. 
(b) NR band populated by elastic neutron scattering from a 
D-D pulsed neutron source; the median (solid) and 
\SI{80}{\percent} band width (dotted) are indicated in blue 
and red, respectively. The mostly vertical gray lines are 
contours of constant energy deposition. For more information, 
see Chapter~\protect\ref{chap:CAS}.}
\end{figure}

The question of discrimination is of great importance in LZ, 
as its dominant background is ERs from 
\IRnttt and \IRnttz daughters as well 
as from solar neutrinos. 
To predict the LZ sensitivity, the electric field strength 
and the light-collection efficiency in the WIMP target
are the main ingredients required. These are key performance 
parameters and we motivate their choice in two separate 
sections below; we also describe the steps we are taking 
to achieve the required performance. We use these parameters 
in conjunction with a full \geantFour Monte Carlo simulation 
\cite{Agostinelli:2002hh} based on the LUXSim package 
\cite{Akerib:2011ec} and incorporating NEST.

The adopted values
---a drift field of \SI{310}{\V\per\cm} and an S1 photon 
detection efficiency (PDE) of \SI{7.5}{\percent}
---motivate an average nominal discrimination of
\SI{99.5}{\percent} for a flat ER spectrum such as that 
from \IRnttt and \IRnttz as well as 
from \Ipp neutrinos (an ER leakage past the NR median 
of 1:200). This is supported by both NEST-based 
simulations and by XENON10 \cite{Aprile:2010bt}, which 
achieved that level of discrimination at similar field 
and light collection as proposed here. PANDA-X recorded 
\SI{99.7}{\percent} at \SI{667}{\V\per\cm}, for a PDE of 
\SI{10.5}{\percent} \cite{Xiao:2014xyn}.
Significantly, LUX initially reported \SI{99.6}{\percent}
discrimination at only \SI{180}{\V\per\cm} in Run 3 
\cite{Akerib:2013tjd}, increasing to \SI{99.8}{\percent} 
in a subsequent reanalysis with improved algorithms
\protect\cite{Akerib:2015wdi}.
Therefore, we are confident of reaching the 
\SI{99.5}{\percent} value assumed in this report.

In fact, Figure~\ref{IDOf:ER-leakage-fraction} shows that the discrimination 
generally improves for smaller S1 signals, as measured 
in LUX. Further, the signal from a light 
WIMP is not distributed symmetrically 
around the NR median: 
Upward fluctuations in S1 that cause 
low-energy events to exceed a 
given analysis threshold also induce
a distribution in $\log_{10}$(S2/S1) that
is systematically lower than the NR median~\cite{Sorensen:2012ts}, 
as shown in Figure~\ref{SIPf:solarCNSfig},
Figure~\ref{SRDf:PLRhowa}, and
Figure~\ref{SRDf:PLRMass}.
The acceptance of the ER rejection cut for light WIMPs is higher 
than the \SI{50}{\percent} one might
surmise from the median of the average
NR distribution. 

Phenomenon like those discussed in the
previous paragraph, combined with the
high-statistics calibrations recently
completed by the LUX 
collaboration~\cite{Akerib:2015wdi,Akerib:2016mzi},
have convinced us to replace the simple
`cut-and-count' technique, and a 
simple leakage figure
of merit, with the profile likelihood 
ratio (PLR) technique described in
Section~\ref{SRDSs:PLR}.

\begin{figure}[htbp]
\begin{center}
\includegraphics[width=0.8\textwidth]{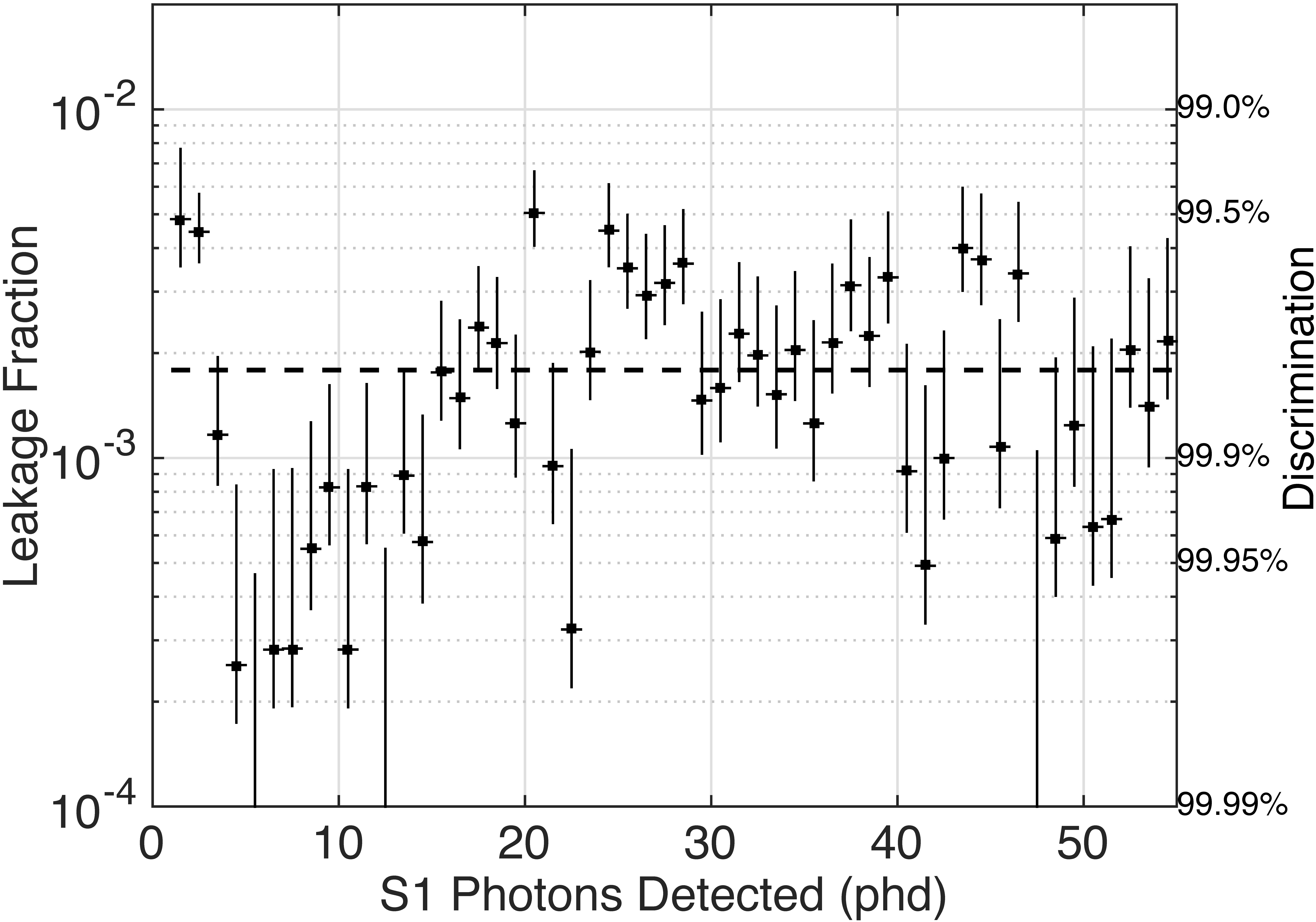}
\end{center}
\tdrfcaption[ER-leakage-fraction]{ER leakage into the NR band}
{ER leakage fraction past the NR median line measured 
with tritium data in LUX. The dashed 
line indicates the average leakage of 0.002 
(\SI{99.8}{\percent} discrimination) in the S1 range
\SIrange{2}{50}{\phe}. The general improvement of 
discrimination at low energies can be clearly seen, with 
the exception of the very lowest S1 data points where the ER
band starts to flare up due to photoelectron statistics.}
\end{figure}

%
%

\tdrsubsec[OuterDetector]{Outer Detector Systems}
A WIMP scatter would deposit a 
few \si{\keV} in the central 
volume of the LXe TPC, with 
no simultaneous energy 
deposit in surrounding materials.
Neutrons caused by radioactivity, which
can fake WIMP interactions when they scatter 
elastically, are likely to interact again
either within the TPC or nearby, and so 
it is broadly desirable to replace as 
much of the neighboring material as possible 
with additional radiation detectors. 
It also helps to minimize intervening 
material between the active LXe in 
the TPC and any such ancillary 
detectors, namely by decreasing the 
thickness of the field-cage and of the 
cryostat vessels. Active material 
surrounding the central LXe volume 
also permits assessment of the local 
radioactivity environment, and thus to 
infer additional information on the
backgrounds in the WIMP search region. 
A persuasive WIMP discovery will require
excellent understanding of all 
background sources, which is best 
done through the characterization of 
those sources in situ.

The LZ apparatus will feature two 
distinct regions where active material
surrounds the LZ TPC. The first is a ``skin'' 
of LXe, formed by liquid between the field
cage and the inner vessel of the cryostat. 
The second is surrounding detectors 
of liquid scintillator (LS), which is 
doped with a small amount of gadolinium, 
to enhance its capability for neutron 
detection. We expect a threshold of 
\SIrange{100}{200}{\keVee} for both 
systems, with the gadolinium-doped LS
detecting neutrons more effectively 
and the skin detector performing best 
for internal gamma rays.

\tdrsubsubsec[Skin]{Xenon Skin Veto}
The side skin region consists of 
an unavoidable \SIrange{4}{8}{\cm} of 
LXe between the outer boundary of the 
field cage and the inner boundary of 
the cryostat. An even thicker LXe region,
known as the dome skin,
exists below the bottom PMT 
array. It is highly desirable to read 
out scintillation light generated in 
these regions for two main reasons:
\begin{enumerate}
\item They comprise anticoincidence 
  detectors, where presence of a signal
  indicates that an interaction in the
  central LXe TPC is not from a WIMP.
\item  They will also veto external LXe
       interactions where VUV photons could
       leak into the TPC and fake S1 light 
       there.
\end{enumerate} 
Our approach is to maximize the optical
isolation between inner and outer
LXe volumes as far as practicable, 
and to instrument as much outer LXe 
as possible. Nevertheless,
the sensitivity to deposited
energy in the skin region will be far 
less than that of the central Xe TPC and, 
no ionization can be detected in the skin.
Reasonable utility for the vetoing 
of gamma rays in the skin results when
the threshold is a small fraction of 
the typical energy of an environmental gamma.
To achieve this threshold, the side skin is
equipped with 90 Hamamatsu R8520-406 PMT 
looking downwards and another 90 looking
upwards to monitor the cylindrical shell between the sides of the TPC and 
the cryostat wall. The dome skin is
viewed by 12 Hamamatsu R11410-22 PMT.

The PMT system achieves a
threshold of about \SI{100}{\keVee},
sufficient to detect Compton recoils 
from MeV gamma rays from radiogenic
backgrounds. The skin detector will have 
low sensitivity for neutron detection 
via elastic scattering due to the mismatch 
in mass between neutrons and Xe nuclei, but
inelastic neutron interactions can be detected in the skin, 
as can gamma rays produced by neutron capture.
The LXe skin veto has some advantages 
over the outer detector for gamma detection
because some gamma rays do not penetrate 
the various vessels all the way to the OD.

\tdrsubsubsec[ScintillatorOD]{Scintillator Outer Detector}
The goal of the outer detector is 
to surround the LZ cryostat with a 
near-hermetic gamma-ray and neutron
anticoincidence system. LZ will employ 
linear alkyl benzene (LAB), an LS solvent 
developed by the reactor neutrino 
community in the past
decade~\cite{Guo:2007ug,DayaBay:2012aa}.
Small quantities of a standard fluor and
wavelength shifter will be added to 
the solvent to provide the scintillation
signal. A PMT system located in the water 
space outside of the clear acrylic tanks
containing the LS will view this 
scintillation light.

To enhance neutron detection, 
\SI{0.1}{\percent} by weight of natural
gadolinium will be dissolved in the LAB 
with a  chelating agent to form 
Gd-loaded LS, or GdLS. Two isotopes of Gd,
\IGdoFF and \IGdoFS, have neutron-capture 
cross sections that are 61 
and \SI{254}{\kilo\barn}, 
respectively. Each of these 
isotopes constitutes about 
\SI{15}{\percent} of natural Gd and, 
at \SI{0.1}{\percent} concentration by 
weight, capture on Gd is about 1
order of magnitude more probable than 
is capture on hydrogen.

A neutron that can cause a Xe NR in 
the same energy range as the recoil from 
a WIMP will have an energy between about 
\num{0.5} and \SI{5}{\MeV}. The source 
of most of these neutrons will be from 
the (\Pga, \Pn) process from material 
around the edges of the Xe, and their 
energy spectrum will be toward the low 
end of this interval. A dangerous neutron 
will enter the LXe TPC and then scatter 
back out after one interaction. Many of 
those neutrons will traverse the 
intervening material and then thermalize 
and capture in the Gd in the GdLS. The 
length scale for thermalization and 
capture is a few centimeters and the 
typical capture time is \SI{\sim30}{\us},
which is small compared with the 
\SI{670}{\us} maximum drift time of the TPC.

After capture on Gd, a total energy 
of about \SI{8}{\MeV} is emitted as a burst 
of several gamma rays, which then 
interact in the LS (or the skin, or the TPC).
This large energy release separates 
neutron captures from the \Pgg rays from
natural radioactivity, which die out above 
\SI{3}{\MeV}.  The thickness of the GdLS 
layer is \SI{61}{\cm}, determined by
detailed \geantFour-based simulations and
mechanical consderations.

For a fraction of \Pgg-rays that 
deposit energy in the 
central LXe TPC, the outer detector 
system also functions as a \Pgg-ray 
veto, for those that propagate through the 
LXe skin and cryostat; these share the 
same detection requirements as the 
capture gammas. 
To achieve good efficiency as 
a \Pgg-ray veto, we require a
threshold to Compton electrons 
near \SI{200}{\keVee}, and we set
a threshold goal of \SI{100}{\keVee}

In reactor neutrino experiments, 
typically a single acrylic 
cylinder contains the Gd-loaded 
scintillator. Transport 
logistics preclude this solution for LZ, 
and instead segmentation of the volume 
into nine smaller tanks that can 
be transported through the shafts 
and drifts that lead to the 
Davis Cavern is the adopted solution. 
To enhance light collection, the outer 
surface of the LZ cryostat will be 
affixed with a diffuse white reflective 
layer of Tyvek\textregistered\ to reflect 
light into the 8-inch Hamamatsu R5912 
PMTs that will surround the tanks. 
The \SI{600}{\um} multilayer
Tyvek\textregistered\ 
we plan to use has a reflectivity in excess of 
\SI{95}{\percent}.

The goal is a sufficient PMT collecting power
to achieve \SI{100}{\keVee} threshold. 
Our preliminary studies indicate 
that, by covering the cryostat in
Tyvek\textregistered\ and employing
another reflective cylinder outside of 
the PMT system, the collecting power of 
120 8-inch PMTs is sufficient to achieve the
required threshold.

The GdLS tanks will be surrounded by 
ultrapure water, and the distance to the
water-space PMT system that detects the 
LS scintillation light must be sufficient 
to attenuate gamma rays from PMT radioactivity 
(these are not low-background models). 
A distance of \SI{80}{cm} from the
scintillator tanks to the PMTs reduces 
the rate from this source to less than 
\SI{5}{\Hz}.

The tightest specifications on 
radioactive impurities in 
the GdLS arise from considerations of 
deadtime caused by the outer detector system.
Our goal is a false veto probability that
does not exceed \SI{5}{\percent} over a \SI{500}{\us}
time window, which results in goals for radioactive
impurities of \SI{<1.3}{\ppt} U,
\SI{<4.5}{\ppt} Th, \SI{<0.8}{\ppt} \IKfz, 
and \SI{<13e-18}{g/g} of \ICof.  Our requirements
on impurities are \SI{<10}{\ppt} U,
\SI{<20}{\ppt} Th, \SI{<3}{\ppt} \IKfz, 
and \SI{<15e-18}{g/g} of \ICof, which permit
a \SI{5}{\percent} inefficiency at
\SI{5}{\percent} false veto probability for a
shorter \SI{170}{\us} time window.

While the LAB itself generally exceeds 
the impurity goals, the 
additives must be purified somewhat more
completely than has been achieved in the
reactor neutrino experiments. 
The purity achieved by the Borexino
Collaboration greatly exceeds the LZ 
requirements for U~/~Th~/~K. Borexino
demonstrated a \ICof impurity slightly 
better than that needed for LZ~\cite{Alimonti:2008gc}.  Should
the \ICof level be higher than our goal
or requirement, we are able to set
a threshold of \SI{200}{\keVee}, above
the endpoint of \ICof \Pgb decay.

Chapter~\ref{chap:CAS} details the implementation and performance 
of the outer detector more fully.

\tdrsec[IntCal]{Internal Calibration with Dispersed Sources}

The physics of self-shielding allows 
LZ to achieve its unprecedented sensitivity 
by reducing the rate of 
\Pgg-ray scatters in the energy range 
of interest to  a level of secondary
importance. This is the central feature 
of the LZ detector design. Conversely, 
the same effect presents a challenge 
for a calibration  program based solely 
on external \Pgg-sources such as \ICsoTS, 
which range out in the first several
centimeters.

LZ will use a suite of gaseous sources
introduced into the xenon circulation 
path that will mix throughout the TPC 
active region and the xenon skin. In 
the active region we will rely primarily on 
the two sources that were developed 
and successfully used by LUX. 
\IKreTm, a source of \SI{9.4}{\keVee} 
and \SI{32.1}{\keVee} conversion electrons,
separated in time by an average of 
\SI{154}{\ns}, will be used to frequently
calibrate spatial effects over the course 
of the run. However, owing to its 
complex decay structure and relatively high
energy compared with the WIMP-search 
energy range, it is not suitable for 
calibrating the ER background. 

The \IKreTm has a 1.8 hour half-life and 
will decay away quickly enough that no 
removal is required. However, because 
of the slow mixing of the xenon in LZ, 
that isotope may not populate the active 
region uniformly. Therefore, LZ will also 
use  \IXeoTom, which has a \num{164}-\si{\keV}
x-ray and an \num{11}-day half-life, to
occasionally make a homogeneous 
distribution of events to uniformly 
calibrate position-dependent 
effects. Its higher energy is also well 
suited to calibrating the xenon skin at 
the expected threshold 
of about \SI{100}{\keVee}. For calibrating the ER background in the
WIMP-search energy range, we will use tritium 
(\IHT), a \Pgbm-emitter with a Q-value of
\SI{18.6}{\keVee}, in the form of tritiated
methane. As we demonstrated in LUX, 
this molecule can be effectively removed 
by the purification  system without leaving
residual activity and its energy is 
ideally suited for modeling the ER band 
down to threshold. 

Further calibration of the skin will use
higher-energy events from \IRnttz, a source 
of \num{6}-\si{\MeV} alpha particles. 
Unlike \IRnttt, this isotope has no 
long-lived progeny and presents no 
background contamination concerns. 
Another benefit of \IRnttz is that some 
of its progeny should live long enough 
to deposit on the detector walls and produce 
a useful calibration  of the S2 response 
there, which as we have seen in LUX, is 
different than in the bulk. 

\tdrsec[XePurity]{Xenon Purity for Detector Performance}

\begin{wrapfigure}{O}[0pt]{0.60\textwidth}
\begin{center}
\includegraphics[width=0.55\textwidth]{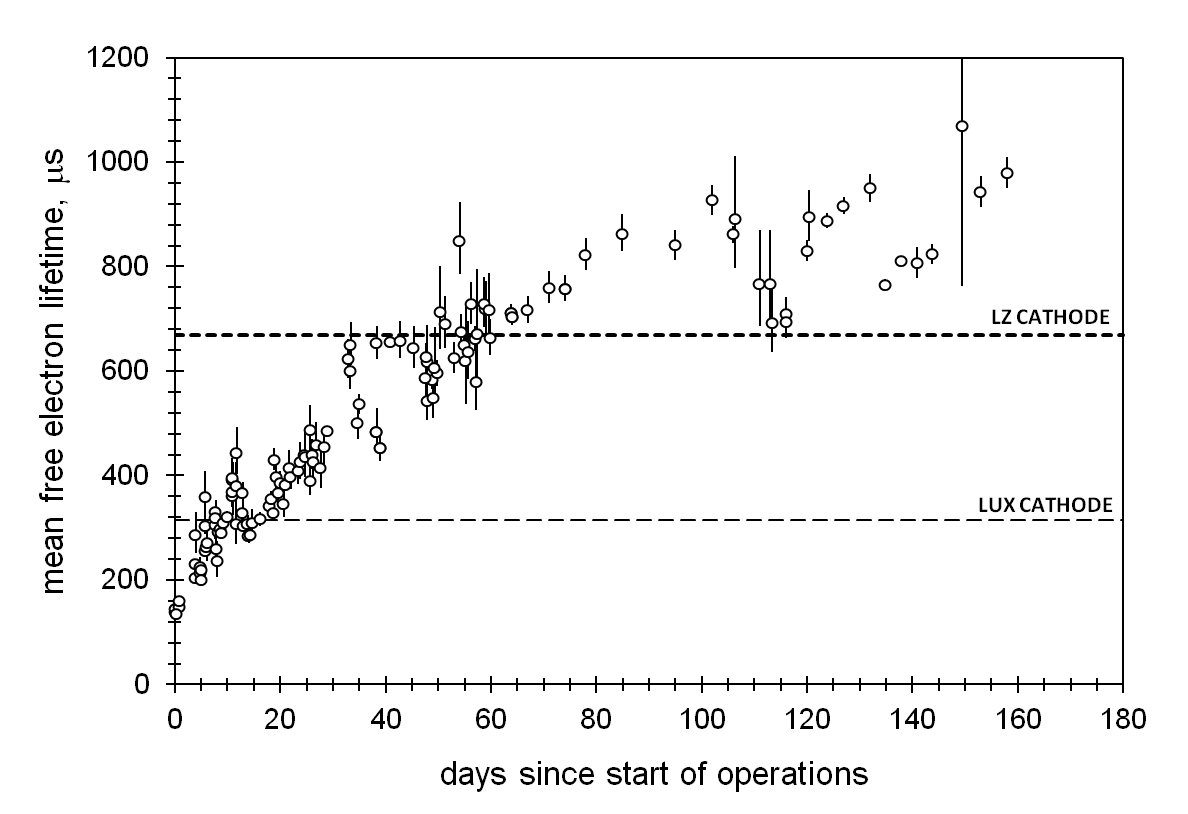}
\end{center}
\tdrfcaption[electron-lifetime-LUX]{Evolution
of electron lifetime in LUX}{Evolution of the mean free electron
lifetime in LUX during Run 3. The electronegative purity required to
observe signals from the LZ cathode (\SI{\sim670}{\mus} for the 
\num{147}-\si{\cm} long TPC) has already been achieved in LUX.}
\end{wrapfigure}

The fluid nature of LXe provides an 
opportunity to manipulate 
the purity of the LZ target material. 
The previous section 
describes how this allows internal calibration sources to 
be temporarily introduced into the LXe. This section considers 
how the purity can be maximized for low-background physics 
running. We consider two classes of impurities: the 
radioactive noble gases \IKreF and \IRnttt -- although 
the latter is addressed fully only in Chapter~\ref{chap:MAS} 
-- and electronegative contaminants such as oxygen and water.

Electronegative impurities are introduced during operations 
by the outgassing of detector materials, and they must be 
continuously suppressed to the level of \SI{\sim0.1}{\ppb} 
by the purification system to ensure good charge and photon 
transport. Previous detectors such as ZEPLIN-III achieved 
this through clean construction techniques with 
low-outgassing materials (i.e., no plastics), which allowed 
these systems to maintain purity without 
recirculation~\cite{Majewski:2011st}. 
In larger detectors such as LUX, the need to utilize large 
volumes of PTFE (see Chapter~\ref{chap:XDS} for detail) 
demands active recirculation in the gas phase. The efficient 
removal of electronegative contaminants is made possible by 
purification technology developed for the semiconductor 
industry. The central elements of this technology are a 
heated zirconium getter for the removal of non-noble 
species, UHV-compatible plumbing and instrumentation, 
and a gas-circulation pump.

Despite the availability of this technology, until recently 
the achievement of good electronegative purity in a LXe TPC 
was considered a significant technical challenge. 
What was lacking was an economical and sensitive monitoring 
technique to allow the purification technology to be fully 
exploited. While free electron lifetime monitors for LXe 
were developed over 20 years ago, these devices cannot 
identify individual impurity species, are insensitive to 
noble gas impurities, and provide little guidance on the 
origin of any impurities.

LZ scientists have developed a mass spectrometry method 
that allows most electronegative and noble gas impurities 
to be individually monitored in real time \cite{Leonard:2010zt}. 
Most crucially, the method provided, for the first time, 
information on the impurity source. For example, the 
presence of an air leak introduces \CNt, \COt, and Ar 
in a characteristic ratio, while a large excess of \CNt 
is a signature of a saturated purifier. Outgassing, on 
the other hand, leads to a uniform rate of increase of all 
common impurity species.

Experience  with the mass spectrometry method, relevant 
to LZ, has been gained by applying it not only to LUX, 
but also to the EXO-200 \Dznbb-decay experiment 
\cite{Dobi:2011zx}. Both experiments achieved 
their purity goals with relative ease due in part to the 
effectiveness of this program, and this valuable experience 
can be brought to bear on LZ. We have learned, for example, 
that vendor-supplied Xe is often relatively pure of 
electro-negatives; that zirconium getters are effective 
for Xe purification but that their performance degrades 
at higher gas flow rates; and that purifier performance 
improves at elevated temperature. This experience is of 
direct relevance for the design of the LZ purification 
system. In LUX, electron lifetimes of the order of 
\SI{1}{\ms} have been demonstrated, as shown in 
Figure~\ref{IDOf:electron-lifetime-LUX}, which is already 
sufficient to drift charge from interactions near the LZ 
cathode.

An additional benefit is that the method is sensitive 
not only to electronegatives but also to trace quantities 
of noble gas impurities, which is of critical importance 
for control of krypton. Krypton is a particularly dangerous 
impurity for LZ because of the presence of the beta emitter 
\IKreF. This isotope, whose abundance at present is 
\num{\sim2e-11} (\IKreF ~/ \IKrnat), presents the leading 
purification challenge for LZ because its noble nature makes 
it impervious to the zirconium getter technology. 
Vendor-supplied Xe typically contains about \SI{100}{\ppb} 
of \IKrnat, which, if left untreated, would give rise to 
an \IKreF \Pgb decay rate of \SI{29}{\milli\becquerel\per\kg} of Xe.

LZ will capitalize on the success of EXO-200 and LUX by 
integrating sensitive monitoring into its Xe handling 
program from the time of Xe procurement until the conclusion 
of the experiment. In addition, because the origin of 
impurities is now understood to be primarily due to 
outgassing, we have in place a comprehensive and 
extensive materials outgassing screening program.

While sensitive impurity monitoring will lay the groundwork for 
the LZ Xe purification program, the heart of the program to reduce 
long-lived noble-element radioisotopes will be the krypton 
removal program using the chromatographic technique developed 
by LZ scientists now at SLAC \cite{Bolozdynya:2007zz}. 
This program was successfully applied to LUX and will be scaled 
up in mass-throughput by a factor of 20 for LZ. The krypton 
concentration goal is \SI{0.015}{\ppt} (g/g), a factor of \num{300} 
below the LUX goal but only a factor of 10 worse than the upper 
limit of \SI{0.2}{\ppt}  (g/g) demonstrated so far with this method. 
At \SI{0.2}{\ppt} (g/g), the rate of beta decays from \IKreF in the 
region of interest is comparable to the pp neutrinos.  In parallel, 
the krypton detection limit of the mass spectrometry method has been 
improved to \SI{\sim0.015}{\ppt} (g/g). The ultimate source of krypton 
impurities in LZ will be the outgassing of detector materials, which 
we can control through careful materials selection and through 
our outgassing plan. 

Similarly, \IRnttt must be controlled by limiting 
the emanation sources within the detector and the 
gas system via a careful screening program. In fact, 
the need to limit radon and krypton in the LZ Xe is 
a driving consideration in the choice of key Xe 
system components such as the gas-recirculation pump. 
More generally, the materials-screening program is 
occurring hand-in-hand with the design efforts of 
all systems whose materials touch Xe, in order to 
ensure that systems meet both their technical and 
their background requirements on schedule.

\tdrsec[DominantBackgrounds]{Dominant Backgrounds}
LZ has a clear background-control strategy with optimal
exploitation of self-shielding to pursue an unprecedented 
science reach. These are the most salient features of 
this strategy: (1) underground operation within an 
instrumented water tank to mitigate cosmogenic backgrounds; 
(2) deployment of a target mass large enough to 
self-shield external radioactivity backgrounds, 
working in combination with outer, anticoincidence 
detectors; (3) construction from low-activity 
materials and purification of the target medium to 
render intrinsic backgrounds subdominant; and (4) 
rejection of the remaining ER backgrounds by S2/S1
discrimination.

The dominant fixed-contaminant radioactive backgrounds 
come from the Xe-space PMTs. 
This background is small when compared to that 
from radon. Our approach is to self-shield those sources 
over the first few centimeters of active liquid and 
thereby be sensitive to a population of ERs caused 
by elastic scattering of solar pp neutrinos --
which can be further discriminated by their S2/S1 
ratio. An irreducible but very small background of 
NRs will also arise from coherent neutrino-nucleus 
scattering. To achieve this we must reduce any intrinsic 
ER backgrounds contained within the Xe itself to a 
tolerable level, with \IKreF, \IArTn, and Rn progeny being a 
particular challenge. Electron recoils caused by the 
\Dtnbb decay of \IXeoTs, now confirmed by 
the EXO-200~\cite{Ackerman:2011gz} and KamLAND-Zen
~\cite{KamLANDZen:2012aa} experiments, are subdominant 
below about \SI{20}{\keVee}. Cosmogenic (muon-induced) 
backgrounds are not significant during operation due 
to the tagging capability of the instrumented water 
tank and the scintillator veto, but cosmogenic 
activation of detector materials prior to deployment 
(including the Xe) must be addressed.

We describe briefly some of these background 
categories here (radioactivity external to the TPC, 
intrinsic contamination of the LXe, and cosmogenic 
backgrounds), as these are important LZ sensitivity 
drivers, but we offer a full discussion on their 
mitigation to Chapter~\ref{chap:MAS}. Neutrino 
backgrounds are explored further in Chapter
~\ref{chap:SIP}. The solar pp neutrino scattering 
rate of \SI{0.8e-5}{events\per\kg\per\day\per\keVee} 
is the benchmark against which other rates are assessed.

\tdrsubsec[MaterialRadioactivity]{Backgrounds from Material Radioactivity}
Radioactivity backgrounds have limited nearly all 
dark matter experiments so far and, in spite of the 
power of self-shielding, we are not complacent in 
addressing them in LZ. 
They impact the thickness of the sacrificial layer 
of LXe that shields the fiducial mass, and they may 
cause rare-event topologies that may be of consequence 
(from random coincidences, atypical surface interactions, 
or Cherenkov emission in PMT glasses, for example). It is
important, therefore, to minimize the rate of \Pga, \Pgb, 
and \Pgg activity around the active volume, as well 
as neutron production from spontaneous fission of 
\IUtTe and from (\Pga,\Pn) reactions. In addition, the 
rate and spatial distribution of such backgrounds must 
be well characterized to build an accurate background 
model for the experiment. The LZ background model is 
derived from a high-fidelity simulation of the experiment 
in the LUXSim framework, which was successfully used 
for the LUX background model \cite{Akerib:2014rda}.

All materials to be used in LZ will be subject to 
stringent constraints as part of the comprehensive 
screening campaign described in Chapter~\ref{chap:MAS}, 
with \SI{10}{\percent} of the solar pp neutrino 
scattering rate and a maximum of \num{\sim0.2} NRs at 
\SI{50}{\percent} signal acceptance being the target 
for the total contribution from material radioactivity 
within the fiducial volume. The dominant rates come 
from the various PMT systems and the LZ cryostat, based 
on the large masses and close proximity to the active 
region of the detector. Table~\ref{IDOt:bkgTable} 
summarizes the count rates from neutron 
and \Pgg-ray emission expected from detector materials 
and other backgrounds, which are discussed in greater 
detail in Chapter~\ref{chap:MAS}. Here and hereafter count rates are given for an indicative \num{5.6}-tonne fiducial mass, considering only single scatter events with no accompanying signal in either the LXe skin or the Outer detector veto systems. The region of interest for WIMP searches is defined as 1) the number of detected photons in the S1 signal to be greater than 0 and less than 20, 2) 3-fold coincidence in S1 between PMTs in the target volume, 3) the number of detected photons in the S2 signal is greater than 350. This region of interest approximately corresponds to energy deposits of \SIrange{6}{30}{\keV} for nuclear recoils and \SIrange{1.5}{6.5}{\keV} for electron recoils. The range of energy deposits is sometimes used to reference the region of interest throughout the text. A full description of the analysis cuts is given in Chapter~\ref{chap:SRD}.

\begin{table}[ht]
\begin{center}
\tdrtcaption[bkgTable]{Summary of backgrounds in LZ.}{Summary 
of backgrounds in LZ, showing the number of counts expected in \num{1000} live days in an
indicative \num{5.6}-tonne fiducial mass in the region of interest with all cuts applied. A comprehensive set of numbers can be 
found in Table~\ref{MASt:BackgroundsTable}.
}

\renewcommand\arraystretch{1.5}
\sffamily
\begin{tabular}{| c | c | c |}
\hline
\rowcolor{mrocol}
\rule{0pt}{1.4em}Item & ER cts  & NR cts \\[0.5ex]
\hline
\multicolumn{1}{|l|}{Detector Componenents} & 6.2	& 0.07 \\
\hline
\multicolumn{1}{|l|}{Dispersed radionuclides (Rn, Kr,  Ar)} & 911	& - \\
\hline
\multicolumn{1}{|l|}{Laboratory and cosmogenic} & 4.3	& 0.06 \\
\hline
\multicolumn{1}{|l|}{Fixed surface contamination} & 0.19	& 0.37 \\
\hline
\multicolumn{1}{|l|}{\IXeoTs \Dtnbb} & 67 	&	- \\
\hline
\multicolumn{1}{|l|}{Neutrinos (\Pgn-e, \Pgn-A)} & 255 	&	0.72 \\
\hline
\multicolumn{1}{|l|}{Total} & 1244	&	1.22 \\
\hline
\multicolumn{1}{|l|}{Total (with 99.5\% ER discrimination, 50\% NR efficiency)}	& 6.22 & 0.61 \\
\hline
\multicolumn{1}{|l|}{Total ER+NR background events}	  & \multicolumn{2}{|c|}{6.83} \\
\hline
\end{tabular}
\end{center}
\end{table}

The PMTs chosen for the LZ TPC are Hamamatsu R11410s, 
which have achieved very low radioactivity values; 
LZ scientists have been involved in a long campaign to 
establish their performance for dark matter experiments, 
working actively with the manufacturer to enable this
\cite{Akerib:2012da}. The PTFE required to fabricate 
the TPC field cage, skin reflectors, and other internal
components may also be an important source of neutron 
emission from the bulk material. We use upper limits 
on the contamination measured by EXO-200 in calculating 
its impact \cite{Leonard:2007uv}. The cryostat is another
dominant component, mostly owing to its large mass. 
For the titanium baseline design (\SI{2140}{\kg}), the 
total neutron emission rate is estimated at 0.6 n/day 
based on recent titanium samples procured for LZ. As 
a result of a 2-year material search campaign, we were 
able to find titanium with U/Th contamination, which 
is a factor of \num{2} lower than that used in LUX
~\cite{Akerib:2011rr} as explained in 
Section~\ref{CROS:MaterialSearch}.

\tdrsubsec[SurfacePlating]{Surface Plating of Radon Progeny}

The noble gas radon consists solely of radioactive isotopes, 
of which four are found in nature: \IRnttt and \IRntoe 
produced in the \IUtTe decay chain, \IRnttz from the 
\IThtTt decay chain, and \IRnton from the \IActtS series. 
As a result of its chemical inertness, radon exhibits long
diffusion lengths in solids. \IRnttt is the most stable 
isotope (T$_{1/2}$ = 3.82 days), and is present in air at 
levels of about ten to hundreds of \si{\becquerel\per\m\cubed}. 
Charged radon progeny -- especially metallic species 
such as \IPotoe\ -- plate out onto macroscopic surfaces 
that are exposed to radon-laden air. A fraction will 
deposit and even implant into material surfaces during 
detector construction or installation \cite{Guiseppe:2011mj}.

Backgrounds from surface beta and gamma radioactivity, as 
well as recoiling nuclei (e.g., \IPbtzs from the alpha 
decay of \IPotoz), are largely mitigated by short 
half-lives and the self-shielding of LXe. However, 
\Pga-particles released in the decay chain, particularly 
from \IPotoz\ -- a granddaughter of the long-lived \IPbtoz
(T$_{1/2}$ = \num{22.3} years) -- result in neutron emission 
following (\Pga, \Pn) reactions. 
This is problematic for TPC materials with large 
(\Pga, \Pn) yields such as PTFE (\SI{\sim e-5}{\Pn\per\Pga}, 
due to the presence of fluorine). Additionally, because 
PTFE is produced in granular form before being sintered 
in molds, plate-out poses further risk because surface
contamination of the granular form becomes bulk 
contamination when the granules are poured into molds.

A second concern relates to our ability to correctly 
reconstruct events at the TPC inner walls, 
since the imperfect reconstruction of these events leads 
to a background population leaking radially toward the 
fiducial volume \cite{Akerib:2013tjd,Alner:2007ja}. 
This concern drives the design of the top PMT array 
and places tight requirements on the plate-out of radon 
progeny on the TPC walls (see Section~\ref{XDSS:PMTs}).

Controls to mitigate background from radon plate-out 
will include limiting the exposure of detector parts 
to radon-rich air; monitoring from point of production 
through transport and storage in Rn-proof materials; 
and employing surface cleaning techniques, such that 
neutron emission is negligible relative to material
radioactivity from bulk uranium and thorium contamination.

\tdrsubsec[IntrinsicBackground]{Intrinsic Backgrounds}
We are confident that our requirements for intrinsic
radioactive contamination from \IKreF and \IRnttt can 
be met with the Xe-purification techniques described in 
Chapter~\ref{chap:XCS}, coupled with the radon emanation
screening of Xe-wetted materials described in 
Chapter~\ref{chap:MAS}. We note that most of these 
backgrounds can be estimated with low systematic 
uncertainty. In addition to direct sampling, the 
\IKreF \Pgb - decay spectrum is well understood and 
the decay rate can be measured during operation with 
delayed \Pgb -- \Pgg coincidences.

Other delayed coincidence techniques as well as \Pga 
spectroscopy allow precise estimation of radon-induced
backgrounds. In fact, it is possible to follow dynamically 
the spatial distribution of these decays throughout 
the detector, which was done very successfully in 
LUX~\cite{Bradley:2014lga}. The two main concerns in 
this instance are a ``weak'' naked \Pgb-decay from 
\IPbtof in the bulk of the TPC (E$_{max}$ = 
\SI{1019}{\keV}, BR = \SI{9.2}{\percent}), and the 
possibility of \Pgg-ray escape for peripheral events 
from the dominant \IPbtof decay modes.

Our goal is to control each of these two backgrounds 
to \SI{<10}{\percent} of the solar pp neutrino rate, 
limiting the \IRnttt activity to \SI{0.67}{\milli\becquerel} and 
the krypton concentration to \SI{0.02}{\ppt} (g/g). In a more
conservative scenario, as requirements we allow the sum of these two components to be 3 times the ER background from pp neutrinos.

Trace quantities of argon are also a concern due to
\Pgb-emitting \IArTn, with a 269-year half-life and 
\SI{565}{\keV} endpoint energy. This background is 
constrained to be less than \SI{10}{\percent} of 
\IKreF, resulting in a specification of 
\SI{4.5E-10}{(g/g)} or \SI{2.6}{\micro\becquerel}. 
The Kr-removal system, which also removes Ar, 
should easily achieve this.

\tdrsubsec[CosmogenicBackground]{Cosmogenic Backgrounds}

A rock overburden of \num{4300} m~w.~e. above the Davis Cavern 
at SURF reduces the muon flux by about a factor of \num{3e6}
relative to the surface \cite{Mei:2009py,Kudryavtsev:2008qh}. Muons crossing 
the water tank and/or liquid scintillator are readily detected via Cherenkov emission in water and/or scintillations, 
and any coincident energy deposition in LZ is similarly 
easily identified. However, neutron 
production in muon-induced 
electromagnetic and hadronic showers, in particular in high-Z materials, may generate background events
\cite{Reichhart:2013xkd,Gray:2010nc}. The total 
muon-induced neutron flux at SURF from the surrounding 
rock is calculated to be about
\SI{0.5e-9}{\Pn\per\cm\squared\per\second}, with 
approximately half of this flux coming from neutrons 
above \SI{10}{\MeV}, and some \SI{10}{\percent} from 
energies above \SI{100}{\MeV} \cite{Mei:2005gm,Lindote:2008nq}. 

Muon-induced neutrons generated in the water shield and liquid scintillator produce a similarly low rate, despite the several hundred tonnes of target mass, due to the low atomic number of water and scintillator and consequent low neutron yield 
(\SI{\sim 2.5e-4}{\Pn/muon\per(g/cm^2)}), 
translating to a production rate of order of
\SI{e-9}{\Pn\per\kg\per\second}). 

Cosmogenic activation -- radioisotopes production 
within materials, largely through spallation reactions 
of fast nucleons from cosmic rays while on the Earth’s 
surface -- can present electromagnetic background 
in LZ. \IScfs produced in the titanium cryostat decays 
through emission of \SI{889}{\keV} and \SI{1120}{\keV} 
\Pgg-rays (with T$_{1/2}$ = 84 days). Cosmogenic 
\ICosz (T$_{1/2}$ = 5.3 years) in copper components 
will produce gamma rays of \SI{1173}{\keV} and 
\SI{1332}{\keV}.

Activation of the Xe itself during storage or transport 
generates several radionuclides, some of which are 
important, especially in the first few months of 
operation. Tritium (T$_{1/2}$ = 12.3 years and 
production rate of \SI{\sim 15}{/kg/day} at the 
Earth’s surface \cite{Mei:2009vn}) was previously 
a concern; however, this is effectively removed 
through purification during operation. Production of 
Xe radioisotopes, such as 
\IXeotS (T$_{1/2}$ = 36.4 days), 
\IXeotnm (T$_{1/2}$ = 8.9 days), and 
\IXeoTom (T$_{1/2}$ = 11.9 days), are more 
problematic, as they cannot be mitigated through 
self-shielding or purification. In particular, atomic
de-excitation of the 2s and 3s shells in \IXeotS 
generates 5.2 and \SI{\leq 1.2}{\keV} energy deposits, 
respectively, which are an important background 
in the WIMP search energy region for certain event 
topologies \cite{Akerib:2013tjd}. These backgrounds 
soon reach negligible levels once Xe is moved underground.

\tdrsubsec[Fiducialization]{Fiducialization}
The backgrounds from nearby materials cluster
near the edges of the LZ central TPC, and they
fall off according to an exponential distributions
described in Fig.~\ref{IDOf:self-shield-LXe}.
The rapid fall off of backgrounds from material
radioactivity suggests the definition of a fiducial
volume, which is an inner region that is relatively
free of background.  

The fiducial volume is
a simplification that allows intuitive understanding
of the LZ background from nearby materials.
Eventually, LZ will abandon the concept of
a fiducial volume in favor of a PLR maximum likelihood
fit which describes the spatial distributions
of the various background components, like the
LUX collaboration has 
recently done~\cite{Akerib:2015rjg}.

We define an LZ fiducial volume that starts
\SI{1.5}{\cm} above the cathode at the bottom
of the TPC, \SI{4}{\cm} inside of the reflective
walls that surround the TPC, and \SI{13.5}{\cm}
down from the gate grid at the top of the TPC.
This fiducial volume encloses \SI{5.6}{\tonnesl}
of LXe, and the event numbers tabulated in
Table~\ref{IDOt:bkgTable}
occur inside this fiducial volume in a
\SI{1000}{\livedayl} run.

\begin{figure}[htbp]
\begin{center}
\includegraphics[width=0.9\textwidth]{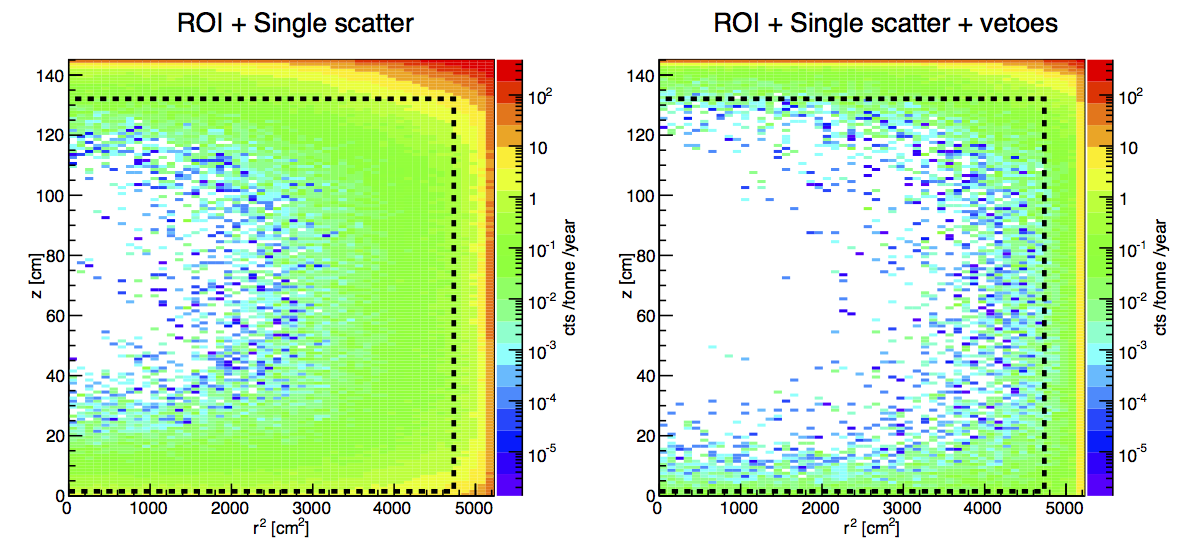}
\end{center}
\tdrfcaption[fiducialization]{Fiducialization}
{Total NR background plus ER leakage from material 
radioactivity for sources external to the LXe in the 
TPC, counted over a \SIrange{6}{30}{\keV} acceptance 
region; a discrimination efficiency of \SI{99.5}{\percent} 
is applied to ERs from beta decays and gamma rays. 
Left: Single scatters only, no 
vetoing by the anti-coincidence systems. 
Right: Adding the combination of 
both the skin veto and the outer detector.
The dashed line shows the boundary of the
\SI{5.6}{\tonnesl} fiducial mass.
}
\end{figure}

Two main factors determine how far 
the fiducial boundary should lie from 
the lateral TPC walls.  The prime
consideration is to ensure a sufficiently 
thick layer of LXe to self-shield against 
the external radioactivity backgrounds.
This is related to the mean 
attenuation length for those particles: 
Figure~\ref{IDOf:self-shield-LXe} 
confirmed that \SI{\sim 2}{\cm} of liquid
decreases the \Pgg-ray background tenfold,
and as much as \SI{\sim 6}{\cm}  
is needed to mitigate neutrons 
by the same factor.  However, the outer
detector is very efficient for 
neutron tagging, which brings these 
two requirements closer together.

Secondly, it is essential that the
reconstructed ($x$,$y$) positions of 
low-energy interactions occurring near 
the TPC walls do not "leak" into 
the fiducial volume. As mentioned above 
(and discussed in Chapter~\ref{chap:XDS})
interactions  from radon progeny plating 
the lateral PTFE are of particular concern:
These can generate events with very small 
S2 signals due to trapping of charge 
drifting too close to the PTFE. If 
allied with poor position resolution, 
this can constitute a very challenging
background \cite{Alner:2007ja}. In the 
vertical direction, only the former
consideration arises. We point out that 
the reverse field region below the cathode
will provide much of the required 
self-shielding (\SI{>14}{\cm}), whereas 
at the top, the small thickness of liquid
above the gate (\SI{0.5}{\cm}) will have 
a limited impact. 

Figure~\ref{IDOf:fiducialization} shows the 
simulated background rate from material
radioactivity in the WIMP region of interest
(\SIrange{6}{30}{\keV}) as a function of
radius squared and height above the 
cathode grid. Nuclear and electron recoil
backgrounds were combined, with \SI{50}
{\percent} acceptance applied to the former
and \SI{99.5}{\percent} discrimination
applied to the latter. The neutrino and
dispersed background contributions listed in
Table~\ref{IDOt:bkgTable} are omitted
from Figure~\ref{IDOf:fiducialization}
because they populate the figures
uniformly.

The left panel of 
Figure~\ref{IDOf:fiducialization} shows
the rate of material backgrounds when
neither of the outer detector systems,
the LXe skin nor the outer detector, is
utilized.  The background-free region
in the central TPC is small, and comprises
about one-half of the \SI{7}{\tonnesl}
of active liquid xenon.  The right panel
shows the background rates after the
application of the two outer detector
systems, which enlarge considerably the
background-free region. The event totals in
Table~\ref{IDOt:bkgTable} are computed
from the events inside the
fiducial volume enclosing 
\SI{5.6}{\tonnesl} delineated by the dashed
line in Fig.~\ref{IDOf:fiducialization}.

\bibliographystyle{apsrev4-2}
\bibliography{LZtdr}

\wlog{In tdrinclude command}
\wlog{\ChapLabel}

\renewcommand{\ChapLabel}{chap:SIP} 
\renewcommand{\BibLabel}{SIPS:bib} 
\renewcommand{\SecPre}{SIP}
\renewcommand{\FloatPre}{SIP}
\tdrchap{Scientific Performance}

The LZ detector system described in the
previous and subsequent chapters is highly
sensitive to a variety of physics signals. 
The principal signal we seek is that
of NRs distributed uniformly throughout the 
LXe TPC volume, in response to an
impinging flux of nonrelativistic WIMPs
that are gravitationally bound to the 
Milky Way galaxy. 
In the first sections of this
chapter, we describe the sensitivity to
various WIMP-particle cross sections. 

The first step in selecting the sample of 
WIMP candidates is to define a 
search region in the two variables: 
S1 (the prompt scintillation light) and 
S2 (the delayed electroluminescence light, 
a measure of primary ionization). 
The use of both
variables allows  the distinction of 
NRs from the much more numerous ERs.

The LUX collaboration has recently
completed extensive calibrations of
the response of liquid xenon to
NRs~\cite{akerib:2016mzi} and
ERs~\cite{akerib:2015wdi}, 
as shown in 
Figure~\ref{IDOf:discrimination}. 
These high-statistics data
permit us to employ the detailed
shapes of the response probability
distribution functions (PDFs) in
a profile likelihood ratio (PLR) fit to
estimate the LZ sensitivity to NRs
from WIMPS.

We first describe our sensitivity 
and discovery potential for the
spin-independent (SI) WIMP-nucleon interaction. We then
discuss interpretations involving more general forms of the WIMP-nucleon
interaction.

Should LZ see a WIMP signal, the 
distribution of that signal in NR 
energy will allow constraints on the 
WIMP-Xe scattering cross section, 
the WIMP-Xe reduced
mass, and on the velocity 
distribution of galactic
WIMPs~\cite{Pato:2012fw}.

A variety of other physics processes can be probed by selective detection of
NRs and ERs as defined with S1 and S2. The central fiducial region of the LZ
detector will be an extraordinarily quiet laboratory for processes that deposit
energy. Among the physics processes that can be probed are:
\begin{enumerate}
 \item Solar neutrino detection.
 \item A neutrino magnetic moment.
 \item Double beta decay.
 \item Neutrinos from supernovae.
 \item Sterile neutrinos.
 \item Interaction of WIMPs with atomic electrons.
 \item Solar and certain dark-matter axion-like particles (ALPs).
 \item Exotic particles that interact in the LZ outer detector.
\end{enumerate}

\tdrsec[WSDP]{WIMP Sensitivity and Discovery Potential}
The principal physics analyses of the LZ experiment will be searches for the
recoils of Xe atoms caused by the interaction of WIMPs with the Xe nucleus.
As discussed above, two types of signal are formed in the LXe response to the
recoils: S1 and S2. In the principal LZ search, the energy of the recoil is
reconstructed from a combination of S1 and S2, and the ratio S2/S1 provides
discrimination of NRs from the background of ERs. The value of the reconstructed
energy depends on whether the event is an NR or an ER.

\tdrsubsec[SSA]{S1+S2 Analysis}

The S1+S2 analysis in LZ will follow the
general framework of the recently published
LUX search for NRs in response to
WIMPs~\cite{Akerib:2015rjg,Akerib:2016vxi}. 
We define a search region in the plane of
$\log_{10}(\textrm{S2})$ 
versus S1, shown in
Figure~\ref{SIPf:LUXSearchN}.\footnote{Sometimes,
the equivalent plane of $\log_{10}(\textrm{S2}/\textrm{S1})$ versus
S1 is utilized to display the same data.}
LUX determines of the sensitivity 
to WIMP-nucleon scattering with
a multi-dimensional PLR fit in that plane,
which is also how the LZ sensitivity is
determined.
 
A comparison of the key performance
assumptions for LZ as well as the
comparable achievements in LUX are
given in Table~\ref{SIPt:CKassump}.
The baseline detector performance 
assumed for LZ is in many aspects more
conservative than that achieved by LUX. 
The most prominent exception in
Table~\ref{SIPt:CKassump} is the 
liquid/gas emission probability, where we
presume that the limitations of the 
LUX electric field will be removed in the
LZ experiment.

\begin{figure}[htbp]
\centering
\includegraphics[width=0.8\textwidth]{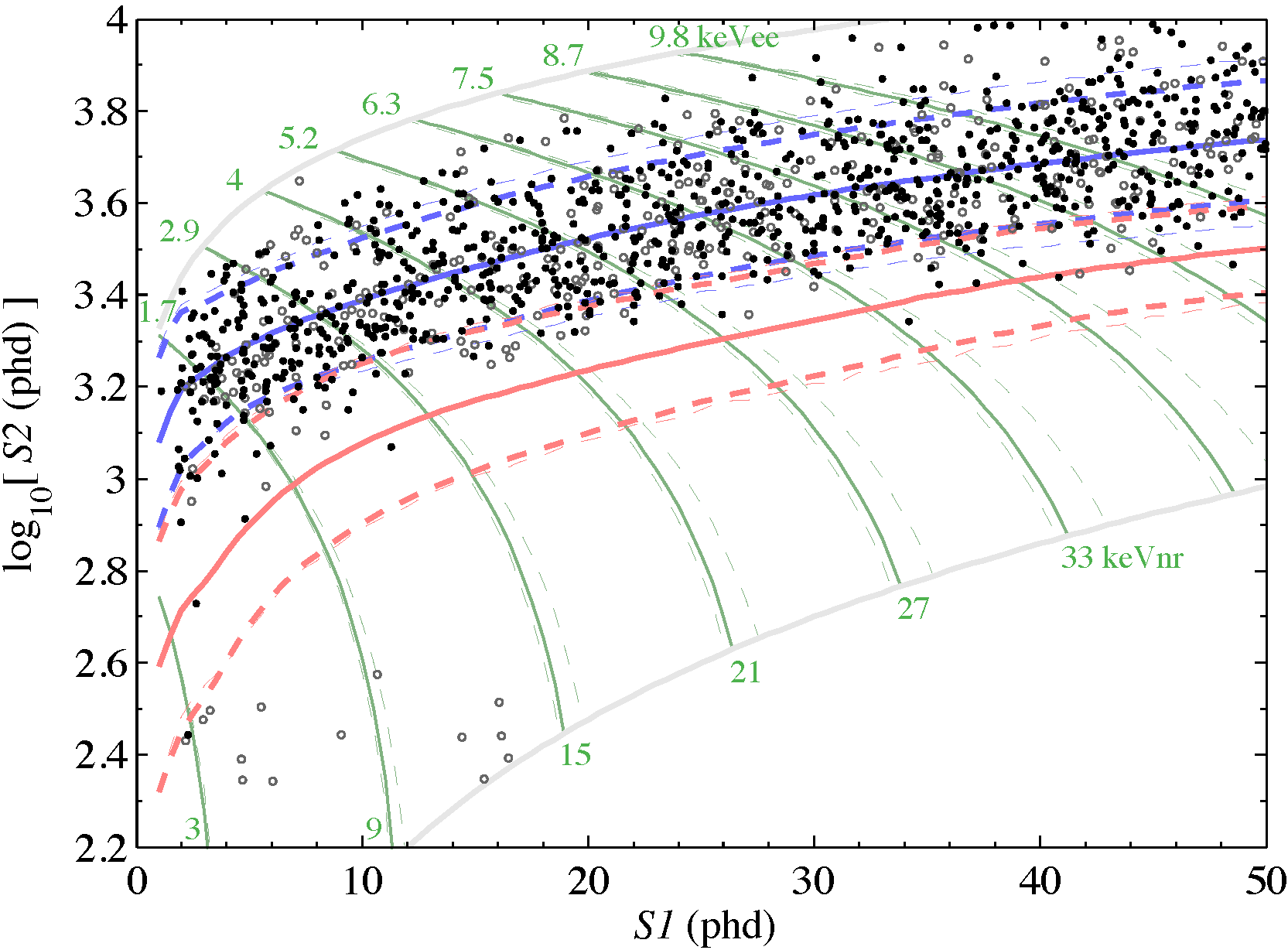}  
\tdrfcaption[LUXSearchN]
{The LUX WIMP search data}
{The LUX WIMP search data~\protect\cite{Akerib:2016vxi}. Shown is
all data after selection criteria for the 332 live days of
the LUX 2014-2016 run.
The logarithm of S2 is plotted versus 
S1, after spatial corrections.  
Filled black circles are in the detector central region
(radius \SI{<18}{cm}) and the
edge of the detector (radius \SIrange{18}{20}{cm}) 
in grey open circles. The centroid (solid) and search
region boundaries (dotted) are red for 
the signal (NR) region or ``band'', and
corresponding lines in blue describe the
primary background (ER) band. The 
dotted lines are \SI{\pm1.28}{$\sigma$} 
around the centroid.  Contours of equal 
recoil energy for NR (\si{\keVnr}) 
and ER (\si{\keVee}) interpretations are 
shown in grey. The unit ``phd'' is photons
detected, and results from correcting the photoelectrons
detected for the probability of one UV
photon inducing two photoelectrons. The data is 
consistent with a background of ERs and 
wall-induced events.}
\end{figure}

\begin{table}
\centering
\tdrtcaption[CKassump]
{Key LZ and LUX Assumptions Compared}
{Key LZ and LUX Assumptions Compared}
\begin{tabular}{|>{\sfs}C{2.5in}|>{\sfs}C{1.0in}|
                 >{\sfs}C{1.0in}|>{\sfs}C{1.0in}|}
\hline
\rowcolor{mrocol}
{\sfbl\Tstrut Quantity} & {\sfbl\Tstrut Units} &
{\sfbl\Tstrut LZ Assumption} & {\sfbl\Tstrut LUX~\cite{Akerib:2016vxi}} \\ \hline
Recoil threshold, \SI{50}{\percent} efficiency & \si{\keVnr} & \num{6}	& \num{3.3} \\ \hline
S1 range & Photons detected & \numrange[range-phrase=~--~]{3}{30} & \numrange[range-phrase=~--~]{2}{50} \\ \hline
S2 range & Photons detected & \num{>350} & \num{>200} \\ \hline
S1 light-collection efficiency & Absolute &
\SI{7.5}{\percent} & \SI{14}{\percent} \\ \hline
Photocathode efficiency	& Absolute &
\SI{25}{\percent} & \SI{30}{\percent} \\ \hline
Liquid/gas emission probability	& Absolute &
\SI{95}{\percent} & \SI{73}{\percent} \\ \hline
ER discrimination & Absolute &
\SI{99.5}{\percent} & \SI{99.8}{\percent} \\ \hline
\end{tabular}
\end{table}

The benchmark process we will use to 
interpret NRs will be the interaction of
WIMPs via an SI process, such as exchange 
of a Higgs particle~\cite{Barbieri:1988zs},
with the gluons in the nucleons in the Xe
nucleus~\cite{Shifman:1978zn}.
This process produces a WIMP-nucleus
scattering rate that is independent of 
the identity, neutron or proton, of 
the nucleon in the nucleus. For the 
low-momentum transfers of typical WIMP
interactions, the scattering amplitude 
is proportional to A, the number of 
nucleons in the nucleus. The scattering 
cross section includes the density of 
states, which also favors larger A, while 
the threshold for energy detection favors
smaller A. The nuclear form factor is 
employed to account for quantum-mechanical
interference attributable to the non-zero
nuclear size~\cite{Lewin:1995rx},
and the standard halo model (SHM) of the
distribution of WIMP velocities in the
Milky Way is used~\cite{McCabe:2010zh}.

The backgrounds expected for LZ are described
in detail in Chapter~\ref{chap:MAS}
and summarized in Table~\ref{IDOt:bkgTable}
and in Table~\ref{SRDt:PLRBG}. In the
LZ Conceptual Design 
Report (CDR)~\cite{Akerib:2015cja}, we performed
a simple cut-and-count estimate of the
LZ sensitivity, under the assumption that
\SI{0.5}{\%} of the ER events would contaminate
an NR signal region defined to retain
\SI{50}{\%} of NR events.  The vastly improved
understanding of the PDFs of ERs and NRs
in S1 and S2 achieved by LUX have
caused us to utilize the more advanced
PLR statistical technique for estimates
of WIMP sensitivity in this report\cite{cowan:2010js}.
In general this technique substantially
reduces the fraction of ER events that
contaminate the NR signal region, as
discussed in Section~\ref{SRDSs:PLR}.
A consequence is the extremely
stringent requirements in the LZ CDR
on the permissible rate of radon decays 
in liquid xenon are substantially relaxed
in this report, to a level of
\SI{\approx20}{\mBq} for the 
\SI{10}{\tonnesl} LZ total volume.  This
rate of radon decays is commensurate with
the achievements of LUX and other existing
liquid xenon experiments.  However, at
a radon decay rate of \SI{\approx20}{\mBq}
ERs from the quiet beta decays of
radon daughters outnumber ERs from
solar \Ipp neutrinos by a factor of
about 3.5.

The resulting sensitivity plot is shown 
in Figure~\ref{SIPf:LZSISens}, along with
LUX and ZEPLIN limits. For the baseline assumptions described in this report, LZ achieves a median
sensitivity at a mass of 
approximately \SI{40}{\GeVpct} of 
\SI{2.3e-48}{cm^2}.

The project LZ sensitivity for low WIMP masses is considerably
improved in this report, compared to the CDR.   The improvement
is due to the utilization in this report of the improved
calibrations reported by LUX\cite{akerib:2015wdi,akerib:2016mzi},
which document higher S1 and S2 responses to low energy NR than were
assumed in the CDR. Further improvements in sensitivity for
low WIMP masses is possible
through an ``S2-only'' analysis\cite{sorensen:2010hv}, or
through the detection of bremmstrahlung
from the nuclear recoil\cite{McCabe:2017rln}.

\begin{figure}[htbp]
\centering
\includegraphics[width=0.95\textwidth]{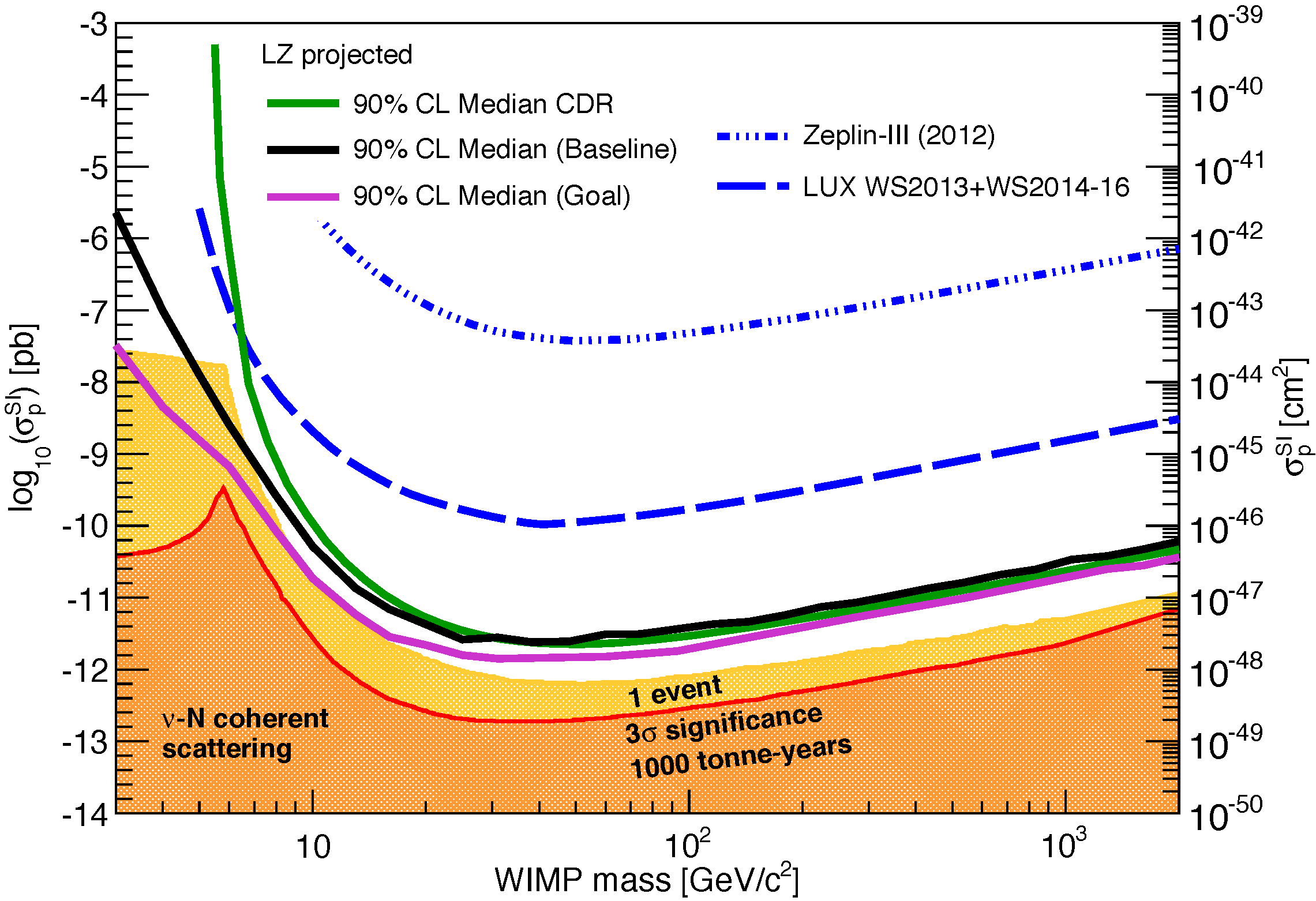}  
\tdrfcaption[LZSISens]
{LZ sensitivity projections}
{LZ sensitivity projections. The
baseline LZ assumptions described in this Technical Design Report 
give
the solid black curve. LUX
and ZEPLIN results are 
shown in broken blue lines.  If LZ achieves the design goals listed
in Table~\ref{SRDt:Dt.1Tab}, the sensitivity would improve, resulting
in the pink sensitivity curve. The gray line shows the projected sensitivity in the LZ Conceptual Design Report (CDR)~\cite{Akerib:2015cja} (see text for details of the changes from the CDR to this report). The shaded regions show 
regions where background NRs from cosmic
neutrinos emerge~\protect\cite{Billard:2013qya,*Ruppin:2014bra}.}
\end{figure}

\tdrsec[NP]{Neutrino Physics}
The LZ detector will have a sufficiently 
large mass and low background that several
types of neutrino interactions will be 
visible. These events will be uniform
throughout the liquid xenon volume and 
cannot be shielded. We have studied the
sensitivity of LZ to solar, atmospheric,
astrophysical, reactor, and geophysical
neutrinos. In particular,
solar neutrinos have been considered as 
both an interesting signal and as 
an irreducible background to a WIMP search. 

LZ will observe the \Ipp
fusion chain of our sun in real time via
elastic \Dnuenue\ scattering, in a lower 
energy regime than the only other real-time
measurement to date, and will most likely
detect neutrinos from \IBe via coherent
nuclear scattering.
The coherent neutrino signal from a 
nearby supernova would be a unique, 
flavor-independent probe of the neutrino flux. 

We have also estimated the
potential of LZ to observe neutrinoless 
double-beta decay (\Dznbb) from \IXeoTs, and
considered the impact on the reactor/source
neutrino anomaly and on searches for a 
neutrino magnetic moment of a prolonged
exposure of LZ to a nearby \ICrFo neutrino
source.

\tdrsubsec[SAN]{Solar and Atmospheric Neutrinos}
\tdrsubsubsec[PP]{Elastic Scattering of Solar Neutrinos}

A prominent background for WIMP dark matter
searches in LZ will come from the elastic
scattering of solar neutrinos from the \Ipp
fusion chain~\cite{Bahcall:2004mz} with
the atomic electrons in xenon.
Our calculations of the rate of these
scatters agree with those of
\cite{Baudis:2013qla} under the same 
assumptions. The calculations in this report,
however, use updated neutrino mixing parameters
\cite{Beringer:1900zz} and solar fluxes
obtained from a luminosity-constrained analysis
of Borexino data (cf. Table 2 of \cite{Robertson:2012ib}). 
Our projections assume the
standard LZ fiducial target 
mass of \SI{5.6}{\tonnesl} and 
an exposure of \SI{1000}{\livedaysl}.
For electron recoil events with energies
between \num{1.5} and \SI{20}{\keVee}, we expect 
\num{838} observable \Ipp events, \num{69} events 
from \IBeS and \num{<10} events from \INoT. 
For electron recoil energies above 
\SI{20}{\keVee}, \Dtnbb\ events from \IXeoTs
are expected to dominate the counting rate. 

In the WIMP dark matter search window between
1.5~and~\SI{6.5}{\keVee}, the corresponding
calculated numbers are 233, 19 and 3 for a
total of 255 electron recoil background events
from solar neutrinos. Our calculation has
neglected atomic binding effects on the
scattering process.
Inclusion these effects will result in
a suppression of result in a suppression
of \SIrange{24}{28}{\%}\cite{Chen:2016eab,*Morgunov:1996jk}.

The LZ experiment would add an interesting 
data point to the existing world experimental
sample on \Ipp solar neutrinos, in the context 
of the MSW-LMA explanation to the observed
solar neutrino flux. Existing data to support
this model in the low-energy regime are from
the SAGE experiment \cite{Abdurashitov:2009tn}
and from Borexino \cite{Bellini:2014uqa}. 
The \SI{50}{\tonnesl} gallium target in SAGE
inferred 854 inverse beta decay events
attributed to \Ipp solar neutrinos over 18
years of operations. The neutrino energy
threshold for that measurement was 
\SI{233}{\keV}.
More recently, the Borexino Collaboration 
made the first real-time detection of
the \Ipp solar neutrinos via the elastic scattering process with atomic electrons.
The Borexino neutrino energy threshold 
was \SI{\approx300}{\keV}, while LZ
will be uniquely sensitive, with a neutrino
energy threshold of a few tens of keV.

Although the LZ experiment will open 
up new experimental territory in the study
of \Ipp solar neutrinos, the current 
consensus in the solar neutrino community
is that the accuracy of an \Ipp solar 
neutrino measurement must be better 
than \SI{1}{\%} to improve understanding of
solar neutrinos~\cite{Robertson:2012ib}. 
To achieve \SI{1}{\%} accuracy, LZ would need
to observe several tens of thousands of \Ipp
neutrino-induced ER events, and also control
systematics at a sub-\SI{1}{\%} level.  The 
elimination of the \IXeoTs isotope and a 
live time of \num{2000} to \num{4000} days would 
allow the accuracy of an LZ measurement 
of \Ipp solar neutrinos to approach \SI{1}{\%}. 

\tdrsubsubsec[CENNS]{Coherent Nuclear Scattering of Solar Neutrinos}
Neutrinos are expected to elastically 
scatter coherently across nucleons in the
nucleus~\cite{Freedman:1973yd,Cabrera:1984rr}.
This process has yet to be observed. Dedicated
experiments aim to measure the process in the
laboratory.
The energy transferred to the nucleus 
from coherent neutrino scattering is 
typically suppressed by $\approx m_e/m_N$
relative to the elastic electron 
scattering process, so
signals from the coherent scattering of 
solar \Ipp neutrinos will fall well below 
the LZ S1+S2 detection threshold.

\begin{figure}[htbp]
\centering
\includegraphics[width=0.49\textwidth]{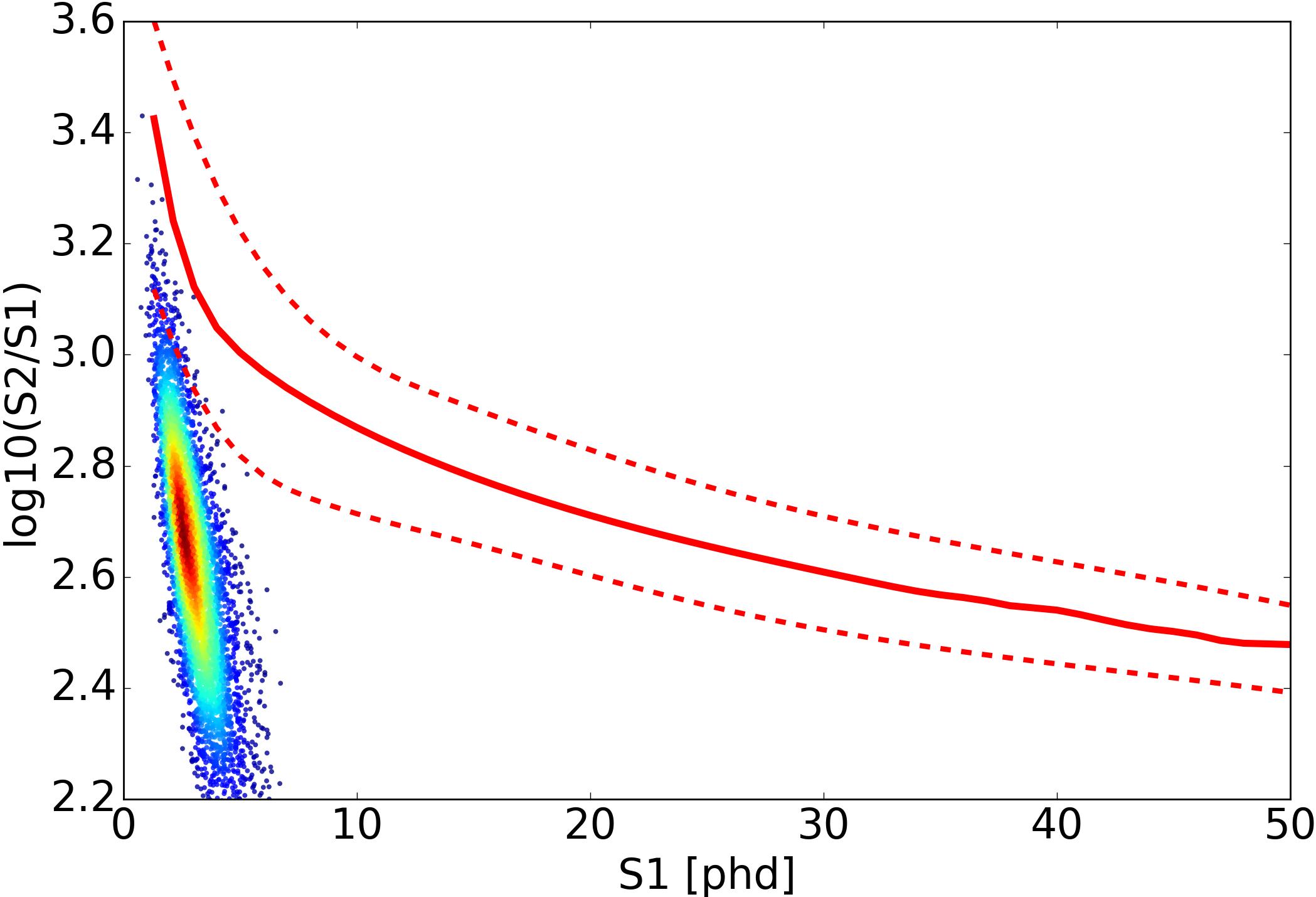}
\includegraphics[width=0.49\textwidth]{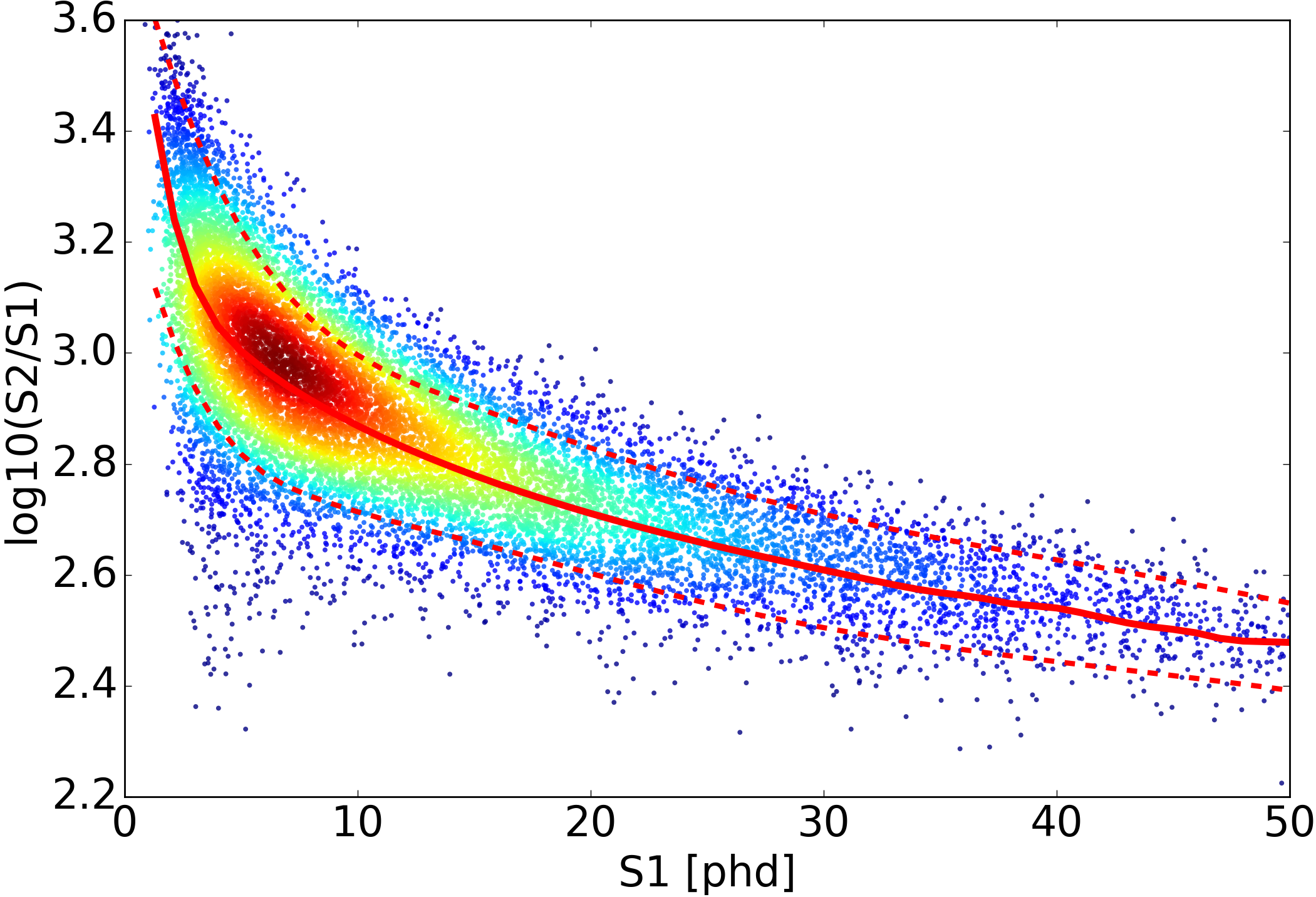}
\tdrfcaption[solarCNSfig]
{Calculated PDFs for NR from neutrinos}
{Calculated high-statistics probability distribution
functions (PDFs) for \IBe
(left panel) and 
atmospheric (right panel) coherent
neutrino-nucleus scattering events.
The solid curve is the centroid of the
nuclear recoil band, and the dotted lines
define a \SI{\pm3}{$\sigma$} band. These lines
are defined for a spectrum
\emph{flat} in nuclear recoil energy.
The recoils from \IBe on the left fall outside
the band because the bulk of the PDF is under the
threshold, and correlated fluctuations must occur
for events to enter this plot.
In \SI{5600}{\tonnedayl}
of LZ exposure, we expect a mean of 7 and 0.5 
NR events from these neutrino sources to meet
selection requirements.}
\end{figure}

Neutrinos from \IBe decay, which occur 
at the end of the \Ipp chain
about 0.1\% of the time, range in 
energy up to \SI{\approx15}{MeV}. This
energy to is sufficient to transfer up 
to a few keV of energy to a xenon nucleus
via coherent scatter, and so these events 
are expected to fall at the threshold
of detectability in LZ. Figure~\ref{SIPf:solarCNSfig} 
(left) shows the expected rate and signal 
distribution of these events. 
Our calculations agree with those of
\cite{Strigari:2009bq} if we make the 
same assumptions. For the calculations in this
report, we assume a \IBe neutrino flux of
\SI{5.25e-6}{cm^{-2}s^{-1}}
as measured by SNO \cite{Aharmim:2011vm}, 
with a total uncertainty of \SI{<5}{\%}.
The largest uncertainty in the number 
of detected \IBe neutrinos is due to
the signal yield in the liquid xenon. 
We assume the latest results obtained
by the LUX Collaboration \cite{Akerib:2015rjg}.
Assuming LZ baseline detector
parameters we expect 7 events from \IBe
coherent neutrino scatter in the full
\SI{5600}{\tonnedayl} LZ exposure. 
Systematic uncertainty, due primarily to 
photon collection and liquid xenon 
signal yields) is comparable to 
statistical uncertainty in this case.

The measurement of the flux of \IBe neutrinos
through coherent neutrino scattering is
sensitive to all neutrino flavors, forming
an interesting result in its own right.
From a dark matter perspective, these 
neutrinos are an irreducible background 
which looks very similar to a \SI{6}{GeV} WIMP. 
The distinct and complimentary calibrations
planned in LZ, described in
Chapter~\ref{chap:CAS} will allow a
thorough mapping of the \IBe neutrino signal
region. Combined with the distinctive,
soft spectrum of \IBe neutrino events, 
LZ will be able to in essence fit and subtract out the \IBe neutrinos from the WIMP search.

In contrast to the atmospheric neutrino
signal (Fig. \ref{SIPf:solarCNSfig}, right
panel), the \IBe signal shown
(Fig. \ref{SIPf:solarCNSfig}, left panel)
appears below the nuclear recoil band.
This is true despite the fact that both
distributions are due to nuclear recoils from
coherent neutrino nucleus scattering. The
primary reason is an artifact of the signal
detection threshold. Because LZ will not be
able to detect fewer than one S1-induced detected photon,
the \IBe signal consists of the tail of upward
fluctuations in the number of scintillation photons
produced. This comes at the expense of the number of
ionized electrons in the S2 signal. Therefore, the
ratio S2/S1 for these events is systematically
biased below the nuclear recoil band, which is
defined for a \emph{flat} distribution in recoil energy.

\tdrsubsubsec[ATM]{Atmospheric Neutrinos}
Atmospheric neutrinos result from muon 
and pion decay in the atmosphere. 
Historically, they were
considered as a background for nucleon 
decay experiments, and then exhibited
a surprising flavor-mixing phenomenon
that has now been verified in accelerator-based
experiments.

Consequently, the literature shows 
measurements or calculations of the atmospheric
neutrino flux for energies \SI{\gtrsim1}{GeV}.
The flux of atmospheric neutrinos is not a
significant background for LZ in the elastic
neutrino-electron scattering channel.
However, coherent neutrino-nucleus 
scattering events present a serious 
background concern. This is because the 
hard energy spectrum of the neutrinos 
results in a recoil spectrum which is
essentially indistinguishable from 
a typical WIMP. The PDF for the
expected spectrum is shown in 
Figure~\ref{SIPf:solarCNSfig} (right).

There is a detector-dependent sweet spot 
in neutrino energy for detection of coherent
neutrino nucleus scattering. For LZ this is 
in the range of a few tens of MeV. Lower
energies cannot register a signal, and 
higher energies begin to suffer a nuclear form
factor suppression from the loss of coherence.
Therefore LZ is singular in its need to
understand the atmospheric neutrino flux 
for energies \SI{\lesssim100}{MeV}.
A single calculation exists for the flux
in this energy region~\cite{Battistoni:2005pd},
and it is tailored to two particular
experimental sites: Kamioka, and Gran Sasso.
The latitude of the site is important 
because the largest uncertainty is
attributed to the geomagnetic cutoff, 
which limits the penetration of the primary
cosmic rays into the atmosphere. 
Previous work \cite{Strigari:2009bq}
assumed the flux values for Kamioka, which 
are about \SI{30}{\%} lower than the flux 
at Gran Sasso.

Our event rate calculations assume the 
Gran Sasso flux values tabulated 
in~\cite{Battistoni:2005pd},
with a single-sided uncertainty 
of \SI{+50}{\%}. This uncertainty was 
estimated by comparing the Kamioka, 
Gran Sasso and SURF locations with a 
vertical cutoff rigidity
map~\cite{Smart20052012}. 
In \SI{5600}{\tonnedaysl}, LZ expects 
to observe 0.5 signal-like
events from atmospheric neutrinos,
distributed in S2 and S1 very much like
the expected signal from a high-mass
WIMP. This PDF shown in Fig.~\ref{SIPf:solarCNSfig} (right panel).

\tdrsubsubsec[NMM]{Neutrino Magnetic Moment}
It is not known if the neutrino has a 
small magnetic moment, and upper limits 
exist on its possible magnitude. The 
strongest direct particle physics upper 
limit is \SI{5.4e-11}{$\mu_B$}
from Borexino \cite{Arpesella:2008mt}, 
while analysis of supernovae provide a 
stronger upper limit of 
\SI{5e-13}{$\mu_B$}~\cite{Lattimer:1988mf}. 
\begin{wrapfigure}{R}{0.65\textwidth}
\centering
\includegraphics[width=0.6\textwidth]{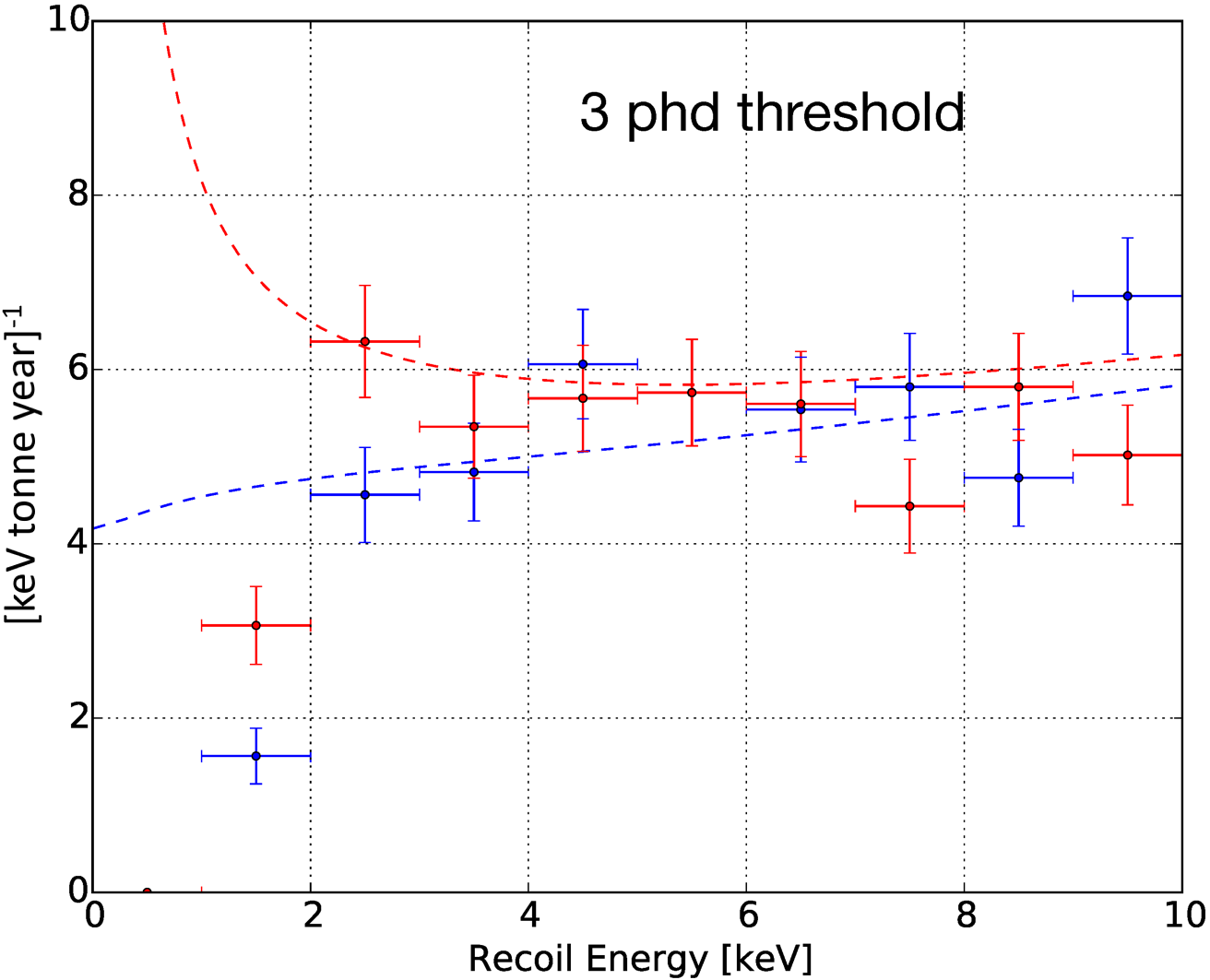}
\tdrfcaption[nuMagMo]{Neutrino magnetic moment signal in LZ}
{Predicted neutrino-dominated electron recoil background rate in LZ, for no
magnetic moment (blue) and a magnetic moment \SI{5e-12}{$\mu_B$} (red).}
\end{wrapfigure}

The effect of a neutrino magnetic moment 
is an increased scattering rate with electrons.
A larger magnetic moment shifts the turn-on 
of the increase to higher energy.
The \SI{\approx1}{keV} energy threshold 
of LZ suggests an order of magnitude
improvement in sensitivity, relative to
Borexino. This is shown
in Fig.~\ref{SIPf:nuMagMo}, assuming
$\mu_{\nu}=$\SI{5e-12}{$\mu_B$}, in
which case an increase in the scattering rate 
at threshold would be just barely observable.

\tdrsubsubsec[ONU]{Other Neutrino Backgrounds}
One other possible source of signal-like 
events arises from coherent neutrino-nucleus
scattering of the diffuse supernova 
neutrino background (DSNB). We estimate this
background in the NR search region to be 
0.05 (DSNB) for the LZ fiducial mass of
\SI{5.6}{\tonnesl} and run duration 
of \num{1000} days.

Geophysical neutrinos from \IUtTe and 
\IThtTt decays have been seen by the
KamLAND~\cite{Sanshiro:2005a,Araki:2005qa,Enomoto:2007eps} 
and Borexino~\cite{Bellini:2010hy} 
detectors. Those detectors have an energy
threshold for neutrinos of about \SI{1.8}{MeV}.
They are unable to detect neutrinos from the
decay of \IKfz, which have an energy 
just below \SI{1.5}{MeV}.
Using the Reference Earth Model and neutrino
flux calculations from the KamLAND work,
we estimate for LZ \SI{1.5}{ER} events/year 
from \IKfz decay, \SI{0.3}{ER} events/year
from \IUtTe decay, and \SI{0.2}{ER} 
events/year from \IThtTt decay. With the
ability to distinguish ER and NR in LZ,
these signals provide negligible backgrounds 
for the dark-matter search.

It is possible for neutrinos to 
capture on xenon nuclei. 
This process, Xe$(\Pgn,\Pem)$Cs,
is analogous to that employed in the 
Davis experiment at Homestake. 
The Feynman diagram for this process 
results from crossing the electron 
capture process, so we expect the electron
capture Q value to set the threshold for
neutrino capture  proceed. 
Only \IXeoTo has a sufficiently
low Q value (\SI{352}{keV}) to exhibit
sensitivity to the \Ipp neutrinos. A single
calculation of the event rate exists
\cite{Georgadze:1997zv}, from which
we estimate \num{<1}~events/year in LZ, 
in the \SIrange{1.5}{6.5}{keV} energy window. 

The nearest power reactors are 
about \SI{800}{km} away, in 
Fort Calhoun, NE (\SI{0.5}{GWe}),
and Cooper, NE (\SI{0.8}{GWe}). The
power/distance$^2$ distribution shows a 
broad peak for reactors in Illinois and
Wisconsin. The net flux is small enough,
however, that we expect negligible 
detected events from power-reactor 
neutrinos in LZ.

\tdrsubsec[BB]{Double Beta Decay} 

\tdrsubsubsec[NBB]{Neutrinoless Double Beta Decay}

If the electron neutrino is its own 
anti-particle, this would allow for the
possibility of a process whereby double beta
decay occurs but the two neutrinos annihilate.
This process is referred to as
neutrinoless double beta decay 
(\Dznbb). In this case, 
all decay energy goes into the two electrons.
Thus, the signature is a mono-energetic, 
single-site event at the Q-value of the decay.
Observation of this process would imply discovery of
\begin{enumerate}
\item fermions which are their own anti-particles (so called Majorana particles).
\item Lepton number violation.
\item Violation of conservation of the net
      difference between baryon and lepton
      number.
\end{enumerate}
Currently the best lower limit on the 
half-life for \Dznbb of \IXeoTs comes 
KamLAND-ZEN results, 
\SI{1.06e26}{y}~\cite{kamland-zen:2016pfg}
at \SI{90}{\%} confidence.  Searches involving
different xenon isotopes and related processes
are discussed in Ref.~\cite{Barros:2014exa}.

Any search for \Dznbb needs low backgrounds, a
large amount of the relevant isotope, and 
good energy resolution. Note that LZ requirement 
R-150004 implies a resolution of better than
\SI{2.0}{\percent} $\sigma/E$ at \SI{2.5}{\MeV}. The criteria 
for a \Dznbb search are very similar to 
those for a competitive dark matter experiment,
however traditionally building an experiment
which is competitive for both tends to be
quite difficult. The large mass and
exceptionally low backgrounds make this search
possible in LZ. Typically searches for \Dznbb 
use enriched \IXeoTs to enhance their signal. 
The \SI{7}{\tonnesl} natural xenon 
implies almost \SI{623}{kg} 
of \IXeoTs, which is more than 
previous \Dznbb searches. 

The main backgrounds at the 
\IXeoTs Q-value are the \SI{2447.7}{\keV} 
\Pgg-line from \IBitof 
in the uranium chain
and the \SI{2614.5}{\keV} 
\Pgg-line from \ITltoz from the thorium
chain. Unlike in the many background that
are distributed uniformly throughout the active
xenon mass in the WIMP-search analysis, these
backgrounds are completely from external
detector components, and the size of LZ
afford substantial screening of these
background.

The same background simulations used to
estimate the sensitivity of LZ to WIMPs
are used to project the sensitivity to
\Dznbb. The analysis for \Dznbb is slightly
different because the result is more 
dependent on the input assumptions. 
The energy resolution at the Q-value affects
the experiment's ability to reject backgrounds
from the penetrating \SI{2614.5}{\keV} 
\ITltoz line. Major background contributions include the TPC PMTs,
the xenon vessel, and the resistors. There is some contribution
from the radioactivity of the Davis cavern walls that is still under
investigation, but at worst some additional shielding may be necessary
above and below the xenon vessel, but not on the sides. The sensitivity estimate also
depends on the minimal vertex separation 
needed to identify a multiple scatter and 
the energy resolution at the Q-value
($\sigma/E$). In previous work by
\cite{Baudis:2013qla}, it was assumed 
multiple scatters could be rejected down 
to \SI{3}{\mm} separations, and we make 
the same assumption. The choice of fiducial volume for the
\Dznbb search is also different than that
for the WIMP search, because of the
penetrating nature of the 
\ITltoz line and the fact that its signal
cannot be distinguished by the S1 and S2
signals. Smaller fiducial volumes have less total background due to the self-shielding effect of xenon.

For the purposes of these projections a fiducial volume of \SI{1000}{\kg}
was chosen as a proper tradeoff between backgrounds and exposure.
A Feldman-Cousins cut-and-count analysis is used with a \num{2}$\sigma$
region-of-interest and Q$_{\beta\beta}$. LZ has the potential to  a
sensitivity to a 90\% median expected C.L. limit on the 0$\nu\beta\beta$
half-life of  of \IXeoTs of \SI{1.2e26}{y} with
an energy resolution 
($\sigma$/E) of \SI{1.0}{\%} 
or better. For comparison, 
at \SI{90}{\%} confidence, the half-life 
limit from EXO 200~\cite{albert:2014awa}
is \SI{1.1e25}{y}, that from GERDA~\cite{Agostini:2017iyd}
is \SI{5.3e25}{y}, 
and KamLAND-Zen has achieved
\SI{1.06e26}{y}~\cite{kamland-zen:2016pfg}.

\tdrsubsubsec[BBNN]{Two Neutrino Double Beta Decay}

Two-neutrino double beta decay
(\Dtnbb) is a standard model decay 
which has been observed in several 
isotopes which occurs when single beta 
decay is energetically forbidden. For 
example, \IXeoTs is lighter than 
\ICsoTs, so conservation of energy 
makes single beta decay of \IXeoTs impossible.
However, \IXeoTs can undergo two simultaneous
beta decays, emitting two electrons and 
two anti-electron neutrinos. This process 
has been observed in many different isotopes
and the half-lives are always greater than \SI{10e18}{y}. 

For example, \IXeoTs has a half-life due
to \Dtnbb of
\SI{2.2e21}{y}~\cite{albert:2014awa} 
and a Q-value of \SI{2456}{\keV}~\cite{Redshaw:2007un}. 
LZ should observed 
\SI{3e6} double beta decays over 
\SI{1000}{\livedaysl}. 

The isotope \IXeoTf is also believed 
to undergo \Dtnbb with a Q-value of  
\SI{826}{\keV}~\cite{Wang:2012ame}.

Although \Dtnbb of \IXeoTs has been observed
by both EXO-200 and KamLAND-ZEN, both had
analysis thresholds at or above the peak of
the spectrum from \Dtnbb of \IXeoTf 
near 800 keV. 
LZ will be in a position to measure the 
full spectrum of \IXeoTf \Dtnbb 
down to \SI{1}{\keVee} and see the turnover 
at the peak of the spectrum. 

\tdrsubsec[SN]{Supernova Neutrinos}

Should a supernova occur in our galaxy 
during LZ operation, neutrinos
emitted from the supernova would be 
detected via coherent neutrino-nucleus
scattering, which is blind with respect to
neutrino flavor. The energy spectrum
of neutrinos emitted from a typical 
supernova peaks near \SI{10}{MeV}, and has
a tail that extends above \SI{50}{MeV}, 
which causes NRs above the LZ
threshold~\cite{Scholberg:2012id}. 
Coherent neutrino-nucleus scattering is
mediated by the weak neutral current, and 
thus provides important information on 
the flux and spectrum of muon and tau
neutrinos from supernovae, complementary 
to the signals that would be seen in other
detectors. From a supernova in our own
galaxy at distance \SI{10}{kpc} from Earth, 
LZ would see \num{\sim50} NR events of 
energy greater than \SI{6}{\keVnr} in 
a rapid 10-sec burst~\cite{Horowitz:2003cz,Chakraborty:2013zua}.

The NR recoil spectrum increases as 
the recoil energy decreases; a threshold
of \SI{3}{\keVnr} would allow detection 
of \num{\sim100} supernova neutrino-induced
NR events. The current world sample of 
19 supernova neutrino-induced events
were detected from supernova 1987a, 
\SI{50}{kpc} from Earth, by detectors with
total mass \num{1200} times greater than LZ. 
A supernova \SI{10}{kpc} from Earth
would cause about \num{7000} 
neutrino-induced events in the \SI{32000}{\tonnesl} 
of water in the Super-Kamiokande detector~\cite{Scholberg:2012id}.

The response of large, liquid xenon detectors to
supernova signals has been recently reviewed\cite{Lang:2016zhv}.

\tdrsubsec[STN]{Sterile Neutrinos}

There are long-standing anomalies 
arising from the detailed study of
antineutrinos from reactors, and from 
source-calibration of solar neutrino
experiments~\cite{Abazajian:2012ys}. 
A recent study has evaluated the
capabilities of deployment of a \SI{5}{\mega\Curie} \ICrFo electron 
neutrino source near to the LZ detector~\cite{Coloma:2014hka}. 
The excellent spatial resolution of 
the LZ liquid xenon TPC allows the 
spatial pattern of electron neutrino
oscillation into a sterile neutrino 
to be detected. A neutrino source experiment
with LZ would not be part of the principal 
LZ science goal, which is the WIMP
search, and could constitute a distinct
follow-on experiment after the WIMP
search had achieved significant results.

The sensitivity achievable by five source
deployments of a \SI{5}{\mega\Curie}
\ICrFo source near LZ is shown in
Figure~\ref{SIPf:SterileNu}. Numerous
proposals are underway to probe the
origin of the reactor/source
anomalies~\cite{Caccianiga:2015ega,*Lhuillier:2015fga},
but the potential LZ advantage is a 
diminished need to control the source
normalization due to LZ's excellent 
spatial resolution. In addition, a source
deployment near LZ will bring sensitivity
to an electron neutrino magnetic
moment that is close to the limits 
deduced from astrophysical
considerations~\cite{Coloma:2014hka}.

\begin{figure}[htbp]
\centering
\includegraphics[width=0.8\textwidth]{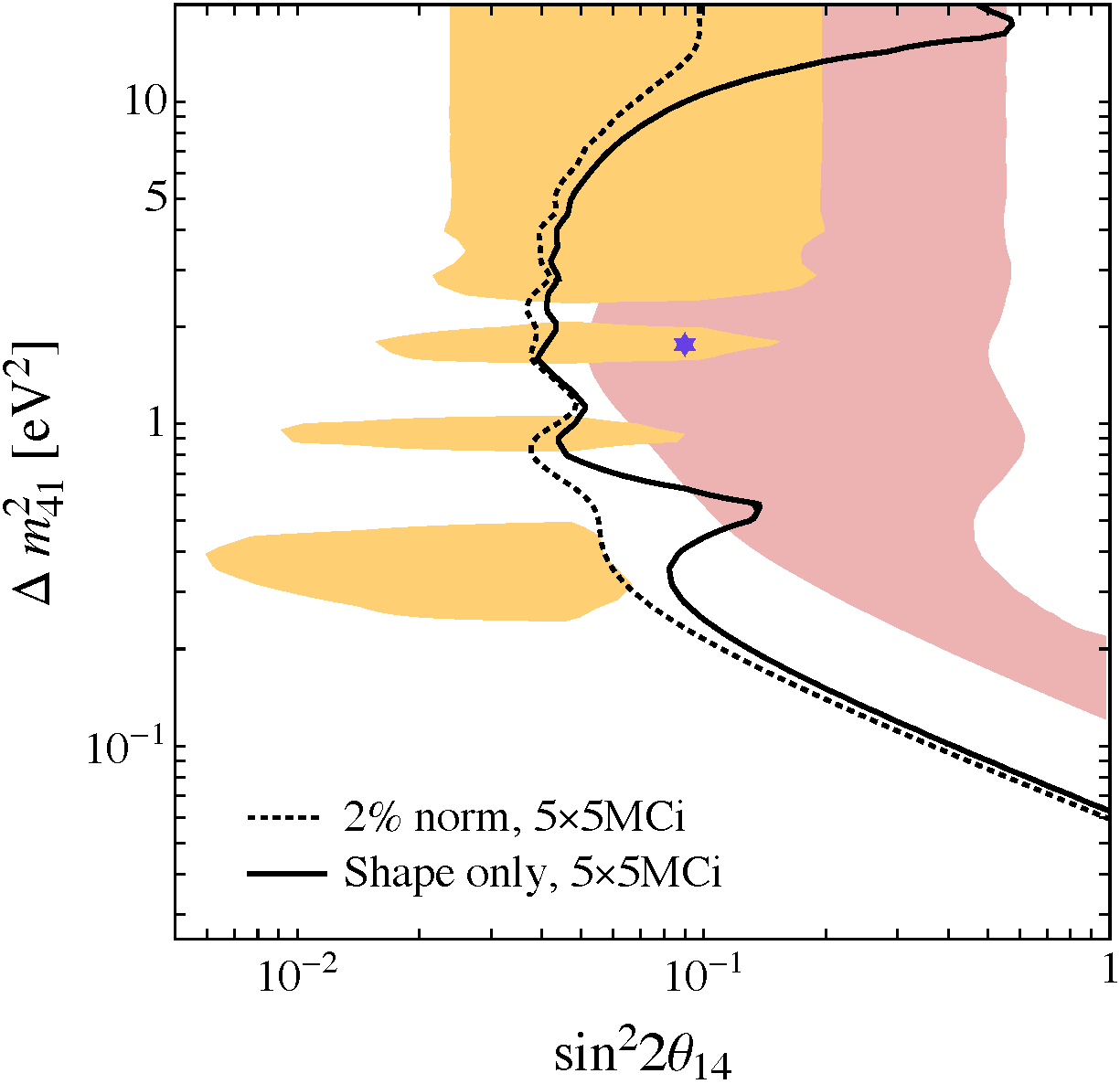}
\tdrfcaption[SterileNu]
{Sensitivity to oscillations to sterile neutrinos}
{Sensitivity to sterile neutrino oscillations as a
function of mass-difference and mixing angle. The
parameter space to the right of each line would be
excluded at \SI{95}{\percent}~CL.
The shaded areas show the \SI{95}{\percent}~CL
allowed regions for source (pink) and reactor (yellow)
anomalies. The blue star is the joint best fit. The black
solid line shows the expected contours for five \SI{100}{\day}
deployments of a \SI{5}{\mega\Curie} \ICrFo source next to LZ, without use
of the source normalization. The dotted line shows the
contour if a \SI{2}{\percent} normalization of the source is available.
From Ref.~\cite{Coloma:2014hka}.}
\end{figure}

\tdrsec[BNRW]{New Physics Beyond Nuclear Recoils from WIMPs}

\tdrsubsec[EW]{Electrophilic WIMPs}

One type of WIMP-matter coupling that 
does not cause NRs, at least at
tree-level, is the coupling of a WIMP 
to a charged lepton. A WIMP-charged
lepton vector coupling induces a 
WIMP-nucleon interaction at one loop in
perturbation theory, where the charged 
lepton loop interacts with the nucleon
via photon exchanges~\cite{Kopp:2009et}. 
This interaction is surprisingly
sensitive. The WIMP-nucleon SI 
cross-section sensitivity of \SI{2.3e-48}{cm^2}
achievable by LZ at a WIMP mass 
of \SI{40}{\GeVpct} corresponds, when
converted via a one-loop calculation, 
to a WIMP-electron cross section of
\SI{1e-50}{cm^2}. Should the interaction 
be exclusively WIMP-muon, the LZ
sensitivity at \SI{40}{\GeVpct} corresponds 
to a vector-mediated WIMP-muon
cross section of \SI{5e-50}{cm^2}; 
for a tau, the corresponding WIMP-tau cross
section is \SI{4e-49}{cm^2}.

If the WIMP is a Majorana particle, all 
its vector couplings vanish, but an
SD axial-vector coupling is still possible.
The axial-vector coupling does not
induce an interaction at higher order in
perturbation theory with the nucleus;
the only observable consequence in LZ of an
axial-vector coupling of a WIMP to
an electron is WIMP-electron scattering.

The electron motion is crucial for
the appropriate treatment of
WIMP-electron scattering. It is 
the very highest momentum tails of 
the electron wavefunction
that determine the cross section for 
an impinging WIMP to ionize a Xe atom.
The resulting events are ERs, and their 
energy spectrum rises very quickly as
the energy deposition falls. Limits on 
axial-vector WIMP-electron scattering
depend critically on the low energy
threshold~\cite{Kopp:2009et}.

Interpretations of the DAMA~\cite{Bernabei:2003ky} 
event excess as axial-vector
WIMP-electron scattering imply a W
IMP-electron cross section of \SI{2e-32}{cm^2}
at a WIMP mass \SI{50}{\GeVpct}. The LZ
experiment is likely to observe an ER
background primarily from \IRnton daughters,
about 4 orders of magnitude lower
than DAMA backgrounds, so LZ should 
achieve a limit, assuming background
subtraction, of approximately \SI{6e-38}{cm^2}.
This sensitivity is comparable
to the indirect astrophysical limits on 
the SD WIMP-electron scattering cross
sections deduced from Super-Kamiokande
data~\cite{Tanaka:2011uf}.

\tdrsubsec[AALP]{Axions and Axion-like Particles}

The axion was introduced to describe the
absence of CP-violation in the strong
interaction. These particles, known as QCD
axions, have a specific relationship
between their mass and their coupling to
fermions~\cite{Peccei:1977hh,Weinberg:1977ma,Wilczek:1977pj}. 
A particle with properties
similar to the axion, but without the
relationship between mass and fermion
coupling, is known as an axion-like particle (ALP)~\cite{RPP:2014rrr,*PhysRevD.86.010001}.

The LZ experiment will be sensitive to 
axions and ALPs via the axioelectric
effect, where an axion is absorbed and 
an atomic electron is
ejected~\cite{Dimopoulos:1985tm}. In 
contrast to the photoelectric effect, the 
mass of the axion or ALP is available 
for transfer to the atomic electron.

Two sources of axions or ALPs contribute 
to a possible signal in
LZ~\cite{Arisaka:2012pb}: 
\begin{enumerate}
	\item Nonrelativistic ALPs that 
might constitute the dark matter of our
galaxy could cause signals in LZ, if 
their masses are sufficient to
provide enough energy to ionize a Xe atom. 
	\item Axions or ALPs with a mass less 
than about \SI{15}{keV} emitted by
bremsstrahlung, Compton scattering, or 
other atomic processes in the
sun also can ionize the Xe atoms 
in LZ~\cite{Redondo:2013wwa}.
\end{enumerate}
Events caused by axions or ALPs in LZ 
would be ERs with energy up to a few
tens of \si{\keVee}. The signal identification 
relies on the distinct shape
of the energy spectrum of the axion or ALP signal.

The signal for a galactic dark-matter ALP
would be a peak in ERs with energy at
the mass of the particle. Our 
studies indicate that the LZ sensitivity 
to the coupling between electrons and 
galactic dark-matter ALPs ranges from a coupling
constant $g_{Ae}$ of \num{e-14} to one of 
\num{e-13} for masses between \SI{1}{\keVpct} 
and \SI{20}{\keVpct}, as shown in
Figure~\ref{SIPf:GalAxion}.

\begin{figure}[htbp]
\centering
\includegraphics[width=0.9\textwidth]{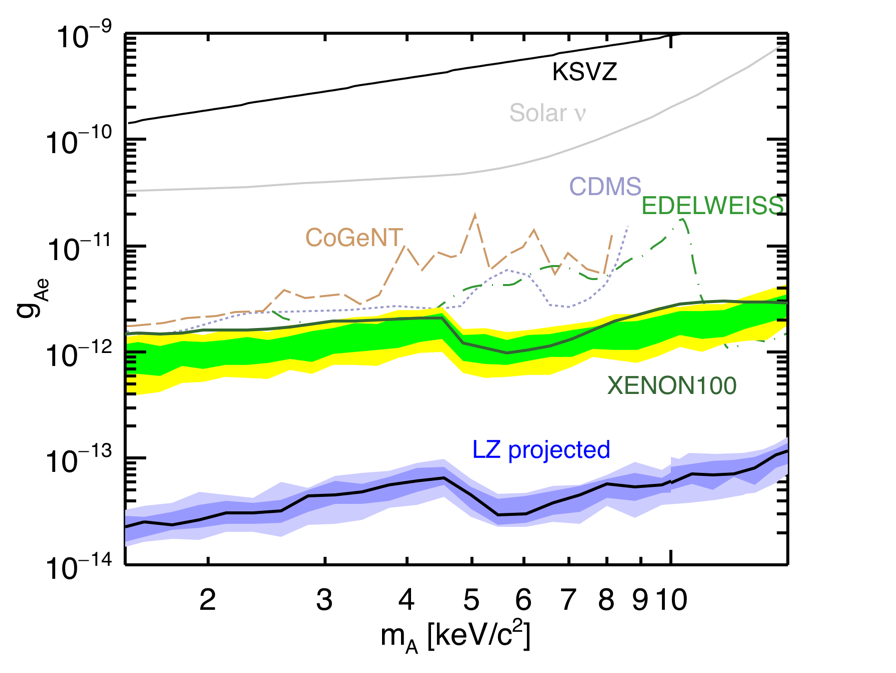}
\tdrfcaption[GalAxion]
{Dark-matter axion-like particle sensitivity}
{Dark-matter axion-like particle sensitivity.
The LZ projected sensitivity for ALPs at \SI{90}{\percent}~CL is
shown by the dark/light blue bands,
which show the \SI{68}{\percent} (\SI{1}{$\sigma$})
and \SI{95}{\percent} (\SI{2}{$\sigma$}) bands for that sensitivity.
The line that defines
KSVZ axions~\protect\cite{Kim:1979if, Shifman:1979if},
an astrophysical upper limit from solar
neutrinos~\protect\cite{Gondolo:2008dd}, is shown. Upper limits by
the experiments CDMS~\protect\cite{Ahmed:2009ht},
EDELWEISS~\protect\cite{Armengaud:2013rta},
CoGeNT~\protect\cite{Aalseth:2008rx}, and XENON100~\protect\cite{Aprile:2014eoa} are
also shown.}
\end{figure}

The signal for solar ALPs is a broad thermal
spectrum caused principally by
bremsstrahlung and the Compton effect in 
the sun convolved with the axioelectric
cross section. Our studies indicate that LZ 
is sensitive to a coupling constant
$g_{Ae}$ between solar ALPs and the electron of about \SI{1.3e-12} for masses
between \SI{0}{\keVpct} and 
approximately \SI{1}{\keVpct}, as shown in
Figure~\ref{SIPf:SolarAxion}.

\begin{figure}[htbp]
\centering
\includegraphics[width=0.9\textwidth]{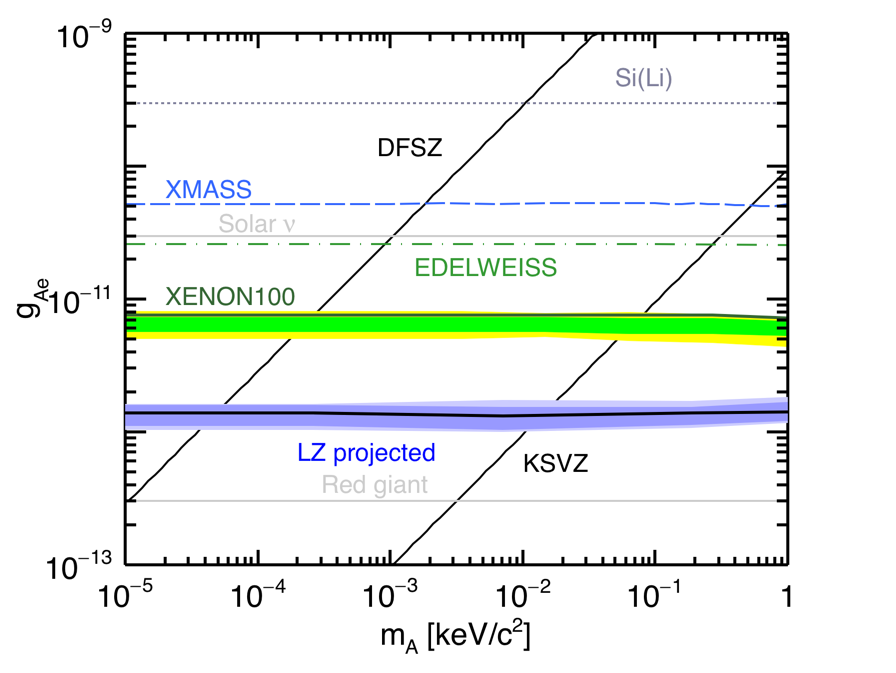}
\tdrfcaption[SolarAxion]
{Solar axion-like particle sensitivity}
{Solar axion-like particle sensitivity.
Horizontal lines all extend down to $m_A=0$. The LZ projected
sensitivity for ALPs at \SI{90}{\percent}~CL is shown by the dark/light
blue bands, which show the \SI{68}{\percent} (\SI{1}{$\sigma$})
and \SI{95}{\percent} (\SI{2}{$\sigma$}) bands for
that sensitivity. The lines that
define DFSZ axions~\protect\cite{Zhitnitsky:1980tq,Dine:1981rt}
and KSVZ axions~\protect\cite{Kim:1979if,Shifman:1979if},
neutrinos~\protect\cite{Gondolo:2008dd}, and from
red giants~\protect\cite{Raffelt:2006cw}, are shown. Upper
limits by the experiments XMASS~\protect\cite{Abe:2012ut},
EDELWEISS~\protect\cite{Armengaud:2013rta}, and
XENON100~\protect\cite{Aprile:2014eoa} are also shown.}
\end{figure}

\tdrsec[POD]{Physics with the Outer Detector}

The primary goal of the LZ Outer Detector 
(OD) consisting of Gd-loaded liquid
scintillator (approximately 20t) 
surrounded by water is to efficiently 
veto events in the LXe TPC which have
additional energy depositions in the OD. 
These events are background to the WIMP
search. However, the OD could
be used for additional physics analyses on 
its own or together with the LXe. A
better understanding of possible 
background induced by muons
requires dedicated studies with the OD. 
The OD is also sensitive to
neutrino and weak signals from exotic
particles which are dark matter candidates. 

\tdrsubsec[MMN]{Muons and muon-induced neutrons}

A few potential physics topics to be 
addressed by the OD of LZ are linked 
to muons and muon-induced neutrons. 
The muon flux at the Davis
Campus at SURF has been calculated as
\SI{6.2e-5}{m^{-2}s^{-1}}, giving the 
muon event rate in the OD (water Cherenkov 
plus liquid organic scintillator) of 
about 300 per day.
Most of these events will be single 
muons with multiple muons contributing a small
fraction (<1\%). 

There will also be a few hundred of 
stopping muons detected in scintillator
and/or LXe in \SI{1000}{days} of running time.
Stopping muons can be identified
by a delayed signal from either muon decay or
absorption on a nucleus
LZ will be able to measure
the rate of stopping muon signals and the 
life-time of muons, although no separation
between positive and negative muons is 
possible, apart from detecting neutron capture
from absorption of negative muons. 

The high
probability of neutron detection in the OD
having a delayed coincidence with a muon 
signal allows the identification of a
negative muon absorption. 
Nuclear recoils in LXe will be delayed by a few
microseconds with respect to the muon 
ionization signal, providing 
a measurements of the
negative muon life time in xenon. 
These measurements require operation of
the OD and LXe in coincidence and an
independent trigger from the OD.
 
The detection of neutron capture events 
will also allow the measurements of neutron
yields from this specific process. 
The accurate interpretation of the results 
will require the
full Monte Carlo simulations of all 
processes involved including detector response.

Probably the most important measurements 
that can be carried out by the OD
(also in combination with LXe target) are
muon-induced neutrons. There
have been a number of measurements of neutron
production by muons including
muon-induced cascades, come carried out 
by the dark matter search experiments. \cite{Araujo:2008ze,Reichhart:2013xkd,Schmidt:2013gdc}

Neutrons are usually identified via their
capture on hydrogen or other elements (for
instance, Gd) in active veto systems 
containing a liquid organic scintillator. 
With a very large OD (water and scintillator),
LZ will detect many thousands of neutrons
within its expected \SI{1000}{days} of running time.
These neutron events will be efficiently
rejected in dark matter analysis using
various cuts but can be studied for the 
purpose of better understanding their
production, transport and detection. 
This is particularly important for designing
3rd generation dark matter experiments,
especially if a dark matter signal is
found, as well as for other rare event
searches.

In a conventional analysis, the 
muon trigger can be provided by either 
the LXe target or the OD system, and neutrons
can be produced in a number of materials
in the LZ setup (xenon, titanium, steel,
scintillator, water), moderated by 
hydrogen in scintillator or water and 
captured predominantly on Gd (or hydrogen) 
in scintillator.

In addition, with a large mass of LXe, 
we expect to have hundreds of events with NRs
without a muon in LXe allowing studies of 
NR rate, multiplicities and separation from
primary muons (also in delayed 
coincidences with neutron capture signals). 

\tdrsubsec[ODN]{Neutrinos}

The possible sources of neutrino signals 
in the LZ OD include solar neutrinos,
geoneutrinos, supernova bursts and neutrinos
from LBNF. In general the LZ OD is
not competitive with dedicated 
neutrino detectors due to its small mass. 
The supernova burst similar to the SN1987A
would give approximately 10 events in
the OD. The rate of geoneutrinos is
approximately \num{1} per year. Our estimates of
signals from the LBNF neutrinos give \num{9} 
events per year for the low energy
beam and 25 events per year for the 
medium energy beam. 

\tdrsubsec[OEX]{Exotic particles}
Models of the exotic candidates to be
the dark matter  include the ones 
where an excitation can happen in the LXe 
and deexcitation in the OD. 
The standard approach would
veto such events. A method the discover 
such interactions would be to measure the
spectrum of energy depositions in the OD as 
it is expected that it should be
monochromatic.  Another exotic 
candidate for DM is a fractionally 
charged particle. The OD could be sensitive 
to particles with charges down to a fraction 
of \num{0.025} of the elementary charge, and
coincidences with the LXe TPC provide an
interesting cross check.
Given the relatively large mass,
quietness of the detector,
and long exposure time the OD can
contribute important capability to
these searches.

\bibliographystyle{apsrev4-2}
\bibliography{LZtdr}

\wlog{In tdrinclude command}
\wlog{\ChapLabel}

\renewcommand{\ChapLabel}{chap:XDS} 
\renewcommand{\BibLabel}{XDSS:bib} 
\renewcommand{\SecPre}{XDS}
\renewcommand{\FloatPre}{XDS}

\tdrchap[XDS]{Xenon Detector System}

\tdrsec[Overview]{Overview}

The direct observation of WIMP scattering within a radiation detector is a significant experimental challenge. Searches for these elusive particles require an extremely sensitive, low-background instrument able to separate NR events at the few-keV energy from a dominant background of ER interactions, some created by particles external to the WIMP target and others arising within it from radioactive contaminants. The LZ experiment addresses part of the background issue by operating deep underground, and by surrounding the instrument with large ultra-pure water and liquid scintillator shields. However, observing these small energy depositions in space and in time in the central LXe target requires a highly instrumented double-phase (liquid/vapor) Time Projection Chamber (TPC) assembled from high-performance, low-radioactivity components operating in a cryogenic environment. An additional constraint is that the detector must operate as conveyed underground, which demands stringent quality requirements of its components.

While LXe is inherently a very radio-quiet detector material, with high enough density and atomic number to very effectively self-shield against external backgrounds, the design of this new detector requires nonetheless that attention be paid to the radiopurity of a number of significant detector elements, such as the PMTs, their voltage-divider bases, cabling, support structures, and reflecting surfaces. This imposes serious constraints on material composition and their location, adding significant complication to the design of the instrument. The details of these developments will be described in the related sections below.

The Xenon Detector System includes the LXe TPC and ancillary systems required for its readout, monitoring and control. An additional anti-coincidence detector is formed by a layer of LXe enveloping the TPC, which we term the ``Xe Skin'' detector. The main components of these two instruments are described in this chapter: the TPC, including HV delivery, PMT systems, and internal liquid flow and monitoring instrumentation; and the Skin detector and its readout. We describe at the end of this Chapter our multi-scale System Test effort which is working to validate the performance of many detector components before their assembly into LZ.

\begin{figure}[ht]
\centering
\includegraphics[width=1.0\textwidth]{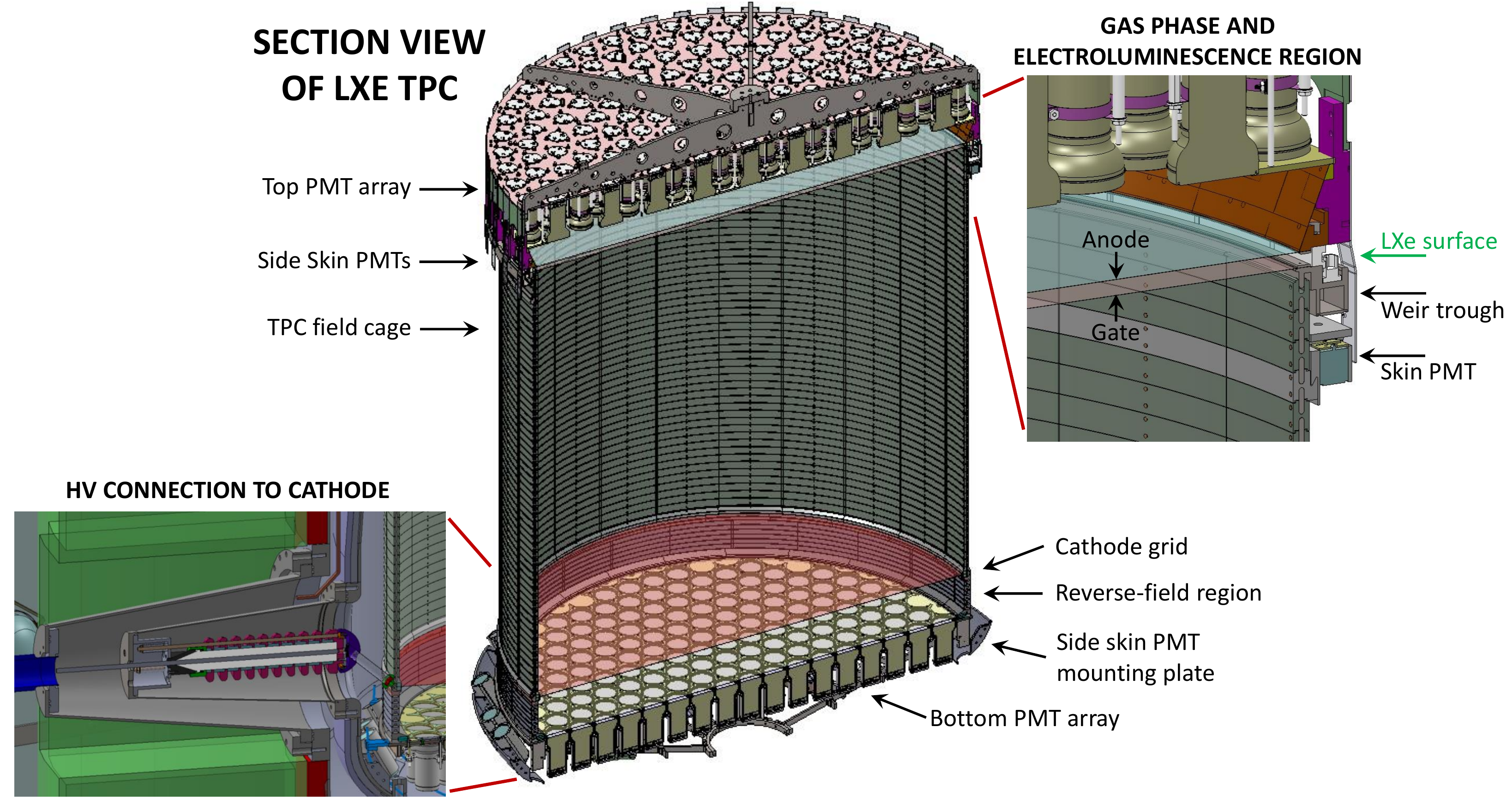}
\tdrfcaption[detector]{Schematic views of the Xenon Detector} {Schematic views of the Xenon Detector. The 7-tonne active region is contained within the TPC field cage between cathode and gate electrodes, viewed by ``top'' and ``bottom'' PMT arrays in the vapor and liquid phases, respectively. S2 signal generation occurs between the liquid surface and the anode (right inset). The HV connection to the cathode (left inset) uses a dedicated conduit leading from outside of the water tank. Below the TPC, a reverse-field region grades the cathode potential to low voltage at the bottom PMT array. The Side Skin PMT readout is shown outside of the TPC field cage.}
\end{figure} 

An overview of the xenon detector system is shown in Figure~\ref{XDSf:detector}. The LZ TPC diameter is determined by the maximum cryostat width that can be conveyed underground and the matching TPC length provides a compact form factor to minimize external radioactivity backgrounds. The TPC itself has a three-electrode configuration: a cathode grid at the bottom, a gate grid just below the liquid surface, and an anode grid just above the liquid surface. It features two arrays of PMTs, one immersed in the LXe viewing up (241~tubes), and the other in the gas phase viewing down (253~tubes). The WIMP target contains \SI{7}{\tonnesl} of active LXe, located vertically between the cathode and gate grids and enclosed laterally by a cylindrical arrangement of PTFE reflector panels. These embed a resistive electrical ladder which grades provides an approximately vertical electric field inside the TPC. Interactions in this region generate prompt vacuum ultraviolet (VUV) scintillation light detected by the PMTs (S1 pulse). The applied electric field sweeps the ionization charge liberated at the interaction site and drifts it upward to the liquid surface past the gate electrode; these electrons are extracted into the vapor phase, where they generate electroluminescence---which is again detected by the same two PMT arrays (S2 pulse). This double-phase TPC technique, which generates two optical pulses per interaction, resolves the energy deposition sites with great spatial accuracy down to very low energies, allowing identification of multiple scatter events and, as described in the previous Chapter, it provides discrimination between ER and NR interactions.

Table~\ref{XDSt:parameters} lists the key design parameters of the Xenon Detector System needed to meet the LZ scientific requirements. An important enhancement beyond LUX is the treatment of the Skin layer of LXe located between the PTFE-clad field cage and the cryostat inner wall, as well as the region beneath the bottom PMT array. A high-quality dielectric standoff is needed between the high electric field regions of the field cage and the grounded vessel wall. A few-cm-thick layer of LXe performs this role, with the added advantage of allowing measurement of any energy deposited in this layer, from which we read out the scintillation light. Operated as a stand-alone veto, this layer is too thin to have high efficiency. However, the combination of this Skin Detector and the liquid scintillator Outer Detector is highly efficient at tagging internal and external backgrounds. The efficiency is further enhanced by the overall minimization of inert materials that can absorb gamma rays and neutrons: The TPC is constructed of the minimum needed mass of PTFE and field-shaping rings, and the vessels and PMT support structures are made of titanium. Both PTFE and Ti have low density and atomic number, and are thus quite transparent to gamma rays. Important design drivers for the Skin are its optical decoupling from the TPC, and compatibility between the Skin readout and the TPC HV design.

\begin{table}[tbh]
\setlength{\extrarowheight}{3pt}
\tdrtcaption[parameters]{Xenon detector parameters}{Main parameters of the Xenon Detector System.}
\centering
\sffamily
\begin{tabular} {|lr|}
\hline
\rowcolor{mrocol}
{\bfseries Parameter} & {\bfseries Value} \\
\hline
\rowcolor{lrocol}
\multicolumn{2}{|l|}{Liquid xenon}\\
\hline
TPC active mass				& \SI{7000}{\kg} \\
Skin mass (side+dome)		& \SI{2000}{\kg} \\
Total mass within cryostat	& \SI{9600}{\kg} \\
\hline
\rowcolor{lrocol}
\multicolumn{2}{|l|}{Photomultipliers} \\
\hline
TPC (Hamamatsu R11410-22)	& \num{253} (top) + \num{241} (bottom) \\
Top Skin (Hamamatsu R8520-406)	& \num{93} \\
Bottom Skin (Hamamatsu R8778)	& \num{20} (side) + \num{18} (dome)  \\
\hline
\rowcolor{lrocol}
\multicolumn{2}{|l|}{Vertical dimensions (cold)} \\
\hline
Electroluminescence region (gate-anode) & \SI{13}{\mm} (\SI{8}{\mm} gas)\\
Drift region (cathode-gate) & \SI{1456}{\mm} \\
Reverse-field region (sub-cathode) & \SI{137.5}{\mm} \\
\hline
\rowcolor{lrocol}
\multicolumn{2}{|l|}{Transverse dimensions (cold)}\\
\hline
TPC inner diameter & \SI{1456}{\mm} \\
Field cage  thickness & \SI{15}{\mm} \\
Skin thickness at surface (at cathode) & \num{40} (\num{80}) \si{\mm} \\
\hline
\rowcolor{lrocol}
\multicolumn{2}{|l|}{Electric fields} \\
\hline
Electroluminescence field (GXe) & \SI{10.2}{\kV\per\cm} \\
Drift field baseline (goal) & \num{0.31} (\num{0.65}) \si{\kV\per\cm} \\
Reverse field baseline (goal) & \num{2.9} (\num{5.9}) \si{\kV\per\cm} \\
Drift (reverse-field) stages & \num{57} (\num{7}) \\
\hline
\rowcolor{lrocol}
\multicolumn{2}{|l|}{Operating conditions} \\
\hline
Operating pressure (range) & \num{1.8} (\numrange{1.6}{2.2}) bar(a) \\
Equilibrium temperature & \SI{175.8}{\K} (\SI{-97.4}{\degree C}) \\
\hline
\end{tabular}
\end{table}

Another area of major difference between the device described here and the previous LUX and ZEPLIN detectors is the side-entry method to deliver the high-voltage connection to the cathode, and the relatively short ``reverse-field'' region (RFR) between the cathode and the lower PMT array. The RFR is especially challenging because of the very high electric field there, which results from balancing a high cathode voltage and minimizing the mass of S2-inactive LXe below the cathode. Our approach to these issues is described below in separate sections on the reverse-field region and cathode HV delivery system.

By design, the structures surrounding the central LXe volume are as lightweight as possible for transparency to gamma rays and neutrons; this also helps to keep their total radioactivity low. The most challenging requirements on the intrinsic radioactivity (i.e., radioactivity per mass or area) are in the largest or most massive components---the PTFE walls and field-shaping rings, and the PMTs with their bases and cables. This Chapter discusses the approach to obtaining the needed radioactivity levels for a number of these major items. However, the absolute level of radioactivity surrounding the detector must be held at acceptable levels, so all components must be carefully selected and screened. The screening program that ensures this is discussed in Chapter~\ref{chap:MAS}.

\tdrsec[TPC]{The Liquid Xenon Time Projection Chamber}

At the heart of the LXe TPC are the field cage embedded in the reflective PTFE panels and the various electrode grids. The grids and field cage create the set of electric fields that drift the electrons to create the S2 signal, and the highly reflective PTFE panels are essential for efficient measurement of the initial S1 scintillation signal. In this section we describe the electrostatic design of the TPC, and then highlight optical and thermal considerations that shape this design.

The electric field configuration inside the TPC volume defines three distinct regions: (1) the drift region, (2) the reverse-field region, and (3) the electroluminescence region. We describe these in turn, with help from Figure~\ref{XDSf:detector} which overviews the detector geometry, and Figure~\ref{XDSf:TPCFields} which illustrates the field distributions.

\begin{figure}[ht]
\centering
\includegraphics[width=0.95\textwidth]{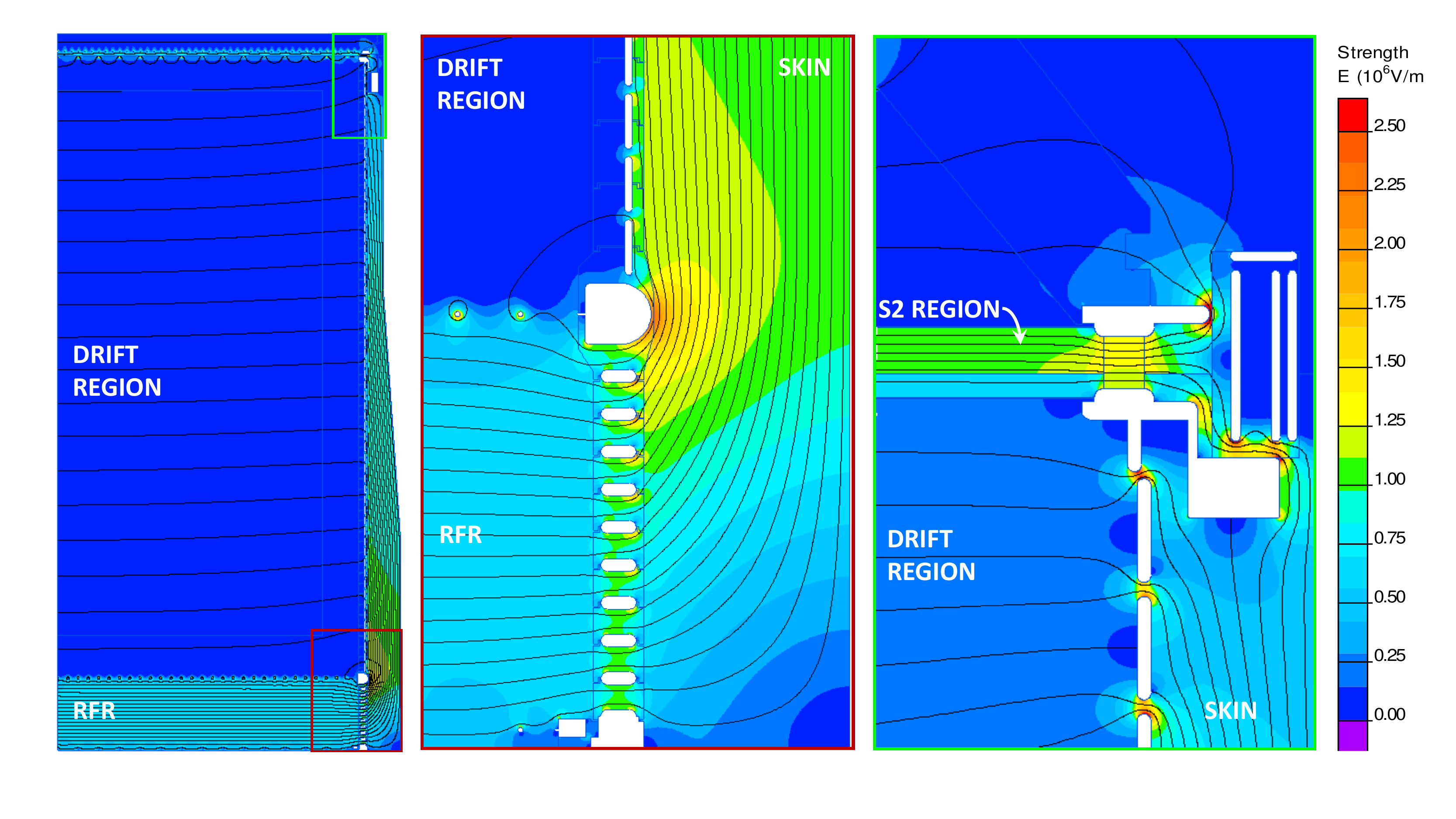}
\tdrfcaption[TPCFields]{TPC electric field regions} {Electrostatic modeling of the various electric field regions of the TPC; applied voltages are design goals rather than nominal in this figure (cathode: \SI{-100}{\kV}, gate: \SI{-7}{\kV}, anode: \SI[retain-explicit-plus]{+7}{\kV}); the field on cathodic surfaces in the liquid phase is kept below the \SI{50}{\kV\per\cm} allowable field. Left: General cross-section showing the drift field in the active region, the side Skin, the reverse-field region below the cathode, and the electroluminescence (S2) region at the top. Center: Detail of the drift and reverse-field regions with field-shaping structures embedded in the PTFE walls (expanding region in red in the left panel). Right: Detail of the electroluminescence region in the TPC top corner (in green in the left panel). In all panels the right-most surface is the grounded inner cryostat wall.}
\end{figure} 

\tdrsubsec[DriftRegion]{Drift Region}

The region between the cathode and gate contains the active volume (WIMP target) and is therefore the most important region of the detector. This is where electrons are removed from the particle interaction site and drifted up to be detected in the vapor layer above the surface. Hence, the electric field strength and uniformity in this region have a major impact on the ability to identify and locate interactions in the detector. It is important that the electric field in this region is as uniform as possible, with field lines parallel to the PTFE reflector walls, so that S2 pulses can be generated for all interactions in this region. The electric field requirement for LZ is \SI{0.3}{\kV\per\cm}, with a design goal of \SI{0.6}{\kV\per\cm}. This range will provide the required level of ER/NR discrimination of \SI{99.5}{\percent} or better.

An additional requirement in our design is that all cathodic surfaces immersed in the LXe must comply with a maximum allowable field of \SI{50}{\kV\per\cm}. As described in the LZ Conceptual Design Report (Section~3.4.2 in~\cite{Akerib:2015cja}], is can be problematic to sustain higher fields in particular on practical stainless steel surfaces.

To produce a uniform electric field between the cathode and the gate electrodes a set of \num{57}~equally-spaced field rings is embedded within the PTFE and connected by pairs of \SI{2}{\giga\Ohm} high-voltage resistors. The rings will be made from titanium from the same source as that used for the cryostat. The details of the design can be seen in Figure~\ref{XDSf:TPCFields}. The rings are ``I''-shaped to help maintain the uniform field pattern needed within the TPC region by keeping the equipotential surfaces nearly normal to the inner surface of the PTFE ring. The field-shaping rings are embedded in vertically-segmented rings of PTFE that have been precision machined and then assembled in a stack to produce the completed field cage. In areas where the large difference in thermal contraction between PTFE and the metal field-shaping rings would cause problems, the PTFE rings are additionally segmented in the circumferential direction to allow the PTFE segments to slide along the field rings when the detector is cooled. The field cage structure will be mounted to the lower reverse-field region and lower PMT support, which in turn is supported from the bottom of the cryostat inner vessel.

All electrodes consist of woven meshes made from thin stainless steel wires---these are described in detail in Section~\ref{XDSSs:GridFab}. The cathode is at the highest potential and the design of its holding ring must be optimized to avoid high-field regions (see Figure~\ref{XDSf:TPCFields} center).

The issue of field uniformity is important. In LUX it was observed that field lines near the lateral edges of the TPC, particularly near the top and bottom, are not fully parallel to the PTFE surfaces. We have come to understand this as being intrinsic to its design: The overall fields resulting from the grids and field cage structure were designed using 2-D electrostatics calculations that treated the grids as continuous conducting sheets. It is well known~\cite{blum2008particle,grids2003} that the 3-D stretched-wire grids have an electrostatic “transparency” such that the bulk electric fields are somewhat (\SI{\sim10}{\percent}) different to the values calculated assuming the grids are conducting planes, and this effect was taken into account in establishing the LZ operating fields. A subtler additional effect happens at the top and bottom of the TPC cylinder, where the transparency of the grids causes some bleed-through of the concentrated fields that terminate on the vessel and other grounded structures just outside the main part of the TPC. A more complete calculation using transparent grids reproduces the observed pattern in LUX. Such an effect was in fact previously observed in XENON100 and understood as described above~\cite{Mei:2011}.

\begin{figure}[ht]
\centering
\includegraphics[width=0.97\textwidth]{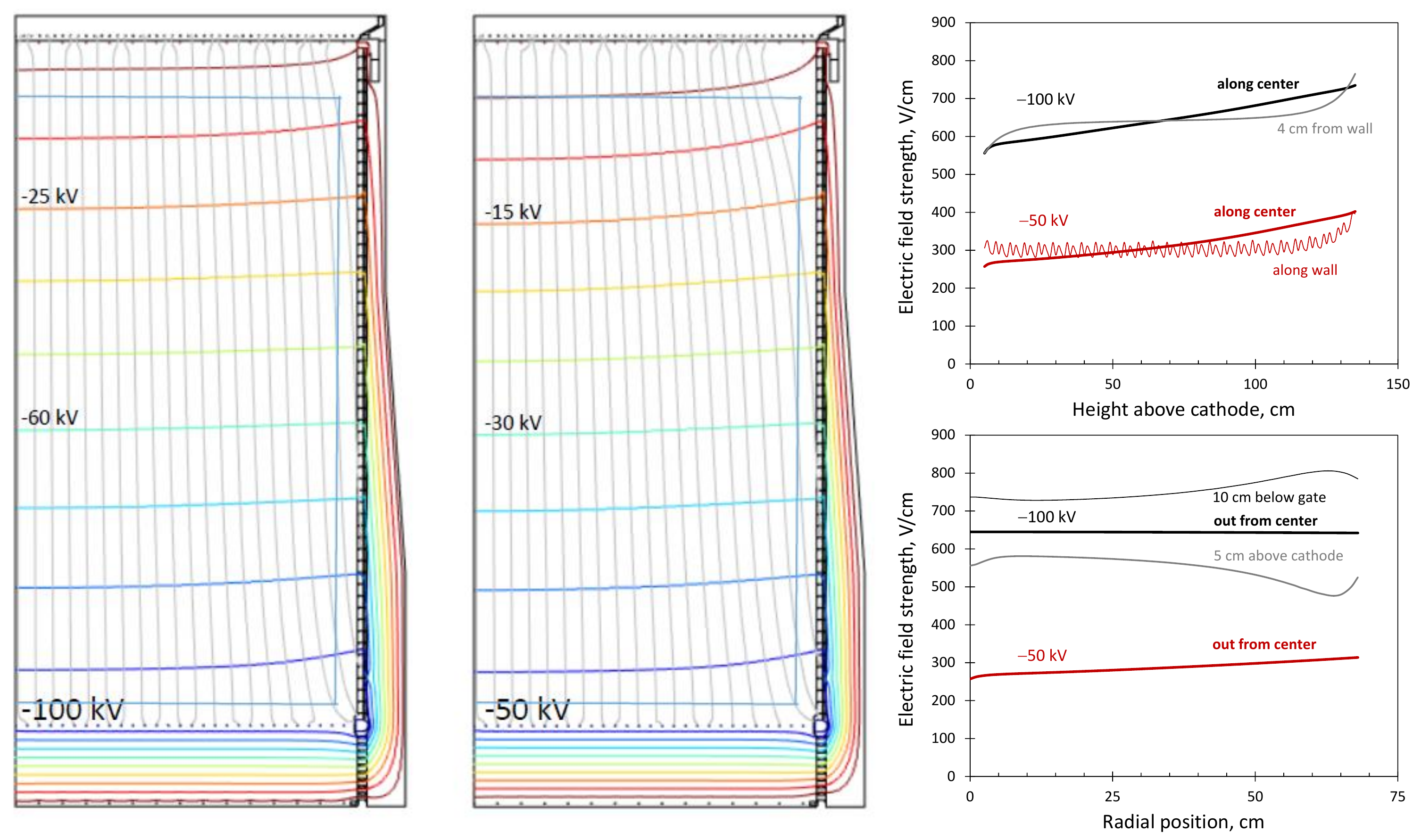}
\tdrfcaption[fieldnonuni]{TPC field non-uniformity} {TPC field uniformity. Left and center: Equipotential lines and electron trajectories for \SI{-100}{\kV} and \SI{-50}{\kV} cathode HV. In both cases a preliminary \num{5.6}-\si{\tonnel} fiducial volume is shown which extends from \SI{5}{\cm} above the cathode to \SI{10}{\cm} below the gate, out to \SI{4}{\cm} from the TPC tall. Right, top: Field magnitude vertically along the center of the fiducial volume as well as along the TPC walls; Right, bottom: Field strength horizontally out from the center at the bottom, middle and top of the fiducial volume.}
\end{figure} 

In LUX, this effect caused electrons at the bottom edge of the detector to deflect \SIrange{2}{3}{\cm} inward as they followed distorted field lines. This did not pose a fundamental problem for the science data, since the effect could be readily corrected for in analysis. Nonetheless, we will seek to better control the fields in LZ. Based on preliminary electrostatic calculations, we believe that this effect can be mitigated, for example by adjusting the values of the last few resistors at the top and bottom of the field cage (this is not assumed in this report). Another design change over LUX is the vertically-segmented design of the PTFE field cage walls. The essentially uninterrupted PTFE surfaces of the field cage are necessary for good light collection, but not ideal from the point of view of good high-voltage design practice, because insulating surfaces can at least in principle accumulate charge that distorts fields. LUX was constructed from single, continuous slabs of PTFE in the vertical direction, whereas the \num{2.5}-\si{\cm}-tall segments in LZ provide much shorter paths to the conducting field rings from any location on the PTFE walls.

\tdrsubsec[ReverseFieldRegion]{Reverse-Field Region} 

Located between the cathode grid and the bottom PMT shield grid, the RFR is a significant challenge for constructing the LZ TPC because of the high fields involved (\SIrange{3}{6}{\kV\per\cm}). The cathode voltage must be graded to near-ground while keeping all surfaces in this region below the \SI{50}{\kV\per\cm} allowable field. At the same time, this space must be kept as small as possible, both to reduce the amount of expensive LXe in this region, and to reduce the rate of events that scatter in both the reverse-field region and the active volume of the TPC. Such events are a class of background that can mimic WIMP signals; however, they have an acceptably low rate for the design presented here. In the LUX detector, due to the much lower cathode voltages and the shorter drift region in the TPC, this was handled with a \SI{4}{\cm} spacing and no field grading between the cathode and PMT shield grids, along with a near-zero field region of \SI{2}{\cm} between the shield and the PMT front surfaces. For the LZ configuration, we have chosen a voltage-grading structure similar to that in the drift region. This better defines the fields, and is a more robust approach to the more challenging LZ voltage requirements.

\begin{figure}[ht]
\centering
\includegraphics[height=5.5cm]{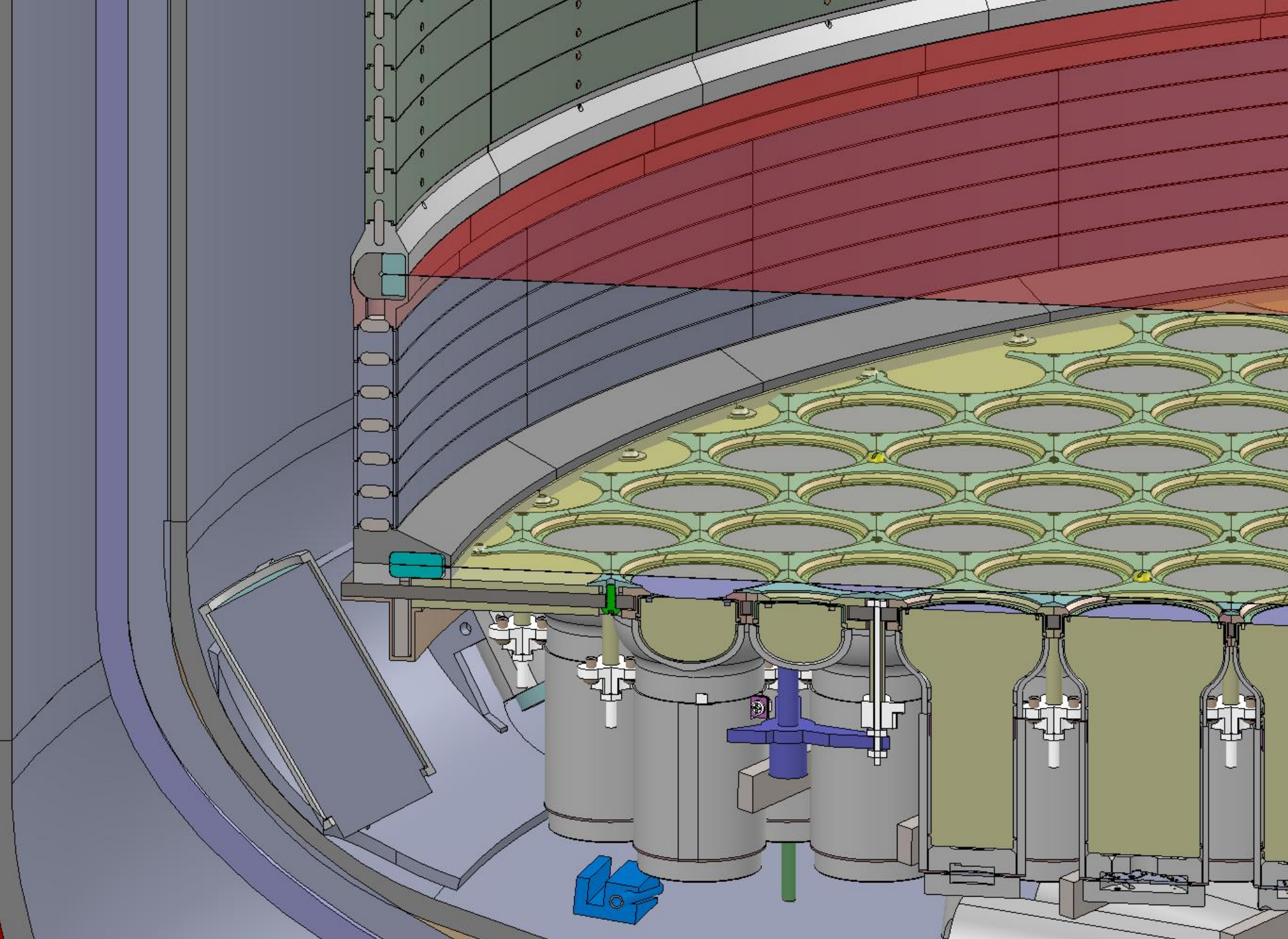}\quad
\includegraphics[height=5.5cm]{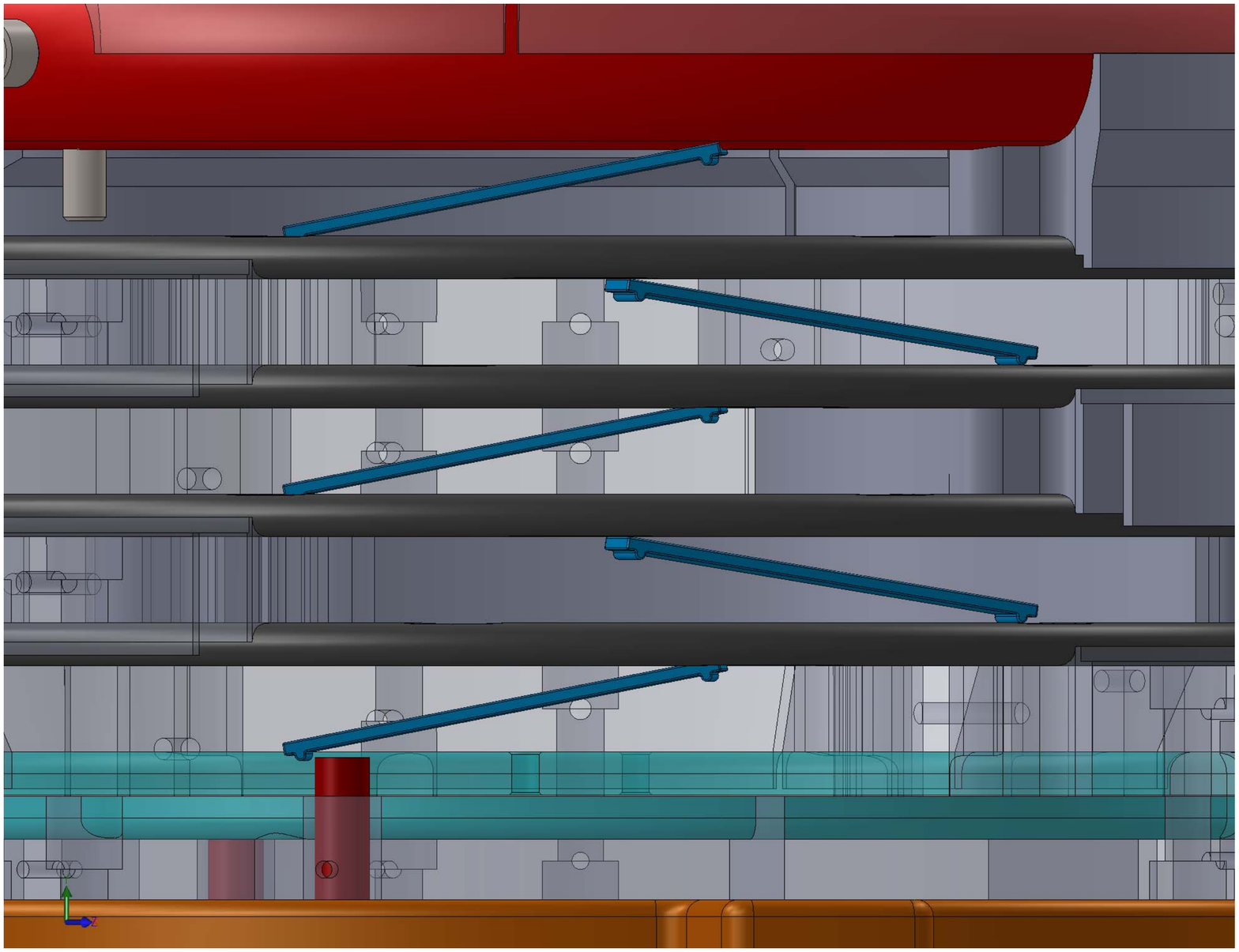}\quad
\tdrfcaption[RFR]{TPC reverse-field region} {Reverse-field region design. Left: The RFR, with the cathode and bottom shield grids visible, and the oval field-shaping rings used here. Right: Placement of the RFR field-grading resistors, which are embedded inside the PTFE ring structure and attached to successive field-shaping rings.}
\end{figure}

The RFR design, shown in Figure~\ref{XDSf:RFR} (left), is composed of a stack of eight PTFE ``rings'', each \SI{\approx1.5}{\cm} high and embedded with titanium field-shaping rings. These conducting rings are optimized for field uniformity above the PMTs---to control the electrical environment near their input optics---while keeping the region between the TPC and the grounded cryostat below \SI{50}{\kV\per\cm}. The central panel in Figure~\ref{XDSf:TPCFields} shows the fields in this region. The smooth oval shape of the RFR rings, compared with the ``I''-shape in the drift region, creates lower surface fields on these elements at the expense of a less-uniform field in the central LXe region. This is not problematic since the uniformity requirement is not strong in the reverse-field region. 

The voltages between each of the field rings are graded down from the cathode potential using a set of series resistors, similar to those used in the drift region, but with a higher resistance value between each ring to accomplish this stronger field grading. The resistors in the reverse-field region are more challenging for radioactivity than those in the drift region because they are larger. The main radioactive challenge in electronic components is alumina ceramics, which in all standard (non-``synthetic'') forms is very high in both gamma activity and neutron yield. We use standard surface-mount resistors with the exception of getting the manufacturer to make them with a thin protective glass coating instead of potting them in epoxy. The resistors have the smallest available ceramic mass for the required voltage rating. Figure~\ref{XDSf:RFR} (right) shows the current design and location of these grading resistors inside the PTFE spacers. The lowest PTFE ``ring'' will be attached to the top of the lower PMT shield grid and this grid will be spaced approximately \SI{2}{\cm} above the PMT surfaces, also using a PTFE spacer ring. The entire assembly will in turn be attached to the lower PMT support structure, which will then be fixed to the cryostat for final mechanical support. Electrostatic and mechanical studies of this region have been carried out, and are part of the System Test program described at the end of this Chapter.

\tdrsubsec[ElumRegion]{Electroluminescence Region} 

In the region above the gate grid, drifting electrons are emitted across the liquid surface and produce electroluminescence (S2) photons while traveling through the gaseous phase on their way to the anode. The separation between these electrodes is merely \SI{13}{\mm}, since only a narrow gas gap (\SI{8}{\mm}) is required to provide enough S2 gain. Fields here are necessarily high (\SI{\approx10}{\kV\per\cm} in the gas phase) demanded by both the cross-phase electron emission process and the S2 photon production. The field in the liquid above the gate is approximately half that in the gas above the surface due to the relative permittivity of the liquid phase ($\epsilon_r$ = 1.96).

The optimization of the grids to create the S2 signal requires care, as is discussed in detail in Section~\ref{XDSS:S2Light}; the mechanics of having gate and anode grids with very low deflection from the large electric fields is challenging; both the Skin PMTs (Section~\ref{XDSS:Skin}) and the weir structure (Section~\ref{XDSS:Fluids}) must be accommodated in a tight space; and the overall structure must maintain a very low level of distortion in the rings supporting those grids, so that a parallel arrangement of electrodes and liquid surface can be obtained (tip-tilt adjustment of the detector to assure parallelism of grids and liquid surface is discussed in Chapter~\ref{chap:CRO}). A close-up view of this region is shown in Figure~\ref{XDSf:elumregion}.

\begin{figure}[ht]
\centering
\includegraphics[width=0.7\textwidth]{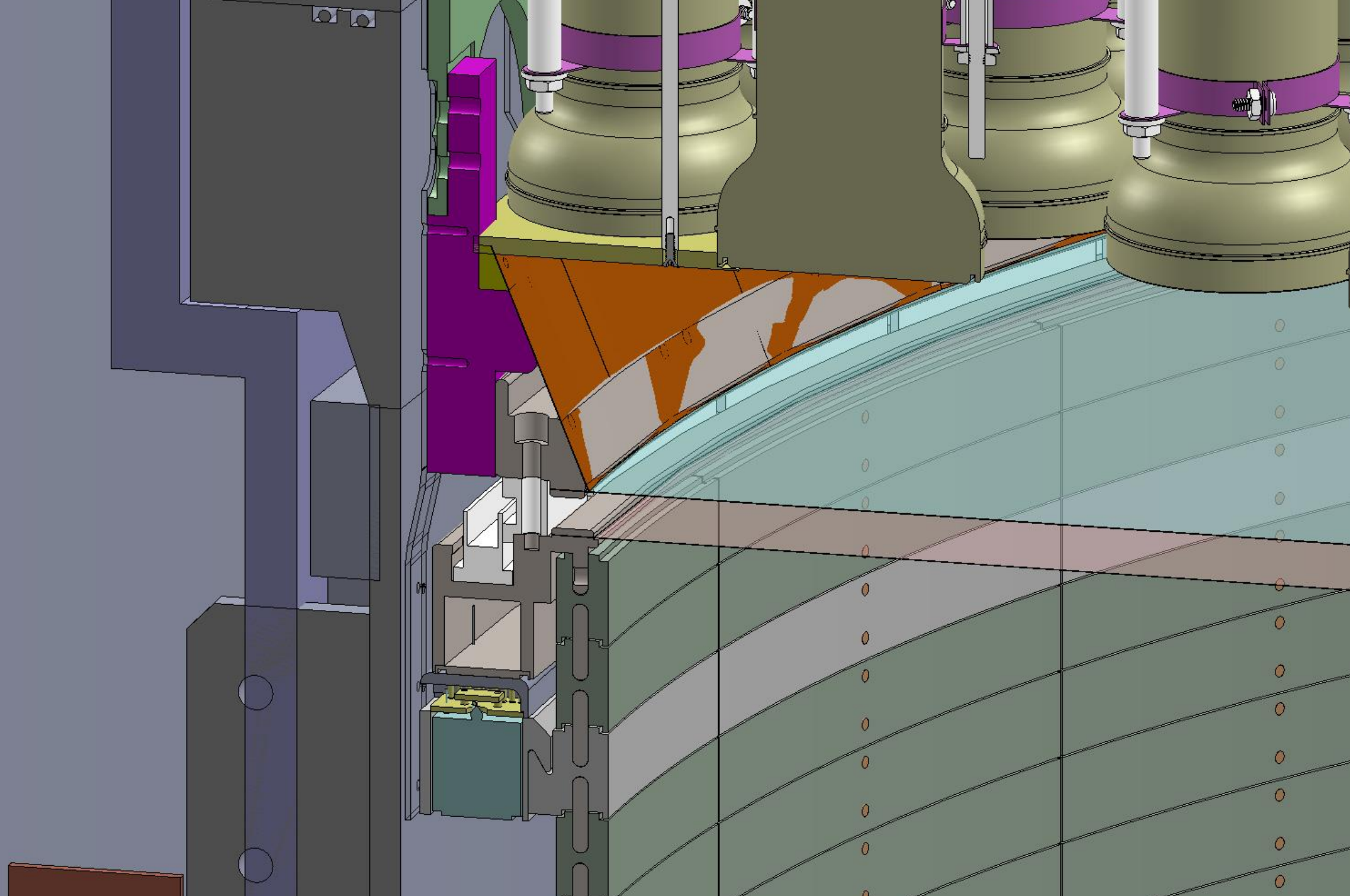}
\tdrfcaption[elumregion]{Electroluminescence region}{The electroluminescence region, with the gate and anode grids shown, along with the weir and top Skin PMTs.}
\end{figure} 

The field in the liquid above the gate must be significantly stronger than the field in the drift region: a \num{\approx5}-\si{\kV\per\cm} ``extraction'' field is needed below the liquid surface in order to give most electrons sufficient kinetic energy to overcome the energy barrier at the liquid surface (\SI{0.61}{\eV}) and be extracted into the gas phase with high probability~\cite{russians1979}. Once electrons enter the gas phase, where the field is approximately twice as strong, they are accelerated and produce electroluminescence photons in the \num{8}-\si{\mm} drift distance until they are collected on the anode grid. The photon yield is \num{\approx800}~photons per emitted electron at \num{1.8}-bar operating pressure, with \SI{10}{\kV\per\cm} in the gas~\cite{Fonseca:2004cd}. For these operating conditions, the electron transit time to the anode is \SI{\approx1.2}{\mus} which, along with diffusion while the electrons travel in the drift region, determines the width of the S2 pulse.

Because the S2 signal develops as the electrons drift from the liquid surface to the anode electrode, it is essential to minimize the variance of S2 photon production for different electron emission points, as this relates directly to the energy resolution achieved in the S2 channel. This imposes requirements on grid deflection as well as micro-uniformity of electric fields. The intricacies of this design are described in Section~\ref{XDSS:S2Light}. Additional constraints come from the need to attach other elements to the side of this region of high field, such as the weirs that control the liquid level and the side Skin PMTs---this can be appreciated from Figure~\ref{XDSf:TPCFields} (right). The ``top corner'' of the TPC is a crowded region and required significant design effort.

\tdrsubsec[Reflectors]{VUV Reflectors}

In the same way that the electrostatic design of the TPC optimizes the detection of ionization electrons extracted from particle interactions, the optical design must accomplish the harder task of maximizing the detection of scintillation photons. Although the numbers of each quanta produced initially are comparable, it is easier to detect electrons efficiently than it is to detect photons. The energy threshold of the TPC is therefore determined primarily by its ability to detect prompt scintillation (S1).

PTFE is the VUV reflector of choice for this purpose, possessing good mechanical properties, low outgassing, and---crucially---achieving hemispherical reflectances in the liquid xenon as high as \SI{\approx97}{\percent} at the xenon scintillation wavelength. The experience of the LUX experiment in optimizing light collection within the TPC volume was very successful, translating to a very low NR detection threshold of below \SI{4}{\keV}. We plan to use machined ``segments'' of high-purity PTFE approximately \SI{1.5}{\cm} thick and \SI{2.5}{\cm} tall to form the inner reflecting surface in the TPC region as well as the outer reflecting surface between the TPC and the cryostat wall, which itself will have a one-mm-thick segmented lining of PTFE.

There are many types of PTFE exhibiting different reflectivity both in the liquid and gas phases~\cite{Silva:2009a,Silva:2009ir}. It is essential that the best material is selected for this purpose since this is a leading parameter that determines the performance of both the TPC and the Skin detectors. The topic of S1 photon detection is discussed in some detail in Section~\ref{XDSS:S1Light}; the PTFE reflectivity is such a critical parameter that we devote part of our System Test program to this topic, as described in Section~\ref{XDSS:SystemTest}.

The radioactivity of this material must be held extremely low both because of direct gamma-ray production but, more importantly, neutron production from (\Pga,\Pn) reactions on F from \Pga decays in the U and Th chains. The raw precursor material for structural PTFE is a powder form produced by only a few suppliers, and is expected to be extremely radio-pure because it is synthesized from gas. A large number of smaller manufacturers produce structural shapes from these powders, and the final material can be very low in radioactivity if there is sufficient care in controlling contamination (e.g., from dust) in this second manufacturing step. In this topic we benefited from the experience of the EXO-200 collaboration, who achieved excellent radiopurity, rendering this critical material sub-dominant in terms of background in LZ, as detailed in Chapter~\ref{chap:MAS}.

\tdrsubsec[Thermal]{Thermal Considerations}

Given that the inner detector region is primarily made from PTFE, stainless steel and titanium parts, attention must be paid to the differential thermal contraction as the detector is cooled to LXe temperatures. The PTFE that makes up the majority of the surface area of the TPC is expected to shrink by \SI{\approx1.5}{\percent} linearly~\cite{White:2002,DuPontPTFE}, or about \SI{7}{\cm} in circumference and \SI{1}{\~cm} in TPC radius when cooled from room temperature to \SI{\approx170}{\K}. Stainless steel, by contrast, contracts only \SI{\approx0.2}{\percent} over the same temperature range, and titanium even less. Where this difference would result in destructive forces in the assembly, the metallic field cage rings will be constructed as solid parts, while the PTFE parts are segmented both horizontally (i.e., into rings) and vertically, so that each ring is itself composed of several segments. These segments contract and slide circumferentially along the solid metal field cage rings. In this way, the overall diameter of the TPC is determined by the metal field cage rings, and thus undergoes a relatively small thermal contraction. As the PTFE shrinks, the seams between the segments open, but the design has overlaps so there will continue to be a reflecting surface in the exposed gaps. In the vertical direction, the dimension of the field cage is determined by a combination of PTFE panels and metal rings. There is an overall height (top PMT array to bottom PMT array) contraction of \SI{\approx1}{\cm}. To minimize the movement in the critical region where the HV connection to the cathode is made (discussed in Section~\ref{XDSS:Cathode}), the entire TPC assembly will be supported from the bottom PMT array, which will be connected to the cryostat vessel. This means that the top PMT array will move downward during cool-down, increasing the clearance in the top dome.

\tdrsec[Cathode]{Cathode HV Delivery System}

\tdrsubsec[CatHVReq]{Cathode HV Requirements}

The cathode HV is a key performance parameter that will directly affect the science reach of the instrument because of its impact on ER rejection and other factors. Table~\ref{XDSt:CathodeHV} summarizes the HV requirement and some detector parameters it determines. The LZ nominal operating voltage is \SI{-50}{\kV}, which establishes a drift field of \SI{\approx300}{\V\per\cm} and allows LZ to meet its baseline sensitivity. The design goal is \SI{-100}{\kV}---the maximum operating voltage for the system. Introduction of HV into the Xe space is challenging because of possible charge buildup and sparking, and since high-field regions can produce spurious electroluminescence that blinds the detector to the minute flashes of light produced by WIMP interactions.

\begin{table}[tbh]
\setlength{\extrarowheight}{3pt}
\tdrtcaption[CathodeHV]{Dependence of TPC parameters on cathode HV}{Dependence of TPC parameters on cathode HV.}
\centering
\sffamily
\begin{tabular} {|lcccl|}
\hline
\rowcolor{mrocol}
Parameter	& \SI{-30}{\kV}	& \SI{-50}{\kV}	& \SI{-100}{\kV} & Comments \\
\rowcolor{mrocol}
& (LUX)		& (Base)	& (Goal)	& \\
\hline
TPC drift field, \si{\kV\per\cm}	& \num{0.17} & \num{0.31} & \num{0.65} & Gate \SI{-5.5}{\kV} \\
\hline
ER/NR discrimination	& \SI{99.6}{\%} & \SI{99.7}{\%} & \SI{99.7}{\%} & NEST LZ04 \\
\hline
Electron drift velocity, \si{\mm\per\mus}& \num{1.5} & \num{1.8} & \num{2.2} & \cite{Miller:1968} \\
Maximum drift time, \si{\mus}& \num{970} & \num{806} & \num{665} & Interactions at cathode \\
\hline
Longitudinal diffusion, \si{\mus}	& \num{2.4} & \num{2.2} & \num{2.0} & FWHM, cathode events\\
Transverse diffusion, \si{\mm} 		& \num{2.4} & \num{1.8} & \num{1.4} & FWHM, cathode events\\
\hline
Gate wire field, \si{\kV\per\cm}			& \num{-64} & \num{-62} & \num{-58} & \\
Cathode wire field,	\si{\kV\per\cm}		& \num{-18} & \num{-31} & \num{-63} & \\
\hline
\end{tabular}
\end{table}

The LZ operating and design voltages were determined through a combination of task-force activity, evaluation of WIMP sensitivity, and project cost and risk. At the nominal voltage, an ER rejection efficiency of \SI{99.5}{\percent} is expected at \SI{50}{\percent} NR acceptance, as demonstrated in previous double-phase Xe detectors and modeled through the Noble Element Simulation Technique (NEST) simulation package. The LZ cathode is required to operate up to \SI{-50}{\kV}; in addition to allowing adequate discrimination power, this voltage further minimizes drift time, event pileup, charge cloud diffusion, and optimized energy resolution. All subsystems in LZ will be designed to withstand higher voltages to help ensure that a \SI{-50}{\kV} operational voltage can be met. High voltage tests of LZ prototypes and final components will also be done to ensure this voltage capability. The planned design safety factors (DSFs) and test safety factors (TSFs) are summarized in Table~\ref{XDSt:SafetyFactors}.

\begin{table}[tbh]
\setlength{\extrarowheight}{3pt}
\tdrtcaption[SafetyFactors]{Safety factors}{Design safety factors (DSFs) and test safety factors (TSFs) for the cathode high voltage delivery system. The DSFs and TSFs are defined as percentages above \SI{-50}{\kV} (e.g.~a \SI{100}{\percent} safety factor implies a voltage of \SI{-100}{\kV}).}
\centering
\sffamily
\begin{tabular} {|lcc|}
\hline
\rowcolor{mrocol}
{\bfseries Item} & {\bfseries Design Safety Factor} & {\bfseries Test Safety Factor} \\
\hline
Cathode power supply & \SI{140}{\%}	& \SI{140}{\%} \\
Warm feedthrough & \SI{200}{\%} & \SI{200}{\%} \\
Cable & \SI{200}{\%} & \SI{200}{\%} \\
Grading region & \SI{100}{\%}	& \SI{140}{\%} \\
Cathode connection region & \SI{100}{\%} & \SI{140}{\%} \\
\hline
\end{tabular}
\end{table}

\tdrsubsec[CatOverview]{HV System Overview}

A schematic overview of the cathode HV system is shown in Figure~\ref{XDSf:cathode}. The baseline LZ design places the cathode HV feedthrough (from air into Xe space) outside the shield at room temperature, at the end of a long vacuum-insulated, Xe-filled umbilical. The cathode high voltage cable has been chosen to be Dielectric Sciences model SK160318. It is rated to 150 kV, and has a conductive polyethylene core, polyethylene insulator, and conductive polyethylene sheath. With the dominant cable material being polyethylene, Rn emanation is minimized. Polyethylene is also known to be a safe material in LXe, mitigating concerns about emanation of electronegative contaminants. With the feedthrough at room temperature and far away from the active LXe, there are no concerns of thermal contraction compromising a leak-tight seal to the Xe space, and no concerns about feedthrough radioactivity. A feedthrough at the warm end of the umbilical allows a commercial polyethylene-insulated cable to pass from a commercial power supply, through a double O-ring seal, and into the gaseous Xe. The cable then travels through the center of the umbilical and routes the HV through LXe and to a field-graded connection to the cathode.

\begin{figure}[ht]
\centering
\includegraphics[width=0.5\textwidth]{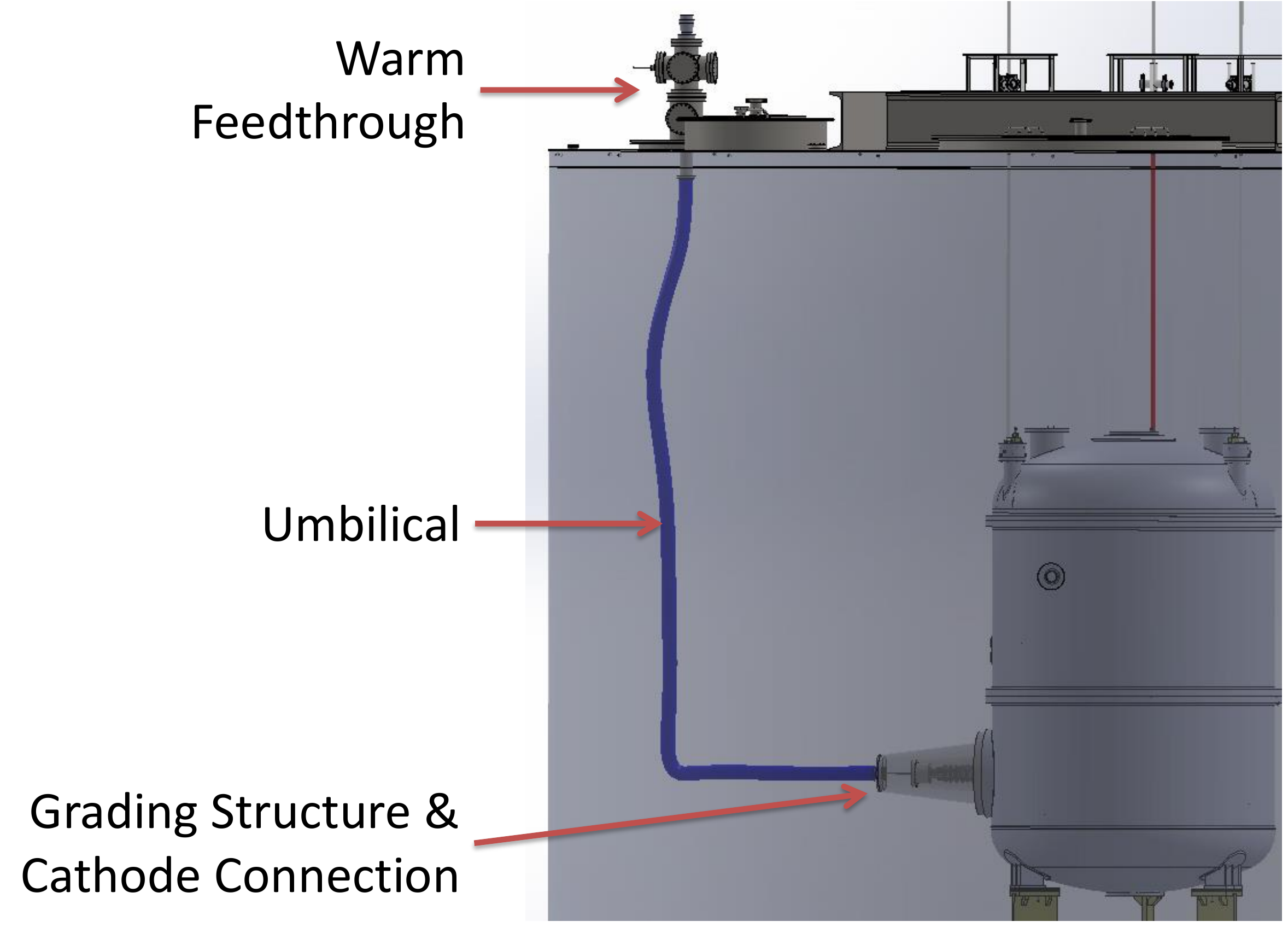}
\tdrfcaption[cathode]{Cathode HV delivery}{Schematic overview of the cathode HV delivery system.}
\end{figure} 

\tdrsubsec[CatSupply]{Cathode Supply and Cable Connection}

The cathode grid power supply (Spellman SL120N10) is rated at \SI{-120}{\kV} and is limited to a maximum current of \SI{100}{\micro\ampere}. This power supply has an adjustable current trip feature which can be used to shut the power supply off in the event of an over-current condition such as HV breakdown to protect the load. The output cable leads directly into the warm feedthrough, and from there into the xenon space and cathode connection region. 

\tdrsubsec[CatFT]{Cathode Feedthrough} 

The warm cathode HV feedthrough, shown in Figure~\ref{XDSf:cathodeFT}, allows high voltage to be passed from outside the LZ detector into the top of the Xe-filled umbilical space. The HV cable passes continuously through the feedthrough, with electric field confined to the cable insulation. The feedthrough includes 2 O-rings that seal to the outside of the conducting outer sheath of the HV cable.  Between the O-rings is a vacuum space, maintained with a turbopump and monitored with a residual gas analyzer.  The first O-ring allows the cable to pass between air and vacuum, and the second O-ring allows the cable to pass from vacuum to xenon gas in the umbilical. The cable is supported mechanically, both in air before the first O-ring, and in the xenon gas space after the second O-ring. 

\begin{figure}[ht]
\centering
\includegraphics[width=0.5\textwidth]{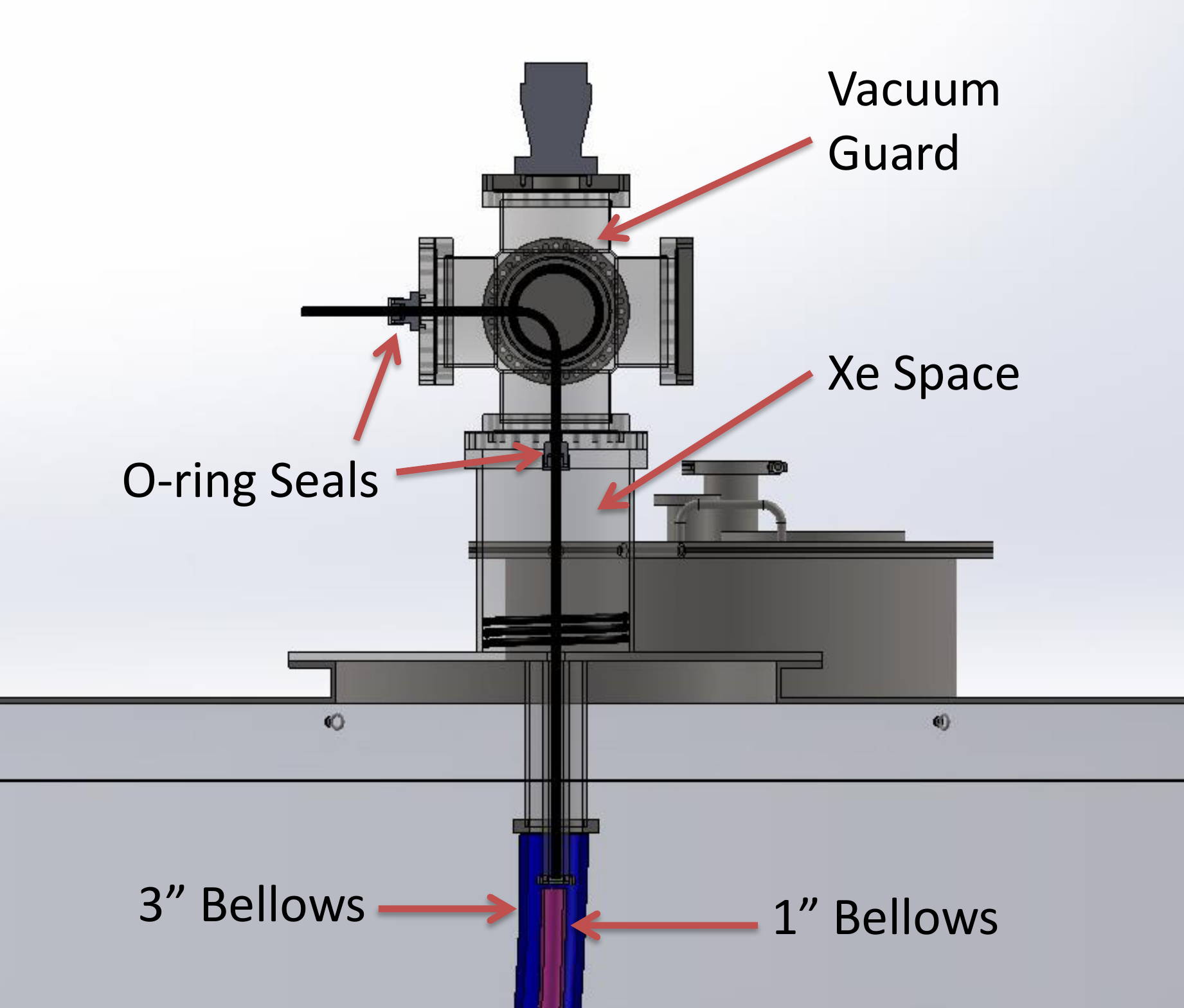}
\tdrfcaption[cathodeFT]{Cathode feedthrough} {Warm feedthrough detail for the HV connection to the cathode.}
\end{figure} 

\tdrsubsec[CatUmb]{Cathode HV Umbilical}

The cathode HV umbilical is designed to carry the Dielectric Sciences HV cable from the warm feedthrough to the cathode of the detector. 
The umbilical is a nested pair of tubes that protrude from the side of the detector at about the height of the cathode. The tubes include formed bellows for flexibility, and after leaving the cryostat horizontally, bend upward to the top of the water tank. The HV cable is also flexible, allowing it to turn with the umbilical. 

The 1-inch-diameter inner tube of the umbilical is connected to the Xe space and is joined to a stainless steel cone that makes a seal to the inner vessel of the detector. To avoid gas bubbles within the cone that might cause electrical discharge, the cone is cooled through a conducting strap in the vacuum space that connects to a detector thermosyphon. The 3-inch-diameter outer tube of the umbilical contains vacuum and is similarly connected to the outer vessel of the detector.  

The outside of the outer tube is immersed in the water of the tank. The evacuated space between the tubes contains super-insulation reflective wrap and acts to thermally isolate the Xe space from the water. This allows LXe to fill the inner tube of the umbilical until it reaches a height equal to the level of the Xe surface inside the detector.  Thus the lower part of the umbilical is filled with LXe, while the upper part contains Xe gas. The long length of LXe is necessary to accommodate the field-grading region of the HV cable. 

A gas return to the circulation pump connects to a tube that leads inside the xenon-filled inner umbilical to just above the liquid surface. This port allows control over the flow of Xe boiling in the lower umbilical. A second port connects to the high end of the umbilical, to slowly purge xenon exposed to the warm cable directly to the Rn scrubber system. Finally, the high end of the umbilical connects to the warm HV feedthrough. The feedthrough and umbilical are supported by the top of the water tank.

\tdrsubsec[CatGrad]{Spark and Discharge Mitigation}

The field-grading structure at the cold end of the HV cable, shown in Figure~\ref{XDSf:cathodeGrad}, allows for the ground braid of the cable to terminate while the polyethylene insulation and conductive center of the cable continue. The departure of the cable ground braid from the cable surface is gradual, and connected to a stress cone made of conductive polyethylene. The stress cone is cryofitted within a xenon-displacing polyethylene wedge, which expands to a polyethylene tube that tightly surrounds the cable insulation. As this tube is larger in outer radius that the commercial cable, the field within the LXe is reduced in comparison with a design that only relied upon the cable insulation. This polyethylene structure is long in order to minimize the electric field parallel to the surface of the cable, and is surrounded by 20 field rings made of conductive plastic. These rings enclose coil springs that grip the cable circumferentially and provide electrical contact to its surface.  The field rings are connected in series by small resistors to establish a uniform voltage grading between them. The highest potential ring is connected to the center conductor of the cable, while the lowest potential ring is connected to the cable ground braid. The surfaces of the rings are heavily rounded, and the resistors are nested between them. This minimizes the field within the LXe that surrounds the grading structure and separates it from the grounded wall of the inner tube of the umbilical. The grading ring structure is supported by three rods entirely from its grounded end, which is attached to the inside of the umbilical. The rods are pulled by springs to the ground end, and the rods compress the ring structure for rigidity. The entire grading structure is immersed within the LXe; all sections of the cable within Xe gas have an intact ground shield. The end of the grading structure connects to a rounded mushroom head, also made of conductive polyethylene, which captures the spring connection to the cathode.

\begin{figure}[htb]
\centering
\includegraphics[width=0.6\textwidth]{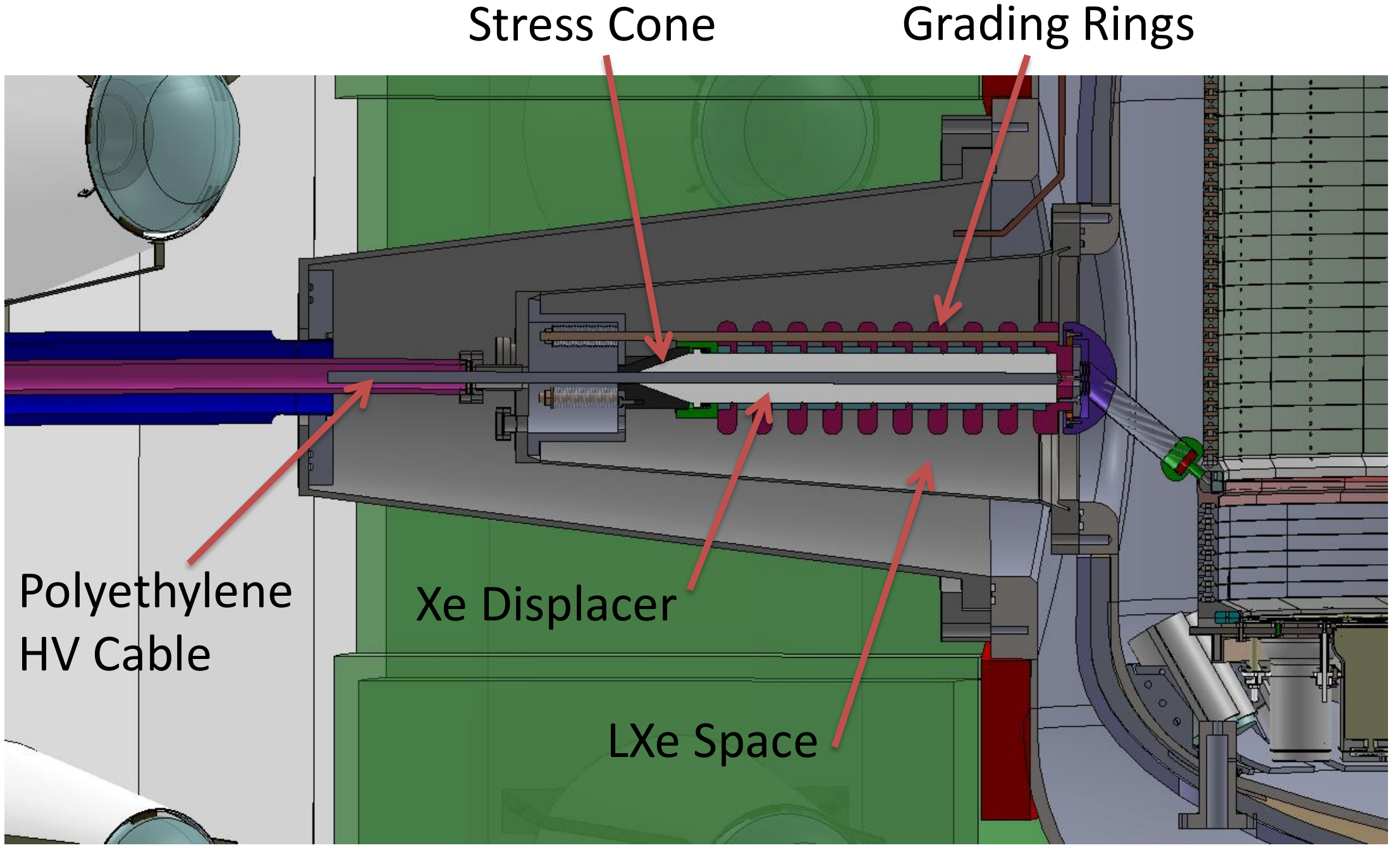}
\tdrfcaption[cathodeGrad]{Cathode field-grading and connection}{Schematic of the flexible HV connection to the cathode grid, showing details of the field-grading structures on the incoming HV cable required to keep the fields in the LXe below \SI{50}{\kV\per\cm}.}
\end{figure}

\tdrsubsec[CatConn]{HV Connection to Cathode Ring}

A schematic of the HV connection to the cathode is shown in Figure~\ref{XDSf:cathodeConnection}, and the simulated electric fields in this region are shown in Figure~\ref{XDSf:cathodeGradingFields}. Because the TPC (including the cathode grid) is supported from the bottom of the vessel, the cathode grid moves down approximately \SI{2}{\mm} as the PTFE TPC components contract when the system is brought from room temperature to operating temperature (\SI{\approx172}{\K}).
To account for this movement, there is a compliant spring connection between the end of the grading structure and the cathode grid ring. 

\begin{figure}[ht]
\centering
\includegraphics[width=0.6\textwidth]{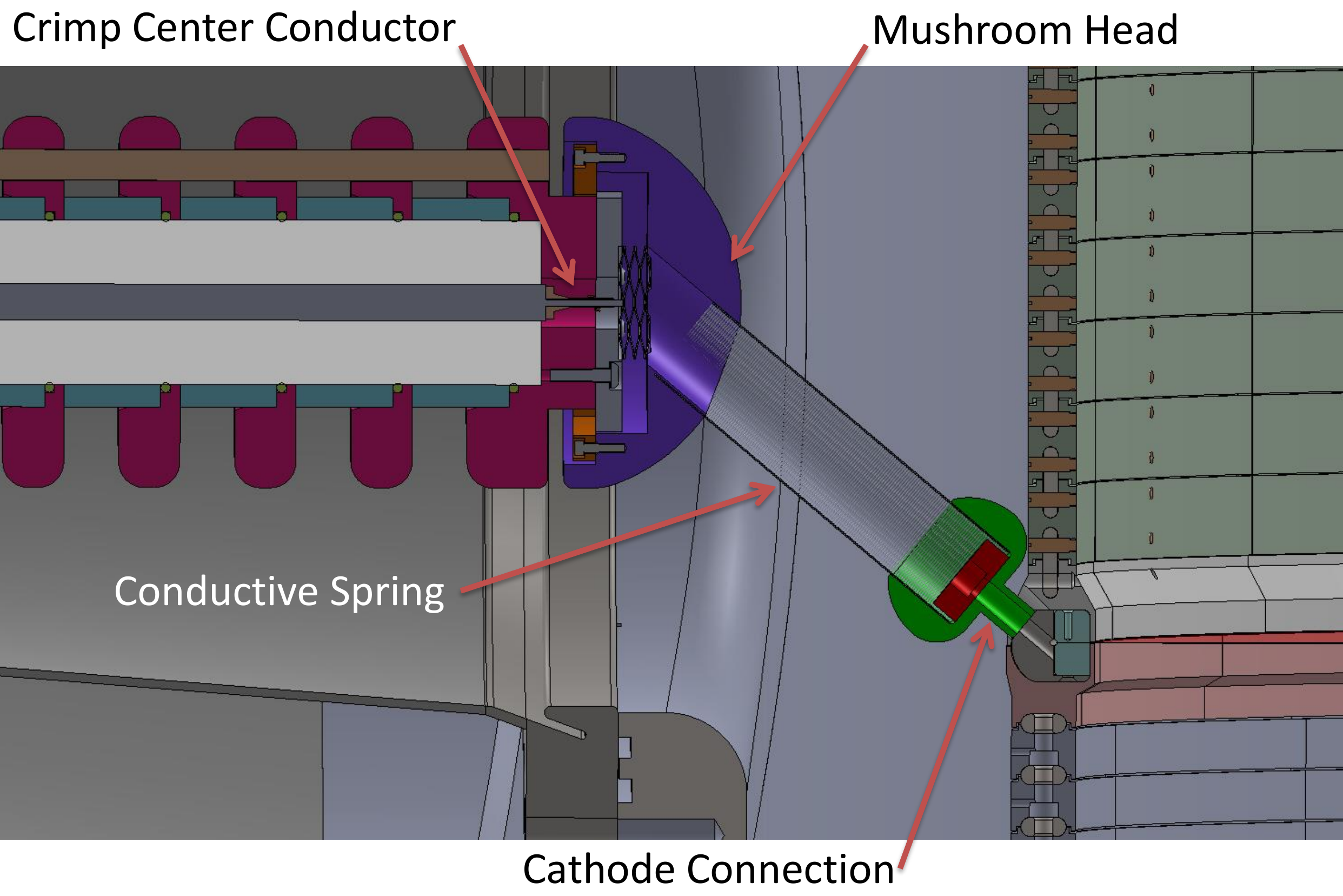}
\tdrfcaption[cathodeConnection]{Cathode connection}{Schematic of the cathode connection region.}
\end{figure}

The grading structure mushroom head is attached to the spring connection prior to installation into the detector.  The lower end of the spring is captured by a rounded cup that has a through hole at its lower end, so that the mushroom head, spring and cup are all one assembly.  A loose screw is captured inside the rounded cup.  This screw will make the connection to the cathode ring.  This screw is captive and cannot fall out of the assembly. The threaded end of this screw sticks out of the lower end of the cup.

During underground installation, the outer and inner cryostat ports are opened.  A long narrow screwdriver is inserted through the mushroom head and into the spring, where it engages the head of the captive screw.  The end of this screw emerges from the lower end of the cup.  The screw inserted into the socket in the cathode ring and tightened.   This mechanically and electrically connects the mushroom head to the cathode ring.  A disk spring held captive under the screw head provides a constant force that prevents unscrewing.

The end of the cold grading structure is then brought close to the mushroom head.  The mushroom head is grasped by hand and pulled toward the end of the cold grading structure, and rotated counter clockwise.  This stretches the spring.  The mushroom head is then pushed into the cold grading structure and rotated clockwise.   This compresses a wave spring that is held captive on the end of the cold grading structure.  The wave spring makes an electrical connection between the cold grading structure and the mushroom head.  It also engages detents in the blue cap and provides a force that prevents the unscrewing of the blue cap.  The mechanical locking of the mushroom head  to the cathode grading structure resembles that of a bayonet mount.

At this point the spring is in its extended position.  The spring is relaxed when the inner CHV cone is brought to meet the face of the cryostat.  Ports in the back of the inner CHV allow for inspection of the spring to make sure it is located correctly.  After the correct position is verified and electrical continuity to the cathode is verified, the bolts connecting inner CHV cone are tightened, making the seal between the inner cone and the inner cryostat.

\begin{figure}[ht]
\centering
\includegraphics[width=0.5\textwidth]{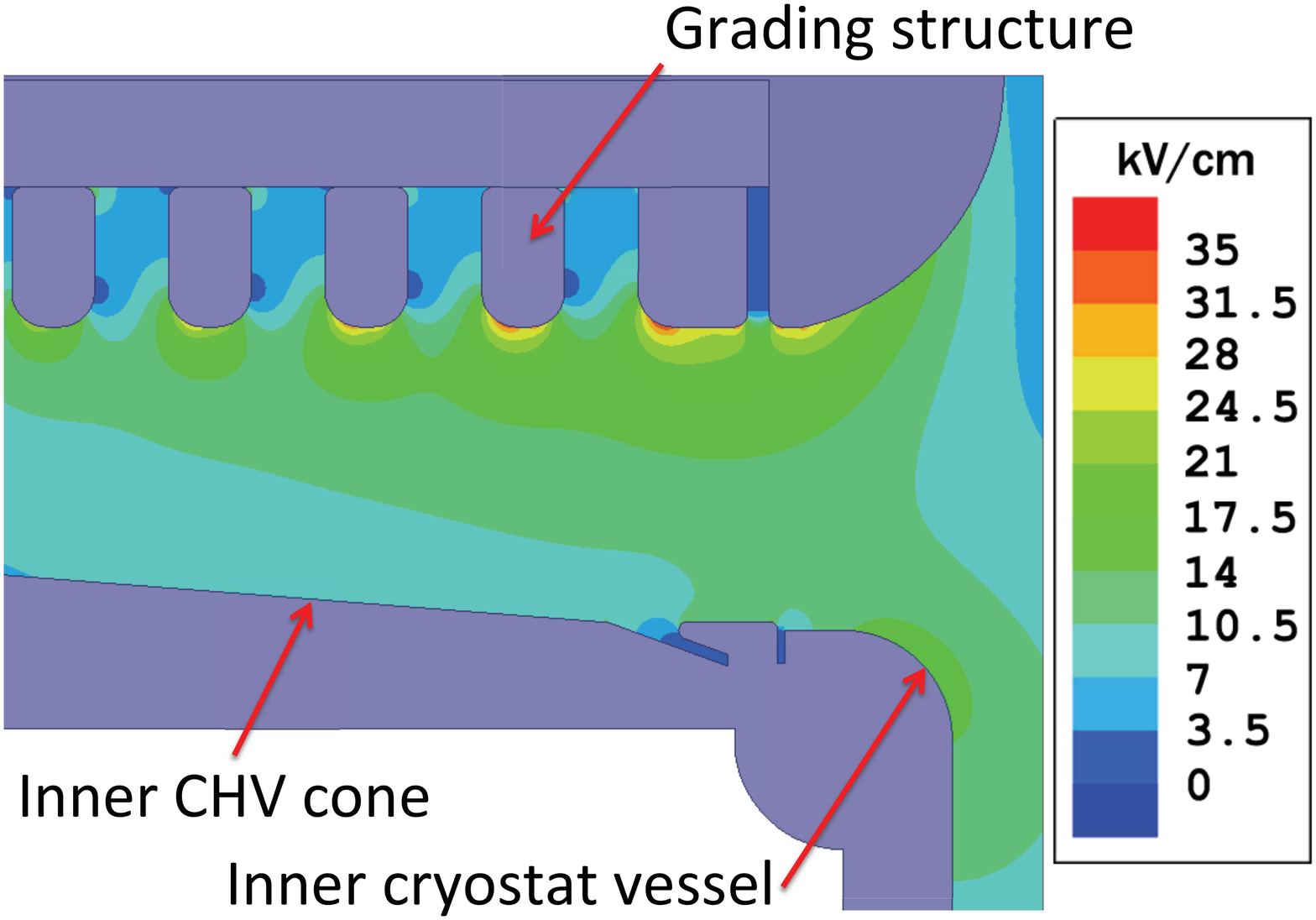}
\tdrfcaption[cathodeGradingFields]{Cathode connection}{Simulation of the electric fields in the cathode connection region, assuming an applied cathode voltage of -100 kV.}
\end{figure} 

\tdrsubsec[CatSafety]{HV Safety Issues}

The combined stored energy from the cathode power supply output capacitance, output cable capacitance, warm feedthrough and umbilical capacitance, and TPC capacitance is \SI{8}{\joule} at the maximum operating voltage of \SI{-100}{\kV} and is classified as a 3.2C~\cite{doe:2013} shock hazard per the DOE Electrical Safety Handbook~\cite{doe:2013}. This shock hazard class indicates that injury or death could occur by contact. To mitigate this hazard, engineering controls are required for operation, and administrative controls are required for electrical work. Specific ``lockout/tagout'' and grounding procedures will be implemented for various operations such as unplugging the output cable and accessing the internals of the warm feedthrough, umbilical, and the TPC. Each worker who is authorized to perform these tasks will have specific high voltage and capacitor safety training.

\tdrsec[PMTs]{Photomultiplier Arrays}

To reach the performance specifications described previously, the Xe detector is equipped with top and bottom arrays of 3-inch-diameter PMTs (Hamamatsu R11410-22) to view the active region of the TPC. A top ring of smaller, 1-inch-square PMTs (R8520) view the scintillation light emitted in the side Xe Skin, with additional 2-inch PMTs (R8778) recycled from LUX complementing the Skin readout at the bottom of the detector. All three PMT types, shown in Figure~\ref{XDSf:PMTs}, have been developed to meet important performance requirements in the field, including good spectral response in the VUV, good single-photoelectron definition, low dark noise, and the ability to operate at LXe temperature---in addition to having ultra-low levels of radioactivity: of order \si{\mBq\per unit} in U / Th / \ICosz~/ \IKfz for the 3-inch and 1-inch tubes, and a few times higher for the 2-inch units. This section details the properties and deployment of the PMTs for the TPC and Skin. Subsequent sections discuss the design and optimization of S1, S2, and Skin optical signals.

\begin{figure}[ht]
\centering
\includegraphics[width=0.75\textwidth]{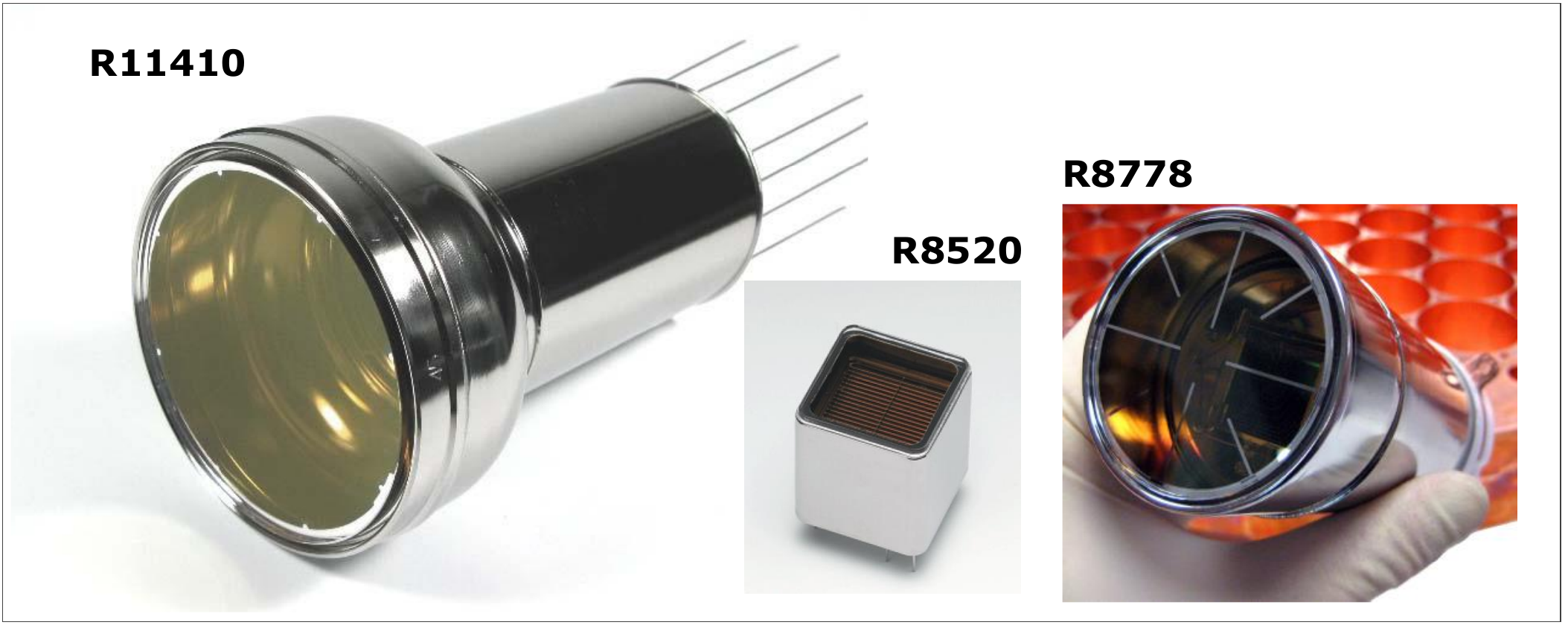}
\tdrfcaption[PMTs]{Xenon-space photomultipliers} {Photographs of xenon-space photomultipliers in LZ.}
\end{figure} 

The LZ Collaboration has been pursuing the development of low-background PMTs tailored specifically for use in LXe with a radioactivity goal of 1/1~mBq per unit in for U/Th and QE \SI{>30}{\percent} at \SI{175}{\nm} wavelength~\cite{Akerib:2012da}. The detector configuration requires \num{\approx500} 3-inch tubes. Because of its outstanding radiopurity, the 3-inch Hamamatsu R11410-22 model has been adopted; this tube contains \num{\sim1000} times less radioactivity than a standard off-the-shelf item and is the result of a coordinated development with the manufacturer and a rigorous screening campaign of sub-components before the items are even manufactured.

The dynode optics in the R11410 are electrically identical to those used in LUX (2-inch R8778), exhibiting similar gain and single photoelectron response. The photocathode diameter is \SI{64}{\mm}. This tube has \num{12} dynodes and provides a minimum gain of \num{2e6} at \SI{1500}{\V} bias voltage. The PMTs are assembled to passive voltage-divider bases and will be negatively biased so that the signal can be collected by directly coupling the front-end electronics at near-ground potential. High peak-to-valley ratios (\num{>2}) are obtained for the single photoelectron response, which is a key parameter to ensure high detection efficiency for the smallest S1 signals that are composed of single photoelectrons appearing in multiple PMTs.

Aside from good VUV sensitivity, these quartz-windowed PMTs are designed to be operated at LXe temperature featuring a special low-temperature bialkali photocathode with low surface resistivity. This obviates the need for metallic underlayers or conductive fingers~\cite{Nakamura:2010zzk}. They also have a pressure rating of \SI{4}{atm} absolute for operation in liquid xenon. Nominal performance will be verified for every unit through a comprehensive low-temperature test program to confirm optical and electrical parameters during and after thermal cycling.

\tdrsubsec[PMTtest]{PMT Test Program}

The full order of the R11410 PMTs is split between the U.S. (2/3) and the U.K. (1/3), delivered in batches of \num{40} per month. It is crucial to characterize every PMT before they are installed, and a detailed testing program is designed to test and radio-assay all the PMTs. This is the testing sequence that will be carried out at Brown University for all tubes (except for the confirmatory radioactivity screening, as explained below):
\begin{enumerate}
\item Warm test \#1
\item Cryogenic test
\item Radioactivity screening 
\item Warm test \#2
\end{enumerate}

In the first warm test the after-pulsing, gain, single photoelectron resolution, dark count rate and linearity will be measured. These measurements are categorized as ``standard testing'' and they will identify any PMT that fails to meet agreed specifications for room temperature operation. In the cryogenic test PMTs will be thermal-cycled from \SI{300}{\K} to \SI{160}{\K} and the standard testing measurements will be repeated, plus some exceptional studies such as photocathode uniformity and long-term stability monitoring. The Brown PMT test chamber known as PATRIC, shown schematically in Figure~\ref{XDSf:PMTtest} (left), is designed to hold \num{14} R11410 PMTs for batch testing. After radioactivity screening (carried out at SURF and Boulby underground laboratories) the PMTs will undergo a second warm test to establish stability over time.

The U.K.-procured PMTs will go through the same test sequence, but a sample of \num{30}~units will undergo characterization of quantum efficiency (QE) and dual-photoelectron emission probability at room temperature and at LXe temperature at Imperial College London. The main goal of the QE test is to establish a statistical correction function to relate the QE at \SI{175}{\nm} obtained by Hamamatsu at room temperature and the relative change of QE upon cooling~\cite{Araujo:2004mj}. The PMT test chamber at Imperial, shown in Figure~\ref{XDSf:PMTtest} (right), uses a xenon scintillation cell maintained at room temperature to shine \SI{175}{\nm} scintillation light into a cryostat containing \num{7} PMTs (\num{6} under test plus a reference tube) which are cooled to LXe temperature. Correlation of QE data with PMT gains, dark counts and the QE values provided by Hamamatsu will also be studied.

\begin{figure}[ht]
\centering
\includegraphics[width=0.7\textwidth]{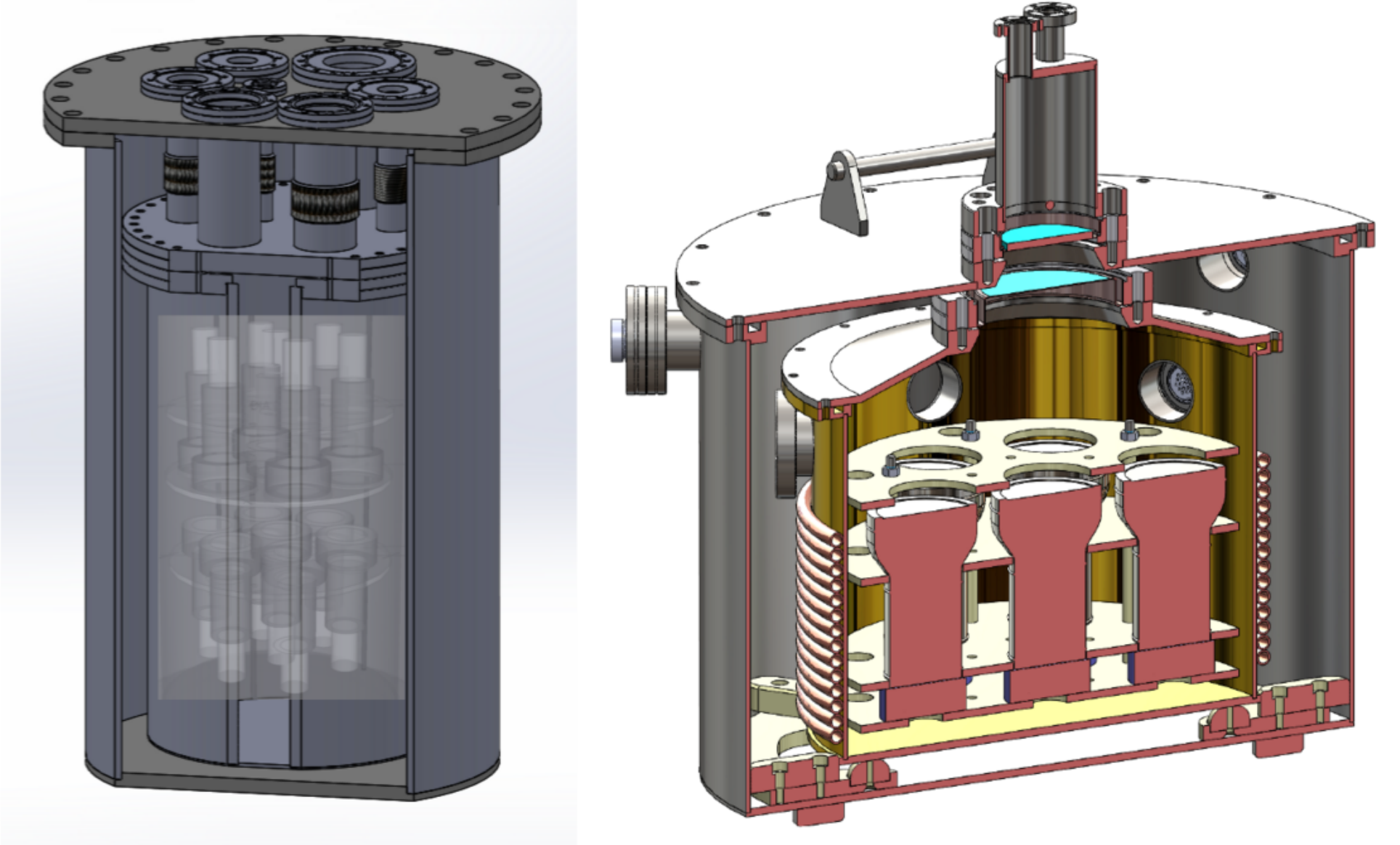}
\tdrfcaption[PMTtest]{Photomultiplier test setups} {PMT test setups. The Brown PMT Array Test Rig In Cryogens (PATRIC), shown on the left, is a pressure vessel that can test \num{14} PMTs at LXe temperature and pressure. After the R11410 PMT testing is completed, the chamber will be modified to test R8520 PMTs. The Imperial setup on the right uses a xenon scintillation cell on top of a vacuum cryostat, shining onto \num{7} PMTs located in cold nitrogen gas.}
\end{figure}

\tdrsubsec[PMTradio]{PMT Radioactivity}

Due to their complexity, total mass (\SI{\approx100}{\kg}), and proximity to the active volume, the Xe-space PMTs are a significant source of radioactivity background in LZ. For this reason, they will be subject to a thorough screening campaign using HPGe detectors (see Chapter~\ref{chap:MAS}). Screening of fabrication materials and sub-components has taken place prior to PMT manufacture, and every assembled PMT will again be screened after delivery from Hamamatsu.  

The 3-inch R11410 has been delivered in part through a 4-year NSF S4 development program with Hamamatsu, which achieved unprecedented radioactivity performance compared with previous generation tubes~\cite{Akerib:2012da,malling2013}. This same model has also been advanced by other collaborations, notably XENON1T with variant R11410-21. Comprehensive radioactive screening results for 216 of these PMTs are publicly available~\cite{Aprile:2015lha}. Screening results from our prototype tubes are in broad agreement, and the pre-screening of construction materials also yielded similar results to those obtained by XENON1T---as detailed in Chapter~\ref{chap:MAS}. LZ will procure the most recent (-22) version of these tubes.

The 1-inch R8520 PMTs used in the top side of the Skin Detector have radioactivity levels that are well understood thanks to their wide use in previous detectors. Nevertheless, they will follow the same screening procedures as the larger R11410 model. The contribution of the 93~Skin PMTs to the background of the instrument is sub-dominant given their comparable specific activity but more peripheral location and smaller number. 

The 2-inch R8778 PMTs used in the bottom of the Skin Detector will be reused from the LUX experiment and are also well understood from both a radioactivity and performance perspective. Although they have somewhat higher radioactivity per PMT relative to the R11410 PMTs, their background contribution is also sub-dominant given their peripheral location and small number.  

\tdrsubsec[PMTbases]{PMT Bases}

Individual voltage-divider bases connected to two coaxial cables (for HV bias and signal) are attached to each PMT. Given their locations, these components are under detailed scrutiny as part of the radioactivity and radon-emanation screening programs, and they must meet also demanding requirements related to operational reliability and xenon poisoning.

\begin{figure}[ht]
\centering
\includegraphics[width=0.8\textwidth]{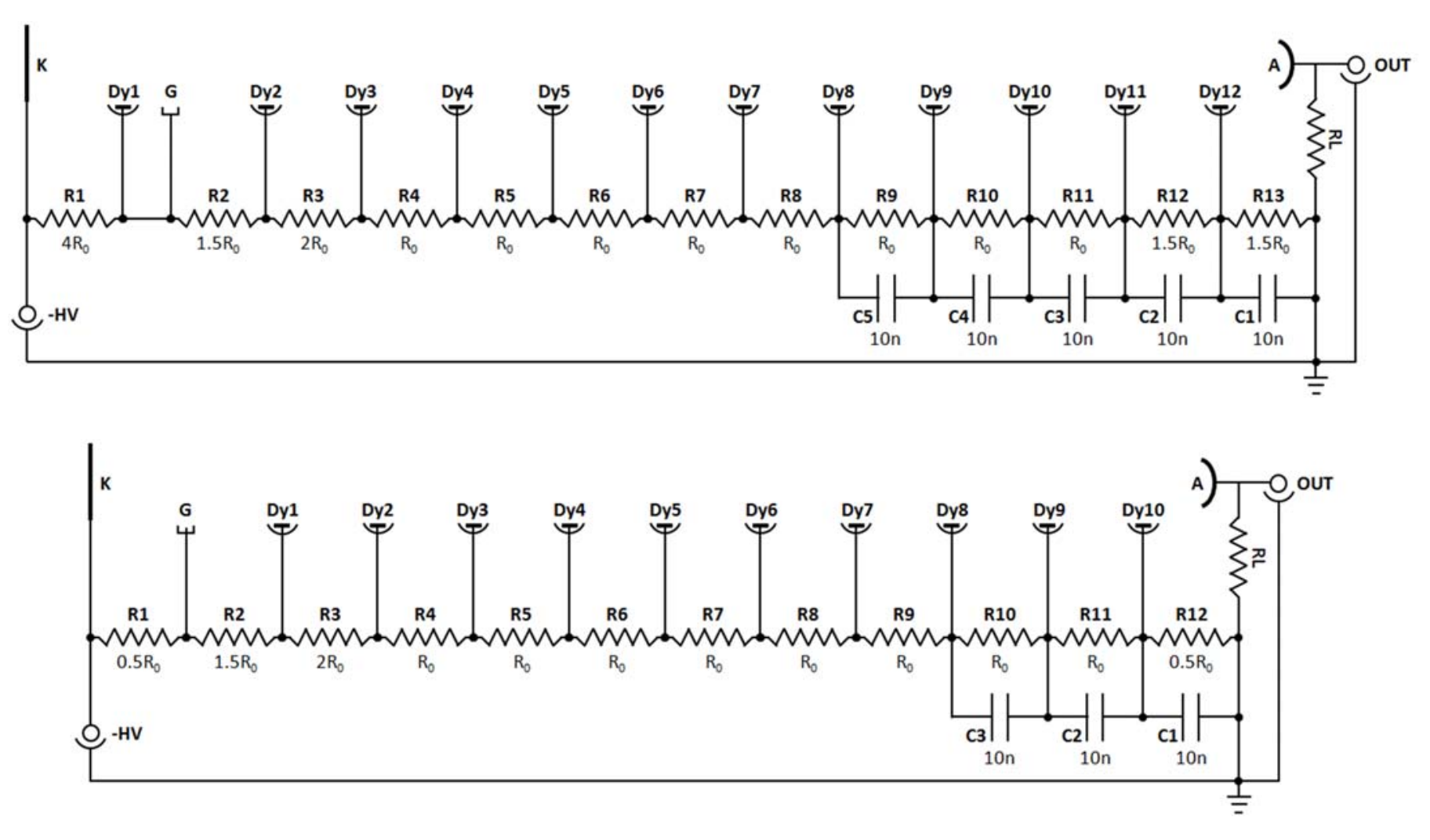}
\tdrfcaption[PMTbases]{Photomultiplier base voltage-divider circuits} {Photomultiplier base voltage-divider circuits for the R11410 PMT (upper panel) and R8520 PMT. In both cases the voltage distribution recommended by Hamamatsu was adopted, with $R_0$=\SI{4.99}{\mega\Ohm}. $R_L$=\SI{49.9}{\Ohm} for the R11410 base and $R_L$=\SI{100}{\kilo\Ohm} for the R8520 base. The 2-inch R8778 Skin PMT bases will employ the same circuit design as the 3-inch R11410 units.}
\end{figure}

The electrical circuits are shown in Figure~\ref{XDSf:PMTbases} for the 3-inch and 1-inch tubes (the 2-inch R8778 base is similar to the former), and a prototype base for the R11410 PMT is shown in Figure~\ref{XDSf:PMTbaseProto}. The voltage distributions are those recommended by Hamamatsu. High-resistance chains are used for low power dissipation in the LXe: \SI{24.3}{\milli\W}/unit at the nominal \SI{1500}{\V} for the 3-inch PMTs, and \SI{10.3}{\milli\W} at \SI{800}{\V} for the 1-inch PMTs. This is required for minimal impact on the thermal design of the detector and to prevent localized bubbling of the liquid. We have confirmed the bubble nucleation threshold of liquid xenon as \SI{\approx20}{\milli\W\per\mm\squared}, in agreement with Ref.~\cite{Haruyama:2002}; all resistors operate far from this value. Still, the divider current is much higher (\num{\approx100}$\times$) than the average PMT anode current expected during calibration, which is required for response linearity. Anode pulse linearity is also addressed by charge supply capacitors (\SI{10}{\nano\farad}, de-rating to \SI{\approx8}{\nano\farad} at LXe temperature) which are added to the last few dynodes. Our main requirement ensures accurate response to \IKreTm calibration events, translating to percent-level non-linearity in top-array channels for S2 signals from this key calibration source. This determines a minimum of five decoupling capacitors for the TPC PMTs. These S2 signals also approach the \SI{\pm 5}{\percent} pulse non-linearity expected in these PMTs above \SI{\approx20}{\milli\A} anode current.
\begin{wrapfigure}{r}{0.45\textwidth}
\centering
\includegraphics[width=0.45\textwidth]{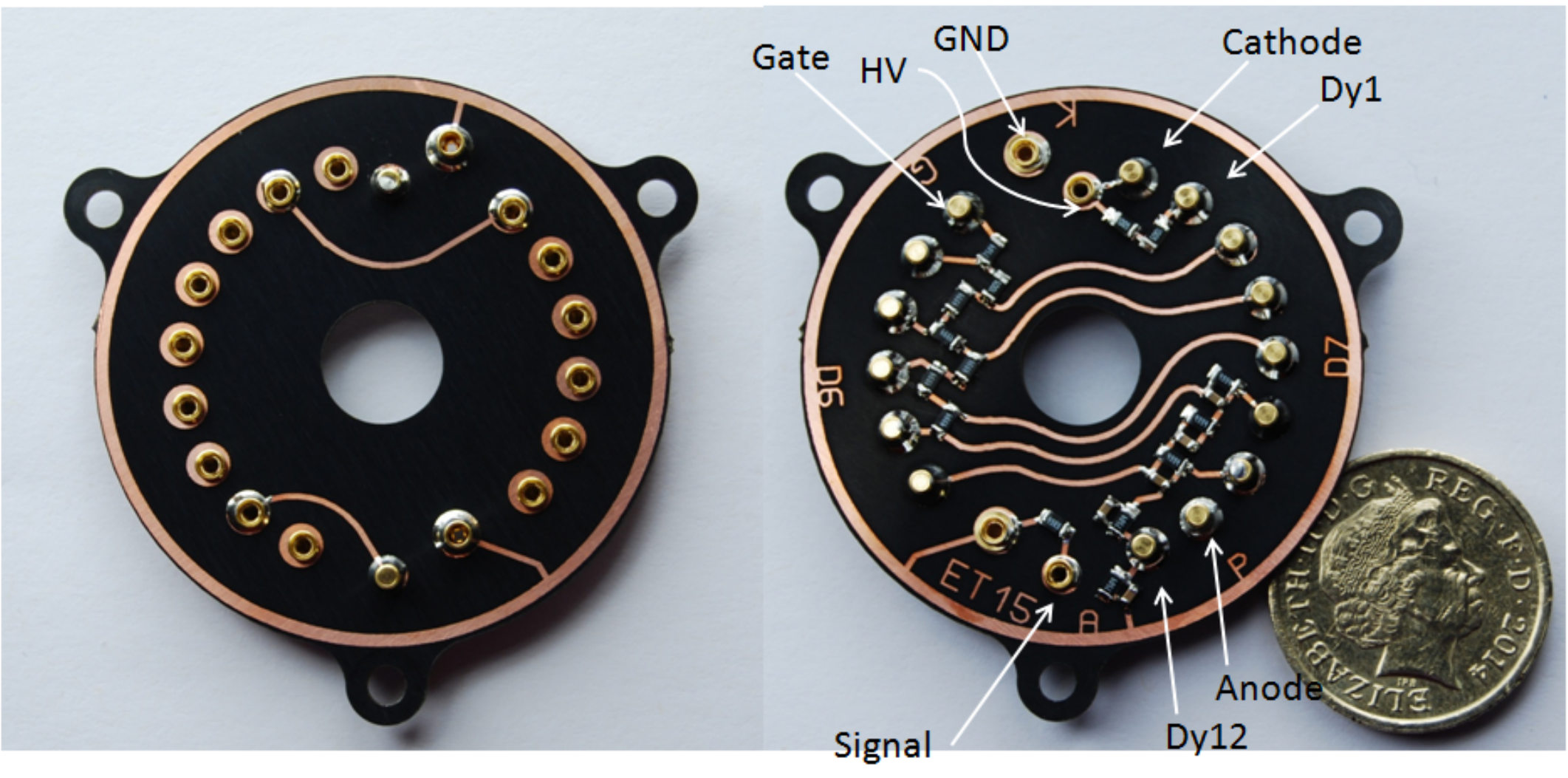}
\tdrfcaption[PMTbaseProto]{R11410 voltage-divider base prototype} {Prototype of the R11410 PMT voltage-divider base.}
\end{wrapfigure}

The bases are made from thick polyimide (Cirlex\texttrademark) PCBs with surface-mounted passive components. Cirlex is a good material for this purpose, having very high dielectric strength, a low thermal expansion coefficient, high tensile strength and low internal stress. It is, however, prone to delamination and fiber release at the edges, demanding careful quality control. The Cirlex is patterned with standard photo-lithography and the cutout and drill-holes are made with a PCB router---all processes are conducted in-house to ensure strict control of radioactivity and cleanliness. Candidate components and full prototypes were tested in gaseous and liquid xenon for electrical resilience and outgassing as well as for spurious optical photon emission. All bases will be thermal-cycled in L\CNt and tested at ambient temperature for 5 days above their maximum allowable voltage. Final inspection, cleaning and radon-tight packaging are conducted in a class \num{1000} cleanroom.

The radioactivity performance of the PMT bases is of special concern, both from the point of view of neutron/gamma emission and from radon emanation. Although Cirlex is intrinsically a radio-clean material and a good Pb-free solder has been identified, the discrete components have significant U/Th contributions in spite of the small masses employed. The X7R ceramic capacitors, the thick-film ceramic-chip resistors, and the PMT-pin receptacles all posed significant challenges and we radio-assayed some \num{25} components in order to select a viable set meeting our radioactivity goal---to be sub-dominant to the PMTs. Ceramic capacitors have typically high uranium content, with those selected (Kemet, 0603 package) showing the lowest levels (\SI{11.5}{\mBqg}); fortunately, the XR7 ceramic is barium titanate (confirmed through SEM/EDX elemental analysis) which has an order of magnitude lower intrinsic neutron yield than alumina, making these components viable---although they pose some radon emanation risk as discussed below. The selected resistors (Vishay, 0805 package) showed acceptable levels of U/Th (\SI{<1}{\mBqg}), but high levels of \IPbtoz (found in all models tested); the resulting (\Pga,\Pn) yield on the alumina becomes a significant contribution to the neutron yield from the bases. Finally, the greatest challenge came from the receptacles that mate to the PMT contact pins and to the cable connectors. These contain BeCu alloy spring clips which are very ``hot'' in early uranium (\num{\approx100}$\times$ out of equilibrium). Avoiding the use of these sockets by directly soldering to the PMT pins and cables would have cleanliness implications that would delay the integration sequence of the PMT arrays, and was thus not considered an attractive option. We identified suitable sockets from Harwin which feature CuNi$_2$Be alloy spring clips within a brass shell; these have lower Be content (\SI{0.4}{\%}) and lower radioactivity than the other models we assayed (note that \SI{\approx50}{\percent} of \IBen(\Pga,\Pn) reactions can be vetoed due to the \SI{4.4}{\MeV} gamma ray~\cite{Liu:2007}; this was considered in the background calculations). In conclusion, we were able to obtain a suitably low radioactivity content, where no single component dominates too significantly, balanced against practical considerations such as robustness and ease of integration. The bottom-up radioactivity estimates are: \SI{390/140}{\micro\becquerel}/unit in mid-U/Th for the R11410 and R8778 base (\num{0.06}~n/yr) and \SI{230/80}{\micro\becquerel}/unit for the R8520 base (\num{0.04}~n/yr), respectively.

Radon emanation from the bases was a concern, prompted especially by the total \IRatts content observed in the capacitors. Fortunately, the \IRnttt emanation rate measured from a very large number of these components was found to be below \SI{1}{\percent} relative to production even at room temperature. The emanation from all bases in the xenon space is expected to be \SI{\approx1}{\milli\becquerel} at ambient conditions; a modest reduction with cooling is expected from the resistors and capacitors, while the Cirlex should exhibit a significant decrease. Therefore, expect the bases to make a limited contribution to the LZ radon budget (and to confortably meet the radon budget requirement).

\tdrsubsec[PMTcabling]{Cabling}

The PMT signals and HV supplies are carried separately between the PMT bases and the warm breakout interface by low-radioactivity coax cables. The baseline design is to use Gore 3007 Coax with no outer jacket, the same cable that was used in this role for LUX. The \SI{50}{\Ohm} characteristic impedance cable uses an AWG~30, silver-plated, Cu-clad steel, surrounded by an AWG~40 stainless steel braid. The cables from the PMTs associated with the upper and lower parts of the TPC are housed in separate conduits, so that no cabling is routed through the side Skin region (these conduits can be seen in Figure~\ref{IDOf:LZSolid} in Chapter~\ref{chap:IDO}). This could interfere with the ability to hold a high voltage on the cathode and it would degrade the light collection efficiency in the Skin.

\begin{table}[tbh]
\setlength{\extrarowheight}{3pt}
\tdrtcaption[Cabling]{Cabling in the xenon space}{Coaxial cables in the xenon space routed via upper and lower conduits.}
\centering
\sffamily
\begin{tabular} {|lrr|}
\hline
\rowcolor{mrocol}
					& Upper & Lower \\
\hline
TPC PMTs			& \num{506} & \num{482} \\
Skin PMTs			& \num{186} & \num{76} \\
LED calibration		& \num{39}  & \num{45}  \\
Dummy channels		& \num{11}  & \num{11}   \\
Monitoring sensors	& \num{59}  & \num{11}  \\
\hline
Total				& \num{801} & \num{625} \\
\hline
\end{tabular}
\end{table}

The lengths of typical top and bottom cables are \SI{12.8}{\m} and \SI{11.6}{\m}, respectively. The upper routing carries \num{801}~cables and the lower routing consists of \num{625}~cables, as listed in Table~\ref{XDSt:Cabling}. The total heat load calculated from the cables is less than \SI{6}{\W}, which is a sub-dominant contribution to the thermal model. The total cable length within the xenon space is \SI{\approx17}{\km}, weighing some \SI{72}{\kg}.

A screening program has been initiated to measure the radon emanation from the baseline cabling, as well as possible alternatives, to ensure that the finally selected model will meet the overall Rn requirements discussed in Chapter~\ref{chap:MAS}. Rn emanation from cables is expected to be dominated by the warm region, which is \SI{\approx8}{\m} long for the upper routing, and only \SI{\approx1}{\m} for the lower routing. To mitigate excessive Rn emanation due both to dust and radioactivity of the cable materials, we are evaluating the benefit of having the cables sheathed in a \SI{150}{\mum} thick coat of FEP Teflon to reduce the rate of Rn reaching the active detector volume. The MJD and GERDA collaborations have observed that including such a FEP jacket provides a suitable Rn barrier in their detectors. Emanation from the feedthroughs (which were previously used in the LUX experiment) will also be measured. Finally, we are planning to add a radon trap to the gas purification system, as discussed in Section~\ref{XCSSs:RnRem}, to address emanation from the cables and room temperature feedthroughs.

The Gore 3007 cable has been tested and shown to support \SI{2}{\kV}, comfortably meeting the HV requirements of all PMTs. The way the cable affects the signal characteristics can be seen in Figure~\ref{EDCf:LZAmpSPHE} in Chapter~\ref{chap:EDC}, which shows the effect of a full cable run (internal and external to the detector) on the single photoelectron response of a R11410 PMT. For these signals there is an amplitude reduction of \SI{47}{\percent} and a pulse area loss of \SI{20}{\percent}.

\tdrsubsec[PMTassy]{Assembly and Integration with TPC}

The \num{253/241} PMTs per top/bottom array will be assembled onto titanium support frames. The PMTs will be held in position using cobalt-free metal belts fabricated by Hamamatsu from the same material used for the PMT body production. Three PTFE columns will then be used to hold the collar to the PMT mounting plates, as shown in Figure~\ref{XDSf:PMTassy}. This mounting system is specified to hold the PMTs in place in the Ti mounting plate both when the plate is in the vertical and horizontal orientations (during assembly and transport).

\begin{figure}[ht]
\centering
\includegraphics[width=0.6\textwidth] {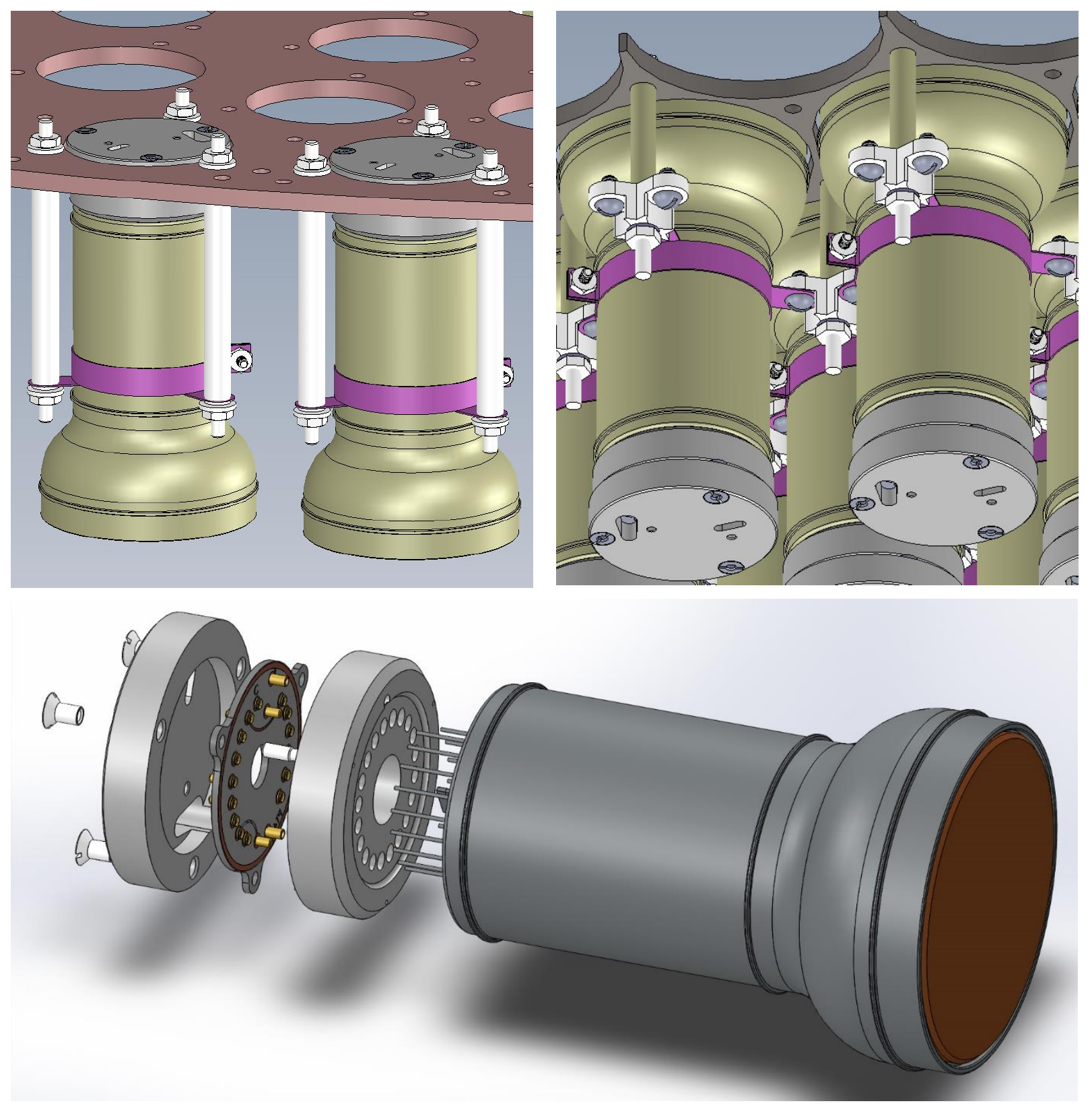}
\tdrfcaption[PMTassy]{PMT assembly} {PMT assembly to the top and bottom array supports(upper right and left, respectively). The lower array PMTs are sleeved in PTFE and attached to the Ti plate with three PTFE rods. The lower panel shows the assembly to the base circuit, which is encased in PTFE caps; that at the back provides high reflectivity for the Skin detector and is used to strain-relive the cables.}
\end{figure} 

\begin{figure}[ht]
\centering
\includegraphics[width=0.65\textwidth] {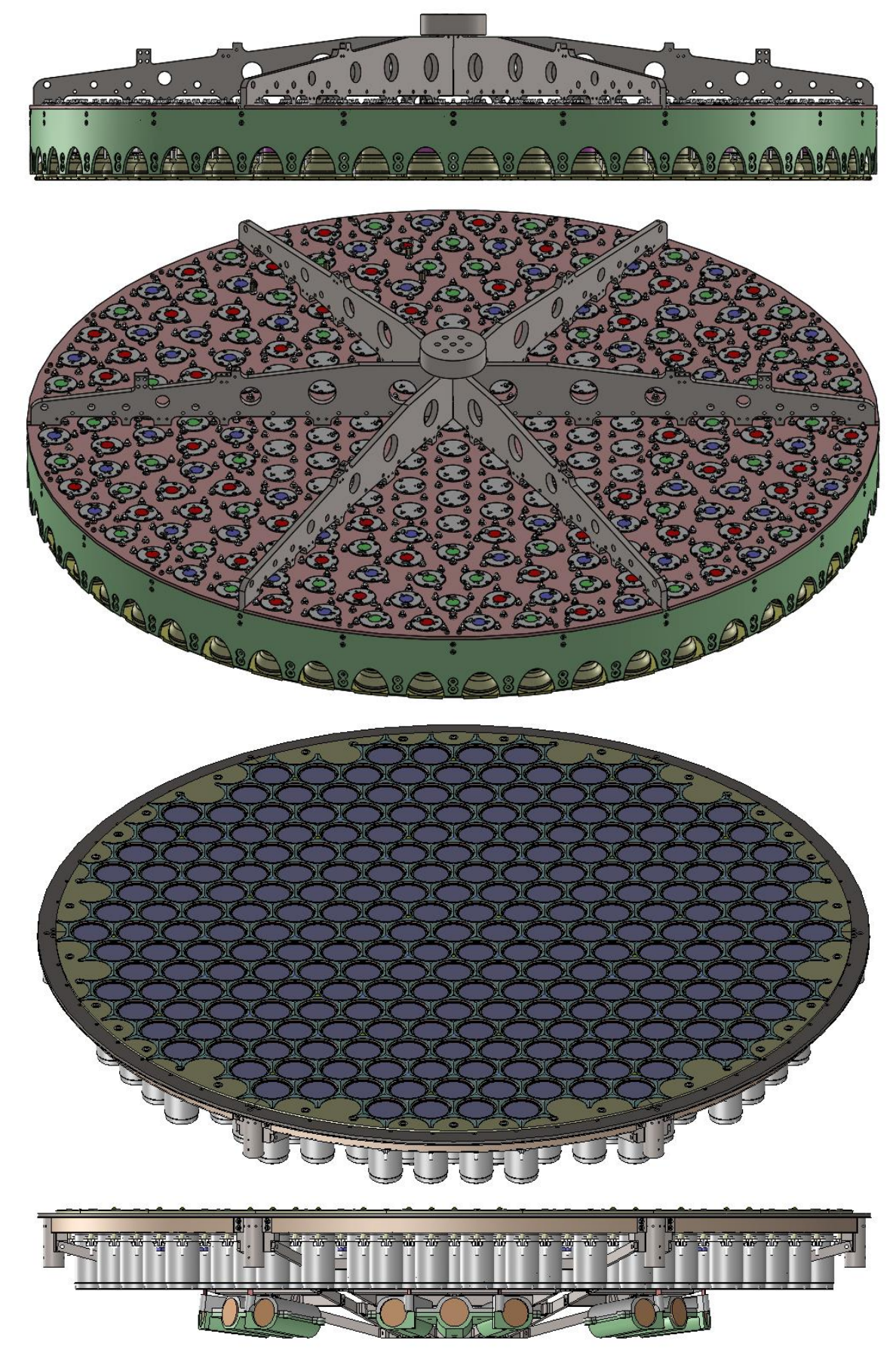}
\tdrfcaption[PMTarrays]{TPC photomultiplier arrays} {The bottom PMT array has \num{241} 3-inch PMTs arranged in hexagonal configuration. The top PMT array accommodates \num{253} PMTs in a hexagonal-circular hybrid configuration.}
\end{figure}

The PMT arrays are shown in Figure~\ref{XDSf:PMTarrays}. The support frames consist of a flat titanium plate with supporting truss-work. The loads on this structure are substantial, particularly in the case of the lower array. For the submerged PMTs, the buoyancy force far exceeds the gravitational force: the net upward load is approximately \SI{8}{\N} per PMT, and collectively the total load for the bottom array is approximately \SI{2200}{\N}. Many configurations of the support frame were considered and simulated using finite element analysis. Starting with a bare plate (no truss-work), a single \numrange{6}{7}-\si{\mm} thick Ti plate deflected upward approximately \SI{19}{\mm}. Other options include successively thicker plates, double plates, curved plates, honeycomb reinforcement, and truss reinforcement. The truss reinforcement had the best overall performance when trying to limit deflection, minimize mass (and therefore background radiation), and provide a relatively open volume for scintillation light in the dome Skin to find its way to the Skin PMTs. The baseline lower PMT support frame is expected to deflect approximately \SI{1}{\mm} upward in operation. The upper PMT support frame (in the gas phase) will have a similar design, but the expected downward deflection is only \SI{\approx0.3}{\mm} for that array.

The Ti surfaces surrounding the front faces of the PMTs, in both the top and bottom arrays, will be covered by PTFE pieces designed for photon recycling, and so increase photon detection efficiency in the main chamber, as discussed in Section~\ref{XDSS:S1Light}. The pieces are designed to provide at least \SI{95}{\percent} coverage of the Ti structural elements, while accommodating the differential thermal contraction coefficients of the PTFE and the Ti mount.

The lower LXe region, below the bottom PMT array, forms part of the Xe Skin detector in which the goal is to maintain over \SI{50}{\percent} detection efficiency for ER events above \SI{100}{\keV} in more than \SI{95}{\percent} of the Skin volume---see Section~\ref{XDSS:Skin}. The rear of the bottom PMTs, which project into this volume, are also sleeved in PTFE in order to increase photon recycling in the LXe below the array---this includes both a PTFE sleeve for the PMT body, and end-caps to cover the PMT bases. The PTFE base covers also prevent stray light leaking into the PMT envelope, and avoid pin short-circuits. The underside of the PMT mounting structure and braces will also be covered in PTFE reflectors where required, to increase the overall photon detection efficiency in the Skin region.

The PTFE components will be fabricated from material that has been pre-screened to achieve the intrinsic activity budget with respect to both gamma and neutron emission, as discussed in Chapter~\ref{chap:MAS}. During machining of the components and the assembly of the PMT arrays, the PTFE components will be maintained in purge boxes to reduce the plating of alpha emitters associated with airborne Rn, and ensure that additional (\Pga,\Pn) neutron generation is significantly below the intrinsic neutron emission goals.

PMTs passing all the testing and screening procedures will be installed onto their arrays inside a dedicated cleanroom at Brown University. Two PMT Array Lifting And Commissioning Enclosures (PALACE) will be built for this purpose. The arrays will be hanged vertically so that all PMT slots can be easily accessed during assembly. PALACE is also designed to be an airtight vessel so that the whole array and PMTs are under nitrogen purge when there is no assembly work on-going. With appropriate high voltage and signal feedthroughs, PALACE will also be used as dark chambers for final testing before TPC integration. Since each PALACE will hold the PMT array and over \num{240}~PMTs, using PALACE as a radon emanation chamber will help understand the radon backgrounds in the detector; we are pursuing the goal of building PALACE as a radon emanation chamber with a target sensitivity of \SI{0.3}{\mBq}.

To meet these design goals, PALACE will consist of a stainless steel frame creating a \num{6 x 6 x 1.5}-\si{\footl\cubed} space to hold the PMT array. The nitrogen purge system will feed a \SI{10}{slpm} flow from L\CNt boil-off into the chamber. All materials will be chosen so they emanate less than \SI{0.1}{\mBq} radon, and the vessel will be welded and electropolished to help achieve this. For the top lid a 1-inch thick plastic plate would be a sufficient radon barrier; we are investigating the best sealing technique for this purpose.

After PMTs are installed and the final electronics checkout is complete, PALACE is ready to be transported to SURF. The chambers will be sealed, purged and triple-bagged. During shipment PALACE will remain airtight, with the PMT arrays sitting in a nitrogen environment. Before the actual shipment happens test shipments with empty crates and accelerometers will be carried out to verify that PALACE is well taken care of under a medical equipment shipping company. After PALACE arrives at SURF it will be unwrapped and placed in the SURF cleanroom. PMTs will also be checked repeatedly to ensure their stability.

\tdrsubsec[PMTcal]{PMT Calibration}

Three calibration techniques will be employed to monitor the PMT performance during LZ operation. The first one entails flashing \SI{470}{\nm} LEDs producing O(100) photons per pulse to measure after-pulsing, providing a direct indication of vacuum integrity of each tube. It is caused by residual gas ions inside the PMT body hitting the photocathode, creating a secondary signal appearing at a certain time after the main pulse. The after-pulse from Xe$^+$ ions would appear around \SI{3}{\mus} after the main pulse~\cite{Faham:2013}. The second calibration is of the single photoelectron response, i.e.~the absolute gain. The LEDs can be pulsed at a very low voltage such that signals seen by the PMTs come from single photons with high probability. The LZ LED calibration system is an evolution on the system developed for LUX. Calibration LEDs will be mounted on the face of each array to shine onto the PMTs opposite under individual DAQ control. A synchronized trigger is also fed into the DAQ system so the LED pulse timing is recorded along with the calibration data.

While the LED measurements monitor PMT gain (at \SI{470}{\nm}) and after-pulsing stability, the VUV response will be measured in-situ as the third calibration technique. In LUX this was achieved by extracting single photoelectron signals from tritium beta decay calibration data. The same calibration will be done in LZ (see Chapter~\ref{chap:CAS}). It was understood from LUX experience that when a PMT sees a 175~nm VUV photon, there is a \SI{\approx20}{\percent} probability that a second photoelectron is emitted in coincidence~\cite{Faham:2015kqa}. This double-photoelectron emission phenomenon has very important consequences---for understanding the optical performance of the detector, and on its energy resolution; therefore, it must be well characterized for all PMTs in operating conditions. The combination of the LED calibration, which produces the true single photoelectron response, with analysis of the response induced by xenon scintillation photons will allow us to obtain this information.

\tdrsec[S1Light]{S1 Light Collection Design}

\tdrsubsec[S1PDE]{Overview of Design and Optical Performance of the TPC}

Achieving the highest possible collection of scintillation photons is a design priority: It leads to the lowest energy thresholds in the S1 and S2 channels and, by reducing statistical fluctuations in the S1 signal, it improves ER/NR discrimination. It also improves energy resolution, which is important for identifying gamma-ray background lines and for $0\nu\beta\beta$ sensitivity. The scintillation yields for electron and nuclear recoils depend on both energy and electric field. For reference, at the LZ field design goal \num{445} scintillation photons are expected to be emitted from a \SI{10}{\keV} ER track and \num{84}~photons from a \SI{10}{\keV} NR interaction. The experimental challenge is to maximize how many are recorded as photoelectrons (phe) in the PMT arrays. In comparison, the high electroluminescence yield in saturated Xe vapor (typically $\approx$\num{1000}~photons/cm per emitted electron~\cite{Fonseca:2004cd}) leads to very large S2 signals such that single electrons are readily measured and sub-keV detection thresholds are obtained. The LXe scintillation emission is centered at \SI{178}{\nm}, with FWHM = \SI{14}{\nm}~\cite{Jortner:1965hi}. Light from electroluminescence in Xe gas has a similar spectrum, but not quite identical \cite{Chepel:2012sj}. Wavelength shifting is not required since the Xe luminescence spectrum is compatible with quartz-windowed photomultipliers. The basic optical properties of LXe are established: The refractive index for scintillation light is $n$ = \num{1.67}~\cite{Solovov:2003ax}, which is well matched to that of quartz ($n$ = \num{1.57}). This allows good optical coupling to the PMTs immersed in the liquid phase. The Rayleigh scattering length is \SIrange{30}{50}{\cm} (see \cite{Chepel:2012sj} and references therein), which must be considered in optical simulations.

The key issue is then to maximize the S1 photon-detection efficiency, $\alpha_1$, defined as the number of detected photoelectrons per emitted scintillation photon from the event site. The main factors affecting $\alpha_1$ in LZ are: (1) the VUV reflectivity of internal surfaces made from PTFE, especially those in the liquid; (2) the photon absorption length in the liquid bulk; (3) the geometric transparency and reflectivity of all grids; (4) the PMT photocathode coverage fraction; and (5) the PMT optical performance.

Most of these VUV optical properties have uncertainties that can result in significant differences in overall light collection. The factors affecting each of these properties are discussed in some detail below. In order to assess the detector performance we have adopted three sets of properties for light collection simulations which should bracket the final LZ response. The first is a realistic ``baseline'' set, which in most cases matches or only slightly exceeds the design requirements for individual optical parameters. The model which feeds into the estimation of the baseline LZ performance is slightly more conservative (\SI{\approx10}{\percent} lower $\alpha_1$) than this baseline optical model, to ensure that the baseline cross-section sensitivity is achieved with high probability. There is also a more optimistic (but still plausible) set of values which feed into the more ambitious sensitivity goal. Finally, for reference, there is a set of pessimistic values which are each somewhat unlikely, and highly unlikely to all occur together. These three sets of properties are listed in Table~\ref{XDSt:OpticalParameters}.

\begin{table}[tbh]
\setlength{\extrarowheight}{3pt}
\tdrtcaption[OpticalParameters]{Optical parameters}{Baseline as well as optimistic and pessimistic sets of optical properties considered for the LZ detector. The baseline LZ sensitivity is calculated using an optical model which is slightly more conservative than the optical baseline model (yielding a fiducial volume (FV) averaged $\alpha_1$ of \SI{7.5}{\percent}). The LZ sensitivity goal assumes the optimistic model.}
\centering
\sffamily
\begin{tabular} {|lccl|}
\hline
\rowcolor{mrocol}
Property & Pessimistic & Baseline & Optimistic \\
\hline
PTFE -- in liquid				& \SI{93}{\%}  & \SI{95}{\%} & \SI{97}{\%}  \\
PTFE -- in gas					& \SI{75}{\%}  & \SI{80}{\%} & \SI{85}{\%}  \\
Average PMT QE					& \SI{22}{\%}  & \SI{25}{\%} & \SI{28}{\%}  \\
Grid reflectivity (liquid and gas)	& \SI{0}{\%}  & \SI{20}{\%} & \SI{40}{\%}  \\
Absorption length in liquid			& \SI{15}{\ m}  & \SI{30}{\m}	& \SI{100}{\m} \\
\hline
FV-averaged S1 PDE ($\alpha_1$)		& \SI{5.5}{\%}	& \SI{8.5}{\%} & \SI{13.3}{\%} \\
\hline
\end{tabular}
\end{table}

In Figure \ref{XDSf:S1PDEScenarios} we show the photon detection efficiency (PDE, or $\alpha_1$) as a function of depth for the three sets of properties, and its partitioning into the top and bottom PMT arrays for the baseline properties. This is defined as the probability of a photon from the interaction site reaching a PMT and creating a photoelectron. The resulting S1 PDE averaged over a preliminary \num{5.6}-\si{\tonnel} fiducial volume is also given in Table~\ref{XDSt:OpticalParameters}. 

We effectively assume here that there is zero probability for single photons to create two photoelectrons. However, for the baseline we adopt an average PMT QE of \SI{25}{\percent}; while typical QEs reported by the manufacturer for LXe scintillation are more like \SI{30}{\percent}, this conservative QE matches the manufacturer specification---and it also allows for the \SI{20}{\percent} fraction of double-photoelectron emission, not captured by the QE specification, that has been observed in these tubes for LXe scintillation \cite{Faham:2015kqa}. Note that there will be an additional \SI{92}{\percent} efficiency for triggering on single photoelectrons, as discussed in Chapter~\ref{chap:EDC}. This is applied for LZ performance estimates, but is not included here.

\begin{figure}[ht]
\centering
\includegraphics[width=0.45\textwidth]{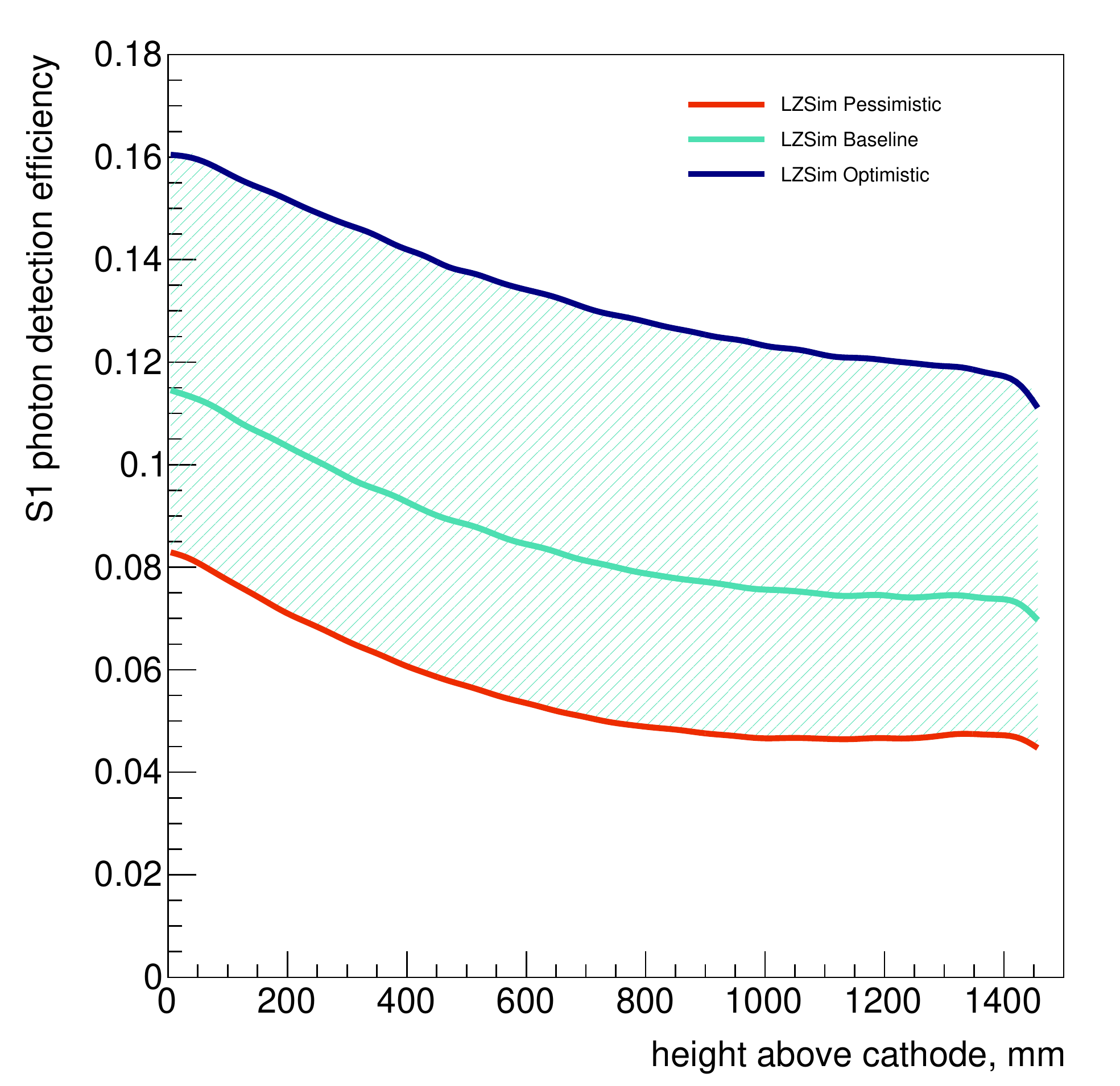}\quad
\includegraphics[width=0.45\textwidth]{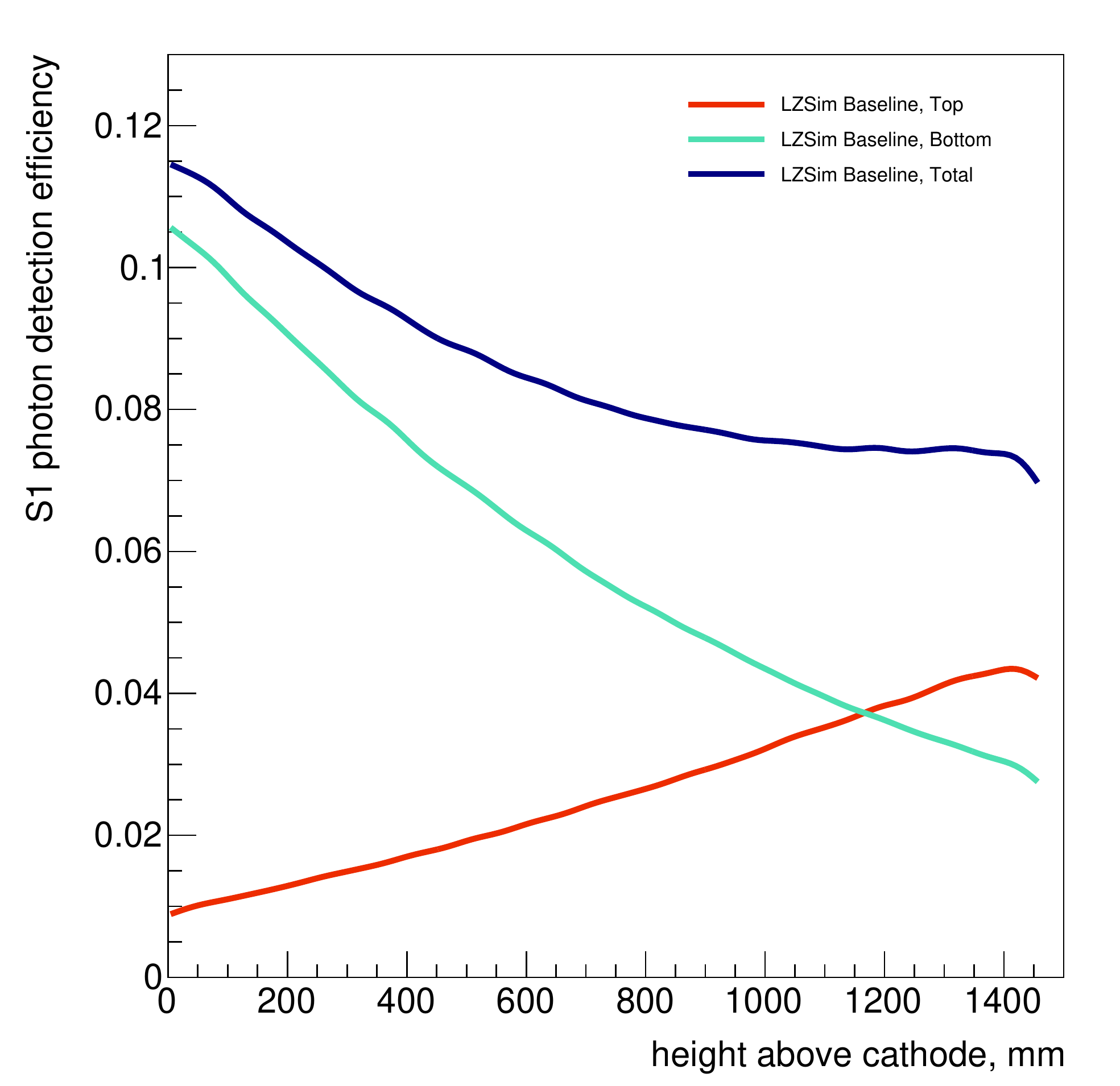}
\tdrfcaption[S1PDEScenarios]{S1 photon detection efficiency scenarios}{Left: Photon detection efficiency ($\alpha_1$) versus interaction height above the cathode for the three optical scenarios listed in Table~\ref{XDSt:OpticalParameters} (fiducial volume averages are also given in the table). Right: contributions from top and bottom arrays for the baseline optical model.}
\end{figure}

The baseline PDE of \SI{8.5}{\percent} translates to an S1 response of \SI{5.3}{phe\per\keV} at zero field for \ICoFS \Pgg-rays (a traditional measure of light yield in LXe chambers)---cf. \SI{8.8}{phe\per\keV} in LUX~\cite{Akerib:2013tjd}, \num{6.6} in XENON10~\cite{Aprile:2010bt,Sorensen:2008a}, \num{6.0} in PandaX-I~\cite{Xiao:2015psa}, \num{5.0} in ZEPLIN-III~\cite{Lebedenko:2008gb}, \num{4.3} in XENON100~\cite{Aprile:2011dd}, and \num{1.1} in ZEPLIN-II~\cite{Alner:2007ja}. According to NEST, the corresponding NR energy threshold is \SI{\approx5.3}{\keV} for a 3-phe coincidence requirement (while we adopt \SI{5.8}{\keV} for sensitivity estimates). Note that a lower, 2-fold coincidence may be possible (as in LUX) that would lower this threshold.

There are several basic features of the light collection response shown in Figure~\ref{XDSf:S1PDEScenarios}. The first, obvious from the figure, is that for most depths more light is collected in the bottom array that the top. This is due to the strong total internal reflection at the liquid surface due to the mismatch in refractive index, giving a critical angle for total internal reflection of \SI{36}{\degree}, and also the good match in VUV refractive indices between the quartz in the PMT windows and LXe. The second is the high degree of scattering from PFTE surfaces. This can be seen by considering the \SI{8.5}{\percent} mean PDE (baseline scenario), which, with a \SI{25}{\percent} PMT QE, implies a \SI{34}{\percent} probability of a photon striking a photocathode, while the photocathode coverage is only \SI{\approx14}{\percent} of the total TPC internal surface. In the simulations the average number of scatters on PTFE is a depth dependent value between about \num{3} and \num{5}. Thus, the value of PTFE reflectivity is quite important: the detector is effectively a ``mirrored box'' in which the value of light collection is the result of a competition between a photon being detected at a photocathode, and absorption that occurs with low probability but a high number of chances as the photon scatters around the detector. Finally, though not obvious in the figure, is that the mean path length traveled is a depth-dependent value between \num{3} and \SI{5}{\m}. Rayleigh scattering is also strong (with \SI{30}{\cm} length, the mean number of Rayleigh scatters is between \num{10} and \num{20}), but does not significantly alter either the pattern or amount of light collection.

\tdrsubsec[S1OpticalProperties]{TPC Optical Properties}

Here we discuss the reflectivity of PTFE in LXe, the absorption length in LXe, and the optical properties of the grids. The values shown in Table~\ref{XDSt:OpticalParameters} are based on our best understanding, but, remarkably, they also happen to result in roughly equal loss of photons to absorption in the PTFE, LXe, and the grids. Somewhat modest gains (or losses) can be achieved by maximizing (or doing less well in) any one of these parameters, while the combined effect of improving (or doing worse in) all three could result in perhaps doubling (or near halving) of the overall light collection.

The photon absorption length in the bulk LXe depends on the purity of the liquid with respect to trace amounts of contaminants with absorption bands overlapping the LXe scintillation spectrum, mostly \CHtO and \COt. For the tight purity requirements for those electronegative species (\SI{0.1}{\ppb}, see Chapter~\ref{chap:XCS}), values in excess of \SI{100}{\m} can be expected (see Figure~15 in \cite{Chepel:2012sj}). We have adopted a somewhat conservative \SI{30}{\m} baseline, along with \SI{15}{\m} and \SI{100}{\m} as pessimistic and optimistic values, respectively. The S1 PDE depth profile is shown in Figure~\ref{XDSf:S1PDEEffects} (left) for several values of this parameter. In the baseline and pessimistic scenarios, absorption is a comparable loss term to absorption on PTFE and grids, but becomes sub-dominant with the optimistic \SI{100}{\m} length and achieving even longer lengths gives diminishing benefit.

\begin{figure}[ht]
\centering
\includegraphics[width=0.45\textwidth]{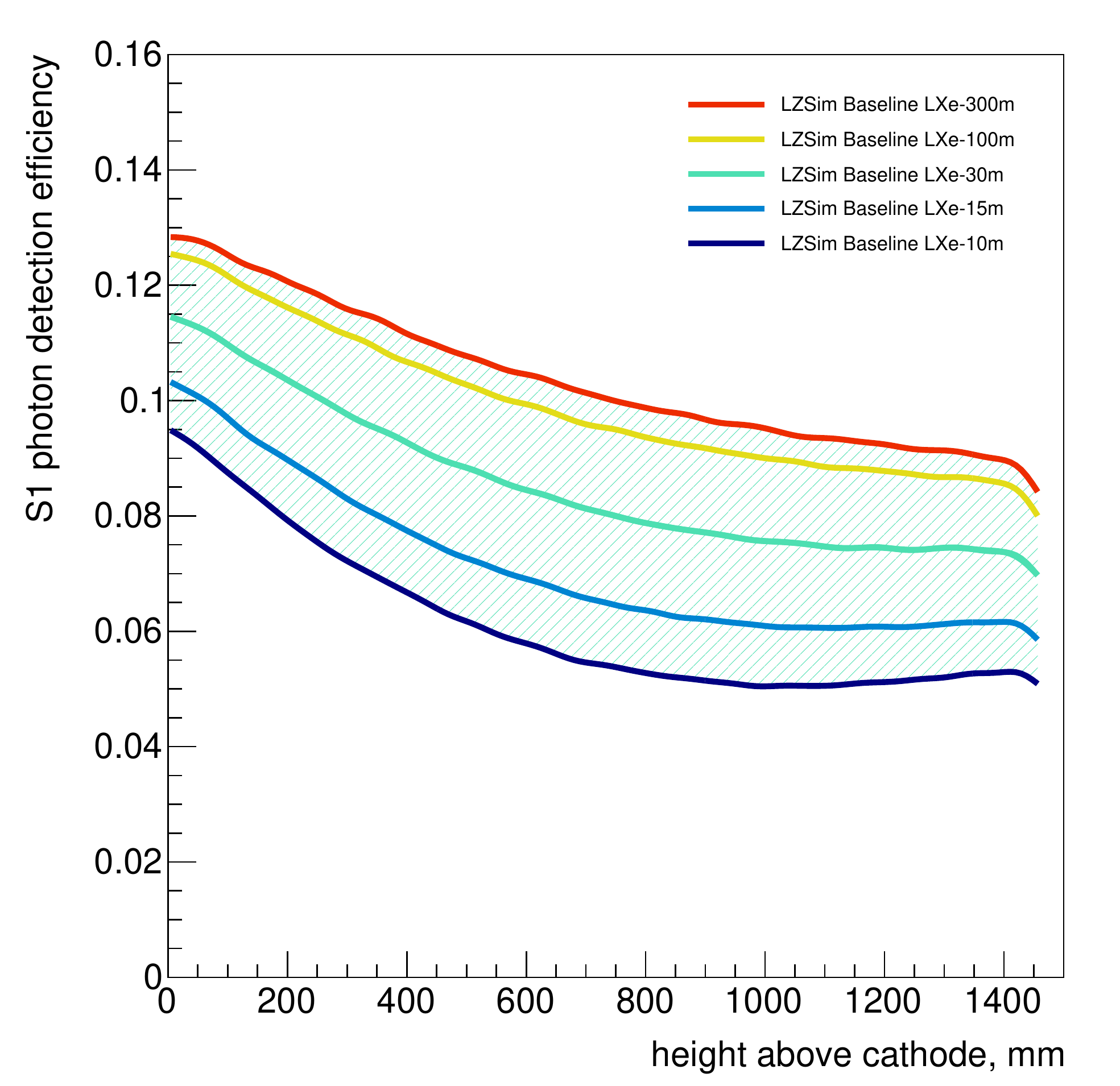}\quad
\includegraphics[width=0.45\textwidth]{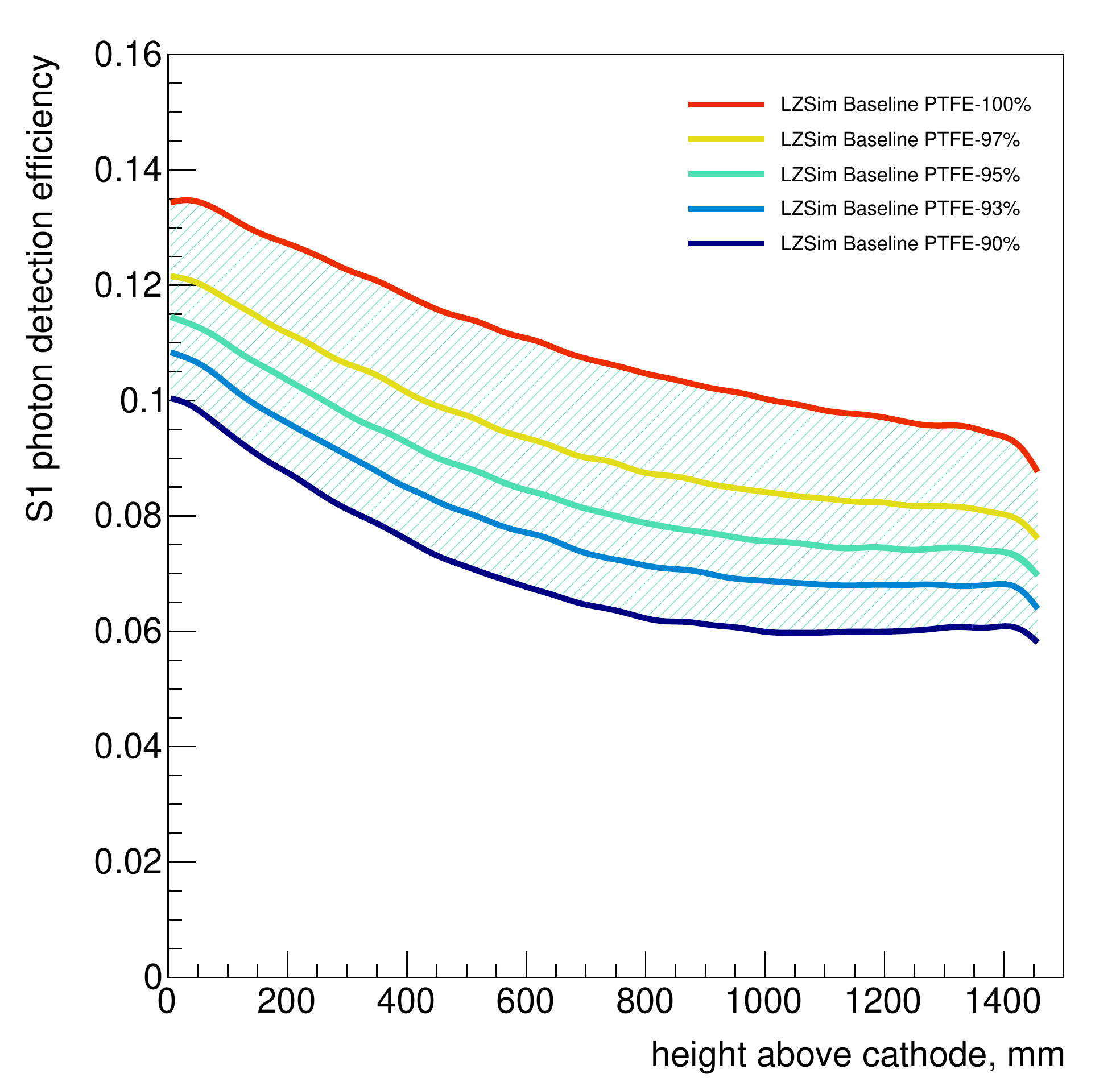}
\tdrfcaption[S1PDEEffects]{S1 PDE as a function of photon absorption length and PTFE reflectivity}{Left: S1 PDE ($\alpha_1$) as a function of the photon absorption length in the liquid; Right: S1 PDE as a function of the PTFE/LXe interface reflectivity.}
\end{figure}

The four electrode grids in the LZ TPC all affect light collection through obscuration and wire reflectivity and, in a chamber where other sources of optical extinction have been minimized, these grids can have a significant effect on the light yield. The limited literature on reflectivity for metals at \SI{178}{\nm} indicates that very high values are unlikely, hence we have adopted values between \SI{0}{\percent} and \SI{40}{\percent}. This range is motivated by expectation that the specific surface treatment of the stainless steel grid material might have an effect on reflectivity, and that the final electropolishing step that we take in grid production could be beneficial. It is conceivable that a (conductive) reflective coating could be deployed (e.g., Al). The grid opacities shown (for normal incidence) in Table~\ref{XDSt:ElectrodeGrids} are selected as a compromise between minimal opacity and minimizing the electric field at the wire surfaces, especially for the cathode and the gate grid, which is also cathodic. It is possible that the ongoing HV tests of grids (Section~\ref{XDSS:SystemTest}) may allow to deploy grids with higher geometrical transparency, which has a similar effect as increasing the reflectivity.  Because light is partially trapped in the liquid phase by total internal reflection at the liquid surface, the grids in liquid have a relatively larger effect than the anode in the gas. This is fortunate, because considerations of generating uniform S2 light, discussed in Section~\ref{XDSS:S2Light}, lead to the anode being the least transparent of the grids.

Perhaps the most important optical parameter is the reflector used in constructing the TPC. Based on a decade of experience, PTFE is indeed the best reflector for LXe scintillation, and also for properties other than optical: The manufacturing process yields very radiopure material (\SI{\sim1}{\ppt} in U/Th); it has good mechanical properties (despite the \SIrange{1.4}{1.5}{\percent} thermal contraction to LXe temperatures~\cite{White:2002,DuPontPTFE}); and outgassing rates are relatively low. The reflectivity values we have adopted are based on optical simulations of the performance of LUX, similar analysis of other detectors, and recent results from small chamber measurements ongoing at LIP-Coimbra and recently starting up at Michigan. These test studies are discussed in Section~\ref{XDSS:SystemTest}. The effect of this parameter on the S1 PDE is shown in Figure~\ref{XDSf:S1PDEEffects} (right).

To reduce the dead volume around the active LXe, as well as outgassing and potential backgrounds, it is desirable to minimize the thickness of the PTFE walls of the TPC and Skin detectors. A lower limit is established by the transmittance of PTFE to Xe scintillation light, and the need for optical isolation between the TPC and Skin regions as well as between these and any dead regions containing LXe. The transmittance of the PTFE used in LUX was measured at LIP-Coimbra as a function of thickness and for different wavelengths: Xe gas scintillation (\SI{178}{\nm}) as well as \SIlist{255; 340; 470}{\nm}. Results for Xe scintillation light show a transmittance \SI{<0.1}{\percent} for \SI{1.5}{\mm} PTFE thickness (but rising significantly to as much as \SI{10}{\percent} for \SI{5}{\mm} in the case of \SI{470}{\nm} blue light). The TPC wall thickness in LZ is \SI{15}{\mm}, while the covering tiles of the vessel in the skin region is \SI{1}{\mm} and a similar though non-uniform thickness is used on the PMT arrays.

\tdrsubsec[OptArray]{PMT Arrays}

The two PMT arrays, shown in Figures~\ref{XDSf:PMTarrays} an \ref{XDSf:PMTArrayLayout}, are similar, but not identical. Since most light is collected in the bottom array, it is a closed-packed hexagonal array of \num{241}~PMTs, optimized purely for maximum photocathode coverage within the mechanical constraints of the low-mass structure that holds the PMTs. The mechanics of the array and PMT mounting are discussed in Section~\ref{XDSSs:PMTassy}. In the bottom array the \SI{76}{\mm} diameter tubes are housed in \SI{80}{\mm}-diameter holes on an \SI{82.5}{\mm} center-to-center spacing, resulting in a \SI{54}{\percent} coverage fraction of the \SI{64}{\mm} diameter photocathode surfaces. The spaces between the photocathode surfaces are fully covered with PTFE.

\begin{figure}[ht]
\centering
\includegraphics[width=0.45\textwidth]{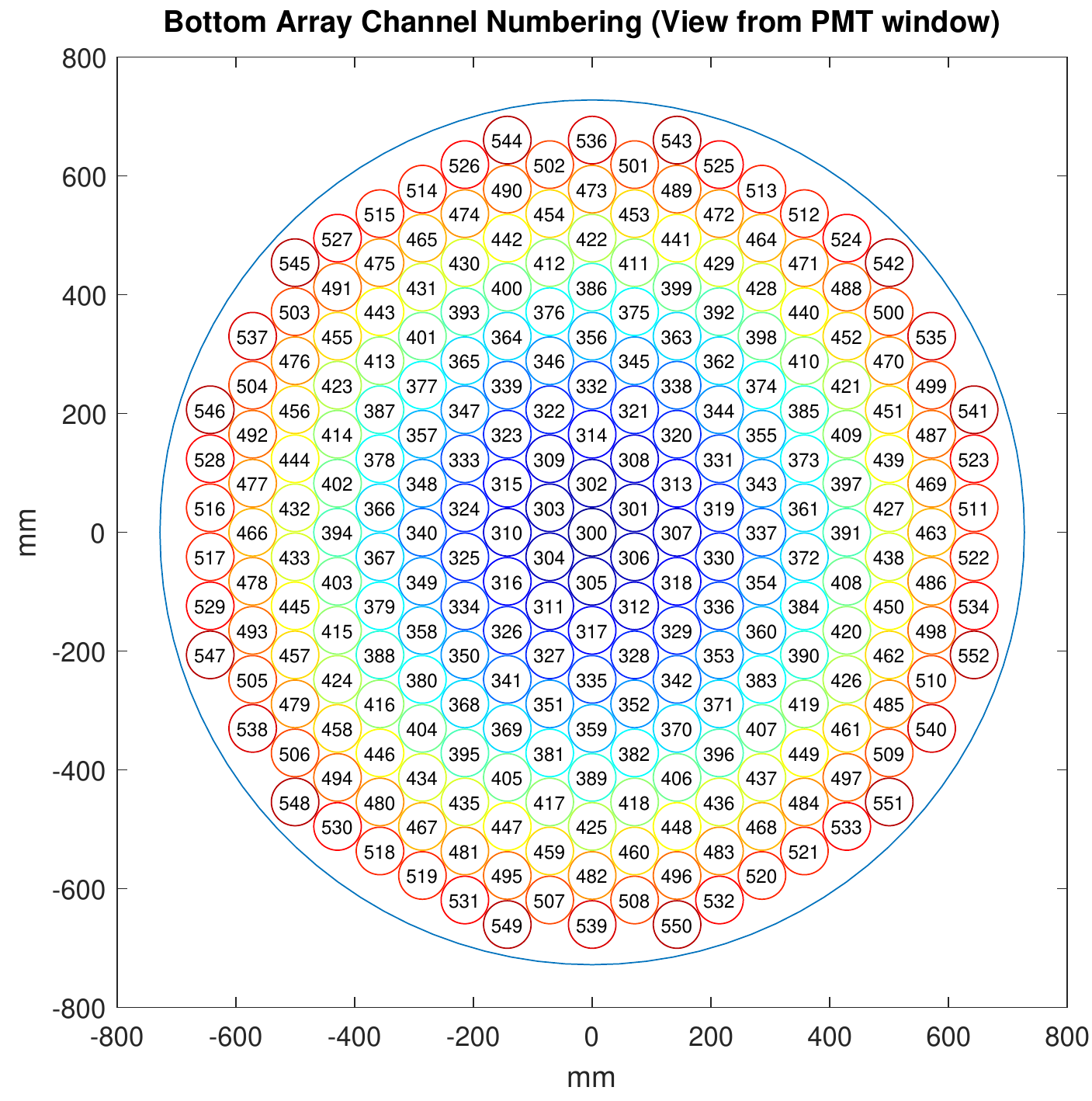} \quad
\includegraphics[width=0.45\textwidth]{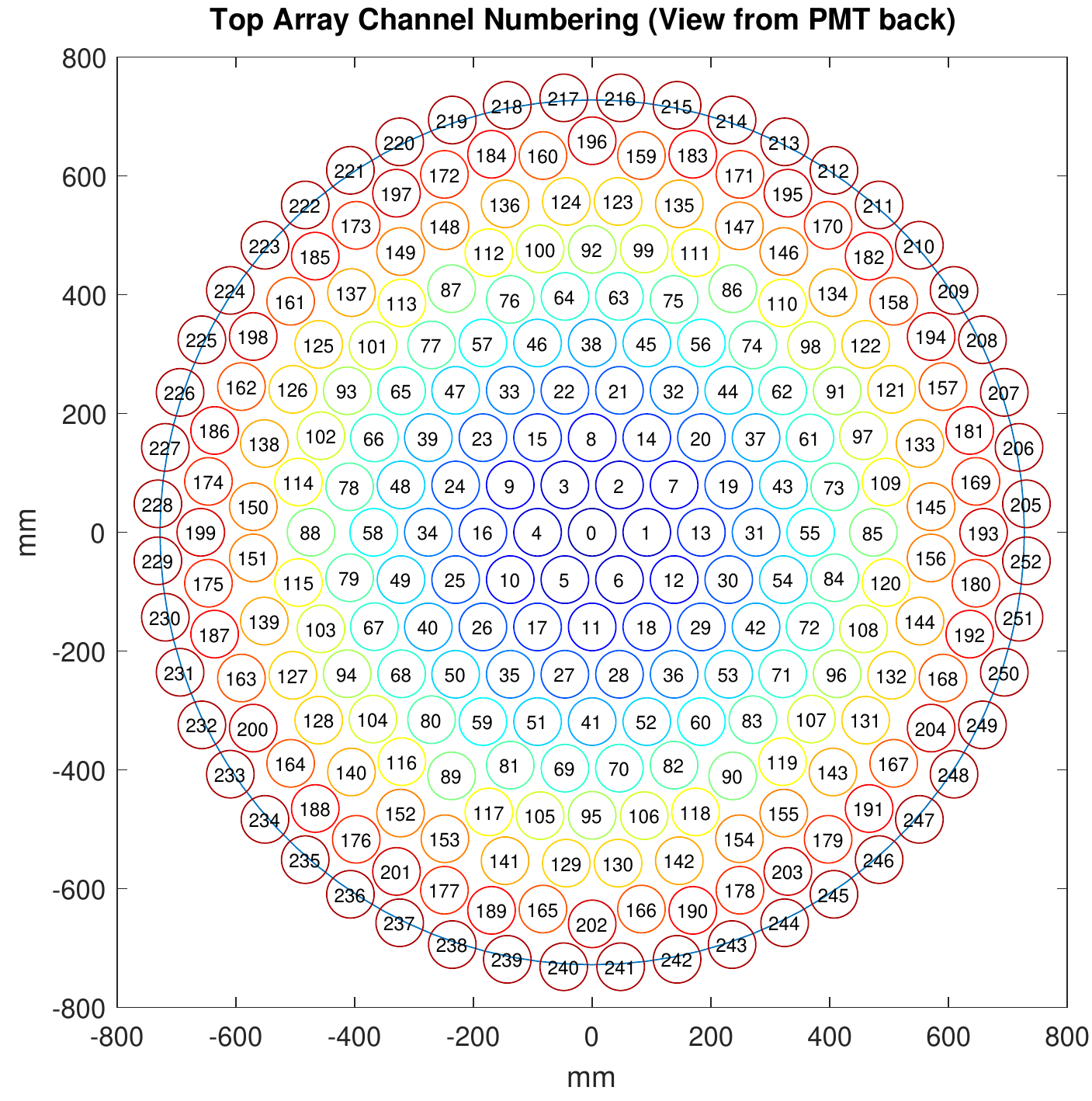}
\tdrfcaption[PMTArrayLayout]{Layout of bottom and top PMT arrays}{Layout of the bottom (left) and top (right) PMT arrays viewed from above. The channel numbering is done by radial position, with similar distances to the center indicated by color.}
\end{figure}

By contrast, while the top array collects a significant fraction of S1 light for events near the top of the detector, this array has the crucial role of  reconstructing the $x$,$y$ location of events from the S2 signal. Especially critical is the accuracy of reconstructing the position of ``wall events'' that result from interactions near the vertical cylindrical surface of the PTFE that defines the TPC. A particular concern is the population of decays from radon progeny plated out on the TPC walls. These consist of ER interactions from the beta decays and gamma-ray emission from the \IPbtoz sub-chain, and from alpha particles and \SI{\approx100}{\keV} \IPbtzs nuclear recoils from the alpha decay of \IPotoz. All of these, due to both energy and charge loss at the wall, lead to a broad distribution of low-energy signals that in part overlaps with the NR signal region in S2-S1 space. Therefore, it is critical to minimize the leakage of these events towards the center of the detector from errors in the reconstruction algorithm in order not to compromise the fiducial mass at low energy. Here the placement of the outer few PMT rows is critical. In contrast to the bottom array design, which is fully contained within the TPC diameter, the top array must overhang the edge of the TPC or all the reconstruction bias will point inward. Ideally, at least a full row of tubes would be located beyond the inner radius of the chamber. This is not possible due to the proximity of the inner cryostat vessel, and instead we locate the outermost circle of tubes at the largest-possible radius, which aligns the PMT centers above the TPC wall. From among a study of several layouts we adopted a ``hybrid'' array of \num{253} PMTs which has two nearly circular rows of PMTs at the perimeter but transitions to an hexagonal pattern in the center. The two outer circular rows maximize uniformity at the edge of the detector, which improves the uniformity of the $x$,$y$ response near the wall and minimizes inward leakage. The larger overall diameter leads to the larger number of tubes, and the array is slightly less compact it the central hexagonal region, with a \SI{93}{\mm} center-to-center spacing (with the same \SI{80}{\mm} holes as the bottom array). We return to this optimization in Section~\ref{XDSSs:WallEvents}.

\tdrsec[S2Light]{S2 Production and Detection}

Electrons escaping the interaction site are drifted under the influence of the ``drift'' electric field in the central part of the TPC to the electroluminescence region where they create the S2 signal. In this section we discuss the design of this region, including that of the electrode grids that create the drift and extraction/electroluminescence fields. The field cage that works in conjunction with the grids to generate the drift field is described in Section~\ref{XDSS:TPC}.

The electroluminescence region of the TPC is located at the top of the field cage, with the gate and anode electrodes (nominally \SI{13}{\mm} apart) straddling the liquid surface. The liquid level is controlled by a weir system at the edge of the TPC, as detailed in Section~\ref{XDSS:Fluids}. This region controls the emission of the drifting electrons into the vapor phase and the subsequent production of electroluminescence photons, in proportion to the number of ionization electrons drifted away from the interaction site. This response channel readily provides sensitivity to single ionization electrons emitted from the liquid.

The S2 signal is also used for spatial localization of the interactions. In particular, the accurate reconstruction of ``wall events'' drives both the gain of the S2 response (involving the optimization of both photon production and collection efficiency) and of the layout of the top PMT array; this optimization is discussed below too.

Three main parameters characterize the S2 response: (1) the photoelectron yield, which depends itself on the cross-surface extraction probability for ionization electrons, the electroluminescence gain, and the efficiency of light collection for photons generated in the electroluminescence region ($\alpha_2$); (2) the S2 pulse width, which is proportional to the electron transit time in the gas phase to first order; and (3) the resolution of the S2 signal, which depends on the detailed electric field distribution near the grid wires. Once the gate-anode separation is fixed (and therefore the gas gap $L_g$), these characteristics depend on operating parameters such as the xenon vapor pressure $P$ and the voltage applied to those electrodes $\Delta V$.

The S2 performance is affected by other electrostatic considerations (such as the maximum fields that can be sustained at the wire surfaces), and by mechanical considerations, such as the manufacture feasibility of large and densely-meshed wire grids, grid deflection and non-parallelism, etc. In the following sections we highlight the baseline design and the design goal, and describe how the S2 response depends on these operating conditions.

\tdrsubsec[S2production]{S2 Photon Production}

For the smallest S2 signals, generated by one to a few ionization electrons, the main S2 requirements are: (1) definition of the single-electron response with a high signal-to-noise ratio, to allow absolute calibration of the ionization channel and to enable physics searches down to S2 signals as small as a few electrons; and (2) sufficiently large S2 signal for accurate reconstruction of the ($x$,$y$) location of peripheral interactions, such as those arising from contamination on the TPC walls. This motivates a photon yield of at least \num{\approx50}~photoelectrons per emitted electron, especially at the edge of the TPC. More details on how the S2 pulse size affects the reconstruction of wall events was given in LZ CDR~\cite{Akerib:2015cja}. Key parameters related to S2 photon production in LZ are presented in Table~\ref{XDSt:S2parameters}.

\begin{figure}[ht]
\centering
\includegraphics[width=0.7\textwidth]{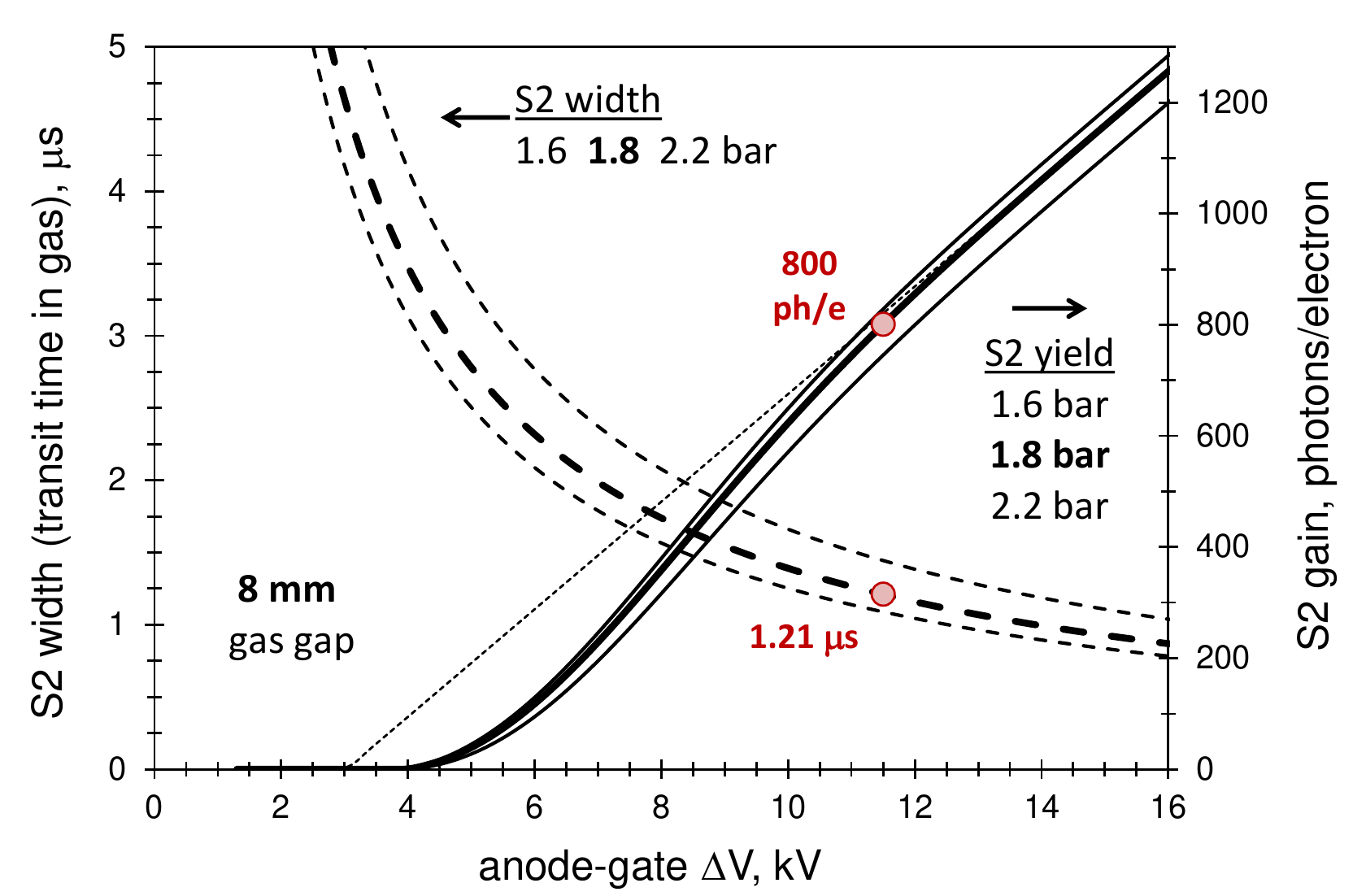}
\tdrfcaption[S2yield]{S2 photon yield and pulse width} {Dependence of the S2 photon yield and mean S2 pulse width (ignoring longitudinal diffusion in the liquid) on the voltage between anode and gate electrodes. The nominal photon yield~\cite{Fonseca:2004cd}, including the electron emission probability~\cite{russians1979}, and the electron transit time in the gas phase (S2 pulse width)~\cite{Santos:1994}, are indicated at the nominal $\Delta V$=\SI{11.5}{\kV}, for operating pressures around the \SI{1.8}{\bar} nominal and a gas gap of \SI{8}{\mm}).}
\end{figure}

Considering an S2 photon detection efficiency of \SI{\approx5}{\percent} for the top array for peripheral interactions, predicted by simulation as presented below, the above photoelectron yield implies a minimum of \num{800} photons generated per emitted electron. For a gate-anode distance of \SI{13}{\mm} with $L_g$=\SI{8}{\mm}, this is achieved with $\Delta V$=\SI{11.5}{\kV} at the operating pressure $P$=\SI{1.8}{\bar}, as shown in Figure~\ref{XDSf:S2yield}. A voltage of \SI[retain-explicit-plus]{+5.75}{\kV} will be applied to the anode and \SI{-5.75}{\kV} to the gate, leaving the liquid surface near \SI{-2.7}{\kV}. Note, however, that the total gate-anode voltage is the real requirement, which gives the flexibility to trade-off between anode and gate voltages in case the nominal division becomes problematic (the anode sits in the gas phase, which is less resilient electrically and can give rise to spurious electroluminescence, but the gate is a cathodic electrode, and therefore susceptible to spurious electron emission).

All of these parameters are intimately connected to S2 light production: Both the electroluminescence yield and the electron drift velocity in the gas are determined by the reduced electric field in that region, $E/P$; in addition to the applied voltages, the electric field depends on both $L$ and $L_g$. Therefore, these parameters must be studied together and their optimization is subtle. We described two viable configurations in the CDR and the one described here is intermediate between them in terms of gas gap and photon yield. The gas gap is sufficiently large to provide high gain, low variance of S2 photon production, and allowance for electrostatic deflection during operation (which will force the grids closer together at the detector center), but small enough to preserve the required dynamic range of the optical readout.

\begin{table}[tbh]
\setlength{\extrarowheight}{3pt}
\tdrtcaption[S2parameters]{Main S2 parameters}{Main S2-related parameters. The predicted S2 response is indicated for the parallel-field model assumed in Figure~\ref{XDSf:S2yield}, as well as for the more detailed modeling illustrated in Figure~\ref{XDSf:S2sims}. A nominal yield of \num{800}~ph/e is used for sensitivity calculations (this includes a non-unity emission probability.}
\centering
\sffamily
\begin{tabular}{|lr|}
\hline
\rowcolor{mrocol}
Parameter & value \\
\hline
Gate-Anode separation (and tolerance)	& \SI{13.0}{\mm} (\SI{\pm 0.2}{\mm}) \\
Gas gap (and tolerance) & \SI{8.0}{\mm} (\SI{\pm 0.2}{\mm}) \\
Field in LXe (GXe)		& \SI{5.2}{\kV\per\cm} (\SI{10.2}{\kV\per\cm}) \\
\hline
Electron emission probability	& \SI{97.6}{\%} \\
\hline
S2 photon yield		& \SI{820}{ph/e} \\
S2 width FWHM		& \SI{1.2}{\mus} \\
\hline
\multicolumn{2}{|l|}{Detailed modeling}\\
\hline
S2 photon yield  		& \SI{910}{ph/e} \\
S2 photon {\em rms}		& \SI{2.0}{\%} \\
S2 width FWHM			& \SIrange{1.0}{2.0}{\mus}$^a$ \\
\hline
\multicolumn{2}{l}{\begin{minipage}{0.5\textwidth}
\vspace{10pt}
\small $^a$ The larger value is for diffusion-broadened S2 pulses from interactions near the cathode (see Figure~\ref{XDSf:S2sims}).
\end{minipage}}\\
\end{tabular}
\end{table}

Nonetheless, the choice of higher voltages across a longer gap has some disadvantages which must be mitigated. The electron emission probability at the liquid surface decreases rapidly when the field in the gas drops below \SI{10}{\kV\per\cm}~\cite{russians1979}. In Figure~\ref{XDSf:S2yield}, the S2 yield assuming full extraction efficiency is represented by the dotted line, while the continuous lines include the field-dependent extraction probability. For our nominal parameters, that probability is close to unity, but poor extraction efficiency may result if nominal voltages fail to be achieved. We mitigate this with Phase-II of our System Test program described in Section~\ref{XDSSs:SLACPhaseIIST}---this includes testing the final gate-anode assembly to ensure that the design voltages can be realized.

Longer electron transit times in the gas also hide the effect of electron diffusion in the liquid, which encodes (modest) interaction-depth information on the S2 pulse shape. This information allows some coarse fiducialization, which is important for an ``S2-only'' analysis. The mean S2 pulse width is \SI{1.2}{\mus}, which will make the precise measurement of diffusion-broadening of the S2 response more difficult. If required, this may be mitigated with operation at lower drift field (see Table~\ref{XDSt:CathodeHV}).

\tdrsubsec[S2detection]{S2 Photon Detection}

The LZSim model was used to obtain the baseline photon detection efficiency for the S2 response ($\alpha_2$), representing the fraction of photons which are detected by the PMT arrays. This parameter multiplies the photon yield per electron discussed in Section~\ref{XDSSs:S2detection} and the number of ionization electrons drifted away from the interaction site to yield the total S2 response (assuming no loss to electronegative impurities). The dependence of $\alpha_2$ on radius is shown in Figure~\ref{XDSf:S2PDEScenarios} and values for central and peripheral locations are summarized in Table~\ref{XDSt:S2PDE}.

\begin{figure}[ht]
\centering
\includegraphics[width=0.45\textwidth]{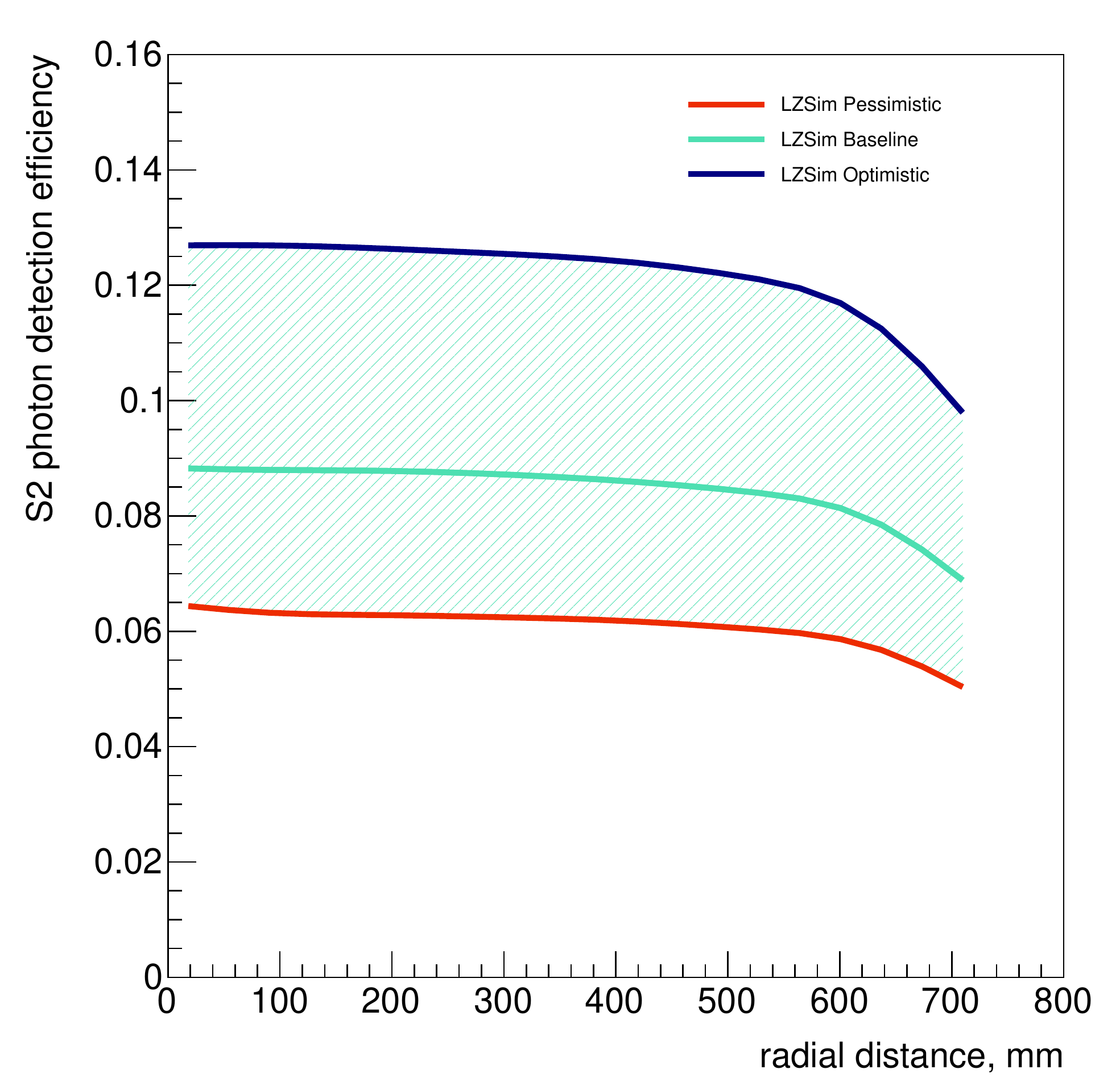}\quad
\includegraphics[width=0.45\textwidth]{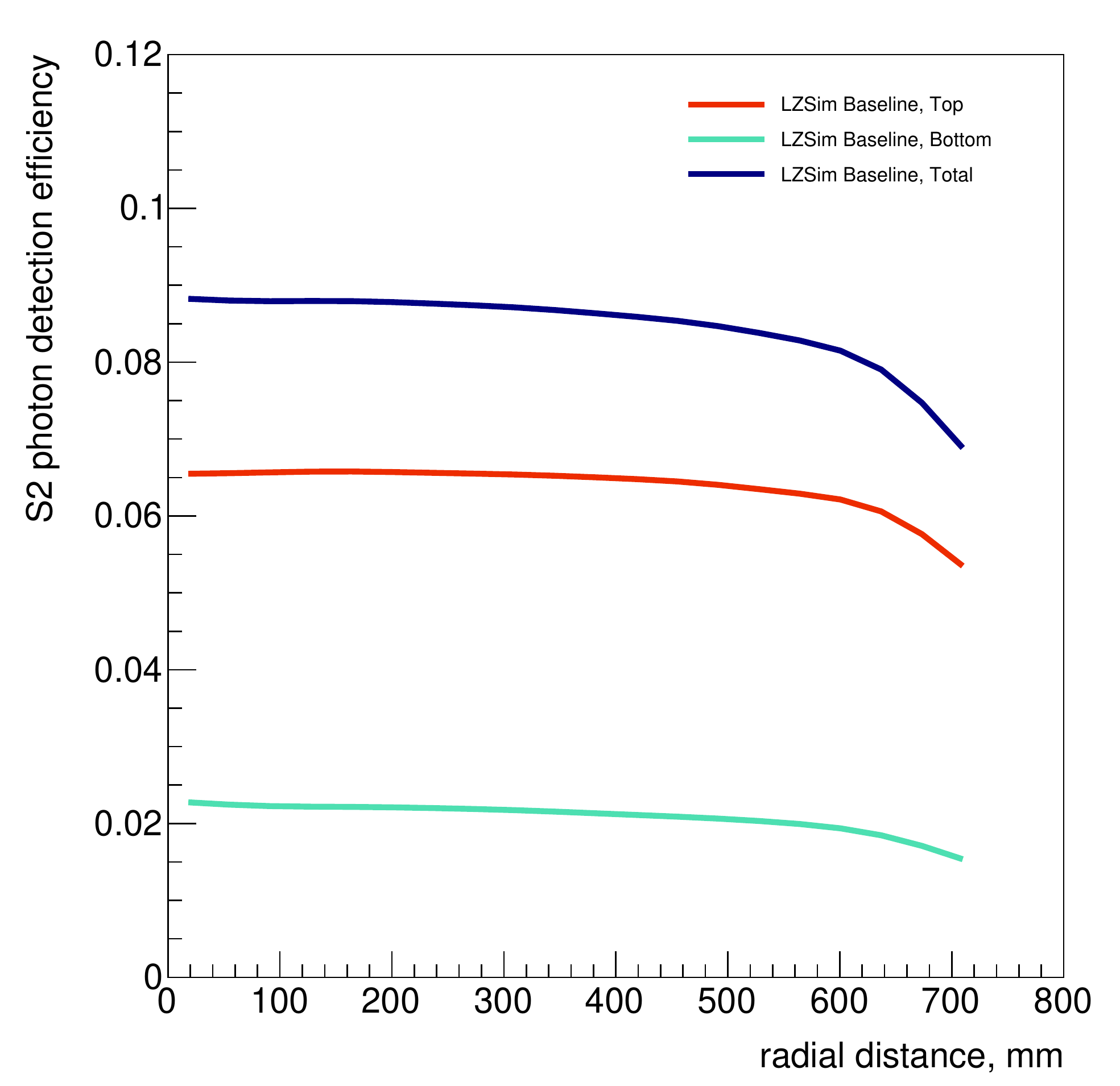}
\tdrfcaption[S2PDEScenarios]{S2 photon detection efficiency}{Left: Radial dependence of S2 photon detection efficiency for several scenarios; the nominal optical model averages $\alpha_2$=\SI{12.0}{\percent} and is used to calculate the LZ baseline performance; Right: Contributions from top and bottom arrays for baseline optical model.}
\end{figure}

The optical parameters are as assumed for the $\alpha_1$ simulation described in Section~\ref{XDSSs:S1OpticalProperties}. In this instance, the top array makes the largest contribution to $\alpha_2$ as expected, partly due to the optical mismatch between the gas and liquid phases. Other driving parameters are the anode transparency and the reflectivity of PTFE in the gas phase (trefoil structures surrounding the PMT windows).

\begin{table}[tbh]
\setlength{\extrarowheight}{3pt}
\tdrtcaption[S2PDE]{S2 photon detection efficiency}{S2 photon detection efficiency ($\alpha_2$) and photoelectron yield for single electron signals (in brackets, assuming \SI{800}{ph/e}) for central and TPC-wall interactions.}
\centering
\sffamily
\begin{tabular} {|lcc|}
\hline
\rowcolor{mrocol}
PMT array & Center & Edge\\
\hline
Top		& \SI{6.6}{\%} (\SI{52}{phe/e}) & \SI{5.4}{\%} (\SI{43}{phe}) \\
Bottom	& \SI{2.2}{\%} (\SI{18}{phe/e}) & \SI{1.5}{\%} (\SI{12}{phe}) \\
\hline
Top+Bottom & \SI{8.8}{\%} (\SI{70}{phe/e})& \SI{6.9}{\%} (\SI{55}{phe}) \\
\hline
\end{tabular}
\end{table}

Towards the center of the TPC $\alpha_2$ exceeds \SI{8}{\percent}, decreasing to below \SI{7}{\percent} for wall events, which enables a single electron response comfortably above \SI{50}{phe/e} everywhere in the detector. For wall events the top array detects some \SI{40}{phe/e}, which is sufficient for the effective reconstruction of plateout backgrounds as discussed later in this section.

The bottom array converts just over \SI{2}{\percent} of the S2 photons, with each PMT contributing a fairly constant \SI{0.01}{\percent} to the overall detection efficiency in that array. This allows LZ to reconstruct large S2 pulses which will saturate many channels in the top array, where an individual PMT located just above the S2 vertex can convert \SI{2}{\percent} of the S2 photons---as much as the whole of the bottom array.

Using instead the optimistic optical parameters presented in Section~\ref{XDSSs:S1OpticalProperties}---which bring the baseline $\alpha_1$ from \SI{8.5}{\percent} to \SI{13.3}{\percent}---the effect on $\alpha_2$ is comparable: the baseline S2 detection efficiency increases from \SI{8.8}{\percent} to \SI{12.7}{\percent} for central interactions, with roughly equal gains in both arrays in absolute terms.

\tdrsubsec[S2resolution]{S2 Resolution and Electrode Configuration}

In addition to appropriate S2 gain and pulse width, we must ensure that the overall energy resolution of the TPC is good. At low recoils energies (ER and NR) this is dominated by statistical fluctuations on the small number of detected S1 and S2 quanta (including significant recombination fluctuations) and the quality of the S2 design is unlikely to be a dominant factor for WIMP searches. However, for \si{\MeV} electron-recoil energies, instrumental effects eventually dominate the overall resolution since a combined energy scale using both S1 and S2 can eliminate the recombination fluctuations, and the number of S1 and S2 quanta is large.

This demands a low dispersion of photon production from the electroluminescence region, which in turn drives the detailed design of the gate and especially the anode grids. This is intimately related to the characterization of detector backgrounds, in particular enabling high-quality spectroscopy of radioactivity gamma-rays generating up to \num{\sim e5} electrons---see Figure~\ref{XDSf:EReres}. These are important to understand low-energy, external ER backgrounds, but especially to constrain radioactivity neutrons. Clearly, fine energy resolution is also required if LZ is to attempt the detection of $0 \nu\beta\beta$-decay in \IXeoTs. We quantify our resolution requirement at the respective Q-value of \SI{2458}{\keV}, and we designed to achieve \SI{2}{\percent} in the combined energy scale (\SI{1.5}{\percent} goal).

\begin{figure}[ht]
\centering
\includegraphics[width=0.8\textwidth]{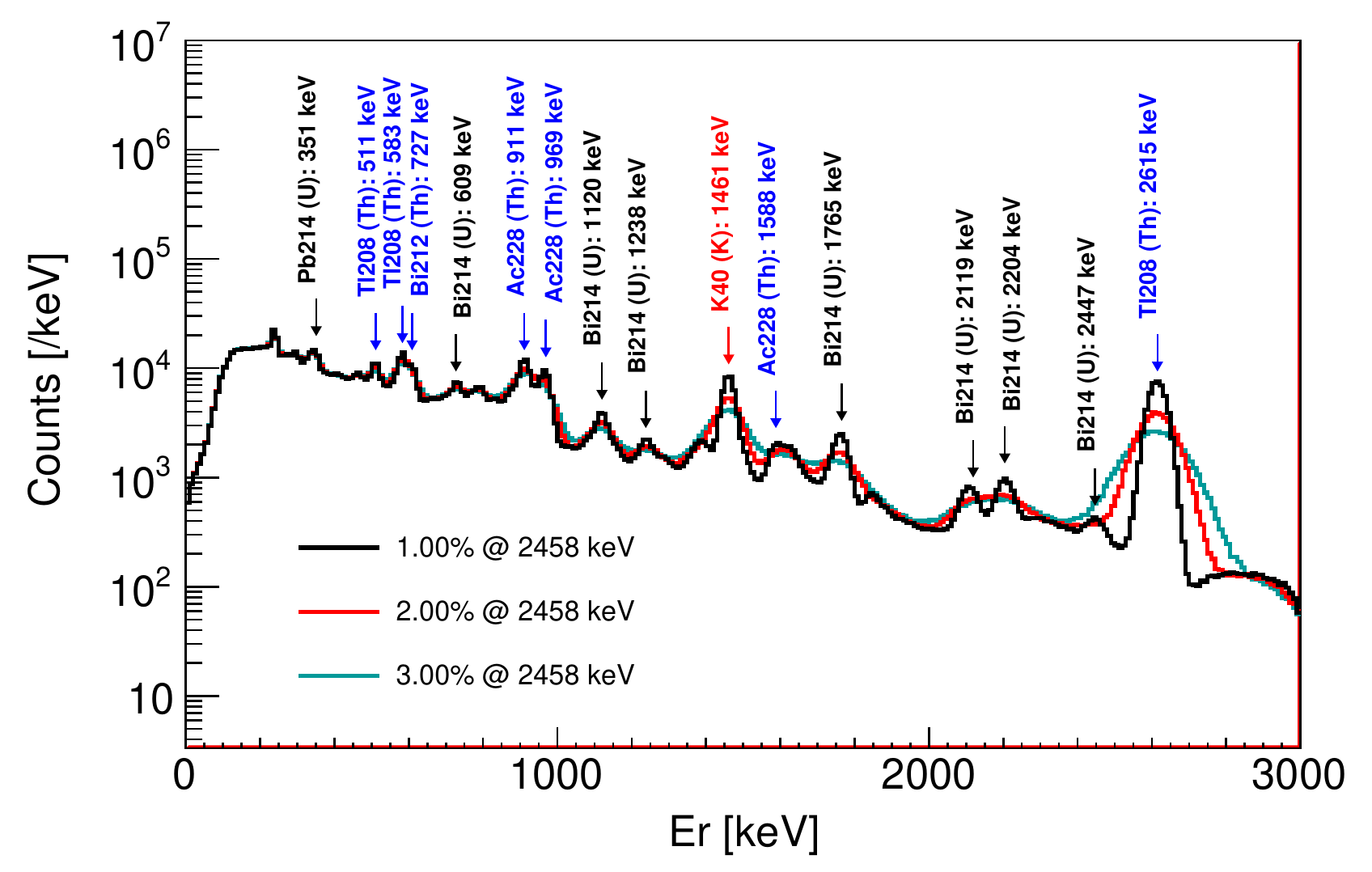}
\tdrfcaption[EReres]{Energy-resolution for high energy electron recoils}{Energy-resolution for high energy electron recoils (S1-S2 combined energy scale). Spectrum is simulated ER background from the titanium cryostat in the full active volume in \num{1000} days.}
\end{figure}

Therefore, fluctuations related to S2 photon production and detection must remain small, at \SI{\sim1}{\percent} level. This motivated a detailed study of electroluminescence and electrode grid configuration presented in the LZ CDR, and subsequently the selection of the electrode grids listed in Table~\ref{XDSt:ElectrodeGrids}. Other factors contributing to the S2 resolution are the uniformity of response in the horizontal plane over the whole TPC diameter (e.g., wire sagging and electrostatic deflection), although those can be calibrated and hence removed to first order.

Aside from diffusion (in both gas and liquid phases), drifting electrons follow electric field lines and therefore their length and the field strength close to the wires must be carefully controlled to avoid substantial dispersion or even the possibility of significant charge multiplication (the first Townsend coefficient for cold Xe vapor at the \SI{1.8}{\bar} operating pressure reaches \SI{1}{e\per\mm} at \SI{35}{\kV\per\cm}~\cite{Santos:1994,Biagi:1999nwa}). Two additional concerns, which are intimately related, are the ability of the electrodes to hold HV and their VUV reflectivity, which depend strongly on the wire material and surface properties; we have investigated these issues through the dedicated R\&D activities described in Section~\ref{XDSSs:SmallWireST}.

\begin{table}[tbh]
\setlength{\extrarowheight}{3pt}
\tdrtcaption[ElectrodeGrids]{TPC electrode grid parameters}{TPC electrode grid parameters (all \SI{90}{\degree} woven meshes). Surface fields are wire-average for \SI{-100}{\kV} cathode voltage. The geometric opacity is given at normal incidence.}
\centering
\sffamily
\begin{tabular} {|lrrcrc|}
\hline
\rowcolor{mrocol}
Electrode & Voltage & Wire diameter/pitch & Number & Wire field & Opacity \\
\hline
Anode   &  \SI[retain-explicit-plus]{+5.5}{\kV}	& \SI{100}{\mum} / \SI{2.5}{\mm} & \num{1184}& \SI[retain-explicit-plus]{+55}{\kV\per\cm} & \SI{8.0}{\%} \\
Gate    &  \SI{-5.5}{\kV}	&  \SI{75}{\mum} / \SI{5.0}{\mm} & \num{592} & \SI{-62}{\kV\per\cm} & \SI{3.0}{\%}  \\
Cathode &  \SI{-50}{\kV}	& \SI{100}{\mum} / \SI{5.0}{\mm} & \num{592} & \SI{-31}{\kV\per\cm} & \SI{4.0}{\%}  \\
Bottom  &  \SI{-1.5}{\kV}	&  \SI{75}{\mum} / \SI{5.0}{\mm} & \num{592} & \SI[retain-explicit-plus]{+34}{\kV\per\cm} & \SI{3.0}{\%}  \\
\hline
\end{tabular}
\end{table}

The optimization of the anode geometry involves a compromise between optical, electrostatic, mechanical, and electroluminescence properties. The latter were assessed through full electron transport modeling, in particular examining the S2 photon production statistics from single electron drifts in the gas phase of several candidate geometries. This method has been validated by comparison of simulated and actual signals in the LUX detector~\cite{Bailey:2016pnn}. Subsequently we applied this methodology to the baseline gate-anode electrode configuration, and some results are shown in Figure~\ref{XDSf:S2sims}. This involved the detailed modeling of electrostatic fields in the electroluminescence region using the Elmer solver~\cite{ELMER:1999} and the meshing tool Gmsh~\cite{Geuzaine:2009}, followed by simulation of electron transport and photon production with Garfield++~\cite{Veenhof:1993hz} using xenon transport parameters from Magboltz~\cite{Biagi:1999nwa}.

As discussed below, all of the grids are woven meshes. Increasing the density (smaller pitch and/or larger diameter wires) decreases the electric field near the wire surfaces, reducing spurious electron emission from the gate and charge multiplication near the anode. It also provides more uniform field lines, especially near the anode, which is important for reducing dispersion of the produced S2 signal. However, decreasing the opacity increases $\alpha_1$ and $\alpha_2$, and increases the ease of manufacture. The gate pitch is half that of the anode pitch, and the two grids are aligned such that an anode wire crossing sits directly above the center of each open square in the gate grid. As the electrons pass through the gate they are focused towards the center of each square, and hence towards this crossing. This alignment gives smallest spread in path lengths traversed from the liquid to the anode, and hence smallest dispersion in the generated S2 signal.

Details of the simulated performance of the baseline gate-anode design are shown in Figure~\ref{XDSf:S2sims}. This study assumes that electrons start evenly spread in ($x$,$y$) below the gate grid. The S2 signal resolution and the timing performance are those reported in Table~\ref{XDSt:S2parameters} above. Without diffusion the pulse would have a ``box'' shape if the anode were a plane; the spike at the end of the pulse (shown in d) is from electrons moving through the short region of enhanced field near the anode wires. The predicted photon yield is slightly higher than suggested by the parallel-plate calculation presented in Figure~\ref{XDSf:S2yield} for this reason. Conversely, some electrons move near the center of the unit cell, experiencing lower average fields, and this motivates the occasional lower yield visible in panel c) of the same figure. Those electrons also arrive slightly later, and form the tail above \SI{1}{\mus} shown in d). Sparser grids aggravate both of these effects very quickly.

\begin{figure}[ht]
\centering
\includegraphics[width=1.0\textwidth]{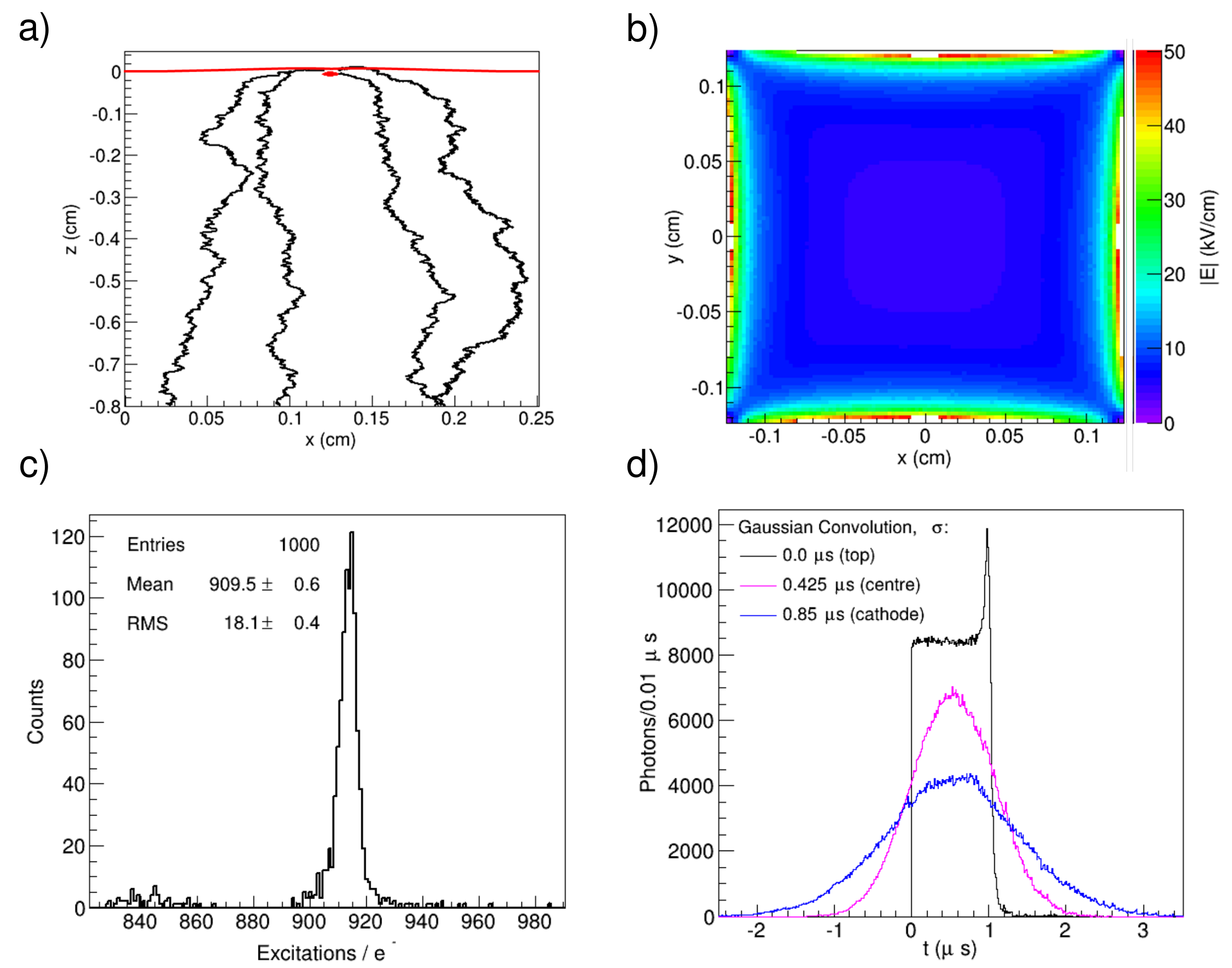}
\tdrfcaption[S2sims]{Simulation of S2 electroluminescence response}{Simulated S2 electroluminescence response of \SI{1000}~electrons for the baseline gate-anode design~\cite{Bailey:2016pnn}. a) Examples of electron drifts from the liquid surface (at $z$=\SI{-0.8}{\cm}) until their collection at the anode mesh wires (shown in red); b) Electrostatic field strength through the anode plane; c) Photon yield (number of excitations), confirming \SI{2}{\percent} {\em rms} for electrons distributed evenly in the liquid xenon bulk; d) S2 simulated pulse shape of the electron signal, without longitudinal electron diffusion in the liquid (black), and the same distribution convolved with the amount of Gaussian diffusion expected for drift from the center (magenta) and bottom of the detector (blue).}
\end{figure}

\tdrsubsec[GridFab]{Design and Fabrication of the Grids}

The four grids, whose basic parameters are list in Table~\ref{XDSt:ElectrodeGrids}, are all of a crossed mesh design. They are fabricated by weaving and individually tensioning a set of series \num{300} stainless steel wires, and held by a set of low mass grid stainless steel rings housed within the field cage. The crossed mesh design has several benefits. Compared to a parallel set of wires, it presents a more uniform load on the ring, allowing it to be smaller in mass. The crossed wire mesh is also more mechanically robust than free-standing wires, and the fields are more uniform. Note that electroformed grids are not readily available at the size of LZ, and also do not naturally have the minimum-surface field producing the profile of a round wire. The wires are chosen to have the highest available surface smoothness.

Large, attractive electrostatic forces exist between the gate and anode grids because of the strong electroluminescence field and, to a lesser extent, between the cathode and bottom grids due to the high field in the reverse field region. The tension on the wires in these grids, nominally \SI{0.25}{\kg} on the anode and \SI{0.5}{\kg} on the others, should achieve less than \SI{2}{\mm} combined maximum deflection between the gate and anode, when the grids are treated as individual wires. These loads are well within the yield strength of available stainless steel wires, but nonetheless represents an important mechanical requirement on the assembly. Evaluation of the mesh grid is much more difficult, though approximate treatments may be possible. We expect that the resulting deflection will be smaller than that estimated from the individual wire calculation, possibly allowing a reduction in tension. The actual tension in full-size prototype grids will be directly measured under field. Note the tension needed to minimize this overall deflection is larger than the minimum tension needed to prevent the well-known wire-to-wire ``saw-tooth'' instability encountered in a single plane of wires in a wire chamber.

The technique for weaving the wires is analogous to the process used in the textile industry for making woven cloth with a loom. Initially, the wire forces are supported by the loom. The technique for anchoring the grid wires is to capture them between two support rings that are epoxied together. The epoxy locks the grid wires in the space between the support rings. Once the epoxy is fully cured, grid wires outside the support rings are trimmed off, and the wire load is solely  carried by the rings. The process for making all the grids is the same as the cathode, albeit with different diameter wires and spacings. After production, the entire grid assemblies will be electropolished and passivated chemically to achieve the highest possible surface quality. Such a treatment has been shown to be beneficial on single wire samples as described in Section~\ref{XDSSs:SmallWireST}. We have demonstrated the technique of electropolishing completed grids for the Phase-I  System Test TPC prototype which is described in Section~\ref{XDSSs:SLACPhaseIST}.

\tdrsubsec[WallEvents]{Reconstruction of Peripheral Interactions}

The mis-reconstruction to smaller radii of peripheral background events---such as those arising from radon progeny plateout on the inner field cage walls---can be a leading source of background in double-phase xenon TPCs (e.g.~ZEPLIN-II~\cite{Alner:2007ja} and LUX~\cite{Akerib:2013tjd} were so affected). This must be addressed by both lowering plateout rates and ensuring good quality spatial reconstruction for the remaining decays. The layout of the top PMT array, and in particular of the peripheral tubes, is therefore of great importance. We considered five array configurations in our initial optimization and this was discussed in the LZ CDR (Section 6.5.3 in Ref.~\cite{Akerib:2015cja}). An additional study was carried out to optimize the hybrid configuration selected previously, based on further considerations including electric field between PMTs, S2 light collection uniformity, and mechanical feasibility.

Based on this new study we adopted for the top array a circular/hexagonal hybrid layout containing \num{253}~PMTs, transitioning from a close-packed hexagonal core to a 48-unit circular outer row (see Figure~\ref{XDSf:PMTArrayLayout}). The minimum PMT separation is \SI{86.3}{\mm}. Our methodology involved extensive optical Monte Carlo using the code ANTS2~\cite{Morozov:2016ywi} coupled to the Mercury vertex reconstruction algorithm~\cite{Solovov:2011bb}. This provided a realistic assessment of the position resolution of the chamber for small S2 signals and, in particular, the fraction of peripheral events that is misreconstructed into the TPC volume. This ``leakage'' fraction was the main design criterion used to select the best array configuration. In addition to the position of the outer PMTs, two other design parameters influence the peripheral position resolution: the distance between the anode grid and the PMT windows, and the reflectivity of the lateral wall in the gas. Regarding the latter issue, a low-reflectance material is desirable so as not to overly distort the spatial response of the outer PMTs. Titanium has \SI{16}{\percent} reflectance at \SI{178}{\nm}, but its oxides can be more reflective in the VUV~\cite{Lynch1997233}, and a non-conducting material is preferred to minimize fields. Thus, that region will be covered by Kapton\texttrademark~foil instead.

\begin{figure}[ht]
\centering
\includegraphics[width=0.6\textwidth]{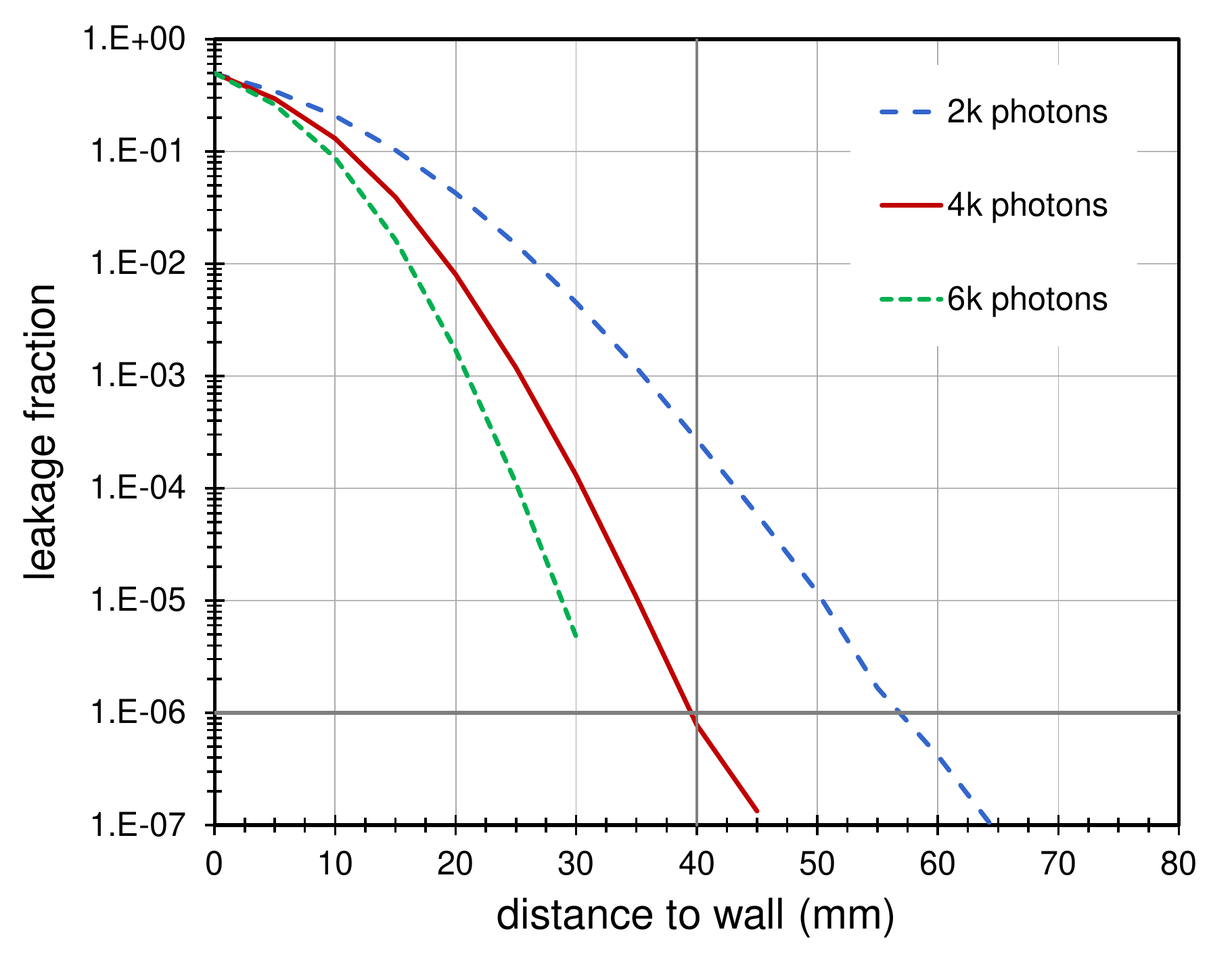}
\tdrfcaption[WallReconstruction]{Reconstruction of wall events}{Simulated reconstruction of wall events as a function of signal size. For \num{4000}-photon S2 signals ($\approx$\num{5}~electrons emitted) a leakage of \num{e-6} is achievable at the edge of our preliminary fiducial volume which is located at \SI{40}{\mm} from the wall.}
\end{figure}

Figure~\ref{XDSf:WallReconstruction} summarizes the results from a high-statistics study of the leakage past various reconstructed radii, with \num{40}{-\si{\mm} being the nominal distance to our preliminary \num{5.6}-\si{\tonnel} fiducial volume. This leakage is indicated as a function of S2 signal size (with \num{4000} S2 photons corresponding to about 5 emitted electrons). This confirms that we can achieve a very small leakage into the fiducial volume above the nominal S2 threshold.

It should be noted that the ionization removed from wall interactions tends to be pushed into the TPC as it drifts and ends up being emitted up to several centimeters away from the TPC edge (as shown in Fig.~\ref{XDSf:fieldnonuni}), which will help mitigate this background. We do not take this field distortion into account in this calculation.

\tdrsec[Skin]{The Xe Skin Detector}

The region between the outer walls of the TPC and the inner cryostat vessel is called the ``Xe Skin''. This region provides necessary mechanical clearance to allow for detector assembly, houses instrumentation including PMTs and other sensors, and provides a standoff between biased TPC components and the electrically-grounded inner vessel. The Xe Skin contains more than \SI{2}{\tonnesl} of LXe, and it is divided into two primary functional regions: a cylindrical, ``side Skin'' region outside of the main TPC field rings, and a ``dome Skin'' region underneath the TPC.

Because of the high density of LXe, gamma rays (and, to a lesser extent, neutrons) have a high probability of interacting or being absorbed in the Skin region. This poses a problem for the efficiency of the outer LS veto detector, a key element in reducing backgrounds in LZ, as secondary scatters that would otherwise be tagged in the Outer Detector are lost in the Skin. Additionally, for events that deposit energy in both the main TPC and the Xe Skin, leakage of light from the Skin to the TPC can compromise the rejection of the dominant ER background by increasing the observed S1 signal and lowering the S2/S1 ratio used for particle discrimination. To counter these effects, we instrument the Xe Skin region with \num{93} dedicated 1-inch R8520 PMTs in the top half of the side Skin, \num{20} 2-inch R8778 PMTs in bottom half of the side skin, and \num{18} more 2-inch R8778 PMTs in the dome, turning the Xe Skin into a second component of the LZ veto strategy to clearly identify events with scattering vertices in the Skin. Instrumenting the Xe Skin also increases the amount of information available about the background environment of the WIMP target, improving the background model and the efficiency of the data analysis. The goal for the Xe Skin detector design is to achieve an energy threshold for observing ER scatters produced by gamma-ray backgrounds or radiative neutron capture on detector materials of \SI{100}{\keVee} in over \SI{95}{\percent} of the volume of the Xe Skin. A second goal is to minimize light leakage between the Xe Skin and the TPC. These are related goals: Since optical isolation is never perfectly realized, any scintillation emitted in ``S2-inactive'' regions of the detector must be read out to prevent the dangerous background topologies described above.

The side Skin is \SI{4}{\cm} wide near the top of the TPC, increasing to \SI{8}{\cm} in the lower half due to the tapered vessel shape. This region is instrumented with \num{93} 1-inch R8520 PMTs viewing down located just below the LXe surface and a further \num{20} 2-inch R8778 PMTs looking up. \num{18} of the R8778 tubes are arranged symmetrically around the bottom of the TPC, with 2 extra PMTs looking specifically at the region where the cathode HV feedthrough enters the detector. Light emission from surfaces is often the first sign of issues maintaining high voltage, and photons produced near the cathode region, whether from radioactivity or from voltage breakdown, have a high probability of entering the feedthrough umbilical and being absorbed. The additional PMTs increase the photocathode coverage in the area to mitigate such photon losses.

The Skin PMTs mount to the weir system as shown in Figure~\ref{XDSf:detector} on page~\pageref{XDSf:detector}, and to the bottom array as depicted in Figure~\ref{XDSf:SkinPMTs}. The inside surface of the inner cryostat vessel is lined with thin PTFE sheets for improved light collection, and the outer surface of the TPC walls also provides a PTFE reflector. The top side Skin region uses 1-inch PMTs, rather than the larger model in the TPC due to mechanical constraints. The Hamamatsu R8520 is specifically designed for LXe operation and was the primary PMT used, for example, in the XENON10 and XENON100 detectors~\cite{Aprile:2010bt,Aprile:2011dd}. The R8520 is a compact 1-inch, square PMT with quartz window and bialkali photocathode with a typical QE of \SI{30}{\percent} at \SI{175}{\nm}. A gain of \num{e6} is provided by an 11-stage metal channel dynode chain. We use a smaller number of 2-inch R8778 tubes in the bottom side skin and in the dome also for mechanical constraints, as R11410 PMTs would not fit in either region. The R8778 PMTs will be recovered from the LUX experiment, where they have operated for several years in a low-background, LXe environment and have been well characterized.  

\begin{figure}[ht]
\centering
\includegraphics[width=0.9\textwidth]{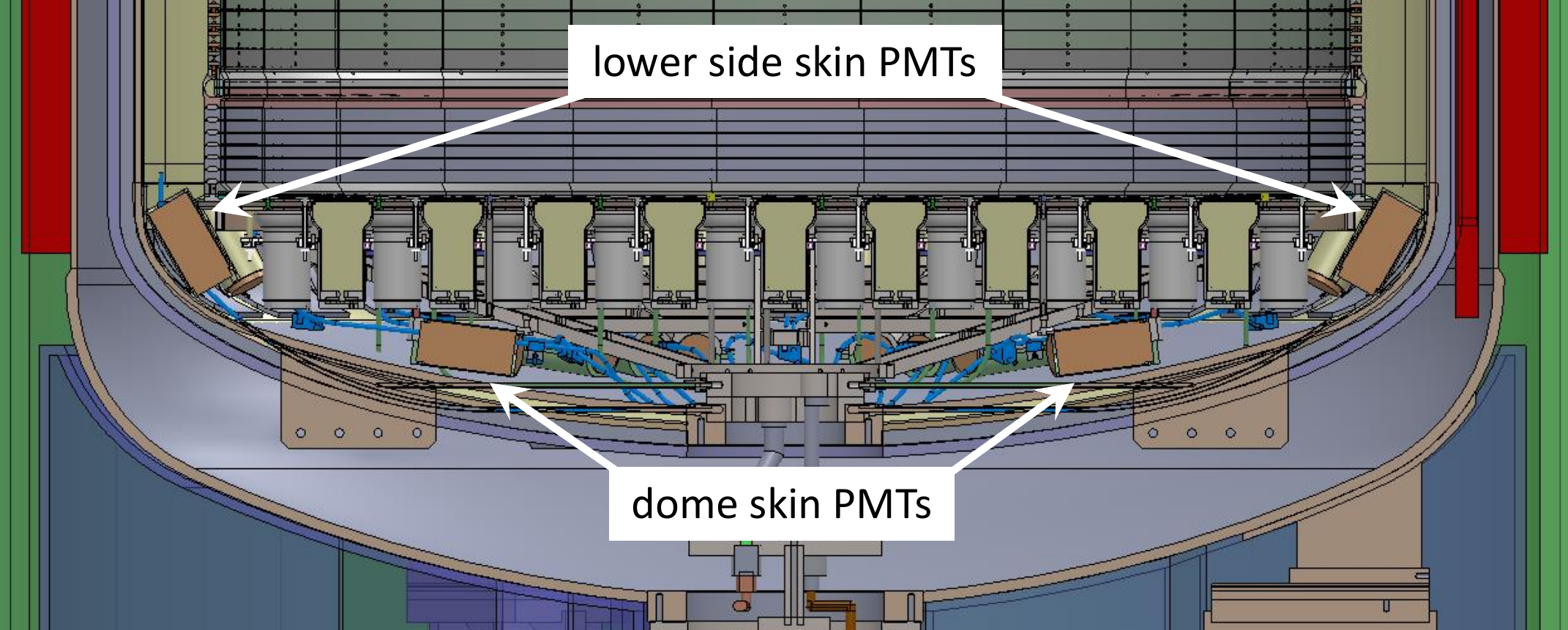}
\tdrfcaption[SkinPMTs]{Mounting of dome and lower side Skin PMTs}{Mounting of the Skin PMTs in the dome and lower side regions. The upper Skin PMTs mount to the weir system as shown in Figure~\ref{XDSf:detector}.}
\end{figure} 

Similar to the inside surface of the TPC, PTFE will be used to cover the underside of the TPC PMT support structure facing the Skin. Both the main TPC PMTs and the R8778 PMTs will have reflective PTFE sleeves to reduce photon absorption on their metal envelopes. The dome Skin PMTs will be mounted to to the lower truss of the TPC structure as depicted in Figure~\ref{XDSf:SkinPMTs}.

\tdrsubsec[SkinOptics]{Skin performance}

Design studies for this region begin by assessing the combined Xe Skin and Outer Detector veto performance as a function of threshold in the Skin. As described in Chapter~\ref{chap:SRD}, a comprehensive model of the LZ detector is implemented in \geantFour, and the effect of radioactivity in various detector components is simulated. The key performance metric of the combined anti-coincidence system is the veto inefficiency, defined as the number of unvetoed single scatter events in a detector divided by the total number of single scatters in that volume. This inefficiency varies with threshold in both veto regions. Figure~\ref{XDSf:SkinThreshold} shows the veto inefficiency for single scatters caused by gamma-rays as a function of Skin threshold, showing a \SI{50}{\percent} increase as the threshold varies from \SI{100}{\keVee} to \SI{200}{\keVee}. In other words, the un-rejected background rate increases by that factor as the Skin threshold is increased. The requirement that the number of electron recoil events produced by detector materials be less than \SI{10}{\percent} of the predicted rate of neutrino-electron scatters therefore drives the requirement that the energy threshold in the Skin must be \SI{100}{\keVee}. As the light collection efficiency in the Xe Skin is very non-uniform (comparing, for example, the side and dome regions), we include in the requirement the demand that \SI{95}{\percent} of the volume in the Skin meets the \SI{100}{\keVee} threshold target.

\begin{figure}[ht]
\centering
\includegraphics[width=0.6\textwidth]{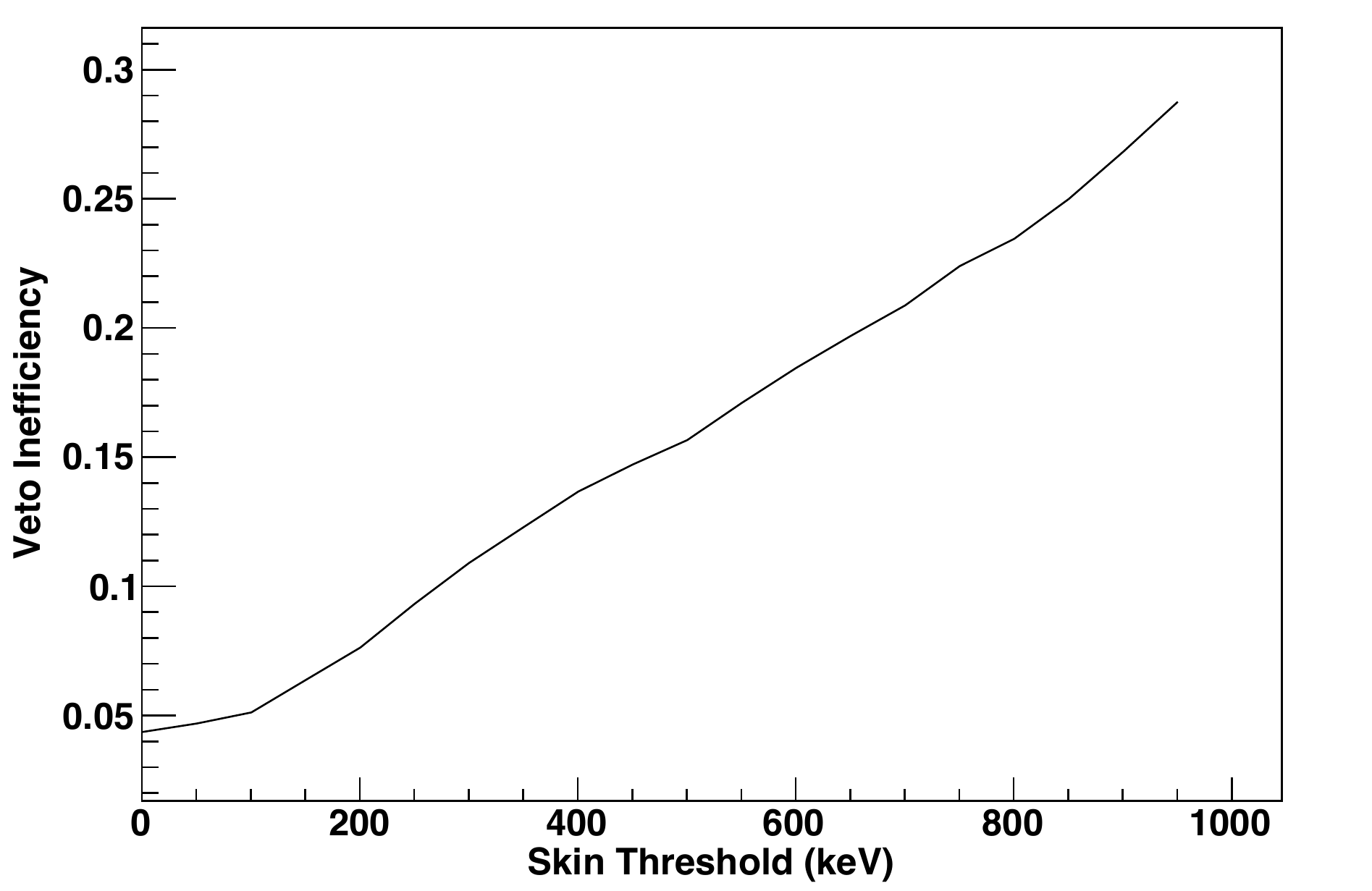}
\tdrfcaption[SkinThreshold]{Xe Skin inefficiency} {Skin veto inefficiency for gamma backgrounds, defined as the number of unvetoed single-scatter events divided by the total number of single-scatter events as a function of energy threshold in the Xe Skin. The inefficiency increases from \SI{5.1}{\percent} to \SI{7.6}{\percent} as the threshold rises from \SI{100}{\keVee} to \SI{200}{\keVee}, representing a \SI{50}{\percent} increase in gamma backgrounds from internal radioactivity. The background requirements for LZ demand that an energy threshold of \SI{100}{\keVee} be achieved in over \SI{95}{\percent} of the Xe Skin volume.}
\end{figure} 

The next step of the design is to create an optical model for the Skin to understand how much photocathode coverage and surface reflectivity are needed. We use the same simulation geometry developed for the background studies, taking care to define the optical interfaces to understand the effect of surface reflectivity. We first divide the Xe Skin into \SI{1 x 1 x 1}{\cm} pixels. The electric field in each pixel is calculated from an electrostatic model, and we use the NEST package to find the mean number of photons produced by a \SI{100}{\keV} electron recoil in each pixel, accounting for suppression in scintillation yield caused by the high electric field. Photons are then generated in each pixel and propagated through the optical model, and the photon collection efficiency (PDE) is defined as the number of observed photoelectrons per generated photon, including PMT effects such as quantum efficiency (as for the TPC). Finally, a pixel is deemed ``good'' if the PDE is high enough that \SI{50}{\percent} of \SI{100}{\keVee} events produce at least \num{3} detected photoelectrons. Figure~\ref{XDSf:SkinOptics} shows the PDE in the Skin as a function of position (averaged over azimuth) for the design described here. Over \SI{97}{\percent} of the Skin region achieves an energy threshold of \SI{100}{\keVee}, meeting the requirement.

\begin{figure}[ht]
\centering
\includegraphics[width=0.5\textwidth]{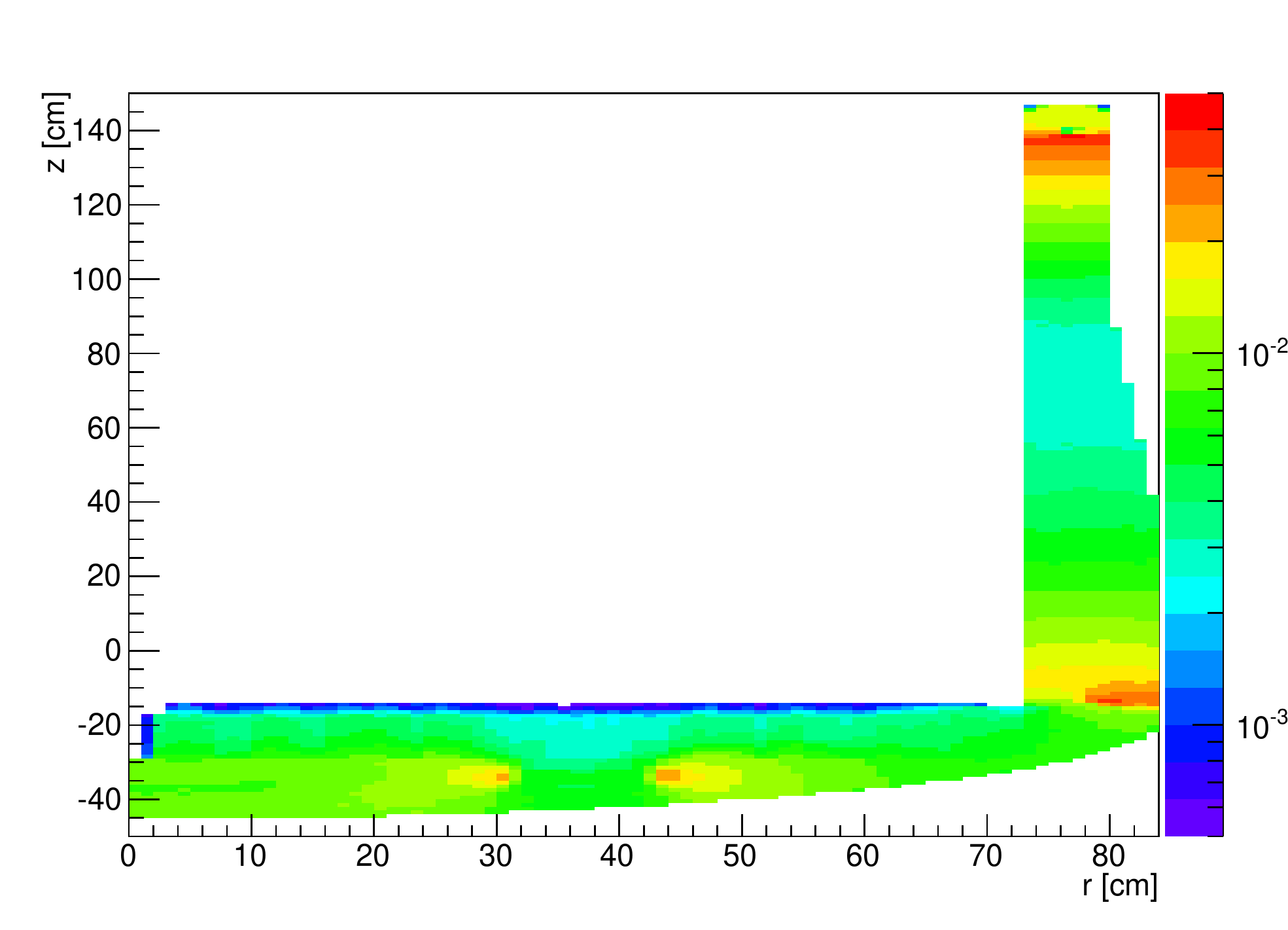}
\tdrfcaption[SkinOptics]{Light collection efficiency of Xe Skin detector} {Photon detection efficiency (PDE) in the Skin region as a function of position (averaged over the azimuth) for the design described in the text. Over \SI{97}{\percent} of the Skin region achieves an energy threshold of \SI{100}{\keVee}, meeting the requirement.}
\end{figure} 

The Skin PDE is highly dependent on the wall reflectivity, especially in the side Skin, and the left panel of Figure~\ref{XDSf:PTFESkin} shows how this parameter varies as a function of $z$ in the side Skin for different reflectivity assumptions. Given the strong dependence, we set a requirement on the PTFE reflectivity of the cryostat liner of \SI{95}{\percent}, the same as for the TPC walls. The final model described above assumes that only \SI{98}{\percent} of the wall is covered, allowing for joints or other imperfections in the liner. The threshold requirements are met in the dome region without any additional PTFE beyond that covering the PMTs.

The PTFE liner will be supported by metal ``buttons'' that are bonded to the cryostat walls using cryogenic epoxy. Simulations of the field in the skin show that metal buttons can be used without causing excessive surface fields, even near the cathode ring. A schematic showing how the buttons will hold PTFE tiles in place can be seen in the right panel of Figure~\ref{XDSf:PTFESkin}. Holes will be cut into the liner to go around the buttons, and PTFE screws will hold PTFE washers above the liner, capturing it in place. 

\begin{figure}[ht]
\centering
\includegraphics[width=0.55\textwidth]{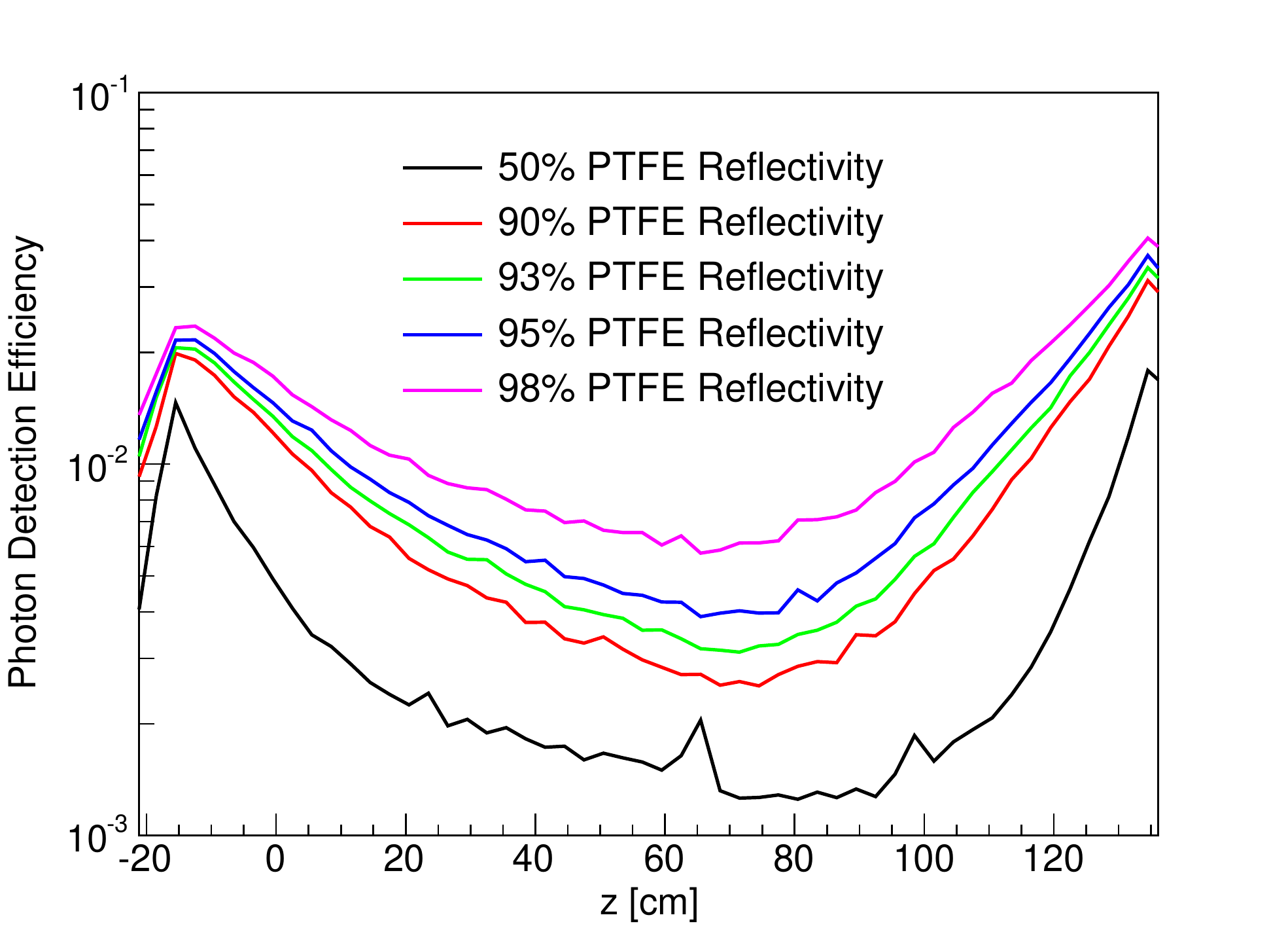}\quad
\includegraphics[width=0.40\textwidth,trim=0 -40 0 0]{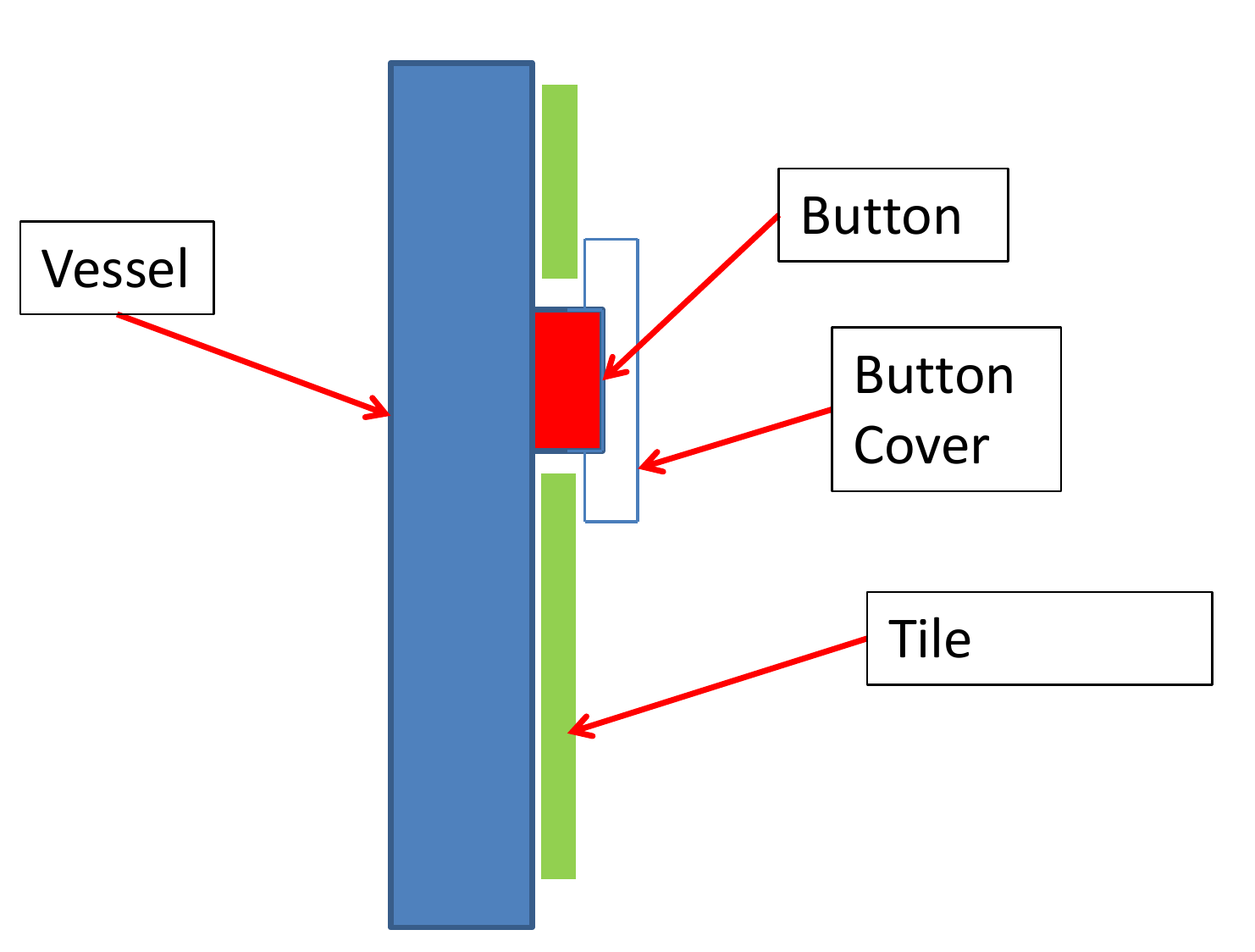}
\tdrfcaption[PTFESkin]{Skin performance as a function of PTFE reflectivity} {Left: Photon detection efficiency as a function of $z$ in the side Skin region for different values of PTFE reflectivity (in percent), where the inner wall of the cryostat is lined with PTFE and the outer wall of the TPC is made of PTFE. The LZ requirement is \SI{95}{\percent} reflectivity for the PTFE/LXe interface. Right: A schematic showing how buttons bonded to the cryostat wall will hold PTFE tiles in place to line the cryostat with reflective material.}
\end{figure}

\tdrsec[Fluids]{Internal Fluid System}

Efficient purification of LXe is a significant challenge, and is especially important given the large size of LZ and the resultant long electron drift lengths and long photon path lengths involved. Purification is discussed in detail in Chapter~\ref{chap:XCS}, and the overall internal flow diagram is shown schematically in Figure~\ref{XDSf:FlowOverview}. Liquid in the detector is continuously circulated to a purification tower located outside of the water tank, where it is evaporated in a two-phase heat exchanger and passed to a gas purification system. There, it is purified by a commercial heated getter. While the getter is highly efficient in a single pass, continuous purification has proved necessary in most previous such detectors, primarily because of the large amount of PTFE and other plastics especially in cables that serve as a long-term source of outgassing. After passing through the getter, the Xe returns to the liquid tower, where it is recondensed in the two-phase heat exchanger, degassed and subcooled, and then passed back to the detector. In addition, separate gas flow through the external purification system purges the space above the liquid, and the conduits. The overall flow rate is 500~slpm of gas, which is \SI{16}{\cm\cubed\per\second} of liquid flow, \SI{2.8}{\kg\per\minute}, and fully processes 10~tonnes of Xe in \SI{2.5}{\day}.

While much of the functionality and complexity of the system shown in Figure~\ref{XCSf:TeaPotDiagramFig} is external to the detector and is described in Chapter~\ref{chap:XCS}, the ``internal circulation system'' that resides within the detector has important design considerations too. Primarily, the system must purify the liquid while sweeping all regions of LXe to avoid stagnant fluid, which is especially important when using the dissolved tritium calibration source. It must do this while maintaining a stable liquid surface with no boiling throughout the volume, and allowing operation within the \SIrange{1.6}{2.2}{\bar} (a) pressure range. Along with the cryogenics system, it provides cooling to counter the various heat loads in the system.  Finally, it should allow, to the extent possible, operation in two modes: one where the liquid is convectively mixed on a time scale that is much shorter than the \num{2.5}-day circulation time, and one where the fluid has a stable thermal gradient with minimal mixing.

\begin{figure}[ht]
\centering
\includegraphics[width=0.6\textwidth]{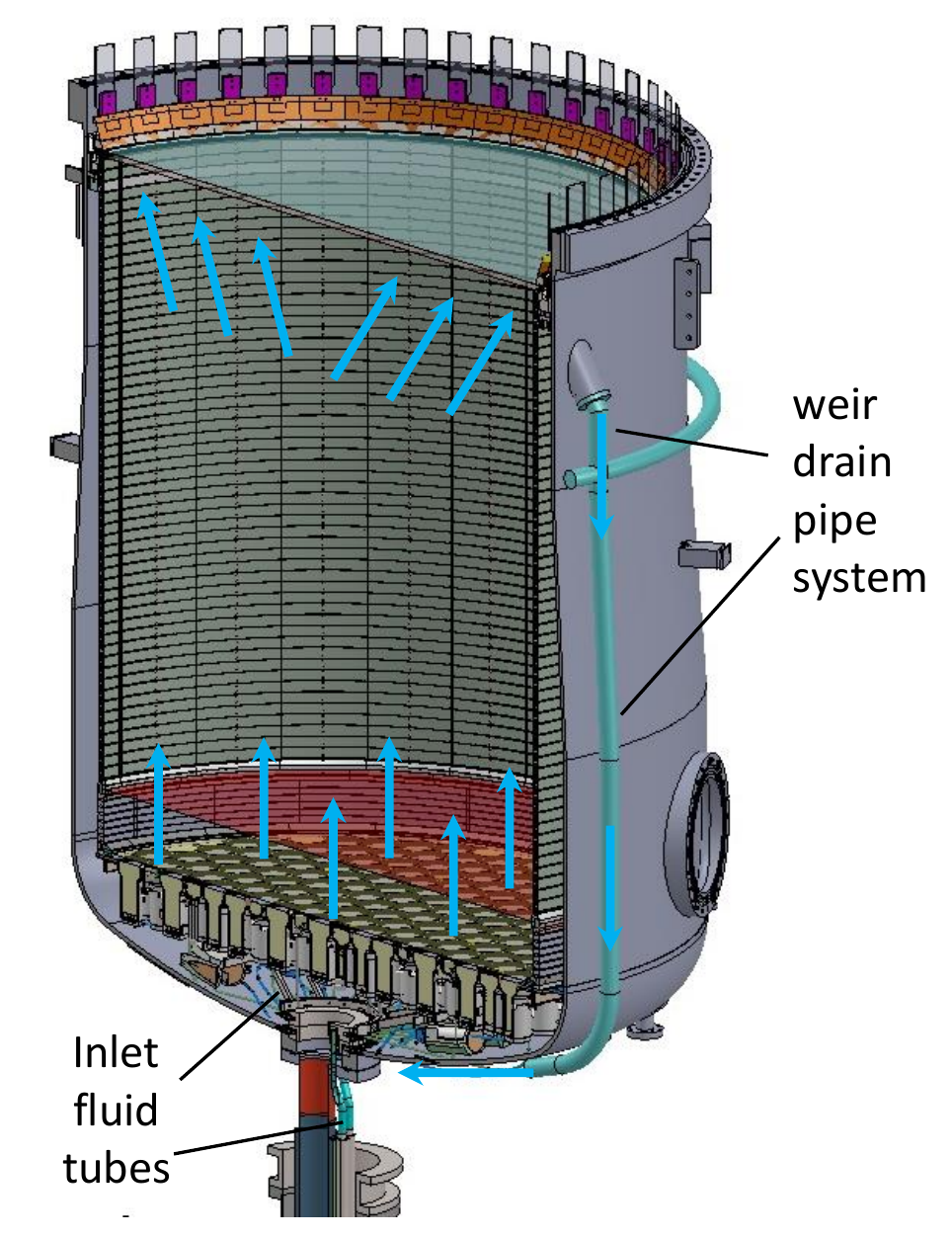}
\tdrfcaption[FlowOverview]{Overview of internal circulation flow}{Cutaway of the Xe detector system, with the internal circulation system elements and flow paths indicated. Purified LXe in two separate tubing sets is distributed into the bottom of the TPC and the bottom of the dome Skin region. The flow is then up to the liquid surface and over a set of liquid level-maintaining weirs, into a set of thee collection troughs and drain pipes, and then return via the bottom conduit to the external system.}
\end{figure}

The overall flow path is shown in Figure~\ref{XDSf:FlowOverview}. No plumbing is routed near or through any of the challenging high-field regions of the detector. This limits locations where fluid lines can access the central TPC volume to the bottom PMT array, and to the perimeter of the TPC near the liquid surface, both of which regions have voltages relatively close to ground. Accordingly two streams of liquid enters the detector from the bottom conduit, one of which is distributed into the  dome Skin, and one of which is distributed through the bottom PMT plate and into the TPC. The flow rates into the Skin and TPC are separately controlled. The overall movement is then upward, with LXe spilling over a set of six evenly spaced level-defining weirs for each of the skin and TPC regions. These drain into a triply-segmented drain collection trough, and out through drain tubes. These penetrate the vessel walls to avoid the high field region of the skin near the cathode, and then back to the bottom conduit and over to the purification tower. All tubing and associated fittings housed in the skin will be made from high-reflectivity PTFE so as to minimize absorption of scintillation photons.

The entering fluid is distributed in the dome region, and through the bottom of the TPC through a set of some \num{20} each small diameter distribution tubes, as seen in Figure~\ref{XDSf:FluidTubeDetail}. The weirs are housed just outside the gate anode rings, a challenging region with confined mechanical space and relatively high fields created by the the two different voltages of the gate and anode grids rings. Details of the weir assembly are shown in Figure~\ref{XDSf:elumregion}. When the liquid flows over the weir, there is a small height of liquid that depends on the flow rate. This design allows a long weir length, and the weir location to be well registered to the grids. This is important because the S2 light production depends directly on the relative levels of the gate and anode grids and the liquid height.

LXe in the bottom conduit and the HV feedthrough conduit is slowly circulated outward by pumping on the gas space at the ends of these conduits while also applying heat to the LXe.  The conduits must be purged because of the outgassing from the plastics in the cables in them, especially at their warm ends for which outgassing is orders of magnitude higher than from cold plastics. A second potential issue is Rn emanation, especially from the cables in the conduits, some portion of which is warm, especially for the top conduit. The flow from the conduits is thus passed through a Rn removal system described in Section~\ref{XCSSs:RnRem}, before it is passed through the main purification system. None of these flows are routed through the heat exchange system.  Finally, there is separate capability to pump the gas directly above the TPC (i.e., in the ``S2'' space) at a low rate, using a set of tubes distributed in the top TPC array much like those in the bottom array.

A central goal in all of these flows is to eliminate as far as practical any stagnant ``dead'' regions---the prime example of which would be the conduits if they were not purged. Such dead spaces, once impure, serve as a slow source of diffusively-driven impurities that can greatly complicate purification. This is an issue not only for purity that affects charge and light collection, but, critically, following the use of radioactive tritium introduced as a calibration source (see Chapter~\ref{chap:CAS}), which must be subsequently removed.

\begin{figure}[ht]
\centering
\includegraphics[width=0.7\textwidth]{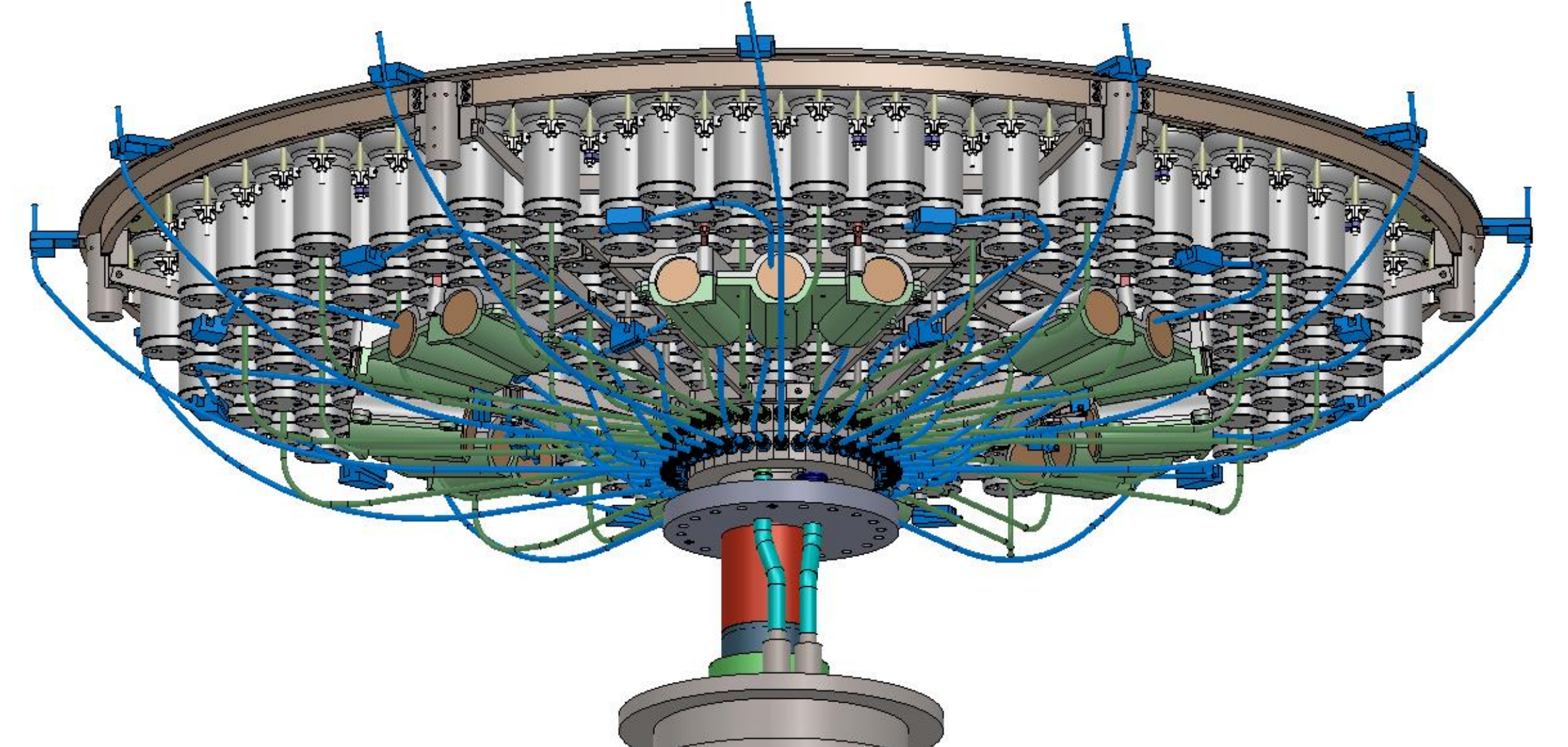}
\tdrfcaption[FluidTubeDetail]{Details of fluid distribution pipes at the bottom of the vessel} {Fluid distribution in the bottom dome region. Fluid tubes in blue distribute flow into the dome and wall skin regions, while tubes in green distribute fluid into the TPC.}
\end{figure}

Both the LXe circulation system and the cryogenic system (Section~\ref{XCSS:Cryo}) must work together to control the overall thermal environment of the detector and also the degree of convection, if any, of the fluid in the TPC. Most cooling of the detector while running will be provided by a set of six thermosyphons attached to the vessel at the level of the liquid surface. The total heat load internal to the vessel (not including cables, whose load is in the conduits) is roughly \SI{40}{\W}, and we plan to deploy up to \SI{30}{\W} from heaters in the bottom PMT array (see below). Most of this heat is applied to the Skin, and so a set of six thermosyphon heads will be deployed around the vessel at the level of the liquid surface. Liquid cooled by these heads and liquid warmed by the radiative load on the wall and the bottom PMT bases in the Skin should set up convective currents that lead to a fairly uniform temperature in the Skin.

The interior of the TPC has a much smaller heat load than the Skin, with only a fraction of the \SI{8}{\W} from the bottom PMTs being transmitted into the TPC (ditto for the \SI{1.4}{\W} from the resistor chains embedded in the TPC field cage). The temperature of the fluids returning to the detector will be set in the purification tower, and so with the fluid entering the TPC can be set to a lower temperature than that in the TPC; this fluid should counter the heat loads from the PMT bases.

In this mode a fairly stable fluid should be achieved in the TPC. Running with a stable, non-mixing fluid has several advantages. It should allow much faster purification because impurities are mostly directly swept out of the detector in the \num{2.5}-day time scale for circulating the fluid. This is termed ``batch mode'' purification in chemical engineering, as opposed to the ``mixed tank'' mode obtained with strong mixing in which purification reduces only exponentially with the circulation time as the time constant. Another advantage of a nearly static fluids is that it may be possible to reduce the effect of Rn events by tagging the position of the various daughters and in particular \IBitof, the decay of which produces an important ER background.

However, the use of the dissolved source \IKreTm argues for convective mixing that is at least as fast as its \SI{1.83}{\hour} half-life. This source provided the main calibration of position response of S1 and S2 signals in LUX, including the crucial measurement of the electron drift length, which varies over time. Because it is outside (but relatively near) the WIMP-search energy window it was used on a frequent basis, and it would be highly desirable to do the same for LZ. The basic plan for achieving mixing in the TPC is to use heaters installed just inside the outlets of the fluid tubes at the bottom of the TPC, so that fluid enters in a heated state. The amount of power planned to use is up to \SI{30}{\W}. We plan to remove this heat by evaporating liquid from the LXe surface via a set of tubes in the top array that mirror those in the bottom array, and a flow that is limited to rates that do not induce boiling. The combination of cooling at the top and heating below should induce convection motion. It is challenging to fully predict the motion of the fluids, though we plan some studies to analyze design using computational fluid dynamics. The goal of the system described here is to support operations with the fluid in either static or convective states.

Initial cooling of the detector must be done with care in order to avoid large thermal gradients in the TPC structure, which has no obvious thermal anchor point. We thus plan to circulate Xe gas, cooled in the purification tower, to slowly cool the entire system at a controlled rate. We will make use of the separate flow streams in the TPC and Skin, as well as the gas purge in the top dome area, and are using computational fluid dynamics (CFD) to plan the rate of cooling. Sensors described in the next section will be used to monitor this process. We will also attach a thermosyphon head to the bottom of the vessel to allow us to cool the bottom of the detector.

\tdrsec[Sensors]{Xenon System Monitoring}

Monitoring sensors are required for controllable and reliable operation of LZ; this Section describes the monitoring systems contained in the xenon space. Some of the data generated by these devices can also be used to identify temporary departures from optimal operating conditions and thus provide additional information for the science analyses. For example, achieving good resolution of the S2 signal relies on a quiet liquid surface, which we will monitor through precision level sensors and acoustic bubble sensors. The thermal profile of the detector is an important aspect of liquid circulation and the stability of the liquid surface, and this is measured by an array of thermometers. The ability of the system to sustain high voltages is very important and a set of loop antennae will measure discharges but may also detect any precursor signals and thus enable mitigation. Bubbles encountering a high-field surface can also lead to discharge and spurious light emission, and thus detection of bubbles is an important aspect of achieving high voltages. Finally, it is important to confirm that the very significant thermal contraction of the plastic field-cage system behaves as expected, and so this motion will be monitored by a set of position sensors. This section discusses the monitoring systems in detail, which are summarized in Table~\ref{XDSt:Sensors}.

\begin{figure}
\centering
\includegraphics[width=0.9\textwidth]{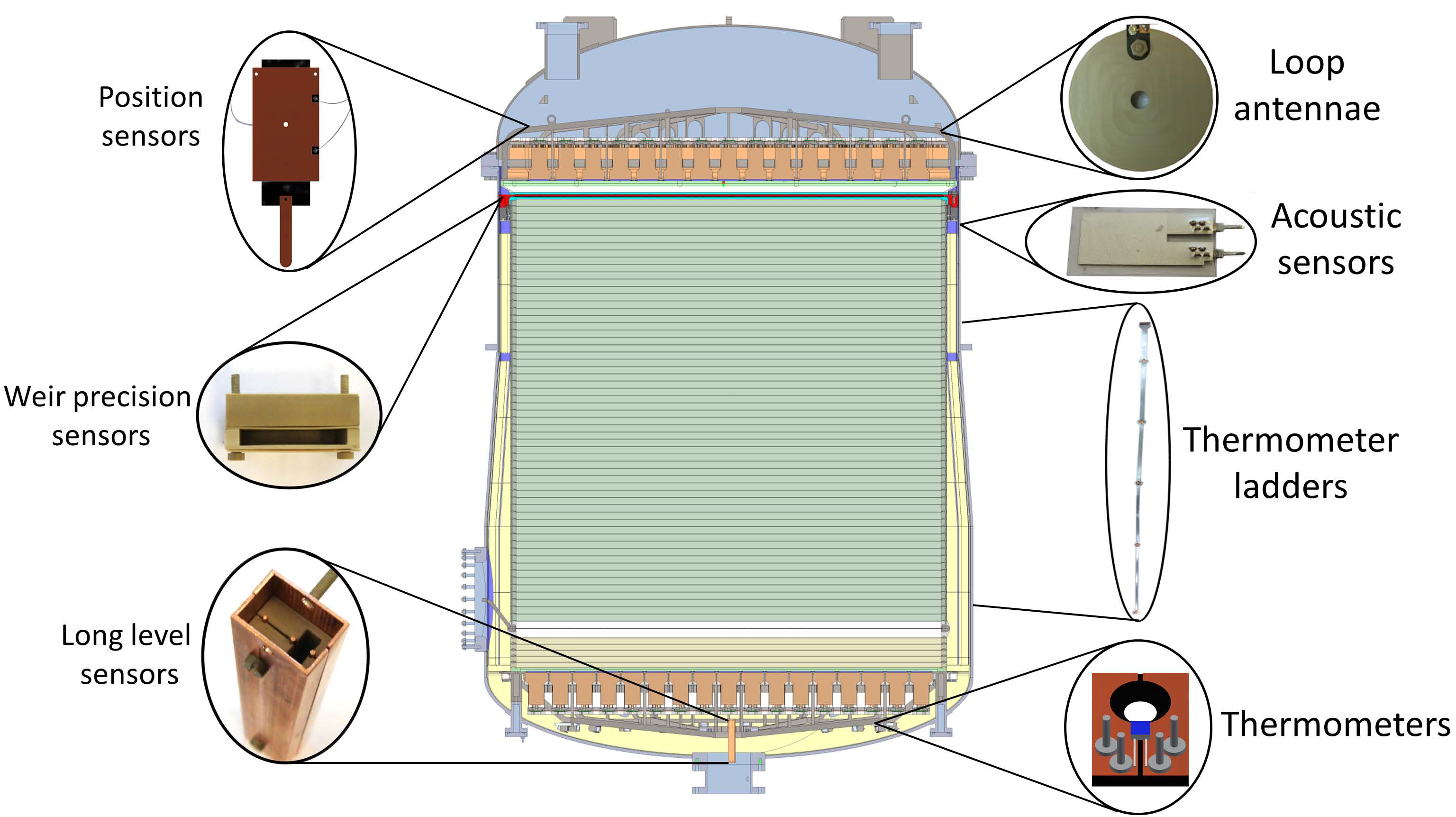}
\tdrfcaption[Sensors]{Internal monitoring sensors}{Schematic diagram of the internal monitoring sensors in the xenon detector.}
\end{figure}

\begin{table}[tbh]
\setlength{\extrarowheight}{3pt}
\tdrtcaption[Sensors]{Monitoring sensors in the xenon space}{Monitoring sensors located in the LZ xenon space.}
\centering
\sffamily
\begin{tabular} {|lccl|}
\hline
\rowcolor{mrocol}
Sensor & Acronym & Number & Location \\
\hline
Thermometers 			& --  & \num{101} &	66 in Xe space, 35 in vacuum space \\
Weir Precision Sensors	& WPS & \num{6}	&	Surface level\\
Long Level Sensors		& LLS & \num{4}	&	Surface level, dome Skin \\
Position Sensors		& POS & \num{6}	&	On top PMT truss \\
Loop Antennae			& LA  & \num{8}	&	On PMT trusses and near HV feedthrough \\
Acoustic Sensors		& AS  & \num{8}	&	Attached to IVC outer wall \\
\hline
\end{tabular}
\end{table}

\tdrsubsec[Thermometers]{Thermometers}

Temperatures need to be monitored at approximately 100 locations throughout the Xe detector and nearby. The chosen temperature sensor is a platinum (PT100-type) resistor. The readout method is 4-wire throughout and the cabling used is twisted wire bundles in the xenon space, while in the vacuum space of the LZ cryostat it is a semi-rigid polyimide-stainless steel layered composite structure; the latter features parallel strip-line pairs sandwiched between shielding/ground planes on the outside. The design of the semi-rigid cabling is individual to each group of sensors, with a maximum length of a single section being \SI{1.5}{\m}, which is sufficient to cover the height of the ICV. There is a connector block at one end of the semi-rigid cabling at which the signals transition to conventional shielded 4-core wiring to cover the long stretch towards the electronics ``breakout boxes'', where mechanical robustness and threading capability are important. At the vacuum barriers of the breakout boxes, standard DB25 connectors are used, with \num{100}~thermometers (\num{400}~wires) requiring at least \num{16} such connectors. The provision of DB25 connectors marks the interface to the LZ Slow Control system (Section~\ref{EDCS:SC}). Calibration of the platinum resistors is via a generic table available for these, where precision requirements are less stringent, or from individual calibrations and corresponding look-up tables or polynomial corrections (within the Slow Control), where higher precision is required. At positions where the potential effects of intrinsic component radioactivity is minor, commercial pin headers and sockets can be used. Closer to the center of the detector, a combination of pins and clean PTFE, PEEK, or Delrin connector bodies will be used. All materials used are checked for compatible radiopurity (see Chapter~\ref{chap:MAS}).

Cryogenic, laminated, layered semi-rigid cabling will be used for reading out the bulk of the thermometers in the vacuum. This cabling and the low-radioactivity connectors (where needed) will be made in-house to individual designs for easy installation on site. The cable is based on polyimide/Kapton, to which stainless steel is laminated. These raw materials are etched to produce individual cabling and are laminated to receive shielding and ground layers on either side. A minimum track width of \SI{150}{\mum} and pitch of \SI{300}{\mum} have been demonstrated. The electronic capacitance between wire pairs is \SI{\approx80}{\pico\farad\per\m}, depending on detailed geometry and operating temperature (via the temperature-dependent permittivity of Kapton). Thermometer mounting blocks in the xenon space will be based on Cirlex boards.

\tdrsubsec[LevelSensors]{Liquid Level Sensors}

For level sensors, two main designs are needed: a parallel-plate type for precision surface sensing, and long coaxial types. The precision surface sensor (or Weir Precision Sensor, WPS) has the plates installed horizontally, straddling the boundary between the LXe and the electroluminescence region to measure the liquid level with high precision, allowing any tilt of the detector to be readily measured, and seeing variations in the liquid surface---for example from bubbles or surface waves. Six WPS sensors are installed at the weir overflow openings. The coaxial Long Level Sensors (LLS) monitor primarily the liquid levels during filling and emptying of the TPC. There are three such sensors, that will span the height of the weir trough and the full \num{13}-\si{\mm} gate-anode distance.

The sensors will be read out with high bandwidth (up to \SI{200}{\kHz}) aiming to monitor the condition of the Xe surface (level, ripples, waves) and, as such, will have a precision of \SI{\sim10}{\mum}.  Further sensors will monitor the bottom Skin region during filling and emptying, and a LLS will be used inside the PMT cabling standpipe to monitor the filling process. A pressure sensor at the bottom of the cryostat vessel measures the head of the liquid and provides further information about the Xe level during filling and emptying. The readout method is via determination of capacitance with respect to a reference capacitance (all sensors employ three electrodes and a feedback readout circuit). This arrangement greatly reduces systematic effects arising from the long cabling in LZ and its variable capacitance (mechanical and thermal effects). The feedback readout circuit is based on  modulated readout with a minimum number of analogue components and the bulk of front-end complexity absorbed into the firmware of a field-programmable gate array (FPGA). For feedthroughs at  the vacuum barrier, a standard flange with DB25 connectors is foreseen into which the 8-channel readout boards plug directly without need for external cabling. Figure~\ref{XDSf:LevelSensors} shows results obtained in a liquid xenon test chamber. The achieved precision satisfied the requirements and features visible in the data are understood.
\begin{figure}
\begin{center}
\includegraphics[width=15cm]{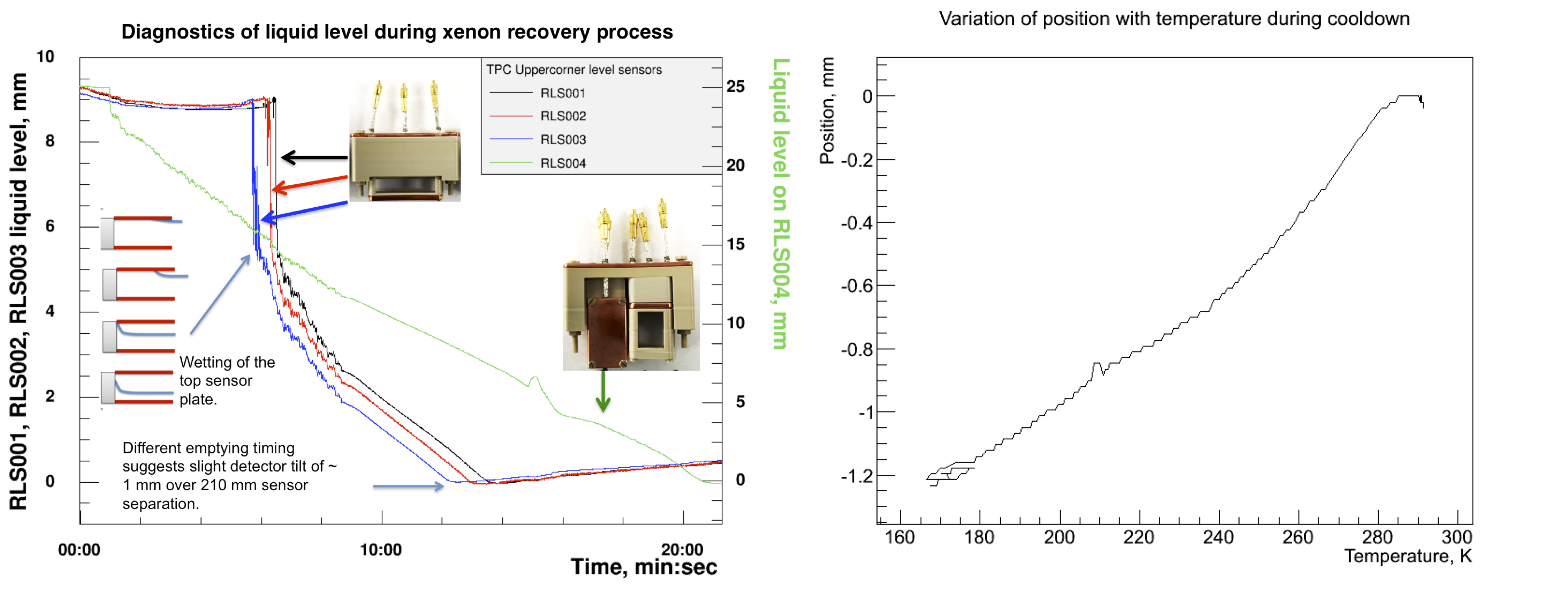}
\tdrfcaption[LevelSensors]{Xenon level sensors data}{Left: Data from a Weir Precision Sensor (WPS) and a Long Level Sensor (LLS), installed  inside a liquid xenon test chamber, showing xenon being emptied. The  shapes of the graphs are well understood; for example, regions pointed  to by the arrows can be linked to xenon gradually overflowing level sensor plates, capillary effect, or known fringe fields. Right: Data from a position sensor, showing the contraction of the TPC as it cools.}
\end{center}
\end{figure}

\tdrsubsec[AcousticSensors]{Acoustic Detection}

Acoustic sensor (AS) heads based on a polymer film (PVDF) have been chosen due to constraints on radioactivity levels. The sensors will be installed in direct contact with the outside of the inner cryostat vessel to pick up internal sound. Eight sensors are foreseen, with three at \SI{120}{\degree} spacing near the lower cylindrical section of the cryostat, three near the top of that section with the same spacing, and one each on the top and bottom domes. The vacuum-barrier electrical connection is via a standard flange with a DB25 connector, interfacing to a dedicated 8-channel differential readout front-end board. Connection to Slow Control is via a standard ModBus interface, allowing to query minimal information such as when an acoustic signal (above trigger) occurred, along with its magnitude and a crude classification of the likely nature of the sound. Simultaneously, via a different Ethernet readout channel, complete acoustic recordings from the sensors, sampled at rates up to \num{200} kS/s will be available for feeding into the fast data stream and for subsequent off-line analysis.

\tdrsubsec[LoopAntennas]{Loop Antennae for Discharge Detection} 

To monitor for onsets or occurrence of HV breakdown events, loop antennae (LA) capable of picking these up will be installed in critical positions. Eight such antennae will be assembled on the top and bottom PMT-array trusses, sufficiently far from high-field regions to ensure that the presence of the metal aerials does not interfere. With such a set of LA sensors, insight into locations of possible HV breakdowns and types should be possible. The feedthrough at the xenon space to the external environment will be through a DB25 connector, similar to the PMT readout. Two sets of fast 4-channel readout electronics with a sampling rate of up to \num{200}~MS/s will be used. The readout board can be polled by the Slow Control through a standard ModBus interface. It will provide a fast digital signal for alerting the HV power supplies to situations when a quick shut-down is required, and there will be capability for reading out triggered fast signals for inclusion in the data stream and further offline analysis.

\tdrsubsec[PositionSensors]{Position Sensors}

Position Sensors (POS) capable of measuring linear displacements of up to about \SI{2}{\cm} will be fitted on and around the top PMT array to monitor the thermal contraction and expansion of the TPC during cooling and warming, allowing to apply countermeasures, if needed, to ensure uniform cooling or warming of the TPC. These sensors will also  give important information on any lateral displacements that would alter the Skin region gap, and the alignment of the top PMT array with respect to the TPC anchor points. The operating principle  of the displacement sensors is based on picking up linear travel via a slider that moves a dielectric between two adjoining parallel plate capacitors. The position sensors are of simple and robust design and made from radiopure materials. The electronic readout is based on the same feedback circuit used in the level sensors, which acts to minimize the effect of cable capacitance on the sensor output. Eight sensors  (three for vertical movement, three for horizontal movement, and two for helical movement) will be installed.

\tdrsec[SystemTest]{Integrated System Testing}

A vital part of the preparation for LZ is the testing of key design aspects and critical components of the Xe Detector and integrated ``System Testing'' of the TPC and associated systems. Several small LXe test chambers, with less than \SI{10}{\kg} of LXe, are available at LZ member institutions, targeting studies in specific areas (spurious electron and photon emission from grid wires, VUV reflectivity measurements, small-component QA testing). Additionally, a larger System Test platform has been developed at SLAC which can operate with over \SI{100}{\kg} of LXe, conducting integrated tests to answer broader HV performance questions, and involving full-scale sub-systems in some cases. This coordinated, multi-scale approach is a key part of the risk mitigation strategy in LZ. We outline the tests being performed at collaborating institutions in Table~\ref{XDSt:SystemTests}.

\begin{table}[ht]
\begin{center}
\tdrtcaption[SystemTests]{LZ System Test planning}{High priority studies and tests to be performed with small LXe chambers or the SLAC System Test facility. Note that SLAC I (II) denotes SLAC (Phase-I) or (Phase-II).  Timing of results is indicated (calender years).}
\setlength{\extrarowheight}{3pt}
\sffamily
\begin{tabular}{|lllc|}
\hline
\rowcolor{mrocol}
Studies/Tests	& Topics 	& Groups Involved & Results \\
\hline
Reflectivity	& PTFE reflectivity in LXe       & Coimbra  /  U-M  & 2016  \\
				& PTFE reflectivity vs thickness & Coimbra  /  U-M  & 2016--2017 \\
\hline
Wire tests	& Electron/photon emission studies	& Imperial & 2015  \\
			& Small wire-grid testing     		& LBNL     & 2016  \\ 
			& Effect of electropolishing/passivation & Imperial & 2016  \\ 
\hline
 HV studies	& Cathode ring: large area at HV & UC Berkeley & 2016 \\
			& Effects of surface treatment   & SLAC I & 2015--2017\\
\hline
TPC prototype tests	& Reverse field region (RFR)  & Yale/SLAC I & 2014--2015\\
					& HV performance & SLAC I  & 2016 \\
                    & Skin optics and performance    & SLAC I & 2016 \\
					& Handling/cleanliness protocols & SLAC I/SLAC II & 2016--2017\\
\hline
Full scale grid tests	& Electroluminescence region & SLAC II  & 2017 \\
                        & RFR with full surface fields & SLAC II & 2017 \\
\hline
\end{tabular}
\end{center}
\end{table}

\tdrsubsec[ReflectivityST]{Reflectivity Measurements}

The performance of both the TPC and the Skin detectors is strongly dependent on their ability to collect xenon scintillation light, and this is in great measure driven by the reflectance of the surfaces that surround these LXe volumes. The optical performance of these systems has been described in Sections~\ref{XDSS:S1Light} and~\ref{XDSS:Skin}. PTFE is the material of choice in LZ and past LXe detectors, used due to its surprisingly high bi-hemispherical reflectance. However, there remains some uncertainty as to the precise value and angular distribution of this reflectance, and its dependence on precise material provenance and surface properties. In fact, high-quality optical data is lacking in the VUV range for many materials of interest. As was shown in Figure~\ref{XDSf:S1PDEEffects}, a few percent difference in PTFE reflectivity has a noticeable impact on the performance of the detector.

At LIP-Coimbra, Portugal, a dedicated chamber is operated for the measurement of PTFE reflectance in LXe, which follows on from earlier studies of angular profile and total reflectance of different PTFE samples for the xenon scintillation wavelength at room temperature \cite{Silva:2009ir,Silva:2009a}. The chamber used for the new measurements is represented schematically in Figure~\ref{XDSf:Coimbra} (left and center). A volume of LXe enclosed by PTFE walls is topped by a PTFE ``plunger''; the chamber is \SI{10 x 10}{\mm} in cross section and up to \SI{150}{\mm} in height. The scintillation light is provided by an \IAmtfo source deposited onto a small stainless steel holder and attached to inner surface of the movable wall. After traveling across the LXe volume and bouncing between the PTFE walls, the scintillation light is detected by a single PMT which makes up the bottom of the chamber.

\begin{figure}
\centering
\includegraphics[width=1.0\textwidth]{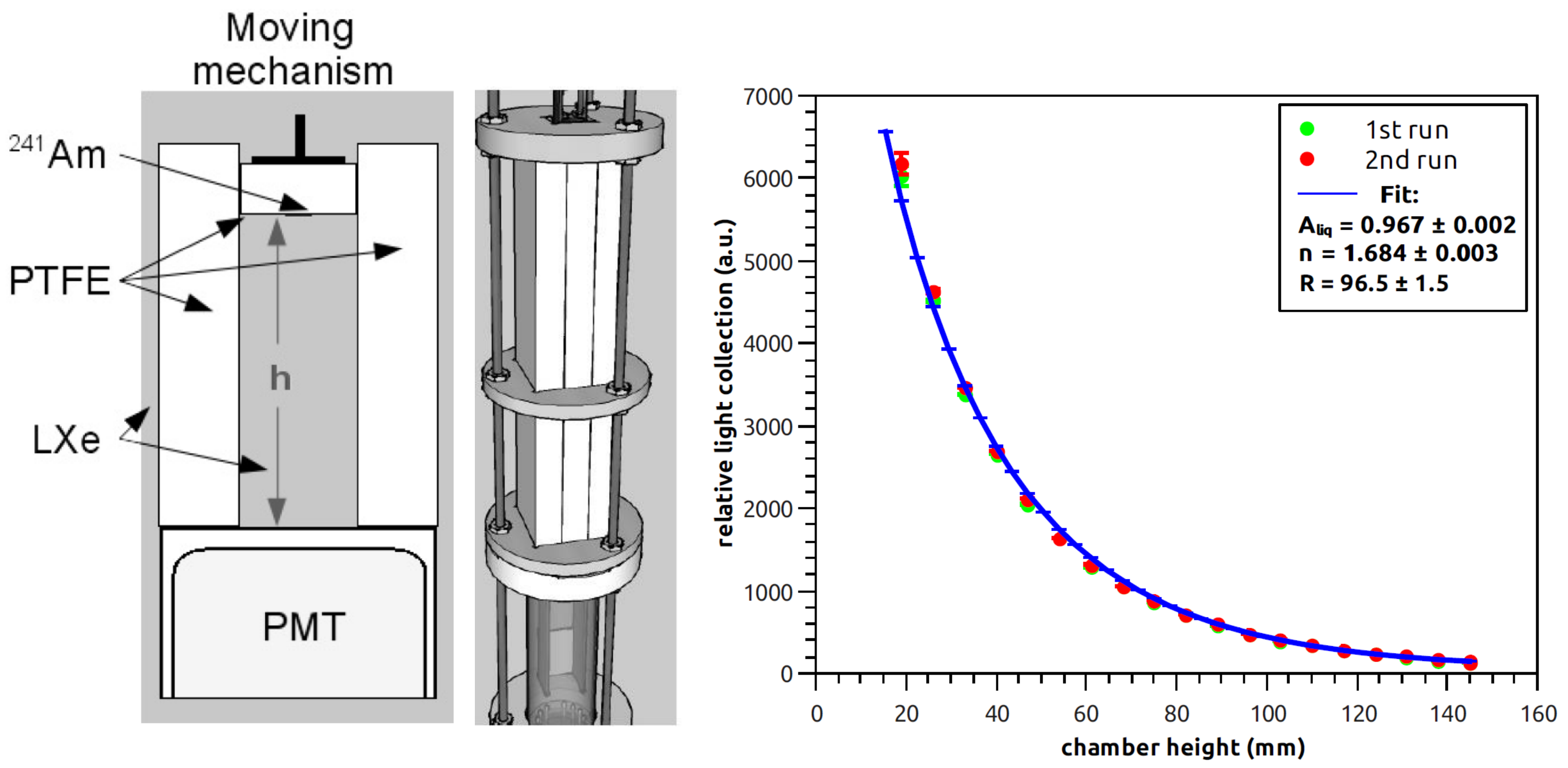}
\tdrfcaption[Coimbra]{PTFE reflectivity measurements at LIP-Coimbra}{Measurements of PTFE/LXe reflectivity at LIP-Coimbra. The PTFE-built chamber is shown schematically on the left and center. The height of the chamber ($h$) is varied with a ``plunger'' mechanism as an alpha source generates scintillation in the LXe; this is detected with a PMT at the bottom. The relative optical response as a function of the chamber height is fitted by the optical model described in Ref.~\cite{Silva:2009a} with three free parameters, from which the reflectivity, $R$, is derived. Preliminary results are shown on the right for two independent runs with the same PTFE sample (the error bars are smaller than the markers); the fit result indicates $R$=\SI{96.5\pm 1.5}{\percent} for this sample.}
\end{figure}

Preliminary experimental results obtained for one PTFE sample (APT \#807NX, a possible material for LZ) are shown in Figure~\ref{XDSf:Coimbra} (right), overlaid  with the corresponding fit. The PTFE/LXe reflectivity ($R$) is obtained by fitting the measured relative light collection at different LXe heights with the prediction of the detailed simulation model described in Ref.~\cite{Silva:2009a}. The Monte Carlo framework for this model is the ANTS2 package~\cite{Morozov:2016ywi}. The free fitting parameters were the PTFE albedo in LXe ($A_{liq}$), the PTFE refractive index ($n$), and the attenuation length ($\lambda$) for the scintillation light in LXe (these are indicated in the figure, and $\lambda$\SI{>2}{\m}). The LXe refractive index and the mean free path for Rayleigh scattering were fixed and set to \SI{1.69} and \SI{29}{\cm}, respectively.

These optical measurements are challenging and prone to systematic uncertainty and, due to the importance of PTFE to our design, additional reflectivity studies are being conducted with the MiX detector at the University of Michigan. These utilize a cylindrical PTFE structure with a PMT on the bottom, a circular PTFE disc on the top that floats on the LXe, and a $^{210}$Po alpha source on the circular disc facing down, as shown schematically in Figure~\ref{XDSf:MiX} (left). As the chamber is filled, the floating circular PTFE disc rises and the fractional area, defined as the PMT photocathode area divided by the total surface area of the chamber, diminishes. 

\begin{figure}
\centering
\includegraphics[width=5.5cm]{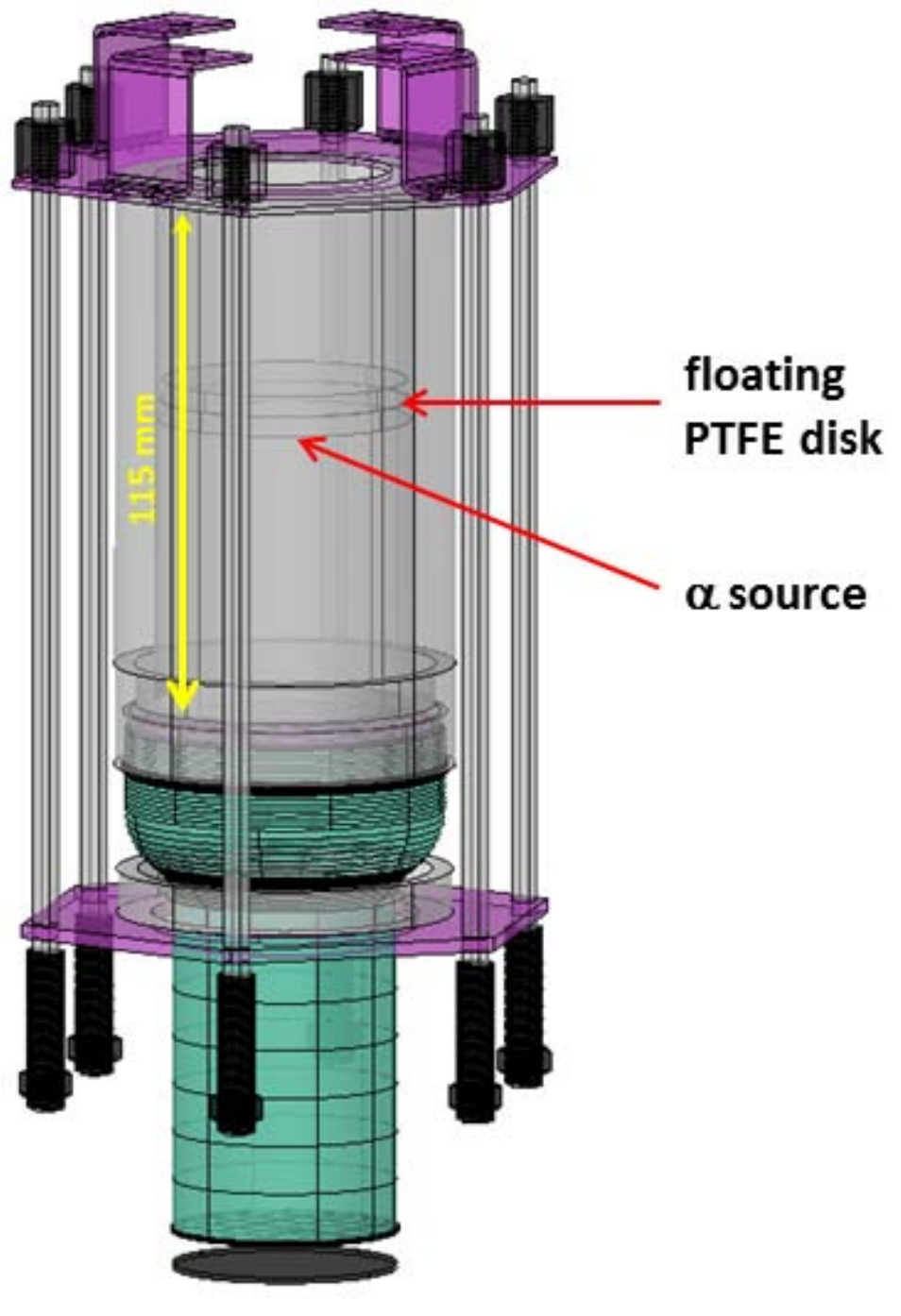} \hspace*{5mm}
\includegraphics[width=10cm]{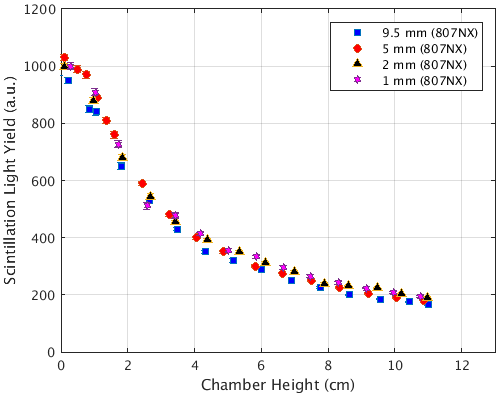}
\tdrfcaption[MiX]{U. Michigan chamber and results for PTFE reflectivity measurements}{Left: Schematic view of the U-M PTFE reflectivity setup which is based on the MiX chamber~\cite{Stephenson:2015qta}. It consists of a 3-inch Hamamatsu R11410 PMT which covers the bottom part of the 4.6-inch high and 2.45-inch ID PTFE cylinder, and a Po-210 source to provide a mono-energetic light source. The fractional area in the setup is varied as the floating disc raises or descends inside the PTFE cylinder.
Right: Preliminary results for light yield as a function of PTFE wall thickness for APT \#807NX PTFE. \textbf{Draft figure, will be updated soon}}
\end{figure}

It is desirable to minimize the thickness of any PTFE panels in the LZ TPC and Skin detectors, in order to reduce dead volumes around the active LXe, outgassing and potential backgrounds. A lower limit on this thickness is established by the PTFE transmittance to xenon scintillation light, and the need for optical isolation between the TPC and Skin regions---as well as between these and any dead regions containing LXe. Preliminary measurements have been  carried out for APT \#807NX PTFE that cover chamber wall thicknesses from \SIrange{1}{9.5}{\mm}, as shown in Figure~\ref{XDSf:MiX} (right), to verify the minimum panel thickness required in LZ. These data suggest that decreasing the PTFE wall thickness in this range does not appear to change reflectance. Further measurements will be carried out for various surface treatments and PTFE materials at thicknesses from \SIrange{1}{9.5}{\mm}. Following the PTFE measurements, these chambers will go on to study the optical properties of grid wires and other materials (e.g., the non-reflective PEEK used at the perimeter of the gas region).

\tdrsubsec[SmallWireST]{HV Studies in Small Two-Phase Chambers}

To ensure the successful delivery of HV to the LZ TPC we take a comprehensive approach, beginning with an experimental study of the physics processes involved in the electric breakdown of individual cathode wires at a microscopic (quantum) level. Chambers at Imperial College London and Lawrence Berkeley National Lab have been built and operated for this purpose.

Electron and photon emission are observed from cathodic surfaces in LXe at relatively low electric fields: \SI{\sim10}{\kV\per\cm}, a factor of \numrange{100}{1000} lower than expected for standard (cold cathode) field emission from metals. These have caused serious limitations to the operating voltage of previous LXe TPCs, almost without exception. These emissions can be caused by local enhancement of the electric field combined with the presence of thin insulating layers, surface defects, or other effects that result in a lower effective work function. This may be accompanied by simultaneous photon emission. We must understand the physical origin of these emissions and then develop effective mitigation in time to inform the production of the LZ wire grids. The extraordinary sensitivity of the S2 response, discussed in Section~\ref{XDSS:S2Light}, means that electron emission must be prevented at all cost up to the \SI{50}{\kV\per\cm} maximum field which we allow for cathodic surfaces in LXe (e.g.,~the cathode and gate grids and their connections).

At Imperial, a small two-phase chamber was developed to study individual cathode wire samples installed at the bottom of the liquid region, where a cathode grid would normally exist---see Figure~\ref{XDSf:ImperialWire}. This configuration ensures that high electric fields can be achieved at the wire surface by applying modest voltages (e.g.~\SI{\approx 150}{\kV\per\cm} for a \num{100}-\si{\mum} sample and \SI{\approx 400}{\kV\per\cm} for \SI{40}{\mum} wire for a cathode voltage of \SI{5.5}{\kV}). This study focuses on the phenomenology associated with the onset of electron and/or photon emission. In particular, we are exploring its dependence on electric-field strength, wire material, surface quality, history, as well as the effectiveness of mitigation steps such as conditioning, electropolishing, and chemical passivation.

\begin{figure}
\centering
\includegraphics[width=1.0\textwidth]{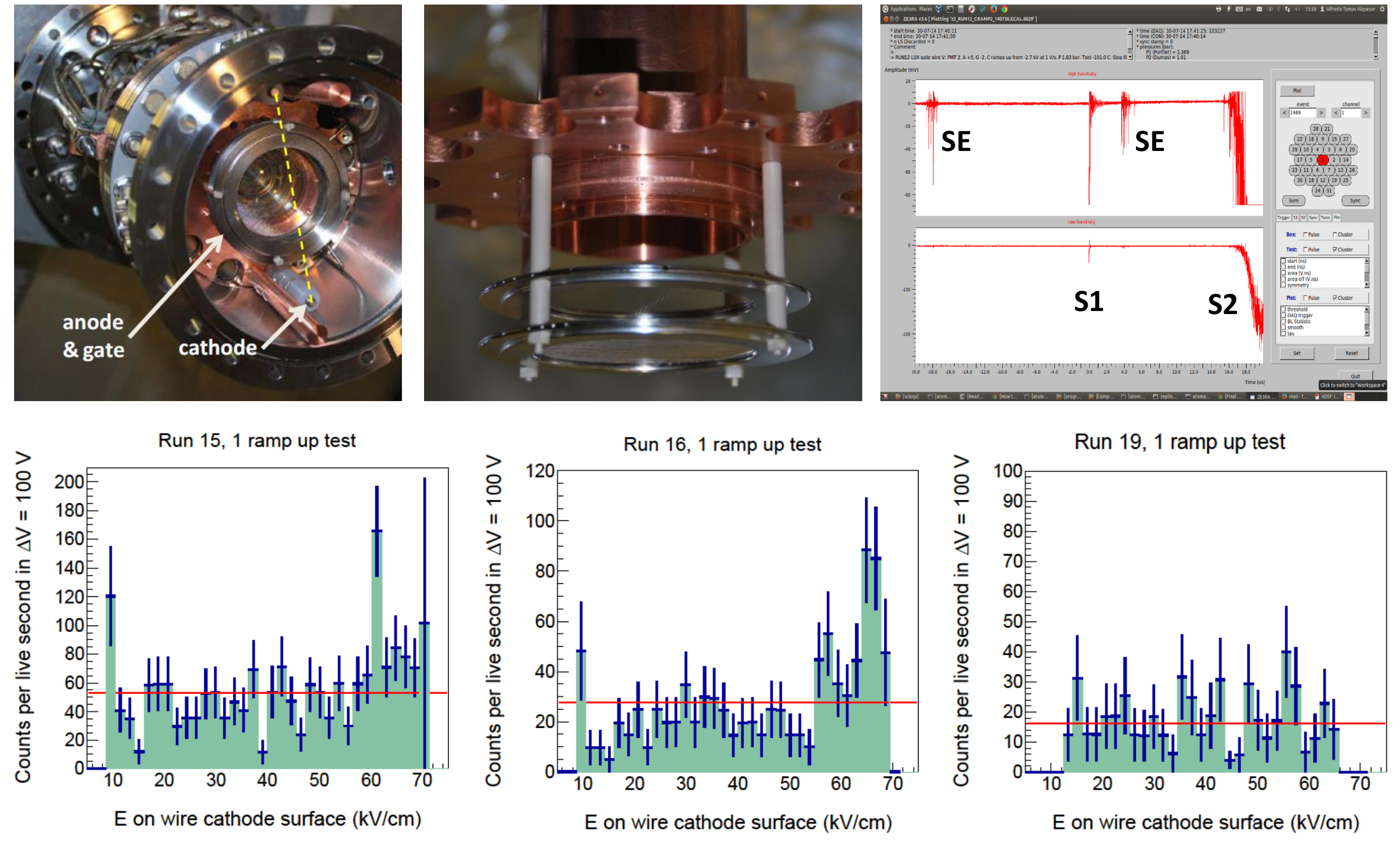}
\tdrfcaption[ImperialWire]{Imperial chamber to study high-field emission processes in thin wires}{Imperial chamber to study electron and photon emission from thin wires; Top from left: Photograph of upward view of the chamber from the bottom, indicating the location of the cathode wire and the gate-anode system (a PMT window can be seen through those grids); Photograph of side view of the gate and anode grids, \SI{14}{\mm} apart; An example event showing S1 and S2 pulses (in dual dynamic range) as well as two single electron (SE) pulses; we search for candidate SE pulses unrelated to any particle interactions producing S1 and S2. Bottom: Evidence for electron emission from a \SI{200}{\mum} stainless steel wire (same type as used in LUX) before treatment (two samples shown left and center) and after electropolishing (right).}
\end{figure}

The chamber has a single internal PMT viewing down from the gas phase to detect both photon and electron emission from the wire sample. The electroluminescence response has single-electron (SE) sensitivity, allowing to measure minute currents (\SI{\sim100}{e\per\second}, or \SI{e-17}{\ampere}). The gate and anode grids, \SI{14}{\mm} apart, straddle the liquid surface, ensuring S2 yields that are mostly independent of the cathode voltage and sufficiently high for efficient cross-phase extraction and SE detection. The \num{130}-\si{\mm}-long cathode wire is mounted \SI{25}{\mm} below the gate, stretched between two feedthroughs that deliver up to \SI{-5.5}{\kV} to the liquid. Electric field lines radiating from the upper surface of the sample guide emitted electrons to the electroluminescence region in the gas where they are detected with high efficiency. Most subsystems required to operate the chamber were inherited from ZEPLIN-III. The ZE3RA analysis software allows full exploitation of the \num{2}-\si{\ns}-sampled waveforms~\cite{Neves:2011ed}, which are recorded in high- and low-sensitivity channels to cover both very small signals and larger S1 and S2 pulses (Figure~\ref{XDSf:ImperialWire}, top right). The voltage applied to the wire sample is ramped up slowly (\SI{1}{\V\per\second}) to several \si{\kV}, with PMT signals being digitized simultaneously. Electron emission from the cathode can, if accompanied by prompt light, be reconstructed to the cathode depth by electron drift time.

We find such emission occurring from fields as low as \SI{\approx10}{\kV\per\cm} in some wires (e.g.~\SI{200}{\mum} stainless steel), while at least one other could withstand hundreds of kV/cm without any discernible effect (\SI{40}{\mum} BeCu). Emission episodes are consistent with average SE rates of \SIrange{100}{1000}{\Hz} (faster, ``impulsive'' structure can be observed within these). Two examples are shown in Figure~\ref{XDSf:ImperialWire} (bottom left and center) for samples from the \SI{200}{\mum} LUX cathode wire, the least resilient tested so far. In some instances we have evidence of simultaneous photon emission, which suggest that the local electric field is higher than the electroluminescence threshold of LXe~\cite{Aprile:2014ila}. Emitters are (probably) localized in field, and they are rarer and fainter on a second ramping test, suggesting a partial conditioning effect. Significantly, they subside at higher fields and are not reproducible subsequently. We tested electropolished samples of the same wire and obtained significantly improved results, as shown in the same figure (bottom right). Finally, we will explore chemical passivation. Both of these measures can be applied to the LZ grids.

At LBNL, the test bed pictured in Figure~\ref{XDSf:Colonial} is operated which is functionally similar to that at Imperial, but devoted to evaluate small cathode wire planes consisting of a stainless steel frame and stretched wires. A key question is whether this performance can be simply inferred from the single-wire studies, or if the length of wire is in fact a critical parameter. A complete cathode grid frame will allow an approximate 10-fold increase in wire length compared with a single wire. This chamber has also improved electric field uniformity as well as higher light collection, allowing a better understanding of the correlation between electron and photon emission. The baseline gate-anode design can be studied in detail at this scale. Other studies with this chamber will include understanding more rigorously the cross-phase electron emission process at the liquid surface.  

\begin{figure}[htb]
\centering
\includegraphics[width=0.7\textwidth]{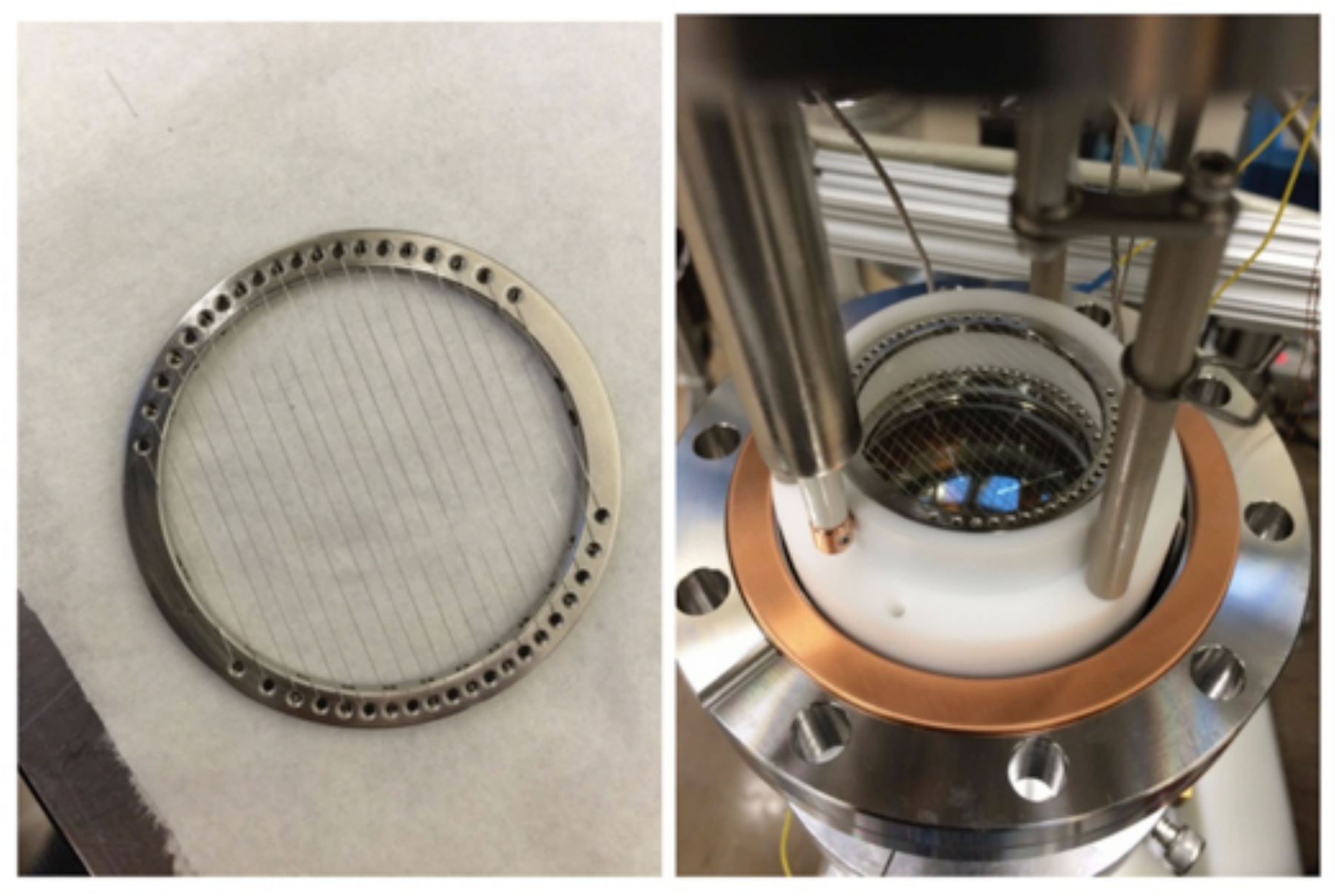}
\tdrfcaption[Colonial]{LBNL test chamber for small wire grids}{Left: A wire grid that will be tested in the small LBNL test chamber. Right: Photograph of the LBNL small grid test chamber.}
\end{figure}

\tdrsubsec[SLACPhaseIST]{TPC Design Testing}

Larger, scaled components of the LZ TPC must also be tested to assure high voltage performance. It is critical that these assemblies sustain the applied cathode voltage without producing light from electrical discharges across their components or to the inner wall of the cryostat. The scaled Phase-I TPC design for testing has a \num{6}-inch inner diameter, and with appropriate scaling of the vessel, achieves the same surface and bulk fields as LZ with the application of half the desired cathode voltage, i.e. \SI{-50}{\kV} in LZ is equivalent to \SI{-25}{\kV} in the System Test.

\begin{figure}[htb]
\centering
\includegraphics[height=0.3\paperheight]{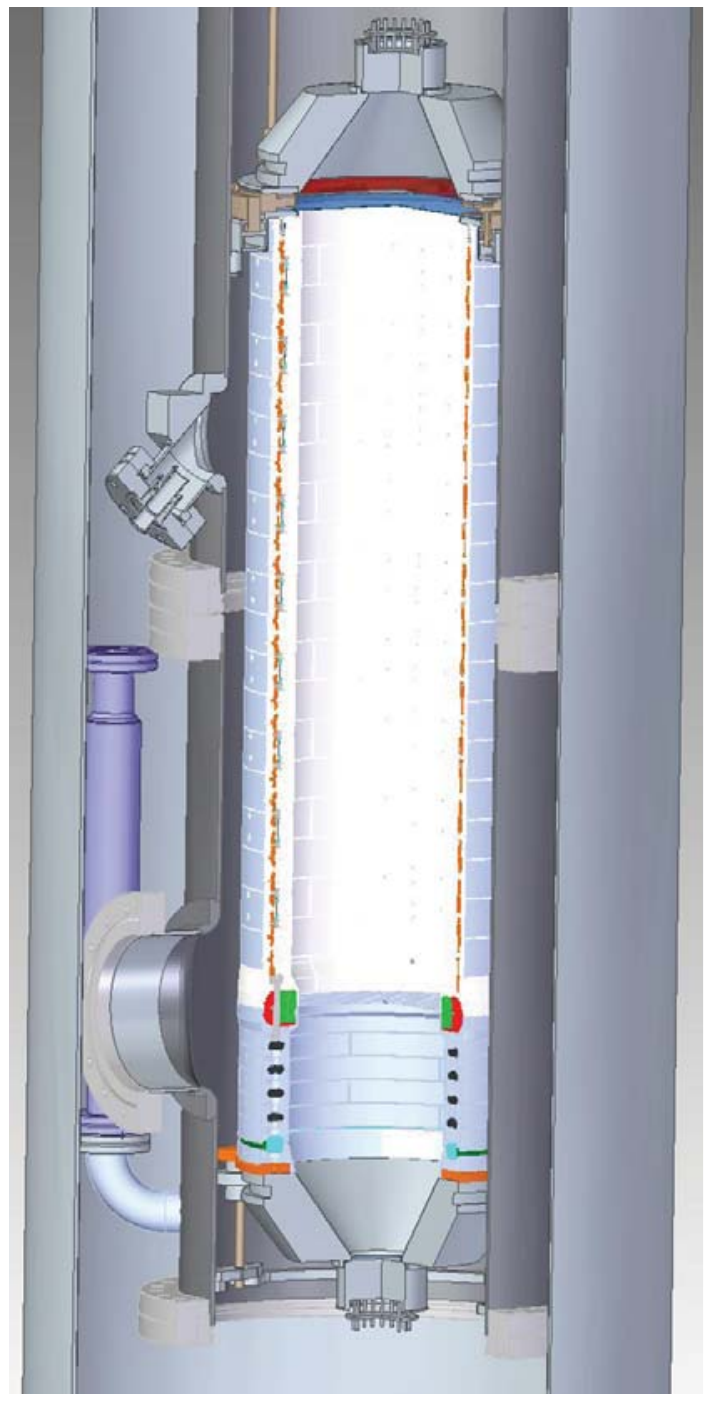}\quad
\includegraphics[height=0.3\paperheight]{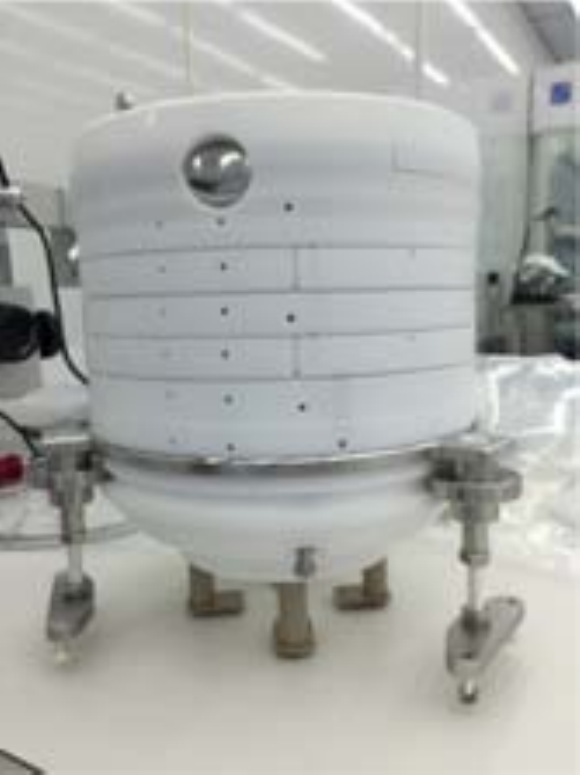}
\tdrfcaption[PhaseITPCCAD]{SLAC System Test Phase-I TPC}{Left: Schematic CAD representation of the full SLAC Phase-I TPC; Right: Photograph of the Phase-I reverse-field region.}
\end{figure}

The design of the Phase-I TPC is shown in Figure~\ref{XDSf:PhaseITPCCAD}. It follows closely that of the LZ TPC, being roughly divided into three functional sections. The first is the reverse-field region between the cathode and bottom grids, with five pairs of resistors grading down to ground with a total \SI{12.5}{\giga\Ohm} impedance set between four field-shaping rings enclosed in PTFE. Above that, the drift field region consists of \num{20} resistor pairs running from the cathode to the gate grid between field shaping electrodes encased in PTFE rings within the LXe. The final section is the electroluminescence region, which consists of the gate grid, weir manifold, and anode grid, and spans the liquid surface. PMTs, located at the bottom and top of the TPC, provide the primary data on emission, supplemented by optical fibers running to an external camera for data on electrical breakdown.

A scaled reverse field region, from the cathode to the bottom grid, was first tested in liquid argon (LAr) at Yale University in 2014. A schematic view of the setup is shown in Figure~\ref{XDSf:YaleAr2} (left) and dewar and HV connection (right).

\begin{figure}
\centering
\includegraphics[height=0.4\paperheight]{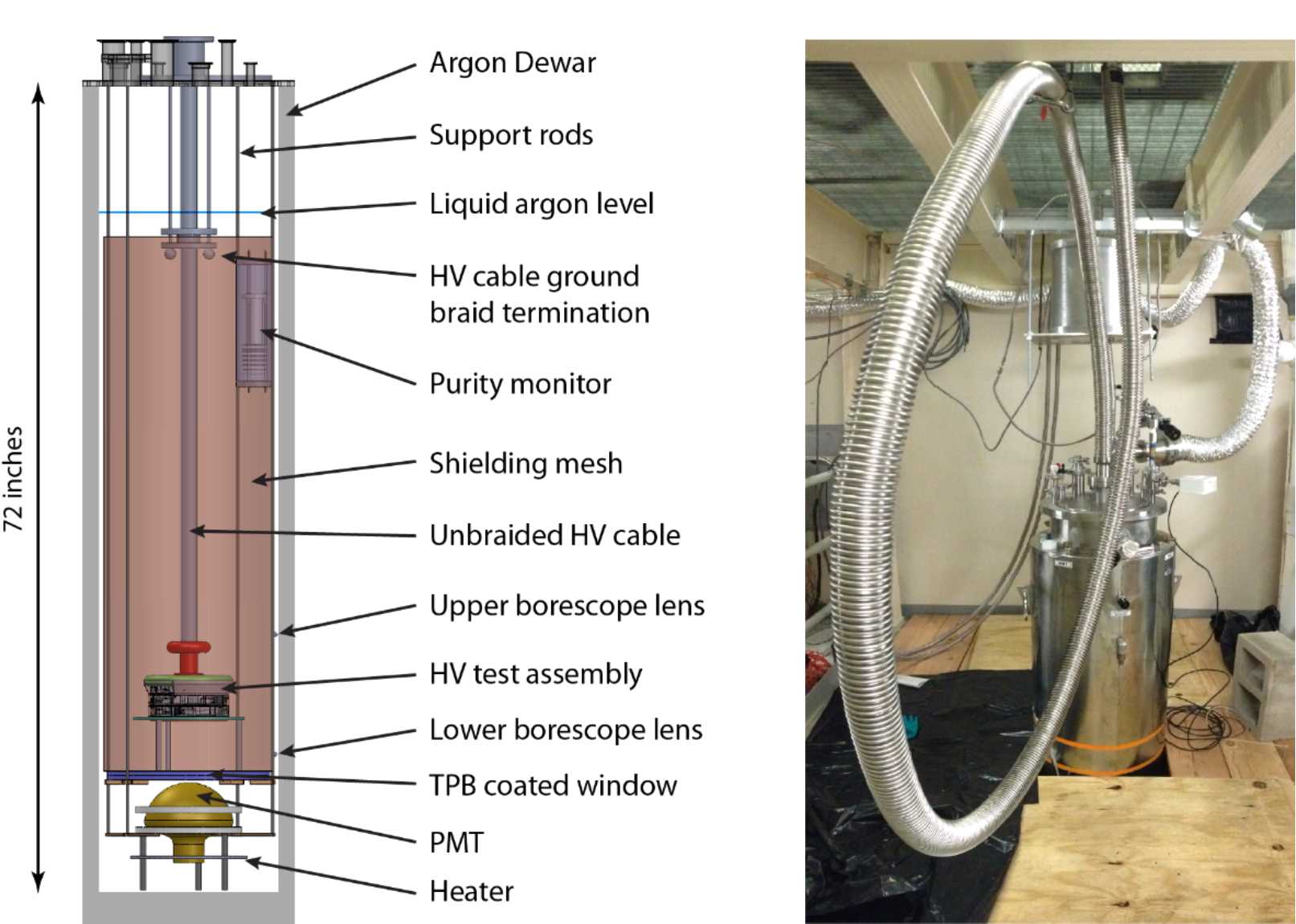}
\tdrfcaption[YaleAr2]{LAr System Test at Yale}{Left: Side view showing the internal components of the liquid argon test system. Note that the vertical support rods (shown out-of-plane in this view) are positioned outside the shielding mesh. For clarity, only two of the seven borescope lenses are shown. The HV test assembly shown here is a portion of the grading structure of the reverse- field-region of the detector. Right: The liquid argon Dewar and HV cable conduit. The HV feedthrough sits above the square hole in the steel grating at the top of the image.}
\end{figure}

The HV tests were performed within a \num{240}-liter cryogenic dewar of \num{16}-inch bore. The bottom of the tested assembly was grounded by the platform; the top was connected to HV that can be ramped to \SI{200}{\kV} to simulate the LZ cathode. The HV is delivered through a polyethylene cable that originates at a feedthrough located \SI{8}{\footl} above the top of the dewar. The feedthrough is connected to a DC power supply from Glassman High Voltage. A controlled electrostatic environment was maintained around the test assembly by surrounding it with a highly transparent grounded metal mesh that shields any nearby structures. Seven lenses at various locations viewed the assembly from just outside the mesh. These were connected to fiber bundles that routed the images to a charge-coupled device (CCD) camera located just above the top flange of the LAr dewar, providing a real-time view of any electrical discharges that occurred during testing. Below the tested assembly was a quartz window coated with fluorescent tetraphenyl butadiene (TPB) wavelength shifter. This window was viewed by an \num{8}-inch PMT, giving efficient detection of ultraviolet light with single-photon sensitivity.

\begin{figure}[htb]
\centering
\includegraphics[width=0.4\textwidth]{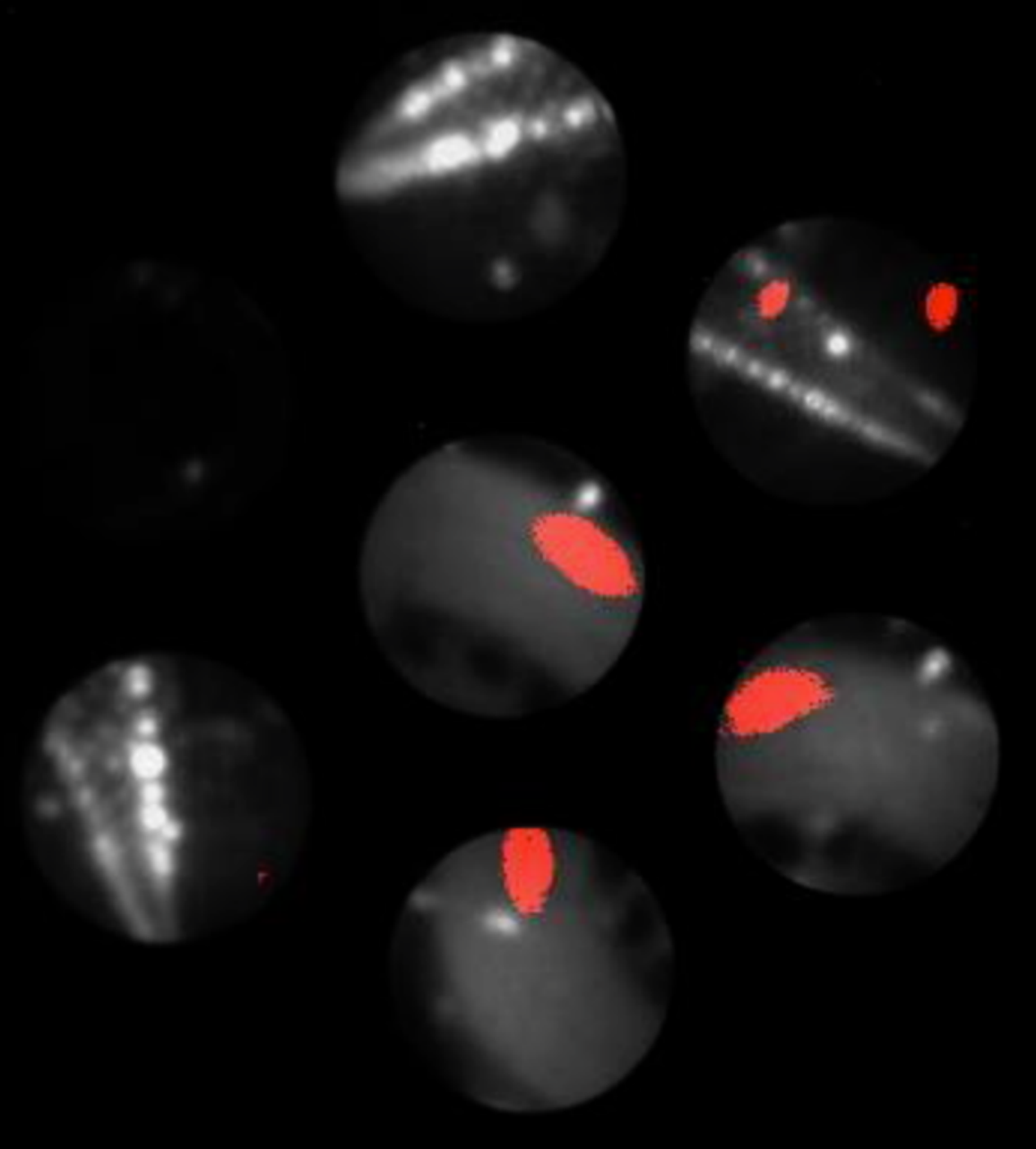}
\tdrfcaption[YaleCameraGlow]{Glow detected from Reverse Field Region tested in LAr}{Image from \num{7} fibers from the RFR tests at Yale. Light preceding breakdown at the locations of the resistors is shown in red, overlaid on the images of the RFR under LED illumination.}
\end{figure}

\begin{figure}[ht]
\centering
\includegraphics[width=0.4\textwidth]{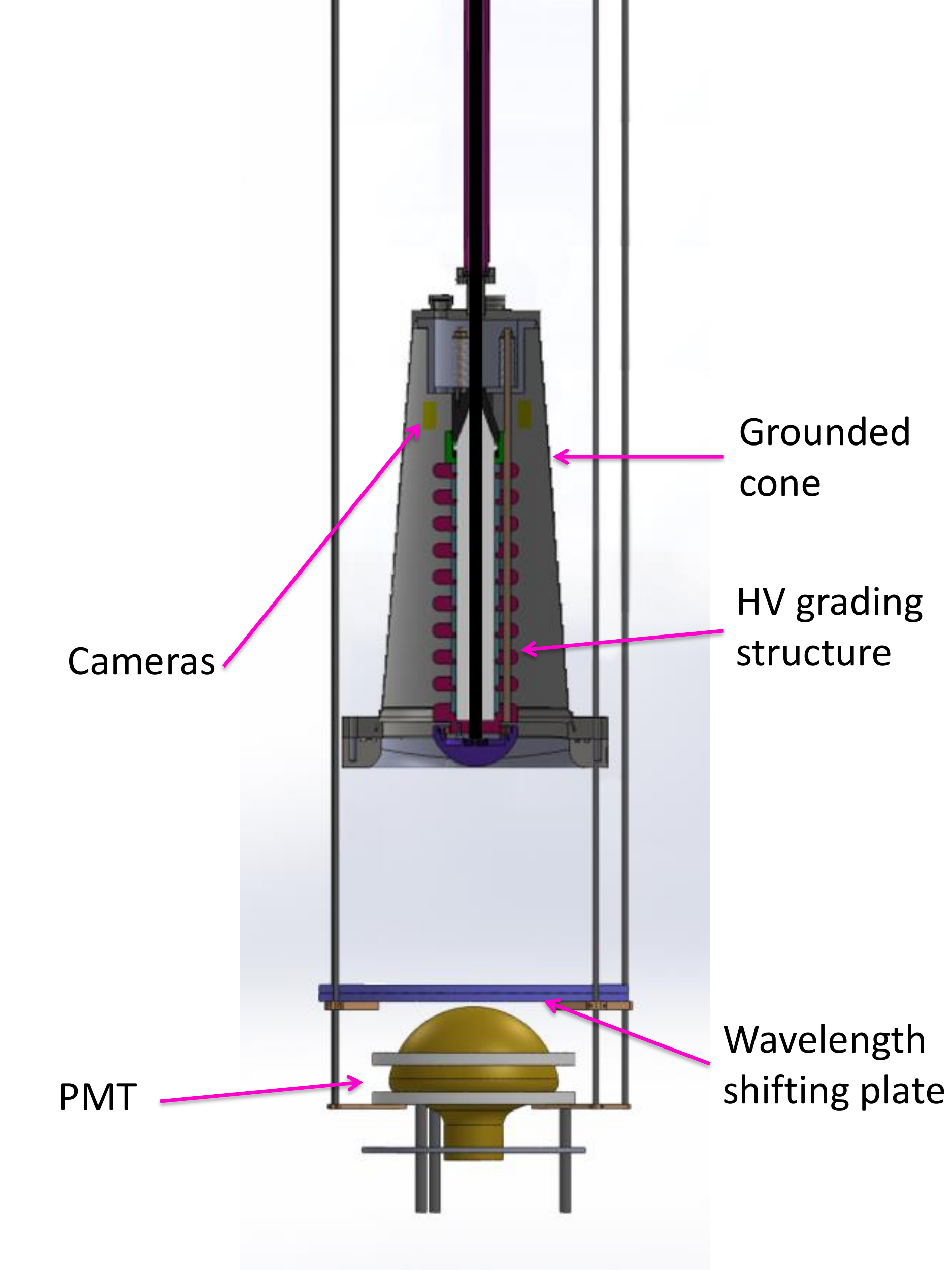}
\tdrfcaption[LArTest]{LArTest}{Schematic of the planned HV test of the cathode connection region immersed in LAr.}
\end{figure}

The tests completed at Yale identified the installed RFR resistors as a weak point when oriented vertically (required by the small radius of curvature of the Phase-I TPC) and with resistors connected in series together between field shaping rings. The camera system identified light produced at the resistor locations during breakdown, as shown in Figure~\ref{XDSf:YaleCameraGlow} at fields equivalent to \SI{200}{\kV} applied in LZ, and this guided us to the current design of more field shaping rings in the reverse field region with resistors oriented horizontally and not connected directly to each other.

Using this system, future HV tests are planned for the cathode connection region. The cathode connection grading structure and the connection itself will undergo testing at voltages up to -150 kV, while viewed by a PMT and CCD camera as in previous tests. Any glow of these components under applied voltage, or electrical discharges from high voltage to ground will be immediately visible. Purity of the LAr will be monitored using a small TPC with Au photocathode to ensure that electron lifetime is sufficiently high that impurities  do not suppress electrical breakdown. A schematic of this test setup may be seen in Figure~\ref{XDSf:LArTest}.

\begin{figure}[htb]
\centering
\includegraphics[width=0.90\textwidth]{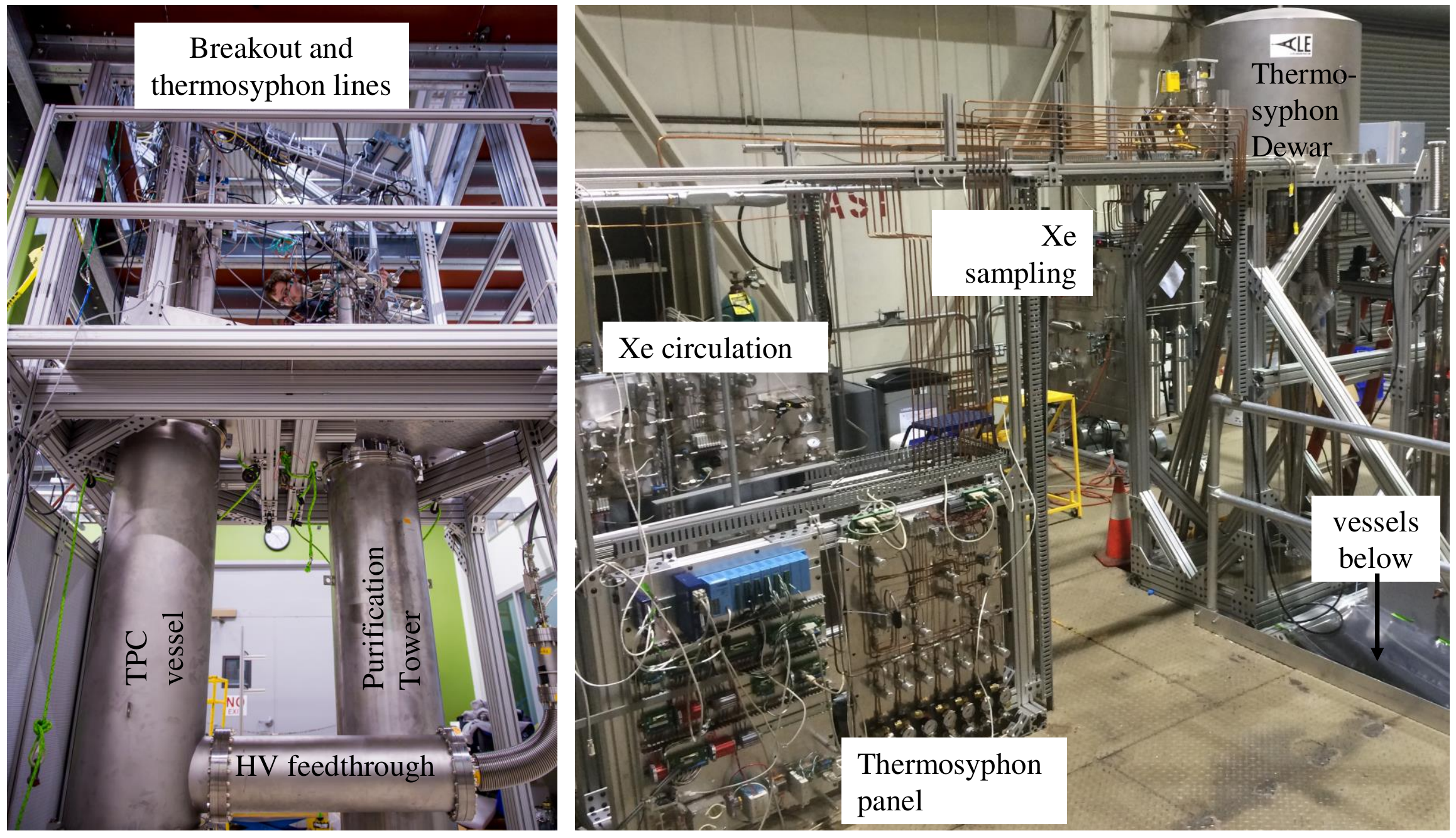}
\tdrfcaption[SLACPlatform]{SLAC System Test platform at SLAC}{The LZ System Test platform at SLAC. Left: Phase-I system, consisting of the TPC vessel on the left with a side-entry cathode HV feedthrough visible at the bottom, purification tower on the right, and various PMT, other grid HV, plumbing and cryogenic breakouts above. Right: Support infrastructure housed above the vessels, consisting of thermosyphon system gas panel at front left, Xe circulation panel behind that, and the multi-port thermosyphon dewar at upper right. The cold-trap Xe assay system developed by the Maryland group is in the back right.}
\end{figure}

The Phase-I TPC will undergo testing in liquid xenon at the SLAC System test platform. The staging at SLAC is shown in Figure~\ref{XDSf:SLACPlatform}. The former BaBar counting room and surrounding space in the IR2 experimental hall was renovated for the purpose of hosting this system, with ample room for expansion. A soft-wall clean area with HEPA units surrounds the vessels. The support systems for these test vessels, partially visible in Figure~\ref{XDSf:SLACPlatform}, are extensive. They build on developments from LUX and ZEPLIN and serve as prototypes of what will be used on LZ. Cryogenics for both phases are supplied by a thermosyphon system, which consists of a multi-port L\CNt dewar capable of providing more than \num{12} separate PID-controlled cooling heads.

The system for online purification through a hot getter uses highly automated gas-handling panels based on 1/2-inch-diameter tubing that will accommodate flow rates well in excess of \SI{100}{slpm}. The system also features a high-flow capacity metal diaphragm compressor as the circulation pump. This technology allows very high flow rates and has been identified (Chapter~\ref{chap:XCS}) as the technology for LZ, but to our knowledge it has not been used in any previous similar Xe experiment. Thus, the system will provide an important test of these pumps.

Critical elements of control and fail-safe recovery of the Xe have also been developed as part of the system test platform, including integration of process loop controllers (PLCs) for essential systems and integration with larger slow-control system development. Phase-I Xe recovery uses a thermosyphon-driven storage and recovery vessel patterned on a similar device used for LUX.  Elements of the planned LZ online and slow-control systems have been developed for the system test to allow a high degree of test automation. The gas system is also designed to be closely integrated with an automated, high-sensitivity purity-monitoring system initially developed by the Maryland group. This will be important in order to achieve purity for successful HV testing; it also allows us to check our understanding of various parameters related to purification that will be important for LZ. Finally, the system is designed to accommodate the range of gaseous radioactive calibration sources deployed in LUX and planned for LZ.

\begin{figure}[htb]
\centering
\includegraphics[height=7.2cm]{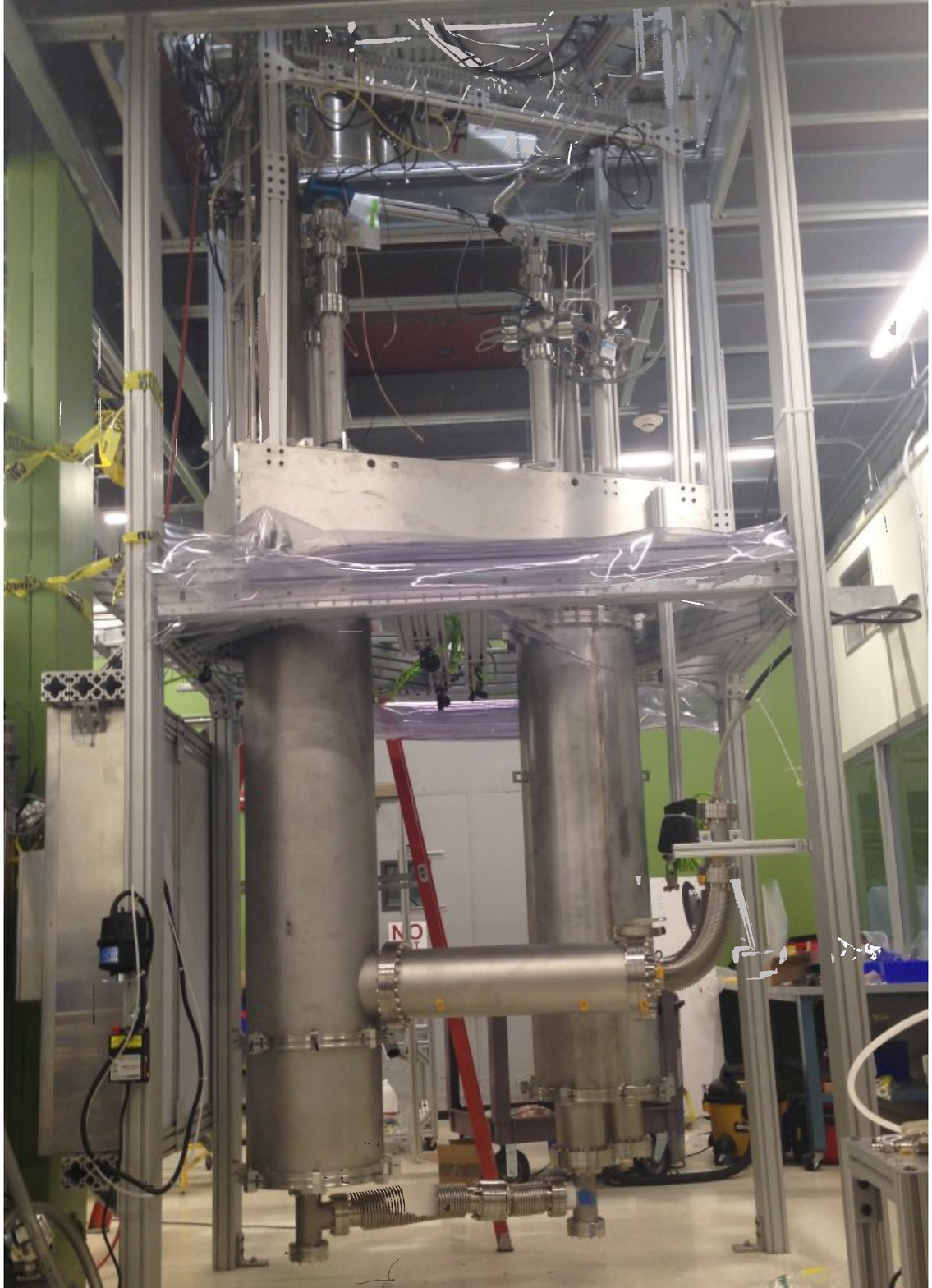}\quad
\includegraphics[height=7.2cm]{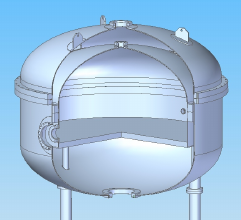}
\tdrfcaption[SLACVessels]{SLAC Phase-I and Phase-II vessels}{Left: SLAC Phase-I vessel set;  On the left is the primary Xe vessel containing the TPC, and the HV feedthrough extends horizontally to the right. The purification tower is the vessel on the right, containing the weir reservoir, heat exchangers and sub-cooler. The HEPA units are visible on the left of the stand, and the cleanroom curtains are rolled up on the detector stand for the photograph. Right: Phase-II Xe vessel design, to test full-scale LZ grids. At the full diameter, \SI{60}{\kg} of xenon will be needed per cm of liquid height.}
\end{figure}

The SLAC Phase-I vessel set is shown in Figure~\ref{XDSf:SLACVessels} (left) with the primary Xe vessel on the left, and the purification tower on the right of the image. The purification tower includes prototypes of the elements of the circulation system planned for LZ: a weir reservoir, two-phase heat exchanger, gas-phase heat exchanger, and a ``sub-cooling'' thermosyphon head on the condensing stream, along with an extensive set of fluid level sensors and thermometers.

The Phase-I tests will continue at SLAC through the spring of 2017 and will allow the assurance of HV operation of the TPC under controlled conditions as well as other LZ subsystems.

\tdrsubsec[SLACPhaseIIST]{Full Scale Grid Testing at SLAC}

A second stage of testing will occur at the SLAC platform to test full LZ-scale grids in liquid xenon. Initially, prototype full scale grids will be tested to confirm the cleaning and handling procedures developed through the Phase-I tests, prior to the testing of the final LZ grids, including the full gate-anode assembly.

The cryostat for Phase-II is shown in Figure~\ref{XDSf:SLACVessels} (right). It will hold \SI{60}{\kg} of LXe per cm of liquid height, and is expected to operate at approximately \SI{6}{\cm} of depth. The extraction region of the gate and anode grids will be tested as a full assembly to full operational voltage. A shortened reverse field region with the cathode and bottom grid will, as in the Phase-I tests, scale the surface fields to LZ values with a reduced cathode voltage.

Optical cameras and a limited number of PMTs will instrument the Phase-II vessel set to look for emission at operational fields. Other instrumentation in the xenon will also combine LZ prototypes, final hardware (such as temperature and level sensors), and some hardware used specifically in the SLAC tests (cathode feedthrough and digitizers).

For Phase-II the purification and circulation system developed for Phase-I will be used, with the exception of requiring a compressor-based recovery system for the larger quantity of xenon. We will recover into standard storage cylinders, as is planned for LZ. This requires a highly reliable system with generator-based backup power. 

The Phase-II testing of the final LZ grids in liquid xenon will be the culmination of the system test program and the final assurance of the TPC meeting both the HV performance requirements and goals.

\clearpage
\bibliographystyle{apsrev4-2}
\bibliography{LZtdr}

\wlog{In tdrinclude command}
\wlog{\ChapLabel}

\renewcommand{\ChapLabel}{chap:OD} 
\renewcommand{\BibLabel}{ODS:bib} 
\renewcommand{\SecPre}{OD}
\renewcommand{\FloatPre}{OD}
\tdrchap[OD]{Outer Detector}

\tdrsec[INT]{Introduction}

In this chapter, we describe the performance and design of the outer detector
system for LZ. The principal signal we seek, that of a WIMP scatter depositing
\SIrange{5}{50}{\keVnr} of energy in the central \SI{5.6}{\tonne}
volume of LXe, will never be accompanied by deposited energy in the
surrounding detector components. In 
contrast, many dominant backgrounds that might fake a WIMP signal will deposit 
energy not only in the central Xe detector but also in the material surrounding 
it. If we are able to detect these secondary interactions, we can veto the 
background event. Table~\ref{IDOt:bkgTable} shows the major backgrounds in LZ, 
which include signals from gammas with energies in the few-MeV range and
neutrons  from (\Pga,\Pn) reactions or created by cosmic-ray interactions. 

To reduce these backgrounds, we surround the large active
Xe volume with an integrated detector capable of tagging gamma rays and
neutrons with high efficiency. Three detector elements are used to achieve this
performance:
\begin{itemize}
\setlength\itemsep{0em}
\item The instrumented ``skin'' of the Xe, the region outside the LXe TPC,
\item The gadolinium-loaded liquid scintillator (Gd-LS), and
\item The portion of the surrounding water that is instrumented as a muon veto.
\end{itemize}
The outer detector system comprises the scintillator and water systems. In
addition to the performance of the integrated veto system, this chapter
describes the design of the outer detector system and modifications needed in
the water tank to accommodate the LZ experiment.
\tdrsec[FP]{Function and Performance}
The outer detector serves two critical functions: 
\begin{enumerate}
\item {\bfseries To veto neutron and gamma backgrounds with high efficiency.}
Although the outer region of the LXe shields the inner region very
efficiently, the outer half of the Xe could not be used as part of the
fiducial mass without an external veto.
By instrumenting the outer skin of the LXe and adding the outer
liquid scintillator detector, we are able to nearly double the
fraction of the Xe in which very-low backgrounds are
achievable. The outer liquid scintillator detector is
essential for vetoing neutrons,
the background that most closely mimics dark-matter scattering.

One risk to the performance of the LZ detector is that some material very close
to the LXe could have a concentration of radioactive impurities higher than expected.
The combination of the LXe skin and the outer detector serves
to mitigate this risk. The integrated veto can suppress most backgrounds even
if they are significantly higher than the design goals, with only a slight
reduction in fiducial volume. 
\item {\bfseries To help characterize and measure the background.} A claim of a WIMP signal
would require extraordinary supporting evidence. The outer detector will
provide crucial supporting evidence necessary to establish a discovery.
In particular, the NR background to a WIMP signal can be estimated by
measuring the number of low-energy deposit scatters in the TPC which are
followed by neutron captures in the outer detector.
\end{enumerate}
The major non-neutrino background sources in LZ are neutrons and gammas from
components within the cryostat and beta decays from radon and krypton distributed
throughout the Xe.  The principal goal of 
the integrated veto system is to reduce the effect of neutron and gamma 
backgrounds to a level smaller than that caused by radon and krypton over a 
very large fraction of the active LXe.

Neutrons represent a particularly troublesome background in the absence of an
external veto. A neutron scatter produces a nuclear recoil (NR), as does a
WIMP scatter, and after scattering they can escape the TPC and skin more
easily than a gamma. The neutron background, which is principally produced
in (\Pga,\Pn) reactions in materials near the LXe, is more difficult to 
predict than the gamma background. If a possible WIMP signal is seen, the 
outer detector will contribute to the identification
and estimation of the neutron background.

The design requirement for NR background is stringent (see Chapter~\ref{chap:MAS}). Without the outer detector, the neutron background would exceed our requirements, so meeting the sensitivity target requires a veto efficiency of
greater than \SI{90}{\percent} for neutrons escaping the TPC. 

\begin{figure}[htbp]
\centering
\includegraphics[width=0.95\textwidth] {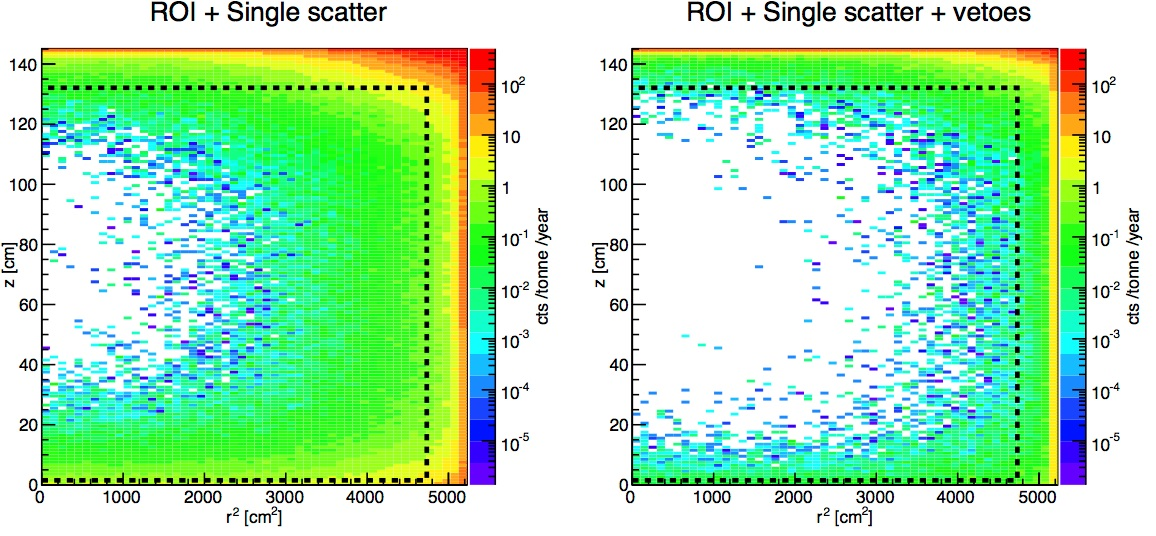}
\tdrfcaption[NRback]{NR Background}
{Total NR background plus ER leakage from material 
radioactivity for sources external to the LXe in the 
TPC, counted over a \SIrange{6}{30}{\keV} acceptance 
region; a discrimination efficiency of \SI{99.5}{\percent} 
is applied to ERs from beta decays and gamma rays. 
Left: Single scatters only, no 
vetoing by the anti-coincidence systems. 
Right: Adding the combination of 
both the skin veto and the outer detector.
The dashed line shows the boundary of the
\SI{5.6}{\tonnesl} fiducial mass.}
\end{figure} 

The principal sources of gamma background are components in direct contact with 
the LXe volume, such as the PMTs, the PTFE reflectors, and the titanium inner
vessel. Gammas in the few-MeV range can scatter at small angles in the outer
region of the TPC, depositing \SIrange{0.5}{6.5}{\keV} of energy corresponding
to the WIMP signal region, and then exit the TPC
without a second scatter. 
Meeting our requirement (see Chapter~\ref{chap:SRD}) requires a veto efficiency of \SI{>70}{\percent} for
such gammas.

We have carried out simulations to characterize the impact of the outer detector
on the background characteristics of the detector.  The results of these simulations
are captured in Figure~\ref{ODf:NRback}, which shows the spatial distribution of the
neutron backgrounds that scatter only once in the LXe volume.

The left panel of Figure~\ref{ODf:NRback} shows the distribution of single
scatters from backgrounds originating from
fixed contamination, such as uranium and thorium,
which generate neutrons that scatter
in the Xe TPC, as well as gamma rays which produce
ERs that leak into the NR band.
The figures plot depth ($z$)
versus radius-squared,
so that the area on the plot is proportional to the volume of LXe. The central
region is, as expected, extraordinarily free of background. But the
neutron background is much higher within \SI{\approx20}{\cm} of
the outer structures. The dashed black line indicates the \SI{5.6}{\tonne} fiducial volume.
Without using information from the LXe skin or the
outer detector, the fiducial mass is a bit less than half
of the active LXe. Most of the active LXe itself in this case
is used as a veto rather than as target
material for WIMPs. If one of the component materials in the cryostat were to
have a larger amount of radioactivity than the design target, even less of the
active LXe would be in the fiducial volume.

The right panel of Figure~\ref{ODf:NRback} shows the performance when vetoing
events that also deposit energy in either the instrumented LXe skin or the outer
detector.  The dashed line shows that the fiducial volume can be extended to
within a few centimeters of the edge of the active LXe. The integrated veto system
enables a fiducial volume of
\SI{5.6}{\tonne}, almost twice as large as for a
stand-alone LXe TPC. Even if the neutron backgrounds
were significantly higher than assumed in this study, the very-low background
needed for effective operation of LZ could be maintained by reducing the \SI{5.6}{\tonne}
fiducial volume by only a very small amount.

Because of the large surface area of the LXe vessel,
the fiducial volume increases
by about \SI{270}{\kg} for every additional centimeter
thickness of LXe at the boundary. To
meet the LZ background requirements for \SI{>5.6}{\tonne}
fiducial volume without
using an external veto would require a TPC containing
\SI{11}{\tonne}, \SI{4}{\tonne} more
than the LZ design value.

\tdrsec[OV]{Outer Detector Overview}

\begin{figure}[htbp]
\vspace{-3pt}
{\centering
\includegraphics[width=0.49\textwidth]{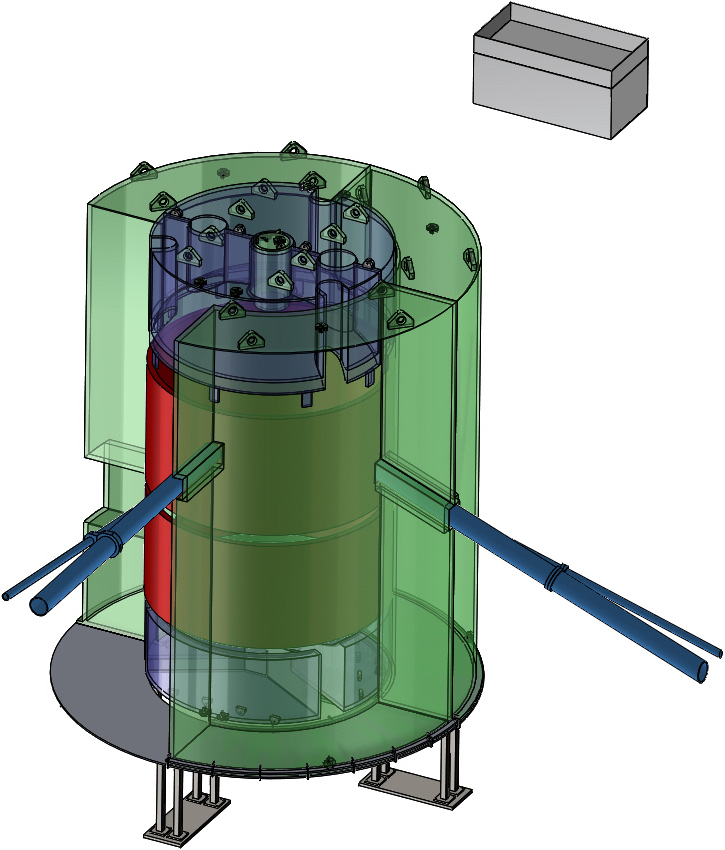}
\includegraphics[width=0.31\textwidth]{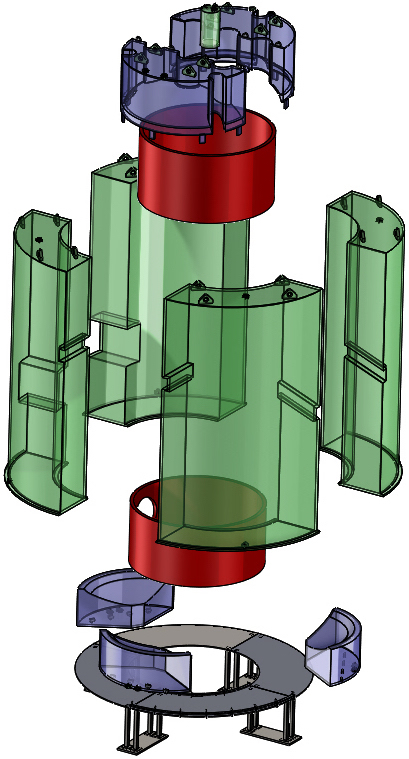}
\tdrfcaption[Detector]{Layout of the LZ outer detector system}
{Layout of the LZ outer detector system, assembled on the left, and
exploded to display the ten acrylic tanks, water displacers (in red),
and stand, on the right.  The tanks will be filled with Gd-LS, and
the largest are the four quarter-tanks
on the sides. Two tanks cover the top, and three the bottom. A small cylindrical 
vessel fits into a cutout at the center of the top tanks, replacing a YBe source 
for most running. The grey box is a reservoir for liquid scintillator.}}
\end{figure}

The proposed layout of the LZ outer detector is shown in
Figure~\ref{ODf:Detector}. A hermetic detector is built from ten vessels
fabricated from UVT acrylic. The use of segmented vessels allows fabrication to
take place at the manufacturer's facility at considerable cost savings. The
sizes of the vessels are chosen to allow straightforward insertion into the
water tank and assembly of the full detector
inside the water tank. Structural finite element analyses (FEAs) of the vessels
have been performed to validate the design without introducing more inert
material than is needed for safe operation.
The acrylic for the side vessels is \SI{1}{\inchl} thick; for the top and bottom
vessels, the acrylic is \SI{0.5}{\inchl} thick except for the top wall of the top
vessel and the bottom wall of the bottom vessel.

The vessels will be viewed by \num{120} \SI{8}{\inchl} Hamamatsu R5912 PMTs. The PMTs are
mounted on stainless steel frames in the water tank, separated from the Gd-LS
vessels by \SI{84}{\cm} of ultrapure water. This arrangement
gives a light-collection efficiency of about \SI{7}{\percent} averaged over the volume of
the outer detector, corresponding to a light yield of about 130~photo-electrons
for a \SI{1}{\MeV} energy deposit. The
water shields the Gd-LS from gammas that originate in the R5912 tubes. A
low-density water displacer will be used to fill in the gaps between the
cryostat and the acrylic vessels and the gaps around
the penetrations, to reduce the probability of absorption in water.
White diffuse reflector will be placed on all available surfaces to
improve collection of the scintillation light.

\begin{figure}[hbtp]
{\centering
\includegraphics[width=0.75\textwidth]{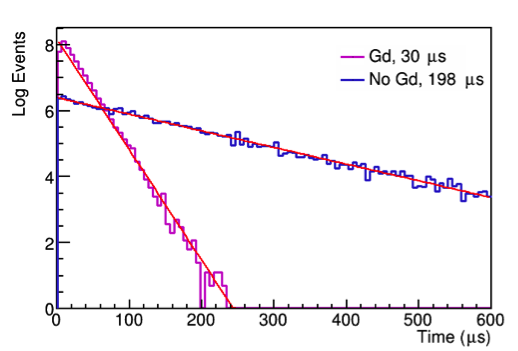}
\tdrfcaption[Capture]{Neutron Capture Times}{The simulated
distribution of capture times for thermal neutrons in
the outer detector for liquid scintillator with and without
\SI{0.1}{\percent} Gd.}
}
\end{figure}

Simulation of the veto performance showed that the veto efficiency varies slowly
with the thickness of the scintillator in the outer detector over the range 
\SIrange[range-units=single]{50}{80}{\cm}. 
This thickness is therefore chosen to reduce the risk of problems
during fabrication and assembly. To insert the side vessels into the water tank
easily, the
scintillator thickness needs to be no greater than \SI{70}{\cm}. Cleaning the side
vessels during fabrication requires that the thickness be no less than \SI{61}{\cm},
which fixed the side tank scintillator thickness.

The radioactivity in the mine walls causes a flux of gamma rays to impinge
on the water tank.  The thinnest portions of water are at the top and bottom
of the tank.  The top and bottom tanks, as well as the vertical extent of the
side tanks, were adjusted to reduce the rate caused by penetrating gamma
rays at the top and bottom, without significant loss in neutron veto
efficiency.   Our simulation indicates a surviving rate
of order \SI{100}{Hz}, however, simulation of the tiny fraction of surviving
penetrating gammas is challenging.  The ``LS screener'' campaign in late
2016, described briefly in Section~\ref{ODS:LS}, will move a small LS
detector through a variety of vertical positions in the water, to test
the simulation.

The liquid scintillator is based upon a linear-alkylbenzene (LAB) solvent,
which is a hydrocarbon chain with one benzene ring attached. LAB has a flashpoint
that exceeds that of diesel fuel, and the safety aspects of diesel fuel in an
underground facility have been explored and defined. The LAB is loaded with Gd,
\SI{0.1}{\percent} by mass, via an organic chelating agent, trimethyl hexanoic acid (TMHA).
This scintillator mix with \SI{0.1}{\percent} Gd doping was used by Chooz
\cite{Apollonio:2002gd}, Palo Verde \cite{Boehm:1999gk,*Piepke:1999db},
and Daya Bay \cite{DayaBay:2012aa,*Guo:2007ug}.
The specific approach adopted by LZ is very similar to that used in the Daya
Bay neutrino experiment, but with additional purification to achieve a lower
radioactivity levels from uranium/thorium/potassium (U/Th/K) impurities.

Gadolinium is added to the scintillator to increase the efficiency for tagging
neutrons while maintaining low veto  deadtime. The benefit of using Gd and
scintillator for this purpose was demonstrated in the ZEPLIN series of experiments.
About \SI{90}{\percent} of the neutrons which moderate to thermal energies
in the scintillator are captured on \IGdoFS or \IGdoFF, releasing 3-4 gammas
with total energy of \SI{7.9}{\MeV} (\IGdoFS) or \SI{8.5}{\MeV} (\IGdoFF);
the remaining \SI{10}{\percent} of these neutrons are captured on hydrogen,
producing a single \SI{2.2}{\MeV} gamma. The Gd captures are tagged with
higher efficiency because of the multiple gammas produced and
the high energy of those gammas. The Gd capture also reduces the neutron capture time to
about \SI{30}{\us}, compared with about \SI{200}{\us} in scintillator without Gd.
Figure~\ref{ODf:Capture} shows the simulated capture
time for low-energy neutrons entering the outer detector. The fast capture
helps reduce deadtime for the veto system.

A veto window of \SI{125}{\us} after a WIMP candidate
is matched to the capture time with Gd.  However,
our detailed simulations have shown that \SIrange{5}{10}{\%}
of neutrons actually capture in the acrylic, and can be accomodated
by a longer veto window.  Our goal is to use a veto window
of \SI{500}{\us}, which we use in our
simulation studies. To maintain a deadtime \SI{<5}{\%} for
a veto window \SI{500}{\us}, we must achieve our goal
of a total OD rate below \SI{100}{Hz}.  We can simultaneously
achieve an inefficiency \SI{<5}{\%} and deadtime \SI{<5}{\%}
with if the OD rate is \SI{300}{Hz}, and we reduce our
veto window to \SI{170}{\us}.

\tdrsec[MS]{Mechanical Systems}
\tdrsubsec[AV]{Acrylic Vessels}
The \SI{17.2}{\tonne} of scintillator liquid are contained in nine permanent acrylic
vessels, as shown in Figure~\ref{ODf:Detector}:
four tall vessels on the sides, two vessels that form a plug on the top, and
three vessels that form a plug at the bottom. A small tenth vessel is used when
the YBe calibration source is not in use.  Taken together, the LS system
forms a \SI{\approx 61}{\cm} thickness of Gd-LS surrounding
the LXe vessel, with several penetrations for connections to the Xe detector and
for calibration systems. Similar acrylic vessels were used for the Daya Bay
Antineutrino Detectors \cite{Guo:2007ug}. 

The masses and volumes of the ten vessels are shown in Table~\ref{ODt:AVmass}.
The side vessels represent the largest part of the veto mass, holding about
\SI{89}{\percent} of the scintillator. Each of the four side
vessels is \SI{375}{\cm} high, extends in radius from 
\SIrange[range-units=single]{97.5}{163.5}{\cm}, and covers
one-quarter of the full azimuth. A vessel is
supported and anchored to a stainless steel base frame, which is in turn
anchored to a base plate installed on the floor of the water tank. The net
upward force on each side vessel when filled is \SI{4940}{\N}.

\begin{table} [H]
\setlength{\extrarowheight}{5pt}
\tdrtcaption[AVmass]{The volume and masses of the scintillator vessels}
{The volume and masses of the scintillator vessels}
\centering
\sffamily
\begin{tabular} 
{|l|c|c|c|c|} 
\hline  
\rowcolor{mrocol}
\multirow{2}{*}
    & {\bfseries Acrylic Mass   } & 
    {\bfseries LS Volume  } & 
    {\bfseries LS Mass } & {\bfseries Net buoyant force } \\ 
	\rowcolor{mrocol}
	&\si{\kg} & \si{\cubic\m} & \si{\kg} & \si{\N}  \\
     \hline
{\bfseries YBe source plug (1) } & \quad \enskip \num{8} & \enskip \num{0.025} & \quad \enskip \num{22} &  \quad \enskip \num{21}\\
 \hline
{\bfseries Side Tank (4) } & \enskip \num{684} & \enskip \num{4.431} & \enskip \num{3823} & \enskip \num{4940} \\
\hline
{\bfseries Top Tank (2)  } & \enskip \num{116} & \enskip \num{0.648} & \quad \num{559} & \quad \num{699} \\
\hline
{\bfseries Bottom Tank (3) } & \quad \num{59} & \enskip \num{0.313} & \quad \num{270} & \quad \num{333} \\
\hline
{\bfseries TOTAL}  & \num{3154} & \num{19.984} & \num{17242} & \num{22178} \\
\hline
\end{tabular}
\end{table}

The two top vessels and the YBe plug form cylinder that fits 
inside the side vessels.  At the thinnest point, the top(bottom) LS
thickness is \SI{40}{\cm}(\SI{34.5}{\cm})-thick, and at the thickest, the
top(bottom) LS thickness is \SI{62}{\cm}(\SI{57}{\cm})-thick.
These tanks will be anchored from the top of
the outer cryostat vessel. The three bottom vessels form a cylinder of the similar
thickness at the bottom. The penetrations
for services and calibrations are positioned within the gaps between the
acrylic vessel and in cutouts in the acrylic vessels.

Filling the acrylic vessel with LS and the water tank with water at the same
time makes it possible to engineer the vessels with acrylic \SI{1}{\inchl} thick,
much thinner than the \SI{4}{\inchl} thickness that
would be required if the vessels were ever filled with LS
and just air surrounding them.
The level of scintillator in each side tank  will be kept near the optimum level for
minimum stress, about \SI{10}{\inchl} above the height of water in the tank. Figure~\ref{ODf:Stress}
shows the results of a finite element analysis of the stress on the side vessels during
the filling process. During filling the stresses will be kept below \SI{1000}{\psil}, well below
the short-term maximum stress of \SI{3000}{psi}. Once the final state has been reached, the
maximum stress is \SI{153}{\psil}, well below the long-term crazing limit of 700 psi. The maximum
deformation at that stage will be \SI{0.05}{\inchl}.

\begin{figure}[htbp]
\centering
\includegraphics[width=0.85\textwidth]{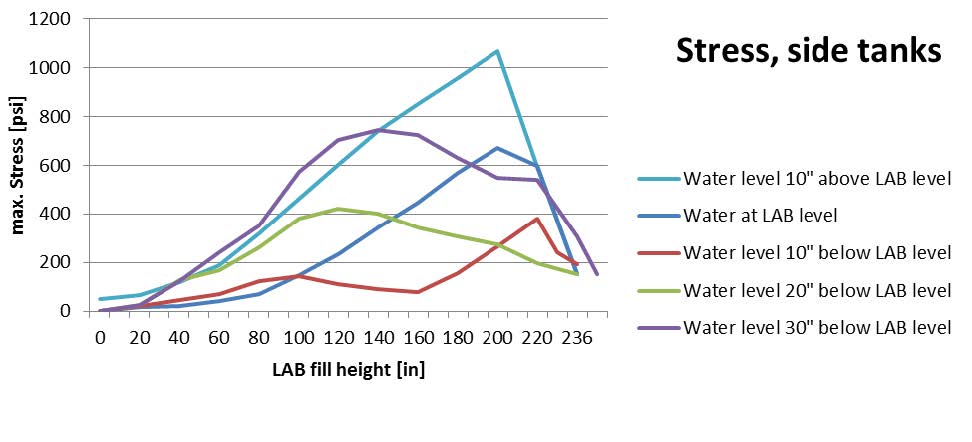}
\tdrfcaption[Stress]{Stress on vessels during filling}{Stress on the side scintillator tanks during filling, from a finite element analysis. The scintillator height in each vessel will be maintained within the range shown in these five plots. The maximum stress will be maintained well below the short term maximum stress of \SI{3000}{psi}.}
\end{figure}

The flanges on the detector cryostat protrude about \SI{2}{\inchl} outside the
cylindrical surface. To avoid building recesses in 
the acrylic vessels to accommodate these protrusions, a low-density closed-cell foam will
be installed as a water displacer around the
outer vessel of the cryostat. This maintains low absorption of gammas between
the scintillator and the LXe skin detector.

The vessels will be cleaned inside and leak-checked at the fabrication vendor.
They will be wrapped in protective sheets at
that time, and placed in double bags before being crated for shipping. The
protective sheets will be removed after they are
installed in the water tank. The final cleaning of the outside of the vessel
will be done at that time. 

\begin{figure}[htbp]
\centering
\includegraphics[width=0.45\textwidth]{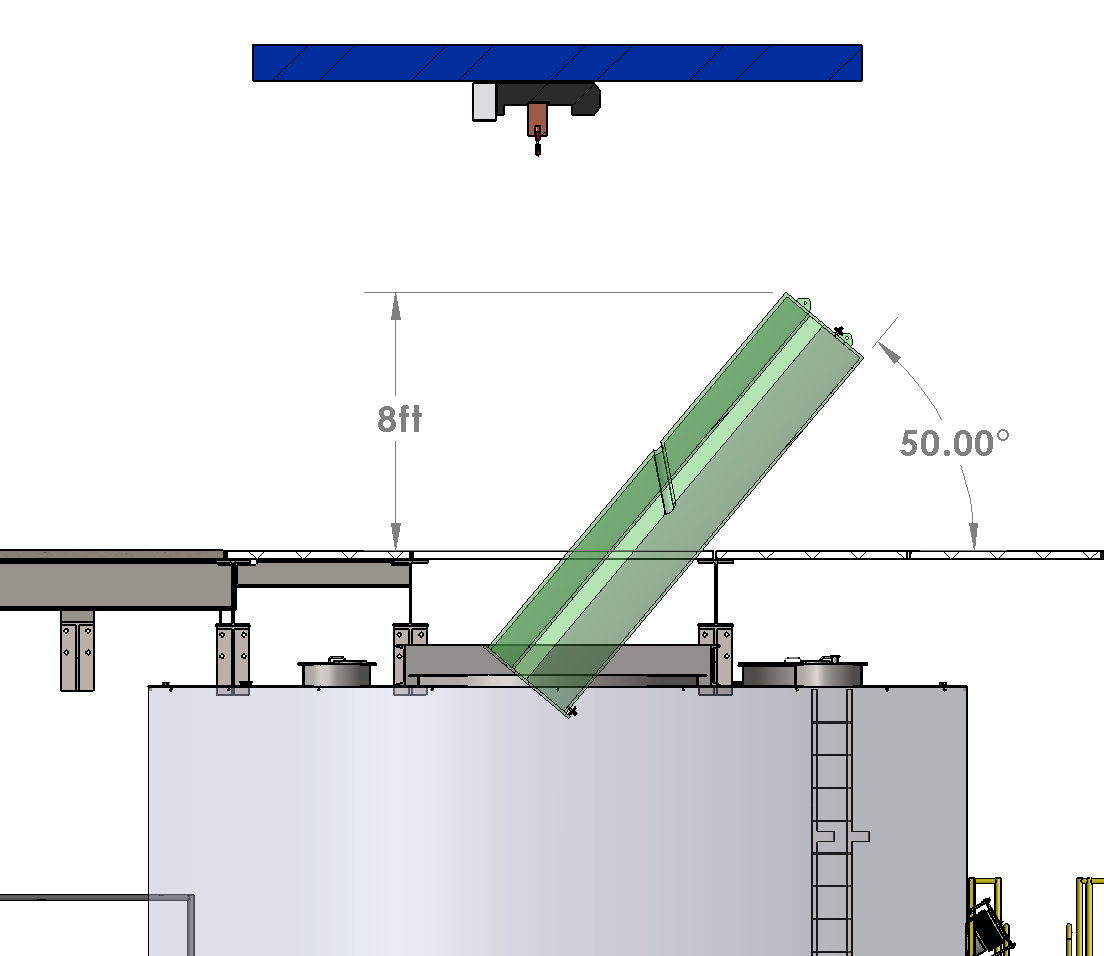}
\includegraphics[width=0.45\textwidth]{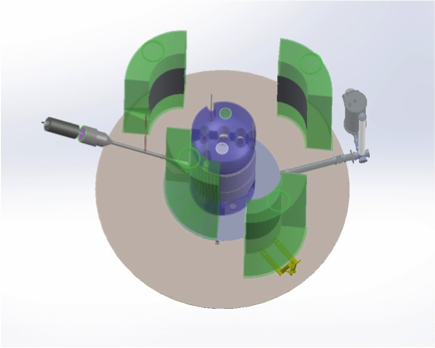}
\tdrfcaption[Assembly]{Two steps in the assembly sequence}
{Two steps in the assembly sequence. The figure on the left shows one
of the quadrant vessels at the point of
maximum height above the water tank. The one on the right slows the four
quadrant vessels in the tank, with one already moved into final location.}
\end{figure}

As a feasibility study, a mock side vessel was slung under the Yates cage,
taken down the shaft, and transported to the cart-wash area just
outside the LUX experimental hall. We have studied the process of installing
the acrylic vessels into the LUX/LZ water tank using a detailed
computer model. The acrylic vessel will be transported in a horizontal position
to the deck immediately above the water tank. The vessel will
then be rotated using lifting eyes at the top and bottom. The left-hand drawing
in Figure~\ref{ODf:Assembly} demonstrates one step of this process,
near the point that requires maximum clearance above the deck. The vessel is
lowered in vertical position into the water tank and then transported radially
outward to near the wall of the tank. 

The right-hand drawing in Figure~\ref{ODf:Assembly} shows the assembly step at
which all of the quadrant vessels are in the tank, and the first one is being
brought into place around the cryostat.
A white diffuse reflector, Tyvek, is strapped with the foam around the cryostat.

The vessels will be fabricated by Reynolds Polymer Technology of Grand Junction, Colorado.
Fabrication will take place during calendar 2017.

\tdrsubsec[SUP]{PMT Supports}

The LAB scintillation light is viewed by 120~\SI{8}{\inchl} PMTs in a cylindrical array
of 20 ladders with six PMTs each. Figure~\ref{ODf:PMTsupport} shows the plan
view of the PMT support system.
The PMT faces are positioned \SI{84}{\cm} from the outer-detector tank wall. The water
between the PMTs and the scintillator vessel shields the active detector
elements from radioactivity in the
PMT assemblies. In this location, the PMTs also see the Cherenkov light from
cosmic-ray muons passing through the water.  

The PMT ladders are attached to the top and bottom of the water tank,
at a radius of \SI{282}{cm}.
The PMT frame has been adapted from from the Daya Bay desin.
A FINEMET magnetic shield to isolate the performance from the local magnetic
field can be deployed The PMT cables run directly
from the PMTs through one of the ports in the top of the water tank to an
electronics rack outside. Strain relief is applied at the port and on the
ladders.

\begin{figure}[htbp]
 \centering
 \includegraphics[width=0.7\textwidth]{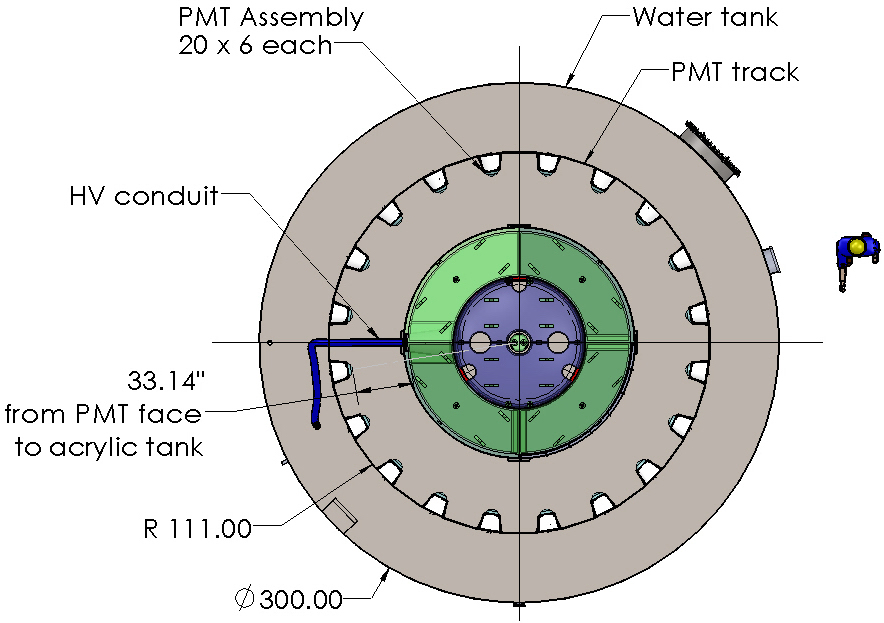}
 \tdrfcaption[PMTsupport]
 {Plan view of the PMT support system}
  {Plan view of the PMT support system. The 20 PMT ladders are
mounted to a circular track on the roof of the water tank.}
\end{figure}

An optical calibration system (OCS) will be deployed on the
support structure, and consists of 35 duplex optical fibers driven
by LEDs.  This system will enable monitoring of the response of the PMT system,
and is described in Section~\ref{ODS:OCS}.

\tdrsubsec[SDS]{Scintillator Distribution System}

Each of the scintillator vessels has one input and one output line.
The lines are Teflon\textsuperscript{\textregistered} tubing, with
strain relief to the PMT ladders. All lines terminate at a feedthrough panel
in the \SI{2}{foot} flange on the north side of the lid
to the water tank. To reduce
pressure on the acrylic
tanks, a \SI{100}{gallon} reservoir will be suspended
from the floor beams above and
next to the north flange of the water tank.

The Gd-LS will be taken underground in \SI{55}{gallon} drums. Secondary
containment will be provided for all of the lines carrying scintillator.  All
volumes to be filled with Gd-LS will first be purged with dry nitrogen,
and all unfilled gas volumes will be kept purged with dry nitrogen to
reduce contamination from atmospheric \IKreF.

To reduce the differential pressure inside and outside the vessels, they will
be co-filled with the water tank that encloses them. The hydrostatic pressure
on the outer surface of the side
vessels varies during the filling process from \SIrange{1}{7.8}{psi}, while the pressure
on the inside from the scintillator varies
from \SIrange{1.7}{7.6}{psi}. The \SI{100}{gallon}
reservoir is connected to each vessel and filled to match the internal pressure to the outside pressure.  

\tdrsec[LS]{Liquid Scintillator}
We chose Gd-LS for the detection medium to achieve excellent efficiency for
neutrons and gammas that reach the outer detector. Various organic liquid
scintillators have been used for neutron tagging due to their production of
relatively large numbers of photons at low energies of a few MeV. The
neutron-capture reaction occurs on the hydrogen in the organic scintillator,
$\Pn+\Pp\to\Pd+\Pgg$, but the cross section is only
\SI{0.33}{\barn}, with a neutron-capture time of about \SI{200}{\us}. The 
single \SI{2.2}{\MeV} \Pgg is also in the energy range of natural radioactivity.

Gadolinium-loaded scintillators have been used in several experiments designed
to detect neutrons produced by inverse beta decay from reactor antineutrinos.
There are several compelling advantages of adding Gd to the liquid scintillator
(LS): 
\begin{itemize}
\item The (\Pn,\Pgg ) cross section for natural Gd is very high, \SI{49}{\kilo\barn}, with
major contributions from \IGdoFF and \IGdoFS isotopes. Because of this
high cross section, only a small concentration of Gd, \SI{0.1}{\percent} by mass, is needed
to dominate over neutron capture on protons.
\item The neutron-capture reaction on Gd releases \SI{8}{\MeV} of energy in a cascade
of 3--4 \Pgg rays. The efficiency for detecting at least one of these gammas is
very high.
\item The time delay for the neutron capture is also significantly shortened to
\SI{30}{\us} in scintillator doped with \SI{0.1}{\percent} Gd, compared to a time of \SI{200}{\us}
in undoped scintillator. This means the veto window can be reduced by a factor
of 7, reducing the deadtime of the outer detector by the same factor. 
\end{itemize}

To detect the low-neutron/gamma backgrounds with high efficiency, the Gd-LS
must have the following key properties: 
\begin{enumerate}
\item Long optical attenuation length, \textgreater \SI{10}{\m} at \SI{430}{\nm};
\item  High light yield, \SI{9000}{\per\MeV};
\item Ultralow impurity content, mainly of the natural radioactive
contaminants, such as U, Th, and K; and
\item  Long-term chemical stability, over the lifetime of the experiment. 
\end{enumerate}
All of these properties have been nearly achieved on a large scale for the Daya Bay
experiment. We are adopting the same formulation and fabrication techniques to
take advantage of this proven performance

It is necessary to avoid any chemical decomposition, hydrolysis, formation of
colloids, or polymerization, which over time can lead to development of color,
cloudy suspensions, or formation of gels or precipitates in the scintillator,
all of which can degrade the scintillator. Recent successful demonstration of
the above-mentioned key items has been done by reactor electron antineutrino
experiments using LAB-based, \SI{0.1}{\percent} Gd-loaded scintillator. 

Assessments of the U/Th contaminations in the Daya Bay scintillator show that
Daya Bay has already achieved levels nearly acceptable for the LZ experiment.
The LZ goal is based upon limiting rate from gamma rays, betas, and alphas in the LXe fiducial volume
to a rate lower by a factor of 8 than that expected from other sources,
to achieve deadtime below \SI{5}{percent}.  The dominant source of
rate in the OD system is the leakage of gamma rays from the walls of the
SURF cavern through the top and bottom of the water shield, and will contribute
of order \SI{100}{Hz} to the OD rates.

The U/Th contamination goals for the scintillator are
\SI{<1.3}{\ppt} and \SI{<4.5}{\ppt} by mass, respectively. Four of the five scintillator
ingredients--\CGdClT, LAB, PPO, and bis-MSB--of the scintillator mixture have
already been analyzed for U/Th contamination by LZ, and met the goals
for Gd-LS.  Since the measurements exceeded the original goals, we have
changed the goals to be equal to the measurements where measurements have been performed.

These goals for each component are summarized in
Table~\ref{ODt:contam}. The goal for \IUtTe is a factor of 15 below
the level achieved by the Daya Bay collaboration. In addition, a
level of \SI{0.8}{ppt} of \IKfz is the required,
which is a factor of 9 lower than achieved by Daya Bay. KAMLAND and Borexino reached
contamination levels orders of magnitude lower by filtering and stripping the
scintillator solvent. We plan to meet the targets by inserting a second
pass of purification into the production process, one step beyond that applied
by the Daya Bay collaboration.

The goal for \ICof is aggressive, and to meet it, underground sources must be used for all organics.
Should the \ICof level be higher than expected, the scintillator threshold will
be raised from \SI{100}{\keVee} to above the \ICof endpoint of \SI{156}{\keVee}.
The scintillator must be kept out of contact with the atmosphere to avoid \IKreF,
and the levels of \IKreF observed in both KAMLAND and Borexino would contribute
a negligible rate to LZ.

\begin{table} [H]
\setlength{\extrarowheight}{5pt}
 \tdrtcaption[contam]{Radioactive impurity goals in LAB-based Gd-LS}
 {Radioactive impurities in LAB-based Gd-LS. The left side
contamination goals for various components of the Gd-LS.  Cells
shaded yellow are based on screening measurements.
Propagation of these contributions to the proposed Gd-LS mixture
is summarized in the last four columns.  The goals (requirements)
keep the rate contribution to the OD below \SI{15}{Hz} (\SI{80}{Hz})
from these sources. }
\centering
\sffamily
\begin{tabular} 
{|l|c|c|c|c|c|c|c|c|c|} 
\hline 
\rowcolor{mrocol}
&  \multicolumn{4}{|c|}{ \textbf{Raw values (ppt)}}  &
{ \bfseries{Gms/liter} } & \multicolumn{4}{|c|}{ \bfseries{Values in
0.1\% Gd-LS (ppt)}} \\
\rowcolor{mrocol} {\bfseries Item}
 & \IUtTe & \IThtTt & \IKfz & \ICof & {\bfseries in Gd-LS} &
\IUtTe & \IThtTt & \IKfz & \ICof \\ 
 \hline
{\bfseries LAB} & \cellcolor{yellow}0.004 & \cellcolor{yellow}0.007 & 0.5 & \num{12e-6} & 860 &
\cellcolor{yellow}0.004 & \cellcolor{yellow}0.007 & 0.5 & \num{12e-6} \\
\hline
{\bfseries Gd} & \cellcolor{yellow}100 & \cellcolor{yellow}100 & 52 & & 0.86 &
\cellcolor{yellow}0.17 & \cellcolor{yellow}0.17 & 0.09 & \\
\hline
{\bfseries PPO} & \cellcolor{yellow}150 & \cellcolor{yellow}640 & \cellcolor{yellow}27 & \num{110e-6} & 3 &
 \cellcolor{yellow}0.5 &  \cellcolor{yellow}2.2 &  \cellcolor{yellow}0.09 & \num{0.37e-6} \\
\hline
{\bfseries TMHA} & 180 & 650 & 27 & \num{140e-6} & 3 & 0.6 & 2.1 & 0.09 & \num{0.44e-6} \\
\hline
{\bfseries bis-MSB} & \cellcolor{yellow}210 & \cellcolor{yellow}190 & \cellcolor{yellow}50 & \num{19e-3} & 0.015
& \cellcolor{yellow}0.004 & \cellcolor{yellow}0.003 & \cellcolor{yellow}0.001 & \num{0.32e-6} \\
\hline
\rowcolor{lrocol}\multicolumn{5}{|l|}{\bfseries Total} & 867.2
& {\bfseries 1.3} & {\bfseries 4.5} & {\bfseries 0.8} & {\bfseries \num{13e-6}} \\
\hline
 \rowcolor{lrocol} \multicolumn{6}{|l|}{\bfseries Requirement} &
 {\bfseries 10} & {\bfseries 20} & {\bfseries 3} & {\bfseries \num{15e-6}} \\
\hline
\rowcolor{lrocol}\multicolumn{6}{|l|}{\bfseries Daya Bay} &
\cellcolor{yellow}{\bfseries 20} & \cellcolor{yellow}{\bfseries 4} & \cellcolor{yellow}{\bfseries 7} & {\bfseries } \\
\hline
\end{tabular}
\end{table}

The production of the Gd-LS is being led by the Brookhaven
National Laboratory (BNL) group. The manager of the LS production effort has
considerable relevant experience, including supervision of the development and
production of the same scintillator for the Daya Bay experiment. The development of
LZ scintillator will be undertaken in two phases: a demonstration phase (to
reach ppt levels), followed by a production phase for deployment. The BNL group
has a state-of-the-art Liquid Scintillator Development Facility equipped with a
variety of instruments (e.g., UV, IR, XRF, LC-MS, medium-scale mixing reactor,
thin-film distillatory, \SI{2}{\m} attenuation length system, etc.) for quality
assurance that are essential to quality control of the scintillator.

To test whether Gd-LS from the demonstration phase meets all the goals in
Table~\ref{ODt:contam}, we have built a ``LS screener,'' consisting of
an acrylic vessel that can hold about \SI{24}{\kg} of Gd-LS viewed by
three very low-background PMTs. We deploy the LS screener in the LZ water tank
in November, 2016 through January 2017.

The most
cost-effective plan for the \SI{17.2}{\tonne} of LZ low-background Gd-LS production
is to carry out the production at the BNL facility and ship the synthesized
scintillator to SURF for filling. The scintillator will be produced at a rate
of \SI{0.5}{\tonne} per week and will be stored in 55-gallon PTFE-lined drums, which
are then shipped to SURF for storage. The purification methods for all
components of the Gd-doped scintillator are developed and will be applied to
each component before synthesis. 

\tdrsec[LD]{Light Detection}
\tdrsubsec[PMT]{Photomultipliers}

Building on the successful use of the Hamamatsu R5912 PMTs in experiments such
as MILAGRO, AMANDA, and most recently Daya Bay, the LZ outer detector will use
120 of these same PMTs. The two existing models for this PMT are R5912 and
R5912-02. The latter possesses four more dynode stages and provides higher
gain, at the cost of higher dark current and slightly degraded timing
characteristics. As LZ does not need the additional gain, the less-expensive
model R5912 was chosen. Even for this model, several subcategories exist that
represent various quality levels of glass window and photocathode materials.
Our assessment of radioactivity levels concluded that the basic model was
sufficient for LZ needs and requirements.

\begin{table} [H]
\setlength{\extrarowheight}{5pt}
\tdrtcaption[PMTs]{Characteristics of the R5912 PMTs.}
{Characteristics of the R5912 PMTs, from the Hamamatsu data sheets.}
\centering
\sffamily
\begin{tabular} 
{|l|c|} 
\hline
\rowcolor{mrocol}
{\bfseries Characteristic} &{\bfseries Value} \\
\hline
Number of dynode stages & 10 \\
\hline
Window material & \qquad Borosilicate glass \qquad \qquad \\
\hline
Photocathode material & Bialkali \\
\hline
Minimum photocathode effective area & \SI{284}{\square\cm} \\
\hline
Typical bias voltage for gain of \num{1e7} &	\SI{1500}{\V} \\
\hline
Maximum voltage & \SI{2000}{\V} \\
\hline
Mean QE at \SI{390}{\nm} &	\SI{25}{\percent} \\
\hline
Pressure rating	& \SI{0.7}{\mega\Pa} \\
\hline
\multirow{3}{*}{Radioactivity levels per PMT (all components)} & \IUtTe: \SI{1200}{\milli\becquerel} \\
 & \IThtTt: \SI{680}{\milli\becquerel} \\
  & \IKfz: \SI{920}{\milli\becquerel} \\
\hline
\end{tabular}
\end{table}

The R5912 PMTs are suitable for the outer detector for the following reasons:
\begin{enumerate}
\item	The spectral response ranges from \SI{300}{\nm} to \SI{650}{\nm}, with a peak
wavelength at \SI{420}{\nm}. This matches well the scintillation light from the LAB
mix between \SIrange{390}{440}{\nm}. The quantum efficiency also covers the relevant
range, with an average expected value of \SI{\sim 25}{\percent} at \SI{430}{\nm}.
\item The PMTs will be submerged in up to 6~m water and must be able to operate
in this environment. Daya Bay has been able to demonstrate successful operation
of the R5912 assembly at higher pressures than those required for LZ. This
experiment used the same assembly that LZ will use.  
\item The radioactivity levels of the PMTs and assembly are a fairly weak
constraint, thanks to the minimum \SI{84}{\cm} of water that separate them from any
active detector volume. The water buffer typically reduces the integrated
flux of incoming gammas by more than 2 orders of magnitude, before taking into
account geometric effects. For the event rate in the Xe target, the \SI{80}{\cm} of
water plus the thickness of the scintillator make the R5912 contribution
largely subdominant to internal sources, for both gammas and neutrons. In the
scintillator itself, the simulated event rate from PMT radioactivity is \SI{<4}{\Hz}
(\SI{<4}{\percent} deadtime would be caused by \SI{100}{\Hz}).
\end{enumerate}

The R5912 PMTs and waterproof assemblies will undergo rigorous individual
testing to fully characterize the response and to validate uniform operation
and long-term stability for the lifetime of the detector. These tests will
include individual electrical behavior, gain measurements, linearity,
after-pulsing, dark current, and dark count. In addition, the PMTs will be
radioactively screened to make sure the activities are consistent with the
values listed in Table~\ref{ODt:PMTs}.

\tdrsec[OCS]{Optical Calibration System}
An optical calibration system (OCS) will be used to monitor the performance of the
outer detector and to maintain calibration. The light will be emitted from 30
LED-driven duplex optical fibers mounted on the PMT support system walls, and
5 duplex fibers routed underneath the bottom tanks.  The system on the PMT
walls will monitor overall system response, including attenuation in the Gd-LS and
in the water. Some fibers under the bottom tanks will use light wavelengths tuned to monitor
acrylic transmission.

The OCS will enable the monitoring of the response of the OD system at the level
of \SI{3}{percent}, in time and in space.  Systems will monitor the stability
of the OCS light sources and supporting light detectors.  The range of light
outputs covered by the OCS focus on the threshold of \SI{\approx100}{\keVee}
up to the expectation of the largest signals from a neutron capture on gadolinium,
about \SI{10}{\MeV\ee}.  The OCS will also enable the monitoring and calibration
of large signals from penetrating muons.

\tdrsec[TSD]{Threshold, singles rate, deadtime}

\begin{wrapfigure}{r}{0.65\textwidth}
 \centering
 \includegraphics[width=0.61\textwidth]{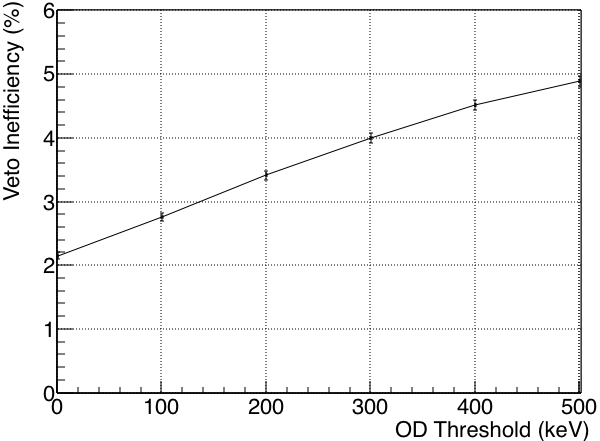}
 \tdrfcaption[Inefficiency]
{Inefficiency for vetoing neutrons}
{Inefficiency for vetoing neutrons from
fixed contamination in detector components
that scatter once in the xenon TPC, 
plotted as a function of threshold in the OD. For this simulation, the 
skin veto threshold is taken to be \SI{100}{\keV}.}
\end{wrapfigure}

As described above, a neutron capture in the Gd-LS releases either a few gammas
of total energy \SI{8}{\MeV} or a single gamma of energy \SI{2.2}{\MeV}. The
neutron-detection efficiency is therefore quite high with the \SI{100}{\keVee} threshold
planned for the outer detector, which corresponds to about 13 photoelectrons
observed. Figure~\ref{ODf:Inefficiency} shows the simulated inefficiency for
vetoing \SI{1}{\MeV} background neutrons that scatter once in the LXe TPC as 
a function of threshold in \si{\keVee}. These simulations indicate that the 
efficiency of the combined veto for neutrons escaping the LXe TPC is about 
\SI{97}{\percent}.

We will read out the outer-detector PMTs each time the LXe TPC produces a
trigger, without need of a hardware threshold. We will therefore be able to
apply the veto offline, with parameters carefully selected to optimize the veto
efficiency. Our goal is to apply neutron veto over a time window of \SI{500}{\us} with
a threshold of \SI{100}{\keVee}. In addition, we will apply a gamma veto
over a much tighter time window of \SI{1}{\us} with a threshold of \SI{100}{\keVee}. There
will be an outer-detector hardware trigger to calibrate and monitor the
background environment.

The sensitivity of the outer detector is estimated to be about 130
photoelectrons/\si{\MeV} from simulation of LZ and this agrees with benchmarks
against similar detectors. Thus, a \SI{100}{\keVee} threshold corresponds to about 
\num{13} photoelectrons, and the trigger rate will be dominated by alphas, gammas, and
betas from the U and Th chains. 

Because the time of the low-energy scatter in Xe is determined well by
measuring the prompt S1 light, the neutron veto window will be determined by
the capture time for neutrons. Our goal is a \SI{500}{\us} veto window.
Our formal requirement is on deadtime, of \SI{<5}{\percent}. This 
requirement can be met with a variety of combinations of
thresholds and veto windows.  Our goal is to
keep the rate in the outer detector below \SI{100}{\Hz},
and use a veto window of \SI{500}{\us}, but currently our
simulations show that we could achieve our deadtime and
ineffiency requirements with an outer detector rate as
high as \SI{300}{\Hz} and a veto window of \SI{170}{\us}.
The radiopurity goals in 
Table~\ref{ODt:contam} keep the overall
background rate in the OD below
\SI{100}{\Hz}.

Ultimately, the of the veto window and threshold needed for
high-efficiency neutron tagging will be determined 
by studying it with calibration data.

\tdrsec[EHS]{Environment, Safety, and Health Considerations}

The most important safety issue to consider underground is flammability. The
flashpoint for LAB is \SIrange{120}{140}{\celsius}, which makes it a Class
IIIB liquid, the lowest hazard category. A Class IIIB liquid is a combustible
liquid with a flashpoint at or above \SI{200}{\celsius}
(\SI{93}{\celsius}). OSHA flammable and combustible liquid regulations
do not apply to IIIB liquids. The boiling point is greater than
\SI{250}{\celsius} and the melting point is below
\SI{-70}{\celsius}, so it is very stable as a liquid. The density of LAB
is \SI{0.86}{gm/cm^3}, significantly less than water. In addition, it has very low
solubility in water. 

The primary risk to consider is a crack in one of the acrylic vessels. The
first preventive step is to thoroughly check the integrity of the vessel before
introducing the scintillator liquid into it. The second is to monitor the
vessel carefully during the filling process. The water tank serves as a
secondary containment system for the scintillator.  We
will also be able to separate LAB from the water in the water treatment. If a
significant leak is observed, we will skim the surface of the water to recover
most of the scintillator, and then remove the residual scintillator by
distillation. 

\clearpage
\bibliographystyle{apsrev4-2}
\bibliography{LZtdr}

\wlog{In tdrinclude command}
\wlog{\ChapLabel}

\renewcommand{\ChapLabel}{chap:CRO} 
\renewcommand{\BibLabel}{CROS:bib} 
\renewcommand{\SecPre}{CRO}
\renewcommand{\FloatPre}{CRO}
\tdrchap{Cryostat}
This chapter discusses the cryostat of the LZ detector. In particular it describes: the results of an extensive screening campaign and the simulations of associated backgrounds that lead to the cryostat material choice, cryostat technical specifications, design with key features, fabrication and tests requirements, and finally transportation and installation underground at SURF.  

The cryostat will be fabricated by a specializing in titanium U-stamped pressure vessel manufacturer selected in the European Union tender process. The fabricator will be in charge of the supplied raw material processing including rolling and forging, cryostat manufacture, tests and certifications, cleaning, packaging and transportation to SURF. 

\tdrsec[MaterialSearch]{Material Search Campaign}
The LZ cryostat is one of the major contributors to the experiment’s background budget due to the 
mass of material required and the vicinity to the LXe target. This imposes stringent 
radiopurity requirements on its materials, and essentially limits the choices to copper, stainless 
steel and titanium – although copper is rejected due to mechanical considerations. Titanium (Ti) has 
a high strength-to-weight ratio, low density and atomic number, and was previously sourced with low 
radioactivity by LUX \cite{Akerib:2011rr}. The level of this radioactivity content would meet LZ 
requirements and the lower particle stopping power combined with considerably lower \ICosz content 
make it advantageous over stainless steel. The material studies and searches have therefore been 
focused on titanium. As a mitigation strategy against failing to source the material with adequate 
radiopurity and in sufficient quantity, on a sensible timescale, parallel studies have been 
conducted of stainless steel activity and a full cryostat design in that material has been 
developed.    

Neutron and gamma-ray radioactivity from the cryostat is due to largely to uranium and thorium 
content in the construction material, as well as \IKfz and \ICosz. To assess the typical 
concentration of these isotopes in titanium and stainless steel the complementary techniques of 
direct gamma-ray spectroscopy and and mass-spectrometry were employed. Samples were procured from 
various stages of production to inform typical activity, reproducibility of particular suppliers and 
points of inclusion of contamination - particularly for titanium.  

\tdrsubsec[Titanium]{Titanium}
 The production of Ti metal is a complex procedure that involves a number of stages in which 
additives and inclusions are deliberately introduced. Several such points in the production cycle 
may contribute to contamination of the final product with elements containing high concentration of 
radioactive isotopes such as \IUtTe, \IThtTt, \IKfz and \ICosz which are of particular concern. 
The refinement of the mineral concentrates, particularly for ilmenite, involves the addition of or 
exposure to coke, coal, oil, and tar prior to the chlorination process. It is not uncommon for such 
materials to contain relatively high levels of U and Th. However, the TiCl$_4$ produced at this 
stage undergoes chemical treatment and filtering to remove chlorides and sludge before pure liquid 
TiCl$_4$ is created, carrying away most impurities, including U and Th. Ultra-pure TiCl$_4$ is 
commercially available, as are titanium hydride and titanium nitride powders that are produced 
through plasmochemical processing from the TiCl$_4$. Beyond this stage, lack of sufficient contact 
controls with surfaces during the Kroll process and metallothermy present other potential sources of 
U and Th. However, the elements introduced, largely Mg and Ar, will probably not be problematic. The 
Ti sponge post-Kroll processing is exposed to several stages in which U and Th can enter the chain. 
Ti ingots and slabs are produced by pressing and melting the Ti sponge, yet often Ti alloy and Ti 
scrap is added at this stage. Other alloys such as aluminum and vanadium may also be included. 
Radioactive contamination contained within the scrap and alloys is then carried through to the Ti 
ingot and into the roll stock. The major stages of this production process, indicating inclusion 
points, are depicted in Figure ~\ref{CROf:titaniumprocess} \cite{Chepurnov:2013}.

\begin{figure}[hbtp]
\begin{center}
\includegraphics[width=0.8\textwidth]{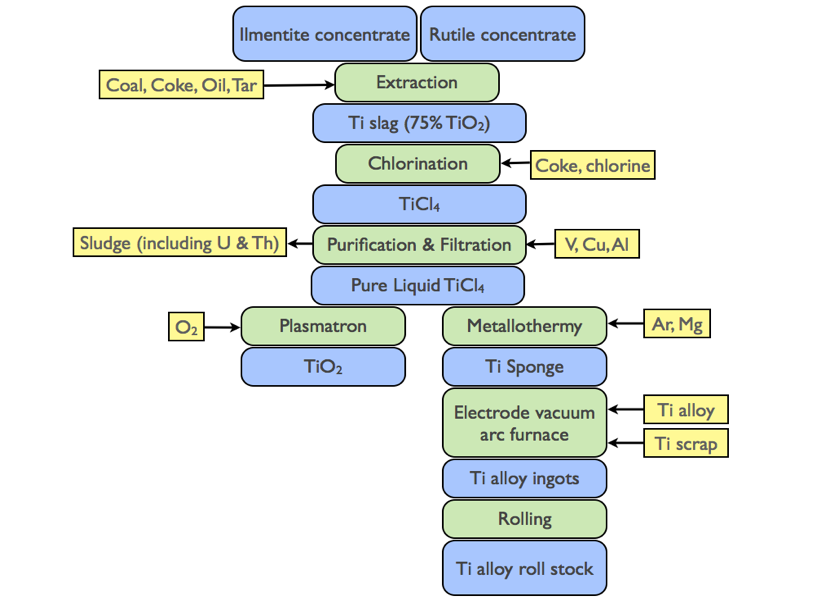}
\end{center}
\tdrfcaption[titaniumprocess]{Stages in commercial production of Ti metal}
{The commercial production of Ti metal, indicating the major stages (green boxes), the post-
processing products (blue boxes), and the additives, as well as reductions during the procedure 
(yellow boxes). Figure adapted from \cite{Chepurnov:2013}}
\end{figure}

In the material search campaign we have engaged several titanium providers including VSMPO 
\cite{VSMPO:2015}, TIMET \cite{TIMET:2015}, Supra Alloy \cite{SupraAlloys:2015}, Honeywell 
\cite{HoneywellEM:2015} and PTG \cite{PTG:2015} to provide sample material, taken from various 
stages along the production process, in order to determine where radioactivity, particularly U and 
Th, enters the chain. In our campaign we have received 23 samples including: eight sponges (TIMET), 
one sample of very high purity Ti (Honeywell), one Gr-1 with 10\% scrap (VSMPO), two Gr-2 sheets 
(Supra Alloy and PTG), seven Gr-1 sheets (Supra Alloy, PTG and TIMET) – and also Ti bolts and nuts. 
Table~\ref{CROt:CryostatTitanium} summarizes all the Ti samples that have been radio-assayed in this 
campaign. Upper screening limits are indicated in italics.   

\begin{table}[htbp]
\begin{center}
\tdrtcaption[CryostatTitanium]{Summary of titanium samples.}{Summary of the 22 titanium samples assayed for LZ cryostat, 
including various grades and types from multiple suppliers.}
\setlength{\extrarowheight}{2pt}
\setlength{\tabcolsep}{2pt}
\sffamily
\begin{tabular}{|c|c|c|c|c|c|c|c|c|c|}  
 \hline
 \rowcolor{mrocol}
 \rule{0pt}{3.5ex}\textbf{\#} & \textbf{Supplier} & \textbf{Sample name} & \multicolumn{2}{c|}{\textbf{\IUtTe 
 [mBq/kg]}} & \multicolumn{2}{c|}{\textbf{\IThtTt [mBq/kg]}} & \textbf{\IKfz [mBq/kg]} & 
 \textbf{Titanium} \\
 \hline
 \rowcolor{lrocol}
 & & & early & late & early & late & & grade/type\\
 \hline
1 & Supra Alloy & Carlson 8J10 & 31.00 & 4.10 & 0 & 2.80  & \itshape{1.80} & Gr-1 Sheet \\
\hline
2 & TIMET & Osaka  & \itshape{2.50} & 248 & 0 & \itshape{4.10}  & \itshape{12.0} & Sponge \\
  &  & 26-29461 &  &  &  &  &  &  \\
\hline
3 & TIMET & Tanghsan & \itshape{2.50} &6200 &0  &\itshape{2.50}   & \itshape{15.0} & Sponge \\
 &  & TX027594&  & &  &  &  &  \\
\hline
4 & TIMET & Toho      & \itshape{2.50} &\itshape{62} & 0 & \itshape{1.60}  & \itshape{12.10} & Sponge \\
  &       &  C 12009C & & &  &  &  &  \\
\hline
5 & TIMET & Toho  & \itshape{2.50} &124 & 0 & \itshape{1.60}  & \itshape{12.00} & Sponge \\
&   & W 112266W &  & &  &   &  & \\
\hline
6 & TIMET & Zaporozhye & \itshape{2.50} & 744 & 0 & \itshape{1.60}  & \itshape{12.00} & Sponge \\
  &       &  6680-12 &  &  &  &   &  & \\
\hline
7 & TIMET & Zuny & \itshape{25.00} & 2480 & 0 & 4.1 & \itshape{12.00} & Sponge \\
 &  & TX027641 &  &  &  &  &  &  \\
\hline
8 & TIMET & HN0021-B/1 & 11.00  & \itshape{0.60} & 0 & \itshape{0.60}  & \itshape{2.50} & Gr-1 Sheet \\
\hline
9 & TIMET & HN0021-B/2 & 4.90 & 3.33 & 2.85 & 0.80 & \itshape{1.50} & Gr-1 Sheet \\
\hline
10 & PTG & ATI W74M & 46.00 & 2.80 & 0 & 2.80  & \itshape{1.80} & Gr-1 Sheet \\
\hline
11 & Supra Alloy & Timet & 110.00 & \itshape{2.40} & 0 & 170.00  & 2.40 & Gr-2 \\
   &             & BN3672(2) RMI &  & &  &   &            &         \\
   &             & 404666 (9) &  & &  &   &            &         \\  
\hline
12 & PTG & Thyssen Krupp & \itshape{9.60}  & 3.60 & 0 & 2.40  & \itshape{2.10} & Gr-2 \\
   &             & 611292 &  & &  &   &            &         \\
\hline
13 & TIMET & Henderson & \itshape{3.70}  & 2480 & 0 & 12.30  & \itshape{18.00} & Sponge \\
   &             & 22-49312 &  & &  &   &            &         \\
\hline
14 & S6MB annulus & Bolts & 13000.00  & \itshape{6.00} & 0 & 160.00  & \itshape{60.00} & Bolts \\
\hline
15 & EE-33 full & Nuts & 500.00  & \itshape{8.40} & 0 & 80.00  & \itshape{60.00} & Nuts \\
\hline
16 & Honeywell & T149858991 & 3.70  & 4.69 & 0 & 1.63  & \itshape{1.50} & Gr-1 Sheet \\
\hline
17 & VSMPO & 528 g & \itshape{61.70}  & \itshape{6.20} & 0 & \itshape{4.10} & \itshape{31.00} & Gr-1 Metal \\
 &  &  &   &  &  &  &  & 10\% scrap \\
\hline
18 & VSMPO & 996 g & 17.28  & 12.35 & 0 & \itshape{4.10} & \itshape{6.20} & Gr-1 Sponge \\
\hline
19 & TIMET & HN2470 & 8.51 & 0.37 & 0 & 0.61  & 0.52  & Gr-1 Sheet \\
\hline
20 & TIMET & Master ID \#46& 8.00 & \itshape{0.124}  & 0 & \itshape{0.12}  & \itshape{0.62}  & Gr-1 Sheet \\
\hline
21 & TIMET & HN3469-T & \itshape{1.6} & \itshape{0.1}  & 0.31 & 0.30  & \itshape{0.62}  & Gr-1 Slab \\
\hline
22 & TIMET & HN3469-M & 2.90 & \itshape{0.10} & \itshape{0.2} & 0.25  & \itshape{0.68}  & Gr-1 Slab \\
\hline
\end{tabular} 
\end{center}
\end{table}

\tdrsubsec[Stainless Steel]{Stainless Steel}
Low background experiments searching for dark matter or neutrino-less double beta decay, such as 
XENON 1T \cite{Aprile:2012zx} and PANDA-X \cite{Cao:2014jsa}, or GERDA \cite{Ackermann:2012xja} and 
NEXT \cite{Gomez-Cadenas:2014dxa},  respectively, use stainless steel for their cryostats, with the majority 
of materials coming from German stockholder NIRONIT \cite{NIRONIT:2015}. 13 samples were procured 
from NIRONIT for radio-assay. Independent assays of samples received directly from the GERDA and 
NEXT experiments were conducted to cross-check published results \cite{Maneschg:2008zz,Alvarez:2013rp} and to measure radioisotopes 
not reported, as well as early U and Th activity. The \num{13} samples (a total of \SI{152}{\kg}) originated from 
different heats and were made by different mills: \num{7} samples at Thyssen Krupp Nirosta (Germany) and \num{6} 
samples at Aperam (Belgium). All samples were electro-polished at LBNL and pre-screened with a 
surface HPGe counter (MERLIN), particularly for excessive \ICosz. Samples with \SI{<20}{\mBqkg} of \ICosz 
were forwarded for more sensitive tests underground at SURF and at the University of Alabama. 
Stainless steel radio-assays are summarized in Table~\ref{CROt:Cryostatstainless}. 

\begin{table}[h]
\begin{center}
\tdrtcaption[Cryostatstainless]{Summary of stainless steel samples.}{Summary of the \num{13} stainless steel samples radio-assayed for LZ. Due to a high \ICosz content detected in samples \numrange{10}{13} during pre-screening, these were not assayed further and as such early-chain contents were not measured}
\sffamily
 \setlength{\extrarowheight}{3pt}
\begin{tabular}{|c|c|c|c|c|c|c|c|c|}  
 \hline
 \rowcolor{mrocol}
 \rule{0pt}{3.5ex}\textbf{\#} & \textbf{Sample name} & \multicolumn{2}{c|}{\textbf{\IUtTe[mBq/kg]}} & \multicolumn{2}{c|}{\textbf{\IThtTt[mBq/kg]}} & \textbf{\ICosz} & \textbf{\IKfz} \\
 \hline
 & & early & late & early & late & & \\
 \hline
1 & NIRONIT 311113 & 7.3 & 0.35 & 1.1 & 4  & 14.5 & \itshape{0.53} \\
\hline
2 & NIRONIT 511803 & 1.2 & 0.27 & 0.33 & 0.49  & 1.6 & \itshape{0.4} \\
\hline
3 & NIRONIT 512006 & \itshape{1} & 0.54 & 0.49 & 1.1  & 1.7 & \itshape{0.59} \\
\hline
4 & NIRONIT 512844 & 1.4 & 0.5 & 0.5 & 0.32 & 2.6 & \itshape{0.5} \\
\hline
5 & NIRONIT 521663 & 1.9 & 0.38 & 0.81 & 0.73  & 5.6 & \itshape{0.46} \\
\hline
6 & NIRONIT 521994 & 0.5 & 1.9 & 1.7 & 1.5  & 4.5 & \itshape{0.5} \\
\hline
7 & NIRONIT 124113 & 0  & 1.1 & 0 & 4.1  & 8.2 & \itshape{3.0} \\
\hline
 & NIRONIT (Alab) 124113 & 0$\pm$22 & 4.89 & 0  & 5.37  & 14.6 & \ 1.7 \\
\hline
8 & NIRONIT 211093 & 0 & \itshape{0.6} & 0 & \itshape{0.8}  & 7.4 & \itshape{3} \\
\hline
 & NIRONIT (Alab) 211093 & 0$\pm$11 & 2.46 & 0.0 & 0.37  & 14.0 & 0 \\
\hline
9 & NIRONIT 528292 & 0 & \itshape{0.6} & 0.0 & \itshape{0.9}  & 6.5 & \itshape{3} \\
\hline
 & NIRONIT (Alab) 528292 & 0$\pm$22 & 2.22 & 0 & 0.67  & 9.69 & 0 \\
\hline
10 & NIRONIT 832090 & 0 & 4 & 0 & 2.2 & 26 & \itshape{4} \\
\hline
11 & NIRONIT 407156 & 0 & \itshape{0.6} & 0 & 4.8 & 32 & \itshape{2} \\
\hline
12 & NIRONIT 528194 & 0 & \itshape{0.8} & 0 & 2.1  & 32 & \ 5 \\
\hline
13 & NIRONIT 828660 & 0 & \itshape{1.4} & 0 & \itshape{1.5} & 335 & \itshape{4} \\
\hline
\end{tabular} 
\end{center}
\end{table}

After screening \num{23} Ti and \num{13} Stainless Steel samples in the material search campaign for the LZ 
cryostat, we have concluded that the highest and reproducible radiopurity is achieved with  
Commercially Pure Grade-1, as per ASTM B256, Ti without added scrap material and using Cold Hearth Electron Beam (CHEB) 
refining technology. CHEB used for production of commercially pure Ti provides an important 
purification mechanism in which high density contaminants are removed by gravity separation. By 
contrast Vacuum Remelting Technology (VAR) does not have such material refining 
capabilities. All samples of the finished material supplied by TIMET were produced with such recipe. 
Their radioactivity levels were consistent and always below the LZ acceptance criteria. Samples 
HN3469-T and –M were from a \SI{15000}{\kg} Ti jumbo slab that TIMET produced at its mill in 
Morgantown (Pennsylvania) and made available for the LZ cryostat. We note that the screening results of those 
samples represent the lowest radio-impurity contamination ever reported for a titanium 
sample.   

\tdrsec[Background]{Monte Carlo simulations of background from the cryostat}
Radioactive background due to the cryostat vessels has been evaluated by means of the detailed Monte Carlo 
simulations as described in~\ref{SRDS:Sims}, which is an evolution of the simulation developed for the LUX ~\cite{Akerib:2011ec}, based on the GEANT4 simulation 
toolkit~\cite{Agostinelli:2002hh}. Whilst sharing the same physics models and software structure, the package developed for LZ has been implemented with the detailed geometry of the LZ detector 
including shielding, cryostat (inner and outer vessels, flanges and bolts), inner detector, photomultiplier tubes (PMTs), PTFE reflectors, the 
cathode and anode grids, field shaping rings, and outer detectors comprising the LXe skin, and 
the Gd-doped liquid scintillator. The outer detectors are implemented appropriately as veto 
systems in these simulations. 

For neutrons, (\Pga,\Pn) reactions and spontaneous fission neutron energy spectra were generated 
using the SOURCE software~\cite{Wilson:2009}. These spectra were then embedded into the simulation framework, which propagates neutrons isotropically emitted from the cryostat. Only neutrons from (\Pga,\Pn) reactions contribute to the nuclear recoil (NR) events background, thanks to the the very high efficiency of the LZ detector in vetoing neutrons from the spontaneous fission process, as described in Section~\ref{SRDSss:SNeutron}. We evaluate separately the early and late part of the \IUtTe enabling us to properly consider samples in which the secular equilibrium is broken. \\
For electron recoil (ER) events from \IUtTe and \IThtTt decay chains, and \IKfz and \ICosz, we used a new particle generator, described in Section~\ref{SRDSss:SGammas}, based on the standard \geantFour process. 

To achieve the required sensitivity of \SI{3e-48}{\cm\squared}, 
we set as a goal the total ER event rate from detector materials to be below 10\% of the rate from pp solar 
neutrinos in the WIMP search region, \SIrange{1.5}{6.5}{\keVee}, and the NR event counts to be below \num{0.2} 
events after S2/S1 discrimination in the full run time of \num{1000} days, within \SIrange{6}{30}{\keVnr}. 

The impact of the cryostat on the background counts has been estimated by the full detector and physics simulation, and is presented in Figure~\ref{MASf:TitaniumAssays}. The plot shows the results for Ti (TIMET) and stainless steel (NIRONIT) samples for LZ in the \SI{1000}{\day} exposure with a \SI{5.6}{\tonnel} LXe fiducial volume and after all the veto systems are applied, for ER events within \SIrange{1.5}{6.5}{\keVee}, with \SI{99.5}{\percent} rejection, and within \SIrange[range-phrase = --,range-units=brackets]{6}{30}{\keVnr}, and \SI{50}{\percent} acceptance, for NR events. The Ti identified by the assay program, indicated by the star corresponding to \num{0.53} counts for ER and \num{0.01} counts for NR, is well below the requirements for LZ, for a specific detector component.

\tdrsec[Cryostat Specification]{Cryostat technical specification}
The intended purpose of the cryostat is to hold \SI{10000}{\kg} of liquid xenon at \SI{-100}{\degree C} 
with an immersed Time Projection Chamber in it. The cryostat consists of three main 
sub-assemblies: an Inner Cryostat Vessel (ICV), an Outer Cryostat Vessel (OCV) 
separated by a vacuum space and a Cryostat Support (CS). The cryostat is submerged in a water 
tank and surrounded by the outer detector acrylic vessels filled with a liquid 
scintillator. A general view of the cryostat with the acrylic vessels below 
(supported by the CS shelves) and above (supported by the OCV head and flange) is shown in Figure~\ref{CROf:cryostatoverview}.

\begin{figure}[hbtp]
\begin{center}
\includegraphics[width=0.5\textwidth]{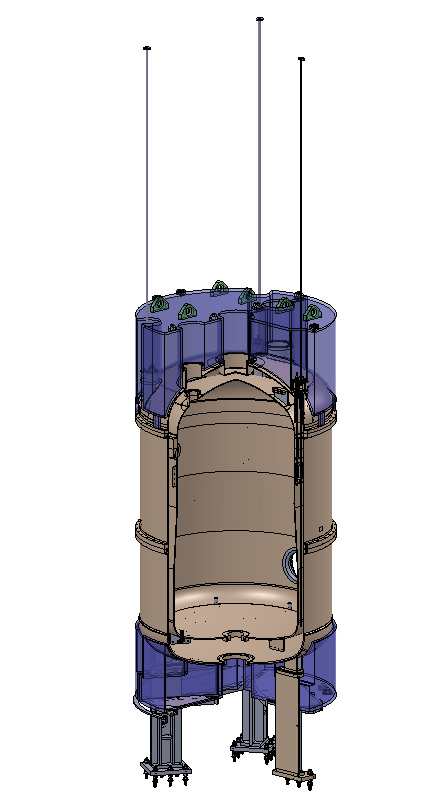}
\end{center}
\tdrfcaption[cryostatoverview]{LZ cryostat overview}
{LZ cryostat assembly with the ICV nested in the OCV supported by the three legs of 
the CS. Acrylic vessels of the Outer Detector are shown at the top and bottom of 
the OCV. Three equally spaced vertical tubes will be used to deliver calibration 
sources into the vacuum space between the vessels.}
\end{figure}

The material for the LZ cryostat is commercially pure Ti, Grade 1 per ASME SB-265, with additional 
low-radioactivity background requirements as presented in~\ref{CROS:MaterialSearch}. The design of 
the vessels complies with the following codes: ASME BPVC \cite{ASME:2015}, 2012 Int. Building Code,
and ASCE 7 \cite{USGSSeismic:2015}, with site soil classification Class B (Rock) for seismic 
conditions. The 2008 U.S. Geological Survey hazard data for this location are: S$_\textrm{S}$ = 
0.121 g, S$_\textrm{MS}$ = 0.121 g, and S$_\textrm{DS}$ = 0.081 g. The Seismic Design Force for 
LZ is 0.054 g, which imposes a force of 6350 N at the center of mass of the cryostat during a 
seismic event. The LZ cryostat operational temperatures (\si{\degree C}) and pressures (bar 
absolute) are summarized in Table~\ref{CROt:Cryostatpresstemp}

\begin{table}[h]
\begin{center}
\tdrtcaption[Cryostatpresstemp]{Pressures and temperatures.}{Pressures and temperatures for cryostat operational conditions: normal, bake out and most severe failures.}
\sffamily
\setlength{\extrarowheight}{3pt}
\begin{tabular}{|c|c|c|c|c|}  
 \hline
 \rowcolor{mrocol}
\rule{0pt}{3ex}\textbf{Vessel} & \multicolumn{2}{c|}{\textbf{Pressure}} & \textbf{Temperature} & \textbf{Condition}\\
 \rowcolor{mrocol}
& \multicolumn{2}{c|}{\textbf{[bar absolute]}}& \textbf{[$^{\circ}$C]} &\\
\cline{2-3}
 \rowcolor{mrocol}
              & \textbf{Internal} & \textbf{External} & & \\
\hline
  & \num{\leq 4.0} & Vacuum & \numrange{-112}{37} & Normal \\
\cline{2-5}  
 & Vacuum & \num{1.01} & \num{\leq 100} & Bake out \\
\textbf{Inner} &  &  &  & dry, no water in water tank \\
\cline{2-5}
 & Vacuum & \num{1.48} & \numrange{-112}{37} & Failure mode  \\ 
 &  &  &  & water flooded between inner and outer vessels \\ 
\hline
 & Vacuum & \num{1.48} & \numrange{0}{37} & Normal \\
\cline{2-5}  
 & Vacuum & \num{1.01} & \num{\leq 100} & Bake out \\
\textbf{Outer} &  &  &  & dry, no water in water tank \\
\cline{2-5}
 & \num{1.48} & \num{1.01} & \numrange{0}{37} & Failure mode  \\ 
 &  &  &  & Xe gas leak between inner and outer vessels\\ 
 &  &  &  & and no water in water tank\\ 
\hline
\end{tabular} 
\end{center}
\end{table}

\tdrsec[Inner Cryostat Vessel]{Inner Cryostat Vessel}

The ICV has conventional cylindrical geometry with ellipsoidal heads, as shown in 
Figure~\ref{CROf:cryostattech}. The inner vessel is split once near the top head with a flange pair. To minimize the passive volume filled with LXe, the diameter of the inner vessel is tapered near its half-height and the bottom head has an ellipsoidal shape with a 3:1 aspect ratio (the top head has the 
more common 2:1 aspect ratio).  The 3:1 aspect ratio head requires greater material thickness 
compared to the 2:1 head, but this trade-off is well worth the savings in LXe.

\begin{figure}[hbtp]
\begin{center}
\includegraphics[width=0.8\textwidth]{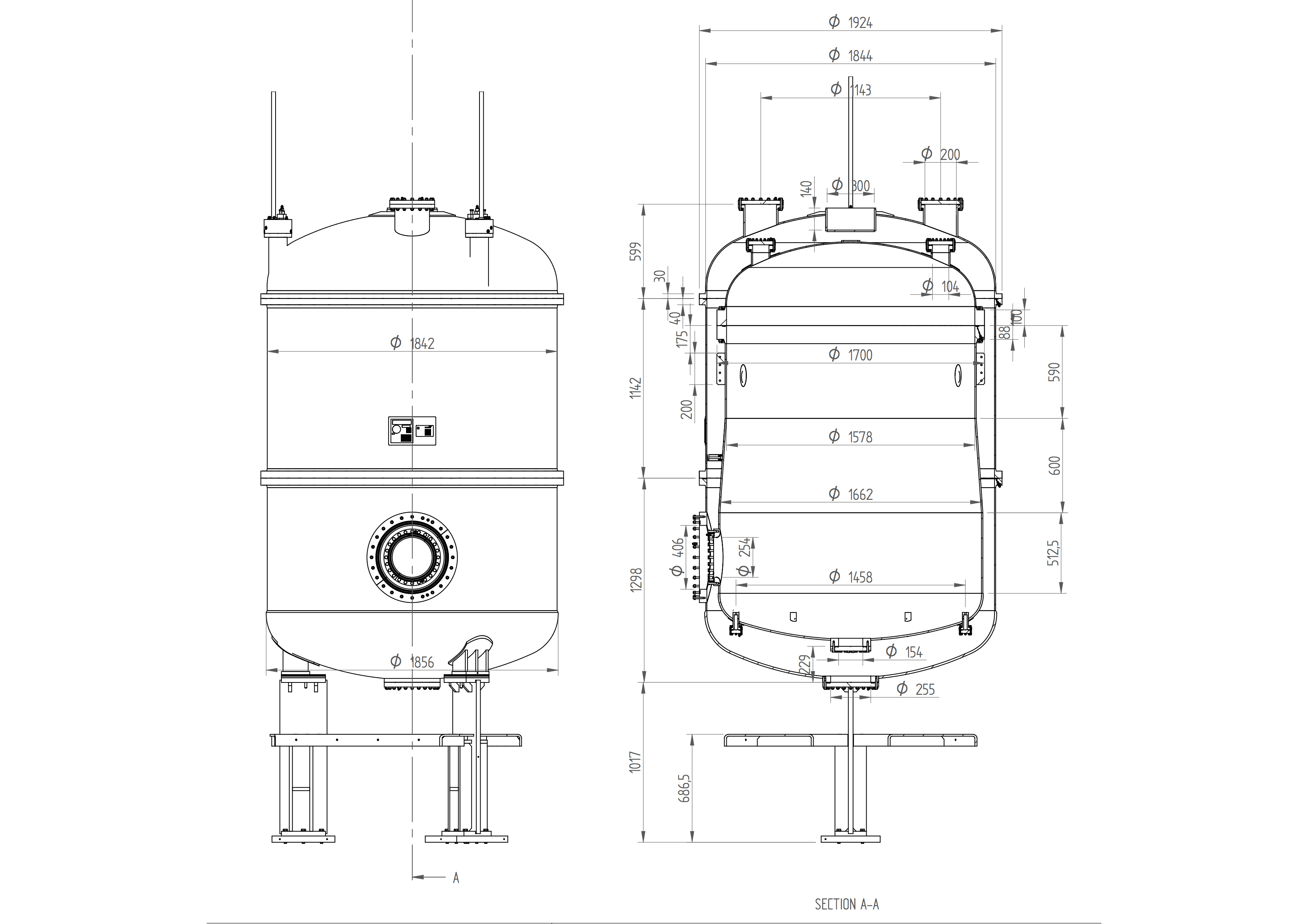}
\end{center}
\tdrfcaption[cryostattech]{LZ cryostat assembly}
{LZ cryostat assembly. Dimensions of major components are shown.}
\end{figure}

For internal pressure, the thickness of the vessel walls in the cylindrical section is governed by a straightforward formula in the pressure vessel code. The minimum wall thickness is a function of the material, vessel diameter, pressure, and quality control measures. At the required internal pressure, the minimum wall thicknesses in the cylindrical section is \SI{5.5}{\mm}. A number of ports are necessary to carry fluids and electrical signals to and from the inner detector. These are added to the top and bottom heads, as well as a side penetration for cathode high voltage. The latter is discussed further in Sections~\ref{XDSS:Cathode}.  Buckling is an important failure mode to consider for vessels that see external pressure (vacuum in this case). The ASME BPVC specifies safe external working pressures based on material, temperature, wall thickness, diameter, and length. If the allowable external pressure is insufficient, a vessel designer has a couple of options. The first is to increase the wall thickness. In the case of LZ, this is undesirable for a number of reasons: Most notably, it creates more background radiation and reduces veto efficiency. The other option is to add reinforcing rings. Reinforcing rings essentially shorten the length of the vessel from a buckling perspective.  To comply with the ASME code, a stiffening ring is located at the top of the tapered (conical) section of the inner vessel. This ring acts as a line of support to increase buckling resistance, and therefore allows a thinner wall to be used, e.g. \SI{6}{\mm} instead of \SI{8}{\mm}.   Comparing the values for internal pressure 4 bar versus external pressure 1.48 bar, it is evident that the vessel design is driven by external pressure. It should also be noted that the minimum wall thickness is the minimum as-built, not the nominal. During the head-forming process, for instance, flat material is drawn or spun into shape, and in that process thinned from its original thickness. It should also be remembered that material is commercially available in discrete increments as opposed to infinitely variable thickness. Total mass and minimum thicknesses required by the ASME code for each segment of the inner and outer vessel made of Ti are summarized in Table~\ref{CROt:Cryostatthick}.

\begin{table}[h]
\begin{center}
\tdrtcaption[Cryostatthick]{Vessel-wall thicknesses.}{Vessel-wall thicknesses [mm] and total mass [kg] imposed by the external pressure at the normal, bakeout and failure conditions.}
\sffamily
\setlength{\extrarowheight}{3pt}
\begin{tabular}{|c|c|c|c|c|c|c|c|c|c|}  
\hline
\rowcolor{mrocol}
\multicolumn{6}{|c|}{\textbf{Inner Vessel}} & \multicolumn{4}{c|}{\textbf{Outer Vessel}} \\
\hline
Top & Upper & Conical & Lower & Dished & Total & Top & Side & Dished & Total\\
head & wall & section & wall & end & mass & head & wall  & end & mass \\ 
\hline
7 & 9 & 9 & 9 & 11 & 950 & 8 & 7 & 14 & 1115 \\
\hline
\end{tabular} 
\end{center}
\end{table}

To minimize the amount of LXe between the TPC and the inner cryostat, the shape of the inner vessel is tapered at its half-height. Studies of the electric-field 
distribution show that the electric field is below the maximum allowed value of \SI{50}{\kV\per\cm}. The inner vessel will be sealed with a sprung metal C-seal because at the 
experimental temperature, O-ring seals with typical elastomeric materials are not 
suitable. Additionally, the large diameter prohibits the use of a knife-edge flange. 
Smaller ports on the vessels will be sealed with sprung metal C-seals as well.  
Inner-vessel flanges with C-seal gaskets for cryogenic service will also feature a 
secondary O-ring seal to facilitate room-temperature leak detection.

\tdrsec[Outer Cryostat Vessel]{Outer Cryostat Vessel}
The OCV supports the ICV through three tie bar assemblies situated in the top head. The ICV contains the TPC and LXe. The ICV can be leveled by adjustment of the tie bars externally from above the water tank.
            
The OCV also provides support for top and bottom Liquid Scintillator tanks, access to the vacuum space for calibration purposes and ports in the top head for cryogenic and cable services. At the bottom a LXe recirculating system of pipes connects to the lower port. The OCV base interfaces with the CS and its position in the water tank determines the orientation of the HV port in the base of the vessel. The HV ports of both vessels are aligned for the umbilical connection.

The OCV, as shown in Figure~\ref{CROf:cryostattech}, has been designed in three pieces for ease of transport to the Davis cavern and assembly in the water tank it is joined together around the circumference by two flange pairs. The top head comprises a flange, 
2:1 ellipsoidal head, three ports for the ICV suspension, two conduit 
ports for cryogenic and cabling services and a large central recess for 
a calibration source called ``YBe''. 
The middle section of the OCV is a straight cylinder with two flanges 
that match the head and base flanges. The flanges are double-sealed with 
two elastomer O-rings and a pump path between the two to facilitate leak 
detection. Both O-rings are expected to seal at experimental 
temperatures since the OCV operates close to room temperature. The 
bottom head comprises a flange that matches the middle section, a HV 
port that enables the umbilical to be connected to the vessel, three 
support feet welded to the 2:1 ellipsoidal head and a heat exchange port 
flange centrally situated in the head. Each foot has a \SI{6.3}{\mm} diameter 
survey hole in the \SI{20}{\mm}-thick plate at the outer surface. The flanges 
are joined with stainless steel bolts and nuts (M12, 48) (same grade as 
for the ICV and supplied by the LZ project). All flanges must be 
integral design. To facilitate assembly, tapered pins have been 
incorporated into the design. During assembly and leak testing, the 
conduit ports, tie bar ports, HV and HX ports will be sealed with 
blanking plates. The sealing plate for the tie bar ports on the OCV head 
has the feature of a hoist ring to lift the head or vessel 
assembly. The tie bar assemblies themselves are used for lowering the 
ICV into the OCV during final assembly.

The ICV is suspended from three tie bar assemblies situated in the OCV 
head. Cross section of the tie bar is shown in Figure~\ref{CROf:cryostattiebar}. 
They incorporate an adjustment feature for leveling the ICV and 
the tie bar is sealed into the assembly with a metal bellows that 
facilitates vertical movement with a small angular offset. 

\begin{figure}[hbtp]
\begin{center}
\includegraphics[width=0.8\textwidth]{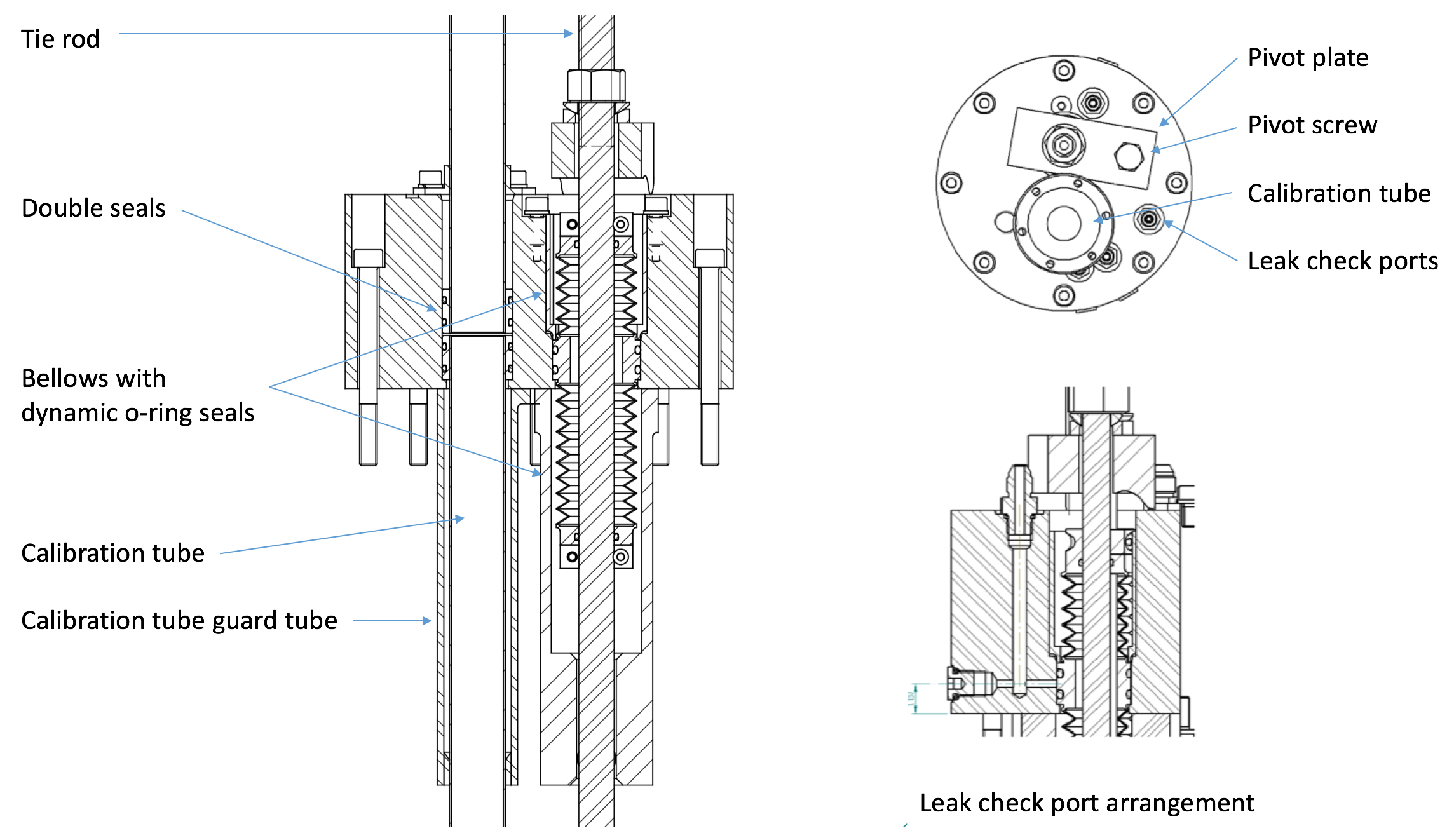}
\end{center}
\tdrfcaption[cryostattiebar]{OCV tie rod port assembly}
{OCV tie rod port assembly}
\end{figure}

There is a coarse and fine adjustment the fine adjustment uses a pivot plate giving 
a 3:1 advantage. Also housed within the assembly is a calibration tube 
that facilitates access to the vacuum space between the vessels and 
enables a calibration source to be lowered down to within a few \si{\mm} of 
the ICV lower head. The inner diameter and the wall thickness of the tubes 
are \num{25.4} and \SI{0.94}{\mm}, respectively. The length of the tube spans 
from the deck above the cryostat to the tangent line of the OCV dished end.  

Double seals are used throughout the LZ cryostat and 
the volume between the grooves is accessed with a sealed nipple that is 
used to detect leaks between seals. The tie bar assemblies have four of 
these items to access the seals. The tie bars are manufactured from high 
strength stainless steel giving a factor of safety of 3:1.

\tdrsec[Cryostat Support]{Cryostat Support}
The CS structure has been designed to support the operating load and 
resist a horizontal force due to a seismic event. The CS interfaces 
with three base plates mounted on studs protruding from the water 
tank base these plates will be surveyed and adjusted flat and level at 
the correct height for the umbilical. Holes for survey targets have 
been included in the relevant faces of the legs for this purpose. The 
support has been designed with stability, strength and installation in 
mind. The cryostat interfaces with flange plates welded to the top of 
the legs and is shimmed to the correct height. The support has also 
been designed with the requirements of the veto counter in mind and 
maximum coverage by the LS tanks has been achieved with a three leg 
flat plate design. The three segmented tanks fit between the legs and 
are supported on shelves between the legs. 

\tdrsec[Interfaces]{Cryostat Interfaces}
The cryostat interfaces with many other subsystems of the experiment. Some of them 
are physically connected to its ports creating additional loads which must be considered in the cryostat design. External loads imposed by the other subsystems on the cryostat ports are listed in Table~\ref{CROt:Cryostatloads}. In addition, the feet of the ICV and OCV shall withstand the load of the vessels when filled with water for the hydrostatic tests.      

\begin{table}[htbp]
\begin{center}
\tdrtcaption[Cryostatloads]{External loads.}{External loads to the cryostat in addition to any loads exerted by vessel internal and external pressures.$^*$Combined load from cryostat, Outer Detector tanks and seismic load.$^{**}$The moment due to the seismic load at 2/3 height.}
\sffamily
\setlength{\extrarowheight}{3pt}
\begin{tabular}{|c|c|c|c|}  
\hline
\rowcolor{mrocol}
\textbf{Cryostat Item} & \textbf{Number} & \textbf {Maximum allowable } & \textbf{Maximum allowable moment}\\
\rowcolor{mrocol}
 & \textbf{of items} & \textbf {load on a single item [N]} & \textbf{in any direction to the flange face [Nm]} \\
\hline
\rowcolor{lrocol}
\multicolumn{4}{|c|}{\textbf{Inner Cryostat Vessel}} \\
\hline
Top head port & 2 & 200 & 200 \\
Weir port & 3 & 100 & 100\\
HV port   & 1 & 2000 & 2000 \\
TPC port  & 6 & 3000 & 200  \\
Dished end port & 1 & 1000 &  1000 \\
\hline
\rowcolor{lrocol}
\multicolumn{4}{|c|}{\textbf{Outer Cryostat Vessel}} \\
\hline
Top head port & 2 & 1000 & 1000 \\
YBe source port & 1 & 3200 & N/A \\
Top flange & 2 & 10000 & N/A \\
Tie bar port  & 3 & 33000 & N/A \\
HV port & 1 & 2000 & 1000 \\
Bottom port & 1 & 1000 & 1000\\
\hline
\rowcolor{lrocol}
\multicolumn{4}{|c|}{\textbf{Cryostat Support}} \\
\hline
Leg & 3 & 50507$^*$ & 15265$^{**}$ \\
\hline
Shelf & 3 5000 & N/A & \\
\hline
\end{tabular} 
\end{center}
\end{table}

\tdrsec[Thermal insulation]{Thermal Insulation}
It is important to limit the heat transfer between the inner vessel and the environment to minimize the amount of refrigeration needed underground. In addition, reducing heat transfer helps to prevent unwanted convection currents in the LXe fluid. The vacuum between nested inner and outer vessels essentially eliminates thermal conduction and convection, in typical Dewar fashion. The dominant mode of heat transfer is therefore radiation, and with a large surface area (\SI{\sim16}{\m\squared}) and a large temperature difference, it is potentially substantial. Multilayer insulation (MLI or superinsulation) is proposed as the baseline solution to reduce this thermal load during normal operation. MLI is a well-known 
insulating material for in-vacuum cryogenic service. The thermal load with and without this material varies by approximately an order of magnitude. In this case, the expected heat load with bare vessels would be several hundred watts, and with MLI several tens of watts. In LZ, the proposed amount of total refrigeration is about a kilowatt, so MLI is the clear choice. A 3-parts MLI blanket design for the ICV is shown in Figure~\ref{CROf:cryostmli}.  

MLI works well during normal operation, but is ineffective in the event of a failure 
in which a gross amount of liquid water enters the volume normally occupied by 
vacuum between the vessels. To mitigate this failure mode, a closed-cell 
polyurethane foam will be applied to the outer surface of the inner vessel anywhere 
it is in contact with LXe (basically everywhere below the main seal flange). The 
proposed foam thickness is \SI{2}{\cm} over the bottom head, and \SI{1}{\cm} over the remainder. 
MLI will be wrapped over the foam, and covers the entire inner vessel and its cold 
appendages. With the foam in place, the maximum heat transfer in this failure mode is expected to be \SI{3600}{\W}, which corresponds to a Xe boil-off of 450 standard liters per minute (slpm). This rate is within the Xe-recovery capacity of the system.

A dedicated thermal analysis has been carried out in order to investigate possible distortions in the calibration tubes from asymmetric thermal gradients caused by thermal radiation from the inner face of the tube onto the cooler cryostat. The steady state thermal simulation was carried out in Ansys CFX~\cite{ANSYS:2015}, and consisted of basic tubular geometries representing the outer and the inner vessels, with a nitrogen filled calibration tube geometry located between the two vessels in a vacuum interspace. Thermal radiation was modeled between the warmer OCV (\SI{20}{\degree C}) and the cooler ICV (\SI{-100}{\degree C}).
For our preferred MLI product, (Coolcat 2 NW) the supplier - Ruag~\cite{RUAG:2015} specify that 10 layers at \SIrange{300}{77}{\K} give heat loss of \SI{<1}{\W\per\m\squared}. Using the equation E$_{eff}$ = P/S x (T$_h^4$ – T$_c^4$) the average effective emissivity over this range is \SI{0.0022}{\W\per\milli\kelvin}.  A more conservative emissivity value of \SI{0.025}{\W\per\milli\kelvin} has been used for the simulation and this value was backed up by wider research of similar MLI products, and also by data used for the cryostat vessel heat transfer calculations. Based on these boundary conditions, the maximum thermal gradient over the length of the calibration tubes was found to be negligible, resulting in a maximum deflection at the end of the tube of less than \SI{0.1}{\mm}.  

\begin{figure}[hbtp]
\begin{center}
\includegraphics[width=0.8\textwidth]{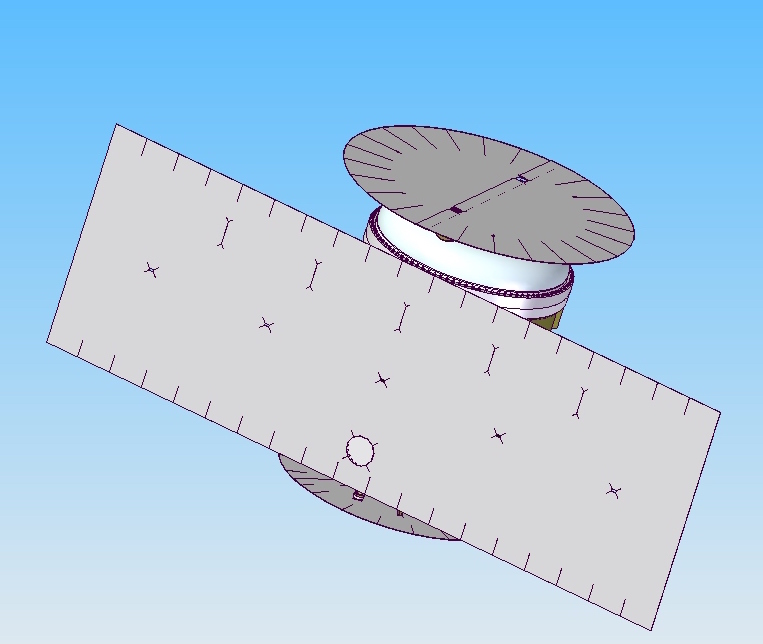}
\end{center}
\tdrfcaption[cryostmli]{Inner cryostat vessel thermal insulation}
{Three parts of the MLI blanket for the ICV thermal insulation.The design with only three parts has been chosen to effectively cover the whole ICV surface with a minimum work required for its assembly. Only final sewing will be needed. Several holes have been provisioned for ICV features such as: ports, fins, tie rods and seismic limiters. Cuts along the blanket's edges are to facilitate wrapping on curved surfaces of the vessel heads.}
\end{figure}
\tdrsec[Fabrication]{Fabrication, Cleanliness, Tests and Certifications}
The titanium for the cryostat is extremely rare and difficult to obtain and as such contingency provision shall be made for spare material.  Potential fabricators are required to demonstrate that they have built in an acceptable level of contingency and that they will take the necessary manufacturing precautions in order to minimize waste and ensure that they do no not exceed their allocation.

Together with the cryostat design, a comprehensive set of drawings has been made which contain toleranced geometric specifications. The general tolerance on machined parts of the cryostat such as flanges and mating surfaces is \SI{1}{\mm}. This tolerance has been relaxed on the form of the barrels in most cases, and wherever possible in order to aid the ease of manufacture.  Where squareness, flatness, position of holes and other features are important for full assembly these are generally tightened to \SI{0.5}{\mm}.  Several key features of the cryostat assembly are highlighted below: 
\begin{itemize}
\item The ICV top head conduit ports must align with those in the OCV with a tolerance of 
\SI{2}{\mm}. 
\item The HV port in the ICV has a squareness tolerance of 1mm relative to top flange and can be positioned \SI{\pm 1}{\mm} from that flange. There is also a \SI{\pm 1}{\mm} tolerance applied to the face from the vessel wall. The face has a positional tolerance of \SI{0.8}{\mm} from the tie rod holes. The required flatness is \SI{0.5}{\mm} and the positional tolerance of the tie rod holes is \SI{\pm 0.8}{\mm}. 
\item The HV port in the OCV has a squareness tolerance of \SI{1}{\mm} relative to the flange in the base and a linear tolerance of \SI{\pm 1}{\mm} to the flange tangent line. The required flatness is \SI{0.5}{\mm}. The tie rod ports in the OCV head have a positional tolerance of \SI{\pm 1}{\mm}. 
\item For helicoflex seals, the manufacturer’s \cite{Technetics:2015} specifications have been used, and for the O-ring grooves, BS1806 standard tolerances have been used.
\end{itemize}
Meeting the geometric requirements of form and alignment of key features can be particularly challenging when producing titanium weldments, where components are prone to distortion during machining and welding procedures. The cryostat shall be welded according to ASME IX BPVC \cite{ASMEweld:2015}. GMAW or MIG welding should be used and the consumable electrode/welding rod shall be the same titanium material as the vessel raw material.  Welding of titanium is a specialist process and the manufacturing contractor is required to demonstrate they are titanium welding specialists with particular emphasis on pre-weld preparation/cleaning, and on protection of the heat affected zone from oxidation during welding. They are required to outline their weld acceptance criteria such that it may be assessed by LZ.

A very high level of cleanliness is required for any component of the cryostat vessels throughout production. Every reasonable effort shall be made to minimize cross-contamination with swarf and loose dust of foreign material as well as environmental radon plate-out on the cryostat surfaces.  The project goal is to achieve a dust level for the internal surface of the cleaned and packaged ICV of \SI{10}{\nano\gram\per\cm\squared}. For all other surfaces the dust level shall not exceed \SI{1}{\micro\gram\per\cm\squared} at delivery. The following procedures shall be applied:
the cryostat raw material shall be stored separately from other material in the workshop; parts shall be individually bagged in the bags provided by LZ to avoid any dust accumulation on the surfaces. Bagged components shall be placed on wooden pallets inside the pallet footprint, safe from the risk of impact or damage to the bag.
The cleaning process for stock material at the mill shall follow the ASTM B 600 standard.  Grease, oil and lubricants are to be removed with alkaline or emulsion type cleaners only.  No mechanical abrasion type cleaning is allowed.  A surface layer of at least 5 microns shall be removed from all stock material by pickling/etching with virgin chemicals.  Chemicals shall be rinsed off after etching with deionized water. 
Prior to machining, rolling and spinning all surfaces in direct contact with cryostat components shall be cleared of dust and swarf with a stiff non-metallic brush, and cleaned with solvent using a lint-free cloth.  All machines shall be flushed of coolant, and coolant replaced with fresh, water soluble coolant, approved by LZ.  During machining, cryostat components shall be separated from other materials in the workshop, and the work area kept clean and tidy.  Post manufacture, all components awaiting further processing shall be hand de-greased with solvent and a lint-free cloth before being bagged using bags provided by LZ, and stored in a clean place.

Prior to welding, components shall be comprehensively cleaned.  Dirt, oil and dust shall be removed from the weld area, and oxide/scale removed by etching.  All surface tables and tools in contact with any part of the cryostat shall be cleaned by hand with fresh solvents.  Upon completion of welding, components shall be bagged using LZ provided bags and stored in a clean place.  Heavy abrading, grinding or wire brushing is prohibited both before and after welding.
For final cleaning, all faces of the cryostat vessels shall be etched using LZ approved virgin chemical solution, and in accordance with ASTM B 600 standard, removing at least 5 microns from all surfaces.  Parts shall then be rinsed with deionized water and thoroughly dried.  All parts shall then be bagged, purged with nitrogen and sealed.

With cooperation from the manufacturing contractor, LZ will carry out the following cleanliness and radioactivity testing: after etching of raw material, a representative 10 kg coupon shall be cut from the 8mm stock plate along with all off-cuts from vessel ports and reinforcement pads, and swarf from bolt holes in large flanges shall be retained and radio-assayed at SURF.  In addition, \SI{10}{\kg} of Ti welding rods and \SI{10}{\kg} of sample welds shall be sent to SURF for radio-assay. The impact of etching on the dimensions of critical features such as grooves shall be assessed prior to machining the real grooves.  LZ shall measure the levels of radon present at various fabrication and cleaning locations.

The following tests will be conducted by the manufacturing contractor and witnessed by designated persons from RAL/STFC:
\begin{itemize}
\item A dimensional check of all components of the cryostat as well as its assemblies, including a localization of the survey markers in 3D space relative to the key features. 
\item Provide proof by mechanical assembly and documentation that the sequence of assembly at SURF shown in Figure~\ref{CROf:cryostatassembly} is achievable.
\item Test and record ICV and OCV leak rate – demonstrate that the leak rate does not exceed \SI{1E-6}{\milli\bar\litre\per\second} (helium). 
\item Conduct a pressure test ICV and OCV for the internal pressure case per the ASME BPVC, as required to achieve certification. 
\item Conduct a suspension full load/leak test of ICV with tie bars assembled.  The ICV shall be loaded to 1.5 x 33,000N = 49,500N and the upper part external to the vessel shall be exposed to water at the operating pressure of \SI{1.48}{\bar} absolute.  The lower part of the vessel shall be evacuated and vacuum leak rate measured and certified to be below \SI{1E-7}{\milli\bar\litre\per\second}.  Further details are contained in the cryostat specification document. 
\item The cryostat support shall be full load tested with 12T applied in increments of 4T, with lateral deflection of the legs as well as vertical displacement of the support recorded.  Measured deflection for each leg shall be smaller than \SI{1}{\mm}. 
\item The vessel shall be stamped/marked relating to the testing and approval. This shall not impart dirt into the system or be etched away during cleaning.
\end{itemize}

\tdrsec[Transporation]{Transportation and Installation}
The cryostat inner vessel along with its contents will be moved as an assembly down the Yates shaft at SURF. The width of the shaft is nominally \SI{1.85}{\m}, and the maximum payload width is \SI{1.70}{\m} to clear features in the shaft cross section and provide some margin of safety. The outer vessel in contrast will be moved down the Yates shaft in three pieces, and assembled once in the Davis Cavern.

\begin{figure}[hbtp]
\begin{center}
\includegraphics[width=0.9\textwidth]{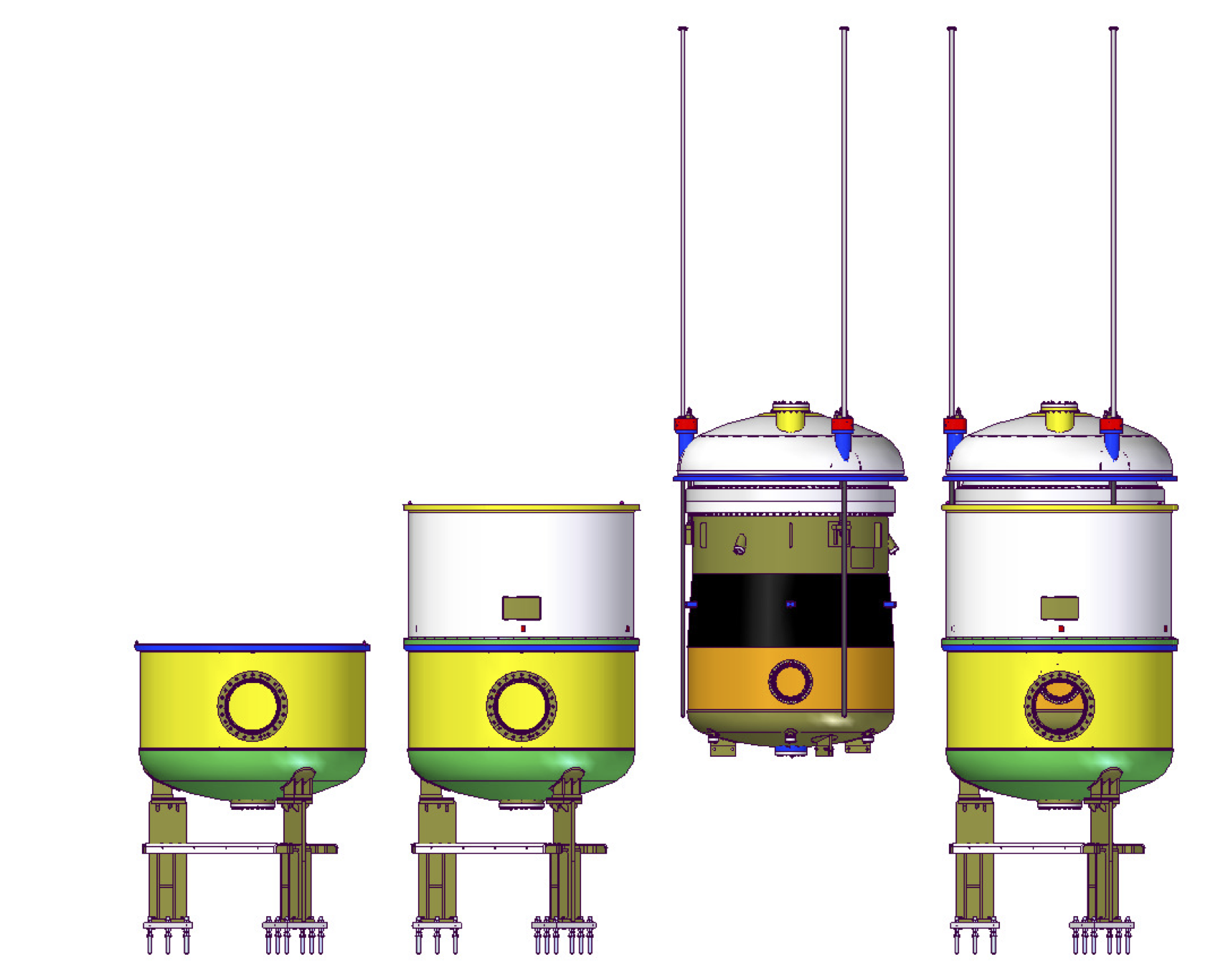}
\end{center}
\tdrfcaption[cryostatassembly]{Cryostat assembly}
{Major steps of the cryostat assembly underground.}
\end{figure}
The CS will be placed onto three base plates that are independently supported inside the water tank. The plates will have been leveled and surveyed prior to the LZ installation. An assembly procedure will be carried out at the manufacturers which will as closely as possible reproduce what will be carried out in the tank at SURF. Having placed the CS onto the plates and loosely assembled the fixing bolts the base of the outer vessel will be assembled onto the top supporting flanges with bolts and shims in place. The shelves between the legs for supporting the LS tanks must be included in this assembly procedure. A preliminary inspection will be carried out to check that there are no gaps between the  mating interfaces greater than 0.5mm. Any gaps greater than \SI{0.5}{\mm} will be shimmed with appropriate thicknesses of shim stock. When this has been completed and a dimensional check has established that the CS support and base are in the correct position the fixing bolts will be tightened sequentially to ensure the geometry is preserved. A trial assembly can now proceed  following the sequence shown in Figure~\ref{CROf:cryostatassembly}. Care must be taken at all times to protect sealing surfaces, the seals will will not be in place when this procedure is carried out at the manufacturer. The middle section of the outer vessel is assembled next onto the base flange using the assembly dowels to locate it in the correct orientation at this stage a minimum of six bolts can be used to hold the assembly together. Next in the sequence is the assembly of the ICV into the OCV using a combined lift with the OCV head, tie bar assemblies and ICV. The OCV head and tie bars will have been assembled prior to the start of this procedure, a convenient supporting frame or gantry will be required to set up this lifting operation. Prior to the tie bars being used they must have been assembled and tested using the test equipment. With the ICV in a position accessible for the crane the OCV head complete with tie bars will be lifted over the ICV and the tie bars aligned with the three access holes in the ICV top head flange. Orientation is important as the HV ports of both vessels have to align with adjustment of the tie bars. There is a supporting CS leg at the 0 degree position and a tie bar assembly aligned with this. The HV ports are at \SI{90}{\degree} to this position. Lower the head and tie bars through the access holes, make sure the top M16 nut and washer is in place before carrying this out. When the tie bar is through its support on the upper part of the ICV assemble the lower M16 nut and washer ensuring that at least twice the diameter is protruding through the nut. Take up the slack with the crane and check dimension of HV port to head flanges and parallelism adjust as necessary. Tighten the M16 nuts and lock tie bars they must not rotate. Lift ICV into OCV with crane, there should be a minimum of three people ensuring that the ICV enters the OCV without hitting the sides, when the head approaches the middle section upper flange place three spacers between the flanges and bring the head into contact with them. Rest the head at this position and check that the HV ports are aligned and that when the spacers are removed the ICV will not hit the bottom of the OCV vessel adjust as necessary. Remove the spacers and lower the OCV head and ICV until the flange engages with the location dowels and mates with the upper flange of the middle section. Secure with 6$\times$~M12 bolts and check alignment of HV ports. At this stage the alignment check can be visual if adjustment is needed then use the tie bars but no more than 10mm of adjustment should be used. This assembly procedure at the manufacturers is an assessment of how easy or difficult it is to assemble the vessels and achieve HV port and axis alignment, the foam insulation, MLI, weir piping and bump stops will not be in place but will be for the assembly at SURF.

\bibliographystyle{apsrev4-2}
\bibliography{LZtdr}

\wlog{In tdrinclude command}
\wlog{\ChapLabel}

\renewcommand{\ChapLabel}{chap:XCS} 
\renewcommand{\BibLabel}{XCSS:bib} 
\renewcommand{\SecPre}{XCS}
\renewcommand{\FloatPre}{XCS}
\tdrchap{Xenon and Cryogenics Systems}

\tdrsec[ov]{Introduction and overview of requirements}

The functions of the xenon handling and cryogenics 
systems are to purify the xenon of the relevant 
contaminants, to condense it and to maintain it the 
liquefied state for the duration of the experiment, 
and to safely recover the xenon to permanent storage 
at the conclusion of the experiment. 

Two classes of impurities drive the purification 
strategy.

\begin{enumerate} 

\item Electronegatives such as oxygen or water 
limit the free electron lifetime and degrade the 
operation of the TPC. We require that the charge attenuation 
length of the LXe be \SI{1.46}{meters} or larger, which is the 
maximum drift length of the detector. This corresponds to a
free electron lifetime of \SI{806}{\mus} at 
the drift field baseline of \SI{310}{\V\per\cm}
(see Table~\ref{XDSt:CathodeHV}), 
or an oxygen-equivalent concentration of \SI{0.37}{\ppb}.
Electronegatives are removed from the xenon by 
continuously circulating the xenon in gas phase through 
a hot zirconium getter during operations.

\item Radioactive noble gases require special measures because 
they are not removed from the getter and because they create ER background
events which can not be mitigated with self-shielding. Our primary 
concerns are \IRnttt and \IKreF, but \IArTn, \IArTS, \IRnttz, 
and \IXeotS also play a role.

\end{enumerate}

As detailed in Sections~\ref{MASSs:Kr} and \ref{MASSs:Ar} 
background considerations imply that the vendor supplied xenon
should be purified of \IKrnat and \IArnat to the levels of \SI{0.015}{\ppt}~(g/g)
and \SI{0.45}{\ppb}~(g/g)\footnote{Unless otherwise specified, throughout 
this chapter we report concentration in units of (g/g). Concentrations of
krypton and argon always refer to \IKrnat and \IArnat.} 
respectively in order to suppress ER events from 
\IKreF and \IArTn. The krypton requirement is particularly 
demanding, and the need to control this species drives much of 
the xenon handling strategy. 

At \SI{0.015}{\ppt}, the amount of krypton
in the ten-tonne xenon stockpile is equivalent to \num{40}~std$\cdot$cc of 
air\footnote{The krypton concentration of air is \num{2.9e-6}~(g/g))}.
To meet this requirement the xenon will be purified of 
both krypton and argon with a custom gas 
chromatography system at SLAC (see Section~\ref{XCSS:KrRem}). 
Because no krypton removal technology will be employed 
during detector operations, it is critically important that 
krypton not be re-introduced once it has been removed. 
Air leaks and detector outgassing are two mechanisms of particular concern.
This places additional burdens on the performance
of the gas handling infrastructure and the operational protocols. 

This chapter is organized as follows. In Section~\ref{XCSS:PurSpec} we review
the air leaks specifications of the purification and storage systems.
In Section~\ref{XCSS:KrRem} we describe the SLAC chromatography system
for removing krypton and argon. The online purification system and the 
xenon recovery systems are described in Section~\ref{XCSS:Online} and 
Section~\ref{XCSS:XeRec}. The long-term xenon storage system is
described in Section~\ref{XCSS:XeStor}. The purity monitoring program
is described in Section~\ref{XCSS:XeSamp}, and the cryogenics systems
are described in Section~\ref{XCSS:Cryo}. Xe procurement is 
described in Section~\ref{XCSS:XeProc}, and some gas 
species diffusion and solubility measurements are reported 
in Section~\ref{XCSS:DiffusionSolubility}.

\tdrsec[PurSpec]{ Xenon purification and storage specifications}

In this section we describe the purification and storage
specifications that bare upon the designs of the various xenon 
handling systems. Many of these specifications are driven by the 
the need to control the krypton concentration in the xenon. We adopt
the following policy: 

\begin{itemize}
\item the krypton removal system should produce xenon
that has no more than \SI{0.015}{\ppt}~(g/g) of krypton; 
\item the concentration should not increase more 
than \SI{0.005}{\ppt}/year while the xenon is
contained in the storage system; 
\item the time-averaged increase in 
the concentration during five years of 
detector operations at SURF should be no more
than \SI{0.015}{\ppt} (equivalent to \SI{0.006}{\ppt}/year).
\end{itemize}

\tdrsubsec[leaksStorage]{Air leaks during storage}

We estimate that the 
xenon will spend on average about six months in storage between krypton
removal and delivery into the detector. If the above requirement
is met, then  the krypton content will not rise by more 
than about \SI[group-separator={}]{0.0025}{\ppt}
from its initial value during this time. 
From the krypton concentration in air, we calculate that the allowed
air leak rate into \SI{10}{\tonnesl} is \SI{0.0173}{\g\per\year}, 
equivalent to a helium leak rate of 
\SI{1.15e-6}{\milli\bar\litre\per\second} (helium).
\footnote{We have used a conversion
factor of one~std~cc$\cdot$atm/sec~(air) = 
\num[group-separator={}]{0.3722}~std~cc$\cdot$atm/sec~(helium).}   
\IArTn will also contribute beta decay backgrounds
in this scenario, however, due to its 
small abundance\footnote{(\IArTn/\IArnat) = 
\num{8e-16} (g/g))~\cite{Loosli:2000}, 
while (\IKreF/\IKrnat) = \num{4} -- \num{23e-12}~\cite{GRL:GRL17458}.}
and longer half-life (\num{269} vs. \num{10.7}~years),
its background burden is only \SI{1.5}{\percent} of 
that of \IKreF and is not
constraining. Isotopes such as \IRnttt and \IArTS are 
relatively short-lived
(\SI{3.8}{\day} and \SI{35}{\day} half-lives, 
respectively) and have a small enough 
concentration in air such that they are not a concern 
during long-term storage. 

\tdrsubsec[leaksOpsOnce]{One-time air contamination during operations.}

If a one-time xenon handling mistake introduces a 
small volume of air into the detector while running,
the \IKreF and \IArTn decay rates will permanently increase, while
\IRnttt and  \IArTS will also contribute unsupported ER backgrounds. 
The background burden from these species is as follows.
We take the nominal \IArTS concentration in air to be 
\SI{1.2}{\milli\becquerel\per\m\cubed}~\cite{Riedmann:2011rp}, 
and we take the Davis campus radon concentration
as \SI{300}{\becquerel\per\m\cubed}. 
We assume that the quantity of dissolved air is 40~cc, 
and that the air is introduced near the beginning of the run. 
Then the number of
ER events between \num{1.5} and \SI{6.5}{\keV} in 
the \SI{5.6}{\tonnesl} fiducial volume in
\num{1000}~days is \numlist{24.3; 2.3; 0.36; 0.11} 
for \IKreF, \IRnttt, \IArTn, and \IArTS,
respectively. Here we have taken
the branching fraction of the \IPbtof beta decay in 
the \IRnttt decay 
chain as being \SI{9.2}{\percent}. \IArTS contributes 
x-rays and Auger electrons to
the ER band between \num{2} and \SI{3}{\keV} at a 
branching fraction of \SI{90}{\percent}.

\tdrsubsec[leaksOpsCont]{Continuous air leak during operations.}

Here we assume that a continuous air leak exists in the Xe handling 
system during five years of underground operations, such that the 
long-lived isotopes \IKreF and \IArTn integrate continuously into 
the ten tonnes of Xe, and such that activity of \IRnttt and  \IArTS is
now supported. We require that the average concentration of \IKreF 
due to the leak lead to no more than \num{24.3} \IKreF events, equivalent 
to the requirements on the krypton removal system. The average 
\IKreF concentration is half the final concentration, and under 
these circumstances we find that the equivalent helium leak rate 
is \SI{1.4e-6}{\milli\bar\litre\per\second} (helium), and the number of 
events due to \IKreF, \IRnttt, \IArTn, and \IArTS is 
\numlist{24.3; 0.51; 0.36; 0.13} respectively (again taking the 
\IRnttt and \IArTS concentrations in air as above).

\tdrsubsec[cosmogenics]{Cosmogenic activation of the Xe}

As detailed in Section~\ref{MASSs:Cosmo}, cosmogenic activation 
of the xenon during surface handling will result in tritium and 
\IXeotS activity in the Xe. After infinite exposure at sea level, 
the tritium decay rate would be about \SI{30}{decays\per\kgdayl}, 
or \num{1.7e8} decays in  \SI{5600}{\tonnedaysl}. 
Tritium is efficiently removed by the hot zirconium getter and 
is not expected to be problematic. \geantFour predicts that the 
\IXeotS activity will be \SI{2.6}{\milli\becquerel\per\kg} in 
equilibrium at sea level, and detector simulations show that the 
rejection factor of all selection cuts will be \num{1.1e-5}. 
Therefore we require that the initial activity be reduced to 
\SI{88}{\milli\becquerel\per\kg} at the start of physics data 
taking so that no more than \num{23} ER events remain after cuts. 
This implies a \num{178} day cool-down period underground, 
which is comparable to the detector commissioning time and 
is acceptable.

\tdrsec[KrRem]{Krypton Removal via Chromatography}

\begin{figure}[htbp]
\centering
\includegraphics[height=2.8in]{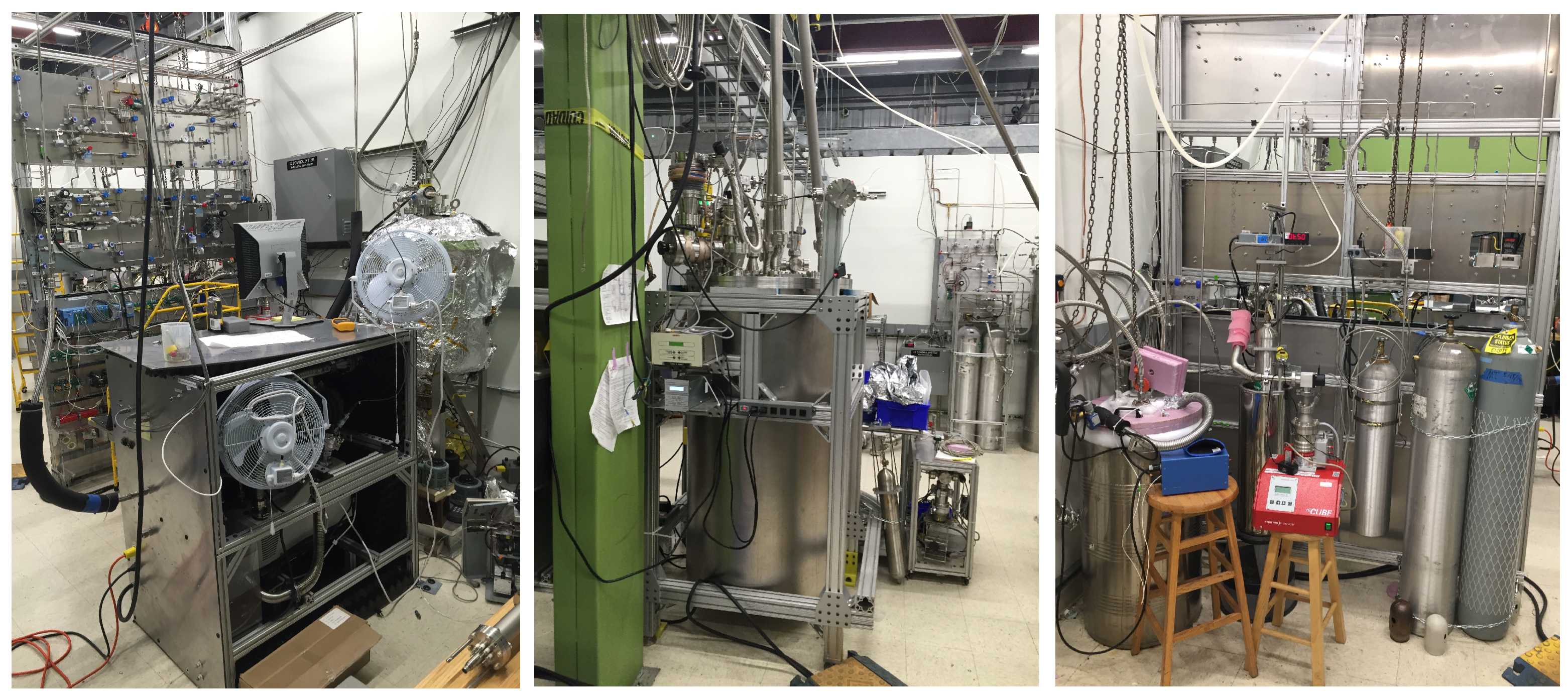}
\tdrfcaption[KrRemovalRandDSystem]{Kr removal R\&D system}
{The three panels show the Kr removal development platform. 
Left: the gas control manifold, pumps, and charcoal column. 
Center: LN thermosyphon cooled condenser for xenon recovery. 
Right: krypton traps and xenon feed and storage located on 
the back side of the manifold.}
\end{figure}

Commercially available research-grade Xe typically contains 
trace Kr at a concentration up to \SI{100}{\ppb} (parts per billion). 
\IKreF is a \Pgb-emitter 
with an isotopic abundance of \num{\sim 2e-11}  in natural 
Kr~\cite{Loosli:2000}, an endpoint energy of \SI{687}{\keV}, and 
a half-life of \SI{10.8}{years}. To reduce the rate of \IKreF 
ER backgrounds to \SI{10}{\percent} of the $pp$ solar neutrino 
ER rate, we require that the total concentration of Kr 
be no more than \SI{0.015}{\ppt} (parts per trillion), equal to a reduction
of up to \num{\sim e7} below that of research-grade 
Xe. Since the Kr is dissolved throughout the Xe, \IKreF 
decays cannot be rejected via self-shielding, nor does the 
getter remove Kr during \emph{in situ} purification ,
due to its inert nature. Gas vendors have indicated that 
they could guarantee Xe with a Kr concentration as small 
as \SI{1}{\ppb} at additional cost, however this would still far exceed the LZ 
background goal. As noted in Section~\ref{XCSS:XeProc}, 20\% of the Xe for LZ 
has been received and assayed, and all of it meets or exceeds spec. 

\begin{figure}[t]
\centering
\includegraphics[height=2.2in]{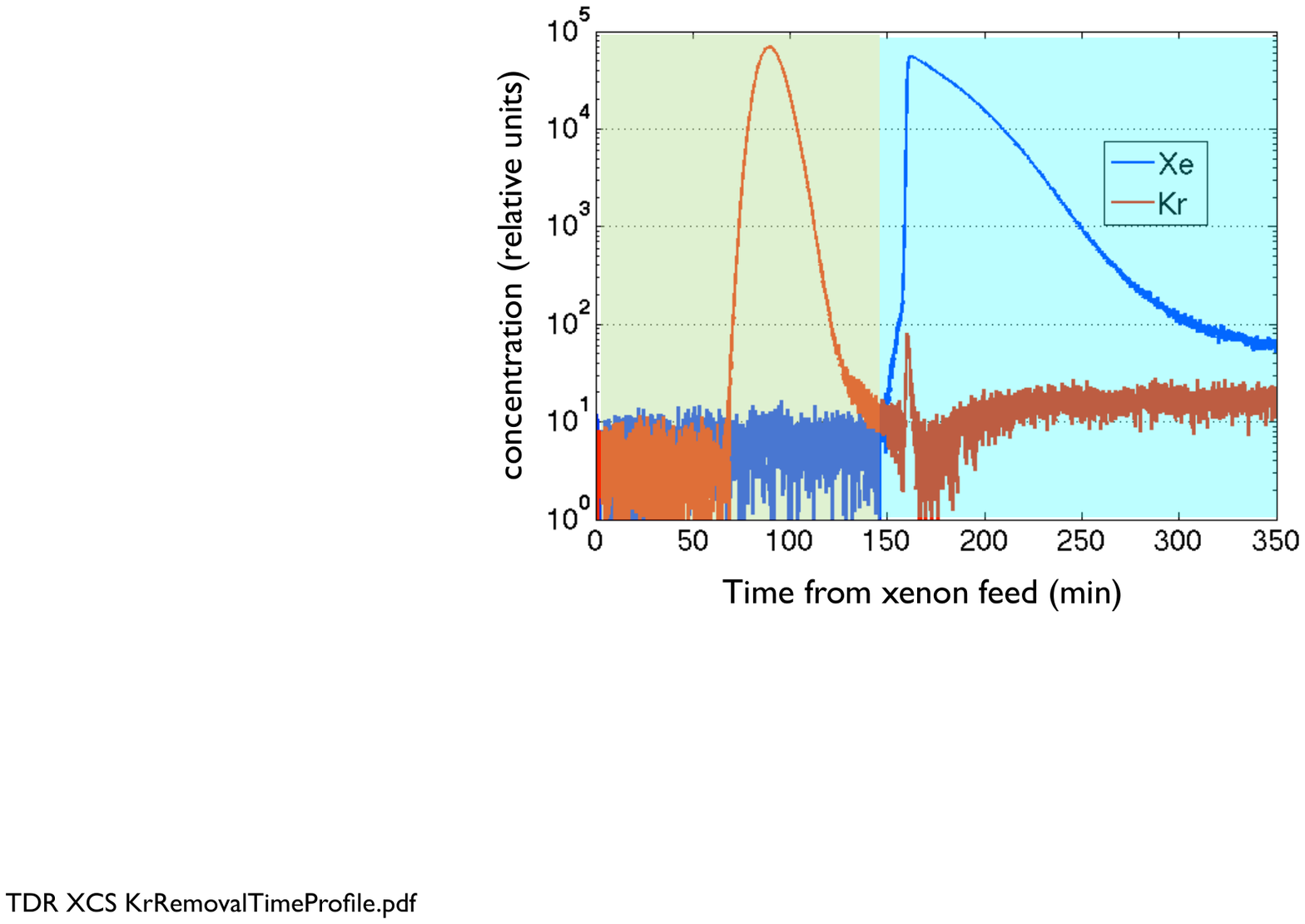}
\tdrfcaption[KrRemovalTimeProfile]{Kr removal time profiles of 
Kr and Xe}{Test data taken with the LUX Kr removal system 
illustrate the time profiles for Kr and Xe as they 
exit the column for a sample of xenon spiked with approximately 
1\% krypton. Data are acquired with a sampling RGA. During 
production, the \SI{\sim 100}{\ppb} Kr in the raw Xe is 
below the RGA sensitivity so spiked samples are used to tune 
the chromatography parameters.}
\end{figure} 

Trace Ar is also a concern due to the presence of the 
\Pgb-emitter \IArTn (endpoint energy of \SI{565}{\keV} and 
half-life of \SI{269}{years}), and we require that the 
background rate due to \IArTn be no more than \SI{10}{\percent}
of that of \IKreF. Due to the low isotopic abundance of \IArTn 
(\num{8e-16})~\cite{Loosli:2000}, this implies an Ar 
concentration requirement of \num{4.5e-10} (g/g), 
substantially less demanding than the krypton requirement. 
Furthermore, because Ar is amenable to the same removal 
techniques as Kr, no special measures are required to 
satisfy the Ar goal. Likewise, no special assays are required, since our 
technique simultaneously measures the residual Ar concentration along with 
the Kr.

LZ will employ gas charcoal chromatography to remove Kr 
and Ar from Xe using the process that we successfully 
developed and executed for XENON-10~\cite{Bolozdynya:2007zz} 
and LUX~\cite{Akerib:2016hcd}. 
The method employs a charcoal column with a continuously 
circulating He carrier gas and exploits the different 
transit time through the column for the various gas species. 
The weaker van der Waals 
binding of Kr to activated charcoal relative to Xe causes 
Kr atoms to flow through the charcoal column more rapidly 
than Xe. This property allows for a separation cycle by 
feeding Xe (with trace Kr) into the column at a fixed 
rate and duration under the influence of the fixed carrier flow. 
The rate and duration are tuned so that 
the last of the Kr exits the column prior to the earliest 
Xe. This permits the Kr to be purged by directing the 
Kr-He stream to a coldtrap that captures the Kr. 
This feed-purge cycle is followed by the Xe recovery cycle, 
in which pumping parameters are altered to accelerate the 
rate at which the Xe exits the column.

Figure~\ref{XCSf:KrRemovalTimeProfile} illustrates this 
process by showing the time profiles for the two species 
as they exit the column. The Kr-He exits first during the chromatography 
cycle to achieve Xe-Kr separation.
Then, as the purified Xe-He stream 
exits the column, the Xe is frozen out from the stream using 
a cryogenic condenser, while the He circulates until recovery is complete. 
After a number of feed-purge-recovery 
cycles, the He is pumped away, the condenser is warmed, 
and the Xe is transferred to a storage cylinder. The purity 
of the Xe is measured using the coldtrap/mass-spectrometry 
technique developed at Maryland~\cite{Dobi:2011vc}  
and described in Section~\ref{XCSS:XeSamp}.

Using this method, the SLAC group, while still at Case Western, processed the \SI{400}{\kg} of 
LUX Xe at a rate of \SI{50}{\kg}/week down to a Kr 
concentration of \SI{4}{\ppt}, exceeding the LUX goal of 
\SI{5}{\ppt} and rendering this background sub-dominant to 
the leading background by a factor of \num{10} during the 
2013 LUX WIMP search~\cite{Akerib:2013tjd,Akerib:2016hcd}. 
The starting concentration was \SI{130}{\ppb} of Kr and 
the typical batch reduction factor was \num{3e4}. In addition, 
one \SI{50}{\kg} batch of Xe that was ``spiked'' to approximately 
\SI{0.01}{\percent} Kr was processed twice, resulting in an upper 
limit on the concentration of \SI{0.2}{\ppt}, which was our assay 
sensitivity at that time. 
This powerful reduction 
of \num{9} orders of magnitude resulted in Xe with a Kr concentration only a 
factor of \num{13} higher than the ultimate LZ goal, and shows that 
sequential processing gives multiplicative 
reduction and is a suitable method to achieve clean Xe 
for LZ. 

\begin{figure}[t]
\centering
\includegraphics[width=5.5in]{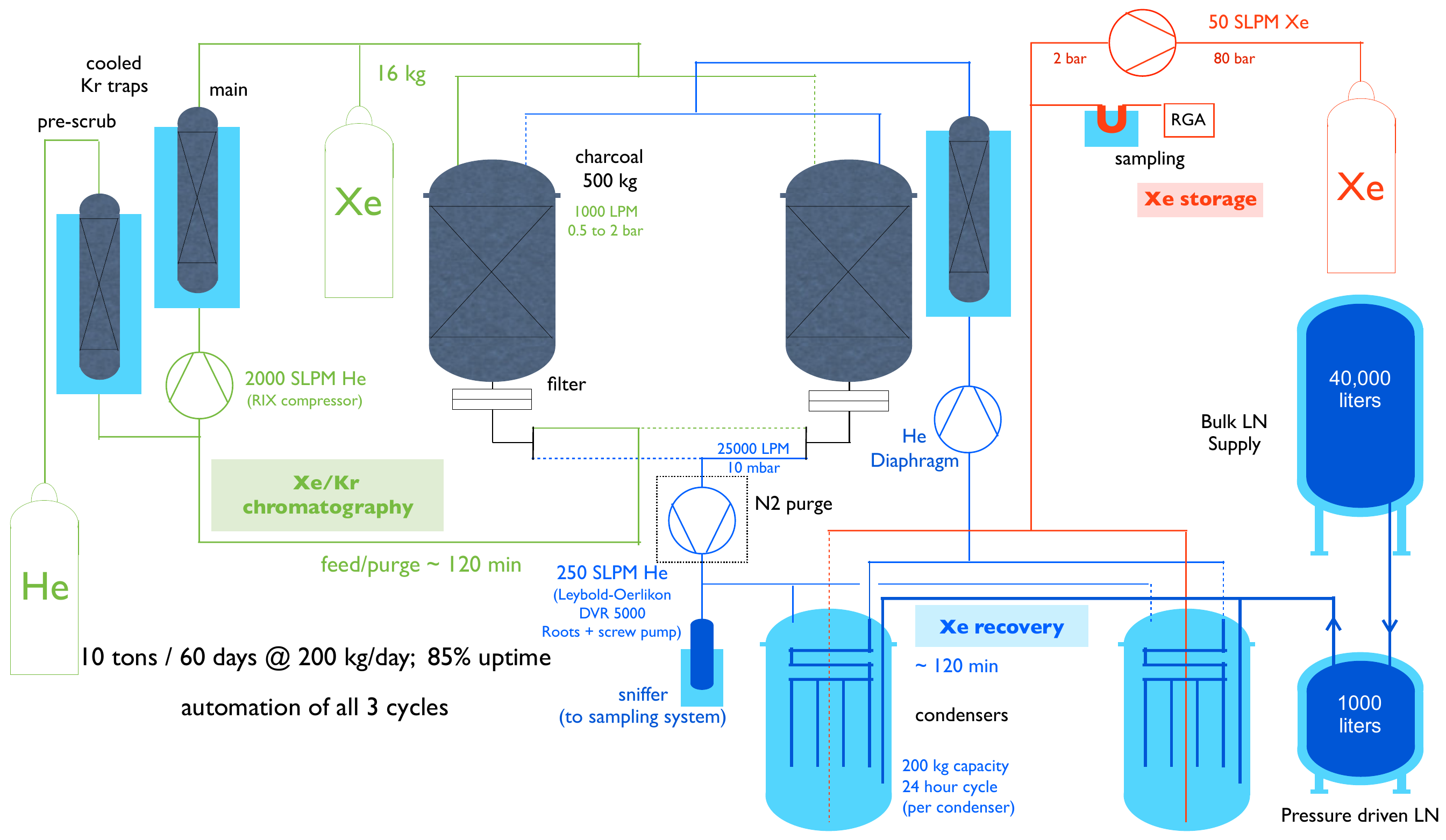}
\tdrfcaption[KrRemovalScaleUpSchematic]{Kr removal simplified 
schematic}{This figure shows the main components and flow paths 
of the LZ Kr removal system. Colored lines indicate the path 
for a given process. The green line shows the left-hand charcoal 
column in the chromatography feed-purge separation cycle with Kr 
being collected in the right trap; a second trap is valved out for 
cleaning (dashed green line). The blue line shows the subsequent 
part of the cycle in which Xe is being recovered from the column 
and collected in the condenser. The red line shows Xe from the 
condenser being stored and analyzed. Key pressures and flow rates 
are indicated.}
\end{figure}

While this  result is encouraging, it demonstrates only 
that cross-contamination of feedstock to recovered Xe can be mitigated 
by multi-pass processing. However, reaching the LZ goal also requires 
that sources of contamination and ingress of Kr-laden air also be 
controlled down to the ultimate required concentration. To ensure 
that our method is sound, we reconfigured the LUX production system 
as a development platform, which is shown in 
Figure~\ref{XCSf:KrRemovalRandDSystem}. We have been using that 
system to refine our process and have learned several valuable 
lessons that are being incorporated into the production system 
including: the pre-scrub of the He carrier gas through a cold-trap 
to remove trace Kr prior to use for purging of the primary traps; 
the placement of the Kr trap as close as possible to the charcoal 
column to mitigate contamination from upstream components; the 
gradual turn-on of Xe into the feed cycle to avoid pressure spikes 
and backflow into the output side of the Kr trap, which otherwise 
results in cross-contamination; the importance of all-metal seals 
throughout the system; and the critical nature of the Xe recovery 
pump, which is the most complex component  through which the 
purified Xe passes. In addition, we have used the assay system 
to prepare 100-gram samples of highly-purified xenon that we backflow 
into various parts of the system, recover, and assay to see if the 
xenon remains pure. We expect to make significant use of this 
powerful diagnostic tool to isolate, identify and mitigate 
deficiencies in the production system, as we have done in the 
R\&D system. Each of the lessons learned has resulted in 
progress in bringing down the level of the Kr, which in our best 
batches to date is 0.075\,ppt. At the time of this 
writing, we await the arrival of an improved Xe recovery pump 
with better isolation between the process space, the 
internal gearbox, and the external environment.

In addition to purity requirements, the processing rate and 
duration of a production run also influence the design of the 
full LZ system. Figure~\ref{XCSf:KrRemovalScaleUpSchematic}
provides a simplified schematic of our design for the production 
system, which will process \SI{200}{\kg} of 
Xe per day, or \SI{10}{\tonnesl} 
in \SI{60}{days} at \SI{85}{\percent} uptime. This design 
has better isolation between the chromatography loop and the 
recovery loop than did the LUX design, which achieved 
\num{3e4} reduction per pass and was likely limited by 
cross-contamination. While we may expect some improvement of 
the Kr rejection factor as the system is tuned during 
initial operations, we conservatively plan for two complete 
passes of the Xe (\SI{120}{days}), which is more than 
sufficient to reduce the Kr concentration in the raw 
stock by a factor of \num{e7}. Fast 
processing also provides schedule insurance in the event 
of an unintentional contamination event. We note that the 
introduction of just \SI{40}{\cm\cubed} of air at standard 
temperature and pressure (STP) corresponds to \SI{0.015}{\ppt} 
of Kr in \SI{10}{\tonnesl} of Xe. A single pass through 
the chromatographic system could easily correct an error as 
large as hundreds of liters. (The corresponding requirement 
for argon is less demanding: \SI{270}{\cm\cubed} of air at STP.)

As illustrated in Figure~\ref{XCSf:KrRemovalScaleUpSchematic}, the 
three principal processing stages are: (1) chromatography to separate 
Xe and Kr, and capture the Kr in a cooled charcoal trap; 
(2) recovery of the Xe from the column into a cooled condenser to 
strip the Xe from the carrier gas; and (3) storage and analytical 
sampling of the purified Xe into LZ storage packs for  
transport to the experiment. The system architecture follows two 
general principles. First, we minimize the components exposed to 
both the raw and purified Xe streams to minimize cross-contamination, 
which leads to the two mostly separate loops in the figure (green 
and blue). Second, the arrangement matches the feed-purge cycle time 
to the Xe recovery time, so that by alternating between the two 
columns, we achieve twice the duty cycle of single-column operation. 
Two condensers are also used so that one is always available in the 
recovery path when the other is in the storage phase. The double-swing 
system with two columns and two condensers gives a factor-of-4 
increase in production rate relative to a single-column-single-condenser 
system at modest additional cost and complexity. Moreover, reduced 
labor for the shorter production run more than offsets the 
incremental hardware cost of the second condenser and column.

The overall processing rate is governed by the differential transit 
time of the Kr and Xe in the column, whereas all other 
components are sized to match this rate. In the LUX system, each feed 
cycle consisted of a \SI{2}{\kg} slug of Xe into a column of \SI{60}{\kg} 
of charcoal at \SI{50}{\slpm}\footnote{standard liters per minute} He 
flow rate. The Kr purge was completed \num{120} minutes after the 
start of the feed, at which time recovery commenced. Straightforward 
scaling to reach \SI{200}{\kg} per day calls for a \SI{16}{\kg}-slug 
every two hours, which in turn requires an approximate eight-fold scaling 
in charcoal mass and flow rate. Some additional tuning will be required 
based on further studies of chromatography and cross-contamination 
using the LUX system. The remaining parameters, as shown in 
Figure~\ref{XCSf:KrRemovalScaleUpSchematic}, are appropriately 
scaled from the existing system. 

The system architecture uses computer-controlled pneumatic valves 
for changing all of the process states between the various cycles. 
Pumps, mass flow controllers, heaters, and diagnostic sensors (pressure, 
temperature, etc.) are controlled and/or read out by this system as 
well. We will use 
programmable logic controllers (PLCs) for process control since they 
are a standard industrial solution that is robust and gives low 
latency for reliable control and system interlocks. The PLC system 
will interface with the SCADA software package Ignition, by Inductive 
Automation, which we have been using in the R\&D system. The three 
primary cycles (chromatography, recovery, storage), as well as Kr 
trap cleaning and analytic sampling, will be fully 
automated, minimizing the chance of 
operator error.

\begin{figure}[h!]
\centering
\includegraphics[width=4.5in]{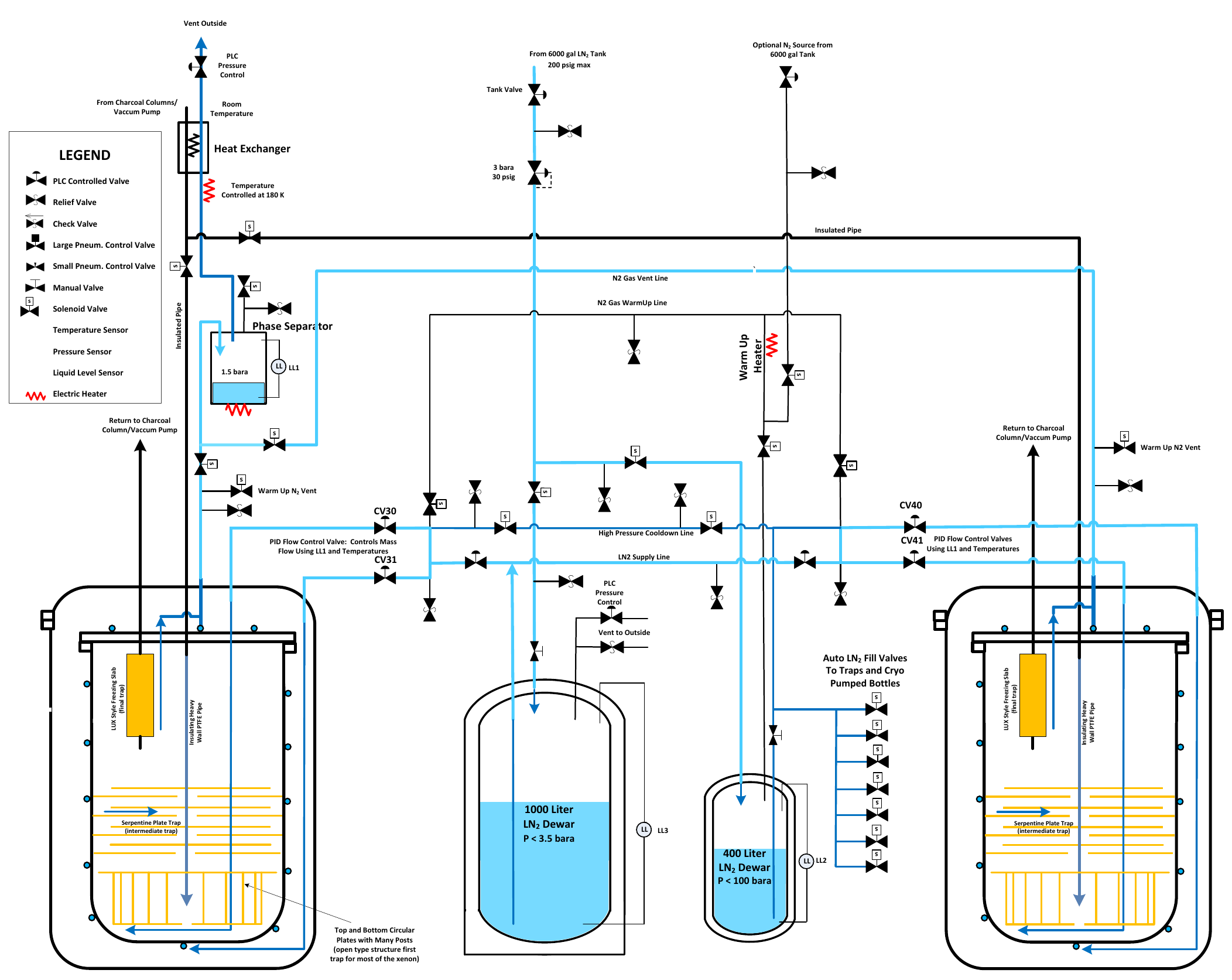}
\tdrfcaption[KrRemovalCondenser]{Kr removal condenser and LN system}
{A schematic of the condenser and LN distribution system. One 
condenser is alternately coupled to the two charcoal columns, 
with LN supplied by the 1000-liter pressurized storage dewar, 
1-2 days of recovery cycles. During that time, the other condenser 
is in the warming, storage, and re-cooling cycle. The 400-liter 
LN dewar supplies cryogens for rapid recooling as well as the 
various LN-cooled traps. The two LN dewars are resupplied by 
the 9,000 gallon tank at IR2 (not shown).}
\end{figure}

The Kr-removal system uses a cryogenic condenser system to freeze 
the purified Xe out of the Xe-He stream at LN temperature. A 
schematic of the combined condenser and LN system is shown in 
Figure~\ref{XCSf:KrRemovalCondenser}. The design of the condenser 
is based on the LUX approach with several new features that will 
simplify construction and operation, and reduce cost. A set of 
LN-cooled surfaces mounted in a pressure vessel accumulates the 
purified xenon. After a number of runs, the He is evacuated, the 
system is warmed to LXe temperature, and the xenon is compressed 
into LZ storage packs for either a second pass or transport to the 
experiment. The Kr removal system will use one of the storage 
compressor skids described in Section~\ref{XCSS:XeRec}, which will 
then travel to South Dakota with the processed xenon. Since we don't 
require precision temperature control as in a LXe detector, rather 
than using a more complex thermosyphon system to provide cooling 
we have designed a simpler force-flow system. This system will 
deliver LN through a system of copper tubes that are brazed to 
the vessel walls and internal structures. The direct LN cooling 
will easily meet peak load and rapidly recool the vessels after 
storage of a previous batch. The pressure-driven LN loops will 
consist of long circuits of copper piping that will be coupled 
to a series of copper plates in the condenser space and to the 
outer wall of the vessel. Because cooling power can be better 
distributed than with individual thermosyphon cold heads, the 
design will have a favorable ratio of xenon capacity to vessel 
size. Finally, the pressure-driven LN system has fewer control 
points than the thermosyphon and will be more straightforward 
to program and operate. A detailed engineering note contains a 
through analysis of the time-dependent heatload defined by the 
variable rate of the Xe recovery 
(Figure~\ref{XCSf:KrRemovalTimeProfile}), as well as the 
reduction in cooling power as frozen Xe accumulates (Xe ice 
is a poor thermal conductor).

\begin{figure}[h!]
\centering
\includegraphics[width=4.0in]{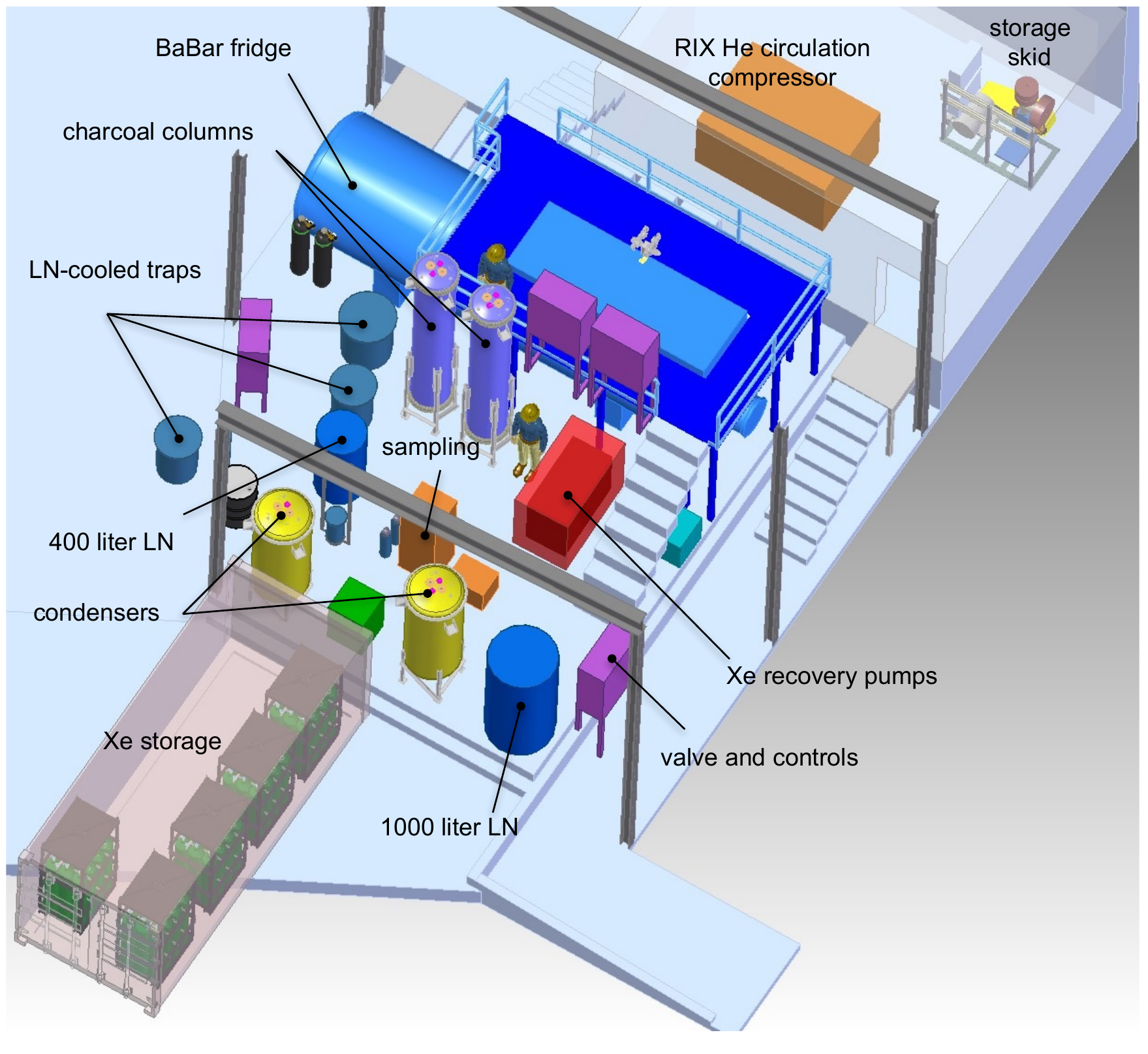}
\tdrfcaption[KrRemovalCryoPadLayout]{Kr removal physical layout}
{A layout of the LZ Kr removal system shows all major 
components located at the former BaBar LHe cryopad in 
building 624 at the SLAC IR2 complex. If a buyer for 
the refrigerator is not found, this layout shows the 
adequacy of the remaining space. The partial soft-wall 
exterior is well suited to mitigate oxygen deficiency 
hazard with the large quantity of cryogens in use. A 
20-foot sea container at the front of pad will provide 
secure storage for Xe cylinder packs and ease in handling.}
\end{figure} 

In Figure~\ref{XCSf:KrRemovalCryoPadLayout}, we show a physical 
layout of the major system components, including  
pumps, columns, condensers, traps, pressure-driven LN supply, and sampling 
subsystems. The system will be installed in Building 624 at the 
IR2 complex at SLAC. The Xe, which is a high-value asset, will 
be stored in a locked "sea container" in front of the pad. Not 
shown is the nearby LN storage tank, which is refilled by truck 
under SLAC contract.  

A high-flow-rate RIX 2-stage piston pump for circulating the carrier 
gas during chromatography has been identified at SLAC and is available 
for LZ. The purity requirements of this pump are modest because the 
He will pass through the LN-cooled charcoal trap prior to the 
Xe feed branch and the chromatography columns. 

As noted in the R\&D discussion, the Xe recovery pump system is 
the most complex component that sees the purified xenon. We have 
worked closely with Leybold-Oerlikon to evaluate commercially 
available dry pumping systems.  We provided the manufacturer with 
a fully-detailed data set rescaled from our LUX slow control data, 
including He and Xe pressures and mass flow rates for the full 
recovery cycle. The manufacturer has used their proprietary 
software to make a detailed a assessment of pump performance, 
and determined that a stock version of the DVR 5000, which 
combines a dry Roots booster backed by a dry screw pump will 
meet our requirements.

During processing, several hundred kilograms of Xe will be in 
various parts of the system. To protect against loss of Xe we 
will rely on backup systems to maintain cooling power. Short-term 
outages are protected by UPS units. For longer outages, the SLAC 
operations team has deployed a standby generator that will come 
online within \SIrange[range-units=single]{1}{2}{minutes}. 
These capabilities will allow the control systems to replenish 
the L\CNt reservoirs in the Kr removal system from the nearby 
\num{9000} gallon LN storage tank at the IR2 complex.

\begin{figure}[t!]
\centering
\includegraphics[height=8.0in]{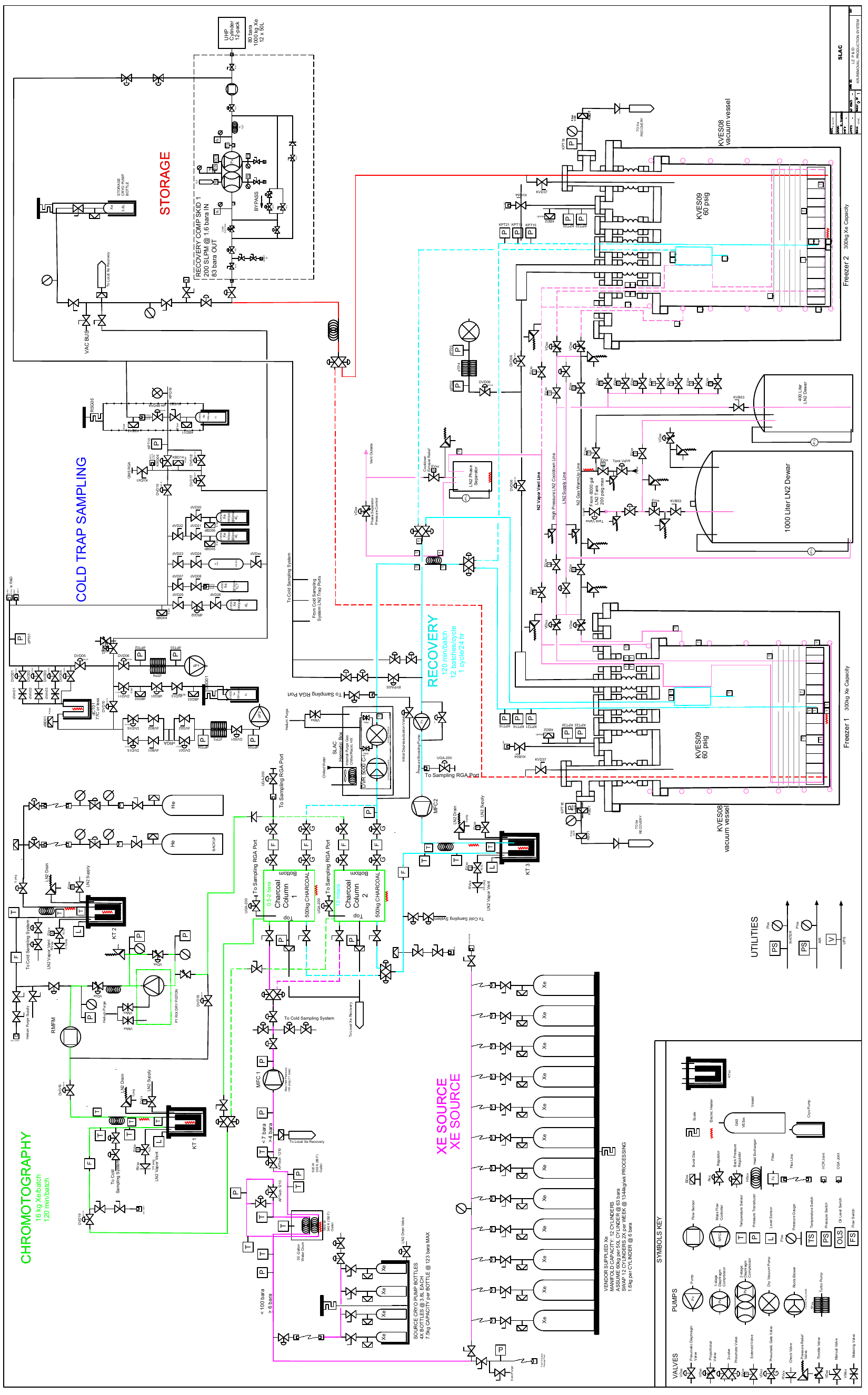}
\tdrfcaption[KrRemovalProductionSystemPID]{Kr removal P\&ID}
{The P\&ID for the LZ Kr removal system.}
\end{figure} 

A fully-detailed P\&ID\footnote{Piping and Instrumentation Diagram} 
has been developed for the production system. 
The design is based on the principles of the LUX system that was 
reconfigured for R\&D, while following the overall architecture 
shown in Figure~\ref{XCSf:KrRemovalScaleUpSchematic}. The production 
system P\&ID is shown in Figure~\ref{XCSf:KrRemovalProductionSystemPID}.

\tdrsec[Online]{Online Xenon Purification System  }

This section describes the online Xe purification system, 
which controls the concentration of 
electronegative impurities such as \COt and \CHtO during 
detector operations through continuous purification. 
The system is shown schematically in 
Figure~\ref{XCSf:TeaPotDiagramFig}. Liquid xenon is 
removed from the detector at the weir 
trough near the top of the TPC. The design
flow rate is \SI{500}{\slpm} (\SI{2.8}{\kg\per\minute}), 
or about \SI{1.0}{\litre\per\minute}. The liquid exits the 
cryostat and flows to the LXe tower through 
vacuum insulated transfer lines that penetrate the water tank. 
In the LXe tower it is evaporated, 
compressed to higher pressure, is purified, and then 
recondensed and returned to the detector. 

A central feature
of this design is a set of two heat exchangers located 
in the LXe tower that make the phase and temperature change 
thermally efficient. Without these heat exchangers, 
\SI{4.3}{\kW} would be required in each  
flow direction at \SI{500}{\slpm} for 
the phase change alone. 

This section is organized as follows. The gaseous portion 
of the circulation system is described in Section~\ref{XCSSs:gXe}.
The LXe tower and the vacuum insulated transfer lines 
are described in Sections~\ref{XCSSs:LXetower} 
and \ref{XCSSs:LXeTransferLines}.
We report on results from a prototype circulation system
that implements the LZ architecture in Section~\ref{XCSSs:SystemTest}.
A radon removal system that treats the gaseous xenon in the 
breakout box and cable conduits is described in Section~\ref{XCSSs:RnRem}, 
and the slow controls is described in Section~\ref{XCSSs:SlowControl}.

\begin{figure}[htbp]
\centering
\includegraphics[height=4.4in]{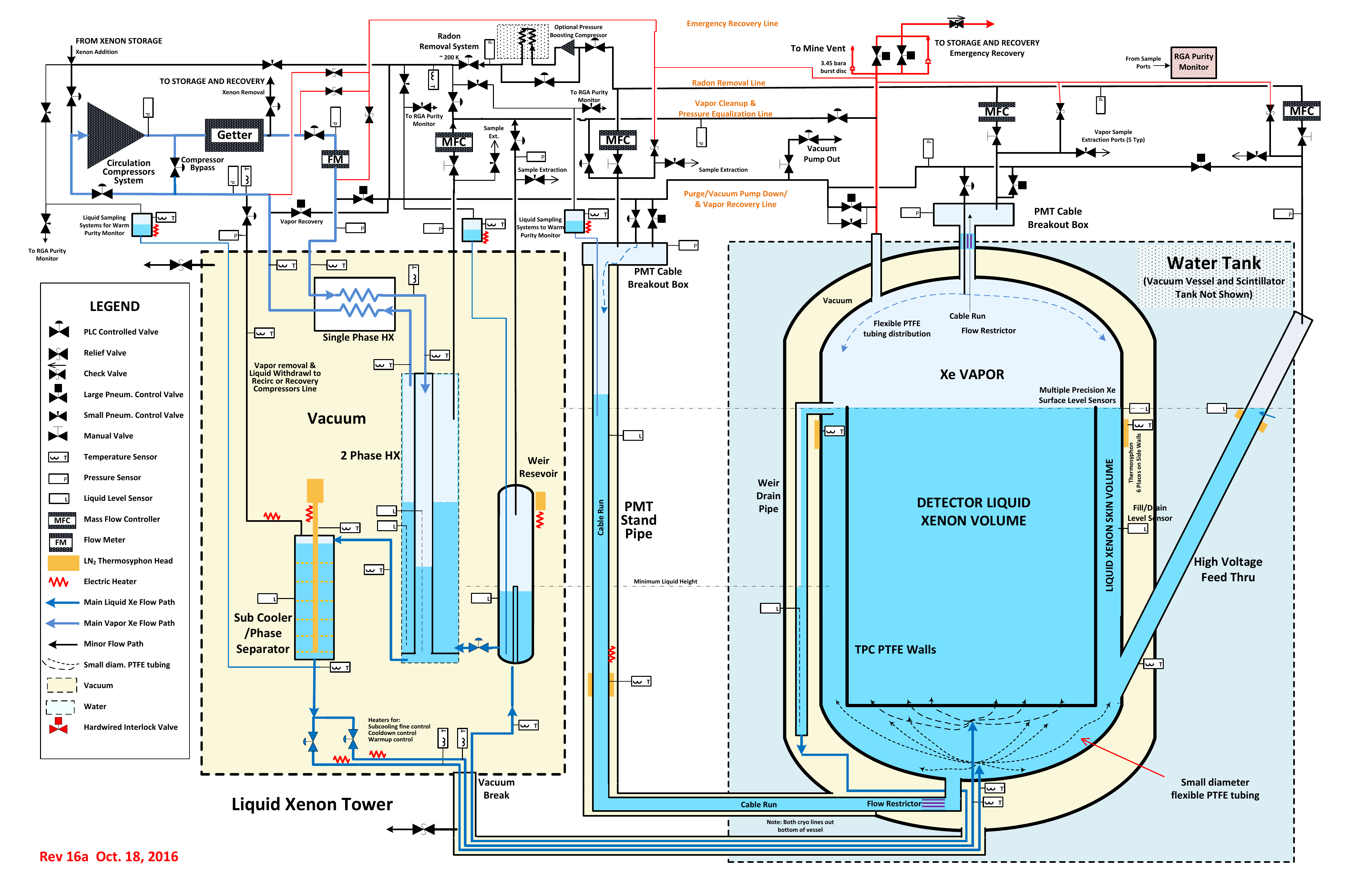}
\tdrfcaption[TeaPotDiagramFig]{Schematic diagram of LZ online 
purification system}{Overall flow diagram of the LZ online 
purification system.  The shaded box on the left represents the LXe 
Tower, where the phase changes between LXe and GXe occurs.  The top 
of the diagram represents the gas recirculation system.  Note that 
this figure captures the design at a high level, but the master P\&ID 
should be relied upon for details.}
\end{figure} 

\tdrsubsec[gXe]{Gas recirculation system}

The Xe gas recirculation system receives gaseous xenon from the 
detector via the LXe Tower (Section~\ref{XCSSs:LXetower}), 
circulates the xenon through the hot zirconium getter to remove 
electronegative impurities, and returns the purified Xe to the 
LXe Tower.  This system also handles Xe purge flow and Rn-Removal 
(Section~\ref{XCSSs:RnRem}) to/from the detector cable 
conduits, as well as calibration source injection. A complete P\&ID for this 
subsystem is provided in Figure~\ref{XCSf:GasRecirculationPID}. 

\begin{figure}[htbp]
\centering
\includegraphics[height=8.25in]{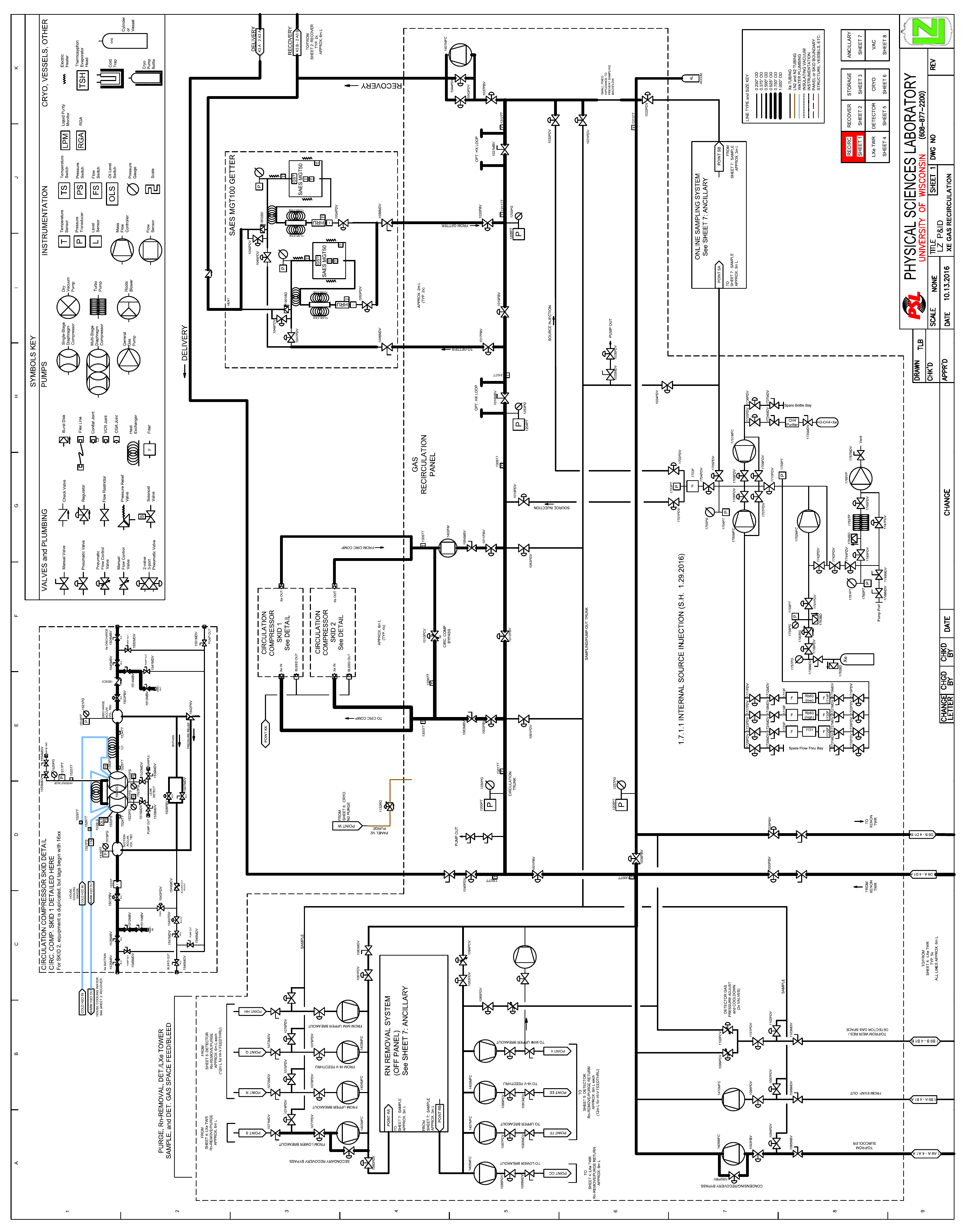}
\tdrfcaption[GasRecirculationPID]{Xe Gas Recirculation 
system P\&ID}{Xe Gas Recirculation system P\&ID.}
\end{figure}

The centerpiece of the LZ online purification system is a hot 
zirconium getter (model PS5-MGT100 from SAES), a purifier 
technology capable of achieving \SI{>99}{\percent} one-pass 
removal efficiency for virtually all non-inert impurities of 
interest, including hard-to-remove species such as \CNt and 
CH$_4$~\cite{Dobi:2010ai}. Impurities 
chemically bond to the surface of zirconium pellets, irreversibly 
removing them from the Xe. The getter operates at elevated 
temperature (\SI{450}{\degree C}) to allow the captured impurities to 
diffuse into the bulk of the getter pellets, leaving the surface 
free for additional gettering. The getter material must be replaced 
when it becomes saturated with impurities. 
The SAES getter comes equipped with valves, plumbing, 
controls, and an engineered integrated safety interlock system 
that will disable the getter power and isolate the getter from 
the process gas in the event that excessive heat is detected.

The Xe circulates through the purification system at a rate of 
\SI{500}{\slpm} (\SI{2.8}{\kg\per\minute}), a value that was 
chosen based upon previous experience with the LUX system and 
based on economic and space constraints. At this flow rate, the 
\SI{10}{\tonnesl} of Xe can be purified in \SI{2.3}{days}, 
comparable to the \num{1.7}-day turnover time of the LUX 
recirculation system. The recirculation rate and the charge 
attenuation goal constrains the allowed outgassing 
rate of the detector, and this drives the outgassing plan of the 
detector. Prior experience with LUX and other liquid noble 
detectors indicates that the scintillation absorption length 
goal (\SI{>15}{\m}) is less demanding than the charge attenuation 
goal and will be satisfied by the same requirements.

Circulation through the purification system is provided by a set 
of all-metal triple-diaphragm two-stage compressors. This type 
of compressor uses an electric motor to drive single-piston 
oil pumps that pressurize the bottom diaphragms of each stage, 
transmitting force to the top diaphragms to compress the gas. 
Middle diaphragms with grooves create a space that is monitored 
for leakage of either oil or Xe. The all-metal design (including 
metal main head seals) minimizes the ingress of krypton and radon 
into the Xe.  A \SI{100}{\slpm} Fluitron 
compressor of this type and utilizing 
this all-metal seal technology has been operating at SLAC as 
part of the System Test (Figure~\ref{XCSf:SLACFluitron}), allowing 
us to begin building a strong understanding of operating performance, 
characteristics, leak-tightness, and maintenance. For LZ, two 
compressors, each rated for \SI{300}{\slpm} at \num{1}~bara
suction pressure and capable of 
\num{10}~bara output pressure, operate in parallel 
to provide the full system flow.  Together these compressors 
nominally deliver \SI{600}{\slpm} of flow, so a proportional 
control valve modulates excess flow back to compressor suction 
to maintain the nominal \SI{500}{\slpm} circulation through the 
purification system.  Scheduled maintenance (every \num{5000} hours) 
and unexpected repairs on these machines will require periodic 
operations at reduced circulation rate (approximately \SI{300}{\slpm} 
from the remaining online compressor).

\begin{figure}[htbp]
\centering
\includegraphics[height=3.0in]{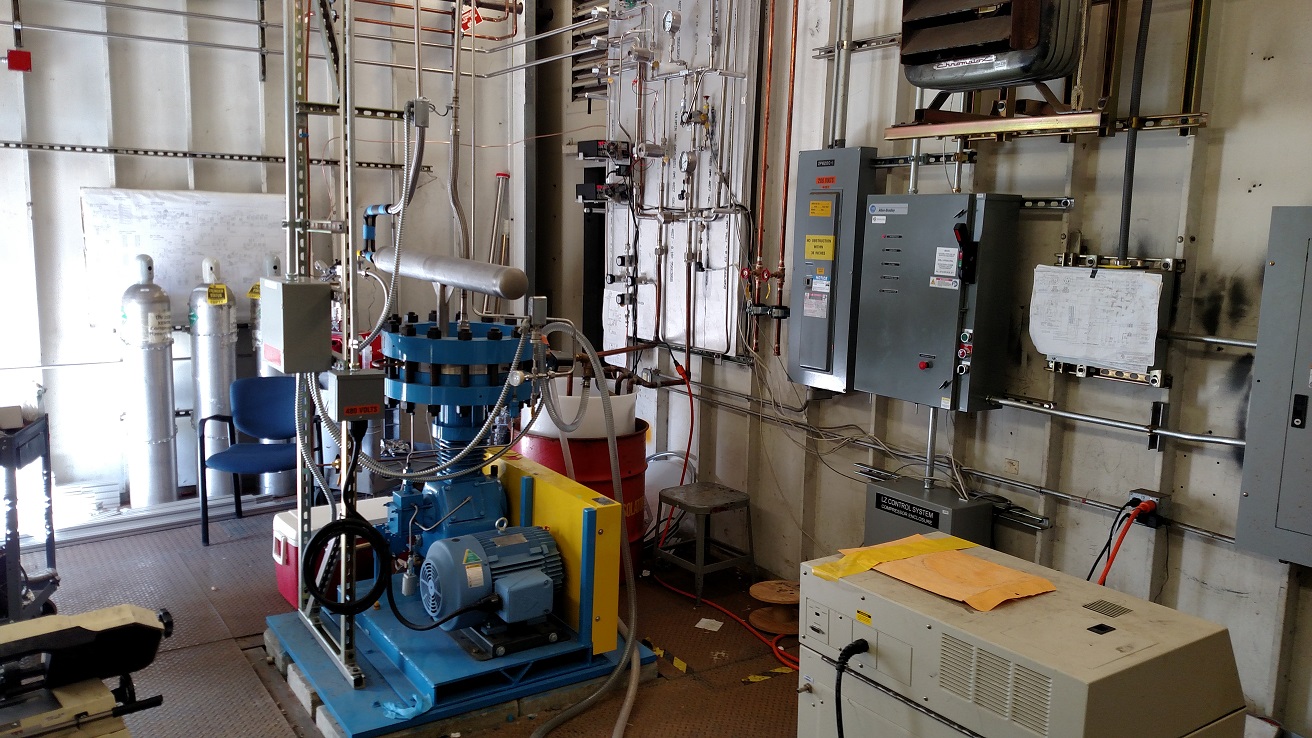}
\tdrfcaption[SLACFluitron]{Fluitron all-metal diaphragm compressor 
at SLAC}{Fluitron \SI{100}{\slpm} all-metal single-stage diaphragm 
compressor operating as part of SLAC System Test.}
\end{figure} 

After-cooling of the compressed gas may be adjusted so 
preheated gas can be delivered to the getter. Hot gas exiting 
the getter is cooled back to room 
temperature by a heat exchanger provided with the getter. 
The purified gas is then 
returned to the LXe Tower.  The online sampling system 
(Section~\ref{XCSS:XeSamp}) has taps to draw samples from 
upstream of the compressor, and upstream/downstream of the 
getter.  Xenon that has already been sampled reenters the 
circulation loop downstream of these taps.

\begin{figure}[htbp]
\centering
\includegraphics[height=8.25in]{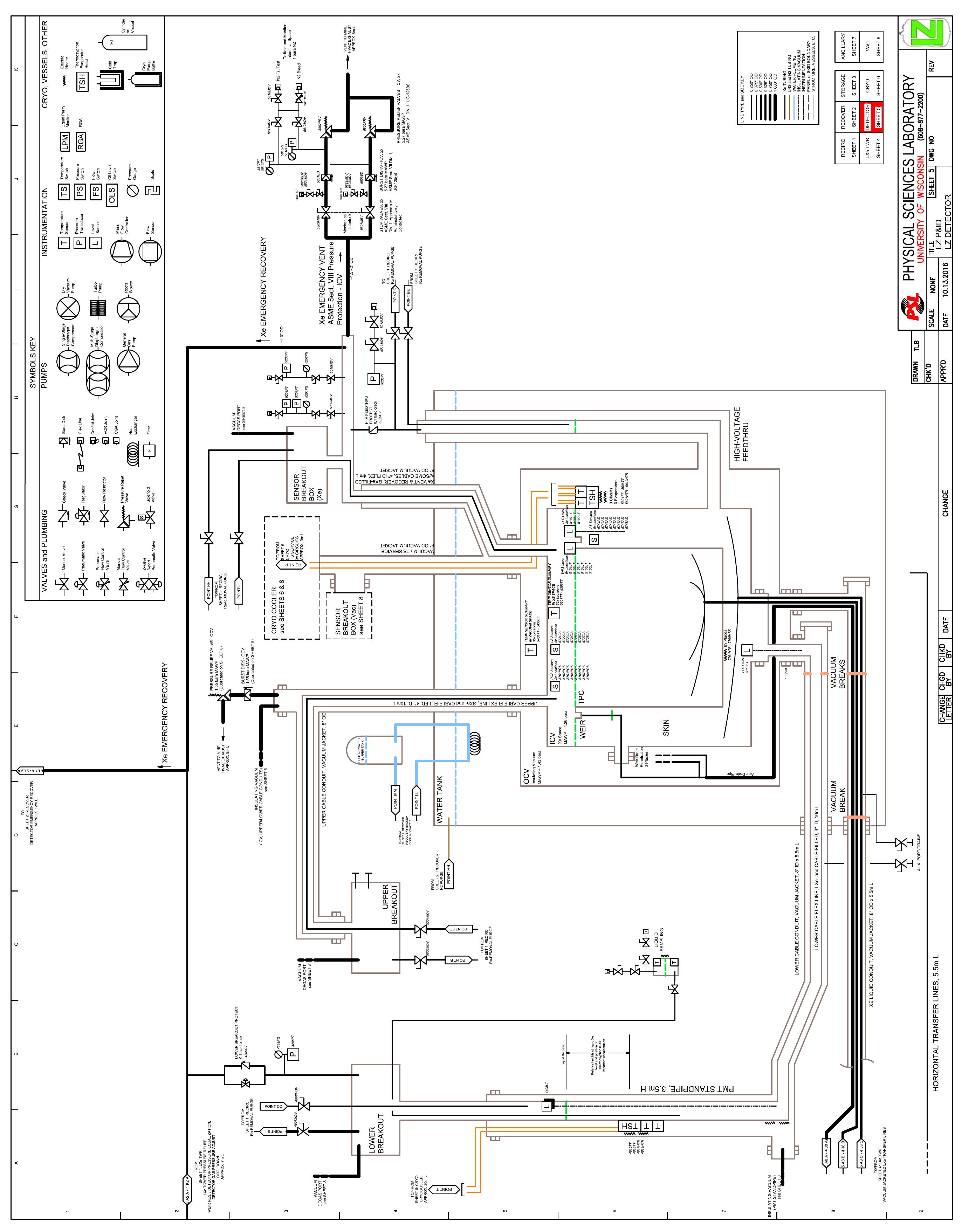}
\tdrfcaption[DetPID]{P\&ID of Detector and Xenon Handling 
Tie-Ins}{P\&ID of Detector and Xenon Handling Tie-Ins. }
\end{figure} 

The gas plumbing for the recirculation system is pre-cleaned 
stainless steel (SS) tubing connected with lanthanated or 
ceriated orbital welds wherever possible. VCR fittings with
metal seals will be used where necessary to open a connection. 
Valves will be of the bellows- or diaphragm-seal type to keep 
the system hermetic.  Gas panels will be sealed to allow for 
panel-wide He leak checking during testing and commissioning, 
as well as continued \CNt purge during operations.

In addition to the primary Xe recirculation path, four secondary 
gas-only recirculation loops act on the 
detector conduits that house the 
cables for the PMTs and instrumentation cables (from Xe vessel 
top and the PMT standpipe). These cables are immersed in LXe 
on one end, but they penetrate the liquid surface and terminate 
at room-temperature feedthroughs filled with Xe gas at the other 
end. These cable bundles are a 
potential source of problematic outgassing, particularly the 
warm ends where the diffusion constants of the insulating 
plastics are the largest. We manage this outgassing by purging 
these conduits with a continuous flow of Xe gas away from the 
TPC at a modest flow rate. This purge gas flow merges at the 
input of the online Rn-removal system (Section~\ref{XCSSs:RnRem}), 
and then returns back to the conduits, bypassing the LXe tower. 
Purge flow is driven by a small independent gas compressor in 
the Rn-Removal system. Plumbing provisions also allow for a 
portion of the Rn-Removal system output flow to be directed 
into the main circulation loop to create a purge flow 
imbalance and provide evaporative cooling to the conduits 
if desired. The purge-gas flow rate required to ensure 
that back-diffusion is negligible is characterized by the 
unitless Peclet number, $P = VL/D$, where $V$ is the linear 
velocity of the gas in the conduit, $L$ is the diffusion 
distance of interest, and $D$ is the diffusion constant of 
the impurity species. For a \SI{4}{\inchl} diameter conduit, a 
gas flow rate of \SI{0.5}{\slpm} per conduit, a diffusion 
distance of \SI{10}{\cm}, and a diffusion constant of 
\SI{0.086}{\cm\squared\per\second} (valid for \COt diffusion 
in Xe gas at room temperature), the Peclet number is \num{8.2},
indicating that back diffusion is negligible. We plan for 
a total of \SI{2.0}{\slpm} of flow across all conduits (three cable 
conduits and one high voltage feedthrough conduit).  
Figure~\ref{XCSf:DetPID} is a portion of the P\&ID detailing 
the detector, which shows how these purge paths are tied into 
the detector conduits.  LXe paths are on bottom, GXe taps are 
above the LXe level.

The circulation system must also accommodate detector 
calibration and purity assays. As discussed in Chapter~\ref{chap:CAS}, 
radioactive sources such as \IKreTm and tritiated methane 
will be introduced into the Xe to calibrate the central 
regions of the TPC, and the online purification system 
has a valve and port system to allow for this. Due to its 
inert nature, \IKreTm may be injected into the circulation 
stream upstream of the getter, whereas tritiated methane 
must be injected downstream. There are also valves to 
allow for the sampling of the LXe in the tower and gaseous 
Xe before and after the getter, as described in 
Section~\ref{XCSS:XeSamp}.  Details of the source injection 
plumbing and how it ties into the gas recirculation system 
can be found lower-center in Figure~\ref{XCSf:GasRecirculationPID}.

\tdrsubsec[LXetower]{Liquid xenon tower}

Because the hot getter in the purification system 
(Section~\ref{XCSSs:gXe}) operates on gaseous Xe (GXe) only, 
the purification system must vaporize and re-condense 
the liquid Xe (LXe) to create a continuous purification circuit. 
This process is made thermally efficient by using counter 
flowing single- and two-phase 
heat exchangers in series with a sub-cooler/phase-separator 
within the LXe Tower assembly. The Xe is transferred 
in liquid form between the LXe Tower and the detector 
through vacuum-insulated transfer lines that penetrate 
through the wall of the water shield and connect to the 
bottom of the TPC vessel.

The LXe Tower is shown in detail in 
Figure~\ref{XCSf:LXeTowerandStandpipe} and contains 
the following major components: 1) a valve block that houses 
LXe flow control and LXe Tower isolation valves; 2) a weir reservoir 
for collecting the liquid returning from the detector 
and preparing it for evaporation; 3) a counter flow serial 
two-phase and single-phase heat exchanger assembly for 
evaporating the LXe and bringing the exiting GXe to room 
temperature while pre-cooling the returning GXe and 
liquefying the xenon gas; 4) a 
subcooler/phase separator for cooling the return liquid 
and removing gas bubbles; and 5) three thermosyphon heads; 
three liquid sampling ports.
The LXe Tower P\&ID is shown in Figure~\ref{XCSf:LXeTowerPID}.

\begin{figure}[htbp]
\centering
\includegraphics[height=3.5in]{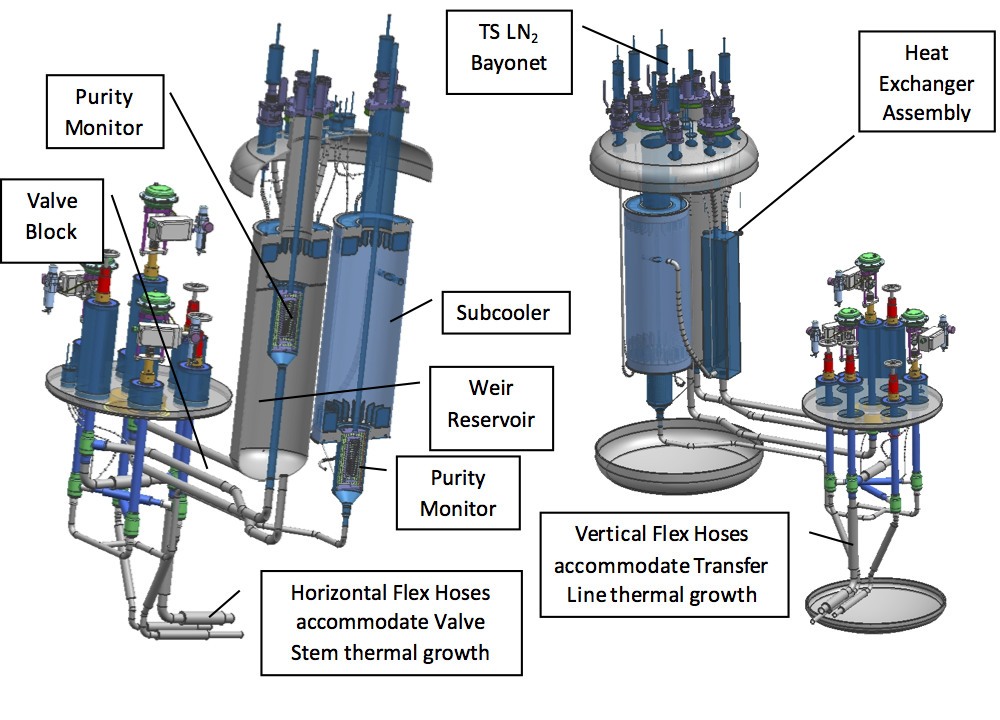}
\tdrfcaption[LXeTowerandStandpipe]{Xe Tower with Internals}
{Two views of the LXe tower assembly. The heat exchanger 
assembly is visible in the right image and the subcooler 
and weir reservoir components can be seen at the left. 
The cryogenic valves are consolidated in a separate valve 
block (visible in both views) to provide easy access to the 
actuators and also to allow for isolation between the LXe 
tower and the detector.}
\end{figure} 

\begin{figure}[htbp]
\centering
\includegraphics[height=8.25in]{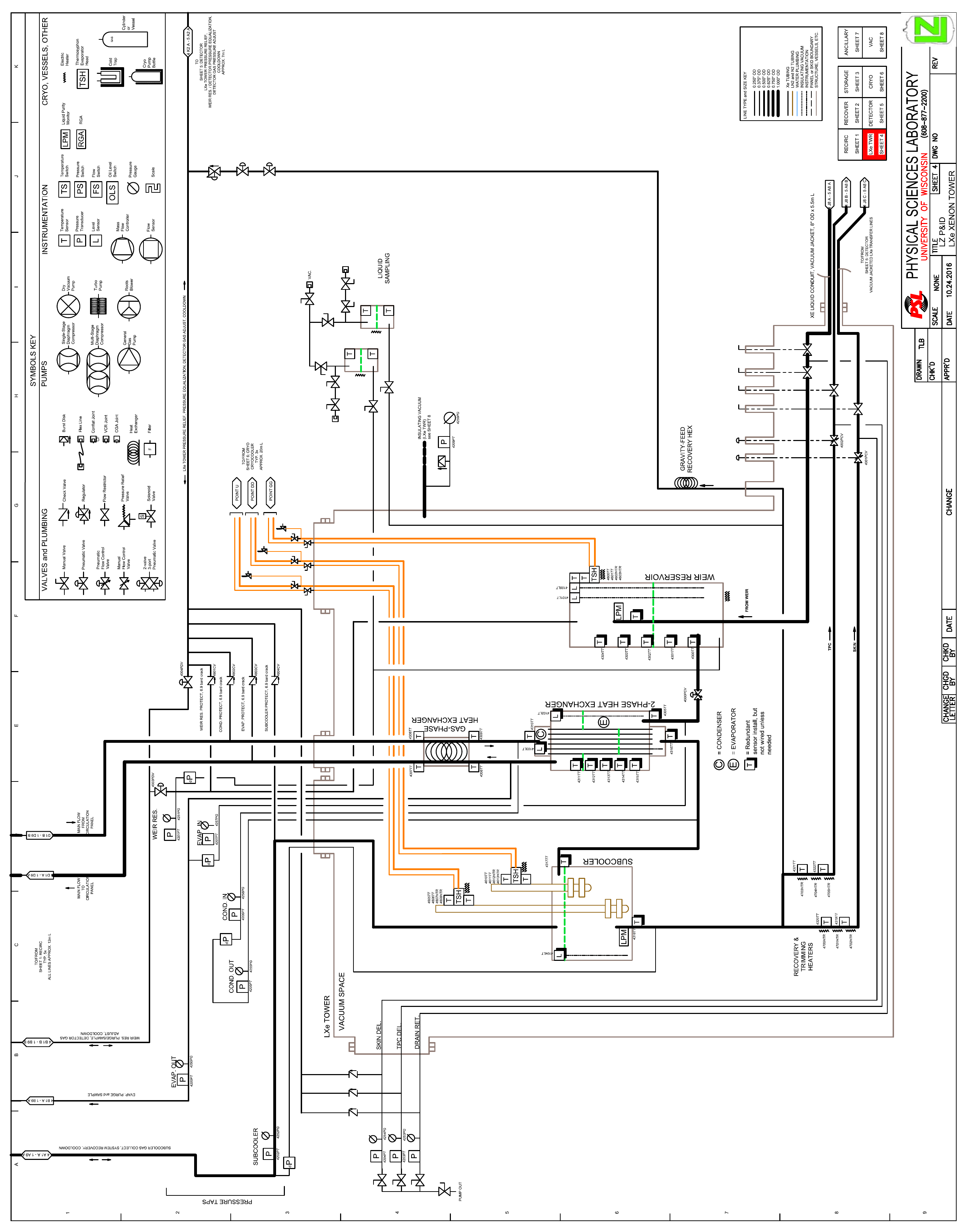}
\tdrfcaption[LXeTowerPID]{P\&ID of the LXe Tower}{P\&ID of 
the LXe Tower.}
\end{figure} 

In addition, the tower design includes space 
for two optional liquid purity 
monitors, one located in the inlet to the weir reservoir, to measure 
the purity of the xenon on the way out of the detector, and the other 
located in the subcooler, to measure the purity of the xenon entering 
the detector.  
Cryogenic components will hang from the top of the vacuum jacket to 
allow for fabrication with the vacuum jacket removed. When the 
internal structure is complete, the vacuum jacket is welded shut 
on the sides and bottom.  There is a vacuum 
break at the bottom of the xenon tower isolating its
vacuum from the detector for safety considerations 
and for LN2 checkout.

Liquid flow from the subcooler passes through two sets of two 
cold control valves that 
partition the flow between the TPC volume and the skin 
volume. Each valve set is paired to allow the LXe tower to be 
isolated from the detector. These valves will be  used during 
normal operation, 
for vessel cool-down, and for changing flow rates and 
patterns during injection of radioactive Kr and methane.  
Both of 
these flow streams have an electric heater with thermometer 
for cool-down and warm-up control.  These valves are also 
used for liquid extraction to compressor suction during 
warm-up.  A third pair of cold control valve is placed between 
the weir reservoir and the evaporator side of the two-phase 
heat exchanger. This valve set works in conjunction with the 
compressor suction control to establish the height of 
the boiling LXe. It will also provide a small amount of 
Joule-Thomson cooling by using the energy supplied from 
the recirculation compressor. All three of these cold 
control valve sets are bellows sealed and designed for use 
in helium liquefiers. 

The weir reservoir is a simple \SI{35.6}{\centi\meter} 
diameter ASME stamped vessel that acts as a 
volume buffer between the weir drainpipe and the two-phase 
heat exchanger. LXe flows down the weir drainpipe in the 
detector, through the transfer line, and up the weir 
reservoir standpipe. Besides establishing the liquid level in 
the weir drainpipe in the detector, the weir reservoir 
also provides a 
location to collect nonvolatile impurity species that may 
be present in the LXe.  Such species, if they exist, would 
become harmlessly trapped in the weir reservoir.  A liquid 
sample tube from the weir reservoir liquid sampling system 
to the bottom of the weir reservoir makes it possible in 
principle to remove contamination.

Liquid flows from the bottom of the weir reservoir into 
the bottom (evaporator) of the two-phase heat exchanger where 
it is vaporized, while Xe from the getter is condensed.  
The bottom of the weir 
reservoir is placed at a relative height above the bottom of 
the two-phase heat exchanger to provide a small liquid 
pressure head to push fluid into the heat exchanger. 
Liquid is also drawn up into the 
evaporator side of the two-phase exchanger by the lower 
pressure of the suction side of the gas 
compressor. As LXe flows from the bottom of the weir 
reservoir into the bottom of the heat exchanger assembly, 
the LXe undergoes a phase change (vaporization) into GXe 
on the evaporator side while GXe from the getter changes 
to LXe in the condenser side of the assembly. The mass 
flow rates are the same for each circuit of the heat 
exchange assembly, during steady state operation. The 
condensing side operates at a higher pressure then the 
evaporator side, and because the heat of vaporization 
is higher at higher pressure, any excess energy released 
during the phase change evaporation process  can be 
used to sub-cool the LXe after condensation  and before 
entering the sub-cooler/phase separator. 

Both the single-phase and two-phase heat exchangers will be 
commercial plate heat exchangers which have a high heat 
transfer efficiency relative to their physical footprint.  
Because the single-phase and two-phase heat exchangers 
are of similar style, they are combined into a single 
heat exchanger package which houses both the gas to gas 
and phase change heat transfer.  As discussed in 
Section~\ref{XCSSs:SystemTest}, data from the LZ System Test 
are being used to ensure that the system conditions that 
affect heat exchanger design are understood.  These system 
conditions include the relative liquid heights within the 
system and the inlet and outlet temperatures and pressures.  

Any additional heat exchange assembly inefficiencies and 
thermal loads on the LXe are removed before the liquid 
enters the detector in the subcooler, the last major 
component of the LXe tower. The subcooler, or phase 
separator, is made with perforated copper plates 
attached to a central copper rod with thermosyphon 
heads at top or bottom. Copper mesh or similar will 
be placed between the copper plates (depending upon 
analysis) to ensure a more than adequate heat transfer 
area. During normal detector operation 
the subcooler serves two purposes.  Any xenon vapor 
not condensed by the two phase heat exchanger can 
separate from the liquid flow in the ullage of the 
subcooler and be condensed, and the liquid itself can 
be sub-cooled before it enters the detector cryostat 
as required.  The refrigeration of the cryostat can be 
optimized by adjusting the power of the subcooler 
thermosyphons and the detector cryostat thermosyphons. 
Using the nominal detector 
vapor pressure of \num{1.8}~bara, the bottom of the detector has 
a pressure of \num{2.25}~bara, translating to a saturation 
temperature of \SI{180.3}{\K}.  A thermosyphon is able to reduce 
the liquid xenon temperature close to the freezing point 
\SI{161.5}{\K}. The temperature of the sub-cooled xenon can 
therefore be adjusted to make up any heat loads on the TPC space. 
The skin volume is heated from 
radiation heat through the vessel wall, detector support 
structure and PMT heat loads. Sub-cooling of the xenon returning 
to the skin will therefore need to
account for the increased heat loads in the skin region.

The subcooler also plays an important role during cooldown of the 
detector. During cooldown operations, xenon gas will be circulated 
through the system and the subcooler thermosyphons and their 
copper fins will gradually cool the circulating gas.  Once the 
detector has reached base temperature, the subcooler thermosyphons 
liquefy xenon gas to fill the detector.

\tdrsubsec[LXeTransferLines]{LXe transfer lines}

\begin{figure}[htbp]
\centering
\includegraphics[height=3.5in]{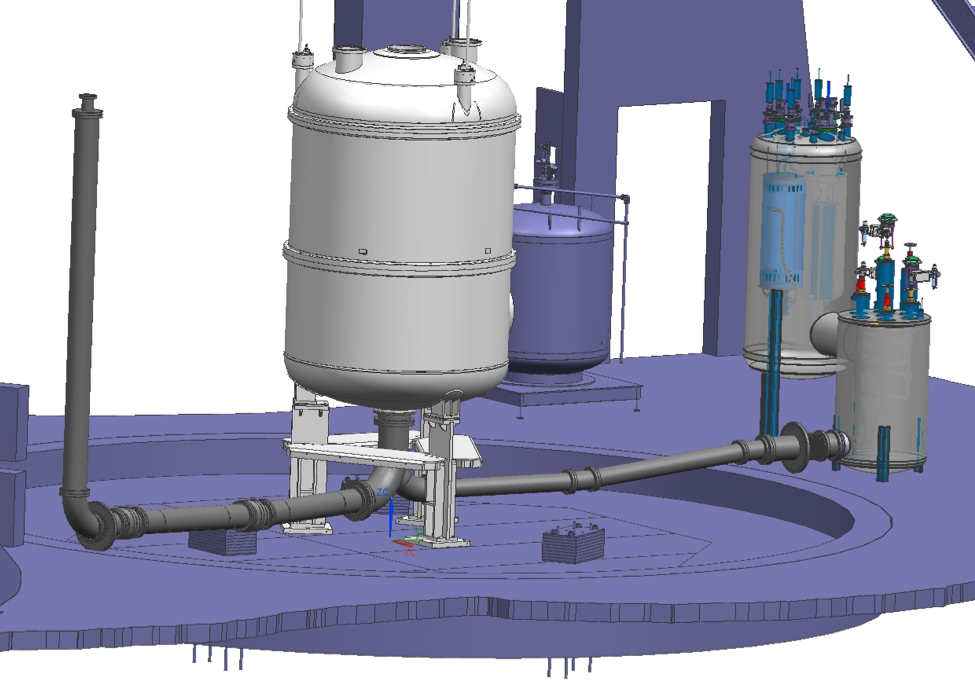}
\tdrfcaption[LXeTowerWithTransferLines]{LXe Tower and LXe 
transfer lines}{The TPC cryostat (center), 
the PMT cable standpipe (left), the LXe Tower (right), and
the two vacuum insulation assemblies that house the LXe
transfer lines.}
\end{figure} 

As shown in Figure~\ref{XCSf:LXeTowerWithTransferLines}, two
vacuum insulated assemblies host cryogenic lines that 
service liquid xenon to and from the cryostat at its
lower flange. These two assemblies meet at a center Tee 
directly underneath the cryostat. Both are 
composed of sections of 8 inch tube and are assembled 
inside the water tank and sealed to its wall with flanged 
connections on opposite sides of the tank. 

One of these assemblies, 
on the right side of Figure~\ref{XCSf:LXeTowerWithTransferLines},
houses the LXe supply and return for the primary 
circulation path of the online purification
system. This assembly hosts two lines that route purified
LXe from the subcooler to the skin and TPC active region, 
and a third line that connects the weir 
drain lines to the weir reservoir.  These three lines are individually 
insulated with multilayer insulation because they can be at 
different temperatures.  Connections will need to be made in 
these lines during assembly.  While it would be ideal to make 
welded connections with an orbital welder, there may need to 
be VCR fittings to allow these connections to be made due to 
space constraints.  A vacuum break is planned just after the 
center Tee below the cryostat.  This helps provide 
protection in case of a leak 
and would also allow softening the vacuum in the transfer line 
region to add heat during Xe recovery.  The weir drain line 
will be actively cooled by a conductive connection to a 
thermosyphon.  This will sub-cool the returning liquid below 
saturation temperature to decrease the possibility of vapor lock.  

The second assembly, seen on the left side of 
Figure~\ref{XCSf:LXeTowerWithTransferLines},
houses the PMT and sensor cables from the TPC bottom PMT 
array. These cables exit the bottom of the detector cryostat, 
and are housed in a 4 inch ID bellows reinforced with a braid.  
The flange at the bottom of the inner cryostat has a connection 
for both the bellows housing the cables and the LXe lines.  
The cables are pulled into the bellows in the reduced radon 
clean room of the surface assembly lab and the bellows is 
sealed for transport underground to keep the cable clean.  
During lowering of the inner cryostat into the outer cryostat 
the lower cable bellows is pulled through the tee in the 
center bottom of the water tank.  The shape of the Tee is 
designed to help guide the cable bellows.   The multilayer insulation 
for this line is wrapped around segments of 180 mm OD 
polyethylene tube that is supported from the inside of the 
8 inch SS tube with plastic thermal standoffs. 

The final section of the cable path is the vertical PMT standpipe 
that is located outside the water tank and 
connects to the bottom of the breakout box. 
The standpipe is shown on the left side of 
Figure~\ref{XCSf:LXeTowerWithTransferLines}. The liquid 
level in the PMT stand pipe is about the same as the liquid 
level in the detector.  The liquid is cooled with evaporation 
by pumping xenon out of the gas space above the liquid surface.  
The xenon gas between the liquid and the bottom of the breakout 
box has a temperature transition to room temperature. 

As noted in Section~\ref{XCSSs:RnRem}, the upper and 
lower PMT conduits and the high voltage conduit will 
extract a total of \SI{2.0}{\slpm} of 
xenon gas through the PMT cable 
breakout boxes. This ensures that flow with the higher 
level of contamination from the plastic is always out 
of the detector where it is cleaned in a radon trap 
before it is returned to compressor suction. Both the 
liquid and vapor PMT cable runs will have a flow 
restrictor as close as possible to the outer flange of 
the cryostat. The restrictor is designed such that the 
Peclet Number (advection/diffusion) \num{>>1}. For the 
liquid line this means a collar around the cables for 
several inches reducing the size of  the big gaps.

\tdrsubsec[SystemTest]{SLAC System Test 
prototype of the circulation architecture}

Previous experience with the LUX cryogenics and its 
xenon handling system is only partially applicable to 
LZ due to the increased complexity of the LZ design. 
To exercise the new architecture, a large System Test platform 
containing over 100 kg of LXe has 
been developed at SLAC, including 
an extensive xenon handling capacity to purify 
LXe to realistic levels. As shown in 
Figure~\ref{XCSf:SLACXeTower}, the platform 
emulates most of the critical design features of the LZ architecture.
It implements a separated LXe tower, containing a weir reservoir, 
a subcooler, and two-phase and single-phase 
heat exchangers, and it is connected to the TPC by 
vacuum insulated LXe transfer lines. These elements perform the 
same functions that their counterparts will in LZ, 
although their designs are not identical. 
Purified liquid is delivered to the bottom
of the TPC and is recovered 
from the top via a weir spillover. That liquid is 
routed to the LXe tower by draining 
into a vertical pipe that penetrates the inner cryostat
and descends to the bottom of the TPC vessel 
in the insulating vacuum space. 

\begin{figure}[htbp]
\centering
\includegraphics[height=2.8in]{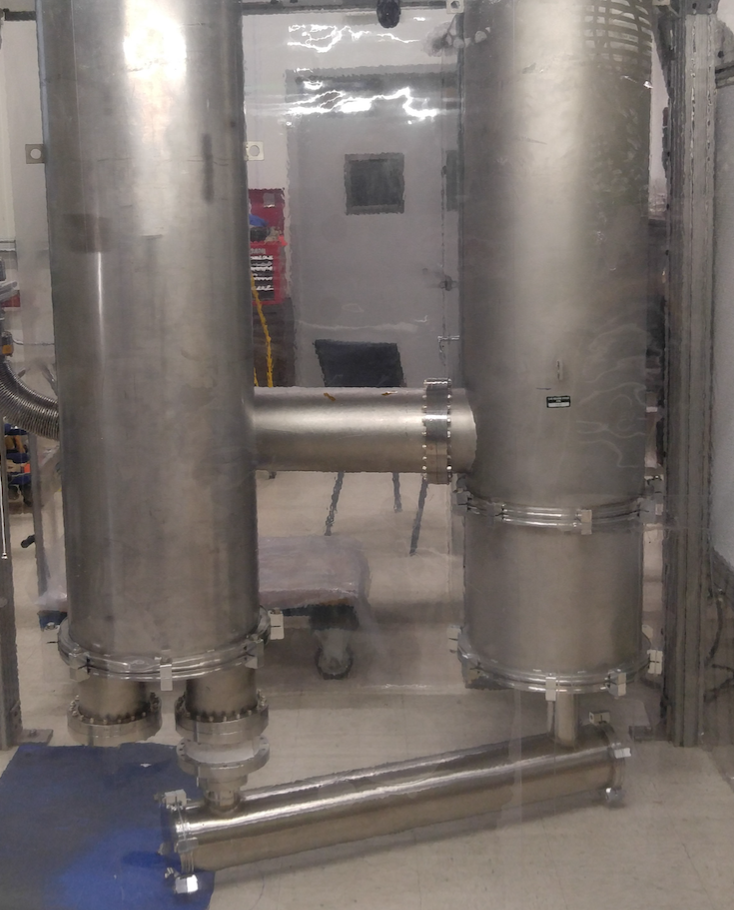}
\includegraphics[height=3.1in]{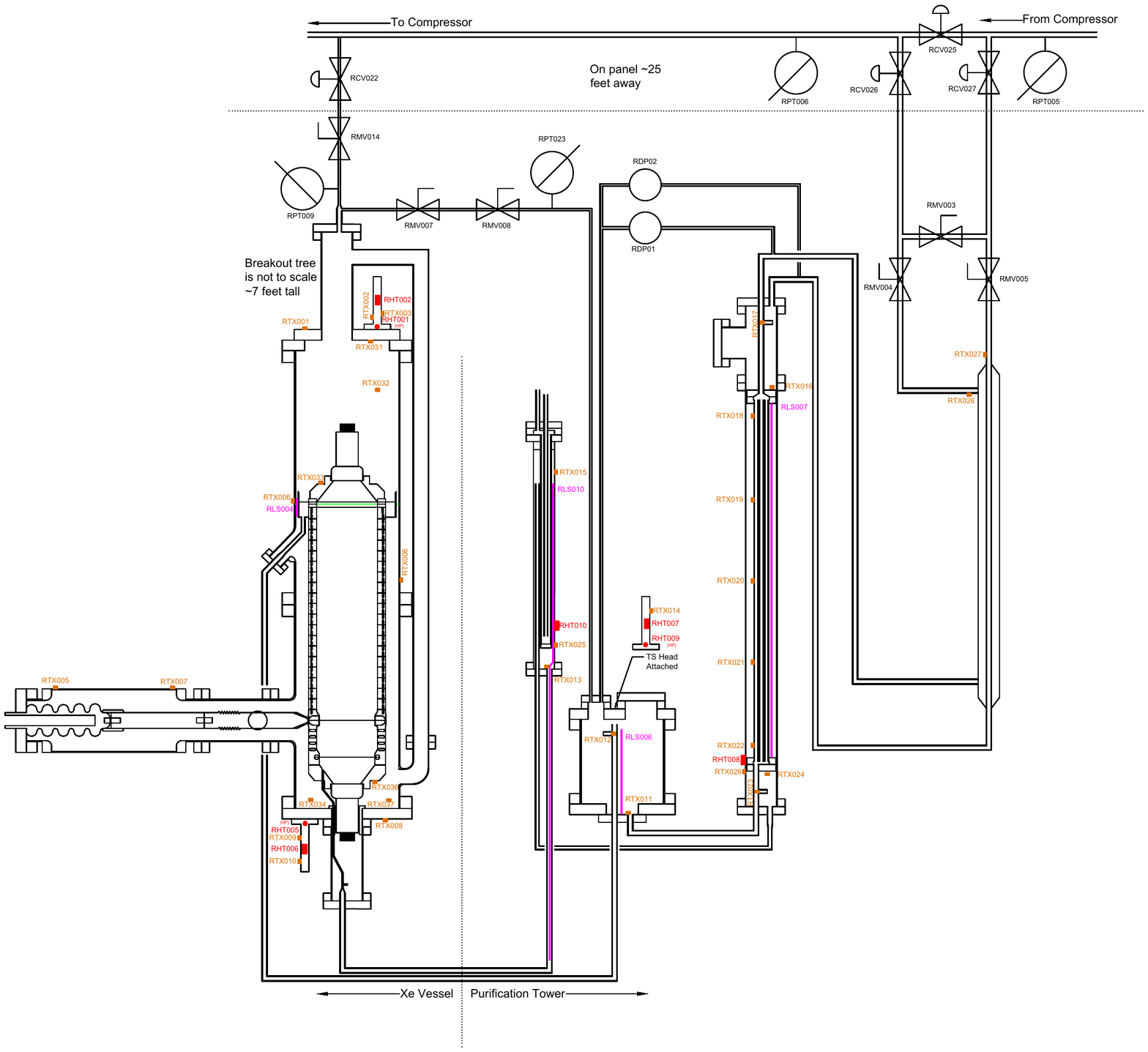}
\tdrfcaption[SLACXeTower]{System Test LXe Tower}{Left:
System Test LXe tower vessel (left vertical vessel) and
TPC cryostat (right vertical vessel), connected by vacuum insulated
LXe transfer lines housed inside the lower horizontal vessel. 
Right: Schematic diagram of the internal elements. At left is the 
TPC, serviced by a large horizontal HV feedthru, and
instrumented with one PMT on top and on bottom. The small vessel 
in the lower middle is the weir reservoir. It is surrounded
by the sub-cooler and the two-phase and single-phase heat
exchangers. The vacuum insulated transfer lines are below.}
\end{figure} 

Several test runs of the circulation system have been
carried out, and after 
some exploration of the thermodynamic parameter 
space and some hardware fixes, uniform
flow at a rate of up to \SI{75}{\slpm} has
been achieved for more than 80 consecutive hours. This provides 
a proof of principle that the LZ circulation architecture is viable.
Under the conditions of uniform flow, the liquid level in the 
TPC is set by the weir spillover as desired, and 
the reservoir and sub-coolers are filled with 
a constant (but flowing) volume of buffer liquid.
Data from these circulation tests provide a 
benchmark for engineering models of the two-phase 
heat exchangers in both the System Test and LZ. 

The System Test has also highlighted some 
challenges presented by the LZ design, particularly 
the vacuum insulated weir drain line. We have found
that, under some conditions, the flow of the 
saturated LXe in this line can be interrupted.
Both the geometry of the line and
the heat leak into it may be responsible for this 
behavior. We are exploring the addition of active cooling
to the line to counter this effect, 
so that uniform circulation can be 
stably achieved under a variety of pressure 
and temperature conditions. 

\tdrsubsec[RnRem]{Radon removal}

Material screening is the primary route to control the 
ER and NR backgrounds resulting from radioactivity in 
the experiment. Recent estimates of radon emanation from 
PMT HV cables, cable feedthroughs and cable conduits, 
shown in Table~\ref{XCSt:Rn-bkgd}, however, suggest 
that \IRnttt backgrounds may be up to \SI{37}{\milli\becquerel}, 
which is substantially above required levels. It is therefore 
required to mitigate radon with a combination of cable 
cladding and radon removal for the xenon gas circulation 
system. Neither action alone is sufficient. Adding 
\SI{150}{\mum} FEP cladding to the braided PMT HV cables 
will greatly reduce the risk of dust or other surface 
contamination in the steel braiding of the cables and 
essentially eliminate emanation from the cables in LXe, 
but will have no effect on PMT feedthroughs. Adding a 
radon trap with a volumetric gas flow of $\phi$=\SI{2}{\slpm}
in the (warm) gas circulation system will 
eliminate \SI{90}{\percent} of radon from cables and feedthrough, 
but will have no effect on the PMT HV cables immersed 
in LXe. 

\begin{table}[ht]
\tdrtcaption[Rn-bkgd]{Radon emanation estimates}{Radon 
emanation estimates for PMT HV cables, cable conduits and 
cable feedthroughs. Also shown are estimates for 
including only cladding or a charcoal trap, as well 
as combining both mitigating methods. With only initial 
measurements of PMT feedthroughs available at this point,
estimates for the other components are based 
on similar items of materials from other experiments. 
All estimates are in \si{\milli\becquerel}.}
\centering
\sffamily
\setlength{\extrarowheight}{3pt}
{\small
  \begin{tabular}{|lr|c|c|c|c|}
  \hline
\rowcolor{mrocol}
& &   \multicolumn{4}{c}{\bfseries Radon Emanation}  \\
   {\bfseries  Item} & {\bfseries Component} & {\bfseries estimated} & 
   {\bfseries w/ cladding} &{\bfseries w/ trap} & {\bfseries w/ cladding \& trap} \\
        & &  &{\bfseries only} &{\bfseries  only} & {\bfseries combined} \\
    \hline
    PMT HV Cables & warm insulation & 0.27  & 0.41  &	0.027 & 0.04 \\
                  & warm braiding	& 1.88  & 0.47  &	0.19  & 0.05 \\
                  & warm dust	    & 13    & 0     &	1.30  & 0 \\
                  & cold insulation & \num{2.7e-4}&\num{4.1e-4}&\num{2.7e-4}&\num{4.1e-4}\\
                  & cold braiding	& 1.88  & 0     &	1.88  & 0 \\
                  & cold dust	    & 13    & 0     &	13    & 0 \\
    \hline
    PMT HV Cables & Subtotal        & 30.0  & 0.88  &  16.4   & 0.09  \\
    Cabling Conduits &  warm \& cold & 0.1   & 0.1   &  0.055 & 0.055 \\
    PMT Feedthroughs & warm        & 7.3  & 7.3  &  0.73 &  0.73 \\
    \hline
    &\textbf{Total}       & {\bfseries37.4}    &{\bfseries 8.3} &  
    {\bfseries17.2} & {\bfseries0.87} \\
    \hline
  \end{tabular}
}
\end{table}

Removal of radon from xenon is challenging due to the similar 
atomic diameters of these two species (\SI{120}{\pm} vs. \SI{108}{\pm}, 
respectively~\cite{Clementi1967}). However a pioneering 
demonstration of such removal at $\phi$=\SI{1}{\slpm} has been 
performed by the XMASS
collaboration using a cooled column of activated 
charcoal~\cite{Abe201250}. This result has been reproduced 
in Ref.~\cite{getterrn} at $\phi$=\SI{0.25}{\slpm}. 

To determine feasibility of radon removal in xenon gas 
for LZ, a series of elution curve measurements were 
performed to determine the breakthrough time, $\tau$, 
(i.e. adsorption coefficient) of radon in xenon carrier 
gas at various temperatures with a \SI{30}{\g} activated charcoal 
trap, since studies~\cite{Munakata:1999yy} of \IRnttt 
adsorption in gases (other than air~\cite{Fukuda:2002uc}) 
are scarce. Breakthrough time is given by $\tau = \frac{m \cdot k_a}{\phi}$, 
where $k_a$ is the dynamic adsorption coefficient, $m$ the mass
of charcoal, and $\phi$ the volumetric gas flow. 

\begin{figure}[htbp]
  \centering
  \includegraphics[height=2.1in]{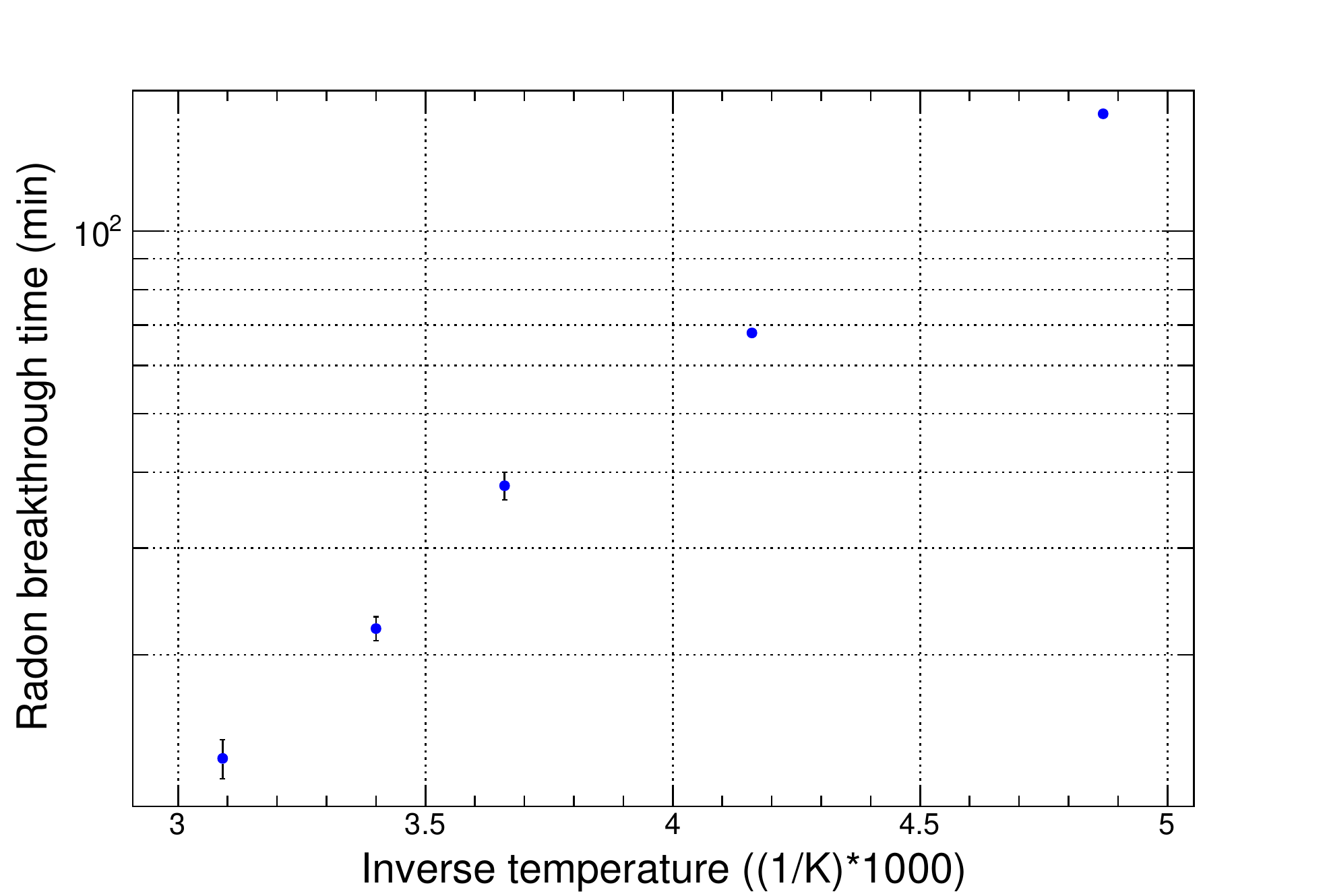}
  \includegraphics[height=2.1in,angle=0,origin=c]{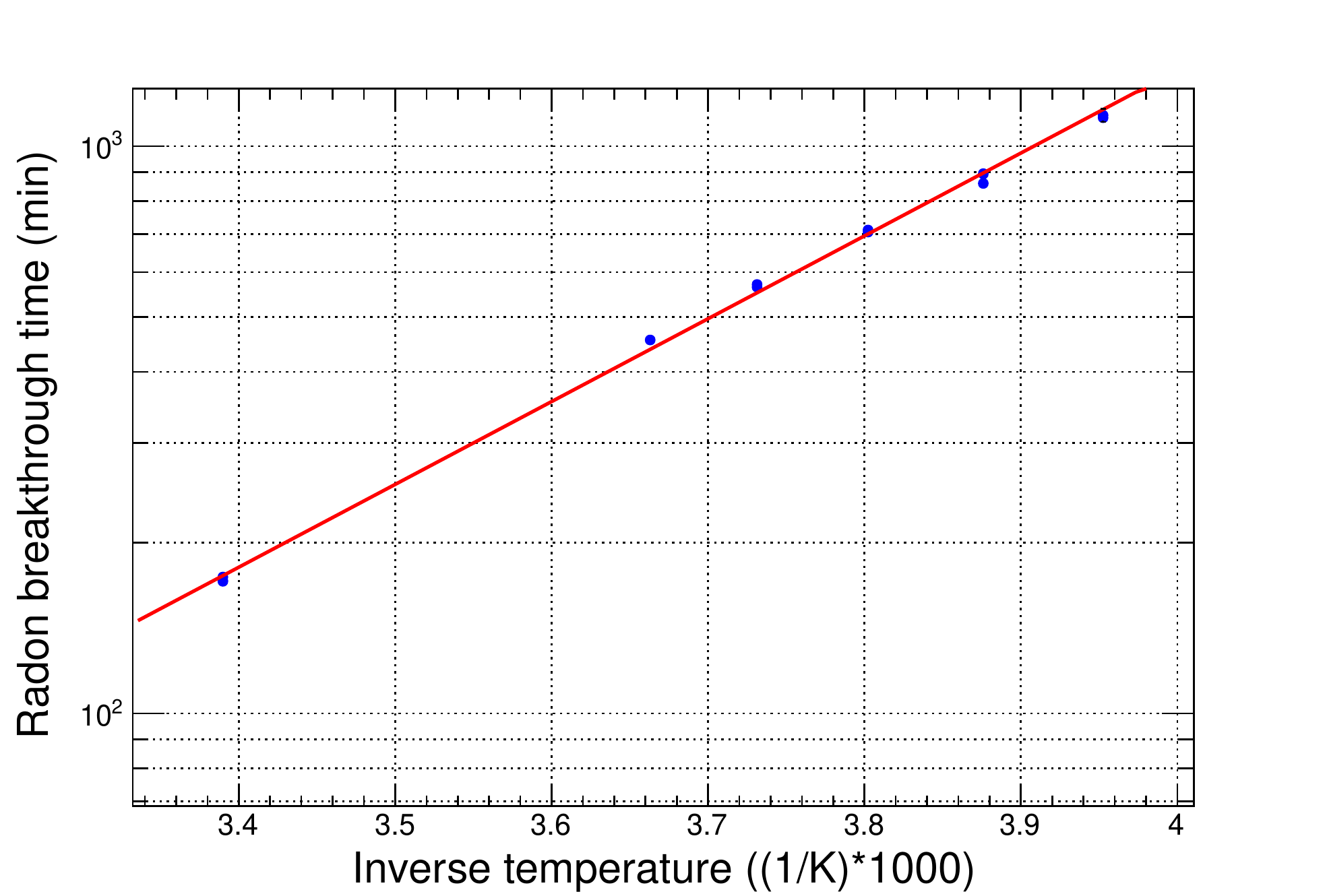}
  \tdrfcaption[temp-dep]{Breakthrough time}{ Radon breakthrough 
  time versus inverse temperature for two different configurations. 
  Left panel:  Breakthrough time in a \SI{30}{\g} charcoal trap with a 
  Xe carrier gas at a flow rate of \SI{0.5}{\slpm}. Right panel: 
  Breakthrough time in a \SI{50}{\g} charcoal trap with nitrogen 
  carrier gas at a flow rate of \SI{2.0}{\slpm}. The red line is 
  a fit of the data to the Arrhenius equation 
  $r=A\exp\left(\frac{E_a}{RT}\right)$, with $r$ the
  inverse breakthrough time, 
  $A$ the collision frequency factor, $T$ the 
  temperature, $E_a$ the activation energy, 
  and $R$ the universal gas constant. The fit values are 
  $A= 0.00187\pm 0.00044$~min$^{-1}$ and $E_a/R = 3375\pm 65$~(K). 
  Note that the temperatures in the traps are known to 
  about $\pm$\SI{1}{\K}.}
\end{figure}

Figure~\ref{XCSf:temp-dep} 
(left panel) shows breakthrough time versus inverse temperature of radon in 
xenon carrier gas with a 30 g activated charcoal trap. Although a
linear relationship is expected over the entire temperature range
in the semi-logarithmic representation shown in
Figure~\ref{XCSf:temp-dep}, the deviation from
linearity may reflect Xe saturation on the charcoal surface. 

Based on the elution curve measurements, a \SI{90}{\percent}-efficient 
radon removal element for the xenon gas circulation system 
was designed as shown in Figure~\ref{XCSf:UM-Rn-removal}. The 
size of the charcoal trap was determined to be \SI{8.6}{\kg} 
(i.e. \SI{19}{liters}) by scaling up the elution curve measurements 
with the small trap ($m$=\SI{30}{\g}, $\tau$=\SI{156}{min}, 
$\phi$=\SI{0.5}{\slpm} at \SI{-68}{\degree C}) to $\tau$=\SI{12.7}{days} 
(i.e. \num{3.3} half-lives), $\phi$=\SI{2}{\slpm} and \SI{-85}{\degree C}. 
This can be accomplished with two \SI{1}{\m}-long columns of \SI{3}{\inchl} 
diameter connected in series.

\begin{figure}[htb]
 \centering
 \includegraphics[height=2.75in,angle=0,origin=c]{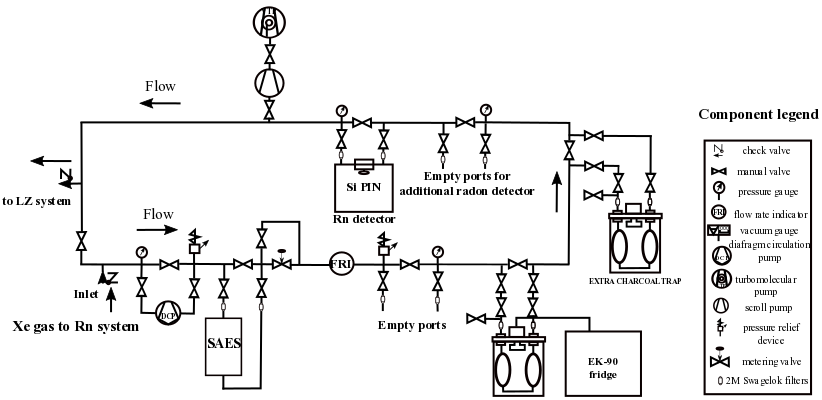}
 \tdrfcaption[UM-Rn-removal]{LZ radon removal system}{Schematic 
 diagram of the radon removal system for LZ.}
\end{figure}

Gaseous xenon from the purge-gas line conduits will be 
introduced into the radon removal system through a SAES 
high temperature getter before it passes through the 
charcoal trap. The charcoal trap will be immersed in a 
cryostat vessel that is filled with HFE-7100 
coolant and kept at a temperature of \SI{-85}{\degree C} using 
a EK-90 refrigerator or a thermosyphon. 
The system will allow 
transmission of xenon gas with flow rates ranging 
from \SIrange{0}{2}{\slpm} using a diaphragm circulation 
pump.  

To measure the amount of radon in the carrier gases, 
a Si PIN photodiode (ORTEC) made from low radioactivity 
components will be used. The electronics readout will 
consist of a 142AH ORTEC preamplifier with low noise 
and fast timing characteristics as well as little 
temperature dependence, and a shaping preamplifier. 
A multichannel analyzer (ORTEC MCA-2K) will process 
the signals from the preamplifier. The radon system 
will be fully automated using high pressure pneumatic 
valves, and pressure and temperature sensors that are 
integrated into the LZ slow control system.

To validate the design and perform studies with 
different trap geometries using various adsorbents 
at various flow rates, a separate full-scale prototype 
system has been designed and built at the University of 
Michigan that makes use of the final cryogenics system. 
Initial measurements with the prototype system are underway. 
The right panel of Figure~\ref{XCSf:temp-dep} shows the breakthrough 
time of radon in nitrogen carrier gas versus inverse 
temperature in a \SI{50}{\g} charcoal trap. The system 
will soon be operated with argon or helium gas before 
expensive xenon gas is used as carrier gas. It contains 
a radon calibration source to study the adsorption 
efficiency of the traps with different adsorbents 
besides natural carbon and different amounts of the 
adsorption materials. The test system can be used to 
study the adsorption efficiency of radon traps in a 
broad range of temperatures and trap geometries. The 
traps can be swapped and regenerated by baking and 
pumping at high temperatures.

\tdrsubsec[SlowControl]{Slow controls}

A set of industrial Programmable Logic Controllers 
(PLCs) will handle instrument read-out and control for 
the xenon purification and recovery systems, as well as 
the cryogenic and systems and TPC high voltages (excluding PMTs).  The PLC 
logic will include interlocks protecting the integrity of 
the xenon system, control loops and automated procedures 
for the various subsystems, and automated emergency response.  
All PLC programming will follow IEC 61131-3 standards.

\begin{figure}
\centering
\includegraphics[width=6in]{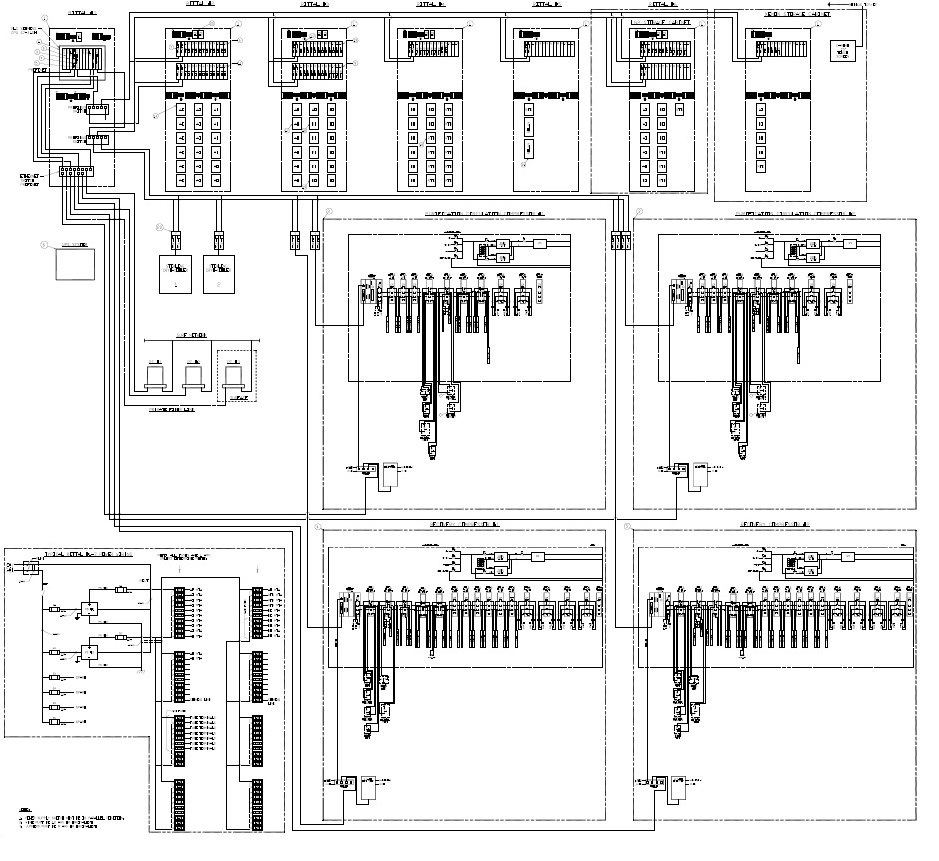}
\tdrfcaption[PLCarchitecture]{PLC Architecture}{Block 
diagram for the PLC System.}
\end{figure}

The overall PLC system architecture includes one top-level 
PLC plus smaller dedicated PLC's for large equipment items, 
namely the cryogenerator system and four xenon compressors 
(2x recovery and 2x circulation), see Figure~\ref{XCSf:PLCarchitecture}.  
The top-level controller is a Siemens S7-410-H Redundant 
Hot Backup PLC, which includes redundant processors with 
independent links to I/O modules.  The redundant processors 
guarantee bumpless switchover in the case of processor 
failure, and also allow for zero-downtime when updating 
PLC logic and I/O mapping.  The Siemens S7 PLC system will 
have a dedicated internally redundant APC Symmetra UPS as 
well as parallel DC power units for the S7 cpu's and I/O 
power, ensuring uninterrupted control when switching to 
generator power.  

All instruments not directly 
related to compressor or cryogenerator control will be 
read-out/controlled-by the S7 PLC.  This amounts to an 
estimated 832~channels of I/O, including 144 thermometers, 
192 other Analog Inputs, 80 Analog Outputs, 80 Discrete 
Inputs, and 336 Discrete Outputs. The S7 logic will handle 
all interlocks and control loops relating to these 
instruments, in addition to
top-level decision making for the compressor and cryogenerator 
systems. The S7 PLC will also be responsible for accomplishing 
xenon recovery in the event of a prolonged power outage 
(see Section~\ref{XCSS:XeRec}).  The S7 PLC will connect to the 
LZ Slow Control system via an Ignition Server on the PLC 
private network (see Section~\ref{EDCS:SC}).  The PLC network 
will receive power from the PLC UPS system and will have a 
direct, redundant fiber link to a backup server in the surface 
facility, ensuring monitoring and control capability in the 
event of power and network failures (see Section~\ref{EDCS:Network}).

Each compressor skid will have a dedicated Beckhoff CX8031 PLC 
that will handle interlocks protecting that compressor, 
automation of compressor start-up and shut-down, and control 
loops to regulate flow rate and pressure at the compressor inlet.  
The Beckhoff PLCs on the recovery compressors will have the 
additional ability to independently initiate emergency xenon 
recovery if the TPC pressure exceeds a set threshold 
(See~\ref{XCSS:XeRec}). 
All variables in the Beckhoff PLC logic will be exposed to 
the Siemens S7 via Profibus, and in normal operations the 
Beckhoff PLCs will receive high-level commands 
(start/load/unload/stop) from the S7 over the Profibus 
connection.  In this way the S7 also serves as the gateway 
between the compressor PLCs and the LZ Slow Control system.  
A touch-screen human-machine-interface (HMI) on 
each skid will allow each compressor 
to be operated as an independent unit when disconnected 
Siemens S7.  

The liquid nitrogen generation system will have a 
vendor-supplied Siemens S7-300 Series PLC with local 
HMI to control the Stirling cryogenerators.  As with 
the compressor PLC's, all variables in the PLC logic will be 
exposed to the main Siemens PLC via Profibus, providing 
the same monitoring and controls as are available through 
the vendor-supplied HMI.

\tdrsec[XeRec]{Xenon Recovery}

The Xe recovery system removes the Xe from the detector 
and returns it to the storage cylinder packs.  
There is also a regulator-based Xe delivery function in 
the opposing direction to support detector cooldown and condensing.

\begin{figure}[htbp]
\centering
\includegraphics[height=5in]{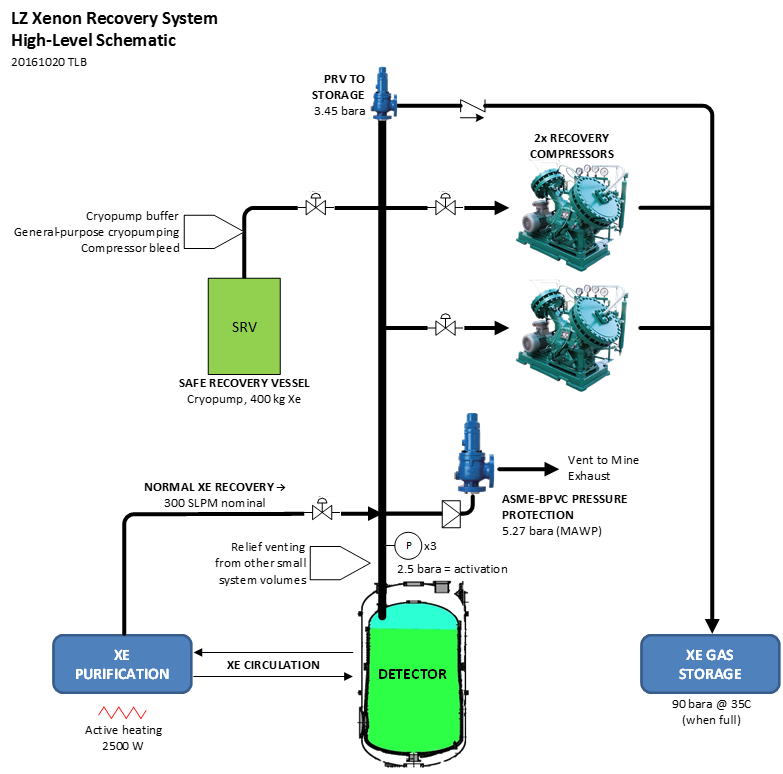}
\tdrfcaption[RecoverySchematicSimple]{Simplified flow 
schematic of LZ Xe Recovery System}
{Simplified flow schematic of LZ Xe Recovery system.}
\end{figure} 

The xenon recovery system must support three functions: 
1) Normal recovery, such as a planned full recovery at 
the end of the experiment or for small operational adjustments 
of xenon quantity in the system, 2) Emergency recovery, 
where the xenon is automatically removed from the system 
and safely stored during an unexpected sustained pressure 
rise or extended power failure, and 3) Venting xenon as 
a last resort to provide detector MAWP\footnote{Max 
Allowable Working pressure} pressure protection. With 
the exception of function 3, in all scenarios the xenon 
asset must be retained and its purity kept intact.  The 
system also offers online cryopumping and cryogenic 
buffer volume that can absorb small pressure spikes, 
xenon feed and bleed to various points throughout the 
system, and redistribution of trace xenon.

At the heart of the recovery system is a pair of recovery 
compressor skids. The skids are redundant replicates, 
each one containing all the hardware necessary for 
autonomous recovery, including valves, sensors and 
interlocks, Beckhoff PLC controller, and the compressor 
itself. Both compressors are commanded to run when 
activated in an emergency.  Compressor suction is 
connected directly to the detector ullage, isolated 
by a pneumatic shutoff valve to allow for independent 
compressor startup routines. A discharge-to-suction 
proportional bypass valve provides variable flow, 
using compressor suction pressure as feedback.  Once 
operating, the compressor is controlled to maintain 
constant suction pressure - this way the compressors 
only transfer the xenon being fed to them.  The 
compressors themselves are all-metal triple-diaphragm 
two-stage compressors, each rated for 350 slpm at 2.5 
bara suction pressure (automatic recovery activation 
pressure), and 100 bara discharge.  A simplified 
flow diagram of the recovery system is shown in 
Figure~\ref{XCSf:RecoverySchematicSimple}.

All pressure relief mechanisms throughout the xenon 
handling system are vented into the detector ullage, 
which acts as a gaseous buffer. Here the vented gas 
either condenses in the detector or can be stored/removed 
by the recovery system. A set of cryo vessels tap off the 
main delivery and recovery gas trunks.  The larger vessel 
is the SRV (Safe Recovery Vessel) repurposed from LUX, 
and has a xenon capacity of 400 kg and maximum pressure 
rating of 83 bara (1190 psig).  This vessel is nominally 
maintained cold at xenon ice vapor pressure, 
\SI{1.0e-3}{\milli\bar}, allowing it to act as an 
online cryopump buffer to absorb small pressure spikes 
within the detector and also bleeding of xenon from 
various points throughout the xenon handling system.  
A smaller cryopump consists of two \SI{4}{\liter} research 
bottles with a combined xenon capacity of \SI{15}{kg}, 
and can be used in conjunction with the SRV to 
redistribute trace xenon by way of volume sharing.

\begin{figure}[htbp]
\centering
\includegraphics[height=4.75in]{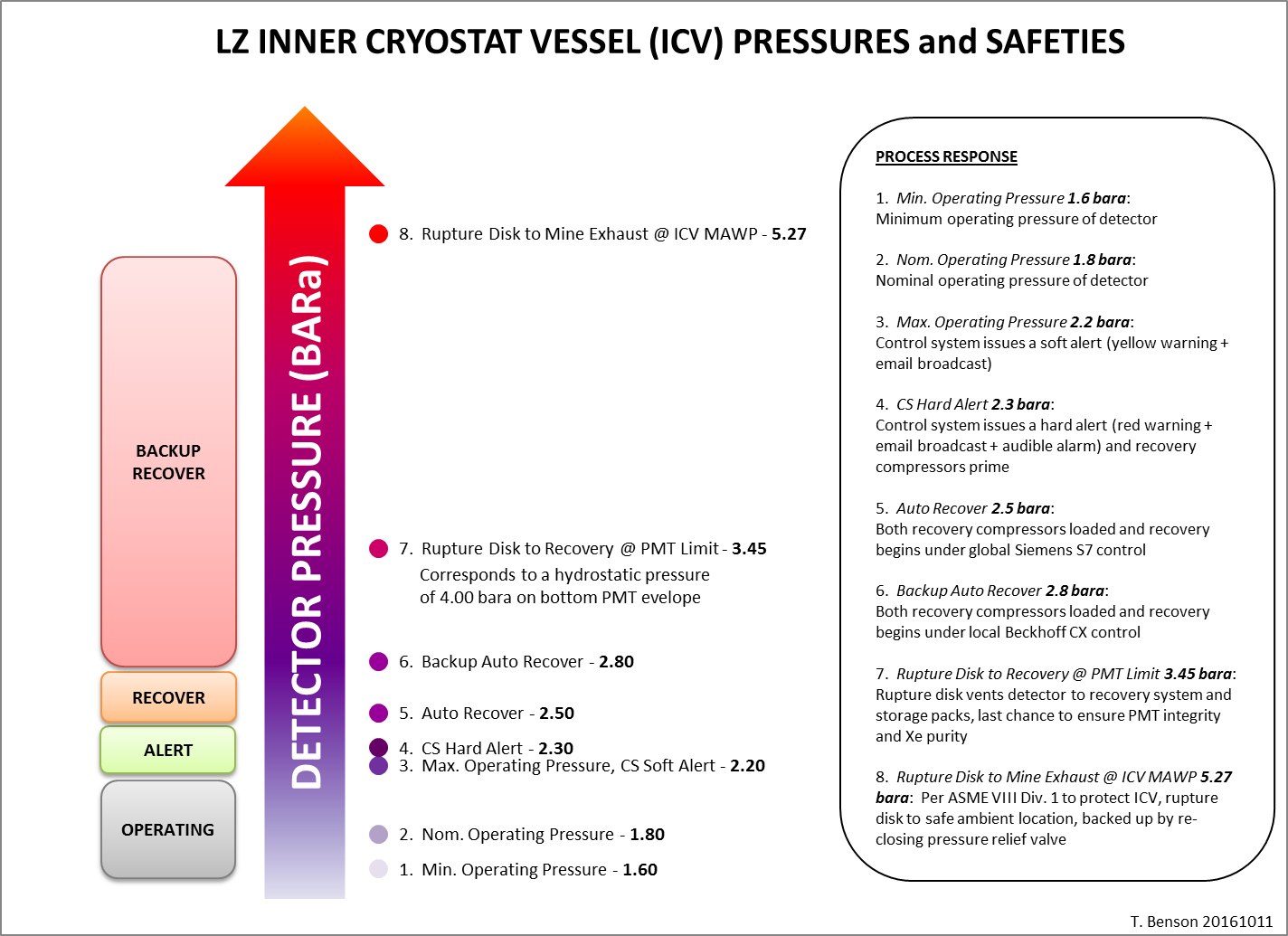}
\tdrfcaption[PressureThresholds]{Xe system pressure 
threshold plot}{Xe system pressure threshold plot.}
\end{figure}

Automatic emergency xenon recovery is a layered approach 
and illustrated in Figure~\ref{XCSf:PressureThresholds}.  
The automatic recovery trigger is keyed to the detector 
ullage pressure, 
which is monitored by three pressure transducers.  All 
three communicate to the main Siemens PLC, two of them 
via the dedicated compressor skid Beckhoff PLCs.  
Detector operating range is \numrange{1.6}
{2.2}~bara.  A set of alerts are issued from the control 
system once detector pressure reaches \num{2.2}~bara.  At 
\num{2.5}~bara, and a 2/3 vote between the three pressure 
transducers, automatic recovery is activated.  If the 
pressure continues to rise to \num{2.8}~bara, each Beckhoff 
controller assumes the main PLC has failed to initiate 
recovery and commands its own skid to begin recovering 
autonomously.  At \num{3.45}~bara the bottom PMTs are at 
risk of failure from hydrostatic pressure.  At this point 
it is assumed the compressors are not functioning, or 
not functioning fast enough, and a pressure relief valve 
vents the detector to the evacuated storage cylinders in 
hopes that this is a transient spike and not a 
continuous rise in pressure.  Finally, as a last resort 
to protect the 
cryostat against rupture, a rupture disk vents 
xenon to mine exhaust at \num{5.27}~bara (detector MAWP).  
This burst disk is backed up by a re-closing pressure 
relief valve so all remaining xenon can be retained after 
the pressure event is resolved.  In this case some 
xenon is lost, the system purity is compromised, and 
some PMTs have possibly mechanically failed.

See Figure~\ref{XCSf:RecoveryDetectorTieIn} for P\&ID 
details of how the recovery system elements and detector tie together.
The full P\&ID for the core system is shown in 
Figure~\ref{XCSf:XeRecoveryPID}.

\begin{figure}[htbp]
\centering
\includegraphics[height=4in]{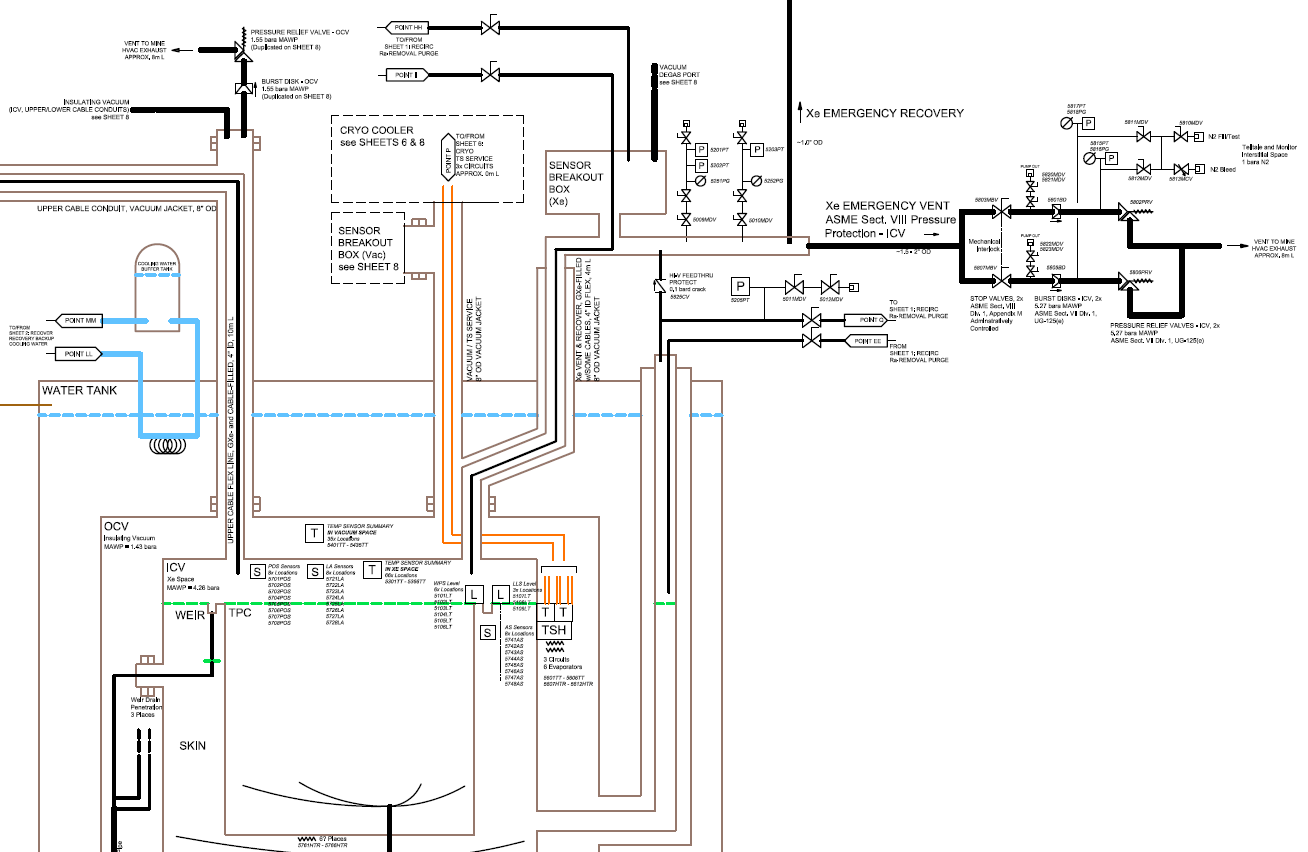}
\tdrfcaption[RecoveryDetectorTieIn]{P\&ID of Recovery/Detector 
Tie-In}{P\&ID detail of how the recovery system ties into 
the detector.  The main recovery path out of the detector 
is via a 4 inch ID bellows containing a small bundle of 
sensor cables.  This path is instrumented with redundant 
pressure transducers, and both the main recovery line and 
the ASME ICV pressure protection tap into this line.}
\end{figure}

\begin{figure}[htbp]
\centering
\includegraphics[height=8.25in]{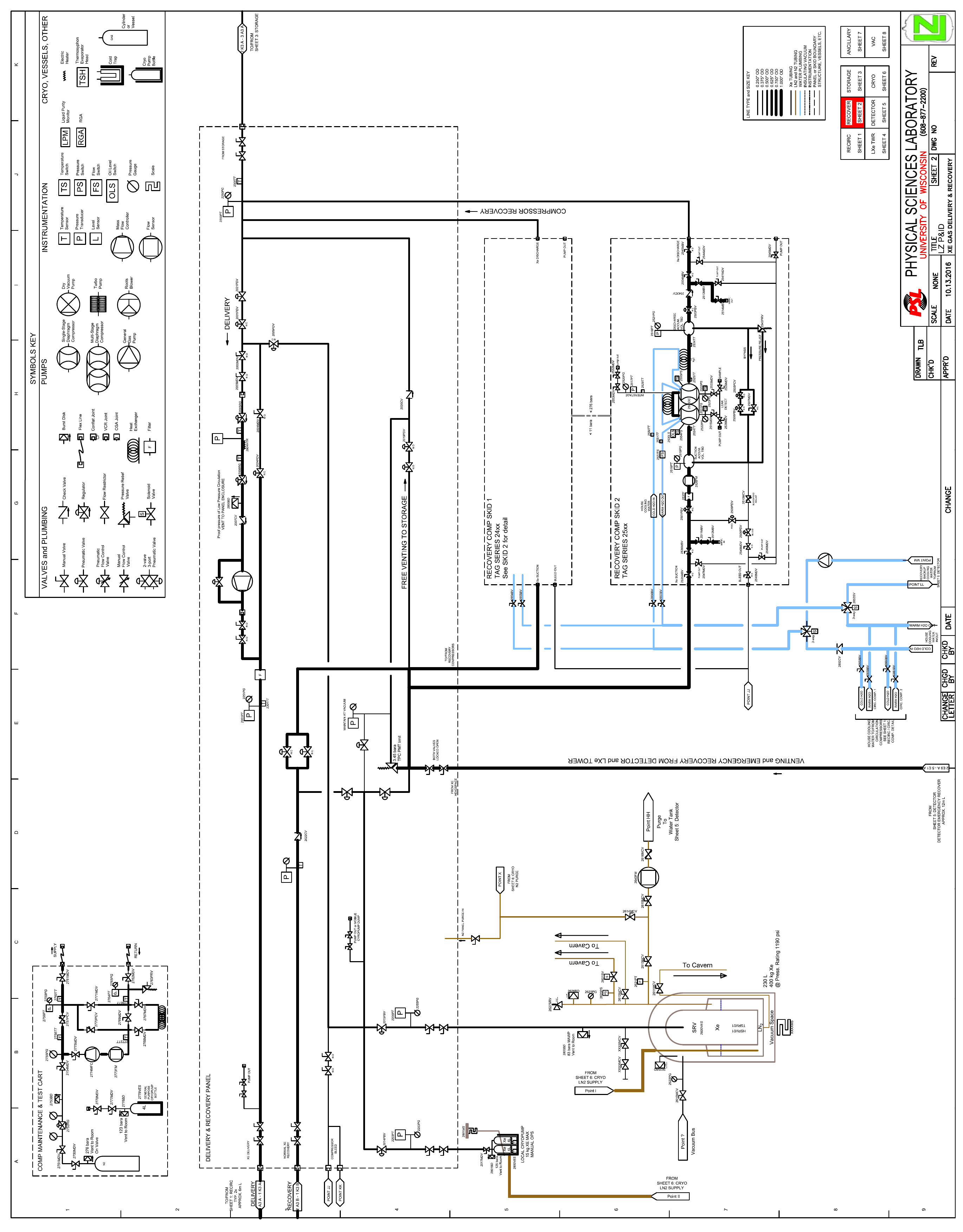}
\tdrfcaption[XeRecoveryPID]{Xe Delivery \& Recovery 
system P\&ID}{Xe Delivery \& Recovery core system P\&ID.}
\end{figure} 

An analysis of various failure scenarios drives the recovery 
system design and sizes the compressors.  Specifically, our 
worst case scenarios would be an air or water breach of the 
vacuum jacket insulating the detector and LXe transfer lines, 
an air-breach being the more probable of the two.  We first 
address these failures by specifying a layer of foam insulation 
to be applied to the outside of the inner cryostat vessel wall 
and bottom head (all vessel surfaces wetted by liquid xenon 
on the other side).  The insulation is specified to have a 
maximum heat transfer coefficient (k/t) of \num{3}~W/m2 and 
\num{1.5}~W/m2 for the wall and bottom, respectively.  
Second, we include vacuum breaks on the lower transfer lines 
to isolate such a failure and minimize the effected heat 
transfer surface area.

With the above design provisions in place, a conservative 
dynamic model was developed to estimate heat loads for the 
air and water breach cases.  From these, initial boiloff 
rates are in the range of \SI{400}{slpm} and \SI{2500}{slpm} 
for the air and water cases, respectively.  Boiloff tapers 
as the LXe level in the detector drops, and significantly 
tapers in the water case as a layer of ice forms and grows.  
Then, by looking at the detector pressure response versus 
time for various sizes of compressors, and making sure it 
always stays below safe limits, we are able to choose an 
optimal compressor size.  Note that compressor throughput 
is a strong function of suction pressure, so at higher 
suction pressure the compressors will move more gas, and 
vice versa.  On a first-order this means the compressors 
self-regulate to match boiloff, and detector pressure is a 
function of how much xenon is being removed, or in other 
words what size the compressors are.  Physical size is 
also a factor since underground space is extremely limited.  
From this analysis, one \SI{350}{slpm} compressor 
(at \num{2.5}~bara suction pressure) can handle the more 
probable air-breach case, and two of these same compressors 
running in parallel can handle the water-breach case.  
Since the analysis is conservative, these compressors 
will be slightly over-sized for the application, which 
makes it important to have variable flow capability by 
way of the proportional bypass valves on the compressor skids.

Normal recovery is driven by electric heaters that actively 
heat and boil the xenon in a controlled manner.  These 
heaters are located in the bottom of the LXe Tower on the 
LXe supply lines leading to the detector 
(Figure~\ref{XCSf:LXeTowerPID}), and are sized to 
deliver a combined \num{2500}~W to boil off xenon at a rate 
of \num{300}~slpm.  Once normal recovery is initiated, main
circulation stops and the recovery heaters begin boiling off 
xenon at the bottom of the LXe Tower.  Flow through the 
supply lines is reversed from normal operations:  LXe is 
draining out of the detector and flowing towards the 
heaters while xenon boiloff travels up through the 
subcooler and out a gas extraction line.  This extraction 
line plugs into the input of the gas recirculation loop.  
From there, the xenon can be routed directly to the 
recovery compressor inputs, or can first be pushed through 
the getter one last time before storage (this requires the 
circulation compressors to be operating to overcome the 
getter's impedance).  Again, the recovery compressors 
operate on constant suction pressure control, and as 
more gas is liberated into the gas plumbing the recovery 
compressors transfer it out.  Assuming a constant 
\num{300}~slpm removal rate, full recovery would take 
approximately 4.2 days.

There are also design provisions for a more passive 
recovery strategy.  Here, a low-elevation cold-to-warm 
liquid tap at the bottom of the LXe Tower is normally 
blocked off by redundant shutoff valves.  Upstream of 
the valves (but at lower elevation) there is a room 
temperature heat exchanger.  With this approach, LXe 
drains out of the detector and into the heat exchanger 
where it is boiled off.  The boiloff gas is routed back 
into the detector ullage, where it both recondenses onto 
the liquid surface and helps raise the bulk xenon 
temperature over time, and raises detector pressure 
as gas builds up.  The pressure buildup will eventually 
trigger the emergency recovery routine.  This backup 
approach may be helpful to supplement and/or kick-start 
a full xenon recovery if there is a need to recover faster.

Two underground generators provide backup emergency 
power to the experiment:  1) A SURF \SI{300}{\kW} 
generator, and 2) An LZ \SI{40}{\kW} generator.  
Within one minute of a power outage, the large generator 
automatically starts and sustains ventilation and critical 
experimental operations in the Davis Campus.  At this 
time, non-critical LZ loads are systematically shed 
while critical loads are maintained, including 
backup cryogenic cooling and any functions relating 
to xenon recovery.  The large generator is coupled 
with a \num{1000} gallon fuel tank and is able to sustain 
ventilation and critical experiment loads for 
approximately \num{3}~days.  During this time, a 
high-level decision can be made whether to initiate 
a full xenon recovery or wait.  If a decision is not 
made, or if outside communications have been severed, 
the recovery system will automatically initiate a 
full recovery after a predetermined amount of time 
after the power loss.  This is to ensure the xenon 
is safely stored before there is no fuel left to do so.
When a low fuel condition is detected 
on the large generator, and power has still not been 
restored, the smaller LZ generator (coupled with a 270~gallon 
fuel tank) begins warming up.  The large generator then 
runs out of fuel, and within 10-20 ms an automatic transfer 
switch transfers the load to the LZ generator.  The 
recovery system plans ahead for the switchover by briefly 
shutting down the compressors and then restarting after 
the switch is complete.  On the other hand, the control 
system is carried through all switchovers by their UPS 
units.  Assuming all recovery loads are active while 
the LZ generator is running, and that it started with 
a full 270~gallons of fuel, it will continue running 
for approximately 4 days.  The combined run time of 
both generators is 7 days, and it takes 4 days for a 
normal full recovery.

\tdrsec[XeStor]{Long-term xenon storage and transportation  }

The LZ Xe will be sourced from multiple gas suppliers 
(Section~\ref{XCSS:XeProc}), shipped to SLAC for krypton 
removal (Section~\ref{XCSS:KrRem}), shipped to SURF to be 
moved underground, and finally liquefied into the LZ 
detector. Vendor-grade Xe arrives at SLAC in vendor-supplied 
gas cylinders.  Once krypton has been removed, the Xe is 
stored and transported in special high-integrity gas cylinder 
packs specifically engineered for LZ.  The Xe storage, 
transport, and transfer must be done without any loss of 
the high-asset xenon, and must also meet or exceed the overall storage 
inbound leak rate requirement of 
\SI{1.15e-6}{\milli\bar\litre\per\second} (helium) as 
specified in Section~\ref{XCSSs:leaksStorage}.

\begin{figure}[htbp]
\centering
\includegraphics[height=4.75in]{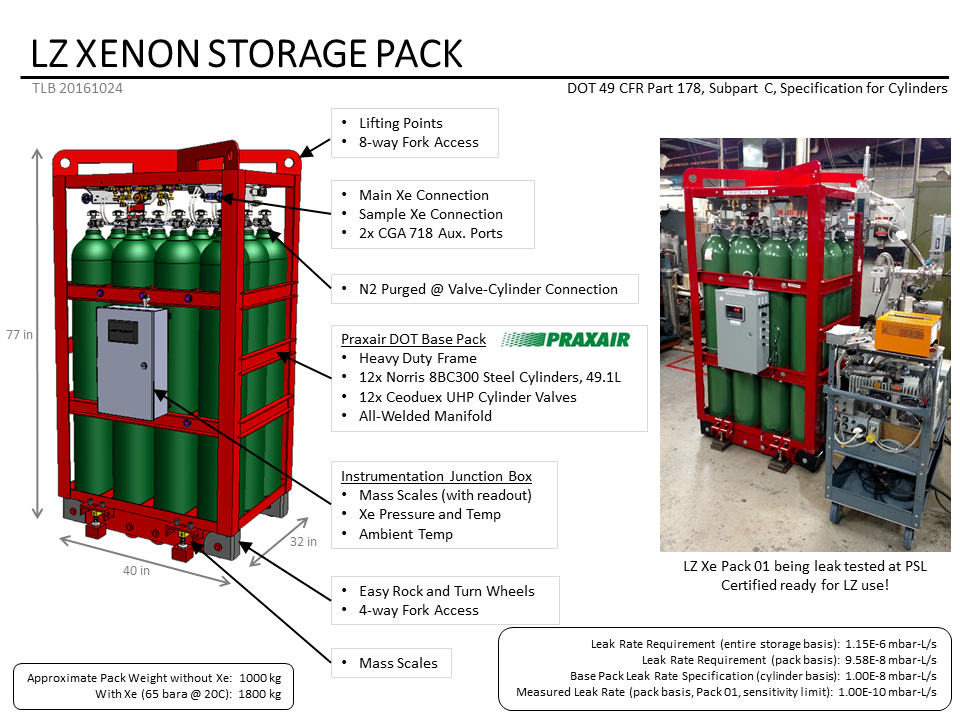}
\tdrfcaption[CylPack]{Overview of LZ cylinder gas pack}
{Overview of LZ cylinder gas pack.}
\end{figure} 

The LZ cylinder packs conform to DOT 49 CFR 
Part 178, Subpart C, Specification 
for Cylinders.  Base packs will be purchased as turnkey 
items from Praxair, with modification and final 
integration and testing occurring at the University of 
Wisconsin's Physical Sciences Laboratory (UW-PSL).  
There will be 12 gas cylinder packs available to contain 
the full \SI{10}{\tonnesl} of Xe for LZ (see Figure~\ref{XCSf:CylPack} 
for some features of the LZ cylinder pack design).  Each 
pack contains 12 DOT-3AA-2400 49.1 liter cylinders, each 
having a working pressure of \num{166}~bara (\num{2400} psig).
The full quantity of Xe can be stored in all \num{144} cylinders 
in a supercritical state at a pressure of \num{65}~bara (\num{928} psig) 
at \SI{20}{\degree C}. Each cylinder then contains approximately 
\SI{69}{\kg} (\num{12700}~standard liters) of Xe and each full pack 
weighs a total of \SI{1800}{\kg}.  The maximum allowable Xe charge 
into these cylinders for transport per DOT regulations 
(pressure must be below 5/4 working pressure at 
\SI{55}{\degree C}) is \SI{91}{\kg} (this 
would require at minimum \num{110} cylinders).  The \num{65}~bara 
target storage pressure offers a favorable packing 
density on the Xe density curve near its critical point 
(\SI{16.6}{\degree C}, \num{58}~bara) without placing excessive 
output pressure requirements on the compressors needed 
to fill the packs.

The cylinder valves are equipped with a pressure burst 
disk rated at \num{277}~bara (\num{4000}~psig) per DOT 
specification.  In the event of a fire, with \SI{69}{\kg}
of Xe charge, this pressure would be reached at a 
temperature of \SI{136}{\degree C}. If the burst disk 
failed to vent, the cylinder burst pressure of \num{398}~bara
(\num{5760} psig) would be reached at a temperature 
of \SI{202}{\degree C}.

Particular attention is given to the seal between the 
valve and the cylinder.  Economics drives us to use a 
standard NGT (National Gas Taper) thread that is 
readily available from DOT cylinder manufacturers.  
This seal can represent the weakest link in the 
system from a leak-tightness perspective.  
The requirement that the total leak rake be no 
more than \SI{1.15e-6}{\milli\bar\litre\per\second} implies
that the average inbound leak should be less than 
\SI{9.6e-8}{\milli\bar\litre\per\second} into each 
of the 12 packs, and less than 
\SI{8.0e-9}{\milli\bar\litre\per\second} 
into each of the 144 cylinders. To
achieve this we set a target inbound helium leak rate of 
\SI{1e-8}{\milli\bar\litre\per\second} for each of the cylinders, 
while we further reduce Kr ingress by flushing
the volume immediately surrounding the NGT seals
with boil-off nitrogen 
from the cryogenic system.  We believe the flushing can 
reduce the Kr ingress rate of Kr by a factor of 10 
over ambient air.  During storage we also 
benefit from a favorable high 
pressure gradient between the stored Xe and outside 
ambient conditions, however this effect is not present 
during operations when the packs are empty.  After 
integrating the base pack into the final LZ design at 
UW-PSL, a leak check is performed on the entire pack, 
including the manifold, to certify it ready for LZ use.  
For these tests, the leak checking system is first 
calibrated with a calibrated helium leak.

Considerable R\&D has been carried out in collaboration 
with Praxair's R\&D Division on achieving our target 
specification.  Initially we tested two valve/cylinder 
pairs, and have since received and tested the first 
complete LZ storage pack.  Initial factory leak rates 
range from 9.0e-8 to 4.1e-9 mbar-L/s, but then after 
pumping down for a few days to reduce helium background 
in the cylinders, measured leak rates drop dramatically.  
In particular, for the first LZ pack, Storage Pack 01, 
the helium leak rate was measured to be less than 
1.0e-10 mbar-L/s, the sensitivity limit of the 
leak test equipment.  Two valve thread preparation 
techniques were compared:  1) standard practice of
PTFE tape with Krytox applied above the 5th thread, 
and 2) thin layer of indium plating, 0.002 inch and 
0.004 inch thicknesses were both trialed.  A 
metal-metal seal must be achieved with the NGT thread 
in order to meet LZ's allowable leak rate requirement, 
and the purpose of the valve thread preparation is 
to allow full engagement and deformation of the NGT 
threads without onset of galling.  The indium seemingly 
has the advantage of flowing to fill any gaps with 
metal, however results of all 14 cylinders (2 R\&D 
cylinders + 12 cylinders in first pack) suggest both 
techniques work, and in fact the PTFE results were 
slightly better.  The most important factors for 
achieving a good seal is careful thread inspection 
and gauging prior to installation, and that 
installation be performed by an experienced 
technician.  All valves for the LZ storage packs 
will be installed by experts at Praxair, and the 
remaining valves will be prepared using the 
standard PTFE tape technique per Praxair's procedures.

We have selected Ceoduex D304 UHP tied diaphragm 
valves with CGA 718 (DISS) output connection type.  
The cylinders in a pack are manifolded together 
with an all-welded stainless steel manifold with 
VCR xenon connection taps and instrumentation 
ports for a pressure gauge and transducer.  Xe 
gas line temperature is monitored and also ambient 
temperature within each pack.  The packs are 
supported by scales that monitor xenon mass to an 
accuracy of +/- \num{0.23}~kg.  Xe temperature 
and pressure, ambient temperature, and weight of 
each pack is read out to the PLC system.  The 
pack frame is a rugged steel frame equipped 
with fork-access both on bottom and top, rigging 
pick points on top, and retractable wheels.

\begin{figure}[htbp]
\centering
\includegraphics[height=3.5in]{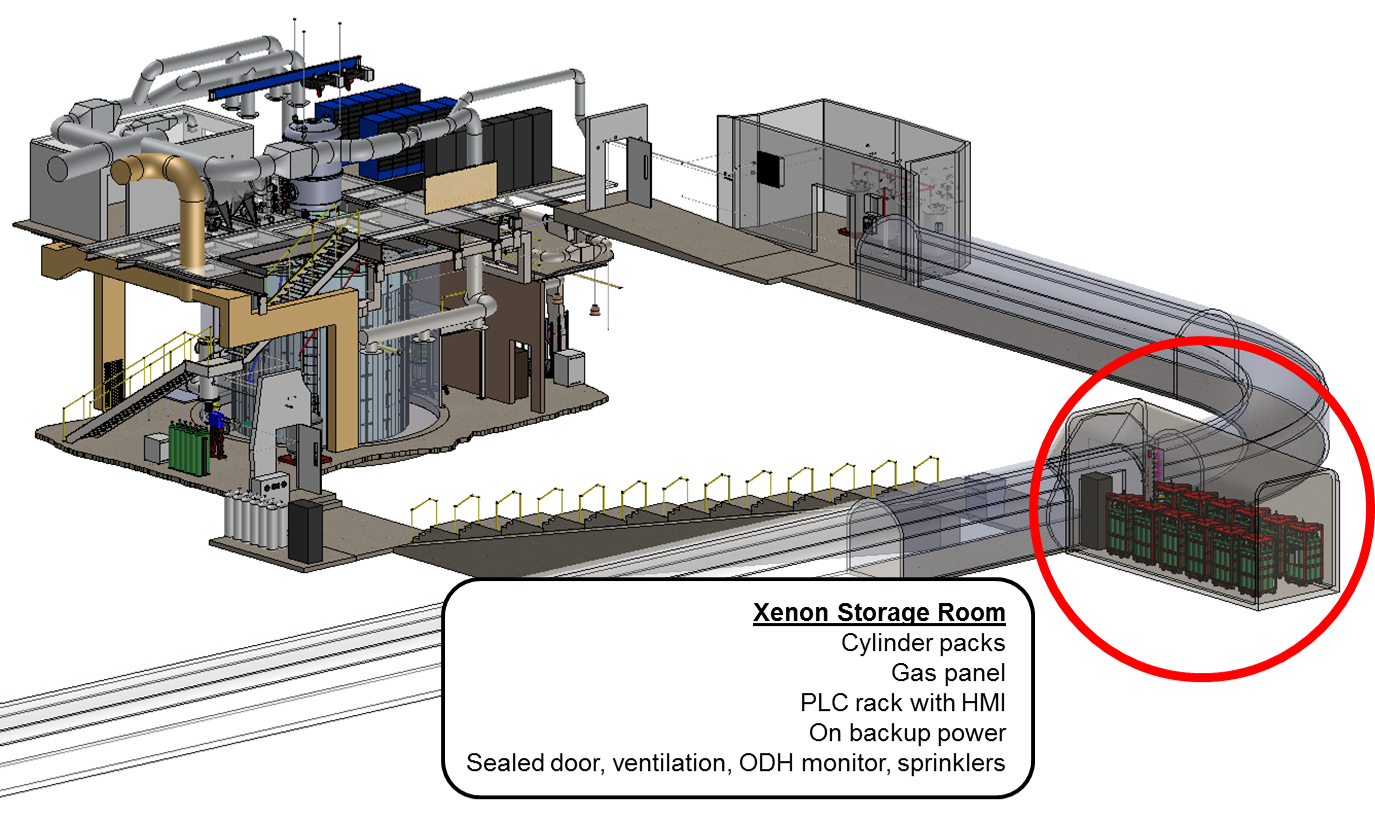}
\tdrfcaption[XeStorageUG]{Location of Xe Storage 
Room in the Davis campus}{Location of the Xe Storage 
Room in the Davis campus, along the access drift to 
the LN2 Storage Room.}
\end{figure}

Transport of the Xe packs will be divided into approximately  
\numrange{3}{4} separate shipments, both to match 
the Kr-removal production schedule and to reduce 
risk of total xenon loss in the event of a serious 
road accident. The shipments will be insured, and 
will travel in secured air-ride trucks.

Once at SURF, the packs will be moved underground 
as soon as possible to limit cosmogenic activation
of \IXeotS. The packs will be loaded on a small rail car in the 
Yates headhouse, loaded into the cage, lowered to the 
Davis Campus level, and rolled into the Davis Campus 
on a rail car. Part of the Davis Campus infrastructure 
work is to prepare a Xe Storage Room for the Xe using 
an existing excavation in the LN storage room access 
drift, as shown in Figure~\ref{XCSf:XeStorageUG}.  In 
order to maintain clearance beneath the HVAC and 
sprinkler utilities in the access drift, the cylinder 
packs may need to travel horizontally.  
In this case, a specially-designed 
pack rigging fixture will assist in the rotation and 
handling of each pack from the Yates 
headhouse to the underground
Xe Storage Room.  We are investigating 
an alternative and preferred 
option to instead modify the utility services 
so that the packs 
may travel in their upright configuration.  
The Xe Storage Room will be 
sealed from the access drift with a block wall 
and secured double doorway.

\begin{figure}[htbp]
\centering
\includegraphics[height=8.25in]{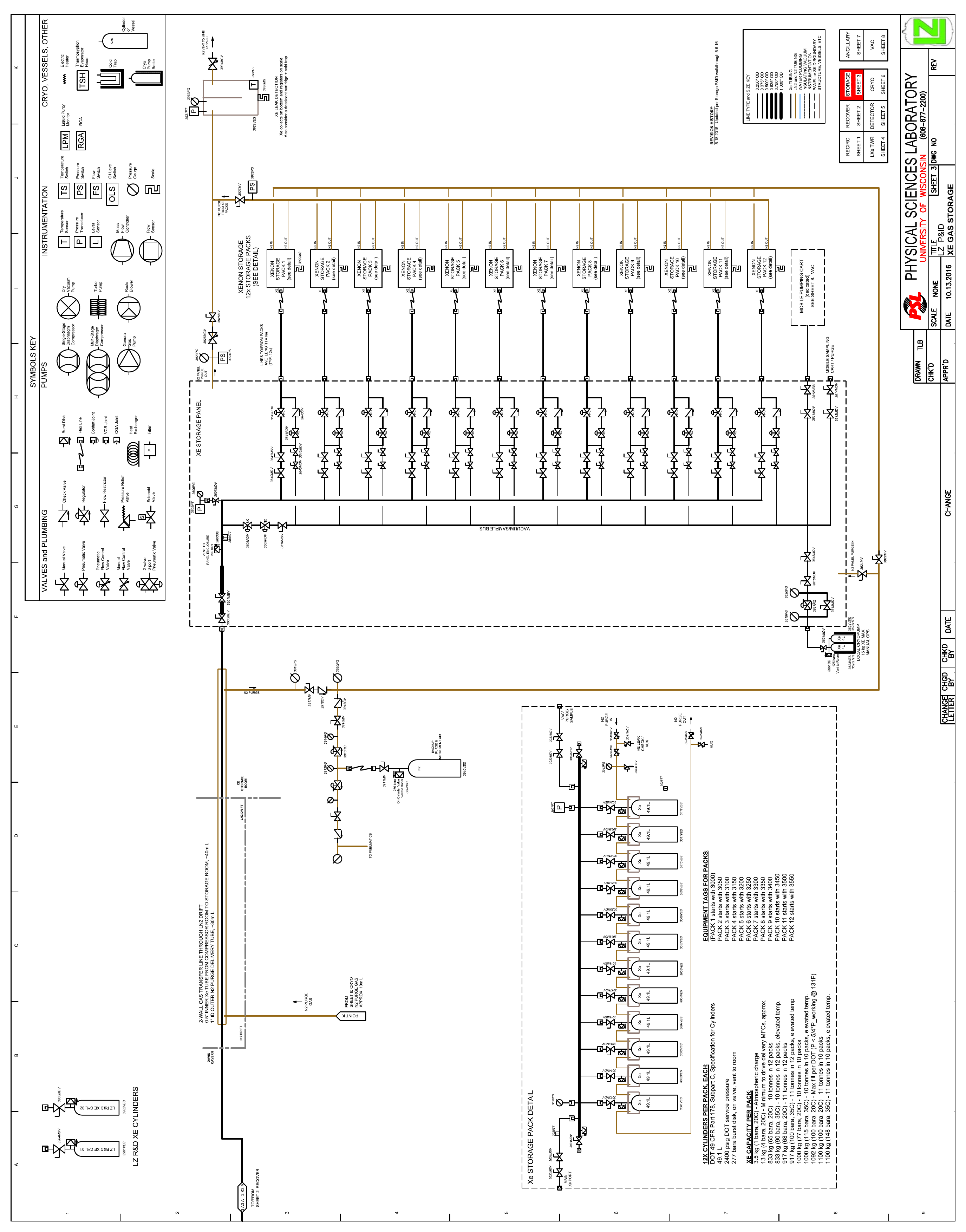}
\tdrfcaption[XeStoragePID]{Xe Storage system P\&ID}
{Xe Storage system P\&ID.}
\end{figure} 

The Xe Storage Room will have a boil-off nitrogen 
supply to purge the valve seal volumes, a sprinkler 
system to control temperature during a fire (activates 
at \SI{68}{\degree C}), and an emergency ventilation and 
oxygen monitoring system in case of an accident. 
Supply and return plumbing from the Xe storage area 
will connect the packs to the LZ detector plumbing 
via a gas panel equipped with a sampling/pump-out bus 
and local cryopump (\num{15}~kg Xe capacity) for 
consolidating xenon locally.  Xe Storage Room controls 
are supported by a local PLC control rack on UPS 
and backup power. Portable 
provisions for clean sampling of various locations 
within the Storage Room will also be provided.  A 
complete P\&ID of the integrated Xe storage plumbing 
is provided in Figure~\ref{XCSf:XeStoragePID}.

Once the Xe has been fully transferred into the LZ 
detector and Xe handling system, any xenon remaining will be 
consolidated to a single pack and the remaining packs will be 
maintained under vacuum and ready to accept incoming 
Xe from any recovery action.

\tdrsec[XeSamp]{Xenon Sampling and Assay }

Sensitive Xe purity monitoring is integrated into the
Xe handling plan. The basic monitoring methodology that 
we employ is the coldtrap/mass-spectrometry method 
developed for LUX and EXO-200~\cite{Leonard:2010zt,Dobi:2011vc}. 
The LUX experience is reviewed in Ref.~\cite{Akerib:2015cja}.
In this section we describe the LZ implementation. 

The method works as follows. A gaseous Xe sample flows 
through a precision vacuum leak valve or MFC to an RGA, 
where the partial pressures of the various species are 
measured. Once a uniform flow is established, these partial 
pressures are proportional both to the abundances in the sample 
and to the gas flow rate, and inversely proportional to 
the volumetric pumping speed of the vacuum system at the RGA. 
The partial pressures can be interpreted as absolute concentrations 
by calibrating the measurement under specified flow and pumping 
conditions with Xe gas samples prepared with known impurity content. 

Good sensitivity may be achieved by maximizing the gas flow 
rate and minimizing the pumping speed, however the total pressure 
cannot exceed about \SI{e-5}{torr} if the RGA and its electron
multiplier are to function well.
Without additional measures the Xe partial pressure would severely 
limit the flow rate, so the Xe is selectively removed by passing 
the gas sample through an LN coldtrap where it forms Xe ice. Under 
these conditions the Xe partial pressure at the coldtrap outlet 
is held at \SI{1.8e-3}{torr}, 
the vapor pressure of Xe ice at \SI{77}{K}. 
This is still too high by a factor of \num{180}, 
however the Xe pressure may be further reduced at the RGA by inserting
an appropriate impedance into the plumbing. 
In the presence of the coldtrap the Xe partial pressure at the RGA 
does not depend on the gas flow rate into the trap.

Some impurity species (such as water) are removed from the 
gas stream by the coldtrap and are thus unobservable at the RGA.
However, many species of interest pass through 
the trap with good efficiency, including \COt, \CNt, CH$_4$, Ar, 
He, and, most importantly, Kr. 
Studies have shown that the partial pressures 
of these species continue to be a good measure of their concentrations 
as long as an appropriate calibration is done at a similar flow rate
and vacuum pumping speed. 

In contrast to Xe, the partial pressure of Kr at the 
RGA is found to be much less sensitive to the vacuum impedance 
between the coldtrap outlet and the RGA. It decreases only moderately 
as the impedance increases, allowing Xe to be selectively suppressed
relative to Kr. The origin of this unusual behavior is
that the Xe signal is a pressure source created by the Xe
ice in the coldtrap, while the Kr signal is a current source
created by the gas flow into the coldtrap.

This distinction in the nature of the Xe and Kr signals
is advantageous because Xe dominates the 
total pressure and thereby limits how
low the pumping speed may be set without
exceeding the RGA's pressure limit.
After Xe suppression, however, the 
pumping speed may be further reduced, increasing the partial pressures of all
species, including Kr and Xe, until Xe once
again reaches the RGA's practical limit of \SI{e-5}{torr}. 
In practice, vacuum parameters such as the impedance and
pumping speed are not modified during a run but are instead 
chosen for optimal sensitivity in prior test runs. 
The resulting system configuration may be calibrated as usual 
with specially prepared Xe gas samples with known krypton content.

\begin{figure}[htbp]
\centering
\includegraphics[height=3.5in]{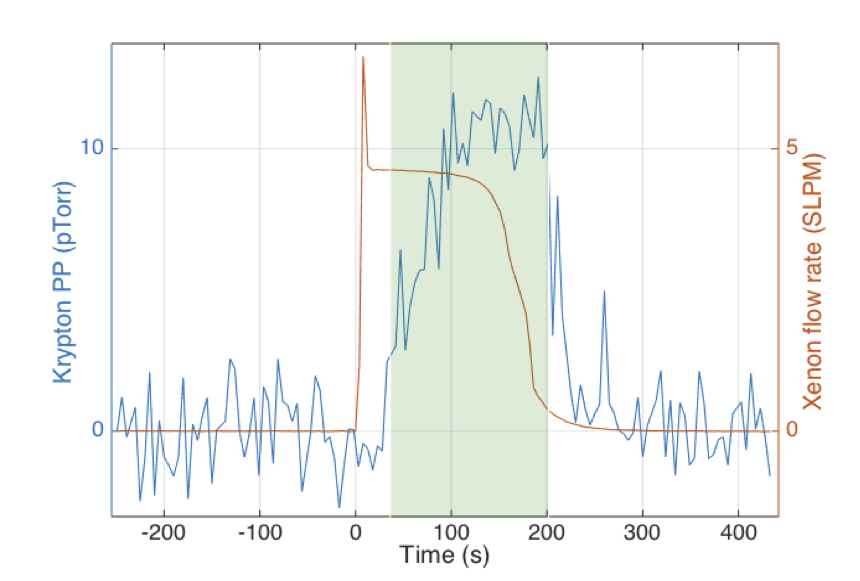}
\tdrfcaption[KrTrace]{Kr pressure trace for \SI{0.34}{\ppt} (g/g)}
{Blue: Kr partial pressure trace at a concentration of 
\SI{0.34}{\ppt} (Kr/Xe) (g/g) (left vertical axis). Orange: Flow rate into
the cold trap in (slpm, right vertical axis). The green 
shaded region is the signal integration window.}
\end{figure} 

Empirical studies have shown that the efficiency to pass Kr 
through the coldtrap depends
upon its tubing diameter, with smaller diameter tubing generally 
performing better. Another important factor is the total amount of 
Xe ice that can be collected before the flow path becomes blocked. 
Larger ice capacity implies that a larger flow rate may be sustained
for the same time period, further enhancing the Kr signal.
For LZ we have adopted a \SI{0.5}{\inchl} OD tube as a baseline compromise
between these competing needs. 

To improve the Kr sensitivity of the method beyond the \SI{0.2}{\ppt}
achieved for the LUX Kr removal campaign, we have increased 
the mass flow rate by a factor of four from \SI{\sim 1.5}{\slpm} to 
\SI{6.5}{\slpm} and decreased the pumping speed by a factor of five
from $\sim$5 to 1~liters/sec. We've also implemented an MFC for better
programmable flow control. Results are presented in 
Figure~\ref{XCSf:KrTrace}, where a prominent Kr partial pressure signal is 
shown for a Kr concentration of \SI{0.34}{\ppt} (Kr/Xe) (g/g). Studies
of the sensitivity of the method, performed with Kr-free xenon gas samples,
have found that the limit of detection is \SI{0.007}{\ppt} (g/g) (Kr/Xe) 
at 90\%C.L.

\begin{figure}[htbp]
\centering
\includegraphics[height=2.7in]{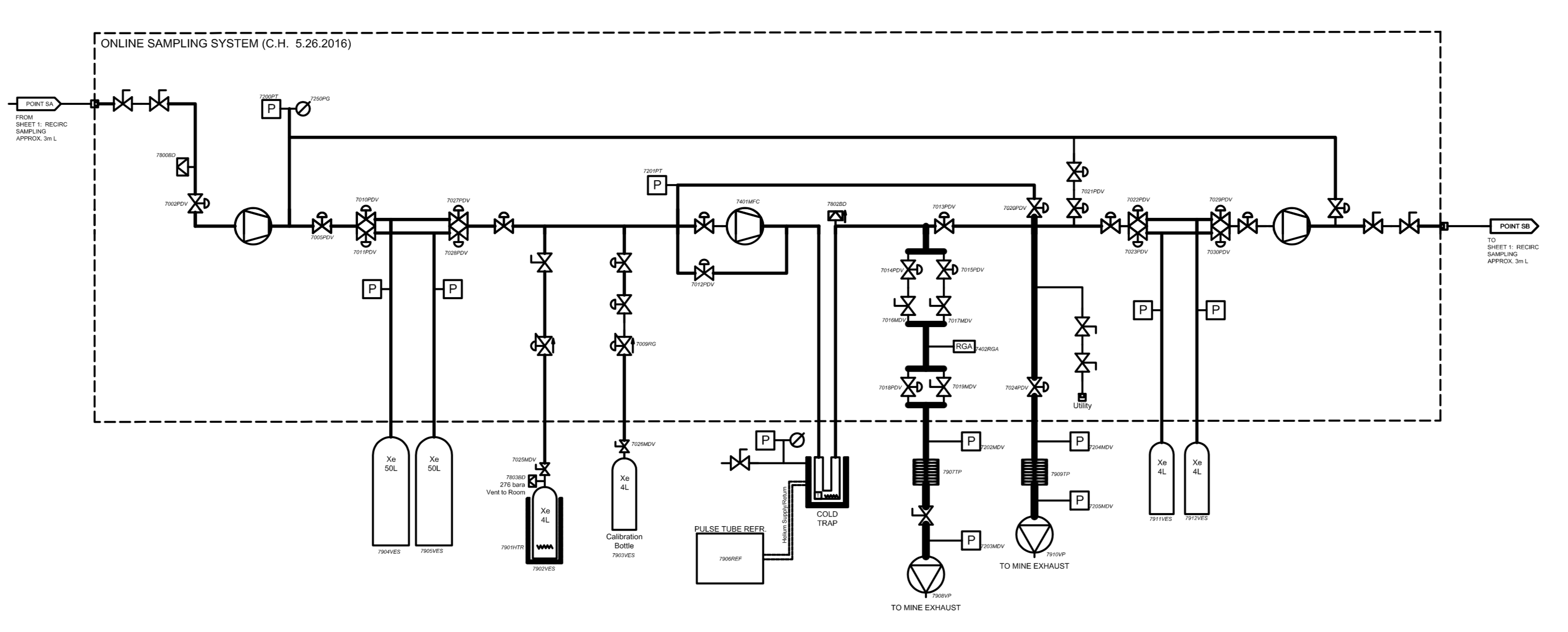}
\tdrfcaption[XeSampPID]{P\&ID of the LZ online Xe sampling system}
{P\&ID of the LZ online Xe sampling system.}
\end{figure} 

\begin{figure}[htbp]
\centering
\includegraphics[height=3.2in]{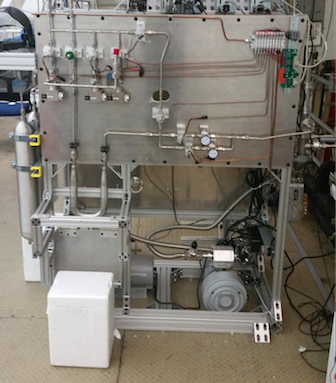}
\includegraphics[height=3.2in]{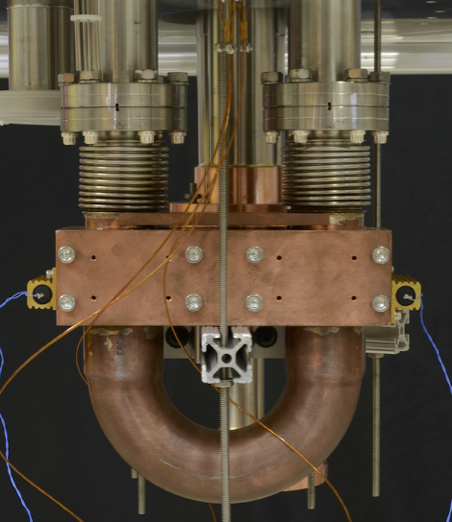}
\tdrfcaption[SLACsampling]{SLAC sampling system and coldtrap}{Left: 
Xe sampling system 
constructed at SLAC in 2015 for monitoring Kr removal activities and 
screening the commercially procured Xe stockpile. 
Right: Prototype copper coldtrap with temperature 
control provided by a pulse tube refrigerator coldhead.}
\end{figure} 

An analytical system incorporating these design features 
has been constructed 
at SLAC to aid the Kr removal campaign described
in Section~\ref{XCSS:KrRem}, and to perform 
quality assurance on the vendor-supplied
xenon as it is acquired. The SLAC system
is currently integrated 
into the R\&D chromatography system, 
and will be re-purposed for use in the production system in 2018. 

At the conclusion of krypton removal operations, the SLAC
sampling system will be shipped to SURF and be permanently
integrated into the LZ xenon circulation system at the Davis campus. 
A P\&ID of this system is shown in Figure~\ref{XCSf:XeSampPID}. 
To simplify continuous operations at SURF, 
the coldtrap will be cooled by a pulse-tube refrigerator
as shown in Figure~\ref{XCSf:SLACsampling}, and accumulation bottles
on the input and output will allow xenon gas to flow continuously
through the system.
The LUX sampling system will also be re-built and re-purposed at SURF 
as a mobile utility sampling system for 
use during detector commissioning
and operations.

\tdrsec[Cryo]{Cryogenics, vacuum services, and breakout boxes}

This section describes the design of the liquid nitrogen cryogenic
systems. We also describe here the vacuum 
pumping systems and the breakout-box feedthroughs
for the internal PMT and instrumentation cables.

Cooling power to maintain Xe in the liquid phase 
is provided by a cryogenic system that distributes 
nitrogen in gas and liquid phases. By utilizing the 
approximate \SI{100}{\K} temperature difference between the 
nitrogen and Xe evaporation temperatures, sufficient 
gradient exists to provide for thermal control and 
temperature modulation. Existing infrastructure from 
LUX, including \num{450}-liter LN storage tanks; 
vacuum-jacketed (VJ) pipe; and miscellaneous valves, 
sensors, and fittings are modified and re-utilized to 
provide a front end to the primary cooling equipment 
for the experiment while providing supplemental LN 
storage. During operations, distribution of LN is 
from a VJ central \num{750}-liter distribution tank that 
is co-located with a Stirling cycle cryocooler. The 
cryocooler liquefies cold evaporated nitrogen gas 
in a closed-loop cycle operating at near atmospheric 
pressure. Multiple thermosyphon heat pipes, cooled 
by the LN in the storage tank, are used as heat sinks
to provide for heat removal from the Xenon system. Primary 
heat removal locations are the detector, high-voltage 
feedthrough, and the LXe tower. The underground
installation is shown in Figure~\ref{XCSf:ug}

\begin{figure}[htbp]
\centering
\includegraphics[height=3.0in]{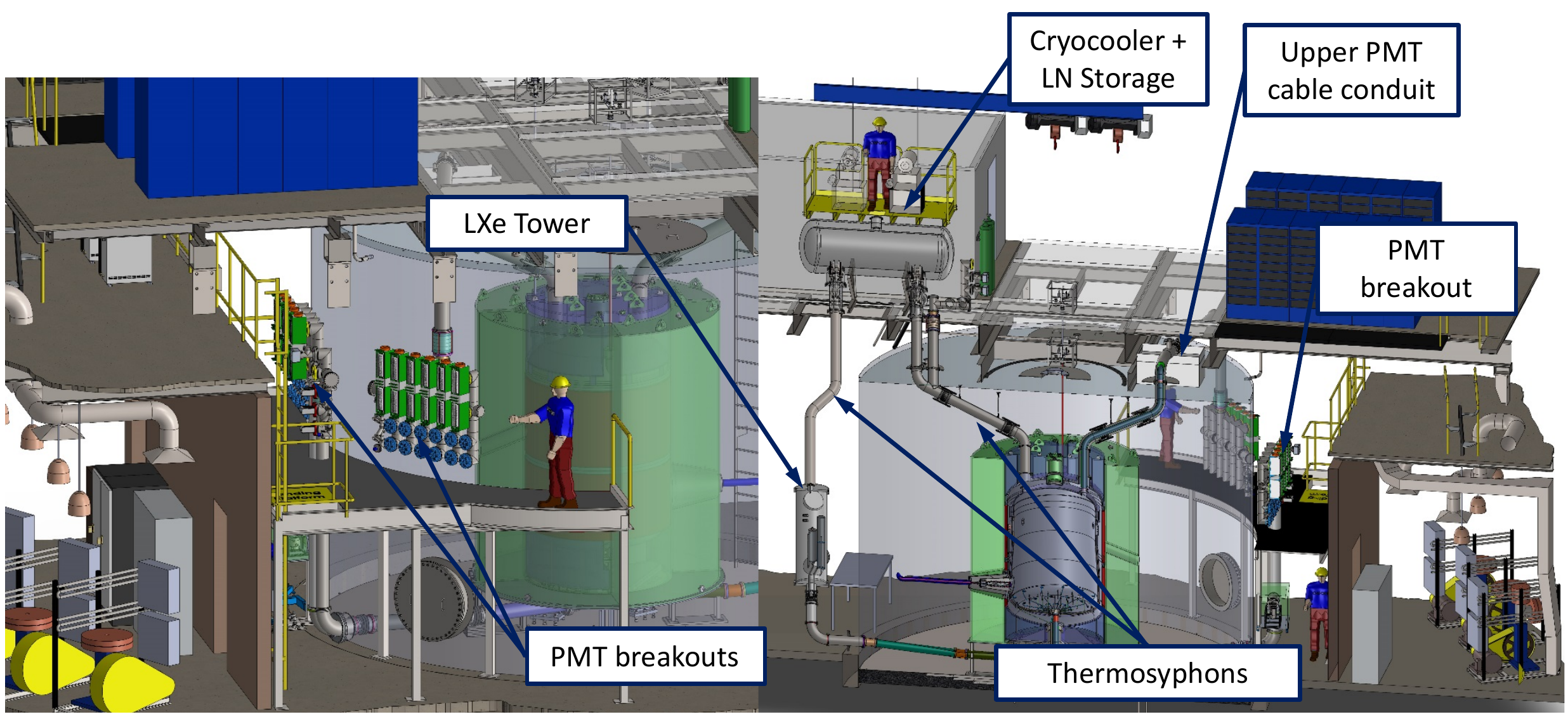}
\tdrfcaption[ug]{Cryogenics installation in the 
Davis Cavern}{Cryogenics installation in the Davis Cavern.}
\end{figure}

Exterior to the Davis Cavern is an LN storage room 
that contains four \num{450}-liter LN storage tanks installed 
on mass scales that monitor LN consumption. A 
commercially purchased vacuum-jacketed (VJ) piping system that has a 
complete implementation of control valves, relief 
valves, and pressure-monitoring equipment connects the 
tanks to equipment in the Davis Cavern. Separate 
small-diameter tubing connected to the storage tanks 
distributes boil-off nitrogen purge gas for the 
freeboard of the water tank, the water-purification 
system vacuum pump, the scintillator reservoir, radioactive
source deployment systems, and Xe equipment protected from 
radon leakage. The four storage tanks are required to
provide an initial liquid volume for startup of the 
cryocooler, meet high load transient conditions, and 
provide backup to the cryocooler until a second cryocooler
procured at a future date. 
Alterations to the VJ piping system are required to 
allow for rerouting of LN to new destinations in the 
Cavern that will differ from the LUX LN use locations. 
The rail-mounted \num{1100}-liter LN storage tank utilized 
by LUX will provide for the resupply of LN from the 
surface during brief cryocooler operation interruptions 
and periodic replenishment of consumed purge gas. 

\begin{figure}[htbp]
\centering
\includegraphics[height=3.5in]{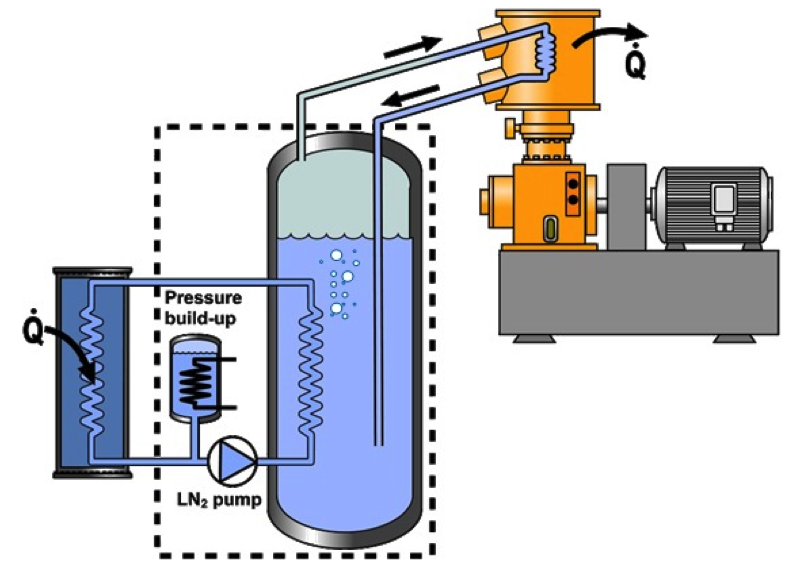}
\tdrfcaption[stirling]{Cryocooler schematic}{Cryocooler 
schematic. A cryogenerator removes heat, Q, from a closed 
liquid nitrogen storage reservoir. Other devices requiring 
heat removal are thermally connected to that reservoir.  
All of the systems are closed-loop.}
\end{figure} 

A cryocooler based on the Stirling thermal cycle is selected 
for cooling of LZ. It is depicted schematically in 
Figure~\ref{XCSf:stirling}. Included in the cryocooler 
design is a cryogenerator, a work platform sized for 
two cryogenerators, a \num{750}-liter LN storage tank to 
provide a distribution reservoir, and a complete 
monitoring and control system. The design of the 
cryocooler is closed cycle. Nitrogen inside the storage 
tank is re-liquefied rather than being vented to the 
cavern as was done during LUX operations. The cryocooler 
is capable of \SI{1000}{\W} of cooling power at the \SI{77}{\K} boiling 
temperature of LN at atmospheric pressure. Preliminary 
estimate of total heat load on the cryogenic system is 
\SI{838}{\W} including \SI{20} {\%} contingency.  
The heat load rollup is shown in Table~\ref{XCSt:heatload}. 
The \SI{1000}{\W} configuration provides a margin against 
unanticipated heat load.

\begin{table}[H]
\tdrtcaption[heatload]{Heat load table.}{Heat load rollup.}
\centering
\sffamily
\setlength{\extrarowheight}{3pt}
\begin{tabular} 
{|c|c|p{6cm}|} 
\hline
\rowcolor{mrocol}
    {\bfseries System/Component  } & 
    {\bfseries Heat (W)  } & 
    {\bfseries Notes  } \\ 
\hline
LZ cryostat &  115  & Assumes 10 liters of MLI; value confirmed 
by independent analysis in U.K. \\
\hline
 PMT Conduit  & 34   &  \\
\hline
 HV Conduit  & 1   &  \\
\hline
 Cryocooler Storage Tank &  28  & Calculated for 1000-liter tank; value diminishes for smaller Dewars.  \\
\hline
 Heat Exchanger  & 349   & Assumed at 94\% efficient (LUX experience) @ \SI{500}{\slpm} Xe flow. \\
\hline
 Xenon Purge  &  56  &  Estimate for 5~\si{\slpm} flow rate. \\
\hline
 Thermosyphons  & 116   &  Dominated by heaters; could go up for higher thermosyphon count. \\
\hline
 Contingency @ 20\%  & 140   &  \\
\hline
{\bfseries Total } & 838   &  \\
\hline
\end{tabular}
\end{table}

Cryocoolers based on the Stirling cycle have the additional 
advantages: quick startup, low energy consumption (each 
unit draws \SI{11}{\kW} of electricity), and variable drive motors 
that allow adjustable cooling levels below \SI{1000}{\W}. These 
units have been deployed at many institutions worldwide, 
including installations at SNOLAB, Gran Sasso (Icarus), 
and multiple university laboratories in the United States, 
Russia, and Asia. Maintenance of the cryocooler is 
required after \num{6000} hours of continuous use, so the 
system is designed for the future addition of a second 
cryocooler (likely purchased as a component of operations). 
Transport of LN from the surface to the Davis Campus can 
sustain operations during short-term maintenance that is 
expected to have a duration not exceeding 8-hours. 

LUX has received deliveries of LN two to three times per week. 
Even though LZ is better insulated, scaling indicates that 
LZ requires one LN delivery every 24 to 36 hours to sustain 
operation. For long-term operations, that would increase 
the inherent transport risk and manpower requirements 
related to LN transfer. Therefore, the addition of a 
second cryogenerator at some point would significantly 
reduce the risk to long-term operations.

Distribution of cooling power from the cryocooler to the 
Xe-containing experimental systems is accomplished by use 
of thermosyphons. A thermosyphon is a type of heat pipe. 
Thermosyphon technology was successfully applied to LUX 
following development at Case Western Reserve University 
(now SLAC). Thermosyphons comprise condenser and evaporator 
heads connected by small-diameter tubing wrapped with 
multilayer insulation (MLI) and charged with nitrogen 
gas at modest pressure. In this application, the 
condenser is immersed in an LN bath, causing the nitrogen 
gas in the thermosyphon to liquefy. The liquid flows in 
the small-diameter tubing by gravity to the evaporator 
that is physically attached to the device that requires 
cooling. Heat from the evaporator causes the LN pooled 
in the evaporator to change phase, removing heat via the 
nitrogen latent heat of vaporization. The warmed nitrogen 
gas then returns to the condenser by buoyancy effect. 

Cooling of the Xe relies upon the temperature difference 
between the LN and Xe vaporization temperatures as well 
as the adjustable cooling power of the thermosyphon. 
Changing the mass of circulating nitrogen can modulate the 
thermosyphon cooling power. As the mass is increased, 
pressure in the condenser rises, effectively increasing 
the boiling temperature in the thermosyphon and inducing 
an increased heat flux to the LN bath that remains at 
atmospheric pressure and 77 K. 

Thermosyphon cooling is completely passive, with a fixed 
amount of nitrogen in the secondary cooling loop. Therefore 
no pumps are required and there is no direct path for large 
quantities of LN (typically stored in Dewars) to get into 
the Xe via a leak. As the thermosyphon transport tubing may 
be \SI{9.5}{\mm} or \SI{12.7}{\mm} OD, a minimal amount of nitrogen can 
be placed into the tubes, on the order of \SI{15}{\g} to reach 
\num{10}~bara. 
This mass gradually increases as the nitrogen 
condenses but the mass remains modest. Over-pressure protection 
is by relief valves and burst disks. The thermosyphons are 
charged with nitrogen via high-pressure cylinders. Control 
of the gas mass (and cooling power) is accomplished with 
control valves, mass flow controllers, sensors, and 
processing devices connected to slow control via Ethernet. 
Thermosyphons deployed in LUX were easily capable of 
delivering more than \SI{200}{\W} cooling for each deployed 
evaporator head. The single largest device needing cooling 
is the LXe tower, in which two-phase, sub-cooler, and gas-gas heat 
exchangers are anticipated to require \SI{349}{\W} 
of cooling to 
recondense Xe returning to the detector. Three thermosyphon 
evaporators are able to provide that energy removal.

Vacuum-pumping systems take advantage of tubing in 
place to route thermosyphons, Xe transfer lines, and PMT cables. 
Packages of combined scroll pumps and turbo-molecular 
pumps are deployed on the decking of upper Davis 
Cavern and near the Xe heat exchanger tower in lower 
Davis Cavern. These pump combinations evacuate: the 
annulus between the inner and outer vessels, detector 
internals, Xe heat-exchanger tower, and the entire Xe 
gas system. Two vacuum leak check carts are provided 
so that there is a minimum of one cart available at any 
time in the Surface Assembly Laboratory and the 
underground Davis Campus. A variety of gauges are 
deployed, including manual gauges (that can show vacuum 
in power outage), thermocouple gauges, and ion gauges.

A critical element of vacuum pumping is removal of residual gases
from detector internals that are either attached to metal surfaces
or dissolved into plastic volumes.  Empirical data developed at U. 
Maryland for gas solubility and diffusion in plastics is used along 
PTFE volume / thickness measurements from SolidWorks CAD models 
to predict the required vacuum pumping to achieve acceptable levels 
of Kr, \CNt, and \COt. Predicted steps for vacuum pumping are as follows: 

\begin{enumerate}
\item Pump detector to \SI{1E-4}{torr} and hold for one day;
\item Backfill with \SI{10}{torr} Xe gas;
\item Circulate Xe gas that is heated to \SI{40}{\degree C} for
45 days;
\item Pump out Xe gas and achieve \SI{1E-6}{torr} predicted
to take 65 days; 
\item Maintain that pressure at room temperature 
for one additional week.  
\end{enumerate}

By that time residual Kr in the 
detector is predicted to be less than \SI{0.03}{\ppt} with a pumping 
duration of 3.3 months.

PMT, sensor, and grid power supply cables from both the inner 
and outer cryostats must be routed to locations where they 
can transition from either a pure Xe or high-vacuum 
environment to atmospheric conditions in the Davis 
Cavern. These cables are routed through VJ conduits. Two conduits 
exit the cryostat from the top and one from the bottom. One upper 
and one lower conduit, primarily carrying PMT cables, are routed 
to a SURF-installed mezzanine level. Sensor and grid cables exiting the 
top of the detector are routed to the main Davis Cavern deck.
The inner tube of the VJ is braided flexible conduit sized to 
allow pulling of all cables during installation. The outer 
jacket has ``tee'' connections to the vacuum-pumping system. 
Stainless steel breakout boxes are installed on the conduits at the
mezzanine and main Davis Cavern deck for installation of hermetic 
feedthroughs at the point where cables must be terminated in order 
to exit the Xe and vacuum spaces. Breakout boxes and inner conduits 
are compatible with the \num{3.4}~bara maximum Xe pressure and meet 
Xe cleanliness requirements. 

Control of temperature, pressure, and liquid levels 
within the cryogen system is needed both for the 
correct operation of detector systems and for safe 
operation. Control of these parameters is via 
connection to the slow control system described in 
Chapter~\ref{chap:EDC}. 
A detailed P\&ID 
along with a related instrument list exists. Each thermosyphon 
utilizes multiple thermometers, pressure transducers, 
a vacuum gauge, a mass flow controller, and controller 
boxes to digitize signals and transmit via Ethernet. 
Similarly, the LN storage room, VJ distribution piping, 
cryocooler, and vacuum pumping system need active 
monitoring and control of temperature and pressure.

\tdrsec[XeProc]{Xenon Procurement}

Approximately \SI{10}{\tonnesl} (\num{1.8e6} gas liters) of Xe 
must be acquired for the LZ detector. The purity requirement 
is an industry standard of \SI{99.999}{\percent} Xe. The Xe currently in 
the LUX detector and from small previous purchases will be 
reused (\num{\sim90000} liters). 
SDSTA initiated a bid and procure process for \num{\sim500000}~liters 
in the second half of 2015. Bids were requested and received 
from all major producers in the world and a few other sources 
with existing Xe stock.  Delivery has started and will continue 
into 2017. A second bid cycle for up to about \num{1.2e6} liters 
was completed in February 2016 and deliveries have started. 
A fixed price and delivery schedule 
consistent with the LZ need to complete delivery by the end of 
2018 was achieved.  The University of Alabama has procured about 
36,000 liters that have been delivered to SLAC. All Xe will be 
delivered to SLAC by the fall of 2018 for later removal 
of Kr. Approximately 20\% of the LZ Xe has been received and 
assayed at SLAC, and all of it has met or exceeded purity specs.

\tdrsec[DiffusionSolubility]{Solubility and diffusion 
constants of common impurity gas species}

To quantify the outgassing burden of the 
LZ TPC materials, the LZ R\&D program produced a set of 
measurements of the diffusion and solubility 
constants of several common impurity species in various plastic and 
elastomer materials. Some result from that campaign were reported  
in Table 9.7.1 of Ref.~\cite{Akerib:2015cja}, however, due to 
an editing error, portions of that table are incorrect.
Table~\ref{XCSt:solubility} presents the 
corrected values.

\begin{table}[H]
\centering
\tdrtcaption[solubility]{Solubility and diffusion
constants of common impurity species.}{Solubility (K) and diffusion
constants (D) of common impurity species. K 
is defined as the ratio of impurity mass per unit volume inside and
outside of the material sample. The PTFE sample is
from the LUX TPC. PE1 is a polyethylene sample from LUX. 
The viton is an off-the-shelf sample from McMaster.
PE2 is a polyethylene sample from a candidate cathode high voltage
cable that was under consideration for LZ during its R\&D phase. All
measurements were performed at room temperature. }
\sffamily
\setlength{\extrarowheight}{3pt}
\begin{tabular} 
{|c|c|c|c|c|} 
\hline
\rowcolor{mrocol}  \multicolumn{5}{|c|}{{ \bfseries Solubility (K) (\%) } } \\
\hline
 & PTFE & PE1 & viton & PE2 \\
\hline
N$_2$ & $10.7\pm0.7$ & $2.1 \pm 0.3$ & $51 \pm 10  $ & \\
O$_2$ & $22\pm2    $ & $1.8 \pm 0.6$ & $22 \pm 1   $ & \\
Kr    & $58\pm5    $ & $9.3 \pm 1.1$ & $23 \pm 2   $ & $11 \pm 2$ \\
Xe    & $89\pm8    $ & $55  \pm 6  $ & $15 \pm 2   $ & \\
Ar    & $8.8\pm0.8 $ & $4.7 \pm 1.5$ & $8.3 \pm 2.6$ & \\
He    & $3.3\pm0.3 $ & $0.64\pm0.07$ & $9.3 \pm 0.8$ & \\
CH$_4$&              & $16  \pm3   $ &               & \\
\hline
\hline
\rowcolor{mrocol}  \multicolumn{5}{|c|}{{ \bfseries Diffusion constant (D) (10$^{-8}$ cm$^2$/s) } } \\
\hline
 & PTFE & PE1 & viton & PE2 \\
\hline
N$_2$ & $ 15.1\pm0.3  $ & $ 16\pm1    $ & $ 2.2\pm0.1  $ & \\
O$_2$ & $ 31.4\pm0.6  $ & $ 39\pm3    $ & $ 6.8\pm0.1 $  & \\
Kr    & $ 5.6\pm0.1   $ & $ 6.4\pm0.4 $ & $ 1.25\pm0.02$ & $ 11.2\pm0.8 $ \\
Xe    & $ 0.80\pm0.02 $ & $ 2.0\pm0.1 $ & $ 1.7\pm0.1  $ & \\
Ar    & $ 16.8\pm0.3  $ & $ 20\pm1    $ & $ 4.0\pm0.1  $ & \\
He    & $ 1270\pm25   $ & $ 435\pm30  $ & $ 436\pm5    $ & \\
CH$_4$&                 & $6.6\pm0.9  $ &                & \\
\hline
\end{tabular}
\end{table}

\clearpage
\bibliographystyle{apsrev4-2}
\bibliography{LZtdr}
\wlog{In tdrinclude command}
\wlog{\ChapLabel}

\renewcommand{\ChapLabel}{chap:CAS} 
\renewcommand{\BibLabel}{CASS:bib} 
\renewcommand{\SecPre}{CAS}
\renewcommand{\FloatPre}{CAS}
\tdrchap[CAS]{Calibration Systems}
A rigorous calibration strategy is a prerequisite for the unambiguous direct detection of hypothetical dark matter interactions in the LZ detector. The basic questions about any event are: (1) how did the particle interact, and (2) how much energy did it deposit? But before these questions can be answered, the ($x$,$y$,$z$,$t$) response of the LZ TPC must be understood.

The LZ calibration strategy is designed to accurately answer these questions, achieve the LZ science goals, and be ready to address the widest possible range of predicted dark-matter signatures. Basic requirements (referred to in this chapter as R-17nnnn for the different requirements) have been defined. The process for capturing requirements is described in Chapter~\ref{chap:SRD}. The principal calibration techniques planned for LZ have been used successfully in previous experiments, especially LUX and Zeplin. A summary of all sources to be used in LZ is included in Table \ref{CASt:sources}.

\begin{table} [h]
\setlength{\extrarowheight}{5pt}
\tdrtcaption[sources]{Baseline calibration sources for LZ.}
{Baseline calibration sources for LZ.}
\centering
\sffamily
\begin{tabular} 
{|c|c|c|c|c|} 
\hline
\rowcolor{mrocol}
Isotope & What & Purpose & Deployment & Custom? \\
\hline
Tritium & beta, Q = \SI{18.6}{\keV} & ER band & Internal & N \\
\IKreTm & beta/gamma, \SIlist{32.1;9.4}{\keV} & TPC $(x,y,z)$ & Internal & Y \\
\IXeoTom & \SI{164}{\keV} \Pgg & TPC $(x,y,z)$, Xe skin & Internal & Y \\
\IRnttz & various \Pga's & xenon skin & Internal & N \\
\hline
AmLi & (\Pga,\Pn) & NR band & CSD & Y \\
\ICftFt & spontaneous fission & NR efficiency & CSD & N \\
\ICoFS & \SI{122}{\keV} \Pgg & Xe skin threshold & CSD & N \\
\IThtte & \SI{2.615}{\MeV} \Pgg, various others & OD energy scale & CSD & N \\
\INatt & back-to-back \SI{511}{\keV} \Pgg's & TPC and OD sync & CSD & N \\
\hline
\IYee Be & \SI{152}{\keV} neutron & low-energy NR response & External & N \\
\IBitzF Be & \SI{88.5}{\keV} neutron & low-energy NR response & External  & Y \\
\IBitzs Be & \SI{47}{\keV} neutron & low-energy NR response & External  & Y \\
\hline
DD & \SI{2450}{\keV}~neutron & NR light and charge yields & External & N \\
DD & \SI{272}{\keV}~neutron & NR light and charge yields & External & Y \\
\hline
\end{tabular}
\end{table}

As described in Chapter~\ref{IDOSs:ExperimentStrategy}, signals in LZ consist of scintillation photons (S1) and ionized electrons, read-out as proportional scintillation (S2). Variation in the ($x$,$y$,$z$) response of S1 and the ($x$,$y$) of S2 arise, for example, from detector geometry, light collection and other factors. Variation in the ($z$,$t$) response of S2 could arise from changes in xenon purity, via attachment of ionized electrons by impurities. These calibrations will be addressed by mono-energetic internal radioisotope sources (Section~\ref{CASS:internalsources}).

Once these variations have been calibrated, it is possible to address the two questions posed in the first paragraph. The answer is contained in a plot like Figure~\ref{IDOf:discrimination} from Section~\ref{IDOSs:Discrimination}, which shows the distribution of background-like events (top) and expected signal-like events (bottom). These bands were obtained from in-situ calibration of the LUX detector. Similar techniques will be used for LZ, as described in detail in the following sections. In Figure~\ref{IDOf:discrimination}, the overlaid curves of approximate event energy are obtained from a knowledge of the absolute photon and electron detection efficiency of the instrument. This in turn is obtained from mono-energetic photon interactions in the detector.

In addition to the broad spectrum nuclear recoil response, LZ will measure the absolute response of LXe to nuclear recoils. It is critical to make this calibration in-situ and  LZ will be able to make more precise measurements than are presently available. This is due to its large target mass and extremely low background count rate. These measurements are discussed in Section~\ref{CASS:photoneutron} and Section~\ref{CASS:ddneutron}.

\tdrsec[internalsources]{Internal Radioisotope Sources}
Liquid Xenon's strong self-shielding ability, of great benefit to the dark matter search by largely eliminating external backgrounds from the central fiducial volume, conversely makes low-energy electron recoil calibrations of this central volume practically impossible using external radioisotope sources.  Rather than rely on extremely long duration (\SI{\gg1}{\day}) exposures to high-rate high-energy (\SI{>1}{\MeV}) sources, gaseous radioisotopes will be mixed into the LXe itself.  This `internal' calibration source strategy has been well-demonstrated in LUX.

\tdrsubsec[Kr-83m]{Metastable Krypton 83 (\IKreTm)}
Metastable \IKreTm is a low-energy (\SI{41}{\keV}) monoenergetic source that is easy to produce and exhibits a conveniently short decay time (half-life of \SI{1.8}{\hour}). The monoenergetic peak enables the production of high-resolution 3-D maps of S1 and S2 detector response, producing a calibration of spatially-varying detector efficiency effects. Electron drift efficiency, extraction efficiency, and S2 light production efficiency are each individually position-dependent, and can combine to produce a larger spatial variation in S2 response, one that can also be time-varying with changes in liquid purity. This motivates requirements R-170002 and R-170007, which necessitate a mono-energetic dispersed throughout the xenon. For all these reasons, regular ($\sim$weekly) 3-D detector response maps are essential to attaining the needs of the WIMP search.

It should be noted that \IKreTm's decay occurs in two steps, separated by a short decay half-life ($\tau$~=\SI{154}{\ns}).  This timing variation from decay to decay produces a variation from event to event in the overall electron-ion recombination probability, meaning that while \IKreTm is quite useful as a monoenergetic peak source, its unusual recombination physics means it is not useful for mimicking typical electron recoil backgrounds.

\IKreTm production and handling for liquid noble detectors is now a mature process: a radon-pure charcoal is infused with a \IRbeT-containing aqueous solution, baked, and then placed in standard VCR plumbing (contained by particulate filters) to produce a \IKreTm-generating source.  To `inject' \IKreTm into the detector, a xenon carrier gas flows over the charcoal, carrying some \IKreTm activity into the main xenon circulation path, with the dosed activity controllable by carrier gas flow rate and flow duration.  As a noble gas, \IKreTm has the practical advantage of first passing through the circulation path getter, for added protection against the incidental injection of impurities.  \IRbeT has a conveniently long half-life of 86.2 days, meaning a single \IKreTm source can sustain a usable activity for approximately one year.

\begin{figure}[h]
\centering
\includegraphics[width=1.0\textwidth]{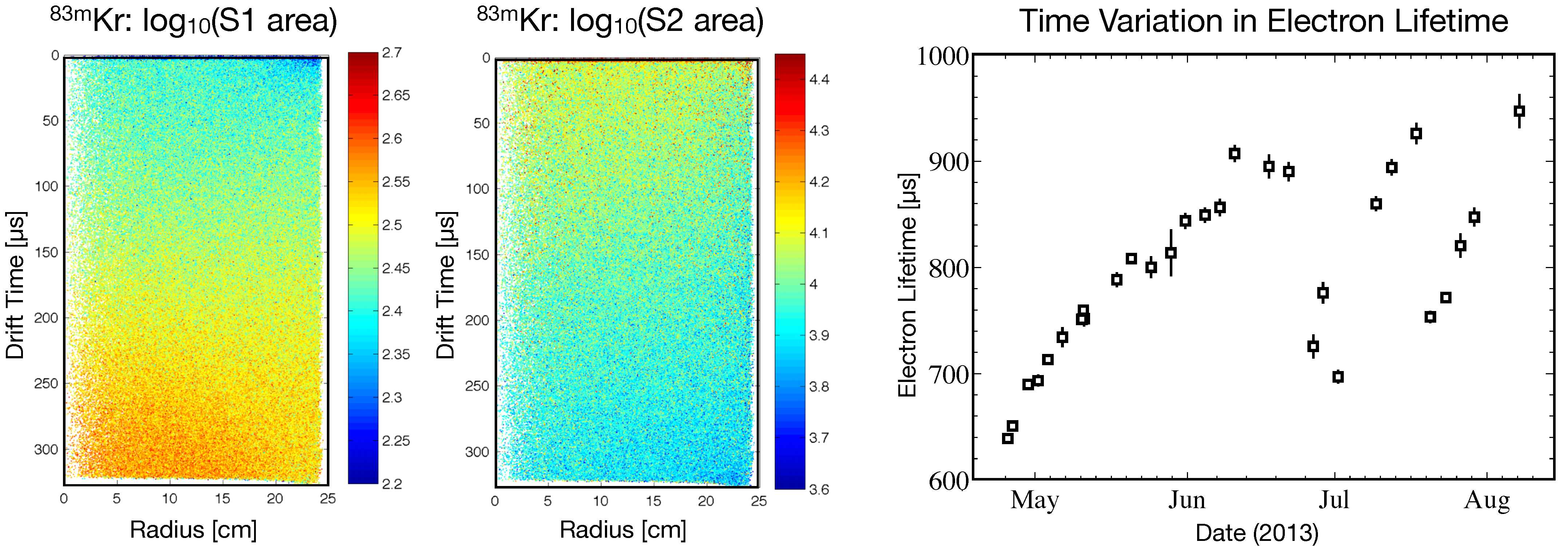}
\tdrfcaption[Kr83m]{Illustrations of \IKreTm calibrations in LUX}{Several illustrations of \IKreTm calibrations in LUX:  (left) S1 area in the radius vs. drift time plane, (middle) S2 area in the same projection (on a particular date), and (right) a time-dependent measurement of electron lifetime in the drift region.}
\vspace{10pt}
\end{figure}

\tdrsubsec[Xe-131m]{Metastable Xenon 131 (\IXeoTom)}
Calibrations of the position reconstruction itself, essential to rejecting spatially-varying external backgrounds and to precisely defining a fiducial volume, are easiest if the calibration source activity is homogeneously distributed throughout the active LXe.  The liquid mixing timescale in LZ may not allow \IKreTm to become fully homogeneous given its short decay timescale, and for this reason a \IXeoTom source is being developed as a long-lived alternative for the specific goal of achieving homogeneity (cf. requirements R-170002 and R-170007). Compared with \IKreTm, \IXeoTom decays at a higher energy (\SI{164}{\keV}) and with a longer half-life (\SI{11.9}{\day}).  The higher energy may result in the partial saturation of the central PMT of each S2 pulse, but the long half-life will guarantee homogeneity for the purposes of calibrating the position reconstruction metrics.  An additional use of this comparatively higher-energy internal source is in calibrating the xenon skin region (cf. requirement R-170008), which will possess a position-dependent detection threshold of roughly the same \SI{\sim100}{\keV} scale.

The hardware for \IXeoTom is nearly identical to the \IKreTm source: xenon carrier gas is flowed over a parent isotope that is emanating the calibration isotope.  In this case, the parent isotope is \IIoTo, which decays entirely to \IXeoTo with a half-life of \SI{8}{days}.  A small fraction of these \IIoTo decays (\SI{\sim0.39}{\percent}) are to the metastable \IXeoTom state rather than the ground state.  \IIoTo is readily available from the medical isotope industry in a variety of forms.  The most convenient form for our purposes is a solid pill, in which \IIoTo is doped into a solid matrix of anhydrous sodium phosphate.

The \SI{164}{\keV} energy of \IXeoTom decays is significantly larger than typical expected dark matter signals, and we estimate that a single PMT in the top array will experience mild analog saturation of its output signal at this energy. The saturation is a combination of exhausting the reserve charge in the coupling capacitors, and of anode saturation \cite{hamamatsu}.
Nevertheless, the result of this saturation is expected to be a negligible impact on the utility of the source for $(x,y,z)$ map-making.

\tdrsubsec[CH3T]{Tritium-Labeled Methane}
In addition to calibrating for position-dependent detector efficiency effects, an internal calibration source is necessary when measuring the low-energy electron recoil physics of the combined S1 and S2 energy scale, and the S2/S1 discrimination ratio (cf requirement R-170003). For this essential calibration, LUX has demonstrated the use of \IHT beta decay, exhibiting a broad energy spectrum with an \SI{18.6}{\keV} endpoint.

\IHT decays with a \num{12.3}-year half-life, this necessitates the ability to remove it from the xenon. This is way tritium-labeled methane is used, as it will be removed by the LZ getter (cf. requirement R-170114). LUX demonstrated that \IHT-labeled CH$_4$ could overcome the paired challenges of a long half life and the polymer diffusion. CH$_4$ exhibits a very small absorption into the PTFE, and is efficiency removed by the getter. The CH$_4$ purification timescale in LZ may be somewhat different from the general Xe re-circulation timescale (\SI{\sim2}{\day}), a result of the methane's differing solubility in the liquid and gas phases of Xe. In LUX, the CH$_4$ purification timescale was shorter than the xenon re-circulation timescale by a significant factor, and we expect the CH$_4$ purification timescale in LZ to be similarly shortened.

\tdrsubsec[RN-222]{Radon 220 (\IRnttz)}
\IRnttz is being considered as a potential large-S1 calibration source for the Xe skin region.  \IRnttz should not be mistaken for \IRnttt; \IRnttz has no long-lived radioactive daughters, thereby making it a compatible with the stringent low-background requirements of LZ.  The longest-lived daughter is \IPbtot, with a half-life of 10.6 hours.

\IRnttz decays through alpha decay at \SI{6.3}{\MeV}, followed shortly after (t$_{1/2}=$\SI{150}{\ms}) by a second alpha at \SI{6.8}{\MeV}.  Alpha emission produces a highly localized and proportionally high-recombination (high-light-yield) Xe response, ideal for calibrating the skin region.  Depending on the liquid mixing timescales, \IRnttz's daughter isotopes (\IPbtot, \IBitot, \IPotot, and \ITltze) are expected to accumulate on surfaces before their decay, providing a unique tool for understanding light and charge yield at xenon-surface interfaces.

Like \IKreTm and \IXeoTom, \IRnttz is most conveniently introduced by flowing a Xe carrier gas over a material containing the parent isotope, in this case \IThtte ($\tau_{1/2}$=\SI{1.9}{\year}).  \IThtte can be purchased in an electroplated form, which both maximizes \IRnttz emanation probability and minimizes the total material (and any possible impurity content).

\tdrsubsec[Int-RSD]{Internal Radioisotope Source Delivery}
An important component of the internal calibration strategy is an effective and precise system for injecting controlled amounts of activity into the Xe circulation flow path (cf requirement R-170102). Fundamentally, the source injection system is a specialized extension of the Xe circulation system, and shares the basic design fundamentals (including the use of stainless steel VCR plumbing standards) and the basic system risks, the most important being the risk of \IRnttt and \IKreF leakage from underground air into the circulating Xe gas.  The mitigation for this radon concern is shared with the circulation system: rigorous leak testing standards during assembly, paired with enveloping the plumbing by a radon-free purge gas rather than underground air.

A schematic of the source injection system is shown in Figure~\ref{CASf:injection-plumbing}. The source injection system will be entirely automated, both to reduce necessary onsite shift burden, and to reduce the risk to the xenon's purity from human operator error. This automation will be made possible through the general slow control system (described in Section~\ref{EDCS:SC}). A connection will be in place between the source injection system and the Xe sampling system (described in Section~\ref{XCSS:XeSamp}), making it possible to perform careful checks of injection material purity before introducing any material to the flow path.

The trace amounts of radioactive gas are transported into the main circulation flow path by a small flow of Xe carrier gas, supplied by a system-specific supply of high-pressure Xe.  This Xe will have gone through the same \IKreF-removal process as the bulk Xe (discussed in Section~\ref{XCSS:KrRem}), introducing no new radiopurity concerns.  This small Xe flow will pass through emanation sources (\IKreTm, \IXeoTom, etc.), regulated using a mass flow controller for a total mass flow roughly proportional to desired injected activity.  \IHT-labeled CH$_4$ is injected according to a slightly different recipe; this activity is stored in a pressurized bottle, which is allowed to fill a small evacuated `dosing region' volume to a precisely-measured pressure.  This dose volume will then be flushed by the Xe carrier gas into circulation, as with the flow-through sources.  The internal source injection system is being designed to be general and flexible, allowing for alternative injection procedures and for the possible use of internal sources not yet developed.

\begin{figure}[h]
\centering
\includegraphics[width=1.0\textwidth]{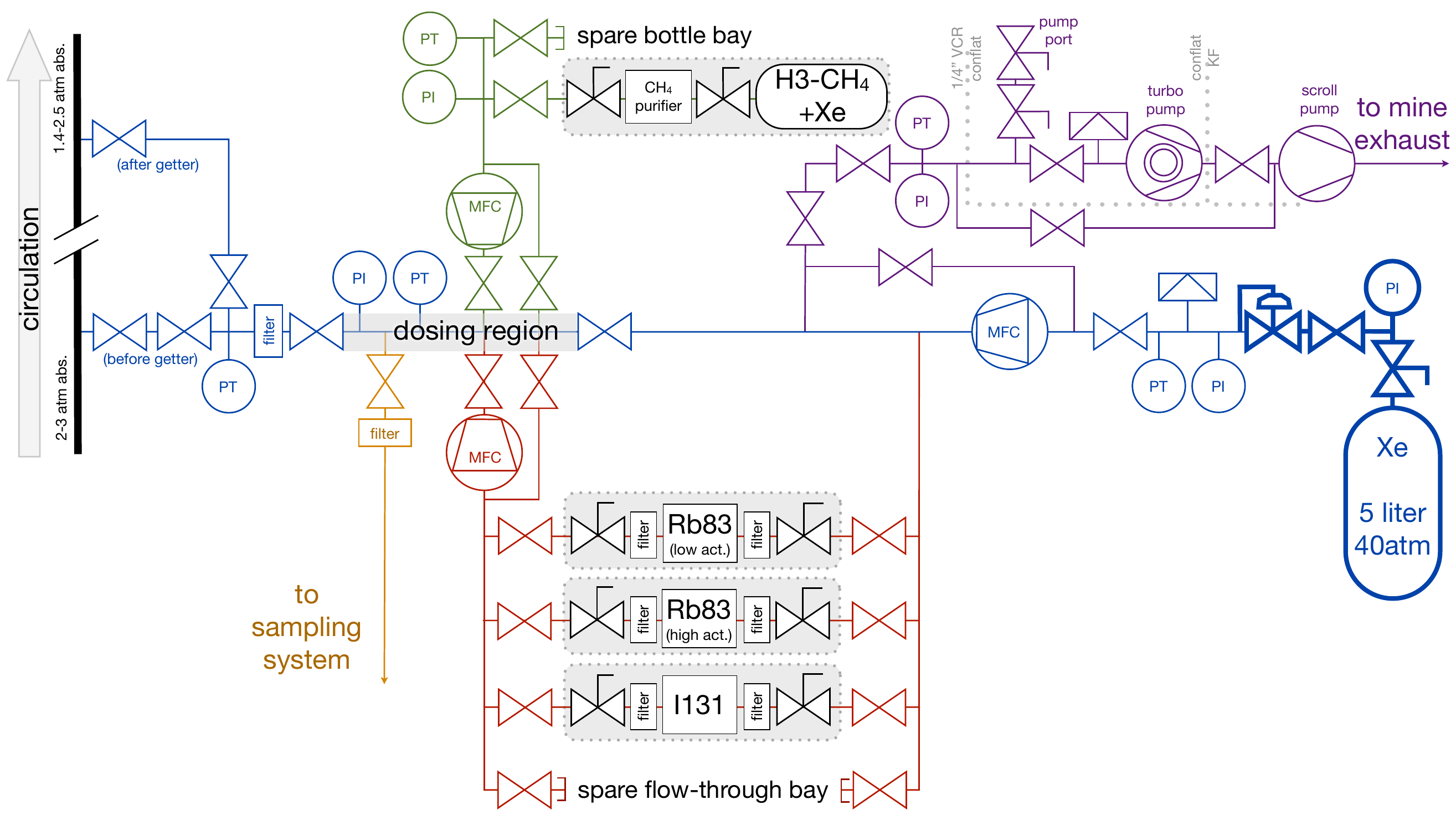}
\tdrfcaption[injection-plumbing]{Internal radioisotope source delivery system.}{The internal radioisotope source delivery system is a portion of Xe plumbing with the goal of releasing a precise portion of various gaseous radioactive isotopes into the main xenon circulation path.  This control is achieved through a combination of Mass Flow Controllers (MFCs) and Pressure Transducers (PTs), with flow from the dosing region into the circulation path (far left) motivated by a high-pressure Xe supply bottle (far right).}
\vspace{10pt}
\end{figure}

\tdrsec[csd]{External Radioisotope Source Delivery}
LZ will have three stainless steel vertical source tubes in the vacuum space between the inner and outer titanium cryostats, as shown in Figure~\ref{CROf:cryostatoverview}. The calibration source deployment (CSD) system will position neutron and gamma calibration sources in the source tubes.  The source tubes will have an inner diameter of \SI{23.6}{\mm}, large enough to accommodate deployment of commercial sources in nearly all cases. The necessary source strengths are well within the available range. Sources that cannot be obtained commercially in our requisite dimensions and rate will be fabricated by LZ (University of Alabama). The source tubes will be sealed at both ends and kept in a nitrogen atmosphere at a pressure of \SI{1.1}{\bar} to suppress contaminants entering the system and prevent the plating of radon daughters. The photoneutron and DD neutron sources require dedicated deployments and are discussed separately below.

\tdrsubsec[nSources]{Neutron sources}
A suite of four neutron sources will provide a broad-spectrum NR calibration, with the additional benefit of four distinct kinematic endpoints, as shown in Figure~\ref{CASf:CSDnsources}. In particular, these broadband sources are useful for measuring the NR band (cf. requirement R-170004). Also shown in the figure are the nuclear recoil detection efficiencies for 2- and 3-fold PMT coincidences. 3-fold coincidence is the baseline for WIMP search analysis, but in calibrations, where the total data taking time is much shorter, 2-fold PMT coincidence thresholds may be more appropriate for understanding low-energy response of the detector.

\begin{figure}[H]
\centering
\includegraphics[width=0.85\textwidth]{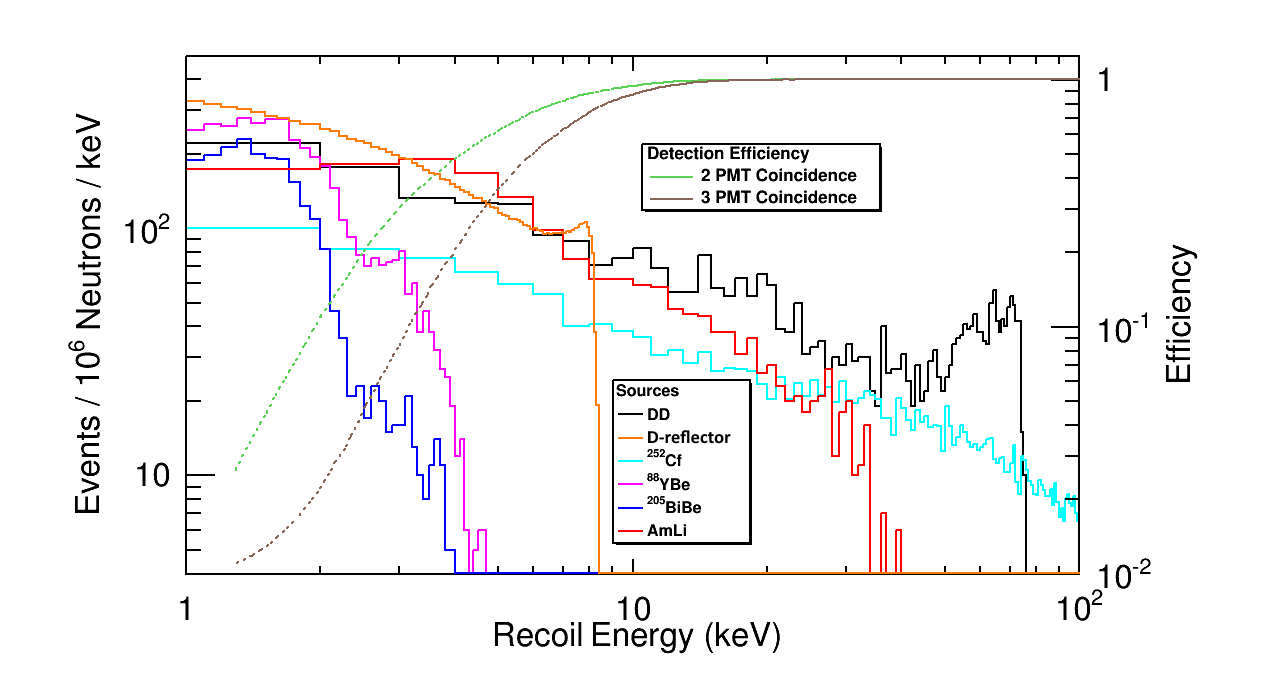}
\tdrfcaption[CSDnsources]{Recoil spectra from neutron sources}{Recoil spectra obtained from each of the primary neutron sources, DD (black), DD back-scatter off deuterium (orange), \ICftFt (cyan), YBe (red), BiBe (blue), and AmLi (magenta). 1 million neutrons corresponds to roughly three hours of calibration with a 100 neutron/s source. The scalings for DD and DD deuterium back-scatter ("D-reflector" above) are arbitrary. These calibrations can easily achieve the single-scatter statistics required and their main goal is to use the double-scatter events. Shown on the y-axis on the right is the nuclear recoil detection efficiency for 2 and 3 fold PMT coincidences.}
\vspace{10pt}
\end{figure}

AmBe $(\alpha,n)$ neutron sources have typically been the broad-spectrum neutron source of choice, as they cover the range from threshold to in excess of \SI{300}{\keV} recoil energy. The motivation to also use an AmLi source is the lower maximum neutron energy of
about \SI{1.5}{\MeV}, which results in a fairly distinct endpoint at about \SI{40}{\keV}.
Simulations show that the yield of single-scatter NR candidates with energy less
than \SI{25}{\keV} is comparable to AmBe but with an enhanced fraction of events at low recoil energy (less than \SI{10}{\keV}). The rates shown in Figure~\ref{CASf:CSDnsources} would be obtained in about three hours of live time, assuming 100 neutron/s source strength. It is notable
that the (\Pga,\Pn) yield is lower for AmLi than for AmBe, so that a higher americium activity of about \SI{2.5}{\milli\Curie} is required to obtain a source strength of 100 neutron/s. The sources will be encapsulated and there are no additional safety concerns.

A generic result of the (\Pga, \Pn) reaction is that the product nucleus may be in an excited state, leading to the release of a gamma. With AmBe, the most common gamma energy is \SI{4.4}{\MeV} and it occurs in 58\% of decays. No literature exists for this process with AmLi sources. Our own preliminary measurements indicate that final state gammas occur at the percent level.

\tdrsubsec[gammaSources-XeTPC]{Gamma Sources for Calibration of the Active Xe TPC}
External gamma sources are not required for any of the primary calibrations of the active Xe TPC. This is by design, as the active region is self-shielded against external gammas. Nevertheless, several important calibrations will be obtained from external gammas. These include studies of higher-energy backgrounds and signal fidelity near the edge of the TPC. A single \IThtte (\SI{2615}{\keV}) source serves this purpose and, in addition, helps provide a calibration of the outer detector. 

\tdrsubsec[gammaSources-SkinOD]{Gamma Sources for Calibration of Xenon Skin and the Scintillator Veto}
The detection thresholds for the Xe skin veto and the organic scintillator veto are \SI{100}{\keV} and \SI{200}{\keV} respectively. Various sources (cf. Table~\ref{CASt:sources}) will be deployed in the source tubes to verify this performance (cf. requirement R-170008 an R-170009). Calibration of the outer detector energy scale (cf. requirement R-170009) is achieved with a \IThtte source as shown in Figure~\ref{CASf:CSDCal}. The simulated response as a function of Z in the Xe skin is also shown in Figure~\ref{CASf:CSDCal}. For the timing synchronicity calibration between the outer detector and the skin, a \INatt source will be deployed (cf. requirement R-170010). The source produces simultaneous back-to-back \SI{511}{\keV} gammas which can be observed in the outer detector and the TPC to look for any difference in timing.

\begin{figure}[h]
\centering
\includegraphics[width=0.45\textwidth]{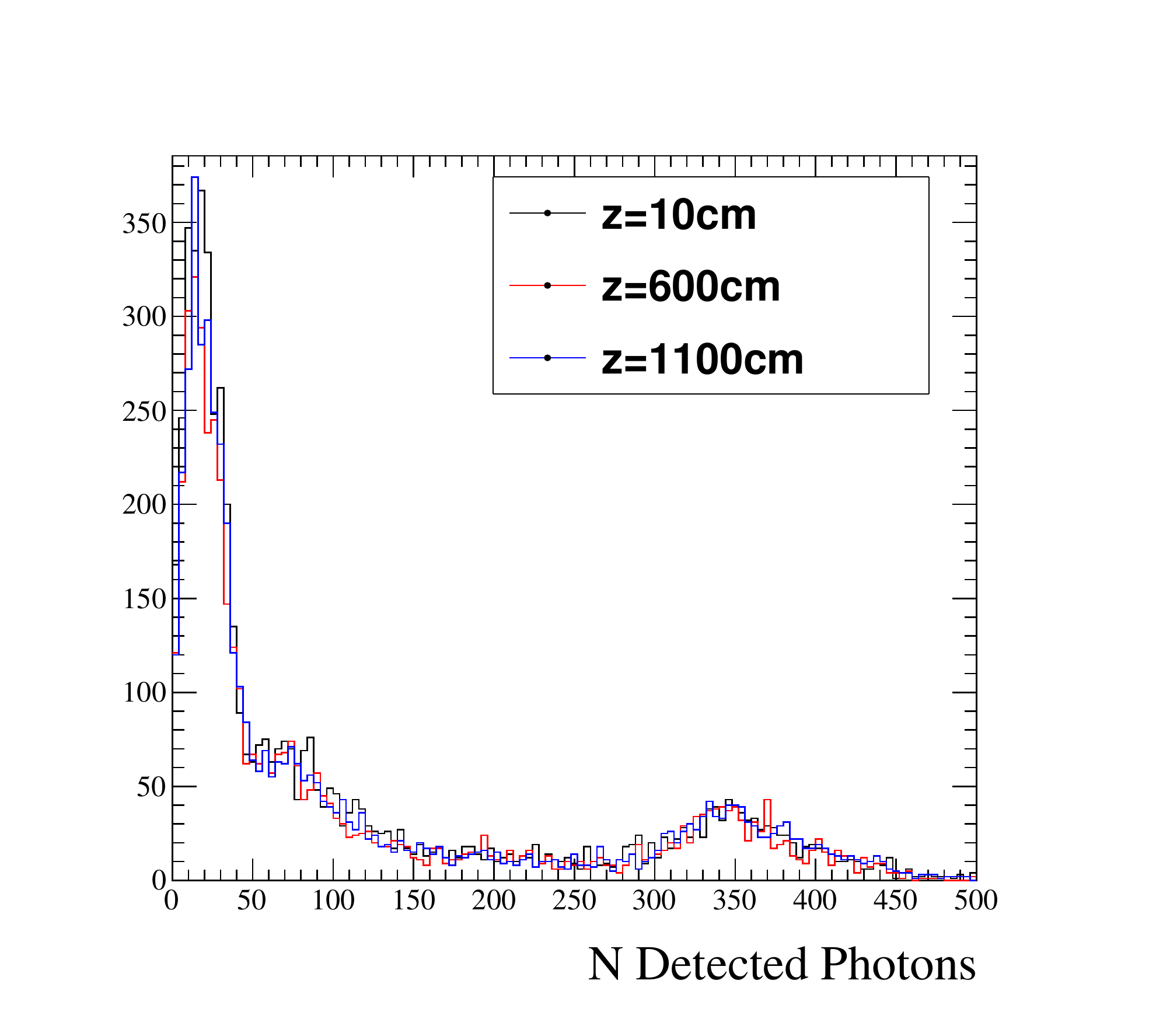}
\includegraphics[width=0.37\textwidth]{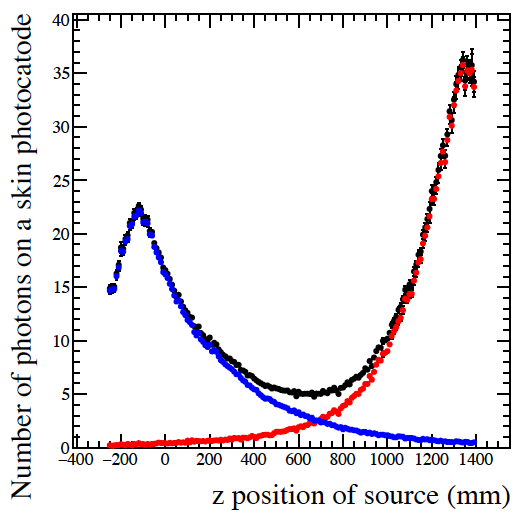}
\tdrfcaption[CSDCal]{Calibration of Xe Skin}{Number of measured photons detected versus Z-position in the outer detector (left panel, \IThtte source) and the xenon skin (right panel, \ICoFS source). The ability to study the performance of the Xe skin as a function of the Z-coordinate motivates the need to be able to locate a source to within \SI{5}{\mm} (cf. requirement R-170201)}
\vspace{10pt}
\end{figure}

\tdrsubsec[CSD]{Calibration Source Deployment (CSD)}
Figure~\ref{CASf:CSD-prototype-TD-drawing} shows the design of a prototype assembly developed to define the core design and components of the CSD. The source deployment system consists of a stepper motor (a SANMOTION F2 2-phase stepping motor model SH2141-5511 with a resolution of \SI{1.8}{\degree} manufactured by SANYO DENKI Co) coupled to the planetary gear-head GP 22A with a 19:1 reduction (manufactured by Maxon). The gear unit couples the stepper motor to a drum holding the deployment filament and source. The deployment filament is a strong thin nylon composite (\SI{\sim0.1}{\mm} diameter, maximum load \SI{12}{\kg}) that carries the load of the source assemblies (\SI{\sim100}{\g}) with a \num{4}$\times$ safety factor. The CSD covers the full \SI{6}{\m} length of the upper and lower calibration tube from the top of the water tank to beyond the cathode level. The CSD mechanics is housed in a K50 Tee-piece that couples via a connection chamber sideways to the calibration source tube making it a very compact system. The stepper motor and gear are held in a support structure clamped to the T-piece. A wheel on a lever feeds the filament from the drum into the calibration tube. Each calibration tube will be fitted with its individual deployment system; the three systems are configured to enable simultaneous operations. A prototype using the same design has been tested at RAL to ensure that thousands of source deployments are possible without failure (cf. requirement R-170202).

As additional control on the z-position of the calibration sources, each deployment system will be fitted with a laser based position monitoring system. For this an ILR1181-30 Micro-Epsilon laser ranger will be coupled to the top of the connection chamber allowing an accurate (better than 2mm precision) reading of the source position. The monitoring system will be incorporated in a feedback protocol that drives the CSD system and thus ensures the z-position accuracy of \SI{\pm 5}{\mm} is met (cf requirement R-170201).

Each system is driven by a ZYBO FPGA board and a PMODSTEP daughter board both manufactured by Digilent. This system allows computer control of the deployment, provides feedback of the operational parameters and will be fully integrated in the Run Control structure.

An unacceptable failure mode of this system would be a source detached from the deployment filament dropped to the bottom of the tube. As a control against this eventuality, the sources are fitted with a ferromagnetic component which allows retrieving sources with a magnet. A full scale mock-up of the design of this system is constructed and extensively tested.
\begin{figure}[h]
\centering
\includegraphics[width=0.8\textwidth]{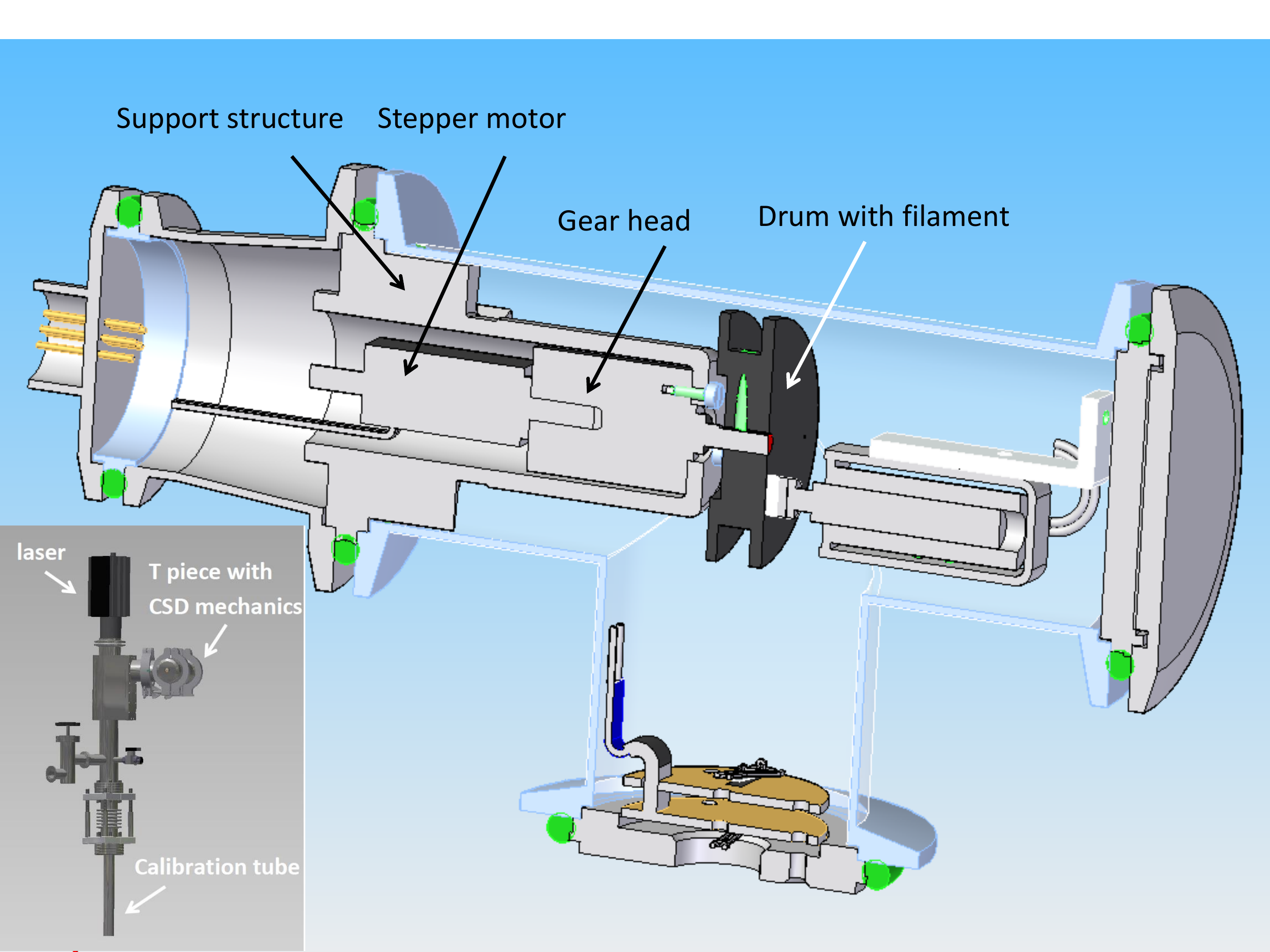}
\tdrfcaption[CSD-prototype-TD-drawing]{Prototype of core of CSD system}{The prototype assembly developed to define the core design of the CSD system: the stepper motor and gear are held in a support structure clamped to the T-piece. The T-piece couples sideways via a connection chamber to the calibration tube as shown in the inset.}
\vspace{10pt}
\end{figure}

\begin{wrapfigure}{R}{0.25\textwidth}
\centering
\includegraphics[width=0.25\textwidth]{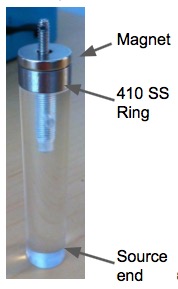}
\tdrfcaption[radioisotope-capsule-prototype]{Prototype dummy source capsule}{Prototype dummy source capsule showing 410SS ring with magnetic recovery system attached.}
\end{wrapfigure}

\tdrsubsec[radioisotopes]{Radioisotope Capsule Design}
Most sources for use in the CSD will be commercial Eckert \& Ziegler (E\&Z) type R sources. The source form factor is a 5" long X 0.625" diameter acrylic cylinders with the source epoxied and encapsulated at one end. The bare acrylic end of the cylinder (the end without the source) will be tapped with an M4 hole, in order to connect to the CSD. The monofilament is captured in a vented screw, which itself secures a martensitic stainless steel (410SS) ring collar to the top of the acrylic cylinder. The use of this ring allows a simple, robust magnetic recovery system can be deployed in the unlikely event of a break in the monofilament which suspends the source.

The AmLi source will be custom fabricated at the University of Alabama since commercial versions are not available in our requisite form factor. Custom sources will have the same form factor and mating structure as commercial sources. Preliminary activities of these sources have been determined. Simulation effort to validate the choices are on-going (cf. requirement R-170301).

\tdrsec[photoneutron]{Photoneutron Sources}
As discussed in Section~\ref{SIPSss:CENNS}, coherent scatters from \IBe neutrinos pose a potential background in LZ as they are difficult to distinguish from a light mass WIMP. However, higher WIMP masses are unaffected by this issue assuming the signal from such recoils is well understood. Having a calibration targeted at understanding the response of LZ to these low-energy scatters is critical. Because of their well-defined kinematic endpoint, photoneutron sources are the most important tool for this purpose and help to fulfill requirement R-170005.

\tdrsubsec[PhotoNeutronSources]{Physics of Photoneutron Sources}
Photoneutron sources exploit (\Pgg,\Pn) reactions on nuclei to produce neutrons. Suppose there is an incident gamma of energy E$_{\gamma}$ incident on a nucleus. That nucleus has some threshold, Q, for emission of a neutron by ($\gamma$, n). If E$_{\gamma}$ $>$ Q there is some finite probability for the nucleus to absorb the gamma and emit a neutron with energy \cite{Knoll1989}
\begin{equation}
E_{n} (\theta) = \frac{M(E_{\gamma} - Q)}{m + M} + E_{\gamma}cos(\theta)\sqrt{\frac{(2mM)(E_{\gamma} - Q)}{(m + M)^{3}}}
\end{equation}
where M is the mass of the nucleus and m is the mass of the neutron. The second term, which has an angular dependence, as $\theta$ is defined as the emission angle of the photoneutron relative to the direction of the incident gamma ray. The angular dependence causes a relative difference of 4.6 \% in the emitted neutron energy between $\theta$ = \SI{0}{\degree} and $\theta$ = \SI{180}{\degree} for \IYee Be source (described below). This energy difference is similar to the end point energy difference caused by isotopic spread (5 \% between \IXeotn and \IXeoTs). Thus, photoneutron sources can be considered to produce mono-energetic neutrons.  One example of a photoneutron source used by \cite{Collar:2013xva} was a \IYee Be source, with E$_{\gamma}$ = \SI{1.836}{\MeV} and Q = \SI{1.666}{\MeV}, producing \SI{153}{\keV} neutrons. A technical challenge of deploying a photoneutron source is that the cross sections for (\Pgg,\Pn) are generally very low. For example, in \cite{Collar:2013xva}, a \SI{1.85}{\mega\becquerel} \IYee source was used and \num{\sim330} neutrons/second were generated. That is \num{>5000} gammas per neutron. For practical purposes such sources need a large amount of lead or tungsten shielding to reduce the gamma rate in the detector, while preserving a useful neutron flux.

\begin{figure}
\centering
\includegraphics[width=0.48\textwidth]{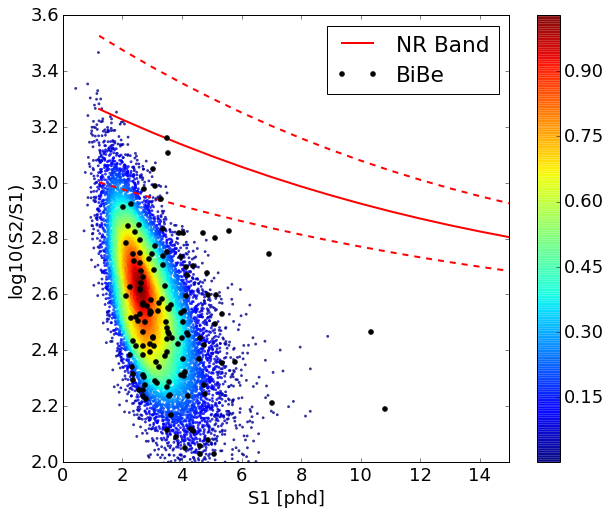}
\includegraphics[width=0.48\textwidth]{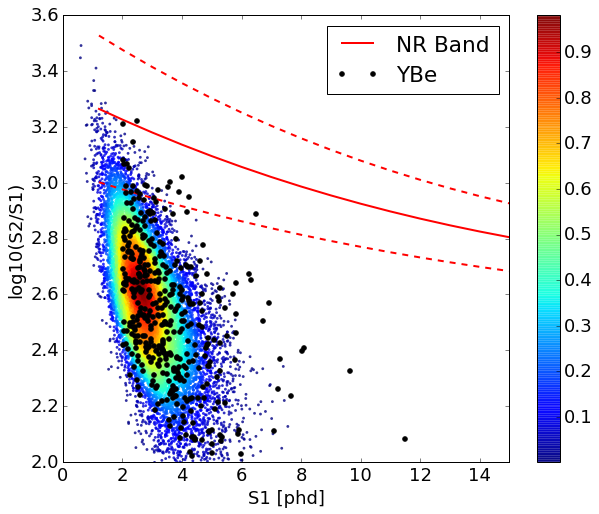}
\tdrfcaption[YBe-BiBe-8B]{S1/S2 spectra for \IBe solar neutrino coherent scattering}{Comparison of S1 versus S2 spectra for \IBe solar neutrino coherent scatters versus \IBitzF Be (top)  and \IYee Be (bottom) in LZ. The \IBe spectrum is a two dimensional PDF represented by the arbitrary color scale. The black dots are the neutron events and the red curves are the nuclear recoil band. The neutron events come from simulation. Events that appear to be at higher energy are the result of multiple scatters, some of which would pass all cuts in a real analysis. Here a 2-fold PMT coincidence cut is assumed, as the live time of these calibrations will be short compared to WIMP search and all the real events will have short drift times. This results in a very low rate of accidental S1+S2 coincidences. The above plot corresponds to 30 hours of calibration with a 100 neutron / s, indicating that with a reasonable activity source, one can perform this calibrations in a few days (cf. requirement R-170404)}
\end{figure}

Two sources of interest for LZ are \IYee Be and \IBitzF Be. \IBitzF Be creates neutrons of three different energies, the dominant one having an energy of  \SI{88.5}{\keV}, with a nuclear recoil end point of \SI{2.7}{\keVnr} in Xe. The nuclear recoil end point from a \SI{152}{\keV} neutron from \IYee Be is \SI{4.6}{\keVnr}. Figure~\ref{CASf:YBe-BiBe-8B} shows a comparison of the signals in LZ from \IBe to the photoneutron sources. \IYee Be bounds the end point for the \IBe spectrum while \IBitzF Be is more similar to the spectrum which would be observed from a \num{1000}-day exposure. Note that because the source is above the detector, most of the neutron scatters occur near the gate and have short drift times. Although the calibrations will contain a large number of gamma interactions, they are easily separated from the nuclear recoils by their apparent energy. The ER events are dominated by neutron capture gammas, as there is enough tungsten shielding in the design to make the gammas directly from the \IYee or \IBitzF decay negligible.

\tdrsubsec[SourceDeployment]{Photoneutron Source Deployment}
Photoneutron sources will be deployed in a \SI{20}{\cm} diameter $\times$ \SI{20}{\cm} tall cylindrical tungsten alloy pig from above LZ. The pig will weigh \SI{116}{\kg} and the cryostat is rated to hold this weight (cf. requirement R-170401). During calibrations the pig will sit on top of the outer cryostat vessel, on axis. Figure~\ref{CASf:Photoneutron-Deployment} shows the individual components and the overall deployment scheme. The pig will lowered through a stainless steel guide tube from three points. The guide tube has a funnel top to accept the pig and holes in the side to displace water. There must be minimal water trapped underneath the tungsten (cf. requirement R-170402) and this is achieved by keeping the relevant surfaces relatively flat. There are two critical safety measures designed into this process to protect the cryostat and the experiment. First, the water above the cryostat acts as a cushion, slowing the descent of the pig. Second, the three lifting points are redundant, so even if one cable breaks, the other two can support the weight of the pig. In the unlikely event that the pig were to get stuck in the guide tube, the source is separately removable. Furthermore, the entire guide tube can be removed with the pig in it.

\begin{figure}
\centering
\includegraphics[width=0.225\textwidth]{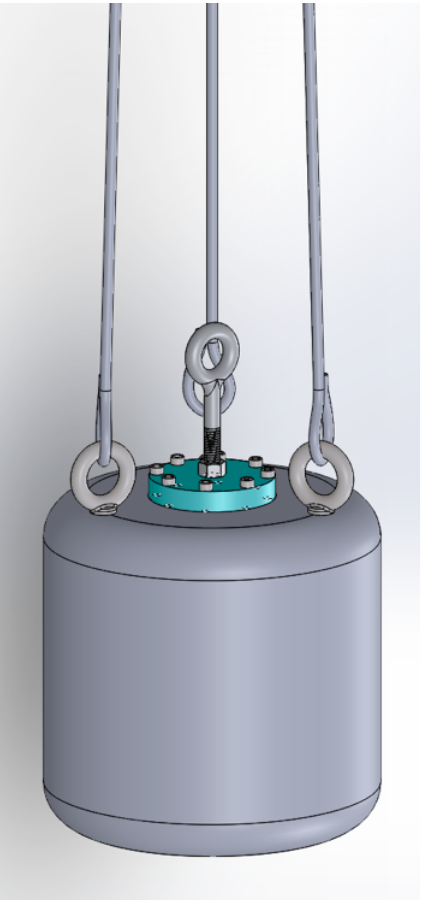}
\includegraphics[width=0.4\textwidth]{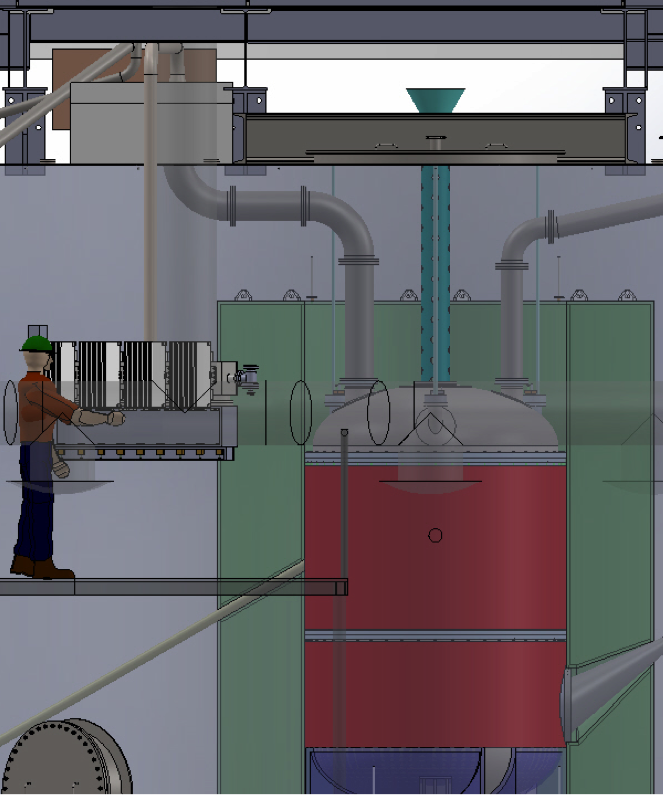}
\includegraphics[width=0.28\textwidth]{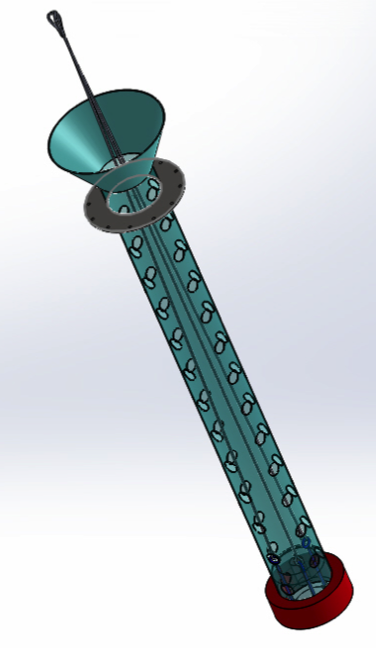}
\tdrfcaption[Photoneutron-Deployment]{Photoneutron pig, various views}{From left to right: Photoneutron pig (\SI{20}{\cm} outer diameter $\times$ \SI{20}{\cm} height tungsten) suspended from three points, side view of the guide tube in-place on top of the cryostat, guide tube (\SI{8.25}{\inchl} outer diameter $\times$ \SI{8.0}{\inchl} inner diameter stainless steel) on its own with the funnel top to accept the pig.}
\end{figure}

\tdrsec[ddneutron]{Deuterium-Deuterium Neutron Source}
A deuterium-deuterium (DD) neutron source, the Adelphi Technologies DD109 neutron generator, will be deployed in the LZ experiment for nuclear recoil calibrations. The generator that will be set up outside of the LZ water tank can produce up to \num{e9} mono-energetic \SI{2.45}{\MeV} neutrons per second into $4\pi$. Conduits in place within the LZ water tank provide a path for neutrons to travel through the water tank and outer detector to reach the Xe volume within the detector (Figure~\ref{CASf:Deut-diag}). So as not to degrade the energy of the neutrons, the amount of water in their path needs to be kept small (cf. requirement R-170501). This novel in-situ NR calibration technique has been successfully implemented in the LUX experiment to calibrate the absolute energy response of NR in LXe down to \SI{1.1}{\keVnr} for light yield (Ly) and \SI{0.7}{\keVnr} for charge yield (Qy) with significant improvements to the calibration uncertainty\cite{akerib:2016mzi} and has helped improve the LUX detector sensitivity to low-mass WIMPs by orders of magnitude\cite{Akerib:2015rjg}.

\tdrsubsec[DD-nSource-Motivation]{Physics motivation of Deuterium-Deuterium Neutron Sources}
This technique will be further upgraded in LZ to explore even lower energy NR response to reduce calibration uncertainty. Four types of measurements are planned to be carried out in LZ using the D-D neutron source as described in the following subsections. 

\tdrsubsubsec[NrQyLyCalib]{NR Qy / Ly calibration of the LZ detector}

In the DD calibrations implemented in LUX, double-scatter events in LXe due to mono-energetic \SI{2.45}{\MeV} neutrons are used to calibrate the NR charge yield (S2) response\cite{akerib:2016mzi}. The absolute deposited energy in the first scatter can be determined by measuring the neutron scattering angle between it and a second scatter. Given the deposited energy, the Qy can be determined by estimating the number of electrons produced at the first scatter. The light yield (S1) can then be inferred by using the calibrated S2 signal to assign an energy deposited to single-scatter events\cite{akerib:2016mzi}. This calibration technique using \SI{2.45}{\MeV} neutrons covers WIMP search energy range (requirement R-170010) from threshold (\SI{6}{\keV}) up to \SI{30}{\keV} (Figure~\ref{CASf:CSDnsources}) with an additional endpoint at \SI{74}{\keV}.

\tdrsubsubsec[D2OCalib]{Calibrating LZ detector using Reflected Neutrons from D2(O) Target} 

As discussed in Section~\ref{CASS:photoneutron}, an understanding of the LZ response to recoil energies at the threshold (requirement R-170005) is critical to probe low mass WIMP and \IBe solar neutrino signals. By placing a deuterium-loaded reflector behind the DD generator and collecting the neutrons that are reflected at a near-\num{180} degree angle (Figure~\ref{CASf:Deut-diag}), the generator's direct \SI{2.45}{\MeV} neutron flux can be converted into a quasi-mono-energetic neutron beam with a minimum energy of \SI{272}{\keV}. These lower energy neutrons can be used to calibrate in a new energy regime. Lower energy neutrons provide smaller uncertainty, because the angles are more favorable. In addition, the recoil spectrum endpoint in Xe is reduced from \SI{74}{\keVnr} to \SI{8.2}{\keVnr}, thus confining the neutron scatters to within this lower energy region of interest (\SIrange[range-units=single]{1}{8}{\keV}). In addition, the slower incident neutron speed would provide greater separation in S1 times for double scatters, which would assist in the direct Ly calibrations planned for LZ (Section~\ref{CASSss:DirectLy}).

\begin{figure}[h]
\centering
\includegraphics[scale=0.09]{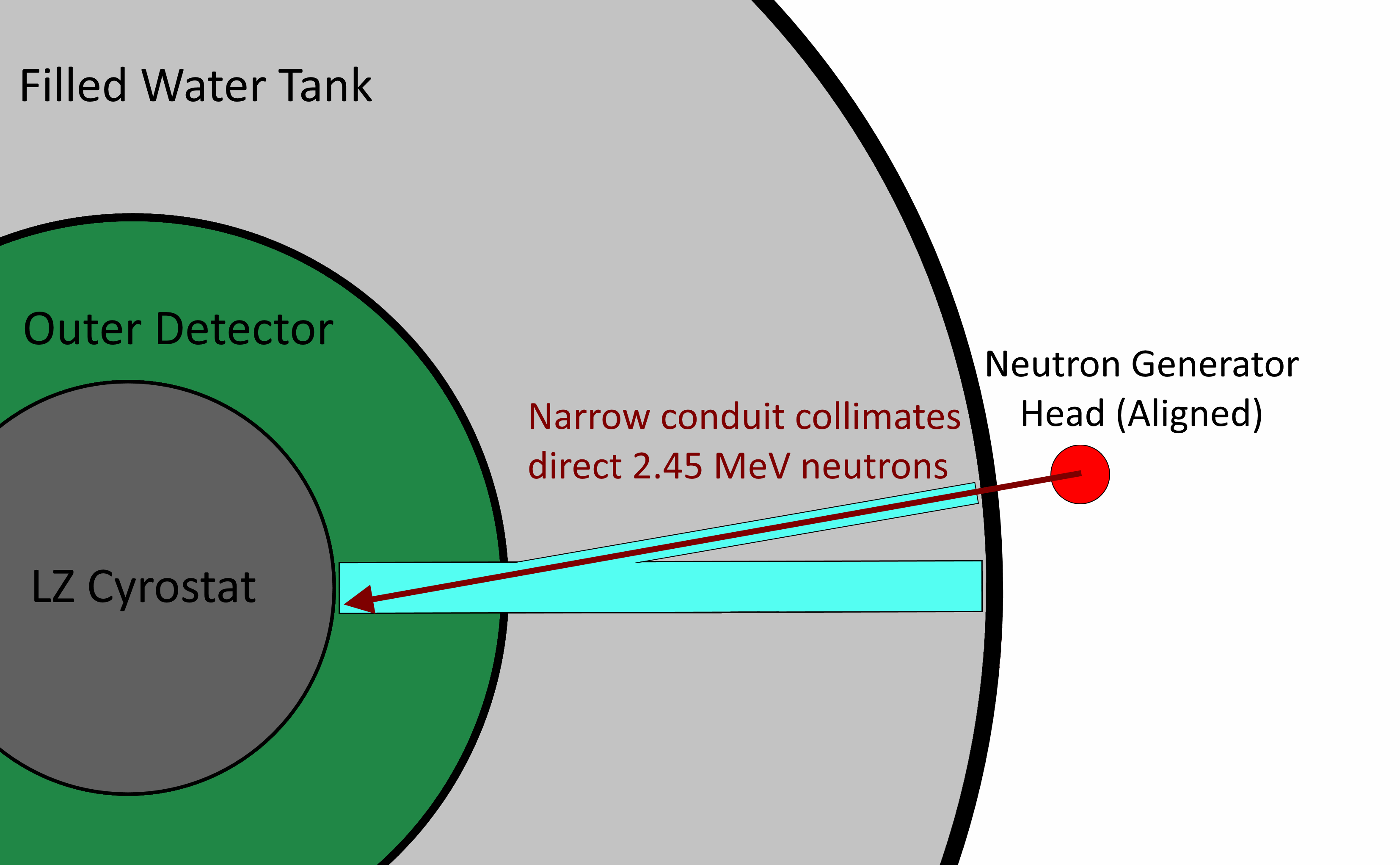}
\includegraphics[scale=0.09]{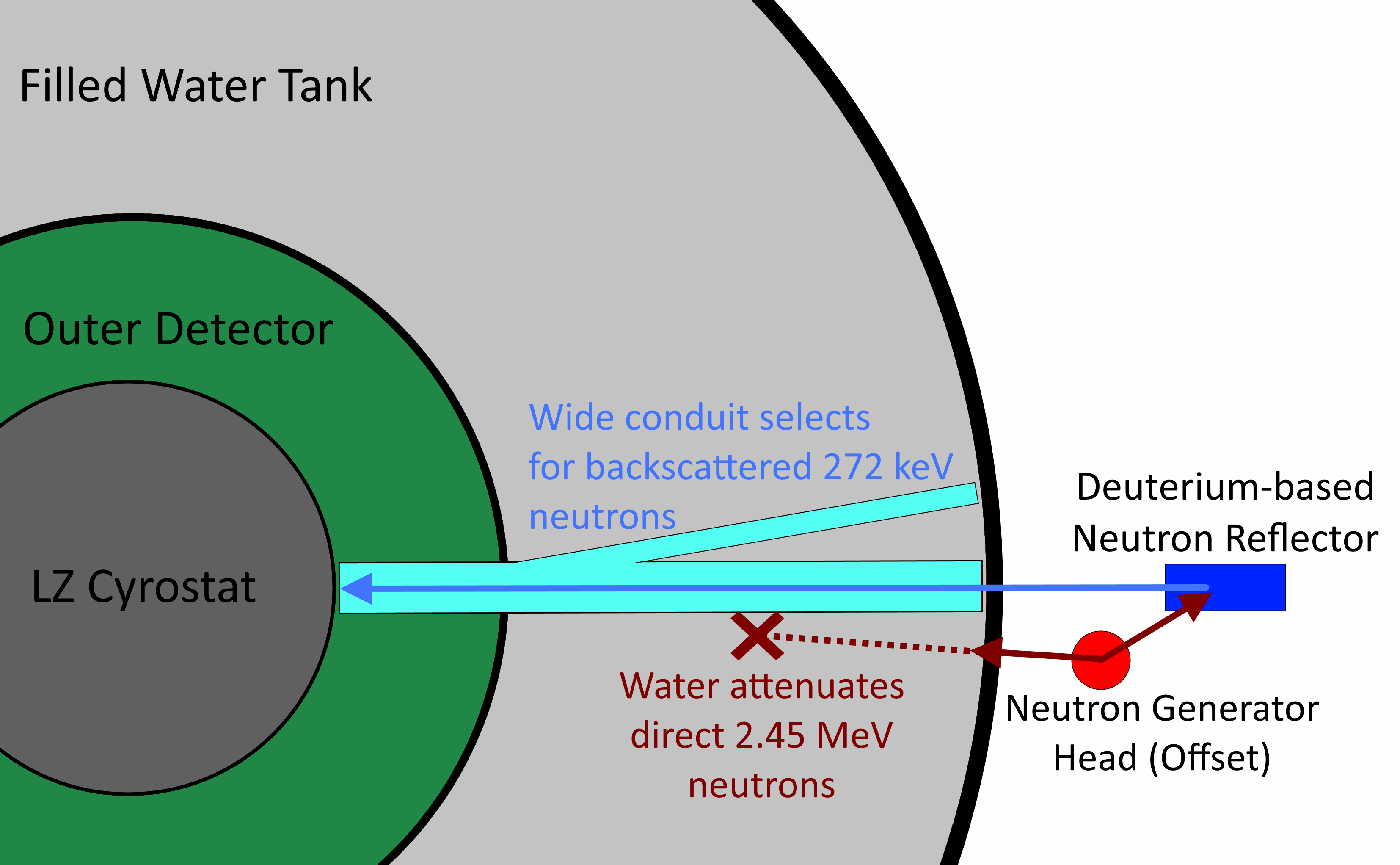}
\tdrfcaption[Deut-diag]{Schematic setup for the utilization of a deuterium-based backscatter reflector}{Approximate setup of the DD generator and neutron conduits. The y-shaped conjoined tube allows for a choice between finer collimation with lower flux by using the smaller, \SI{5.25}{\cm} inner diameter tube and broader collimation but greater flux by using the larger \SI{15.4}{\cm} inner diameter tube. The smaller tube joins the larger one in a bonded joint just outside the outer detector, though the path of the neutrons from the smaller tube continues up to the cryostat wall. (Left) The configuration for \SI{2.45}{\MeV} calibration. The neutrons go directly from the generator to the detector through the narrow conduit. (Right) The configuration for \SI{272}{\keV} calibration. The generator head (red) is offset from the neutron conduit to prevent direct neutron flux into the xenon via water attenuation by the water tank, while neutrons reflected off the deuterium reflector (blue) may enter via the air-filled neutron conduit (cyan). }
\vspace{10pt}
\end{figure}

\tdrsubsubsec[NeutronCalib]{Calibrating LZ Detector with Neutron Events with no S1 Light}
Another technique to calibrate LZ response at threshold (requirement R-170005) is to explore neutron events with no scintillation light. This requires an additional tool to establish t$_{0}$ for the electron drift of a neutron event, which is usually determined by S1. This additional tool is given by shorter neutron production pulses, resulting in tighter bunches of neutrons. In LZ, our plan is to upgrade the Adelphi Neutron Generator such that it is capable of reducing the neutron bunch width down to the few-microsecond level(as shown in Figure~\ref{CASf:shortTTL}). The neutron trigger pulse will be digitized by one of the DAQ channels to synchronize with other channels that digitize PMT signals. This allows identification of t$_{0}$. There are two direct benefits of implementing this technique. First, it provides an additional timing cut that helps significantly to reduce neutron analysis background events. Second, it allows probing no-S1 neutron events; the measured number of DD scatters with a known number of zero-collected-photon S1 pulses provides considerable statistical leverage and provides a stronger constraint on Ly because the number of no-S1 events can serve as a normalization for the observed number of S1>0 events.

\begin{figure}
\centering
\includegraphics[width=0.62\textwidth]{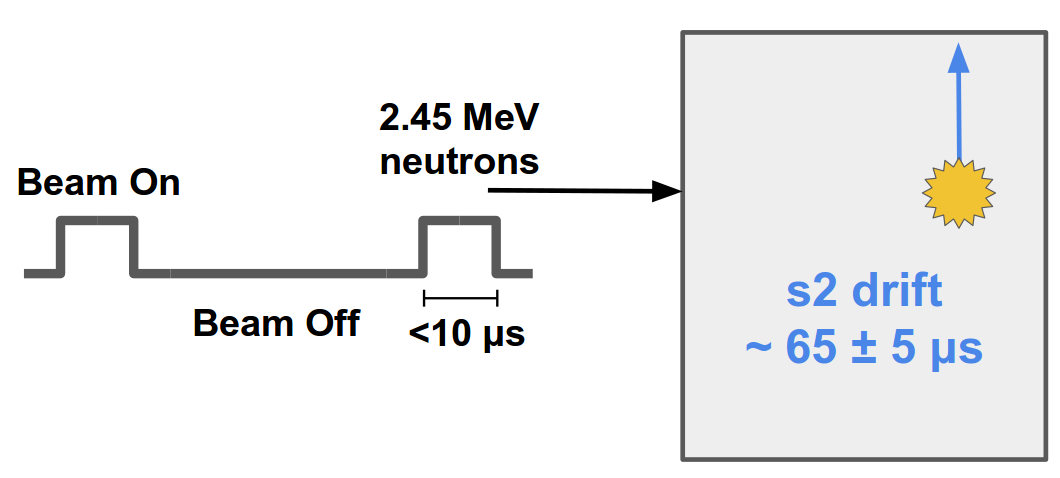}
\tdrfcaption[shortTTL]{LZ DD square wave neutron trigger pulse}{The square wave on the left indicates the DD trigger pulse. The schematic on the right indicates a no-S1 single-scatter neutron event based on a tube centered \SI{12.5}{\cm} from the LXe surface. The $z$ position resolution from the DD trigger is comparable to the $x,y$ position of such events. As described in the text, this technique gives additional capability to probe nuclear recoil response at threshold.}
\vspace{10pt}
\end{figure}

\tdrsubsubsec[DirectLy]{Direct Ly Measurement Using Double (Multiple) Scatter Events}

A direct light yield measurement will significantly reduce the systematics in the existing technique (Section~\ref{CASSss:NrQyLyCalib}). A direct light yield measurement in LUX is not feasible because S1 light from two scatters overlaps in time. However, given that the LZ LXe volume is at the meter scale, it is possible to separate S1 signals from the first two scatters in a double (multiple) scatter D-D neutron event by leveraging the time of flight of neutrons in greater path lengths in the LZ LXe volume, especially for \SI{272}{\keV} neutrons, whose speed is a factor of \num{3} lower than \SI{2.45}{\MeV} neutrons but have similar scattering cross sections, and thus take significantly longer for the second scatter to occur on average Figure~\ref{CASf:s1-sepa}. This will thus allow a direct light yield measurement by using double (multiple) scatter events in a similar way to measuring Qy in the standard calibration, helping eliminate the systematics due to S2-energy mapping in the indirect Ly measurement.

\begin{figure}
\centering
\includegraphics[width=0.62\textwidth]{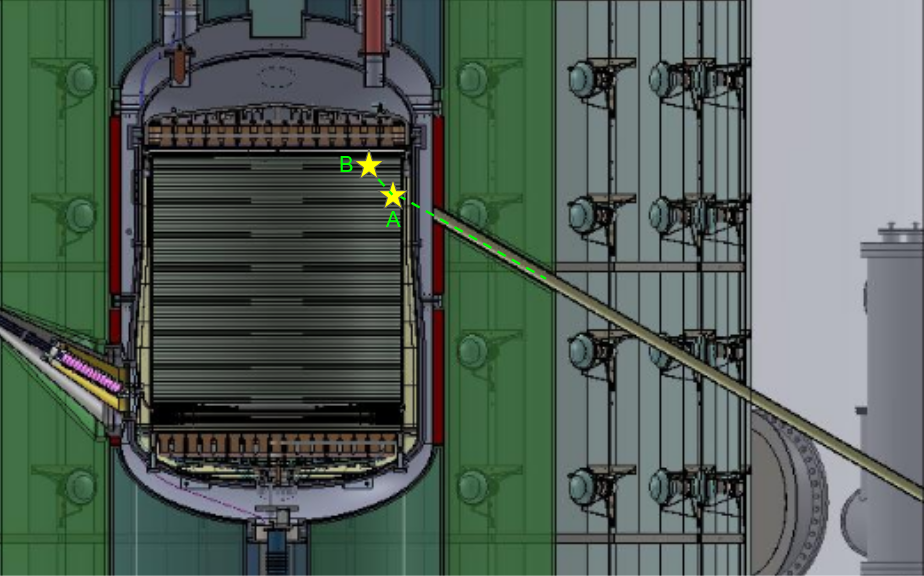}
\includegraphics[width=0.62\textwidth]{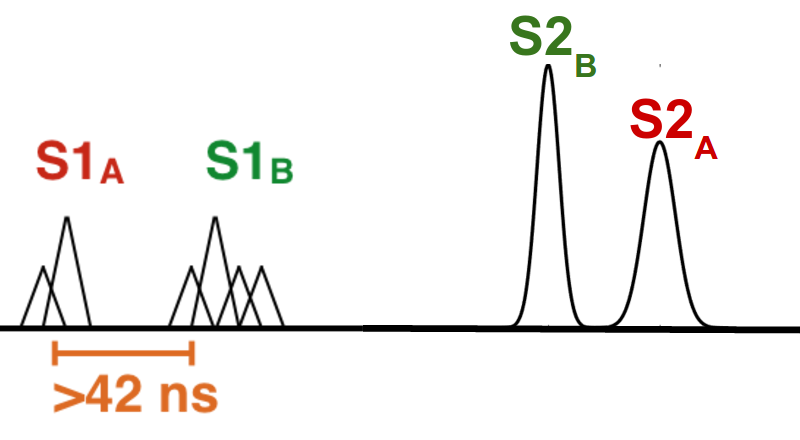}
\tdrfcaption[s1-sepa]{Solid model of LZ detector with angled neutron tube}{Top shows the solid model of LZ detector with angled neutron tube. The leveled tube is not shown in this figure. A cartoon neutron double scatter event in the scenario of separated S1s is also shown in the figure. Bottom shows the expected waveform of a neutron double scatter event with two separated S1s.}
\vspace{10pt}
\end{figure}

\tdrsubsec[DD-Source-Deployment]{DD Source Deployment and Neutron Rates}
The generator, manufactured by Adelphi Technology Inc, uses deuterium-deuterium (DD) fusion to generate neutrons at \SI{2.45}{\MeV}. D2 gas is pumped into a chamber where it is ionized via RF induction discharge. A titanium plate is biased using a negative high voltage between \SIlist[list-units=single]{-80;-125}{\kV}, which draws the D+ ions. The ions embed in the titanium plate, forming titanium hydrate. Subsequent ions strike the embedded ions and eject neutrons via the fusion reaction $\textrm{D} + \textrm{D} \rightarrow \IHeT + \Pn$. At maximum operating parameters, the generator is designed to output \num{e9} n/s into $4\pi$. \cite{adelphi}

The DD generator is located outside the LZ water tank. The neutrons will be collimated via two sets of tubes inside the water tank, each of which has two options for the diameter: a \SI{16.8}{\cm} outer diameter tube and a \SI{6.03}{\cm} outer diameter tube. (See Figure~\ref{CASf:Deut-diag} for a drawing of the two-diameter approach and Figure~\ref{CASf:s1-sepa} for a depiction of the angled tube.) During WIMP search, the tubes are filled with water, thus maintaining the water shielding around the detector. During calibrations, the tubes are drained of water and filled with a nitrogen purge, which allows the neutrons to pass into the detector. One tube is horizontal and will be centered \SI{12.5}{\cm} below the LXe surface, while one is angled 20 degrees to the horizontal, the center of which enters the LXe volume \SI{40.5}{\cm} below the LXe surface. The horizontal tube mimics the implementation of the LUX DD tube, which has been used to great effect, while the angled tube seeks to take advantage of the preferential forward scattering via the higher precision in $z$-coordinate reconstruction.

The generator itself will be mounted on a platform supported by a modified SLA-10 (Super Lift Advantage) Genie lift. This lift can be moved to either neutron conduit site and will be capable of raising the generator head to the entrance of both neutron conduits. Form-fitting generator shielding will adequately reduce neutron flux in the Davis Cavern, and plugs in the shielding at both conduit angles will be added or removed to allow for maximum flux into the desired conduit.

For the DD backscatter calibrations, the generator head is placed off-axis of the neutron conduit while a backscatter target is aligned with the conduit. As the neutrons are produced by the generator head, those hitting the deuterium backscatter target scatter at a near \SI{180}{\degree} angle and are sent down the conduit to the the interior of LZ, while direct neutron flux is attenuated by the water tank. Kinematically, the \SI{180}{\degree} backscatter results in a reduction in neutron energy from \SI{2.45}{\MeV} to \SI{272}{\keV}. Losses in neutron flux due to backscattering will be compensated by a heightened generator neutron flux (up from \num{4e6} n/s to \num{1e9}). The diameter of the neutron conduit (\SI{15.4}{\cm} ID) is driven by the need to get enough flux in deuterium back-scatter calibration.

\tdrsec[calrates]{Calibration Rates}
In general, event rates in the LZ detector of course depend on source location, type and energy spectrum. For neutron calibrations, a number of generally applicable conclusions can be obtained from the representative case of \SI{2.45}{\MeV} neutrons emanating from the wall of the outer Ti cryostat vessel. The scenario is identical to what would be realized during a DD neutron calibration, assuming neutrons that strayed out of the conduit do not contribute appreciably to the event rate in the Xe TPC. The scenario is also qualitatively similar to the foreseen AmLi calibration, obtained from a source located in one of the three external source tubes. 
\tdrsubsec[Gd-Doping]{Effect of Gd doping in liquid scintillator}
A generic concern for neutron calibrations of the Xe TPC is that Gd doping of the liquid scintillator in the outer detector will cause a large rate of gamma events in the Xe TPC. Monte carlo simulations suggest that this concern is not warranted. As shown in Figure~\ref{CASf:ncapGd}, the event rate due to neutron captures on Gd is only slightly larger than it would have been with undoped liquid scintillator, and only in the energy window \SI{2.2}{\MeV} $\lesssim E \lesssim$ \SI{4.0}{\MeV}.

\begin{figure}
\centering
\includegraphics[width=0.62\textwidth]{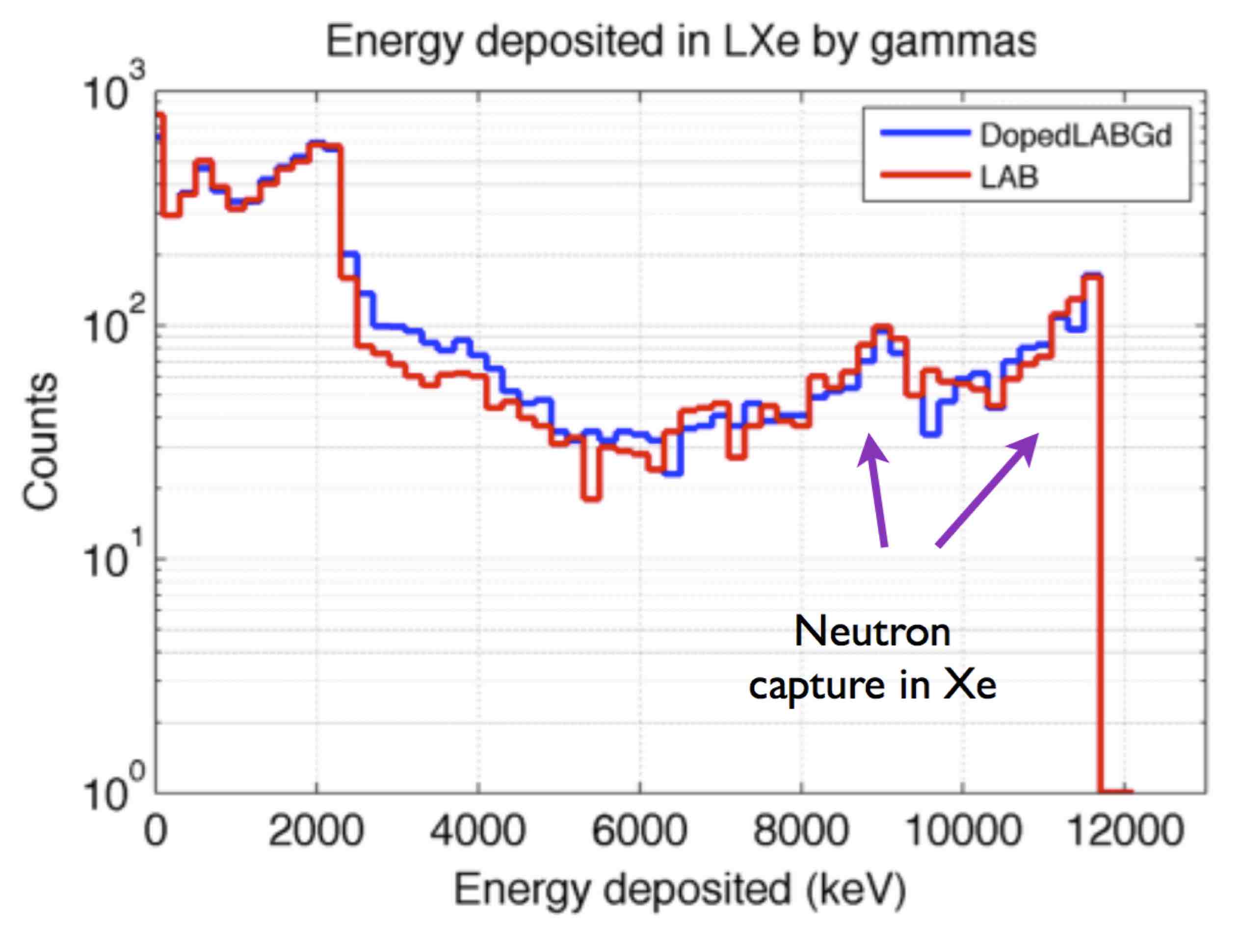}
\tdrfcaption[ncapGd]{Event rate due to all gammas in the active xenon TPC from neutron captures on Gd}{Event rate due to gammas in the xenon TPC from neutron captures on Gd.}
\end{figure}

\tdrsubsec[gammaCaptures]{Effect of gamma captures on useful neutron event rates}
A key number for neutron calibrations is the fraction of useful events. As mentioned above, this will depend on the geometry and energy of the source. Table~\ref{CASt:singlesdoubles} provides numbers for the simplified, representative case described at the beginning of Section~\ref{CASS:calrates}. Shown are fractions of neutrons events for various scenarios. Single scatters with gamma veto provide the relevant useful event fractions for defining the NR band. $\sim$\SI{3}{\percent} of neutrons result in a useful single-scatter in the detector, after removing events that contain any ER component. Double scatters with gamma veto provide the relevant useful event fractions for most DD studies.

\begin{table} [h]
\setlength{\extrarowheight}{5pt}
\tdrtcaption[singlesdoubles]{Fraction of neutron scatters in the Xe TPC.}{Fraction of neutron scatters in the Xe TPC for various assumptions about the outer detector material in the acrylic vessels. Gd doped LAB is the baseline choice. LUX, which has only \CHtO outside the TPC, is shown for reference. Also shown for comparison to make the impact of LAB and Gd doping clear are LZ with LAB only (no Gd doping) and water instead of scintillator.}
\centering
\sffamily
\begin{tabular} 
{|l|c|c|c|c|} 
\hline
\rowcolor{mrocol} & \multicolumn{3}{|c|}{ LZ } & LUX \\
\hline
OD material & Gd-doped LAB & LAB & \CHtO & - \\
\hline
{\bfseries single scatters (no cuts)} & 0.129 & 0.123 & 0.115 & 0.266 \\
\hline
{\bfseries single scatters (\Pgg veto)} & 0.032 & 0.031 & 0.033 & 0.164 \\
\hline
{\bfseries double scatters (no cuts)} & 0.106 & 0.107 & 0.114 & 0.073 \\
\hline
{\bfseries double scatters (\Pgg veto)} & 0.020 & 0.025 & 0.026 & 0.042 \\
\hline
\end{tabular}
\end{table}

\tdrsubsec[MaxRates]{Maximum useful calibration rates}
The maximum useful calibration rate will in most cases be driven by the drift time of the TPC. This limitation can be circumvented by delivering events near the top of the TPC, in order to shorten the drift time. However, in the interest of understanding the most general case first, the projections shown in Figure~\ref{CASf:eventrate} assume an event window defined by twice the maximum drift time.

For an average event rate $\lambda$ in a time window $t_w$, the probability to obtain $n$ or more random events in any particular window is given by
\begin{equation}
P = e^{-\lambda} \sum_{n=1}^{\infty} \frac{\lambda^n}{n!},
\end{equation}
from which Figure~\ref{CASf:eventrate} is derived. A \SI{10}{\percent} pileup rate could be achieved by keeping the calibration rate below \SIrange[range-units=single]{50}{70}{\Hz}, depending on the drift time. 

\begin{figure}[H]
\centering
\includegraphics[width=0.62\textwidth]{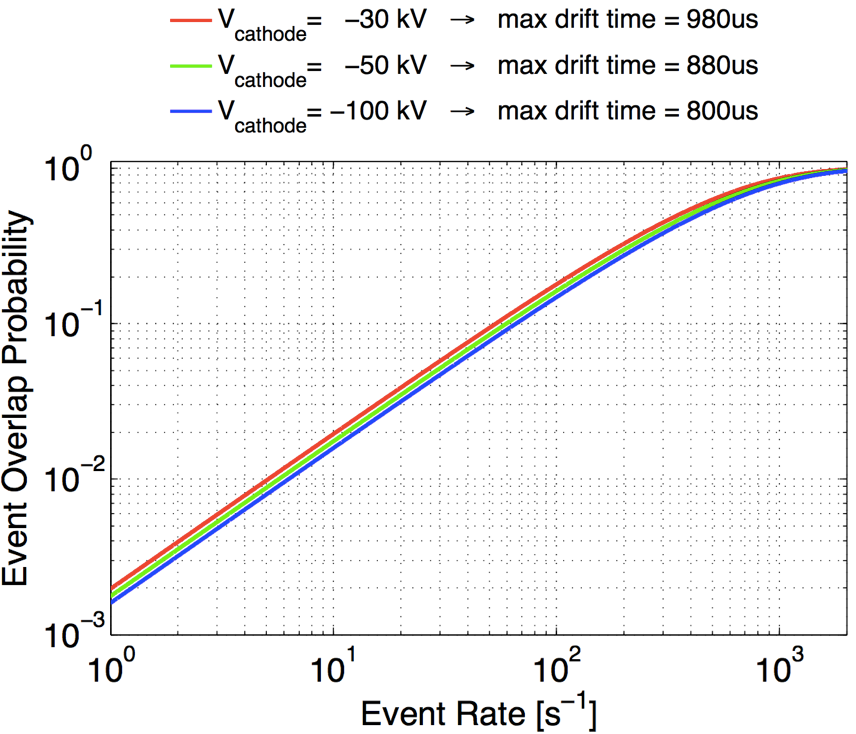}
\tdrfcaption[eventrate]{Event overlap probability as function of event rate}{The electron drift time produces a time window in which an unrelated event could produce S1-S2 pairing confusion.  The Poisson probability of event overlap is plotted here, and provides an estimate as to the maximum calibration rate in LZ for a desired maximum event overlap probability, for a specified field configuration (electron drift velocity), and a specified event distribution in $z$.  Here, we assume all calibration events are at the position of longest drift time, and vary the field configuration according to several possible running modes.  If a \SI{10}{\percent} overlap probability is acceptable, then the bottom of the detector could be calibrated at \SIrange[range-units=single]{50}{70}{\Hz}.}
\vspace{10pt}
\end{figure}

\tdrsec[EHS]{Environment, Health and Safety Concerns}
Safety concerns related to calibrations mainly relate to the use of radioactive sources. Most sources are either internal, and therefore don't expose users to radioactivity, or are NRC exempt. The most active source to be used in LZ is the DD generator, which will be capable of producing \SI{e9} neutrons/second. This source was also used in LUX and the safety concerns were mitigated by having an exclusion zone around the generator while it is operating. Because the generator can be operated remotely, no one needs to be inside this exclusion zone during operation. For LZ there is a planned \SI{9}{\m} exclusion zone. Points inside and outside this exclusion zone will be monitored with neutron detectors to ensure the areas where work will continue to take place are still safe.

Photoneutron calibrations require gamma sources with activities up to \SI{3}{\mega\becquerel} (\SI{81}{\micro\Curie}) which are not NRC-exempt. However because these sources will always be deployed inside of a tungsten shield, there is no reason they can't generally live inside an insert with a small amount of shielding to protect the users. Only briefly upon initial receipt of a source does a user need to handle one outside of such shielding. Specific safe handling procedures are under development.

\clearpage
\bibliographystyle{apsrev4-2}
\bibliography{LZtdr}
\wlog{In tdrinclude command}
\wlog{\ChapLabel}

\renewcommand{\ChapLabel}{chap:EDC} 
\renewcommand{\BibLabel}{EDCS:bib} 
\renewcommand{\SecPre}{EDC}
\renewcommand{\FloatPre}{EDC}
\tdrchap{Electronics, DAQ, Controls, and Online Computing}

This chapter describes the LZ signal processing electronics, data acquisition, detector-control system, and online data processing.

\tdrsec[SigProc]{Signal Processing}

The processing of the signals generated by the TPC PMTs is schematically shown in Figure~\ref{EDCf:SignalFlowXenon}. The TPC PMTs operate at a negative HV supplied by the LZ HV system, described in more detail in Section~\ref{EDCS:HV}. HV filters are installed at the HV flange on the breakout box. The PMT signals leave the breakout box via a different flange and are processed by the analog front-end electronics, described in more detail in Section~\ref{EDCS:Analog}. The amplified and shaped signals are connected to the data acquisition system (DAQ), described in more detail in Section~\ref{EDCS:DAQ}. The digitized data are sent to Data Collectors and stored on local disks. 
The PMTs of the outer-detector system operate at positive HV. The processing of the signals from these PMTs, shown schematically in Figure~\ref{EDCf:SignalFlowOD}. The same type of amplifier used for the Xe PMTs is used for the outer-detector PMTs.  Gain and shaping parameters of these amplifier will be fixed once the operating conditions of the outer-detector PMTs have been finalized.

\begin{figure}[htbp]
{\centering
\includegraphics[width=0.75\textwidth]{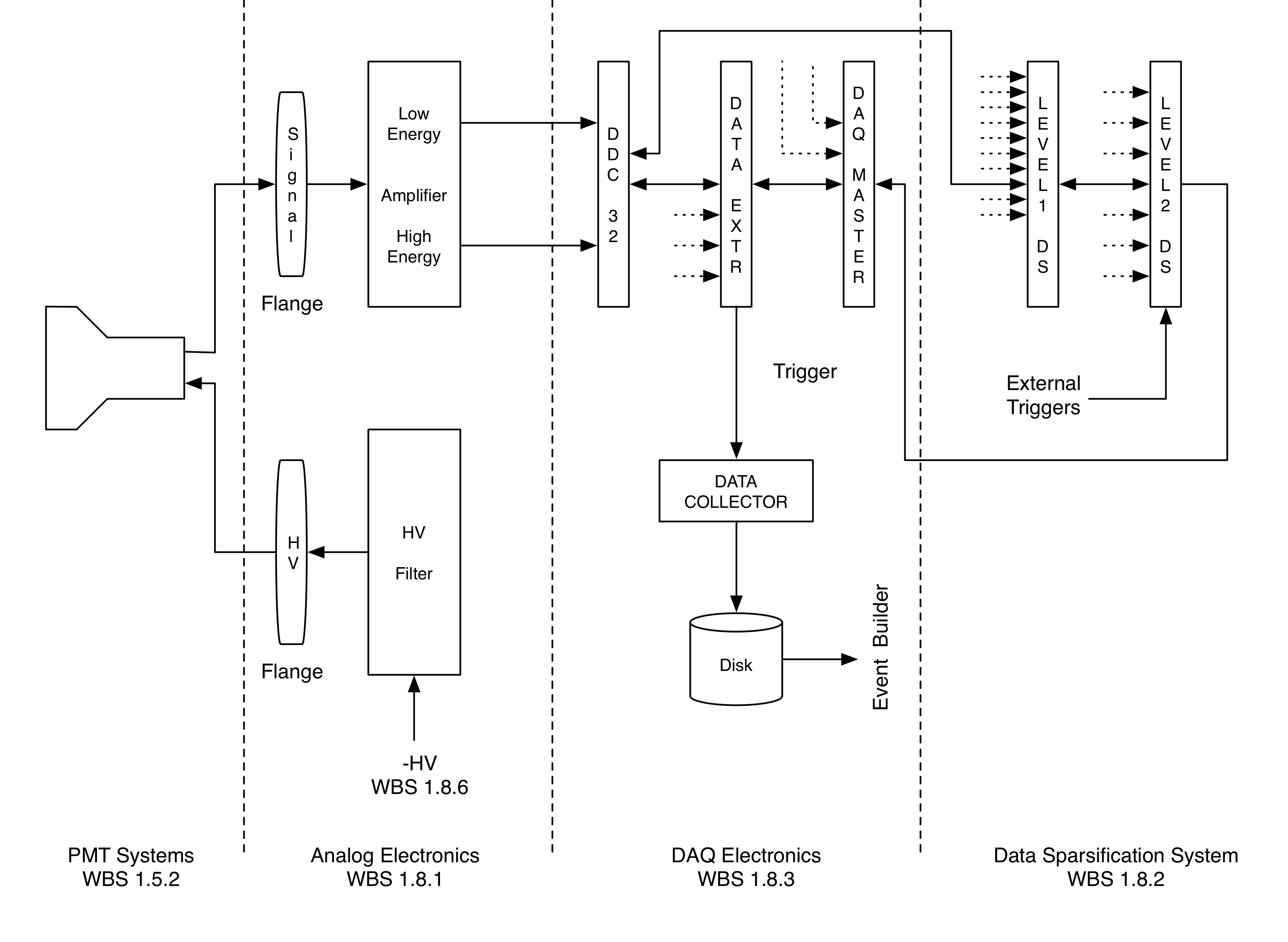}
\tdrfcaption[SignalFlowXenon]{TPC PMT signal processing schematics}{A schematic of the signal processing of the TPC PMTs. The TPC PMTs use dual-gain signal processing. The skin PMTs only utilize the low-energy section of the amplifiers.}
}
\end{figure} 

\begin{figure}[htbp]
{\centering
\includegraphics[width=0.75\textwidth]{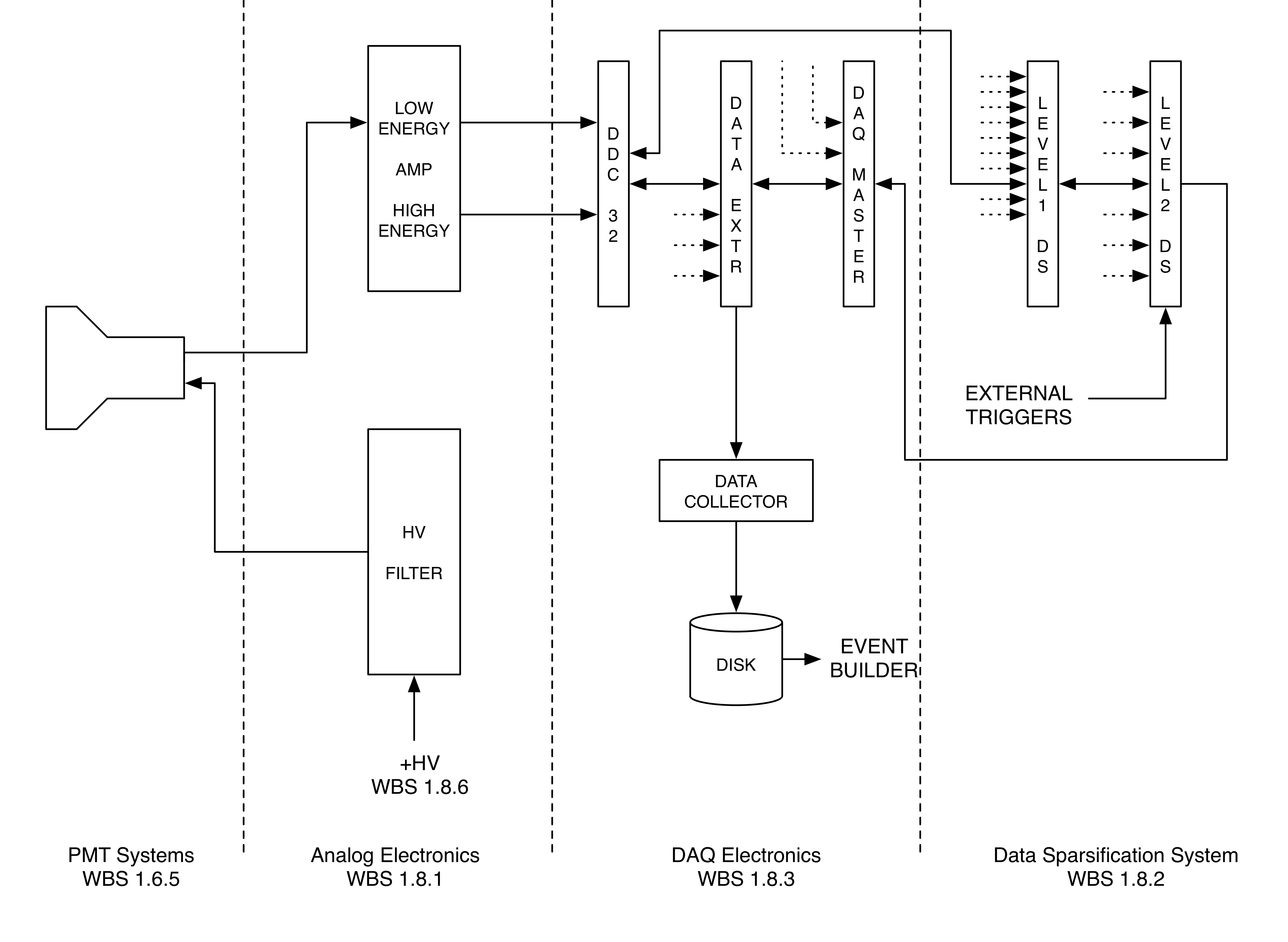}
\tdrfcaption[SignalFlowOD]{OD PMT signal processing schematics}{A schematic of the signal processing of the outer-detector PMTs. The outer detector PMTs use dual-gain signal processing.}
}
\end{figure} 

The data flow is schematically shown in Figure~\ref{EDCf:DataFlow}. The event builder assembles the events by extracting the relevant information from the Data Collector disks, DAQ1~-~15. This step is discussed in more detail in Section~\ref{EDCS:Online}. The event files are stored on local RAID arrays, RAID 1 and RAID 2, before being distributed to the data-processing centers for offline data processing and analysis.

\begin{figure}[htbp]
{\centering
\includegraphics[width=0.75\textwidth]{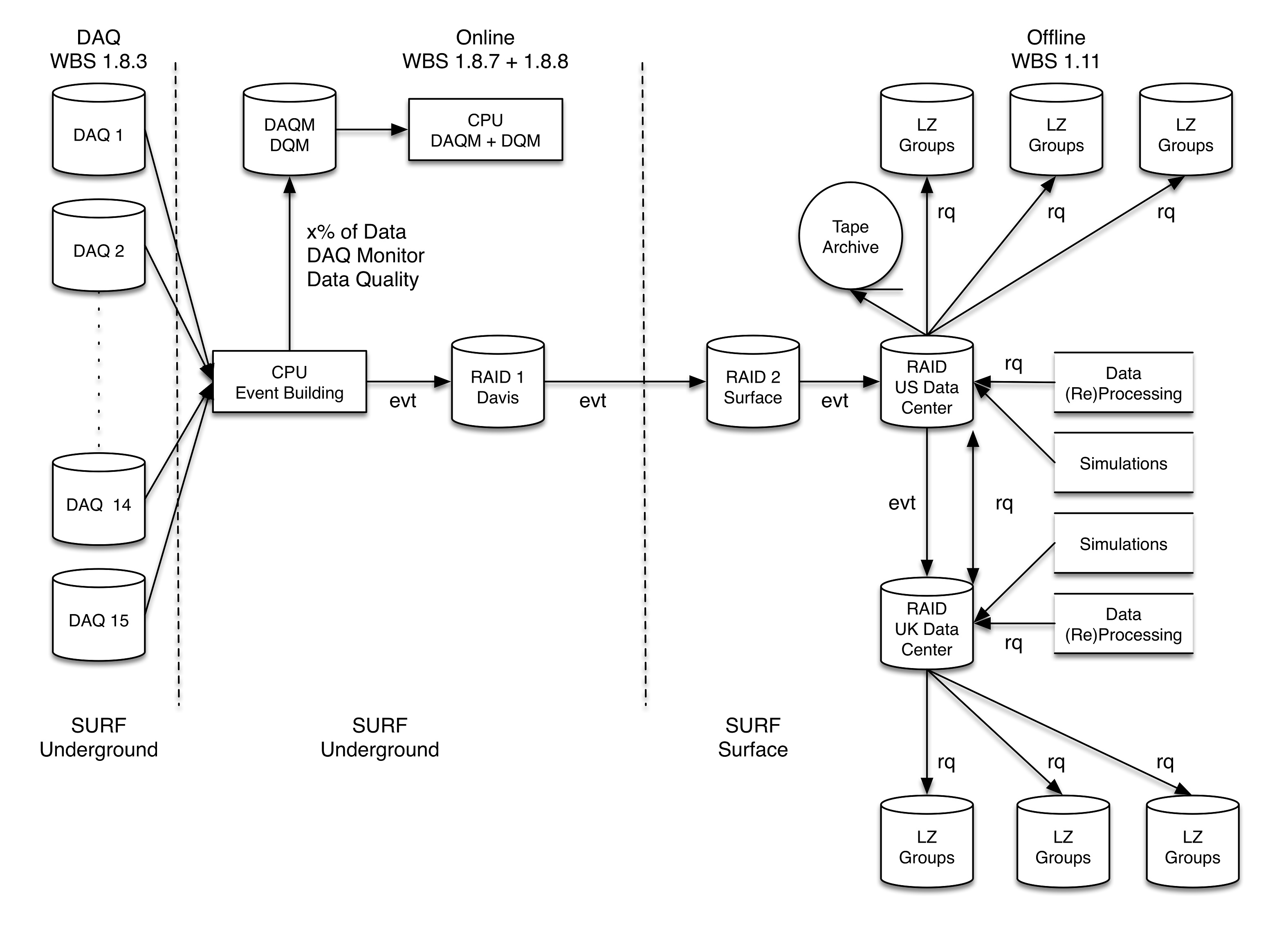}
\tdrfcaption[DataFlow]{A schematic of the data flow}{A schematic of the data flow.}
}
\end{figure} 

\tdrsec[Requirements]{Requirements}
The parameters of the analog and digital electronics are defined based on the properties of the PMT signals, the required dynamic range of the different PMTs, and the expected calibration rates.

Three different PMTs are used in LZ. The TPC PMTs will see S1- and S2-type signals. The skin and outer-detector PMTs will only see S1-type signals. The relevant properties are listed in Table~\ref{EDCt:PMTs}.

The cables that connect the xenon PMTs to the analog electronics are \SIrange[range-units=single]{40}{50}{\foot} long; the actual length depends on the final design of the conduits and the location of the area in which the analog electronics will be installed. The internal cables will be similar to the type used for LUX (Gore). Measurements with \num{45}-ft-long LUX cables have shown an area reduction of S1 signals by approximately \SIrange[range-units=single]{13}{23}{\percent}.

The design of the analog electronics is constrained by the required dynamic range of the LZ signals. Our design relies on the assumption that a single photoelectron (SPHE) detected in a TPC PMT generates a \SI{11.5}{\mV\ns} pulse at the input of the amplifiers. The required dynamic range is defined by the sources used to calibrate LZ and the desire to detect high-energy events for background studies. Chapter~\ref{chap:CAS} provides details on the LZ calibrations. 

The DAQ system design is based on our experience with LUX and the required LZ calibration rates. Typical calibration rates are listed in Table~\ref{EDCt:Calibrations}. The TPC source calibration rates are limited by the maximum drift time of \SI{700}{\mus} in LZ. During source calibrations of the TPC, a \SI{150}{\Hz} calibration rate results in a \SI{10}{\percent} probability of detecting a second calibration event within the drift time of the previous calibration event. Neutron calibrations are carried out to define the NR band in the TPC. These calibrations utilize external neutron sources and a neutron generator. External sources, inserted into the source tubes around the central cryostat, are used to calibrate the skin and the outer-detector PMTs. The count rates for these calibrations are not limited by the drift time in the TPC and they can be carried out at substantially higher rates. Weekly LED calibrations are done to examine the SPHE response of the PMTs and to monitor the PMT response. These calibrations can be carried out with rates as high as \SI{4}{\kHz}.

\begin{table} [H]
\setlength{\extrarowheight}{5pt}
\tdrtcaption[PMTs]{Properties of the PMTs}{Properties of the PMTs}
\centering
\sffamily
\begin{tabular} 
{|l|c|c|c|r|} 
\hline
\rowcolor{mrocol}
    {\bfseries System} & 
    {\bfseries Type PMT}& 
    {\bfseries Number}& 
    {\bfseries Gain}&
    {\bfseries HV}\\ 
\hline
TPC & R11410-20 & 494 & \num{<5e6} & \SI{<1750}{\V} \\
\hline
Skin & R8520 & 93 & \num{<1e6} & \SI{<900}{\V} \\
\hline
Skin & R8778 & 38 & \num{<5e6} & \SI{<1750}{\V} \\
\hline
Outer Detector & R5912  & 120 & \num{<1e7} & \SI{<1750}{\V} \\
\hline
\end{tabular}
\end{table}

The data volume to be handled by the DAQ system can be estimated on the basis of our experience with LUX. In WIMP search mode, we will focus on events with energy depositions below \SI{40}{\keV}. During krypton calibrations, in which the total energy deposition is \SI{41.6}{\keV}, the average LUX event size was \SI{203}{\kilo\byte} (or \SI{1.7}{\kilo\byte\per channel}). The size of each event was dominated by the width of the S2 signals. The event size of LZ can be estimated by scaling the LUX event size by the ratio of the number of PMT channels. Because LZ has four times as many TPC PMTs as LUX, and taking into account the dual-gains, we expect that the event size in LZ will be roughly \num{8} times as large as the LUX event size, or \SI{1.6}{\mega\byte}. This estimate does not include the data volume associated with the outer-detector and the skin PMTs. If each outer-detector and skin PMT detects a single S1-like pulse, that increases the event size by about \SI{45}{\kilo\byte}. With compression, the typical event size for LZ is estimated to be \SI{0.53}{\mega\byte}. Monte Carlo simulations show that the total background rate in LZ will be about \SI{40}{\Hz}. The background rate in the WIMP search region (\SIrange[range-units=single]{0}{40}{\keV}) will be about \SI{0.4}{\Hz}. At \SI{40}{\Hz}, the data rate is 21 MB/s. In \num{1000}~days, LZ will thus collect \SI{1.9}{\peta\byte} of WIMP-search data. Other estimates, including energy depositions above \SI{41.6}{\keV}, result in an estimated total data volume of \SI{2.8}{\peta\byte}. By optimizing the event selection, we expect to be able to reduce the total volume of WIMP-search data by a factor of \num{2} (see Section \ref{EDCSs:DAQDS}).

\begin{table} [H]
\tdrtcaption[Calibrations]{Calibrations and expected count rates.}{Calibrations and expected count rates.}
\setlength{\extrarowheight}{5pt}
\centering
\sffamily
\begin{tabular} 
{|l|c|c|c|} 
\hline
\rowcolor{mrocol}
    {\bfseries Calibration Type} & 
    {\bfseries System}& 
    {\bfseries Typical Count Rate}& 
    {\bfseries Frequency}\\ 
\hline
Internal sources & TPC & \SI{<150}{\Hz} & Twice a week (Kr) \\
\hline
External gamma sources & Skin and outer detector & TBD & TBD \\
\hline
Neutrons & TPC  & \SI{<150}{\Hz} & TBD \\
\hline
LEDs & TPC and outer detector  & \SI{4}{\kHz} & weekly \\
\hline
\end{tabular}
\end{table}

\tdrsec[Analog]{Analog Electronics}

\tdrsubsec[AnalogDesignCriteria]{Design Criteria}

The analog front-end design for the LZ experiment has benefited immensely from the experience with the LUX detector. In what follows, we have retained all the features of LUX electronics that performed well, while improving in some areas. The most important figure of merit of the LUX analog front end was its low noise characteristics, which allowed us to set our thresholds well below the SPHE level. Assuming the LZ noise characteristics will be the same as those for LUX, we expect to be able to run comfortably with a threshold of \SI{0.25}{PHE} (photoelectron) for each PMT. Figure~\ref{EDCf:TwoFoldThreshold} shows a simulated distribution of total S1 pulse area for events in which one PHE each is detected in two PMTs (left) and in three PMTs (right). A threshold of \SI{>0.25}{PHE} is set for each PMT. The distributions are governed primarily by the \SI{\sim 34}{\percent} rms width of the PMT signal (gain variation) and receives almost no contribution from electronic noise. Setting a threshold at \SI{1}{PHE} for the total S1 signal yields an efficiency of \SI{\sim 93}{\percent} for events that produce at least two PHEs in the PMTs. For the case of three PHEs, the corresponding efficiency is approaching \SI{\sim 100}{\percent}. These electronics thresholds correspond to the lowest energy threshold achievable in such a detector, and hence drive the noise specifications for the front end. The choice between two- or three-fold coincidence will be governed by the dark currents in the PMTs, which have been specified to be in the \SIrange[range-units=single]{50}{200}{\Hz} range.

The second key design parameter for the LZ front end is the dynamic range, which is defined by the energy range of the sources used to calibrate LZ. Isotopes such as \IKreTm (\num{32.1} and \SI{9.4}{\keV} transitions), activated Xe (\SI{236}{\keV} and \SI{164}{\keV} transitions), and tritium (endpoint at \SI{18.2}{\keV}) will be present or injected directly into the LXe volume and will be used to calibrate the detector periodically. LZ will retain the ability to detect high-energy events that saturate the PMTs of the top array by using only the light collected by the bottom array to determine the total S2 area.

The LZ electronics will provide excellent resolution for single liquid electrons, which are expected to yield at most \SI{50}{PHEs}, depending on the height of the gas gap and the strength of the electric field in that region of the TPC. The typical duration of such pulses will be about \SIrange[range-units=single]{0.5}{1.1}{\mus}. At the same time, we will need to provide extremely clean measurements of SPHEs in order to have a sharp turn-on of the S1 efficiency. SPHE spectra also help with maintaining an in situ calibration of the PMT gains. To meet these requirements, the analog electronics provides one low-energy (high-gain) and one high-energy (low-gain) output for each PMT.

Figure~\ref{EDCf:LZAmpSchematic} shows this concept. The high-energy channel has low gain and a \num{30}-\si{\ns} full width at tenth maximum (FWTM) shaping-time constant. Its dynamic range is defined by the \num{236}-\si{\keV} Xenon activation line. The low-energy channel has a \num{10} times higher area gain and wider shaping of \SI{60}{\ns} FWTM. It is optimized for an excellent SPHE response and dynamic range. The shaping times and gains are derived from one assumption: the DAQ will have a usable dynamic range of \SI{1.8}{\V} at the input and will sample the pulses at \SI{100}{\MHz} with \num{14}-bit accuracy. A \SI{0.2}{\V} offset is applied to the digitizer channels in order to measure signal undershoots of up to \SI{0.2}{\V}. In summary, the relevant parameters are:
\begin{itemize}
\item Typical SPHE response of the PMTs: \SI{11.5}{\mV\ns} pulse. The performance should also be verified if this value is as low as \SI{6.5}{\mV\ns} for some PMTs
\item The dark current for the PMTs at operating voltage has been specified to be in the \SIrange[range-units=single]{50}{200}{\Hz} range. The distribution within this range is not yet known.
\item S1 light yield at \SI{236}{\keV}: \SI{620}{PHEs} (for \SI{650}{\V\per\cm}).
\item S1 light distribution: evenly distributed over all PMTs. The bottom array receives \SI{80}{\percent} of the total S1 light.
\item S2 light yield for ERs at \SI{236}{\keV}: \num{42} liquid electrons / \si{\keV} and \SI{70}{PHEs} / extracted liquid electron (conservative maximum).
\item S2 light distribution: \SI{22}{\percent} of the S2 light is in a single top PMT. The S2 light is distributed evenly over all bottom PMTs; the bottom array receives \SI{45}{\percent} of the total S2 light.
\end{itemize}

\begin{figure}
		\centering
		\includegraphics[width=0.9\linewidth]{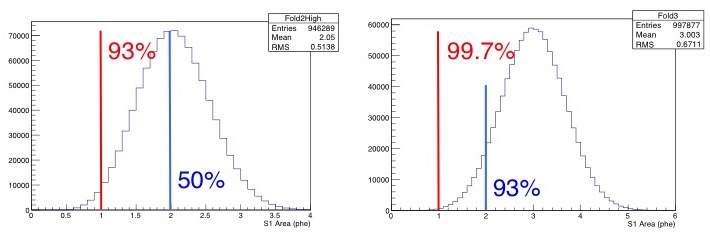}
		\tdrfcaption[TwoFoldThreshold]{Simulation of events with \SI{>0.25}{PHE} in two PMTs}{A simulated distribution of total pulse area for S1 pulses with two (left) and three (right) photoelectrons (left). Requiring a total area larger than 1 PHE provides nearly 100\% efficiency for events with 3 or more S1 photoelectrons detected.}
\end{figure}

\begin{figure}
		\centering
		\includegraphics[width=0.49\linewidth]{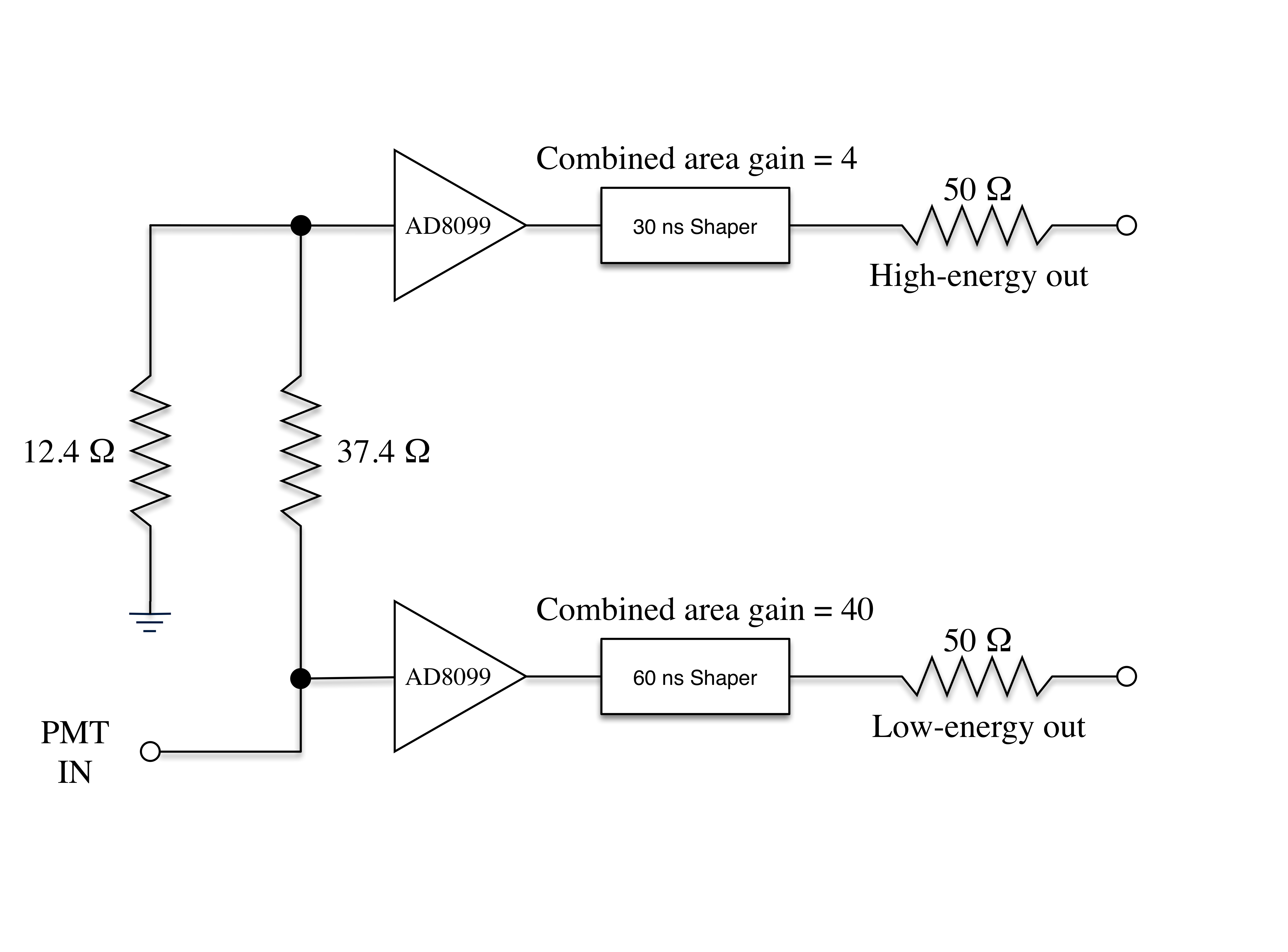}
		\includegraphics[width=0.49\linewidth]{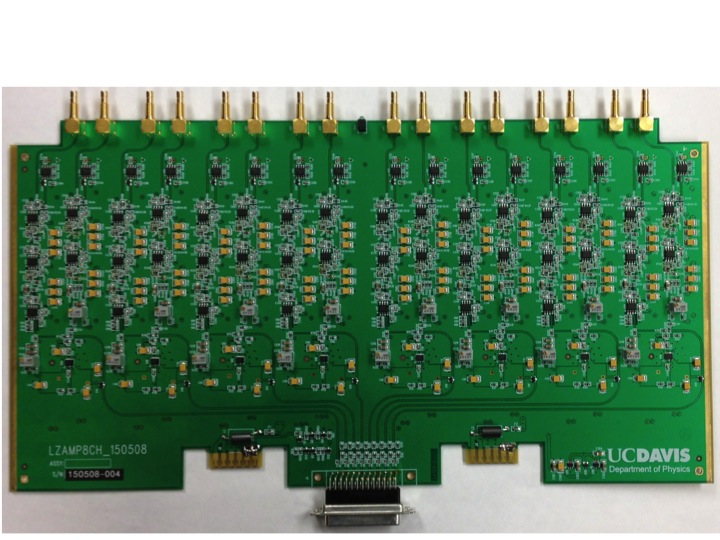}
			\tdrfcaption[LZAmpSchematic]{LZ amplifier}{Left: A schematic diagram of a single channel of the LZ amplifier. Right: A photograph of the 8-channel prototype board.}
\end{figure}

The op-amps indicated in Figure~\ref{EDCf:LZAmpSchematic} (left) are very similar to those used in LUX. The gain and shaping-time constants of the amplifiers were optimized using simulations. Figure~\ref{EDCf:LZAmpSimulationTop} and \ref{EDCf:LZAmpSimulationBottom} shows the results of simulations of S2 pulses associated with the \num{236}-\si{keV} transition in activated Xe (top) and a \num{3}-\si{\mega\eV} energy deposition (bottom). Figure~\ref{EDCf:LZAmpSimulationTop} shows that S2 saturation in the top PMT array for the \num{236}-\si{keV} transition is not a problem if the amplitude of a single PHE is less than \SI{1.5}{\mV} (\num{12} ADCC). If we allow one PMT to saturate, single PHEs of up to \SI{3.0}{\mV} can be accommodated. Figure~\ref{EDCf:LZAmpSimulationBottom} shows that the S2 associated with larger energy depositions will not start to saturate the PMTs of the bottom array if the amplitude of a single PHE is less than \SI{15}{\mV} (120 ADCC). The two outputs of the amplifiers have the following properties (assuming a \SI{11.5}{\mV\ns} SPHE response of the PMT):
\begin{itemize}
\item Low-energy output: Area gain = \num{40}, \SI{1}{PHE} = \SI{105}{ADCC} (amplitude), S1 dynamic range: \SI{140}{PHEs} (\SI{1700}{\keV}), S2 dynamic range \SIrange[range-units=single]{12}{26}{\keV} (top) and \SIrange[range-units=single]{1300}{2800}{\keV} (bottom) for \SIrange[range-units=single]{0.5}{1.1}{\mus} wide pulses ($1\sigma$) with no saturation.
\item High-energy output: Area gain = \num{4}, \SI{1}{PHE} = \SI{21}{ADCC} (amplitude), S1 dynamic range: \SI{700}{PHEs}, S2 dynamic range \SIrange[range-units=single]{120}{260}{\keV} (top) and \SIrange[range-units=single]{13000}{27000}{\keV} (bottom) for \SIrange[range-units=single]{0.5}{1.1}{\mus} wide pulses ($1\sigma$) with no saturation.
\end{itemize}

\begin{figure}
\begin{minipage}[t]{.49\textwidth}
	\centering
	\includegraphics[width=0.9\linewidth]{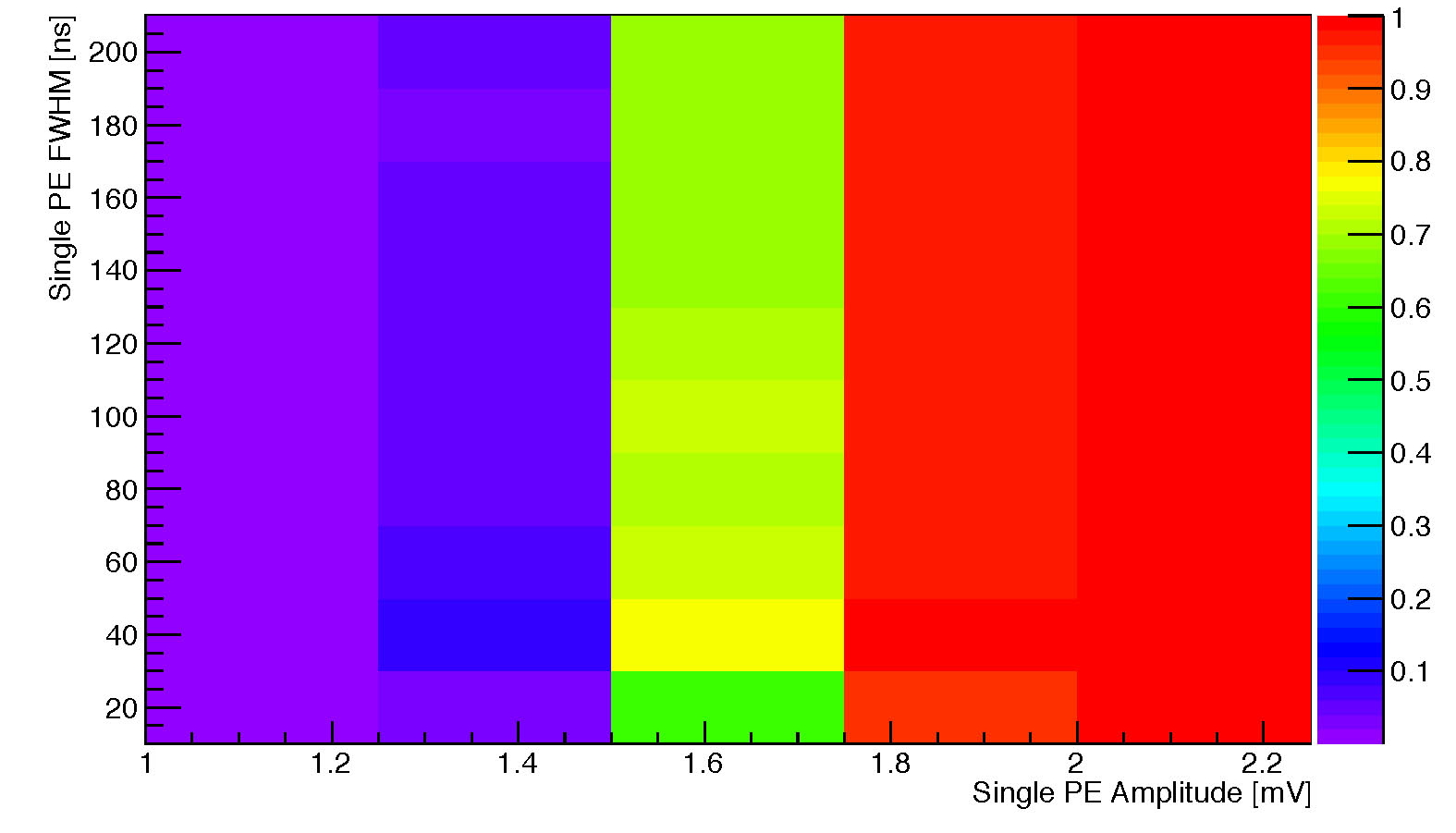}
	\tdrfcaption[LZAmpSimulationTop]{Simulation of S2 response for a \num{236}-\si{\keV} Xe transition}{A simulation study of the S2 response for a \num{236}-\si{\keV} Xe transition in the center of the detector, as seen by the top PMTs. The gain and the shaping width of the SPHE response are varied. The color code shows the fraction of events in which the peak PMT saturates.}
\end{minipage}
\hfill
\begin{minipage}[t]{.49\textwidth}
	\centering
	\includegraphics[width=0.9\linewidth]{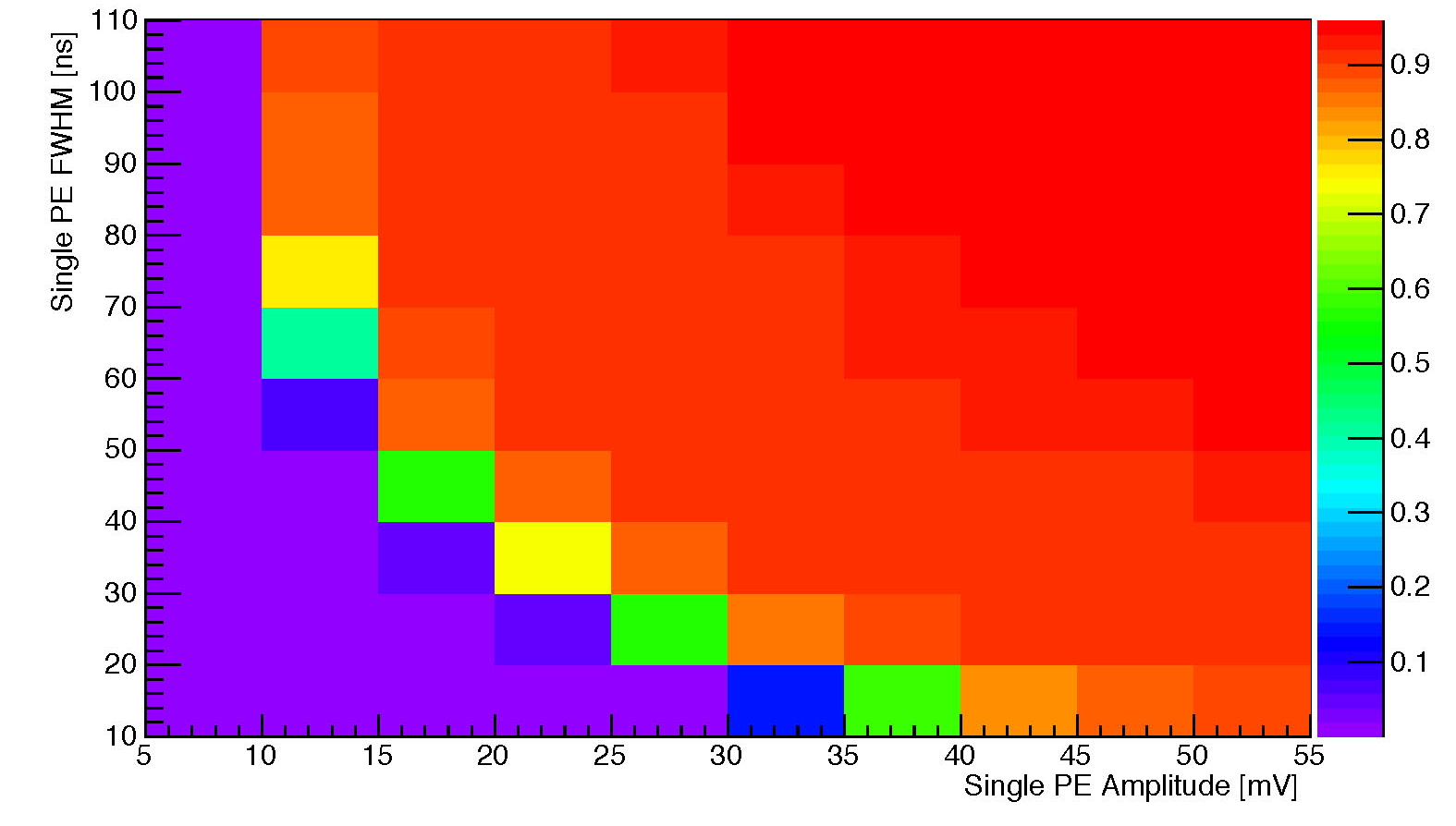}
	\tdrfcaption[LZAmpSimulationBottom]{Simulation of S2 response for a \num{3}-\si{\mega\eV} Xe energy deposition}{A simulation study of the S2 response for a \num{3}-\si{\mega\eV} energy deposition in the center of the detector, as seen by the bottom PMTs. The color code shows the fraction of PMTs of the bottom array that saturate.}
\end{minipage}
\end{figure}

The dynamic range for S2 signals is shown in Figure~\ref{EDCf:S2DR} \cite{Yin:2015jtk}. The low-energy channel of the amplifier provides the dynamic range required for the tritium and krypton calibrations. The high-energy channel is required to provide the dynamic range required to measure the activated Xe lines. S2 signals due to $0\nu\beta\beta$ will saturate one or more channels of the top array, but the S2 pulse area can still be reconstructed using the low-energy channels of the bottom PMTs.

\begin{figure}
\begin{minipage}[t]{.49\textwidth}
	\centering
	\includegraphics[width=0.9\linewidth]{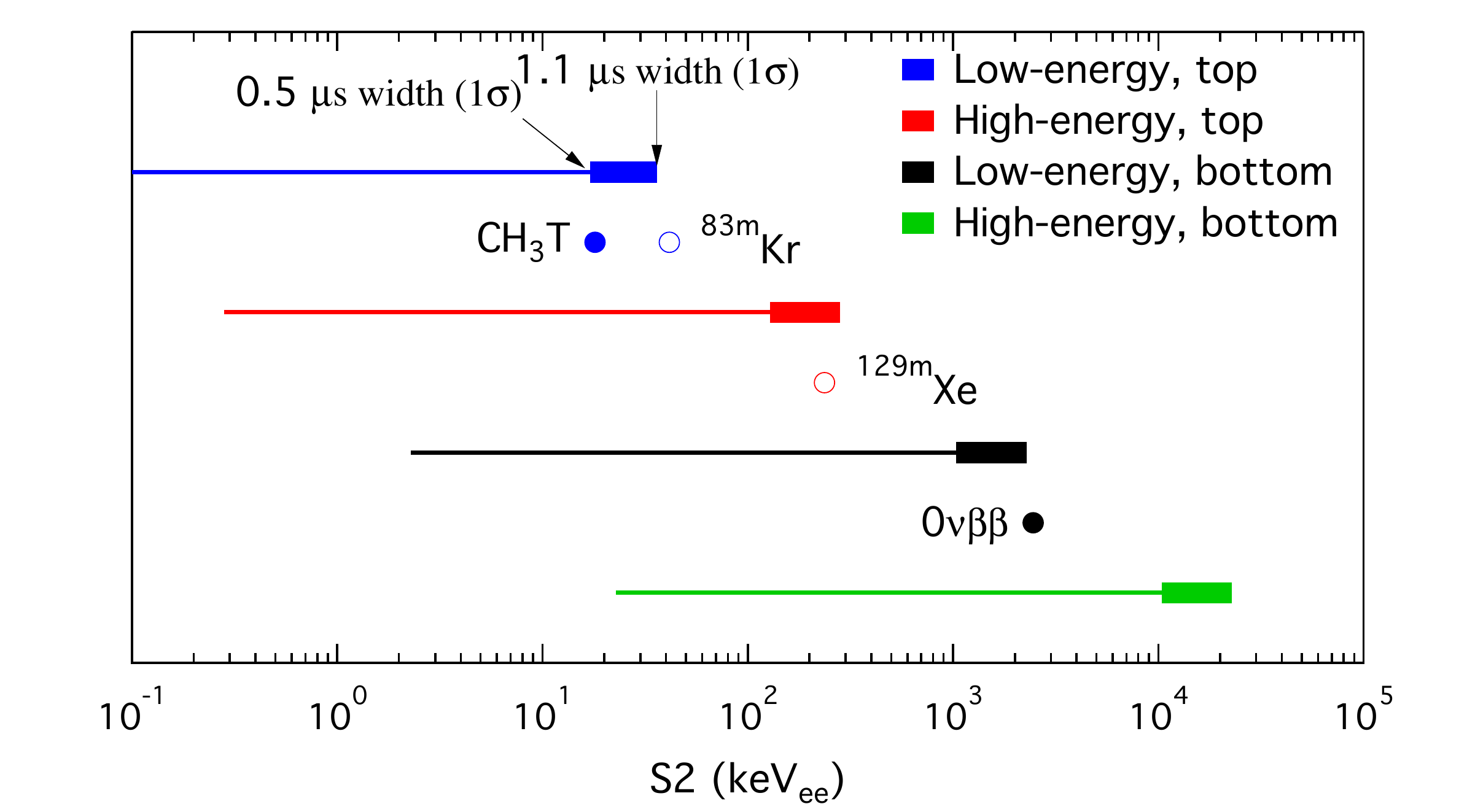}
	\tdrfcaption[S2DR]{Dynamic range for S2 signals detected in the top and bottom PMTs}{S2 dynamic range, expressed in terms of electron-recoil energy depositions for the low- and high-energy channels of the top and bottom PMT arrays \cite{Yin:2015jtk}. The bars indicate variations in the upper level of the dynamic range due to the variations in the width of the S2. The lower and upper ends of these bars show the dynamic range for \num{0.5}-\si{\mus}-wide and \num{1.1}-\si{\mus}-wide pulses, respectively. Energy depositions for various calibration sources are indicated by the circles.}
\end{minipage}
\hfill
\begin{minipage}[t]{.49\textwidth}
	\centering
	\includegraphics[width=0.9\linewidth]{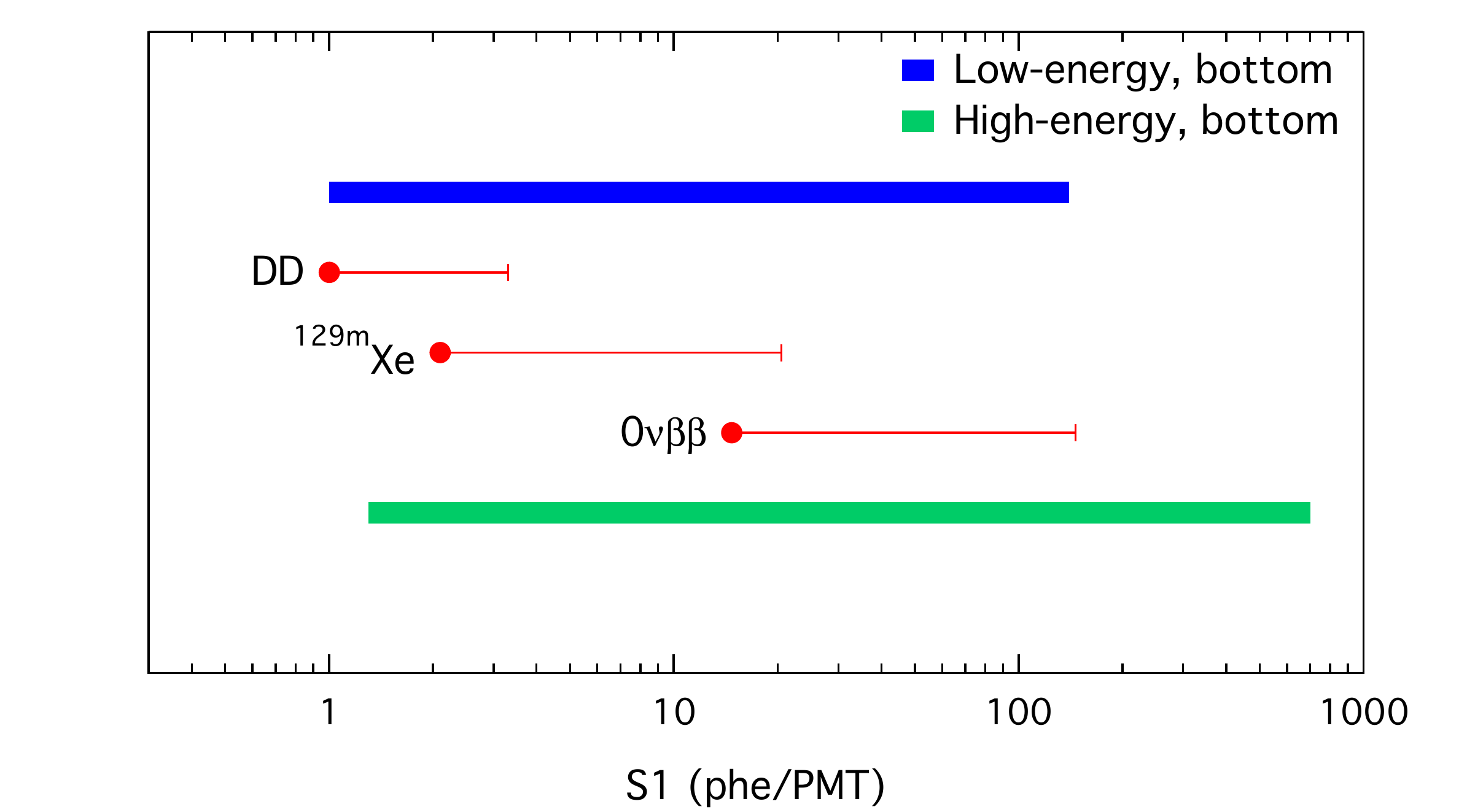}
	\tdrfcaption[S1DR]{Dynamic range for S1 signals detected in the bottom PMTs}{Dynamic range for S1 signals detected in the bottom PMTs \cite{Yin:2015jtk}. The range required for DD neutrons, the \IXeotnm activation line, and $0\nu\beta\beta$ decay of \IXeoTs are also shown. The left-end of each source line corresponds to an energy deposition in the center of the TPC; the right-end corresponds to an energy deposition \SI{1}{\cm} above the cathode.}
\end{minipage}
\end{figure}

The dynamic range for S1 signals is shown in Figure~\ref{EDCf:S1DR} \cite{Yin:2015jtk}. The figure shows the dynamic range of a bottom PMT. Also shown are the number of PHEs associated with the full-energy deposition of the \num{236}-\si{\keV} Xe activation line and $0\nu\beta\beta$ decay of \IXeoTs. The dynamic range provided by the dual-gain channels is sufficient for all LZ calibrations. 

Minor changes to the gain and shaping parameters of the amplifiers can be easily accommodated via changes in resistor and capacitor values. This will be done if more detailed simulations, including the electronics response and the noise of all components of the electronics chain, indicate a need for modifications.

The same amplifier design will be used for the skin and outer-detector PMTs although the gain and shaping may be adjusted slightly, if needed. Dual gain amplifiers will be used for the outer-detector PMTs, in order to make measurements of cosmic muons.  For the skin PMTs, only the low-energy channel will be instrumented. A summary of the number and type of analog signals is shown in Table~\ref{EDCt:GainAndShaping}. 

\begin{table} [H]
\tdrtcaption[GainAndShaping]{Summary of the number and type of the \num{1359} analog signals.}{Summary of the number and type of the \num{1359} analog signals.}
\setlength{\extrarowheight}{5pt}
\centering
\sffamily
\begin{tabular} 
{|l|c|c|c|c|c|c|} 
\hline
\rowcolor{mrocol}
	\multicolumn{1}{|l|}{}  &
	\multicolumn{3}{|c|}{\bfseries High-gain Signals} &
	\multicolumn{3}{|c|}{\bfseries Low-gain Signals} \\
\hline
	{\bfseries PMT Type} & 
	{\bfseries \#} & 
    {\bfseries Area Gain}& 
    {\bfseries Shaping}& 
    {\bfseries \#}&
    {\bfseries Area Gain}& 
    {\bfseries Shaping}\\ 
    {\bfseries } & 
	{\bfseries } & 
    {\bfseries }& 
    {\bfseries (FWTM)}& 
    {\bfseries }&
    {\bfseries }& 
    {\bfseries (FWTM)}\\ 
\hline
Top TPC & 253 & 40 & \SI{60}{\ns} & 253 & 4 & \SI{30}{\ns} \\
\hline
Bottom TPC & 241 & 40 & \SI{60}{\ns} & 241 & 4 & \SI{30}{\ns} \\
\hline
Skin (1")& 93 & 100 & \SI{60}{\ns} & 0 & NA & NA \\
\hline
Skin (2")& 38 & 40 & \SI{60}{\ns} & 0 & NA & NA \\
\hline
Outer Detector & 120 & 40 & \SI{60}{\ns} & 120 & 4 & \SI{30}{\ns} \\
\hline
Total & 745 &  &  & 614 &  &  \\
\hline
\end{tabular}
\end{table}

\begin{figure}
\begin{center}
\includegraphics[width=1.0\textwidth]{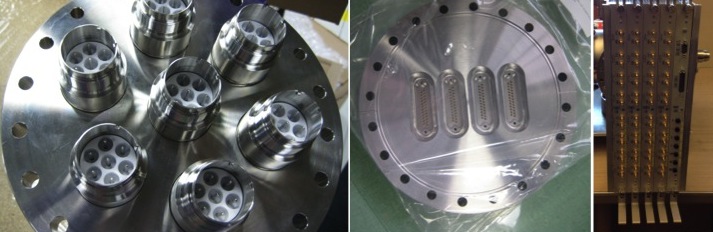}
\end{center}
\tdrfcaption[LZAmpPrototype]{Flanges and amplifier crate prototype}{Photographs of the PMT HV flange (left), the \num{32}-channel PMT signal flange (middle) and a \num{32}-channel crate that mounts on the signal flange. }
\end{figure}

\tdrsubsec[AmpPrototype]{LZ Amplifier Prototype}

Figure~\ref{EDCf:LZAmpSchematic} (right) shows a photograph of the first amplifier prototype with eight input channels. The amplifiers connect to the PMT signal lines using the DB-25 connector visible at the bottom of the figure. The DB-25 connector allows the signal lines to be interleaved with two ground lines for minimizing cross-talk. The 32-channel flange is shown in Figure~\ref{EDCf:LZAmpPrototype} (middle). The amplifiers will be housed a 5-card mini crate that mounts on the signal flange, as shown in Figure~\ref{EDCf:LZAmpPrototype} (right). It houses four amplifier cards and a fifth card for power distribution and monitoring.

\begin{figure}
	\centering
	\includegraphics[width=0.8\linewidth]{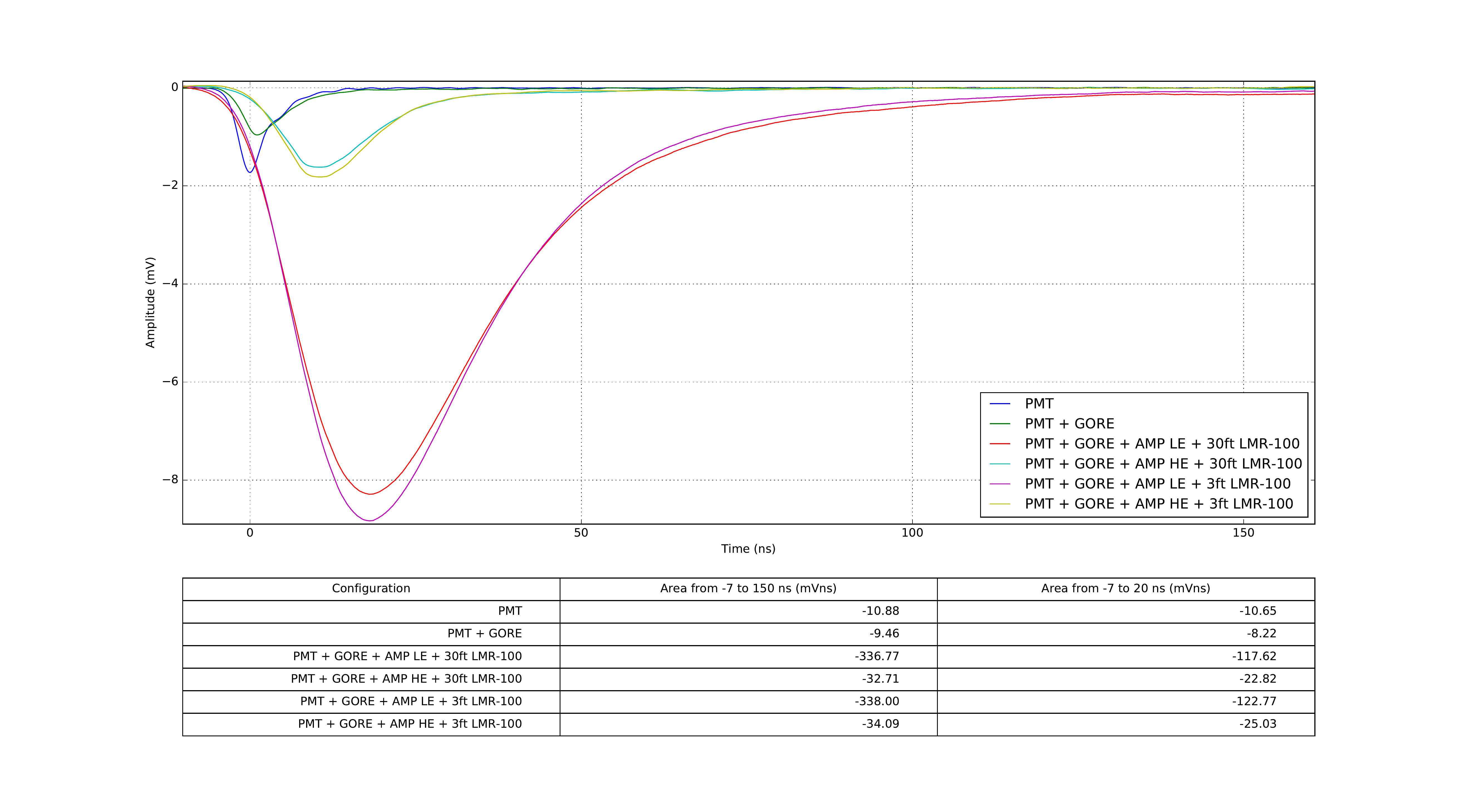}
	\tdrfcaption[LZAmpSPHE]{SPHE Waveforms}{Waveforms for an SPHE at various stages of the analog chain. The PMT was operated at \SI{-1300}{\V}. The effect of the internal Gore cable and the external LMR-100 cables is shown. The desired shaping and gain specifications have been achieved.}
\end{figure}

Waveforms captured with an oscilloscope at various stages of the analog chain are shown in Figure~\ref{EDCf:LZAmpSPHE}. The response of a PMT, operating at \SI{1300}{\V}, to an SPHE at the start and end of the cable in the vacuum space is shown.  Also shown are the outputs of the LE and HE channels of the amplifier, at its output and at the end of external co-axial LMR-100 cables. The table at the bottom of the figure shows the losses and gains at each stage.

\begin{figure}
\centering
\includegraphics[width=0.8\textwidth]{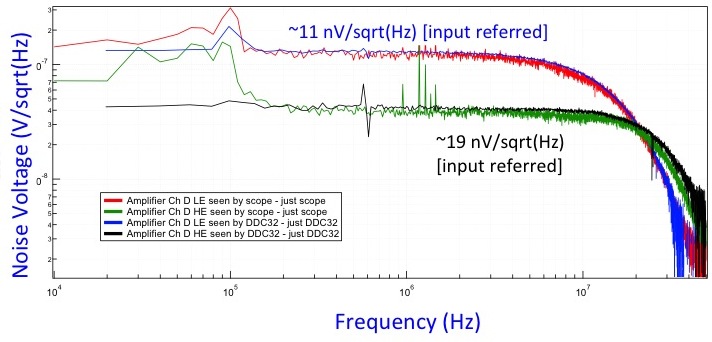}
\tdrfcaption[NoisePower]{Amplifier Noise}{The measured spectrum of noise power in the amplifier channels. }
\end{figure}

Figure~\ref{EDCf:NoisePower} shows the measurements of the noise power in both LE and HE channels. The RMS ADC noise of the free-running DDC-32 digitizer channels was measured to be \SI{1.19+-0.01}{ADCC}. In the spectra shown, the measurements were made using a DDC-32 and also using an oscilloscope. In both cases the contributions from the digitization have been subtracted in quadrature. The resulting input referred noise voltage in the ``white'' part of the spectrum were measured to be \SI{11}{\nV}/$\sqrt{\si{\Hz}}$ and \SI{19}{\nV}/$\sqrt{\si{Hz}}$ respectively, for the LE and HE channels. These values are in excellent agreement with the circuit simulation model.

\begin{figure}
\centering
\includegraphics[width=0.8\textwidth]{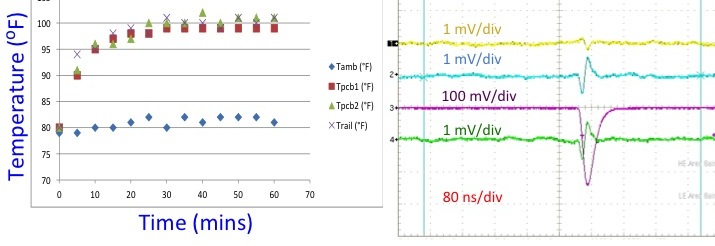}
\tdrfcaption[PowerCrosstalk]{Power and Crosstalk}{Left: The power dissipation and equilibrium operating temperature of a crate without any active cooling. Right: Measurements of the cross-talk for two neighboring and one next-neighbor channels.}
\end{figure}

Figure~\ref{EDCf:PowerCrosstalk} (left) shows measurements of the power dissipation. A fully loaded crate was powered up in a lab with an ambient temperature of \SI{80}{\farad}, without any active cooling or forced air through the crate.  Within \num{30} minutes the crate reached a temperature of \SI{100}{\farad} and stabilized at that value. This is well within the operating range of the electronics.

Figure~\ref{EDCf:PowerCrosstalk} (right) shows measurements of crosstalk between the individual amplifier channels. The central channel was pulsed such that it had an output of \SI{240}{\mV}. The neighboring channels shown in cyan and green had  bipolar induced pulses with a pulse height of \SI{0.6}{\mV}, or \SI{0.25}{\percent} crosstalk. The next-neighbor channel, shown in yellow, had a pulse height of \SI{0.3}{\mV}. These small values of crosstalk are completely acceptable. Moreover, bipolar pulses contribute very little area when integrated.

\begin{figure}[htbp]
{\centering
\includegraphics[width=0.45\textwidth]{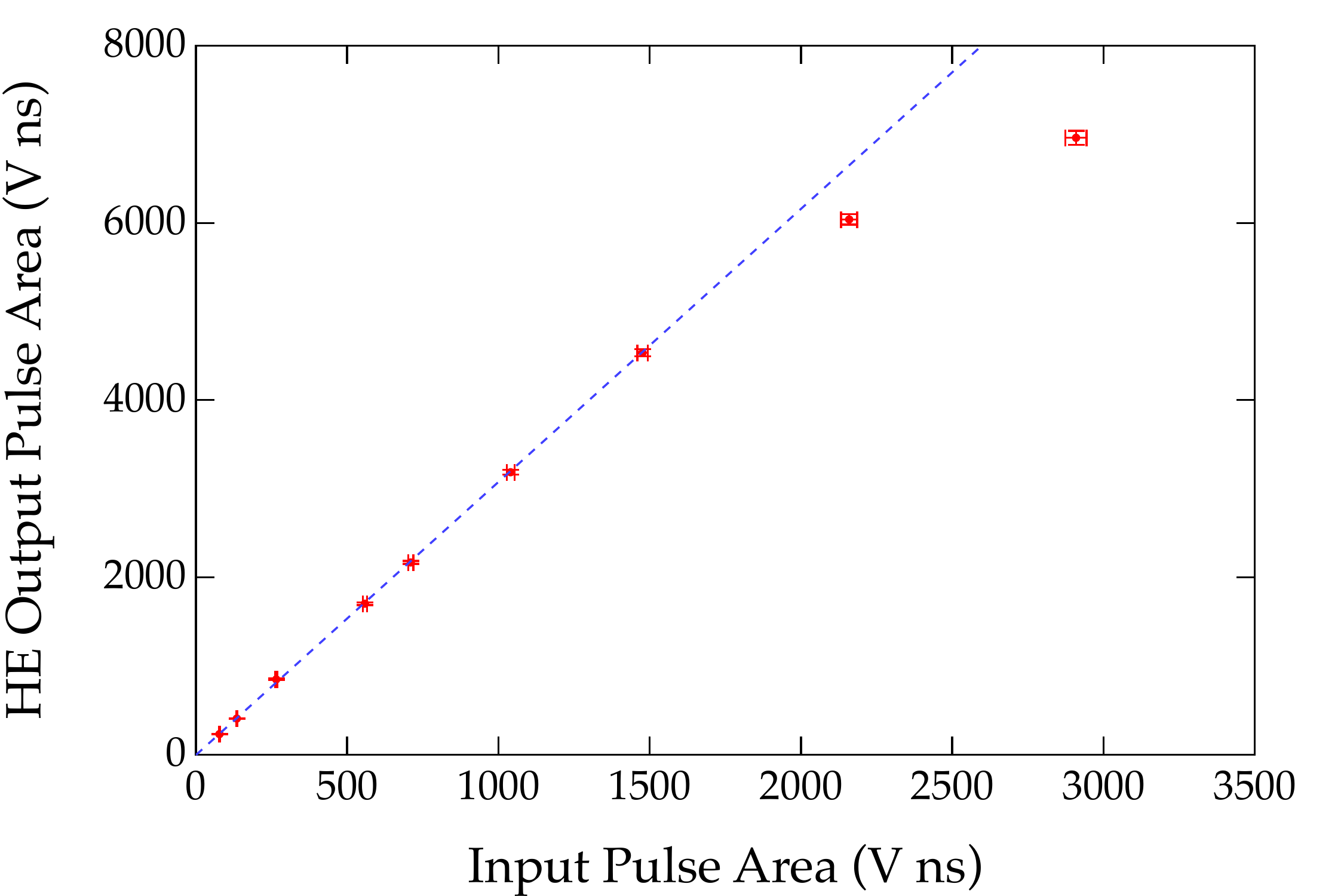}
\includegraphics[width=0.45\textwidth]{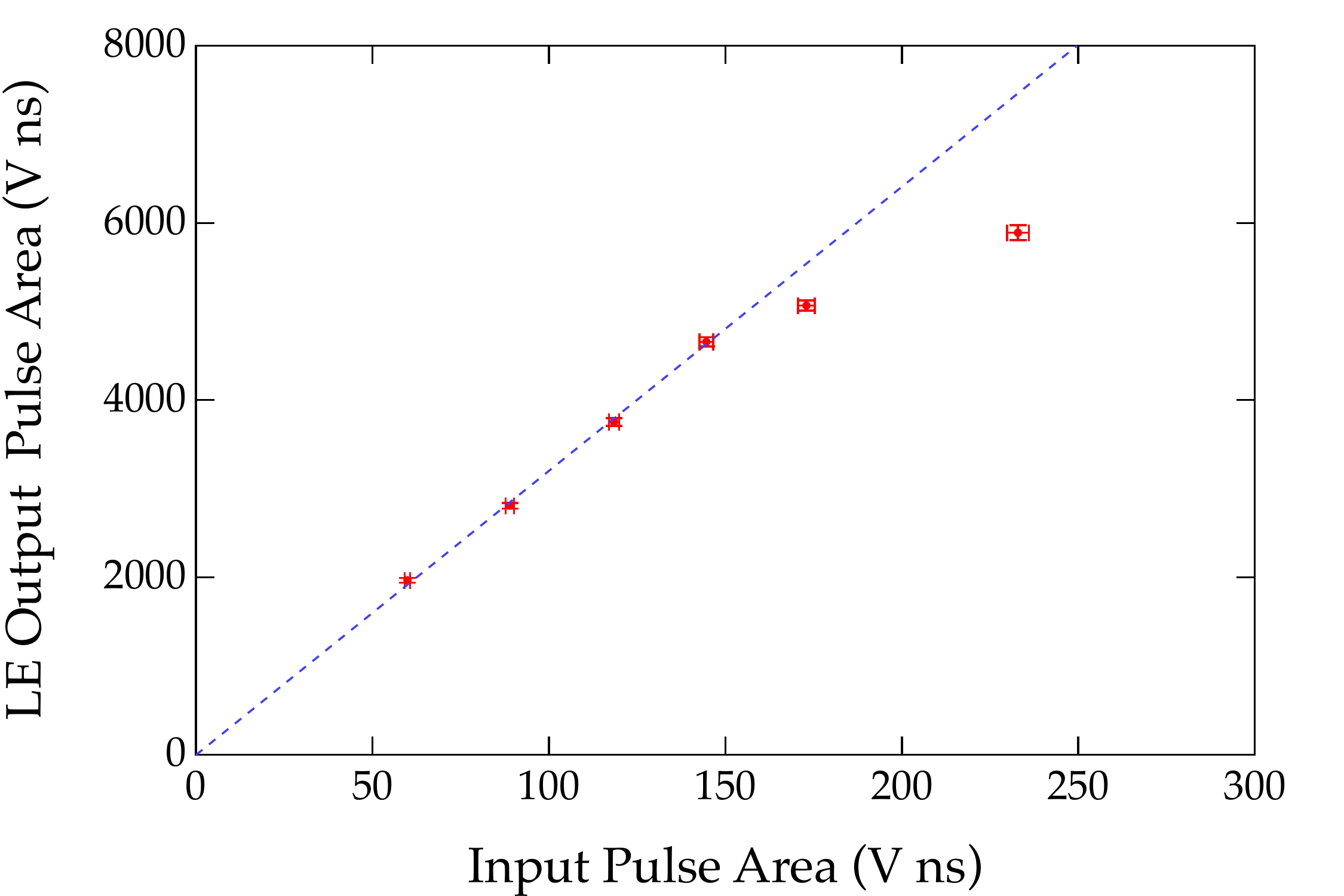}
\tdrfcaption[LinearityHighLowGain]{Results of linearity measurements}{Results of linearity measurements. The area of the output pulse is plotted as function of the area on the input pulse. The results obtained for the high-energy (low-gain) channel are shown on the left while the data collected for the low-energy (high-gain) channel are shown on the right.}
}
\end{figure} 

The linearity of the amplifier was studied using S2-like test pulses with a width of \num{1}~\si{\mus}~(FWHM). The test pulses were propagated through \SI{45}{ft} of gore cable before reaching the amplifier.  The output signals saturate when their amplitude exceeds \SI{2.6}{\V}. Examples of the results of these linearity studies with S2-like pulses are shown in Figure~\ref{EDCf:LinearityHighLowGain} The low-energy and high-energy outputs becomes nonlinear when the area of the input signals exceed \SI{150}{\V\ns} and \SI{1500}{\V\ns}, respectively. Although the response is nonlinear in this region, there is still a one-to-one correlation between the area of the output pulse and the area of the input pulse. 

\begin{figure}
\centering
\includegraphics[width=0.5\textwidth]{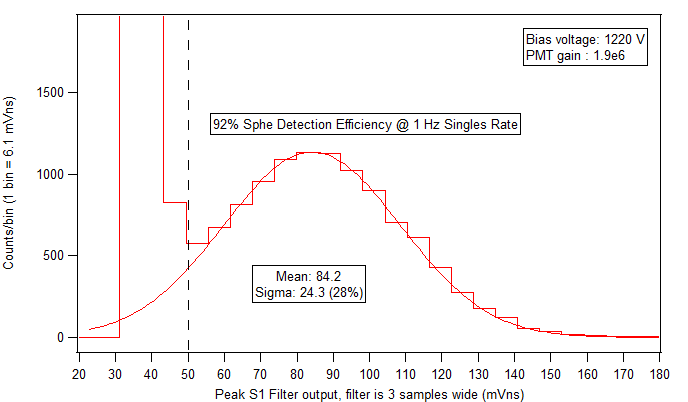}
\tdrfcaption[LowGainResponse]{Lower Gain}{Measurements of the PMT response to SPHE at a gain much lower than the nominal value.}
\end{figure}

Finally, the response of the amplifiers was measured for the case of PMTs that fail to provide the nominal gain. For this test, the HV on the PMT was lowered to \SI{-1220}{\V}, corresponding to a gain of \num{1.9e6}. Figure~\ref{EDCf:LowGainResponse} shows a histogram of the S1 filter output for a three-sample wide filter for signals from the LE channel. The SPHE peak can be seen to have a peak value of \SI{84.2}{\mV\ns} and an rms width of \SI{28}{\%}. The noise peak is also shown. Placing a cut at \SI{50}{\mV\ns} results in an SPHE detection efficiency of \SI{92}{\%}, which meets specifications.  The singles rate of 1 Hz due to noise at this cut is negligible compared to the \SIrange[range-units=single]{50}{200}{\Hz} dark noise expected from the PMTs. 

\tdrsec[Digital]{Digital Electronics}

The LZ digital electronics is based on a digital platform (a motherboard); a prototype of this platform is shown in Figure~\ref{EDCf:DigitalBaseboard}. The final LZ motherboard will be based on this design, but will operate with a more powerful Series-7 Kintex field-programmable gate array (FPGA) from Xilinx ~\cite{Xilinx:2016}. The final motherboard will provide gigabit Ethernet, RS-232, and low-voltage differential signaling (LVDS) interfaces, and four logic outputs, either TTL or NIM. Waveform memory (\SI{3578}{\kilo\byte}) will be provided by the FPGA. A large event-buffer memory of up to \SI{128}{\mega\byte} will be provided by the dual-core processor. The onboard clock can be driven externally in order to synchronize multiple boards to the same clock source. Very high processing power, nominally \num{52} giga-operations per second, will be provided by the onboard FPGA. Two daughter card connectors can host two separate daughter cards, or one daughter card of twice the size. The I/O pins of these connectors are arranged as differential pairs, supporting either the differential or single-ended signals. A dual-core processor will be connected to the FPGA with the 32-bit memory bus, as well as two dedicated 16-bit-wide FIFOs. Readout of the FPGA data can be performed either via the memory bus or via the FIFOs, depending on the application. The board can be hosted in a 6U VME crate, or it can be powered with a tabletop power supply. Power consumption is minimized by using low-voltage chips. 

\begin{figure}[htbp]
{\centering
\includegraphics[width=0.4\textwidth]{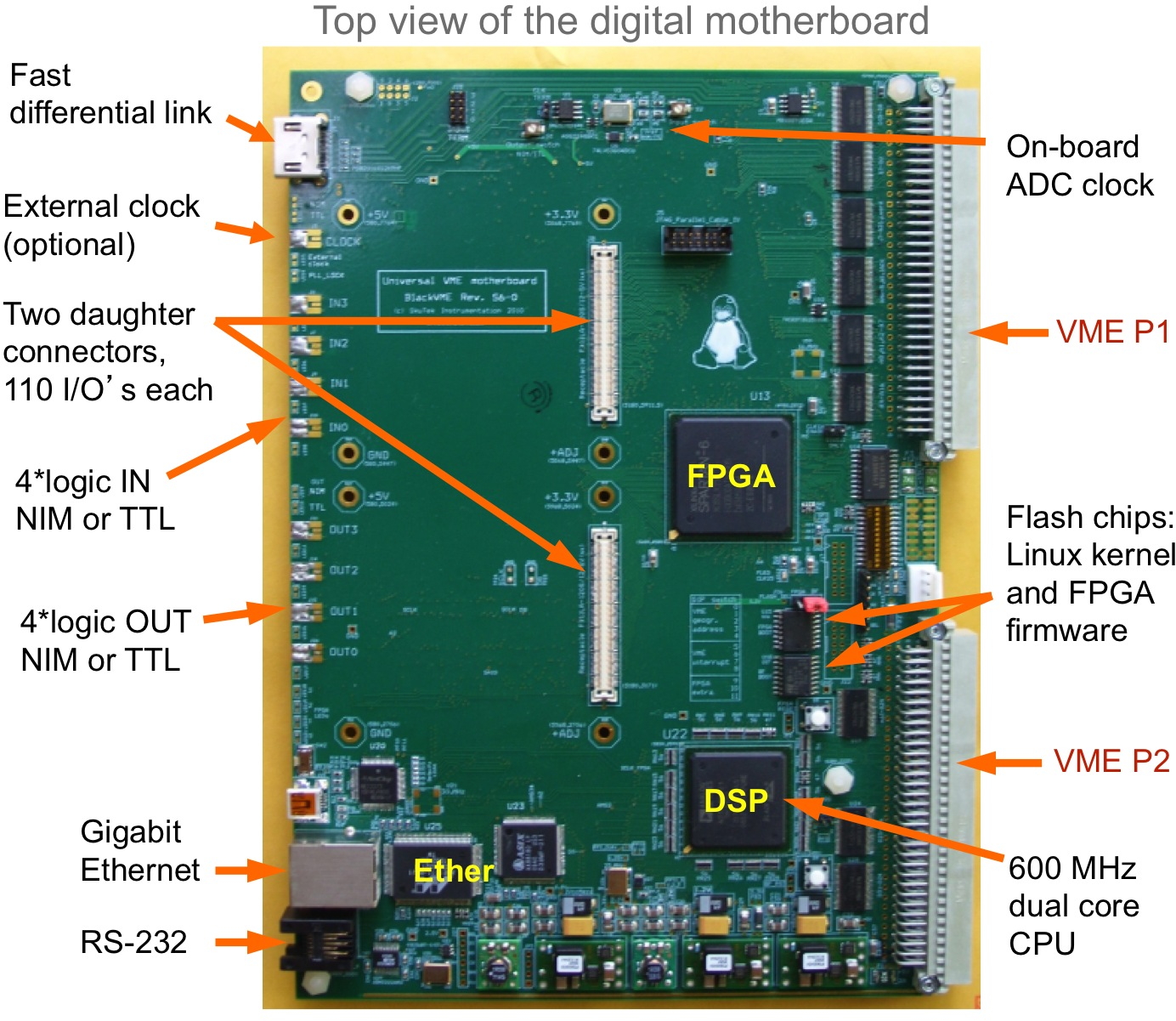}
\includegraphics[width=0.4\textwidth]{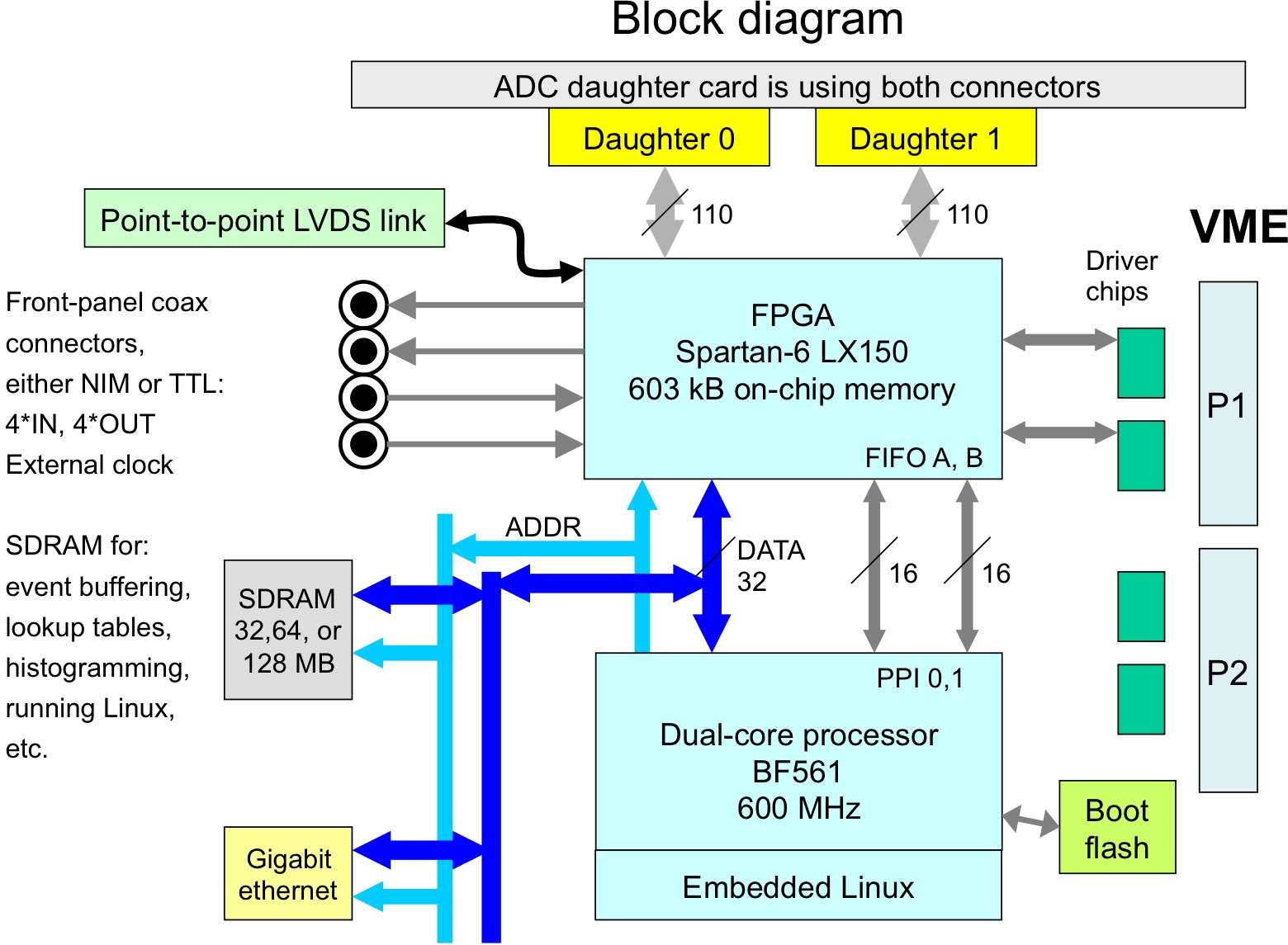}
\tdrfcaption[DigitalBaseboard]{Digital motherboard used to develop LZ digitizers}{The digital motherboard used to develop the LZ digitizers. It provides gigabit Ethernet, RS-232, USB-2, VME, and LVDS interfaces. The FPGA and the dual-core processor are rated at \num{52} and \num{2.4} giga-operations per second, respectively.}
}
\end{figure} 

The onboard processing power and multiple interfaces provide flexibility that can be applied to almost any project. The LVDS links enable custom communication architectures. The Ethernet provides support for distributed experiments and/or standalone remote applications. The processor is running Linux, which is popular, free, and fully customizable, allowing each board to perform on-the-fly data processing and online diagnostics.

\begin{figure}[htbp]
{\centering
\includegraphics[width=0.75\textwidth]{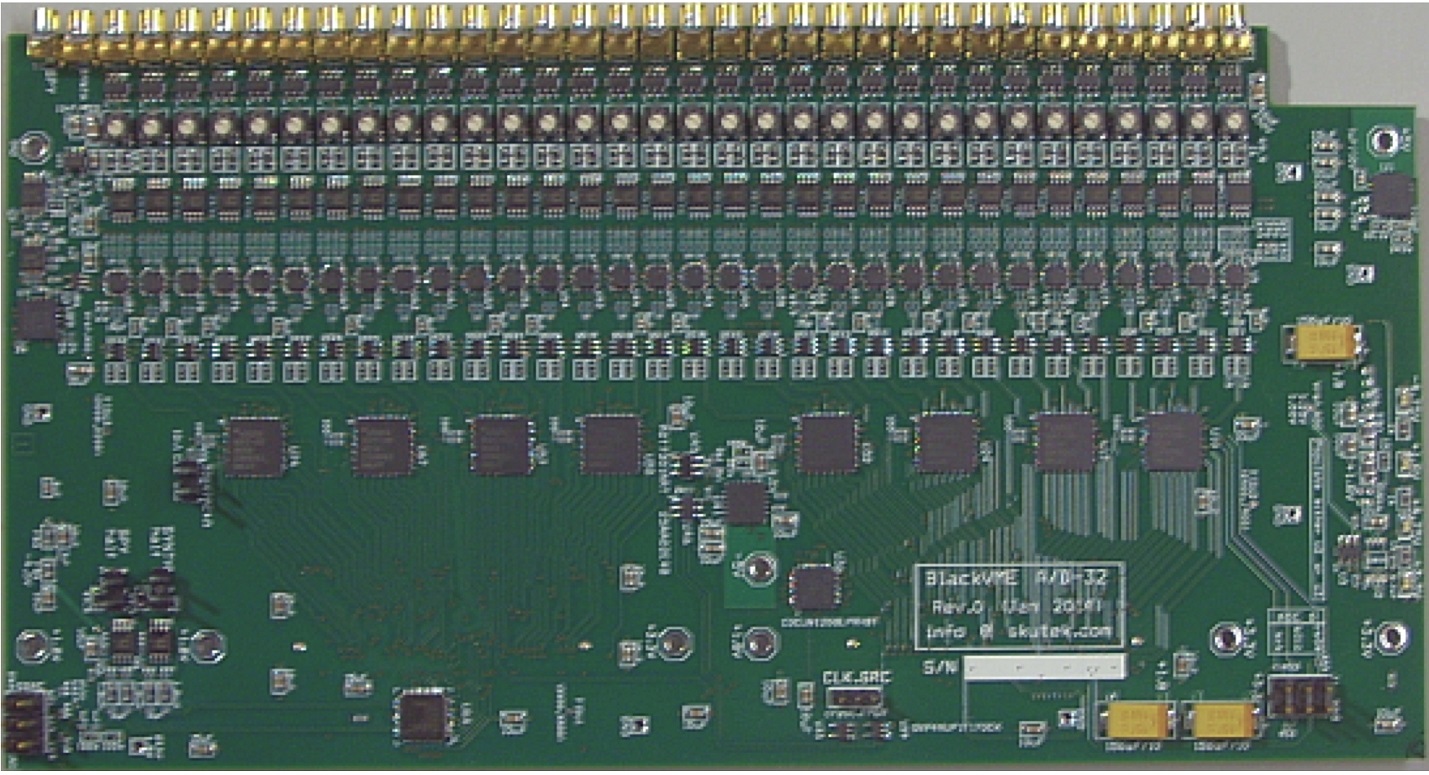}
\tdrfcaption[DDC32DaughterCard]{Prototype ADC daughter card with 32 channels}
{Prototype ADC daughter card with 32 channels.}
}
\end{figure} 

The 32-channel ADC card shown in Figure~\ref{EDCf:DDC32DaughterCard} implements the digitizer front end. It provides \num{32} channels of digitization and two waveform reconstruction outputs for diagnostic. The ADC channels feature remote DC offset control. The card is connected to the two daughter connectors of the digital baseboard that provide the control signals and power. The printed circuit-board layout can accommodate quad A/D chips with sampling frequency up to \SI{125}{\MHz}. This card, installed on the digital motherboard, is referred to as the DDC-32 in the remainder of this chapter.

\tdrsec[DAQ]{DAQ}

The top-level architecture of the LZ DAQ system is shown schematically in Figure~\ref{EDCf:DAQTopLEvel} \cite{Druszkiewicz:2015pcl}. The DDC-32s continuously digitize the incoming PMT signals and store them in circular buffers. When an interesting event is detected, the Data Extractor (DE) collects the information of interest from the DDC-32s. The DEs compress and stack the extracted data using their FPGAs and send the data to Data Collectors (DCs) for temporary storage. The Event Builder (EB) takes the data organized by channels and assembles the buffers into full event structures for online and offline analysis. The DAQ operation is controlled by the DAQ Master (DM) for high-speed operations such as system synchronization and waveform selection, and by the DAQ Expert Control/Monitoring (DECM) system for slow operations such as running setup/control and operator diagnostics. The entire system runs synchronously with one global clock. 

\begin{figure}[htbp]
{\centering
\includegraphics[width=0.75\textwidth]{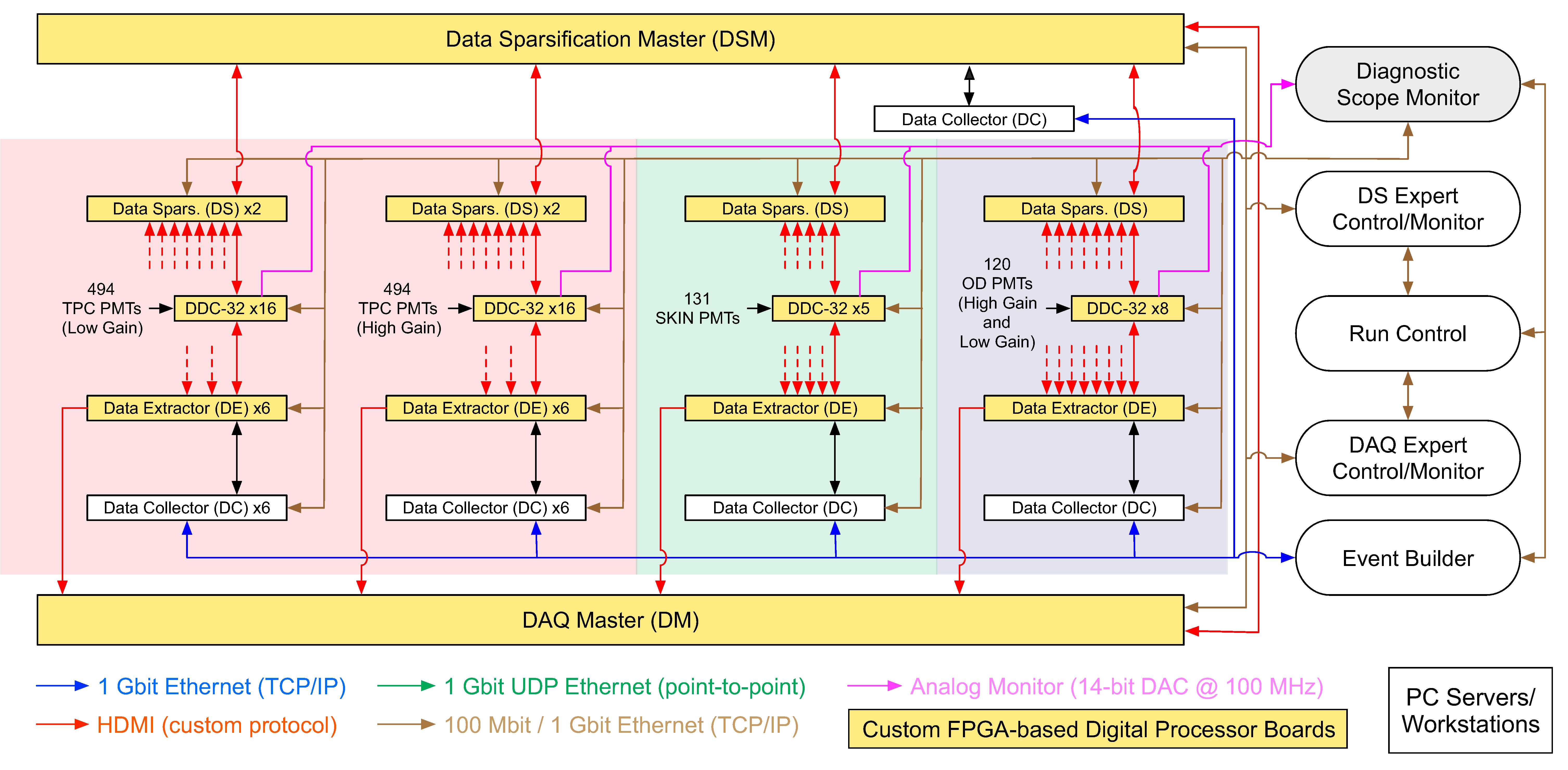}
\tdrfcaption[DAQTopLEvel]{Diagram of the DAQ architecture}{Diagram of the DAQ architecture \cite{Druszkiewicz:2015pcl}. Groups of digitizers (DDC-32) capture the amplified and shaped signals from the Xe, skin, and outer-detector PMTs. The waveforms of interest are extracted from the DDC-32s and compressed by the Data Extractors (DEs) before they are passed to Data Collectors (DCs) for temporary storage. One additional Data Collector will capture reduced sparsification quantities to be merged by the Event Builder with the waveform data in full event files. The DAQ Master Board (DM) coordinates the high-speed operation of the entire DAQ system when the Data Sparsification Master (DSM) signals the detection of waveforms to be preserved. The global clock will be distributed over the shown HDMI links or dedicated NIM clock inputs that are not shown in this diagram.}
}
\end{figure} 

\tdrsubsec[DAQDE]{Data Extraction}

Figure~\ref{EDCf:DAQDetails} shows a more detailed overview of the different key elements of the DAQ system. The digitizers are sampling at \SI{100}{\MHz} with \num{14}-bit resolution over a \num{2}-\si{V} range. During normal operation, the boards will collect waveforms in a Pulse Only Digitization (POD) mode, which is expected to effectively reduce the raw waveform volume by a factor of 50~\cite{akerib:2011ix}. The amount of memory assigned to each channel is set so that no data truncation is expected even if the POD mode reduces the data volume by only a factor of \num{20}. The POD waveforms are stored in dual-buffered memory that is divided into sections that hold the header information and the actual POD samples, as shown in Figure~\ref{EDCf:DAQMemory}. Separate POD header and payload memories will improve the performance of extracting waveform data when the Data Sparsification Master (DSM) detects an event of interest~\cite{Druszkiewicz:2015}. 

\begin{figure}[htbp]
{\centering
\includegraphics[width=0.75\textwidth]{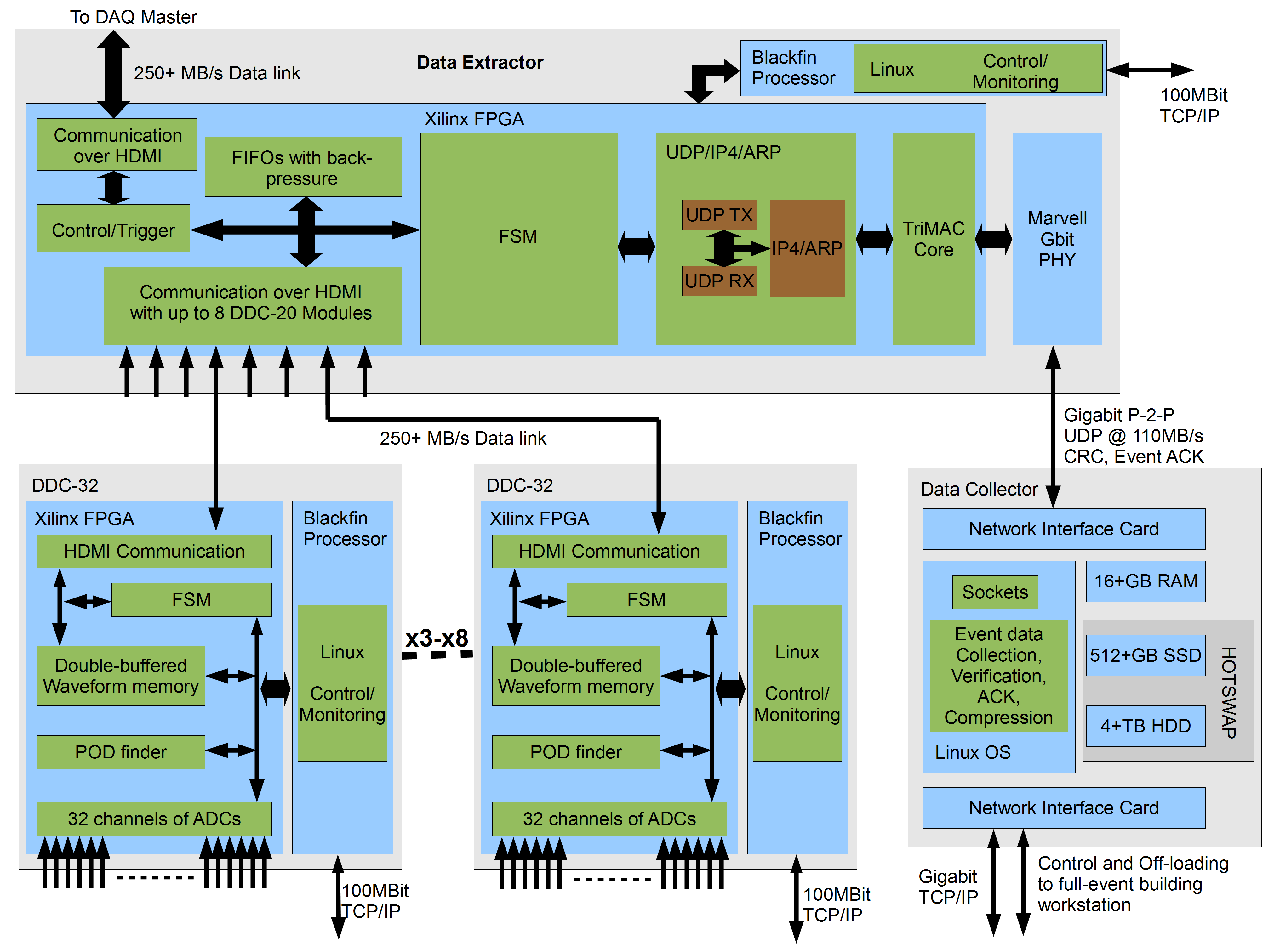}
\tdrfcaption[DAQDetails]{Detailed depiction of interaction between DAQ elements}{Detailed depiction of the inside of and interaction between the key elements of the proposed DAQ system.}
}
\end{figure}

\begin{figure}[htbp]
{\centering
\includegraphics[width=0.75\textwidth]{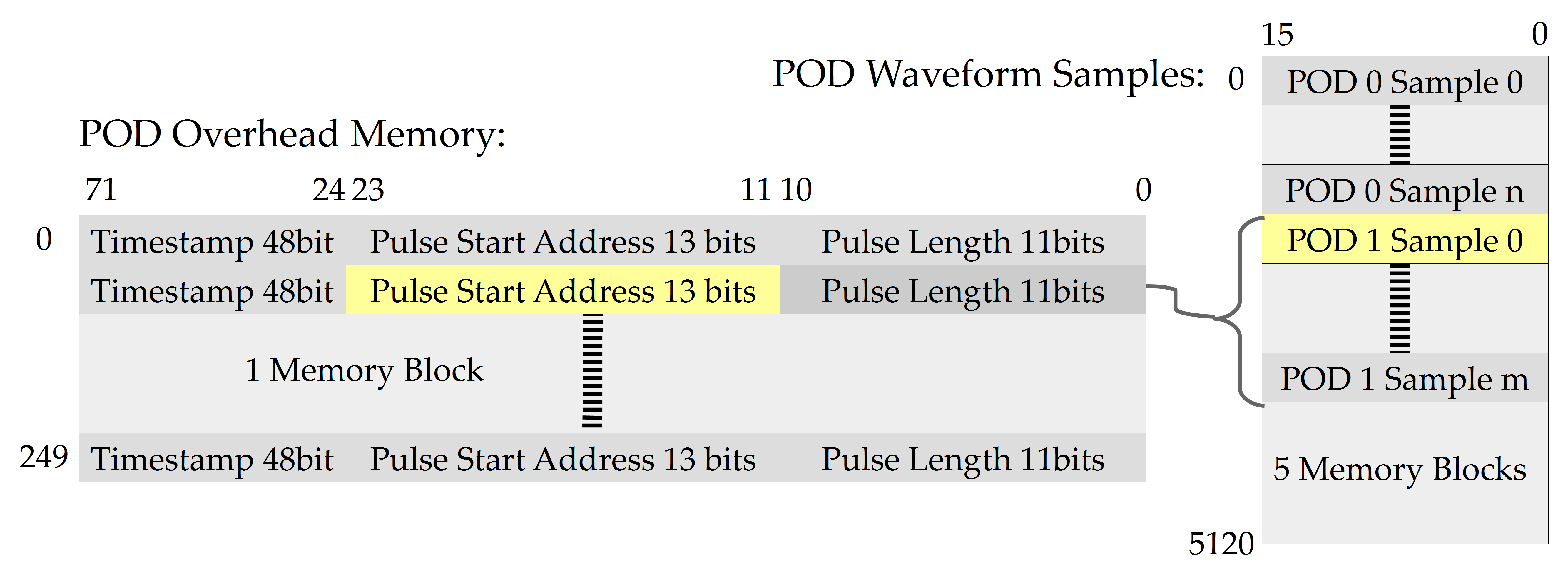}
\tdrfcaption[DAQMemory]{Proposed memory organization of POD waveform storage}{Depiction of the proposed memory organization of POD waveform storage in the FPGA for a single buffer (out of two) of a single channel. Separation of POD overhead and POD samples will improve the performance and ease of extracting information from memory.}
}
\end{figure}

The extreme flexibility that comes with using FPGAs and their internal memories allows us to assign the entire on-chip memory to just one specific channel when needed. This feature will be used for system diagnostics and noise measurements where capturing long, continuous (non-POD) waveforms is important. In such a mode, the DDC-32 will be able to capture \num{10}-\si{ms}-long waveforms, suitable for power-spectral-density analysis. 

The DEs and the DM use the same hardware but different firmware. They use the same motherboard as the digitizer boards, with different daughter cards that enable communication with multiple DDC-32 modules and the DCs. Each daughter card can serve up to \num{14} DDC-32s over HDMI links and one DC over a dedicated gigabit Ethernet connection. 

The HDMI link has seven single-ended lanes used for communicating states of the finite state machines (FSMs) between boards and four LVDS lanes used for fast offloading of the waveforms from the digitizers. On the current motherboard, we have used input/output serial/deserializer (IOSERDES) elements, offered in the Spartan-6 FPGA series, and have confirmed the advertised \num{1}-\si{\giga\bit} throughput per LVDS lane.

\begin{table} [H]
\tdrtcaption[DCSpecs]{Key parameters of the prototype Data Collector.}
{Key parameters of the prototype Data Collector.}
\setlength{\extrarowheight}{5pt}
\centering
\sffamily
\begin{tabular} 
{|m{3cm}|m{4cm}|m{3cm}|m{4cm}|} 
\hline
{\bfseries Processor:} & Intel Xeon E3-1270V3 3.5GHz Quad-Core & {\bfseries HDD 1:} & SAMSUNG 840 Pro Series 256GB SSD \\
\hline
{\bfseries Motherboard:} & ASUS P9D-V ATX & {\bfseries HDD 2:} & Western Digital RE 4TB 7200 RPM \\
\hline
{\bfseries Memory:}	& 16GB Kingston DDR3 SDRAM ECC & {\bfseries Case:}Case: & NORCO RPC-270 2U Server Case \\
\hline
{\bfseries NIC:} & Intel Ethernet Server Adapter I350-T2 & {\bfseries Hot Swap:} & ICY DOCK 3.5" and 2.5" SATAIII 6Gps HDD Rack Tray \\
\hline
\end{tabular}
\end{table}

\begin{table} [H]
\tdrtcaption[DAQLinks]{Summary of the performances of the DAQ links}{Summary of the performances of the DAQ links and their expected utilization levels.}
\setlength{\extrarowheight}{5pt}
\centering
\sffamily
\begin{tabular} {|c|c|c|c|}
\hline
\rowcolor{mrocol}
    {\bfseries Link} & 
    {\bfseries Expected}& 
    {\bfseries Maximum}& 
    {\bfseries Usage} \\
\rowcolor{mrocol}
    & 
    {\bfseries Performance}& 
    {\bfseries Expected Usage}& 
    \\
\hline
LVDS over HDMI & \num{250}+\si{\mega\byte\per\second} & \SI{8.6}{\mega\byte\per\second} & \SI{<3.5}{\%} \\
\hline
DE $\rightarrow$ Gigabit UDP $\rightarrow$ DC & \SI{109}{\mega\byte\per\second} & \SI{34.4}{\mega\byte\per\second} & \SI{<33}{\%} \\
\hline
\end{tabular}
\end{table}

\begin{table} [H]
\tdrtcaption[DAQRates]{Summary of data collectors expected peak storage and buffering capabilities}{Summary of the expected storage and buffering capabilities on the Data Collectors during calibrations.}
\setlength{\extrarowheight}{5pt}
\centering
\sffamily
\begin{tabular} {|c|c|c|c|}
\hline
\rowcolor{mrocol}
    {\bfseries Source} & 
    {\bfseries Data Rate per Data Collector}& 
    {\bfseries Solid-state Drive} &
    {\bfseries Hard-disk Drive} \\
\rowcolor{mrocol}
    & 
    {\bfseries (uncompressed)}& 
    {\bfseries (512 GB)}&
    {\bfseries (1 TB)} \\
\hline
\IKreT & \SI{\sim5.2}{\mega\byte\per\second} & \SI{\sim1.1}{days} (raw) & \SI{\sim9}{days} (raw) \\
      &           & \SI{\sim4.4}{days} (7z) & \SI{\sim35}{days} (7z) \\
\hline
LED & \SI{\sim34.4}{\mega\byte\per\second} & \SI{\sim4}{hours} (raw) & \SI{\sim1.4}{days} (raw) \\
      &          & \SI{\sim16}{hours} (7z) & \SI{\sim5.3}{days} (7z) \\
\hline
\end{tabular}

\end{table}

The gigabit Ethernet link between the DE and the DC utilizes the User Datagram Protocol (UDP). To mitigate the limitations of UDP, we will add for each event a cyclic redundancy check (CRC) and acknowledgment, and also use the dedicated Ethernet link in a point-to-point topology to eliminate congestion and packet loss. All UDP packets formed by the DE’s FPGA will be sent over a single gigabit Ethernet cable to a dedicated network port on the DC.

The DCs will be implemented using server-grade, rack-mountable (2U) workstations. We have tested a prototype DC with the parameters shown in Table~\ref{EDCt:DCSpecs}. We were able to confirm reliable data transfer from the DE to the solid-state drive (SSD) in the DC. With appropriate modifications to the network interface card (NIC) drivers, we were able to transfer data at a constant rate of \SI{109.8}{\mega\byte\per\second}, with no data loss or corruption. Table~\ref{EDCt:DAQLinks} summarizes the tested link performances and their expected utilization.

The data stored on the DCs are processed by the Event Builder (EB). The EB builds events by sampling the different DC disks and collecting all information associated with a given event. The resulting event (EVT) files are compressed in order to achieve the smallest possible file size for storage and transmission. The EVT files are written to RAID array \#1, located in the Davis Cavern, before being transferred to the RAID array \#2, located on the surface at SURF. The RAID arrays have sufficient storage capacity for a full month of data taking.

The DAQ system is designed to be able to allow LED calibrations of the TPC PMTs in about \num{10} minutes. This requires an event rate of \SI{4}{\kHz}, resulting in a \SI{\sim340}{\mega\byte\per\second} total waveform data rate. The fast (\num{400}+\si{\mega\byte\per\second} for SSDs) storage drives in the individual DCs allow for sustaining such collection rates, and the amount of space offered by each DC permits significant buffering in case of connection problems to the off-site permanent storage, as shown in Table~\ref{EDCt:DAQRates}.

We are planning to use 7z compression. Nominally, 7z uses the LZMA algorithm, but we have found that the PPMd algorithm~\cite{Salomon:2010} is better suited for the waveforms we are going to collect. Based on comparisons made using LUX data, the 7z PPMd compression offered an additional \SI{33}{\percent} data reduction for waveforms and a \SI{15}{\percent} data reduction for reduced quantities over gzip, which was used in LUX. 

Because the POD mode relies on time stamping, the entire DAQ system will run with a global \SI{100}{\MHz} clock. 
We are considering two ways of distributing the clock, either by using a proven method of dedicated NIM logic clock FAN-IN/OUTs or using clock recovery from the serial links of the HDMI cables; the latter method is currently being evaluated. We are also looking at the possibility of using one of the single-ended channels of the HDMI cable for a dedicated clock signal.

The configuration, acquisition control, and monitoring will be done using the DECM workstation. For improved robustness, the DECM will communicate with all DAQ elements using an isolated 100/1000 LAN; the run-control (RC) system, described in Section~\ref{EDCS:Online}, will not have direct access to the front-end digitizers. We have successfully tested and are planning to use ICE middleware as our communication framework~\cite{Henning:2004}.

\tdrsubsec[DAQDS]{Data Sparsification}

The design of the LZ data sparsification system is based on our experience with event selection in the LUX trigger system. The top architecture of the LZ data sparsification system is shown schematically in Figure~\ref{EDCf:DAQTopLEvel}. The DAQ and data sparsification firmware operates in parallel on the FPGAs of the DDC-32 digitizers.

The DDC-32s continually process the incoming pulses and extract specific required quantities such as pulse area, pulse height, and time of occurrence. By using digital filters with various filter lengths, the system can distinguish S1 and S2 pulses. The parameters extracted from the waveforms by the DDC-32s are sent to the Data Sparsifiers (DSs) for further processing and generation of secondary quantities such as the multiplicity vectors of groups of PMTs. These secondary quantities are sent to the Data Sparsification Master (DSM), where the final waveform selection decision is made. The decision to preserve the current waveforms is sent to the DAQ Master. It is important to reiterate that much of the information based on which a given event selection was made is going to be off-loaded alongside event waveform data and stored in the full event files. This will allow for simpler cross-checking and verification of the system performance off-line.

A new feature under development is the creation of a total waveform digital sum that can be created at the DSM from all or subset of DDC-32 channels as shown in Figure~\ref{EDCf:DAQSumCreation}. This allows for including total area cuts in the event selection decision process.

\begin{figure}[htbp]
{\centering
\includegraphics[width=0.65\textwidth]{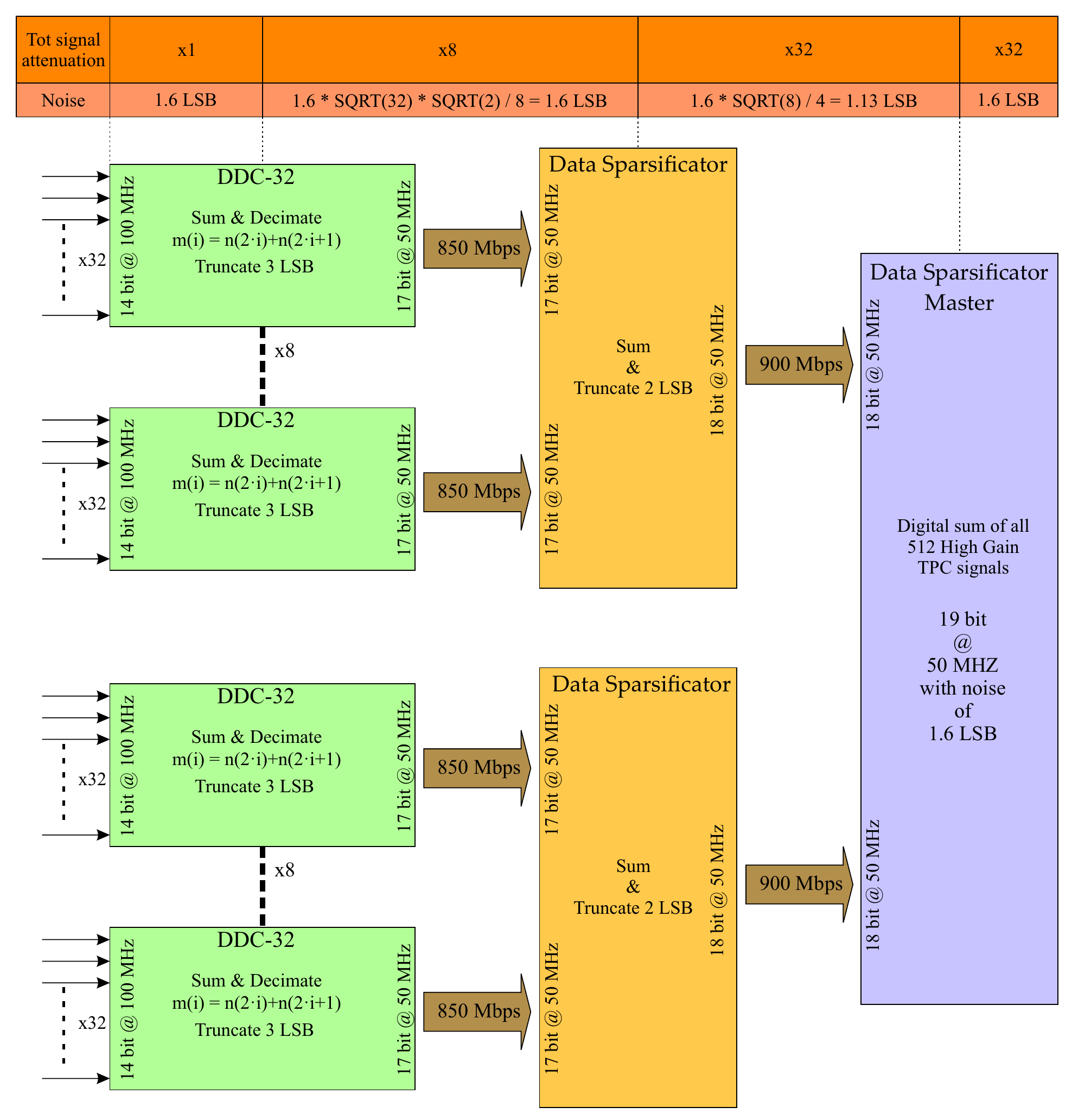}
\tdrfcaption[DAQSumCreation]{Full digital waveform sum creation}{Schematic of digital waveform sum creation from all or subset of DDC-32 channels that allows the total pulse area to be a part of the sparsification decision making.}
}
\end{figure}

The LZ DAQ/Sparsification system is capable of handling event rates up to \SI{\sim250}{\kHz}. This is much higher than the highest event rates expected in the LZ detector, which will occur during LED calibrations (\SI{\sim4}{\kHz}).

As in LUX, the PMT signals will be processed by parallel digital integrating filters, allowing for discrimination against area of the incoming pulses. The filter with an integration width of about \SIrange{60}{100}{\ns} is tailored to S1-like pulses; the second filter with a few-microsecond width is optimal for wider S2-like pulses. The filters are designed to also perform automatic baseline subtraction.

As shown in Figure~\ref{EDCf:DAQTopLEvel}, the TPC, skin, and outer-detector PMTs will use separate DDC-32s and DSs because these PMTs have different purposes and the firmware will be tailored to their specific needs. Such separation also makes it easier to apply different scaling factors for the digitization of signals from different PMT groups. The three major waveform-selection modes for the central TPC PMTs are summarized in Table~\ref{EDCt:DSModes}.

\begin{table} [H]
\tdrtcaption[DSModes]{Summary of three major waveform-selection modes for the central TPC PMTs.}{Summary of three major waveform-selection modes for the central TPC PMTs.}
\setlength{\extrarowheight}{5pt}
\centering
\sffamily
\begin{tabular} {|m{1.5in}|m{4.5in}|}
\hline
\rowcolor{mrocol}
    {\bfseries Trigger Mode} & 
    {\bfseries Summary} \\
\hline
S1 Mode	& Detection of coincident S1-like signals across selected channels. No fiducialization. \\
\hline
S2 Mode & Detection of coincident S2-like signals across selected channels. Fiducialization in the $x$,$y$ plane. \\
\hline
S1 and S2 Mode & Detection of S1-like signals followed by S2-like signals within a selected drift time range. Fiducialization in $x$, $y$ and $z$ planes. \\
\hline
\end{tabular}
\end{table}

If a waveform selection condition is met, the DSM sends a signal to the DM. At the same time, a packet containing the selection parameters used to make the decision is merged with the captured waveform data. This allows for offline evaluation of the selection decision for every individual event. This feature has proved extremely valuable in monitoring LUX data quality.

Similarly to the DAQ system, the DS system will be controlled and monitored by a dedicated DS expert control/monitoring (DSECM) workstation over an isolated 100/1000 LAN network. The DS system will operate on the same \SI{100}{\MHz} global clock as the DAQ.

The LZ DS system will adopt and expand on the monitoring capabilities of the LUX system. Capabilities such as performing continuous noise sweeps (monitoring S1 and S2 filter crossing rates as a function of threshold) or monitoring channel hit distributions of the selected events, all in parallel to regular uninterrupted data-sparsification operation, have been invaluable in LUX and surely will be in LZ.

Each of the DDC-32s has a dual \num{14}-bit, \SI{100}{\MHz} analog reconstruction output (SPY) allowing for diagnostics and scope monitoring. The SPYs can be sourced with individual incoming channels or their digital sum. They can also be sourced with S1 and S2 filter outputs, individual or summed, or any other signal internal to the FPGA. This feature will be very useful, especially in development and deployment stages, as proven in the LUX project.

\tdrsec[HV]{PMT HV}
LZ will use the Wiener Mpod LX HV system to bias the xenon and outer-detector PMTs. This same system is currently being used for LUX. The system will provide negative HV to the xenon PMTs and positive HV to the outer-detector PMTs. The HV modules are part of the EDS 201-30x 504 series. Each module provides \num{32} HV channels and uses a common floating ground. The voltage ripple is less than \SI{5}{\mV}. The HV can be set with a resolution of \SI{10}{\mV}, and the current on individual channels can be measured with a resolution of \SI{50}{\nA}.

HV connections to the HV filters are made using Kerpen cable with Redel connectors on both ends. Each Kerpen cable carries HV for \num{32} channels. Important properties of the HV system are listed in Table~\ref{EDCt:LZHVSystem}.  

\begin{table} [H]
\tdrtcaption[LZHVSystem]{Details of the PMT HV system.}{Details of the PMT HV system.}
\setlength{\extrarowheight}{5pt}
\centering
\sffamily
\begin{tabular} {|l|c|c|c|c|c|}
\hline
\rowcolor{mrocol}
    {\bfseries HV Module} & 
    {\bfseries Maximum HV}& 
    {\bfseries Maximum}& 
    {\bfseries PMTs}&
    {\bfseries Channels}&
    {\bfseries Number} \\
\rowcolor{mrocol}
	& 
    {\bfseries HV}& 
    {\bfseries Current}& 
    &
    &
	{\bfseries of Modules} \\
\hline
EDS 201 30n & \SI{-3000}{\V} & \SI{500}{\micro\ampere\per ch} & TPC/Skin & \num{625} & \num{21} \\
\hline
EDS 201 30p & \SI[retain-explicit-plus]{+3000}{\V} & \SI{500}{\micro\ampere\per ch} & Outer-detector & \num{120} & \num{4} \\
\hline
\end{tabular}
\end{table}

\tdrsec[Cables]{Cables}
All network, signal, and HV cables are low smoke zero halogen (LSZH) cables. LZ uses the same cables that have been approved for LUX use in the Davis Laboratory. The exact lengths of the HV and the signal cables will be fixed once the location of the analog amplifiers is fixed and the route of the cable trays has been finalized.

The types of cable and their lengths are listed in Table~\ref{EDCt:LZCables}. The networking cables will have various lengths; individual lengths will vary based on the location of the network switched and the location of the devices to be connected. About half of the network cables are used by the slow-control system; the other half will be used for the networks associated with the DAQ, trigger, and online systems.

We have not been able to identify a suitable HDMI cable that provides good performance and uses LSZH materials. The total length of these cables is small and the electronics racks are enclosed and vented directly into the exhaust system of the Davis Campus. This arrangement will be assessed by the SURF safety team. Note that rack enclosures and connections to the exhaust system are also used in LUX for this same reason.

\begin{table} [H]
\tdrtcaption[LZCables]{Information on LZ signal, logic, HV, power, and network cables.}{Information on LZ signal, logic, HV, power, and network cables.}
\setlength{\extrarowheight}{5pt}
\centering
\sffamily
\begin{tabular} 
{|l|c|c|c|c|} 
\hline
\rowcolor{mrocol}
    {\bfseries Cable Type} & 
    {\bfseries Type}& 
    {\bfseries Length (ft)}& 
    {\bfseries Number of Cables}&
    {\bfseries Notes}\\ 
\hline
Network	& Belden 7936A & \num{10000} & \num{750} & Cat 6, LSZH \\
 & & (total) &  & Various lengths \\
\hline
Signal & LMR-100A-FR & \num{56} & \num{143} bundles & LSZH \\
 &  &  & (\num{8} cables/bundle) &  \\
\hline
Logic & LMR-100A-FR & \num{2} & \num{500} & LSZH \\
\hline
HV & Kerpen & \num{56} & \num{25} & LSZH \\
 & & &  & \num{32} channels per cable \\
\hline
Power cords & CordMaster & \num{6} & \num{100} & LSZH \\
\hline
HDMI & TBD & TBD & \num{99} & \\
\hline
\end{tabular}
\end{table}

\tdrsec[SC]{Slow Control}

\tdrsubsec[SCFunctions]{Functions of the slow control}

The slow control system performs supervisory control and monitoring of all the major subsystems of the experiment. The functions of slow control can be classified in several categories as described below: 

\begin{itemize}
\item \textbf{Experiment parameter monitoring and logging.} This includes periodic readout of various sensors distributed throughout the xenon circulation, storage and recovery systems (temperature, pressure, gas flow etc.) and of advanced sensors installed in the detector (capacitive level and distance sensors, acoustic sensors and loop antennas). For the electronic racks the status of the uninterruptible power supplies (UPS) as well as temperature and current consumption of the front-end amplifiers will be monitored. Also, the environmental data, such as ambient temperature and pressure and radon concentration in the air, will be collected. 
\item \textbf{Control over experiment subsystems.} This includes controlling (by means of electrically-triggered pneumatic valves) pathways for performing operations of xenon circulation, storage and recovery, controlling the detector temperature via thermosiphons and heaters, and controlling distribution of liquid nitrogen required for thermosiphon operation. The slow control will also serve as a front-end to the calibration systems (krypton injection, radioactive source delivery, optical calibration) to guarantee that they are not interfering with normal detector operation. Finally, high voltage power supplies for the field-shaping grids and the PMTs in the detector will be programmed and monitored through slow control.
\item \textbf{Safety.} One of the most important functions of the slow control system is to ensure safe operation of the detector and all the experiment subsystems in order to prevent equipment damage and xenon loss. This is achieved by implementing a set of interlocks that would protect the system from erroneous operator command. Additionally, the alarm system will be programmed to alert on-site crew and/or system experts when certain key parameters are out of the predefined range. The emergency xenon recovery, activated by pressure rise in the detector, will safely transfer xenon to the storage facility if the thermal control of the detector fails.
\item \textbf{User interface.} The slow control system will provide the GUI to display all relevant telemetry information and controls organized in pages by subsystem. The plots of parameter history with be available as well. The controls available to the user will depend on privilege and expertise levels as determined by querying the centralized access control and privilege management facility provided by the experiment IT infrastructure. All control changes performed by users will be logged.
\item \textbf{Automation and interfaces with run control and offline systems.} In addition to the set of scripts (procedures) directly accessible to slow control, a set of standard high-level operating procedures will be made compatible with remote procedure calls from run-control or to the offline systems.   Programming these scripts on the slow-control (server side) will  make these sequences faster, more reliable, and will eliminate possibility of an operator error.   In some cases these scripts will include break points requiring explicit feedback from the run control client to advance to the next step, further ensuring system safety and accuracy.
\end{itemize}

\tdrsubsec[SCRequirements]{Requirements}

The main requirement to the slow control system is to provide the monitoring and control infrastructure that will guarantee safety of the experiment subsystems and xenon supply in all modes of operation. The Level 2 requirement R-180004 states: \emph{Guarantee the safety of xenon supply and the xenon circulation system. Use Programmable Logic Controllers (PLCs) to control and monitor the xenon purification, circulation, and storage systems.} The corresponding Level 3 requirements specify the necessary parameters of the monitoring and control infrastructure:
\begin{itemize}
\item \textbf{R-180501:} Monitor state of detector. Monitor up to 2,500 channels.
\item \textbf{R-180502:} Provide access to detector controls. Control up to 1,400 channels.
\item \textbf{R-180503:} Provide an alarm system. Allow for alarm priorities, alarm profiles, and alarm pipelines.  
\item \textbf{R-180504:} Record system operations and system state. Provide a history of all monitored parameters and changes in the status of controls as well as a log of user access and actions.
\end{itemize}

\tdrsubsec[SCOverview]{Overview}

\begin{figure}[htbp]
{\centering
\includegraphics[width=0.75\textwidth]{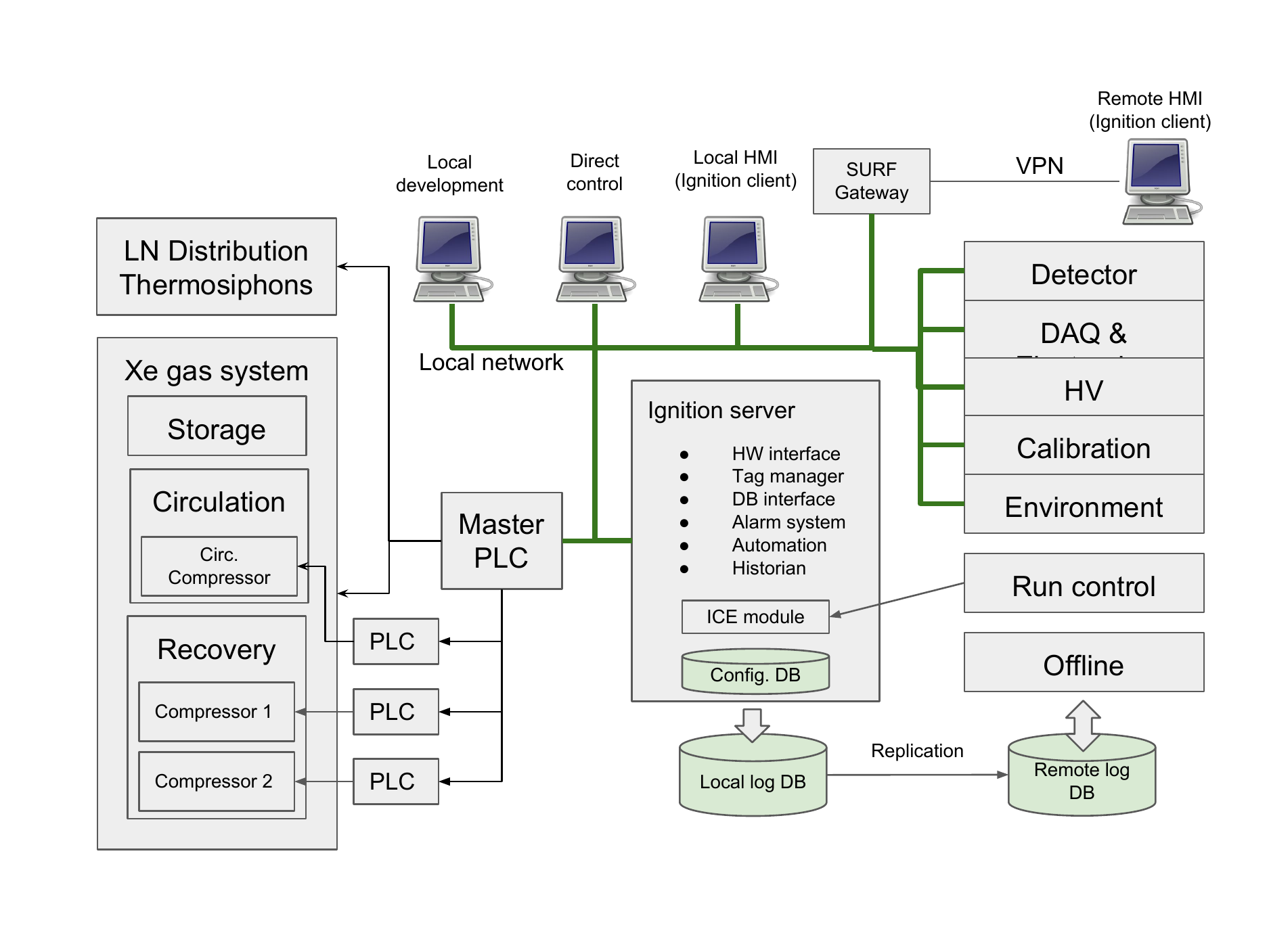}
\tdrfcaption[SlowControlFunctionalDiagram]{Slow control functional diagram}
{Slow control functional diagram.}
}
\end{figure}

The functional diagram of the slow control system and its interaction with the experiment subsystems and infrastructure is shown in Figure~\ref{EDCf:SlowControlFunctionalDiagram}. The core of the slow control system is composed of two components: (1) the integrated supervisory control and data acquisition (SCADA) software platform \emph{Ignition} from Inductive Automation~\cite{sqlbridge:2015}, and (2) the Siemens SIMATIC S7-410H programmable logic controller (PLC) with associated I/O modules~\cite{Siemens410-5h}. The Ignition server works as the main hub of the slow control system and is responsible for collecting and archiving the sensor data and providing the authorized users with access to the controls. Additionally, the Ignition server runs the alarm system, provides the scripting engine for experiment automation and provides the ICE server for interfacing with run control. It also provides a GUI for accessing historical data in the form of plots by local and remote clients. The configuration data (properties of sensors and controls, alarms, user preferences etc.) are stored in the local configuration database. 

The PLC component is designed to guarantee safety of the critical subsystems (xenon circulation, storage and recovery, detector temperature control and liquid nitrogen distribution) by implementing necessary interlocks to safeguard their correct operation. It is also programmed to ensure safety of the xenon supply even in the case of a major infrastructure failure, for example a prolonged power or network outage with no access underground. The data flow between the Ignition server and the PLC passes through a dedicated TCP/IP network. The Ignition server directly communicates with the non-critical experiment subsystems through the same network. The computers used for local programming and control of both PLC and Ignition server are connected to this network as well. Communication between the Ignition server, the PLC, other non-critical experiment subsystems, operator consoles, and PLC software development workstations uses Ethernet. Multiple TCP/IP subnets in conjunction with router access control lists will be used to separate the slow control network into security zones with appropriate access restrictions.

The historical data log is stored in a local MySQL slow control database mirrored to a replica database from which these data can be accessed by clients on the LZ experiment network including the offline system. 

\tdrsubsec[SCPLCSystem]{PLC System}

The PLC system is the core component of the slow control which is expected to maintain the detector and xenon system in a safe state even in the case of failure or shutdown of the rest of the slow control infrastructure. It is designed to provide continuous real time monitoring and control of the critical subsystems, including: 

\begin{itemize}
\item Xenon purification, circulation and storage systems,
\item recovery compressors,
\item detector temperature control and liquid nitrogen distribution systems,
\item vacuum pumps. 
\end{itemize}

The master control of the the critical subsystems is performed by a Siemens S7-410H Redundant Hot Backup PLC with S7-300 associated I/O modules. This PLC allows bumpless transfer from one active CPU to a backup CPU, with synchronization and self-testing of both CPUs. Each CPU is connected to a separate interface module (IM), allowing fault tolerant process control and bumpless programming and hardware changes in run mode.  Members of the LZ run control team have extensive positive experience with the Siemens S7-410.  This PLC system will be programmed using the Siemens PCS7 engineering programming software (meeting IEC 61131-3 standard).

Four compressor units (primary and secondary recovery compressors and two circulation compressors) are controlled by local Beckhoff PLCs, sharing data with the master PLC via Profibus. The Master PLC provides system data and interlocks to these subordinate PLC systems. Compressor-related instruments and valves are connected directly to Beckhoff modules. In the unlikely event of the failure of the master PLC, local PLCs can operate independently. Control logic resides in both master and locals PLCs. Local control of PLCs is also provided via touch screen so the autonomous units can be controlled directly by the on-site operator in case of an emergency. All PLCs are connected to a single dedicated private network with TCP/IP connectivity to 3 development machines (2 underground and one at the surface). The development machines will be remotely accessible through a secured connection (e.g. VPN).

More information on the PLC system and its support of xenon recovery operations is given in section~\ref{XCSSs:SlowControl}.

\tdrsubsec[SCIgnition]{Ignition}

The choice of Ignition as the software platform for LZ slow control is based on the wide range of highly customizable, easy to use core functions and tools for development of efficient and robust SCADA systems. This set of tools includes: the tag management system, scripting and automation facilities, a versatile alarm system, a user management interface, a historian function, a database interface, a versatile toolbox for GUI desgin as well as a number of other functions and utilities.  Ignition comes with a comprehensive library of device drivers supporting most of the hardware used in LZ.

Ignition follows a web-based client-server model where the server is deployed on site and takes care of communication with all the hardware while GUI clients can run both locally and remotely. It has a modular, easily expandable architecture: an extensive SDK allows developers to build their own custom modules and device drivers in Java or Python. At a higher level, Sequential Function Charts (SFC) or Python scripts can be used for development of automation procedures.

A high level of system availability is maintained even in the case of a system fault through the automatic swap to a mirror server to which configuration is constantly backed up (only a few minutes are lost in such an event).  If even higher system availability is determined to be necessary, virtually zero downtime can be achieved by running two redundant servers in parallel with automatic hot swap (for the cost of the second server license). Security is provided by support of secure communication with clients, user authentication (site-wide authentication is possible through an Active Directory), user role management and action logging.

Ignition is also expected to be more cost-efficient for scientific automation than most commercial SCADA solutions as it is licensed by server with no limitation imposed on number of devices, tags or clients. Also, due to its multi-platform nature, it does not limit the developers or users to a single operating system.

\tdrsubsec[SCTagsInterfaces]{Tags and device interfaces}

A \emph{tag} is the term used for an elementary unit of information in industrial automation. A tag can be bound to a device (made readable for monitoring or writable for control), a record in a database or a variable in control software. Ignition tags follow an object oriented philosophy. \emph{User Defined Tags} (UDT), equivalent to classes used in programming, allow the user to define a set of generic tags each associated with a certain device type from which all the tags for individual devices are later derived. A well-structured UDT scheme speeds up the setup, configuration and modification of large groups of sensors. UDTs can be associated with GUI templates with positive impact on the GUI usability as well as a reduction in the development time and effort. A UDT library developed for the System Test slow control system will be used as the base for LZ slow control UDT.

The information transfer between the tags and the associated devices can be handled in several different ways. The preferred one is by using the OPC-UA protocol supported by many PLCs either directly or through middleware solutions (for example the Kepware server). In addition, Ignition has its own OPC-UA server with drivers for many well-known PLC brands, including Siemens PLCs. For the rest of the devices the preferred protocol will be MODBUS which is also supported by the Ignition OPC-UA server. The devices not supporting MODBUS (e.g. RGA units) will be interfaced via custom made Java drivers written in the Ignition SDK. All these interfacing schemes were successfully implemented and field-tested for the System Test slow control system.

\tdrsubsec[SCHistorianAlarmsAutomation]{Historian, alarms and automation}

Ignition offers highly customizable historian features for logging the values of the selected tags in a database. There are options for both periodic logs and those on an evaluate-on-change basis. It is also possible to have different timing parameters for different tag classes or temporarily increase the readout rate for a particular tag. The GUI history explorer module developed for the System Test provides tools for rapid historical data visualization and analysis.

Ignition also provides a powerful alarm framework whereby each tag can have multiple alarm levels associated with it. Alarms can be configured with different profiles that include priority, escalation scheme and notification pipeline. Notifications can be both local (client GUI, speaker, siren) and remote (email, VoIP or text message). The configuration of these alarm profiles is done by the developers in the Designer. Individual alarm activation/deactivation, setpoint configuration and acknowledgments can be done in the client GUI by a user with adequate privilege level. This allows developing a unified interface for setting alarms and storing pre-defined alarm configurations on a per-user basis; for example, a gas system expert may have a particular set of alarms that they want to be active during a troubleshooting exercise.

For further customization, Ignition provides a Python scripting engine. Python scripts can run both in the client or gateway and can be attached to event handlers (e.g., mouse clicked, key pressed, a tag changing state events), a timer or run inside an SFC. Since Python scripts running on the client block the client execution, these are only suited for GUI customization or very simple automation routines. On the other hand, scripts running in the SFC which are executed in the gateway, are intended to control complex automation routines and procedures. SFCs use the IEC61131-3 programming specification (also used for PLCs) and can be developed in the designer using a drag and drop programming tool. SFCs can be started and monitored both from gateway or Clients GUI but they always run on the gateway. Several SFCs can run in parallel and can be programmed to run for extended periods (even surviving a gateway restart).

\tdrsubsec[SCGUI]{Graphical User Interface}

Ignition offers an efficient and intuitive environment for high performance GUI development. The GUI design concepts were extensively tested and refined  during the development of slow control for the System Test until the GUI configuration met with the approval of both project scientists and engineers. The LZ slow control GUI will be based on this configuration and will offer different levels of supervision and control of the system.

At a higher level, several panels will show the real-time status of the system and subsystems from which an operator will be able to tell the overall health of the system. These panels will allow the authorized user to verify and control key system parameters and assess high level information, for example alarm status, current operation mode, status of automation scripts, etc. These panels will be designed so that even a novice operator could recognize an abnormal situation. Furthermore, access to the solutions for the common problems and tasks will be provided.

At a lower GUI level, detailed information about every sensor and actuator in the system will be provided for debugging and development purposes. The panels will be organized in form of complete but simple P\&IDs, as shown in the example in Figure~\ref{EDCf:SlowControlGUI}. Upon a click on a P\&ID element, a new window will open and the system experts will be able to see and change the sensor configuration or status of the control, allowing complete access and control for the run control system experts.

At a very high level, a single summary plot of the entire system can be prepared by slow control and sent to run control for display in a single status pane on the run control GUI.   This high level display can ensure that key information is available to any shift personnel regardless of run-control expertise, and regardless of the specific arrangement of windows, dialog boxes or other information on the slow-control system.

\begin{figure}[htbp]
{\centering
\includegraphics[width=0.75\textwidth]{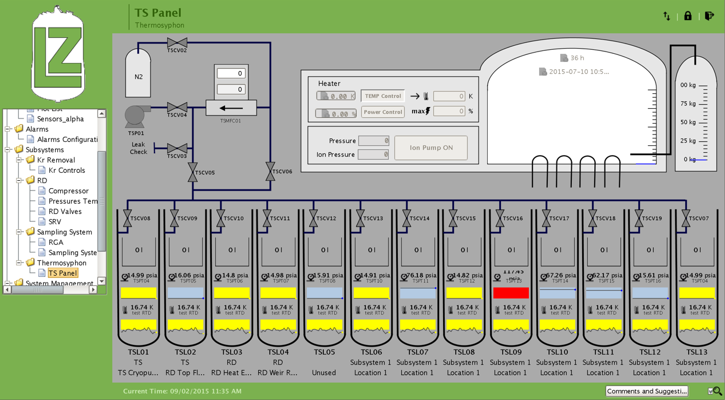}
\tdrfcaption[SlowControlGUI]{An example of a low-level panel in the System Test slow control GUI}
{An example of a low-level panel in the System Test slow control GUI.}
}
\end{figure}

In a addition, the GUI will provide panels for historical data plotting (individual, multi- and scatter-plots) and analysis.  The lists of sensors, controls and alarms similar to those used in the LUX slow control system will be provided as well. For increased usability, all the viewing panels will be fully customizable so that each user can save personalized views between usages. Finally, administrative panels will give access to advanced system configuration such as tag metadata, user management, etc.

\tdrsubsec[SCIntegration]{Integration into experiment online IT}

The slow control system will seamlessly integrate into the LZ experiment online IT and take advantage of the network security architecture and the authentication/authorization infrastructure provided for LZ online systems (see \ref{EDCS:Network}). The main servers for slow control (Ignition and the MySQL database) will be run as virtual machines, allowing for an additional layer of redundancy, fault tolerance and easy backup/restore.

With the possible exception of operator consoles where physical access is restricted, all users will authenticate to the slow control system with their own credentials (no shared accounts). All roles and permissions for slow control are maintained in the central directory and made available to Ignition through LDAP. 

\tdrsec[Online]{Online System}

The core of the online system is the run-control system (RC), schematically shown in Figure~\ref{EDCf:OnlineBlockDiagram}. It comprises the software and hardware required to allow the operator to define and initiate data collection runs of different types, control and monitor subsystems (DAQ, slow control, the event builder (EB), and offline software), and log key information to the database. The physical system is hosted on a single rack-mount computer, identical to the hardware used for the event builder. The use of identical hardware with an on-site spare minimizes any downtime of this critical system. RC software is committed to the LZ GIT repository and is maintained on a mirror system throughout the project and operation periods to allow off-line debugging and code development.

\begin{figure}[htbp]
{\centering
\includegraphics[width=0.5\textwidth]{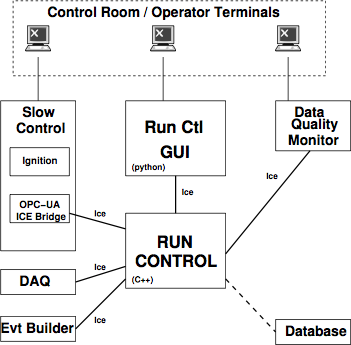}
\tdrfcaption[OnlineBlockDiagram]{Primary interfaces between RC and other LZ subsystems}{Block diagram showing the primary interfaces between the RC system and other LZ subsystems. The users use a GUI to interact with the RC system.}
}
\end{figure}

The work on the run-control software development started with a definition of the architecture (adopting various software standards), the development of detailed specifications for the software and the identification of key interfaces with the various subsystems. The interplay of run control with slow control is of central importance:  Where slow control provides detailed control and monitoring of the detector hardware, run control both initiates high-level slow control scripts (to configure the instrument for various run types) and is the sole entity that configures and controls the data acquisition subsystems (DAQ, EB and DQM).   OPC/UA (and similar) standards are used by much of the hardware for slow control, but the middleware (communication protocol) needed for coordinating other systems has different requirements.  Thus one critical decision in determining the RC architecture was the selection of the middleware to be used for communications between RC and the subsystems. We have past experience with a number of middleware packages, including the industry standard CORBA (Common Object Request Broker Architecture) system. Given the limitations of CORBA (e.g., scalability, throughput, reliability, documentation and thread-safety) a new solution was considered with similar functionality but improved performance. A recent study ~\cite{Dworak:2011} to identify middleware to run CERN accelerators ranked ICE (Internet Communications Engine) and ZeroMQ as the top two evaluated systems. The criteria for this evaluation included reliability and speed, and the ability of the system to handle a large number of messages per second (both small and large messages) and publish to a large number of clients in minimal time. Based on our comparisons of different Middleware options, ICE has been identified as the best choice for the LZ run control system. ICE is a remote-procedure-call (RPC) system with open source C++ and Python implementations.  In ICE there is no clear distinction of client or server, but it is convenient to think of the subsystem ICE nodes as acting primarily as servers, while the RC process is the main ICE Client.  The RC process also acts as a server for the RC GUI. The low-level RC program is implemented in C++, while the RC GUI is implemented in Python.   An API is being developed to allow LZ subsystems to implement ICE communications either through Python or C++.   The RC Gui is being developed in PyQt~\cite{riverbank:2016} with Qwt for real-time displays.  

To provide a seamless interface to the DAQ system, each data collection mode (e.g., calibration runs, data runs) is associated with a corresponding set of command sequences to the various subsystems. The list of data collection modes includes normal data-taking modes (S1 only, S2 only, S1 and S2; see Table~\ref{EDCt:DSModes} for more details) and a number of calibration modes, including the LED, krypton injection, tritium, and neutron calibration (see Chapter~\ref{chap:CAS} for more details). The RC group interacts closely with the groups responsible for the various subsystems to determine the required interfaces, including defining what status information is shown, what commands are provided, and how status information should be displayed.   Based on this information, the RC group has been developing skeleton (server/daemon) program for each subsystem, including the appropriate handles for communication with RC, covering commands and sharing of status information. These skeleton server programs provide the starting point for further development by each group for their specific subsystem, working in close consultation with the RC programmer.  This model has already been used to develop the DAQ interface, where the Washington University RC engineer provided a simple DAQ ICE server skeleton program to the DAQ system developer, who subsequently used this code to quickly develop the DAQ server side communications.   The initial RC DAQ system is now complete, and was used for instantiating runs in the LZ electronics chain test.   The run control/slow control interface is currently under development.   As before, the RC engineer is providing code to the SC system programmer to be used in the development of an ICE/OPC-UA bridge to be used for the RC-SC command and communication link.   

Throughout the course of the development, the run-control and slow-control groups have worked together closely and try to adhere to common standards for coding, code-management, and common libraries for communication and interfaces. Most commands and status data communication will be handled by the ICE/OPC-UA bridge, providing a bridge between the RC ICE RPC mechanism (optimized for data acquisition), and a tag-based interface primarily using OPC-UA (the standard for slow hardware interface) for Ignition.   As with other subsystems, data sharing between run-control and various subsystems can occur through the database, but commands and critical data communication will occur only through the ICE connection between slow-control server computers and the run-control system.

The final RC system provides the user with a simple GUI to the DAQ, data quality monitoring, and the slow-control systems. The GUI starts and stops all subsystems, and displays key status information and alerts if communication links are lost.   The system will guide the user through setting up and starting/stopping various data-collection modes and starting runs, making sure all settings, data, and operator comments are logged into the database. The GUI includes key alerts for out-of-bound values, as determined by the slow-control system or other subsystems, and provides a small number of user-selected plots that allow the user to quickly monitor the health of the LZ system.

A skeleton version of the RC software is already complete and was used for the electronics chain test.  See below for a screen shot of the RC GUI as of mid-January 2016. During this initial development period, the options for the basic system framework were determined.   In the next R\& D stage, system hardware will be purchased. The RC group will work with software developers for the other subsystems to develop interface control documents (ICDs) specifying command sequences, settings, and critical data to be exchanged directly through an ICE link or indirectly through the database. Lessons learned from the electronic system test and the feedback from early users will provide input for the final design phase, during which ICDs will be finalized and the full set of interfaces between RC and the subsystem implemented.
\begin{figure}[htbp]
{\centering
\includegraphics[width=0.5\textwidth]{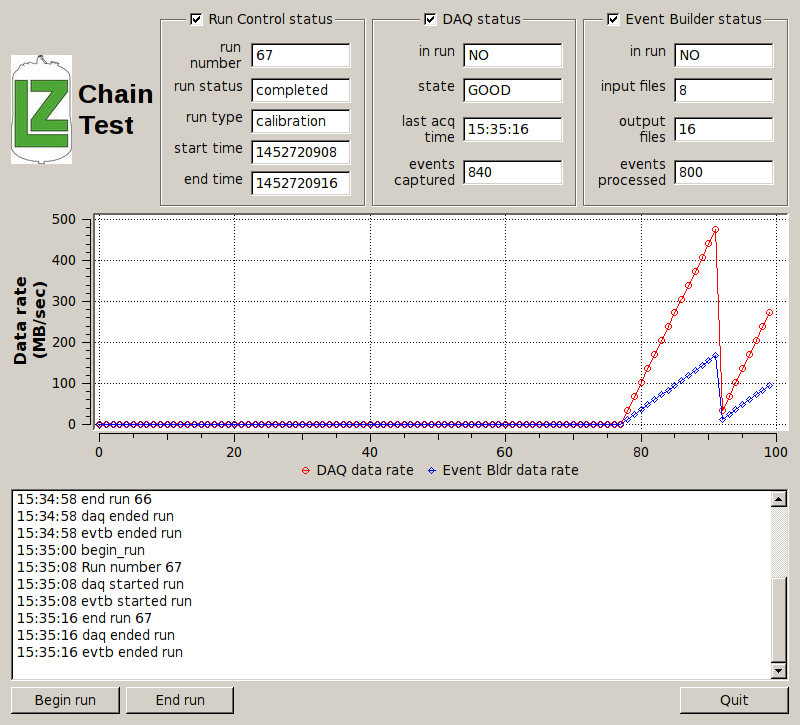}
\tdrfcaption[RCGUI]{Run Control GUI screenshot}{Run Control GUI for for electronics chain test.}
}
\end{figure}

The other major component of the online system is the Event Builder (EB).  
The Event Builder reads the binary files from the 15 individual data
collectors (14 for the TPC, skin and outer detector channels, and one for the
sparsification/trigger system), assembles the events, and writes out the raw
data files on the underground RAID array.

The format and naming convention for both binary input and raw output files is
documented in an ICD between WBS 1.8 and 1.11,
i.e., between the online and offline systems.
Based on the experience from the Daya Bay experiment, it has been decided that
the size of the raw output files should be kept at about \SI{1}{\giga\byte}/file.
The choice of the raw data format was based on three factors, namely
a self-defining data structure, machine independence, and the availability of
a generic event viewer (which allows a quick inspection of the data without
relying on an experiment-specific framework).
This narrowed down the choices to HDF5 and {\normalfont\ttfamily root}.
The raw output format was chosen to be based on {\normalfont\ttfamily root}.
An HDF5-based EB was also developed, but its performance (both in speed and
internal compression) was shown to be inferior to that of the
{\normalfont\ttfamily root}-based EB.
The EB needs to be able to keep up with the highest data rate expected in
the LZ detector, e.g., about \SI{140}{\Hz} during \IKreT calibrations.
Therefore, assuming a safety factor of \numrange{2}{3}, the EB must be able to operate at
\SIrange{280}{420}{\Hz}.
The internal compression is defined to be acceptable if it is comparable to the
level achieved by the standard gzip compression algorithm on the binary files.

The EB is controlled by Run Control, which initiates the start of the EB for
a particular run.
The EB reports back to Run Control the current event and file number with
a frequency that can be adjusted through a parameter.
The end of run (of predetermined duration or number of events) is propagated to
the EB directly through a flag included in the binary files, but can also be
initiated directly by Run Control if the DAQ is non-responsive.
Once a run has been successfully completed by the EB, Run Control is notified by
the EB, which in turn releases the binary files on the data collectors to be
deleted.
All communications between Run Control and the EB occur through ICE.

Because the RC code will be developed and maintained by an experienced professional programmer, it is expected that the RC system (and other subsystem communication links) will be well maintained throughout the LZ project and operation periods.   The RC software engineer will ensure that other online systems adhere to the ICDs, and use the software development standards decided by the WBS 1.8 group.   The RC software engineer will assist with other online programming tasks as needed, lending support and responding to user requests for changes in the online system as needed.

\tdrsec[ChainTest]{Electronics Chain Test Facility}

\tdrsubsec[ECTGeneralDescription]{Description and goals of the Electronics chain test facility}

To test all the components of the signal processing chain, an electronics chain test facility (ECT) has been set up at the University of Rochester \cite{Khaitan:2015slt}.  The main goals of the ECT are:

\begin{enumerate}

\item \textbf{Evaluate Analog Electronics:}
\begin{itemize}
\item  \textbf{Noise and cross-talk.} To measure the noise and cross-talk of the amplifiers, digitizers, and PMT bases.
\item  \textbf{Signal propagation.} To measure and optimize shaping, gain, linearity, and saturation.
\end{itemize}

\item \textbf{Evaluate Digital Electronics:}
\begin{itemize}
\item  \textbf{Firmware and software development.} The ECT provides a direct hardware testing platform minimizing the need for time-consuming simulations.
\item  \textbf{Clock distribution.} Determine if we can distribute the clock over the HDMI links or if we have to use dedicated clock distribution resources.
\item \textbf{System performance.} Verify that the system is capable of handling the expected data rates:
	\begin{itemize}
		\item SerDes links utilization.
		\item P-2-P Gigabit UDP link utilization.
		\item Data Collector CPU/disk speed utilization (storing on local disks and providing the data to the event builder).
	\end{itemize}
\end{itemize}

\item \textbf{Develop Online Systems:}
\begin{itemize}
\item \textbf{Run Control.} Optimize interactions between run control and DAQ and DS systems.
\item \textbf{Event Building.} Optimize event building with data collected with the entire electronics chain.
\item \textbf{Data Quality Monitor.} Develop software tools to monitor the quality of the data collected.
\end{itemize}

\end{enumerate}

In the first phase of the ECT, we have a single chain setup shown in Figure~\ref{EDCf:ChainTestFirstStage}. This stage allows us to concentrate on the quality of signal propagation from PMT to disk.  The first results from phase 1 are described in more detail in the remainder of this Section.

Figure~\ref{EDCf:ChainTestFullStage} shows the chain test setup in its second phase. The second phase is geared towards developing and scaling the DAQ system.

\begin{figure}[htbp]
{\centering
\includegraphics[width=0.75\textwidth]{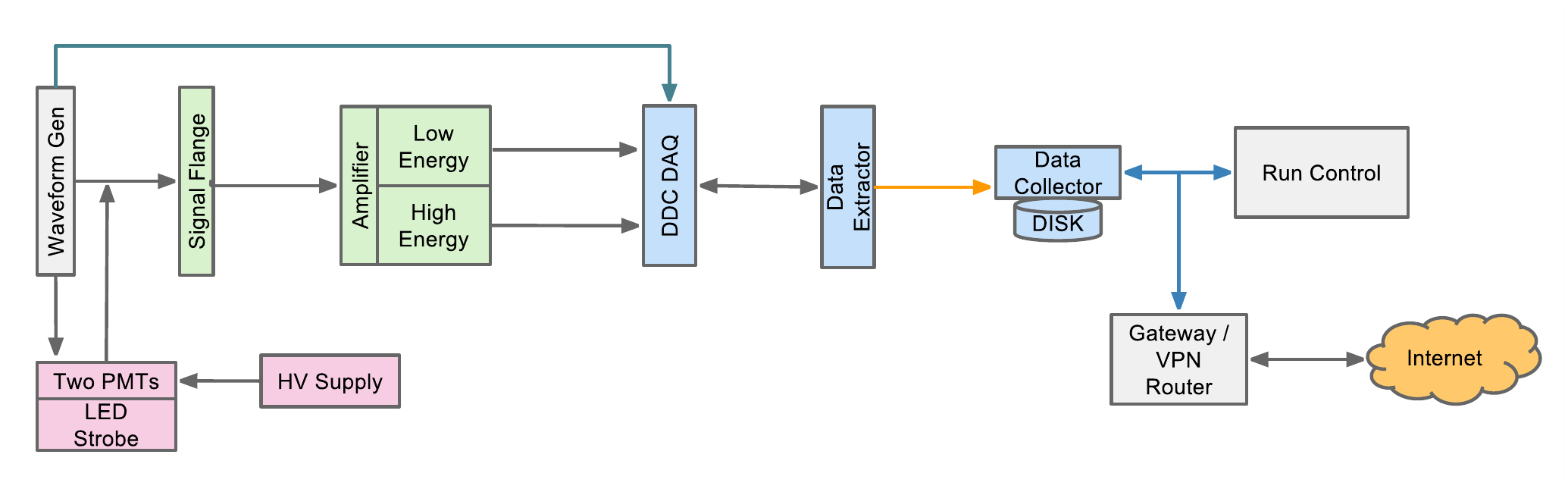}
\tdrfcaption[ChainTestFirstStage]{Diagram of the first phase of the Electronics Chain Test Facility}{Diagram of the first stage of the Chain Test setup. The setup includes: two LZ PMTs, LZ internal signal and HV cable, LZ signal flange, LZ amplifier cage with one LZ amplifier board, one digitization chain, Run Control, and OpenVPN remote access.}
}
\end{figure}

\begin{figure}[htbp]
{\centering
\includegraphics[width=0.75\textwidth]{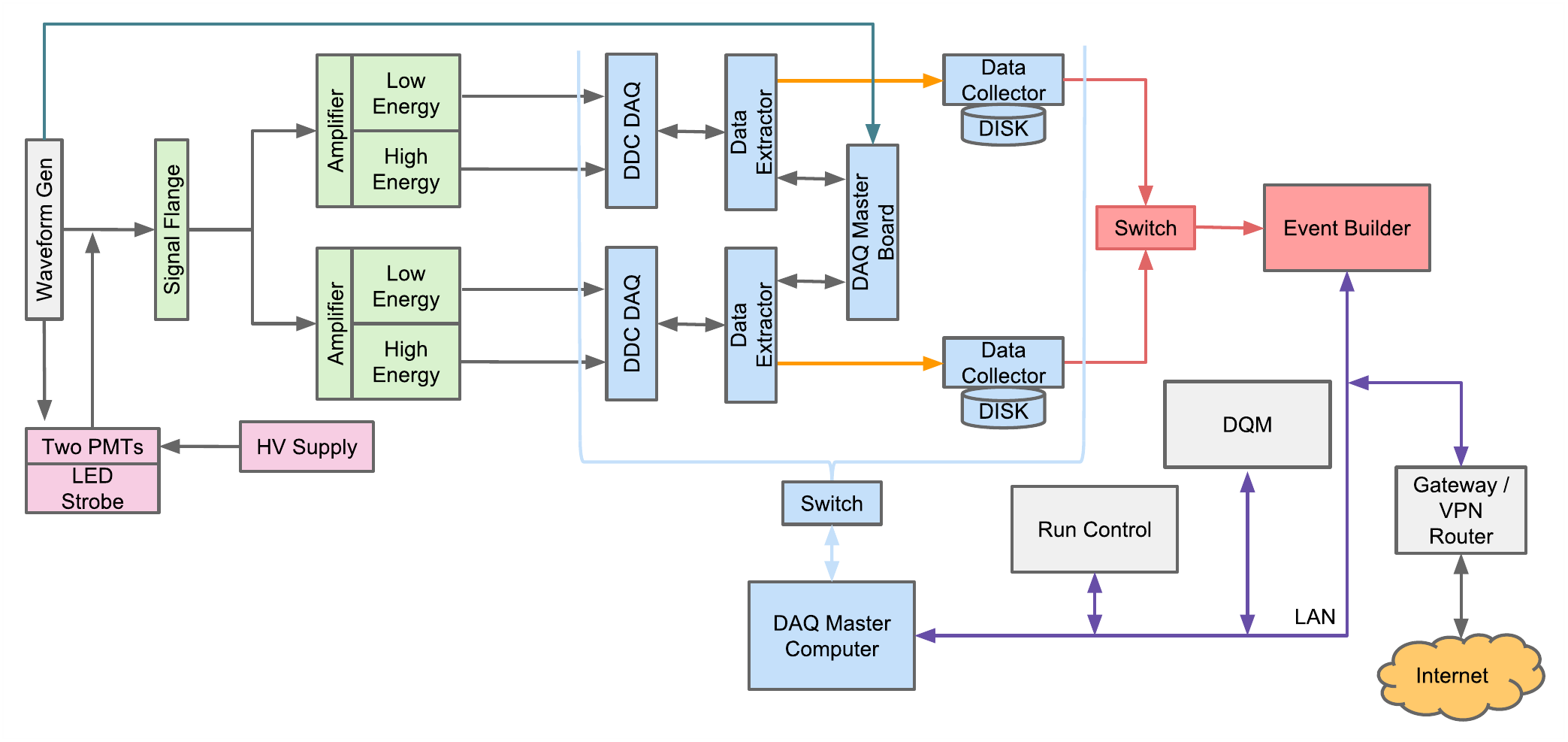}
\tdrfcaption[ChainTestFullStage]{Diagram of the Chain Test setup}{Diagram of the second phase of the Electronics Chain Test Facility \cite{Khaitan:2015slt}. The setup includes: two LZ PMTs, LZ internal signal and HV cable, LZ signal flange, LZ amplifier cage with two LZ amplifier boards, two parallel digitization chains, Event Builder, Run Control, DQM, DAQ Master Computer and OpenVPN remote access.}
}
\end{figure}

\tdrsubsec[ECTSphePropagation]{Single photoelectron pulse propagation}

The initial measurements carried out at the ECT focused on SPHE pulses from a single R11410-20 PMT.  These studies allowed use characterize signal-to-noise ratios and timing resolution of the electronics chain.

Figure~\ref{EDCf:PMTSPHEnoAmpandLE}a shows an intensity map and an average of 1000 SPHE pulses captured at the signal flange (before amplifiers), using a Tektronix MSO4104 oscilloscope with \SI{250}{\MHz} set bandwidth and \SI{2.5}{\GHz} sampling.
Figure~\ref{EDCf:PMTSPHEnoAmpandLE}b shows a corresponding SPHE pulse captured at the output of the LE amplifier channel.

\begin{figure}[htbp]
{\centering
\includegraphics[width=0.75\textwidth]{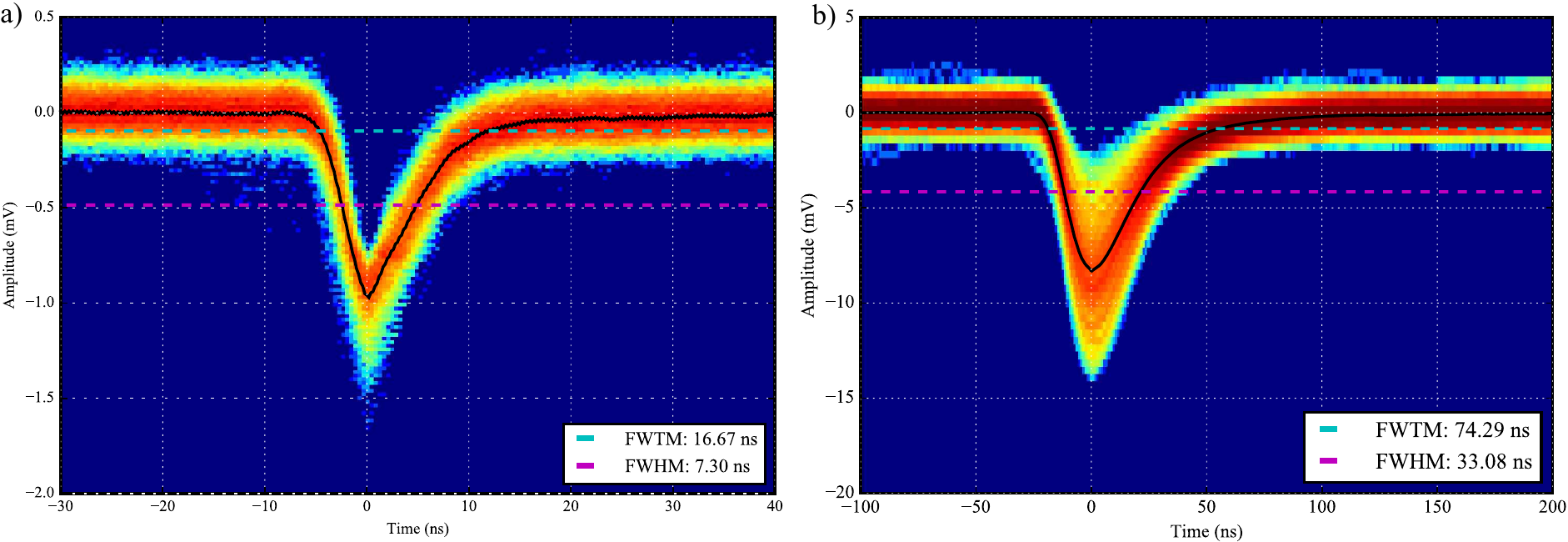}
\tdrfcaption[PMTSPHEnoAmpandLE]{Single photoelectron intensity plots}{Single photoelectron intensity plots. (a) LZ PMT @ \SI{1300}{\V} $\rightarrow$ \SI{45}{\foot} GORE cable $\rightarrow$ Signal Flange $\rightarrow$ \SI{6}{\foot} LMR-100 $\rightarrow$ Scope, (b) LZ PMT @ \SI{1300}{\V} $\rightarrow$ \SI{45}{\foot} GORE cable $\rightarrow$ Signal Flange $\rightarrow$ LZ Amplifier LE output $\rightarrow$ \SI{30}{\foot} LMR-100 cable $\rightarrow$ Scope}
}
\end{figure}

The R11410-20 are expected to operate at \num{3.5e6} gain, but since the manufacturer guarantees them to \num{2e6}, we measured the SPHE response for both scenarios.
Figure~\ref{EDCf:SinglePheLikePulseEntireChain} shows two SPHE spectra obtained at PMT gains of \num{3.25e6} and \num{1.9e6}. The signals were propagated through the entire analog chain composed of 45 feet of gore cable, the amplifiers, and 30 feet of LMR-100 cable, before being digitized on the DDC-32. The waveforms were passed through an S1 filter and the filters' peak output (proportional to area) was histogrammed. 
The filters' width was set to 3 samples to optimize the peak to baseline noise ratio. The measured detection efficiency of \SI{98}{\percent} (at \num{3.25e6}) and \SI{92}{\percent} (at \num{1.9e6}) are obtained with only \SI{1}{\Hz} of singles rate (where the S1 filter captures a fluctuation in the baseline noise).

\begin{figure}[htbp]
{\centering
\includegraphics[width=0.75\textwidth]{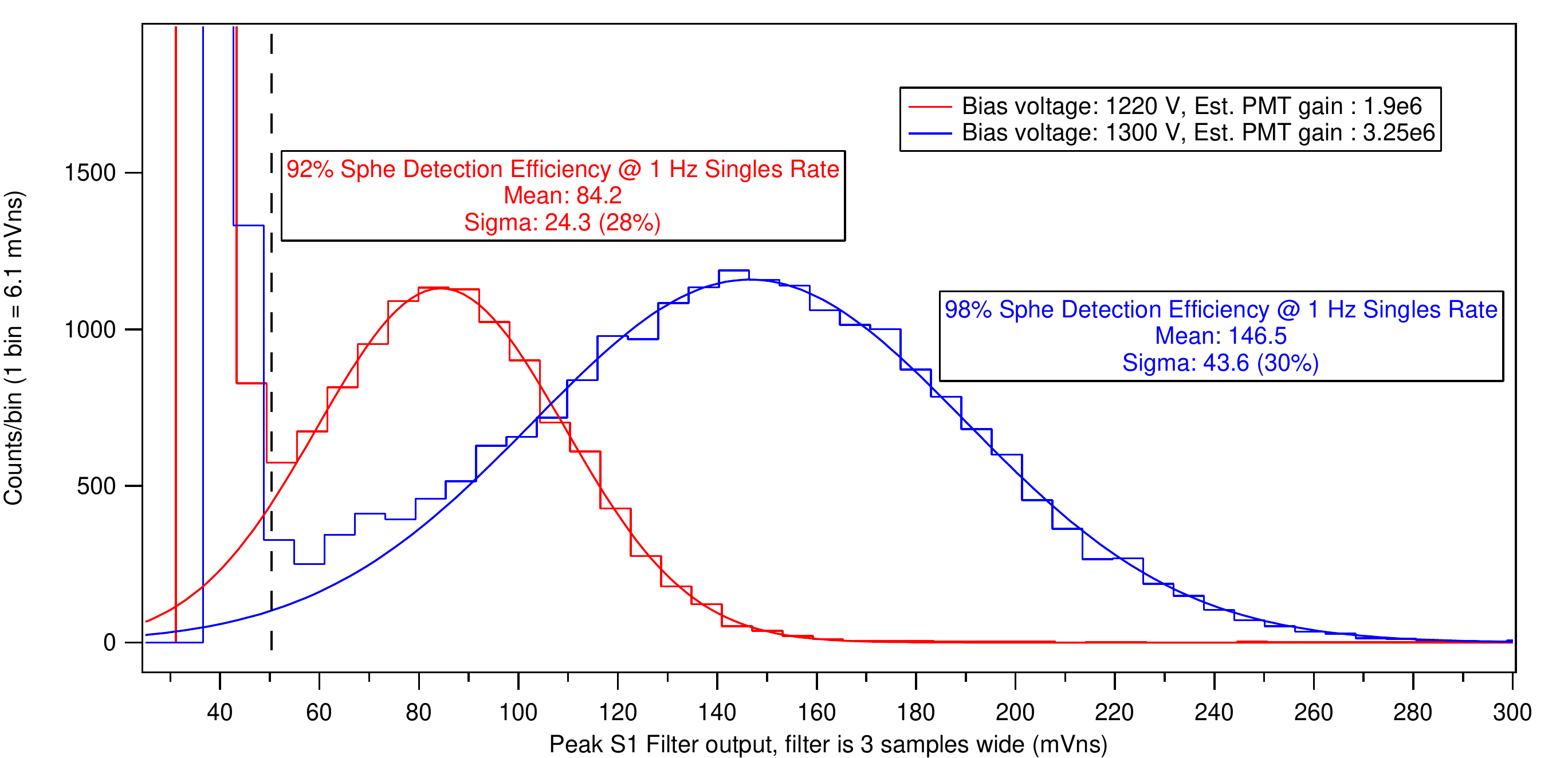}
\tdrfcaption[SinglePheLikePulseEntireChain]{Single photoelectron spectrum at estimated \num{1.9e6} and \num{3.25e6} gain}{Single photoelectron spectrum at estimated \num{1.9e6} and \num{3.25e6} gain. Detection efficiency at \SI{1}{\Hz} singles rate is \SI{92}{\percent} and \SI{98}{\percent}, respectively.}
}
\end{figure}

In order to determine how well we can reconstruct the relative timing between SPHE signals with a \SI{60}{\ns} shaping time (FWTM), digitized at \SI{100}{\MHz}, we carried our a series of measurements with short light pulses from an LED. As light source we used the CAEN SP5601 LED pulser, which has a measured resolution of \SI{0.44}{\ns}~(1$\sigma$)~\cite{CAENLedPulser}. To limit the impact of the PMT's transit time spread (TTS) which has been measured to be \SI{2.55}{\ns}~(1$\sigma$)~\cite{Lung:2012pi} for photons distributed across the entire photocathode, an aperture (\SI{1}{\cm} diameter) was placed in front of the PMT. The PMT was operated with a gain of \num{3.25e6}. Figure~\ref{EDCf:LZSpheTiming}a shows that measured reconstructed time difference between the LE SPHE signals and the LED pulser start signal has a resolution of \SI{1.65}{\ns} (1$\sigma$). It should be noted that the resolution obtained between two LED pulser start signals, also digitized with the DDC-32, is \SI{54}{\ps} (1$\sigma$) (Fig.~\ref{EDCf:LZSpheTiming}b).

\begin{figure}[htbp]
{\centering
\includegraphics[width=0.75\textwidth]{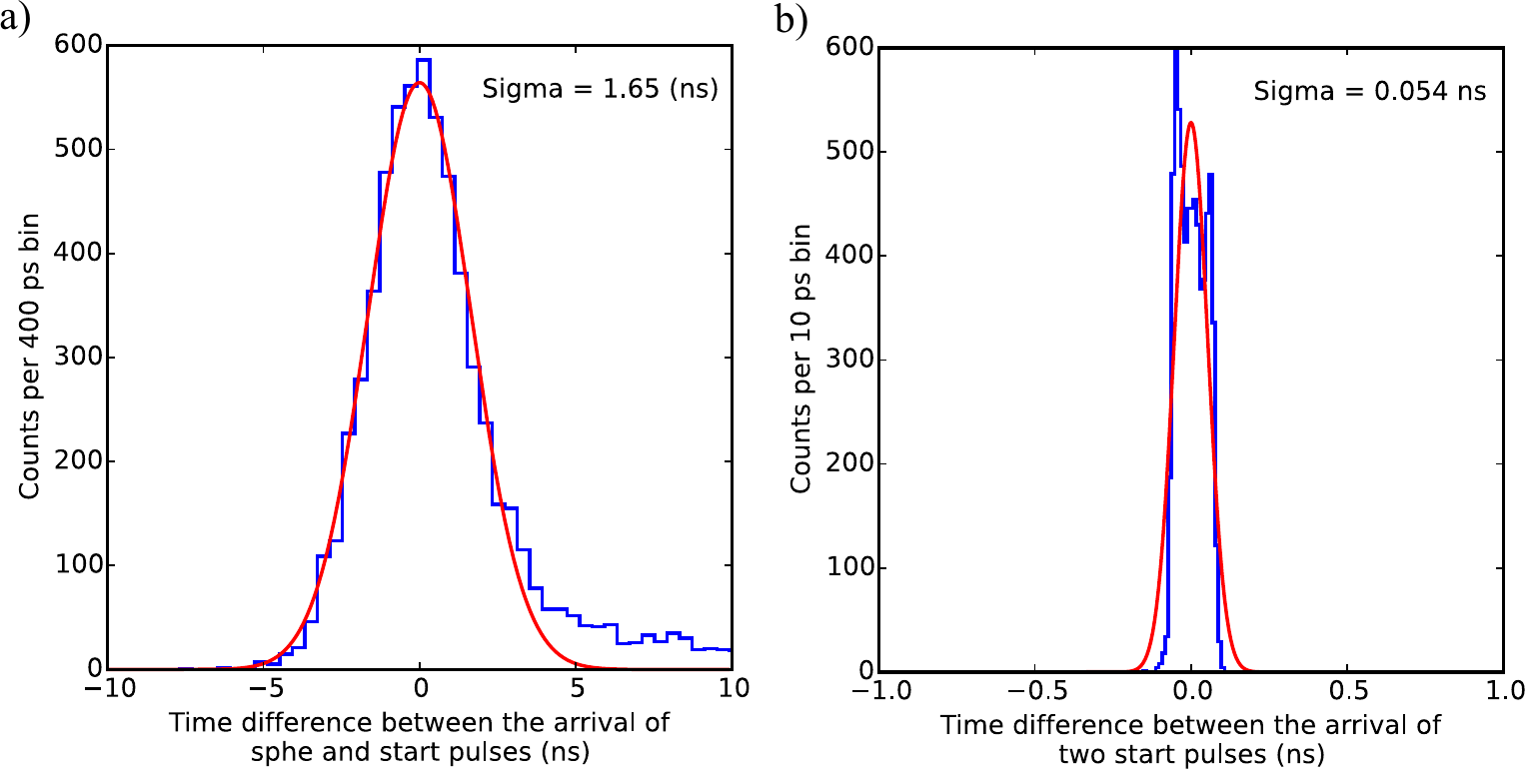}
\tdrfcaption[LZSpheTiming]{SPHE timing measurements}{Time difference distributions for: a) LE SPHE signals obtained with a gain of \num{3.25e6}, b) two LED start pulses. The time range in b) is ten times smaller than the time range in a).}
}
\end{figure}

\tdrsubsec[ECTPMTBaseSaturation]{PMT Base Saturation}

The PMT base was evaluated for saturation with S2-like pulses. Wide pulses were sent to a \SI{425}{\nm} LED and varied in amplitude to produce S2-like responses in the PMT with a width of \SI{0.5}{\mus} (1$\sigma$). To determine the amount of light emitted by the LED, we collected data in two modes: 1) with a \SI{1}{\cm} diameter aperture in front of the PMT to limit the amount of incident light in order to calibrate the LED, and 2) without the aperture where we expect PMT saturation for large LED signals. Signals from the PMT are propagated through \num{45} feet of Gore cable to the DB25 connector on the signal flange and digitized on the Tektronix MSO4104 oscilloscope along with the signals sent to the LED. Figure ~\ref{EDCf:LZEctSaturationAvgPulses} shows the averaged PMT pulses collected without the aperture for different LED pulses.

With the aperture, the PMT pulse area increases linearly with the LED pulse area, demonstrating that as the amplitude of the pulse to the LED increases the amount of light being generated increases. This allows us to calibrate the amount of light incident onto the PMT. In the measurements without the aperture, we know that the LED pulse area is proportional to the light emitted by the LED.  Figure ~\ref{EDCf:LZEctSaturationArea} shows a plot of the measured PMT pulse area when the aperture is removed versus the PMT pulse area we would have obtained if the aperture would have been in front of the PMT. This figure shows that for the two largest LED amplitudes, the PMT base saturates.

In the region of saturation, the shape of the response of the PMT changes, as can be see in Figure ~\ref{EDCf:LZEctSaturationAvgPulses}. The saturation manifests itself in the form of narrowing the shape of the pulse for larger light input to the PMT and no increase in pulse area. Saturation occurs when the amplitude of the signal from the PMT at the input of the oscilloscope exceeds \SI{0.6}{\V}.

\begin{figure}
\begin{minipage}[t]{.49\textwidth}
	\centering
	\includegraphics[width=0.75\textwidth]{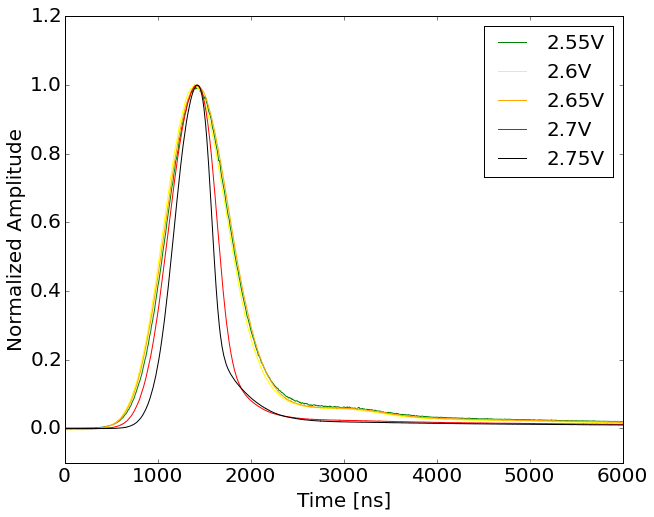}
	\tdrfcaption[LZEctSaturationAvgPulses]{PMT Base Pulse Shape Saturation}{ Average peak aligned and normalized pulses showing the S2-like response of the PMT. The legend corresponds to the peak amplitude of the pulse used to drive the LED. The first signs of saturation begin to manifest itself for an LED amplitude of \SI{2.65}{\V}.}
\end{minipage}
\hfill
\begin{minipage}[t]{.49\textwidth}
	\centering
	\includegraphics[width=0.75\textwidth]{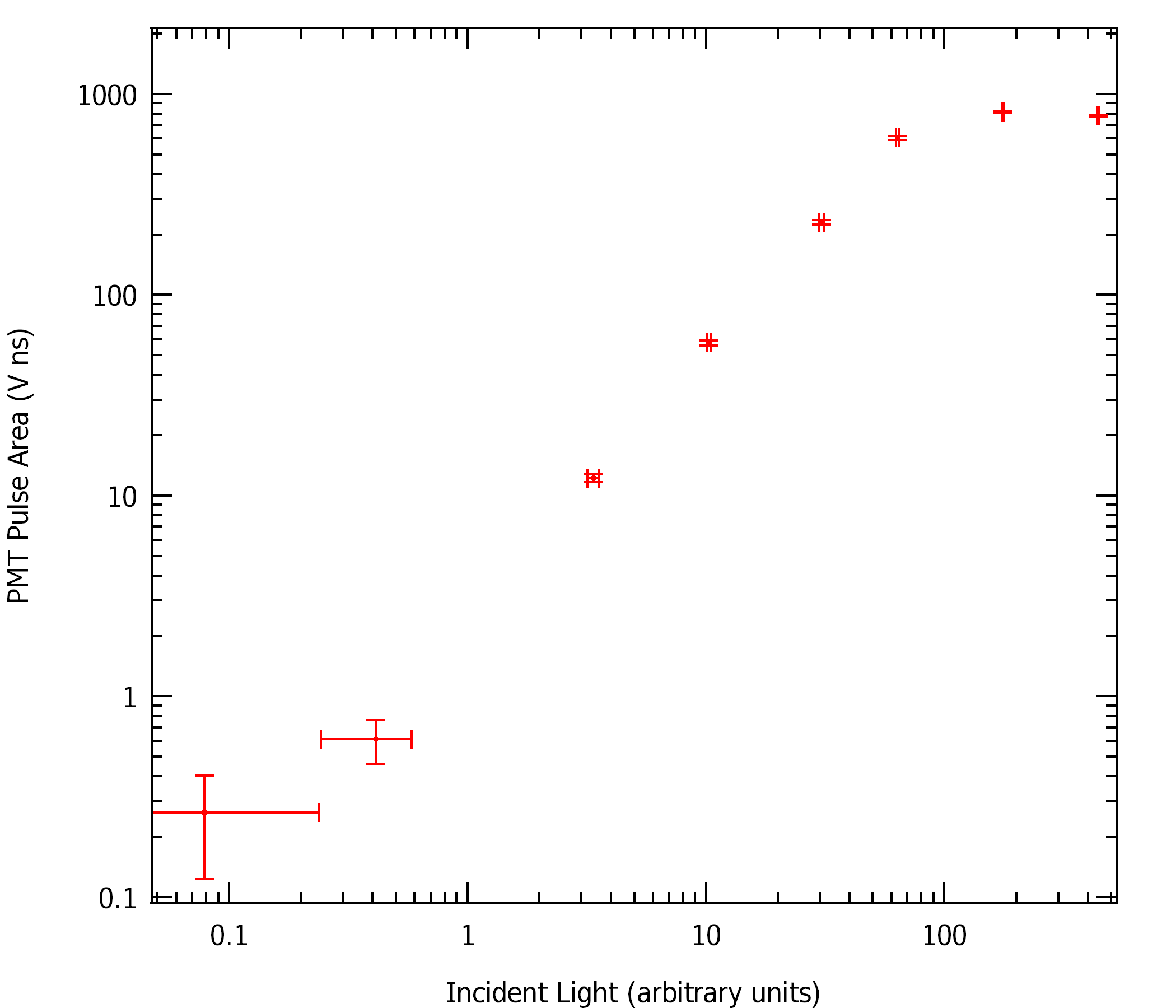}
	\tdrfcaption[LZEctSaturationArea]{PMT Base Area Saturation}{Result of measured pulse area detected by the PMT without a hole versus the expected pulse area of the PMT with a hole. As the expected amount of light increase the PMT's measured response increases till it begins to saturate for the last two points.}
\end{minipage}
\end{figure}

\tdrsubsec[ECTDAQPerformance]{DAQ Code and Performance}
The main Data Collector acquisition code is named DCCore and has been rewritten in C++ using Qt based multithreading framework. It is responsible for receiving UDP datagrams from the Data Extractors, performing consistency checks on it and storing it to a local SSD drive. The DCCore successfully communicates with the Run Control over ICE by responding to run start/stop commands and reporting the acquisition status and progress.

During active data collection, at the rate of \SI{70}{\mega\byte\per\second}, the DCCore has been shown to utilize a single CPU core at only \SI{\sim15}{\percent} level. Currently the measurement was done with the peak rate of \SI{70}{\mega\byte\per\second}, because the FPGA waveform capture firmware is implemented in a single-buffering mode and thus not push the Gigabit UDP link to its maximum of \SI{109}{\mega\byte\per\second}. The single-buffered rate is understood and agrees perfectly with what we estimate in this scenario.

Additionally a Python based DAQScope code has been developed. It communicates with the DCCore using ICE. It is a GUI tool that allows for real-time preview of the waveforms incoming into the Data Collector. It also allows for continuous calculation of the waveforms' power spectral density (PSD) plots.
The screenshot of the tool in operation is shown in Figure~\ref{EDCf:DAQScopeGUI}

\begin{figure}[htbp]
{\centering
\includegraphics[width=0.75\textwidth]{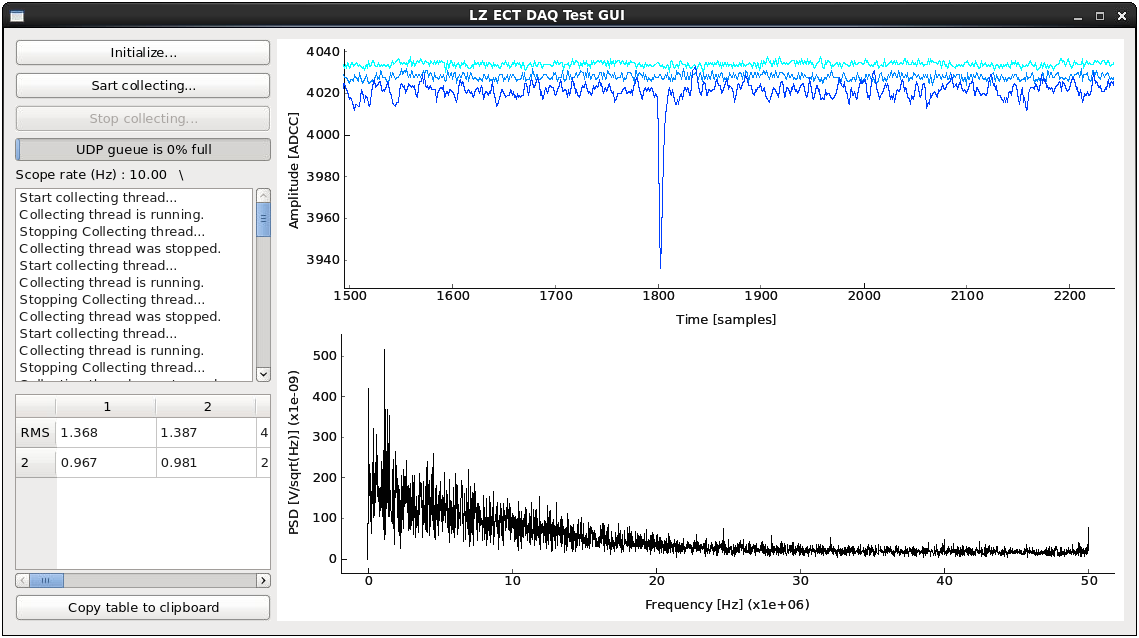}
\tdrfcaption[DAQScopeGUI]{DAQScope tool screenshot}{A screenshot of the DAQScope tool displaying a subsample of the waveforms incoming into a given Data Collector and showing a continuously calculated PSD spectrum for one of the channels.}
}
\end{figure}

\tdrsubsec[ECTatLUX]{Real Signals from LUX}

At the end of LUX, we had the opportunity to replace the existing signal chain with the LZ signal chain from amplifier to disk. This allowed us to benchmark the performance of our electronics against those used in an active dark matter detector and to look at real signals. For these tests, two LZ amplifiers were mounted on a LUX signal flange that had signals originating from both the top and bottom PMT arrays. Dual gain signals from the amplifier were routed through 45 feet of LMR-100 cable to be digitized on a prototype DDC32. Data was extracted from the DDC32 via and HDMI link to a DE board and then pushed to a DC where it was stored in a partial event file format. The LUX grounding scheme was used for the electronics chain. 

\begin{figure}[htbp]
{\centering
\includegraphics[width=0.75\textwidth]{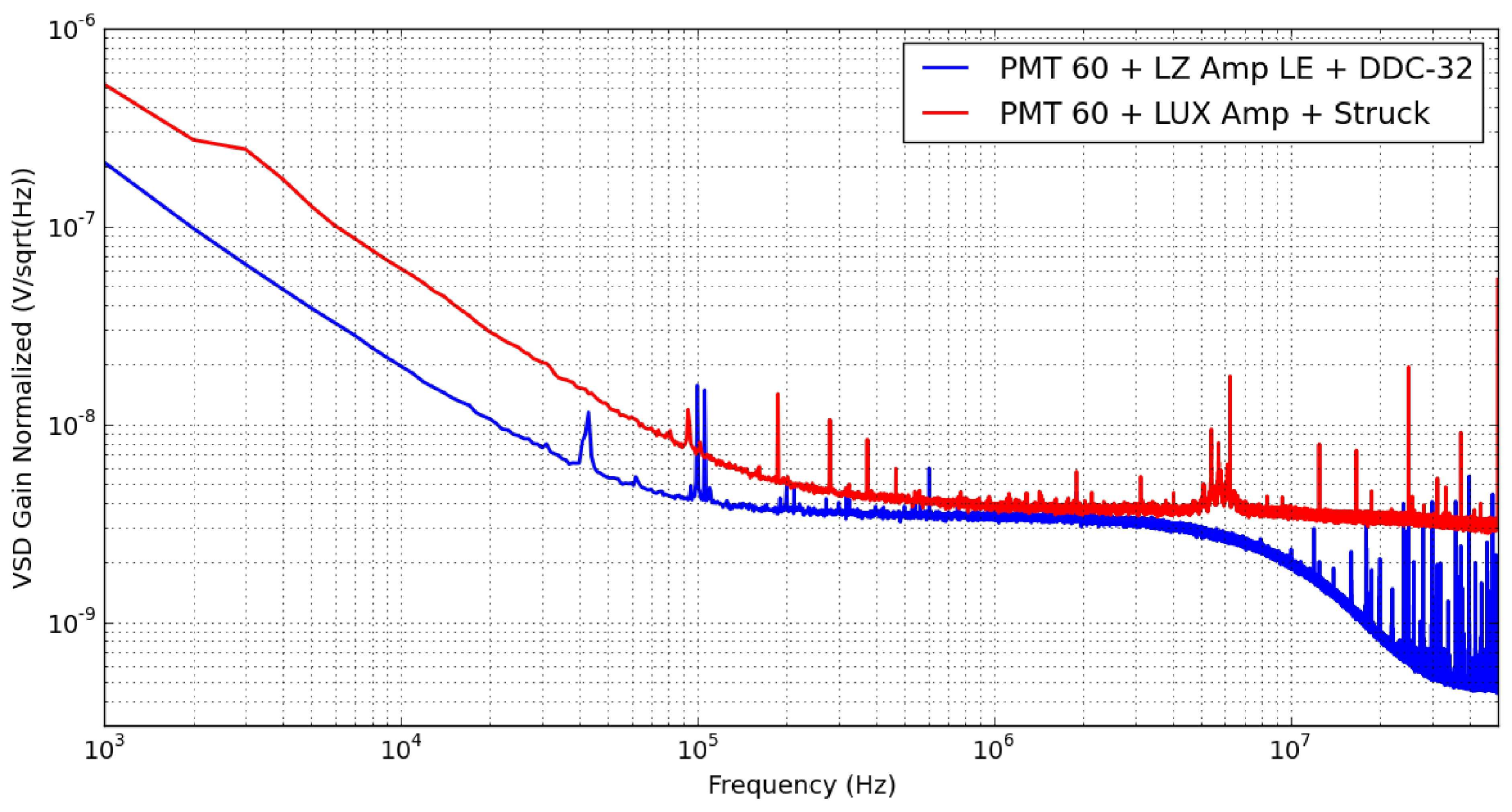}
\tdrfcaption[LZatLUXForNormalizedNoiseSpectrum]{Noise spectrum from LUX}{Gain normalized noise spectra compared between the LUX and LZ DAQ chains.}
}
\end{figure}

Noise measurements, SPHE calibrations, and real event signals from LUX were collected during these tests. The gain corrected noise measurements are shown in Figure~\ref{EDCf:LZatLUXForNormalizedNoiseSpectrum}. These measurements show that, even though no special consideration was given to optimize the grounding scheme, the LZ system performs as well as the LUX systems. From SPHE calibrations we measure a SPHE detection efficiency greater than 90\%. An example of the S1 filter output for SPHEs is shown in Figure~\ref{EDCf:LZatLUXForSampleSpheSpectrum}. 

\begin{figure}[htbp]
{\centering
\includegraphics[width=0.75\textwidth]{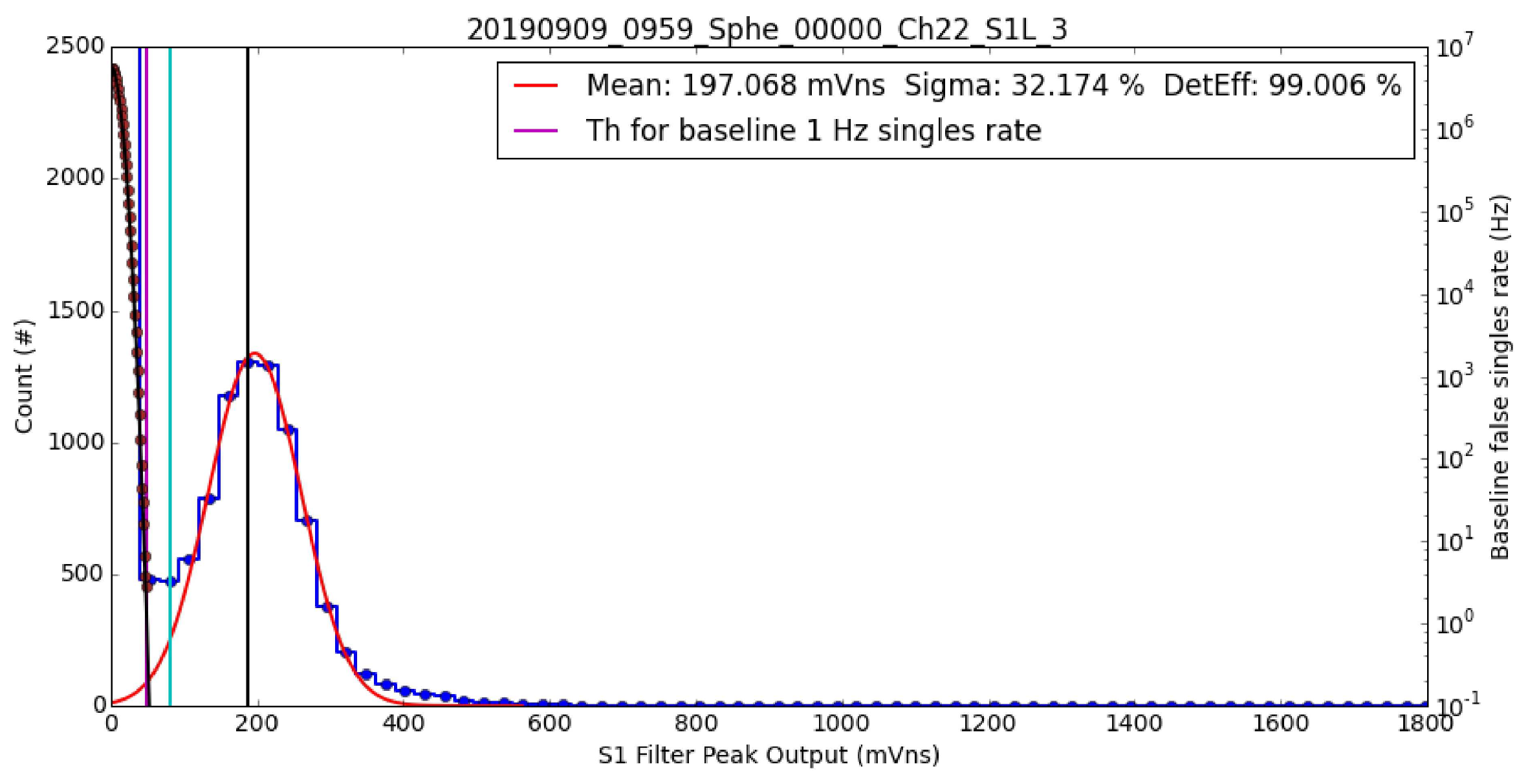}
\tdrfcaption[LZatLUXForSampleSpheSpectrum]{SPHE spectrum obtained at LUX}{Maximum S1 filter output for SPHEs obtained with the LZ DAQ electronics installed on the LUX detector.}
}
\end{figure}

\tdrsec[Grounding]{Grounding}
The LZ experiment comprises many electrical and electronic sub-systems that may be adversely affected by electrical noise. Each sub-system in the LZ experiment will be guarded from noise induced by other sub-systems. Guarding will also reduce noise introduced into other sub-systems. The PMT signal collection chain is one of the most sensitive and most critical sub-systems in LZ. PMT signals are in the sub-\si{\mV} range before amplification so even a small amount of noise has a potentially large impact. PMT signal quality directly affects the quality of the physics data therefore signal integrity is of great concern.

Unwanted signals or noise can be coupled from an ``aggressor'' circuit into a ``victim'' circuit through capacitive coupling, inductive coupling and conduction. Proper grounding can be critical to reducing interference through these mechanisms. Grounding plans for experiments of the size of LZ are complex and require significant engineering effort. 

The LUX grounding system serves as the starting point for the LZ grounding scheme. The main lesson learned in implementing the LUX grounding system was the importance of a single-point or ``star'' ground configuration (Figure~\ref{EDCf:LUXStarGndRackGnd} a). All of the equipment in the electronics racks was grounded to the rack grounding bars using flat braided cable (Figure~\ref{EDCf:LUXStarGndRackGnd} b) and all of the grounding rods have dedicated grounding paths to the central star point using 2 AWG copper welding cable.

\begin{figure}[htbp]
{\centering
\includegraphics[width=0.9\textwidth]{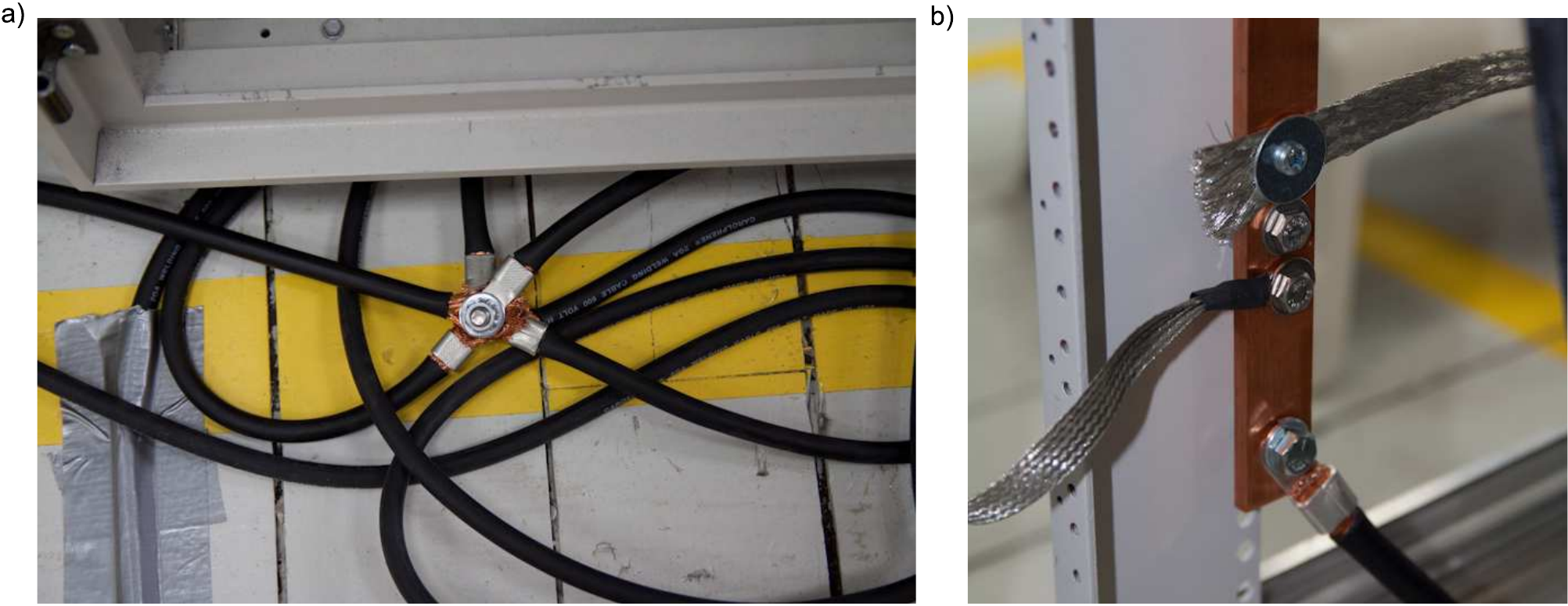}
\tdrfcaption[LUXStarGndRackGnd]{Grounding}{a) Star ground central point to all the rack grounds. b) Sample grounding connections within one of the racks, where individual equipment and the rack door are grounded with steel braid to the rack grounding rod. The grounding rod is connected to the center star grounding point using a high gauge copper welding cable.}
}
\end{figure}

For many of the circuits in LZ, the signal cable shield also acts as the signal return path. In order for the shield to be effective, it must not conduct significant amounts of current. If excess current does flow in the shield, it will show up as noise in the signal. This can occur when the ``ground'' potential is different at each end of the cable. In situations such as this, galvanic isolation can break the flow of current and eliminate the induced noise. We are currently testing the feasibility of using isolation transformers on the signal lines between the amplifier outputs (at the breakout) and the digitizer inputs (\SI{30}{\foot} of cable away in the electronic racks above).
We are testing two candidates from Mini-Circuits: FTB-1-6+ and FTB-1-1+.
As these are typically meant for RF applications, we need to quantify how these transformers impact the pulse time characteristics of typical amplified LZ PMT signals. If the transformers do not significantly degrade the signal quality they should allow for minimizing potential ground loops between the breakout area and the upper electronics racks. 

Significant noise can also be generated by electrical equipment such as pumps, chillers, and blowers that draw significant current or have switched loads that might cause electrical transients. Interference from these sources will be eliminated or greatly reduced by isolating them on separate electrical circuits and by physically locating them as far from the sensitive PMT signal cables as practical. Capacitive and inductive coupling to PMT signal cables can occur if cables from other sub-systems run in close proximity to the PMT cables for significant distances. Again, by placing the PMT signal cables as far as is practical from other sub-system cables, we will greatly reduce the effects of capacitive and inductive coupling.

Ideally we will be able to identify all sources of potential noise and eliminate them during the design stages. But experience shows that it is unlikely that every noise source will be identified prior to the installation and commissioning of the system. During deployment/construction of the LZ experiment we will, early on, have a subset of the DAQ system installed such that we can constantly monitor PMT signals for any noise pickup as equipment is installed. If we see any noise being picked up from other systems during the installation of equipment for the project, the data we gather from the DAQ system will greatly simplify pinpointing the source and allow us to put in place fixes to ensure excellent signal integrity.

\tdrsec[Network]{Network Infrastructure, Security, Remote Access}

The LZ network and computing infrastructure will be designed according to best practices to meet the LZ requirements for performance, security and availability / operational reliability and uptime. A prototype has been developed and successfully demonstrated at the SLAC LZ System Test Facility, discussed in Section~\ref{XDSS:SystemTest}.

\tdrsubsec{Local Area Network}

\begin{figure}[htbp]
{\centering
\includegraphics[width=0.75\textwidth]{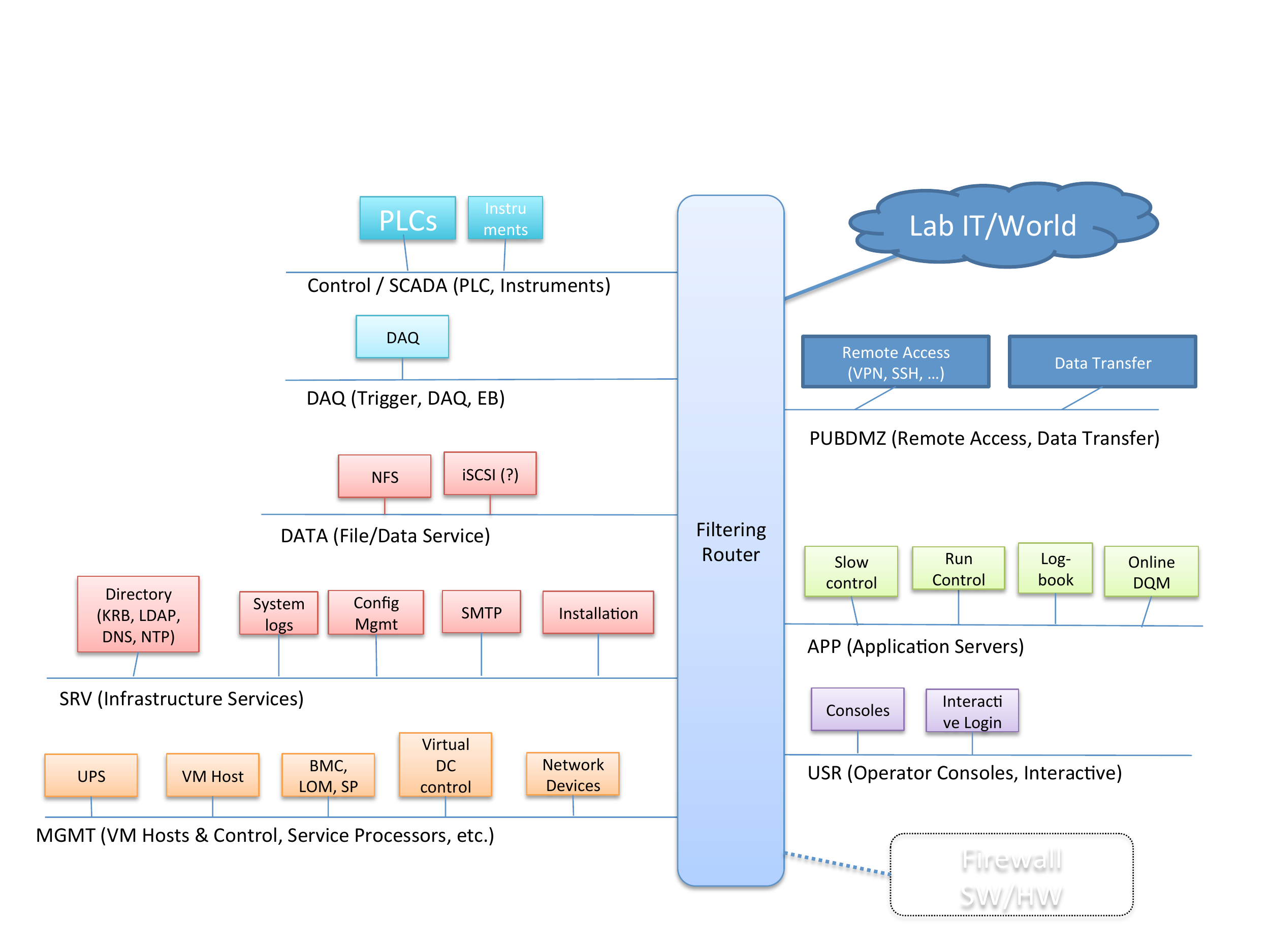}
\tdrfcaption[LZOnlineNetwork]{The LZ Online network}{The LZ Online Network}
}
\end{figure}

The LZ experiment LAN in the Davis Campus will be implemented with a single central router/switch that acts as communication hub for the experiment, provides separate network zones using Virtual LANs (VLANs) and IP subnets and enforces inter-zone network restrictions using simple access control lists (ACLs). Network zoning will be used to implement a standard network security architecture with ``Demilitarized Zones'' (DMZ) and restricted networks. If necessary, enhanced network security controls can be implemented with host-based firewalls or a dedicated firewall appliance. Figure~\ref{EDCf:LZOnlineNetwork} shows a logical diagram of the LZ experiment LAN.

\tdrsubsec{Reliability and Availability}

Connection to the surface will use dedicated and redundant fiber connections on physically separate paths (one fiber path through Ross shaft, one fiber path through Yates shaft). The central LZ router/switch located in the Davis Campus will be implemented as redundant stack of two switches with full power, uplink and control redundancy. All critical devices such as main PLCs and critical computing infrastructure will have redundant Ethernet connections to both stack elements. A separate, dedicated network connection to the surface will allow emergency out-of-band access to the PLCs and the network and IT infrastructure underground.

\tdrsubsec{Information Security}

The LZ computing architecture will implement a defense-in-depth security design based on a zoned network, central authentication and authorization (Kerberos and LDAP) with appropriate account-lockout policies, a configuration management system for provisioning, managing and patching hosts, central logging and monitoring, and frequent backups for recovery from a wide class of problems. Advanced firewalls, network- or host-based intrusion detection, or other security controls can be easily added if deemed necessary. 

Network zoning will be used to create a layered configuration where devices with weak security controls or network stacks (such as embedded systems, sensors, PLCs) are only allowed to communicate with trusted and managed systems.

Central management of authentication and authorization in a ``Directory'' based on Kerberos and LDAP will be used to manage user accounts and all privileges in the LZ system, including host access privileges and slow control roles. This will allow for easy enforcement of credential requirements (such as minimum password complexity or password expiration), secure auditing and logging of user privileges and lockout of user accounts if necessary. It will also allow to securely implement user self-service (e.g. for password reset or contact information change) and delegation of privilege management to subsystem managers (e.g. the slow control manager can assign roles to users without having to ask a system administrator to make the changes).

A configuration management system (such as e.g. SaltStack~\cite{SaltStack}) will be used to provision hosts with a secure baseline configuration and to maintain and update the host configuration as necessary. Using such a system minimizes the amount of effort required for system management and provides a way to detect unauthorized changes relative to the baseline configuration. 

Central logging and monitoring provides another avenue to detect unauthorized activities, especially when combined with simple keyword-based automated log analysis, but is also a valuable tool for general non-security related health monitoring, diagnosis and debugging of the system.

Where possible, frequent downtime-less file system snapshots and backups of all important system configuration and user/application data will be performed. Access to backup files will be restricted so that even a system compromise will not allow the attacker to modify historical backup data. 

Virtualization technologies may be used to reduce the effort required for system management and to provide virtual machines that can be easily controlled remotely by authorized users and that can be easily backed up.

\tdrsubsec{Remote Access}

The implementation of secure remote access will depend on the applications that need to be accessed and the client devices (desktop, laptop, tablet/phone) accessing them. All remote access protocols must use secure and encrypted communication and strong authentication mechanisms. While the details are still under study, we foresee some combination of web, command line, remote desktop and application client access over encrypted communication channels provided by a VPN and/or SSH.

\tdrsec[Installation]{Installation and Commissioning}

The electronics will be installed in two locations. The analog electronics for the Xe PMTs will be installed on the mezzanine level, as shown schematically in Figure~\ref{EDCf:DavisCavernView}. The amplifiers will be arranged in \num{23} racks, with \num{4} amplifier cards installed in each rack. The racks will be located horizontally on the wall of the water tank in order to improve cooling and provide better access to each amplifier crate. Exhaust ducts may must be installed at this location to facilitate cooling of the electronics. Because it will be difficult to enclose the electronics crates at this location, it is important the air flowing over the electronics is directed into the exhaust system. This will prevent any smoke generation as a result of an electrical problem, e.g., overheating of components, from spreading into the Davis Campus. The outputs of the amplifiers are routed in cable trays to the DAQ and trigger systems installed in the electronics racks installed at deck level (see Figures~\ref{EDCf:DavisCavernView} and \ref{EDCf:BreakoutBox}).

\begin{figure}[htbp]
{\centering
\includegraphics[width=0.75\textwidth]{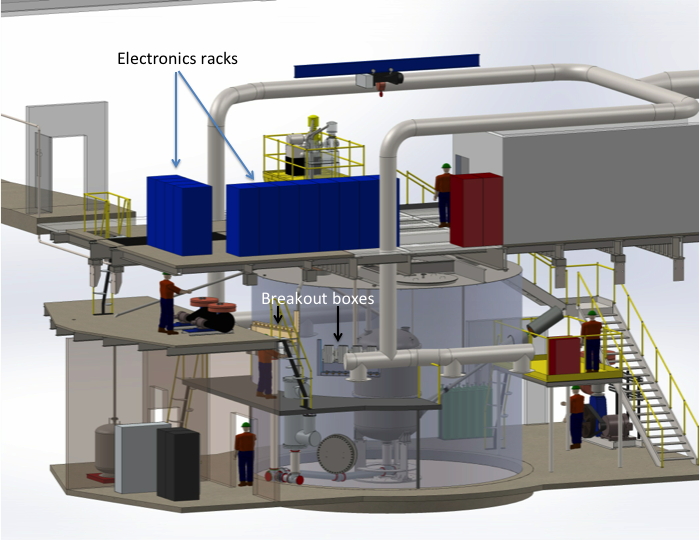}
\tdrfcaption[DavisCavernView]{Location of analog and digital electronics}{Installation of the analog electronics for the xenon PMTs at the mezzanine level and the electronics racks at the deck level.}
}
\end{figure} 

\begin{figure}[htbp]
{\centering
\includegraphics[width=0.75\textwidth]{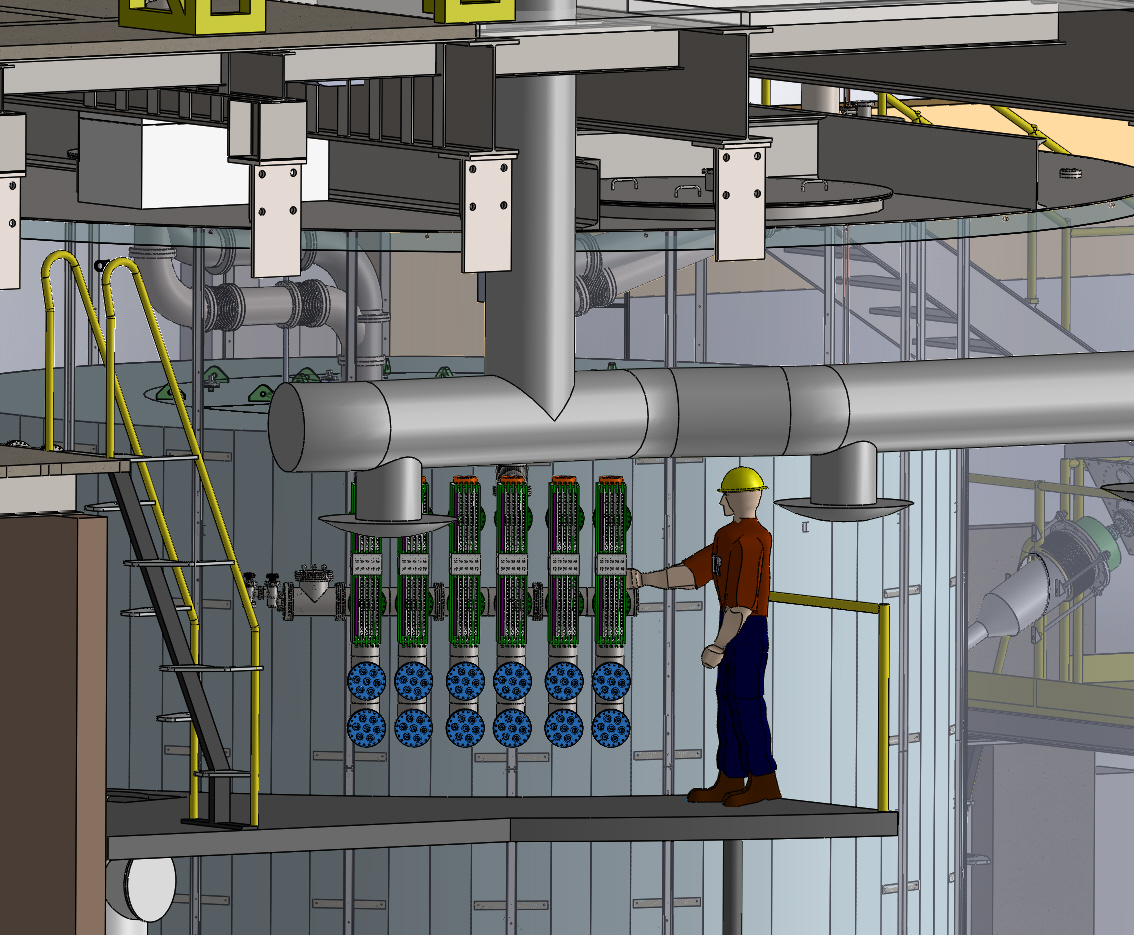}
\tdrfcaption[BreakoutBox]{The top breakout box at the mezzanine level}{Close-up of the mezzanine level showing the amplifier racks installed on the top breakout box.  The HV filter boxes are not shown.}
}
\end{figure}

A schematic of one element of the breakout box is shown in Figure \ref{EDCf:CrossElbowSections}. The internal signal and HV cables enter this section of the breakout box through the ports in he center. The signal cables are routed to the two flanges, shown in yellow, with four DB-25 connectors on which the amplifier crates mount. The HV cables are routed to the HV flange, shown in blue, visible at the bottom of this breakout section.

The amplifiers and the HV pickoffs for the outer-detector system will be installed in the electronics racks shown in Figure~\ref{EDCf:DavisCavernView}. The HV cables from the outer-detector PMTs are routed via one of the flanges on top of the water tank to this location. The design of the flange is such that no connectors are required to ensure a light-tight and radon-tight connection. Not using connectors on the flange improves signal quality.

\begin{figure}[htbp]
{\centering
\includegraphics[width=0.25\textwidth]{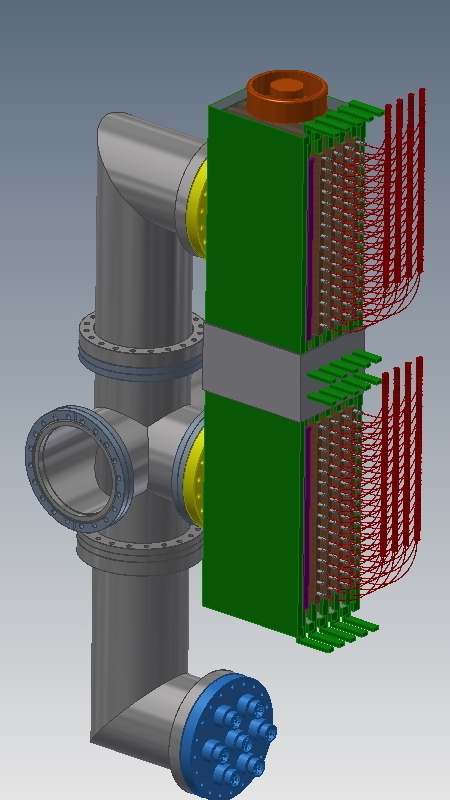}
\includegraphics[width=0.25\textwidth]{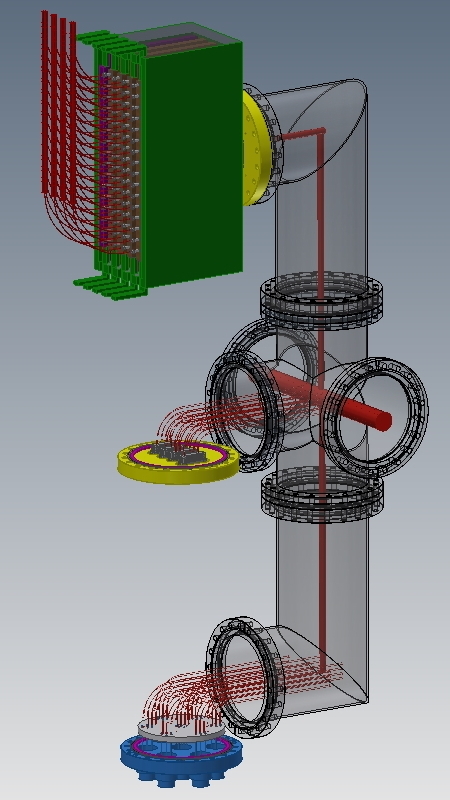}
\tdrfcaption[CrossElbowSections]{Details of one elements of the breakout system}{Close-up of one element of the breakout box.  The figure on the left shows two amplifier cages installed on the top two signal flanges (yellow) of this section.  The HV flange is visible at the bottom (blue). The figure on the right shows the routing of the signal and HV cables in this breakout section. }
}
\end{figure}

\begin{figure}[htbp]
{\centering
\includegraphics[width=1.0\textwidth]{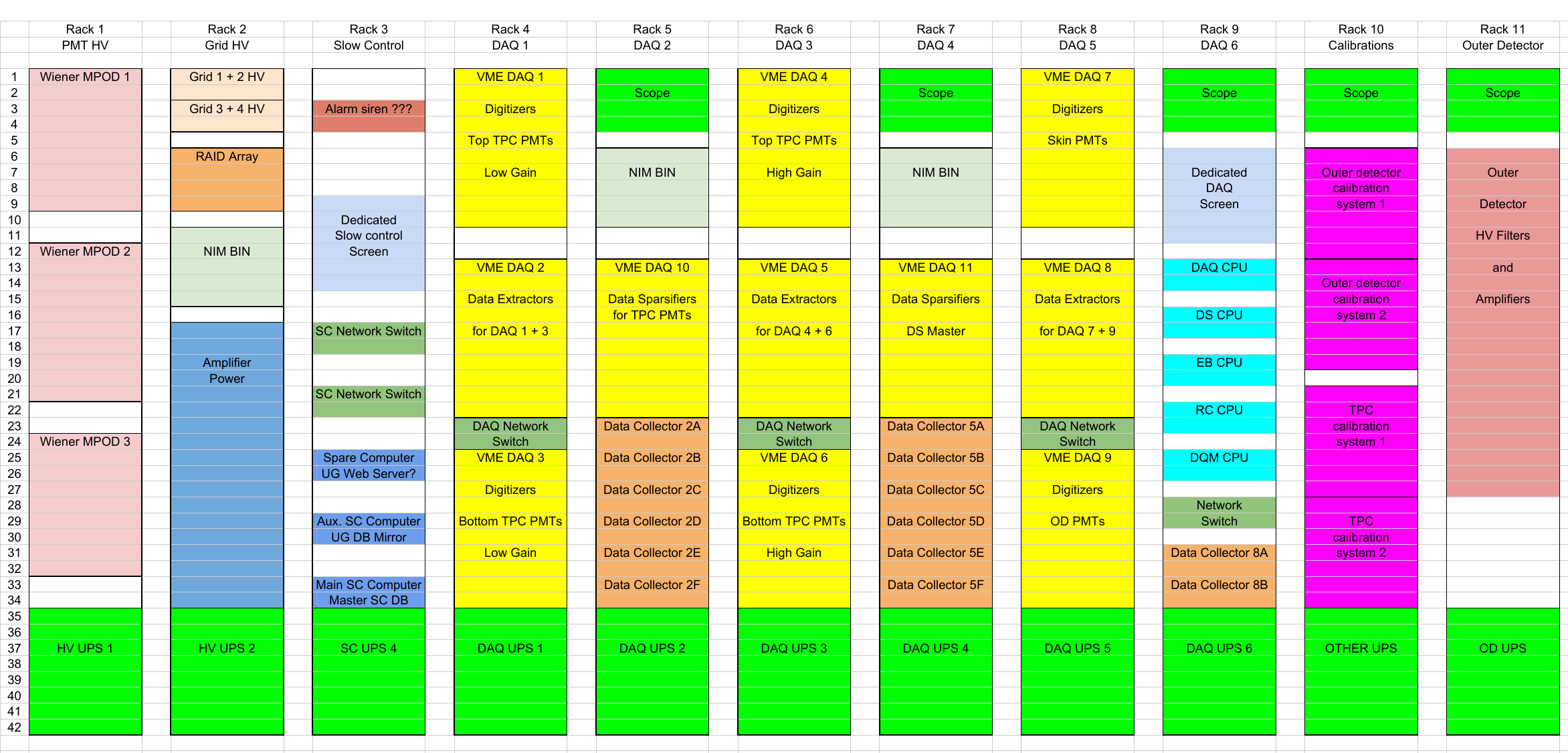}
\tdrfcaption[RackLayout]{Layout of the LZ electronics racks}{Detailed layout of the LZ electronics racks, installed at the deck level.}
}
\end{figure}

A total of \num{11} electronics racks are installed at the deck level, as shown in Figure~\ref{EDCf:DavisCavernView}, to provide space for all electronics systems, except the xenon amplifiers. The racks are fully enclosed and the forced airflow across the electronics is directed into the exhaust system of the Davis Laboratory.
Figure~\ref{EDCf:RackLayout} shows the layout of the components installed in these \num{11} racks.

The power requirements of the electronics system in the Davis Laboratory have been estimated based on the measured power consumption of the various components of the system. The power required by the amplifiers, the DAQ, the computer systems, and the PMT HV systems is \SI{26}{\kW}.  The power required for the various calibration systems for the outer detector and the xenon PMTs remains to be determined.  The power required by the electronics is provided by uninterruptible power supplies, installed at the bottom of each electronics rack. The total maximum load of the electronics is \SI{94}{\kW}. This is the load of the equipment at startup.

\clearpage
\bibliographystyle{apsrev4-2}
\bibliography{LZtdr}

\wlog{In tdrinclude command}
\wlog{\ChapLabel}

\renewcommand{\ChapLabel}{chap:MAS} 
\renewcommand{\BibLabel}{MASS:bib} 
\renewcommand{\SecPre}{MAS}
\renewcommand{\FloatPre}{MAS}
\tdrchap[MAS]{Material Screening}

\tdrsec[Intro]{Introduction}

The LZ sensitivity is assessed using a profile likelihood ratio (PLR) method, as described in Section~\ref{chap:SRD}. For material screening purposes, however, we evaluate conservative tolerable count rates by considering the residual events in a WIMP search region of about 0--20 detected photons (phd), equivalent to approximately \SIrange{1.5}{6.5}{\keVee} or \SIrange{6}{30}{\keVnr}. This energy window is functionally equivalent to the one employed in the full PLR method when considering the experiment's optimum sensitivity at about \SI{40}{\GeVpct} WIMP mass. Events are considered within the \SI{5.6}{\tonnel} fiducial mass, with vetoes from the liquid xenon (LXe) skin and outer detector (OD) applied, but before discrimination through S2/S1 and NR selection efficiency. In order for LZ to realize a sensitivity to a WIMP-nucleon cross section below \SI{3e-48}{\cm\squared} within three years of data taking, the maximum tolerable electron recoil (ER) background from non-astrophysical sources is approximately \SI{37e-6}{events\perkkd} (\SI{37}{\mudru}). The nuclear recoil (NR) background must also be low, with \num{\sim1}~count in the exposure. 

Although this represents an unprecedented low background rate for dark-matter detectors, it can be achieved through the use of several proven low-background assay techniques that have been successfully employed in recent rare-event searches for dark matter, as well as in neutrinoless double-beta decay and neutrino experiments~\cite{Leonard:2007uv,Akerib:2012ys,Araujo:2011as,Aprile:2011vb,Aprile:2013tov,Ghag:2011jd,Abgrall:2013rze,KamLANDZen:2012aa,An:2013zwz}.  The collaboration maintains significantly stricter goals for all the detector elements to reduce the risk of any one element violating the required levels: \SI{\sim37}{\mudru} and an NR rate of \SI{\sim1}{count} in the exposure. In this section we frequently refer to the goals, which are between 1 - 10{\%} of the levels specified as project requirements. The techniques LZ uses to monitor and control these backgrounds include:

\begin{itemize}
\itemsep0em
\item A comprehensive material-screening campaign to select components that satisfy stringent intrinsic radioactivity goals such that the single scatter rate within the fiducial volume and WIMP search energy range is less than \SI{0.4}{\NR} counts and below \SI{1}{\mudru} of ER background resulting from fixed contamination in the detector components; 
\item Direct measurements of radon emanation from construction materials for a maximum activity of \SI{10}{\mBq} throughout the liquid xenon. The goals for emanation are \SI{1}{\mBq}; 
\item Adherence to cleanliness protocols for control of airborne radioactivity and particulates to limit background at the level of that from materials to contact the LXe, contributing \SI{10}{\mBq} from radon emanation. This same level of surface contamination, in addition, contributes ER and NR backgrounds from the radioactive decays in the dust in the TPC and (\Pga,\Pn) reactions on the TPC components of  \SI{1}{\mudru} ER and \SI{0.4}{\NR} counts, respectively. The goals for emanation from dust are <\SI{1}{\mBq} with a corresponding reduction in ER and NR;
\item Removal of radioactive elements such as Ar and Kr from the liquid xenon to limit their single scatter ER backgrounds in the WIMP search energy range to below \SI{1}{\mudru}.
\end{itemize}

Material screening is the primary route to controlling the ER and NR backgrounds resulting from radioactivity in the experiment. The measurement of radioactive isotopes in and on detector materials is required. These are primarily the \Pgg-ray emitting isotopes \IKfz; \ICsoTS; and \ICosz, as well as \IUtTe, \IUtTF, \IThtTt, and their progeny. The U and Th chains are also responsible for neutron production following spontaneous fission and (\Pga,\Pn) reactions. Kr and Rn outgassing from materials into the Xe also results in ER backgrounds, and \Pga-emitting Rn daughters can contribute to neutron backgrounds when deposited on certain materials.

Generally, radioactive contaminants in massive components or those closer to or within the central volume of Xe present more stringent cleanliness and radio-purity requirements. These include the PMTs, PMT bases and cables; the TPC components, including the PTFE sheets and support structures; and the Ti cryostat. Our screening campaign includes several mature techniques for the identification and characterization of radioactive species within these bulk detector materials, namely \Pgg-ray spectroscopy with High Purity Germanium detectors (HPGe), Inductively-Coupled Plasma Mass Spectrometry (ICP-MS), Neutron Activation Analysis (NAA), and measurement of Rn emanation from components before their integration into LZ. These complementary techniques collectively produce a complete picture of the radiological contaminants. LZ has all of these available with sufficient sensitivity and throughput to meet the project goals, described in Section~\ref{MASS:TechSens}. 

Detector components are matched with the appropriate assay technique depending on the material and requisite sensitivity, defined by Monte Carlo simulations described in Section~\ref{OCSS:Sims}. In some cases, the final detector material or components are assayed. When manufacturing or fabrication processes are complex, raw components as well as final components will be assayed to assist in maintaining purity through the manufacturing process and to assist in selecting those processes that do not introduce additional contamination.

Stringent constraints are also applied to ``intrinsic'' contamination of the Xe by radon, argon and krypton. The LZ Xe purification program will remove \IKreF from the Xe down to the level of \SI{0.015}{\ppt} using chromatographic techniques (where concentration is for g/g here and throughput this section). The Xe purification program is detailed in Section~\ref{chap:XCS}. All components that come into contact with any Xe in the experiment, whether within the primary instrument or in the gas storage, circulation, or recovery pipework, are screened for fixed contaminants, Rn emanation, and Kr outgassing to ensure that the intrinsic background remains within defined limits. These emanation and outgassing measurements are performed in dedicated chambers built and operated by LZ institutions. Similarly, techniques to measure bulk contamination of materials with alpha-emitting radon progeny that are not readily identified using traditional HPGe, ICP-MS, or NAA have been developed by the collaboration. This is particularly important for the PTFE reflector panels within the TPC where prolonged exposure to radon-contaminated air during manufacture will result in the presence of alpha-emitting daughter of \IPbtoz. The high cross section for (\Pga,\Pn) reactions on the fluorine in the PTFE will result in neutron emission, setting stringent constraints on \IPbtoz content in the bulk material.   

The results of screening measurements are entered into an LZ materials database and build on the existing LUX screening campaign data~\cite{Akerib:2014rda}. The database collates assay results from materials selected for use and identifies the components that contain them. These are referenced to results from Monte Carlo simulations that detail the background from the components in LZ. The contributions from several other sources are included in the database, where each contributes no more than materials radioactivity ER and NR goals. The first is the contribution from dust and radon-daughter plate-out on components, especially during component transport, storage, and assembly. This is controlled through use of dedicated cleanrooms available to the project at SURF (especially the radon-reduced Surface Assembly Laboratory), active monitoring of the environment for radon, and following established cleanliness, handling, and storage procedures. Selected lightweight plastics and rubbers with low radon diffusion coefficients are used to enclose materials in transit and during temporary storage. Coupons and witness plates are used for measurements of surface contamination with high sensitivity alpha-screening devices and cameras. The second  contribution comes from cosmogenic activation of components before they are moved underground, such as \IScfs production from Ti activation or \IXeotS and radioactivity from the laboratory environment. While contribution to the background from local radioactivity and potential activation products including radioactive isotopes of Xe are assessed with a number of simulation toolkits, data from the LUX experiment in particular is able to provide considerable input to reduce the systematic uncertainty for such calculations. 

Although the sensitivity goals for LZ require unparalleled low-background contamination control for dark matter experiments and consequent severe constraints on material contamination, the screening campaign outlined here builds on the demonstrated substantial experience of the collaboration and established procedures or techniques employed by rare-event searches for background mitigation to meet these challenges with confidence. The LZ collaboration has accomplished important steps in developing necessary infrastructure for our assay and cleanliness program, and in identifying appropriately low-activity materials for the five most critical detector components or materials including Ti for the cryostat, PTFE used in several locations in the inner TPC, PMT base materials, PMT raw materials, and stainless steel for large mass support structures. In the following subsections we define the goals and requirements for our screening program, present the details of assays associated with the major components together with our collected data for all materials, present the impact of the background sources in LZ as determined through Monte Carlo simulations, and summarize the techniques and resources available to the project to meet our assay and cleanliness requirements on schedule. 

\tdrsec[goalsReqs]{Goals, Requirements and Estimates}

The LZSim Monte Carlo simulation package, described in detail in Section~\ref{OCSS:Sims}, has been constructed to model the experiment and inform the design, determine optimal performance parameters, and define tolerable rates from background sources. Developed using the \geantFour toolkit~\cite{Agostinelli:2002hh}, the framework inherits from, and closely follows, the successful implementation of the LUX model~\cite{Akerib:2011ec}, with evolving design of all parts of the experiment, including the inner detector, Xe skin, cryostat, and the OD, reflected in appropriate changes to the model geometry.  

Simulations are performed to assess the contribution from all expected background sources, including intrinsic radioactivity in the Xe, emission from every component in the experiment, as well as estimates of neutrino interactions. This extends to subcomponents, with the model accurately representing the physical distribution of contaminants, particularly since (\Pga,\Pn) neutron yields vary by many orders of magnitude depending on the primary constituents of the materials containing the alpha-emitting uranium and thorium and decay products. Similarly, the physical distribution of \Pgg-ray, alpha, and beta particle emitters are modeled, as electrons created by these may produce detectable photons through Cherenkov and Bremsstrahlung processes, particularly in quartz or plastics close to or in contact with Xe. Evaluations of material radioactivity include contributions from spontaneous fission of heavy nuclei. Spontaneous fission of \IUtTe, \IUtTF, \IUtTf, \IPatTo, \IThtTz, or \IThtTt, present in the U and Th decay chains generates multiple prompt and delayed neutrons and  \Pgg-rays per decay. LZSim models these products with accurate energy, multiplicity, timing, and angular distributions. The single scatter identification capability in the TPC coupled to the high vetoing efficiency of the Xe skin and OD systems allow \SI{>99.99}{\percent} rejection of all spontaneous fission events, and effective removal from our background models. The performance of the veto skin and OD veto systems are presented in Section~\ref{XDSS:Skin} and Chapter~\ref{chap:OD}, respectively. Events in the TPC with associated energy depositions identified in the Xe skin or OD with signal equivalent to approximately equal to or more than \SI{100}{\keV} in Xe skin and \SI{200}{\keV} in OD are rejected. 

All energy depositions from interactions in the Xe and OD are recorded in LZSim. Where necessary, optical tracking is performed following scintillation and ionization generation implemented using NEST~\cite{Szydagis:2011tk}. LZSim models photon hit patterns and timing to mimic S1 and S2 signal generation in LZ, and allows for accurate studies of rare mechanisms that might produce backgrounds such as MSSI (multiple-scintillation single-ionization) events or background pile-up. Such detailed characterization and quantification of all background sources and their impacts are necessary to assign confidence to expected background event rates, their spectra, and their physical distribution in the detector. As a discovery instrument, the expected background in LZ must be well understood and quantified before significance can be ascribed to observation of any potential signal and WIMP discovery.  

Version controlled release management of LZSim includes validation of materials and geometries against CAD drawings, physics lists implementation and detector response using pre-defined macros, and impact tests from radio-activity released from major components as well as from specific benchmark locations distributed throughout the detector volumes. Details of LZSim and the simulations are given in Section~\ref{OCSS:Sims}. 

With the exception of astrophysical neutrinos discussed in Section~\ref{SIPS:NP}, the major sources of background in LZ will be radioactivity from construction materials surrounding the central fiducial volume, and radon daughters and krypton distributed throughout the xenon. The goal for the maximum unvetoed differential single-scatter ER rate from each of these non-astrophysical sources after cuts in the WIMP search energy range has been set to about \SI{1}{\mudru}. This is approximately \SI{10}{\percent} of that expected from irreducible \Ipp solar neutrinos deduced from our Monte Carlo simulations of the detector. Similarly, an upper limit of \num{0.4} unvetoed WIMP-like single-scatter NRs due to neutron emission from material radioactivity is the goal for a \num{1000}-day exposure, reduced to \num{0.2} after a \SI{50}{\percent} NR efficiency is applied. Goals for maintaining cleanliness throughout the experiment construction, for parts manufacture, transport, storage, assembly and integration, allow for contribution to ER and NR background equivalent to that from materials. Table~\ref{MASt:BackgroundsTable} presents the estimate of the backgrounds in the experiment. Material assay results in hand and achieved by similar collaborations form the input for material activity, and these meet project goals. Values for radon emanation are based on conservative estimates, as described in Section~\ref{MASS:RnEm}, that match our requirements, but are currently estimated to exceed our goals. As we assay our materials we anticipate a reduction from this contribution to Table~\ref{MASt:BackgroundsTable}. The project requirements are discussed in Section~\ref{chap:SRD}, where the sensitivity of LZ to WIMP dark matter is presented. The LZ collaboration maintains more aggressive goals for its radioactive backgrounds for several reasons. The first is to provide contingency for measured radio-content in materials and for intrinsic contributions, invoking safety factors that mitigate large variations in expected and final radioactivity. Second, the goals facilitate simplified analyses, with total material and intrinsic background contributions sub-dominant to the ER backgrounds from astrophysical neutrinos that will be uniformly distributed across the active volume. Finally, our goals enhance physics capability beyond the primary WIMP search, as described in Section~\ref{chap:SIP}, particularly for neutrino interactions, neutrinoless double-beta decay, and alternative dark matter models including axion-like particles (ALPs).  

Acceptance of screened materials for use in LZ depends on the Monte Carlo simulations and the overall radioactive background budget. When a component is identified as required in LZ, it is incorporated into the LZSim model and a preliminary estimate of maximum tolerable activity from that component is calculated. This requirement necessarily depends on activity from other components and the overall budget, and initial inputs to LZSim for detector-related backgrounds are based on measured values, or from screening results from previous experiments. Initially the background budget is evenly divided among the major components, such that if there are ten components, the allowable background for each component is assigned 1/10 of the allowable contamination. The maximum tolerable activity for the new component, including contingency for dominant materials such as the PMTs, is then translated to a required screening sensitivity for radio-assaying a material sample. This in turn informs the screening technique and facility that will be employed for the assay. Screening results are then fed back into LZSim to produce an accurate assessment of ER and NR background and overall impact. The acceptance of the component depends on whether it meets requirements, or if it can be accommodated given related constraints and achieved radio-purity in other components and materials. In some cases, and as is justified by our assay experience, we may employ sampling of complete components. As the materials are assayed, this screening provides ``as-built'' input for the LZ background model.  

The ER and NR background calculations take into account radioactivity from all the components. Requirements on different materials and components vary, with their impact depending on material and position in the experiment. The simulations inform the necessity for screening U and Th in materials at the order of tens of ppt levels, tens to hundreds of ppb for K, and \SI{0.05}{\mBqkg} for \ICosz. Sensitivity at \SI{10}{\mBqkg} to \IPbtoz in bulk materials is required. Radon emanation systems are required to meet sensitivity to \IRnttt at the level of \SI{0.3}{\mBq} from materials. Cleanliness-maintenance procedures that will be applied to all materials to guarantee adequate background control and accurate modeling set requirements on maximum dust or radon exposure, calculated taking into account assembly and integration duration. We require less than a \SI{10}{\mBq} contribution to ER background from dust, conservatively equivalent to less than \SI{1}{\gram} of dust in total on all wetted surfaces. We require a maximum \IPbtoz activity from radon progeny on surfaces of \SI{10}{\mBqmt}, reduced to \SI{0.5}{\mBqmt} for the inner surfaces of the TPC reflectors. 

Materials of sufficient radio-purity have been successfully deployed in rare-event search experiments and will be procured and incorporated in the LZ project following sample/component measurements with available technology and facilities that incorporate screening, cleanliness maintenance, and outputs from the completed R\&D program. These measurements and procedures are well on the way to reliably identify clean materials and maintain their purity throughout the chain, from fabrication to installation and operation. The assaying program to achieve these minimum limits and estimates adopted are detailed below.

\tdrsubsec[matsTab]{Materials Table}

Table~\ref{MASt:MaterialsTable} presents the justification for the assumed values of \IUtTe, \IThtTt, \ICosz, and \IKfz content used as initial input to LZSim for the major components. 
Throughout this document, we refer to early and late values for \IUtTe and \IThtTt. In the case of \IUtTe, we define the early part of the chain (\IUtTeE) as containing any isotopes above \IRatts since chemical processes may induce a break of secular equilibrium at this point and it will take many years to be restored.  The late part of the chain (\IUtTeL) is counted from \IRatts and below. 
In practice, standard HPGe detectors are not sensitive to the low-energy \Pgg-ray lines from \IPbtoz at the bottom of the chain but the BEGe and well-type detectors available to LZ are (see Section~\ref{MASSss:BUGS}). 
We populate the table with values for \IPbtoz where appropriate for our materials and assays. 
For the \IThtTt chain, we define the early part of the chain (\IThtTtE) as coming from isotopes above \IRattf and the late part of the chain (\IThtTtL) as coming from isotopes below as chemical processes may, again, break the secular equilibrium. 
The relatively long half-life of \IRatte (\SI{5.7}{\year}) means that, for example, with \SI{100}{\percent} removal of Ra isotopes it would take \SI{\sim50}{\year} for 
\IRatte and \IRattf to re-equilibrate.
LZ and LUX assays are maintained in our dedicated database but for those items which we are yet to screen, XENON100 assays are taken from \cite{Aprile:2011ru}; EXO-200 from \cite{Leonard:2007uv}; XENON1T ~\cite{Aprile:2015lha}; \textsc{Majorana}, GERDA, and SuperNEMO by private communication or conference presentations; SNO from ~\cite{Boger:1999bb}. 

\afterpage{
\begin{center}
\sffamily
\renewcommand{\ltbcp}{MaterialsTable}
\renewcommand*{\arraystretch}{1.}
\begin{longtable}{| >{\centering\arraybackslash}m{4.0cm} | >{\centering\arraybackslash}m{1.3cm} | >{\centering\arraybackslash}m{1.3cm} | >{\centering\arraybackslash}m{1.3cm} | >{\centering\arraybackslash}m{1.3cm} | >{\centering\arraybackslash}m{1.1cm} | >{\centering\arraybackslash}m{1.3cm} | >{\centering\arraybackslash}m{1.1cm} |}
\caption[Table of material radioactivity \ifKL\space\texttt{\FloatPre t:\ltbcp}\fi]{Materials in the LZ design listed
with radioactivity \si{\milli\becquerel\per\kg}
as determined by direct assay data from the LZ screening program, and
from other published experimental results. 
Reference values are from LZ$^{[1]}$, EXO-200$^{[2]}$, XENON100$^{[3]}$, SuperNEMO$^{[4]}$, SNO$^{[5]}$ and GERDA$^{[6]}$.
The activities shown represent error weighted averaged values for \Pgg-ray detection or \SI{68}{\percent} C.L. upper limits (italicized). \ifKL\space\texttt{\FloatPre t:\ltbcp}\fi\label{\FloatPre t:\ltbcp}}\\
\hline
\rowcolor{mrocol}
{\rule[-0.7em]{0pt}{2.0em}\textbf{Material}} & 
{\rule[-0.7em]{0pt}{2.0em}\textbf{\IUtTeE}} &
{\rule[-0.7em]{0pt}{2.0em}\textbf{\IUtTeL}} & 
{\rule[-0.7em]{0pt}{2.0em}\textbf{\IThtTtE}} &
{\rule[-0.7em]{0pt}{2.0em}\textbf{\IThtTtL}} & 
{\rule[-0.7em]{0pt}{2.0em}\textbf{\ICosz}} &
{\rule[-0.7em]{0pt}{2.0em}\textbf{\IKfz}} & 
{\rule[-0.7em]{0pt}{2.0em}\textbf{\IPbtoz}}
\\ 
\hline
\rowcolor{mrocol}
\multicolumn{8}{|c|}{\textbf{Values in (\si{\mBqkg})}}\\
\endfirsthead
\hline
\rowcolor{mrocol}
{\rule[-0.7em]{0pt}{2.0em}\textbf{Material}} & 
{\rule[-0.7em]{0pt}{2.0em}\textbf{\IUtTeE}} &
{\rule[-0.7em]{0pt}{2.0em}\textbf{\IUtTeL}} & 
{\rule[-0.7em]{0pt}{2.0em}\textbf{\IThtTtE}} &
{\rule[-0.7em]{0pt}{2.0em}\textbf{\IThtTtL}} & 
{\rule[-0.7em]{0pt}{2.0em}\textbf{\ICosz}} &
{\rule[-0.7em]{0pt}{2.0em}\textbf{\IKfz}} & 
{\rule[-0.7em]{0pt}{2.0em}\textbf{\IPbtoz}}
\\
\hline
\rowcolor{mrocol}
\multicolumn{8}{|c|}{\textbf{Values in (\si{\mBqkg})}}\\
\hline
\endhead
\hline
\rowcolor{lrocol}
\multicolumn{8}{|c|}{(continued on next page)}\\
\hline
\endfoot
\endlastfoot
\hline 
\rowcolor{lrocol}
\multicolumn{8}{| c |}{\textbf{General Materials}} \\ \hline 
Titanium$^{1}$ & \textit{1.60} & \textit{0.09} & 0.28 & 0.23 & \textit{0.02} & \textit{0.54} & -   \\
PTFE$^{2}$ & \textit{0.02} & \textit{0.02} & \textit{0.03} & \textit{0.03} & - & \textit{0.12} & -   \\
PEEK$^{1}$ & 17.0 & 16.6 & 16.1 & 8.50 & \textit{0.52} & 47.8 & -   \\
LED$^{1}$ & 2000 & \textit{100.0} & \textit{200} & \textit{100.0} & \textit{-} & \textit{1000} & -   \\
Copper$^{1}$ & \textit{1.30} & \textit{1.20} & \textit{14.5} & \textit{1.70} & \textit{0.40} & \textit{7.00} & -   \\
Cable (RG174)$^{3}$ & 29.8 & 1.47 & 3.31 & 3.15 & 0.65 & 33.1 & -   \\
Stainless Steel$^{1}$ & 1.20 & 0.27 & 0.33 & 0.49 & 1.60 & 0.40 & -   \\
Epoxy$^{2}$ & 0.55 & 0.55 & 0.10 & 0.10 & - & 0.63 & -   \\
Viton O-ring$^{1}$ & 2630 & 2490 & 220 & 220 & 10.0 & 2150 & -   \\
Tyvek$^{4}$ & 6.00 & 6.00 & 2.20 & 2.20 & - & 5100 & -   \\
HDPE$^{5}$ & 5.96 & 0.37 & 0.63 & 0.62 & - & 3.40 & -   \\
Rubber O-ring$^{2}$ & 124 & 124 & 41.0 & 41.0 & - & 24.5 & -   \\
Mini-HV cable$^{1}$ & 40.0 & 2.10 & 1.50 & 1.30 & - & 0.17 & -   \\
\hline 
\rowcolor{lrocol} 
\multicolumn{8}{| c |}{\textbf{Liquid Scintillator}} \\ \hline 
LAB$^{1}$ & 0.00005 & 0.00005 & 0.00003 & 0.00003 & - & 0.00001 & -   \\
\CGdClT.6\CHtO$^{1}$ & 1.24 & 1.24 & 0.41 & 0.41 & - & 0.0006 & -   \\
PPO$^{1}$ & \textit{1.85} & \textit{1.85} & \textit{2.60} & \textit{2.60} & - & \textit{0.0008} & -   \\
TMHA$^{1}$ & \textit{0.25} & \textit{0.25} & \textit{0.29} & \textit{0.29} & - & \textit{0.0003} & -   \\
bis-MSB$^{1}$ & \textit{2.60} & \textit{2.60} & \textit{0.78} & \textit{0.78} & - & \textit{0.0009} & -   \\
\hline 
\rowcolor{lrocol}
\multicolumn{8}{| c |}{\textbf{Outer Detector}} \\ \hline 
Glass bulb$^{1}$ & 1507 & 1507 & 1065 & 1065 & - & 3900 & -   \\
LS Tanks$^{2,5}$ & \textit{0.01} & \textit{0.01} & \textit{0.005} & \textit{0.005} & - & \textit{0.07} & -   \\
Displacement foam$^{1}$ & 20.0 & 57.0 & \textit{2.60} & \textit{9.00} & \textit{6.00} & \textit{80.0} & -   \\
\hline 
\rowcolor{lrocol}
\multicolumn{8}{| c |}{\textbf{PMT Base Materials}} \\ \hline 
Resistors (4.99M)$^{1}$ & 965 & 284 & 240 & 172 & \textit{18.0} & 3249 & 35475   \\
Capacitors$^{1}$ & 4540 & 11715 & 4035 & 4035 & \textit{10.0} & 350 & 11590   \\
Cirlex Board$^{1}$ & 23.9 & 19.1 & 3.19 & 3.19 & \textit{0.63} & \textit{15.1} & 25.5   \\
Solder$^{1}$ & \textit{58.2} & \textit{11.8} & \textit{10.7} & \textit{10.7} & \textit{2.24} & \textit{31.8} & -   \\
Receptacles: PMT pins$^{1}$ & 1178 & \textit{7.00} & 22.0 & 15.0 & - & 110 & 22393   \\
Receptacles: Signal/HV$^{1}$ & 833 & 7.50 & 13.2 & 15.4 & - & 77.3 & 21171   \\
Receptacles: Ground$^{1}$ & 568 & 20.5 & 16.7 & 8.20 & - & \textit{19.0} & 29112   \\
Resistors (7.5M)$^{1}$ & 1097 & 389 & 106 & 106 & \textit{16.3} & 3082 & 41600   \\
Resistors (10.0M)$^{1}$ & 2420 & 414 & 226 & 226 & \textit{11.6} & 987 & 21613   \\
Resistors (2.49M)$^{1}$ & 1787 & 571 & 645 & 150 & \textit{65.0} & 7069 & 52702   \\
Resistors (100k)$^{1}$ & 3460 & 1036 & 510 & 144 & \textit{169} & 6118 & 97516   \\
Resistors (50)$^{1}$ & 3430 & 670 & 314 & 264 & \textit{29.0} & 5425 & 61140   \\
Receptacles: Spring$^{1}$ & 4758 & 38.0 & 50.0 & 50.0 & \textit{7.00} & 131 & 1364   \\
\hline 
\rowcolor{lrocol}
\multicolumn{8}{| c |}{\textbf{R11410 PMTs}} \\ \hline 
Metal Bulb$^{1}$ & \textit{17.9} & \textit{0.90} & \textit{1.67} & \textit{1.28} & - & \textit{6.41} & - \\
Dynode$^{1}$ & \textit{216} & \textit{2.02} & \textit{4.10} & \textit{3.40} & 4.00 & \textit{4.60} & - \\
Shield Plate$^{1}$ & \textit{77.0} & \textit{3.10} & \textit{5.00} & \textit{3.20} & 4.60 & \textit{6.00} & - \\
Faceplate$^{1}$ & \textit{11.0} & \textit{0.67} & \textit{1.00} & \textit{0.80} & - & \textit{4.00} & - \\
Insulator plate$^{1}$ & \textit{20.9} & \textit{1.05} & \textit{1.63} & \textit{1.16} & - & \textit{6.28} & - \\
Electrode Disk$^{1}$ & \textit{203} & 9.50 & \textit{4.30} & 14.0 & 8.50 & \textit{9.60} & - \\
Faceplate Flange$^{1}$ & \textit{162} & \textit{2.75} & \textit{3.80} & \textit{4.20} & 12.5 & \textit{14.4} & - \\
Ceramic Stem$^{1}$ & 105 & 20.0 & 12.9 & 9.60 & - & 110 & 5.60 \\
Ceramic Stem Flange$^{1}$ & \textit{198} & \textit{0.63} & \textit{2.32} & \textit{0.84} & 12.0 & \textit{3.30} & - \\
Aluminium Ring$^{1}$ & \textit{62.0} & \textit{1.23} & 2.33 & \textit{0.94} & \textit{0.34} & \textit{8.50} & - \\
Getter$^{1}$ & \textit{2508} & \textit{39.0} & \textit{133} & \textit{102} & \textit{9.40} & \textit{173} & - \\
Stem Coating$^{1}$ & \textit{22.0} & 178 & \textit{9.00} & 7.50 & - & 61.0 & \textit{539} \\
\hline  
\rowcolor{lrocol}
\multicolumn{8}{| c |}{\textbf{R8520 Skin PMTs}} \\ \hline 
Metal Package$^{3}$ & 19.0 & 19.0 & \textit{13.0} & \textit{13.0} & 40.0 & 90.0 & -   \\
Glass in Stem$^{3}$ & 970 & 970 & 340 & 340 & \textit{10.0} & 2300 & -   \\
Spacer$^{3}$ & 780 & 780 & 260 & 260 & \textit{12.0} & 800 & -   \\
Seal$^{3}$ & 17.0 & 17.0 & 370 & 370 & \textit{27.0} & 5.00 & -   \\
Electrodes$^{3}$ & 19.0 & 19.0 & 18.0 & 18.0 & 12.0 & 0.15 & -   \\
Window$^{3}$ & \textit{0.50} & \textit{0.50} & \textit{1.80} & \textit{1.80} & \textit{0.10} & 18.0 & -   \\
\hline  
\rowcolor{lrocol}
\multicolumn{8}{| c |}{\textbf{TPC Items}} \\ \hline 
Loop Antenna Wire$^{2}$ & 0.20 & 0.20 & 0.12 & 0.12 & - & \textit{1.86} & -   \\
Thermometers$^{1}$ & 4038 & 3433 & 1439 & 1042 & \textit{6.6} & 156416 & 1364   \\
Acoustic Sensor PVDF$^{1}$ & 1710 & 595 & 2179 & 1951 & - & 1389 & -   \\
Field Shaping Resistor$^{1}$ & 5679 & 1.73 & 0.57 & 0.57 & - & 0.19 & -   \\
\hline  
\rowcolor{lrocol}
\multicolumn{8}{| c |}{\textbf{HV Conduits}} \\ \hline 
Tivar$^{1}$ & \textit{6.20} & 22.2 & \textit{1.22} & \textit{1.22} & - & \textit{9.30} & -   \\
Delrin$^{4}$ & \textit{4.00} & \textit{0.70} & \textit{0.18} & \textit{0.18} & \textit{0.30} & 18.0 & -   \\
\hline  
\rowcolor{lrocol}
\multicolumn{8}{| c |}{\textbf{Cryostat Insulation}} \\ \hline 
SI Aluminium$^{6}$ & 1.13 & 1.13 & 0.37 & 0.37 & - & 25.5 & -   \\
SI PTFE$^{6}$ & 0.34 & 0.34 & 0.75 & 0.75 & - & 38.0 & -   \\
Foam-insulation$^{1}$ & 57.0 & 57.0 & 9.00 & 9.00 & \textit{6.00} & \textit{80.0} & -   \\
Helicoil$^{2}$ & \textit{2.40} & \textit{2.40} & \textit{1.30} & \textit{1.30} & 41.0 & \textit{13.5} & -   \\
\hline
\end{longtable}
\end{center}
}

Our assay campaign has made good progress in measuring the main contributors to the experiment's ER and NR backgrounds and those that represent challenging requirements. LZ has performed over 250 material assays from April 2015 to October 2016, and have already identified five critical components or materials that collectively account for \SI{60}{\percent} of the NR and \SI{50}{\percent} of ER background from intrinsic material radioactivity in LZ. These are the Ti used for the cryostat, field shaping rings, and PMT support structures; PMTs including all PMT raw materials; PMT base components; stainless steel for OD supports; and PTFE. All these materials satisfy the LZ goals that are well below the project requirements. Below we summarize the assay campaigns and results from these components and materials. 

\par

\textbf{Titanium:} The two cryostat vessels and flanges will be constructed with CP-1 grade titanium.  
The selected material has measured \IUtTe and \IThtTt chain activities of \SI{<0.09}{\mBqkg} and \SI{0.23}{\mBqkg}, respectively, representing the lowest reported to-date worldwide and substantially below the LZ goals. Two billets of Ti from the same production lot have been assayed and material procured to provide the raw material for all Ti components. The field-shaping rings, constructed from \SI{260}{\kg} of Ti, and the PMT support structures, constructed from \SI{104}{\kg} of Ti, will use the same raw material as the cryostat.

A long campaign in assaying titanium from various sources and states of processing has been underway for a couple years, including investigation of alternative options such as stainless steel. The cryostat material assays have primarily been carried out by the Berkeley Low Background Facility using its surface screening detectors at LBNL and the underground HPGe system \textsc{Maeve} both before and after its relocation from the Oroville Dam to the 4850L of SURF (see Section~\ref{MASSss:BHUC}). Moreover, there is consistency in multiple assays of similar geometries with this HPGe counter which was also used to assay titanium samples ultimately selected for use in LUX. Dozens of titanium and stainless steel samples were assayed for the LZ experiment. As seen in Figure~\ref{MASf:TitaniumAssays} a wide range of radioactivity values have been covered and that, in general, titanium has been more consistently clean as compared to stainless steel, which varies in both the natural radionuclides as well as \ICosz.

The selected titanium sample from TIMET, Heat Number (HN) 3469 was initially screened at the Berkeley Low Background Facility using \textsc{Maeve} in May of 2015. This initial sample consisted of \SI{10.1}{\kg} of plates selected from the top portion of the single slab that comprised this heat (HN3460-T). A second sample from this slab, taken from the middle (HN3469-M), was acquired and assayed in September of 2015. Both samples of this titanium stock were found to be consistent with one another with their contained radioactivity. The samples were analyzed using a \geantFour model of the \textsc{Maeve} detector with an exact sample geometry of the titanium placed around the detector on four sides and the top to simulate detection efficiency of \Pgg-rays in the regions of interest for the detector. The simulation was first compared to calibrated samples of known activity to benchmark the detector geometry, then modeled with the titanium geometry. As a check, the titanium sample was also directly compared to a calibration sample in a similar geometry, which agreed well with the analysis performed with a simulated efficiency.

In both of the samples, each counted for approximately three weeks, the early uranium chain was non-detectable or barely detectable within the limited abilities of HPGe detectors to assay this portion of the decay chain via \Pgg-ray spectroscopy. The late portion of the chain (at \IRatts and below) however, is quite accessible via \Pgg-ray spectroscopy due to both the branching ratio and detection efficiency for \Pgg-rays emitted during the decay of constituent isotopes, registered no detectable activity above background down to the few ppt level. The late uranium value is based upon the \SI{609}{\keV} peak from \IBitof and is consistent when compared to upper limits from other useful peaks in the late uranium chain (such as \num{295}, \num{352}, and \SI{1764}{\keV}). Both samples had detectable levels of the thorium series in both the early and late portions of the chain which implied equilibrium in both samples. The late series measurement is based upon the \SI{238}{\keV} \Pgg-ray from \IPbtot, which is the strongest peak based upon the product of the detection efficiency of that \Pgg-ray line and its branching ratio in the thorium chain for assay with \textsc{Maeve}. The contributions of the uranium, thorium, and potassium concentrations varies across the chains and their corresponding decays. The assays are summarized in Table~\ref{MASt:titaniumHN3469}. The late uranium chain in the titanium cryostat, for example, is a larger overall source of NR backgrounds within the detector than the early uranium chain because there is a rapid succession of several high energy alpha decays which induce (\Pga,\Pn) reactions. These various contributions from the chains towards the detector background are totaled and accounted for within Figure~\ref{MASf:TitaniumAssays}.

In terms of cosmogenic activation there were several isotopes of scandium present, most of which are the result of cosmic ray-induced reactions with the several stable isotopes of titanium. Detected in the sample was \IScfs (\num{889}, \SI{1120}{\keV}); as well as small amounts of  \IScfS (\SI{159}{\keV}),  \IScfe (\num{983}, \num{1037}, \SI{1312}{\keV}), and \IScffm (\num{271}, \SI{1157}{\keV}).   None of the scandium isotopes are a cause of concern as they are all short-lived, the longest of which is \IScfs with a half live of \SI{84}{\day} is listed in Table~\ref{MASt:titaniumHN3469}. The \IScfS,  \IScfe, and \IScffm activities were not listed as they have short half lives of a few days or less so essentially disappear over the course of the measurement. The reported value for \IScfs was decay corrected to the start of counting for each of the samples. All limits are one sigma upper limits and uncertainties are statistical only. A conservative systematic uncertainty of up to \SI{10}{\percent} should also be assumed.  
The assays have being confirmed by ICP-MS and with additional direct counting assays with the Chaloner detector at Boulby. 

\begin{table}
\tdrtcaption[titaniumHN3469]{Results from the radio-assay of titanium sample TIMET HN3469}{Results from \Pgg-ray spectroscopy of TIMET titanium HN3469 for both Top and Middle samples. All limits are one sigma upper limits and uncertainties are statistical only. A systematic uncertainty of up to \SI{10}{\percent} should also be assumed. The results are shown below as  activities in the \IUtTe, \IThtTt, and \IKfz chains --- all in units of \si{\mBqkg}. Results were obtained with the \textsc{Maeve} detector and confirmed with ICP-MS and the Chaloner detector.}
\begin{center}
\sffamily
\begin{tabular}{|l|c|c|}  
 \hline
 \rowcolor{mrocol}
 & \textbf{Top} & \textbf{Middle} \\
 \hline
date & 5-2015 & 9-2015 \\
sample mass & \SI{10.07}{\kg} & \SI{8.58}{\kg} \\
livetime & \SI{23.9}{\day} & \SI{20.8}{\day} \\
\IUtTeE  & < 1.6 & 2.9(15) \\
\IUtTeL & < 0.09 & < 0.10 \\
\IThtTtE & 0.28(3) & < 0.20 \\
\IThtTtL & 0.23(2) & 0.25(2) \\
\IKfz  & < 0.54 & < 0.68 \\
\ICosz & < 0.02 & < 0.03 \\
\IScfs & 2.0(1) & 2.7(1) \\
\hline
\end{tabular} 
\end{center}
\end{table}

\begin{figure}
\includegraphics[scale=0.6]{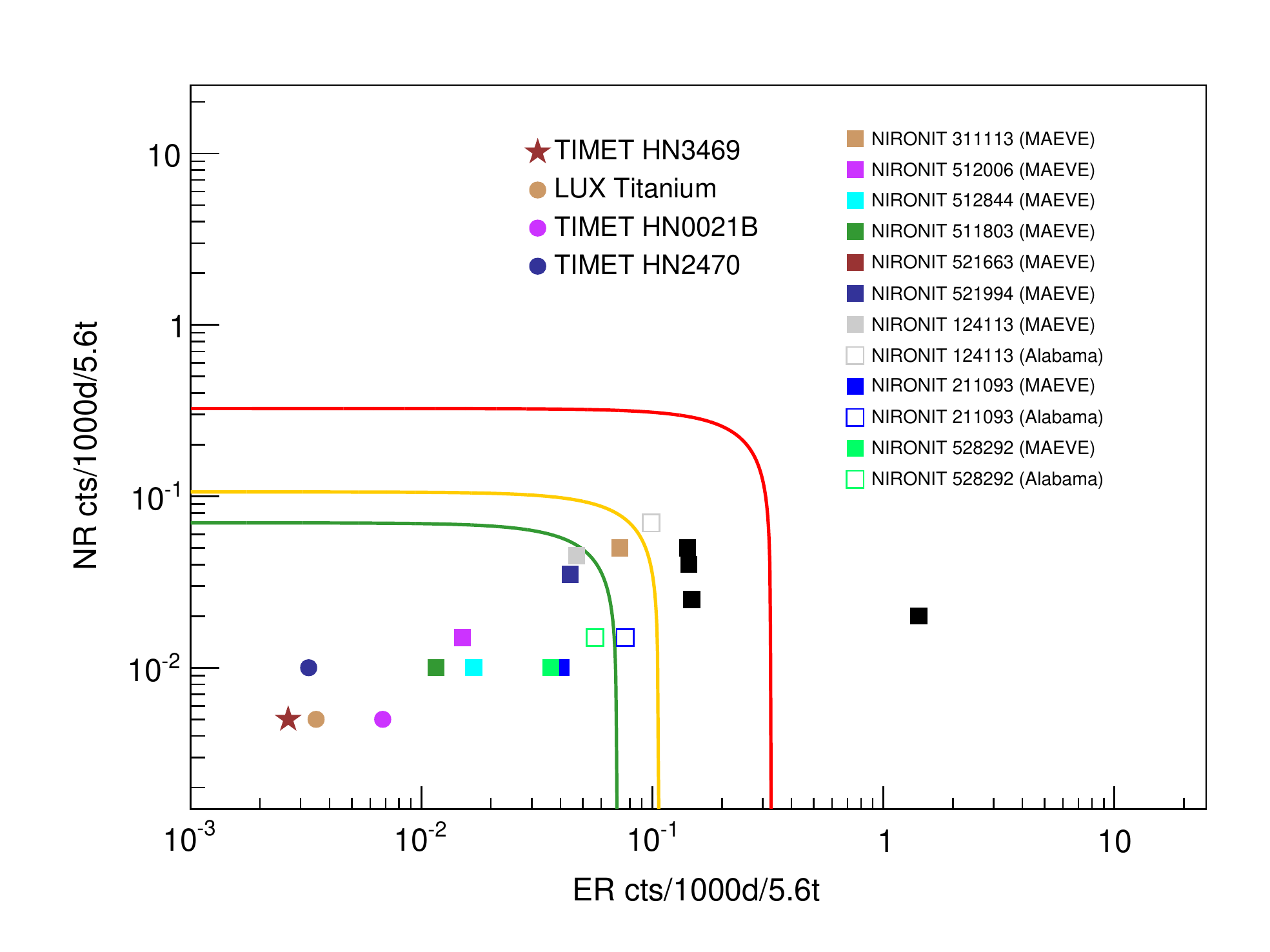} 
\centering
\tdrfcaption[TitaniumAssays]{ER and NR background in the cryostat}{The plot shows the background counts resulting from LZ's Ti  (TIMET) and stainless steel (NIRONIT) samples from the LXe cryostat in the \SI{1000}{\day}-exposure with a \SI{5.6}{\tonnel}-fiducial volume and after all the veto systems are applied, for ER events within \SIrange{1.5}{6.5}{\keVee}, with \SI{99.5}{\percent} rejection, and within \SIrange{6}{30}{\keVnr}, and \SI{50}{\percent} acceptance, for NR events. 
The red curve corresponds to the sum of \SI{10}{\percent} of the \Ipp solar neutrinos background for ERs and of \SI{0.2}{\NR} events. The yellow curve is for the sum of \SI{5}{\percent} of the \Ipp neutrino background and \SI{0.05}{\NR} events. The green curve corresponds to the sum of \SI{3.3}{\percent} of the \Ipp neutron background and \SI{0.03}{\NR} events, the requirement for the LZ cryostat. The Ti identified by the assay program, indicated by the star, is well below requirements of LZ.}
\end{figure}

\textbf{R11410 PMTs:} The activity for early production models of the 3-inch R11410 PMTs has been measured by the LZ collaboration~\cite{Akerib:2012da}. Multiple batches of PMTs (25 total) were assayed in groups of five. The raw material used to produce the experiment's PMTs were also assayed. The component materials used to produce the final LZ PMTs have now been assessed for radio-purity. The results are presented in Table~\ref{MASt:PMTCompskg}, and are generally consistent with results from the XENON1T collaboration (albeit using \SI{68}{\percent} rather than \SI{90}{\percent} CL upper limits) following a similar campaign of screening of their R11410 PMTs and constituent materials~\cite{Aprile:2015lha}. The LZ assays of PMT materials have been performed with \Pgg-ray spectroscopy principally with the SOLO detector and coordinated by the Brown University group, with the \textsc{Maeve} detector at SURF and Chaloner detector at Boulby providing additional throughput and low energy sensitivity, particularly for early U chain content. Details of these detectors are given in Section~\ref{MASSs:HPGe}. The materials have been returned to Hamamatsu and PMT production is underway. Screening of tubes that will be installed in LZ has begun at Boulby using the Chaloner and Lunehead detectors and will continue through 2017 when the screening will also be supported by the BHUC detectors.

\begin{table}
\tdrtcaption[PMTCompskg]{Assay results from R11410-20 PMT component materials}{Intrinsic radioactivity of the component materials which will be used in the manufacture of the R11410-20 PMTs to be used in the LZ experiment. All values are presented in \si{\mBqkg} of material. Measurements are performed using the SOLO detector~\textsuperscript{[1]}, \textsc{Maeve}~\textsuperscript{[2]}, and Chaloner~\textsuperscript{[3]}. Upper limits are italicized}
\begin{center}
\sffamily
\begin{tabular}{|>{\centering\arraybackslash}m{2.8cm}| >{\centering\arraybackslash}m{1.4cm} | >{\centering\arraybackslash}m{1.3cm} | >{\centering\arraybackslash}m{0.9cm} | >{\centering\arraybackslash}m{0.9cm} | >{\centering\arraybackslash}m{1.1cm} | >{\centering\arraybackslash}m{1.1cm} | >{\centering\arraybackslash}m{0.9cm} | >{\centering\arraybackslash}m{0.8cm} | >{\centering\arraybackslash}m{0.9cm} |}
\hline
\rowcolor{mrocol}
\multicolumn{10}{|c|}{\textbf{Values in (\si{\mBqkg})}}\\ 
\hline
\rowcolor{mrocol}
\rule[-0.7em]{0pt}{2.0em}\textbf{Material} & 
\rule[-0.7em]{0pt}{2.0em}\textbf{Mass screened (g)} & 
\rule[-0.7em]{0pt}{2.0em}\textbf{Mass /PMT (g)} & 
\rule[-0.7em]{0pt}{2.0em}\textbf{\IUtTeE} & 
\rule[-0.7em]{0pt}{2.0em}\textbf{\IUtTeL} & 
\rule[-0.7em]{0pt}{2.0em}\textbf{\IThtTtE} & 
\rule[-0.7em]{0pt}{2.0em}\textbf{\IThtTtL} & 
\rule[-0.7em]{0pt}{2.0em}\textbf{{\ICosz}} & 
\rule[-0.7em]{0pt}{2.0em}\textbf{{\IKfz}} & \textbf{{\IPbtoz}}\\
\hline
Metal Bulb\textsuperscript{2} & 506 & 78 & \textit{17.9} & \textit{0.90} & \textit{1.67} & \textit{1.28} & - & \textit{6.41} & - \\
Dynode\textsuperscript{1} & 530 & 7.2 & \textit{216} & \textit{2.02} & \textit{4.10} & \textit{3.40} & 4.00 & \textit{4.60} & - \\
Shield Plate\textsuperscript{1} & 519 & 4 & \textit{77.0} & \textit{3.10} & \textit{5.00} & \textit{3.20} & 4.60 & \textit{6.00} & - \\
Faceplate\textsuperscript{2} & 1168 & 30 & \textit{11.0} & \textit{0.67} & \textit{1.00} & \textit{0.80} & - & \textit{4.00} & - \\
Insulator plate\textsuperscript{2} & 838 & 8.6 & \textit{20.9} & \textit{1.05} & \textit{1.63} & \textit{1.16} & - & \textit{6.28} & - \\
Electrode Disk\textsuperscript{1} & 517 & 9.9 & \textit{203} & 9.50 & \textit{4.30} & 14.0 & 8.50 & \textit{9.60} & - \\
Faceplate Flange\textsuperscript{1} & 532 & 18 & \textit{162} & \textit{2.75} & \textit{3.80} & \textit{4.20} & 12.5 & \textit{14.4} & - \\
Ceramic Stem\textsuperscript{1,3} & 757.5 & 15.7 & 105 & 20.0 & 12.9 & 9.60 & - & 110 & 5.60 \\
Ceramic Stem Flange\textsuperscript{1} & 1568 & 14 & \textit{198} & \textit{0.63} & \textit{2.32} & \textit{0.84} & 12.0 & \textit{3.30} & - \\
Aluminium Ring\textsuperscript{1} & 506 & 0.6 & \textit{62.0} & \textit{1.23} & 2.33 & \textit{0.94} & \textit{0.34} & \textit{8.50} & - \\
Getter\textsuperscript{1} & 7 & 0.07 & \textit{2508} & \textit{39.0} & \textit{133} & \textit{102} & \textit{9.40} & \textit{173} & - \\
Stem Coating & 100 & 0.00012 & \textit{22.0} & 178 & \textit{9.00} & 7.50 & - & 61.0 & \textit{539} \\
\hline
\multicolumn{2}{|c|}{\textbf{Mass Weighted Ave}} & \textbf{186.1} & \textbf{\textit{71.6}} & \textbf{\textit{3.20}} & \textbf{\textit{3.12}} & \textbf{\textit{2.99}} & \textbf{\textit{2.82}} & \textbf{\textit{15.4}} & \textbf{\textit{0.47}} \\
\multicolumn{2}{|c|}{\textbf{Total (mBq/PMT)}} & \textbf{186.1} & \textbf{\textit{13.3}} & \textbf{\textit{0.60}} & \textbf{\textit{0.58}} & \textbf{\textit{0.56}} & \textbf{\textit{0.53}} & \textbf{\textit{2.87}} & \textbf{\textit{0.09}} \\
\hline
\end{tabular}
\end{center}
\end{table}

\textbf{PMT Bases:} The activity of the components used to produce the bases for the R11410 TPC and R8520 Xe skin PMTs have been assayed at Boulby, principally with the Chaloner detector. These items include the resistors, capacitors, connectors and substrate material. We have also performed assays to measure \IPbtoz content in these materials, exploiting the capability of Boulby's Lumpsey detector (described in Section~\ref{MASSss:BUGS}). This contamination is challenging to assay and usually overlooked, despite it being present in large quantity in components such as resistors and capacitors, with the potential to contribute significantly to neutron production. The \IPbtoz is accounted for in our background model. The radioactivity of the base components meet LZ goals and are presented for the TPC PMTs in Table~\ref{MASt:PMTBase}, and for the Xe skin in Table~\ref{MASt:PMTSkinBase}.

\begin{table}
\tdrtcaption[PMTBase]{Assay results from 3-inch R11410 PMT base component materials}{Assay results from 3-inch R11410 PMT base component materials. All values are presented in mBq/base.}
\begin{center}
\sffamily
\begin{tabular}{|>{\centering\arraybackslash}m{1.8cm}| >{\centering\arraybackslash}m{1.8cm} | >{\centering\arraybackslash}m{1.5cm} | >{\centering\arraybackslash}m{0.9cm} | >{\centering\arraybackslash}m{0.9cm} | >{\centering\arraybackslash}m{1.1cm} | >{\centering\arraybackslash}m{1.1cm} | >{\centering\arraybackslash}m{0.9cm} | >{\centering\arraybackslash}m{0.8cm} | >{\centering\arraybackslash}m{0.9cm} |}
\hline
\rowcolor{mrocol}
\multicolumn{10}{|c|}{\textbf{Values in (\si{\muBq}/base)}}\\ 
\hline
\rowcolor{mrocol}
\multicolumn{2}{|c|}{{\rule[-0.7em]{0pt}{2.0em}\textbf{Material}}} & 
\rule[-0.7em]{0pt}{2.0em}\textbf{Mass per base (mg)} &
\rule[-0.7em]{0pt}{2.0em}\textbf{\IUtTeE} & 
\rule[-0.7em]{0pt}{2.0em}\textbf{\IUtTeL} & 
\rule[-0.7em]{0pt}{2.0em}\textbf{\IThtTtE} & 
\rule[-0.7em]{0pt}{2.0em}\textbf{\IThtTtL} & 
\rule[-0.7em]{0pt}{2.0em}\textbf{{\ICosz}} & 
\rule[-0.7em]{0pt}{2.0em}\textbf{{\IKfz}} & 
\rule[-0.7em]{0pt}{2.0em}\textbf{{\IPbtoz}}\\
\hline
\multirow{4}{*}{Resistors} & 4.99M & 34.4 & 32.8 & 14.1 & 7.3 & 5.9 & \textit{0.6} & 130 & 1290 \\
 & 7.5M & 14.1 & 15.3 & 5.4 & 1.5 & 1.5 & \textit{0.3} & 43.2 & 584 \\
 & 10.0M & 12.9 & 31.5 & 5.3 & 2.9 & 2.9 & \textit{0.2} & 12.8 & 280 \\
 & 50R & 4.3 & 12.9 & 2.9 & 1.4 & 1.1 & \textit{0.1} & 23.5 & 265 \\\hline
Capacitors & 10nF & 24.5 & 111 & 286 & 98.5 & 98.5 & \textit{0.1} & 8.5 & 283 \\\hline
\multirow{3}{*}{Receptacles} & PMT Pins & 1170 & 775 & \textit{6.5} & 25.8 & 12.8 & \textit{6.0} & 129.0 & 26200 \\
 & Signal/HV & 156 & 103.4 & \textit{0.9} & 3.4 & 1.7 & \textit{0.8} & 17.2 & 3495 \\
 & Ground & 244.2 & 103.4 & \textit{0.9} & 2.9 & 1.7 & \textit{0.8} & 15.2 & 5780 \\\hline
Base & Cirlex & 3300 & 94.8 & 66.8 & 13.8 & 13.8 & \textit{2.0} & \textit{40.5} & 56.8 \\\hline
Solder & Pb Free & 180 & \textit{10.5} & \textit{2.1} & \textit{1.9} & \textit{1.9} & \textit{0.4} & \textit{5.7} & \textit{63.0} \\
\hline
\multicolumn{2}{|c|}{\textbf{Total}}  & \textbf{5140} & \textbf{\textit{1291}} & \textbf{\textit{390}} & \textbf{\textit{159}} & \textbf{\textit{143}} & \textbf{\textit{11}} & \textbf{\textit{426}} & \textbf{\textit{38300}} \\
\hline
\end{tabular}
\end{center}
\end{table}

\begin{table}
\tdrtcaption[PMTSkinBase]{Assay results from 1-inch R8520 skin PMT base component materials}{Assay results from 1-inch R8520 skin component materials. All values are presented in mBq/base.}
\begin{center}
\sffamily
\begin{tabular}{|>{\centering\arraybackslash}m{1.8cm}| >{\centering\arraybackslash}m{1.8cm} | >{\centering\arraybackslash}m{1.5cm} | >{\centering\arraybackslash}m{0.9cm} | >{\centering\arraybackslash}m{0.9cm} | >{\centering\arraybackslash}m{1.1cm} | >{\centering\arraybackslash}m{1.1cm} | >{\centering\arraybackslash}m{0.9cm} | >{\centering\arraybackslash}m{0.8cm} | >{\centering\arraybackslash}m{0.9cm} |}
\hline
\rowcolor{mrocol}
\multicolumn{10}{|c|}{\textbf{Values in (\si{\muBq}/base)}}\\ 
\hline
\rowcolor{mrocol}
\multicolumn{2}{|c|}{{\rule[-0.7em]{0pt}{2.0em}\textbf{Material}}} & 
\rule[-0.7em]{0pt}{2.0em}\textbf{Mass per base (mg)} &
\rule[-0.7em]{0pt}{2.0em}\textbf{\IUtTeE} & 
\rule[-0.7em]{0pt}{2.0em}\textbf{\IUtTeL} & 
\rule[-0.7em]{0pt}{2.0em}\textbf{\IThtTtE} & 
\rule[-0.7em]{0pt}{2.0em}\textbf{\IThtTtL} & 
\rule[-0.7em]{0pt}{2.0em}\textbf{{\ICosz}} & 
\rule[-0.7em]{0pt}{2.0em}\textbf{{\IKfz}} & 
\rule[-0.7em]{0pt}{2.0em}\textbf{{\IPbtoz}}\\
\hline
\multirow{5}{*}{Resistors} 
& 4.99M & 34.4 & 32.8 & 14.1 & 7.3 & 7.3 & \textit{0.6} & 130 & 1292\\
& 7.5M & 4.7 & 5.1 & 1.8 & 0.5 & 0.5 & \textit{0.1} & 14.4 & 195\\
& 10.0M & 4.3 & 10.5 & 1.8 & 0.1 & 0.1 & \textit{0.1} & 4.28 & 93.4\\
& 2.49M & 8.4 & 15 & 4.8 & 1.2 & 1.2 & \textit{0.5} & 59.4 & 443\\
& 100k & 3.6 & 12.3 & 3.7 & 0.5 & 0.5 & \textit{0.6} & 21.8 & 347\\\hline
Capacitors 
& 10nF & 14.7 & 66.3 & 171 & 59.1 & 59.1 & \textit{0.2} & 5.1 & 170\\\hline
\multirow{2}{*}{Receptacles} 
& Signal/HV & 156 & 103 & 0.9 & 1.7 & 1.7 & \textit{0.8} & 17.2 & 3495\\
& Ground & 244 & 103 & 0.9 & 1.7 & 1.7 & \textit{0.8} & 15.2 & 5780\\\hline
Base & Cirlex & 1400 & 40.2 & 28.3 & 5.8 & 5.8 & \textit{0.8} & \textit{17.2} & 24.1\\
Solder & Pb Free & 250 & \textit{10.5} & \textit{2.1} & \textit{1.9} & \textit{1.9} & \textit{0.4} & \textit{5.7} & \textit{63.0} \\
\hline
\multicolumn{2}{|c|}{\textbf{Total}}  & \textbf{2120} & \textbf{\textit{400}} & \textbf{\textit{230}} & \textbf{\textit{81}} & \textbf{\textit{81}} & \textbf{\textit{5}} & \textbf{\textit{291}} & \textbf{\textit{11900}} \\
\hline
\end{tabular}
\end{center}
\end{table}

\textbf{Stainless Steel:} The base of the OD support stand and water PMT stands total some \SI{770}{\kg} of material and are constructed from stainless steel. The assay values are taken from the LZ determination of NIRONIT raw material using the \textsc{Maeve} detector. These samples were identified as part of our R\&D campaign to select and procure sufficiently low background material for the detector cryostat. Though there is large variation across the sample activities, activities as low as \SI{<1.60}{\mBqkg} \IUtTe and \SI{0.23}{\mBqkg} \IThtTt have been realized. Goals for OD support radioactivity are easily met by all of the samples assayed. In addition to the OD supports stainless steel will be used for several internal structures such as grid supports and pipework. Several NIRONIT samples meet the radioactivity goals and all meet the requirements for these components. Candidate materials for procurement and the materials used for the final fabrication will be assayed. 

\textbf{PTFE:} The LZ collaboration has assayed four samples of PTFE in September 2015 using NAA at the MIT Research Reactor with surface \Pgg-ray counters at Alabama. The NAA technique is described in Section~\ref{MASSs:NAA}. The samples were prepared at Alabama together with NIST certified fly ash used to calibrate the neutron flux, and measured in port 2PH1 with \SI{5.5}{\mega\watt} of thermal reactor power. The samples include material from Flontech and DuPont, where the latter is the supplier used by the EXO experiment, employing sintering techniques developed with Applied Plastics Technology Inc. (APT) in Bristol, Rhode Island. The assayed contamination levels are presented in Table~\ref{MASt:ptfe results}. While these are slightly higher than those obtained by EXO, they satisfy the radioactive contamination goals of LZ.

\begin{table}
\tdrtcaption[ptfe results]{NAA measurements of PTFE}{NAA measurements of \IKfz, \IUtTe, \IThtTt, and \ILaoTe activities in PTFE samples.} 
\begin{center}
\sffamily
\begin{tabular}{|>{\centering\arraybackslash}m{3.5cm} | 
>{\centering\arraybackslash}m{1.5cm} | 
>{\centering\arraybackslash}m{1.5cm} | 
>{\centering\arraybackslash}m{1.5cm} | 
>{\centering\arraybackslash}m{1.5cm} |} 
\hline 
\rowcolor{mrocol}
\multicolumn{5}{|c|}{\textbf{Values in (\si{\mBqkg})}} \\ 
\hline 

\rowcolor{mrocol}
& \rule[-0.7em]{0pt}{2.0em}\textbf{\IUtTe ($\times 10^{-3}$)} 
& \rule[-0.7em]{0pt}{2.0em}\textbf{\IThtTt ($\times 10^{-3}$)} 
& \rule[-0.7em]{0pt}{2.0em}\textbf{\IKfz ($\times 10^{-3}$)} 
& \rule[-0.7em]{0pt}{2.0em}\textbf{\ILaoTe ($\times 10^{-6}$)}\\
\hline
PTFE8764 & $<$42 & 13 $\pm$ 3 & 66 $\pm$ 4 & 2.8 $\pm$ 0.2 \\
DuPont 807NX & 38 $\pm$ 9 & 29 $\pm$ 2 & 96 $\pm$ 5 & 9.4 $\pm$ 0.5 \\ 
DuPont NXT85 & $<$21 & 28 $\pm$ 2 & 122 $\pm$ 10 & 6.4 $\pm$ 0.3 \\
Flontech FLON008 & $<$27 & 51 $\pm$ 3 & 329 $\pm$ 17 & 25 $\pm$ 1 \\
PTFE8764 sheet & 18 $\pm$ 4 & 29 $\pm$ 1 & 76 $\pm$ 1 & 8.5 $\pm$ 0.2 \\

\hline 
\end{tabular}
\end{center}
\end{table}

LZ has developed assay capability to determine the \IPbtoz content of the bulk materials used to fabricate the PTFE. The Lumpsey detector at Boulby, described in Section~\ref{MASSss:BUGS}, has sensitivity meeting the requirement of \SI{10}{\mBqkg} to \IPbtoz and will be used to measure the PTFE. 

The collaboration will continue assaying all materials used in the detector as they are designed and procured. We anticipate completing assays for all these items with LZ-specified materials and assembling a detailed background model prior to the start of integration and assembly. The impact of these backgrounds is presented in Table~\ref{MASt:BackgroundsTable}.

\tdrsubsec[BackTab]{Backgrounds Table}

Table~\ref{MASt:BackgroundsTable} presents the impact of the background sources assembled in Table~\ref{MASt:MaterialsTable} as determined through Monte Carlo simulations with LZSim and by adopting the radio-activities as assayed by LZ or as expected based on literature values. The activities for the material components easily satisfies the LZ requirements and are also well below the goals, with a total contribution to background of under \SI{0.1}{\NR} events and \SI{0.3}{\mudru} in \SI{1000}{\day} with a fiducial volume of \SI{5.6}{\tonnel} of Xe. This is for a WIMP search region defined by a 3-fold PMT coincidence lower bound and 20 phd upper bound,  
equivalent to approximately \SIrange{1.5}{6.5}{\keVee}, before discrimination or NR efficiency is applied, but employing the spontaneous fission rejection power of LZ, OD and Xe skin vetoes, and single scatter requirements. This energy region, selected for evaluation of trial components, is functionally equivalent to that employed in the full PLR when considering the experiment's optimum sensitivity. 

Most of the entries in Table~\ref{MASt:BackgroundsTable} are formed from composite materials, where more than 130 components or subcomponents contribute to make up the different elements. A comprehensive table that includes separate sections for each detector element detailing each subcomponent is maintained in a version-controlled archive; Table~\ref{MASt:BackgroundsTable} is the summary from the complete table. The assayed contamination values are used to generate mass-weighted average activities for Monte Carlo simulations for ER backgrounds, and in major components all the individual materials are used to generate NR rates from (\Pga,\Pn) and spontaneous fission neutrons. Neutron emission rates in the simulations include contributions from \IPbtoz at measured values for both the TPC PMT bases and skin PMT bases. For several components \SI{90}{\percent} C.L. upper limits have been set for ER rates where simulations are statistics limited given the lack of any recoils in the xenon that pass selection criteria. Where background is generated by some isotopes, such as  \IUtTe or \IUtTF, but not others, such as \IKfz, mean values and upper limits are combined.
This occurs in cases where either components are located far from the xenon target, are situated where radiation released is vetoed with very high efficiency, or where contamination levels are so low that no background is generated on timescales several times the LZ exposure. The components in Table~\ref{MASt:BackgroundsTable} where this is the case, and reported ER counts include a significant rate from an upper limit, are the R8520 skin PMT bases and the HV Conduit and Cables. 

\afterpage{
\begin{center}
\sffamily
\renewcommand{\ltbcp}{BackgroundsTable}
\renewcommand*{\arraystretch}{1.}
\begin{longtable}{| 
>{\centering\arraybackslash}m{3.55cm} | 
>{\centering\arraybackslash}m{0.8cm} |
>{\centering\arraybackslash}m{0.8cm} |
>{\centering\arraybackslash}m{0.8cm} | 
>{\centering\arraybackslash}m{1.0cm} | 
>{\centering\arraybackslash}m{1.0cm} | 
>{\centering\arraybackslash}m{0.7cm} | 
>{\centering\arraybackslash}m{0.6cm} |
>{\centering\arraybackslash}m{0.9cm} | 
>{\centering\arraybackslash}m{0.8cm} | 
>{\centering\arraybackslash}m{0.8cm} |
}

\caption[{Table of Backgrounds in LZ} \ifKL\space\texttt{\FloatPre t:\ltbcp}\fi]{Estimated background counts in the WIMP search region of interest, as discussed in the text, from all significant sources in the LZ \SI{1000}{\day} exposure. Mass-weighted average activities obtained from Table~\ref{MASt:MaterialsTable} are shown for composite materials. Solar \IBe neutrinos are expected to contribute \SI{7\pm3}{\NR} but only at very low energies and are excluded from the table. \ifKL\space\texttt{\FloatPre t:\ltbcp}\fi\label{\FloatPre t:\ltbcp}}\\
\hline
\rowcolor{mrocol}
\rule[-0.7em]{0pt}{2.0em}\textbf{} & 
\rule[-0.7em]{0pt}{2.0em}\textbf{Mass} & 
\rule[-0.7em]{0pt}{2.0em}\textbf{\IUtTeE} & 
\rule[-0.7em]{0pt}{2.0em}\textbf{\IUtTeL} & 
\rule[-0.7em]{0pt}{2.0em}\textbf{\IThtTtE} &
\rule[-0.7em]{0pt}{2.0em}\textbf{\IThtTtL} &
\rule[-0.7em]{0pt}{2.0em}\textbf{\ICosz} & 
\rule[-0.7em]{0pt}{2.0em}\textbf{\IKfz} & 
\rule[-0.7em]{0pt}{2.0em}\textbf{n/yr} & 
\rule[-0.7em]{0pt}{2.0em}\textbf{ER} & 
\rule[-0.7em]{0pt}{2.0em}\textbf{NR} \\
\cline{3-8}
\rowcolor{mrocol} & 
\textbf{(kg)} & 
\multicolumn{6}{|c|}{\textbf{\si{\mBqkg}}} 
& 
& 
\textbf{(cts)} & 
\textbf{(cts)}
\\ 
\endfirsthead
\hline
\rowcolor{mrocol}
\rule[-0.7em]{0pt}{2.0em}\textbf{} & 
\rule[-0.7em]{0pt}{2.0em}\textbf{Mass} & 
\rule[-0.7em]{0pt}{2.0em}\textbf{\IUtTeE} & 
\rule[-0.7em]{0pt}{2.0em}\textbf{\IUtTeL} & 
\rule[-0.7em]{0pt}{2.0em}\textbf{\IThtTtE} &
\rule[-0.7em]{0pt}{2.0em}\textbf{\IThtTtL} &
\rule[-0.7em]{0pt}{2.0em}\textbf{\ICosz} & 
\rule[-0.7em]{0pt}{2.0em}\textbf{\IKfz} & 
\rule[-0.7em]{0pt}{2.0em}\textbf{n/yr} & 
\rule[-0.7em]{0pt}{2.0em}\textbf{ER} & 
\rule[-0.7em]{0pt}{2.0em}\textbf{NR} \\ 
\cline{3-8}
\rowcolor{mrocol} & 
\textbf{(kg)} & 
\multicolumn{6}{|c|}{\textbf{\si{\mBqkg}}}
& 
& 
\textbf{(cts)} & 
\textbf{(cts)}
\\
\hline
\endhead
\hline
\rowcolor{lrocol}
\multicolumn{11}{|c|}{(continued on next page)}\\
\hline
\endfoot
\endlastfoot
\hline 
Upper PMT Structure     & 40.5 & 3.90 & 0.23 & 0.49 & 0.38 & 0.00 & 1.46 & 2.53 & 0.05 & 0.000 \\
Lower PMT Structure     & 69.9 & 2.40 & 0.13 & 0.30 & 0.24 & 0.00 & 0.91 & 6.06 & 0.05 & 0.001 \\
R11410 3" PMTs          & 91.9 & 71.6 & 3.20 & 3.12	& 2.99 & 2.82 & 15.4 & 81.8 & 1.46 & 0.013 \\
R11410 PMT Bases        & 2.8  & 288  & 75.8 & 28.4 & 27.9 & 1.43 & 69.4 & 34.7 & 0.36 & 0.004 \\
R8778 2" Skin PMTs		& 6.1  & 138  & 59.4 & 16.9 & 16.9 & 16.3 & 413  & 52.8 & 0.13 & 0.008 \\
R8520 Skin 1" PMTs      & 2.2  & 60.5 & 5.19 & 4.75 & 4.75 & 24.2 & 333  & 4.60 & 0.02 & 0.001 \\
R8520 PMT Bases    		& 0.2  & 213  & 108  & 42.2 & 37.6 & 2.23 & 124  & 3.62 & 0.00 & 0.000 \\
PMT Cabling             & 104  & 29.8 & 1.47 & 3.31 & 3.15 & 0.65 & 33.1 & 2.65 & 1.43 & 0.000 \\
\hline
TPC PTFE                & 184  & 0.02 & 0.02 & 0.03 & 0.03 & 0.00 & 0.12 & 22.5 & 0.06 & 0.008 \\
Grid Wires              & 0.8  & 1.20 & 0.27 & 0.33 & 0.49 & 1.60 & 0.40 & 0.02 & 0.00 & 0.000 \\
Grid Holders            & 62.2 & 1.20 & 0.27 & 0.33 & 0.49 & 1.60 & 0.40 & 6.33 & 0.27 & 0.002 \\
Field Shaping Rings     & 91.6 & 5.41 & 0.09 & 0.28 & 0.23 & 0.00 & 0.54 & 10.8 & 0.23 & 0.004 \\
TPC Sensors             & 0.90 & 21.1 & 13.5 & 22.9 & 14.2 & 0.50 & 26.3 & 24.8 & 0.01 & 0.002 \\
TPC Thermometers        & 0.06 & 336  & 90.5 & 38.5 & 25.0 & 7.26 & 3360 & 1.49 & 0.05 & 0.000 \\
Xe Tubing 				& 15.1 & 0.79 & 0.18 & 0.23 & 0.33 & 1.05 & 0.30 & 0.64 & 0.00 & 0.000 \\
HV Components  			& 138  & 1.90 & 2.00 & 0.50 & 0.60 & 1.40 & 1.20 & 4.90 & 0.04 & 0.001 \\
Conduits     			& 200  & 1.25 & 0.40 & 2.59 & 0.66 & 1.24 & 1.47 & 5.33 & 0.06 & 0.001 \\
\hline
Cryostat Vessel         & 2410 & 1.59 & 0.11 & 0.29 & 0.25 & 0.07 & 0.56 & 124  & 0.63 & 0.013 \\
Cryostat Seals          & 33.7 & 73.9 & 26.2 & 3.22 & 4.24 & 10.0 & 69.1 & 38.8 & 0.45 & 0.002 \\
Cryostat Insulation     & 23.8 & 18.9 & 18.9 & 3.45 & 3.45 & 1.97 & 51.7 & 69.8 & 0.43 & 0.007 \\
Cryostat Teflon Liner   & 26   & 0.02 & 0.02 & 0.03 & 0.03 & 0.00 & 0.12 & 3.18 & 0.00 & 0.000 \\
\hline
Outer Detector Tanks    & 3200 & 0.16 & 0.39 & 0.02 & 0.06 & 0.04 & 5.36 & 78.0 & 0.45 & 0.001 \\
Liquid Scintillator     & 17600& 0.01 & 0.01 & 0.01 & 0.01 & 0.00 & 0.00 & 14.3 & 0.03 & 0.000 \\
Outer Detector PMTs     & 205  & 570  & 470  & 395  & 388  & 0.00 & 534  & 7590 & 0.01 & 0.000 \\
OD PMT Supports         & 770  & 1.20 & 0.27 & 0.33 & 0.49 & 1.60 & 0.40 & 14.3 & 0.00 & 0.000 \\
\hline
\multicolumn{9}{|l|}{\textbf{Subtotal (Detector Components)}} & 6.20 & 0.070 \\
\hline
\multicolumn{9}{|l|}{\IRnttt (\SI{2.0}{\muBqkg})}   	& 722  & -    \\
\multicolumn{9}{|l|}{\IRnttz (\SI{0.1}{\muBqkg})}    	& 122  & -    \\
\multicolumn{9}{|l|}{\IKrnat (\SI{0.015}{\pptgpg})}  	& 24.5 & -    \\
\multicolumn{9}{|l|}{\IArnat (\SI{0.45}{\pptgpg})}   	& 2.47 & -    \\
\multicolumn{9}{|l|}{\IBitoz (\SI{0.1}{\muBqkg})}   	& 40 & -    \\
\hline
\multicolumn{9}{|l|}{Laboratory and Cosmogenics}  		& 4.3   & 0.06 \\
\multicolumn{9}{|l|}{Fixed Surface Contamination}       & 0.19   & 0.37 \\
\hline
\multicolumn{9}{|l|}{\textbf{Subtotal (Non-\Pgn counts)}} & 922 & 0.50 \\
\hline
\multicolumn{9}{|l|}{\IXeoTs \Dtnbb}                               & 67    & 0.00 \\
\multicolumn{9}{|l|}{Astrophysical \Pgn counts (\Ipp+\IBeS+\INoT)} & 255   & 0.00 \\
\multicolumn{9}{|l|}{Astrophysical \Pgn counts (\IBe)}             & 0.00  & 0.00 \\
\multicolumn{9}{|l|}{Astrophysical \Pgn counts (\Ihep)}            & 0.00  & 0.21 \\
\multicolumn{9}{|l|}{Astrophysical \Pgn counts (diffuse supernova)}& 0.00  & 0.05 \\
\multicolumn{9}{|l|}{Astrophysical \Pgn counts (atmospheric)}      & 0.00  & 0.46 \\
\hline
\multicolumn{9}{|l|}{\textbf{Subtotal (Physics backgrounds)}}      & 322   & 0.72 \\
\hline
\multicolumn{9}{|l|}{Total}                                        & 1,244 & 1.22 \\
\multicolumn{9}{|l|}{Total (with \SI{99.5}{\percent} ER discrimination, \SI{50}{\percent} NR efficiency)} & 6.22 & 0.61 \\
\hline
\multicolumn{9}{|l|}{\textbf{Sum of ER and NR in LZ for \SI{1000}{\day}, \SI{5.6}{\tonnel} FV, with all analysis cuts}} & \multicolumn{2}{|c|}{\textbf{6.83}} \\
\hline
\end{longtable}
\end{center}
}

There are several contributors to background in addition
to the ER and NR counts induced by fixed radioactivity
in materials. Radioactivity associated with radon emanated
into the instrument is expected to contribute approximately
\num{840} counts to the ER events, where half of this comes
from materials and the rest from radon released from dust on
material surfaces. This level of background, discussed in
Section~\ref{MASS:RnEm}, is based on conservative estimates
for emanation from materials and from dust. The activity,
\SI{2}{\muBq\per\kg}, corresponds to a total of approximately
\SI{20.2}{\mBq} from \IRnttt dispersed throughout the xenon
target, which is at the level of our requirements (\SI{20}{\mBq}).
A total of \SI{20}{\mBq} in the xenon corresponds to
\SI{13.4}{\mBq} from \IRnttt in the active volume and
\SI{11.2}{\mBq} in the fiducial target. We do not expect
a contribution from \IRnttz in LZ. However, due to the
observation of a population of alphas attributed to \IRnttz
in LUX, where radon mitigation was minimal, we nonetheless
conservatively include a rate from a \SI{0.1}{\muBq\per\kg}
concentration of \IRnttz in our background estimates.

We anticipate the decay-daughters from \IRnttt to accumulate
on the surfaces of the detector during manufacturing and assembly.
LZ has taken steps to minimize this plate-out deploying clean
rooms and Rn-reduced air for some of the critical assembly tasks.
Several large volume liquid scintillator
experiments~\cite{KamLAND-Zen:2016pfg, Alimonti:2008gc}
reported observing mobility of these products, in
particular \IBitoz.  The untagged beta decay would present
an ER background similar to the \IRnttt. Studies were conducted
to look for the mobility of \IBitoz in LUX data.  Severe limits
obtained from LUX, when applied to the LZ plate out requirement,
place an upper limit on the \IBitoz of \SI{0.1}{\muBq\per\kg},
resulting in a limit of \SI{40} ER counts.  We have explicitly
added an entry in the table for this background source. 

As discussed in Section~\ref{MASS:goalsReqs}, we maintain
a goal for total activity from radon emanation in the xenon
of \SI{1}{\mBq}. This would lower the contribution from
radon emanation to \SI{1}{\mudru}, matching goals for
krypton and from fixed material radioactivity. Contributions
from Kr and Ar, with 27 counts, are discussed in
Section~\ref{MASS:LXeCont}. Laboratory and cosmogenic
backgrounds  contribute just \SI{4.3}{\ER} and \SI{0.06}{\NR}
background events, well below goals for fixed material
radioactivity; these backgrounds are discussed in
Section~\ref{MASS:labCosmo}. Finally, surface contamination
on materials is described in Section~\ref{MASS:Cleanliness}.
This generates both ER and NR counts at the level of \num{0.19}
and \num{0.37}, respectively, from intrinsic radioactivity
in the dust and backgrounds due to radon plate-out.  

Contributions from the irreducible physics sources
are discussed in Section~\ref{chap:SIP}. These
include ER contributions from the double beta decay
of \IXeoTs and astrophysical neutrinos, in
particular \Ipp and \IBeS solar neutrinos. Solar \IBe
and \Ihep neutrinos, together with atmospheric and
diffuse supernovae neutrinos contribute to NR
background through coherent neutrino-nucleus scattering.
As described in Section~\ref{chap:SIP}, \SI{7\pm3}{\NR}
are expected from Solar \IBe neutrinos. However, these
are at very low energies and are therefore excluded
from Table~\ref{MASt:BackgroundsTable}. 

The total expected background to the WIMP search in LZ under
the assumption of \SI{99.5}{\percent} discrimination against ERs and
\SI{50}{\percent} NR acceptance is \num{6.8} counts. 

\tdrsec[TechSens]{Techniques and Sensitivities}

The LZ project will require approximately 500 materials assays, determined by counting materials, parts and components detailed in the project's CAD drawings. This is consistent with expectations based on our experience from LUX, EXO, and other similar low-background experiments when including multiple techniques and confirmation measurements. As already discussed we have  achieved good success for major items through completed assays. ICP-MS, HPGe, NAA and Radon Emanation systems will all be utilized, as no single technique has sensitivity to all radioactive isotopes within all materials, nor can any single technique at present provide sensitivity to the full \IUtTe and \IThtTt decay chains. Use of all four techniques will provide the required accurate model of a material's full \Pgg-ray, neutron and beta emission, and the subsequent impact on the radiation budget and sensitivity of the experiment. Any rare-event search would benefit from the availability of all techniques, which vary in their sample throughput and screening duration, requirements for sample size, access to instruments, ability to do bulk screening of complete components, and preservation or destruction of samples in the assaying process. Table~\ref{MASt:techniques} provides a summary of these assay techniques. 

\begin{table}
\tdrtcaption[techniques]{Primary material radio-assay techniques}{Primary material radio-assay techniques, indicating isotopic sensitivity and detection limits, as well as typical throughput or single-sample measurement duration.}
\begin{center}
\sffamily
\begin{tabular}{| >{\centering\arraybackslash}m{1.9cm} | >{\centering\arraybackslash}m{2.1cm} | >{\centering\arraybackslash}m{2.0cm} | >{\centering\arraybackslash}m{1.3cm} | >{\centering\arraybackslash}m{1.7cm} | m{4.8cm} | }
\hline
\rowcolor{mrocol}
\textbf{Technique} & \textbf{Isotopic Sensitivity} & \textbf{Typical Sensitivity Limits} & \textbf{Sample Mass} & \textbf{Sampling Duration} & \textbf{Destructive/Non-destructive and Notes}\\
\hline
\rule[-0.7em]{0pt}{2.0em}\textbf{HPGe} & 
\rule[-0.7em]{0pt}{2.0em} \IUtTe, \IUtTF, \IThtTt chains, \IKfz, \ICosz, \ICsoTS any \Pgg-ray emitter  & 
\rule[-0.7em]{0pt}{2.0em}\SI{50}{\ppt} U, \SI{100}{\ppt} Th &
\rule[-0.7em]{0pt}{2.0em}\si{\kg} & 
\rule[-0.7em]{0pt}{2.0em}Up to 2 weeks &
\rule[-0.7em]{0pt}{2.0em}Non-destructive, very versatile, not as sensitive as other techniques, large samples \\
\hline
\rule[-0.7em]{0pt}{2.0em}\textbf{ICP-MS} & 
\rule[-0.7em]{0pt}{2.0em} \IUtTe, \IUtTF, \IThtTt (top of chain) & 
\rule[-0.7em]{0pt}{2.0em}\SI{e-12}{\gpg}& \si{\mg} to \si{\g} & 
\rule[-0.7em]{0pt}{2.0em}Days &
\rule[-0.7em]{0pt}{2.0em}Destructive, requires sample digestion, preparation critical \\
\hline
\rule[-0.7em]{0pt}{2.0em}\textbf{NAA} & 
\rule[-0.7em]{0pt}{2.0em} \IUtTe, \IUtTF, \IThtTt (top of chain), K & 
\rule[-0.7em]{0pt}{2.0em}\SIrange{e-12}{e-14}{\gpg} & 
\rule[-0.7em]{0pt}{2.0em}\si{\g} &
\rule[-0.7em]{0pt}{2.0em}Days to weeks & 
\rule[-0.7em]{0pt}{2.0em}Destructive, sensitive to some contaminants\\
\hline
\rule[-0.7em]{0pt}{2.0em}\textbf{GD-MS} & 
\rule[-0.7em]{0pt}{2.0em} \IUtTe, \IUtTF, \IThtTt (top of chain) & \rule[-0.7em]{0pt}{2.0em} \SI{e-10}{\gpg} & 
\rule[-0.7em]{0pt}{2.0em}\si{\mg} to \si{\g} &
\rule[-0.7em]{0pt}{2.0em}Days & 
\rule[-0.7em]{0pt}{2.0em}Destructive, minimal matrix effects, cannot analyze ceramics and other insulators\\
\hline
\rule[-0.7em]{0pt}{2.0em}\textbf{Radon Emanation} & 
\rule[-0.7em]{0pt}{2.0em}\IRnttt, \IRnttz & 
\rule[-0.7em]{0pt}{2.0em}\SI{0.1}{\mBq} & 
\rule[-0.7em]{0pt}{2.0em}\si{\kg} & 
\rule[-0.7em]{0pt}{2.0em}1 to 3 weeks &
\rule[-0.7em]{0pt}{2.0em}Non-destructive, large samples, limited by size of emanation chamber \\
\hline
\end{tabular}
\end{center}
\end{table}

Sensitivity to U and Th decay chain species down to \SI{\approx10}{\ppt} has been demonstrated using ultralow-background HPGe detectors. HPGe can also assay \ICosz, \IKfz, and other radioactive species emitting \Pgg-rays. This technique is nondestructive and, in addition to sample screening, finished components can be assayed prior to installation. Under the assumption of secular equilibrium, with all isotopic decays remaining within the same volume, the activity and concentration for any particular isotope in the chain may be inferred from the measured U and Th content, assuming natural terrestrial abundance ratios. However, secular equilibrium can be broken through removal of radioactive  isotopes during chemical processing or through emanation and outgassing. HPGe readily identifies the concentrations of isotopes from mid- to late-chain isotopes from the early chain decays of \IUtTe and \IThtTt, particularly those with energies in excess of several hundred keV. Background-subtracted \Pgg-ray counting is performed around specific energy ranges to identify radioactive isotopes. Taking into account the detector efficiency at that energy for the specific sample geometry allows calculation of isotopic concentrations. A typical assay lasts \numrange{1}{2} weeks per sample to accrue statistics at the sensitivities required for the LZ assays. These direct \Pgg-ray assays probe the bulk of the radioactivity from any material, including identification of equilibrium states. The instruments available to the project are described in Section~\ref{MASSs:HPGe}, and will be used to assay all materials. The HPGe technique is less sensitive, however, to the progenitor isotopes or the low-energy or low-probability \Pgg-ray emission from decays in the early chain~\cite{Malling:2013jya}.

ICP-MS offers very precise determination of elemental contamination with potentially up to \num{100}$\times$ better sensitivity for the progenitor U and Th concentrations compared to \Pgg-ray spectroscopy. Since ICP-MS directly assays the \IUtTe, \IUtTF, and \IThtTt progenitor activity it informs the contribution to neutron flux from (\Pga,\Pn) in low-Z materials, but also the contribution from spontaneous fission, which in specific materials can dominate. However, it is limited in identifying daughter isotopes in the U and Th decay chains that are better probed by HPGe. The ICP-MS technique assays very small samples that are atomized and measured with a mass spectrometer. As a destructive technique, it is not used on finished components. The limitation of ICP-MS is that the sample must be soluble - typically in mixtures of \CHF, \CHCl, and \CHNOT - and that several samples from materials must be screened to probe contamination distribution and homogeneity. Assays take \SIrange{1}{2}{\day} per material, dominated by the sample preparation time, where extreme care must be taken to avoid contamination of solvents and reactants. ICP-MS will be used to provide confirmation of the direct \Pgg-ray counting assays, and inform the background model with potentially significant contributions from the tops of the U/Th chain inaccessible to \Pgg-ray measurements; particularly important for constraining systematic uncertainties in the PLR for signal identification. It will also provide chemical compositions to improve calculations of neutron emission yields from materials. Finally, ICP-MS available to the LZ project mitigates potential throughput limitations in long duration underground assays, with rapid direct U/Th measurements that can be used to establish background contributions under assumptions of secular equilibrium through the decay chains. This may be particularly useful for small/low-mass components that need rapid assessment, since they can be performed and material cleared for use before the \Pgg-ray assays that will determine the bulk of the radio-content with sensitivity to equilibrium states are performed. The LZ ICP-MS systems are presented in Section~\ref{MASSs:ICP-MS}.

Neutron activation analysis (NAA) achieves sensitivities up to \num{1000}$\times$ better than direct counting. Samples are irradiated with neutrons from the reactor to activate some of the stable isotopes, which subsequently emit \Pgg-rays of well-known energy that are detected through \Pgg-ray spectroscopy. Elemental concentrations are then inferred, using tabulated neutron-capture cross sections convoluted with the reactor neutron spectra. Samples must be specially prepared and compatible with neutron irradiation in a reactor, and then measured with surface \Pgg-ray counters. NAA probes the bulk contamination simultaneously and is not limited by the composition of the material since no sample digestions or ablations are required. Indeed, of all known techniques, NAA can provide the best sensitivity to U and Th concentration. However, concurrent activation of trace contaminants of little interest or from the primary constituents of the sample can produce high \Pgg-ray fluxes that present a background to the U and Th measurements, severely compromising sensitivity.  Counting can be timed to allow interfering species to decay, allowing identification of the isotopes of interest with high accuracy. As with ICP-MS, this technique requires small sample masses, does not assay finished components, and assumptions of secular equilibrium need to be made since this technique measures the top of the U and Th chains. NAA is particularly useful for materials with very low activities difficult to assay directly with \Pgg-ray counting, and also difficult to digest for ICP-MS. The PTFE in the LZ detector represents one such material. The proximity of the PTFE to the active Xe, both in defining the TPC and around the PMTs, coupled to a high (\Pga,\Pn) cross section, sets stringent constraints on the acceptable U and Th content. As described in the previous subsection, NAA has been successfully utilized to identify suitable material for the LZ project, with details of the facility presented in Section~\ref{MASSs:NAA}.     

Particular attention must be paid to radon, as it is a noble gas consisting solely of radioactive isotopes, is produced in the decay chains of uranium and thorium, and has the ability to enter Xe volumes due to its chemical inertness and subsequent long diffusion lengths through solids. Outgassing of radon from a material in which it has been produced is commonly termed ``radon emanation''. Especially for materials in contact with or in close proximity to Xe, radon emanation must be taken into account in setting the levels of U/Th that can be tolerated (with stringent limits particularly on U) due to the presence of \IRnttt in the \IUtTe decay chain, as well as \IRnttz from \IThtTt decay. LXe cannot provide self-shielding against the dispersed Rn, unlike radioactivity from fixed contaminants.  Radon emanation from the bulk of a material may be estimated once its U/Th decay chain content (in particular its \IRatts content) has been assayed using HPGe detectors. However, such estimates must be supplemented by direct screening for radon emanation for critical materials due to limited sensitivity in HPGe, systematic error from assumptions on equilibrium chain states, uncertainties in describing temperature-dependent radon transport in materials, and uncertainties on the amount of contaminants near surfaces. Direct measurements are performed by allowing bulk materials, either large samples or finished components, to emanate radon within a sealed chamber for a period of one or more weeks. The radon is then transferred using a carrier gas such as He or \CNt with high efficiency to a detector system, where radon progeny is detected through its radioactive decay, either using alpha counting or \Pgg-rays. The four radon emanation systems available to LZ all meet requirements, with sensitivity ranging from \SIrange{0.1}{0.3}{\mBq}. These systems are presented in Section~\ref{MASS:RnEm}. 

Most of the facilities for the direct assaying measurements are operated directly by LZ groups or exist at LZ institutes, allowing us to maintain control of sample preparation, measurements, analysis, and interpretation of data necessary to ensure sufficient sensitivity with reliable reproducibility and control of systematics. Commercial facilities that provide Glow Discharge Mass Spectrometry (GD-MS) are available to the collaboration. GD-MS has poorer sensitivity (\SI{\approx0.1}{\ppb} U/Th) than ICP-MS, can only be used with conductive or semiconductive solids, and commercial service providers are typically limited in sensitivity due to regular exposure of their instruments and sample preparation infrastructure to materials with high concentrations of contaminants. However, it may be exploited for additional throughput or rapid confirmation of measurements if necessary.

\tdrsec[Intrinsic]{Intrinsic Contamination Techniques and Devices}

Identification of key materials presented in Section~\ref{MASSs:matsTab} with ultra-low activity assays for intrinsic contamination demonstrate existing instrument sensitivity and capabilities within the project. Here we present the instruments and facilities used to perform the fixed contaminant assays used already and available to the project to meet the both throughput and sensitivity requirements to schedule. The materials assay schedule includes all instruments described below, and is managed through the project to allow flexibility by assigning resources depending on sample type, size, required sensitivity, timescale, and risk. The project has performed cross-calibration campaigns of our instruments that have included circulating standard sources, blind samples, and exchanging data and analysis routines, to ensure reliability across our assays.

\tdrsubsec[HPGe]{HPGe}

Eleven HPGe detectors located in facilities both above- and underground are available to the LZ collaboration, with differences in detector types and shielding configuration providing useful dynamic range both in terms of sensitivity to particular isotopes and physical sample geometries. All of the instruments, with several previously used for LUX or ZEPLIN, are managed and operated by LZ collaborating institutes and are already in use for the LZ material screening campaign. The detectors are typically several hundreds of grams to several kilograms in mass, with a mixture of n-type, p-type, and broad energy Ge (BEGe) crystals, providing relative efficiencies at the tens of percent through to in excess of \SI{100}{\percent} (as compared to the detection efficiency of a ($3 \times 3$)-inch \CNaI crystal for \SI{1.33}{\MeV} \Pgg-rays from a \ICosz source placed \SI{25}{\cm} from the detector face). While p-type crystals can be grown to larger sizes and hence require less counting time due to their high efficiency, the low energy performance of the n-type and broad energy crystals is superior due to less intervening material between source and active Ge. Clean samples are placed close to the Ge crystal and sealed for several days to weeks in order to accrue sufficient statistics, depending on the minimum detectable activity (MDA). The detectors are generally shielded with low-activity Pb and Cu, flushed with dry nitrogen to displace the Rn-carrying air, and sometimes are surrounded by veto detectors to suppress background from Compton scattering that dominates the MDA for low-energy \Pgg-rays. To reduce backgrounds further, most of the detectors are operated in underground sites, at SURF within the Black Hills State University Underground Campus (BHUC), and at the U.K. Boulby Underground Laboratory within the Boulby Germanium Suite (BUGS), lowering the muon flux by several orders of magnitude. We also retain a number of surface counters that are particularly useful for pre-screenings before more sensitive underground assays. All of the HPGe detectors available to LZ are shown in Table~\ref{MASt:Gamma Counters}.

\begin{table}
\tdrtcaption[Gamma Counters]{The LZ \Pgg-ray spectroscopy detectors}{\Pgg-ray-counting facilities available for LZ material radio-assays. Sensitivities shown are approximate detectable activities after 2 weeks of counting and samples of order-kg mass. Typical cavity size within the shielding of these detectors within which samples may be placed is \SI{0.03}{\m\cubed}.}
\begin{center}
\sffamily
\begin{tabular}{|c|c|c|c|c|c|c|c|}
\hline
\rowcolor{mrocol}
\textbf{Detector} & \textbf{Site} & \begin{tabular}[c]{@{}c@{}}\textbf{Site Depth} \\ \textbf{(mwe)}\end{tabular} & \begin{tabular}[c]{@{}c@{}}\textbf{Crystal} \\ \textbf{Type}\end{tabular} & \begin{tabular}[c]{@{}c@{}}\textbf{Crystal Mass} \\ \textbf{(Relative Eff)}\end{tabular} & \begin{tabular}[c]{@{}c@{}}\textbf{Sensitivity,} \\ \textbf{U (mBq/kg)}\end{tabular} & \begin{tabular}[c]{@{}c@{}}\textbf{Sensitivity,} \\ \textbf{Th (mBq/kg)} \end{tabular} \\ \hline
\textbf{Chaloner} & Boulby & 2805 & BEGe & \SI{0.8}{\kg} (\SI{48}{\percent}) & 0.6 & 0.2 \\
\textbf{Ge-II} & Alabama & 0 & p-type & \SI{1.4}{\kg} (\SI{60}{\percent}) & 4.0 & 1.2 \\
\textbf{Ge-III} & Alabama & 0 & p-type & \SI{2.2}{\kg} (\SI{100}{\percent}) & 4.0 & 1.2 \\
\textbf{Lumpsey} & Boulby & 2805 & Well & \SI{1.5}{\kg} (\SI{67}{\percent}) & 0.4 & 0.3 \\
\textbf{Lunehead} & Boulby & 2805 & p-type & \SI{2.0}{\kg} (\SI{92}{\percent}) & 0.7 & 0.2 \\
\textbf{\textsc{Maeve}} & SURF & 4300 & p-type & \SI{1.7}{\kg} (\SI{85}{\percent}) & 0.1 & 0.1 \\
\textbf{\textsc{Merlin}} & LBNL & 180 & n-type & \SI{2.3}{\kg} (\SI{115}{\percent}) & 6.0 & 8.0 \\
\textbf{\textsc{Mordred}} & SURF & 4300 & n-type & \SI{1.2}{\kg} (\SI{60}{\percent}) & 0.7 & 0.7 \\
\textbf{\textsc{Morgan}} & SURF & 4300 & p-type & \SI{2.1}{\kg} (\SI{85}{\percent}) & 0.2 & 0.2 \\
\textbf{SOLO} & SURF & 4300 & p-type & \SI{0.6}{\kg} (\SI{30}{\percent}) & 0.5 & 0.2 \\
\textbf{Wilton} & Boulby & 2805 & BEGe & \SI{0.4}{\kg} (\SI{18}{\percent}) & 7.0 & 4.0 \\
\hline
\end{tabular}
\end{center}
\end{table}

\tdrsubsubsec[BHUC]{BHUC}

The BHUC is a dedicated facility for low-background counting at SURF and was completed in late 2015. Four ultra-low background HPGe detectors are operational: \textsc{Maeve}, \textsc{Morgan}, \textsc{Mordred}, and SOLO. \textsc{Maeve}, formerly situated at Oroville before operations in the Davis campus at SURF and final re-location in the BHUC, is an \SI{85}{\percent} p-type HPGe detector in a low-activity Pb- and Cu-shielded and Rn-flushed chamber. The \textsc{Maeve} shield was recently upgraded with the addition of a layer of very old Pb shielding. For several \si{\kg} sized samples, the sensitivity after a week of counting reaches approximately \SI{10}{\ppt}  U 
(\SI{\approx0.1}{\mBqkg} \IUtTe) and \SI{25}{\ppt} Th (\SI{\approx0.1}{\mBqkg} \IThtTt), \SI{20}{\ppb} for K (\SI{\approx0.7}{\mBqkg} \IKfz)  and \SI{\approx0.03}{\mBqkg}  for \ICosz. Nearly an identical detector, \textsc{Morgan}, was paired with \textsc{Maeve} in 2015, and is now fully operational.

LBNL upgraded the CUBED detector, now called \textsc{Mordred} with a low background cryostat and improved design. \textsc{Mordred} has a \SI{1.2}{\kg} ORTEC n-type coaxial HPGe detector with a \SI{254}{\cm\cubed} active volume and a relative efficiency of \SI{60}{\percent}. The sample chamber, with a dimension of \SI{8000}{\cm\cubed}, is surrounded by a \SI{10}{\cm}-thick \SI{99.9}{\percent} OFHC copper shield, enclosed in a stainless steel box that is itself sealed by \SI{10}{\cm} of lead. \textsc{Mordred} achieves similar performance to \textsc{Maeve} and \textsc{Morgan}, and as an n-type detector will aid in measuring early U chain and \IPbtoz. 

The SOLO detector, formerly operated at Soudan Mine, has also been moved to BHUC. The SOLO shielding houses a nitrogen-flushed Pb shield that has a minimum thickness of \SI{30}{\cm} \SI{50}{\Bqkg} \IPbtoz activity) with a \SI{5}{\cm} inner liner of \SI{150}{\year}-old low-activity Pb (\SI{50}{\mBqkg} \IPbtoz activity). A counting chamber of \SI{8000}{\cm\cubed} contains the \SI{0.6}{\kg} ``Diode M'' HPGe detector. This detector approaches sensitivities at the \SI{50}{\ppt} level for \IUtTe and \IThtTt and \SI{25}{\ppb} for \IKfz for multi-\si{\kg} samples. 

The LBNL group manage and operate the BHUC counters for the LZ project. \textsc{Maeve} delivers the best sensitivity of all the project's detectors, and \textsc{Mordred} will extend the facilities low-energy sensitivity as an n-type detector. The SOLO and \textsc{Morgan} detectors are reserved with \SI{100}{\percent} live-time screening of LZ PMTs following delivery in 2016. The BHUC detectors, together with surface and Boulby counters, meet the LZ project's expected throughput demands with MDA's sufficient for most materials used in the construction of LZ. However, it is anticipated that additional counters will be added to the array in 2016 from the South Dakota School of Mines and Technology (SDSMT), UC Berkeley, and the University of South Dakota (USD).

\tdrsubsubsec[BUGS]{BUGS}

The Boulby Underground Facility, at \SI{2805}{mwe}, has been upgraded through 2015 to now include a dedicated low-background counting area that is operated as an ISO Class 6 cleanroom~\cite{Ghag:2015fla}. The Boulby Underground Germanium Suite (BUGS) is housed in this area and includes three primary high sensitivity ultra-low background detectors: Chaloner, Lunehead, and Lumpsey, and a fourth pre-screener: Wilton. The primary detectors are each housed in custom-built lead and copper shields using existing Boulby stock material. The shields feature a retractable roof to aid the reproducibility of backgrounds. The shield cavities range between \SIrange{30}{40}{\litre}, and are purged with \CNt gas fed directly into the inner cavity to reduce the ambient radon activity from \SI{3}{\BqmT} to negligible levels. The shields allow detectors to be retracted and interchanged, allowing detector maintenance without dismantling, and matching of instruments to castles to optimize sensitivity across the suite.

The Chaloner BEGe detector is a Canberra BE5030 (\SI{0.8}{\kg} Ge, \SI{48}{\percent} relative efficiency) installed in 2014. The crystal is configured in a unique planar geometry, yielding greater peak-to-Compton ratios at the \Pgg-ray energies of interest, and improved energy resolution (by \SI{\approx30}{\percent} over typical p-type detectors at \SI{122}{\keV}). BEGe detectors also have considerably lower energy thresholds due to a factor-70 reduction in dead layer thickness on the front face of the Ge crystal and the utilization of a carbon-fiber end-cap window. These factors provide useful efficiency down to \SI{10}{\keV} (as opposed to \SI{\approx80}{\keV} for p-type HPGe detectors) and consequently can directly measure \IPbtoz — a problematic source of background that is particularly difficult to quantify with other techniques. The BE5030 achieves \SI{<50}{\ppt} sensitivity to \IUtTe and \IThtTt for typical samples despite the low relative efficiency given the geometry and high resolution of the crystal. The energy range of the BEGe also gives us (unique to such a screening program) information about the elemental content of some materials through x-ray fluorescence. This has proved particularly useful in verifying composition of sample of candidate capacitors, found to contain \CBaTiOT, as the detector is sensitive to fluorescence x-rays from barium, rather than \CAltOT.

The Lunehead detector is a GEM-XX240-S p-type HPGe of \SI{92}{\percent} relative efficiency. Used extensively for the ZEPLIN-III experiment~\cite{Akimov:2010hr}, this detector has undergone complete refurbishment. With the exception of the Ge crystal, the detector has been overhauled and retrofitted with ultra-low-background components in 2014 to become a GEMXX-95-LB-C-HJ model with J-type neck and carbon fiber entrance window. Lunehead achieves sensitivity to about \SI{50}{\ppt} of \IUtTe and \IThtTt. 

The Lumpsey detector is a Canberra SAGe Well-type ultra-low-background GSW275L7950-30U-ULB instrument. Lumpsey has a \SI{1.5}{\kg} crystal that operates as a conventional co-axial (p-type) detector for large samples, with approximately \SI{40}{\keV} threshold, and sensitivity equivalent to the Chaloner and Lunehead detectors. However, the crystal includes a \SI{28}{\mm} diameter, \SI{40}{\mm} cavity (`well') where small samples can be placed. This provides unique capability with high sensitivity rapid screening of small components or materials by enveloping the sample with $>3\pi$ coverage. The thin lithium contact inside the well allows low energy threshold at \SI{20}{\keV}, similar to the BEGe detectors. This is particularly important for measurement of the \SI{46.5}{\keV} \Pgg-ray from the decay of \IPbtoz. As described earlier, detection and mitigation of \IPbtoz in bulk PTFE is required to mitigate (\Pga,\Pn) production on fluorine by \IPotoz, a \IPbtoz daughter nucleus. Traditional HPGe \Pgg-ray counting, NAA, and ICP-MS cannot readily be applied to detect \IPbtoz or its daughters at the \si{\mBqkg} level. Lumpsey demonstrates sensitivity of \SI{35}{\mBqkg} for a small (\SI{30}{\g}) sample after \SI{21}{\day} of counting and \SI{10}{\mBqkg} for more massive samples after \SI{50}{\day}, meeting the project requirement for \IPbtoz in bulk material screening. The Lumpsey well-detector will be used for the assays of PTFE for \IPbtoz content. 

The Wilton detector is a Canberra BE2825 with low mass (\SI{0.4}{\kg}) BEGe crystal. Though Wilton is an ultralow-background instrument it has a straight neck from dewar to crystal, and is housed in a smaller flip-lid lead shield with tin and copper liner. Wilton is used as a pre-screener, where samples are first assessed in a \numrange{1}{2} day measurement. Wilton's sensitivity is at the level of \SI{7}{\mBqkg} to \IUtTe and \SI{4} {\mBqkg} to \IThtTt. This allows materials with high radioactivity to be immediately rejected if requirements are not met, without disturbing the primary high-sensitivity instruments. Where samples are to be passed to the primary detectors, the pre-screener can inform screening times required to meet sensitivity for the assay.

BUGS is managed by the LZ groups at Oxford and UCL, through the DMUK consortium, with dedicated effort from Boulby providing significant operational and infrastructure support. The instruments have already been used extensively through the project's R\&D phase and beyond to perform critical assays in particular to identify PMT base materials, cryostat materials, and support PMT construction material assays with SOLO and \textsc{Maeve}. The Chaloner and Lunehead detectors are reserved exclusively for screening of PMTs, begun in mid-2016. 

\tdrsubsubsec[SurfaceHPGe]{Surface HPGe Detectors}

The \textsc{Merlin} detector, as with Wilton at Boulby, is used to pre-screen materials and determine suitability for underground assaying and required sampling livetime. Materials identified as exceeding requirements by \textsc{Merlin} are not assayed with the high sensitivity counters in the BHUC. \textsc{Merlin} is a \SI{115}{\percent} relative efficiency n-type low-background HPGe operated at the surface Berkeley Low Background Facility (BLBF) within a $4\pi$ shielded room with \SI{1.5}{\m}-thick low-activity serpentine rock concrete walls. The HPGe detector head is mounted on a J-hook to reduce line-of-sight for background from electronics and the cryostat, and is shielded in a Pb and OFHC Cu castle. It has an MDA of approximately \SI{0.5}{\ppb} (\SI{6}{\mBqkg}) to \IUtTe for O(\si{\kg}) samples from \SI{1}{\day} of counting, and \SI{2}{\ppb} (\SI{8}{\mBqkg}) for \IThtTt. Sensitivity to K and \ICosz is at the level of \SI{1}{\ppm} and \SI{0.04}{\pCikg}, respectively. 

Finally, the University of Alabama operates two high sensitivity low-background HPGe detectors at its surface screening facility: the Ge-II and Ge-III detectors. Each detector is housed in a lead-copper-shield equipped with an active cosmic-ray veto systems. The shield cavities are suitable for large samples. These devices can reach \SI{0.3}{\ppb} MDA for \IUtTe and \IThtTt with two weeks of counting, useful for many LZ components, and also provide pre-screening. These detectors are essential for the NAA screening program for LZ, described in Section~\ref{MASSs:NAA}. 

\tdrsubsec[ICP-MS]{ICP-MS}

The UCL group operate an ICP-MS facility dedicated to the LZ project~\cite{Ghag:2015fla}.  At UCL, An ISO Class 6 cleanroom, erected in late 2015, contains all instruments and the sample preparation areas. The primary instrument is an Agilent 7900 ICP-MS, with sensitivity to U and Th below \SI{e-12}{\gpg}, allowing assay of materials with U and Th content at the level of a several ppt. The ICP-MS has an integrated auto-sampler for high speed discrete sample uptake with low-flow, Peltier-cooled sample introduction system. The system octopole can provide species discrimination through interference removal either with kinetic energy discrimination (KED) in helium collision mode, or in reaction mode using \CHt, in addition to running with no gas at all, which in some cases is sufficient for U and Th measurements. High-purity Ar, He, and \CHt gases (5N grade) are introduced to the system as carrier gas, collision cell gas, or reaction gas, respectively. The system has been fitted with an inert sample introduction kit such that a micro flow nebuliser and platinum skimmer and sampling cones are used to allow up to \SI{20}{\percent} concentration of acid in samples introduced to the ICP-MS, including \CHF. Calibration of the system is performed using tuning solutions containing Li, Y, Tl, Co, and Ce, in \SI{2}{\percent} \CHNOT. The mass numbers 7 (Li), 89 (Y) and 205 (Tl) are tracked for stability, sensitivity, resolution and linearity for the integrated system across the mass range. The backgrounds at these masses allow sub-ppt sensitivity, \SI{<3}{\percent} relative standard deviation (RSD) in count rate over a \SI{20}{\minute} stability measurement period, and detector resolutions of \SI{<1}{\percent}.

The limiting factors in realizing reproducible high throughput ppt sensitivity is clean sample preparation, requiring digestion apparatus and procedures, as well as cleanliness of acids, that do not contaminate samples. Sample preparation infrastructure has been procured to address these issues and mitigate risks. A Milestone EthosUP microwave digestion system is able to digest solid samples at high temperatures in closed containers in minutes; this cannot readily be achieved using open hot-plate technology since volatiles are lost, the open vials may be contaminated depending on the ambient environment and handling, and low temperatures limit the digestion efficiency, further increasing exposure to background. The digestion oven also allows for the use of high quantities of \CHF that greatly simplifies digestion and, crucially, allows measurement of virtually all materials that are being considered for use in LZ. The quality of the reagents used in the sample material and, more importantly, analytical blank digestions further limit sensitivity through variations in contamination levels, even for ultra-pure commercial products. Milestone acid distillation and reflux systems have been installed for control of acid purity and reproducibility, and the laboratory includes a Veloia 18~M$\Omega$ water purifier. Finally, a Pyro-260 microwave ashing system allows the possibility of digestion of materials such as PTFE and acrylics that are not otherwise digested easily, and are required with low radioactivity levels inaccessible to HPGe, necessitating the need for NAA. 

Protocols and methodologies for operations, sample handling, and digestions are largely based on established protocols and methodologies developed that demonstrate sub-ppt measurements~\cite{Leonard:2007uv,LaFerriere:2014rva,Grinberg:2005} The cleanroom contains an ISO Class 4 laminar flow unit, and a fume cupboard with H+ filtration for sample handling. Digestion protocols with specific acid chemistries and heating profiles with microwave digestion on a material-by-material basis are also available for the EthosUP system following our R\&D phase and microwave digestion routines developed in partnership with Milestone and U.K. operators Analytix Limited. Microwave energy couples directly to ions, rotating around the dipole to cause friction and release heat, such that acids with higher dipole moments absorb microwaves readily for fast and even heating of reactant solutions. The vessels used in these high pressure reactors, in contrast, are constructed from materials with low or no dipole moment, such as tetra-fluoromethoxy (TFM), making them transparent to microwaves. High pressures in these closed vessels allow acids to be heated beyond their boiling points, further aiding material sample digestion. Protocols for digestion of all materials within the R11410 PMT, for example, have been developed and successfully tested already. A mixture of \CHNOT, \CHCl and \CHF at 220 degrees Celsius (reaching 30 bar) over 45 minutes is sufficient for complete dissolution of all component materials. The nitric acid is commonly used to digest organic material, \CHCl for Fe-based alloys due to ability to hold chloro-complex in solution, and \CHF used for decomposing silicates.

The UCL ICP-MS facility became fully operational in December 2015, and material assays began in early 2016 following the establishment of QC and QA procedures for consistency checks, calibrations, optimizing of cleaning procedures, and assessment of systematics. Sample assays for the LZ project have begun with initial assays of plastics and foams for the cryostat sub-system, achieving results consistent with HPGe assays with the BUGS detectors. 

The Center for Underground Physics (CUP), S. Korea, operates an ICP-MS laboratory supporting a number low-background experiments, including LZ.  The facility and operations are largely the same as those described for UCL above, including ISO 6 cleanroom, Agilent 7900, and identical microwave digestion system now on order.  The instrument was made operational and staffed its 20~m$^{2}$ unidirectional-flow cleanroom (about 180 air changes per hour) in October of 2015.  The ICP-MS is equipped for He collision mode with potential to add a Hydrogen line at a later date if a strong need arises.  It is has a UHMI (gas-diltuion) introduction system which can help to optimize sample introduction load into the plasma and reduce water and reagent backgrounds. It also was purchased with cold-plasma capability allowing removal of particular interferences with high ionization potentials.  This is particularly useful for measurement of $^{39}$K which is severely limited by interference from $^{38}$ArH$^{+}$.  The lab contains an in-house acid distillation system and water is supplied from a 500L/hr 18~M$\Omega$ building-wide supply as input to a Millipore Advantage A10 water-purifier. 

Protocols for cleaning are generally very similar to those stated above, derived largely from the same sources of experience, but sample preparation has so far been restricted only to un-assisted open digestion.  Particular attention has been paid to quantifying and understanding the ultimate limits of measurement backgrounds, based largely on previous work performed at the University of Seoul.  Using Savillex brand digestion vessels, cleaning procedures consistent with descriptions above,and statistical subtraction of sample-dependent continuous backgrounds described by Ref.~\cite{Leonard:2014jha}, CUP achieves blank contaminants of U and Th statistically consistent with zero (in the blanks this level can be at or below a few fg/g but increases for high concentration samples), and with sample limits significantly below the background equivalent level. A figure of merit using the calibration response, the measured background rate, and simple propagation of statistical errors is used to accurately predict sensitivities in a particular sample based on results of initial in-matrix tuning. Three sigma detection capability below 2~ppt has been demonstrated, but not yet pursued for a sufficiently pure material.   Details depend on the sample chemistry, but similar levels are reasonable to expect for most easily acid-dissolvable samples such as metals. 

These measurements take significant time owing to the need for high statistics, standard addition calibration of high-concentration samples, and frequent machine cleaning. For more routine measurements, the backgrounds are simply monitored and reported as an equivalent concentration, and lower concentration samples allow for external calibrations and a significantly relaxed cleaning schedule, allowing to measure less demanding samples at the rate of several per day with sensitivities to U and Th on the order of a few hundred ppt.  Finally some initial work has been done to develop measurements of potassium using the cold plasma configuration, leading to measurements of a few hundred ppb of K, and to indications of potential sensitivity much lower, but not yet explored.

ICP-MS facilities are also available Black Hills University Campus, and at the University of Alabama. BHUC also operate an Agilent 7900, with additional laser ablation sample introduction capability with an ESI NWR 213 system. The system is housed in an ISO 7 clean area within a dedicated ICP-MS laboratory. The University of Alabama group have access to a Perkin-Elmer SCIEX-ELAN 6000 ICP-MS within the Geology department that has demonstrated capability to detect U/Th down to the level of tens of ppt and has availability to screen tens of samples on the timescale of a few months. This facility does not, however, provide services for sample digestion and separation/concentration of U/Th content. The Alabama group has carried out a systematic program to certify the capability of the facility, develop protocols to prepare samples for analysis without risk of cross-contamination, and set up a laboratory where common digestion and separation/concentration procedures can be performed.   

\tdrsubsec[NAA]{NAA}

The University of Alabama group within the LZ collaboration utilizes the 5.5~MW$_{th}$ MIT Reactor II (MITR-II) to perform neutron activation of samples and subsequent measurements with the Ge-II and Ge-III surface HPGe detectors described in Section~\ref{MASSss:SurfaceHPGe}. MITR-II is a double-tank reactor with an inner tank for light-water coolant moderator and an outer one serving as heavy-water reflector~\cite{MITReactor:2015}. Two pneumatic sample insertion facilities are available. Steady-state thermal neutron fluxes of up to \SI{5.5e13}{\nsct} can be achieved. The sample insertion facilities can accommodate multiple samples that range in size but are typically a few \si{\mm} in diameter and several \si{\cm} in length. The two sample insertion facilities offer differing thermal over fast neutron flux ratios. Sample irradiations ranging from minutes to days can be performed, allowing accumulation of very large neutron fluences, which is key for reaching high analysis sensitivity. 

LZ samples are prepared, cleaned, and hermetically sealed at the University of Alabama in a cleanroom prior to activation. Polyethylene irradiation vials and samples are separately soaked in ultrapure \CHNOT, rinsed and dried in a vacuum oven. Vials are welded shut with the clean samples within, and leak tested in a heated water bath. Irradiation at MITR-II is typically for about \SI{10}{\hour}, with a NIST certified fly ash sample used to calibrate the neutron flux also irradiated for about \SI{5}{\minute}. After activation, the samples can be recovered cleanly and safely to avoid any carry-over contamination problems from the activation vials or other instruments. The counting of the activated samples at Alabama utilizes a double differential time-energy analysis over a period of about 2~weeks. The typical shipping delay of \SI{24}{\hour} is acceptable compared with the half-lives of the activation products of interest (such as \IKft, \IPatTT, and \INptTn). An exponential decay is fit to each time series of activity, for each radionuclide over multiple time intervals. The activity is reported relative to an earlier reference time, by fitting the decay of each radionuclide with a known half-life for determining the composition of the sample. Elemental concentrations are inferred using tabulated energy dependent radiative neutron capture cross sections folded with a standard reactor neutron spectrum, where the model of the neutron flux assumes a thermal Maxwell-Boltzmann neutron energy distribution plus an epi-thermal tail. 

The Alabama group routinely achieved \SI{e-12}{\gpg} sensitivity for Th and U using these techniques, as reported in~\cite{Leonard:2007uv}, appropriate for the LZ material screening campaign. Indeed, NAA results obtained by the same group for the KamLAND experiment reached sensitivity to U and Th at the \SIrange{e-14}{e-15}{\gpg} level~\cite{Leonard:2007uv} for liquid scintillator. Requirements for assaying PTFE for the LZ project have already been met using NAA, as described in Section~\ref{MASSs:matsTab}.

\tdrsec[RnEm]{Radon Emanation}

The background from radon emanation is dominated by the ``naked'' beta emission from \IPbtof in the \IRnttt sub-chain as it decays to \IBitof, whereas the \IBitof beta decay itself is readily identified by the subsequent \IPotof alpha decay that would be observed within an LZ event timeline (\thlf=\SI{160}{\mus}). Similar coincidence rejection also occurs where beta decay is accompanied by a high-energy \Pgg-ray, which may still be tagged by the LXe skin or external Gd-LS vetoes even if it leaves the active Xe volume. Radon-220 generates \IPbtot, which decays with a short-timescale Bi-Po (beta-alpha delayed coincidence) scheme similar to \IPbtof. Radon daughters are readily identified through their alpha decay signatures, as demonstrated in LUX, and can be used to characterize the \IRnttt and \IRnttz decay chain rates and distributions in the active region, providing a useful complement to estimating radon concentration from the beta decay contribution to the ER background. Indeed, these isotopes were the only sources of alpha decay identified in LUX~\cite{Akerib:2014rda}. 
As detailed in Table~\ref{MASt:RnBudget}, there are multiple potential sources of radon emanation (e.g., PTFE reflectors, PTFE skin, PMT glass, PMT and HV cables, grid resistors, components in the circulation system), and radon emanation screening must be sensitive to sources that individually sustain smaller populations. We use \SI{0.67}{\mBq} in the active target within the TPC as the goal for \IRnttt, which equates to a steady-state population of approximately 300 atoms. This activity corresponds to \SI{0.56}{\mBq} in the fiducial LXe, and \SI{1}{\mBq} in the total amount of LXe.  This activity results in a background contribution that matches those from Kr and from intrinsic material radioactivity, and is about \SI{10}{\percent} of the irreducible \Ipp solar neutrino background. 
Requirements for LZ are twenty times higher than these goals, with \SI{20}{\mBq} \IRnttt total, of which \SI{13.4}{\mBq} is in the active LXe, and \SI{11.2}{\mBq} in the fiducial LXe.  This requirement ensures that the background due to radon does not dominate significantly over the irreducible \Ipp solar neutrino background, so that the WIMP-search reach of the experiment is not significantly reduced. Background from \IRnttz is not expected given its very short half-life. Due to observation of an alpha population ascribed to \IRnttz in LUX, we conservatively include a contribution in our estimates. \IRnttz is required to contribute no more than than \SI{20}{\percent} of the ER counts from \IRnttt.  

\begin{table}
\tdrtcaption[RnFacilities]{Radon emanation screening facilities available to LZ}{Radon-emanation facilities available for LZ material radio-assays. Sensitivities shown are approximate detectable activities after 2 weeks of emanation and counting. With the exception of the UCL system, all are managed by LZ groups.  The UMd and Alabama systems are dedicated to LZ, while the UCL system is primarily for SuperNEMO, and the SDSM\&T system will be shared (at \SI{<25}{\percent} time) with SuperCDMS.}
\label{MASt:RnEmanationStations}
\begin{center}
\sffamily
\begin{tabular}{| 
>{\centering\arraybackslash}m{1.8cm} |
>{\centering\arraybackslash}m{1.7cm} | 
>{\centering\arraybackslash}m{1.7cm} | 
>{\centering\arraybackslash}m{1.7cm} | 
>{\centering\arraybackslash}m{1.4cm} | 
>{\centering\arraybackslash}m{1.7cm} | 
>{\centering\arraybackslash}m{1.5cm} | 
>{\centering\arraybackslash}m{1.4cm} | 
}
\hline
\rowcolor{mrocol}
\textbf{Detector} & \textbf{Type} & \textbf{Detector} \textbf{Efficiency} & \textbf{Detector} \textbf{Background} \textbf{(mBq)} & \textbf{Samples/} \textbf{Year}& \textbf{Chamber  Volume (l)} & \textbf{Transfer} \textbf{Efficiency} &\textbf{Blank} \textbf{Rate} \textbf{(\si{\mBq})} \\ \hline
\multirow{2}{*}{\textbf{UCL}}      & \multirow{2}{*}{PIN-diode}  & \multirow{2}{*}{\SI{30}{\percent}}& \multirow{2}{*}{0.2} & \multirow{2}{*}{6} & 2.6  & \SI{97}{\percent}~~ & 0.2\\
& & & &  & 2.6 & \SI{97}{\percent} & $0.4$ \\
\hline
\textbf{Maryland} & PIN-diode & \SI{24}{\percent} & 0.2 & 12 & 4.7 &  \SI{96}{\percent}~~ &  0.2\\
\hline
\multirow{2}{*}{\textbf{SDSM\&T}}  & \multirow{2}{*}{PIN-diode} & \multirow{2}{*}{\SI{20}{\percent}} & \multirow{2}{*}{0.15}& 12 & 13 & \SI{94}{\percent} & <0.3\\
& & & & 12 & 300 & \SI{80}{\percent} & 0.3 \\
\hline
\multirow{2}{*}{\textbf{Alabama}} & Liquid & \multirow{2}{*}{\SI{40}{\percent}} & \multirow{2}{*}{<0.15}  & 12 & 2.6 & \multirow{2}{*}{\SI{30}{\percent}}~~ & \multirow{2}{*}{<0.4} \\
& Scint. & & & 12  & 2.6 &  &  \\
\hline
\end{tabular}
\end{center}
\end{table}

All components that may contribute to the radon load within the LXe will be screened for radon emanation.  These components include all that are within the inner cryostat or that come into direct contact with Xe during experimental operation. 
Some materials are screened to inform material selection.  These assays are scheduled carefully in conjunction with material procurements. In other cases (such as the PMTs), the materials cannot be changed due to finite resources and schedule, but the assay is performed in order to inform the background model. 

The LZ collaboration has extensive access to four radon-emanation screening stations that meet the sensitivity requirement for our assays. These are summarized in Table~\ref{MASt:RnEmanationStations}. 
In one of the stations, at Alabama, the radon atoms are collected by passing the radon-bearing gas through liquid scintillator, with the Bi-Po coincidence detected through gated coincidence logic using one PMT viewing the scintillator. 
In the other three stations, radon atoms and daughters are collected electrostatically onto silicon PIN diode detectors to detect \IPotoe and \IPotof alpha decays.  One of these stations was developed at Case Western Reserve University and has been commissioned at the University of Maryland.  Work at this system is focusing on emanation of systems that act as their own emanation chambers, such as plumbing for the LZ Xe recirculation system.  The third station, at SDSM\&T, has two emanation chambers including one large, \num{300}-liter chamber that will be used to emanate large amounts of materials to achieve the best sensitivity.
Finally, there is additional capability and throughput available in the U.K. using the system employed by the SuperNEMO group at UCL~\cite{Mott:2013}. This system was used extensively in the R\&D phase and LZ is expected to have access sufficient to measure emanation from six samples per year with it.

All stations were initially evaluated using calibrated sources of radon, with a cross-calibration program performed to ensure the accuracy of each system's overall efficiency and ability to estimate and subtract backgrounds.  The first cross-calibration sample had a relatively high emanation rate, so that system efficiencies could be determined without possible interference from backgrounds.  The second cross-calibration sample has a rate close to the quoted system sensitivities, to check the accuracy of background subtraction with the systems.

Screening a single sample for LZ takes about two weeks, including emanation and collection/detection times.  Taking into consideration the desire for repeated measurements to check reproducibility and improve sensitivity, as well as runs to ensure stability of background rates and transfer efficiencies, no more than one sample per month per emanation chamber is planned for the scheduling of radon emanation screening. 

\begin{table}
\tdrtcaption[RnBudget]{Materials in LZ to be assayed for radon emanation}{List of materials in contact with Xe, indicating the quantity of the material and the requirement for radon emanation from the material.  This requirement is set by dividing the maximum activity from the materials, 10 mBq in the full 10 tons of LXe, evenly among the 9 major systems. A further 10 mBq from dust makes up the full requirement of 20 mBq from radon. The estimates of radon emanation are based either on direct assays (listed in boldface) or on the most similar object or material for which emanation rates are available in the literature. Some materials (labeled with $^\ast$) are expected to emanate less radon when cold, but the estimate listed conservatively does not take such reduction into account.  Only the estimate for PMT cables, as described in the text, takes this into account.  We also list the quantity of material planned for screening.  
Expected reduction of radon by the carbon trap described in Section~\ref{XCSSs:RnRem} is included in estimates for those components affected (labeled with $^\dagger$).}
\begin{center}
\sffamily
\begin{tabular}{| 
>{\centering\arraybackslash}m{3.1cm} |
>{\centering\arraybackslash}m{3.0cm} | 
>{\centering\arraybackslash}m{1.6cm} | 
>{\centering\arraybackslash}m{0.9cm} | 
>{\centering\arraybackslash}m{1.5cm} | 
>{\centering\arraybackslash}m{1.5cm} | 
>{\centering\arraybackslash}m{1.8cm} | 
}
\hline
\rowcolor{mrocol}
\textbf{Material} & \textbf{Component(s)} & \textbf{Quantity} & \textbf{Unit} & \begin{tabular}[c]{@{}c@{}} \textbf{Require-} \\ \textbf{ment} \\ \textbf{(mBq)}\end{tabular}  & \begin{tabular}[c]{@{}c@{}} \textbf{Estimate} \\ \textbf{(mBq)}\end{tabular} & \begin{tabular}[c]{@{}c@{}}\textbf{Screening} \\ \textbf{Quantity}\end{tabular}\\ 
\hline
Al$_2$O$_3$ resistor & PMT Bases & 9790 & \# & 0.66 & \textbf{ 0.58$^\ast$} &  3,650\\ 
BaTiO$_3$ capacitor & PMT Bases & 3010 & \# & 0.66 & \textbf{0.016$^\ast$} &  100,000\\ 
Cirlex & PMT Bases & 6000 & \si{\cm\squared} & 0.11 & \textbf{0.37$^\ast$} &  668\\ 
\hline
Titanium & Cryostat, PMT Mounts, Field Rings, Grid Supports  & 412,000 & \si{\cm\squared} &  
1.70 & 0.41 &  550\\ 
\hline
PTFE &Reflectors, HV Umbilical & 840,000 & \si{\cm\squared} & 0.66 & \textbf{<1.3$^\ast$} &  205,000\\ 
\hline
PMT Cabling$^\dagger$ & PMT Cabling & 17,000 & \si{\m} & 0.55 & 0.09 &  3,000\\ 
PMT Feedthrough$^\dagger$ & PMT HV Flange & 122 & \# & 0.11 & \textbf{0.49} &  5\\ 
PMT Feedthrough$^\dagger$ & Signal Flange & 88 & \# & 0.11 & \textbf{<0.24} &  5\\ 
Steel Conduit$^\dagger$ & Cabling Conduit & 100,000 & \si{\cm\squared} & 0.22 & 0.055 &  100,000\\ 
R11410 PMT & R11410 PMT & 488 & \# & 1.10 & 1.26 &  488\\ 
R8520 PMT & R8520 PMT & 90 & \# & 0.47 & 0.15 &  90\\ 
R8778 PMT & R8778 PMT & 36 & \# & 0.08 & 0.09 & 36\\
Polyethylene & HV Umbilical & 4200 & \si{\cm\squared} & 0.11 & 0.10 &  42,000\\ 
Tin-coated copper & HV Umbilical & 11,000 & \si{\cm\squared} & 0.11 & 0.002 &  110,000\\ 
Tivar & HV Umbilical & 3894 & \si{\cm\squared} & 0.22 & 0.004$^\ast$ &  20,000\\ 
Acetal & HV Umbilical & 195 & \si{\cm\squared} & 0.11 & 0.0002$^\ast$ &  2000\\ 
Copper & HV Umbilical & 39 & \si{\cm\squared} & 0.11 & 0.000007 &  400\\ 
Epoxy & HV Umbilical & 1000 & \si{\cm\squared} & 0.11 & 0.0001$^\ast$ &  10,000\\ 
\hline
Steel & Cryostat Seals, Xe Recirculation  & 135,000 & \si{\cm\squared} & 0.77 & 0.104 &  135,000\\ 
\hline
Recirculation Pump & Xe Recirculation & 1 & \# & 0.22 & 0.1 &  1\\ 
\hline
Purification Getter & Xe Recirculation & 2.5 & \si{\kg} & 1.10 & 1.34 &  2.5\\ 
\hline
Transducers \& Valves & Xe Recirculation & 30 & \# & 0.44 & 0.17 &  30\\ 
\hline
Welds & Recirculation System, Cryostat & 32.3 & m & 0.22 & \textbf{0.11} &  18.3\\ 
\hline
Dust &  &  &  & 10.0 & 10.0 & \\ 
\hline
\hline
Total &  &  &  & 20.0 & <16.9 & \\ 
\hline
\end{tabular}
\end{center}
\end{table}

In advance of the radon emanation screening, estimates were made of expected radon emanation from all materials in contact with the Xe, based on previous measurements of radon emanation from similar materials, or based on the combination of HPGe measurements of the bulk \IRatts contamination together with measurements or estimates of Rn diffusion in the material.  It must be noted that surface contamination may cause a higher rate of radon emanation from a material than is estimated from its bulk contamination, especially if the material's bulk is relatively radiopure or if the material has a very low radon diffusion constant.  Table~\ref{MASt:RnBudget} summarizes the resulting radon budget from the critical materials.
In cases where the dominant radon emanation is expected due to diffusion of radon from the material bulk, the expected radon emanation is based on room-temperature assays, conservatively ignoring the expected reduction of radon diffusion at LXe temperatures for all materials except the cables.
For materials with low radon diffusion, such as metals, whose radon emanation is expected to be dominated by recoil punch-out, no emanation suppression is expected at reduced temperatures. 

Without remediation, the cables would have been  expected to be a major source of radon, emanating up to \SI{30}{\mBq} based on previous measurements~\cite{GERDAcablesRn}.  This total would likely be dominated by the large surface area of the stainless steel braiding and contamination, such as dust, that is caught in this braiding.  
 Mitigation of this large radon source will be achieved by adding a thin FEP cladding to the outside of the cables and by the installation of a carbon trap (described in Section~\ref{XCSSs:RnRem}) to filter radon out of the Xe gas from the conduits before it enters the main re-circulation.   
 The cladding should make it easier to minimize contamination of the cables with dust.
Furthermore, radon diffuses very slowly through the cold FEP cladding, nearly eliminating the emanation into the liquid Xe by the cables, while the  trap will reduce the radon emanation into the gas portion of the conduits by at least \SI{90}{\percent}. 
 The combination of these two strategies reduces the expected emanation by the cables to \SI{0.09}{\mBq}.  The carbon trap should significantly reduce the radon load from other room-temperature materials such as the PMT cable feedthroughs and conduits.  Table~\ref{MASt:RnBudget} includes the expected reduction from the carbon trap.
 
Other potentially significant contributors were identified.  
Early assays of the components of the PMT bases indicate sufficiently low emanation even at room temperature that mitigation strategies, such as potting, need not be explored. 
The Xe purification getter material has been identified 
as a source of radon~\cite{getterRn}. In order to minimize the radon emanation from the getter while allowing the high throughput needed by LZ, the baseline plan is to purchase a full-size getter cabinet with a partially-loaded getter cartridge (see Section~\ref{XCSSs:gXe}).  In addition, a screening program to identify a clean Zr substitute is underway, along with an investigation of other material options.  

The largest single contributor to radon backgrounds in the experiment is expected to be due to dust on material surfaces.  As described in Section~\ref{MASSs:Dust}, certified protocols for assembly and cleaning will provide the means to achieve a low enough dust concentration on materials in LZ. The dust is conservatively estimated to generate \SI{10}{\mBq} of activity. We assume that \SI{25}{\percent} of \IRnttt from the dust on material surfaces escapes into the xenon; however preliminary measurements of of radon emanation from dust support fractions considerably lower (see Section~\ref{MASSs:Dust}).

The radon-emanation screening campaign, coordinated through dedicated management in the screening working group, extends beyond initial material selection. The system from SDSM\&T will be relocated underground to SURF in order to screen large-scale assembled detector elements and plumbing lines. As pieces or sections are completed during installation of gas pipework for the LZ experiment, they will be isolated and assessed for Rn emanation and outgassing for early identification of problematic seals or components that require replacement, cleaning, or correction. 

\tdrsec[Cleanliness]{Surface Cleanliness}

Once materials and components have met screening requirements and components have been procured, they will be kept clean during fabrication, storage, transport, and final assembly and integration into the experiment. We refer to this task as cleanliness. The major sources of contamination that must be addressed by the LZ cleanliness program are radon diffusion and daughter-nuclei plate-out, and dust and debris. Radon-daughter plate-out onto material surfaces may generate NR backgrounds through two mechanisms: (\Pga,\Pn) processes that release neutrons into the xenon; and ions from the \IPbtoz sub-chain originating at the edge of the TPC being mis-reconstructed as NRs within the fiducial volume. \IPbtoz on surfaces may also present ER background if it or daughters become mobile and enter the fiducial volume, with \IBitoz of note as a beta-emitter. Dust on component surfaces carried into the xenon may produce ER and NR from intrinsic activity, as well as contributing significantly to the radon emanation background. We require that the effects of radon plate-out and dust deposition do not contribute to the background in LZ any more than the material components. Other sources of contamination must also be addressed, for example residual chemicals from fabrication processes, removal of which should be effected at the same time as cleaning to remove dust. 

LZ draws on extensive cleanliness experience from earlier involvement in LUX, EXO, SNO, KamLAND, \textsc{Majorana} Demonstrator, and other low-background experiments to develop cleanliness protocols and assay techniques for dust and surface contamination. LZ is developing experiment-specific techniques for validating protocols and establishing a rigorous and comprehensive Quality Control (QC) and Quality Assurance (QA) program.  
All protocols and documentation are stored in the LZ information repository where they are linked to the specific material or component. 
No material or component will be integrated into the experiment without adequate documentation demonstrating handling following these protocols and successfully passing quality control tests.  

In the following subsections we present details on the origins of these background sources; requirements and goals for both radon plate-out and dust depositions and equivalent maximum exposure times; controls and mitigation protocols; and assay techniques to ensure QC and QA certification.

\tdrsubsec[Dust]{Dust}

Dust is typically characterized as fine-grained (\SI{<100}{\mum}) material that results from the breaking/grinding of materials in the local environment. The specific content will reflect the location, but in general will include minerals from rock and soil, small amounts of plant pollen, human and animal hairs, textile fibers, paper fibers, human skin cells and even burnt meteorite particles. The contents and their relative proportions will determine the type and level of radioactivity, but in general the major contributors are \IKfz, \IUtTe and \IThtTt, all with activity levels around \SI{\sim10}{\mBqg}. Fine grains may easily become airborne, to be transported over significant distances, and then deposit on local surfaces. Electrostatic attraction allows vertical as well as horizontal surfaces to become contaminated.  We expect to encounter dust in all locations where work is conducted, including the sourcing of materials where dust may become internalized to components, fabrication and construction phases, and in assembly and deployment.  A detailed understanding of the radiological content of the dusts to be encountered has been developed, primarily through HPGe screening of floor sweepings and HEPA filter samples from the SURF underground areas. Measurements of the dust-particle count density, size, and deposition rate at SURF have also been completed~\cite{Tiedt:2013} and will be extended to other assembly locations at LZ institutes and external sites where appropriate, such as the Ti-cryostat manufacturer. 

The quantity of airborne dust in a specific location is found to vary significantly, with strong dependence on the local environment, geometry of any buildings, rate of airflow, and the use of air filtration to reduce the presence of dust, for example through the use of cleanrooms. The effectiveness of a cleanroom is highly dependent on associated procedures, for example the level of use, gowning procedures of users, and supplementary cleaning activities. The rate at which dust settles out on surfaces depends on the dust level in the air, the particle size distribution and density, and the rate of airflow.  Where possible dust levels and dust deposition rates have been measured directly, but also modeling of these factors has been performed under a number of reasonable assumptions and model bases. We have developed two web--based calculators, denoted ``SNO'' and ``ASML-Delft'', respectively, that give allowed exposure times based on model and environmental inputs. The ``SNO'' model is based on the mass fallout model developed by R.~Stokstad for SNO~\cite{stokstad:1991}.  Input parameters include fresh air dust content, volume exchange rate, recirculation fraction, filter efficiency, and mass carry-in rate. ``ASML-Delft''~\cite{asml-delft:2012} is an empirical model derived from test measurements and validated by monitoring particle fallout rates as a function of air particle concentrations in operational cleanrooms in the semiconductor industry.  Test measurements and operational cleanroom observations are consistent with parameterization to within an order of magnitude. 
As an example, the SNO model predicts that a volume of fresh air (ISO class~8) that has 120~volume exchanges per hour including  \SI{0.8}{\percent} fresh air, and a carry-in rate of dust of \SI{0.25}{\gday}, would accumulate \SI{500}{\ngct} of dust in a period of 2.3~days. The dust fallout rate is sensitive to the carry-in rate: for example, reducing the carry-in rate to \SI{0.025}{\gday} increases the accumulation time to \SI{22}{\day}. The carry-in rate is a factor that we  control by implementation of cleanroom protocols. We are conducting a program of dust fall-out measurements within the LZ cleanrooms to determine actually realized rates under particular LZ conditions and locations. Our results indicate that fall-out rates are below those expected from the ASML-Delft models. 

There are a number of mechanisms by which the radiological content of dust that is present on the surface of a component may be problematic. The relative importance of these varies depending on the properties and location of the component in the experiment. Contributions to ER and NR rates have been estimated based on the estimated levels of dust, known nuclear properties, and simulations propagating the decay products to the LZ experiment.

The most significant radiological concern for LZ from dust, and the one that drives our requirements, is radon emanation from dust within the inner cryostat and conduits. As previously described in Section~\ref{MASS:RnEm}, this contributes ERs via the naked beta decay of \IPbtof. The requirement for the total activity from radon emanation in the xenon is \SI{20}{\mBq}, with a goal of \SI{1}{\mBq}. Dust contributes half the rate in the requirement. Our present estimate is that the radon emanation from dust is equivalent to \SI{10}{\mBqg}. This estimate is based on measured rates of radon emanation from naturally--occurring materials~\cite{schumann:1996} and SNO reports on radon emanation from mine dust~\cite{rnemanation:sno}. This sets the requirement on maximum amount of dust within the xenon at \SI{1}{\g}. We anticipate the contribution from dust will be significantly less given conservative estimates that define our requirements. Direct HPGe measurements we have performed to infer radon emanation from debris collected at SURF has indicated a \SI{12}{\percent} radon emanation fraction, suppressed to \SI{4}{\percent} at cryogenic temperatures due to reduced radon mobility. This is considerably lower than the \SI{25}{\percent} we assume in defining our requirements. Direct radon emanation measurements of dust at room temperature will be performed end-2016. The goal for dust is a total of \SI{10}{\mg} within the xenon, corresponding to an activity of \SI{0.1}{\mBq} (based on the assumed \SI{25}{\percent} emanation fraction). The total surface area from all components within and including the ICV is about \SI{1.6e6}{\cm\squared} and we conservatively set the required limit on dust surface mass density at \SI{0.5}{\mugct}. 
Our goal defines a density limit of \SI{5}{\ngct}. Dust on all other components is constrained to be at the same level, driven by the need to avoid cross contamination during assembly, however radon emitted from dust on non-wetted surfaces not in contact with any xenon is not a concern. 

Beyond radon emanation, the intrinsic activity of dust generates both ER and NR background, and this contribution has also been evaluated in detail. Dust concentrations that meet requirements are assumed for the surfaces of all components within the ICV, taking into account the appropriate surface areas from the LZ CAD model. LZSim is used to generate radioactivity from these surfaces by propagating \Pgg-rays and neutrons and estimating impact in LZ. These calculations account for (\Pga,\Pn) reactions on the appropriate targets. 
The range of few-MeV alpha particles in dust is on the order of \SI{100}{\mum} and thus in most decays it is reasonable to expect the alpha particle to escape. For a surface deposit, a simple geometric consideration predicts half of these to then impact the surface. Of particular concern is dust upon the surface of PTFE components, since the number of neutrons emitted for each incident alpha particle is particularly high for fluorine, \num{9.48}~neutrons per \num{e6} \Pga-particles~\cite{Heaton:1989vbt}. Aluminum also has a moderately high (\Pga,\Pn) cross section resulting in \num{0.63}~neutrons per \num{e6} \Pga-particles. The (\Pga,\Pn) cross sections of other elements used in LZ all have significantly smaller cross sections. Of note, the design for the internal PTFE reflectors has them attached by pins to the inner cryostat, deliberately allowing a thin layer of Xe to penetrate between the titanium and the PTFE, thus mitigating against alpha particles emitted from U and Th contaminants in dust on the cryostat surface from inducing (\Pga,\Pn) reactions on the adjacent PTFE reflector. Between the inner and outer cryostats is the vacuum region. Dust here has been considered on the titanium surfaces, the MLI superinsulation, which is aluminized mylar, and displacer foam. Beyond the cryostats, there is a layer of displacer foam for mechanical stability, the outer detector comprised of acrylic vessels holding gadolinium loaded scintillator, mechanical supports, the water tank and its PMTs and reflective liners. While some of these have large areas, our calculations suggest that their distance from the fiducial volume prevent them from posing significant concern under any reasonable assumption of dust level on their surfaces. 
Dust contributes a negligible \SI{0.19}{\ER} counts to the background in LZ before any discrimination is applied. The total NR contribution from dust is similarly low, at \num{0.05}~counts before application of NR efficiency. These contributions are included in Table~\ref{MASt:BackgroundsTable}.   

Detailed cleanliness protocols to clean components and maintain purity are developed for each subsystem, specific to the materials and adopting industry standard cleaning procedures. Each protocol is defined to satisfy the cleanliness goal rather than just the requirement. Each subsystem develops its own cleanliness protocols that are reviewed by the cleanliness working group before being put into effect. 
The protocols describe how the subsystem components are to be cleaned and kept clean through fabrication, assembly, integration and installation, and describe the validation of cleaning and assay methods. Documentation with each component demonstrates that protocols have been followed step-by-step, and includes results from assay of witness plates and coupons, and from environmental monitoring during fabrication and storage. Upon delivery to SURF each component's documentation pack will include a requirement for signoff for dust cleanliness at the required level.
The protocols and documentation are stored in the LZ Information Repository. No part is to be accepted for final assembly and integration on-site without QA and requisite documentation that protocols having been successfully followed. 

Final shipping of clean components will involve the use of sealed bags made from films of certified cleanliness. Vendors have been identified for procurement of 50 micron thick Nylon bags certified to be dust free at Level 50 of U.S. MIL-SPEC 1246C, equivalent to there being less than one \SI{50}{\mum} sized particle per square foot. For large components, custom--fabricated bags meeting cleanliness requirements will be used. The standard packaging procedure will employ three layers of packaging: 
outermost will be ``bug bag'' which will be removed just before transferring the package from the loading dock to an inside climate--controlled area; the next layer of packaging will be removed in an anteroom to the cleanroom assembly/integration area; and the innermost package will be removed inside the cleanroom just before the part is integrated. Dust deposition rates of \SI{1}{\ngct\per\hour} have been achieved previously by the SNO experiment, and by the LZ SDSM\&T group. 

Essential to evaluating cleanliness is the ability to assay at the required levels to maintain QC and perform QA before components are accepted for integration. Several methods for the assay of dust have been developed for the LZ project. At SDSM\&T, an optical system has been used. Glass witness slides are imaged, revealing dust, fibers and hairs that have been deposited onto the slides. An automated software process analyses these images and determines the size and mass of each particle. Sensitivity at the level of \SI{5}{\ngct} has been demonstrated, sufficient not only for our requirement but also for our goal. In parallel, tape lifts are also being used on non-transparent surfaces, such as Cirlex that is to be used in the PMT bases. Acetate tape is first immersed in acetone and then placed on the surface of the material to be assayed. The tape secures the dust and, once dry, is removed along with the dust. This tape is then optically imaged using the same method as for the glass witness slides. The size distributions for all surfaces assayed broadly follows the expected logarithmic behavior, with the majority of particles measuring below \SI{10}{\mum} and only a few larger. However, these few larger particles dominate the mass of the sample. For large areas of titanium sweeping of dust from the surface and then measurement of the weight of that material is possible with sensitive scales. An area of \SI{1}{\square\m} with dust deposition at the level of \SI{5}{\ngct} equates to a mass of \SI{50}{\mug}, an amount that may be collected and measured with a standard microbalance. 

\tdrsubsec[Plate-out]{Radon Plate--out}

Radon plate-out is a phenomenon in which charged radon progeny are deposited onto the surfaces of materials exposed to air that typically contains concentrations of \IRnttt (\thlf=\SI{3.82}{\day}) ranging from tens to hundreds of \si{\BqmT}~\cite{Guiseppe:2011mj,Arpesella:2001iz}. The decay daughters can be embedded into material as they recoil due to subsequent decays. Beyond radon concentration and surface area, the susceptibility to plate-out depends on the material and factors such as air-flow rates, which are difficult to predict and therefore must be measured. In the cases where measurements are not available, conservative estimates are used to predict contamination risk from plate-out. Plate-out may be further enhanced in the presence of an electric field, since positively charged radon daughters are deposited on negatively charged surfaces such as electrodes within the TPC. The background due to radon daughters on the surfaces arises predominantly from neutron production. In particular the long-lived \IPbtoz (\thlf=\SI{22.3}{\year}) in the decay series decays to \IPotoz (via \IBitoz), which emits an \Pga that feeds (\Pga,\Pn) reactions. In addition to neutron background, progeny from Rn plated onto the inner surfaces of the TPC, particularly \IPbtzs, can lead to spatial leakage of mis-reconstructed events at the TPC walls, rapidly reducing the fiducial mass. Furthermore, incomplete charge collection of these recoils at the edges of the TPC can cause them to overlap with the low-energy NR band. We have conducted detailed simulations, utilizing the position reconstruction algorithm successfully deployed in both LUX and ZEPLIN-III and adapted for the TPC, extraction electrodes, and top PMT array configurations of LZ, to study position reconstruction of such edge events. In the following  we describe in detail our estimate of the level of background produced by plate--out, our approach to 
estimating the rate at which plate--out could occur in LZ, and the measures we will take to adequately mitigate background risk.

The first background due to (\Pga,\Pn) depends on the area and material composition of component surfaces.   It also depends on the degree of surface contact between two components. As described for \Pga's from dust above, if two components are in close contact, \Pga's emitted from the surface of one material could induce (\Pga,\Pn)  in the other material. Of the LZ detector components, the
only ones with a significant product of surface area and (\Pga,\Pn) reaction yield are the teflon components, due to the fluorine, and the \numrange{10}{20} layers of superinsulation containing aluminum between the inner and outer vessels of the cryostat. 
Using results on the simulation of the detector response to neutrons and application of  analysis cuts, the expected NR background counts from the (\Pga,\Pn) reaction on teflon and aluminum has been estimated, not only for plate--out activity directly on these surfaces but also for plate-out activity on surfaces of other components (e.g. PMTs) in tight contact with teflon or aluminum.  The results are shown in Table~\ref{MASt:plateoutAlphaN}, with a low total of only 0.05 NR counts before NR efficiency is applied. This is based on a required surface activity density of \SI{10}{\mBqmt}. 

The second background is the mis--identification of ion daughters which recoil into the TPC active volume from the decays of radon progeny on the surface of the teflon cylindrical reflectors.  In the long--lived \IPbtoz sub--chain, the
only decay capable of producing an ion recoil of sufficient energy is the \IPotoz decay into an alpha and \IPbtzs ion. Using a fiducial volume enclosing \SI{5.6}{\tonnel} on LXe, the probability that an ion recoil is mis-identified as an NR event has been estimated to be \num{e-6}. At the minimum  \IPotoz activity LZ expects to be able to detect, namely, \SI{0.5}{\mBqmt}, the expected NR background is \num{0.16}~counts. This background is expected to be mitigated with straightforward measures since it affects only small S2 signals, and its rate falls as a steep function of the distance of the fiducial volume boundary from the physical reflector surface. It therefore does not affect the peak sensitivity of LZ, and may in any case be removed efficiently through fiducialization. Nonetheless, we include this background in Table~\ref{MASt:BackgroundsTable}, and derive a requirement on plate--out activity on the inside of the TPC walls at \SI{0.5}{\mBqmt}.

Radon plate-out may also generate ER background if the \IPbtoz or any of its daughters become mobile and detach from surfaces to enter the fiducial volume. \IBitoz in particular would generate single site ER background if present in the target. As discussed in Section~\ref{MASSs:BackTab}, LUX data has been assessed to place upper limits on the mobility of \IPbtoz daughters from the TPC walls. For an activity of \SI{0.5}{\mBqmt} on the PTFE panels of the TPC in LZ, an upper limit of 40 ER counts may be expected; a contribution similar to the fixed radioactivity or intrinsic contamination goals.

\begin{table}
\tdrtcaption[plateoutAlphaN]{Expected NR background due to (\Pga,\Pn) reaction from plate--out activity.}{Expected NR background due to (\Pga,\Pn) reaction from plate--out activity at the level of \SI{10}{\mBqmt}.}
\begin{center}
\sffamily
\begin{tabular}{|l|c|}
\hline
\rowcolor{mrocol}
 \textbf{Component}  & \textbf{NR Background} \\
 \hline
 TPC PMTs (3--inch) & 0.002 \\
 Teflon & 0.021 \\
 PMT cables (teflon jacket) & 0.017 \\
Field-shaping rings & 0.004 \\
Superinsulation & 0.007 \\
\hline
Total & 0.051 \\
\hline
\end{tabular}
\end{center}
\end{table}

For evaluating measures to control backgrounds due to plate-out, one must estimate the rate at which plate-out occurs.  As already mentioned, plate-out is a complex process depending not only on the concentration of radon in the air and the area of exposed surfaces but also on environmental conditions and surface properties and
conditions. We have developed three rate calculators, one (``Guiseppe'') that parameterizes the daughter deposition rate in terms of the radon concentration in air, surface area, and a ``deposition velocity'' that has been measured for different radon daughters under different conditions~\cite{Guiseppe:2011mj}; a second (``Borexino'') based on a model used by the Borexino experiment~\cite{borexino:shutt} that parameterizes the daughter deposition rate in terms
of radon concentration, a column height (e.g. the height of a cleanroom above a work surface), and plate-out fraction; and a third (``dead--air'') based on a pessimistic model in which all of the radon decays in a defined volume result in \IPbtoz daughters deposited on the enclosing surface of the volume. Plate-out measurements carried out within the LZ collaboration to validate this model 
under conditions in which radon-laden air was circulated very slowly (one volume exchange every 12 hours) over aluminum, teflon, and glass plates set at the bottom of 
a \num{8}-liter rectangular purge box gave results consistent with the dead--air model. 

Numerical results for the estimated exposure time required for \IPbtoz
activity to reach a level of \SI{10}{\mBqmt} are presented in Table~\ref{MASt:plateoutExposureTime}.  For the first calculator, the largest measured value for the 
deposition velocity was used as a conservative value, corresponding to the measured plate-out of \IPotoe on acrylic; for the second calculator a column height of \SI{2.5}{\m} and a 
plate-out fraction of \num{0.01} was used, compatible with radon daughter plate-out onto nylon as measured by Borexino; for the third calculator, the area of the enclosing surface was set to
\SI{8.0}{\m\squared} and the volume was set to \SI{3.5}{\m\cubed}, corresponding approximately to the volume of the TPC.   

\begin{table}
\tdrtcaption[plateoutExposureTime]{Allowable exposure times (days) to reach radon daughter plate-out requirements}{Allowable exposure times (days) to reach radon daughter plate-out requirements}
\begin{center}
\sffamily
\begin{tabular}{|c|c|c|c|c|}
\hline
\rowcolor{mrocol}
  \textbf{Maximum Activity} & \textbf{Radon Concentration} & & & \\
\rowcolor{mrocol}  \textbf{(\si{\mBqmt})} &\textbf{(\si{\BqmT})}  & \textbf{Guiseppe} & \textbf{Borexino} & \textbf{Dead-Air} \\
 \hline
 10 & 20 & 235 & 168 & 13.4 \\
 \hline
 10 & 0.3 & 15700 & 11200 & 895 \\
 \hline
 0.5 & 20 & 11.8 & 8.4 & 0.7 \\
 \hline
 0.5 & 0.3 & 783 & 560 & 44.8 \\
 \hline
\end{tabular}
\end{center}
\end{table}

The estimates from the Guiseppe and Borexino models, in qualitative agreement with each other, indicate that detector fabrication and 
assembly in well-ventilated areas significantly reduces the plate-out rate.   Even with good ventilation, however, the allowed exposure times under
typical laboratory conditions (Rn concentration on the level of \SI{20}{\BqmT}) are short 
for a maximum  surface activity of \SI{0.5}{\mBqmt} allowed for the TPC cylindrical reflectors.   To assure that radon plate-out is controlled
during fabrication, assembly, and integration, the following measures will be taken:
\begin{enumerate}
\item Final assembly and integration at SURF of the inner cryostat vessel and all interior components, 
projected to take approximately 6~months, will be carried out in radon-scrubbed clean room with a design
radon concentration of \SI{0.3}{\BqmT}. This dramatically increases the acceptable exposure for the TPC reflectors to meet installation and integration timescales, as shown in Table~\ref{MASt:plateoutExposureTime}.
\item During storage and transport, parts will be sealed in at least three layers of nylon or metalized mylar bags backfilled with nitrogen or argon.  Metallized
mylar and nylon are known to be excellent radon barriers although use of nylon in  high-humidity environments significantly degrades its effectiveness.
Tests within the LZ collaboration using commercially-supplied 2-mil nylon bags and 2.5-mil aluminized mylar bags have shown 
that the radon concentration reached inside a single bag after one month is at least two orders of magnitude below external levels in the case of nylon
and at least three orders of magnitude less for the case of aluminized mylar.   
\item In the case of critical parts for which the storage time is long compared to a month, nitrogen purge boxes will be used.
\item Assembly and integration procedures for the TPC cylindrical reflectors will assure good ventilation to avoid higher plate-out rates which
would be expected to occur in a dead-air environment.
\item Witness plates and coupons will be assayed at critical points to assure that plate-out is being controlled at the required level. 
\item The detailed prescription for implementing the above measures and documenting that they have been successfully followed will be incorporated into the cleanliness protocol for each subsystem.
\end{enumerate}

Within LZ, two sensitive detectors will be used to carry out the assays to determine surface activity and plate-out.  The first is the commercial XIA Ultralow 1800 surface alpha detector system, suitable for routine screening of small samples including our witness plates and coupons. Recently installed and commissioned at Brown University, the detector has already achieved a sensitivity to 
\IPotoz at the level of \SI{1}{\mBqmt}, easily meeting the requirements for the materials contributing to the (\Pga,\Pn) backgrounds. Based on the experience of operating this instrument at other experiments, we expect to ultimately achieve a sensitivity below \SI{0.5}{\mBqmt} that will match requirements for assaying the inner reflector panels of the TPC to determine an upper limit on ion recoil mis-reconstruction backgrounds. The second detector has been developed by the LZ group at SDSM\&T, deploying a panel of large-area Si detectors installed in a large vacuum
chamber.  This detector allows assays of bulky detector components as well as planar witness plates and coupons.  With a single Si detector, a sensitivity to \IPotoz surface activity at the level of a few \si{\mBqmt} has already been achieved, meeting requirements for the materials generating (\Pga,\Pn) backgrounds. The full system will be online for routine LZ assays by early-2017. For materials such as Teflon, which are produced in granular form before being sintered in molds, plate-out comprises an additional dimension of risk because surface contamination of the granular form becomes contamination in bulk when the granules are poured into molds. Measurements of ~\IPbtoz~/~\IPotoz in bulk PTFE to the required sensitivity of \SI{10}{\mBqkg} will be performed using the Lumpsey detector as previously described in Section~\ref{MASSss:BUGS}. At this activity, the \IPbtoz in the bulk PTFE will contribute a background of 0.1 NR counts and is included in Tab.~\ref{MASt:BackgroundsTable}.

\tdrsec[LXeCont]{Liquid Xenon Contamination}

Krypton and argon present in the xenon will generate backgrounds that, as with radon emanation, cannot be self-shielded against. Here we outline the strategies for limiting the total background contribution from these elements to a relatively low level of less than \SI{1}{\mudru} total activity from both, equivalent to the ER background from material radioactivity. 

\tdrsubsec[Kr]{Krypton}

Krypton contamination dispersed throughout the fiducial 
volume can generate ER background. \IKreF is a beta-emitter with a half-life of \SI{10.8}{\year} 
and a dominant (\SI{99.6}{\percent} branching ratio) bare beta 
decay mode of endpoint energy \SI{687}{\keV}. Its presence 
in the atmosphere is largely anthropogenic, resulting 
from nuclear fuel reprocessing and testing of 
nuclear weapons~\cite{Ahlswede201334,Collon:2004xs}. 
Coupled with a long half-life and diffusion properties 
as a noble gas, it can become a significant contaminant 
in the course of the production and storage of Xe. 

The isotopic abundance of \IKreF
in air is nominally \num{\sim2e-11}
~\cite{Collon:2004xs,GRL:GRL17458}.
It is present in some air samples at levels that 
are \SIrange{10}{20}{\percent} higher than this. The XMASS experiment found the \IKreF 
isotopic abundance in its distilled xenon to be 
\num{0.6+-0.2 e-11}~\cite{Abe:2013tc}. For the purpose of calculating
the LZ requirement we use an abundance of \num{2e-11}.

The research-grade Xe used in LUX contained an average 
\SI{130}{\ppb} \IKrnat/Xe upon procurement and was 
reduced to \SI{3.5+-1.0}{\ppt} in LUX, resulting in a measured 
event rate of \SI{0.17+-0.1}{\mdru}~\cite{Akerib:2014rda}. Recent
measurements of the \IKrnat content of the first 100,000 liters of
Xe purchased for LZ from Praxair found an average concentration of \SI{1.2}{\ppb}.

The allowed \IKrnat concentration in LZ is determined by
requiring that ER backgrounds from \IKreF 
contribute no more than \SI{10}{\percent} of the solar \Ipp neutrino rate.
To achieve this, the \IKrnat concentration in the 
LXe must be less than \SI{0.015}{\ppt}, accounting for the \IKreF 
isotopic abundance and the beta decay spectrum and branching ratio.
This concentration will be accomplished 
with chromatographic separation at SLAC prior 
to physics operations (see Section~\ref{XCSS:KrRem}), followed by a
comprehensive program to limit the ingress of air 
from leaks during storage (see Section~\ref{XCSSs:leaksStorage}).
Additional requirements are placed on the acceptable leak rate of the 
Xe handling system during operations (see Section~\ref{XCSS:Online}), 
and on the acceptable level of Kr
outgassing from detector components (see Chapter 9 of Ref.~\cite{Akerib:2015cja}).
There is a corollary requirement of \SI{0.015}{\ppt} \IKrnat detection 
sensitivity to confirm that the removal and storage programs 
have been successful (see Section~\ref{XCSS:XeSamp}.) 

\IKrnat concentration levels of \SI{<0.2}{\ppt} 
have been demonstrated during the LUX production run by double-processing a \SI{50}{\kg} 
LXe batch. 

\tdrsubsec[Ar]{Argon}

Trace quantities of argon are also a concern due to beta-emitting \IArTn, 
with a \SI{269}{\year} half-life, a \SI{565}{\keV} endpoint energy, and a \SI{100}{\percent} branching 
fraction to the ground state. 
The isotopic abundance is \num{8e-16}~\cite{Loosli:2000}.
This background is constrained to be less 
than \SI{10}{\percent} of \IKreF, resulting in a specification of 
\SI{4.5e-10}{\gpg} or \SI{2.6}{\muBq}. The Kr removal system, 
which also removes Ar, should easily achieve this goal. 

\IArTS
is another isotope of concern. It decays via electron capture, producing 
bare x-rays and Auger electrons between 2 and 3 keV with a \SI{90}{\percent} branching ratio.
Its half-life is 35 days, similar to \IXeotS, 
but since it is not cosmogenically 
produced while the Xe is in storage, it is not expected to be present in Xe
stockpile at the start of physics operations. Its concentration in 
air is typically \SI{1.2}{\mBqmT}~\cite{Riedmann:2011rp}, and at this level
it is of secondary importance to the leak rate tolerance of the experiment compared
to \IKreF
(see Section~\ref{XCSS:Online}). However its local concentration in air 
may be elevated depending on soil conditions, being produced via neutron 
capture on calcium. We are pursuing a measurement of the \IArTS
activity of the Davis campus air in collaboration with the Pacific 
Northwest National Laboratory to confirm that it is not a driving factor for the 
leak rate tolerance of the experiment.

\tdrsec[labCosmo]{Laboratory and Cosmogenic Backgrounds}

Mitigation of external radioactivity arising from the laboratory environment or from cosmic-ray muons is largely derived from the LZ water shielding and OD. Use of the LXe skin and fiducial volume definition in the TPC for single scatter events also reduces this background, as well as radioactivity arising from cosmogenic activation of materials in the LZ project during manufacture, transport, or storage. Cosmogenic activation of the Xe target also produces radioisotopes but these are intrinsic to the LXe target and cannot be shielded against. While such cosmogenic production of radioisotopes underground in LXe may be neglected due to the low cosmic-ray flux, it is surface activation during production and storage that will generate background in LZ. These backgrounds are shown in Table~\ref{MASt:externalBackgrounds}, and are discussed in the following subsections.

\begin{table}
\tdrtcaption[externalBackgrounds]{Background from laboratory, muon-induced neutrons, and cosmogenic activation}{Events expected from different external or activation sources in \SI{1000}{\day} live exposure of \SI{5.6}{\tonnel} fiducial mass in LZ. Rates are given before expected \SI{99.5}{\percent} discrimination for ER events or \SI{50}{\percent} efficiency for NRs, with the exception of the last row where the expected count rate in the WIMP search region is quoted.}
\begin{center}
\sffamily
\begin{tabular}{|l|l|c|l|}
\hline
\rowcolor{mrocol}
Source & Recoils & Number of events & Systematic uncertainty \\ 
\hline
Walls & ER & 4.1 $\pm$ 0.8 & Factor of 2 \\ 
Walls & NR & $<$0.001 & Included in the limit \\
Muons & NR & $<$0.056 & \SI{30}{\percent}  \\ 
\IXeotS & ER & 0.11 & Factor of 5 \\
\IScfs & ER & $<$0.1 & Included in the limit \\
\hline
Total & (\SI{99.5}{\percent} discr., \SI{50}{\percent} NR) & 0.01 -- 0.09 & Included in the range \\
\hline
\end{tabular}
\end{center}
\end{table}

\tdrsubsec[labBack]{Laboratory Backgrounds}

The background in the underground SURF facility outside the water tank is dominated by \Pgg-rays and neutrons produced in the cavern walls.  The sources of these \Pgg-rays and neutrons are the uranium and thorium decay chains with a contribution of \IKfz decay to high-energy \Pgg-rays. Many samples of rock have been screened to get U and Th contamination varying from \SIrange{0.1}{7.5}{\ppm} U, from \SIrange{1.9}{47}{\ppm} Th and from \SIrange{0.04}{3.3}{\percent} of natural potassium, depending on a specific sample and rock formation~\cite{Lesko:2011qk,Roggenthen:2008}. At the 4850 level, the \Pgg-ray flux has been measured as \SI{2.16+-0.06}{\per\square\cm\per\s} above \SI{0.1}{\MeV} and \SI{0.632+-0.019}{\per\square\cm\per\s} above \SI{1}{\MeV}~\cite{Mei:2009py}. This agrees well with the simulations of \Pgg-ray flux based on the average values of contamination in the Homestake formation: \SI{1.51}{\ppm} U, \SI{7.38}{\ppm} Th and \SI{0.96}{\percent} of natural potassium~\cite{Mei:2009py}. Some other sources indicate a lower contamination of U/Th in rock around the lab (down to \SI{0.22}{\ppm} U and \SI{0.33}{\ppm} Th) but the shotcrete and concrete on the walls may give the radioactivity levels an order of magnitude higher and similar to (or even higher than) those quoted above for the Homestake formation (see for instance \cite{Lesko:2015sma} and references therein). Recently the flux and spectrum of \Pgg-rays in the laboratory hall (East Counting Room) have been measured with a bare HPGe detector~\cite{Thomas:2014}. The intensities of \Pgg-ray lines from U, Th and K have been found to be higher than previously reported, giving the integrated flux of \SI{2.19}{\per\square\cm\per\s} for energies \SI{>1}{\MeV} with a large uncertainty (about a factor of 2 towards lower flux values) due to the fact that the flux in the continuum comprises Compton scattering events in the detector itself, with no on-site calibration of the detector. Recent measurements of a sample of gravel from beneath the LZ water tank~\cite{Harvey:2016} have revealed concentrations of \SI{1.65}{\ppm} U, \SI{0.30}{\ppm} Th and \SI{0.066}{\percent} of natural potassium, significantly smaller than previous reports. In our estimates we have used the results from~\cite{Thomas:2014} as a conservative estimate.

The \Pgg-ray and neutron fluxes from rock have been shown to be attenuated by many orders of magnitude by the water shielding and then by the LUX background rejection strategy so they do not contribute to the measured number of events in LUX. 
LZ, however, is expected to achieve \num{100}$\times$ better sensitivity to WIMPs and the thickness of shielding is slightly reduced compared to LUX due to a larger cryostat. 
We have simulated \Pgg-ray production in U and Th decay chains and from \IKfz, and transport of these \Pgg-rays from lab walls through the shielding down the LXe detector in several steps and applied the same event selection algorithms as for other background sources. 
The decay chain of Th should give the highest contribution because of the presence of the most energetic \SI{2.62}{\MeV} \Pgg-ray line in the Th decay chain. 
The simulation results have been normalized to high-energy line intensities as measured in \cite{Thomas:2014} which suggested the concentrations of \SI{6.42}{\ppm} (\SI{26.1}{\Bqkg}) Th, \SI{5.95}{\ppm} U (\SI{73.4}{\Bqkg}) and \SI{2.31}{\percent} of natural K (\SI{716}{\Bqkg} of \IKfz).
Our simulations have shown that the number of single ER events in \SI{5.6}{\tonnel} of fiducial volume and a live-time of \SI{1000}{\day} after all cuts is \num{4.1\pm0.8} dominated by the 2.6 MeV gammas from Th decay chain. 
The errors are purely statistical and have been calculated from the number of observed events in simulations. 
The systematic uncertainty is about a factor of 2.
Assuming discrimination power of \SI{99.5}{\percent} the limit on the number of events in the NR band will be <0.05 events, much less than the background from internal sources and neutrinos. 
Neutrons from the laboratory walls are attenuated efficiently by water and scintillator that will surround the LZ cryostat with a minimum thickness of hydrogenous shielding to be \SI{70}{\cm} and a reduction of the neutron flux by more than 6 orders of magnitude.
We have carried out simulations of neutron transport through rock and shielding beneath the detector where one of the conduits is located and have shown that no neutrons are likely to pass through this bottom conduit to the LXe target. Other conduits are smaller in diameter and longer which, together with a big thickness of water and scintillator make other sides of the detector better protected against neutrons from rock.
This will bring the event rate from rock neutrons to less than 0.001 events in \SI{5.6}{\tonnel} fiducial mass for a live time of \SI{1000}{\day}, well below the contribution from detector components. 

Muon-induced neutron production and impact in LZ has been assessed using full Monte Carlo. Muon simulations for LZ were carried out using accurate surface profile for the Davis campus at the 4850 level at SURF. Initially, muons with different energies at the surface were transported through various thicknesses of rock using MUSIC~\cite{Kudryavtsev:2008qh,Antonioli:1997qw}. The resulting spectra of surviving muons have been convolved with muon spectra at the surface for different zenith angles and slant depths. Underground muons have been sampled on the surface of a box which contained the cavern and \SIrange{5}{7}{\m} of rock around the cavern using the MUSUN code~\cite{Kudryavtsev:2008qh} and transported through the rock and the detector using LZSim. All particles have been tracked and their interactions in LXe and OD recorded. Selection of surviving events has been based on energy deposition, hit multiplicity, fiducial volume and the anti-coincidence with LXe skin and the Outer Detector; identical to all other event selection criteria in LZSim background simulations. No candidate events has survived the cuts following \SI{118.5}{\year} of equivalent live time. The rejection efficiency of muon-induced neutron background benefits from the large multiplicity of particles produced in cascades initiated by muons. The absence of observed events after cuts allows us to set a \SI{90}{\percent} CL upper limit on the muon-induced background rate in LZ of 0.056 NR events for a live-time of \SI{1000}{\day}, reduced to 0.023 events after applying the \SI{50}{\percent} NR efficiency.

\tdrsubsec[Cosmo]{Cosmogenic Activation}

The irradiation of xenon by cosmic rays at the surface produces a number of radioactive isotopes including those of Xe, tritium, \ICsoTf, \IIotF, \ITeotom, \ITeotTm. These cannot be shielded against but non-xenon isotopes will be efficiently removed from the xenon during purification. Radioisotopes of Xe that can be produced through cosmogenic or neutron activation are \IXeotS (\thlf=\SI{36.4}{\day}), \IXeotnm (\thlf=\SI{8.9}{\day}), \IXeoTom (\thlf=\SI{11.9}{\day}), and \IXeoTT (\thlf=\SI{5.3}{\day}). Most of these will decay quickly and will not pose any threat after a few months of commissioning and calibration runs. Decaying xenon isotopes, in particular \IXeoTom, and \IXeotnm, will also serve as important calibration sources in these early runs. The exception will be \IXeotS, with a half-life of \SI{36.4}{\day}. LUX measurements~\cite{Akerib:2014rda} showed a decay rate of \SI{2.7+-0.5}{\mBqkg} of \IXeotS after xenon was exposed to cosmic rays at the Earth's surface. This can be a factor of 3-4 higher if xenon is stored on the surface at SURF (at about \SI{1600}{\m} elevation). 

\IXeotS produces energy depositions within the WIMP search region of interest and also poses a background for axion searches. \IXeotS undergoes electron capture that results in an orbital vacancy that is filled by electron transitions from higher orbitals, resulting in an X-ray or Auger electron cascade. An \SI{85}{\percent} probability for the capture electron coming from the K shell results in a cascade with a total energy of \SI{33}{\keV}. A further \SI{12}{\percent} of captures from the L shell generate cascades of \SI{5.2}{\keV} total energy deposition, and the remaining \SI{3}{\percent} of decays come from higher shells (M and N) to deposit up to \SI{1.2}{\keV}. For a WIMP search energy window of \SIrange{1.5}{6.5}{\keVee}, significant numbers of L, M, and N shell decays will generate background. However, the daughter \IIotS nucleus is left in either a \SI{619}{\keV}, \SI{375}{\keV}, or \SI{203}{\keV} excited state (with no direct population of the ground state as part of the electron capture). The subsequent decay of the \IIotS to the ground state emits internal conversion electrons or \Pgg-rays that permit almost all of the \IXeotS background to be rejected by coincidence tagging — only those X-ray/Auger events for which the associated \Pgg-rays are not detected contribute to the low-energy ER background in the Xe active region. This effect is expected predominantly at the edge of the Xe target. For example, with a mean free path of \SI{2.6}{\cm} in LXe, the \SI{375}{\keV} \Pgg-ray can potentially escape the active region, reducing the efficiency of coincidence rejection further for events at the edges of the LXe, as seen in LUX~\cite{Akerib:2014rda}. LUX data showed that, after a cooling down period, \SI{0.115}{\mBqkg} of \IXeotS produced a background single hit rate of \SI{0.5e-3}{events\perkkd} in the ROI. In LZ, the skin and external veto systems significantly aid rejection and characterization of this background. Our simulations with LZSim show that the initial ER background in LZ after cuts at \SIrange{1.5}{6.5}{\keVee} is 0.22 events per day (assuming no exposure to cosmic rays at \SI{1600}{\m} elevation). After 8 months of cooling down time underground this rate drops by two orders of magnitude to 0.0022 events/day or about 0.11 events for a live time of \SI{1000}{\day} (before discrimination and taking into account the continuous decay of the isotope). The early storage of xenon underground mitigates this background further. 

The most dangerous isotope due to activation of titanium, used to manufacture the cryostat and many internal detector components, is \IScfs (\thlf = \SI{83.8}{\day}). Using \geantFour and ACTIVIA~\cite{Back:2007kk} codes, we have determined that after 6 months of activation of titanium at sea level, the decay rate of \IScfs will be \SI{2.4}{\mBqkg}. This is consistent with measurements from LUX~\cite{Akerib:2014rda}. Assembling the TPC within the inner vessel at the surface at SURF will increase this rate by a factor of 2. The beta-decay of \IScfs is almost always followed by the emission of two \Pgg-rays, of energies \SI{1120}{\keV} and \SI{889}{\keV}. We expect that, after 8 months of cooling, activated \IScfs will contribute less than 0.1 event for a live-time of \SI{1000}{\day} of LZ operation before \SI{99.5}{\percent} discrimination, rendering this background negligible. Effects from activation of other materials are small compared to possible background rates from activated xenon and titanium due to either smaller masses of materials in the vicinity of the LXe target (copper, stainless steel) or absence of long-lived 'dangerous' radioactive isotopes (PTFE). Activation of various materials underground has also been evaluated using \geantFour and found to be negligible.

\clearpage
\bibliographystyle{apsrev4-2}
\bibliography{LZtdr}

\wlog{In tdrinclude command}
\wlog{\ChapLabel}

\renewcommand{\ChapLabel}{chap:IAI} 
\renewcommand{\BibLabel}{IAIS:bib} 
\renewcommand{\SecPre}{IAI}
\renewcommand{\FloatPre}{IAI}
\tdrchap[IAI]{SURF Infrastructure, Assembly, and Integration}

This section describes the surface and underground infrastructure 
improvements and additions needed at SURF in order to facilitate 
LZ assembly and installation. A detailed assembly and 
installation sequence for LZ is also presented here. 
Infrastructure at SURF includes the following: (1) surface 
laboratory space for assembly of the Xe detector (WBS 1.5) into 
the inner cryostat vessel (WBS 1.2); (2) staging space for 
scintillator veto tanks and scintillator (WBS 1.6), for the outer 
cryostat subcomponents (WBS 1.2), and other detector components; 
(3) secure storage space for Xe; (4) custom tooling for lowering 
the fully assembled Xe detector sealed inside the inner cryostat 
down the Yates shaft; and (5) modifications to the Davis Campus 
at the 4850L of SURF. These infrastructure elements are described 
below. Surface and infrastructure improvements are being funded 
primarily by the South Dakota Science and Technology Authority 
(SDSTA). The design of the infrastructure improvements has been a 
joint effort between the SDSTA engineering and technical staff, an external contractor 
and the LZ collaboration. The firm Leo A. Daly~\cite{LeoADaly} 
is under contract to SDSTA for the detailed design of the surface 
and underground infrastructure improvements. The final design is complete.

\tdrsec[Surface-Infrastructure]{Surface Infrastructure}

\begin{figure}[H]
\centering
\includegraphics[width=0.6\textwidth]{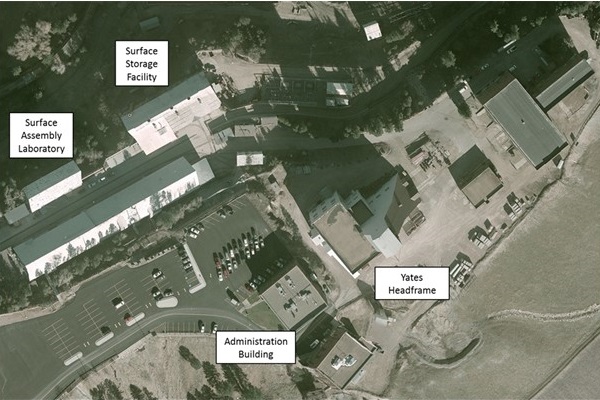}
\tdrfcaption[AerialViewSURF]{SURF Aerial View}
{Aerial view of the SURF site showing the locations of the
Surface Assembly Laboratory and the Surface Storage Facility.}
\end{figure}

The principal components of the surface infrastructure specific 
to LZ are: (a) the Surface Assembly Laboratory (SAL), (b) the 
Surface Storage Facility (SSF), and (c) office and meeting space 
at the SDSTA administration and education buildings. Office and 
meeting space already exists for ongoing experiments at SDSTA; 
general improvements to these capabilities will serve the broader 
experimental community at SURF, and are not described here. The 
locations of the SAL and the SSF at SURF are shown in Figure
~\ref{IAIf:AerialViewSURF}

The SAL was utilized for the assembly of the LUX detector and for 
operation of LUX in a water tank located in the SAL. Significant 
upgrades to the SAL are planned for assembly of the LZ detector 
and are described later. The most significant improvement to the 
SAL is the addition of radon-reduced air-handling capability. A 
new building to house a system that provides reduced-radon air to 
the SAL will be constructed between the SAL and the SSF. The SSF 
is currently used for general storage and modest interior
improvements were made in order to be used for storage and disposition of 
components during LUX decommissioning, which began in September 2016. These improvements will also provide staging space for LZ components.

\tdrsubsec[SAL]{Surface Assembly Laboratory (SAL)}
The SAL building is a wood-frame structure with four levels: the 
surface level and levels $-1$, $-2$, and $-3$, which are successively 
deeper below the surface. This building was renovated for LUX 
assembly and testing and has a sprinkler system, HVAC, and 
\SI{220}{kVA} installed electrical capacity. An existing cleanroom of 
\SI{2900}{\foot\cubed} will be used for cleaning and staging metal parts 
and bagged PTFE parts prior to Xe detector assembly in the 
low-radon cleanroom, and for other assembly tasks not requiring 
a low-radon environment. A new low-radon cleanroom will provide 
workspace beneath an existing monorail and over a pit in Level~$-1$. 
The new cleanroom will be connected, but separated by doors, 
from the existing cleanroom and will have a separate 
air-handling system. Transport of the detector will be facilitated by 
rolling it to the building shipping dock. The layout of the SAL 
is shown in Figure~\ref{IAIf:SAL}. The design of the new 
cleanroom is based in part on the design of a low-radon 
cleanroom in place at SDSM\&T~\cite{Street:2015nwa}. 
The design goal for this new cleanroom is \SI{<1}{\BqmT}
(unoccupied). 
The cleanroom is under fabrication by a commercial vendor to 
specifications set by SDSTA and the LZ Project.

\begin{figure}[H]
\centering
\includegraphics[width=0.8\textwidth]{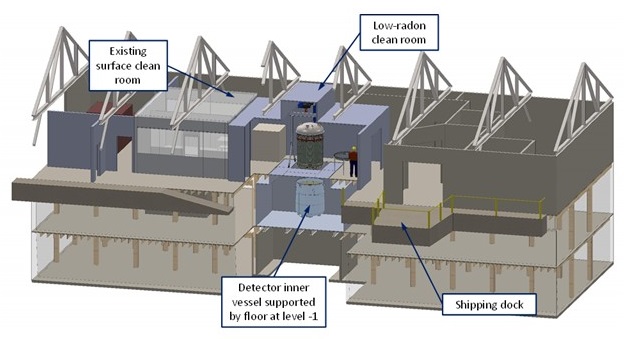}
\tdrfcaption[SAL]{The Surface Assembly Laboratory}
{The Surface Assembly Laboratory.}
\end{figure}

\tdrsubsec[Rn-redAirSys]{Radon-reduced Air System}
Radon is one the highest-risk contaminants for low-background 
experiments because it can easily escape from bulk material and 
quickly diffuse into active parts of the detectors. The innermost 
materials and parts of ultralow-background experiments must be 
manufactured or assembled in a radon-suppressed atmosphere. Final 
assembly of the LZ time projection chamber (TPC) will occur in a 
cleanroom in the SAL, which has a radon-reduced air system.

\begin{figure}[H]
\centering
\includegraphics[width=0.4\textheight]{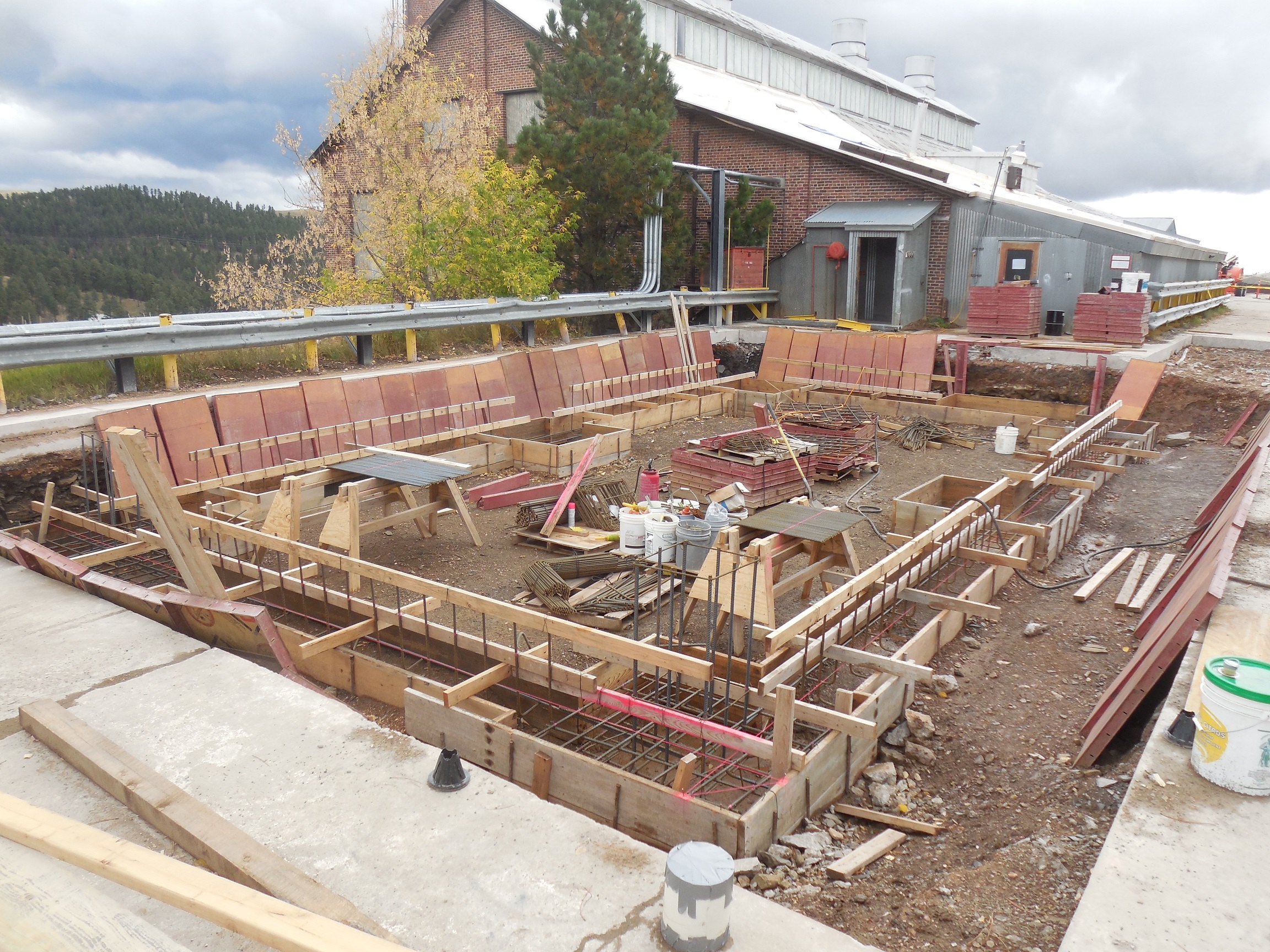}
\tdrfcaption[AdditionalBuilding]{Plan view of additional 
building to be constructed}{A picture (October 2016) of the additional 
building being constructed to house the radon-removal system (RRS).}
\end{figure}

\begin{wrapfigure}{o}[0pt]{0.65\textwidth}
\centering
\includegraphics[width=0.65\textwidth]{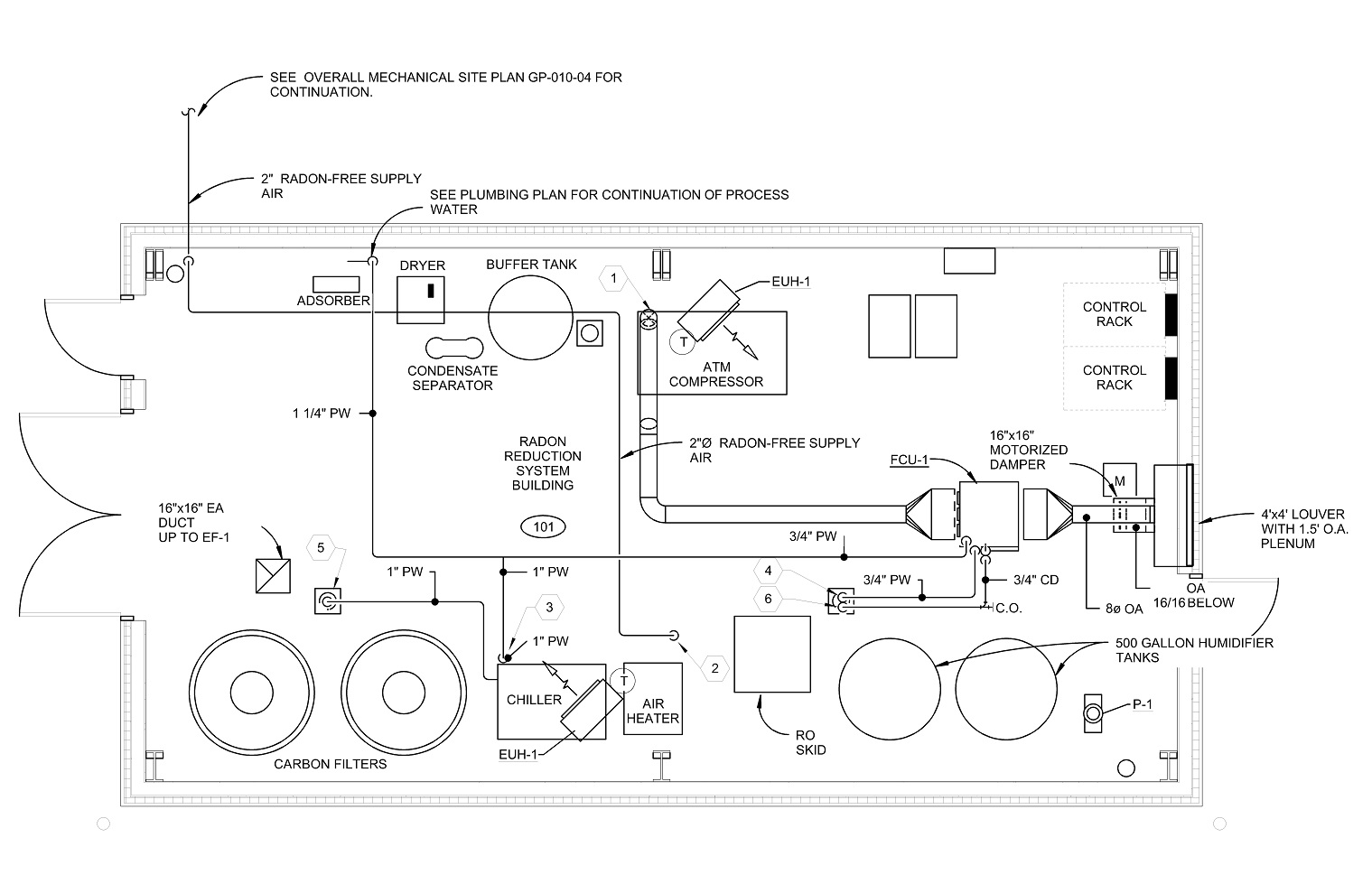}
\tdrfcaption[RRSBuilding]{Plan view of RRS Building}
{Plan view of RRS Building.}
\end{wrapfigure}

Carbon adsorption is the preferred technique for removing radon 
from air and has been effective for varying degrees of radon 
contamination, from as low as a few \si{\mBqmT} for low-
background experiments to levels as high as \SI{\sim100}{\BqmT}.
The surface air radon level at SURF is expected to be in the 
range of \SI{\sim20}{\BqmT}. The LZ detector assembly cleanroom 
will be designed with a radon-reduction factor of greater 
than \num{1000}. It will be flushed with radon-reduced air at a flow-
rate of about \SI{250}{\m\cubed\per\hour}, which is produced by 
compressing, drying, cooling, and pushing the air through two 
\SI{1600}{\kg} activated-carbon towers. Our plan is 
to use a commercial unit provided by ATEKO, which has developed 
and built large commercial radon-removal systems for other 
sensitive underground experiments. These units are capable of 
reducing radon concentration in the air by a factor of greater 
than \num{1000}. The fabrication of the LZ unit has begun for delivery in Spring 2017.

The process is 
based on compression, cleaning, and drying (dew point 
\SI{-70}{\degree C} min.) of air, cooling to 
\SI{-55}{\degree C}, adsorption of radon from air 
on activated carbon at \SI{-50}{\degree C} at approximately 
\SI{8}{bars} of pressure, followed by heating and pressure 
reduction of air to ambient pressure and temperature.  
A new building is being constructed to specifically house 
this system, and will be located between the SAL and SSF 
on an existing concrete pad illustrated in 
Figure~\ref{IAIf:AdditionalBuilding}. 

A plan view of the building, showing the location of key 
components, is illustrated in Figure~\ref{IAIf:RRSBuilding}.

\tdrsubsec{Surface Storage Facility (SSF)}
Short-term storage and staging of equipment needed for experiment 
assembly – including the large scintillator tanks - will be 
facilitated by upgrades to an existing building next to the SAL. 
The SSF has a large high bay with approximately \SI{445}{\square\m} 
(\SI{4800}{\square\foot}) of lay-down space under a 40-ft-wide 10-ton 
bridge crane. Hook height on the bridge crane reaches \SI{7.3}{\m}
(\SI{24}{\foot}). Handling of equipment will also be facilitated 
by three wall-mounted jib cranes rated at \num{1} ton, \num{2} tons, 
and \num{3} tons, respectively. Truck entry to the building is via 
a \num{4}-m (\num{13}-\si{\foot})-wide $\times$ \num{5}-m 
(\num{16}-\si{\foot})-tall entry rollup door. A structure within 
the SSF has been built to first house materials and equipment 
resulting from the removal of LUX from the Davis Cavern and 
decommissioning, and then LZ components. A view of this building 
showing scintillator barrels (about \SI{50}{\percent} of the total) and 
transport boxes containing the outer detector acrylic vessels is 
shown in Figure~\ref{IAIf:SSFBuilding}.

\begin{figure}[H]
\centering
\includegraphics[width=0.75\textwidth]{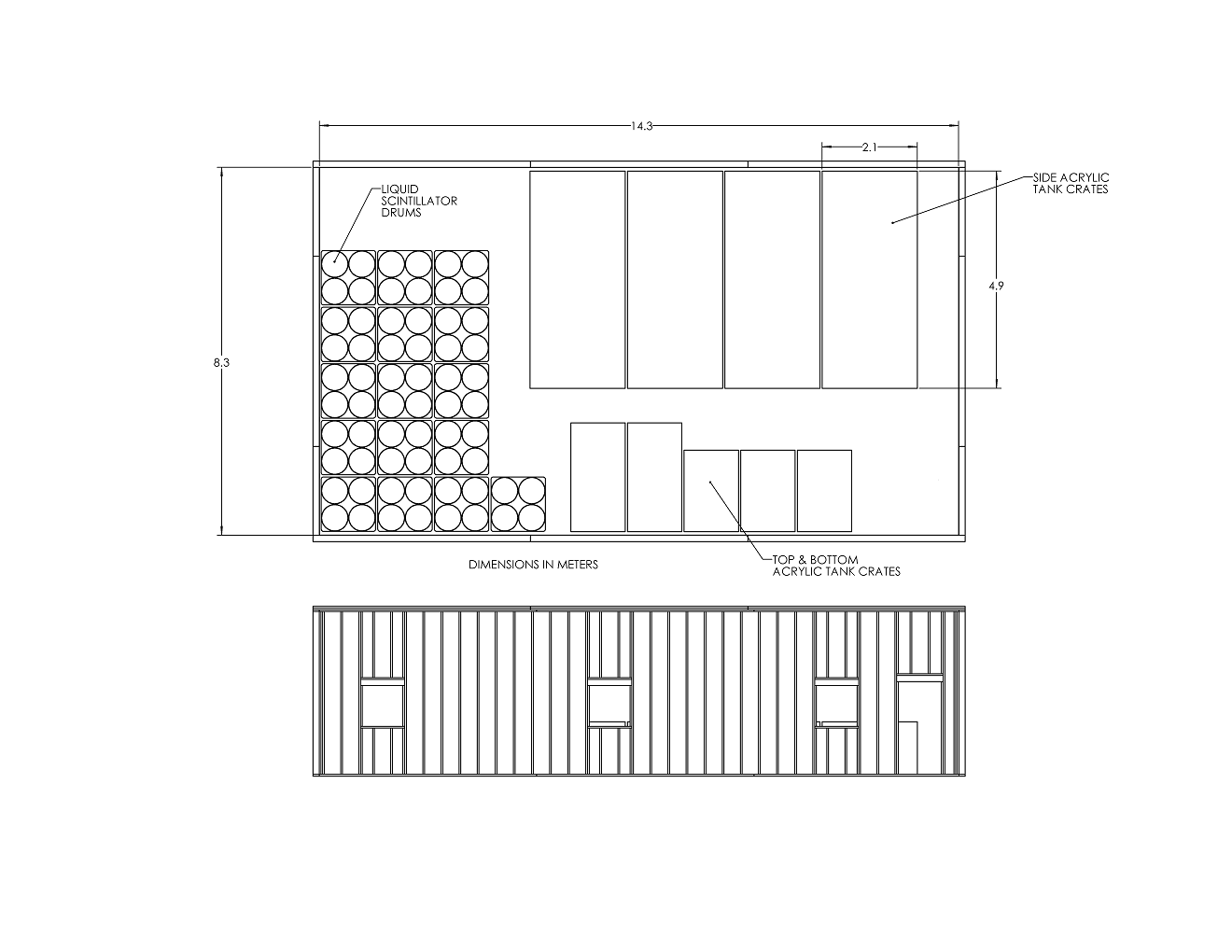}
\tdrfcaption[SSFBuilding]{Building constructed for staging 
LZ detector components}{Building constructed for staging 
LZ detector components.}
\end{figure}

\tdrsec[YatesShaft]{Yates Shaft Infrastructure and Custom Transport}
The LZ TPC inside the inner cryostat will be transported from the 
SAL to the Yates headframe in a horizontal orientation. The 
transport method will be similar to the successful transport of 
LUX from the same building to the headframe. A large telehandler 
was used, along with continuous monitor of orientation and 
G-forces. LUX was transported in the Yates cage but the LZ 
assembly will be slung under the cage, given its larger size. 
The LZ detector/inner cryostat will be placed on a custom 
transport frame. 

\begin{figure}
\begin{tabular}{ll}
	\begin{minipage}[b]{.49\textwidth}
		\includegraphics[width=0.99\textwidth]{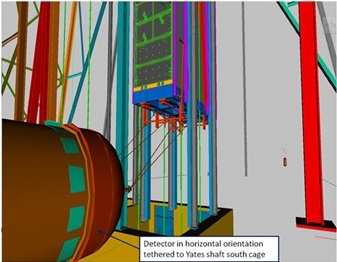}
	\end{minipage}
	&
	\begin{minipage}[b]{.49\textwidth}
		In the first image, 
		the LZ detector lies horizontally at the shaft entrance and 
		the cage bottom is 6 to 7 feet above the floor. Crews will 
		cover the shaft by installing a work platform and attach the 
		slings to the bottom of the cage. These are also attached 
		to the top of the LZ detector. The platform will be removed 
		and the cage hoisted to remove slack.  Another wire rope 
		control line from a winch is attached to the bottom of the 
		detector.
	\end{minipage}
\\
	\begin{minipage}[b]{.49\textwidth}
		\includegraphics[width=0.99\textwidth]{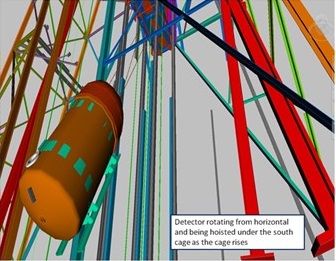}
	\end{minipage}
	&
	\begin{minipage}[b]{.49\textwidth}
		In the second image, 
		the hoist is lifting the LZ detector in the shaft to its vertical 
		position. The control line mounted on the detector bottom 
		will maintain tension in order to control the swing into the 
		shaft. Even though not fully in the shaft, the bottom of 
		the custom transport frame is off the ground. Once 
		hoisting is in process, it will not be stopped unless a 
		difficulty arises. The entire procedure is short-lived.
	\end{minipage}
\\
	\begin{minipage}[b]{.49\textwidth}
		\includegraphics[width=0.99\textwidth]{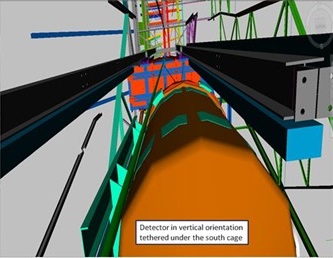}
	\end{minipage}
	&
	\begin{minipage}[b]{.49\textwidth}
		In the third image, 
		looking almost vertically. Here, the detector is hanging 
		vertically. It will be positioned properly in order for the 
		crew to attach the bumpers or guide shoes for clearance 
		and centering during shaft transit. After inspections, a crew 
		will man the cage and watch as it travels underground, 
		being lowered at a rate of \SI{0.7}{\foot\per\second}.
	\end{minipage}
\\
\end{tabular}
\tdrfcaption[TransportSequence]{LZ transport method from building 
to headframe}{The transport method of the LZ TPC inside the 
inner cryostat will be similar to the successful transport of LUX 
from the same building to the headframe.}
\end{figure}

Once the detector/inner cryostat assembly arrives at the Yates 
headframe, the assembly on the transport frame will be lowered in 
the Yates shaft, as shown in Figure~\ref{IAIf:TransportSequence}. 
Lowering the assembly on its transport cart slung under the Yates 
cage will take some hours. A crew member will accompany the 
assembly in the Yates conveyance, as was done for LUX. The 
assembly and transport cart will be extracted from the Yates 
shaft at the 4850L and rotated back to the horizontal 
orientation. A trial lowering in the Yates shaft of a mockup of 
the transport frame and the cryostat has been completed 
successfully.

\tdrsec[Underground]{Underground Infrastructure}
A plan view of the Davis Campus is shown in Figure~\ref{IAIf:DavisPlan}.
After arriving on the 4850L, the detector will be transported 
to the Davis Cavern via the Primary Access Drift, passing by the 
room housing the MAJORANA detector, as depicted in Figure~
\ref{IAIf:DavisPlan}. That is the same access-way utilized for transport of 
the LUX detector. The larger LZ inner cryostat can be moved 
through the same space by temporarily removing short segments of 
the Davis Campus HVAC air-supply ducts to allow for sufficient 
clearance. All access doorways are sufficiently large to allow 
for detector passage. Entry to the Davis Cavern via the Primary 
Access Drift allows the transporter design to take advantage of 
the 8,100-lb maximum floor loading afforded by design of the 
Davis Cavern structural steel. It would also be possible to 
transport the inner detector in the drift that passes by the LN 
storage room, but this would require the removal (and later 
replacement) of a wall between the drift and the Davis Cavern.

\begin{figure}[H]
\centering
\includegraphics[width=0.85\textwidth]{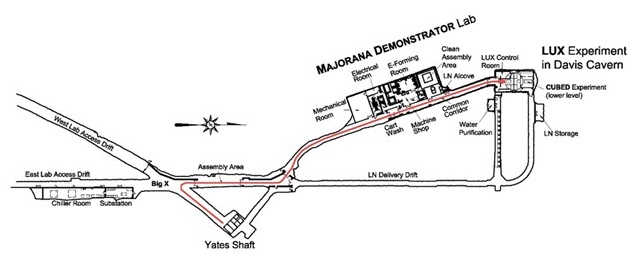}
\tdrfcaption[DavisPlan]{Plan view of the Davis Campus at the 
4850L}{Plan view of the Davis Campus at the 4850L.}
\end{figure}

LZ will take advantage of design features built into the Davis 
Cavern to allow for deployment of a larger detector than LUX. A 
\num{102}-inch-diameter flange is built into the \num{70000}-gallon water 
tank. Additionally, the redundant \num{50}-ton water chillers and the 
Davis Campus \SI{1500}{kVA} substation were implemented with a large 
future detector in mind.

Figure~\ref{IAIf:DavisCampusLayout} shows an isometric view of the LZ 
deployment in the Davis Campus. Xenon storage cylinders will be 
securely deployed in a newly upgraded portion of the present LN 
storage room access drift (shown in the lower-right corner). 
Currently, this is an unfinished area that has been used for rock-
moving equipment and storage. SURF will upgrade this area by 
providing concrete floors and covered walls while also 
closing off the space to allow for flow-through ventilation. 
Ventilation will be adequate to mitigate potential oxygen -
deficiency hazards that could occur in the event of a release 
from one of the Xe storage cylinders. The Xe cylinders are more 
thoroughly described in Chapter~\ref{chap:XCS}. 

The LN storage rooms and control rooms used for LUX will be 
largely reused. LZ will utilize a similar scheme for LN storage 
tanks until a second cryocooler is secured for long-term 
operational flexibility. These tanks will supply makeup nitrogen 
to the cryocooler, assist with transient startup effects, purge 
gas for the water purification system, and provide some buffer 
against short-term power interruptions. The control room is 
expected to remain unchanged, providing minimal office space for 
underground workers during assembly, commissioning, and data 
taking.

The ventilation duct that exhausts the LN storage room will be 
modified to also ventilate the Xe storage room. New supplemental 
ducting and a fan added to the existing ventilation will provide 
fresh air flow.

\begin{figure}[H]
\centering
\includegraphics[width=0.9\textwidth]{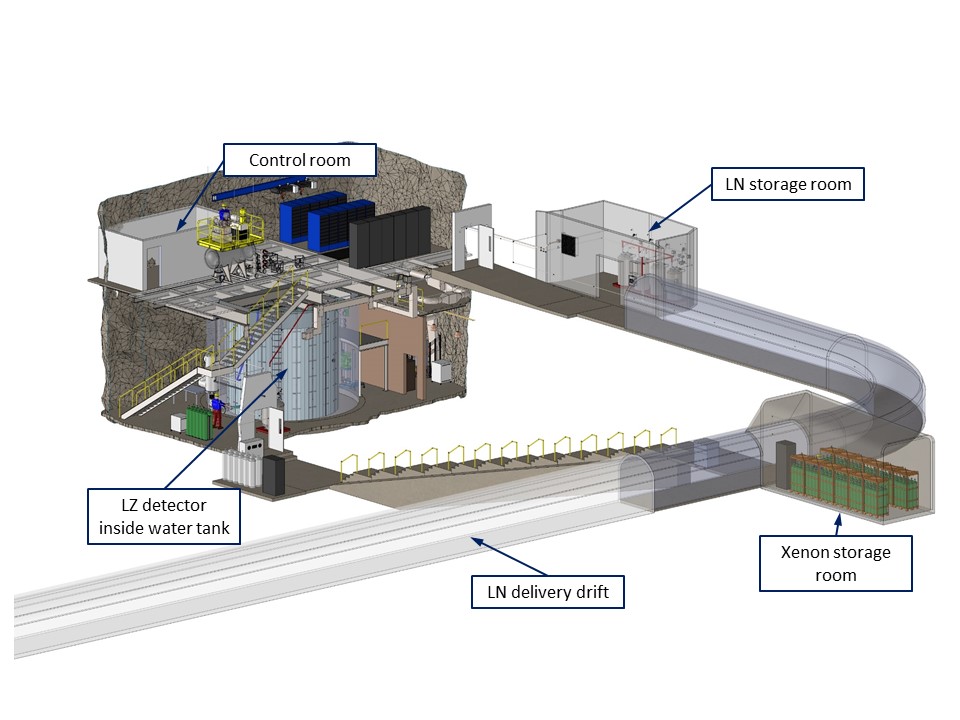}
\tdrfcaption[DavisCampusLayout]{Overall layout of LZ in the Davis 
Campus}{Overall layout of LZ in the Davis Campus.}
\end{figure}

Figure~\ref{IAIf:LZ-Davis} depicts a close-up view of the Davis Cavern 
showing supporting infrastructure installed. Key elements include 
data acquisition (DAQ) cabinets, a cryocooler to support cooling 
thermosyphon lines, a platform that interfaces to the PMT and sensor 
cable breakout, a XE purification tower for circulation heat exchange, 
and Xe circulation/storage equipment deployed both in and above the 
current low-background counting rooms in the lower Davis Cavern.

To facilitate deployment of a larger number of DAQ racks and 
cryocoolers on the deck of the upper Davis Cavern, SURF will 
decommission and remove the existing cleanroom to open up floor 
space. A new platform deployed near the PMT cable breakouts will also
serve to significantly increase the effective floor space in the cavern.

All other Xe circulation, storage, and recovery equipment is co-
located in space behind block walls previously utilized for low-
background experiments. Included are the Xe recovery compressors. 
These large \SI{480}{\V} units are positioned near existing in-cavern 
power panels and also the low-pressure/higher-loss suction side 
of the Xe gas system. 

The Davis Campus is fed from a \SI{1500}{kVA} substation. The 
substation is presently loaded at approximately \SI{60}{\percent} 
during worst-case conditions, and has ample reserve capacity to 
accommodate the projected LZ electrical load of about \SI{115}{\kW} (an 
increase of about \SI{\sim75}{\kW} from LUX). The campus 
has a \SI{300}{\kW} backup generator that supplies the air-
handling systems, communications, facility control and alarm 
systems, and egress lighting for up to 48 hours. An additional \SI{40}{\kW} 
generator will be installed to provide backup power for LZ.

A system to provide reduced-radon air during critical assembly 
steps inside the water tank will be located underground just 
outside the entrance to the clean area of the Davis Campus in an 
existing space. This system will provide reduced-radon air for 
the few months of critical connections to the Xe detector and possibly later. Air 
will be piped from this system to the water tank. The baseline 
plan for this system follows closely from a similar system 
constructed at the South Dakota School of Mines and Technology
\cite{Street:2015nwa}. 

Because the water tank installed for the LUX experiment was 
designed to be able to accommodate a much larger experiment, the 
modifications required for LZ are modest. 

\begin{figure}[H]
\centering
\includegraphics[width=0.85\textwidth]{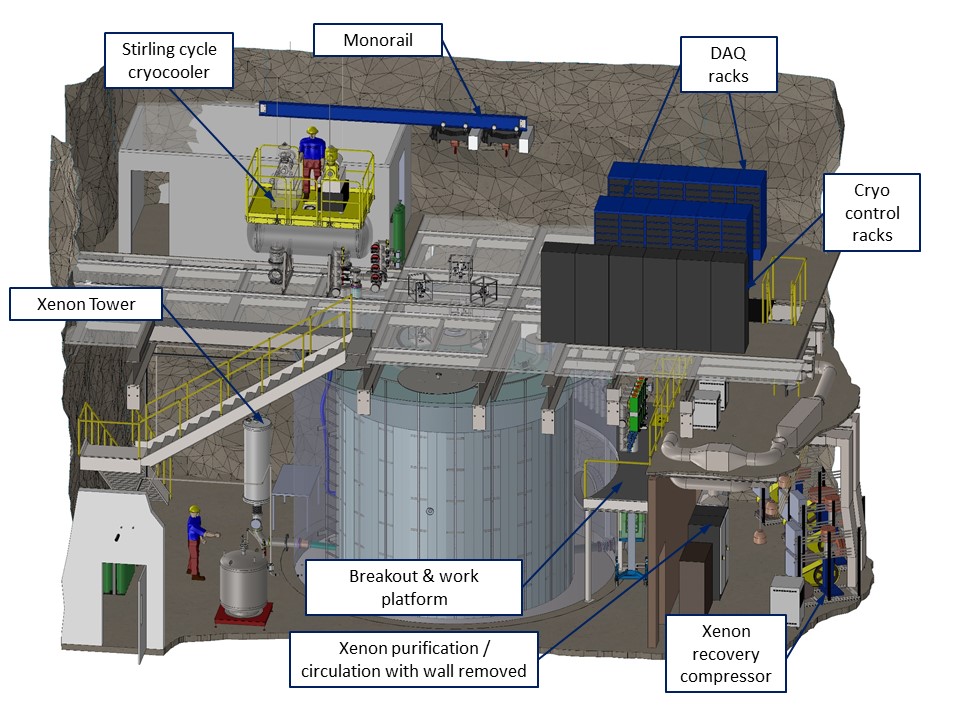}
\tdrfcaption[LZ-Davis]{LZ and related support systems in the 
Davis Cavern}{LZ and related support systems in the Davis Cavern.}
\end{figure}

Mounting plates will be added to the tank floor for the LZ 
detector along with mounting points for the four side 
scintillator vessels. At larger radius, mounting points will be 
installed for the outer detector PMT ladders. In addition, a support for the 
lower conduit that connects the detector to the LXe tower and PMT 
cable breakout will be installed. 

The detector HV system for LZ requires a feedthrough in the top 
of the water tank. In addition, two penetrations of the tank wall are 
needed for routing the conduits that connect the detector to the 
LXe tower and the PMT cable breakout. Two neutron tubes will be 
installed for the neutron calibration system, and that tube will 
be filled with water for normal running and nitrogen for 
calibration. The top of the water tank will also have several 
feedthroughs to accommodate thermosyphons, detector cables, 
source tubes, scintillator lines, water PMT cables, LED flasher 
cables, and vacuum pumping.

\tdrsec[IntegrationAssembly]{Integration and Assembly}
This section describes the effort to integrate work at the 
subsystem level into a coherent design that meets the science 
requirements and that will result in an operational detector at 
SURF. A detailed assembly sequence is presented. While the 
overall scope of integration and assembly is contained in WBS 
1.9, significant resources from every subsystem are required. 
Much of this effort comes from the distributed pool of engineers 
working for subsystems at many institutions. Coordination of this 
engineering effort is part of the integration task. 

\tdrsubsec[Integration]{Integration}
For LZ to achieve its science goals, it uses primary requirements 
that define what the subsystems must do; subsystem requirements 
drive more specific design specifications (see Chapter~\ref{chap:SRD}). 
The subsystems must be safe, affordable, 
timely, compatible with the other subsystems, and possible to 
assemble and operate with the available infrastructure. The 
Integration Group provides the necessary management, engineering, 
design, organizational tools, and administrative effort to assist 
all subsystems in completing the LZ design. Phone meetings and 
technical workshops help identify interface issues and hidden 
constraints created by design choices in other subsystems. An 
interface matrix is used to record and highlight interfaces 
between WBS areas. Interface Control Documents (ICDs) are 
authored by the key driver of the interface and reviewed and 
approved by all stakeholders. The Integration Group maintains CAD 
models of the overall LZ detector and of the Davis Campus. This 
helps define physical interferences and design gaps that are not 
adequately covered. At integration meetings, engineers from each 
subsystem, and many scientists, share ideas to help with some of 
the more difficult design challenges.

The Integration Group develops general standards and controls, 
such as engineering document standards to be used project-wide 
for CAD file exchange, engineering drawings and design notes, 
specifications and procedures, controled Technical Documents (CTDs), technical change (TCs) documents, and a document numbering 
and organization system. The Integration Group defines component 
reference names and an overall coordinate system for the 
experiment. The group maintains a cable and feed-through list 
with the help of the subsystem management. A key parameters list 
is maintained as a reference for design, modeling, and science.
 
As designs mature, they must be documented and reviewed. The 
Integration Group works with project and subsystem management to 
arrange and execute design reviews at appropriate times. 
Conceptual Design Reviews evaluate whether the design meets the 
requirements, interfaces have been identified and successfully 
coordinated, and engineering details are sufficiently developed 
to proceed. Preliminary Design Reviews focus on 
manufacturability, cost, schedule, risk, and safety. Final Design 
Reviews ensure that documentation and drawings are complete and 
the fabrication plan fits within project budgets and timelines. 
All LZ systems have completed Final Design Reviews. Production Readiness Reviews for 
WBS 1.2, WBS 1.5 (PMTs),  
and parts of WBS 1.6 (liquid scintillator production and acrylic vessels) have taken place.

LZ is being assembled and installed in an underground area 
administered by SURF. Integration includes working with SURF to 
be sure the infrastructure is adequate to support assembly, 
installation, and operation of the LZ experiment. SURF engineers 
are tasked with design and execution of infrastructure projects 
to support the LZ project. The Integration Group coordinates 
communication of requirements with SURF and the detector 
subsystems.

The overall planning of the on-site assembly and installation of 
the detector at the Davis Campus is part of the integration 
effort. This work includes defining the sequence of steps to put 
the detector together, creating a schedule for this work, and 
developing an understanding of the resources needed to accomplish 
it. Subsystems support this effort by providing details for 
handling components and aiding in resource and schedule 
development.  Subsystems also develop detailed work procedures 
that technicians can follow for cleaning, assembly, and testing. 

\tdrsubsec[Assembly]{Assembly}
The LZ assembly will happen in three stages---off-site 
subassembly, surface assembly in the Surface Assembly Lab (SAL), 
and underground assembly in the Davis Campus---with 
transportation between stages. The general strategy is to do as 
much work as practical off site at universities and national 
laboratories, where highly skilled, specialized labor can easily 
work with students and scientists. This also reduces travel 
costs. Parts delivered to the site will be tested, clean, and 
ready to use, with a few exceptions. The development of detailed 
delivery requirements and procedures, particularly cleanliness 
relevant to achieve the low-radioactivity needed, is under active 
development. A database and electronics logging tools will be 
used to track the location and status of parts and subassemblies, 
from initial fabrication through delivery to SURF, and up to 
assembly into the final detector. Assembly workers will check 
that the part has been approved for radioactivity, cleanliness, 
and function before using it on the detector.

Xe PMTs will be assembled, tested, and characterized prior to 
delivery to SURF. The PMT vendor will do some QA testing, but burn
-in, final electrical characterization, and cold testing of each 
PMT will be done by LZ. Assembly includes connecting a PMT base 
to the PMT, attaching polytetrafluoroethylene (PTFE) reflectors, 
and final cleaning. Because they have PTFE as a reflector, the 
assemblies must be kept under a nitrogen purge during storage and 
shipping to prevent radon contamination.  Internal PMT cables will have a pin-and-socket 
connection at both ends. They will ship separately and be routed 
to the PMTs after the PMTs are installed on the support arrays. 
The warm end of the cables will be connected to the feed-throughs 
mounted in flanges after the cables are routed. The cables must 
be kept clean and under nitrogen purge during storage and shipping 
because they contain fluorinated ethylene propylene (FEP) as the 
primary insulator. 

The tested PMTs will be placed into the titanium PMT support 
structure, including preliminary placement and routing of the 
cables. The baseline plan is to do this work in a cleanroom at 
Brown University. The titanium arrays will be mounted in the two  
``PMT Array Lifting And Commissioning Enclosures''
 (PALACE -- see Section~\ref{XDSSs:PMTassy}), a multipurpose 
light- and gas-tight enclosure. 
The PALACE will be opened for PMT, reflector, and cable 
placement, and then closed and purged with nitrogen between 
operations and during testing. This work will create an upper and 
lower populated PMT array. The PALACEs will also be used to 
protect the arrays during shipping. The upper skin PMTs will be mounted into teflon trays that attach to the sides of the TPC.  This work will be done in the reduced radon environment of the SAL clean room at SURF.  The lower skin PMTs will be attached to a titanium support that is mounted to the bottom of the ICV.  This work will also be done in the SAL.  

The wire grids for cathode, gate, anode, and PMT shields will 
also be manufactured off site. The grids will be cleaned, 
inspected, packaged, and shipped to SURF. The packaging must keep 
the wires clean from any debris and protect the fragile wires 
from shock during shipping. Boxes of the same design as  
the PALACEs will be used for grid storage, shipment, and testing 
to save on design and manufacturing cost. Inspection procedures 
after fabrication and after arrival at SURF are under development 
and will be guided by experience in the system test at SLAC (see 
Chapter~\ref{chap:XDS}). These could involve 
automated optical inspection and voltage testing in gas.

The field-grading region of the TPC is made from conductive metal 
rings, insulating PTFE spacers, and resistors. These parts will 
 be fabricated by vendors and then inspected and cleaned off site 
at LBNL before shipping to SURF. Radon exposure of completed PTFE 
parts will be minimized after final machining. They can be stored 
and shipped in nitrogen-filled metalized bags.

The cryostat will be manufactured in Italy as fully tested and 
cleaned code-stamped vessels under contract with the Imperial College. 
The fabrication vendor has the responsibility for final design, 
fabrication, testing, cleaning, and shipping to SURF. The inner 
and outer vessel will be shipped separately (not nested) in 
nitrogen-filled sealed bags from the cleaning facility. Receiving 
acceptance for the cryostat will include visual inspection for 
any shipping damage, separate assembly of each vessel, and vacuum 
leak-checking with helium. Rate of rise leak checking of the double O-ring seals will also be performed to get a baseline rate of rise for each joint.  The reflective PTFE liner for the inner 
vessel will be attached to the inner cryostat vessel wall in the 
SAL. The plan is to suspend the bottom of the inner cryostat 
upside down in the vessel-assembly area and to access the inside 
of the vessel from underneath with a manlift. This will allow a 
worker to tile the inside while standing on a stable surface.  The bottom skin PMT array will be installed into the bottom of the ICV.  The ICV will be sealed after the line is installed.  A radon emanation test of the lined ICV can be performed both to discover if there are any large radon issues and to establish the emanation from this part of the assembly so it can be compared to the emanation of the full assembly later. 

The HV umbilical will be delivered from LBNL as a clean, tested, 
and sealed assembly. The heat-exchanger tower subassembly will be 
built, cleaned, and tested at the University of Wisconsin Physical 
Sciences Lab, and transported as a sealed assembly. The Gd-LS 
acrylic tanks will be manufactured, cleaned internally, tested off 
site, and shipped in protective packaging. 
They will be stored in the SSF. The outside of the Gd-LS tanks 
will be cleaned on site, underground. The ladders, PMTs, and 
Tyvek reflectors of the outer detector will be shipped separately. 

It is unlikely that the timing of delivery can match the time 
when each piece is needed. The SSF is the identified storage 
location at SURF. A temperature-controlled storage room has been built in the SSF to store the Gd-LS and the acrylic tanks. 
 
The subassembly work and transportation described above will 
primarily be the responsibility of the subsystems. Once things 
arrive at SURF, primary responsibility shifts to the Integration 
and Installation Team (WBS 1.9). The plan is to work with SURF to 
hire a pool of local technicians to perform much of the work. 
This pool will include experts in cleaning, vacuum, 
rigging, and mechanical assembly. A lead technician will handle 
managerial supervision and work direction, but technical 
supervision will be supplied by engineering and scientific staff. 
An LZ engineer will be on site to coordinate the work and provide 
technical oversight. Experts from subsystems will be on site 
during assembly of their subsystems. Existing SURF engineering 
and technical personnel will also provide support.  Worker training 
and work control will be performed as part of the SURF safety program. 

On-site assembly starts above ground with assembly of the TPC. 
Cleanliness is critical for this assembly work, both to reduce 
radioactivity backgrounds in the detector and to control 
particles that could cause field enhancements and reduce 
operating voltage. This work will be done in the SAL at SURF 
with a large reduced-radon cleanroom, described previously. 
Figure~\ref{IAIf:SurfaceAssemblySequence} shows the major steps 
of assembly in the SAL. The first step is receiving and 
inspection of the parts and subassemblies. Clean parts destined 
to enter the cleanroom will be shipped in triple bags. The first 
bag will be removed in the receiving area to keep dust and dirt 
out of the cleanroom area. The second bag will be removed in a 
soft-walled semi-cleanroom at the entrance to the cleanroom. 
The third bag will be removed inside the cleanroom after the 
particle count in the room has returned to normal after opening 
the door. The goal of the inspection is to ensure that the parts 
have not been damaged in shipment and that they will meet 
functional requirements. This includes cleanliness. The PMTs in 
the arrays will be tested to ensure they were not damaged in 
shipment. The grids will also be inspected.

The next several subsections describe the various stages of 
assembly. Each subsection concludes with a description of the 
suite of checkouts that will be performed prior to declaring the 
stage complete. Final definition of these checkouts is a crucial 
aspect of the assembly and will be made by WBS 1.9 in close 
consultation with the relevant subsystem owners.

\tdrsubsubsec[TPC-Assembly]{TPC Assembly (Steps SA2 -- SA6)}

\begin{figure}[h]
\centering
\includegraphics[height=0.8\textheight]
{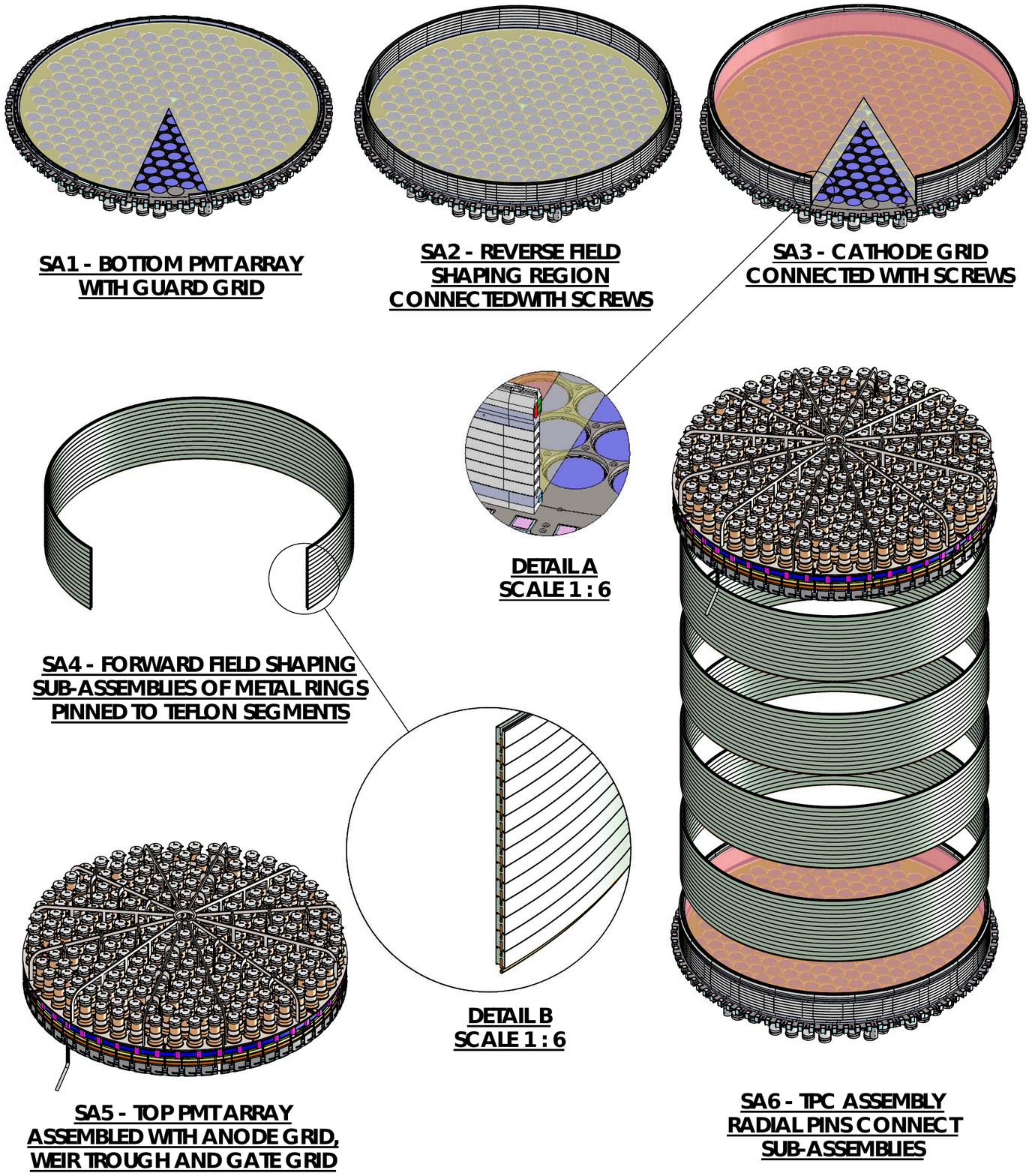}
\tdrfcaption[SurfaceAssemblySequence]{LZ surface assembly 
sequence}{TPC assembly forms subassemblies of the lower PMT array with 
reverse field region and cathode grid; the upper PMT array, anode grid, 
weir trough and gate grid; and four short stacks of field shaping grids.  
These are then combined into a complete assembly}
\end{figure}

The assembly will start with the lower PMT array (see 
Chapter~\ref{chap:XDS} for descriptions of the components 
referenced here) The lower PMT array will be set upside down
on a ring shaped cart for initial assembly. The cart will have 
telescoping legs and locking wheels that allow adjustment to a 
comfortable working height and easy movement into an extra-clean 
storage garage.  Eighteen skin 2-in PMTs will be 
connected to the titanium truss of the lower array. Xe return 
liquid distribution tubing and manifolds will be mounted and 
dressed. Loop antennae and temperature sensors will then be 
placed and cabled. The fully populated bottom subassembly will be 
rotated \SI{180}{\degree} and set back on the ring cart.  A layer 
of arc shaped PTFE segments is screwed to the titanium support 
plate and then the PMT guard grid is placed on top of the PTFE.  
The reverse field-grading assembly composed of arc shaped 
PTFE segments and complete metal rings will be assembled onto the 
PMT array next. The parts are held together with axial PEEK screws that pass through a counterbored hole in the top PTFE part, a through hole in the titanium field shaping ring, and into a PEEK nut press fit into a coutnerbored hole (farside) in the bottom PTFE part.  Leads from resistors that fit into pockets in the PTFE will be 
attached with screws to electrically connect the metal rings as 
they are stacked. When the reverse field region has been stacked to 
the correct height, The cathode grid is installed on the top of the 
stack.  An additional layer of PTFE segments is placed on top of the 
cathode to allow connection to the forward field shaping subassemblies.  
The completed bottom subassembly will be stored in the garage when 
it is completed.  Subassemblies are also stored in the garage whenever 
active work is not being done on them.  This will reduce contamination 
of surfaces with dust. 
Sections of the forward field-grading assembly will be built on 
similar carts with ring shaped tops.  The open support structure is 
designed to allow airflow through the pieces being assembled so less 
dust will accumulate on the work pieces. The forward field PTFE segments 
and rings are held together with radial pins.  This allows the assemblies 
to be joined with only access from the outside.  There will be four 
forward field subassemblies.  The extraction region will be assembled at LBNL and shipped as a subassembly consisting of the weir trough, anode grid, gate grid, and a few layers of field shaping rings.  The extraction region sub assembly will be mounted to a similar cart and the upper PMT array will be rigged in place over it.  The extraction sub-assembly will be by connecting to the titanium plate of the upper array assembly.  The 93 upper skin PMTs will be attached to the outside of this assembly.
The cables from these PMTs will be carefully dressed with the 
other PMT cables. Voltage-control cables will be connected to the 
grids. Loop antennas, temperature sensors, and position sensors 
will then be placed and cabled. The completed extraction region 
subassembly will then be stored in the garage.

The full TPC will be assembled by moving carts holding sections 
under one side of the monorail, lifting the sections off the 
carts with the monorail crane, transporting them to the other 
side of the monorail, and lowering them onto the stack. The 
sections will be connected with PEEK radial pins. Before a 
subassembly is lifted, it will be inspected and tested for dust 
accumulation. This test and possible cleaning mitigations are 
under development. The assembled TPC will undergo a series of 
tests to ensure light-tightness of the field-grading cylinder, 
and function of all PMTs, HV grids, resistor network, and sensors. 
\clearpage

\tdrsubsubsec[TPC-Insertion]{TPC Insertion into Cryostat with Fluid 
and Electrical Final Routing}
The inner cryostat vessel (ICV) will be staged in the lower level 
of the cleanroom under a removable grated floor. The first step 
of insertion is to remove the lid of the ICV and raise it so the 
top flange is a few inches above the floor. The TPC will then be 
lifted with the hoist on the monorail, transported on 
the monorail over the ICV, and lowered into the ICV. Guide bars will be attached to the sides of the ICV and go vertically up to the top of the TPC to help guide the TPC as it is lowered.  During this 
process, the PMT and sensor cables coming from the lower PMT 
array will be routed through the central port in the bottom of 
the inner cryostat. Xe fluid circulation lines also routed 
through this port will have been placed into position earlier, 
but may need adjusting and securing as part of the cable routing. 
Access for this operation will be through the HV connection port 
and the bottom port of the vessel. The three Xe tubes from the 
weir trough must be routed to the ports in the inner cryostat 
wall. These tubes are PTFE bellows that will initially be 
pointing straight down and then guided into the ports as the TPC 
is lowered. The TPC is supported on six posts projecting upward 
from the bottom head of the inner cryostat. Tapered guide pins 
will be installed in the bottom mounting holes of the TPC to 
engage the holes in the posts and guide the TPC into the correct 
position. Once the TPC is in the correct place, the guide pins 
will be removed and bolts will be installed to secure the TPC to 
the posts. The titanium plate for the upper PMT array will be 
guided from the inner cryostat near the main flange with tabs 
that allow vertical motion but constrain radial motion. Access 
for this work is from the top over the main flange. After these 
upper guides are secured, the TPC will be fixed in the vessel. 
The location of the weir surfaces that establish the LXe plane 
will be surveyed relative to known positions on the outside of 
the inner cryostat. This will allow rough leveling of the weir 
surface during future assembly steps. The next step is to stage 
the lid of the inner cryostat over the bottom and install a 
temporary safety support. The PMT cables, grid cables, sensor 
cables, and Xe gas lines coming from the upper PMT array need to 
be routed through a port in the inner cryostat lid. The sensors 
that monitor the position of the TPC relative to the inner 
cryostat wall will be installed and cables dressed. The lid can 
then be lowered onto the bottom of the inner cryostat and the 
large-diameter seal formed. This seal is designed as two seals 
(inner helicoflex metal seal and outer elastomer o-ring seal) to 
facilitate a check for leaks by pumping between 
the seals and looking at the rate of pressure rise. This rate can be comared to the baseline rate established during cryostat acceptance helium leak checking.  All other 
ports on the inner cryostat must be sealed for transportation to 
keep the TPC clean. The cables will be wiped clean and tested for 
dust before they are pulled through a long bellows. The bellows 
will be sealed to the flanges on the ICV and leak-checked. The 
outer end of the bellows will be capped. The sealed inner 
cryostat will be pumped down to vacuum to ensure the seals are 
adequate. Other testing will be done at this point. This could 
include PMT functional tests, sensor tests, light-tightness tests 
between the skin and central TPC volume, and HV tests.  A radon emanation test of the completed and closed vessel can be performed at this point.  Closed-cell foam insulation and superinsulation blankets will 
be fit-tested on the outside of the inner cryostat for later 
underground installation. After these tests, the bellows will 
be dressed to the outside of the ICV and the ICV with TPC 
assembly will be double-bagged so it is ready to be moved 
underground. The special transport frame for this move will be 
brought into the clean room.  The vessel will be rotated from 
vertical to horizontal using two hoists on the monorail and raised 
so the special transport frame can be placed beneath it.  
The vessel will be lowered on the special transport frame and 
secured for transport.

\tdrsubsubsec[Underground-OD]{Underground Outer Detector Tank 
Preparation / Staging (Steps U1 -- U2)}
The first step of underground assembly is to prepare the water 
tank. All LUX components have been removed.  Infrastructure work includes reorienting the overhead monorail crane so it runs North South.  Some welding is needed in the water tank for attachment points for the 
cryostat support, the Gd-LS tanks, and the outer detector PMT 
ladders. New penetrations are needed for the HV umbilical, the 
heat exchanger (HX) conduit through the wall of the tank, and 
calibration tubes.  After 
these welding operations, the tank will be passivated again to 
improve corrosion resistance to the pure water. The water tank 
will then be cleaned and made into a cleanroom with 
reduced-radon air delivered from the system located underground. 
The tank will be kept with slightly positive pressure to reduce 
air infiltration. The access door in the side of the tank will be 
outfitted with a temporary changing room and air lock.

\begin{figure}[H]
\centering
\includegraphics[height=0.77\textheight]{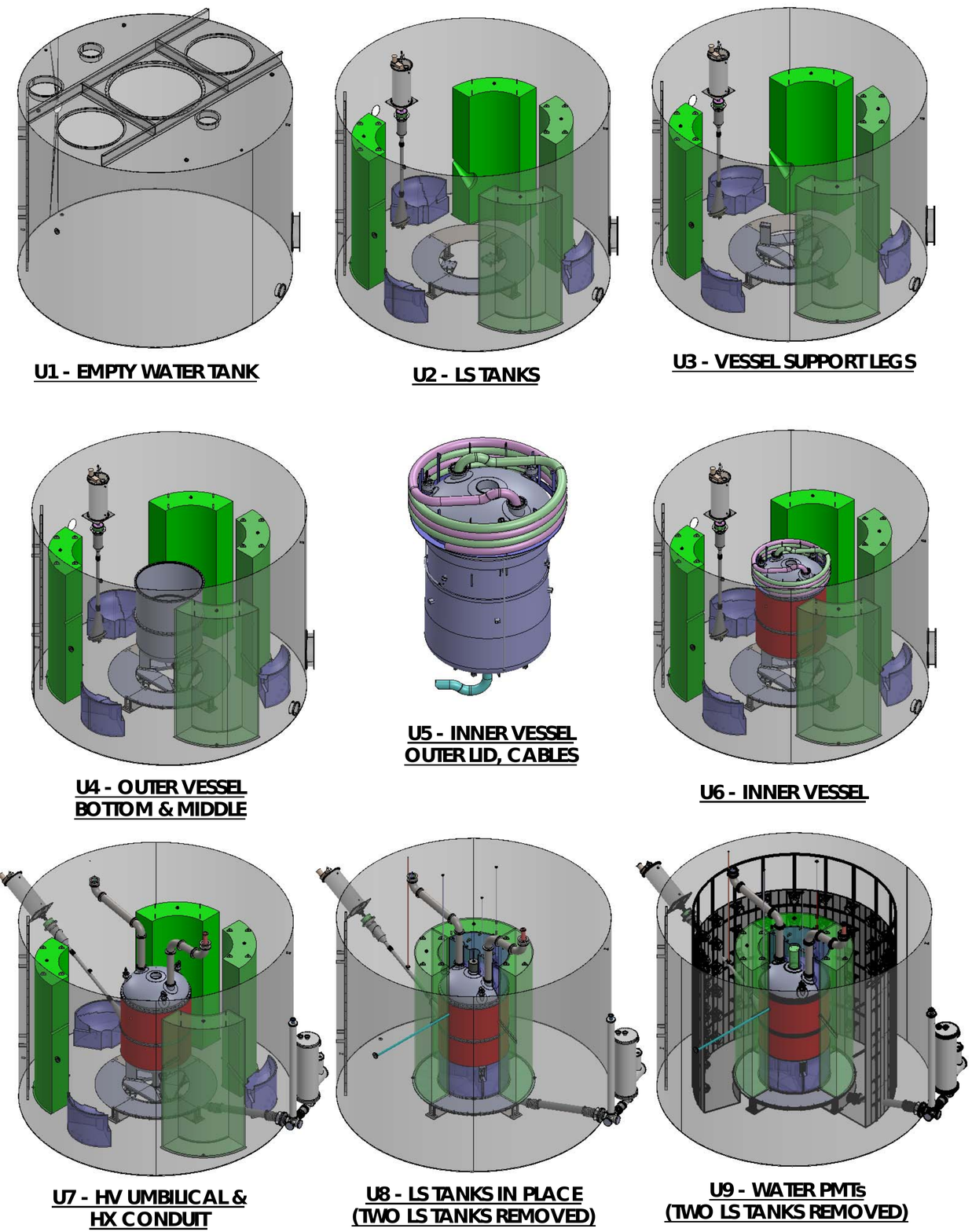}
\tdrfcaption[OD-Installation]{Underground installation sequence 
in water tank}{Underground installation sequence in water tank.}
\end{figure}

The central top port of the water tank is the only port big 
enough to allow installation of the Gd-LS tanks. Once the 
cryostat is installed, this path will be blocked. So the bottom 
and side Gd-LS tanks must be transported underground and staged 
in the water tank. Figure~\ref{IAIf:OD-Installation} shows the 
sequence of installation in the water tank. The Gd-LS tanks will 
arrive clean on the inside and covered with protective plastic 
and rigging frames. The acrylic tanks will be brought into the 
Yates headframe with a telehandler and set on a cart in front of 
the Yates cage. The four large side tanks will be moved first. 
Rigging will be used to attach the steel frame around the tank to 
the underside of the cage. The cage will be raised to lift the 
tank until it is vertically under the cage. A drag line will 
guide the lower end of the tank. The cage will be lowered slowly 
until the bottom of the tank is at the 4850L. A drag line will be 
reconnected to the bottom to pull the tank out as it is lowered 
further. The tank will be placed on a receiving cart, which will 
transport the tank to the entrance to the Davis Campus 
cleaning area (so-called cart wash). 
The external frame and external packaging around the tank will be 
cleaned to remove mine dust from transport. The bottom of the 
tank must enter the Davis area first. One hook from the monorail 
in the Davis Cavern will connect to two points near the bottom of 
one side of the rigging frame. A second hook from the same 
monorail will connect to the top of the tank’s rigging frame. The 
motion of the hooks will be choreographed to lower the acrylic 
tank into the water tank, keeping the rigging vertical over the 
lifting points by moving the hooks along the monorail. Once the 
tank is vertical and set down on the floor of the water tank, it 
will receive a final inspection and leak test. The external 
packaging and protective plastic will be removed from the acrylic 
tank. A temporary beam with hoist will be installed to the inside 
top of the water tank to move the Gd-LS tanks inside the water 
tank and aid with unpackaging.  The outside of the acrylic tanks 
will be cleaned with Alconox and water and visually inspected for 
any cracks. The three bottom tanks and the two top tanks will be 
transported inside the cage and unpackaged and cleaned in the 
cart-wash area. The three bottom tanks will be staged inside the 
water tank. The two top tanks will not be staged, as they will be 
installed from the top after the cryostat and cable-conduit 
installation is completed.  Stainless steel 
support stands that hold the four tall outer Gd-LS tanks will 
also be staged into the water tank. The HV umbilical and parts 
for the HX conduit may also be staged in the water tank. A simple 
mockup that had the appropriate dimensions for the largest 
acrylic vessel transport frame was successfully lowered in the Yates 
shaft and partially transported underground.

\tdrsubsubsec[CryostatTransport]{Cryostat Transportation and 
Underground Assembly (Steps U3 -- U6)}
The installation of the cryostat starts with the cryostat 
support. The survey reference system for the detector will be 
established and a template for drilling the anchor bolts for the cryostat support leg will be located.  Nine anchors will be installed into holes that are drilled and tapped into the thick steel shielding plates beneath the water tank.  Once the anchor rods are installed, nuts will lock them to the floor.  These nuts need to be seal welded to the bottom of the tank and the threaded rod so water can not leak out.  This work will most likely be done during the modification to the water tank.  Three machine leveling jacks will be place on the floor of the water tank temporarily.  Three mounting plates will be installed onto the anchor plates and supported by the leveling jacks.  The cryostat support legs, cleaned and triple-bagged, will be brought down 
in the cage. The outside bag will be removed at the entrance to 
the Davis Campus. The inside 
bag will be removed on the top deck near the opening to the water 
tank. The legs will be lowered into the water tank through the 
central port with a single hoist. The final bag is removed inside 
the water tank. The legs will be set into position on the mounting plated and loosely secured.  
The outer cryostat has been designed as three pieces so each 
piece can fit in the Yates cage. Each piece will be brought down 
in the cage cleaned and triple-bagged. The bottom head of the 
outer cryostat is lowered into the water tank with a single hoist 
and positioned over the three support legs.  Once the bottom head is 
lowered, it will be bolted to the legs. The bottom section is then surveyed and located using the machine leveling jacks.  Once it is in place, the other bolts and nuts are tightened.  The final position is then surveyed for verification.  The middle section will 
be brought in next, rigged into position above the bottom, sealed 
to the bottom, and leak-tested. The middle section will be 
surveyed to ensure it is level. Adjustment will be needed if it 
is out of range. A displacer around the outer cryostat 
cylindrical section reduces the amount of water between the Gd-LS 
tanks and the cryostat. This may be installed around the outer 
cryostat at this point or after the HV connection is completed. 
The top lid will be brought into the Davis Campus and staged on 
the top deck with its inner bag still in place.

The inner cryostat will not fit in the cage, so it must be hung 
under the cage. It will be horizontal for some of its journey 
from the SAL to the Davis Campus, and vertical for other parts. The TPC 
support system will be designed to accommodate support in both 
conditions and the transition between them. An external rigging 
company will be contracted to remove the ICV from the SAL and 
transport to the Yates headframe. It is anticipated that this 
move will also be done with a telehandler. There is a detailed 
plan for transport of the inner cryostat from the surface to the 
4850L, described earlier in this chapter. On the 4850L, a special 
cart with air skates will be used to bring the inner cryostat 
from the Yates shaft to the Davis Cavern (this was the method 
used for LUX). The special transport frame must be cleaned. The outer bag 
around the cryostat will also be removed in the cleaning area. 
The transport frame will move the inner cryostat onto the deck near the 
entry hole to the water tank. The deck plates above the central hole of the water tank will have already been removed.  A temporary platform will be built resting on the lower flanges of the beams that support the Davis deck.  This temporary platform will be about 19 inches below the main floor.  The platform is designed to roll along the flanges of the beam so it can be moved clear of the large port in the water tank.  The two hooks on the monorail will 
be used to lift the inner cryostat off the transport frame, rotate it back 
to vertical, and rest it on the temporary platform. The bellows full of cables exiting the bottom of the cryostat will need to be managed during this lift.  A temporary cleanroom will be built around the inner 
cryostat. Reduced-radon air from the underground system will be 
used to provide a clean atmosphere. The inner bag around the 
inner cryostat will be removed. Closed-cell foam insulation will 
be installed onto the sides and bottom of the cryostat.  The LXe weir drain lines will be connected and routed.  The cryostat cooling thermosyphon evaporators will be attached to the fins on the cryostat walls.  Temporary supports for the outer cryostat lid will be installed 
onto the inner cryostat. The outer cryostat lid that was 
previously staged will be unbagged, rigged over the inner 
cryostat, and set on the temporary supports. The two upper 
bellows will need to be guided through the ports in the outer 
cryostat lid as it is staged. The permanent support rods that 
hold the inner cryostat from the outer cryostat will be 
installed. The supports will be adjusted to position the weir 
surface of the inner cryostat parallel to the sealing surface of 
the outer cryostat lid. Then 
prefabricated superinsulation blankets will be installed. The lower section of the three 
calibration ports will also be installed.  The outer cryostat lid will then be 
lifted and the load from the inner cryostat will transfer from 
the temporary supports on the deck to the permanent supports from 
the lid. The temporary platform will be rolled along the flange of the beam until it is clear of the water tank opening.  The 
assembly of the outer cryostat lid and inner cryostat will be 
lowered into the water tank and into the outer cryostat bottom. 
As the inner cryostat is lowered, the bottom bellows will need to 
be threaded through the central port of the outer cryostat bottom 
head. The reduced-radon air will again be flowing into the water 
tanks for this process. The crane will set the assembly down so 
the outer cryostat lid rests on the outer cryostat middle-section 
top flange. The inner cryostat will still be hanging from the 
lid. This flange can then be assembled and leak-checked.

\tdrsubsubsec[UtilityConnection]{Utility Connection (Step U7)}
The lower cable bellows are routed to the edge of the water tank 
through a vacuum jacket. Xe transport lines have a separate 
vacuum jacket and connections will be made with an orbital welder 
whenever possible. Some connections may be made using VCR 
fittings. These lines continue through the outer wall of the 
water tank to the cryo tower. The lower PMT cable conduit continues vertically after penetration of the water tank wall and connects to the breakout box.  The cables continue into the breakout boxes and connect to an
array of flanges and hermetic feed-throughs. The upper bellows 
will be routed through vacuum jackets to the appropriate flanges 
on the top of the water tank.  One bellows connects to vacuum 
pumps and Xe recovery system plumbing; the other connects to 
another breakout box with an array of flanges and hermetic feed-
throughs. During the connection of the cables in the breakout 
boxes, reduced-radon air will be routed through the ICV to 
prevent back diffusion of mine air into the ICV. This area has 
many details that must be carefully planned.
 
The HV umbilical attaches to the large side port. To reduce 
krypton and radon absorption by the internal plastic components, 
this assembly will also be purged with nitrogen or reduced-radon 
air whenever it is opened. The HV umbilical is a flexible assembly that connects to the top of the water tank and the side of the cryostat.  
The central cable of the umbilical needs to be electrically 
connected to the cathode. The inner tube of the umbilical then 
seals against the inner cryostat. The flange will have a 
double seal (inner helicoflex and outer o-ring) so it can be 
leak-checked at this point. Then 
the outer vacuum jacket will slide down toward the detector and 
make a seal to the outer cryostat. This seal will have a 
double o-ring so it can be leak-checked. Sealing rings at the 
water tank wall are installed to seal the inner tube to the outer 
tube and the outer tube to the vacuum tank. There are no direct 
water-to-Xe seals.

The final step of cryostat installation is positioning and 
leveling. The connections for the conduit and HV umbilical add 
load and positional constraints to the hanging inner cryostat. 
The support rods will be adjusted using feedback from built-in 
electronic level sensors. We have designed in enough compliance 
to these connections so the inner vessel can be moved. The 
cryostat should now be sealed and the reduced-radon air flow can 
be stopped. The inner cryostat will be pumped down to start long-
term outgassing of the internal plastics. 

\tdrsubsubsec[OD-Assembly]{Outer Detector Assembly (Steps U8 -- U9)}
The Gd-LS tanks can now be placed into final position. The three 
bottom tanks are set on platforms connected to the cryostat legs. 
The upper Gd-LS tanks are then installed through the top port of 
the water tank. They are lowered slightly radially outward from 
their final positions to clear the PMT cable conduit and 
thermosiphon conduits and, once they are low enough, translated 
under the conduits to the correct final position. The upper Gd-LS 
tanks are supported by the top flange of the outer cryostat. The 
four side tanks will be moved in adjacent to the displacer around 
the outer cryostat and rest on stainless steel supports. There 
are notches for the HV umbilical so the tanks have to come in 
from the proper direction. After positioning, the tops of the 
tanks are connected for stability. Each tank is secured so it 
will not float in the water or tip over in an earthquake. Each 
tank has a fill line and vent line that come to a common overflow 
reservoir on the top of the water-tank lid. The final system will 
be visually inspected and leak-checked with a low-pressure gas.
  
The outer detector PMTs will be installed onto half-ladders and 
lowered into the water tank through one of the larger off-axis 
ports in the lid. The half-ladders are assembled together and 
secured to the roof and floor of the water tank and cables are 
run up to one of the top water-tank ports. Cables are sealed at 
these ports. The Tyvek reflectors that direct light lie on the 
floor, are hung vertically from the top of the ladders, and are 
stretched across the top of the ladders.

The detector is now ready to be filled with Xe, Gd-LS, and water. 
Xe filling must wait until the inner cryostat has been at vacuum 
long enough to get the residual gas content of the plastics to an 
acceptable level. Warm low-pressure Xe gas may be circulated to 
heat the plastic and enhance diffusion. This Xe would be pumped 
out and repurified or sold. Once plastic outgassing is at an 
acceptable level, the vessel will be filled with Xe gas and 
slowly cooled with LN2 until the Xe starts to condense. Filling 
continues until the Xe liquid is at the desired level. Gd-LS is 
received on site ready to use in 55-gallon drums, temporarily 
stored in the SSF. The Gd-LS and water must be filled at the same 
time to minimize stress on the acrylic walls. The levels do not 
need to match exactly, so one drum of Gd-LS can be added to the 
tanks one at a time as the water level rises. It is added by 
pressurizing the drums with nitrogen gas to force Gd-LS through 
the filling tubes at the overflow reservoir. Water is purified 
before it is added, and covered with a nitrogen head once the 
tank is filled. A flowing nitrogen head is maintained over the Gd-
LS to protect it from both radon and oxygen.

While the main detector installation and assembly sequence 
described above are occurring, the support equipment and 
utilities for the experiment will be installed in Davis. This 
includes cryogenic cooling equipment, vacuum pumps, L\CNt 
thermosiphons, Xe purification and circulation equipment, TPC HV 
supplies, PMT readout electronics, PMT HV supplies, calibration 
source tubes, connections and hardware, the emergency Xe recovery 
system, DAQ, and control systems. Details of these items are 
covered in other chapters.

The duration of the work in the SAL from the start of the 
assembly of TPC to the inner cryostat being sealed and ready to 
move underground is expected to be about seven months. Before 
underground installation work can begin, LUX must be 
decommissioned and removed (by very early 2017) and Davis 
infrastructure work described earlier in this chapter will need 
to be completed. LZ installation underground has an estimated 
duration of seven months from staging of the Gd-LS tanks to being 
ready to fill.  

\clearpage
\bibliographystyle{apsrev4-2}
\bibliography{LZtdr}

\wlog{In tdrinclude command}
\wlog{\ChapLabel}

\renewcommand{\ChapLabel}{chap:OCS} 
\renewcommand{\BibLabel}{OCSS:bib} 
\renewcommand{\SecPre}{OCS}
\renewcommand{\FloatPre}{OCS}
\tdrchap{Offline Computing and Software}

\tdrsec[Intro]{Introduction}

This section describes the LZ offline computing systems, including offline
software for the LZ experiment, the definition of the computing environment, the
provision of hardware and manpower resources, and the eventual operation of the
offline computing systems.

The offline computing organization provides the software framework, computing 
infrastructure, data-management system, and analysis software as well as the
hardware and networking required for offline processing and analysis of LZ data.
The system will be designed to handle the data flow starting from the raw event
data files (the so-called EVT files) on the SURF surface RAID array, all the way
through to the data-analysis framework for physics analyses at collaborating
institutions, as illustrated in Fig.~\ref{OCSf:dataflow}.
\begin{figure}[hbtp]
\begin{center}
\includegraphics[width=0.9\textwidth]{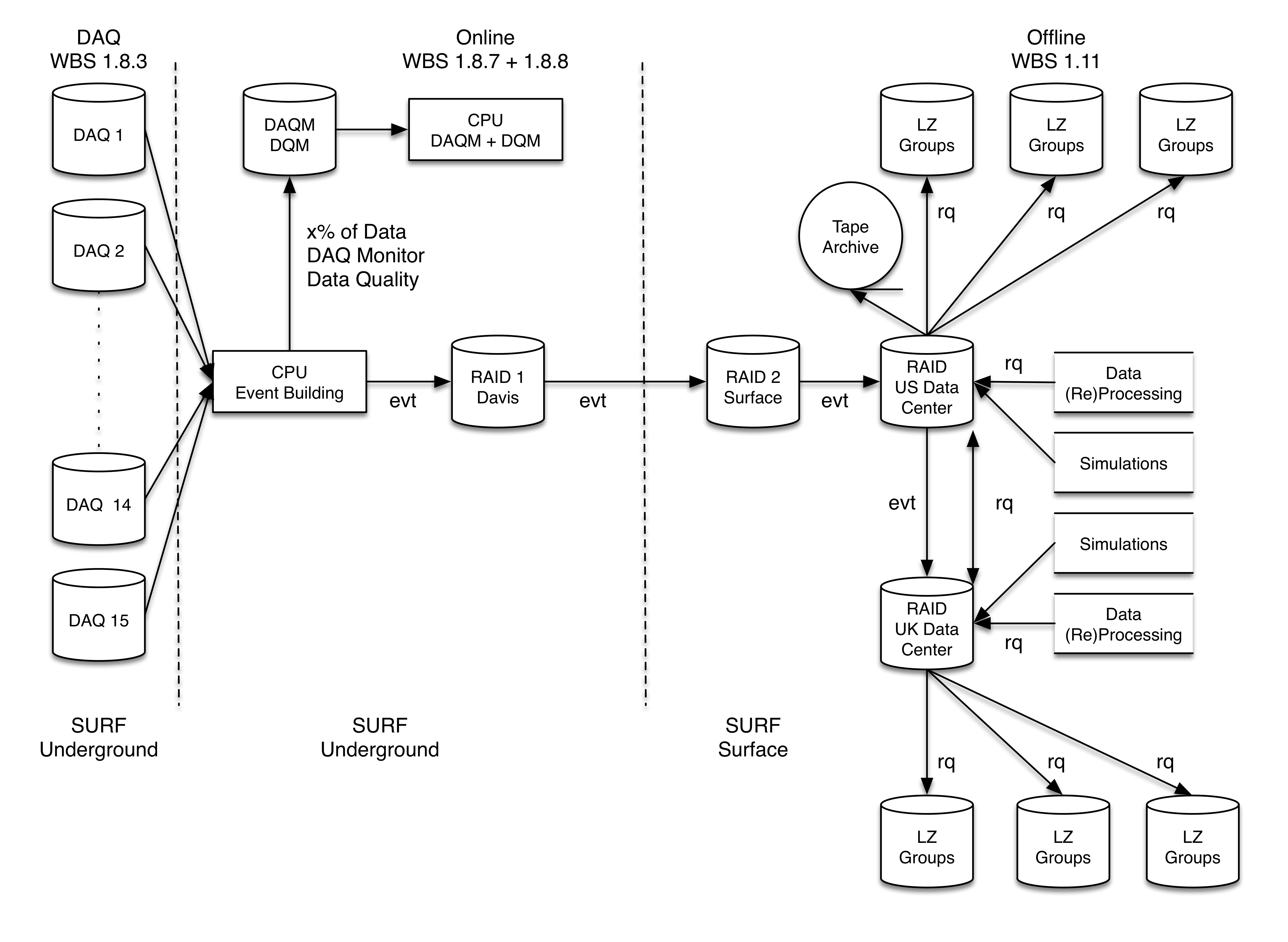}
\end{center}
\tdrfcaption[dataflow]
            {Schematic data-flow diagram for LZ}
            {Schematic data-flow diagram for LZ.}
\end{figure}
%
\tdrsec[DvDpDC]{Data Volume, Data Processing, and Data Centers}
%
The LZ data will be stored, processed and distributed using two data centers,
one in the U.S. and one in the U.K.
Both data centers will be capable of storing, processing, simulating and
analyzing the LZ data in near real-time.
The SURF surface staging computer ships the raw data files (EVT files) to the
U.S. data center, which is expected to have sufficient CPU resources for
initial processing.
The National Energy Research Scientific Computing (NERSC) center at LBNL will
contain the resources to act as the LZ U.S. data center.

The run processing software (LZ analysis package, or LZap for short) extracts 
the PMT charge and time information from the digitized signals, applies the 
calibrations, looks for S1 and S2 candidate events, performs the event reconstruction, 
and produces the so-called reduced quantity (RQ) files. The RQ files will be accessible 
to all groups in the collaboration and represent the primary input for the physics analyses.

The EVT and the RQ files are also mirrored from the U.S. data center to the U.K.
data center (UKDC, located at Imperial College London) partly as a backup, and partly
to share the load of file access/processing, giving better use of resources for
all LZ collaborators.
The EVT file transfer to the U.K. data center is done from the U.S. data center
as opposed to directly from SURF, in order to avoid any reduction of the
bandwidth available to ship the raw data from the experiment.
Subsequent reprocessing of the data (following new calibrations, reconstruction
and identification algorithms, etc.) is expected to take place at one or both
centers, with the newly generated RQ files copied to the other center and made
available to the collaboration.

From the hardware point of view, the system must be able to deal with the LZ
data volume in terms of storage capacity and processing. 
Based on the LUX experience and appropriate 
scaling for LZ (in terms of number of channels, single/dual gains, event rates, etc.), 
the amount of WIMP search data generated in \num{1,000} days of LZ running is estimated to be 
\SI{2.8}{PB}.
Including calibration runs, the total amount of LZ data produced per year is
expected to be \SIrange{1.3}{1.4}{PB}, depending on the amount and type of calibration
data collected during yearly operation. This estimate assumes that about three hours of 
calibration data are collected each week. The breakdown of the different sources of data in
LUX and their scaling to LZ is given in Table~\ref{OCSt:dailyrates}, which
clearly shows that the data volume is dominated by the S2 signals.
\begin{table}[h]
\setlength{\extrarowheight}{5pt}
\sffamily
\tdrtcaption[dailyrates]
            {LZ daily data rates.}
{Daily raw and compressed data rates in LZ based on scaled LUX data.
 The scaling factors have been computed as follows:
 (a) PMT surface area ratio (2) times number of channels ratio — not including
     the low-gain channels (4), $2 \times 4 = 8$;
 (b) number of channels ratio (8) times rate ratio (13), $8 \times 13 = 104$;
 (c) liquid surface area ratio.
 The compression factor is taken to be 3, as described in the text.
 (Abbreviations: PE = photoelectron, SE = single electron.)}
\begin{center}
\begin{tabular}{|l|r|r|r|r|} \hline
\rowcolor{mrocol}
Source         & LUX (GB/d)& Scaling factor& LZ (GB/d)& LZ compressed (GB/d)\\
\hline
Single PE      &  44.00 &   $8^{(a)}$ &   352 &   117\\
S1             &   0.24 & $104^{(b)}$ &    25 &     8\\
S2             &  76.34 & $104^{(b)}$ & 7,939 & 2,646\\
Uncorrelated SE&  20.00 &   $9^{(c)}$ &   180 &    60\\ \hline\hline
Total          & 140.58 &             & 8,496 & 2,832\\ \hline
\end{tabular}
\end{center}
\end{table}

The SURF staging computer will have a disk capacity of \SI{192}{TB}, enough storage
for slightly more than two months (\SI{68}{days}) of LZ running in WIMP-search mode
(at \SI{2.8}{TB/day}), similar to its underground counterpart.
The capacity of the staging arrays was based on the assumption that any network 
problems between SURF underground and the surface, or the surface to the
outside, would take at most several weeks to be fully resolved.
The remaining storage capacity can be used to store additional calibration data.
The anticipated data rates imply that the network must be able to sustain
a transfer rate of about \SI{33}{MB/s}. Such rates do not represent a particular 
challenge for the existing networks between SURF and LBNL or between LBNL and Imperial College.
We note that the LUX experiment currently sends data from SURF to the primary data 
mirror at Brown University with an average throughput of \SI{100}{MB/s}.

From the current LUX experience, we expect that processing one LZ event should
take no more than one second on one core (using a conservative estimate based
on an Intel Xeon ES-2670 at \SI{2.6}{\GHz} and \SI{4}{GB} of RAM per core).
Therefore, assuming a data-collection rate of \SI{40}{\Hz}, LZ needs 40 cores to keep
up with the incoming data stream in WIMP-search mode.
For reprocessing, as analysis software and/or calibrations are refined,
a larger number of cores will be needed to keep the processing time within
reasonable limits (e.g., a factor of 10 more CPU cores allows reprocessing of
a year’s data in approximately one month).

Simulated data will also be created and stored. The top-level estimates in 
terms of storage capacity and CPU power for Monte Carlo simulations based on 
existing simulations are summarized in Table~\ref{OCSt:simscomputing}.
They add up to a total data volume of about \SIrange[range-units=single]{85}{100}{TB}
and require approximately \num{e6} CPU hours per year.

\tdrsubsec[USDC]{The U.S. Data Center}

The U.S. data center will be located at NERSC/LBNL.
Currently NERSC has three main systems: the Parallel Distributed Systems
Facility (PDSF) and two CRAY systems.
PDSF provides approximately \num{3200} cores running Scientific Linux and
is a dedicated system for astrophysics, high-energy and nuclear physics
projects.
The CRAY systems, Cori Phase 1 and Edison, provide approximately 
\num{52000} and \num{153000}
cores, respectively.
All systems can access the Global Parallel File System (GPFS) with a current
capacity of about \SI{8}{PB}, which is coupled to the High Performance Storage
System (HPSS) with a \SI{240}{PB} tape robot archive.

The LZ resources will be incorporated within the PDSF cluster.
Our planning assumes modest needs for data storage and processing power for 
simulations, as described in Section~\ref{OCSS:Sims}, a rapid growth in
preparation for commissioning and first operation, and then a steady growth of
resources during LZ operations.
The planned evolution of data storage and processing power at the U.S. data
center is given in Table~\ref{OCSt:USDCstaging}.
The amounts of raw and calibration data per year are assumed to be \SI{1120}{TB} 
and \SI{320}{TB}, as described in the text, while the Monte Carlo data are 
ramped up to the maximum estimated capacity over the Project period (\SI{85}{TB}
from Table~\ref{OCSt:simscomputing}, increased to \SI{100}{TB} as a safety margin). 
Once the regular data-taking begins, the amount 
of simulations data is doubled in order to be able to accommodate both current 
and previous simulations. The processed data are assumed to be \SI{50}{\percent} 
of the Monte Carlo  simulations and \SI{10}{\percent} of the data (assuming a 
slightly higher percentage of \SI{10}{\percent} in the size of the RQ-files 
compared to the \SI{7}{\percent} in LUX).
The user data are assumed to be \SI{50}{\percent} of the Monte Carlo 
simulations in the years prior to experimental data, and \SI{5}{\percent} 
of the total data once LZ is running. The total disk space allocated includes 
a \SI{20}{\percent} safety margin with respect to the total amount of 
calculated data.
\begin{table}[h]
\sffamily
\setlength{\extrarowheight}{5pt}
\tdrtcaption[USDCstaging]{U.S. and U.K. data center staging.}
{Planned storage (in TB) and processing power by U.S. fiscal year at the U.S. and U.K. data centers.}
\begin{center}
\begin{tabular}{|l|r|r|r|r|r|r|r|r|r|r|r|} \hline
\rowcolor{mrocol}
FY               & 2015& 2016& 2017& 2018& 2019 & 2020 & 2021 & 2022 & 2023 & 2024 & 2025\\
\hline
Raw         data & $-$ & $-$ & $-$ & $-$ &  $-$ &  560 & 1680 & 2800 & 3920 & 5040 & 6160\\
Calibration data & $-$ & $-$ & $-$ & $-$ &  $-$ &  160 &  480 &  800 & 1120 & 1440 & 1760\\
Simulation data  &  40 &  80 &  80 & 100 &  100 &  200 &  200 &  200 &  200 &  200 &  200\\
Processed data   &  20 &  40 &  40 &  50 &   50 &  172 &  316 &  460 &  604 &  748 &  892\\
User data        &  20 &  40 &  40 &  50 &   50 &   55 &  134 &  213 &  292 &  371 &  451\\
Total data       &  80 & 160 & 160 & 200 &  200 & 1147 & 2810 & 4473 & 6136 & 7799 & 9463\\ \hline\hline
USDC: Disk space       &  40 & 220 & 220 & 220 &  220 & 1360 & 3360 & 5360 & 7360 & 9360 & 11360\\ \hline
USDC: CPU cores        & $-$ & $-$ & 175 & 350 &  350 &  390 &  830 & 1270 & 1710 & 2150 & 2590\\
\hline\hline
UKDC: Disk space       &  150 & 220 & 220 & 270 &  650 & 1597 & 3260 & 4923 & 6586 & 8249 & 9913\\ \hline
UKDC: CPU cores        & 150 & 175 & 350 & 350 &  350 &  390 &  830 & 1270 & 1710 & 2150 & 2590\\
\hline
\end{tabular}\end{center}
\end{table}

The CPU power is ramped up to reach the maximum of 350 cores needed by the 
simulations in two years (2016 and 2017) and is increased by 40 cores to 
390 cores in the commissioning year (2020) in order to be able to continue 
Monte Carlo production in parallel with real-time data processing (which 
assumes \SI{1}{\second/event} at \SI{40}{\Hz}). 
In the subsequent years of operation, the CPU power is increased by 440 
cores per year, in order to be able to perform full data reprocessing in 
a reasonable time, in addition to the real-time data processing 
(400 + 40 cores). The CPU estimates for simulations are based on the total 
number of \num{e6} CPU hours from Table~\ref{OCSt:simscomputing},
which yields an average of about 115 cores/year in a steady state operation.
Assuming that the simulations have a duty factor of about 1/6, i.e., 
run for 1 month and analyze/develop for 5 months, the total number of 
cores needed for simulations yields about 700, which is then equally 
divided between the U.S. and U.K. data centers.

Data flow from the surface data cache onsite to NERSC is automated by
use of the Spade system. Spade will transfer raw data files from SURF to
NERSC within 15 minutes of file close by the DAQ. At NERSC, data will be
written to NGF (NERSC Global File system) and automatically archived to
HPSS (High Performance Storage System) tape. Data integrity will be
verified by comparing checksums before and after transfer. From NERSC,
data will automatically be transferred to the UKDC and verified by a 
second Spade data pipeline.
Nominally (when network continuity is complete),
all DAQ data files will be automatically replicated at the USDC and UKDC
and backed up to HPSS within 30 minutes of DAQ file close.
Although the U.S. data center will be located at NERSC, other U.S.
computing resources are likely to be available to the collaboration. We
will utilize resources available to the collaboration as appropriate.

\tdrsubsec[UKDC]{The U.K. Data Center}
The U.K. data center is led by Imperial College, and built on an infrastructure of GridPP ~\cite{Faulkner:2006:gridpp,Britton:2009:gridpp} hardware and software. It currently runs Monte Carlo production and analysis jobs, providing the data products to the collaboration via GridFTP and XROOTD. In the remainder of the construction phase, it will run the end-to-end-simulation, event-reconstruction and analysis applications described in Sections~\ref{OCSS:Sims} and~\ref{OCSS:Framework}. The output will be background and signal models which capture the full physics of detection and measurement with the precision needed to predict dark-matter sensitivity. Tests with LUX data, system-test data, and simulation in the mock data challenges will assure that a validated pipeline is in place ready for first underground data. Timely commissioning, calibration and performance monitoring, as well as evaluation of final physics results, will all be helped by the high throughput, high availability and flexible scaling afforded by a Grid computing model.

UKDC will provide redundancy and parallel capacity for carrying out the 
first level, near real-time processing of all LZ raw data (when needed), 
and carrying out reprocessing of the entire data set on timescales of 
several weeks. The U.S. and U.K. data centers will use the same analysis 
framework, access the same central databases, and run identical software. 
The single, central installation is distributed via CVMFS and the single,
dedicated LZ virtual organization provides user authentication. Both are 
in place since June 2015, served from the University of Wisconsin.

Grid storage at UKDC was initiated in 2015 at \SI{150}{TB}, to store the 
results of background simulations informing this Design Report. As of October 
2016 the storage is \SI{220}{TB} and a gradual increase is planned in 2018 
and 2019 to reach \SI{650}{TB} at the time of detector deployment (see 
Table~\ref{OCSt:USDCstaging}). A further \SIrange{1}{1.5}{PB} per year, 
dependent on realized background rates and trigger strategy, will be added to 
maintain a complete copy of acquired LZ data.

Production using GridPP nodes at Imperial started during the simulation campaign of autumn 2015. Since December 2015, LZ has been using CPU resources at 5 GridPP sites with up to 2,000 jobs running simultaneously. This proves that the UKDC can meet its requirement of reprocessing WIMP search and calibration from scratch in one-tenth of the acquisition time. Transfer rates from UKDC to USDC were tested during the same simulation campaign. Performance using the globus data transfer toolkit with no optimization comfortably exceeded the 80~MBps required for a factor-of-two margin when mirroring acquired data.

The second large scale processing campaign was conducted in September 2016 for the validation of the background model. The simulations were performed for a modified geometry and for a larger number of detector components with increased statistics. A total of \num{5e10} events were generated within 411,000 CPU hours in one month. These files are using \SI{85}{TB} of disk space. In addition, the UKDC has been used on request, for specific studies thorough the year using significant CPU hours (283,000) and disk storage (\SI{>50}{TB}). The requests are currently managed via a custom job submission system using Google sheet documents. This system will be soon superseded by a production job submission system currently in development. 

The DIRAC system of middleware~\cite{Tsaregorodtsev:2008:dirac} is used to manage LZ data and computation on GridPP, providing command line, web portal, and Python API interfaces for submission, monitoring and accounting. Development has begun of tools to streamline building, editing, and merging the output of LZ jobs, using the object-oriented Python package Ganga~\cite{Moscicki:2009:ganga} (the same approach as the LHCb collaboration, which has long experience running the Gaudi framework on the Grid). The LZ-specific aspects of UKDC are handled by a researcher at \SI{50}{\percent} FTE and a programmer at \SI{10}{\percent} FTE. Support of non-LZ-specific systems, which can include development of Grid infrastructure software as required, is provided by the GridPP team at Imperial.

\tdrsec[Framework]{Analysis Framework}
%
The key element for the LZ data processing is the analysis framework
(AF), which will contain some standard processing modules and will 
also allow users to put together modular code for data analysis to 
automatically take care of the basic data handling (I/O, event/run 
selection, etc.).
A dedicated task force was created at the end of July 2014 to 
evaluate various options for LZ. In terms of existing frameworks, 
two {\normalfont \ttfamily ROOT}-based frameworks were considered: 
Gaudi (developed at CERN and used by ATLAS, LHCb, MINERvA, Daya Bay, 
etc.)~\cite{Barrand:2001ny} and {\normalfont\ttfamily art} (developed 
at Fermilab and used by MicroBooNE, NOvA, LBNE, DarkSide-50, 
etc.)~\cite{Green:2012gv}.
In parallel, the possibility of evolving the framework developed 
for LUX, which is based on Python scripts and a MySQL database, 
and supports modules written in Python, C++/{\normalfont \ttfamily ROOT}
or MATLAB was also evaluated.
For completeness, developing a new framework from scratch was also 
considered as another alternative.
However, given the amount of effort this would have required (of 
the order of at least several FTE-years based on estimates from other
experiments such as CMS, Double Chooz, MiniBooNE, T2K, etc.), this 
option was dismissed.

Input from the entire collaboration regarding the desired features 
for the LZ framework was collected and organized by the task force, 
and the three candidate frameworks were evaluated and ranked against 
this list.
In addition, presentations and live demos for each of the three 
contenders were given during the regular task force meetings, while 
core frameworks were also installed on different test platforms to 
evaluate the respective installation processes.

The task force ranked unanimously Gaudi in first place, followed 
by a distant second by {\normalfont \ttfamily art} and LUX in a close 
third position. The recommendation to adopt Gaudi as the LZ analysis 
framework was  approved by the collaboration in April 2015. With the 
adoption of Gaudi as the foundation for the analysis framework, the first 
milestone (see Table~\ref{OCSt:milestones}) was the creation of a First Release,
which was delivered to the collaboration in February 2016.
This release was based on the Gaudi Hive branch of the Gaudi code, which is 
specially designed to support multi-threading within the framework itself. 
Goal of the release was to demonstrate end-to-end processing of mock 
(user-generated) events through the entire framework chain, and to support
the development of different modules from a team of LZ collaborators.

A Physics Integration Release is currently under production and will be
delivered to the collaboration in November 2016. The features of the framework 
needed for this release are:
\begin{itemize}
\item Definition of the transient data model for DAQ and Physics data;
\item Creation of the necessary modules to read and write raw DAQ data based on the above transient model;
\item Provision of a programming and build environment in which the physicists can adapt existing modules and develop new ones to run within the framework;
\item Access to non-DAQ data, such as calibrations, slow control variables, etc., from a conditions database.
\end{itemize}

Before any of those can be achieved it is necessary to be able to build a version of Gaudi within the LZ infrastructure (see~\ref{OCSS:Infrastructure}). This has been done by importing the Gaudi Hive codebase into our own GitLab in order to manage the impact of any changes made to Gaudi Hive in the future and building it from that source.

The definition of the transient data models are well underway. The DAQ one is a straightforward representation of the objects read out by the DAQ. The physics one requires more development as it has to capture the objects and relationships of derived quantities and these have not all yet been defined. The creation of the input and output modules has begun.
Gaudi has a well established mechanism for extending its input and output modules to handle custom formats and we plan to use the experience of other experiments, such as Daya Bay, with this task to speed its development.

The Gaudi codebase presently supports two types of build systems, the legacy one based on CMT~\cite{cmt:web} and a new one base one CMake~\cite{cmake:web}.
LZ has decided to use the CMake-based version for its build and is now using this as a model to develop the programming and build environment for developing modules for its framework.
The creation of the input and output modules is being used as the test bed on which this mechanism can be developed.

Gaudi itself has a fully developed fault handling system so that any problems that arise during processing can be evaluated and either stop the processing altogether, stop the processing for a single event or allow the event to continue processing.
This, together with its comprehensive logging system, means that is straight forward to decide which processings are successfully, which need to be redone once the cause of the fault has been addressed and which are irrecoverable.

Wherever possible, we anticipate that existing code from the successful LUX and ZEPLIN experiments will be adapted and optimized for use within the LZ analysis framework.
The LZ processing and analysis codes will be written to be as portable as possible to ensure straightforward running on both Linux and OSX platforms for those groups who wish to do analysis in-house in addition to (or instead of) running codes on the data centers.

All non-DAQ data, i.e. any data recorded or developed that is not read out by the DAQ with each event, will be stored in a database (DB) known as the ``conditions database''.
The challenges of this database are that it not only needs to understand the interval of validity for each piece of data, but also needs to support versioning of that data for instances such as when better calibrations are available.
This problem is not unique to LZ and therefore it was decided to used the DBI package developed for the MINOS experiment.
Not only does this allow for intervals of validity and versioning, it also supports a hierarchy of data sources.
This means that during development of code or calibrations it is possible to specify an alternate DB to be used for certain values which supersede the values in the main underlying DB.
This also allows for the validation of new entries before they are inserted into the main conditions DB.
\tdrsec[Sims]{Simulations}
Detailed, accurate simulations of the LZ detector response and backgrounds are
necessary, both at the detector design phase and during data analysis.
Current LZ simulations use the LZSim package, which in turn is based on the existing LUXSim software package~\cite{Akerib:2011ec}, originally developed for the LUX experiment.
The LZSim codebase is entirely separated from LUXSim, although several of the
developers contribute to both.
It is managed through the collaboration's git repository.
This software provides object-oriented coding capability specifically tuned for
noble liquid detectors, working on top of the Geant4 engine~\cite{Agostinelli:2002hh}.
LZ intends to further update its simulation software in 2017 with two
significant changes: first a switch to the most recent version of Geant4, 10.2,
in order to take advantage of a number of critical improvements to the code and
physics lists.
Second, a recast of the LZSim framework into the more generalized BACCARAT
framework, which is also based on LUXSim but not tied to any detector-specific
or legacy code.

All LZ simulations will be integrated into the broader LZ analysis framework,
from production to validation and analysis.
The framework will in fact largely be developed using simulation output until
detector data is available.
Two output formats are supported, a raw simulation output at the GEANT
interaction level, and a reduced tree format at the event level.
Both use the {\normalfont\ttfamily ROOT} format.

The simulations group is organized into several distinct areas of technical
expertise, a structure reflected in the organization of this task.
In addition to the computing-centric approach described here, physics output
coordination is managed within a working group of the entire scientific
collaboration. Tasks include:
\begin{itemize}
\setlength\itemsep{0em}
\item[(a)] Simulation software packages maintenance, development, and collaboration
           support;
\item[(b)] Definition, maintenance, and implementation of an accurate detector
           geometry;
\item[(c)] Maintenance and continued improvement of the micro-physics model of
           particle interactions in liquid xenon, as captured in the NEST
           package~\cite{Szydagis:2013sih};
\item[(d)] Detector response implementation — which transforms the ensemble of
           individual GEANT4 photon hits at the PMTs to produce an event file of
           the same format and structure as in the data.
\item[(e)] Generators for relevant event sources in LZ for both backgrounds and
           signal;
\end{itemize}
Table~\ref{OCSt:simscomputing} shows the computing needs estimates for all simulation tasks,
based on recent production data.
These needs are on the order of about \SI{100}{TB} of disk storage and a total of 
\num{e6} CPU hours per year, mostly concentrated in short 1-to-4-week periods of burst activity.
Both hard disk storage and CPU needs are dominated by background simulations, including
a \SI{50}{\percent} contingency to account for the need to repeat and/or compare some studies.
There is also a provision for very-high statistics data sets for three major detector
components, in order to produce a high-granularity map of the background spatial profile
inside the TPC.
These data sets are only kept in their most reduced format in order to conserve storage
resources.
For all other simulations, both reduced files and raw GEANT4 output files (translated to
{\normalfont \ttfamily ROOT} format) are kept.
The total also includes optical simulations, which only provide photon collection
results and are less demanding in terms of resources, calibration simulations, necessary
to validate calibration data and early R\&D efforts, and a number of ad-hoc small
simulation tasks all grouped together.
Resource time profiles for different simulation categories feed into the higher level computing estimates.
\begin{table}[h]
\setlength{\extrarowheight}{5pt}
\tdrtcaption[simscomputing]{LZ simulations computing resources needs}
{High-level summary table of LZ simulations projections for computing power 
and storage needs, based on the simulation campaign of autumn 2015. The 
dominant part of the resources is required for detailed background simulations 
of detector components, including \SI{50}{\percent} contingency for updates and 
comparative studies. Storage needs are contained within a \SI{100}{TB} envelope 
for the duration of the project through commissioning (another \SI{100}{TB} 
is allocated for processed and user data). Corresponding computing power is 
relatively modest as a yearly average, however most of the needs occur in short 
bursts with a 1-4 week timescale.}
\begin{center}
\sffamily
\begin{tabular}{|l|r|r|r|} \hline
\rowcolor{mrocol}
Simulation type & Raw {\normalfont \ttfamily ROOT} & Reduced {\normalfont \ttfamily ROOT} & CPU needs \\
\rowcolor{mrocol}
                & files (TB) & files (TB) & (core-hours) \\
\hline
Background simulations			& 30	& 2		& $1.0 \times 10^5$ \\
High-statistics background map	& 0		& 17	& $8.0 \times 10^5$ \\
Contingency / repeats			& 15	& 2		& $5.0 \times 10^4$ \\
\hline
Optical simulations				& 2		& 0		& $5.0 \times 10^3$ \\
Calibration simulations			& 10	& 1		& $3.3 \times 10^4$ \\
Other misc. simulations			& 5		& 0.5	& $1.7 \times 10^4$ \\
\hline\hline
Totals							& 62	& 23	& $1.0 \times 10^6$ \\
\hline
\end{tabular}
\end{center}
\end{table}
%
\tdrsec[Infrastructure]{Software Infrastructure}
%
All LZ software is centrally maintained through a software repository based on GitLab~\cite{gitlab:web}, which is currently operating at the University of Alabama. The repository is backed up daily and the snapshots are retained for 15 days. GitLab implements excellent tools for release management and code review. A GitLab snapshot from a recent developement cycle of LZSim is shown in Figure~\ref{OCSf:gitLab}: different developers were working simultaneously on a round of updates to the detector geometry. After extensive testing, the updates were eventually merged into the master branch and folded into a tagged release. GitLab also offers a continuous integration tool, allowing for automatic testing and installation of the offline codebase on the U.S. and U.K. data center servers. Build automation is inherited from the Gaudi infrastructure and supported via CMake and cmt. 

\begin{figure}[bt]
\begin{center}
\includegraphics[width=0.9\textwidth]{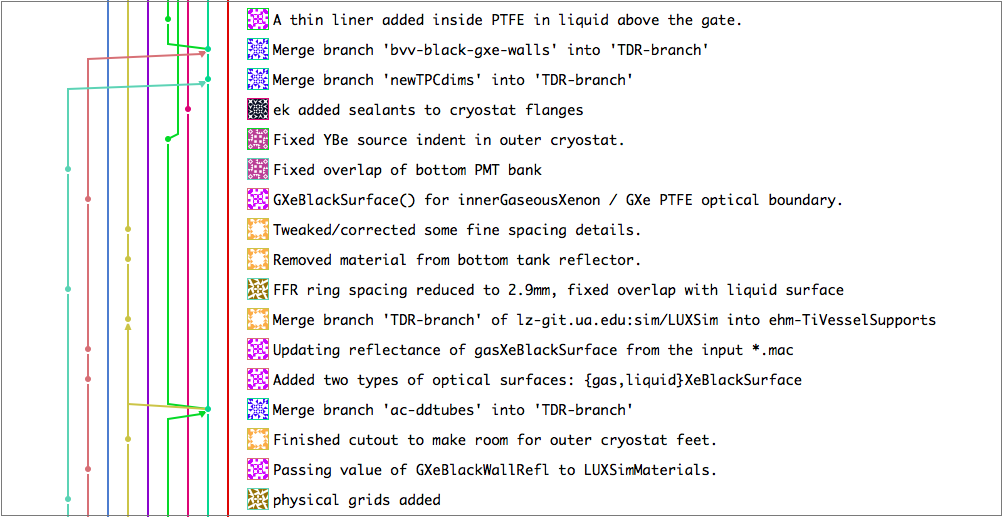}
\end{center}
\tdrfcaption[gitLab]{Parallel development workflow in GitLab}{Parallel development workflow in GitLab: several contributors were updating the LZSim detector geometry simultaneously. Each vertical line corresponds to a different development branch. After extensive testing, every update was eventually merged to the master branch.}
\end{figure}

Release Management and Version Control standards were strictly enforced from a very early stage of the project to ensure sharing, verifiability and reproducibility of the results. Each code release undergoes a battery of tests before being deployed to production. For every update cycle, the code is inspected by a moderator, who checks its integrity and reports possible inconsistencies or conflicts. The package is then installed and executed on each data center and checked for functionality. If one test fails, the code update is rejected. Release management also acts as a bridge between development and production: this ensures that all the changes are properly communicated and documented, to achieve full reproducibility.

A comprehensive validation suite is being developed, and each class of software update has a well-defined set of tests, depending on the nature of the change and on its magnitude. For example, when the detector geometry in LZSim is updated, the standard overlap checks and geantino tests provided by Geant4 are always performed. If the update leads to a tagged release, a large statistics of photons is simulated in every part of the detector and the light collection carefully examined (light collection is the parameter most sensitive to geometry changes). Based on the successful experience from the Fermi-LAT software validation suite~\cite{Atwood:2009ez}, the results from these simulations are compared via a mono-dimensional Kolmogorov–Smirnov test. If discrepancies are found, they are reported back to the developers, who are tasked with explaining the changes or fixing the underlying errors before the code is deployed to production.

Software distribution is achieved via CernVM File System (CVMFS)~\cite{Blomer:2011zz}. CVMFS is a CERN-developed network file system based on HTTP and optimized to deliver experiment software in a fast, scalable, and reliable way. Files and file metadata are aggressively cached and downloaded on demand. Thereby the CernVM-FS decouples the life cycle management of the application software releases from the operating system. The LZ CVMFS server is hosted at University of Wisconsin, Madison and is visible to all the machines in the U.S. and U.K. data centers, and to most computing centers available to the collaboration (Wisconsin, SLAC, Edinburgh, Sheffield, etc.). It can be loaded to each collaborator's personal laptop by installing a FUSE client. All the LZ software releases and external packages are currently delivered via CVMFS: this ensures a unified data production and analysis stream, because the data centers access exactly the same versions of the same executables, removing possible dependencies on platform and configuration. 

CERN software is also made available to the collaboration via CVMFS. Besides the above-mentioned Geant4 and Gaudi, we use several external packages, including CLHEP, ROOT and AIDA. The environment defining a specific combination of external packages is inherited via LCGCMT~\cite{lcgcmt:web} and shares a build automation infrastructure with Gaudi. Again, delivering the external packages via CVMFS removes unwanted dependencies on architecture and environment. However, in order to maximize software availability to each collaborator, we also plan to deliver precompiled tarballs of LZ tagged releases and external packages for individual download.

Cybersecurity risks posed to the offline computing systems relate to the experiment’s data and information systems. Much of the LZ computing and data will be housed at major computing facilities in the United States (NERSC/LBNL) and U.K. (Imperial College), which have excellent cybersecurity experience and records.
Specific risks posed to the LZ project relate to data transfer (in terms of data loss or corruption during transfer) and malicious code insertion. File checksums will mitigate the danger of loss or corruption of data during transfer, while copies at both the U.S. and U.K. data centers provide added redundancy. 
Moreover, CVMFS features robust error handling and secure access over untrusted networks~\cite{Dykstra:2014kea}. By requiring digital signatures and secure hashes on all distributed data, CVMFS also provides a strong security mechanisms for data integrity.
Malicious code insertion can be mitigated by monitoring each commit to the code repository by the offline group and requiring username/password authentication unique to each contributor to the code repository. A comprehensive policy for release management and continuous testing will be the key factor in preventing malicious software from being deployed to production.
\tdrsec[Schedule]{Schedule and Organization}
%
Offline software by its nature is heavily front-loaded in the schedule.
To enable the scientists to commission the LZ detector, the software for reading, 
assembling, transferring, and processing the data must be in place before detector 
installation.
This implies, in particular, that the data transfer, offline framework, and analysis 
tools themselves will have been developed, tested, debugged, and deployed to the 
collaboration.
We rely on the collaboration’s existing experience with the LUX experiment and others 
(Daya Bay, Double Chooz, Fermi, LAT, etc.), which routinely handled similar 
challenges.
Key offline computing milestones are summarized in Table~\ref{OCSt:milestones}.

The decision on the choice of analysis framework for LZ has been taken in 
April 2015. The first version of the framework with a minimal number of 
modules was released in early March 2016. This will be followed 
by the first physics integration release planned for November 2016.
This version includes all necessary modules for real-time processing (i.e., 
hit-finding algorithms, calibration constants modules, S1/S2 identification, 
event reconstruction), as well as a fully integrated simulations package 
(i.e., from event generation through photon hits, digitization, trigger, 
and data-format output).

\begin{table}[htbp]
\sffamily
\setlength{\extrarowheight}{5pt}
\tdrtcaption[milestones]{Key offline computing milestones}{Key offline computing milestones.}
\begin{center}
\begin{tabular}{|r|l|c|}
\hline
\rowcolor{mrocol}
Date      & Milestone                         & Status \\ \hline
Mar. 2015 & Analysis Framework decision       & Done (Apr. 2015)\\ \hline
Feb. 2016 & First Analysis Framework release  & Done (Mar. 2016)\\ \hline
Nov. 2016 & First physics integration release & Done (Nov. 2016)\\ \hline
Sep. 2017 & First mock data challenge         & \\ \hline
Feb. 2018 & OCS FDR - Final Design Review     & \\ \hline
Aug. 2018 & Second mock data challenge        & \\ \hline
Nov. 2019 & Third mock data challenge         & \\ \hline
Dec. 2019 & Full software release for commissioning  & \\ \hline
\end{tabular}
\end{center}
\end{table}

The first mock data challenge (September 2017) will test both the data flow 
(transfers, processing, distribution, and logging), as well as the full physics 
analysis functionality of the framework, separately. The first few weeks of LZ 
commissioning (calibrations included) will be simulated; participants should 
be able to quantify detector response and main backgrounds, based on simulated data.
The second data challenge (July 2018) will be dedicated to testing the entire data 
chain. It will contain 6 months of simulated data with physics signals, including 
calibration data. Participants should be able to establish a detailed background 
model and perform low-energy calibrations (ER/NR).
The third data challenge  (October 2019) will test the complete analysis strategy 
and is expected to validate the readiness of the offline system just before the LZ 
cool-down phase. It will include \num{1000} days of simulated data including physics signals.
No MC truth will be available to participants and physics signals will be known 
only to a subset of organizers. It will simulate the analysis for the first LZ 
science paper, including possible blinding/salting plans.

Offline computing will be led by physicists experienced in software development 
and use and a computing professional from LBNL.
The software professional will also liaise with NERSC for collaboration on providing 
LZ compute resources — in particular, provisioning and/or allocating of network, CPU, 
disk, and tape resources sufficient for LZ collaborators to transfer, manage, 
archive, and analyze all data for the experiment.
The infrastructure software effort will also involve professional software 
engineering from LBNL.
This person will provide technical leadership, oversight, and coordination of LZ 
collaboration efforts on infrastructure software as well as the design, 
implementation, testing, and deployment of critical LZ infrastructure components.
LZ infrastructure software includes data management and processing, offline systems 
and monitoring, offline interfaces to LZ databases, and the analysis framework.
The remainder, and bulk, of the software is a collaboration responsibility.
Software for simulation, analysis, monitoring, and other tasks will be written and 
maintained by collaboration scientists.

\clearpage
\bibliographystyle{apsrev4-2}
\bibliography{LZtdr}
\wlog{In tdrinclude command}
\wlog{\ChapLabel}

\renewcommand{\ChapLabel}{chap:SRD} 
\renewcommand{\BibLabel}{SRDS:bib} 
\renewcommand{\SecPre}{SRD}
\renewcommand{\FloatPre}{SRD}
\tdrchap{Simulations, Requirements, and Detector Performance}

This chapter ties together the
previous descriptive chapters and
describes the performance we expect
from the LZ apparatus.
Our estimates
of performance are based on 
detailed simulations, which have been substantially
overhauled since completion of the LZ
Conceptual Design 
Report~\cite{Akerib:2015cja}.
The LUX collaboration has performed extensive
high-statistics calibrations of 
ER background~\cite{Akerib:2015wdi},
NR signal~\cite{Akerib:2016mzi}, and implemented
these calibrations in an improved analysis using
the profile likelihood 
ratio technique~\cite{Cowan:2010js}
of the LUX WIMP search data~\cite{Akerib:2015rjg}.

We have incorporated the performance improvements
made by LUX, which change the sensitivity
in two principal manners:
\begin{enumerate}
\item The LZ sensitivity to low-mass WIMPs and
to nuclear recoils from solar \IBe neutrinos
is much improved.
\item The LZ sensitivity to sources of ER background,
including, prominently, the "quiet" beta decay
of daughters that trace their parentage to radon
impurities, is much reduced.
\end{enumerate}

\tdrsec[Sims]{Simulations}
In this section we describe 
in detail the simulations used to develop the
background and sensitivity studies presented in 
this report. The simulations were performed using
LZSim, an offshoot of the LUXSim package originally
developed for the LUX experiment and based on the
\geantFour particle physics simulation
software~\cite{Akerib:2011ec,Agostinelli:2002hh}. 
Designed specifically for low-background detector modeling, LZSim generates events and records particle interactions on a detector geometry component-by-component basis, but with an infrastructure independent of the detector geometry. 

The LZSim infrastructure allows the user to define any
detector component as a \geantFour sensitive detector
at run-time with a macro command. It incorporates
internal bookkeeping to automate the generation of
backgrounds arising from a variety of event
generators, each with intensities set by the user at
run time, and with a time-ordered stochastic primary
event record. The geometry components are customized
using coding techniques familiar to users of the base
\geantFour code. The event generators can be based
either on the internal \geantFour classes or created
from scratch by the users. LZSim automatically records
quality control information to a header in each 
output file to establish a record of how the data 
was generated.

We use \geantFour version 4.9.5, the
physics list \texttt{QGSP\_BIC\_HP}, and 
the libraries of CLHEP version 2.1.0.1. 
LZSim provides the option to incorporate 
the NEST model that describes ionization
and scintillation formation for NRs 
and ERs~\cite{Szydagis:2011tk,Szydagis:2013sih}.
Currently we instead pass the output from LZSim 
to a standalone version of the NEST model that
incorporates the latest results from the LUX
experiment~\cite{Akerib:2015rjg,Akerib:2015wdi}.

\clearpage
\tdrsubsec[SGeometry]{Geometry construction}
A detailed model of the LZ detector geometry was
created within the LZSim framework according to CAD
drawings of the detector. Components with significant
mass or high amounts of radio-impurities as well as
those located very close to the active xenon target
are included. Elements that influence light 
collection and their respective optical 
properties are also described in the model.

Our model describes the detector from the outside in,
with a few exceptions, nesting each successive volume
or component within the one preceding it. The
outermost volume is the steel water shielding tank.
Placed within the water of the shielding tank are 
the major components of the outer detector of
Chapter~\ref{chap:OD}, including the segmented 
acrylic vessels, liquid scintillator, foam 
displacer, reflectors and R5912 PMTs. Services 
for the TPC such as the cryostat support stand,
cathode high voltage conduit and thermosyphon 
and PMT cabling conduits require penetrations
in the acrylic tanks that can impact veto 
performance and are therefore implemented in the
detector model. Conduits that contain multiple 
or complex materials such as coaxial cable, 
gaseous or liquid xenon, and vacuum are modeled 
by a single material which represents the average
density and composition of all materials in that
conduit. Angled and horizontal neutron calibration
tubes and a port for the YBe source are included 
for calibration source studies.

Located within the outer detector is the titanium
cryostat, built according to the engineering models,
as shown in Fig.~\ref{SRDf:CryostatComp}. An inner
cryostat vessel is contained within the outer 
cryostat vessel, also made of titanium and built
to engineering specifications. The model includes multi-layer insulation between these two
volumes, in addition to vacuum. No optical
properties are defined between these regions
as no photons are expected to be produced here.

\begin{wrapfigure}{o}[0pt]{0.60\textwidth}
\centering
\includegraphics[width=0.5\textwidth]{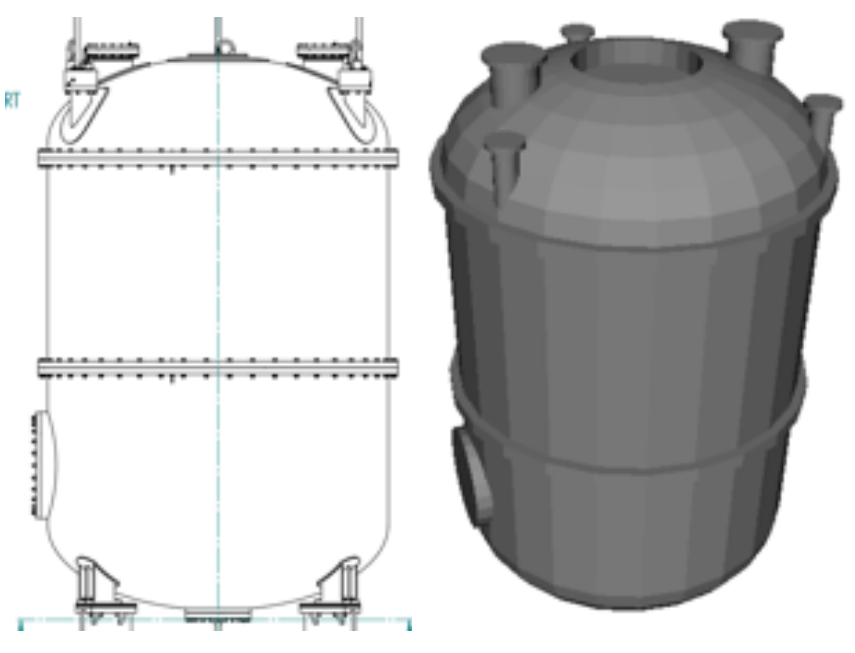}
\tdrfcaption[CryostatComp]
{Engineering drawing and simulation geometry of the outer cryostat}
{Engineering drawing and simulation geometry of the outer cryostat.}
\end{wrapfigure}

The first liquid xenon volume, known as the xenon
skin, is just inside the inner cryostat vessel and
continues inward to the outer surface of the TPC.
The skin also  extends below the bottom of the TPC
through the bottom dome~(see Chapter~\ref{chap:XDS}
for more details). The inner wall of the cryostat 
in the simulation is covered in a thin PTFE liner 
to aid light collection. 

Optical boundaries, which can be defined at run 
time, are used between the liquid xenon skin and 
the components within to study the effects of 
changes in the assumed reflectivities. 

The heart of the experiment, the liquid xenon TPC
where the primary scintillation (S1) and secondary
scintillation from ionization (S2) signals would
be caused by a WIMP, is nested inside the skin volume.
A rendering of the TPC is shown
Fig.~\ref{SRDf:TPCComp}. The liquid xenon in the 
TPC is divided into two volumes. The active volume
contains all the liquid xenon above the cathode, 
where well-reconstructed S1/S2 events occur. The 
second volume is the reverse field region (RFR), 
below the TPC cathode.  Energy depositions in this
volume cause an S1 signal, but no S2.  Nevertheless,
sometimes an RFR S1 signal becomes associated with 
an unrelated S2 signal, resulting in a class of
events known as ``gamma-X''.
 A third xenon volume, the gaseous xenon above the
liquid where electroluminescence develops for
the S2 signal, is also defined.

A cylinder of PTFE with extremely high
diffuse reflectivity forms the TPC wall.
Field shaping rings and grading resistors in
the walls are represented. TPC components
including the bottom shield grid, cathode grid,
gate grid, and anode grid, according to the 
design described in
Table~\ref{XDSt:ElectrodeGrids}, are
represented.
Optical boundaries between the liquid xenon 
and the PTFE walls as well as between the 
liquid xenon and the stainless steel of the 
grid wires are included to describe the light collection properties of the detector.

\begin{wrapfigure}{o}[0pt]{0.60\textwidth}
\centering
\includegraphics[width=0.5\textwidth]{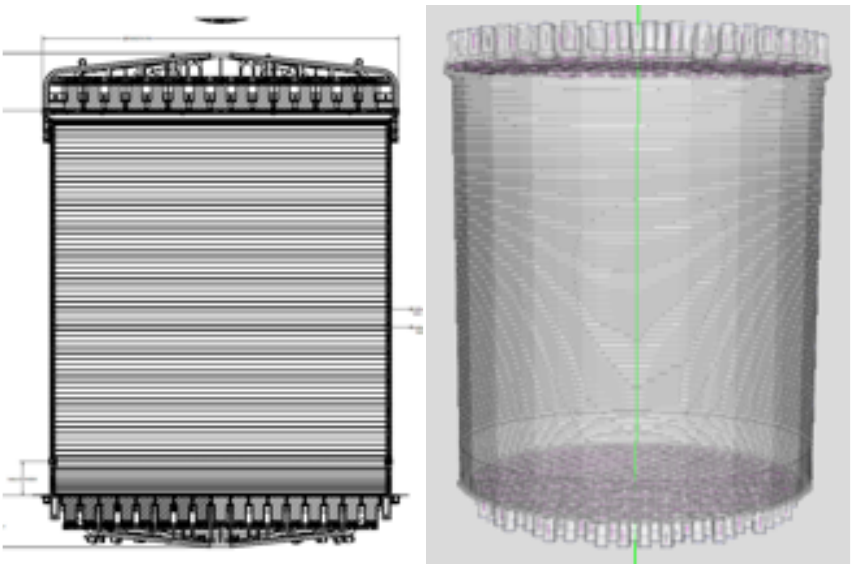}
\tdrfcaption[TPCComp]
{Engineering drawing and simulation geometry of the TPC}
{Engineering drawing and simulation geometry of the TPC}
\end{wrapfigure} 

At the top and bottom end of the TPC cylinder 
are the PMT arrays, containing representations
of each R11410 PMT that will be used in the 
LZ experiment. The PMTs are described as 
a stainless steel shell with vacuum inside. 
Materials representing the dynode chain are
included to model the radioactivity of those
components. 

The face of the PMT is a quartz window with 
an additional thin layer of quartz buried 
within to represent the photocathode. The 
window and photocathode volumes are divided 
to model cases where a photon reflects off 
the front face of the PMT and fails to penetrate
to the photocathode. 

Optical interfaces at the top of the TPC 
between the PMT windows and gaseous xenon 
in the extraction region, and at the bottom 
array between the PMT windows in liquid xenon,
are described so \geantFour appropriately 
handles the reflection and transmission of photons. 
The PMT photocathode is defined as a \geantFour 
sensitive detector to collect optical photons.
The PMT array volumes also contain
the support plates and support trusses that 
provide the mechanical support for the PMTs.
Figure~\ref{SRDf:PMTarray} shows a comparison
between the current engineering design and the
simulated geometry of the bottom PMT array. 

\begin{figure}[H]
\centering
\includegraphics[width=0.75\textwidth]{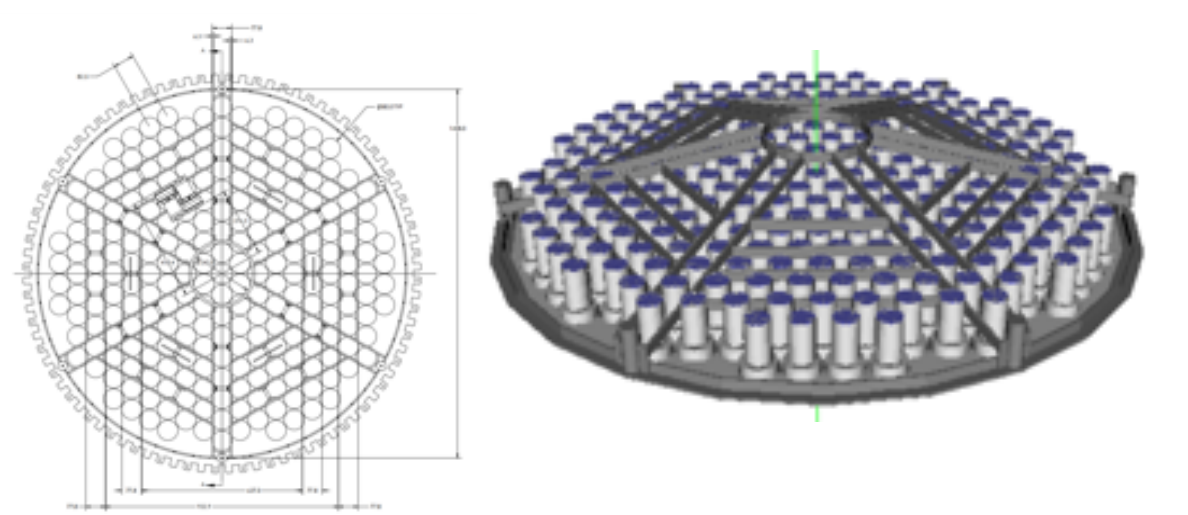}
\tdrfcaption[PMTarray]
{Engineering drawing and simulation geometry of the bottom PMT array}
{Engineering drawing and simulation geometry of the bottom PMT array and support truss. The skin PMTs that mount to this truss are not shown.}
\end{figure} 

\tdrsubsec[SEGens]{Event Generators}
A variety of software packages are employed
to simulate the physics of signals and backgrounds
that induce responses in the LZ detector.
The event generators which describe the WIMP
signal and neutrino physics are derived from the
references cited in Chapter~\ref{chap:SIP}.
Here we discuss the event generators employed
to describe various types of background phenomena
in the LZ detectors.

\tdrsubsubsec[SNeutron]{Neutron production in detector materials}
Neutrons emitted from radioactive processes
in the material near to the LZ liquid xenon
detector can create isolated nuclear recoils
that might fake the recoils expected from WIMPs.
To simulate neutron backgrounds from radioactivity
(the \IUtTe, \IUtTF and \IThtTz decay chains),
LZSim uses input neutron spectra as calculated 
with the SOURCES-4A~\cite{Wilson:1999a} package.

The SOURCES-4A code calculates neutron yields 
and spectra from spontaneous fission, (\Pga,\Pn)
reactions and delayed neutron emission due to 
the decay of radionuclides. Its library contains
all alpha emission lines from known radioactive
isotopes. The code takes into account the energy
losses of alphas, cross-sections of (\Pga,\Pn)
reactions and the probabilities of nuclear
transition to different excited states. 
We use an option for a thick target neutron 
yield allowing for calculation of neutron yields
and spectra under the assumption that the size of 
a material sample exceeds significantly the 
range of alphas. 

The original SOURCES-4A code has been
modified~\cite{Carson:2004cb,Lemrani:2006dq,Tomasello:2010zz}
to extend the energy range of alpha particles 
to \SI{10}{MeV} and to include (\Pga,\Pn) 
cross-sections and transition probabilities 
to excited states for most isotopes relevant 
to underground rare event experiments, based 
either on measurements or on
EMPIRE1.19~\cite{Herman:2007a}. 

The neutron spectra from SOURCES-4A are 
implemented as generators in LZSim, allowing 
any detector component to become a source of
neutrons. For each component of interest, 
we simulate concentrations of \SI{10}{\ppb} of
uranium or \SI{10}{\ppb} of thorium; these
concentrations are then scaled to match the 
results of the materials screening. The 
uranium decay chains are split into early and 
late branches, and the \IPbtoz sub-chain is
calculated separately. 

The spontaneous fission (s.f.) process is 
treated separately to exploit the ability 
of LZ to reject decays producing multiple 
neutrons and gammas, such as those produced 
in s.f. events. The SOURCES-4A package
generates individual neutron spectra without
accounting for simultaneous multiple neutrons
and gamma rays.  A special generator was
developed to accurately simulate multiple 
neutrons (2.01 on average for \IUtTe) and 
gammas (6.44 on average for \IUtTe) in s.f. 
events. Accounting for multiple neutrons and
gammas permits accurate description of 
multiple simultaneous signals in the LZ detector,
and accurate accounting of the rejection of events
with multiple signals.  Most spontaneous fission
events, particularly those in materials close to
the LXe target, are rejected through their 
tendency to produce multiple hits.

\tdrsubsubsec[SMuon]{Muon-induced neutrons}
Energetic neutrons can be produced by
atmospheric muons that penetrate to the rock
around the Davis Cavern. Simulations of 
this process commence with the selection of
muons with positions, directions, and energies
sampled according to the MUSUN
code~\cite{Kudryavtsev:2008qh} after 
transport using standalone MUSIC code~\cite{Antonioli:1997qw}. 
MUSUN has been integrated into LZSim as a 
particle generator in such a way that events 
are generated from the surface of a cuboid
surrounding the detector.

Muons sampled with the MUSUN code are then passed to \geantFour which transports them and their secondaries including neutrons to the detector. All processes relevant to muon, photon, electron, and hadron interactions are included and the models are equivalent to those used in the \geantFour physics list called Shielding, recommended by the \geantFour developers for this application.

\tdrsubsubsec[SGammas]{Gamma activity in large detector components}
Background ER events from \IKfz, \ICosz, 
and the \IUtTe and \IThtTt decay chains are 
also modeled using  a particle generator 
developed in LZSim. Ions of the parent isotopes 
are positioned in a component and the full 
decay chain is simulated according to the 
physics described by the \geantFour radioactive
decay data libraries. Crucially, the decay 
chain is produced in equilibrium, which 
allows splitting of the chain by individual
isotopes during analysis. Thus, only one 
simulation is required, independent of the 
relative decay rates of individual radioactive
isotopes within the chain. As for the neutron
studies, fixed activities are simulated for each
component, which are then renormalized based on
materials screening results. 

\tdrsubsubsec[SGammaRock]{Gammas from the cavern rock}
The external background from gammas in the 
rock walls of the cavern are also assessed 
using the radioactive decay chain generators
described in the previous section. However, 
due to the large number of primary decays 
required to accumulate events in the active 
xenon, event biasing is used to boost statistics.
This involves saving events at decreasing 
distances from the target and feeding them back
into LZSim multiple times as primary particles. 
By default, LZSim records all hits within a
detector component with a defined record level,
but for these simulations the structure has been
modified so that the output is recorded for 
tracks according to the distance from the 
center of the TPC. Additionally, modifications
were made to allow the LZSim binary output files
to be input as primary particles.

\tdrsubsubsec[SBench]{Benchmark points}
A new generator has been developed to allow the
assessment of the impact of the radioactivity
of small size components (e.g. sensors) quickly.
LZSim was designed to primarily distribute
primary events throughout a given volume.
Our new generator can associate radioactive
events with a specific geometric location 
(the `Benchmark Point') within the LZSim geometry. 

\tdrsubsec[SEDep]{From energy deposition to signals}
The event generators described in the previous
section are used to generate different 
radioactive decay products. These particles
are tracked by \geantFour as they deposit their
energy in different volumes of the geometry. 
LZSim allows any part of the volume to be a
sensitive detector, and for these simulations, 
we record all depositions in the TPC, as well 
as associated energy depositions in the skin 
region and the outer detector.  Among the data
saved in each volume for each event are the
locations, times, and magnitudes of energy 
deposits made by various particle types.

In the skin and TPC liquid xenon volumes, 
energy depositions within \SI{400}{\micro\meter} are clustered (to encompass the largest possible ER and NR track sizes) and categorized as 
either ER or NR depending on the interaction. The result is a list 
of energy-deposition clusters from a given 
particle type (ER or NR). The energy and 
associated particle type of each cluster 
as well as the local electric field are then fed
into the NEST (Noble Element Simulation Technique)~\cite{Lenardo:2014cva,Szydagis:2013sih,Szydagis:2011tk,Lenardo:2014cva} 
package which stochastically computes 
the number of expected photons and electrons
produced at each cluster. 

NEST models the scintillation light and
ionization charge yields of nuclear and electron
recoils as a function of electric 
field and energy or $\textrm{d}E/\textrm{d}x$. NEST also models the drift, diffusion, absorption, extraction, and electroluminescence of the electrons as they move through the liquid and gas. 
``NEST'' refers both to a collection 
of microscopic models for energy deposition in
noble elements and to the Monte Carlo
simulation code that implements these models. NEST
provides mean yields and intrinsic fluctuations 
due to the physics of  excitation, ionization, 
and recombination, including both Gaussian and 
non-Gaussian components of the energy resolution.
 
The NEST methodology was initially trained on data 
from a small dual-phase detector from 
Case Western Reserve University (Xed), which
yielded comprehensive data sets in
terms of energy range and field 
sweep~\cite{Dahl:2009nta}. The NEST model 
used in this simulation has been updated to
incorporate the latest calibration results 
from the LUX
experiment~\cite{Akerib:2015rjg,Akerib:2015wdi,Akerib:2016mzi}.

After the clustered energy depositions
have been converted from energy to quanta (scintillation photons and ionization
electrons) via the NEST models, a detector 
model is applied to convert these raw quanta 
into the detector observables, that is,
into S1 and S2. 

The primary scintillation light from the NEST
models is propagated to the faces of the PMTs, 
accounting for binomial fluctuations, using 
the light collection model, including Fresnel
transmission and reflection, described in
Section~\ref{XDSS:S1Light}.  We refer to the
average light collection efficiency as
$\alpha_1$, which the optical model currently predicts to be
\SI{8.5}{\%} (better than our baseline value and requirement of \SI{7.5}{\%}),
and we correct the raw signal (denoted S1) for the variation
of light collection with position in the detector (denoted S2).
Photoelectron production at the photocathode accounts for the
double-photoelectron phenomenon described
in~\cite{Faham:2015kqa}.  A trigger
requirement of three-fold coincidence of PMT
hits is applied prior to subsequent analysis. Both S1 and S1c are reported to the user.  

The S2 signal is produced from the number of
ionization electrons that drift away from the
interaction site. The LZ extraction efficiency
(see Table~\ref{XDSt:S2parameters}) is applied to
determine how many electrons are extracted to the
gas phase. The extracted electrons are converted 
to a number of luminescent photons with 
NEST, accounting for LZ parameters, and these photons 
are then converted into photoelectrons at the PMTs.
The S2 signal
is also corrected for the position of the event in the detector, including the effect of non-infinite electron lifetime
presumed in LZ to be \SI{850}{\mus} (corresponding to an absorption length of greater than \SI{1.5}{\meter}). The raw and 
corrected signals, denoted S2 and S2c,
are reported to the user.

In the central TPC the energy-weighted position and variance of all energy clusters is computed. 
The total S1 signal in the central xenon 
volume includes the S1 signal created by 
energy deposits in the reverse field 
region (RFR) below the cathode. Here, the 
electric field drifts electrons downward,
making no contribution to the S2 signal.

A similar procedure is followed to model 
the S1 signal observed in the skin region
(excluding S2 as no charge is collected in 
the skin). A light collection model is 
determined by simulating photons throughout 
the skin. Each energy deposit in the skin 
is converted into photon quanta using an 
electric field model of the skin and the 
NEST package. These photons are then 
converted to detected photoelectrons, and the final skin S1 is reported to the user. 

The process that translates energy 
depositions into observed S1c and S2c 
signals does not currently account for 
PMT or DAQ electronics noise. A more 
complete model incorporating all known 
sources of  noise leading to the generation 
of simulated waveforms is under development. 

A substantial effort has gone into simulating
the light collection for the outer detector
liquid scintillator system.  Currently, however,
for the primary TDR physics simulations, only
energy depositions from \geantFour in the outer
detector are used.  The transport, capture,
and conversion to gamma rays and nuclear recoils
of neutrons are simulated in \geantFour.

\tdrsubsec[SAnalysis]{Analysis cuts}
For each event, we have a tree containing 
the energy and location of interactions in 
the TPC, 
the skin region, and the outer detector, 
and energy depositions in liquid xenon have 
been translated into raw and corrected 
S1 and S2 as described in the previous section. 
We apply a set of cuts to the simulation data 
to determine the backgrounds produced by a 
given radioactive decay chain in a given 
detector component. The baseline cuts that 
are used to produce the numbers in Table~\ref{MASt:BackgroundsTable} are as follows:
\begin{itemize}
\item{Region of Interest: $0 < \mathrm{S1c} < 20$
detected photoelectrons, but assuming 
3-fold coincidence in the TPC PMTs. In other 
words, three PMTs have to have observed light, 
but the total sum of the signal can be 
arbitrarily small. For ER, this range 
is approximately \SIrange[range-units=single]{1.5}{6.5}{\keVee}, and 
for NR, this range is approximately \SIrange[range-units=single]{6}{30}{\keVnr}. 
In addition, the uncorrected S2 signal is required to be \num{>350} detected photoelectrons (5 emitted electrons) ensuring adequate signal size for position reconstruction.}
\item{Single scatter event: $\sigma_Z <$ \SI{0.2}{\cm} 
and $\sigma_r <$ \SI{3.0}{\cm}. The energy-weighted
variance in position must be less than the 
expected spatial resolution of the detector 
for an event to be classified as a single
scattering event. The given values are based on the LUX performance with an estimated scaling to LZ. Gamma-X events are 
treated as single scatters, as no S2 can be
observed from the second interaction vertex. }
\item{Skin cut: \SI{<3} detected photoelectrons 
in the skin veto region, to ensure that no 
visible energy is deposited in the skin within the \SI{800}{\mus} coincidence window.}
\item{Outer detector cut: \SI{<200}{\keVee}
deposited in the outer detector, to ensure that 
no visible energy is observed in the outer detector within the \SI{500}{\mus}  coincidence window.}
\item{Fiducial volume cut: the fiducial volume 
is defined as \SI{4}{\centi\meter} from the TPC
walls, \SI{1.5}{\centi\meter} from the cathode 
grid and \SI{13.5}{\centi\meter} from the gate
grid, corresponding to \SI{5.6}{\tonnesl} of LXe.}
\end{itemize}

In some cases, particularly for the simulation 
of gamma backgrounds, adequate statistics could 
not be generated due to limitations in time and
available disk space. In these cases, the upper
bound of the region of interest was increased 
to \SI{100}{\keVee} to increase the statistics 
in the analysis. This scaling is only valid if 
the spectrum of background events is roughly 
flat below \SI{100}{\keVee}; in the fiducial 
volume used here, this condition was met.  

\tdrsubsec[SValidation]{Validation}
To ensure that the simulations are an accurate reflection 
of the detector design, the simulation code 
and outputs are validated in several ways. 
First, because the LZSim simulation package 
shares a code base with LUXSim, the extensive
validations of LUXSim that have been performed
using LUX data can be incorporated directly 
into LZSim. Therefore, the \geantFour physics 
list, event generators, and the NEST models 
are vetted against LUX data before being applied 
to the LZ detector model. 

Where possible, specific predictions of LZSim 
are validated against external, independent 
models. For example, the light collection 
studies described in Sec.~\ref{XDSSs:S1PDE} 
are produced using the full LZSim model, but 
are checked against  an independent, 
MATLAB-based ray tracing code package developed 
by collaborators within LZ. Similarly, 
the parameters that drive the S2 photon 
detection described in Sec~\ref{XDSS:S2Light} 
are compared with results from independent 
electron transport models that are validated
against LUX S2 pulses.  
 
To validate the detector geometry, all 
components are checked against engineering 
drawings by at least two people. Given the 
high level of confidence in the optical 
model, based on the agreement with LUX data 
and the independent checks, the light 
collection models are used as a second 
validation step. The majority of mistakes 
in implementing the geometry become 
immediately apparent when looking at the 
predicted light collection. Before a 
modification to the geometry can be accepted 
into the repository, any changes to the 
light collection output that result from 
that modification are studied and understood.

Finally, a set of high-level cross-checks 
provides additional quality assurance for the 
key simulation results used for background rate 
estimates. These include comparisons to 
back-of-the-envelope calculations for the 
dominant sources of background, comparison 
of the SOURCES-4A neutron yields to those 
from an alternative simulation package, and 
sanity checks of neutron and gamma attenuation 
lengths along critical paths throughout the 
LZ geometry. All cross-checks are consistent 
with the full simulations output at a level 
that does not impact the LZ sensitivity
requirement.

\clearpage
\tdrsec[Reqs]{Requirements}

In this section, we summarize the 
key requirements for the LZ experiment. 
The LZ collaboration has established a 
small number of requirements to guide and 
evaluate the design and fabrication of the 
detector systems. 

The top-level scientific requirement is the
sensitivity to WIMP dark matter, via the 
spin-independent scattering process. Subsidiary
high-level science requirements and the flow-down
from the overall sensitivity are shown in
Fig.~\ref{SRDf:LZRequirementsFlow}. The 
high-level requirements, including the 
key infrastructure requirements, are 
summarized in Table~\ref{SRDt:RTopTab}. 
These requirements flow down to the 
detector subsystems and are captured 
in a concise form available to the 
collaboration. There are two practical 
high-level requirements. First, all equipment 
and subassemblies must be transported via 
the Yates shaft (see Chapter~\ref{chap:IAI}), 
which imposes dimensional and weight limits. Second, the existing water tank now housing 
the LUX detector will be reused. 
The collaboration has also captured the
requirements for detector subsystems at 
WBS Level 2. 

\begin{figure}[htbp]
\centering
\includegraphics[trim={0 1cm 0 0},clip,width=1.0\textwidth]{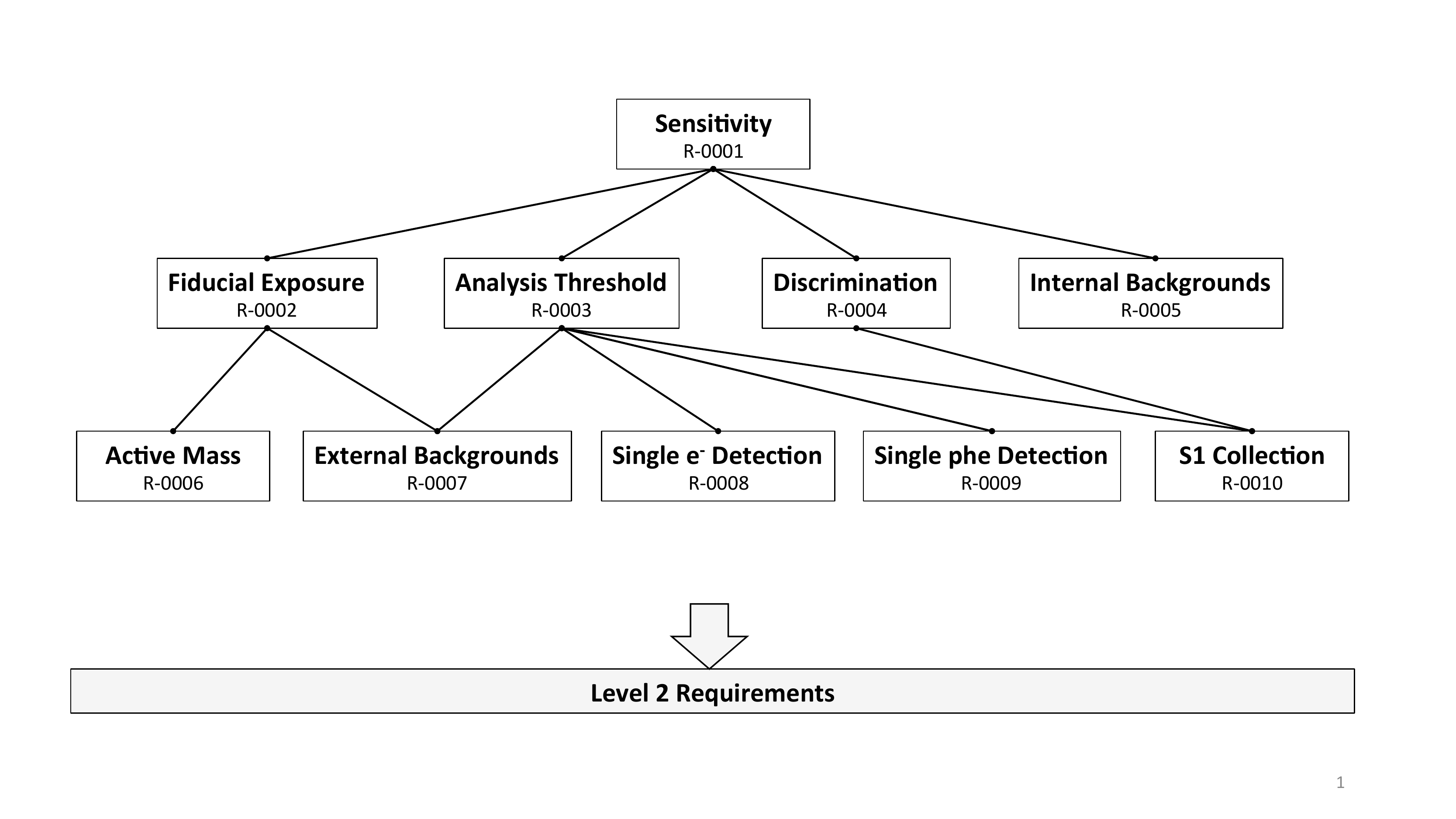}
\tdrfcaption[LZRequirementsFlow]
{Top Level LZ Science Requirements Flow}
{Flow down of the top level scientific requirements.}
\end{figure} 

Requirements development and explication have 
been key elements of internal reviews of LZ
detector systems and will be an important 
aspect of configuration control. All top-level 
and Level-2 requirements have been developed 
and reviewed in dedicated meetings. The
requirements are captured in a Google 
document, along with additional material 
and documentation for each relevant WBS 
Level-2 element. The LZ instrument scientist 
and chief and deputy chief engineers (see
Chapter~\ref{chap:MCS}) are primarily 
responsible for the maintenance of the 
requirements and their further development, 
if required, in close collaboration with the 
Level 2 managers. 

\newpage
\begin{center}
\sffamily
\renewcommand{\ltbcp}{RTopTab}
\begin{longtable}{ | m{1.7cm} | m{1.2cm} | m{2.55cm} | m{3.4cm} | m{5.55cm} | }
 \caption[Top Level Science Requirements
       \ifKL\space\texttt{\FloatPre t:\ltbcp}\fi]
       {LZ Top Level Science Requirements
       \ifKL\space\texttt{\FloatPre t:\ltbcp}\fi\label{\FloatPre t:\ltbcp}}\\
\hline
\rowcolor{mrocol}
{ Number} & { WBS} & {Requirement Name} & {Requirement Description} & { Rationale} \\ \hline
\endfirsthead
\caption[]{\sfs (continued)}\\
\hline
\rowcolor{mrocol}
{ Number} & { WBS} & {Requirement Name} & {Requirement Description} & { Rationale} \\ \hline
\hline
\endhead
\rowcolor{lrocol}
\multicolumn{5}{|c|}{\sfs\Tstrut (continued on next page)}\\
\hline
\endfoot
\endlastfoot
\rowcolor{lrocol} \multicolumn{5}{|c|}{Primary}  \\ \hline
	R-0001 & Science & WIMP Sensitivity & Sensitivity to \SI{40}{\GeVpct} WIMPs is \SI{3e-48}{cm^2} or better & Probe limit of liquid xenon technology set by solar neutrino background. Approach sensitivity to atmospheric neutrinos. Test supersymmetric and extra-dimension models of dark matter  \\ \hline
\rowcolor{lrocol} \multicolumn{5}{|c|}{Secondary}  \\ \hline
	R-0002 & Science & Fiducial Exposure & Minimum of \SI{5600}{\tonnedaysl} & Needed to achieve sensitivity requirement. Achievable with fiducial mass of \SI{5.6}{\tonnesl} and assumed running period of \SI{1000}{\livedaysl} or less fiducial mass and longer running time \\ \hline
	R-0003 & Science & Analysis threshold & \SI{50}{\percent} efficiency at \SI{6}{\keVnr} & Probe WIMP mass range down to \SI{5}{\GeVpct} with non-negligible sensitivity \\ \hline
	R-0004 & Science & ER Discrimination & \SI{99.5}{\percent} ER discrimination for \SI{50}{\percent} NR acceptance & Limit background from ERs so as to reach WIMP sensitivity requirement \\ \hline
	R-0005 & Science & Internal backgrounds & Internal backgrounds from radioactive noble gases (Rn, Kr, Ar) not to exceed four times the solar neutrino ER background & Limit ERs from internal backgrounds to an acceptable level. Solar neutrino rate does not include \IBe \\ \hline
\rowcolor{lrocol} \multicolumn{5}{|c|}{Tertiary}  \\ \hline
	R-0006 & Science & Active mass & \SI{7.0}{\tonnesl} & Required to reach fiducial exposure \\ \hline
	R-0007 & Science & External backgrounds & Backgrounds from radioactivity of the detector components (not including internal backgrounds, R-0005).
ER counts before discrimination \SI{<25} and
NR counts before discrimination \SI{< 0.7} & ER counts constrained to be \SI{<10}{\percent} of ERs from solar neutrinos (not including \IBe), including uncertainty in this rate. NR events to be constrained to be comparable to neutrino rate. We rely on veto efficiency to reduce the NR rate contribution. This rate, and to a lesser extent external ER contributions, define the fiducial mass. Analysis threshold also depends on size of these backgrounds. \\ \hline
	R-0008 & Science & Single electron detection & 50 S2 photoelectrons detected per emitted electron & Sufficiently large S2 signal for accurate reconstruction of peripheral interactions, such as those arising from contamination on the TPC walls.  \\ \hline
	R-0009 & Science & Single photoelectron detection & Single S1 photoelectron detection with \SI{>90}{\percent} efficiency, so as to reach \SI{>70}{\percent} efficiency for \SI{3}{\phe} & Main determinant of analysis threshold. \\ \hline
	R-0010 & Science & S1 light collection & Volume-averaged S1 photon detection efficiency (geometric light collection times effective PMT quantum efficiency) of \SI{>7.5}{\percent}  & Good discrimination and low energy threshold, equal or better than past Xe experiments. Exponentially falling (in recoil energy) WIMP spectrum means more recoils at lower energies, and low-energy recoils produce less S1 (both total and per-unit-energy) driving the S1 light collection efficiency requirement. \\ \hline 
\rowcolor{lrocol} \multicolumn{5}{|c|}{Infrastructure}  \\ \hline
	R-0100 & General & All parts fit down Yates shaft & All detector elements must be sized so that they can be lowered via the Yates shaft & Yates shaft is primary access to the Davis campus \\ \hline
	R-0110 & General & Reuse Davis water tank & Existing Davis water tank is reused. Include minor modifications and refurbishment. & Not practical or cost effective to replace water tank. Insufficient underground space to make larger tank. \\ \hline

\end{longtable}
\end{center}

The baseline design described in this Technical Design Report meets or exceeds all requirements. We briefly summarize the key requirements by WBS element in the subsections below. Linkage among the various requirements has been part of the requirements review process.
\vspace{1.5cm}

\tdrsubsec[R1.1]{WBS 1.1 Xenon Procurement}
There is a single requirement for WBS 1.1 Xenon Procurement, that being to procure 10 total tonnes of xenon. This requirement flows down from R-0006, as 10 total tonnes are adequate to result in 7 active tonnes in the detector. 
\clearpage
\tdrsubsec[R1.2]{WBS 1.2 Xenon Vessel}
The primary requirements for the cryostat vessels are shown in Table~\ref{SRDt:R1.2Tab}. The size of the cryostat determines the amount of target that can be deployed and flows down from the top level science requirement but also the infrastructure requirement that all parts must fit in the Yates shaft. The cryostat vessels must also be low radioactivity. Safety considerations dictate that the vessels are compliant with standard engineering codes. 

\begin{center}
\sffamily
\renewcommand{\ltbcp}{R1.2Tab}
\begin{longtable}{ | m{1.7cm} | m{0.95cm} | m{3.1cm} | m{4.5cm} | m{3.9cm} | }
 \caption[Level 2 Requirements for WBS 1.2 Xenon Vessel
       \ifKL\space\texttt{\FloatPre t:\ltbcp}\fi]
       {Level 2 Requirements for WBS 1.2 Xenon Vessel
       \ifKL\space\texttt{\FloatPre t:\ltbcp}\fi\label{\FloatPre t:\ltbcp}}\\
\hline
\rowcolor{mrocol}
{ Number} & { WBS} & {Requirement Name} & {Requirement Description} & { Rationale} \\ \hline
\endfirsthead
\caption[]{\sfs (continued)}\\
\hline
\rowcolor{mrocol}
{ Number} & { WBS} & {Requirement Name} & {Requirement Description} & { Rationale} \\ \hline
\hline
\endhead
\rowcolor{lrocol}
\multicolumn{5}{|c|}{\sfs\Tstrut (continued on next page)}\\
\hline
\endfoot
\endlastfoot
	R-120001 & 1.2 & Inner Cryostat Vessel (ICV) Maximum Outer Diameter & Inner cryostat outside diameter less than \SI{1.702}{\meter} (\SI{67.0}{\inchl}).  Ports must be at angles where there is sufficient clearance. & Must fit within Yates shaft. \\ \hline
	R-120002 & 1.2 & Inner cryostat vessel (ICV) compact geometry & Design compact ICV to minimize use of passive xenon: ICV conical section and ellipsoidal 3:1 dished end & Saving xenon \\ \hline
	R-120003 & 1.2 & Outer Cryostat Vessel segmented & No fabrication underground & Segmented OCV fits into Yates shaft \\ \hline
	R-120004 & 1.2 & Low radioactivity of the cryostat material  & Radioactivity budget from background simulations. Maximum contribution to the overall background: \SI{3.3}{\%} of pp solar neutrinos and 0.03 NR event  & Low radioactivity, low density minimizing, efficiency of Outer Detector \\ \hline
	R-120005 & 1.2 & Outer Cryostat Vessel (OCV) and Inner Cryostat Vessel (ICV) designed for \SI{1.48}{bar} external pressure and vacuum internal & Working conditions for OCV and most severe failure mode for ICV  & Vacuum inside and water outside    \\ \hline
	R-120006 & 1.2 & Inner Cryostat Vessel (ICV) designed for \SI{4}{bar} inner pressure and vacuum external  & ICV working conditions & Xe gas inside with maximum pressure - including hydrostatic pressure at the bottom dished end; \SI{3.4}{bar} top head \\ \hline
	R-120007 & 1.2 & Design compliance to codes & Compliance to ASME BPVC code VIII, 2012 Int. Building Code, ASCE 7 with soil classification Cals B for seismic conditions, Fabricator holds U-stamp certificate & Required by SURF SD regulations   \\ \hline
\end{longtable}
\end{center}

\tdrsubsec[R1.3]{WBS 1.3 Cryogenic System}
The primary requirements for the Cryogenic System are shown in Table~\ref{SRDt:R1.3Tab}. The main objective of this system is to enable safe and efficient cooling of the LXe target volume. In particular, the cooling power must be adequate to enable rapid circulation of the Xe volume for purification to satisfy the Level-2 requirements in WBS 1.4, which is an example of linkage between WBS requirements. 

\begin{center}
\sffamily
\renewcommand{\ltbcp}{R1.3Tab}
\begin{longtable}{ | m{1.7cm} | m{0.95cm} | m{2.5cm} | m{5.0cm} | m{3.9cm} | }
 \caption[Level 2 Requirements for WBS 1.3 Cryogenic System
       \ifKL\space\texttt{\FloatPre t:\ltbcp}\fi]
       {Level 2 Requirements for WBS 1.3 Cryogenic System
       \ifKL\space\texttt{\FloatPre t:\ltbcp}\fi\label{\FloatPre t:\ltbcp}}\\
\hline
\rowcolor{mrocol}
{ Number} & { WBS} & {Requirement Name} & {Requirement Description} & { Rationale} \\ \hline
\endfirsthead
\caption[]{\sfs (continued)}\\
\hline
\rowcolor{mrocol}
{ Number} & { WBS} & {Requirement Name} & {Requirement Description} & { Rationale} \\ 
\hline
\endhead
\rowcolor{lrocol}
\multicolumn{5}{|c|}{\sfs\Tstrut (continued on next page)}\\
\hline
\endfoot
\endlastfoot
	R-130001 & 1.3 & Sufficient cooling power & Cryogen cooling systems shall be sufficient to remove heat for \SI{500}{\slpm} of Xenon circulation; purge of the detector Xenon gas space; and thermal losses of all the system components. & There must be adequate cooling to liquefy Xenon within the parameters of flow required to attain purity. \\ \hline
	R-130002 & 1.3 & Oxygen deficiency hazard safety & Engineered controls shall be implemented to achieve ODH Class 2 or better. & SURF requires ODH Class 2 or better for all experimental implementations \\ \hline
	R-130003 & 1.3 & Pressure safety - valves & Redundant relief devices shall be employed for pressure safety.  Relief devices shall be sized per CGA S-1.3. & Properly sized redundant relief devices are a primary engineered control for pressure systems. \\ \hline
	R-130004 & 1.3 & Pressure safety - monitoring & Active pressure monitoring shall be utilized such that alarms provide warning of pressure going higher than the planned operating pressure range. & Monitoring of pressure may provide an opportunity to act on elevated pressure before one of the primary safety devices is activated. \\ \hline
	R-130005 & 1.3 & Materials & Materials exposed to LN temperatures shall be: a) low carbon stainless steel; b) aluminum based alloys; c) nickel based alloys; d) copper / copper based alloys; or e) pure titanium. & These materials will remain ductile at \SI{77}{K}. \\ \hline
	R-130006 & 1.3 & Thermal insulation & Equipment at \SI{175}{K} shall have a minimum of 10 layers and equipment at \SI{77}{K} shall have a minimum of 25 layers of multi-layer insulation (MLI). & MLI is the most effective method of minimizing radiative heat load.  Lower operating temperatures require additional layers of MLI. \\ \hline
\end{longtable}
\end{center}

\clearpage
\tdrsubsec[R1.4]{WBS 1.4 Xenon Purification}
The primary requirements for WBS 1.4 Xenon Purification are shown in Table~\ref{SRDt:R1.4Tab}. These requirements primarily flow down from R-0003 and R-0005, as the Xe must be pure enough to enable efficient extraction of signals to satisfy the analysis threshold requirement but also low enough in internal radioactive sources like Kr and Rn to satisfy the internal backgrounds requirement.  

\begin{center}
\sffamily
\renewcommand{\ltbcp}{R1.4Tab}
\begin{longtable}{ | m{1.7cm} | m{0.95cm} | m{3.1cm} | m{4.5cm} | m{3.9cm} | }
 \caption[Level 2 Requirements for WBS 1.4 Xenon Purification
       \ifKL\space\texttt{\FloatPre t:\ltbcp}\fi]
       {Level 2 requirements for WBS 1.4 Xenon Purification
       \ifKL\space\texttt{\FloatPre t:\ltbcp}\fi\label{\FloatPre t:\ltbcp}}\\
\hline
\rowcolor{mrocol}
{ Number} & { WBS} & {Requirement Name} & {Requirement Description} & { Rationale} \\ \hline
\endfirsthead
\caption[]{\sfs (continued)}\\
\hline
\rowcolor{mrocol}
{ Number} & { WBS} & {Requirement Name} & {Requirement Description} & { Rationale} \\ \hline
\hline
\endhead
\rowcolor{lrocol}
\multicolumn{5}{|c|}{\sfs\Tstrut (continued on next page)}\\
\hline
\endfoot
\endlastfoot
	R-140001 & 1.4 & Xe electronegative purity & Charge absorption length \SI{> 1.5}{\meter}, \COt equivalent \SI{=0.4}{\ppb}. & Collect charge and scintillation throughout the entire volume of the TPC. \\ \hline
	R-140002 & 1.4 & Removal of Kr and Ar from Xe. & Ability to remove natural Kr from the Xe to a concentration of  \SI{<0.015}{\ppt\g\per\g}. Ability to remove natural Ar from the Xe to a concentration of \SI{<0.45}{\ppb\g\per\g}. & Limit ER Background from \IKreF and \IArTn.  \\ \hline
	R-140003 & 1.4 & Control the ingress of \IRnttt into the Xe & Control the decay rate of \IRnttt in the Xe. & ER background from \IPbtof. Limit the \IRnttt decay rate in the active Xe to \SI{13.4}{\milli\becquerel}. \\ \hline
	R-140004 & 1.4 & \IXeotS activity & \SI{88}{\micro\becquerel\per\kg} activity in the \SI{5.6}{\tonnesl} fiducial volume at the start of physics data taking. & No more than 23 single-scatter events below \SI{6}{\keVee} from \IXeotS in the lifespan of the experiment.  \\ \hline
	R-140005 & 1.4  & Safe Xe recovery & Safe recovery of the Xe during normal operations and during an emergency & Protection of the Xe investment. \\ \hline
\end{longtable}
\end{center}

\clearpage
\tdrsubsec[R1.5]{WBS 1.5 Xenon Detector System}
There are several requirements in WBS 1.5 Xenon Detector System. 
For example, the TPC must be large enough to accommodate the required target mass, the electric fields need to be adequate to achieve efficient single electron detection and to achieve the required \SI{99.5}{\percent} ER/NR discrimination, the TPC must have adequate light collection, and the detector components must be made from clean materials to limit the external backgrounds seen by the LXe. R-150009 is another good example of the linkages between different WBS systems, as the power dissipated in the detector must be low enough that the cooling power specified in R-130001 is adequate.  The primary requirements for WBS 1.5 Xenon Detector System are listed in Table~\ref{SRDt:R1.5Tab}. 
  
\begin{center}
\sffamily
\renewcommand{\ltbcp}{R1.5Tab}
\begin{longtable}{ | m{1.7cm} | m{0.95cm} | m{3.1cm} | m{3.0cm} | m{5.4cm} | }
 \caption[Level 2 Requirements for WBS 1.5 Xenon Detector System
       \ifKL\space\texttt{\FloatPre t:\ltbcp}\fi]
       {Level 2 Requirements for WBS 1.5 Xenon Detector System
       \ifKL\space\texttt{\FloatPre t:\ltbcp}\fi\label{\FloatPre t:\ltbcp}}\\
\hline
\rowcolor{mrocol}
{ Number} & { WBS} & {Requirement Name} & {Requirement Description} & { Rationale} \\ \hline
\endfirsthead
\caption[]{\sfs (continued)}\\
\hline
\rowcolor{mrocol}
{ Number} & { WBS} & {Requirement Name} & {Requirement Description} & { Rationale} \\ \hline
\hline
\endhead
\rowcolor{lrocol}
\multicolumn{5}{|c|}{\sfs\Tstrut (continued on next page)}\\
\hline
\endfoot
\endlastfoot
	R-150001 & 1.5 & TPC inner dimensions & Dia. \SI{=1456}{\milli\meter} / length  \SI{=1456}{\milli\meter} & \SI{7}{\tonnel} active mass, optimal self-shielding \\ \hline
	R-150002 & 1.5 & TPC drift field & \SI{300}{\volt\per\centi\meter} & \SI{99.5}{\%} discrimination; drift time \SI{<1}{\milli\second} \\ \hline
	R-150003 & 1.5 & Electroluminescence field (GXe) & \SI{10}{\kilo\volt\per\centi\meter} & \SI{95}{\%} emission; \SI{50}{\phe\per\elementarycharge^-}; e-trains; wall events \\ \hline
	R-150004 & 1.5 & Energy resolution & \SI{2.0}{\%} at \SI{2.5}{\MeV} (S1+S2) & Gamma spectroscopy for background model \\ \hline
	R-150005 & 1.5 & LXe Skin threshold & \SI{100}{\keVee} with \SI{3}{\phe} in \SI{>95}{\%} of skin volume & Veto efficiency required for fiducial mass \\ \hline
	R-150006 & 1.5 & Component radioactivities & $<250/240/240/540$ \si{\milli\becquerel} U/Th/Co/K & \SI{<0.4} NR cts, \SI{<10}{\%} pp neutrino ER cts (matches R-200114 to R-200117) \\ \hline
	R-150007 & 1.5 & Photocathode coverage & \SI{38}{\%} top and bottom arrays & S1 Photon detection efficiency, vertex resolution, discrimination \\ \hline
	R-150008 & 1.5 & PTFE reflectivity & \SI{>95}{\%} in LXe & Light collection efficiency in TPC and Skin \\ \hline
	R-150009 & 1.5 & Total heat in or on cryostat & 140 W & Cooling power requirement, fluid model \\ \hline
	R-150010 & 1.5 & Max field for cathodic surfaces & \SI{50}{\kilo\volt\per\centi\meter} & Limit spurious photon/electron emission \\ \hline
\end{longtable}
\end{center}

\clearpage
\tdrsubsec[R1.6]{WBS 1.6 Outer Detector}
WBS 1.6, the Outer Detector, acts as the main neutron veto, and therefore most requirements in this WBS flow down from R-0007. The Level 2 Requirements for WBS 1.6 Outer Detector are shown in Table~\ref{SRDt:R1.6Tab}.   

\begin{center}
\sffamily
\renewcommand{\ltbcp}{R1.6Tab}
\begin{longtable}{ | m{1.7cm} | m{0.95cm} | m{3.1cm} | m{4.5cm} | m{3.9cm} | }  
 \caption[Level 2 Requirements for WBS 1.6 Outer Detector
       \ifKL\space\texttt{\FloatPre t:\ltbcp}\fi]
       {Level 2 Requirements for WBS 1.6 Outer Detector
       \ifKL\space\texttt{\FloatPre t:\ltbcp}\fi\label{\FloatPre t:\ltbcp}}\\
\hline
\rowcolor{mrocol}
{ Number} & { WBS} & {Requirement Name} & {Requirement Description} & { Rationale} \\ \hline
\endfirsthead
\caption[]{\sfs (continued)}\\
\hline
\rowcolor{mrocol}
{ Number} & { WBS} & {Requirement Name} & {Requirement Description} & { Rationale} \\ \hline
\hline
\endhead
\rowcolor{lrocol}
\multicolumn{5}{|c|}{\sfs\Tstrut (continued on next page)}\\
\hline
\endfoot
\endlastfoot
	R-160001 & 1.6 & Neutron veto efficiency & Detection efficiency of \SI{95}{\%} for a \SI{1}{\MeV} neutron that scatters once in the xenon & Needed to meet the neutron part of requirement on external backgrounds \\ \hline
	R-160002 & 1.6 & Outer Detector threshold & Threshold of \SI{200}{\keV} (\SI{50}{\%} efficiency) on energy deposit in liquid scintillator & Highest achievable veto efficiency with acceptable deadtime \\ \hline
	R-160003 & 1.6 & Number of photoelectrons detected per energy deposit &  \SI{>80}{\phe\per\MeV}  & Corresponds to about \SI{15}{\phe} at threshold of \SI{200}{\keV} \\ \hline
	R-160004 & 1.6 & Light collection efficiency & Efficiency of \SI{>5}{\%} for having VUV photons  strike a PMT & Light detection is proportional to light collection efficiency. \\ \hline
	R-160005 & 1.6 & Deadtime & OD must not veto more than \SI{5}{\%} of the WIMP search livetime  & Keep overall activity low to limit deadtime imposed by OD veto triggers \\ \hline
\end{longtable}
\end{center}

\tdrsubsec[R1.7]{WBS 1.7 Calibration System}
The Calibration System must provide accurate calibrations of the light and charge yields, position dependencies, time variations, energy threshold, resolution, and discrimination parameters of the central TPC without taking an undue amount of running time away from the primary dark matter search. The Calibration System must also provide an accurate picture of the performance of the veto systems: the Xe skin and Outer Detector.  The L2 requirements for WBS 1.7 Calibration System are listed in Table~\ref{SRDt:R1.7Tab}. 

\begin{center}
\sffamily
\renewcommand{\ltbcp}{R1.7Tab}
\begin{longtable}{ | m{1.7cm} | m{0.95cm} | m{3.1cm} | m{4.5cm} | m{3.9cm} | }  
 \caption[Level 2 Requirements for WBS 1.7 Calibration System
       \ifKL\space\texttt{\FloatPre t:\ltbcp}\fi]
       {Level 2 Requirements for WBS 1.7 Calibration System
       \ifKL\space\texttt{\FloatPre t:\ltbcp}\fi\label{\FloatPre t:\ltbcp}}\\
\hline
\rowcolor{mrocol}
{ Number} & { WBS} & {Requirement Name} & {Requirement Description} & { Rationale} \\ \hline
\endfirsthead
\caption[]{\sfs (continued)}\\
\hline
\rowcolor{mrocol}
{ Number} & { WBS} & {Requirement Name} & {Requirement Description} & { Rationale} \\ \hline
\hline
\endhead
\rowcolor{lrocol}
\multicolumn{5}{|c|}{\sfs\Tstrut (continued on next page)}\\
\hline
\endfoot
\endlastfoot
	R-170001 & 1.7 & Calibration Times & \SI{<12} hrs/week for periodic calibrations and \SI{<100}{\day} total for infrequent calibrations & 12 hrs can fit in a single day of SURF operations (2 shifts). 100 days is \SI{10}{\percent} of total exposure target. \\ \hline
	R-170002 & 1.7 & Calibrate TPC (x,y,z) variation & \SI{<2}{\percent} uncertainty on S1 and S2 mean area (of a monoenergetic peak) in bins of 5x5x5 cm & Understand detection efficiencies and gains at scales relevant to variation \\ \hline
	R-170003 & 1.7 & Calibrate ER band & \SI{<2}{\percent} uncertainty on the measured ER band mean and $\pm1\sigma$ contours (in 1 phd bins of S1) & Clearly define ER background region \\ \hline
	R-170004 & 1.7 & Calibrate NR band  & \SI{<2}{\percent} uncertainty on the measured NR band mean and $\pm1\sigma$ contours (in 1 phd bins of S1) & Clearly define signal region \\ \hline
	R-170005 & 1.7 & Calibrate NR response at threshold &  \SI{<5}{\percent} uncertainty on energy at \SI{50}{\percent} acceptance	&Clearly define threshold and map out \IBe neutrino response  \\ \hline
	R-170006 & 1.7 & Calibrate TPC, skin and LS signal time offsets & $<\pm1$ sample uncertainty in primary scintillation rise time jitter across the three analog electronics chains & Necessary for optimal veto efficiency \\ \hline
	R-170007 & 1.7 & Calibrate temporal response &  \SI{<2}{\percent} uncertainty in S2 mean area (of a monoenergetic peak) in z-direction bins of \SI{5}{\cm} & To ensure stable energy reconstruction \\ \hline
	R-170008 & 1.7 & Calibrate Xe skin energy response & \SI{<2}{\percent} uncertainty on ER phd/keV (\SI{1}{\cm} z bins) in Xe side skin &	Understand skin threshold   \\ \hline
	R-170009 & 1.7 & Calibrate energy scale of LS & \SI{<2}{\percent} uncertainty on ER phd/keV (volume-averaged) at \SI{2.2}{\MeV} &	Verify outer detector performance
  \\ \hline
  R-170010 & 1.7 & Calibrate NR response & \SI{<5}{\percent} uncertainty on energy at about \SI{30}{\keV}	& Bound the upper end of the WIMP-search range
 \\ \hline
    R-170011 & 1.7 & Calibrate fiducial volume fraction & \SI{<5}{\percent} uncertainty on fraction of TPC LXe mass satisfying fiducial selection criteria	& Directly scales exposure and sensitivity \\ \hline
\end{longtable}
\end{center}

\newpage
\tdrsubsec[R1.8]{WBS 1.8 Electronics, DAQ, Controls, and Computing}
The primary requirements for WBS 1.8 Electronics, DAQ, Controls, and Computing are shown in Table~\ref{SRDt:R1.8Tab}.  The main requirement here is to enable a low energy threshold in the detector. However, there are two requirements that flow across from WBS 1.7, as the data acquisition must be robust enough to handle the event rates listed in Table~\ref{SRDt:R1.7Tab} needed to effectively calibrate the detector. On the controls side, this subsystem is responsible for providing monitoring and control during emergencies, and in particular the control of the xenon recovery system of WBS~1.4. 

\begin{center}
\sffamily
\renewcommand{\ltbcp}{R1.8Tab}
\begin{longtable}{ | m{1.7cm} | m{0.95cm} | m{3.1cm} | m{3.0cm} | m{5.4cm} | } 
 \caption[Level 2 Requirements for WBS 1.8  Electronics, DAQ, Controls, and Computing
       \ifKL\space\texttt{\FloatPre t:\ltbcp}\fi]
       {Level 2 Requirements for WBS 1.8  Electronics, DAQ, Controls, and Computing
       \ifKL\space\texttt{\FloatPre t:\ltbcp}\fi\label{\FloatPre t:\ltbcp}}\\
\hline
\rowcolor{mrocol}
{ Number} & { WBS} & {Requirement Name} & {Requirement Description} & { Rationale} \\ \hline
\endfirsthead
\caption[]{\sfs (continued)}\\
\hline
\rowcolor{mrocol}
{ Number} & { WBS} & {Requirement Name} & {Requirement Description} & { Rationale} \\ \hline
\hline
\endhead
\rowcolor{lrocol}
\multicolumn{5}{|c|}{\sfs\Tstrut (continued on next page)}\\
\hline
\endfoot
\endlastfoot
	R-180001 & 1.8 & Energy threshold & \SI{90}{\%} efficiency for a single \si{phe}  & The S1 analysis threshold is defined by our efficiency to capture single-photoelectron signals.  For a specific gain, this requirement fixes the noise requirements of the electronic chain.  This calculation depends sensitively on the assumed \SI{35}{\%} variations in the PMT response. \\ \hline
	R-180002 & 1.8 & Source count rate & \SI{150}{\Hz} & Able to handle source calibrations.  Rates are limited by the drift time. \\ \hline
	R-180003 & 1.8 & LED count rate & \SI{4}{\kilo\Hz} & Able to handle LED calibration rates. \\ \hline
    	R-180004 & 1.8 & Guarantee the safety of xenon supply and the xenon circulation system & Use Programmable Logic Controllers (PLCs) to control and monitor the xenon purification, circulation, and storage systems & PLCs are used to ensure that the detector and xenon system are maintained in a safe state during emergencies that result in a shutdown of the slow-control infrastructure.  The PLC system provides continuous real-time monitoring and control of the critical subsystems and will initiate automatic recovery of the xenon to the storage facility during an emergency. \\ \hline
\end{longtable}
\end{center}

\newpage
\tdrsubsec[R1.9]{WBS 1.9 Integration and Installation}
The primary requirements for WBS 1.9 Integration and Installation are shown in Table~\ref{SRDt:R1.9Tab}. There are two main requirements in this WBS. The first is that all parts must be installed correctly, and the second is that the detector must be installed with clean parts in such a way to minimize exposure to radon and dust. These requirements flow down from the top level science requirements, particularly R-0005 and R-0007. 

\begin{center}
\sffamily
\renewcommand{\ltbcp}{R1.9Tab}
\begin{longtable}{ | m{1.7cm} | m{0.95cm} | m{2.6cm} | m{5.6cm} | m{3.4cm} | }
 \caption[Level 2 Requirements for WBS 1.9 Integration and Installation
       \ifKL\space\texttt{\FloatPre t:\ltbcp}\fi]
       {Level 2 Requirements for WBS 1.9 Integration and Installation
       \ifKL\space\texttt{\FloatPre t:\ltbcp}\fi\label{\FloatPre t:\ltbcp}}\\
\hline
\rowcolor{mrocol}
{ Number} & { WBS} & {Requirement Name} & {Requirement Description} & { Rationale} \\ \hline
\endfirsthead
\caption[]{\sfs (continued)}\\
\hline
\rowcolor{mrocol}
{ Number} & { WBS} & {Requirement Name} & {Requirement Description} & { Rationale} \\ \hline
\hline
\endhead
\rowcolor{lrocol}
\multicolumn{5}{|c|}{\sfs\Tstrut (continued on next page)}\\
\hline
\endfoot
\endlastfoot
R-190001	&1.9	&Parts used have acceptable radioactivity	&All parts have a traceable history that shows they have low enough radioactivity to be used.  This will be documented in a database and there will be an acceptance sheet with sign off that the part is usable.	&Control background level in the experiment \\ \hline
R-190002	&1.9	&Parts used are clean	&All parts have a cleaning procedure to ensure surface contamination is at an acceptable level.  This will be documented in a database and there will be an acceptance sheet with sign off that the part is usable.	&Control background in the experiment.  Control contamination of Xenon that would reduce electron drift length and/or increase light absorption \\ \hline
R-190003	&1.9	&Parts are moved and lifted without damage	&Moving and lifting will be done to procedures written and approved by subsystem experts.  This will include analysis of rigging attachment and support loads where applicable.  Workers will review the procedures and have rigging experience	&Detector is assembled without damaging parts\\ \hline
R-190004	&1.9	&Parts are correctly assembled	&Assembly work will be done to procedure written and approved by subsystem experts.  Workers will review the procedure and have training where necessary.  	&Detector is assembled correctly\\ \hline
R-190005	&1.9	&Control Part exposure to Radon	&Parts have an allowable total exposure to Radon, based on material and location.  Monitored time of exposure and Radon level of air are recorded in a database.  For critical parts, may use samples.	&Control background level in the experiment\\ \hline

\end{longtable}
\end{center}

\newpage
\tdrsubsec[R1.10]{WBS 1.10 Cleanliness and Screening}
The primary requirements for WBS 1.10 Cleanliness and Screening are shown in Table~\ref{SRDt:R1.10Tab}. At Level 2, the WBS 1.10 requirements dictate that the project has the sensitivity and capacity to screen all detector components at the level needed to ensure that the radioactivity requirements are satisfied. WBS 1.10 also keeps track of allowed radioactivity levels for individual components, and more details can be found in Chapter~\ref{chap:MAS}.

\begin{center}
\sffamily
\renewcommand{\ltbcp}{R1.10Tab}
\begin{longtable}{ | m{1.7cm} | m{0.95cm} | m{4.4cm} | m{2.3cm} | m{4.8cm} | }
 \caption[Level 2 Requirements for WBS 1.10 Cleanliness and Screening
       \ifKL\space\texttt{\FloatPre t:\ltbcp}\fi]
       {Level 2 Requirements for WBS 1.10 Cleanliness and Screening 
       \ifKL\space\texttt{\FloatPre t:\ltbcp}\fi\label{\FloatPre t:\ltbcp}}\\
\hline
\rowcolor{mrocol}
{ Number} & { WBS} & {Requirement Name} & {Requirement Description} & { Rationale} \\ \hline
\endfirsthead
\caption[]{\sfs (continued)}\\
\hline
\rowcolor{mrocol}
{ Number} & { WBS} & {Requirement Name} & {Requirement Description} & { Rationale} \\ \hline
\hline
\endhead
\rowcolor{lrocol}
\multicolumn{5}{|c|}{\sfs\Tstrut (continued on next page)}\\
\hline
\endfoot
\endlastfoot
R-200001 &	1.10 &	Assay Sensitivity for U and Th by Direct Counting	& \SI{10}{\ppt} & Sensitivity needed to assess total radioactive background \\ \hline
R-200002&	1.10	&Assay Sensitivity for U and Th by Neutron Activation Analysis	&  \SI{1.5}{\ppt} & Sensitivity needed to assess total radioactive background \\ \hline
R-200003&	1.10&	Assay Sensitivity for U and Th by ICP-MS &	\SI{10}{\ppt} & Sensitivity needed to assess total radioactive background\\\hline
R-200004	&1.10	& Assay Sensitivity for Radon Emanation 	& \SI{0.3}{\milli\becquerel} &  Sensitivity needed to assess total radioactive background \\\hline
R-200005 &	1.10 &	Assay Sensitivity for \IPbtoz in the bulk	& \SI{10}{\milli\becquerel\per\kilo\g} & Sensitivity needed to assess total radioactive background \\ \hline
R-200006&	1.10	&Assay Sensitivity for \IPbtoz Plateout	&  \SI{60}{\nano\becquerel\per\cm^2} & Sensitivity needed to assess total radioactive background \\ \hline
R-200007&	1.10&	Assay Sensitivity for Dust Accumulation &	\SI{10}{\nano\gram\per\cm^2} & Sensitivity needed to assess total radioactive background\\\hline
\end{longtable}
\end{center}

\newpage
\tdrsubsec[R1.11]{WBS 1.11 Offline Computing}
The primary requirements for WBS 1.11 Offline Computing are shown in Table~\ref{SRDt:R1.11Tab}, and are primarily the amount of disk and CPU to enable adequately detailed studies of the detector and background levels to be performed. Without sufficient computing resources, the project is unable to determine whether it is satisfying the top level science requirements. WBS 1.11 is also critically important in coordinating computing and software tasks with other WBS elements, particularly in simulations and analysis tools. 

\begin{center}
\sffamily
\renewcommand{\ltbcp}{R1.11Tab}
\begin{longtable}{ | m{1.7cm} | m{0.95cm} | m{3.1cm} | m{3.0cm} | m{5.4cm} | }
 \caption[Level 2 Requirements for WBS 1.11 Offline Computing
       \ifKL\space\texttt{\FloatPre t:\ltbcp}\fi]
       {Level 2 Requirements for WBS 1.11 Offline Computing 
       \ifKL\space\texttt{\FloatPre t:\ltbcp}\fi\label{\FloatPre t:\ltbcp}}\\
\hline
\rowcolor{mrocol}
{ Number} & { WBS} & {Requirement Name} & {Requirement Description} & { Rationale} \\ \hline
\endfirsthead
\caption[]{\sfs (continued)}\\
\hline
\rowcolor{mrocol}
{ Number} & { WBS} & {Requirement Name} & {Requirement Description} & { Rationale} \\ \hline
\hline
\endhead
\rowcolor{lrocol}
\multicolumn{5}{|c|}{\sfs\Tstrut (continued on next page)}\\
\hline
\endfoot
\endlastfoot
R-210001 &	1.11 &	US Disk Costs FY15-FY18	&Procure NERSC GPFS disk for simulation, development, and initial data taking	& Handle simulation data, derived data, and user data volumes as defined in Table~\ref{OCSt:USDCstaging} \\ \hline
R-210002&	1.11	&US CPU Costs FY15-FY18	&Procure NERSC PDSF CPU for simulation, development, and initial data taking	& Handle simulation data, derived data, and user data volumes as defined in Table~\ref{OCSt:USDCstaging}  \\ \hline
R-210003&	1.11&	US Server Costs FY15-FY18&	Procure and maintain servers for simulation and development &	Handle simulation data, derived data, and user data volumes as defined in Table~\ref{OCSt:USDCstaging}  \\\hline
R-210004	&1.11	&UK Data Center	&Design, deploy, test, and maintain UK Data Center at Imperial&	Handle simulation data, derived data, and user data volumes as described in Sec.~\ref{OCSSs:UKDC}  \\\hline
\end{longtable}
\end{center}

\clearpage
\tdrsec[SensDetPerformance]{Sensitivity and Detector Performance}

We employ the simulation tools
described in Section~\ref{SRDS:Sims} to evaluate
the sensitivity of LZ, as defined by
the requirements in Section~\ref{SRDS:Reqs}, 
to a WIMP signal.  Additionally, we evaluate
the impact of variations in the parameters that
define the LZ apparatus on its sensitivity. 

The high-statistics calibrations
performed by LUX and their incorporation into
NEST have driven a conversion from the simple
cut-and-count statistical methods used in the CDR
to more advanced likelihood methods.  The use
of likelihood methods has markedly reduced the 
sensitivity of LZ to the dominant electron recoil
backgrounds. Table~\ref{SRDt:PLRBG} shows the number of counts with S1 signals between 0 and 20 photoelectrons coming from different background sources in a \SI{1000}{\day} run with no discrimination applied,
analogous to Table~\ref{MASt:BackgroundsTable} (see Sec.~\ref{SRDSs:SAnalysis} for further discussion of the cuts used to generate these numbers).  For the majority of the chapter, we use the PLR method described in Sec.~\ref{SRDSs:PLR} to exploit the ER/NR discrimination power of liquid xenon to calculate our sensitivity to dark matter. 

\begin{table}[htbp]
\centering
 \tdrtcaption[PLRBG]{Backgrounds used in the profile
likelihood analysis}{Backgrounds described by
PDFs in the profile likelihood analysis, with
the counts expected with S1 signal between
0 and 20 photoelectrons in a \SI{1000}{\day} run, with no discrimination applied,
analogous to Table~\ref{MASt:BackgroundsTable} (see Sec.~\ref{SRDSs:SAnalysis} for further discussion of the cuts used to generate these numbers). }
\centering
\sffamily
\begin{tabular}{ | m{6.7cm}|c|m{1.5cm}| }   
\hline
\rowcolor{mrocol} Background  & Type & 
Counts  \\ \hline
\IBe solar \Pgn & NR & \num{7}  \\ \hline
\Ihep \Pgn & NR & \num{0.21} \\ \hline
DSN \Pgn & NR & \num{0.05}  \\ \hline
ATM \Pgn & NR & \num{0.46}  \\ \hline
\Ipp+\IBeS+\INof solar \Pgn &ER & \num{255} \\ \hline
\IXeoTs (\Dtnbb) &ER & \num{67} \\ \hline
\IKreF &ER & \num{24.5}  \\ \hline
\IRnttt &ER & \num{722}  \\ \hline
\IRnttz &ER & \num{122}  \\ \hline
Detector components + Environmental &ER & \num{11.3}  \\ \hline
Detector components + Environmental &NR & \num{0.5}  \\ \hline
\end{tabular}
\end{table}

Our evaluation of the LZ sensitivity should be considered
as a snapshot in time. The field of direct dark matter
detection with liquid xenon TPCs continues to
achieve increasing sensitivity.
Since the completion of the
LZ CDR~\cite{Akerib:2015cja},
the LUX Collaboration has substantially advanced the
understanding of the response of liquid xenon
to both background 
and signal~\cite{Akerib:2015rjg,Akerib:2016mzi}, knowledge which
has been incorporated in this report.  As the
LZ construction project progresses, key response
and radioactive background properties will be
measured and our sensitivity evaluation will
adapt.  We endeavor to bound this evolution
by providing a "baseline" parameter set as
well as "goal" and "reduced" sets, summarized
in Table~\ref{SRDt:Dt.1Tab}.   The variation
of sensitivity with excursions in many LZ
parameters is captured in
Section~\ref{SRDSs:ParScans}.

\tdrsubsec[PLR]{Profile Likelihood Ratio Method}

The sensitivity projections in this report are 
based on a profile likelihood 
ratio (PLR) method~\cite{Cowan:2010js},
which allows near-optimum exploitation of the differences
between signal and background in the key parameters
that are reconstructed by the LZ apparatus.
The parameters with the highest sensitivity
to these differences are the position-corrected 
S1 (primary scintillation light) and
S2 (secondary luminescence from ionization) signals.
The position of events in radius $r$ and height $z$
also allows distinction between signal and 
the background events which originate from
radioactive impurities
in the material in the vicinity of the liquid
xenon TPC of LZ; these background events cluster
near the edges of the TPC, while the
WIMP signal is uniform in the liquid xenon
mass.  In this report,
we apply a simple fiducial volume cut, where
\SI{5.6}{\tonnesl} of the inner liquid xenon
is retained as the sensitive volume.  In the
future, our sensitivity studies will exploit
the distinct spatial distributions of signal
and background just as the LUX experiment
has done, and the fiducial requirement
will become unnecessary.

\begin{figure}[htbp]
\centering
\includegraphics[width=0.7\textwidth]{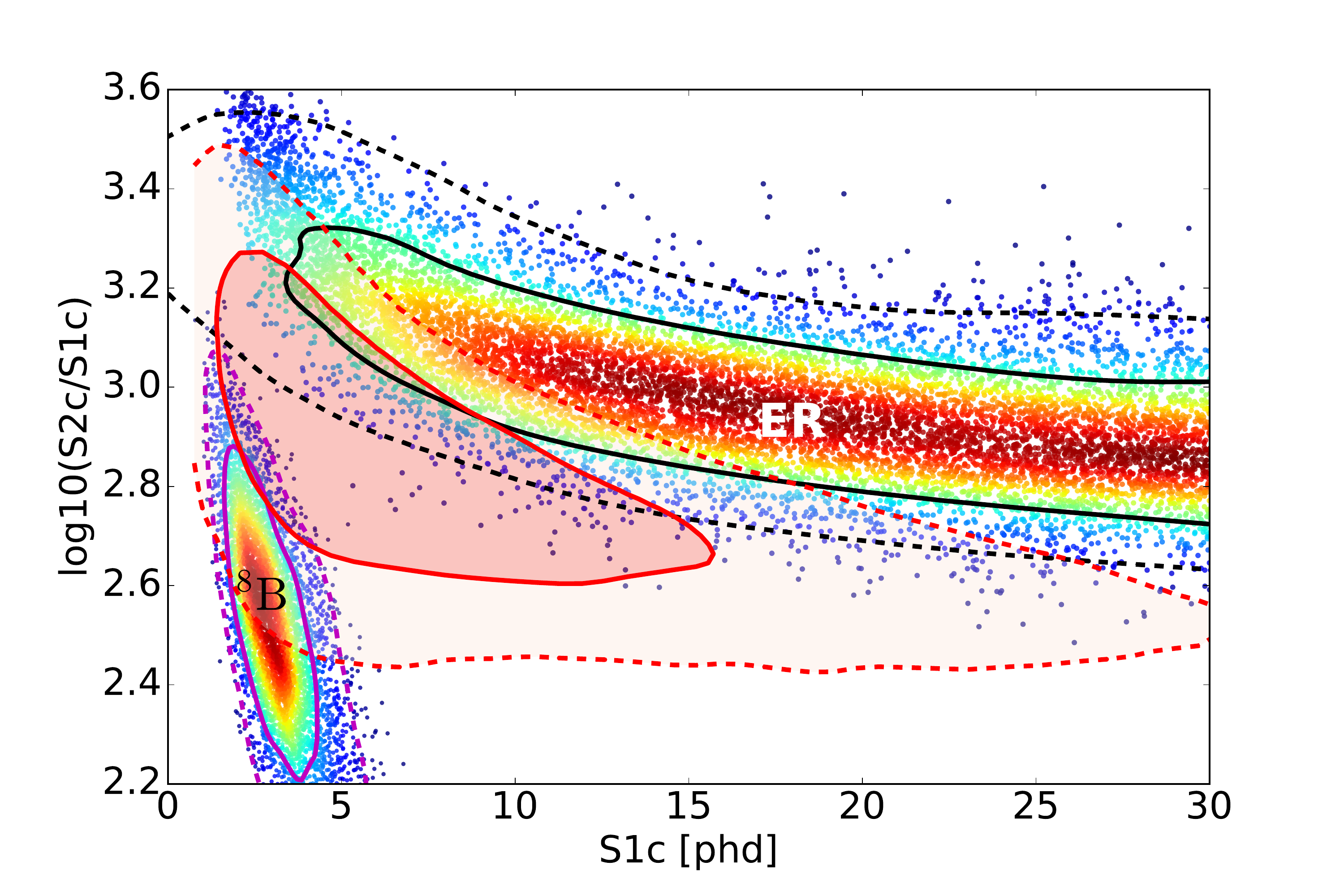}
\tdrfcaption[PLRhowa]{Profiles of background and signal in LZ}{Simulations of the most
prominent ER and NR (from \IBe) backgrounds are plotted in the 
$\log_{10}(\textrm{S2c}/\textrm{S1c})$-S1c
plane.  The statistics shown represent 5x the expected ER background and 500x the expected NR background in the nominal LZ exposure. 
The red tinted area shows the expectation for events from a
\SI{40}{\GeVpct}-mass WIMP, falling between the two background
populations with the region enclosed by the solid(dashed) line
representing the \SI{1}{\sigm}(\SI{2}{\sigm}) band.}
\end{figure} 

To execute the PLR sensitivity estimate,
signal and background probability distribution
functions (PDFs) are created in S1 and S2 after application of 
the fiducial cut in $r$ and $z$.  The signal
PDF for each WIMP mass is generated by converting
the differential energy spectrum calculated
from~\cite{McCabe:2010zh} to S1 and S2 signals in
the LZ detector using NEST and the parameterization
of detector response described in
Section~\ref{SRDSs:SEDep}.

The background PDFs are broken into the 
eleven individual components listed in
Table~\ref{SRDt:PLRBG}. 
The simulations for detector components and environmental backgrounds are 
summed together into a single PDF each for ER 
and NR events.
For each WIMP mass, we scan over the 
cross section to set a \SI{90}{\percent} 
confidence interval (CI) for the expected 
number of signal events, evaluated 
using RooStats~\cite{Moneta:2010pm}. In the PLR technique we use the
unbinned likelihood computed in the plane
of 
$\log_{10}(\textrm{S2}/\textrm{S1})$ 
versus S1.  Poisson
fluctuations are innate to the technique.

The power of the PLR technique 
arises from an optimal weighting of
the background-free and background-rich
regions in the
$\log_{10}(\textrm{S2}/\textrm{S1})$-S1
plane.
Figure~\ref{SRDf:PLRhowa}
 shows high-statistics 
simulations of the
most prominent backgrounds 
(ER events from \Ipp solar neutrinos,
\IRnttt, and \IRnttz, and NR events 
from \IBe neutrinos) in the 
$\log_{10}(\textrm{S2}/\textrm{S1})$-S1
plane, representing 5x and 500x the count rates expected in the nominal LZ exposure for ER and NR, respectively. Also shown is the
region that would be populated by
events from a \SI{40}{\GeVpct} WIMP signal, which  falls between the
\IBe and ER background areas. The PLR
technique optimally combines the 
background-free
and background-rich regions.  
The intense calibrations
recently conducted by the LUX 
collaboration provide
confidence in the knowledge of the 
shapes of the backgrounds and the signal.

\begin{figure}[htbp]
\centering
\includegraphics[width=0.7\textwidth]{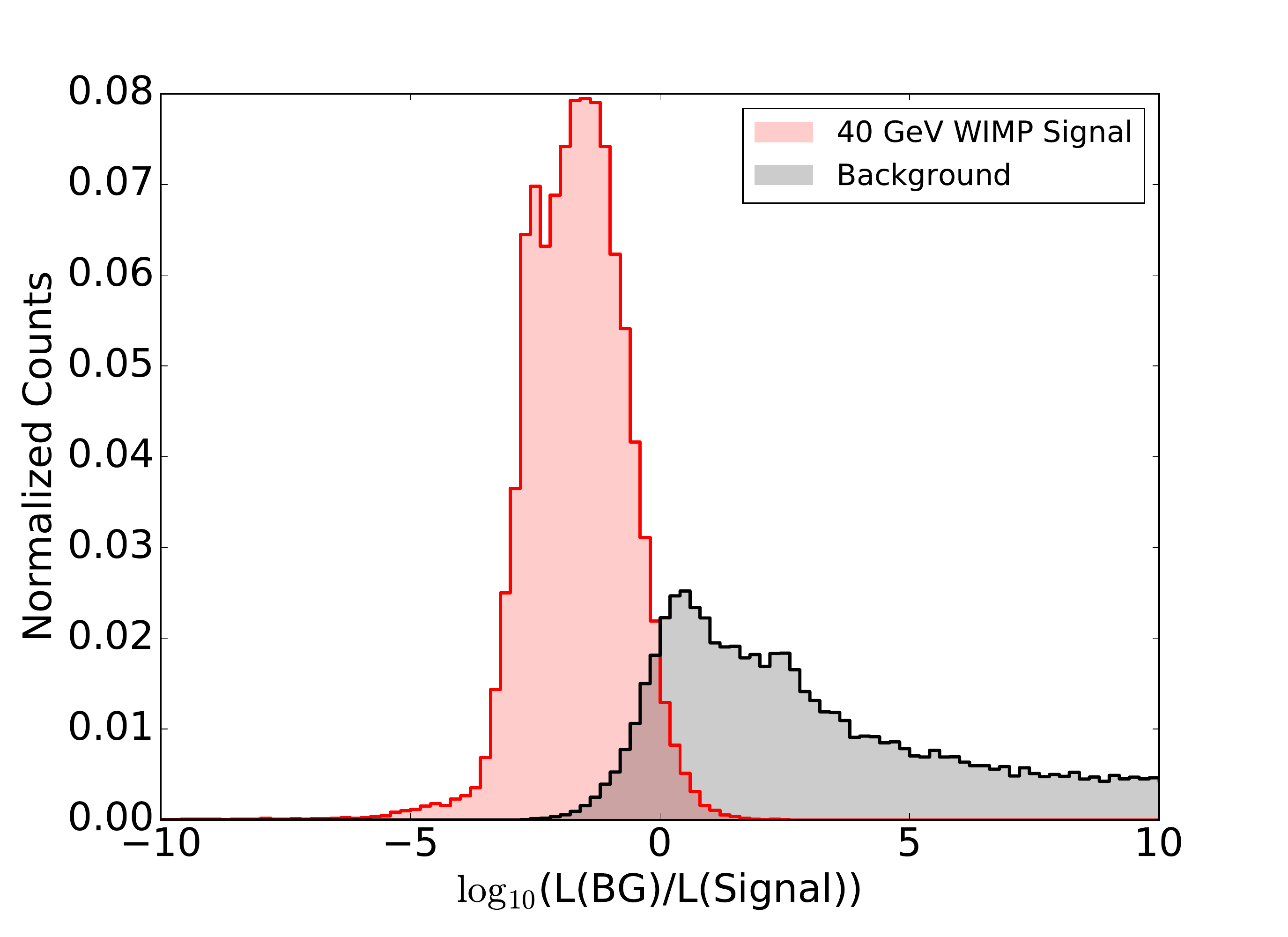}
\tdrfcaption[PLRhow]{PLR discrimination statistic}{A one-dimensional projection
of the PLR discrimination statistic. 
Two ensembles of points in the 
$\log_{10}(\textrm{S2}/\textrm{S1})$-S1
plane are considered, one distributed like
a \SI{40}{\GeVpct} WIMP signal, 
and the other like the expected background (combining both the ER and \IBe bands of Fig.~\ref{SRDf:PLRhowa}).}
\end{figure} 

\begin{figure}[htbp]
\centering
\includegraphics[width=0.49\textwidth]{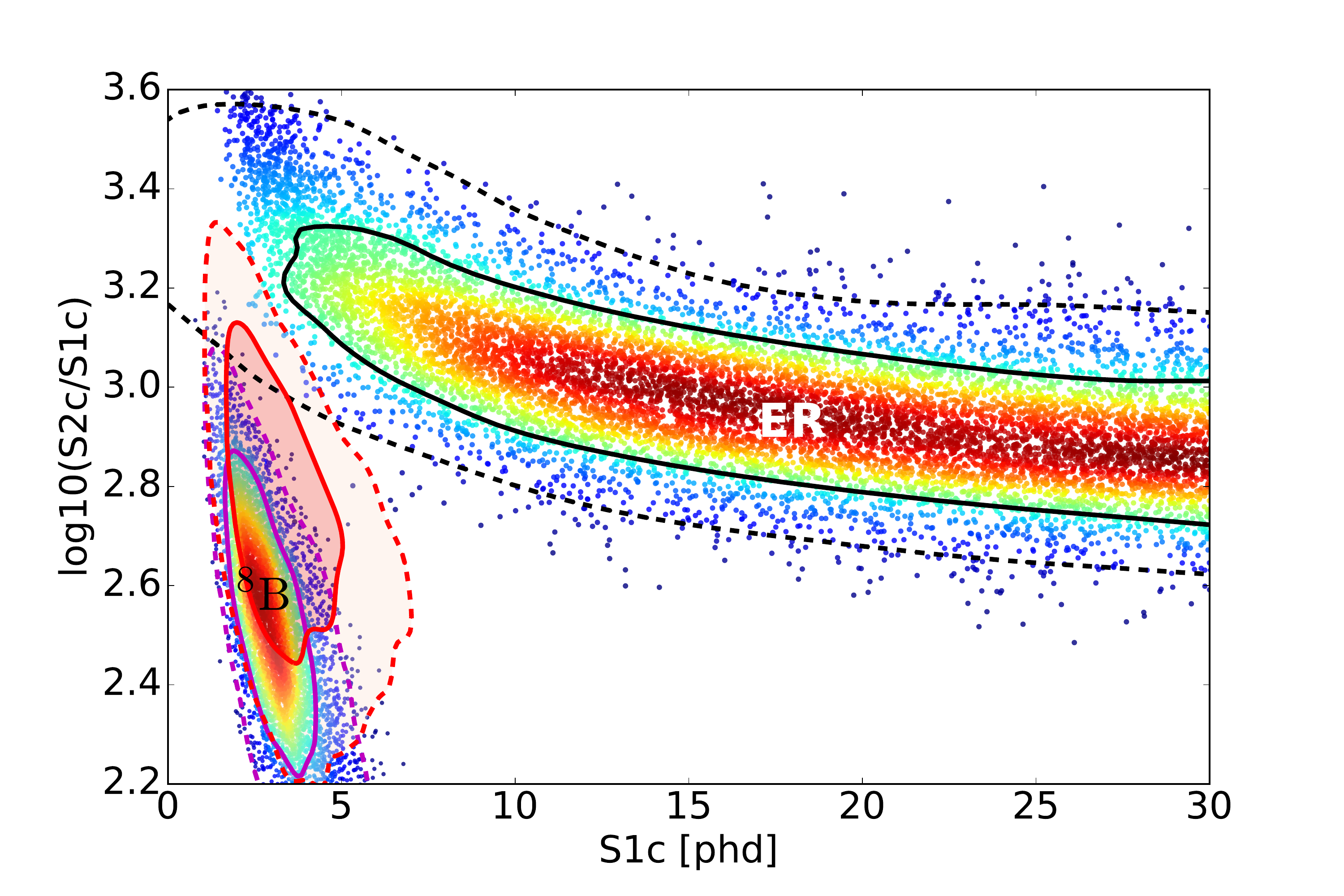}
\includegraphics[width=0.49\textwidth]{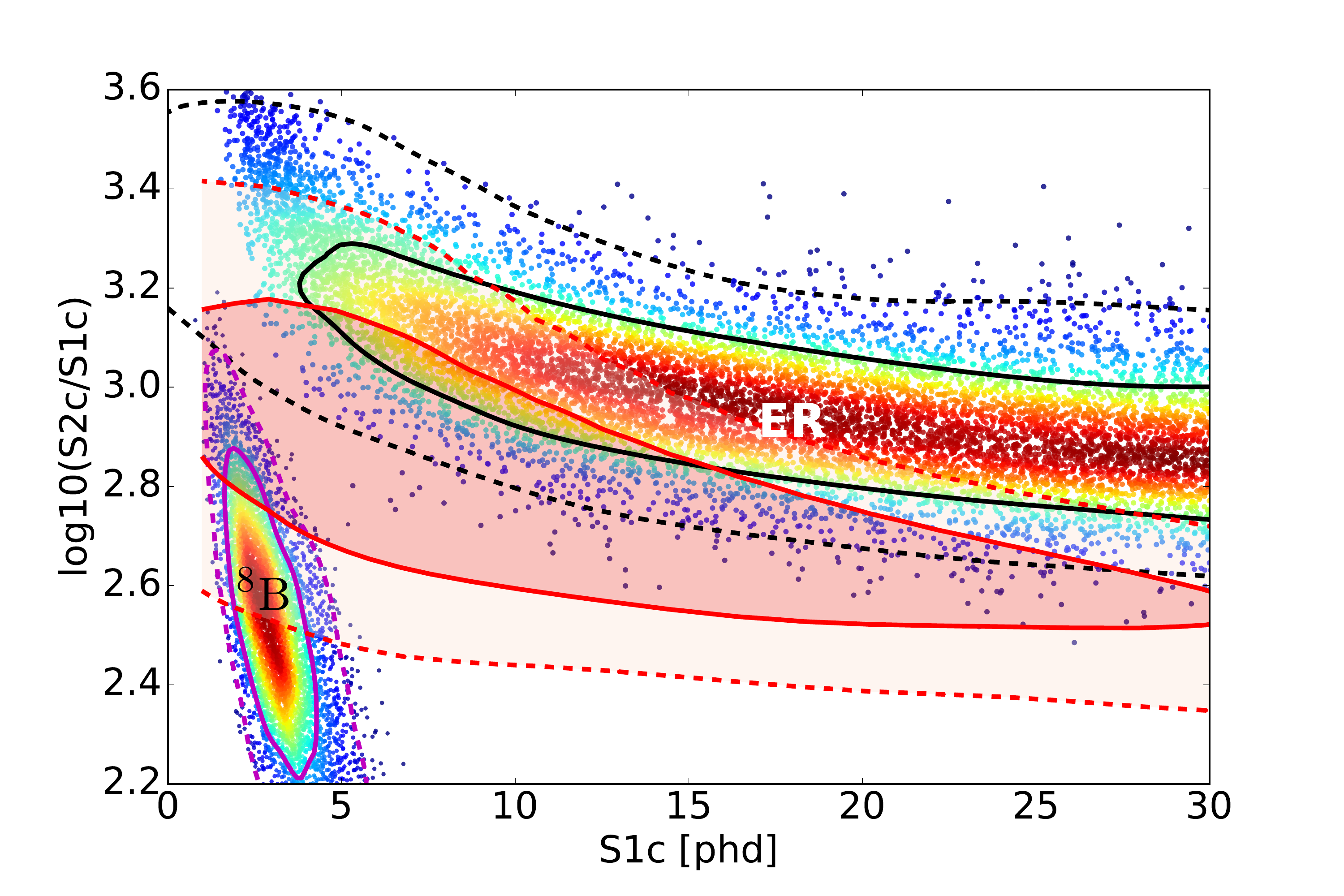}
\tdrfcaption[PLRMass]{PLR technique for
 different masses}{PLR technique for 
different masses.  The background
distributions are the same as in Figure~\ref{SRDf:PLRhowa}, but the
expected signal regions are for a
\SI{10}{\GeVpct} WIMP (left) and a
\SI{1000}{\GeVpct} WIMP (right).
The signal regions merge, respectively,
into the \IBe and ER background regions.
The expected signal regions are
tinted red, with the darker(lighter)
color \SI{1}{\sigm}(\SI{2}{\sigm}).
}
\end{figure}

The discrimination statistic that quantifies 
whether each point in the respective
ensemble more resembles background or 
signal is evaluated in one dimension
as the difference between
logarithms of the likelihoods that
a point in the ensemble is background
or signal.  Low values correspond to
poor likelihood to be background, and
high values to a good likelihood to be
background.  
Figure~\ref{SRDf:PLRhow} shows 
 the PLR discrimination statistic integrated over the full energy range for 
a \SI{40}{\GeVpct} WIMP signal
and the expected background (combining both the ER and \IBe bands of Fig.~\ref{SRDf:PLRhowa}). Comparison of the two
ensembles shows the considerable separation
between signal and background available using this method.
This plot
and similar plots constructed for different
WIMP masses are used to describe
the discrimination power of the LZ apparatus
against background.

\begin{figure}[htbp]
\centering
\includegraphics[width=0.8\textwidth]{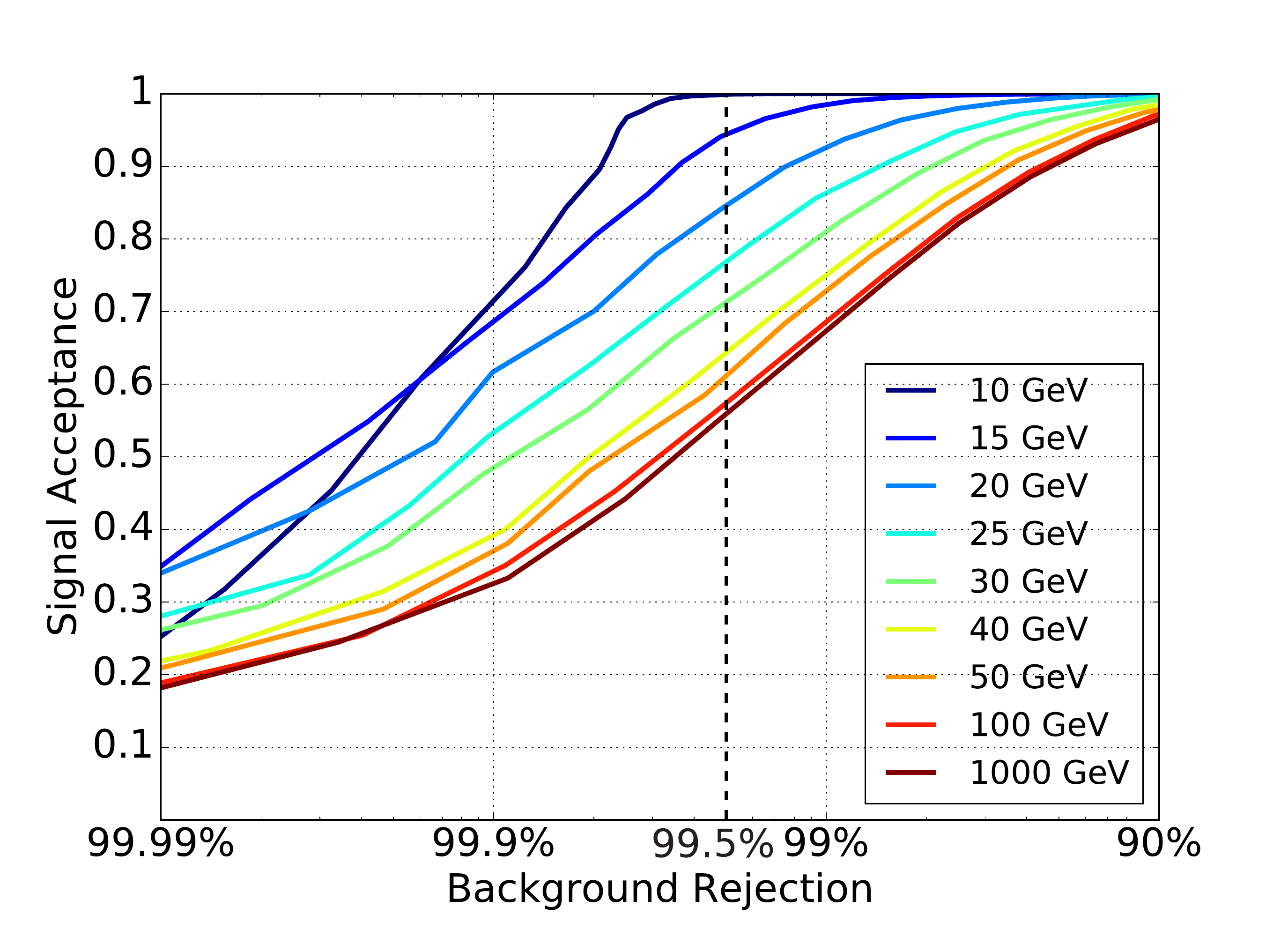}
\tdrfcaption[ROC]{Acceptance and rejection for WIMP signals in LZ }{Acceptance and rejection for WIMP signals in LZ.  For a variety of
WIMP masses, histograms like that shown in the
Fig.~\ref{SRDf:PLRhow} are
integrated to derive
the curves shown. Backgrounds from both
\IBe and ER events are included.  The
requirement of \SI{99.5}{\%} rejection
at \SI{50}{\%} acceptance is projected for all WIMP
masses.}
\end{figure}

The expected signal region varies according to the WIMP mass, and Figure~\ref{SRDf:PLRMass} shows the signal regions for WIMP masses of \SI{10}{\GeVpct} and \SI{1000}{\GeVpct} compared to the same ER and NR simulations of Fig.~\ref{SRDf:PLRhowa}. 
The PLR method naturally takes into account the shape
of the expected WIMP signal, and the achievable 
discrimination at all WIMP masses is thus significantly better than 
that attainable in the `cut and count'
technique utilized in the LZ CDR.  Figure~\ref{SRDf:ROC} shows the
signal acceptance for WIMPs of a variety of
masses as a function of the fraction
of the background rejected.  
For all WIMP masses, a background
rejection that exceeds \SI{99.5}{\%} for
signal acceptance of \SI{50}{\%} is
projected.

\tdrsubsec[LZSen]{LZ Sensitivity Projection}

To evaluate the projected sensitivity
for LZ, we consider a run
of \SI{1000}{\livedaysl} and a
\SI{5.6}{\tonnel} fiducial mass. We assume the same background models of Table~\ref{SRDt:PLRBG}, 
where background counts were shown for a signal region encompassing 0~to~\SI{20}{\phd}, effectively the search region for \SI{100}{\GeVpct} WIMP masses and below. To increase the sensitivity over the entire WIMP mass range, we consider
an expanded S1 signal region of 0~to~\SI{50}{\phd},
corresponding to 1.5~to~\SI{16}{\keVee}
for ER and 6~to~\SI{60}{\keVnr} for NR. Although the expanded search region brings with it a higher absolute count of backgrounds than those listed in the table, the use of the PLR method smoothly accounts for the profiles of these backgrounds relative to each WIMP mass as described in the previous section. We include the requirement of a 3-fold coincidence for the PMTs.
\begin{figure}[htbp]
\centering
\includegraphics[width=0.9\textwidth]{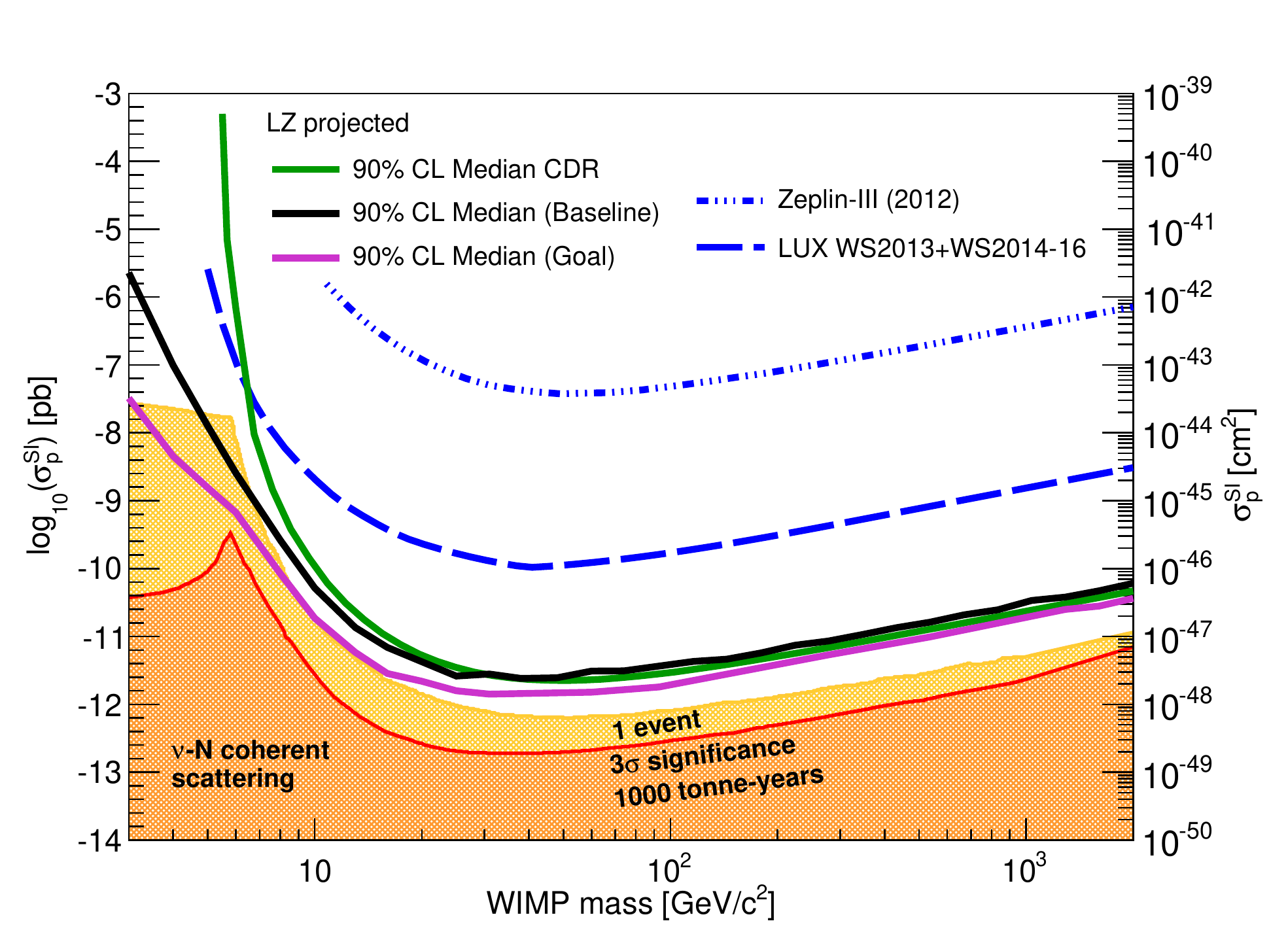}
\tdrfcaption[LZProjection]{LZ sensitivity
projections}{LZ sensitivity projection. The
baseline LZ assumptions in this report 
give
the solid black curve. LUX
and ZEPLIN results are 
shown in broken blue lines.  If LZ achieves the design goals listed
in Table~\ref{SRDt:Dt.1Tab}, the sensitivity would improve, resulting
in the pink sensitivity curve. The green line shows the projected sensitivity in the LZ Conceptual Design Report (CDR)~\cite{Akerib:2015cja} (see text for details of the changes from the CDR to this report). Lastly,
the shaded regions show where coherent scattering neutrino
backgrounds emerge.}
\end{figure} 

The projected sensitivity curve for 
LZ is shown in Figure~\ref{SRDf:LZProjection}.
The best sensitivity is 
\SI{2.3e-48}{\cm^2} for a \SI{40}{\GeVpct} 
WIMP mass, which satisfies our top 
level science requirement. 

Figure~\ref{SRDf:LZProjection} also shows the
projected sensitivity from the LZ Conceptual
Design Report (CDR)~\cite{Akerib:2015cja}. There are three main
differences in this projection with respect to
the CDR:
\begin{enumerate}
\item We have increased the assumed level of
\IRnttt and \IRnttz by a factor of twenty, so
radon is now the dominant source of 
ER events in the detector. 
\item The PLR technique as described in the
previous section is used to evaluate the dark 
matter sensitivity, instead of a simple 
cut and count approach. These two changes
effectively offset: there is a higher 
overall background level but those
backgrounds are more effectively rejected by
the analysis technique. Our confidence in
the ER background shape, most important
for using the PLR technique, rests on the
high-statistics LUX tritium ER
calibration~\cite{Akerib:2015wdi}.
\item The LUX neutron calibration results,
which allow LZ to more confidently project
the response of the detector to very low 
energy nuclear recoils. Use of the neutron
calibration provides the greatly increased
sensitivity to very low-mass WIMPs in the new
projection. 
\end{enumerate}

Figure~\ref{SRDf:LZDiscovery} shows the discovery potential for LZ under the baseline assumptions, calculated using a cut-and-count technique via the TRolke package~\cite{Lundberg:2009iu,Rolke:2004mj}. With the baseline parameters, LZ will have $3\sigma$ significance for \SI{40}{\GeVpct} WIMP mass at a cross section of \SI{6.0e-48}{\centi\meter^2}. We also show the $5\sigma$ significance curve, which falls below the projections from the XENON1T experiment at all WIMP masses.  Figure~\ref{SRDf:LZ3sig} shows the background and signal events populating the $\log_{10}(\textrm{S2}/\textrm{S1})$-S1
plane in an example LZ experiment with 1000 days and 5600 kg exposure for the $3\sigma$ example combination of WIMP parameters. 
\begin{figure}[htbp]
\centering
\includegraphics[width=0.8\textwidth]{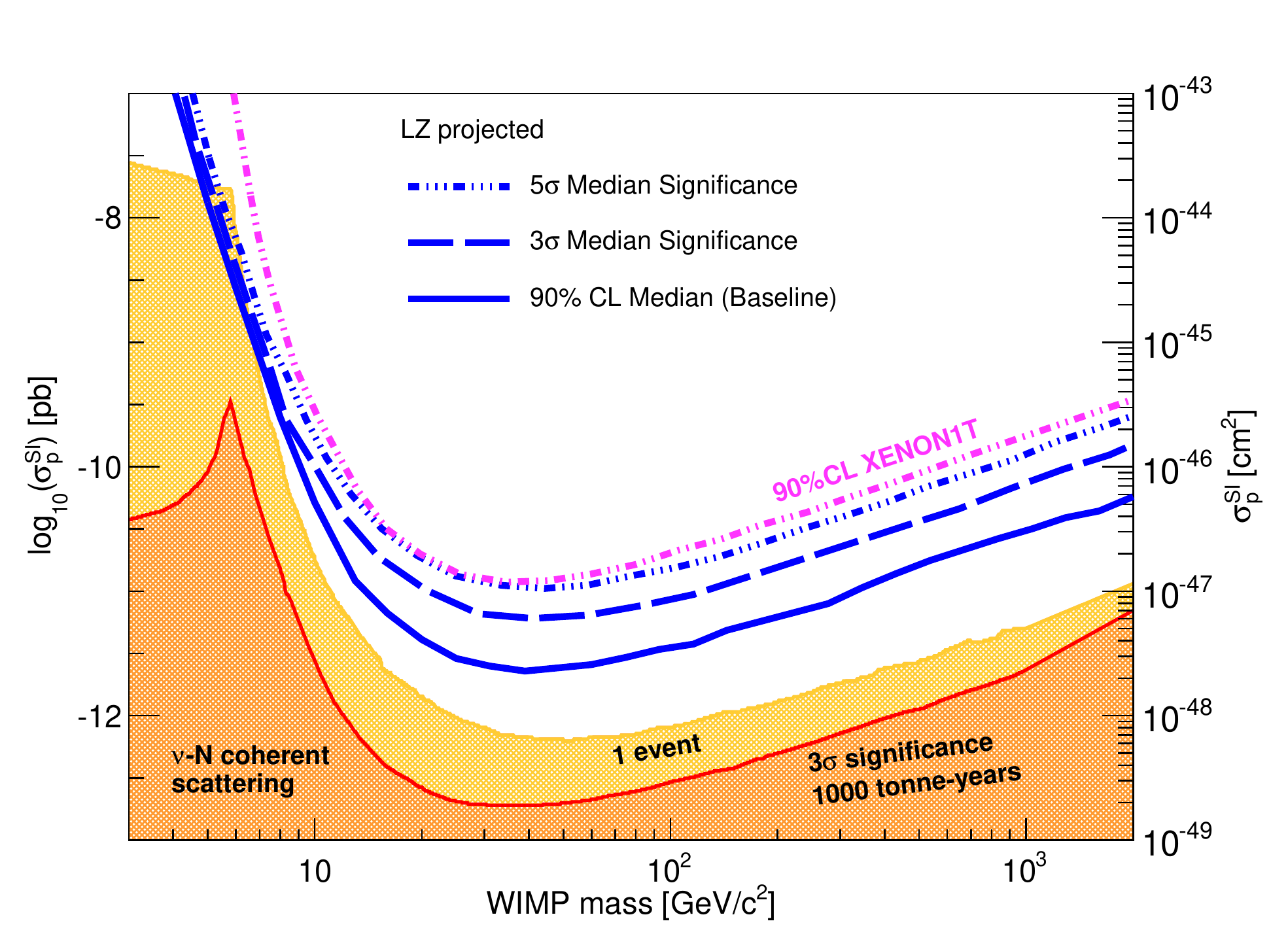}
\tdrfcaption[LZDiscovery]{LZ discovery potential}{The discovery potential for LZ under the baseline assumptions, calculated using a cut-and-count technique via the TRolke package~\cite{Lundberg:2009iu,Rolke:2004mj}. With the baseline parameters, LZ will have $3\sigma$ significance for \SI{40}{\GeVpct} WIMP mass at a cross section of \SI{6.0e-48}{\centi\meter^2}. The $5\sigma$ significance expectation is just below the expected \SI{90}{\percent} CL limit from a two year run of XENON1T at all WIMP masses.}
\end{figure} 

\begin{figure}[htbp]
\centering
\includegraphics[width=0.8\textwidth]{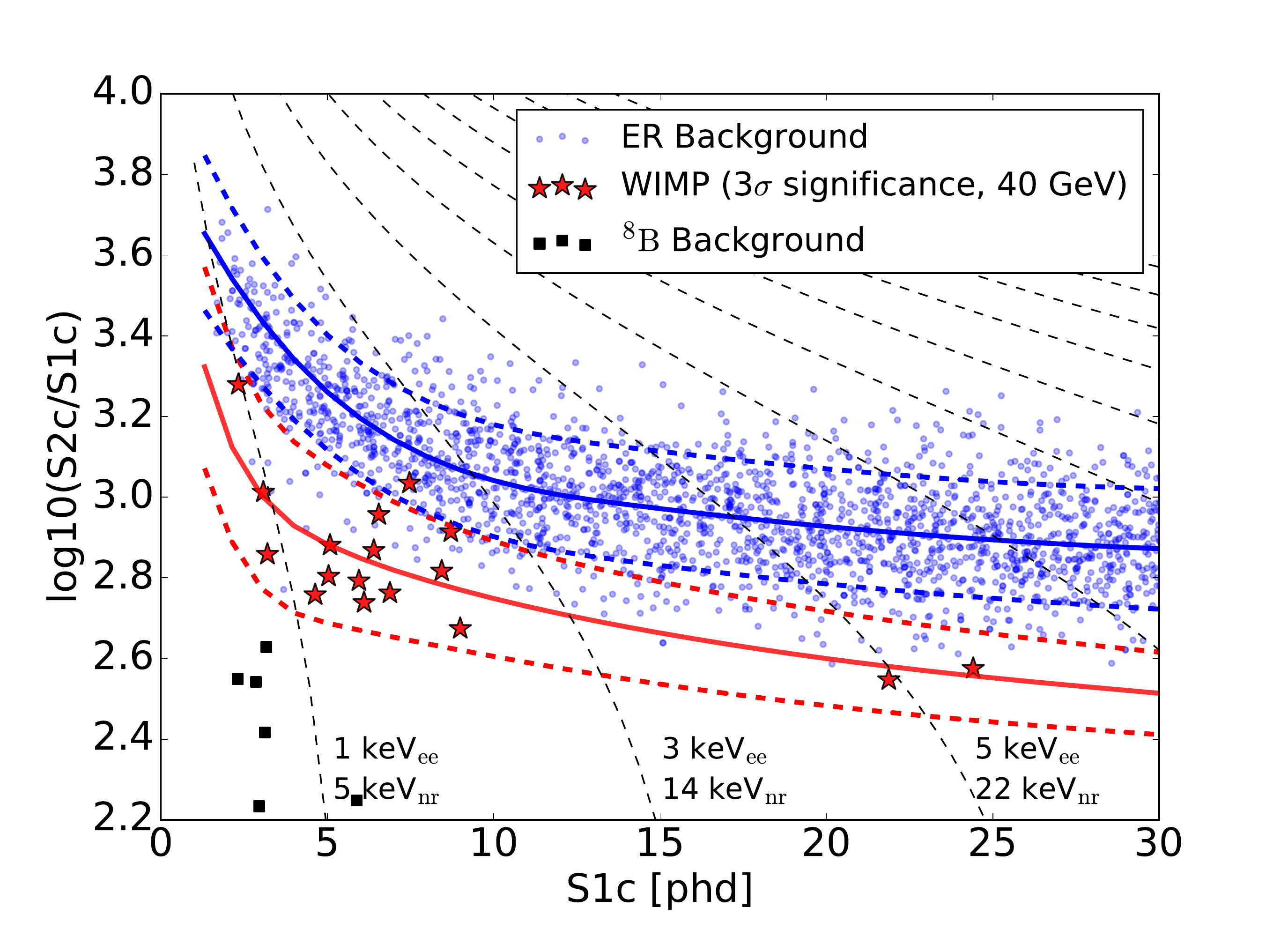}
\tdrfcaption[LZ3sig]{Example LZ exposure}{A sample LZ exposure under the baseline assumptions where the signal represents a \SI{40}{\GeVpct} WIMP mass with a cross section of \SI{6.0e-48} {\centi\meter^2}. LZ will observe several \IBe events under these assumptions, along with the nominal ER backgrounds. For these WIMP parameters, LZ will have a $3\sigma$ median significance for WIMP discovery. The solid blue and red lines represent the median of the ER and NR bands, the dashed blue and red lines indicate the \SI{10}{\%} to \SI{90}{\%} intervals for each population. The dashed lines running from the top left corner down to the x-axis show lines of constant recoil energy.}
\end{figure} 

\tdrsubsec[ParScans]{Parameter Scans of LZ Sensitivity}
We have explored the dependency of the LZ
sensitivity to spin-independent
interactions of WIMPs upon critical
detector performance assumptions.  Our
baseline parameters lead to the sensitivity
shown by the solid blue curve in
Fig.~\ref{SRDf:LZProjection}

One such study defines ``goals'' as a set
of parameters that are likely to be 
achieved during the fabrication
of LZ, but which are somewhat less
conservative than the baseline.  We
define a "reduced" set of parameters
as an unlikely worst-case. Both the
goal and baseline parameter sets meet
all requirements in Section~\ref{SRDS:Reqs}.
These three parameters sets are listed in
Table~\ref{SRDt:Dt.1Tab}, and their projected
performance  is shown in
Figure~\ref{SRDf:LZSenvsDetPerf}. 
Reaching the LZ goals achieves a 
sensitivity of \SI{1.1e-48}{\cm^2} 
at \SI{40}{\GeVpct}. The reduced parameter 
set degrades the sensitivity to 
\SI{5.1e-48}{\cm^2} at \SI{40}{\GeVpct}.

\begin{table}[H]
\centering
\sffamily
 \tdrtcaption[Dt.1Tab]{Key parameters for reduced, baseline, and goal detector performance}{Key parameters for reduced, baseline and goal detector performance as explained in the text.}
\begin{tabular}{ | m{4.5cm} | c | c | c | }   
\hline
\rowcolor{mrocol}	Detector Parameter & Reduced & Baseline & Goal \\ \hline
 Light collection (PDE) & 0.05 & 0.075& 0.12 \\ \hline
 Drift field (\si{\volt\per\centi\meter}) & 160 & 310 & 650 \\ \hline
 Electron lifetime (\si{\micro\second}) & 850 & 850 & 2800 \\ \hline
 PMT phe detection & 0.8 & 0.9& 1.0 \\ \hline
 N-fold trigger coincidence & 4 & 3 & 2 \\ \hline
 \IRnttt (\si{\milli\becquerel} in active region) & 13.4 & 13.4 & 0.67 \\ \hline
 Live days & 1000 & 1000 & 1000 \\ \hline

\end{tabular}
\end{table}

A notable observation that would accompany
the achievement of the ``goal'' set of
parameters: over 300 \IBe neutrino
events would be observed in a
\SI{1000}{\livedaysl} LZ run.  While
these events would slow the discovery
of a WIMP in the \SI{7}{\GeVpct} mass range,
they would also demonstrate a physical process
not yet observed in nature, the coherent
scattering of neutrinos from nuclei.

\begin{figure}[htbp]
\centering
\includegraphics[width=0.75\textwidth]{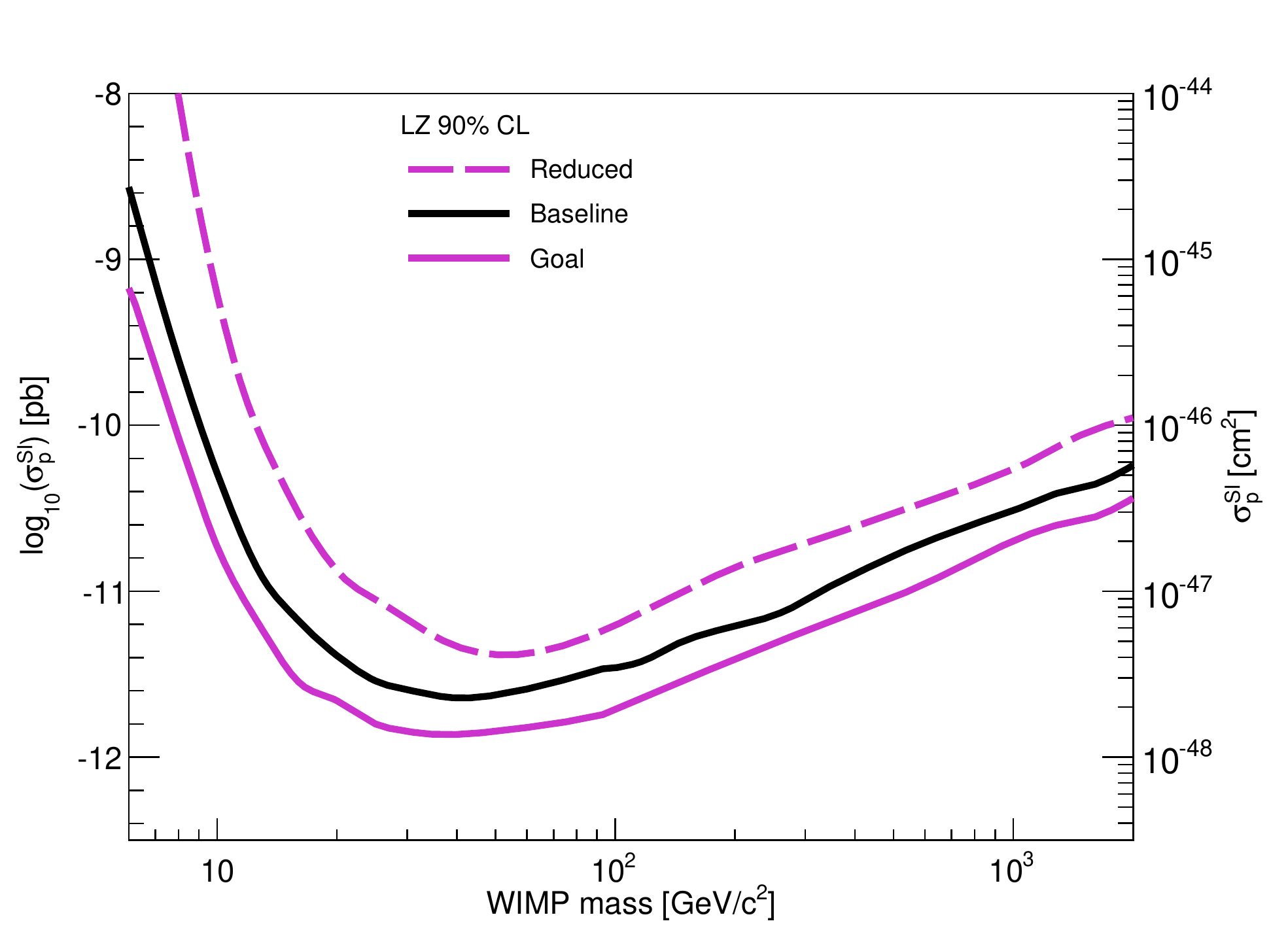}
\tdrfcaption[LZSenvsDetPerf]{LZ sensitivity
projections for goal, baseline, and
reduced parameters}{LZ sensitivity 
projections for goal, baseline, and
reduced parameters. The respective best
sensitivities are \SI{1.1e-48}{\cm^2},
\SI{2.3e-48}{\cm^2} (which both satisfy
the primary science requirement), 
and \SI{5.1e-48}{\cm^2} at a WIMP
mass of \SI{40}{\GeVpct}.}
\end{figure} 
In the next several figures, we vary key detector performance parameters to gauge their impact on the median \SI{90}{\%} CL of the upper limit for WIMP cross-section. Figure~\ref{SRDf:LZSenvsRn} shows the impact of \IRnttt, the dominant ER background, on LZ sensitivity. Assuming that the goal of \SI{0.67}{\milli\becquerel} is achieved, the sensitivity (to a \SI{40}{\GeVpct} WIMP) improves by \SI{20}{\%} over the baseline case. If the Rn background is increased another 10x over the nominal case then the sensitivity would degrade \SI{20}{\%}. The \IRnttt rate is representative of a generic flat ER background in LZ. Even a factor of 10 increase in the radon has marginal impact on sensitivity, demonstrating the power of using the PLR to separate ER backgrounds from signal like events. The effect of radon on the discovery potential is more significant, as shown in Fig.~\ref{SRDf:LZDiscvsRn}, where a 10x increase in the radon rate degrades the discovery potential of the detector by a factor of two. 

Figure~\ref{SRDf:LZSenvsATM} shows the impact on sensitivity of scaling the atmospheric (ATM) coherent scattering neutrino background. ATM neutrinos produce nuclear recoils similar to that of a \SI{40}{\GeVpct} WIMP, and their PDF has a high degree of overlap with the WIMP signal. Adjusting the ATM rate serves as a proxy for changing the overall NR count. When the ATM rate is turned up by a factor of 10, it is equivalent to having roughly 5 extra NR counts, degrading the sensitivity (to a \SI{40}{\GeVpct} WIMP) by \SI{25}{\%}.

Figure~\ref{SRDf:SenvsLightColl} shows the effect of changes in the S1 photon detection efficiency alone. Greater light collection leads to better S1 resolution, tightening the distribution of NR and ER events and leading to improved discrimination. Better light collection also improves the low energy threshold of LZ, enhancing sensitivity to low-mass dark matter, although this effect is somewhat countered by also seeing a higher \IBe background at low energy.  Figure~\ref{SRDf:LZSenvsDrift} shows the effect of changes in the purity of the LXe, as represented by the electron lifetime. There is significant margin until the drift time drops below half the baseline value. Figure~\ref{SRDf:LZSenvsEEE} gives the dependence on electron extraction efficiency. With only \SI{50}{\percent} extraction efficiency, larger fluctuations in the S2 signal lead to a reduction in discrimination power and a corresponding loss of sensitivity. Figure~\ref{SRDf:LZSenvsEF} shows that the sensitivity depends weakly on the drift field.     
Figure~\ref{SRDf:LZSenvsNfold} shows the dependence of the sensitivity on the coincidence trigger requirement, where the baseline design assumes a 3-PMT coincidence trigger. Going to 2-PMT coincidence reduces the threshold and makes a significant impact for low-mass WIMPs and similarly the \IBe neutrino signal.

Lastly, Figure~\ref{SRDf:LZSenvsExp} shows how extending the run from 1,000 to \SI{3000}{\livedaysl} would improve the sensitivity of the experiment. The plot shows the median \SI{90}{\%} confidence level upper limit on the cross section for a \SI{40}{\GeVpct} WIMP. In the baseline case, the sensitivity can be improved from \SI{2.3e-48}{\cm^2} to \SI{1.3e-48}{\cm^2}. If all the design goals are achieved, the sensitivity can be improved from \SI{1.1e-48}{\cm^2} to \SI{6e-49}{\cm^2} with \SI{3000}{\livedaysl}. Figure~\ref{SRDf:LZSenvsExp} also shows the $1\sigma$ bands on the expected sensitivity for both the baseline and goal parameter sets in a given exposure of LZ.

\begin{figure}[htbp]
\centering
\includegraphics[width=0.7\textwidth]{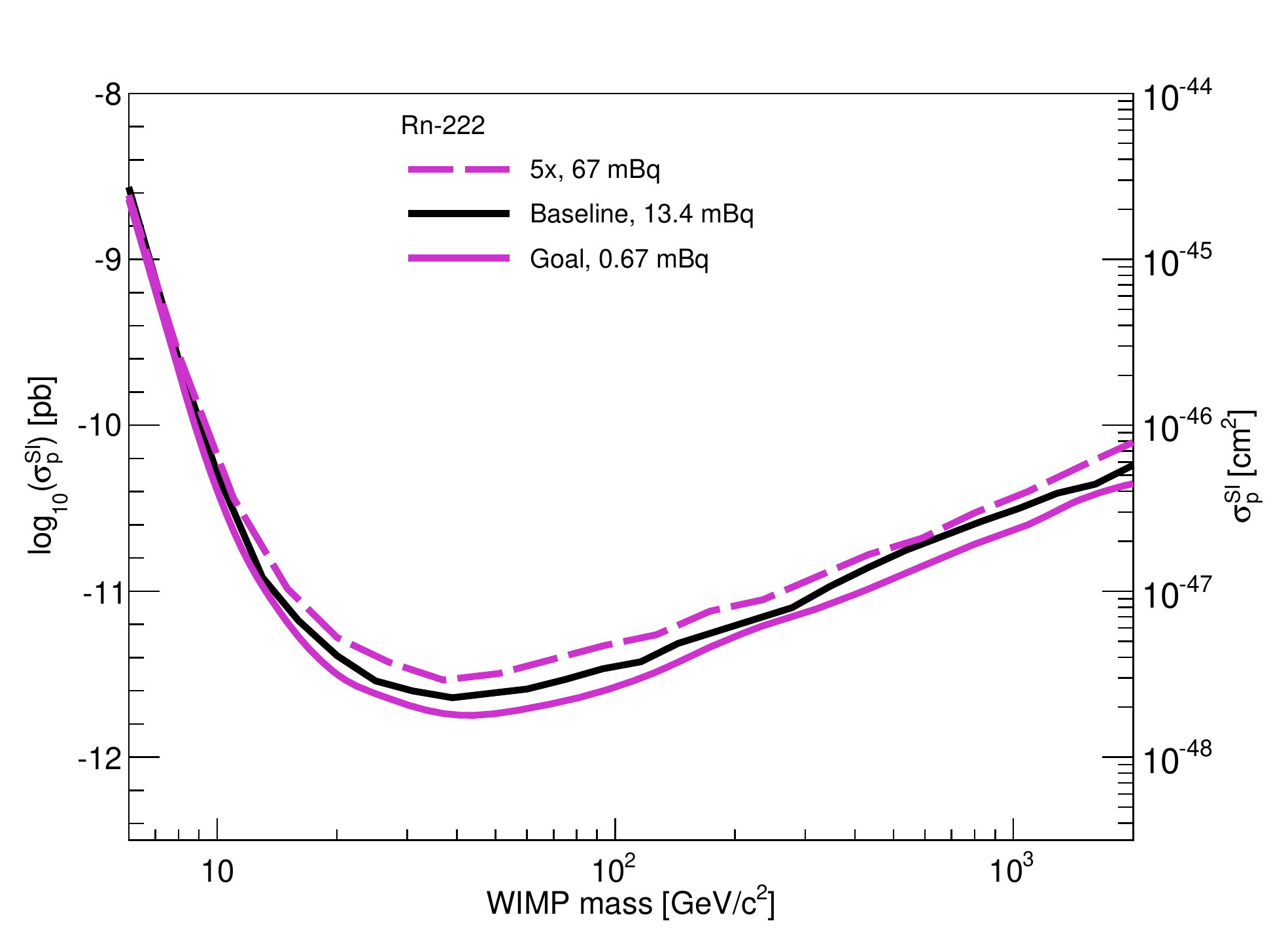}
\tdrfcaption[LZSenvsRn]{LZ sensitivity projections vs. radon concentration}{LZ sensitivity projections for three different assumptions on the concentration of radon in the active volume.}
\end{figure} 

\begin{figure}[htbp]
\centering
\includegraphics[width=0.7\textwidth]{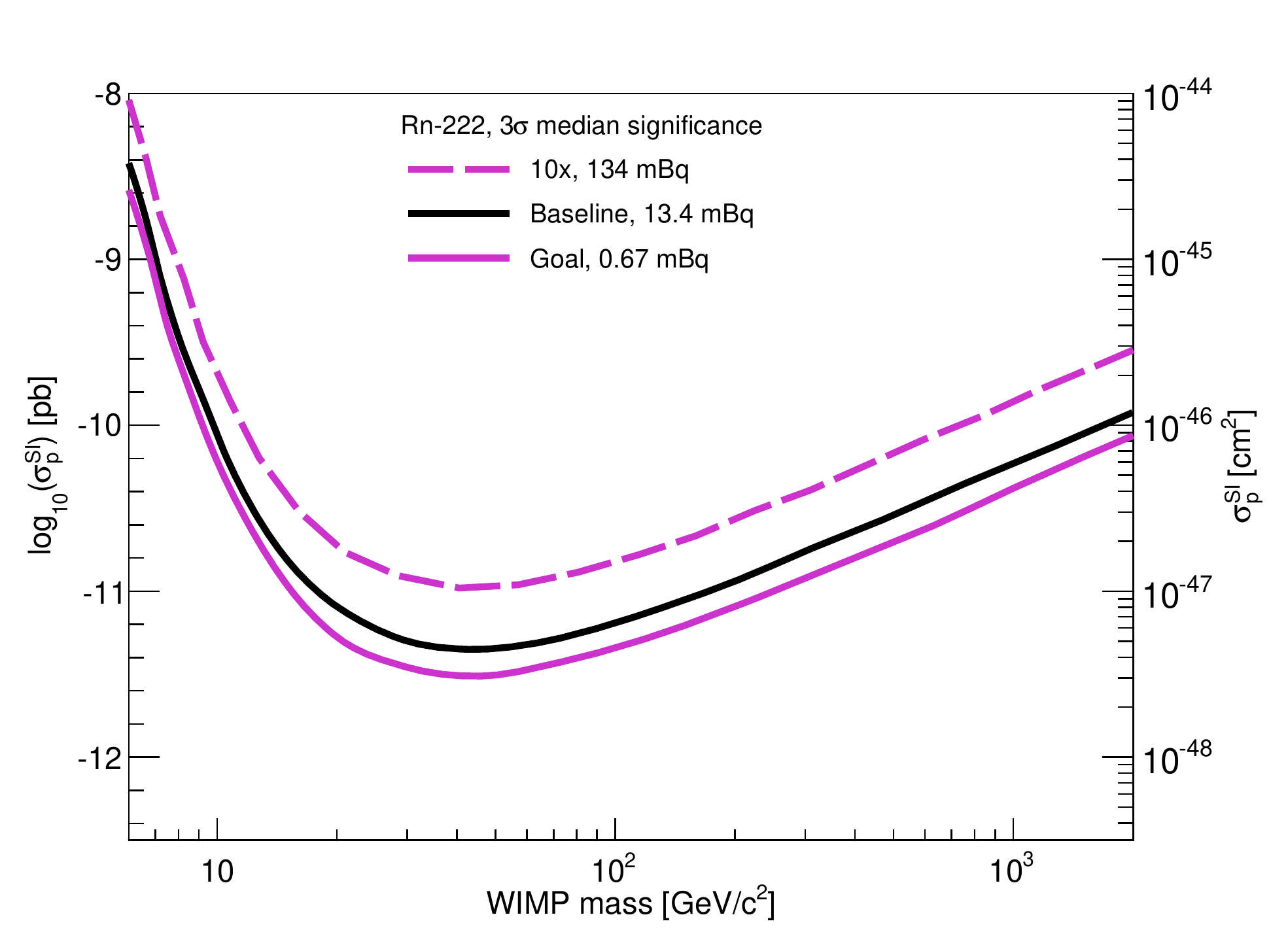}
\tdrfcaption[LZDiscvsRn]{LZ sensitivity projections vs. radon concentration}{LZ 3$\sigma$ median significance projections for three different assumptions on the concentration of radon in the active volume.  }
\end{figure} 

\begin{figure}[htbp]
\centering
\includegraphics[width=0.7\textwidth]{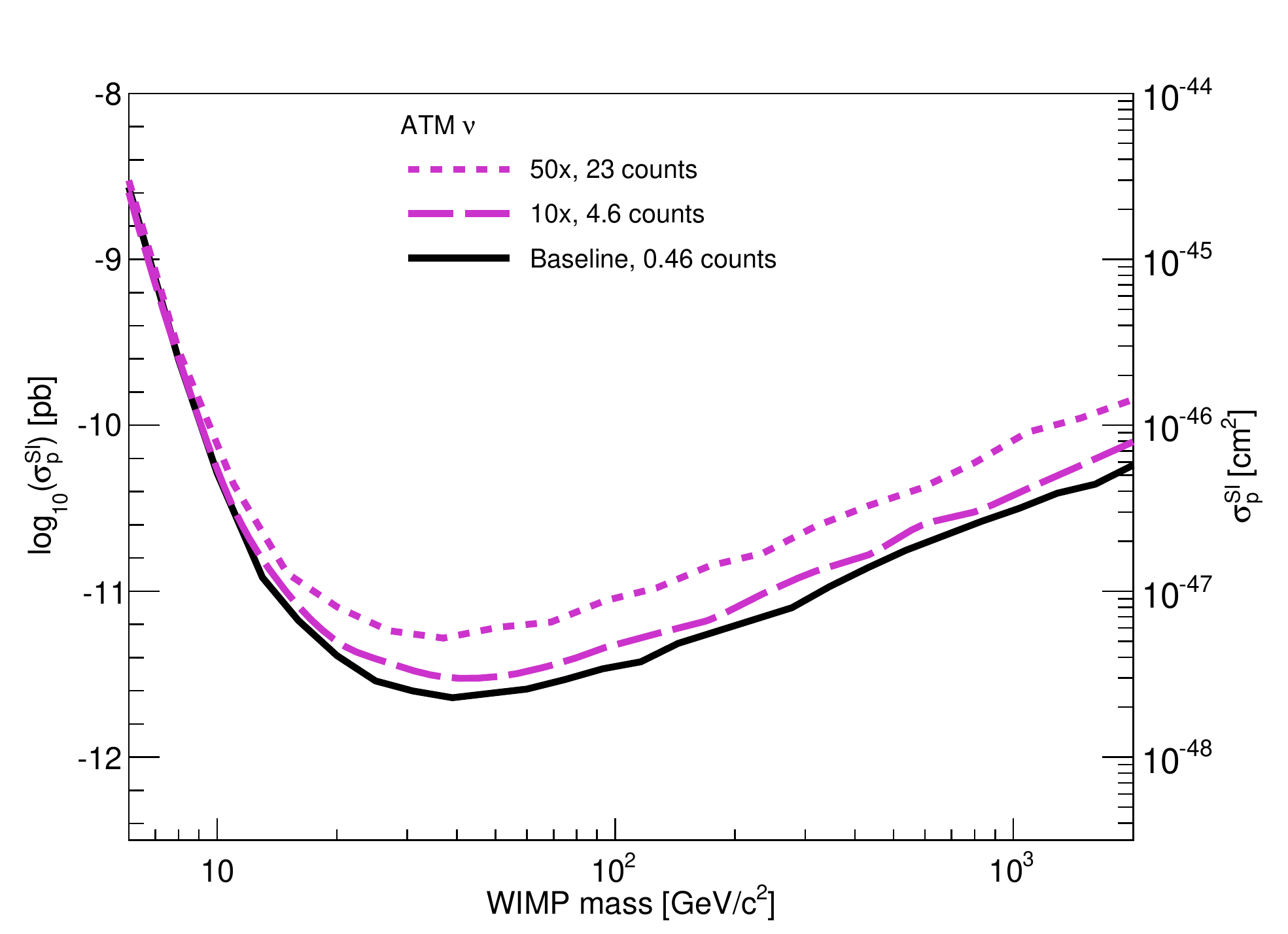}
\tdrfcaption[LZSenvsATM]{LZ sensitivity projections vs. atmospheric neutrino rate}{LZ sensitivity projections vs. scaled atmospheric neutrino flux. Scaling the atmospheric neutrino rate is a proxy for scaling the overall NR backgrounds in LZ.}
\end{figure} 

\begin{figure}[htbp]
\centering
\includegraphics[width=0.7\textwidth]{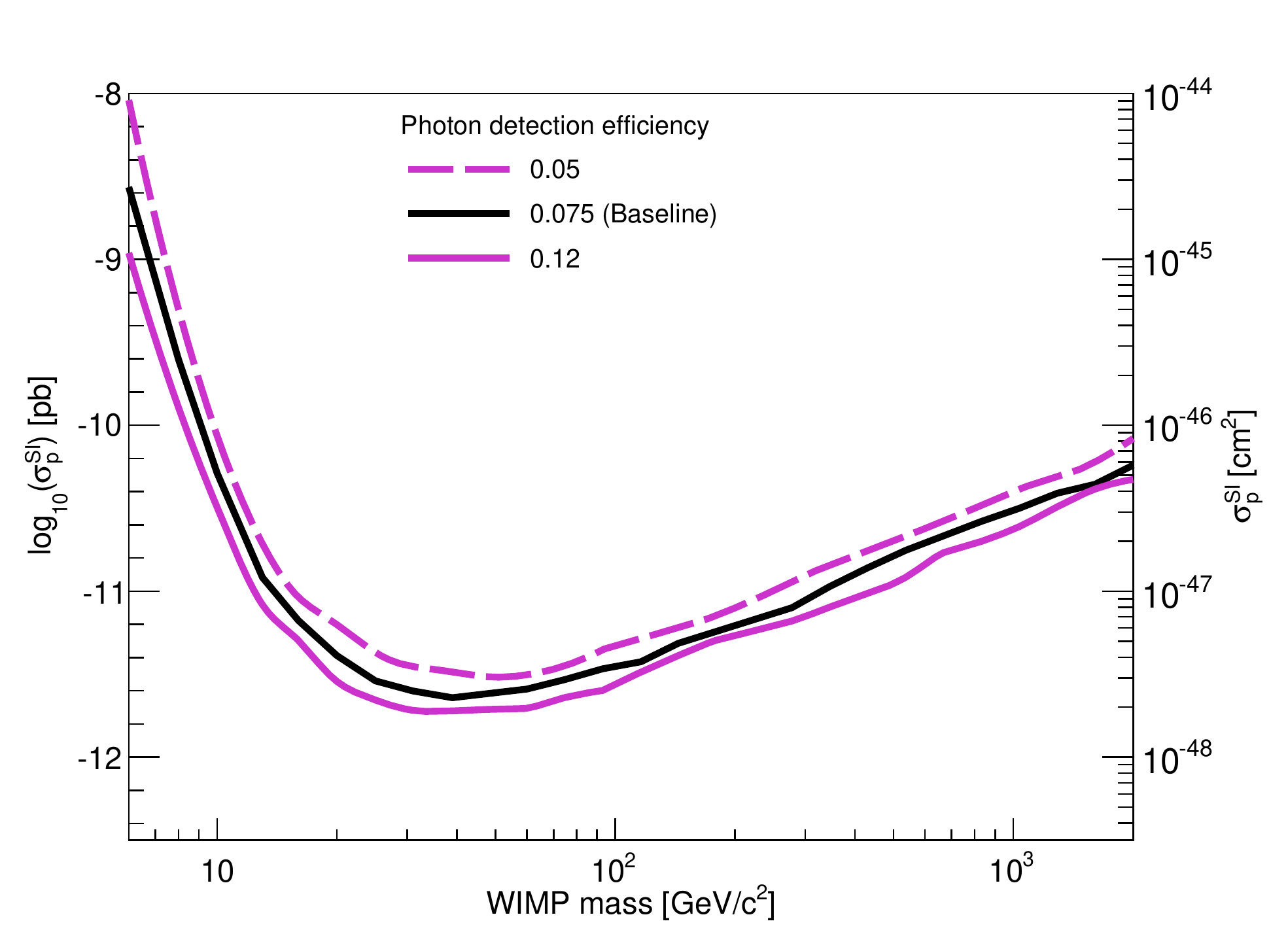}
\tdrfcaption[SenvsLightColl]{LZ sensitivity projections vs. S1 photon detection efficiency}{LZ sensitivity projections vs. S1 photon detection efficiency. We assume a photon detection efficiency of \num{0.075} for the baseline sensitivity, matching the requirement. The current model of the detector predicts a value of \num{0.085} (see Sec.~\ref{XDSSs:S1PDE}).}
\end{figure}

\begin{figure}[htbp]
\centering
\includegraphics[width=0.7\textwidth]{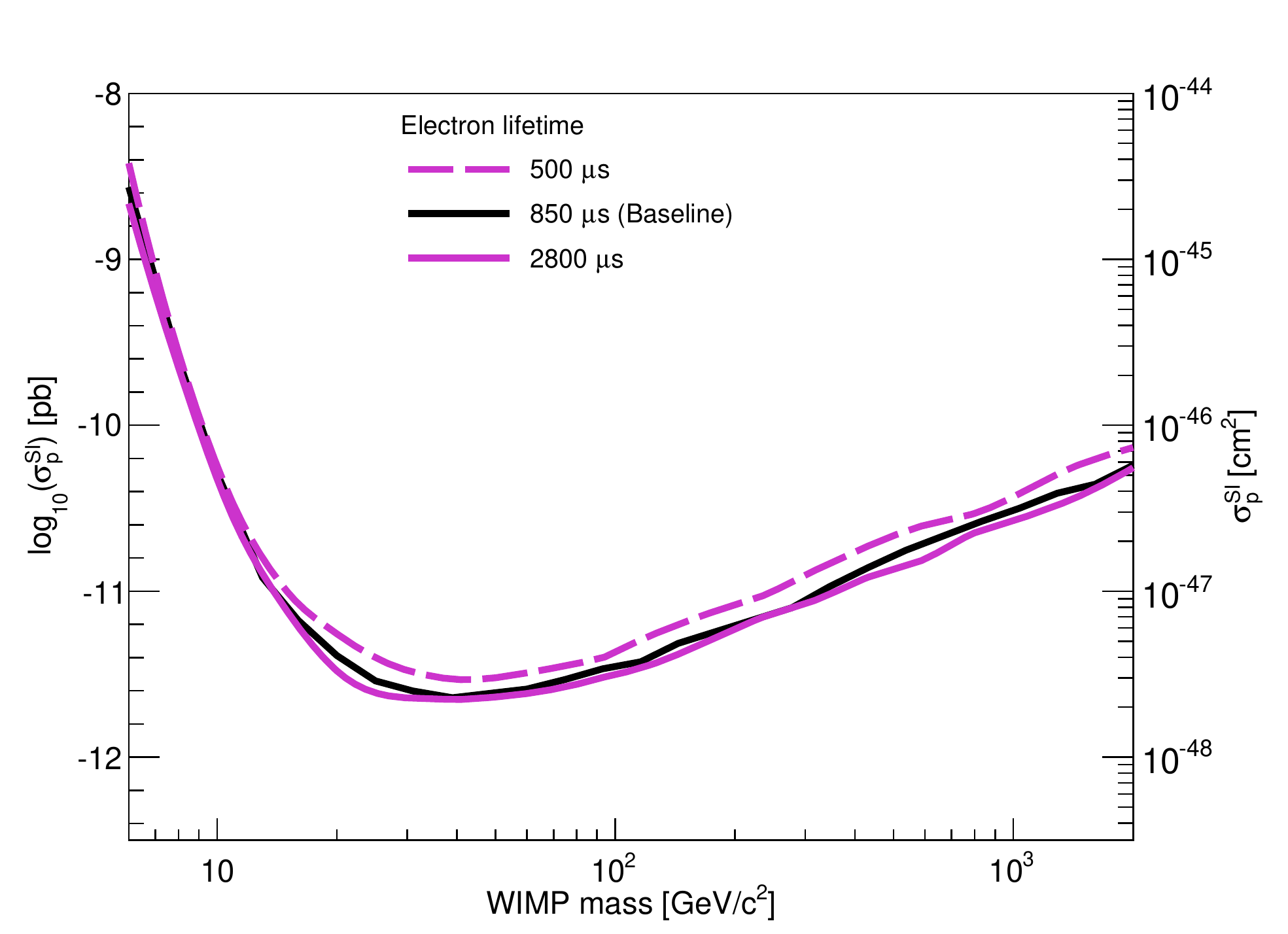}
\tdrfcaption[LZSenvsDrift]{LZ sensitivity projections vs. electron lifetime}{LZ sensitivity projections vs. electron lifetime (for the nominal electric field value of 310 \si{\volt\per\centi\meter}). LZ does not lose significant sensitivity until the lifetime drops below half the nominal value of \SI{850}{\micro\second}.}
\end{figure} 

\begin{figure}[htbp]
\centering
\includegraphics[width=0.7\textwidth]{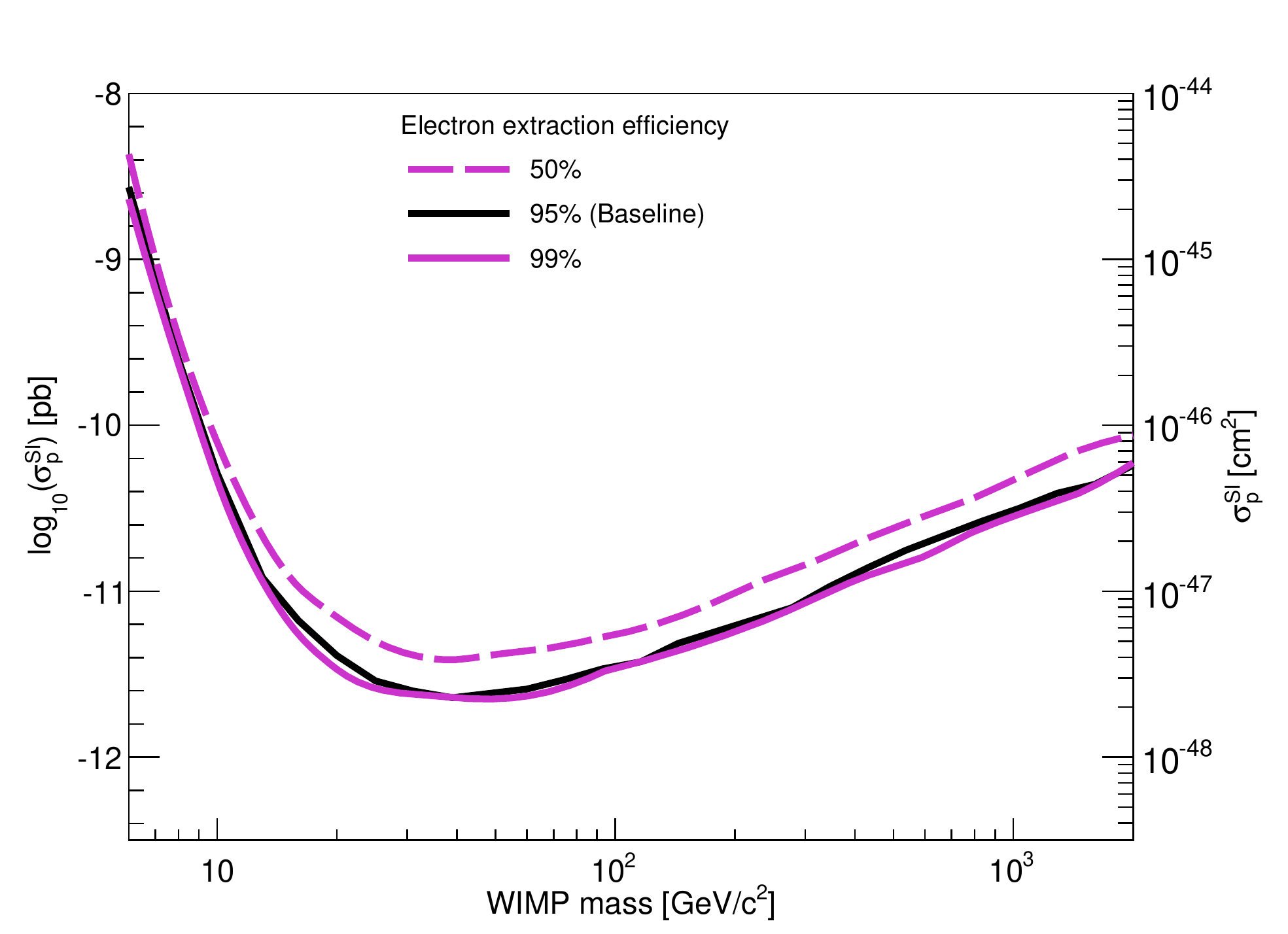}
\tdrfcaption[LZSenvsEEE]{LZ sensitivity projections vs. electron extraction efficiency}{LZ sensitivity projections vs. electron extraction efficiency. With only \SI{50}{\percent} extraction efficiency, larger fluctuations in the S2 signal lead to a reduction in discrimination power and a corresponding loss of sensitivity.}
\end{figure} 

\begin{figure}[htbp]
\centering
\includegraphics[width=0.7\textwidth]{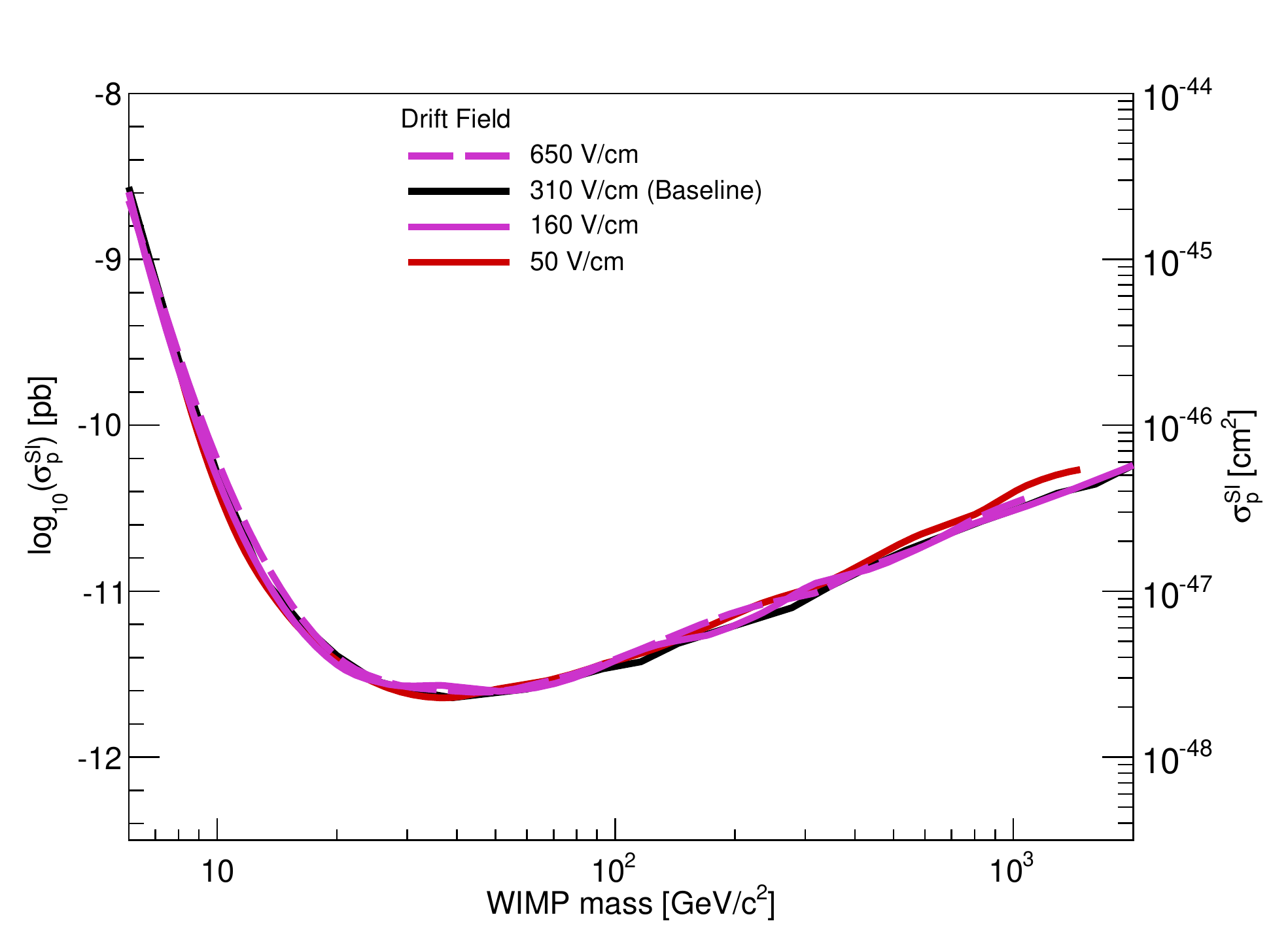}
\tdrfcaption[LZSenvsEF]{LZ sensitivity projections vs. electric field}{LZ sensitivity projections vs. electric field. In this regime, the ER/NR discrimination is robust to changes in the drift field, and the effect on sensitivity is minor.}
\end{figure} 

\begin{figure}[htbp]
\centering
\includegraphics[width=0.7\textwidth]{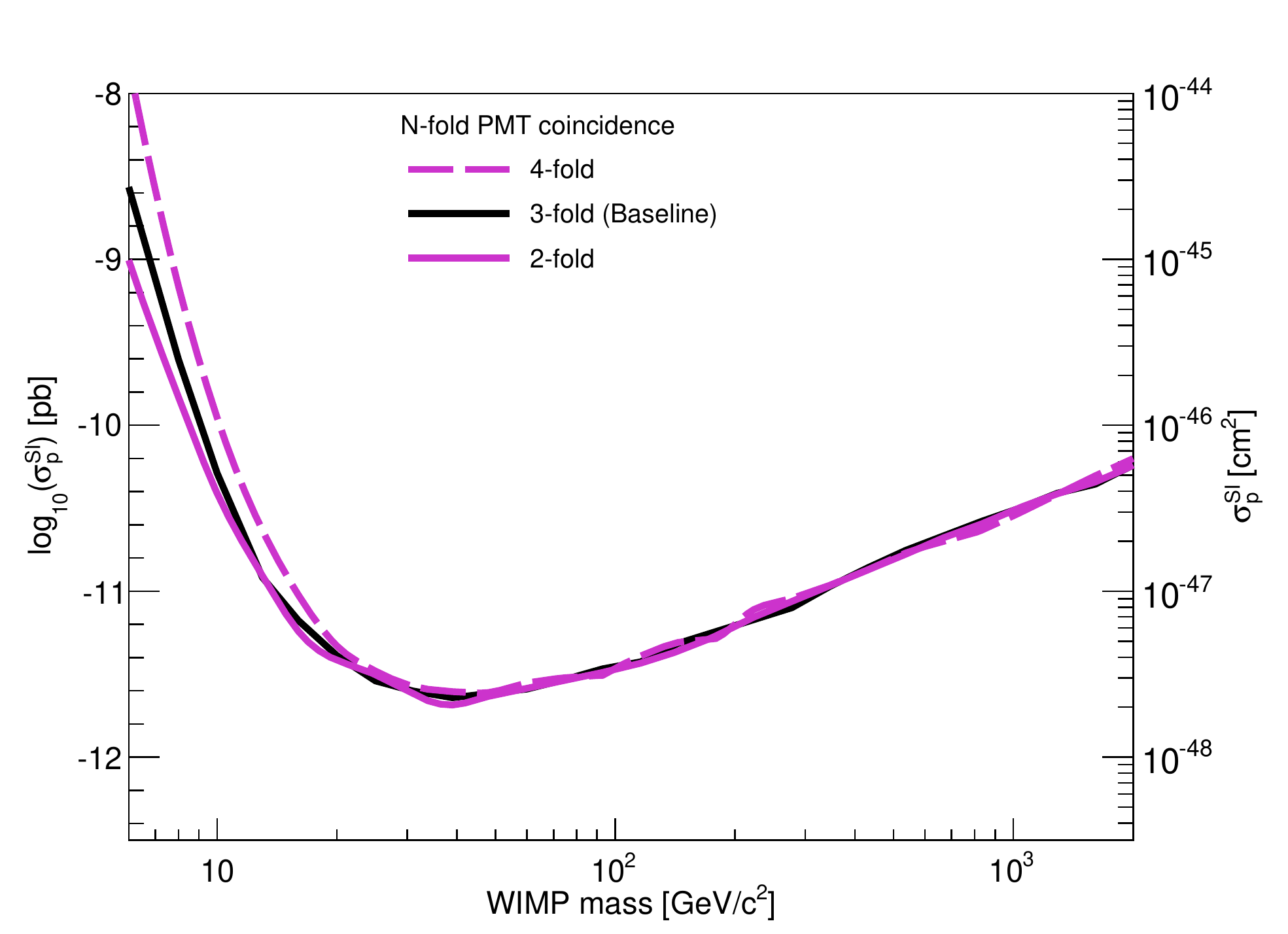}
\tdrfcaption[LZSenvsNfold]{LZ sensitivity projections vs. trigger coincidence level}{LZ sensitivity projections vs. trigger coincidence level. The primary effect of the coincidence requirement is on the detection of very low energy events, with direct consequences for sensitivity to low WIMP masses. For comparison, the LUX detector operates with a 2-fold trigger coincidence.}
\end{figure} 

\begin{figure}[htbp]
\centering
\includegraphics[width=0.7\textwidth]{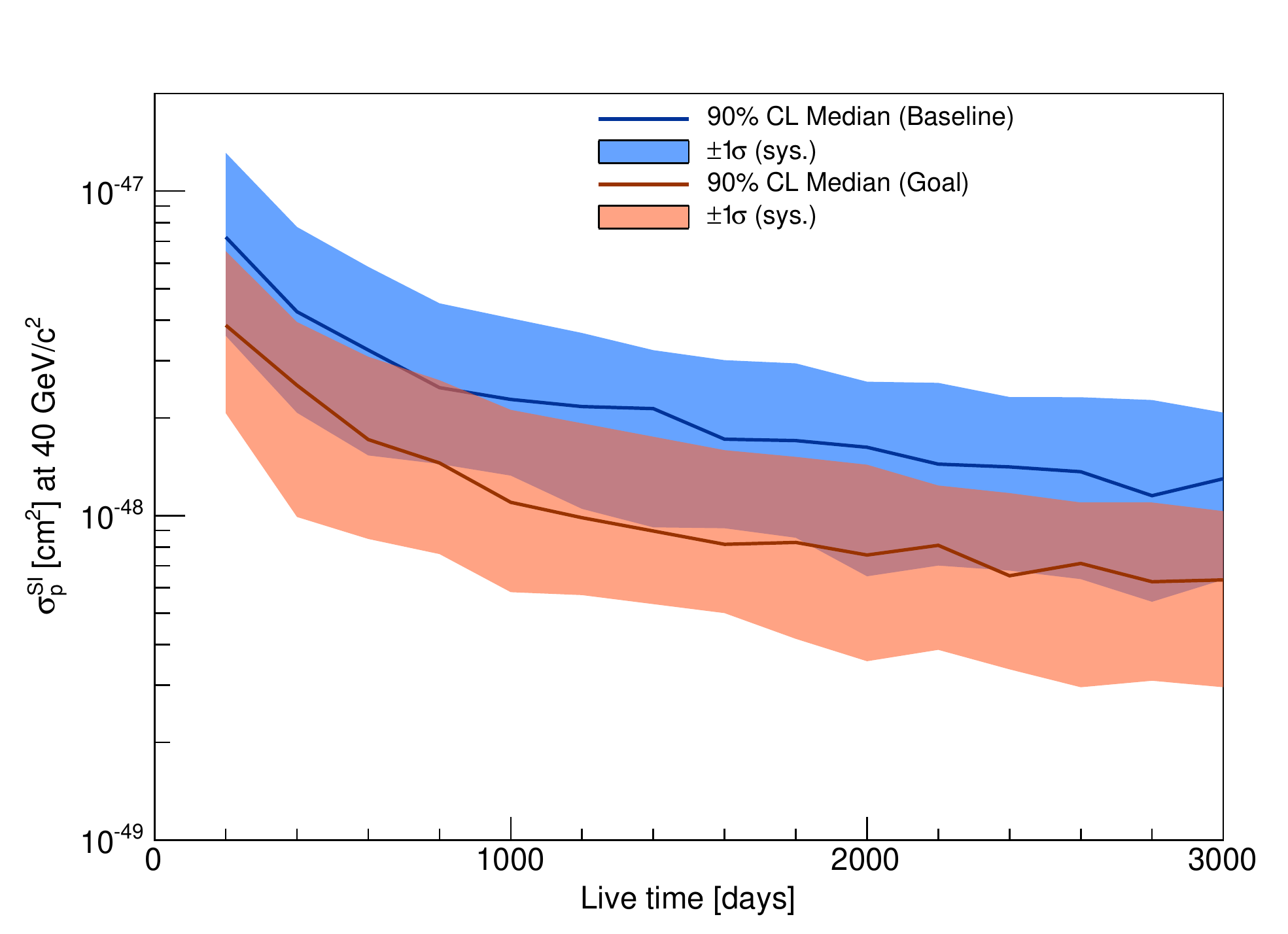}
\tdrfcaption[LZSenvsExp]{LZ sensitivity projections vs. exposure in days}{LZ sensitivity projections vs. exposure for the baseline (blue) and goal (orange). The impact of running the experiment for an additional 2000 days is to improve the sensitivity to \SI{1.2e-48}{\cm^2} and \SI{6e-49}{\cm^2} for the baseline and goal scenarios, respectively. Also shown are the $1\sigma$ bands showing the range of possible sensitivities for a given exposure. }
\end{figure} 

\clearpage
\bibliographystyle{apsrev4-2}
\bibliography{LZtdr}

\wlog{In tdrinclude command}
\wlog{\ChapLabel}

\renewcommand{\ChapLabel}{chap:MCS} 
\renewcommand{\BibLabel}{MCSS:bib} 
\renewcommand{\SecPre}{MCS}
\renewcommand{\FloatPre}{MCS}
\tdrchap{Management, Cost, and Schedule Summary}

The LZ Project (the Project) is international in scope, funding, and organization. 
This chapter presents an overview of the overall Project organization and a summary 
of the Project cost and schedule. The integrated overall Project Management 
organization is also described here. This Project organization has authority and 
responsibility over all aspects of the Project, including those funded by DOE, 
SDSTA, and non-U.S. agencies: the U.K.'s Science \& Technology Facilities Council 
(STFC); Portugal's Funda\c{c}\~{a}o para a Ci\^encia e a Tecnologia (FCT); and Korean-funded scope. The functions of the Project Advisory Board 
(PAB) are also described. A detailed discussion of Project Management, management 
systems, and approaches is described separately in the \textsl{LZ Project Execution Plan} (PEP) 
document. 

\tdrsec[org]{LZ Project Organization}
\begin{figure}[H]
\centering
\includegraphics[width=0.8\textwidth]{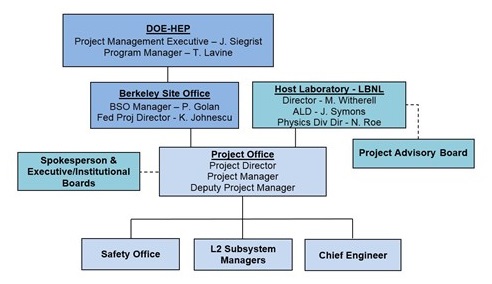}
\tdrfcaption[Reporting]
{LZ Project reporting}
{LZ Project reporting and responsibility organization chart, with an emphasis on the relationship to DOE.}
\end{figure} 
The Project’s organization from the perspective of DOE is summarized in Figure~\ref{MCSf:Reporting}.
LBNL is the DOE lead laboratory for LZ. As lead laboratory, LBNL will be responsible for Project
management and funding from DOE and for ensuring that essential manpower and necessary
infrastructure are provided to the Project during the R\&D, construction, and operations phases. 
The Project Director and Project Manager will be from the lead laboratory and will report to 
the Physics Division Director of LBNL. These two Project positions must be jointly approved 
by LBNL and by the collaboration Executive Board (EB). The LBNL Project Management 
Office (PMO) will review and provide oversight of the Project and its management systems 
to ensure that all DOE project guidelines and procedures are followed.

Figure~\ref{MCSf:Organization} presents the internal organization of the Project. The 
Spokesperson is elected by the collaborating institutes to represent the scientific interests 
of the collaboration. The roles and term of the Spokesperson are defined in a governance 
document. The current Spokesperson is Prof. Harry Nelson (University of California, Santa 
Barbara). The Spokesperson chairs the EB, which is a representative, elected body of 
senior collaboration members. The EB will help guide the Project organization in its goal of 
delivering the experimental apparatus and software that will meet the scientific requirements 
of the LZ collaboration. An Institutional Board (IB), with representatives from each 
collaborating institution, meets regularly with the Spokesperson and Project team.
\begin{figure}[H]
\centering
\includegraphics[width=0.8\textwidth]{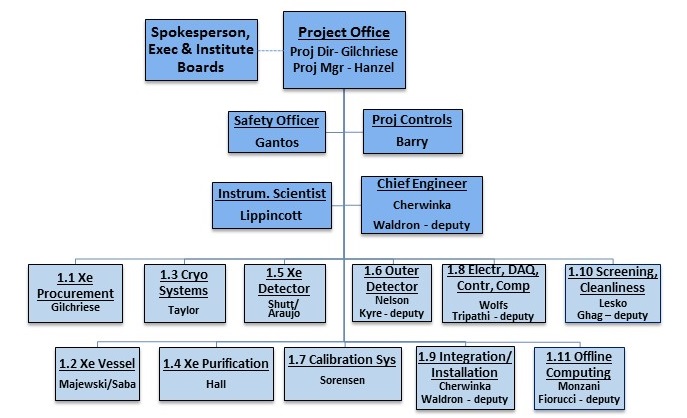}
\tdrfcaption[Organization]
{LZ Project organization}
{LZ Project organization.}
\end{figure} 

\tdrsubsec[PAB]{Project Advisory Board}
The PAB is an external board, gathered from the U.S. and non-U.S. scientific communities, 
that has expertise in large scientific projects. This board will provide valuable guidance and 
advice to the Project over the course of the construction life cycle. The PAB is charged by, 
and reports to, the Physics Division Director of LBNL. The current members are: 
David McFarlane (SLAC-chair), Jay Marx (Caltech-retired), Chris Bebek (LBNL), 
Elaine McCluskey (Fermi National Accelerator Laboratory [FNAL]), 
Bob McKeown (Jefferson National Accelerator Facility [J-LAB]), Mark Thomson (Cambridge), 
and Dan Dwyer (LBNL). The PAB will be supplemented as required to provide advice on a 
specific subject and for specific reviews.

\tdrsubsec[PMO]{Project Management Office}
PMO personnel include Project Director M.~Gilchriese (LBNL), Project Manager K. Hanzel (LBNL), 
Deputy Project Manager W. Edwards (LBNL), Instrument Scientist H.~Lippincott (FNAL), 
Chief Engineer J.~Cherwinka (Physical Sciences Laboratory, University of Wisconsin), 
Deputy Chief Engineer W.~Waldron (LBNL), Safety Officer 
J.~Gantos (LBNL), and Project 
Controls Officer M.~Barry (LBNL). Systems engineering and QC/QA functions are also 
under the direction of the PMO. 

\tdrsubsec[WBS]{Project Work Breakdown Structure}
The LZ Work Breakdown Structure (WBS) has 12 major categories, as shown in Table~\ref{MCSt:WBS}.

\begin{small}
\renewcommand{\ltbcp}{WBS}
\renewcommand*{\arraystretch}{1.}
\setlength{\extrarowheight}{5pt}
\sffamily
\begin{longtable}{ | m{0.6cm} | L{2.1cm} | m{12.3cm} | }
\caption[LZ Work Breakdown Structure \ifKL\space\texttt{\FloatPre t:\ltbcp}\fi]{LZ Work Breakdown Structure (WBS) shown at L2 and description of what each element covers \ifKL\space\texttt{\FloatPre t:\ltbcp}\fi\label{\FloatPre t:\ltbcp}}\\
\hline
\rowcolor{mrocol}
{WBS} & {WBS Title} & {WBS Description} \\ \hline
\endfirsthead
\caption[]{\sfs (continued)}\\
\hline
\rowcolor{mrocol}
{WBS} & {WBS Title} & {WBS Description} \\ \hline
\hline
\endhead
\rowcolor{lrocol}
\multicolumn{3}{|c|}{\sfs\Tstrut (continued on next page)}\\
\hline
\endfoot
\endlastfoot
1.1 & Xenon Procurement & Specification and procurement of the Xe necessary for the LZ 
experiment. Xe storage \& transportation vessels are covered in WBS 1.4, Xe Purification 
\& Handling.\\
\hline
1.2 & Xenon Vessel (Cryostat) & Labor, materials, and equipment associated with the design, 
prototyping, materials selection, construction, certification, and delivery, as well as planning 
and oversight of assembly and testing efforts on site, for the cryostat vessel system, its 
tanks, connecting flanges, insulation, and support structures.\\
\hline
1.3 & Cryogenic Systems & Labor, materials, and equipment associated with the design, 
prototyping, procurement, construction, assembly, testing, and delivery of the liquid nitrogen 
cryogenic system and nitrogen purge system.\\
\hline
1.4 & Xenon Purification \& Handling & Labor, materials, and equipment associated with 
the production of high-purity LXe, its storage, delivery to, and recovery from the TPC. 
This element covers the online purification system, the Xe purity analysis systems, the 
automated fail-safe Xe recovery system, and selected radon-reduction systems. A major 
subcomponent of this element is the stand-alone krypton-removal system, which will be used 
to purify the Xe prior to experimental operations.\\
\hline
1.5 & Xenon Detector & Labor, materials, and equipment associated with the design, 
prototyping, fabrication, testing, and assembly planning for the central Xe detector. 
This element covers the central detector region with its PMTs and the accompanying 
field-shaping electrodes and reflecting walls. It includes the “skin” veto region outside the 
main TPC volume and its PMTs. Included are the cathode, anode, and gate HV power 
supplies and the cathode HV umbilical connection to the TPC cathode and the grid structures, 
as well as the internal Xe liquid fluid system that brings liquid into the TPC region, providing 
cooling surfaces for temperature control. Also included is monitoring equipment for temperature, 
pressure, fluid flow, and other necessary measurements.\\
\hline
1.6 & Outer Detector System & Labor, materials, and equipment associated with the design, 
fabrication, testing, and assembly planning for the outer detector system. This includes 
the acquisition of the acrylic vessels, construction of the scintillator filling system, the 
acquisition and testing of the outer detector PMTs, the mixing and handling of the 
gadolinium-loaded liquid scintillator, procurement of reflector materials, as well as all the 
support infrastructure required. It also includes the planning, procedures, and oversight, 
plus the installation tooling required during the assembly of the system inside the water tank.\\
\hline
1.7 & Calibration & Labor, materials, and equipment associated with the design, prototyping, 
construction, delivery, assembly, and testing of the calibration system for the Xe detector and 
the outer detector system, along with the mechanisms, plumbing, valves, and radiation sources 
required to implement the calibration systems. Included are safety and administrative custodial 
requirements for source security, handling, and shipping.\\
\hline
1.8 & Electronics, DAQ, Controls, Computing & Labor, materials, and equipment associated 
with the design, prototyping, construction, delivery, assembly, and testing of the analog 
and digital electronics for the Xe and outer detector PMTs, the DAQ and trigger systems, 
the PMT HV system, the detector control system, and the online and offline hardware and 
software. This element includes the external signal, PMT HV, and network cables. Not included 
are the internal HV and signal cables for the PMTs (covered by WBS 1.5) and the detector 
sensors/instruments required for detector control. This element provides standard interfaces 
for detector sensors/instruments; custom interfaces required to connect custom 
sensors/instruments to the detector control system will not be provided by this element.\\
\hline
1.9 & Integration \& Installation & Labor and materials necessary to integrate the design 
effort of the subsystems into an overall detector design, maintain CAD models of the LZ 
detector and Davis Campus, upgrade the SURF infrastructure to support the detector 
assembly and operation, and perform on-site surface-level assembly of the detector and 
installation into the Davis Campus underground. Other subsystem elements maintain the 
responsibility to support integration by communicating design requirements, interface issues, 
subsystem CAD models, infrastructure needs, and assembly and operation needs. 
WBS 1.9 supplies planning, management, and skilled labor for assembly and installation, 
and the subsystems supplies experts on site to support this as needed.\\
\hline
1.10 & Cleanliness \& Screening & Labor, materials, and equipment associated with 
specification of radioactive background-level tolerances in the experiment; material 
radioassaying and control of radioactive background contaminants in the Xe resulting 
from component outgassing; control of ambient radioactivity; and establishing cleanliness 
controls, monitoring, and maintenance procedures for manufacture, transport, storage, 
handling, assembly, and integration of detector components.\\
\hline
1.11 & Offline Computing & Software professional labor and computing hardware needed to 
begin operations of the LZ experiment. Interface to collaboration responsibilities for data 
processing, analysis, and simulation software.\\
\hline
1.12 & Project Management & The cost of labor, travel, and materials necessary to plan, 
track, organize, manage, maintain communications, conduct reviews, and perform necessary 
safety, risk, and QA tasks during all phases of the Project. Subsystem-related management 
and support activities for planning, estimating, tracking, and reporting as well as their specific 
EH\&S and QA tasks are included in each of the subsystems.\\
\hline
\end{longtable}
\end{small}

\tdrsubsec[PSO]{Project Subsystem Organization}
The current Subsystem Managers (at Level 2 of the WBS and selected Level 3 Managers) 
and lead engineers are listed in Table~\ref{MCSt:Managers}. The LZ Technical Board comprises 
the WBS Level 2 Managers (\textbf{Bold}), their deputies, and the Project Office.

The Level 2 or Subsystem Managers, in addition to being members of the LZ Technical Board, 
are responsible for overseeing the development of the Project baseline with regard to their 
subsystems. They work with the Project Office to establish a level (L3) organization, helping 
to ensure that adequate technical resources have been identified, and defining the 
subsystem-specific requirements as they flow down from the overall Project. The L2 
Managers oversee the development of the technical design as well as the schedule and 
cost estimates associated with design, fabrication/execution, assembly, and test of their 
subsystems. They are also responsible for producing design reports and internal and 
external reviews. Ultimately, they are responsible for executing the Project Plan with 
respect to their subsystems. 

Design reviews are held for all relevant subsystems and are organized by the PMO. 
Each major subsystem and procurement undergoes multiple reviews (typically preliminary, 
final, and production readiness) as the design of the particular subsystem matures and 
reaches readiness for construction. 

\begin{small}
\renewcommand{\ltbcp}{Managers}
\renewcommand*{\arraystretch}{1.}
\setlength{\extrarowheight}{2pt}
\sffamily
\begin{longtable}{ | R{1.0cm} | L{3.0cm} | L{4.1cm} | L{3.0cm} | L{3.2cm} | }
\caption[LZ Work Breakdown Structure (WBS) shown at L2 \ifKL\space\texttt{\FloatPre t:\ltbcp}\fi]{The LZ Project Level 2 and 3 managers and lead engineers. \ifKL\space\texttt{\FloatPre t:\ltbcp}\fi\label{\FloatPre t:\ltbcp}}\\
\hline
\rowcolor{mrocol}
{WBS} & {Description} & {L2/3 Manager} & {Deputy or Co-mgr.} & {Lead engineer}\\ \hline
\endfirsthead
\caption[]{\sfs (continued)}\\
\hline
\rowcolor{mrocol}
{WBS} & {Description} & {L2/3 Manager} & {Deputy or Co-mgr.} & {Lead engineer}\\ \hline
\hline
\endhead
\rowcolor{lrocol}
\multicolumn{5}{|c|}{\sfs\Tstrut (continued on next page)}\\
\hline
\endfoot
\endlastfoot
\multicolumn{1}{|l|}{\textbf{1.1}}&\textbf{Xe Procurement}			&\textbf{M. Gilchriese (LBNL)}	&&\\	
\hline
\multicolumn{1}{|l|}{\textbf{1.2}}		&\textbf{Xe Vessel}				&\textbf{P. Majewski (RAL)}	& \textbf{J. Saba (LBNL)}	&\textbf{E. Holtom (RAL) / J. ODell (RAL)}\\
\hline
1.2.1   &Design                 &E. Holtom (RAL) &&\\
1.2.2   &Material Selection     &P. Majewski (RAL) &&\\
1.2.3   &Fabrication            &J. ODell (RAL) &&\\
1.2.4   &Cleaning               &J. ODell (RAL) &&\\
1.2.5   &Transportation         &J. ODell (RAL) &&\\
1.2.6   &Acceptance, Assembly, Installation  &J. ODell (RAL) &&\\
1.2.7   &Subsystem Management    &P. Majewski (RAL)/J. Saba (LBNL) &&\\
\hline
\multicolumn{1}{|l|}{\textbf{1.3}}		&\textbf{Cryogenic Systems}		&\textbf{D. Taylor (SDSTA)}	&&	\\
\hline
1.3.1   &Nitrogen Distribution  &D. Taylor (SDSTA) &&\\
1.3.2   &Thermosyphons          &D. Taylor (SDSTA) &&\\
1.3.3   &Vacuum System          &C. Maupin (SDSTA) &&\\
1.3.4   &Breakout               &C. Maupin (SDSTA) &&\\
1.3.5   &Controls and Power     &C. Maupin (SDSTA) &&\\
1.3.6   &Subsystem Management   &D. Taylor (SDSTA) &&\\
\hline
\multicolumn{1}{|l|}{\textbf{1.4}}		&\textbf{Xe Purification/Handling}		&\textbf{C. Hall (UMd)}	&&	\\
\hline
1.4.1	&Xenon Sampling			&C. Hall (UMd)	&&\\
1.4.2	&Kr Removal				&D. Akerib (SLAC)	&&	\\
1.4.3	&Xe Storage  \& Transport	&T. Benson (UW-PSL)	&&	\\
1.4.4	&Xe Gas Delivery \& Recovery&T. Benson (UW-PSL)	&&	\\
1.4.5	&Xe Gas Recirculation		&J. Cherwinka (UW-PSL) &&	\\
1.4.6	&Liquid Xe Tower			&H. Lippincott (FNAL)	&&	\\
1.4.7	&LXe Transfer Lines		&J. Cherwinka (UW-PSL)	&&	\\
1.4.8	&Undergnd Instal. Planning	&C. Hall (UMd)		&			&J. Cherwinka (UW-PSL)\\
1.4.9	&Subsystem Management	&C. Hall (UMd) &&		\\
1.4.10	&Xe Recovery \& Cont. Man.&E. Dahl (Northwestern)	&& \\	
\hline
\multicolumn{1}{|l|}{\textbf{1.5}}	  	&\textbf{Detector} 				&\textbf{T. Shutt (SLAC)}		&\textbf{H. Araujo (Imperial)}	&\textbf{J. Saba (LBNL)}\\
\hline
1.5.1	&Cathode High Voltage		&D. McKinsey (UCBerkeley)&		&W. Waldron (LBNL)\\
1.5.2    	&U.S. PMT Systems		&R. Gaitskell (Brown)	&&	\\
1.5.3    	&U.K. PMT Systems		&H. Araujo (Imperial)	&&	\\
1.5.4    	&Field Cage				&J. Saba (LBNL)	&&	\\
1.5.5	&Grids					&W. Wisniewski (SLAC)			&			&K. Skarpaas\\
1.5.6    	&Xe Monitoring System		&H. Kraus (Oxford)	&&	\\
1.5.7    	&Internal Fluid System		&T. Shutt (SLAC)	&&	\\
1.5.8    	&Skin System				&H. Lippincott (FNAL)	&&	\\
1.5.9	&Assembly and Installation	&J. Saba (LBNL)	&&	\\
1.5.10	&Screening Coordination		&R. Webb (TAMU)	&&	\\
1.5.11	&System Test	 			&K. Palladino (SLAC)&&		\\
1.5.12	&Subsystem Management	&T. Shutt			&H. Araujo	&\\
\hline
\multicolumn{1}{|l|}{\textbf{1.6}}		&\textbf{Outer Detector System} 	&\textbf{H. Nelson (UCSB)}	&\textbf{S. Kyre (UCSB)}		&\textbf{D.White (UCSB)}\\
\hline
1.6.1	&Scint. Vessels			&S. Kyre (UCSB)	&&	\\
1.6.2	&Reflector System			&S. Kyre (UCSB)	&&	\\
1.6.3	&LS Filling System			&D. White (UCSB)	&&	\\
1.6.4	&Liquid Scint.				&M. Yeh (BNL)	&&	\\
1.6.5	&Water Tank				&D. White (UCSB)	&&	\\
1.6.6	&PMT Supports			&S. Burdin (Liverpool)	&		&D. White (UCSB)\\
1.6.7	&PMTs					&S. Fiorucci (LBNL)	&&		\\
\hline
\multicolumn{1}{|l|}{\textbf{1.7}}		&\textbf{Calibration System} &\textbf{P. Sorensen (LBNL)}	&&	\\
\hline
1.7.1	&Internal Radioact. Sources	&S. Hertel (UCBerkeley)	&&	\\
1.7.2	&Calibration Source Delivery	&M. van der Grinten (RAL) &&		\\
1.7.3	&Radioisotope Sources	&P.~Sorensen (LBNL)	&&	\\
1.7.4	&Photoneutron Sources	&P.~Sorensen (LBNL)	&&	\\
1.7.5   &DD Neutron Source      &D. Huang (BrownU)  && \\
1.7.6   &Assembly and Installation Planning  &P.~Sorensen (LBNL)  &&  \\
1.7.7   &Subsystem Management   &P. Sorensen (LBNL)  &&  \\
\hline
\multicolumn{1}{|l|}{\textbf{1.8}}		&\textbf{Electr., DAQ, Controls, Computing}&\textbf{F. Wolfs (URochester)} &\textbf{M. Tripathi (UCDavis)}&	\\
\hline
1.8.1	&Analog Electronics	 		&M. Tripathi (UCDavis) 		&&R. Gerhard (UCDavis)\\
1.8.2	&Data Sparsification System	&E. Druszkiewicz  (URochester)	&&E. Druszkiewicz  (URoch)\\
1.8.3	&Data Acquisition System	&W. Skulski (URochester)		&&W. Skulski (URoch)\\
1.8.4	&External PMT HV, Signal	&F. Wolfs (URochester)	&&	\\
1.8.5	&Slow Control				&V. Solovov (Coimbra)		&&TBD (FNAL)\\
1.8.6	&PMT HV Supplies			&M. Tripathi (UCDavis)		&&R. Gerhard (UDavis)\\
1.8.7	&Online HW				&J. Buckley (Washington U)	&&M. Olevitch (WashingtonU), P. Zarzhitsky (UAlabama)\\
1.8.8	&Online SW				&J. Buckley (Washington U)	&&	\\
1.8.9	&Assembly \& Installation Planning&F. Wolfs (URochester)	&&R. Gerhard (UCDavis), E. Druszkiewicz  (URoch)\\
1.8.10  &Subsystem Management   &F. Wolfs (URoch)           &&  \\
1.8.11  &Detector Response Simulations  &H. Kraus (Oxford)      &&  \\
\hline
\multicolumn{1}{|l|}{\textbf{1.9}}		&\textbf{Integration \& Installation}	&\textbf{J. Cherwinka (UW-PSL)}&&		\\
\hline
1.9.1   &SURF Infrastructure	&S. DeVries (LBNL)			&&	\\
\hline
\multicolumn{1}{|l|}{\textbf{1.10}}		&C\textbf{leanliness \& Screening}	&\textbf{K. Lesko (LBNL)}	&\textbf{C. Ghag (UCL)	}&\\
\hline
1.10.1	&Screening for Fixed Contaminants			&P. Scovell (Oxford)/A. Cole (LBNL)	& &	\\
1.10.2	&Radon Emanation Screening	&R. Schnee (SDSMT)&&		\\
1.10.3	&Cleanliness Maintenance& A. Manalaysay (UCDavis) &&		\\
1.10.4	&Background Simulations	&A. Lindote (Coimbra)	&&	\\
1.10.5	&Information Repository				&J. Busenitz (UAlabama)	&&	\\
\hline
\multicolumn{1}{|l|}{\textbf{1.11}}		&\textbf{Offline Computing}	&\textbf{M.E. Monzani (SLAC)}	&\textbf{S. Fiorucci (LBNL)} &	\\
\hline
1.11.1	&U.S. Data Center			&C. Tull (LBNL)	&&	\\
1.11.2	&U.K. Data Center			&A. Vacheret (Imperial)	&&	\\
1.11.3	&Infrastructure Software	&S. Patton (LBNL)	&&	\\
1.11.4	&Simulations				&J. Dobson (UCL)	&&	\\
1.11.5	&Analysis Software			&C. Carmona (Penn State)	&&	\\
1.11.6	&Subsystem Integration and Validation&M.E. Monzani (SLAC)	&&	\\
1.11.7	&Subsystem Management	&M.E. Monzani (SLAC) & S. Fiorucci (LBNL)&	\\
\hline
\multicolumn{1}{|l|}{\textbf{1.12}}		&\textbf{Project Management}		&\textbf{M. Gilchriese (LBNL)}		&\textbf{K. Hanzel (LBNL)}	&\textbf{J. Cherwinka (UW-PSL)}\\
\hline
\end{longtable}
\end{small}

\tdrsec[Safety]{Safety}
Personnel safety, protecting the environment, and equipment safety are high priorities for 
the LZ Project. Its scientific goals cannot be achieved without an effective safety and 
environmental protection program that is integrated into the overall management of the 
experiment. The details of the EH\&S organization are described in an \textit{Integrated 
Safety Management} document. A separate \textit{Hazard Analysis Report} (HAR) 
describes the hazards that will be encountered and their associated controls during the 
execution of the Project. 
The HAR received significant input from the L2 Subsystem Managers who are, and will 
remain, closely involved in identifying and mitigating these hazards. Many hazards will be 
similar to those found in past operation of similar experiments (e.g., LUX and Daya Bay). 
The LZ Project work will take place at multiple institutions in addition to LBNL. Safety of the 
work at each institution will be the responsibility of the institution and work will be performed 
in accordance with the requirements and management systems of the home institutions. 
A sharing of lessons learned for the various locations is expected. Additionally, the LZ 
Project team will assist collaborating institutions as requested to address any hazard 
concerns.  

Final assembly of the LZ experiment and its operation will take place at SURF, where 
integrated safety management is well established and will be employed in all phases. 
SURF EH\&S rules and responsibilities will apply to all LZ activity at the SURF site, and 
SURF will provide relevant safety training for all members of LZ who work on the site. SURF has 
established an external EH\&S panel that will review the LZ Project during the construction 
period and prior to commissioning and operations. This panel has completed the first 
review of LZ and provided review of the HAR.

\tdrsec[Risk]{Risk}
The LZ risk program has several key aspects. The first is the early identification of 
potential risks in each of the detector elements as well as the system as a whole. 
Second, an early R\&D program focuses on understanding, reducing, or eliminating 
the identified risks. Third is the formal tracking of the remaining risks and mitigation 
strategies throughout the life of the experiment’s construction phase. Last is an 
accounting for technical, cost, and schedule risk in developing the contingency analysis 
for the cost of the experiment. These first three components (ID, R\&D, tracking) 
will be discussed in this chapter. A \textit{Risk Registry} for LZ has been assembled 
and is updated and reviewed regularly. A Risk Management Plan has also been completed.

Subsystem Managers have performed a risk assessment of their technical systems. 
These have been gathered by the Project Office and disseminated back out to the 
Subsystem Managers, key engineering leads, and the rest of the Project leadership team. 
The \textit{Risk Registry} will be reviewed and discussed regularly in subsystem and 
overall Project meetings. Updates to the \textit{Risk Registry} are considered monthly as the Project proceeds, as more information and experience are 
gathered and risk status changes.

\tdrsec[Operations]{Operations}
Experiment Operations will be managed centrally from an Operations Project office in much the same 
way the construction phase of the Project is managed. Our plan is based on the 
successful experience operating the LUX detector at SURF and other projects. The 
collaboration will provide much of the necessary resources for shifts and on-call experts. 
A small engineering and technical group will provide maintenance planning, oversight, 
and the resources for achieving them. A small computing and software maintenance 
group will ensure high availability of computing hardware and software and will support 
the collaboration’s data production and analysis activities.

The elements of the LZ operations support are:

\begin{itemize}
\item \textbf{LBNL operations manager.} Provides oversight of budget and EH\&S matters.
 
\item \textbf{Engineering support.} Provides engineering oversight during operations, 
particularly during the early phases of operations.
\item \textbf{On-site EH\&S officer.} Provides on-site EH\&S oversight. This is an 
SDSTA employee under contract to LBNL.
\item \textbf{On-site operations technical staff.} Provides on-site maintenance, support, 
and interface to SURF staff. This will involve both operations aspects and technical support. 
Multiple positions are planned and these will be SDSTA staff under contract to LBNL.
\item \textbf{Procurement.} Obtains materials, supplies, consumables (e.g. liquid 
nitrogen) and equipment necessary for operations and maintenance. Under the direction 
of the LBNL operations manager.
\item \textbf{University travel support.} Provides support of travel to the site. Under 
the direction of the LBNL operations manager.
\item \textbf{Computing support.} Provides support for professional services for 
computing hardware and software.
\end{itemize}

Transition-to-operations support for LZ-related travel of scientific staff began in FY 2016 and will
necessarily expand as work on site increases.

\tdrsec[Cost]{Cost and Schedule Summary}
The overall LZ Project plan is summarized in this chapter, including an overview of the 
Project schedule and the concept for the division of scope and cost among the various 
funding sources. The planned contributions supported by DOE, SDSTA, the U.K.'s 
Science \& Technology Facilities Council (STFC), Korea, and Portugal 
are outlined. This is a joint project with an international collaboration, and the 
cost-accounting approaches differ. Therefore, we have attempted to utilize the U.S. 
DOE cost-accounting approach for the costs summarized in this chapter.

\tdrsubsec[Schedule]{Project Schedule}
\begin{figure}[H]
\centering
\includegraphics[width=0.95\textwidth]{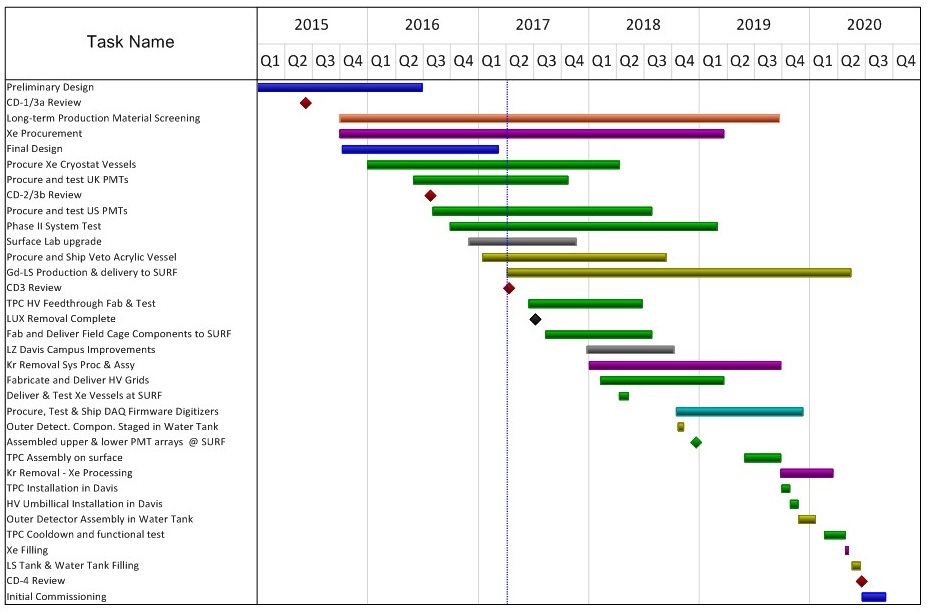}
\tdrfcaption[Schedule]
{Summary of LZ schedule}
{Summary of LZ schedule. The dates are in U.S. fiscal year quarters.}
\end{figure} 

The goal is to begin commissioning and early operations by early 2020. Infrastructure 
improvements (to surface laboratories) began in September 2016. Components begin to arrive 
at SURF in mid-2017. Underground installation begins in mid-2018. The DOE critical 
decision milestones in this plan are CD-1/3a obtained in April 2015, 
CD-2/3b obtained in August 2016 and a CD-3(all remaining items) review in January 2017. The procurement of long lead items (Xe, 
the cryostat vessels, PMTs, outer detector acrylic vessels and others) has already started. A more complete view of the 
Project schedule is shown in Figure~\ref{MCSf:Schedule}. The dates shown in this 
schedule correspond to early-finish milestones. 

\clearpage
\tdrsec[Scope]{Project Scope}
The Project’s technical scope has been described in previous chapters. The complete LZ 
Project includes the detector elements -- purified Xe, cryogenic systems, Xe detector, 
cryostat, veto system, calibration system, electronics, DAQ, trigger, online and offline 
software -- as well as all the integrating activities -- system tests, system integration, 
assembly/installation, on-site infrastructure, and project management.

The planned scope division among the various U.S. and non-U.S. agencies is summarized 
briefly here.
\begin{itemize}
\item The major elements of U.K. scope deliverables include about one-third of the 
low-background PMTs for the Xe detector, the cryostat set (inner and outer), elements 
of the low-background counting capability, contributions to integrated system tests, the 
source calibration delivery mechanism, and extensive contributions to computing and software.
\item Korea is expected to contribute PMTs for the outer detector, low-background assay 
capability, and software.
\item Portugal is expected to contribute to control systems, software, and a measurement 
system for TPC quality control. 
 
\item The SDSTA scope includes above- and belowground modifications to required facilities 
and much of the Xe needed for the experiment. 
\item The NSF is assumed to support scientific efforts for those U.S. collaborating institutions 
funded by NSF but will not contribute to the Project scope.
\item The DOE is assumed to fund all remaining Project scope.
\end{itemize}

\clearpage
\tdrsec[CostSummary]{Cost Summary}
The U.S.-based cost estimate associated with the above scope is shown in 
Table~\ref{MCSt:CostSummary} in at-year dollars without contingency. U.S. contingency amounts to about 25\% of the DOE - funded amount. 

\begin{center}
\renewcommand{\ltbcp}{CostSummary}
\renewcommand*{\arraystretch}{1.}
\setlength{\extrarowheight}{4pt}
\setlength{\tabcolsep}{4pt}
\sffamily
\begin{longtable}{ | L{1cm} L{8cm} | R{2.5cm} | R{3cm} |}
\caption[LZ Project cost summary \ifKL\space\texttt{\FloatPre t:\ltbcp}\fi]{LZ Project U.S.-equivalent base cost (without contingency) in at-year kilo-dollars. The equivalent U.S. costs for non - DOE items are also shown. \ifKL\space\texttt{\FloatPre t:\ltbcp}\fi\label{\FloatPre t:\ltbcp}}\\
\hline
\endfirsthead
\caption[]{\sfs (continued)}\\
\hline
\endhead
\rowcolor{lrocol}
\multicolumn{4}{|c|}{\sfs\Tstrut (continued on next page)}\\
\hline
\endfoot
\endlastfoot
\rowcolor{mrocol}
\multicolumn{2}{|c|}{WBS (2)} & DOE Funded & NONDOE Funded\\
\hline
1.01 - &Xe PROCUREMENT	 		& \$k\hfill 420 	 	& \$k\hfill 9846 \\ 
1.02 - &Xe VESSEL SUBSYSTEM	 	& \$k\hfill 5 	 	& \$k\hfill 1466 \\
1.03 - &CRYOGENIC SUBSYSTEM	 	& \$k\hfill 1881 	& \$k\hfill -   \\
1.04 - &Xe PURIFICATION SUBSYSTEM & \$k\hfill 7399 	& \$k\hfill - \\  
1.05 - &Xe DETECTOR SUBSYSTEM	& \$k\hfill 10332 	& \$k\hfill 2838 \\
1.06 - &OUTER DETECTOR SUBSYSTEM& \$k\hfill 4191 	& \$k\hfill 661 \\
1.07 - &CALIBRATION SUBSYSTEM	& \$k\hfill 727 	 	& \$k\hfill 98 \\
1.08 - &ELECTRONICS, DAQ, CONTROLS \& COMPUTING SUBSYSTEM & \$k\hfill 5274 	 & \$k\hfill 34 \\
1.09 - &INTEGRATION \& INSTALLATION SUBSYSTEM & \$k\hfill 5379 	 & \$k\hfill 1971 \\
1.10 - &CLEANLINESS AND SCREENING SUBSYSTEM	 & \$k\hfill 1153 	 & \$k\hfill 650 \\
1.11 - &OFFLINE COMPUTING SUBSYSTEM	 & \$k\hfill 2322 	 	& \$k\hfill - \\
1.12 - &PROJECT MANAGEMENT	 & \$k\hfill 4733 	 &\$k\hfill -   \\
\hline
\multicolumn{2}{|c|}{Total} & \$k\hfill 44,817 & \$k\hfill 17,661 \\
\hline
\end{longtable}
\end{center}

\clearpage

\wlog{In tdrinclude command}
\wlog{\ChapLabel}

\end{document}